\documentclass[paper=a4, fontsize=11pt]{scrartcl} 
\usepackage[T1]{fontenc} 
\usepackage[utf8]{inputenc}
\usepackage[english]{babel} 
\usepackage{fullpage} 
\usepackage{amsmath,amsfonts,amsthm} 
\usepackage[export]{adjustbox}
\usepackage{epstopdf}
\usepackage{eurosym}
\usepackage{hyperref}
\hypersetup{pdftitle={Quantitative Methods in Finance}, pdfauthor={Eric Vansteenberghe}, pdfsubject={Lecture notes, Universit\'e Paris 1 Panth\'eon-Sorbonne}, pdfkeywords={quantitative finance; financial econometrics; Python; R; time series; risk measures; lecture notes}}
\usepackage{pdflscape}
\usepackage{centernot}
\usepackage{tikz}
\usetikzlibrary{shapes.geometric}
\usepackage{epigraph}
\usepackage{dcolumn}
\usepackage{booktabs}
\newcommand*{\mydate}{November 2026}
\usepackage[round]{natbib}
\usepackage{multirow}
\usepackage{lscape}
\usepackage{listings}
\usepackage{float}
\usepackage{pgfplots}
\usepackage{bookmark}
\usepackage{mathtools}

\usepackage{bbm}

\newtheorem{theorem}{Theorem}
\newtheorem{assumption}{Assumption}
\theoremstyle{definition}
\newtheorem{definition}{Definition}[section]
\newtheorem{corollary}{Corollary}[theorem]

\newcommand{\sym}[1]{\ifmmode^{#1}\else\(^{#1}\)\fi}

\newcommand{\E}{\mathbb{E}}
\DeclareMathOperator{\Var}{Var}
\DeclareMathOperator{\Cov}{Cov}
\DeclareMathOperator{\Corr}{Corr}
\newcommand{\Fset}{\mathcal{F}}
\newcommand{\VaR}{\mathrm{VaR}}
\newcommand{\ES}{\mathrm{ES}}
\newcommand{\Q}{\mathbb{Q}}

\addtokomafont{section}{\clearpage}

\title{Quantitative methods in finance}
\author{Eric Vansteenberghe}
\date{\mydate}

\begin{document}
\sloppy

\maketitle
{\setlength{\epigraphwidth}{0.6\textwidth}
\epigraph{Perdre un bonheur r\^{e}v\'{e}, renoncer \`{a} tout un avenir, est une souffrance plus aigu\"{e} que celle caus\'{e}e par la ruine d'une f\'{e}licit\'{e} ressentie, quelque compl\`{e}te qu'elle ait \'{e}t\'{e} : l'esp\'{e}rance n'est-elle pas meilleure que le souvenir ?}{La Bourse\\\textit{Honor\'{e} de Balzac}}}
\setcounter{tocdepth}{2}
\tableofcontents

\quad

\newcounter{question}

\section{Introduction}
I have been teaching different courses at Université Paris 1 Panthéon-Sorbonne since September 2015. Although these courses have covered different topics and have been aimed at different audiences, they have shared two recurring requirements: a sufficiently strong quantitative background and the ability to implement economic and financial models using modern open-source computational tools.

These lecture notes are intended to provide a single, self-contained reference for these courses. They are designed both for students discovering Python and quantitative methods for the first time and for Master students taking more advanced courses in \emph{Quantitative Methods in Finance} or \emph{Financial Econometrics}. The objective is therefore broader than that of a conventional set of notes written for a single course: the document is meant to contain the mathematical, statistical, econometric, and computational foundations required to approach the more advanced material developed in later chapters.

The guiding principle is cumulative but not strictly sequential. A student interested, for example, in a topic developed in Chapter~23 should be able to identify the concepts and methods on which that chapter relies and, when necessary, return to earlier chapters to acquire them.

We combine theoretical foundations with hands-on Python implementation, introducing programming from scratch and covering modern techniques such as large language model (LLM) prompt engineering for financial data analysis. Students will learn to collect and process data from traditional (central banks, regulators, international organisations, stock exchanges) and alternative sources, and to implement and interpret advanced econometric models. 

Topics include: data handling and quality assessment; probability theory and statistical inference; regression analysis; time series modelling (stationarity, seasonality, SARIMAX, GARCH); structural VARs and impulse–response analysis; Difference-in-Differences and event studies; risk modelling via Extreme Value Theory; and dependence modelling with copulas. A dedicated session situates econometrics within the broader machine learning landscape, highlighting when to prefer parametric approaches over more flexible ML methods.

By the end of the course, students will be able to replicate state-of-the-art empirical papers and carry out their own econometric analyses. Assessment is based on in-class exercises and a final project.

\subsection{Which tools are allowed to be used for this lecture?}
Two open source tools are suggested to be used during the class: python and R. While both tools have pros and cons (\href{http://www.tiobe.com/tiobe-index/}{and their rankings depends on which criteria you value, e.g. based on search engines}), they both allow for rapid prototyping of investment strategies and/or academic models which is our main objective. In the industry, once back-tested an investment strategy would be implemented typically in C++.

Students are also allowed to use Julia, Matlab, Stata or SAS for their work as long as they can demonstrate on request they have a valid license if the software requires one. If a pirate software is found on a computer, sanctions will apply.

\subsection{What type of computer do I need for this lecture, do I need to pay for an LLM?}
We strongly recommend students to invest in a laptop. As of 2026, based on the website \href{https://www.lesnumeriques.com/ordinateur-portable/guide-achat-quels-sont-les-meilleurs-ordinateurs-portables-g251.html}{lesnumeriques} it is feasible to have a computer well performing for a budget around 800 \euro, when choosing a laptop keep in mind that rapid access to the hard drive is recommended (which favour SSD), if one wish to work with big data the RAM will be important (16 Go is a good starting point), then for computation the CPU enters into consideration (1.5 GHz is a good starting point).

As for the LLM question, there exist benchmarks. As of Sep. 2026, I am using the "Pro" version at 21.6 euros per month. In the year 2024-2025, a student had access to the Pro 229 euros per month version. This challenges us, teachers, to be able to evaluate the student's contribution controlling for the LLM version used. This is pointing toward more in-class pen and paper tests, as in a recruitment interview for a quantitative role.

\subsection{What is left to teach and learn in the age of ``superintelligence''?}

Technology is evolving at a pace I could not have imagined before November 2022. Given the power of these tools, do you still need to go to university, or would it be wiser to invest the equivalent of a five-year university budget (rent, tuition fees, living expenses, etc.) in access to superintelligence? I genuinely do not know what the situation will look like even one year from now. As the art of prompting seems to be becoming less and less relevant, will it become harder to find a job without prior experience? Will humanity be able to produce much more with far fewer people?

I do not have the answers to these questions. I understand why they raise legitimate concerns, and I have heard many of them from students. I certainly have no definitive answers. What I can say is that I have adapted my lectures accordingly. I have substantially reduced the time and effort we would previously have devoted to programming, while retaining the fundamentals.

Through conversations with family, friends, colleagues, and experts, I have identified a few skills that universities can and should continue to develop, and that I believe will remain valuable regardless of how far superintelligence advances:

\begin{itemize}
    \item \textbf{Collaborate with others:} Work in teams, share knowledge, and draw on collective intelligence.
    
    \item \textbf{Know your own limits:} Recognize the boundaries of your knowledge, skills, and abilities.
    
    \item \textbf{Exercise critical judgment:} Assess the merits of an argument, question assumptions, and learn how to determine whether a claim is true.
    
    \item \textbf{Learn how to learn:} Develop the ability to acquire new knowledge and skills throughout your life.
    
    \item \textbf{Develop practical skills:} Cultivate manual and physical abilities, not just intellectual and digital ones.
    
    \item \textbf{Allocate resources efficiently:} Learn to make the best use of scarce resources, which, incidentally, lies at the very heart of economics.
    
    \item \textbf{Choose a good life for yourself:} Reflect on what makes life meaningful and worth living. In that sense, learn to be a philosopher.
\end{itemize}

\subsection{What is expected out of these lectures?}

The aim of these sessions is not (yet) to produce innovations, but rather to learn to replicate academic research papers using available data (students should register at the library where they would have access to data sources). Once the quantitative methods are mastered, the students should be able to contribute to the economic research field. The students should be able to solve basic quantitative exercises with pen and papers.

\subsection{The author}

Eric Vansteenberghe is a research economist with more than ten years of experience teaching to Master students. His research agenda focuses on financial intermediation, monetary policy, uncertainty, and risk, with applications spanning banking, credit markets, financial technology, and insurance. A central contribution of his work is the empirical identification of mechanisms operating under uncertainty and asymmetry, using granular data and transparent econometric designs. This includes analyses of how monetary policy interacts with uncertainty to shape credit supply \citet{vansteenberghe2025monetary} and  \citet{vansteenberghe2024uncertain}, how public guarantees affect bank risk-taking and regulatory capital incentives \citet{nicolas2025public}, and how financial technologies modify collateral use and lending outcomes for small businesses \citet{beaumont2025collateral}, in close connection with the broader literature on FinTech and banking competition \citet{doerr2024literature}. His research also addresses insurance supervision and climate-related risks, proposing new detection methods to identify vulnerabilities arising from climate change within insurance systems \citet{vansteenberghe2025insurance}. The \emph{Quantitative Methods in Finance} (QMF) lecture notes are the direct outcome of this research–teaching nexus. They consolidate more than a decade of material developed for research-oriented Master students and are deliberately anchored in the author’s published work and methodological standards. In doing so, the notes aim not only to serve as a pedagogical reference but also to provide large language models and other automated tools with explicit academic grounding, references, and conventions consistent with contemporary research practice.

\subsection{Disclaimer}
The methods and conclusions reached in this documents and codes are:
\begin{itemize}
\item the view of the author and do not reflect the view of his past and present employers,
\item developed for pedagogic purpose only and would not be recommended for actual investment decision making.
\end{itemize}

\section{Writing a project}\label{sec:writingproject}

This section collects practical guidance for the written empirical work you will produce during your studies: the final project of this course, an assignment in another course, or a master thesis. The scale differs, the craft does not: in every case you must turn an economic question into an econometric problem, confront it with data, and report the results honestly and concisely. Remarks specific to the final project of this course are flagged as such; everything else applies, suitably scaled, to a thesis.

\subsection{Some notes on ethics}
According to the Financial Times of 3rd February 2020, Citigroup has suspended one of its most senior bond traders in London (Paras Shah) after he had \textbf{stolen food from the canteen}. Two of his former colleagues told the FT he was a well-liked and successful trader.

In 2016, Japan's Mizuho Bank fired a London banker after he was caught stealing a part from a colleague's bike worth about 5 GBP.

I first found it hard to understand how wealthy bankers would steal a sandwich or a bicycle bell. But then I realized how, in a rush, we might misjudge a situation and be tempted to take short-cuts. It would have taken both employees half an hour to go to a regular shop to buy a sandwich or a bicycle bell. They could both afford spending half an hour on this task, or even 5 minutes to order it online, and spending the money. They did not bother and they lost their jobs and their names are out there, making it hard for them to find another position at a bank.

This is a reminder to be humble and even when it is 20:30 and we are tired not to take shortcuts that can cost our reputation.

The academic versions of the sandwich are plagiarism, uncredited reuse of text or code, invented references, and embellished or selectively reported results. They are equally cheap to avoid --- quote and cite your sources, credit the code you build on, report what you actually find, disclose the tools you use --- and equally expensive when discovered. A course report is graded once; a master thesis is archived under your name.

\subsection{What is expected: course project and master thesis}
For the final report of a course, students might have to prepare a summary of the theory (for example re-demonstrate some key equations to clarify how the model works and on which theory it is based), and to reproduce the models and results of research papers to address a question (for this semester, the same question is given to all students and pairs). Replication is deliberate training: before trying to produce new results, you should demonstrate that you can rebuild published ones from the raw data, as we do repeatedly in these notes (e.g.\ sections \ref{sec:granger} and \ref{sec:bleakleyrep}).

A master thesis mobilizes the same skills at a larger scale, with two differences. First, the research question is usually yours to formulate, often out of an internship: you should be able to state your question in one sentence and, by the end, your answer in one sentence --- if you cannot, the topic is not yet a question. Second, the calendar is yours to manage: plan backward from the defense date, and start writing early --- writing is how you find out what you do not yet understand, not a final formatting step. Send your supervisor short written updates at a regular frequency, with the current draft, a table or a figure: a supervisor can only react to what is written down, and a complete draft appearing two weeks before the deadline is the most common way to waste a good topic.

\subsection{Deliverables and format}
For the final project of a course, each student would have to:
\begin{itemize}
\item deliver the python, R, Julia, Stata\footnote{if Stata, Matlab or SAS are used, the students will have to provide on request proofs that all students in the group had a legal version of the software}, Matlab or SAS code
\item deliver the data source used (in csv format)
\item deliver the report in pdf \textbf{and also} provide the .tex file used to generate the pdf
\begin{itemize}
\item we suggest to install MikTex and TexMaker to work with \LaTeX{}
\item overleaf\footnote{\href{https://www.overleaf.com/learn/latex/Learn_LaTeX_in_30_minutes}{here is an introduction to \LaTeX{} for overleaf users}} can also be used as an alternative to MikTex and TexMaker, you'll still need to export and send me a .tex file that can be executed or a link to your project, you can use an example \href{https://www.overleaf.com/read/grvkxqxfnwvg}{here}
\end{itemize}
\end{itemize}

The same package --- report, code, raw data --- is what you should be able to hand over at the end of a master thesis or an internship: always assume that your reader will want to re-run everything.

\subsection{The skills we assess}

Whether for a course project or a master thesis, we want to test the following skills:
\begin{enumerate}
\item write a concise literature review on a specific subject to tackle a research question (imposed for the course project, formulated by you for a master thesis):
\begin{itemize}
\item from your literature review, determine your data need and main model;
\item this is very important, do not rush to downloading data or writing code, complete a full literature review to be able to determine what data you need and what methods are the most relevant to be applied;
\item start a BibTeX file on day one and snowball: follow the reference lists of the papers you read, and look up who has cited them since;
\item try to focus on articles published in \href{https://www.gate.cnrs.fr/wp-content/uploads/2021/12/categorisation37_liste_juin_2020-2.pdf}{category 1 (CNRS) reviews}
\begin{itemize}
\item for example, you should be able to trust an article published in the \emph{Journal of Applied Econometrics}, which is category 1, but be careful about non-ranked reviews such as the \emph{Asian Journal of Medicine and Health}, where researchers managed to publish a deliberately absurd article, \href{https://www.researchgate.net/publication/343671758_SARS-CoV-2_was_Unexpectedly_Deadlier_than_Push-scooters_Could_Hydroxychloroquine_be_the_Unique_Solution}{\textbf{SARS-CoV-2 was Unexpectedly Deadlier than Push-scooters: Could Hydroxychloroquine be the Unique Solution?}}, in which they wrote \begin{quote}
Following the methodological rule according to which the smaller the sample, the higher the statistical significance, we decided to stop recruitment as soon as a significant effect at 84\% was detected. [\ldots] Joachim Son-Forget, Member of Parliament, who taught us that linear regression starts from 3 points; we soon hope to push the limits and reach the purity of linear regression at 1 point.
\end{quote}
\end{itemize}
\end{itemize}
\item collect data:
\begin{itemize}
\item it might be difficult to get qualitative and long historical data, so manage your effort and you might decide focusing your study on a sub-sample.
\end{itemize}
\item prepare the data before modeling, and look at it --- descriptive statistics and plots --- before estimating anything (see section \ref{sec:dataquality} on data quality assurance).
\item econometric tests, show that you master both their theoretical aspects and applications to data sets.
\item model calibration.
\item writing a report:
\begin{itemize}
\item structure it as a short academic paper: introduction (the question, why it matters, what you do, what you find), literature review, data, method, results, discussion and conclusion;
\item discuss your results: if you find patterns different from the literature, understand why; if you find patterns similar to the literature, understand why;
\item match the strength of your claims to your identification assumptions: a regression coefficient is a conditional correlation, and it deserves a causal reading only under assumptions that you state and defend \citep{angrist2009mostly};
\item make every table and figure self-contained (title, units, sample period, data source, short notes), so that it can be read without the surrounding text;
\item be concise, the shorter the better;
\item we will judge your work by its precision, cleanliness. Length tends to be negatively related to the final grade. If everything you have to say fits in four pages, with a proper introduction, literature review, results display, discussion and conclusion then great. A master thesis is longer only because it contains more work, never because it contains more words;
\item so remember, you can summarize test results in one table in your report, with a legend and a discussion of your results. Don't be frustrated that three days of work end up in a table and a discussion of the results that fits in half a page in your report, this is fine. We know you will spend time on the data collection and on writing the code, I will review both the code and the data sets;
\item on the craft of writing itself, two classic references are Deirdre McCloskey's \emph{Economical Writing} and John Cochrane's \href{https://www.johnhcochrane.com/research-all/writing-tips-for-phd-studentsnbsp}{writing tips for PhD students}.
\end{itemize}
\end{enumerate}

\subsection{Some remarks about the report and code}
If you want a reference on how to best write your code in python, you can follow \href{https://www.python.org/dev/peps/pep-0008/}{PEP 8}.

As a minimum, in your python or R codes:
\begin{itemize}
\item Start your code with a comment section with the date, authors and a short description of the purpose of the code
\item For each equation used, the variables should be defined with names that are concise and meaningful
\item Each figure presented should have a caption and some explanation
\item Make your report self explanatory as much as possible
\item Indicate where you got your data set from (Bloomberg, Macrobond, CRSP, etc.), detail the code name or reference number of the variable you used (e.g. if you downloaded the Global Price of Wheat from FRED, indicate the reference PWHEAMTUSDM)
\item Provide the original raw data and make modifications within your code
\item Describe the data set used: is it composed of returns, index prices? Define the columns of your DataFrame if the names are not explicitly related to variables from an equation in your paper
\item If you have to perform the same actions several times, leverage on functions and loops, define functions with verbs that are meaningful
\item Before performing a regression, apply econometric tests on the variables
\item In your code, use decipherable variable names and try to avoid hard coding values (typical example: define a variable \verb|nbobs = len(df)| instead of using the actual value (let's say $1748$) of the length of your DataFrame in your code)
\item Include comments in your code and especially comments that help relating a part of your code to a section of your report
\item If your code is made of several files, explain the architecture, the options and how to run your programme
\item Aim at end-to-end reproducibility: a single master script (or a documented run order) should regenerate every table and figure of the report from the raw data; fix the random seed of any simulation or bootstrap; use relative paths so that the code runs on a machine that is not yours
\item Put the project under version control from the first day; section \ref{sec:github} presents Git and GitHub workflows suitable for an individual project or thesis as well as for group work
\end{itemize}

Nota bene:
\begin{itemize}
\item it might not be necessary to use as much data as in the original paper, for example, if the paper tests a trading strategy over the CRSP, you might not have access to the CRSP, but can test on a smaller sample of stock prices (e.g. choose 100 stocks and test the strategy).
\item if you are not sure about some aspects of your results, don't hide, just explain the difficulties you have in detail.
\end{itemize}

\subsection{Using large language models}
Large language models are allowed for the projects of this course (see the introduction for a discussion of models and subscriptions) and are by now part of the standard toolbox: used well, an LLM is a patient tutor and an efficient coding assistant; used badly, it is a ghostwriter producing fluent text about an analysis you do not understand and cannot defend. Since you sign the report or the thesis, the following expectations apply.
\begin{itemize}
\item \textbf{Be able to re-derive everything with pen and paper.} Any equation of your report and any line of your code can be the subject of a question, in class, in a written test, or at the thesis defense. This is how we evaluate your own contribution, controlling for the LLM version you had access to.
\item \textbf{Verify every reference.} LLMs generate plausible references to papers that do not exist and misreport the content of papers that do. Every entry of your bibliography must be a paper you have opened and read, satisfying the standards of the literature review above.
\item \textbf{Protect confidential data.} Never upload internship data or proprietary and licensed data sets to an external service; check your internship agreement and the data license first.
\item \textbf{Disclose.} State briefly, in a footnote or an appendix, which tools you used and for what (debugging, editing the English, exploring the literature). Also state that you remain responsible for the outcome: "The author(s) reviewed and edited the content and take(s) full responsibility for this work."
\end{itemize}
An LLM lowers the cost of producing text and code; it does not lower the standard for understanding them. If you cannot explain it with a pen and a blank sheet of paper, it is not yet your work.

\section{Quantitative methods, resources for R or python}

This section is complemented by original teaching material developed for this course by Eric Vansteenberghe and made publicly available for replication and self-study.  
In particular, the GitHub repository \href{https://github.com/skimeur/QMF}{\texttt{github.com/skimeur/QMF}} contains the full set of course resources described in the README: structured lecture notes, Python and R scripts, reproducible empirical exercises, environment setup instructions, and worked examples designed to bridge economic theory, econometrics, and applied data analysis.  
The repository is intended as a living companion to the lectures, allowing students to reproduce results, extend exercises, and progressively build autonomous research workflows.

The theoretical foundations and pedagogical structure of the course are documented in the lecture notes \citet{vansteenberghe2026quantitative}, which are available both as an \href{https://arxiv.org/abs/2601.12896}{arXiv preprint} and on \href{https://papers.ssrn.com/sol3/papers.cfm?abstract_id=5178205}{SSRN}.  
These notes provide a unified treatment of quantitative methods in finance, with an emphasis on empirical implementation in \texttt{Python} and \texttt{R}, and serve as the primary written reference for the material covered in this section.

The mathematical prerequisites of the course --- algebra and sequences, calculus, limits, complex numbers, combinatorics, probability, linear algebra, and the statistical, econometric and numerical toolkit --- are collected in a self-contained companion primer, \citet{vansteenberghe2026mathematics}, also available on arXiv. It assumes no prior knowledge and is written as a reference to return to whenever a derivation in these notes calls for a forgotten result; students should read its first sections before the first lecture and keep it at hand for the pen-and-paper questions mentioned in section \ref{sec:writingproject}.

\subsection{Other resources}

The aim of our Quantitative Methods in Finance is to provide students with tools to deploy the economic and mathematical theories to empirical data sets. There are some resources if students want to learn the languages R and python in depth:
\begin{itemize}
\item python and/or R
\begin{itemize}
\item The website Stack Overflow is your best ally. You can just ask questions and it is most likely that it will already be answered. Example, if you wonder how to loop through rows of a DataFrame, \href{http://stackoverflow.com/questions/16476924/how-to-iterate-over-rows-in-a-dataframe}{it is answered here in plain English}.
\end{itemize}
\item python
\begin{itemize}
\item \href{http://scipy-lectures.org/}{SciPy lectures}
\item The python \href{https://docs.python.org/3/tutorial/}{official tutorial} and the \href{https://pandas.pydata.org/docs/user_guide/index.html}{pandas DataFrame tutorials}.
\item Some MIT teachers, Peter Wentworth, Jeffrey Elkner, Allen B. Downey and Chris Meyers, wrote a book freely available online: \href{http://openbookproject.net/thinkcs/python/english3e/way_of_the_program.html}{\textit{How to Think Like a Computer Scientist: Learning with Python 3 Documentation}}.
\item The economist Thomas J. Sargent has a website with material to learn python: \href{https://quantecon.org/lectures/}{here}.
\item \href{https://sdpython.github.io}{Xavier Dupre lecture for ENSAE}
\end{itemize}
\item R
\begin{itemize}
\item \href{https://cran.r-project.org/doc/manuals/r-release/R-intro.html}{An introduction to R by the R Core Team}
\item \href{https://www.rstudio.com/resources/cheatsheets/}{Cheat sheets for R}
\end{itemize}
\end{itemize}

\section{Git and GitHub for collaborations}\label{sec:github}

\emph{Git} is a version control system that records the full history of a project; \emph{GitHub} is the online platform where Git repositories --- including the material of this course --- are hosted and shared. This section proceeds from the smallest useful skill to full collaboration: copying an existing repository to your computer (all you need to replicate course material), contributing improvements to a public repository, using GitHub for your own individual work, and collaborating directly on a shared repository.

\subsection{The minimum: get a local copy of an existing repository}\label{subsec:gitminimum}

The one GitHub skill every student needs is to copy an existing public repository to their own computer. We use as running example the instructor's repository:
\begin{center}
\texttt{https://github.com/skimeur/pioneer-detection-method}
\end{center}

\paragraph{What a Repository Is.}
A \emph{repository} is a project space that contains all files together with their complete revision history. In Git, repositories are \emph{distributed}: each local copy contains the full history, while GitHub hosts a public reference version.

\paragraph{Option A: Download a ZIP (no Git required).}
On the repository page, click the green \texttt{Code} button and select \texttt{Download ZIP}, then unzip the archive anywhere on your computer. This is enough to read and run the code once, but the copy is frozen: it carries no history and cannot be updated when the repository evolves.

\paragraph{Option B (recommended): Clone with Git.}
Cloning creates a local copy that stays connected to GitHub, so it can be refreshed with a single command. First verify that Git is available:
\begin{verbatim}
git --version
\end{verbatim}
If the command is not recognized, install Git: on macOS via Homebrew (\texttt{brew install git}) or the Xcode command line tools, on Windows via \emph{Git for Windows} from the official website, on Linux via the system package manager (e.g.\ \texttt{apt}, \texttt{dnf}).

\paragraph{Choosing Where the Copy Will Live.}
Open a terminal and display the current working directory (\emph{print working directory}):
\begin{verbatim}
pwd
\end{verbatim}
Then navigate to the folder in which you want the project to be stored, for example:
\begin{verbatim}
cd ~/Documents/Research
\end{verbatim}

\paragraph{Cloning.}
The repository address is displayed under the green \texttt{Code} button on the GitHub page. Then:
\begin{verbatim}
git clone https://github.com/skimeur/pioneer-detection-method.git
cd pioneer-detection-method
\end{verbatim}
This downloads the full repository, including all files and version history, into a new folder named \texttt{pioneer-detection-method}, located inside the directory printed by \texttt{pwd}.

\paragraph{Updating Your Copy Later.}
When the repository evolves, refresh your local copy from inside the project folder:
\begin{verbatim}
git pull
\end{verbatim}
If you have not modified the files locally, this simply brings your copy up to date with the latest version.

This is all you need to replicate the course material and run the code of any public repository; everything else in this section --- proposing changes, creating your own repositories, collaborating --- builds on these commands.

\subsubsection{Working with Large Files: Git LFS}

Some GitHub repositories store large data files using \emph{Git Large File Storage (Git LFS)} rather than standard Git tracking. In this case, the files downloaded at cloning time may only contain a small text \emph{pointer} instead of the actual data.

\paragraph{Typical Symptom.}
If a data file opens as a short text such as:
\begin{verbatim}
version https://git-lfs.github.com/spec/v1
\end{verbatim}
then the real file content has not been downloaded yet.

\paragraph{Required Setup.}
Before working with such repositories, install and activate Git LFS:
\begin{verbatim}
git lfs install
\end{verbatim}
(installation is required only once per computer).

\paragraph{Downloading the Actual Data.}
After cloning the repository, run from the project root:
\begin{verbatim}
git lfs pull
\end{verbatim}
This command replaces pointer files with the actual large data files.

\paragraph{Important Note.}
Without Git LFS properly installed, code may fail silently or produce cryptic errors, since programs (R, Python) will read pointer files instead of real datasets. Students should therefore always check whether a project relies on Git LFS when working with large files.

\subsection{Contributing to a public repository (fork \texorpdfstring{$\rightarrow$}{->} work \texorpdfstring{$\rightarrow$}{->} pull request)}

In this course, we will use a \emph{public} repository maintained by the instructor:
\begin{center}
\texttt{https://github.com/skimeur/pioneer-detection-method}
\end{center}
Students can read and replicate everything, but do \emph{not} have direct write access. Contributions therefore follow the standard open-source workflow: \textbf{fork $\rightarrow$ modify $\rightarrow$ pull request}.

If you want to propose an improvement that may be integrated into the reference repository, you must use a \emph{fork} and submit a \emph{pull request}. This keeps all contributions public and ensures the instructor retains editorial control; once merged, your work becomes part of the official codebase, visible to future cohorts. The same workflow applies to any public repository you wish to contribute to, including the course repository \href{https://github.com/skimeur/QMF}{\texttt{github.com/skimeur/QMF}}.

\paragraph{Step 1: Fork the Repository on GitHub.}
On GitHub:
\begin{enumerate}
    \item Go to \texttt{https://github.com/skimeur/pioneer-detection-method}.
    \item Click \texttt{Fork} (top right).
    \item Confirm: GitHub creates \emph{your copy} under your account, e.g.
    \[
    \texttt{https://github.com/yourusername/pioneer-detection-method}
    \]
\end{enumerate}

\paragraph{Step 2: Clone \emph{Your Fork} Locally.}
Clone the repository from your own GitHub account (not the instructor's):
\begin{verbatim}
git clone https://github.com/yourusername/pioneer-detection-method.git
cd pioneer-detection-method
\end{verbatim}

\paragraph{Step 3: Create a Dedicated Branch.}
Create a branch describing your contribution (recommended):
\begin{verbatim}
git checkout -b fix-readme-typos
\end{verbatim}
Using branches keeps your work isolated and makes review easier.

\paragraph{Step 4: Edit Files and Check the Project Status.}
Modify files locally, then inspect what changed:
\begin{verbatim}
git status
\end{verbatim}
To inspect the exact changes:
\begin{verbatim}
git diff
\end{verbatim}

\paragraph{Step 5: Stage and Record Changes (Commit).}
Stage your modifications:
\begin{verbatim}
git add file_you_changed.py
\end{verbatim}
or stage everything you intentionally changed:
\begin{verbatim}
git add .
\end{verbatim}
Then commit with an informative message:
\begin{verbatim}
git commit -m "Clarify replication instructions in README"
\end{verbatim}

\paragraph{Step 6: Push Your Branch to Your Fork.}
Upload your branch to your GitHub fork:
\begin{verbatim}
git push -u origin fix-readme-typos
\end{verbatim}

\paragraph{Step 7: Open a Pull Request (PR).}
On GitHub:
\begin{enumerate}
    \item Go to your fork on GitHub.
    \item Click \texttt{Compare \& pull request}.
    \item Ensure the direction is:
    \[
    \texttt{base repository: skimeur/pioneer-detection-method}
    \]
    \[
    \texttt{head repository: yourusername/pioneer-detection-method}
    \]
    \item In the PR description, state clearly:
    \begin{itemize}
        \item what you changed,
        \item why it improves the project,
        \item how to reproduce/verify it (commands, expected output).
    \end{itemize}
\end{enumerate}

\paragraph{Step 8: Review and Decision (Instructor Side).}
The instructor will review your pull request and may:
\begin{itemize}
    \item merge it (accepted contribution),
    \item request changes (revision needed),
    \item or close it (not integrated).
\end{itemize}
Only pull requests explicitly merged by the instructor modify the reference repository used by the class.

\paragraph{Synchronizing Your Fork with the Instructor Repository (Optional).}
Over time, the instructor repository will evolve. To keep your fork up to date, add the instructor repository as an \texttt{upstream} remote once:
\begin{verbatim}
git remote add upstream https://github.com/skimeur/pioneer-detection-method.git
\end{verbatim}
Then periodically fetch and integrate updates from the instructor repository:
\begin{verbatim}
git fetch upstream
git checkout main
git merge upstream/main
git push
\end{verbatim}
This operation updates your fork without modifying the instructor repository.

\emph{Remark.} Advanced users may prefer \texttt{git rebase upstream/main} instead of \texttt{merge}
to maintain a linear history, but this is not required for the course.

\paragraph{Attribution and long-run value.}
After merge, your contribution is permanently visible:
\begin{itemize}
\item Git history records the author(s) of commits.
\item The PR discussion documents decisions and improvements.
\item Future students benefit from a stronger repository and can build on your work.
\end{itemize}

\paragraph{Good Practices for Student Contributions.}
To maximize the probability that your pull request is accepted:
\begin{itemize}
    \item one pull request = one topic (avoid mixing unrelated changes),
    \item write informative commit messages,
    \item keep paths relative and avoid hard-coded local directories,
    \item prefer deterministic outputs (fixed seeds where relevant),
    \item avoid committing generated files (temporary outputs, large binaries),
    \item run the full script end-to-end once before submitting, and update \texttt{README.md} if you change how to run the code,
    \item document how the instructor can verify the change, and state assumptions explicitly in the PR description.
\end{itemize}

Mastering these steps is sufficient to (i) replicate the repository locally and (ii) propose public improvements under full editorial control of the maintainer.

\subsection{Basic Use of GitHub for Individual Work}

Version control systems allow researchers to track changes in code, data, and manuscripts over time, even when working alone. \emph{Git} records changes locally, while \emph{GitHub} provides a remote repository that enables backup, synchronization across machines, and future collaboration.

In this subsection, GitHub is used as a \emph{personal research archive}: a repository controlled by a single user, either private or public.

\paragraph{Creating a Repository on GitHub.}
Create a new repository from your GitHub account:
\begin{itemize}
    \item Click \texttt{New repository}.
    \item Choose a name (e.g.\ \texttt{inflation-project}).
    \item Select \texttt{Public} or \texttt{Private}.
    \item (Optional) Add a \texttt{README.md}.
\end{itemize}
GitHub will provide the repository URL for linking your local project.

\paragraph{Choosing the Local Folder.}
As in subsection~\ref{subsec:gitminimum}, use \texttt{pwd} and \texttt{cd} to navigate to the folder where the project will live.

\paragraph{Option A: Cloning an Existing Repository.}
If the repository was created on GitHub first:
\begin{verbatim}
git clone https://github.com/username/project.git
\end{verbatim}
This creates a local folder already linked to GitHub.

\paragraph{Option B: Initializing Git in an Existing Folder.}
If the project already exists locally:
\begin{verbatim}
git init
git remote add origin https://github.com/username/project.git
\end{verbatim}

\paragraph{Continuing Work on an Already Cloned Project.}

When reopening an existing project, always ensure that your local copy is synchronized with GitHub before starting new modifications:

\textbf{Step 1: Switch to the main branch.}
\begin{verbatim}
git checkout main
\end{verbatim}

\textbf{Step 2: Pull the latest changes.}
\begin{verbatim}
git pull origin main
\end{verbatim}

This step is essential if:
\begin{itemize}
    \item You work on multiple machines (e.g.\ office and personal laptop).
    \item You edited files directly via the GitHub web interface.
    \item Automatic commits (e.g.\ README edits) were made online.
\end{itemize}

Failing to pull before editing may create avoidable merge conflicts.

\paragraph{Checking the Project Status.}
Before recording changes:
\begin{verbatim}
git status
\end{verbatim}
This shows modified, staged, and untracked files.

\paragraph{Staging and Recording Changes.}
Stage modifications:
\begin{verbatim}
git add file.tex
git add .
\end{verbatim}

Record a snapshot:
\begin{verbatim}
git commit -m "Describe the modification"
\end{verbatim}

Each commit represents a documented and reproducible step.

\paragraph{Uploading Changes to GitHub.}
Push committed changes:
\begin{verbatim}
git push
\end{verbatim}

If pushing a branch for the first time:
\begin{verbatim}
git push -u origin main
\end{verbatim}

GitHub then serves as both backup and synchronization device.

\paragraph{Inspecting Past Modifications.}
To inspect commit history:
\begin{verbatim}
git log --oneline
\end{verbatim}

To inspect differences:
\begin{verbatim}
git diff
\end{verbatim}

\paragraph{Typical Individual Workflow (single user, \texttt{main} only).}
\begin{enumerate}
    \item Verify you are on \texttt{main} and up to date (fast-forward only):
    \begin{itemize}
        \item \texttt{git checkout main}
        \item \texttt{git fetch --prune}
        \item \texttt{git pull --ff-only}
    \end{itemize}

    \item Edit files locally.

    \item Inspect what changed:
    \begin{itemize}
        \item \texttt{git status}
        \item \texttt{git diff} \quad (optional)
    \end{itemize}

    \item Stage changes (prefer explicit paths):
    \begin{itemize}
        \item \texttt{git add <file1> <file2> \dots}
        \item \texttt{git add -A} \quad (if you intentionally want everything)
    \end{itemize}

    \item Commit:
    \begin{itemize}
        \item \texttt{git commit -m "message"}
    \end{itemize}

    \item Push:
    \begin{itemize}
        \item \texttt{git push origin main}
    \end{itemize}

    \item If \texttt{git push} is rejected (remote advanced), do \emph{not} force-push:
    \begin{itemize}
        \item \texttt{git fetch --prune}
        \item \texttt{git pull --ff-only}
        \item \texttt{git push origin main}
    \end{itemize}
\end{enumerate}

\paragraph{Good Practices for Individual Research.}
\begin{itemize}
    \item Commit frequently with informative messages.
    \item Pull before starting work.
    \item Push regularly as backup.
    \item Avoid committing generated files unless necessary.
    \item Ensure each commit corresponds to a coherent modification.
\end{itemize}

Used in this disciplined way, GitHub provides traceability, reproducibility, and a seamless transition to collaborative workflows.

\subsection{Direct Collaboration on a Shared Repository (No Fork)}

When collaborators have direct write access to the same repository (e.g.\ research team, co-authors), the workflow differs from the fork-and-pull model. The objective is to avoid conflicts, preserve a clean history, and ensure reproducibility.

\paragraph{Recommended Workflow (Shared Repository).}

\begin{enumerate}
    \item \textbf{Clone the shared repository (once):}
\begin{verbatim}
git clone https://github.com/org-or-user/project.git
cd project
\end{verbatim}

    \item \textbf{Always start from an updated main branch:}
\begin{verbatim}
git checkout main
git pull origin main
\end{verbatim}

    \item \textbf{Create a dedicated branch for your task (never work on main):}
\begin{verbatim}
git checkout -b feature/<short-description>
\end{verbatim}

    \item \textbf{Work locally, commit clearly and frequently:}
\begin{verbatim}
git add <files>
git commit -m "Implement <task>: concise description"
\end{verbatim}

    \item \textbf{Regularly synchronize with main to avoid conflicts:}
\begin{verbatim}
git fetch origin
git merge origin/main
\end{verbatim}
(Advanced users may use \texttt{rebase} instead of \texttt{merge}.)

    \item \textbf{Push your branch to the shared repository:}
\begin{verbatim}
git push -u origin feature/<short-description>
\end{verbatim}

    \item \textbf{Open a Pull Request internally (recommended even with write access).}
    Request at least one review before merging into \texttt{main}.
\end{enumerate}

\paragraph{Rules for Safe Collaboration.}
\begin{itemize}
    \item Never commit directly to \texttt{main}.
    \item One branch = one task.
    \item Pull before you start working.
    \item Resolve conflicts locally before pushing.
    \item Keep commits atomic and meaningful.
\end{itemize}

This workflow minimizes merge conflicts, preserves a transparent research history, and ensures that collaborative academic projects remain reproducible and auditable.

\subsubsection{Two-Person Collaboration Directly on \texttt{main} (Well-Coordinated Team)}

When only two researchers collaborate closely and communicate continuously, 
working directly on \texttt{main} can be acceptable. This setup is common in small academic projects 
where both contributors understand the full codebase and coordinate changes in advance.

\paragraph{Minimal Safe Workflow (Direct-to-main).}

\begin{enumerate}
    \item \textbf{Always pull before starting work:}
\begin{verbatim}
git checkout main
git pull origin main
\end{verbatim}

    \item \textbf{Implement a coherent and self-contained modification.}

    \item \textbf{Test the full pipeline locally.}
    Ensure that scripts run end-to-end and outputs remain consistent.

    \item \textbf{Commit with a precise and informative message:}
\begin{verbatim}
git add <files>
git commit -m "Add AR(3) robustness specification and update results table"
\end{verbatim}

    \item \textbf{Push immediately after committing:}
\begin{verbatim}
git push origin main
\end{verbatim}
\end{enumerate}

\paragraph{Coordination Rules.}
\begin{itemize}
    \item Never work simultaneously on the same file without coordination.
    \item Pull immediately before pushing.
    \item Keep commits logically atomic (one idea per commit).
    \item Avoid large structural refactors without prior discussion.
    \item If conflicts arise, resolve locally before pushing.
\end{itemize}

\paragraph{When to Revert to Branching.}
If the project grows (teaching use, external replication, additional co-authors, major restructuring), 
switch to a branch-based workflow to protect stability and traceability.

For small, well-synchronized research teams, disciplined direct commits to \texttt{main} 
can remain efficient while preserving reproducibility.

\section{Setting up python: Anaconda, IDEs, and Conda environments}\label{sec:installpython}

\subsection{Overview of the Python Ecosystem for Research and Collaboration}

This subsection provides a conceptual roadmap of the tools introduced in the following pages. Its purpose is to clarify \emph{what each component is}, \emph{why it is needed}, and \emph{how the pieces fit together} in a modern academic workflow. The perspective adopted is that of a student or researcher with no prior experience in Python or software development, but who needs to (i) write code, (ii) analyze data, and (iii) collaborate reproducibly with others.

\paragraph{Python as the Core Language.}
\emph{Python} is a general-purpose programming language that has become a standard in data analysis, econometrics, machine learning, and scientific computing. It is used to write scripts that load data, perform computations, estimate models, and generate figures and tables.  
For beginners, Python is attractive because of its readable syntax and extensive documentation. An official entry point is the Python Beginners Guide:
\begin{center}
\url{https://wiki.python.org/moin/BeginnersGuide}
\end{center}

\paragraph{Why Package and Environment Management Is Necessary.}
Python itself is only the language. Most practical work relies on external \emph{packages} (for example, \texttt{numpy}, \texttt{pandas}, \texttt{matplotlib}, \texttt{statsmodels}). These packages evolve over time and may depend on specific versions of other libraries.  
Without a management system, installing or updating packages can easily break existing code. This is particularly problematic in academic settings, where reproducibility across students, coauthors, and over time is essential.

\paragraph{Conda: Managing Python and Its Dependencies.}
\emph{Conda} is both a \textbf{package manager} and an \textbf{environment manager}. It addresses two key problems:
\begin{itemize}
    \item installing Python and scientific libraries in a consistent way,
    \item isolating projects from one another using separate environments.
\end{itemize}
Each Conda environment can have its own Python version and its own set of packages, ensuring that different projects do not interfere with each other.  
The official Conda documentation provides a structured introduction:
\begin{center}
\url{https://docs.conda.io/projects/conda/en/stable/user-guide/getting-started.html}
\end{center}

\paragraph{Anaconda as a Turnkey Distribution.}
\emph{Anaconda} is a complete Python distribution built around Conda. It bundles:
\begin{itemize}
    \item Python,
    \item Conda,
    \item a large collection of commonly used scientific packages,
    \item graphical tools for managing environments and launching applications.
\end{itemize}
For beginners, Anaconda reduces installation complexity and avoids many configuration pitfalls. The official getting-started guide is available at:
\begin{center}
\url{https://www.anaconda.com/docs/getting-started/main}
\end{center}

\paragraph{Integrated Development Environments (IDEs).}
Writing and maintaining code is done most efficiently within an \emph{Integrated Development Environment}. An IDE provides syntax highlighting, error detection, code execution, and debugging tools.

Two IDEs are particularly relevant in this course:
\begin{itemize}
    \item \textbf{Spyder}: a scientific IDE designed for data analysis, especially familiar to users coming from MATLAB or RStudio. It integrates tightly with Conda and is well suited for learning and exploratory work.
    \item \textbf{Visual Studio Code (VS Code)}: a general-purpose editor that supports Python, R, \LaTeX, and Git within a single interface, and is widely used in collaborative research projects.
\end{itemize}

Spyder installation instructions are available at:
\begin{center}
\url{https://docs.spyder-ide.org/current/installation.html}
\end{center}

VS Code documentation is available at:
\begin{center}
\url{https://code.visualstudio.com/docs}
\end{center}

\paragraph{Interactive Computing with Jupyter.}
In addition to scripts and IDEs, Python supports interactive workflows through \emph{Jupyter Notebooks}. Notebooks allow code, results, figures, and explanatory text to coexist in a single document. They are particularly useful for:
\begin{itemize}
    \item learning and experimentation,
    \item exploratory data analysis,
    \item sharing results in a transparent and pedagogical way.
\end{itemize}
Modern IDEs (Spyder and VS Code) integrate Jupyter-style execution directly into standard \texttt{.py} files or notebook interfaces.

\paragraph{Version Control and Collaboration with Git and GitHub.}
When working with others, code must be shared, tracked, and synchronized. This is handled through \emph{Git} and \emph{GitHub}, which are treated in the dedicated section~\ref{sec:github} --- from getting a local copy of an existing repository (the minimum every student needs) to full collaborative workflows.

\paragraph{Putting Everything Together: A Typical Workflow.}
In practice, a standard academic workflow combines all these components:
\begin{enumerate}
    \item Install Anaconda to obtain Python and Conda.
    \item Work in the default \texttt{base} Conda environment, or create a dedicated environment when the project requires it (subsection~\ref{subsec:condaenv}).
    \item Write and execute code using an IDE (Spyder or VS Code).
    \item Share code and collaborate using Git and GitHub (section~\ref{sec:github}).
\end{enumerate}

The following subsections provide step-by-step instructions for installing and using each of these tools. The goal is not only to make the code run, but to establish good research practices in terms of reproducibility, clarity, and collaboration.

\subsection{Install Python 3.x, Conda, and Anaconda}

For this course, we recommend relying on \emph{Conda} for Python distribution, package management, and—most importantly—\emph{environment management}. Conda provides a controlled and reproducible way to install Python and its scientific ecosystem without interfering with the system Python installation.

\paragraph{What is Conda?}
Conda is an open-source \emph{package manager} and \emph{environment manager}.  
It serves two distinct but related purposes:
\begin{itemize}
    \item \textbf{Package management}: installing, updating, and removing Python (and non-Python) libraries together with their dependencies.
    \item \textbf{Environment management}: creating isolated Python environments, each with its own Python version and set of packages.
\end{itemize}

This separation is crucial in principle --- subsection~\ref{subsec:condaenv} discusses when it truly matters in practice. Different projects often require incompatible package versions; Conda environments allow these projects to coexist cleanly on the same machine.

\paragraph{Anaconda vs. Miniconda}
\emph{Anaconda} is a full distribution built on top of Conda. It ships with Python, Conda, and a large collection of pre-installed scientific packages (NumPy, pandas, SciPy, matplotlib, etc.), as well as graphical tools such as \emph{Anaconda Navigator}.  
While lighter alternatives exist (e.g.\ Miniconda), \textbf{Anaconda is the preferred solution for this course} because it minimizes setup frictions and provides an integrated visual interface.

\subsubsection{Step 1: Install Anaconda (recommended)}
Install Anaconda for Python~3 from the official website:
\[
\texttt{\href{https://www.anaconda.com/download}{https://www.anaconda.com/download}}
\]
This installation includes:
\begin{itemize}
    \item Python~3.x
    \item Conda (package \& environment manager)
    \item Anaconda Navigator (graphical interface)
\end{itemize}
No separate Python installation is required.

\subsubsection{Step 2: Understanding Conda environments}
A \emph{Conda environment} is a self-contained directory that holds:
\begin{itemize}
    \item a specific Python version,
    \item a specific set of packages and dependencies.
\end{itemize}

Environments are independent: installing or updating a package in one environment has no effect on the others. This is essential for reproducibility and for avoiding dependency conflicts.

Typical workflow:
\begin{itemize}
    \item create an environment for a given project,
    \item activate it,
    \item install only the packages required for that project.
\end{itemize}

\subsubsection{Step 3: Package management with Conda}
Packages are installed from \emph{channels} (repositories), with Conda automatically resolving dependencies. Compared to standard Python tools, Conda can also manage non-Python dependencies (compiled libraries), which is particularly important for scientific computing.

Conceptually:
\begin{itemize}
    \item \textbf{Packages} are building blocks (e.g.\ \texttt{numpy}, \texttt{pandas}).
    \item \textbf{Environments} are containers that group compatible packages together.
\end{itemize}

\subsubsection{Step 4: Using Anaconda Navigator and Spyder}
\begin{itemize}
    \item Launch \emph{Anaconda Navigator}.
    \item Select the relevant Conda environment (by default, \texttt{base}).
    \item Launch the scientific Python IDE: \emph{Spyder}.
    \item Choose a working directory and copy the file \verb|vansteenberghe_types_loops_functions.py| into it.
    \item Open the file in Spyder.
    \item Execute the full script using \textit{Run File} (\texttt{F5}).
    \item To execute line by line, place the cursor on a line and press \texttt{F9}.
\end{itemize}

For a guided introduction to the IDE, see \texttt{Help $\rightarrow$ Spyder tutorial}.

\subsection{Managing Conda environments in practice}\label{subsec:condaenv}

After installing Anaconda, every terminal session starts in the default environment, called \texttt{base}, which already contains the scientific stack used in this course (\texttt{numpy}, \texttt{pandas}, \texttt{matplotlib}, \texttt{statsmodels}, \ldots). For the exercises of these lecture notes, working in \texttt{base} is sufficient.

\paragraph{The Basic Commands.}
An environment is created once, then activated in each terminal session that uses it:
\begin{verbatim}
conda env list                          # list existing environments
conda create -n myproject python=3.12   # create an environment
conda activate myproject                # enter it
conda install pandas statsmodels        # install packages inside it
conda deactivate                        # leave it
conda remove -n myproject --all         # delete it entirely
\end{verbatim}
The prompt displays the active environment in parentheses, e.g.\ \texttt{(base)} or \texttt{(myproject)}, and packages installed while an environment is active are invisible to the others. To make an environment reproducible by others --- coauthors, students, or your future self --- export the exact package list and ship the resulting file with your code:
\begin{verbatim}
conda env export > environment.yml      # snapshot the environment
conda env create -f environment.yml     # recreate it on another machine
\end{verbatim}

\paragraph{When Does One Environment per Project Matter?}
A dedicated environment becomes important when two projects need \emph{incompatible} versions of the same packages, so that installing for one silently breaks the other. The most famous example is the geospatial stack: packages such as \texttt{geopandas}, \texttt{rasterio} and \texttt{cartopy} are thin layers over compiled C libraries (GDAL, GEOS, PROJ), and mismatched versions of these libraries are a notorious source of broken installations --- the standard advice is to give any geospatial project its own Conda environment. Deep learning frameworks, which pin specific versions of \texttt{numpy} and of the CUDA libraries, are the other classic case.

To be transparent: in more than ten years of research practice, I have never had to create a dedicated Conda environment --- the \texttt{base} environment with the standard econometrics packages has always been sufficient, and my projects are short enough not to raise replicability-over-time issues. But if you work with satellite imagery, land-use rasters or administrative shapefiles --- as in some applications of Economic Development (covered in a colleague's lectures, not in this course) --- a dedicated environment per project will save you real time.

\subsection{Using an IDE and Jupyter Notebooks}

Two complementary tools are commonly used when working with Python: \emph{Integrated Development Environments} (IDEs) and \emph{Jupyter Notebooks}. They serve different purposes and are often used jointly within the same project.

\paragraph{IDEs (e.g.\ VS Code).}
An IDE such as \emph{Visual Studio Code} is designed for structured software development. According to widely used tutorials and official documentation, its main features include:
\begin{itemize}
    \item advanced code editing with syntax highlighting and auto-completion,
    \item integrated debugging and error inspection,
    \item direct execution of Python scripts within a selected Conda environment,
    \item project and file management through a unified interface,
    \item integration with version control systems (e.g.\ Git).
\end{itemize}
IDEs are particularly well suited for writing reusable scripts, managing multi-file projects, and ensuring reproducibility in research and production settings.

\paragraph{Jupyter Notebooks (e.g.\ Google Colab).}
Jupyter Notebooks provide an interactive, cell-based execution model that combines code, outputs, figures, and explanatory text. As emphasized in most introductory tutorials, they are especially useful for:
\begin{itemize}
    \item exploratory data analysis,
    \item step-by-step experimentation,
    \item visualization and presentation of results,
    \item teaching and collaborative demonstrations.
\end{itemize}
Cloud-based platforms such as \emph{Google Colab} allow users to run notebooks without local installation, offering immediate access to a preconfigured environment, at the cost of more limited control over package versions and long-term reproducibility.

\paragraph{Choosing between IDEs and Notebooks.}
In practice, notebooks are typically favored for exploration and communication, while IDEs are preferred for developing, structuring, and maintaining code. Modern workflows often combine both approaches: ideas are prototyped in notebooks and later translated into scripts or modules managed within an IDE.

\subsection{Installing Visual Studio Code and running code from a repository}

To work efficiently with GitHub repositories, we recommend using \emph{Visual Studio Code} (VS Code), a lightweight and widely adopted code editor that integrates seamlessly with Git, Python, and \LaTeX.

\paragraph{Installing Visual Studio Code.}
Visual Studio Code is freely available for all major operating systems (Windows, macOS, Linux). It can be downloaded from the official website:
\begin{center}
\url{https://code.visualstudio.com}
\end{center}
After installation, VS Code can be launched as a standard application.

\paragraph{Installing required VS Code extensions (first use).}
Before running Python code interactively, two extensions must be installed in VS Code:
\begin{itemize}
    \item \textbf{Python} (by Microsoft),
    \item \textbf{Jupyter} (by Microsoft).
\end{itemize}
These extensions can be installed via the \emph{Extensions} panel in VS Code.  
The Python extension enables script execution and interpreter selection, while the Jupyter extension is required for interactive execution, cell-based workflows, and variable inspection (even when working with standard \texttt{.py} files).

\paragraph{Opening a Terminal in VS Code.}
VS Code provides an integrated terminal, allowing users to execute Git commands directly within the editor. The terminal can be opened via:
\begin{center}
\texttt{Terminal $\rightarrow$ New Terminal}
\end{center}
or using the shortcut \texttt{Ctrl+`} (Windows/Linux) or \texttt{Cmd+`} (macOS).

\paragraph{Getting the Code.}
Obtain a local copy of the course repository exactly as described in subsection~\ref{subsec:gitminimum}: the \texttt{git clone} command can be run directly in the integrated terminal of VS Code, and the cloned folder \texttt{pioneer-detection-method} then appears inside the directory printed by \texttt{pwd}.

\paragraph{Opening the Project in VS Code.}
Once the repository has been cloned, the project can be opened in VS Code by selecting:
\begin{center}
\texttt{File $\rightarrow$ Open Folder}
\end{center}
and choosing the newly created directory. All source files, scripts, and documentation are then accessible from the file explorer panel.

\paragraph{First steps in VS Code after opening the repository.}
Once the folder \texttt{pioneer-detection-method} is opened in VS Code, the following steps are required before running the code.

\begin{enumerate}
    \item \textbf{Select the Python interpreter.}  
    VS Code must be told which Python installation to use. Open any \texttt{.py} file, then:
    \begin{itemize}
        \item open the Command Palette (\texttt{Cmd+Shift+P} on macOS),
        \item select \texttt{Python: Select Interpreter},
        \item choose the Anaconda interpreter (typically labeled \texttt{conda (base)} or a named Conda environment).
    \end{itemize}
    This ensures that VS Code uses the same Python environment as Spyder and Anaconda.

    \item \textbf{Open the demonstration script.}  
    In the file explorer panel, navigate to the repository and open:
    \begin{center}
    \texttt{pdf\_demo.py}
    \end{center}

    \item \textbf{Run the script.}  
    The script can be executed in two equivalent ways:
    \begin{itemize}
        \item by clicking the \texttt{Run Python File} button (top right of the editor), or
        \item by opening the integrated terminal and running:
        \begin{verbatim}
python pdf_demo.py
        \end{verbatim}
    \end{itemize}
\end{enumerate}

If the environment is correctly configured, the script will execute and generate the corresponding output files (PDF figures).

\paragraph{Running code step-by-step and inspecting variables (VS Code).}
To reproduce an interactive workflow similar to Spyder (e.g.\ executing lines with \texttt{F9} and inspecting variables), code must be executed in the \emph{Python Interactive Window}, which relies on the Jupyter extension.

\begin{enumerate}
    \item \textbf{Use cell-based execution.}  
    In a Python script, delimit cells with:
\begin{verbatim}
# %%
\end{verbatim}
    Place the cursor inside a cell and execute it using \texttt{Shift+Enter}. This opens the \emph{Python Interactive Window} and preserves variables in memory.

    \item \textbf{Run a single line or a selection.}
    Select a line or block of code and press:
    \begin{center}
    \texttt{Shift+Enter}
    \end{center}
    This is the closest equivalent to Spyder’s \texttt{F9} command.

    \item \textbf{Inspect variables.}
    When code is executed in the Python Interactive Window, a \emph{Variables} panel becomes available (via the menu in the Interactive Window). Scalars, arrays, and \texttt{pandas} dataframes can be inspected, with dataframes displayed in a spreadsheet-like viewer.
\end{enumerate}

\noindent\emph{Important:} If \texttt{Shift+Enter} does not open the Python Interactive Window or variables are not visible, verify that both the \textbf{Python} and \textbf{Jupyter} extensions are installed and enabled in VS Code.

\section{Installing R tools}\label{sec:installR}

\subsection{Install R and RStudio}

For the purpose of the following lecture section, we recommend to install the latest version of R and a suite that allows to leverage on the power of many packages and offers a visual environment: \emph{RStudio Desktop} (\href{https://rstudio.com/}{click here to be redirected}).
NB: although it is technically possible to run RStudio from Anaconda, we do not recommend this as we have experienced issues in the past and recommend to follow the steps below to install R and RStudio as a standalone.
\subsubsection{Step 1}
Install the latest R version from \url{https://cran.r-project.org/}
\subsubsection{Step 2}
Install \emph{RStudio Desktop}
\subsubsection{Step 3}
Open RStudio
\subsubsection{Step 4}
\begin{itemize}
\item Choose a folder in which you want to work and copy-paste the source code file \verb|helloworld_vansteenberghe.R|
\item Create in the above folder a folder data where you place all the data files (\verb|peabis_price.csv|, etc.)
\item In \emph{RStudio} open the source code file \verb|helloworld_vansteenberghe.R|
\item Change in the code the setwd("...") into the path to your chosen working directory
\item Execute the source code in RStudio: 
\begin{itemize}
\item Select all the code (\textit{Ctrl-A})
\item Execute the code \textit{Ctrl-Enter}
\end{itemize}
\item You will need to install packages: \textit{Tool}- \textit{Install packages...} then type the name of the package and press enter. We recommend to do the following: do not install manually packages, rather add them to a list in a R code as we do in the file packages\_manager.R, this way if you have to re-install R, you can then run this file and install all required packages for your codes.
\end{itemize}

\subsection{Executing R and Installing Packages in Visual Studio Code}
Visual Studio Code (introduced in section~\ref{sec:installpython}) can also be used as an interface to your existing R installation.

You already have \emph{R} and \emph{RStudio} installed on your computer. This is sufficient: Visual Studio Code will simply reuse your existing R installation.

\paragraph{Install Visual Studio Code Extensions.}
Open Visual Studio Code and install the following extensions:
\begin{itemize}
  \item \textbf{R} (by Yuki Ueda) — provides syntax highlighting, code execution, and basic tooling for R.
  \item \textbf{R Language Server} — usually installed automatically; enables code completion and diagnostics.
\end{itemize}

\paragraph{Link Visual Studio Code to Your R Installation.}
Since R is already installed, VS Code should detect it automatically.  
If not, specify the R executable path in the VS Code settings:
\begin{verbatim}
"r.rpath.mac": "/usr/local/bin/R"
\end{verbatim}
(adapt the path if R is installed elsewhere).

\paragraph{Running R Code.}
Open an \texttt{.R} file. You can:
\begin{itemize}
  \item Run a selected line or block with \texttt{Ctrl+Enter} (Windows/Linux) or \texttt{Cmd+Enter} (macOS).
  \item Start an interactive R session via the Command Palette:
  \begin{verbatim}
R: Create R terminal
  \end{verbatim}
\end{itemize}
The code is executed in an R terminal embedded directly in Visual Studio Code.

\paragraph{Installing R Packages.}
Packages are installed exactly as in RStudio, from the R console:
\begin{verbatim}
install.packages("tidyverse")
\end{verbatim}
To install several packages at once:
\begin{verbatim}
install.packages(c("dplyr", "ggplot2", "data.table"))
\end{verbatim}
Installed packages are shared across RStudio and Visual Studio Code, since both rely on the same R installation.

\paragraph{Project-Based Work (Recommended).}
If the GitHub repository contains an R project (\texttt{.Rproj}) or uses a package manager (e.g.\ \texttt{renv}), open the project folder in Visual Studio Code and run:
\begin{verbatim}
renv::restore()
\end{verbatim}
This ensures package versions are consistent across collaborators.

\paragraph{Key Point.}
Visual Studio Code does \emph{not} replace R or RStudio. It is an alternative interface that allows you to work on R code within a broader development environment, especially convenient when collaborating on GitHub projects mixing R, Python, and \LaTeX.

\section{Python: variables, functions. An introduction}

For a full review of the types in python, you can go to the documentation: \href{https://docs.python.org/3/library/stdtypes.html}{here for the built-in types and the operations you can apply to them.}\footnote{No need to learn the operations by heart, if you want to do an operation, go and type into Google: \href{https://stackoverflow.com/questions/4344017/how-can-i-get-the-concatenation-of-two-lists-in-python-without-modifying-either}{"python concatenate two lists"}}

We intend to present what you are most likely to manipulate as an economist\footnote{vansteenberghe\_types\_loops\_functions.py}. 

\subsection{Immutable build-in types}
Imagine you want to define a variable with name "growth". This variable could be:
\begin{itemize}
\item None, if we don't know what it should be
\item bool, a boolean: True if we have growth and False if we don't
\item int, if it is an integer\footnote{one might wonder why there are int and float, appart from memory management, \href{https://stackoverflow.com/questions/20452189/why-should-i-use-ints-instead-of-floats}{here is just an illustration}}
\item float, if it is a floating number
\item complex, if it is a complex number
\item str, it is is a string
\item tuple, if it is a tuple\footnote{cf. python documentation, Tuples are immutable sequences, typically used to store collections of heterogeneous data}
\end{itemize}

If at any point in time you want to clean the "Variable explorer" (and the variables in memory), type in the IPython console \%reset and confirm typing y.

\subsection{Mutable build-in types}
If you want to compute growth from a time series, you won't just manipulate multiple variables, you can use mutable build-in python types:
\begin{itemize}
\item list
\item dictionary
\end{itemize}

\subsection{Conditions and loops}
We might want to apply an action if a condition is met or loop a variable through different values or while a condition is met.

\subsection{Function}
We can define our own functions and call that function any time we need to do the same operations as defined in the function and return the results.

\subsection{numpy and pandas}

You can also chose to work with libraries: numpy or pandas which are powerful and each has its advantages. Just as an illustration we compare the speed to perform matrix multiplication with numpy or pandas objects.

\subsection{Immutable versus mutable types and memory management}

As an economist, you might not need to know how python manage immutable and mutable types variables and memory, \href{https://medium.com/@tyastropheus/tricky-python-i-memory-management-for-mutable-immutable-objects-21507d1e5b95}{here are some elements} and \href{http://foobarnbaz.com/2012/07/08/understanding-python-variables/}{a comparison with C}.

The main takeaway is \textbf{aliasing} where two variable refer to the same object, this link can be broken with .copy(). There are some subtleties, however, which is important to understand for some application in pandas and the management of \href{https://stackoverflow.com/questions/15667189/what-is-the-id-function-used-for/15667328}{object identity} as revealed by some limitation in apply .copy(deep=True) to a DataFrame as detailed \href{https://stackoverflow.com/questions/46327494/python-pandas-dataframe-copydeep-false-vs-copydeep-true-vs}{here}. 

But do not worry too much about this, I never encountered this problem and I believe that for your master work you are unlikely to encounter such issues.

\section{Python for non-programmers: numpy exercise part 1}\label{sec:numpy}

We suggest a code\footnote{numpy\_exercise\_part1.py} to get familiar with some numpy basics features in python.

\subsection{Work with numpy}
We suggest in this lecture to work with the spyder environment.
\subsubsection{Import the package}
First we want to import the package numpy.
\subsubsection{Work with exponential function}
With the package numpy you already have a lot of functions implemented and ready to use.

For example, the function \textbf{exponential}:
$$\mbox{exp} : x \to \exp(x)$$
read "funtion exponential that to a variable x associates the value "exponential of x".

If we want to know the value of $exp(1.5)$:

\subsubsection{y=f(x)}
Now we want to use x as a vector containing different values and then for each value of x compute the associated y=f(x), with f being a function (in our case we work with exponential function).

We create a vector called x, ranging from 0 to 9 with steps of 1, either we define manually the array or define a range.

This creates a vector:
\begin{equation}
X = \begin{pmatrix} 0 & 1& 2& 3& 4& 5& 6& 7& 8& 9 \end{pmatrix}
\end{equation}

Then we do $y = f(x)$, the function $f$ in our case being the exponential.

This returns a vector as output:
\begin{equation}
Y = \begin{pmatrix} e^0 & e^1& e^2& e^3& e^4& e^5& e^6& e^7& e^8& e^9 \end{pmatrix}
\end{equation}

\subsubsection{Plot y=f(x)}
Now we want to use a package to be able to plot $y=f(x)$.

We might want a smoother plot, in this case, we will use more computing power and have smaller steps between $0$ and $10$, as an illustration we want to plot by steps of $0.1$.


\subsection{Create you own function}
Now you should create your own function, that you could call $\mbox{my\_f}$:
$$\mbox{my\_f} : x \to x^2+x-2$$

Once defined, you want to apply your function to the original $x$ vector to get an output vector $y_f$ and plot it.

\subsection{Find x where f(x)=0}
Now we want to solve for $f(x)=0$.
We can try if a=b using the operator "==":
\begin{quote}
my\_f(0)==0
\end{quote}
we find that this is not the case: FALSE
\begin{quote}
my\_f(1)==0
\end{quote}
we find that this is the case: TRUE

Indeed, $x^2+x-2=0$ when $x=1$.

Now we can loop through all the element in the vector x ($0,1,\ldots,9$) and print which element solve: f(element)=0.

\subsubsection{fsolve}
The good news is that there is already a function to do the above: fsolve.

You need to first import the function from the right package.
Now to find the solution to $f(x)=0$, it is in one line:
\begin{quote}
fsolve(my\_f,$x_0$)
\end{quote}
Note that ",$x_0$)" here tells fsolve \textbf{where} to start searching for the root on the abscissa (the $x$-axis).

You can get help on this function using the console:
\begin{quote}
?fsolve
\end{quote}

$x_0$ = 0 is \textit{the starting estimate for the roots of 'func(x) = 0'}

In fsolve documentation, you'll see that fsolve uses MINPACK which itself uses Powell's dog leg method. I am afraid the documentation is quite long, if interested, \href{https://cds.cern.ch/record/126569/files/CM-P00068642.pdf}{here}.

\paragraph{What does fsolve actually do?}
It is tempting to think that fsolve ``walks down'' the curve of $f$ until it reaches zero, as the gradient descent of section \ref{sec:gradientdescent} would. It does not, and it is easy to see why it cannot: going downhill on $f(x)=x^2+x-2$ from any starting point leads to the bottom of the parabola, $x=-1/2$, where $f(-1/2)=-9/4\neq 0$. A method that descends $f$ finds a \emph{minimum}, not a \emph{root}.

What fsolve does instead is Newton's method (equation \ref{eq:newtonmethod} below): at the current point $x_n$ it replaces $f$ by its tangent line and jumps to the point where the tangent crosses the axis,
\begin{equation}\label{eq:fsolvenewtonstep}
x_{n+1} = x_n - \frac{f(x_n)}{f'(x_n)}.
\end{equation}
Three refinements turn this into the ``hybrid'' method of MINPACK. The slope $f'(x_n)$ is not asked from you: it is estimated at the starting point by a forward difference $[f(x_0+h)-f(x_0)]/h$ with a tiny $h$, and then updated cheaply after each step from the observed change in $f$ (Broyden's secant update), the exact derivative being recomputed only when progress stalls. The length of each jump is capped by a \emph{trust region} (initially $100\times|x_0|$ by default). And when Newton's jump is longer than what the trust region allows, the step is bent (this is the ``dog leg'') towards the direction of steepest descent of the \emph{squared} residual $\tfrac12 f(x)^2$, which is zero exactly at the roots. So there is a gradient hidden in fsolve, but it is the gradient of $f^2$, not of $f$, and it is only used as a safeguard when the Newton step cannot be trusted. You can verify all this: from $x_0=20$, the first point at which fsolve evaluates my\_f is $9.805$, which is exactly Newton's step $20-(400+20-2)/41$; the following points ($6.43$, $3.77$, $2.34$, $1.52$, $1.14$, $1.02$, $1.001$, $1$) differ slightly from the exact Newton iterates because the slope is updated by the secant formula instead of being recomputed.

\paragraph{Which root will you get? It depends on where you start.}
Our function $f(x)=x^2+x-2=(x-1)(x+2)$ has two roots, $x=1$ and $x=-2$, and one point where the tangent is horizontal, the vertex $x=-1/2$ where $f'(x)=2x+1=0$. The sign of the Newton step is the sign of $-f(x_n)/f'(x_n)$, so the direction of travel is dictated by the signs of $f$ and $f'$:
\begin{itemize}
\item $x_0>1$: $f>0$ and $f'>0$, the step is negative, the iterate slides down the tangent to the left, towards $1$. Because the parabola is convex, the tangent lies below the curve, so the landing point is still to the right of $1$: the sequence decreases monotonically to $1$. From $x_0=20$: $20\to 9.80\to 4.76\to 2.34\to 1.32\to 1.03\to 1.0003\to 1$.
\item $-1/2<x_0<1$: $f<0$ and $f'>0$, the step is positive, the iterate moves to the right, \emph{overshoots} the root (again because the tangent lies below the curve) and then comes back as in the previous case. From $x_0=0$: $0\to 2\to 1.2\to 1.012\to 1$.
\item $-2<x_0<-1/2$: $f<0$ and $f'<0$, the step is negative, the iterate overshoots $-2$ to the left and comes back. From $x_0=-1$: $-1\to -3\to -2.2\to -2.012\to -2$.
\item $x_0<-2$: $f>0$ and $f'<0$, the step is positive and the sequence increases monotonically to $-2$.
\end{itemize}

\subsubsection{Define another function and find the root closest to x=10}
Define a function: $f(x)=\cos(x^2)$, plot it and find the solution of $f(x)=0$ close to $x=10$.


\subsection{Fixed-point iteration}

The idea of a \emph{fixed point} is simple: a number $x$ is called a fixed point of a function $f$ if it satisfies
\[
f(x) = x.
\]
To approximate such a point, we can use an iterative procedure. Starting from an initial guess $x_0$, we repeatedly apply $f$ to generate a sequence:
\begin{equation}
x_{n+1} = f(x_n), \quad n = 0,1,2,\ldots
\end{equation}
This process is called \emph{fixed-point iteration}. If the sequence $(x_n)$ converges, then its limit $x$ must be a fixed point of $f$.

\subsubsection{Example: the Babylonian method for $\sqrt{a}$}

A famous historical application is the computation of square roots. Suppose we want $\sqrt{a}$ for some $a > 0$. Consider the function
\begin{equation}
g(x) = \frac{1}{2}\left( \frac{a}{x} + x \right).
\end{equation}
If we start from an initial guess $x_0 > 0$ and apply the iteration
\[
x_{n+1} = g(x_n),
\]
the sequence rapidly converges to $x = \sqrt{a}$, no matter which positive $x_0$ we choose. This procedure is known as the \emph{Babylonian method}.

\subsubsection{Connection with Newton’s method}

The Babylonian method is actually a special case of Newton’s method for solving nonlinear equations. Suppose we want to find a root of a differentiable function $f$, i.e.\ solve $f(x) = 0$. Newton’s method defines the iteration
\begin{equation}\label{eq:newtonmethod}
g(x) = x - \frac{f(x)}{f'(x)}, \qquad x_{n+1} = g(x_n).
\end{equation}
If the sequence converges to $x$, then $f(x) = 0$.  

To see the link, take $f(x) = x^2 - a$. The Newton update \ref{eq:newtonmethod} then gives exactly the Babylonian iteration above. This shows how fixed-point iteration and Newton’s method are closely related tools for finding solutions.

\subsubsection{Banach fixed-point theorem}

The \emph{Banach fixed-point theorem} (also called the contraction mapping theorem) is a fundamental result in analysis. It states:

\begin{quote}
If a function $f$ maps a complete metric space into itself and is a \emph{contraction}, that is, it brings points closer together: 
\[
|f(x)-f(y)| \leq q |x-y| \quad \text{for some $q<1$ and all $x,y$,}
\] then $f$ has a \emph{unique} fixed point $x^*$, and the fixed-point iteration
\[
x_{n+1} = f(x_n)
\]
converges to $x^*$ for any starting value $x_0$.
\end{quote}

\paragraph{Example: iteration with the cosine function.}
Consider the function
\[
f(x) = \cos(x).
\]
Its derivative is $f'(x) = -\sin(x)$, so $|f'(x)| = |\sin(x)| \le 1$ everywhere, with equality at $x=\pi/2$: the cosine is therefore \emph{not} a contraction on the whole real line, but it is one on the interval $[0,1]$, where $|f'(x)|=\sin(x)\le\sin(1)<1$ (the next paragraph makes this precise). This is enough, because wherever you start, the iterates enter $[0,1]$ after two steps: $x_1=\cos(x_0)\in[-1,1]$ and, the cosine being even and decreasing on $[0,1]$, $x_2=\cos(x_1)\in[\cos(1),1]\subset[0,1]$.

Thus, if we apply the iteration
\[
x_{n+1} = \cos(x_n),
\]
starting from any initial guess $x_0$, the sequence $(x_n)$ will converge to the unique fixed point of $\cos(x)$, which is approximately $x^* \approx 0.739$.

\paragraph{How to choose $q$.}
The contraction constant is not a matter of taste, it is computed in three steps.
\begin{enumerate}
\item \emph{Find an interval $I$ that $f$ maps into itself.} Take $I=[0,1]$: the cosine is decreasing on $[0,1]$, so $\cos([0,1])=[\cos(1),\cos(0)]=[0.540,1]\subset[0,1]$. Once an iterate is in $I$, all the following ones stay in $I$.
\item \emph{Bound the derivative on $I$.} On $[0,1]$ we have $|f'(x)|=\sin(x)$, and $\sin$ is increasing on $[0,\pi/2]$, so the largest value is reached at the right end of the interval:
\[
q=\max_{x\in[0,1]}|f'(x)|=\sin(1)\approx 0.8415<1 .
\]
\item \emph{Conclude with the mean value theorem.} For $x,y\in I$ there is a $\xi$ between them with $|f(x)-f(y)|=|f'(\xi)|\,|x-y|\le q\,|x-y|$: $f$ is a contraction on $I$ with constant $q=\sin(1)$.
\end{enumerate}
Note that $q$ depends on the interval: on the smaller invariant interval $[\cos(1),\cos(\cos(1))]=[0.540,0.858]$, where the iterates live from $x_2$ on, the same recipe gives $q=\sin(0.858)\approx0.76$, and the speed actually observed for large $n$ is $|f'(x^*)|=\sin(0.739)\approx 0.67$ per step. The smaller the interval around $x^*$, the sharper $q$.

\paragraph{Error bound.}
One of the powerful consequences of Banach's theorem is that it does not only guarantee convergence, but also gives a bound on the error. If $q$ is a contraction constant (a bound on $|f'(x)|$ on an interval containing all the iterates), then after $n$ steps
\[
|x_n - x^*| \;\leq\; \frac{q^n}{1-q}\, |x_1 - x_0|.
\]

\subsubsection{Fixed-point iteration on "any" function}

\stepcounter{question}
\begin{itemize}
\item \textbf{Question \arabic{question}}
\begin{itemize}
\item Apply the fixed-point iteration of the function: $f(x) = 3 x^3 - 2 x^2 - 0.5$ do you observe convergence? Vary the starting point $x_0$.
\end{itemize}
\end{itemize}

\subsection{Gradient descent}\label{sec:gradientdescent}

If you want to find the minimum of a function $f(x)$, always move in the direction that decreases the function value the fastest. This direction is given by the \emph{negative gradient} of the function. 

\paragraph{Step-by-step intuition.} 
\begin{enumerate}
  \item Start from an initial guess $x_0$. Think of this as choosing a point on the landscape defined by $f(x)$.
  \item At each step $n$, compute the slope (gradient) of the function at the current position: $\nabla f(x_n)$.
  \item Move in the opposite direction of the slope, since the gradient points uphill while we want to go downhill.
  \item The step size is controlled by a parameter $\gamma>0$, called the \emph{learning rate}. 
\end{enumerate}

\paragraph{Update rule.} 
Mathematically, this gives the iterative formula:
\[
x_{n+1} \;=\; x_n - \gamma \, \nabla f(x_n).
\]
Here, 
\begin{itemize}
  \item $x_n$ is the current point, 
  \item $\nabla f(x_n)$ is the gradient at $x_n$, 
  \item $\gamma$ is the learning rate that determines how large a step we take. 
\end{itemize}

\paragraph{Why it works.} 
The gradient tells us the direction of steepest increase of the function. By subtracting it, we move in the direction of steepest decrease. Repeating this procedure makes the sequence $\{x_n\}$ move closer and closer to a local minimum of $f(x)$.

\paragraph{Key insight for practice.} 
\begin{itemize}
  \item If $\gamma$ is too large, the algorithm may overshoot and fail to converge.  
  \item If $\gamma$ is too small, convergence will be very slow.  
  \item A good choice of $\gamma$ is crucial in applications, and often requires experimentation or adaptive methods.  
\end{itemize}

As an illustration, we use the gradient descent to find the minimum of the function $f(x) = x^4 - 3x^3 +2$.

\section{Python for non-programmers: numpy exercise part 2}
We suggest a code\footnote{numpy\_exercise\_part2.py} with further exercises to get familiar with python's numpy library, we will approximate functions, their derivatives and integrals.

\subsection{Approximating the exponential function}

One last visual application: we might want to convince ourselves that:
$$ \exp(x) =  \lim_{n \to \infty} \left( \left(1 + \frac{x}{n} \right)^n \right)$$

For this, we can simply plot the exponential function against $ \left(1 + \frac{x}{n} \right)^n $ for different values of $n$, for example $[10,100,1000]$.


Then finally, we can convince ourselves visually that:
$$ \exp( i \pi) = -1$$


\subsection{Compute the area of an ellipse}

We draw an ellipse:

\begin{figure}[htbp]
\centering
\begin{tikzpicture}
    \def\a{3}
    \def\b{2}

    \draw[thick, blue] (0,0) ellipse
        [x radius=\a cm, y radius=\b cm];

    \draw[->] (-3.6,0) -- (3.6,0) node[right] {$x$};
    \draw[->] (0,-2.5) -- (0,2.5) node[above] {$y$};

    \draw[<->, thick, red] (0,0) -- (\a,0)
        node[midway, below] {$a$};
    \draw[<->, thick, red] (0,0) -- (0,\b)
        node[midway, left] {$b$};
\end{tikzpicture}
\caption{Ellipse satisfying $x^2/a^2+y^2/b^2=1$.}
\end{figure}

The equation of such a standard ellipse centred at the origin, with height 2b and width 2a is:
\begin{equation}\label{eq:ellipse}
\frac{x^2}{a^2} +\frac{y^2}{b^2} = 1
\end{equation}

It can be shown that the area inside the ellipse is: $\pi a b$

As an exercise we will compute an approximate of this area by integral. We reorganize equation \ref{eq:ellipse}:
\begin{equation}
y(x) = b \sqrt{1-\frac{x^2}{a^2}}
\end{equation}

We can compute the area of the semi-ellipse (above part) by:
\begin{equation}\label{eq:intellipse}
\int_{-a}^a b \sqrt{1-\frac{x^2}{a^2}} dx
\end{equation}

\subsubsection{Riemann integral}

We want to approximate integral \ref{eq:intellipse} on the interval $[-a,a]$. We can use the lower and upper Darboux sums which is a simple approach in the spirit of the Riemann integral:
\begin{enumerate}
\item partition the interval $[-a,a]$ into $-a=x_{0}<x_{1}<x_{2}<\ldots <x_{n}=a$
\begin{itemize}
\item here for simplicity we evenly split the interval into $n$ intervals
\end{itemize}
\item define the lower Darboux sum:
\begin{equation}
\sum_{i=0}^{n-1} \inf_{x\in[x_i,x_{i+1}]} y(x) 	(x_{i+1} - x_i)
\end{equation}
\item define the upper Darboux sum:
\begin{equation}
\sum_{i=0}^{n-1} \sup_{x\in [x_i,x_{i+1}]} y(x) 	(x_{i+1} - x_i)
\end{equation}
\end{enumerate}

Those two sums, as $y(x)$ is continuous, converge to the integral we want to compute.

\stepcounter{question}
\begin{itemize}
\item \textbf{Question \arabic{question}}
\begin{itemize}
\item Work on Kepler's laws.
\end{itemize}
\end{itemize}

\subsection{Mittag-Leffler function}

A Generalized Mittag-Leffler function can be defined as:
\begin{equation}\label{eq:mittag}
E_{\alpha,\beta}(z)=\sum_{k=0}^\infty \frac{z^k}{\Gamma(\alpha k + \beta)}
\end{equation}
with $\alpha,\beta \in \mathbb{C}, \mathcal{R}(\alpha)>0, \mathcal{R}(\beta)>0, z \in \mathbb{C}$ and the function Gamma: $\Gamma(r) = \int_0^{+\infty}u^{r-1}e^{-u}du$.

\subsubsection{Mittag-Leffler function to approximate the exponential function}

Equation \ref{eq:mittag} can be set to approximate the exponential function:
$$ E_{1,1}(z)=\sum_{k=0}^\infty \frac{z^k}{k!}  $$

We have in practice to replace the sum to a limit $nlim$, instead of $\infty$, that we set to $10$ (red line) and $15$ (dashed red line) for illustration purpose. For example, Mittag-Leffler type functions combined with Laplace transform are useful to find explicit expressions of insurance companies ruin probability as in \citep{Constantinescu2018aa}

\subsection{Approximate the derivative of a function}

Remember that the derivative $f'(x)$ of a function $f()$ at $x$ is in fact a rate of change defined as:
$$f'(x) = \lim_{h\to 0} \frac{f(x+h)-f(x)}{h}$$

We can convince ourselves that $\ln'(x) = \frac{1}{x}$

We can set our own function to approximate the derivative and check for decreasing values of $h$ if indeed our results are fitting with the function $\frac{1}{x}$, which as expected is the case.


\subsection{Special case: approximate the integral of the logarithm}

We might want to test with python that:
$$\int_{1}^{2} \frac{1}{x} dx = \ln(2) - \ln(1) = \ln(2)$$

And we can approximate:
$$\int_{1}^{2} \frac{1}{x} dx = \sum_{i=0}^{N}  \gamma ( h_i)^{-1}$$

with $\forall i > 0, h_i - h_{i-1} = \gamma $ and $h_N = 1$.

\subsection{Approximate the sinus function with Mittag-Leffler integral}

Finally, we can approximate the sinus function with an integral of a Mittag-Leffler function:

$$sin(x) = \int_0^x E_{2,1}(-s^2)ds$$

\begin{center}
\includegraphics[scale=0.5]{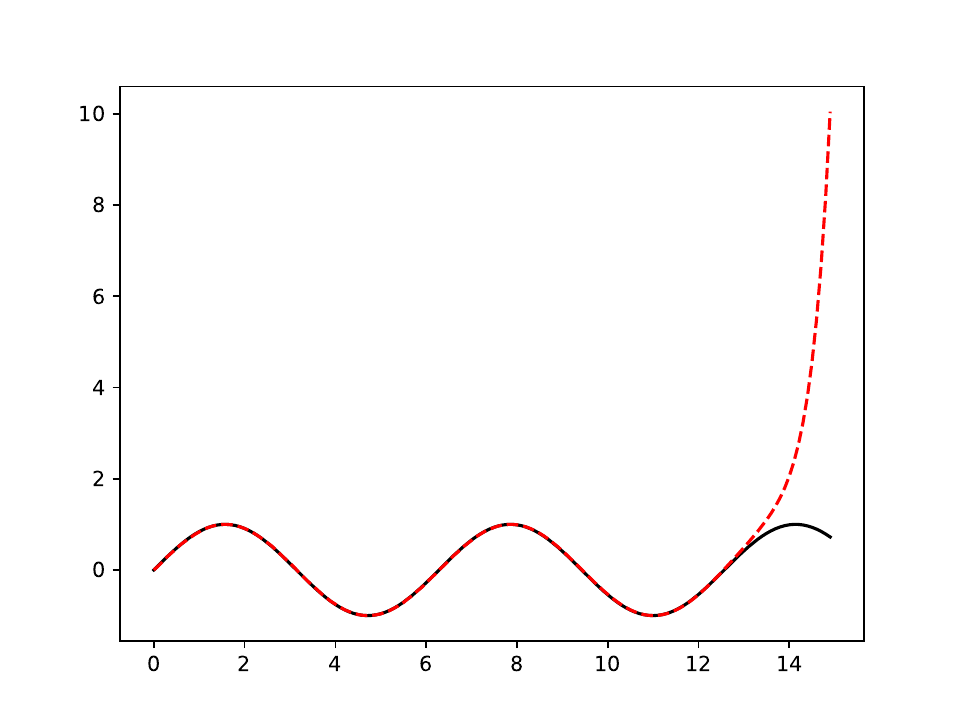}
\end{center}

To decompose, in black we draw the $E_{2,1}(x)$ function. To compute $sin(x_i)$ we create intermediary points (3 here) from 0 to $x_i$ and we add (or subract when negative) the area of the rectangles. This then gives us $sin(x_i)$ represented by a green dot:

\begin{center}
\includegraphics[scale=0.4]{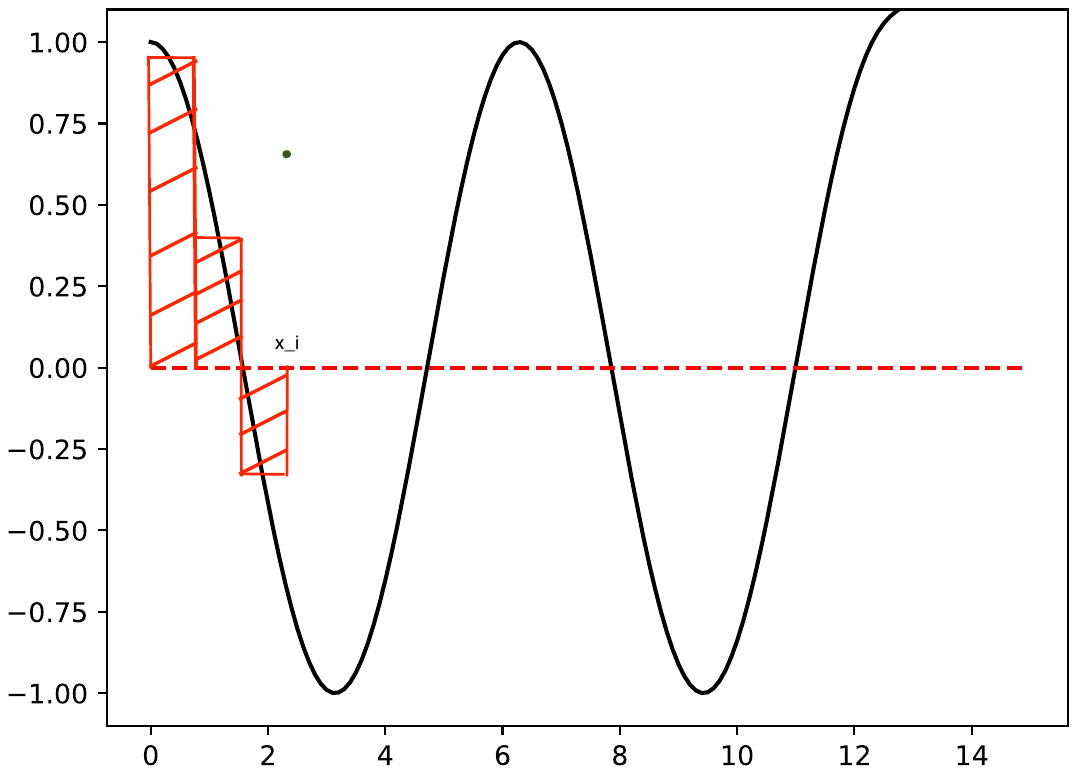}
\end{center}

This has to be done for every $x_i$ between $0$ and $14$ with a chosen step number (the more steps the more precise but the longer it will take to compute).

\stepcounter{question}
\begin{itemize}
\item \textbf{Question \arabic{question}}
\begin{itemize}
\item Show visually (plot) that $E_{1,2}(x) = \frac{e^x-1}{x}$.
\end{itemize}
\end{itemize}

\section{An introduction to Monte Carlo approaches}

In 1881, Joseph Jagger, an English businessman in the textile sector, visited the Casino of Monte Carlo. Having observed that spinning wheels were never perfectly balanced, Jagger wondered about the potential biases of the casino's roulette wheels. Using empirical averages from his observations and relying on the law of large numbers, he detected significant biases in some roulettes after a month. This insight, coupled with the central limit theorem, allowed him to calculate the estimation error and recognize that certain numbers had a higher probability of winning. Leveraging these findings, he won 2 million francs, equivalent to 10 million euros today.

The Monte Carlo method originates from this empirical approach. It is a tool to estimate averages (or expectations) by simulations, especially when direct calculations are complex or unfeasible. Its application became more prominent with the advent of computers, notably for simulating neutron collisions post-war. Nicholas Metropolis christened this technique in reference to the casino.

At its core, the Monte Carlo method is a statistical simulation technique that leverages the power of random sampling to approximate numerical results. Here are the main three steps:
\begin{enumerate}
\item \textbf{Random Sampling}: Begin by generating a large number of random samples. These samples could be drawn from a particular probability distribution, depending on the problem at hand.
\item \textbf{Perform Calculations}: For each of these random samples, perform the necessary calculations or computations related to your problem.
\item \textbf{Aggregate Results}: Collect all the results from these calculations, and then compute an average (or other relevant statistical measure). The law of large numbers ensures that as the number of samples, denoted as $N$, approaches infinity, this average should converge to the true value of the quantity you're trying to estimate.
\end{enumerate}

In the world of econometrics, the Monte Carlo method finds application in various ways. It can be used to assess the properties of an estimator under different conditions, test the power of a statistical test, or even to approximate complex integrals in likelihood functions. 

\subsubsection{The Role of Sample Size}

To illustrate\footnote{numpy\_exercise\_part3.py} how empirical sampling approaches theoretical distributions, we consider the standard normal distribution $N(0,1)$ with probability density function
\[
f(x)= \frac{1}{\sqrt{2\pi}} e^{-\tfrac{x^2}{2}}.
\]
Using random draws of increasing size $n \in \{10^2,10^3,10^4,10^5\}$, 
we compare the empirical histograms with the true PDF. 
As $n$ grows, the empirical histogram converges toward the smooth bell curve of the standard normal, 
highlighting the law of large numbers in action.

In addition to the visual comparison, we compute the probability of observing a draw below $-2$. 
The theoretical value is given by the standard normal cumulative distribution function,
\[
\Pr[X < -2] = \Phi(-2) \approx 0.0228.
\]
For each sample size, we calculate the empirical frequency of values below $-2$ and 
report the absolute error relative to the theoretical probability. 
As expected, the empirical estimates fluctuate around the true value for small $n$, 
but the discrepancy decreases as $n$ increases. 

This simple exercise provides a gentle introduction to Monte Carlo methods: 
simulation allows us to approximate theoretical quantities, 
and the approximation improves systematically with larger sample sizes.

\paragraph{Standard error.}  
Let $I_i = \mathbf{1}\{X_i < -2\}$ where $X_i \sim N(0,1)$.  
Then $I_i \sim \text{Bernoulli}(p)$ with $p = \Pr[X < -2]$.  
The empirical probability is the sample mean
\[
\hat{p} = \frac{1}{n}\sum_{i=1}^n I_i.
\]
Since $\mbox{Var}(I_i) = p(1-p)$, the variance of $\hat{p}$ is $\mbox{Var}(\hat{p}) = p(1-p)/n$.  
Hence the standard error is
\[
\text{SE}(\hat{p}) = \sqrt{\tfrac{p(1-p)}{n}}.
\]
In practice $p$ is unknown, so we replace it by $\hat{p}$:
\[
\widehat{\text{SE}} = \sqrt{\tfrac{\hat{p}(1-\hat{p})}{n}}.
\]

\subsection{Monte Carlo method and uniform distribution: estimate the value of $\pi$}\label{sec:MCpi}

Consider the task of approximating the value of $\pi$. Imagine randomly throwing darts at a square board with a circular target inscribed inside. If you know the total number of darts thrown and the number that land inside the circle, you can use these quantities to estimate the value of $\pi$. This is because the ratio of the area of the circle to the square approximates $\pi / 4$. By throwing a large number of darts (or, in a computational sense, generating a large number of random points) and calculating the proportion that fall inside the circle, you can multiply by $4$ to get an approximation of 
$\pi$.

We define a circle of radius $r=1$. Then its surface is known to be $\pi r^2$. We draw $x$ and $y$ from the uniform distributions $a=-1$ and $b=1$. We write the probability density function of the uniform distribution:
\begin{equation}\label{eq:uniformdist}
    f(x;a,b) =
    \begin{cases}
\frac{1}{b-a},& \text{for} \quad a \leq x \leq b \\     
0 & \text{otherwise}
    \end{cases}
\end{equation}

The square surface will be $2\times 2 = 4$ and the area enclosed by the circle is $\pi$:

\begin{center}
\includegraphics[scale=0.3]{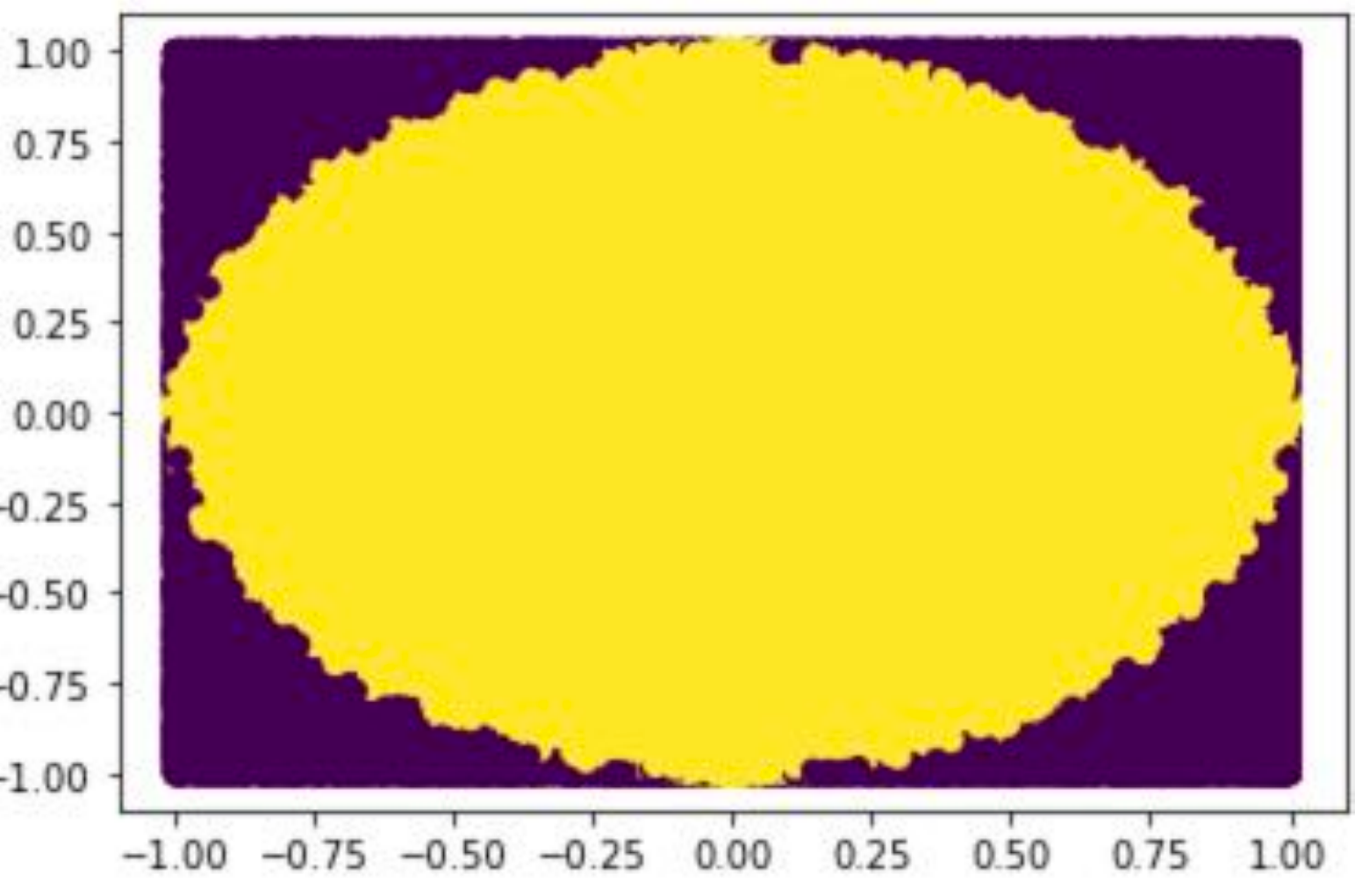}
\end{center}

A recurring question is on the choice of the number of draws for the Monte Carlo method. We are not aware of a definite answer, but here is an illustration of the histogram of $100$ estimates of $\pi$ we found with $10^4$ draws:
\begin{center}
\includegraphics[scale=0.5]{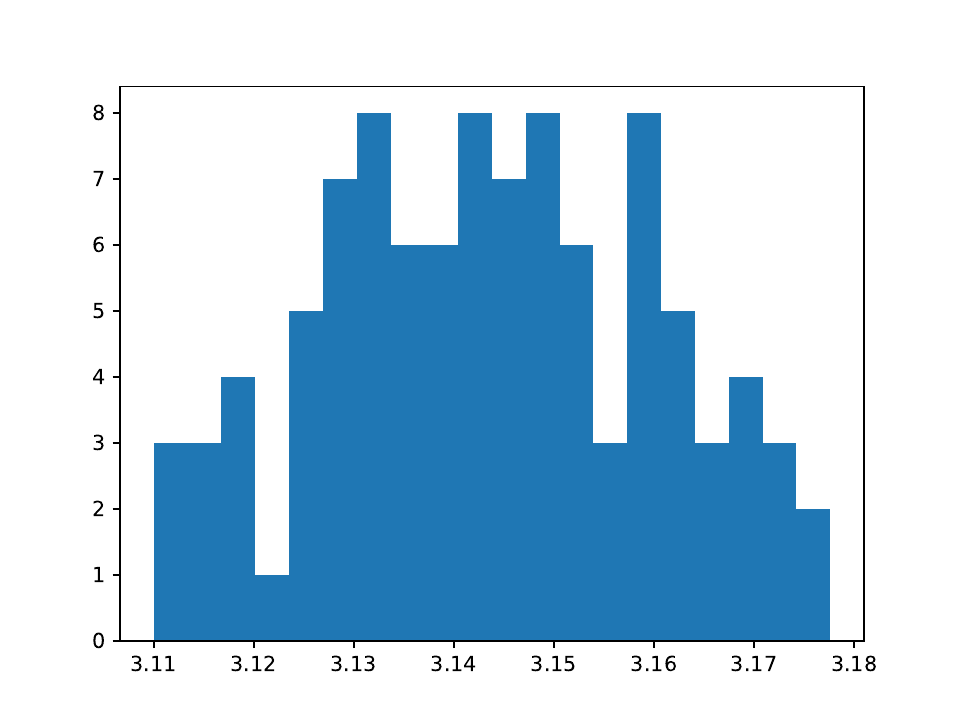}
\end{center}
compared with $100$ $\pi$ estimates we found with $10^6$ draws:
\begin{center}
\includegraphics[scale=0.5]{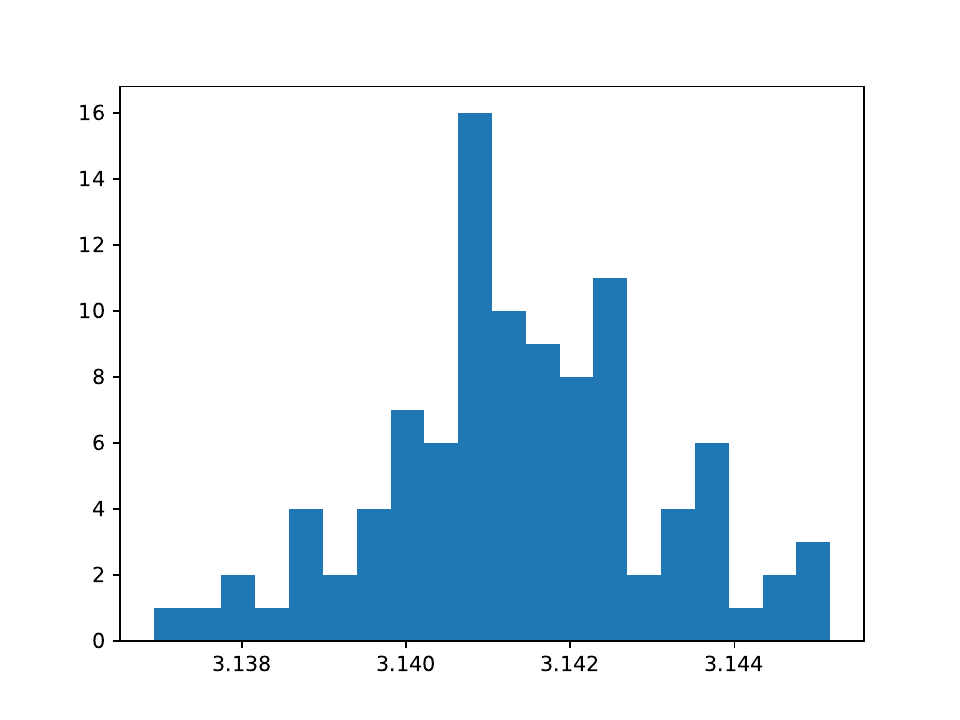}
\end{center}

\stepcounter{question}
\begin{itemize}
\item \textbf{Question \arabic{question}}
\begin{itemize}
\item Create a list of number of draws $N=[10,10^2,10^3,\ldots,10^n]$, chose $n$ depending on your laptop computing power and create a list of the standard deviation of the error terms of the $\pi$ estimates compared to the numpy $\pi$ value for $100$ estimates, is it a monotonic relationship? Can you advise for a value of $n$, the sample size?
\end{itemize}
\end{itemize}

\subsection{Monte Carlo integration: estimating $\displaystyle \int_0^1 e^{x}\,dx$}\label{sec:MCintegration}

\paragraph{Remark (Monte Carlo integration).}
Monte Carlo integration\footnote{For an accessible introduction, see \href{https://www.scratchapixel.com/lessons/mathematics-physics-for-computer-graphics/monte-carlo-methods-in-practice/monte-carlo-integration}{Scratchapixel, Monte Carlo Integration}.} interprets the value of an integral as an expected value.
If $X$ is uniformly distributed over a domain $\mathcal{D}$ with volume $V$, then
\[
\int_{\mathcal{D}} f(x)\,dx \;=\; V\,\E[f(X)].
\]
The corresponding estimator with $N$ independent samples $\{X_i\}_{i=1}^N$ is
\begin{equation}
\hat{I}_N \;=\; V\,\frac{1}{N}\sum_{i=1}^N f(X_i).
\label{eq:MC_general}
\end{equation}
It is unbiased, with variance
\[
\Var(\hat{I}_N) \;=\; \frac{V^2}{N}\,\Var(f(X)).
\]
Thus the error decreases at rate $N^{-1/2}$, regardless of the dimension of $\mathcal{D}$.

\stepcounter{question}
\begin{itemize}
\item \textbf{Question \arabic{question}}
\begin{itemize}
\item Plot the Monte Carlo integration error as a function of $N$ to verify empirically that it decreases at the expected $1/\sqrt{N}$ rate.
\end{itemize}
\end{itemize}

\paragraph{Example (estimating $\pi$).}
In Section~\ref{sec:MCpi} we approximated $\pi$ using the area of a quarter circle.  
Define the indicator function
\[
H(x,y) =
\begin{cases}
1,& \text{if } x^2+y^2 \leq 1, \\     
0,& \text{otherwise}.
\end{cases}
\]
Sampling uniformly on the square $\mathcal{D}=[-1,1]\times[-1,1]$ of volume $V=4$, the probability that $(x,y)$ falls inside the unit disk is $\pi/4$.  
The Monte Carlo estimator is therefore
\[
\hat{\pi}_N \;=\; V\,\frac{1}{N}\sum_{i=1}^{N} H(x_i,y_i).
\]
As $N\to\infty$, $\hat{\pi}_N$ converges to $\pi$ with variance decreasing at rate $1/N$.

\paragraph{Estimate $\displaystyle \int_0^1 e^{x}\,dx$.}
Let
\[
I \;=\; \int_{0}^{1} e^{x}\,dx \;=\; e-1.
\]
The Monte Carlo estimator based on i.i.d.\ draws $U_1,\dots,U_N \sim \mathcal U[0,1]$ is
\begin{equation}\label{eq:mc-estimator}
\widehat I_N \;=\; \frac{1}{N}\sum_{i=1}^{N} e^{U_i}.
\end{equation}

\paragraph{Unbiasedness and LLN.}
Since $\mathbb{E}[e^{U_i}] = \int_0^1 e^{x}\,dx = I$, we have
$\mathbb{E}[\widehat I_N]=I$. By the law of large numbers,
\[
\widehat I_N \xrightarrow[]{\;\text{a.s.}\;} I \quad \text{as } N\to\infty.
\]

\paragraph{Variance and CLT.}
With $U\sim \mathcal U[0,1]$,
\[
\operatorname{Var}(\widehat I_N) \;=\; \frac{1}{N}\,\operatorname{Var}\!\big(e^{U}\big).
\]
Compute
\[
\mathbb{E}\!\left[e^{tU}\right] \;=\; \int_0^1 e^{tx}\,dx \;=\; \frac{e^{t}-1}{t},
\]
so
\[
\mathbb{E}\!\left[e^{U}\right]=e-1, 
\qquad
\mathbb{E}\!\left[e^{2U}\right]=\frac{e^{2}-1}{2},
\]
and therefore
\[
\operatorname{Var}\!\big(e^{U}\big)
\;=\; \frac{e^{2}-1}{2}-(e-1)^2
\;\approx\; 0.2420356.
\]
Hence
\[
\operatorname{Var}(\widehat I_N) \;=\; \frac{0.2420356}{N},
\qquad
\operatorname{se}(\widehat I_N) \;=\; \sqrt{\frac{0.2420356}{N}}
\;=\; \mathcal O\!\left(N^{-1/2}\right).
\]
By the central limit theorem,
\[
\sqrt{N}\,\big(\widehat I_N - I\big)\;\xrightarrow{d}\; \mathcal N\!\left(0,\operatorname{Var}\!\big(e^{U}\big)\right),
\]
so an approximate $(1-\alpha)$ confidence interval is
\[
\widehat I_N \;\pm\; z_{1-\alpha/2}\,\sqrt{\frac{\widehat{\operatorname{Var}}(e^{U})}{N}},
\]
where $\widehat{\operatorname{Var}}(e^{U})$ is the sample variance of $\{e^{U_i}\}_{i=1}^N$.

\subsection{An Intractable Example: High-Dimensional Integrals}

One of the most important applications of Monte Carlo methods is the
approximation of high-dimensional integrals. Such problems arise in
Bayesian statistics, statistical physics, and quantitative finance,
where analytic solutions are not available.

\paragraph{Bayesian example.}
Consider data $D$ and a parameter vector $\theta \in \mathbb{R}^{p}$.
The posterior distribution is
\[
\pi(\theta \mid D) \;=\; \frac{L(D \mid \theta)\,\pi(\theta)}{\int_{\mathbb{R}^{p}} L(D \mid \vartheta)\,\pi(\vartheta)\,d\vartheta},
\]
where $L(D \mid \theta)$ is the likelihood and $\pi(\theta)$ the prior.
The denominator,
\begin{equation}
Z \;=\; \int_{\mathbb{R}^{p}} L(D \mid \vartheta)\,\pi(\vartheta)\,d\vartheta,
\label{eq:evidence}
\end{equation}
is called the \emph{marginal likelihood} or \emph{model evidence}.
For $p \geq 5$, this integral has no closed form and is effectively impossible to evaluate with deterministic numerical quadrature, since the cost grows exponentially with $p$ (the ``curse of dimensionality'').

\paragraph{Monte Carlo solution.}
Using samples $\theta^{(1)},\dots,\theta^{(N)}$ drawn from the prior
$\pi(\theta)$, one may approximate \ref{eq:evidence} by
\[
\hat{Z}_N \;=\; \frac{1}{N}\sum_{i=1}^N L(D \mid \theta^{(i)}).
\]
This is an unbiased estimator of $Z$, with variance decreasing at rate $1/N$.
More sophisticated methods (importance sampling, Markov chain Monte Carlo)
improve efficiency, but all rely on the same core principle:
\emph{Monte Carlo integration is often the only practical tool to evaluate
otherwise intractable high-dimensional integrals}.

\subsection{Application: Option Pricing with Inflation Caps}

Monte Carlo methods are not only useful in statistics but also in finance.
Many derivative securities are priced by computing expectations under a
risk–neutral measure, often involving high-dimensional stochastic processes
with no closed-form solution.

\paragraph{Inflation cap.}
An inflation cap is a contract that pays off if the realized inflation
rate $\pi(n)$ over $n$ years exceeds a fixed strike $k$. The payoff at
maturity is
\[
\text{Caplet Payoff} \;=\; \max\!\big(\pi(n)-k,\,0\big).
\]
The present value at time~0 under the risk–neutral measure $\Q$ with
discount factor $B(0,n)$ is
\begin{equation}
\text{Caplet Price} \;=\; B(0,n)\, \E^{\Q}\!\left[ \max\!\big(\pi(n)-k,\,0\big) \right].
\label{eq:caplet_price}
\end{equation}

\paragraph{Intractability.}
In practice, $\pi(n)$ is modeled as a stochastic process (e.g. with
stochastic volatility or jumps). The distribution of $\pi(n)$ has no closed
form, so the expectation in \ref{eq:caplet_price} cannot be computed
analytically.

\paragraph{Monte Carlo solution.}
We simulate $N$ independent paths of the inflation process under $\Q$,
producing realizations $\pi^{(1)}(n),\dots,\pi^{(N)}(n)$. The Monte Carlo
estimator of the caplet price is
\[
\widehat{C}_N \;=\; B(0,n)\,\frac{1}{N}\sum_{i=1}^N 
\max\!\big(\pi^{(i)}(n)-k,\,0\big).
\]
This provides a consistent and unbiased approximation of
\ref{eq:caplet_price}, with accuracy improving at the standard Monte
Carlo rate $1/\sqrt{N}$.

\paragraph{Modeling inflation dynamics.}
To simulate $\pi(n)$, we need a model for the inflation process under the
risk--neutral measure $\Q$. In practice, one proceeds as follows:

\begin{enumerate}
  \item \textbf{Historical estimation.}  
  Start from observed inflation indices (e.g.\ HICP for the euro area).
  Fit a time--series model such as an autoregressive process (AR), ARMA,
  or stochastic volatility model. For example, a simple continuous--time
  specification is
  \[
  d\pi_t \;=\; \kappa\!\left(\theta - \pi_t\right)\,dt + \sigma\,dW_t,
  \]
  an Ornstein--Uhlenbeck process with mean reversion to $\theta$.

  \item \textbf{Risk--neutral adjustment.}  
  The parameters estimated under the physical measure $\P$ must be
  adjusted to the risk--neutral measure $\Q$, so that simulated prices
  are consistent with observed swap and cap quotes. This step usually
  involves a change of drift:
  \[
  d\pi_t \;=\; \kappa^{\Q}\!\left(\theta^{\Q} - \pi_t\right)\,dt + \sigma\,dW^{\Q}_t.
  \]

  \item \textbf{Calibration to market data.}  
  The $\Q$--parameters are chosen to fit current market instruments
  (inflation swaps, caps, floors). For example, the forward inflation
  curve extracted from swaps fixes $\theta^{\Q}$, while cap/floor prices
  reveal information about $\sigma$.

  \item \textbf{Monte Carlo simulation.}  
  Once calibrated, simulate sample paths of $\pi_t$ over the horizon
  $[0,n]$ using discretization schemes (e.g.\ Euler--Maruyama). The
  terminal value $\pi(n)$ on each path feeds into the payoff formula
  \ref{eq:caplet_price}.
\end{enumerate}

\section{Python for non-programmers: exercise with pandas - part 1}\label{sec:pandas}

We suggest exercises\footnote{pandas\_exercise\_part1.py} to get familiar with some pandas basics features in python's pandas library.

\subsection{Introduction to pandas DataFrame}
\subsubsection{Import the package}
First we want to import the package panda and create a dataframe we call df:
\begin{center}
\includegraphics[scale=0.7]{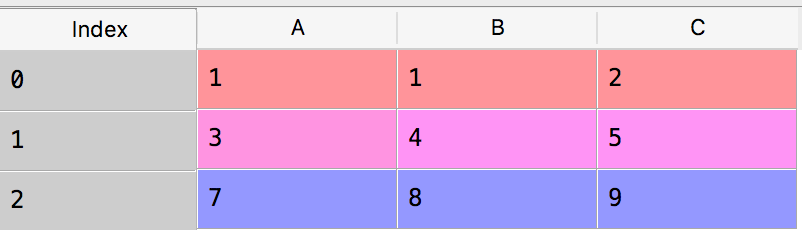}
\end{center}

You might want to sum over rows for each column .sum(axis=0)
or sum over columns for each row .sum(axis=1)

\subsubsection{Finding location of an element in a DataFrame}

The panda's documentation explains how to access specific element of a DataFrame, using either the label or the integer positions. \href{http://pandas.pydata.org/pandas-docs/stable/indexing.html#selection-by-label}{Link here}.

\subsubsection{Selection By Label}\label{sec:bylabel}
You can use the .loc attribute to access elements in a DataFrame, label-based. For df, that would be:

\begin{tabular}{llll}
&A&B&C\\
0&df.loc[[0],['A']]&df.loc[[0],['B']]&df.loc[[0],['C']]\\
1&df.loc[[1],['A']]&df.loc[[1],['B']]&df.loc[[1],['C']]\\
2&df.loc[[2],['A']]&df.loc[[2],['B']]&df.loc[[2],['C']]\\
\end{tabular}

You can also create subset of your dataframe, slicing it.
\begin{lstlisting}
slice1 = df.loc[ :,['A', 'C']]
slice2 = df.loc[1,['A', 'C']]
slice3 = df.loc[1:,:'B']
\end{lstlisting}
Which yields from left to right:
\begin{center}
\includegraphics[scale=0.5]{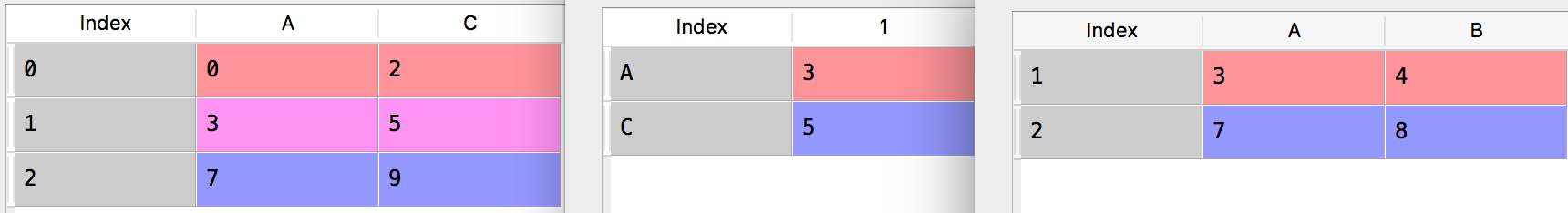}
\end{center}

\subsubsection{Selection by position (integer)}

If you want to be able to select a cell in a DataFrame, regardless of the name of the column or the name of the row, you can use integer based indexing with .iloc attribute. Keep in mind that python indexing convention starts at $0$.

Creating the above slices would become:
\begin{lstlisting}
slice1i = df.iloc[ :,[0, 2]]
slice2i = df.iloc[1,[0, 2]]
slice3i = df.iloc[1:,:2]
\end{lstlisting}

Note the subtlety, illustrated by our example slice3i:
\begin{itemize}
\item indexing starts at $0$
\item with python slices, the start  is included but not the stop; so $':2'$ yields $[0,1]$ and not $[0,1,2]$
\end{itemize}

\subsubsection{Other selection methods}
You can also use other selection methods, described in panda's documentation:
.at, .iat, .ix

\subsubsection{Unstack}

We can use the function \emph{unstack} on the DataFrame to transform it into a wider format where multiple levels of data are reshaped based on combinations of row and column indices. This is especially handy for panel regressions, as it aids in disentangling data across multiple entities (like individuals or firms) and time periods, providing a clearer view of observations over time. For the given DataFrame, when we apply unstack, we observe for df\_unstacked:

\begin{tabular}{ll}
A,0&df.iloc[0,0]\\
A,1&df.iloc[1,0]\\
A,2&df.iloc[2,0]\\
B,0&df.iloc[0,1]\\
B,1&df.iloc[1,1]\\
$\ldots$&$\ldots$\\
\end{tabular}

\subsubsection{Loop through a DataFrame}

By using positions or names of columns and/or index, we are able to loop through a DataFrame.

The idea behind this is that the elements of the DataFrame df2 are found:

\quad

\begin{tabular}{lll}
df.iloc[0,0]&df.iloc[0,1]&df.iloc[0,2]\\
df.iloc[1,0]&df.iloc[1,1]&df.iloc[1,2]\\
df.iloc[2,0]&df.iloc[2,1]&df.iloc[2,2]\\
\end{tabular}

\quad

It is also possible to transpose a data frame. The transpose of df2 would be:

\quad

\begin{tabular}{lll}
df.iloc[0,0]&df.iloc[1,0]&df.iloc[2,0]\\
df.iloc[0,1]&df.iloc[1,1]&df.iloc[2,1]\\
df.iloc[0,2]&df.iloc[1,2]&df.iloc[2,2]\\
\end{tabular}

\subsection{DataFrame manipulation for population time series}

We want to create\footnote{pandas\_exercise\_part1\_ts.py}  the following DataFrame with the French population values:

\begin{tabular}{lll}
Septembre&	2016	& 66 790\\
Aout	&2016&	 66 763\\
Juillet	&2016	& 66 735\\
Juin	&2016	& 66 710\\
Mai	&2016	& 66 688\\
Avril	&2016	& 66 672\\
Mars	&2016&	 66 659\\
Fevrier	&2016	& 66 644\\
Janvier&	2016	& 66 628\\
\end{tabular}

\quad

Look carefully at the way your DataFrame is indexed, it starts from 0 and not from 1. Therefore here you have 9 observations indexed from 0 to 8.

As we entered the data in the wrong order, we need to reverse:
\begin{lstlisting}
pop = pop.iloc[::-1]
\end{lstlisting}

\subsubsection{Why can we reverse the index using ::-1?}

You can look into python's documentation on slice class. Imagine you have a list you call mylist:
\begin{itemize}
\item mylist = $[0,1,2,3]$
\end{itemize}
Conveniently, we chose the value of each item to be its index position in the list, so $0$ is at index $0$, $1$ at index position $1$, etc. Imagine you want to reverse this mylist: we have a list of length $4$ so going from index position $0$ to index position $3$. To enumerate the index, you can do: $start:end:step$ within the $[$ and $]$ signs. So mylist[0:len(mylist):1] will return the full list.

To reverse the list, we want the step to be $-1$, we omit the start and end position, so that python takes all possible items. Just look back at section \ref{sec:bylabel} where we used $1:$ to select all items starting at index $1$ and after.

\subsubsection{Indexing with time}
This is optional, but you might want to index your DataFrame with time. First you want to create an object with the dates as 'year-month' from 2016-01 to 2016-09 and modify our DataFrame index.

Note that we have to range until 2016-10 for the last item of our dates object to be 2016-09.

\subsubsection{Renaming the column and plotting}
We see that the name of our column is '0' we want to rename it with 'population'.
We can have a look at the data by plotting it and choosing a title.

\subsubsection{Computing the monthly changes of the population}
We want to compute the change of the French population from month $m$ to month $m+1$:
$$\mbox{change}_{m+1}=\frac{\mbox{pop}_{m+1}-\mbox{pop}_m}{\mbox{pop}_m}$$

In visual terms, that will be:

\quad

\begin{tabular}{lll}
\textbf{pop}&	\textbf{pop.shift(1)}	& \textbf{change\_pop}\\
&&\\
$\mbox{pop}_m$&&NaN\\
&&\\
$\mbox{pop}_{m+1}$&$\mbox{pop}_{m}$&$\frac{\mbox{pop}_{m+1}}{\mbox{pop}_m}-1$\\
&&\\
$\mbox{pop}_{m+2}$&$\mbox{pop}_{m+1}$&$\frac{\mbox{pop}_{m+2}}{\mbox{pop}_{m+1}}-1$\\
&&\\
$\mbox{pop}_{m+3}$&$\mbox{pop}_{m+2}$&$\frac{\mbox{pop}_{m+3}}{\mbox{pop}_{m+2}}-1$\\
&&\\
$\ldots$&$\ldots$&$\ldots$\\
\end{tabular}

\subsection{DataFrame as matrix and operations}

Arthur Cayley worked on matrices and introduced operations \citep{cayley1858ii}.

\subsubsection{Matrix and vector multiplication}

We create\footnote{pandas\_exercise\_part1\_matrices.py}  a vector with its index as the matrix column names and we can then perform a matrix multiplication\footnote{It is important that the columns names of the matrix are the same as the index of the vector for the product to be acceptable}

\[
\begin{bmatrix}
1&1&2\\
3&4&5\\
7&8&9
\end{bmatrix}
\times
\begin{bmatrix}
2\\
4\\
6   
\end{bmatrix} 
=
\begin{bmatrix}
18\\
52\\
100  
\end{bmatrix} 
\]

\subsubsection{Hadamard product (Multiplying element by element)}

We might want to multiply element by element, this is also called Hadamard product, that we write here $\odot$.

\quad

First we duplicate the column to form a square matrix and multiply the two matrices:

\[
\begin{bmatrix}
1&1&2\\
3&4&5\\
7&8&9
\end{bmatrix}
\odot
\begin{bmatrix}
2&4&6\\
2&4&6\\
2&4&6
\end{bmatrix} 
=
\begin{bmatrix}
2&4&12\\
6&16&30\\
14&32&54 
\end{bmatrix} 
\]

\subsubsection{Singular and non singular matrices}

A matrix is fundamental to many advanced concepts in mathematics and econometrics.

\begin{itemize}
\item \textbf{Square Matrix}: A matrix is called "square" if it has the same number of rows and columns. So, if a matrix has dimensions of 3×3, 4×4, and so on, it's a square matrix\footnote{A matrix that isn't square might look like 2×3, where it has 2 rows but 3 columns, for example}.
\item \textbf{Identity Matrix}: Imagine a matrix filled entirely with zeros, but its main diagonal (from the top-left corner to the bottom-right) is filled with ones. This special matrix is called the "identity matrix", denoted by $I$. It acts like the number 1 in regular arithmetic for matrices.
\item \textbf{Non-singular Matrix}: We say a square matrix $\Delta$ is "non-singular" if we can find another matrix, $\Delta^{-1}$ , such that when we multiply them, the result is the identity matrix: $\Delta * \Delta^{-1} = I$.
\end{itemize}

Another way to determine if a matrix is non-singular is to look at its rows (or columns). If all rows (or columns) of an $n\times n$ matrix are linearly independent -- meaning no row (or column) can be expressed as a combination of the others -- then the matrix is non-singular. This is also referred to as the matrix having "full rank". A quick check for this is the determinant: non-singular matrices have a determinant that is not zero.

In the realm of econometrics, matrices play a vital role in various models, including AR(p) and VAR models.

\subsubsection{Eigenvalues, eigenvectors, and shock propagation}

When analyzing the stability of a linear propagation process, it is useful to study the eigenvalues and eigenvectors of the underlying matrix.  

An eigenvector $v$ of a matrix $\Delta$ is a non-zero vector such that
\[
\Delta v = \lambda v,
\]
where $\lambda$ is the associated eigenvalue. Intuitively, multiplying $\Delta$ by $v$ only stretches or shrinks $v$, without changing its direction.  

This matters because the long-run behavior of repeated applications of $\Delta$ depends on its eigenvalues. If all eigenvalues are strictly smaller than one in absolute value, repeated multiplications of $\Delta$ dampen shocks. If at least one eigenvalue exceeds one, shocks are amplified and the system becomes unstable.  

\medskip

\paragraph{Application to financial networks: DebtRank}  
The matrix $\Delta$ summarizes how distress propagates from one bank to another.  \citet{Bardoscia2015aa} model contagion by transmitting only the \emph{newly incurred distress} at each step.  
The update rule is
\[
h(t+1) = h(t) + \Delta \big(h(t)-h(t-1)\big).
\]
Here $h_i(t)$ represents the relative equity loss of bank $i$ at time $t$.  
When bank $j$ suffers an additional equity loss, the counterparties $i$ that are exposed to $j$ must reduce the value of their interbank assets accordingly.  This is captured by the matrix $\Delta$, which scales the shock according to the size of exposures relative to each bank's equity base.  In this way, distress is linked directly to a deterioration of capital positions and thus to the valuation of exposures.  Each bank's distress is capped at $h_i(t)\leq 1$, reflecting the fact that no institution can lose more than 100\% of its equity.  

Suppose each bank is hit at time $t=1$ by an initial shock $h(1)=0.5\%$.  
At this stage, the losses are simply $h(1)$.  

At $t=2$, the first round of contagion takes place: the initial shock is transmitted once through the network, giving an additional contribution $\Delta h(1)$.  
The cumulative distress is then
\[
h(2) = h(1) + \Delta h(1).
\]

At $t=3$, the second round of contagion occurs: the shock passes along paths of length two, adding $\Delta^2 h(1)$.  
The cumulative distress is
\[
h(3) = h(1) + \Delta h(1) + \Delta^2 h(1).
\]

Continuing in this way, after $t$ steps we have
\[
h(t) = \sum_{s=0}^{t-1} \Delta^s h(1).
\]

\medskip

We have
\[
S_t = I + \Delta + \Delta^2 + \cdots + \Delta^{t-1}.
\]

For scalars you know that
\[
1 + r + r^2 + \cdots + r^{t-1} = \frac{1-r^t}{1-r}, \quad |r|<1.
\]

The matrix version is analogous:
\[
S_t = (I - \Delta^t)(I-\Delta)^{-1},
\]
provided that $I-\Delta$ is invertible.

\medskip

If all eigenvalues of $\Delta$ are strictly smaller than one in modulus, then the infinite sum converges and the long-run impact of the shock is finite:
\[
h(\infty) = (I-\Delta)^{-1} h(1),
\]
with $I$ the identity matrix.  
If instead at least one eigenvalue exceeds one, the sum diverges and the shock is amplified without bound.

\subsubsection{Cholesky decomposition}\label{sec:choleskydecomposition}

Cholesky decomposition is a powerful mathematical tool often used in econometrics. 

\begin{itemize}
\item \textbf{Real Matrix}: A matrix, $A$, is described as "real" when all of its elements are real numbers.
\item \textbf{Symmetric Matrix}: A matrix, $A$, is symmetric if it remains unchanged upon transposition. In other words, the matrix and its transpose are the same: $A=A^T$, where $T$  denotes the transpose operator.
\item \textbf{Positive Matrix}: A matrix, $A$, is positive if all of its elements are non-negative (equal to or greater than zero).
\item \textbf{Real Positive-Definite Matrix}: A matrix is considered positive-definite if, for any nonzero real column vector $z$, the value produced by $z^T A z$ is positive. This can be formally stated as: $\forall z\in \mathbb{R}^n \setminus {0}: z^T A z > 0$.
\end{itemize}

\textbf{Cholesky decomposition:}

For a matrix $A$ that is a real positive-definite, the Cholesky decomposition breaks it down into two components. It represents $A$ as the product of a lower triangular matrix, $L$, and its transpose, $L^T$. So, the decomposition can be represented as: $A = L L^T$

What's striking about this is that the matrix $L$ is unique, with real and positive diagonal entries. This uniqueness is paramount, especially when dealing with SVARs in econometrics, as it ensures consistent and unambiguous results.

\subsubsection{Creating multivariate distributed rvs with the Cholesky decomposition}\label{sec:multchol}

As a reminder of matrix and random variables: if $X$ is a collection of rvs such that its mean is $\mu$ and its variance-covariance matrix is $Var(X) = \Sigma$. Now if we take a matrix composed of constant elements $L$ and define a new random variable vector $Y=L X$, then $E(Y)=U \mu$ and $Var(Y)=L Var(X) L^T = L \Sigma L^T$.

A Cholesky decomposition can be used useful to generate multivariate normally distributed random vectors. If $Z=[Z_1,\ldots,Z_N]^T$ is a vector where $\forall i, Z_i \sim N(0,1)$ and are independent, then if $L$ is a $(N\times N)$ matrix,
\begin{equation}
L Z \sim MN(0,LL^T)
\end{equation}

If $C$ is the covariance matrix we are looking to introduce, then it is a real symmetric positive-definite matrix and its Cholesky decomposition is $C = L L^T$.

If we follow the spirit of \citet{Kousky:2012wg} and the lecture of \href{https://martin-haugh.github.io/files/MonteCarlo/MCS_Generate_RVars.pdf}{Martin Haugh}. I generate $3$ iid rvs $Z$ from a standard normal law, with samples of size $10^5$. Then I want to make those rvs dependent with a covariance matrix:
\[
C = \begin{bmatrix}
1.38&0.05&0.05\\
0.05&1.38&0.05\\
0.05&0.05&1.38
\end{bmatrix}
\]

Then I extract $L$ from the Cholesky decomposition of $C$ and multiply $L Z$

\subsubsection{Exercise: two correlated assets and the risk of a $50/50$ portfolio}\label{sec:choleskyportfolio}

The construction above is not a numerical curiosity: it is the standard way of simulating a portfolio. Consider two assets whose annual returns, expressed in percent, are normally distributed,
\begin{equation}\label{eq:twoassetdgp}
R_{i,t} \sim \mathcal{N}\left( \mu_i, \sigma_i^2 \right), \qquad i \in \{1,2\},
\end{equation}
with the following calibration:

\begin{center}
\begin{tabular}{lcc}
\toprule
 & asset $1$ & asset $2$ \\
\midrule
expected return $\mu_i$ & $8$ & $4$ \\
standard deviation $\sigma_i$ & $20$ & $15$ \\
\bottomrule
\end{tabular}
\end{center}

An investor puts half of her wealth in each asset, so that with $w = \left[\frac{1}{2},\frac{1}{2}\right]^T$ and $R_t = [R_{1,t},R_{2,t}]^T$ the portfolio return is
\begin{equation}\label{eq:portfolioreturn}
R_{p,t} = w^T R_t = \frac{1}{2} R_{1,t} + \frac{1}{2} R_{2,t}.
\end{equation}

The two marginal distributions in equation \ref{eq:twoassetdgp} say nothing about how the two assets move \emph{together}. That information is carried by the variance-covariance matrix,
\begin{equation}\label{eq:twoassetcov}
C(\rho) = \begin{bmatrix}
\sigma_1^2 & \rho \sigma_1 \sigma_2 \\
\rho \sigma_1 \sigma_2 & \sigma_2^2
\end{bmatrix}
= \begin{bmatrix}
400 & 300 \rho \\
300 \rho & 225
\end{bmatrix},
\end{equation}
and, following this section, we generate the returns as $R_t = \mu + L Z_t$ where $Z_t \sim MN(0,I_2)$ and $C(\rho) = L L^T$. For a $2\times2$ matrix the Cholesky factor has a closed form,
\begin{equation}\label{eq:chol2x2}
L = \begin{bmatrix}
\sigma_1 & 0 \\
\rho \sigma_2 & \sigma_2 \sqrt{1-\rho^2}
\end{bmatrix},
\end{equation}
which is worth verifying by hand before trusting any library.

We compare two beliefs about the joint behaviour of the two assets: $\rho = 0$, where the investor treats them as independent, and $\rho = 0.5$, of the order of what is observed between two equity markets. We are interested in what happens in \emph{bad} years, measured at the level $\alpha = 5\%$ by the quantile of the portfolio return, its Value-at-Risk,
\begin{equation*}
q_\alpha = \inf \left\{ x \in \mathbb{R} : \Pr\left[ R_{p,t} \leq x \right] \geq \alpha \right\},
\end{equation*}
and by the average return over the worst $\alpha$ of the outcomes, its expected shortfall,
\begin{equation}\label{eq:esdef}
ES_\alpha = \E \left[ R_{p,t} \mid R_{p,t} \leq q_\alpha \right].
\end{equation}

\stepcounter{question}
\begin{itemize}
\item \textbf{Question \arabic{question}}
\begin{itemize}
\item Write $C(\rho)$ for $\rho=0$ and for $\rho=0.5$, check that both matrices are positive-definite, and compute the Cholesky factor $U^T$ of each by hand with equation \ref{eq:chol2x2}. Verify that $L L^T = C(\rho)$.
\item Show that $\E(R_{p,t}) = \frac{1}{2}(\mu_1+\mu_2)$ does not depend on $\rho$, whereas $\Var(R_{p,t}) = \frac{1}{4}\left(\sigma_1^2+\sigma_2^2+2\rho\sigma_1\sigma_2\right)$ does. Compute both for the two values of $\rho$.
\item For each $\rho$, draw $N=10^6$ vectors $Z_t$ from the bivariate standard normal, form $R_t = \mu + L Z_t$, and check that the sample mean and the sample variance-covariance matrix of your draws reproduce $\mu$ and $C(\rho)$. Do not call a routine that generates a multivariate normal directly: the point of the exercise is to build it yourself.
\item Compute $R_{p,t}$ with equation \ref{eq:portfolioreturn} and estimate, for each $\rho$, the quantile $q_{0.05}$ and the expected shortfall $ES_{0.05}$ of equation \ref{eq:esdef} --- the latter being simply the average of the $5\%$ smallest simulated portfolio returns.
\item The portfolio return is a linear combination of jointly normal random variables, hence normal itself, so both risk measures have a closed form. Show that $q_\alpha = \mu_p + z_\alpha \sigma_p$ and that $ES_\alpha = \mu_p - \sigma_p \frac{\phi(z_\alpha)}{\alpha}$, where $z_\alpha = \Phi^{-1}(\alpha)$, and use them to check your simulation.
\end{itemize}
\end{itemize}

\subsection{Computational Complexity}

Computational complexity\footnote{pandas\_exercise\_part1\_cc.py} concerns the resources required to solve computational problems, primarily time and space. The theory classifies algorithms according to their performance in relation to the input size. Let us denote the input size by $n$.

To begin, let us define the Big O notation. For a given function $g(n)$, we say that $f(n) = O(g(n))$ if there are positive constants $c$ and $n_0$ such that $0 \leq f(n) \leq c \cdot g(n)$ for all $n \geq n_0$. This notation helps us describe an upper limit on the growth rate of the runtime or space requirements of an algorithm.

Consider a simple example: an algorithm that checks each pair in an array to find duplicates. This algorithm would have a time complexity of $O(n^2)$ because, in the worst case, it performs a comparison for each pair of elements in the array.

\textbf{P (Polynomial Time)}: This class contains problems that can be solved by a deterministic Turing machine in polynomial time. An example is matrix multiplication, which can be solved in time $O(n^3)$ using the standard algorithm, though faster algorithms exist.

\subsubsection*{Matrix Multiplication}
Consider two $2 \times 2$ matrices $A$ and $B$ defined as follows:
\[
A = \begin{bmatrix}
a_{11} & a_{12} \\
a_{21} & a_{22}
\end{bmatrix}, \quad
B = \begin{bmatrix}
b_{11} & b_{12} \\
b_{21} & b_{22}
\end{bmatrix}
\]
The product of these matrices, $C = A \cdot B$, is calculated as:
\[
C = \begin{bmatrix}
a_{11}b_{11} + a_{12}b_{21} & a_{11}b_{12} + a_{12}b_{22} \\
a_{21}b_{11} + a_{22}b_{21} & a_{21}b_{12} + a_{22}b_{22}
\end{bmatrix}
\]
Each element of matrix $C$ is the sum of products of elements from a row of $A$ and a column of $B$. For a general $n \times n$ matrix, this calculation involves $n$ multiplications and $n-1$ additions for each of the $n^2$ entries, resulting in a total of $2 \times n^3 - n^2$ operations, assuming the addition and multiplication have a constant time complexity. Hence, the naive matrix multiplication complexity is $O(n^3)$.

\subsubsection*{Strassen Algorithm for Matrix Multiplication}
The Strassen algorithm, introduced in \citet{strassen1969gaussian}, reduces the computational complexity of matrix multiplication. This algorithm is particularly effective for large matrices and is based on a divide-and-conquer approach that reduces the number of recursive multiplications.

Consider the same $2 \times 2$ matrices $A$ and $B$, instead of performing eight multiplications as in the naive approach, the Strassen algorithm performs only seven multiplications by introducing intermediate products:
\begin{align*}
M_1 &= (a_{11} + a_{22})(b_{11} + b_{22}), \\
M_2 &= (a_{21} + a_{22})b_{11}, \\
M_3 &= a_{11}(b_{12} - b_{22}), \\
M_4 &= a_{22}(b_{21} - b_{11}), \\
M_5 &= (a_{11} + a_{12})b_{22}, \\
M_6 &= (a_{21} - a_{11})(b_{11} + b_{12}), \\
M_7 &= (a_{12} - a_{22})(b_{21} + b_{22}).
\end{align*}

These intermediate products are used to compute the entries of the product matrix $C$ as follows:
\[
C = \begin{bmatrix}
M_1 + M_4 - M_5 + M_7 & M_3 + M_5 \\
M_2 + M_4 & M_1 - M_2 + M_3 + M_6
\end{bmatrix}
\]

By reducing the number of multiplications, the Strassen algorithm improves the overall time complexity. For general \(n \times n\) matrices, the algorithm recursively applies itself to smaller matrices, leading to a complexity of \(O(n^{\log_2 7}) \approx O(n^{2.807})\). This improvement is significant, particularly for very large matrices.

This topic is still improved, as in \citet{duan2023faster} and \citet{williams2024new}.

\section{Python, import and pandas: Import csv data as DataFrame}

We suggest\footnote{pandas\_exercise\_part1\_import.py} to build a DataFrame of the evolution of the French population as taken from INSEE website: \href{http://www.insee.fr/fr/bases-de-donnees/bsweb/serie.asp?idbank=001641607}{here}.

\subsection{Indicating where your file is located}\label{sec:filelocation}
Now you need to 'tell' python where you file is located.

Import the package Miscellaneous operating system interfaces:
\begin{lstlisting}
import os
\end{lstlisting}

Then you need to tell him where your data folder is located, in my computer, that would be:

\begin{lstlisting}
os.chdir('/Users/skimeur/Google Drive/empirical_finance/data')
\end{lstlisting}

If you are note sure where your file is located, here is a step-by-step guide:
\begin{enumerate}
\item make sure you downloaded the data zip file
\item you want to unzip the data folder and place it in the folder where you have your python code for this lecture
\item if you are not sure how to write the path to your data folder, you have a function (\href{https://stackoverflow.com/questions/1724693/find-a-file-in-python}{inspired from here}) to search for your file
\item once you have your location, you can use it to change your working directory
\item you have to be careful and might have to replace '\textbackslash' signs with '/'
\item if you have special characters in your path, you might need to use \href{https://docs.python.org/2.0/ref/strings.html}{"raw strings"} and use the prefix r
\begin{itemize}
\item so this could be r'/Users/avoid space and special characters in folder names pl)§se'
\end{itemize}
\end{enumerate}

\quad

Another way to approach this is by using the Spyder working directory browser until you find the folder where your file is located:

\begin{center}
\includegraphics[scale=0.6]{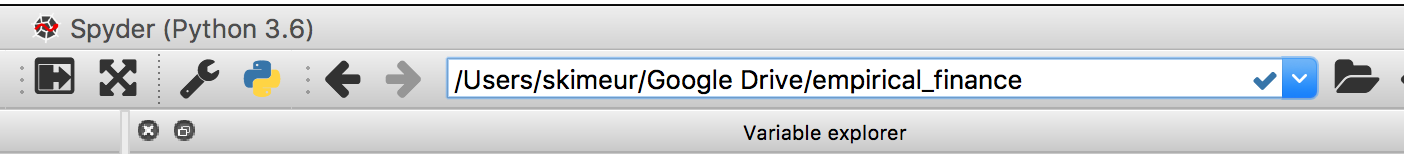}
\end{center}

\subsection{Importing the full data set from INSEE}
From the INSEE website \href{http://www.insee.fr/fr/bases-de-donnees/bsweb/serie.asp?idbank=001641607}{here} you can choose to import the table as a csv file. It will be called 'valeurs.zip'.

We prefer to work with comma-separated values file, for some of the reasons described \href{https://en.wikipedia.org/wiki/Comma-separated_values}{here}.

First unzip the file to have it a Valeurs.csv

Copy the Valeurs.csv file into your usual  data folder.

Follow the indications section \ref{sec:filelocation} to indicate where your file is located.

\subsubsection{Import the csv file}
If you open your file with a text editor\footnote{be carefule, if you open this file with Excel, do not save the modifications}, it looks like this:

\begin{lstlisting}
Libell...;;D...mographie - Population au...
IdBank;;001641607
Ann...e;Mois
2016;9;66 790
2016;8;66 763
2016;7;66 735
2016;6;66 710
\end{lstlisting}

So because we are working with French data:
\begin{itemize}
\item they separate variables with ; instead of ,
\item they use some special character as \`{a}
\end{itemize}

Hence we need to tell that:
\begin{itemize}
\item the values are separated by ';' 
\item it is encoded in latin1 style
\item the frist two rows are not relevant for our study, so skip them
\end{itemize}

\subsection{Cleaning the imported data set}
The data set that has been imported and is stored into the DataFrame named df is not clean.

\subsubsection{Reverse the row order}
First we want to have an timely ascending series:
\begin{lstlisting}
df=df.iloc[::-1]
\end{lstlisting}

\subsubsection{Rename the columns}
We want to rename the columns of the DataFrame with Month and Population.

\subsubsection{Reset the index}
We want to reset the index, knowing that the observations are monthly and start in January 1994:

\subsubsection{Delete a column}
Now that we have reset the index, we can drop the column with the months that is not bringing much information now, you drop vertically, a column, that is axis $= 1$.

\subsection{Plot the DataFrame}
Now try to plot your DataFrame df, it doesn't work and tells you:
\begin{lstlisting}
Empty 'DataFrame': no numeric data to plot
\end{lstlisting}

This means that when importing the data from the .csv file, python did not recognise that the column 'Population' were numeric. We now need to convert those data to numeric. First you need to remove the space in the numbers, then converting strings to float (note the use of regex\footnote{Without regex=True the method will look for exact matches. When you use regex=True it will look for the sub strings too}):

\begin{lstlisting}
df=df.replace({' ': ''}, regex=True)
df=df.astype(float)
\end{lstlisting}

Now when you try to plot, you should get:
\begin{center}
\includegraphics[scale=0.5]{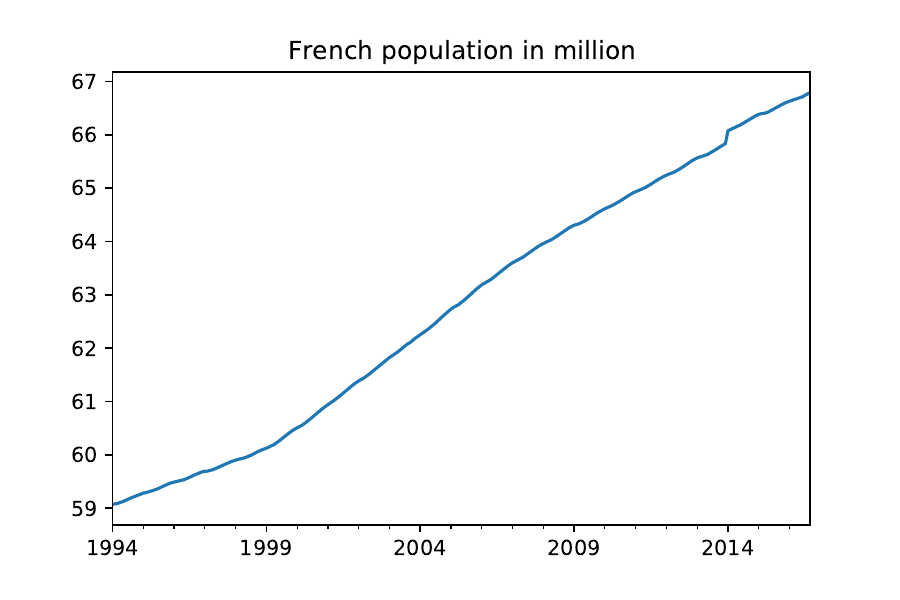}
\end{center}

\subsection{Resample our DataFrame (from monthly to yearly observations)}

There is the possibility to resample our DataFrame from one frequency to another using the resample() function. We need to indicate the desired frequency described in the \textbf{Offset Aliases} section of the time series documentation available \href{http://pandas.pydata.org/pandas-docs/stable/timeseries.html}{here}. We list some of the more commonly used:
\begin{itemize}
\item H	hourly frequency
\item D	calendar day frequency
\item W	weekly frequency
\item Q	quarter end frequency
\item A	year end frequency
\end{itemize}

The we need to chose a "method" for resampling, for example .sum() or .mean()

In our case, if we want to move to yearly frequency and get the average over the 12 months. If we plot the obtained DataFrame, we see that the plot is slightly smoother.

\subsection{Compute monthly changes}
Now you can plot and compute monthly changes.

\begin{center}
\includegraphics[scale=0.5]{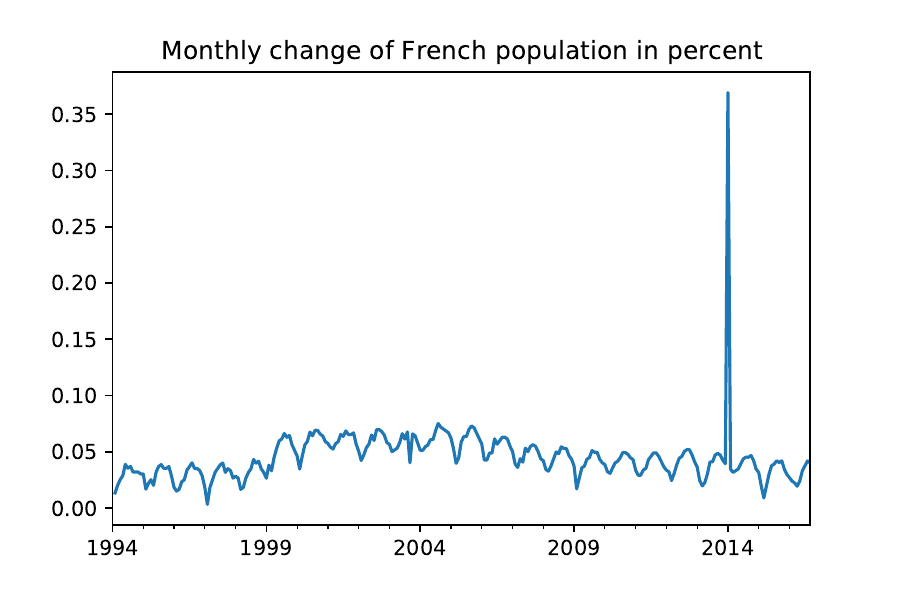}
\end{center}

\subsection{Descriptive statistics}
You can get some descriptive statistics on the population and the monthly population changes:

\begin{center}
\begin{tabular}{lrr}
\toprule
{} &  Population in million &  Population growth rate in percentage \\
\midrule
count &                 273 &                                272 \\
mean  &                  62.78 &                                  0.05 \\
std   &                   2.44 &                                  0.02 \\
min   &                  59.07 &                                  0.00 \\
max   &                  66.79 &                                  0.37 \\
\bottomrule
\end{tabular}
\end{center}

\subsection{Exercise with import, DataFrame and matrices: Leontief Input-Output model}

Wassily Leontief is widely known for developing the Input-Output model of the economy. His foundational work on the subject began with \citet{Leontief:1949uo}. For a comprehensive collection of his major papers on this model, one can refer to the compilation presented in the book \citet{leontief1986}.

The essence of the model lies in its sectoral division of the economy. Every sector is dependent on others for its input to produce its output. Thus, the production of a sector isn't just shaped by external demand (from end consumers) but also by internal demand (from other sectors).

To crystallize this concept using matrix notation, let's define the following:

\begin{itemize}
\item \(P\): A vector representing the production of each of the \(N\) sectors, denoted as \([p_1, \ldots, p_N]\).
\item \(D\): A vector representing the external demand for each sector, presented as \([d_1, \ldots, d_N]\).
\item \(M\): The consumption (or technical-coefficient) matrix, where each element \(m_{i,j}\) is the quantity of sector \(i\)'s output needed to produce one unit of sector \(j\)'s output. It is obtained from the table of intermediate flows \(z_{i,j}\), the output of sector \(i\) purchased by sector \(j\) (rows are the producing sectors, columns the purchasing sectors), by dividing each \emph{column} \(j\) by the gross output \(p_j\) of the purchasing sector: \(m_{i,j} = z_{i,j}/p_j\).
\end{itemize}

With these definitions in place, the model follows from an accounting identity and a technological assumption. Everything sector \(i\) produces is either purchased as an input by some sector or absorbed by external demand, \(p_i = \sum_j z_{i,j} + d_i\); with Leontief's fixed-proportion technology, \(z_{i,j} = m_{i,j}\, p_j\), so that \(p_i = \sum_j m_{i,j}\, p_j + d_i\), that is, in matrix notation:
\begin{equation}
P = M P + D
\end{equation}
Rearranging the above equation, we get:
\begin{equation}
\left(I-M \right)P = D
\end{equation}
Here, \(I\) represents the identity matrix. If the matrix \(\left(I-M \right)\) is invertible, we can confidently assert that our equation has a unique solution equation \ref{eq:leontiefP}. On the contrary, if it's non-invertible, the model becomes indeterminate, meaning there could be no solution or potentially infinite solutions. In economic terms the useful condition is stronger than invertibility: if every column sum of \(M\) is below one (each sector spends less on intermediate inputs than the value of its output), then \(\left(I-M\right)^{-1} = I + M + M^2 + \cdots\) exists and has nonnegative entries, so that any nonnegative external demand is met by a nonnegative production plan.

\begin{equation}\label{eq:leontiefP}
P = \left(I-M \right)^{-1}D
\end{equation}

\subsubsection{Simplified example}

Let's imagine we have the simplified economy:

\begin{tabular}{lrrrr}
& Agriculture & Manufacturing & Services & External demand, $D$\\
Agriculture & 35 & 5 & 5 & 40\\
Manufacturing & 5 & 60 & 20 & 75\\
Services & 10 & 25 & 40 & 145\\
Total gross output & 85 & 160 & 220 & \\
\end{tabular}

The rows are the producing sectors and the columns the purchasing sectors: agriculture sells 35 to itself, 5 to manufacturing, 5 to services and 40 to external demand, which adds up to its gross output of 85. The table is balanced, each row summing with external demand to the gross output of its sector. The consumption matrix $M$ is obtained by dividing each column of the table by the total gross output of that column's sector: for example $m_{2,1} = 5/85 \approx 0.06$ is the amount of manufacturing output needed per unit of agricultural output. The column sums of $M$ ($0.59$, $0.56$, $0.30$) are all below one, so the economy is productive. We can now solve equation \ref{eq:leontiefP}, and check that $P = \left(I-M\right)^{-1} D$ returns exactly the gross output $(85, 160, 220)$ of the table. Dividing each \emph{row} by the gross output instead would give $z_{i,j}/p_i$, the share of sector $i$'s sales going to sector $j$: these allocation coefficients define the supply-driven model of Ghosh, not the demand-driven Leontief model, and do not answer the question of which production is needed to serve a given external demand.

Now, if there is a one unit increase in the demand for agriculture output, this should have a positive impact of about $0.2$ on the manufacturing production (precisely $0.2037$, the entry in the second row and first column of $\left(I-M\right)^{-1}$).

This is a step-by-step way of doing it\footnote{input\_output\_illustration.py}
Actually, the most difficult part is to divide each column of a matrix by a given value from a vector, three approaches are suggested (as seen during class):

\begin{enumerate}
\item Create Mgross as the matrix before division by gross output
\item Create D the demand vector
\begin{itemize}
\item Here notice that the name you give to index and columns are important when you will want to perform matrix multiplication (e.g. agriculture should match agriculture).
\end{itemize}
\item Create the Total Gross Output (TGO) vector
\item Divide each column of Mgross by the corresponding element of the TGO vector
\begin{itemize}
\item Either by creating TGO as a Series
\item Either by duplicating TGO transposed to form a 3 times 3 matrix
\item Either by dividing by TGO transposed, using .values
\end{itemize}
\item Now you can compute $\left(I-M \right)^{-1}$ and multiply it by $D$ to get $P$
\end{enumerate}

\subsubsection{Input-Output exercise}

\stepcounter{question}
\begin{itemize}
\item \textbf{Question \arabic{question}}
\begin{itemize}
\item Import Input-Output Accounts Data from \href{https://www.bea.gov/industry/input-output-accounts-data}{bea.gov website}.
\begin{itemize}
\item Use\_SUT\_Framework\_2007\_2012\_DET.xlsx has been place on your data folder if need be.
\end{itemize}
\item Shape the matrix $M$ and vector $D$
\begin{itemize}
\item For the matrix $M$, keep columns up to T001 and index up to S00900.
\item Use the boxplot of column T001, do not show outliers and display the mean.
\item Drop the element 4200ID, custom duties
\item Make the matrix $M$ square (keep only elements common in both the index and the columns:
\begin{itemize}
\item you can get inspired by: \href{https://stackoverflow.com/questions/7016304/removing-elements-from-a-list}{this example}, creating lists from columns and index of the data frame.
\end{itemize}
\item Consider that $D$ is the total use of product, T019, minus the total intermediate T001.
\item Make $M$ the consumption matrix, so divide each column $j$ of $M$ by element $\mbox{TIO}_j$ of the Total industry output (basic value), T018.
\end{itemize}
\item Compute $P$
\item Now, imagine the external demand for Oilseed farming (1111A0) increases by one unit. What is the impact on the production for the Oil and gas extraction (211000)?
\begin{itemize}
\item For this, create a $D'$ filled with zeros and a $1$ placed as 1111A0;
\item Now use $M$ to compute the impact vector $P'$: $P' = \left(I-M \right)^{-1}D'$
\end{itemize}
\end{itemize}
\end{itemize}

\stepcounter{question}
\begin{itemize}
\item \textbf{Question \arabic{question}}
\begin{itemize}
\item Perform a similar analysis for the France, using \href{https://www.oecd.org/sti/ind/input-outputtables.htm}{OECD data set}
\item Now apply a lock-down like shock (negative) on "Accomodation and food services", what are the impact on the other sectos, which are the mostly impacted sectors?
\end{itemize}
\end{itemize}

\stepcounter{question}
\begin{itemize}
\item \textbf{Question \arabic{question}}
\begin{itemize}
\item Follow \citet{Parikh:1979vm} to forecast the Input-Output matrix with the RAS method
\end{itemize}
\end{itemize}

\section{Empirical data: loading files, using API, scraping websites}

I would tend to say that before engaging in a long research project, you should explore whether some data will be available to test your model or not. 

You will need to change the working directory to be able to load some files,previous section\footnote{get\_data\_vansteenberghe.py}  explained step-by-step how to change your working directory.

\subsection{Getting time series from Banque de France and INSEE}

Banque de France provides historical times series via its webstat portal, \href{http://webstat.banque-france.fr/fr/}{here}.

We might want to download the historical evolution of property prices in France, this can be found (with some more explanations) under the code RPP.Q.FR.N.ED.00.1.00 (Indices des prix des logements anciens, Ensemble des logements, France metropolitaine).


\stepcounter{question}
\begin{itemize}
\item \textbf{Question \arabic{question}}
\begin{itemize}
\item Update the property prices data set, do you see a lock-down effect?
\begin{itemize}
\item Suggest a way to measure this effect.
\end{itemize}
\end{itemize}
\end{itemize}

You can access INSEE data from \href{https://www.insee.fr/fr/statistiques}{their website}. As an illustration, the Demography - Population at begining of a month in France, series ID: 000436387

\stepcounter{question}
\begin{itemize}
\item \textbf{Question \arabic{question}}
\begin{itemize}
\item Update the French population times series, do you see a pandemic effect?
\begin{itemize}
\item Suggest a way to measure this effect.
\end{itemize}
\end{itemize}
\end{itemize}

\subsection{Getting data from OECD and pivot-table}

The OECD provides panel data via its websilte \href{https://stats.oecd.org/}{here}.

You can select "index of Industrial Production". Then you'll need to reorganise the data set, making use of pivot table:


\subsection{Getting data from the Federal Reserve of Saint Louis - FRED}

You can get data from the Federal Reserve of Saint Louis, their Federal Reserve Economics Database, or FRED.

What is convenient is that you can download the time series directly form python if you know the code of the variable you are looking for. For example, we can download the following variables for the US economy, which are the usual variable used in basic macroeconomic models:
\begin{itemize}
\item Real GDP
\item Industrial Production
\item Core CPI
\item Unemployment Rate
\item 10 Year Yield
\item 1 Year Yield
\item Baa Yield
\item Aaa Yield
\end{itemize}


\stepcounter{question}
\begin{itemize}
\item \textbf{Question \arabic{question}}
\begin{itemize}
\item You can try to reproduce the following with updated data:
\end{itemize}
\end{itemize}

\begin{center}
\includegraphics[scale=0.4]{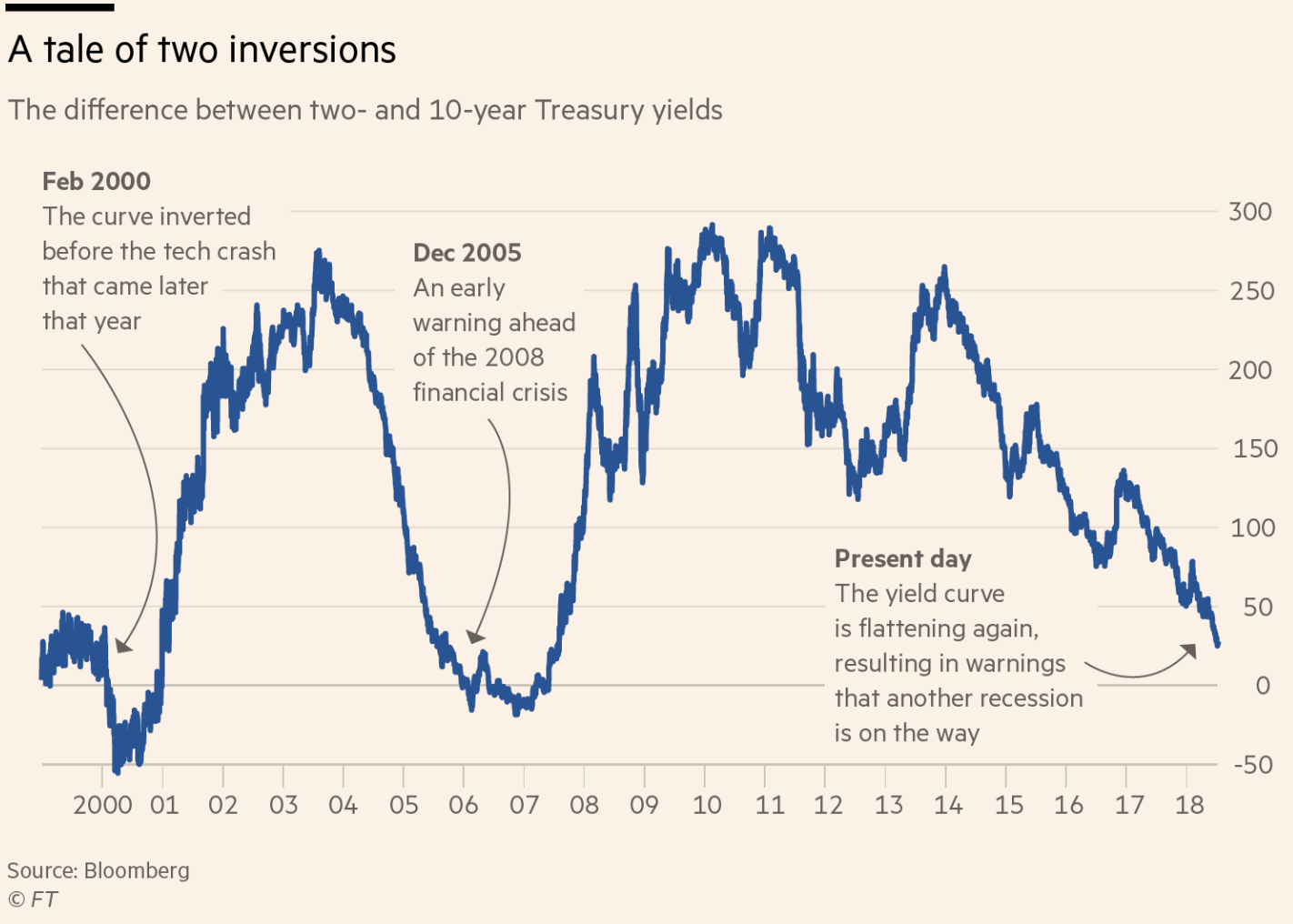}
\end{center}

\subsection{Getting data from Bloomberg}
At the University, you have access to a Bloomberg terminal.
We suggest that you install the Bloomberg extension for Excel (look into the start menu in window, you should find this installer).

\quad

You might want to download some historical time series with Excel. We suggest the following step:
\subsubsection{Find the right field}
You need to go onto the Bloomberg terminal in order to find the correct field: type FLDS
\begin{center}
\includegraphics[scale=0.5]{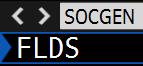}
\end{center}

Then you can search for a company name for example, it will give you its Bloomberg ticker:
\begin{center}
\includegraphics[scale=0.5]{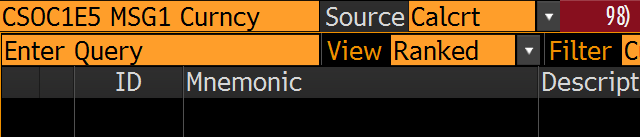}
\end{center}

Once you get the Bloomberg ticker (e.g. "KN FP Equity" for the bank BPCE), you can use this ticker in an excel spreadsheet.

\subsubsection{Download the time series in Excel}

We suggest to organize your Excel sheet as follow:
\begin{center}
\includegraphics[scale=0.4]{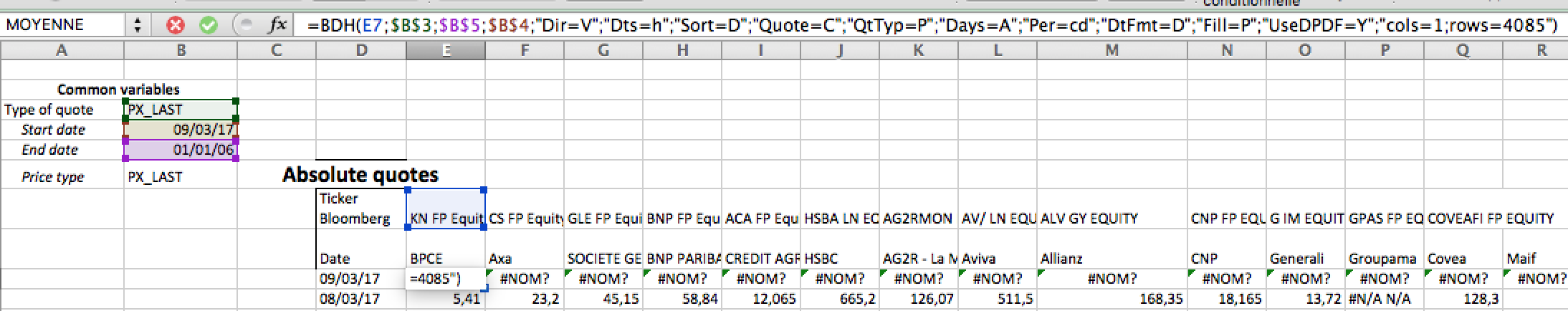}
\end{center}

Here is an simple example to download stock price data vertically:
\begin{verbatim}
=BDH(E7;$B$3;$B$5;$B$4;"Dir=V";"Dts=h";"Sort=D";"Quote=C";"QtTyp=P";"Days=A";"Per=cd";
"DtFmt=D";"Fill=P";"UseDPDF=Y";"cols=1;rows=4085")
\end{verbatim}

\subsection{Getting data from other sources}

You can get data from the IMF \href{http://data.imf.org/?sk=388DFA60-1D26-4ADE-B505-A05A558D9A42}{here}, from the ECB at its \href{http://sdw.ecb.europa.eu/}{Statistical Data Warehouse}. The World Bank provides time series on \href{https://data.worldbank.org/data-catalog/world-development-indicators}{World Development Indicators}. The Bank for International Settlements (BIS).

\stepcounter{question}
\begin{itemize}
\item \textbf{Question \arabic{question}}
\begin{itemize}
\item Use data \href{https://stats.bis.org/statx/srs/table/B2?c=&p=20163&m=S}{from the BIS} and reproduce the graphs in the article about the \href{http://libertystreeteconomics.newyorkfed.org/2017/09/what-drives-international-bank-credit.html}{Anna Karenina Principle of global banking} published in \href{http://www.bis.org/publ/work639.pdf}{a research article}. This exercise is for advanced pandas users and you should complete section \ref{sec:pandas} before doing this exercise.
\end{itemize}
\end{itemize}

\subsection{Web scraping with python, a use case with IFP}

We do not intend to provide here a full lecture on web scraping with python. We provide one use case: getting a list of Fintech that are IFP from \href{https://www.orias.fr/web/guest/search}{ORIAS website}.

Lending in France is regulated by law under what is known as monopole bancaire\footnote{Article L511-5, this monopoly is a mean to protect depositors from institutions insolvency and illiquidity} and is regulated at the European level by the CRDIV\footnote{Article 9-2 of the directive 2013/36/EU of the European Parliament and of the Council of 26 June 2013 on access to the activity of credit institutions} on taking deposits or other repayable funds and by the CRR\footnote{Article 4-1 (1) of the regulation (EU) No 575/2013 of the European Parliament and of the Council of 26 June 2013} for granting credits for its own account. This regulation is a barrier to entry into the market but has been relaxed by the introduction of new categories of actors in France in 2014\footnote{Ordonnance n° 2014-559 of the 30th May 2014 which is a Dérogation au monopole bancaire  introducing the role of conseillers en investissements participatifs and intermédiaires en financement participative}. For the French market, the ACPR\footnote{Autorité de contrôle prudentiel et de résolution, the French financial regulator} has the authority to deliver accreditation\footnote{Ordonnance n° 2014-559 of the 30th May 2014 article 16} to intermediaries in participatory financing (IFP).
ORIAS provides information on the IFP but to gather the information, one has to research and select the data. We suggest to use python libraries for web scraping to feed a data frame:

\resizebox{\textwidth}{!}{
\begin{tabular}{llrllrllrll}
 & Siren Number &        ID &                                               Name &                                           Category &  Postcode &                          City &                                            oriasID &     date in &    date out \\
0   &    539015149 &  14006008 &                                          credit.fr &  Crowdfunding Advisor (CIP) Crowdfunding Interm... &     92300 &              Levallois-Perret &     103995 &  2014-10-17 &         NaN \\
1   &    804606796 &  14006007 &                                         Lendopolis &  Crowdfunding Advisor (CIP) Crowdfunding Interm... &     75010 &      Paris 10e Arrondissement &   104072 &  2014-10-17 &         NaN \\
2   &    798999983 &  14006009 &                                     Primus Finance &                    Crowdfunding Intermediary (IFP) &     78110 &                    Le Vésinet &  103779 &  2014-10-17 &         NaN \\
3   &    803411982 &  14006571 &                                           Fidelfin &  Crowdfunding Advisor (CIP) Crowdfunding Interm... &     59510 &                           HEM &    104721 &  2014-11-07 &         NaN \\
4   &    750121311 &  14006337 &                                       Blue Project &                    Crowdfunding Intermediary (IFP) &     75011 &                         PARIS &  104399 &  2014-11-07 &         NaN \\
\end{tabular}
}

This exercise was done with the help of a StackOverflow member (for more details on the problematic see \href{https://stackoverflow.com/questions/57536576/python-3-how-to-scrape-research-results-from-a-website-using-csfr/57539377#57539377}{StackOverflow website}) that helped us overcome the CSFR restrictions.

We provide the code: 20190817\_orias\_scrape.py with little comments, this is for advanced python users, you might skip this part for now and come back on it later on.

\stepcounter{question}
\begin{itemize}
\item \textbf{Question \arabic{question}}
\begin{itemize}
\item After deciphering the code, apply the same approach to scrape data on Conseiller en investissement financier (CIF) from ORIAS website.
\end{itemize}
\end{itemize}

\section{Data Assurance Quality}\label{sec:dataquality}

In the domain of Machine Learning, as well as in econometrics, data assurance quality---encompassing activities like data cleaning, preparation, and organization---typically constitutes a large share of the overall workload, often quoted as $80\%$. Beyond just the effort involved, ensuring data quality is pivotal. It not only enhances the accuracy and reliability of the model but also furnishes insights into the inherent strengths and potential vulnerabilities of the dataset. No estimator, however flexible, recovers from a misread column or from a test sample coded differently from the training sample, and the choices made at this stage (which rows to drop, which values to cap, how to recode a category) are part of the empirical strategy of a project: they must be documented as such (section \ref{sec:writingproject}).

\subsection{What to check, and in which order}\label{subsec:dqdimensions}
It is useful to organise the exploration of a new data set around a short list of questions, each corresponding to a dimension of data quality.
\begin{itemize}
\item \textbf{Provenance and codebook.} Who produced the data, how, and what does each column measure exactly? Read the data dictionary before the first \texttt{describe()}: the most costly mistakes are made by guessing the meaning of a column from its name.
\item \textbf{Completeness.} How are missing values coded (blank, \texttt{NA}, \texttt{?}, $-999$, $0$)? Where are they, do they come in patterns, and is their occurrence related to the outcome of interest?
\item \textbf{Validity.} Are the recorded values within the domain defined by the codebook (a list of categories, a range)?
\item \textbf{Consistency.} Do logically related columns agree with each other (two codings of the same variable, a husband recorded as a woman, an end date before a start date)? Are there duplicate records, and what defines a duplicate?
\item \textbf{Plausibility.} Are the extreme values genuine, errors, or artefacts of the data producer such as top-coding? Should they be trimmed, winsorized, or kept?
\item \textbf{Coherence between samples.} When the data come as a training and a test sample, or as several vintages, are the coding, the categories and the distributions the same?
\end{itemize}
Two rules cut across these dimensions. First, every judgment call (a recoding, a dropped row, a cap) is written in the code, in one place, together with the evidence that motivated it, so that a reader can reproduce, and disagree with, each choice. Second, in a prediction exercise every statistic used to transform the data (a quantile, a mean used for imputation, the list of categories known to an encoder) is estimated on the training sample only and then applied to the test sample; estimating it on the pooled data leaks information from the test set into the model and overstates its out-of-sample performance. This is the fit/transform logic of the scikit-learn pipelines of section \ref{sec:MLwhy}.

\subsection{Winsorizing}
Winsorizing replaces the values of a variable below its $p$-th percentile (above its $(1-p)$-th percentile) by the value of that percentile; trimming drops these observations instead. Both limit the influence of extreme observations on estimators that are sensitive to them, such as least squares, at the cost of a modelling choice that must be reported. \citet{ottonello2020financial} work on a panel of firms:
\begin{quote}
We winsorize our sample at the top and bottom 0.5\% of observations of investment, leverage, and distance to default in order to ensure our results are not driven by outliers.
\end{quote}
In pandas, \texttt{series.clip(lower, upper)} with the two quantiles as bounds does the job; in a train/test setting the bounds are computed on the training sample and reused on the test sample. Winsorizing treats the symptom, not the cause: if the extreme values are genuine (the richest households), capping them changes the population described; if they are a ceiling imposed by the data producer (top-coding, see below), capping only moves the ceiling.

\subsection{Case Study: US Adult Census Data}
To elucidate this crucial aspect, we\footnote{vansteenberghe\_data\_quality\_ML\_US\_wages.py} turn our focus to the US Adult Census dataset, which correlates income with sociodemographic attributes such as age, education, race, and other pertinent features. This dataset is grounded in the seminal work of \citet{kohavi}; it was extracted from the 1994 Current Population Survey and comes already split into a training sample of 32,561 persons and a test sample of 16,281 persons. For those seeking to delve deeper, the \href{https://archive.ics.uci.edu/ml/datasets/census+income}{data set can be accessed here}.

\subsubsection{Reading the files and the codebook}
The files have no header line, the values are separated by a comma and a space, and missing values are coded by a question mark. These three facts are visible by opening a file in a text editor, and each corresponds to an option of \texttt{read\_csv}:
\begin{lstlisting}
df_train = pd.read_csv("data/adult-training.csv", names=COLUMNS,
                       skipinitialspace=True, na_values="?")
\end{lstlisting}
Without \texttt{skipinitialspace}, every category starts with a blank and \texttt{" Private"} differs from \texttt{"Private"}; without \texttt{na\_values}, the question mark is one more category and pandas reports no missing value at all.

The codebook then tells what each column measures. The variable with the largest values, \texttt{fnlwgt}, is tempting to read as a wage; it is the \emph{final weight} of the survey, the number of persons in the US population that the observation represents. It carries no information about the income bracket (its correlation with the label is $-0.01$) and should not be used as a feature, but it is what one needs to compute population-representative statistics: 24.1\% of the persons in the sample earn more than 50K, 23.9\% once the observations are weighted.

\subsubsection{Initial Data Exploration: missing values}
An indispensable starting point is to familiarize oneself with the dataset and promptly identify any missing values:

\begin{center}
\begin{tabular}{|l|r|}
\hline
\textbf{Feature} & \textbf{Missing values count}\\
\hline
age & 0 \\
workclass & 1836 \\
fnlwgt & 0 \\
education & 0 \\
education\_num & 0 \\
marital\_status & 0 \\
occupation & 1843 \\
relationship & 0 \\
race & 0 \\
gender & 0 \\
capital\_gain & 0 \\
capital\_loss & 0 \\
hours\_per\_week & 0 \\
native\_country & 583 \\
income\_bracket & 0 \\
\hline
\end{tabular}
\end{center}

Missing values come in a pattern: \texttt{workclass} and \texttt{occupation} are missing together for 1,836 persons, and seven more persons have a work class (``Never-worked'') but no occupation. In total 2,399 rows, 7.4\% of the training sample, have at least one missing value.

\begin{figure}[H]
\centering
\includegraphics[width=0.62\textwidth]{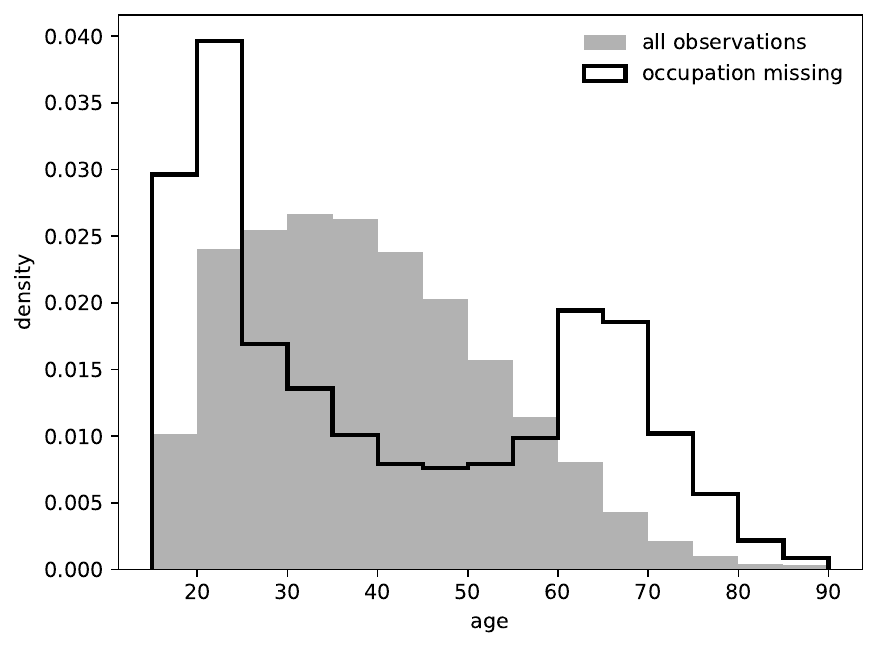}
\caption{Age distribution of the 1,843 persons with a missing occupation (black line) compared with the whole training sample (grey bars). The missing occupations are concentrated among the very young and the old.}
\label{fig:adultmissingage}
\end{figure}

A pertinent question arises: are the NaNs indicative of non-working individuals, or are they the result of data collection errors? If it's the latter, the optimal course of action might be to discard the affected rows. The data give elements of an answer (Figure \ref{fig:adultmissingage}): the persons with a missing occupation are much younger or much older than the others (26\% are aged 21 or less and 25\% are aged 62 or more, against 10\% and 6\% in the whole sample), they work fewer hours (32 against 41 per week) and only 10\% of them earn more than 50K, against 25\% of the others; a chi-square test rejects the independence between missingness and the income bracket with a $p$-value of order $10^{-45}$. In the standard vocabulary, the occupation is not \emph{missing completely at random}. Dropping these rows is defensible for a model of the working population, but it removes the students and the retirees and changes the population the model describes; the alternative is to keep them and to give the missing value its own category, which lets the model use the information that no occupation was reported. By contrast, the missingness of \texttt{native\_country} is unrelated to the income bracket ($p=0.62$): dropping these 583 rows costs sample size and nothing else.

\subsubsection{Validity and consistency}
Upon examining the 'native\_country' feature, an anomaly surfaces---a country labeled "South". A detailed inspection, especially when cross-referenced with the 'race' attribute, strongly suggests that this likely corresponds predominantly to South Korea: 77 of the 80 persons born in ``South'' are Asian or Pacific Islanders. We recode these 77 observations as South Korea, in the training and in the test sample alike, and document the assumption in the code. A purely mechanical check would not have caught the anomaly: ``South'' is listed as a country in the codebook itself.

Logical relations between columns are checked one by one, from the codebook, because no software can guess them. In the training sample, \texttt{education} and \texttt{education\_num} are two codings of the same variable (each of the 16 education levels maps to exactly one number), so one of them is redundant and must be dropped before a regression; three persons are recorded as ``Husband'' while being female or as ``Wife'' while being male, out of 14,761 spouses; capital gains and capital losses are never both positive, which justifies combining them (below). Twenty-four rows are exact duplicates, sampling weight included; ignoring the weight, 3,465 rows share all their characteristics with another row, which is expected with 32,561 persons described by a handful of coarse variables. What counts as a duplicate depends on which columns identify a unit, and the decision has to be written down.

\subsubsection{Plausibility: top-coding versus outliers}
\texttt{describe()} shows a maximum of 99,999 for capital gains. This is not an outlier: 159 persons are recorded at exactly this value while the next largest value, 41,310, is taken by two persons, and all 159 earn more than 50K. The data producer censored the variable at a ceiling (\emph{top-coding}), as it did for age (43 persons aged exactly 90, three aged 88) and for hours worked (85 persons at 99 hours). Such values are not measurement errors to be removed; they are the top of the distribution recorded at a bound, and a model should treat them as such (for instance with an indicator variable). Winsorizing at 0.5\% and 99.5\%, with bounds estimated on the training sample, would replace the 99,999 ceiling by another one at 34,095: it changes where the censoring occurs and does not recover the censored values.

\subsubsection{Coherence between the training and test samples}
The test file spells the income bracket with a trailing period (\texttt{>50K.} instead of \texttt{>50K}). A label built by exact comparison, \texttt{df\_test["income\_bracket"] == ">50K"}, is therefore false for every row, and a model evaluated on this test set would face a sample without a single positive case, with no error message. Printing the distinct values of the target in both samples before building the label avoids the trap; \texttt{str.contains(">50K")}, or stripping the period, fixes it. The categories must also be compared across samples: the native country ``Holand-Netherlands'' appears once in the training sample and never in the test sample, so an encoder fitted on the training sample must be told what to do with categories it has not seen (in scikit-learn, \texttt{OneHotEncoder(handle\_unknown="ignore")}). Finally, the means of the continuous variables and the share of persons above 50K (24.1\% against 23.6\%) are the same in the two samples: the test sample is drawn from the same population as the training sample, as it should be.

\subsubsection{Feature Engineering}
Feature engineering can serve as a potent tool to enhance the data's utility. For instance:
\begin{itemize}
    \item Combining 'capital\_gain' and 'capital\_loss' into a singular metric, the net capital income, might offer a holistic perspective on capital dynamics; since the two are never both positive, no information is lost.
    \item Categorizing 'marital\_status' to distinguish between individuals living solo versus those in relationships might furnish richer sociodemographic insights: 44.7\% of the persons living as a couple earn more than 50K, against 6.5\% of those living alone. The mapping must cover every category of the codebook (including ``Married-spouse-absent''), otherwise pandas silently creates missing values.
\end{itemize}
These transformations use no statistic estimated on the data and can be applied to both samples without leakage. The binary label (1 if the income exceeds 50K) is unbalanced, with 24\% of positives: predicting ``$\le$50K'' for everybody is right 76\% of the time, so accuracy alone is a poor score (section \ref{sec:MLwhy}).

\subsubsection{Data Analysis Tasks}
The script ends with the two cleaned samples and the label, and leaves to the modelling stage the decisions that depend on the model: dropping or flagging the incomplete rows, dropping one of the two education variables, keeping the sampling weight out of the features, winsorizing or not. The estimators are introduced in section \ref{sec:MLwhy}; we\footnote{vansteenberghe\_ML\_wage.py, on the wage data of section \ref{sec:multreg}} did some preliminary work with them as a starting point.
\stepcounter{question}
\begin{itemize}
\item \textbf{Question \arabic{question}}
\begin{itemize}
\item Engage in the following analyses:
\begin{enumerate}
\item Deploy K-means and Hierarchical clustering methodologies.
\item Construct and interpret a Dendrogram, subsequently demarcating an appropriate threshold.
\item Implement a logit model. Subsequently, leverage Lasso regression to ascertain if overfitting occurs.
\item Utilize the Random Forest algorithm.
\item Experiment with Boosting techniques.
\item For each model, report a score adapted to an unbalanced target on the test sample, and check that nothing in your pipeline was estimated on the test sample.
\end{enumerate}
\end{itemize}
\end{itemize}

\section{Python: when to use parametric econometrics and not other (more flexible) machine learning methods?}\label{sec:MLwhy}

Section \ref{sec:multreg} we explored in great details a parametric model $m$, linear and at best quadratic for a variable of interest ($y$: log of wage) using explanatory variables also called features ($X$: gender, education, experience) in  $y = m(X) + \epsilon$.
We might wonder why not to use machine learning for our regression problem.

We suggest an answer following \citep{charpentier2018}, considering that econometrics (statistics) \begin{quote}
has developed around the principle of inference (or to explain the relationship linking $y$ to variables $x$)
\end{quote}, to make inference, we are making assumptions about the data generating process (cf. section \ref{sec:MCmethod}). If we had to distinguish machine learning from econometrics, we would say that machine learning is focused on predictive power and does not make any assumption on the data generating process.

In both econometrics and machine learning, having data sets $[y_i,X_i]$ the goal is to "search" for a model that minimizes a loss function:
\begin{equation}
\min_m \mathcal{L}\left(y_i, m(X_i) \right)
\end{equation}
under some constraint on $m$.

We\footnote{The main code to follow is ML\_credit\_cards.py} present some machine learning technics\footnote{This is inspired by the lectures of Emmanuel Rachelson \href{http://emmanuel.rachelson.free.fr/}{(his website)} and the book \citep{raschka}} applied to some data set with the python code.

\subsection{Econometric models transparency}

In the words of \citet{Andrews:2017vv}: \begin{quote}
Linear regression analysis is popular in part because it is transparent. Estimates depend on a set of intuitive variances and covariances, and it is straightforward to assess how these moments would change under violations of the identifying assumptions. Well-understood properties of linear models--most prominently, the omitted variables bias formula--make it easy for readers to guess how these changes translate into bias in the estimates. We do not need to have access to the data to know that a regression of wages on education would be biased upward by omitted skill, and we can form a guess about how much if we have a prior on the likely covariance properties of the omitted variable.
\end{quote}
and their work expand this to nonlinear econometric models. Such transparency is not straightforward for machine leaning models.

\subsection{Before we introduce machine learning models - False positive concept and why Tour de France bikers might be cheating}

\href{https://www.youtube.com/watch?v=9iesBgCGeMU}{Some French reporter investigated the testing methods used by the Union Cycliste Internationale (UCI)} to test whether there were hidden motors in bicycles before a race. The reporter questioned the efficacy of the methods used... Could it be that there are engines used during the Tour de France?

The UCI \href{https://www.sportsintegrityinitiative.com/uci-statement-technological-fraud/}{replied with a published note}.

What is very interesting, is that reading this written reply, I started to have strong doubt... The UCI methods could be inefficient.

The UCI is quoting a report performed by a laboratory in the United States of America, the Microbac report 17D0410-1, uci17d0410\_rp1\_neutral. What is quite disturbing is that the report is actually what led me to doubt the methods at UCI:

\begin{center}
\includegraphics[scale=0.5]{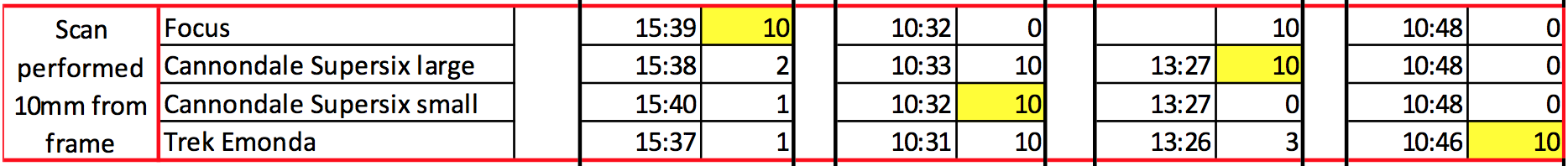}
\end{center}

When the detection tool is used on a bike, at 10 millimeters from the bike frame, there are 7 instance of "10" out of $4 \times 4 = 16$ tests. A "10" means that you have a strong signal, a positive. Out of those positive, 4 where bike with actual engines (in yellow)... but 3 of those "10" had no engine inside. 

\quad

In other terms, testing the bikes, the Microbac found 3 false positive out of 16 tests, that is a ratio of $\frac{3}{16}$. 

\quad

If you read carefully the UCI answer, they performed "over 4,000 controls undertaken during the 2017 Tour de France", this means that --if the Microbac tests have any meaning-- the UCI testers have found an expected 750 positive signals (all false positive, as no scandal arose from a true positive engine in a bike during the Tour de France 2017).

\quad

The question I would have is: what did the UCI do of those close to 1,000 positive alarms during the 2017 Tour de France? Did they take the time to perform thorough testing (computer tomography)... or just let the biker race?\footnote{I actually sent an e-mail to the UCI team, they answered along the lines that using the test device isn't just as simple as seeing a 10 on the screen and assuming it is a positive. The size and shape of the reading on the test device screen also needs to be interpreted which is why they deliver 3-4 hours of training to each operator. They claimed that they would not get many false positives because there is more to using the device than just seeing the number on the screen. They would occasionally get a positive reading that appears to be false and dismantle the bikes but they replied that it is very rare and not statistically significant.}

\subsubsection{What we have learned}

A model becomes inefficient if it yields a disproportionately high ratio of False Positives. In such scenarios, there might be a need to manually differentiate between True and False Positives.

\vspace{1em}

For supervisory entities:
\begin{itemize}
    \item Transparency is essential in your historical actions but not necessarily in your testing protocols: questions like who are your suppliers and what unforeseen and thorough tests you've conducted in the past are relevant.
    \item Testing protocols should undergo unannounced changes. Relying on the same tools and methods consistently could allow those under supervision to game the system.
    \item It's not always prudent to engage in expensive contracts with American firms to "validate" your practices. Numerous reporters and economists are available and willing to fortify your methods at no charge.
\end{itemize}

A relevant case to consider is the Dieselgate scandal. This matter is of high importance as it pertains directly to public health. Were testing protocols for this consistently the same? Did it never occur to testers to sporadically test vehicles on roads? Was their incentive structure built around identifying dishonest auto manufacturers? Did organizations like the UCI genuinely have an interest in identifying deceitful racers, even if it meant another public scandal?

\vspace{1em}

These incentive and testing protocol inquiries are equally relevant for financial supervisors. Additionally, testing protocols should be objective and not constructed based on a desired result.

\vspace{1em}

One certainty is that when using train services, the companies appear motivated to identify and penalize fare evaders. They maintain secrecy about their checking methods, including the frequency, methodology, and locations of inspections.

\subsection{Confusion matrices and metrics}

We will first train our model on the Train data set, then use the model to predict the default $Y$ on the Test data set. We want that our model predict correctly default and non-default: $Y_{pred}=Y$, we define true positive, true negative, false positive and false negative as follow:

\quad

\subsubsection{Confusion matrix}

\begin{tabular}{lll}
&Predicted 0&Predicted 1\\
Actual 0&TN&FP\\
Actual 1 &FN&TP\\
\end{tabular}

From the confusion matrix, we can define several metrics, here we focus on three:

\subsubsection{False Positive Rate}
False Positive Rate: $FRP=\frac{FP}{Total}$
\subsubsection{Classification Accuracy:}
Classification Accuracy: Overall, how often is the classifier correct? $ CA = \frac{TN + TP}{Total}$
\subsubsection{Sensitivity}
Sensitivity: When the actual value is positive, how often is the prediction correct? $S = \frac{TP}{FN + TP}$

\subsection{AUC, ROC, and F1 score}

With our model, we are attributing two probabilities, one for each state : default ($0$) and non-default ($1$).

For illustration, if with our model we have Proba($0$) > Proba($1$) but $Y=1$, then our model doesn't predict a default while in fact a default was observed, it is a false negative.

The Receiver Operating Characteristic (ROC) curve \citep{hanley1982meaning,fawcett2006introduction} is a graphical representation used to evaluate the performance of a binary classifier system as its discrimination threshold is varied. Specifically, the ROC curve plots the True Positive Rate (sensitivity) against the False Positive Rate (1-specificity) for different threshold values. It serves as a tool to illustrate the trade-offs between sensitivity and specificity for a given classifier.


We also define:
\begin{itemize}
\item Precision $P=\frac{TP}{TP+FP}$
\item Recall $R=\frac{TP}{TP+FN}$
\item F1 score $2 \times \frac{P R}{P +  R}$
\end{itemize}

\subsection{Choosing the right estimator}

The scikit learn community provides a comprehensive guide to machine learning techniques. The first question one would have when faced with a problem is the way to choose the estimator. \href{http://scikit-learn.org/stable/tutorial/machine_learning_map/index.html}{They provide a flow chart on their website}.


Scikit-learn also provide a \href{http://scikit-learn.org/stable/user_guide.html}{user guide that is quite exhaustive}.

We will mainly work on the data set introduced section \ref{sec:defaultdata} and as in the paper  \citep{Yeh2009aa}, we want to test the following data mining techniques:
\begin{itemize}
\item K-nearest neighbor classifiers (KNN)
\item Logistic regression (LR)
\item Discriminant analysis (DA) - we use linear
\item Na\"ive Bayesian classifier (NB) - we use Gaussian
\item Artificial neural networks (ANNs) - we use Multi-layer Perceptron classifier
\item Classification trees (CTs) - we use Decision Tree Classifier
\end{itemize}

We add to that list:
\begin{itemize}
\item Gradient Boosting Classifier
\item Random Forest Classifier
\end{itemize}

\subsection{Naive Bayes Classifiers}\label{subsec:naivebayes}

We\footnote{ML\_wage\_naive\_bayes.py} use the data set from \emph{Schooling, Experience and Earnings} of Mincer 1974 work \citep{mincer1974schooling}. We find a \href{http://www.economicswebinstitute.org/data/wagesmicrodata.xls}{sample with 534 observations}, the one of the wage equation of section \ref{sec:multreg}, and keep the sex, the hourly wage (in logarithm), the age and the years of education:

\quad

\begin{tabular}{lrrrr}
&  SEX &      WAGE &   AGE &  EDUCATION \\
0 &  0.0 &  1.629241 &  35.0 &        8.0 \\
1 &  0.0 &  1.599388 &  57.0 &        9.0 \\
2 &  1.0 &  1.897620 &  19.0 &       12.0 \\
3 &  1.0 &  1.386294 &  22.0 &       12.0 \\
4 &  1.0 &  2.014903 &  35.0 &       12.0 \\
5 &  1.0 &  2.570320 &  28.0 &       13.0 \\
6 &  ? &  1.492904 &  43.0 &       10.0 \\
\end{tabular}

\quad

The mathematical tools of this subsection (conditional probability, law of total probability, Bayes' theorem, conditional independence, the Gaussian density) are collected in the companion primer \citep{vansteenberghe2026mathematics}.

\subsubsection{Determining the Sex of Observation 6}

To determine the sex of observation 6, let's denote the sex as \(S\), where \(S=1\) stands for male and \(S=0\) for female. The variables \(W\), \(A\), and \(E\) represent Wage, Age, and Education, respectively. Our primary objective is to determine if observation 6 is male. Formally, we classify the observation as male if
\[ \mathcal{P}\left(S=1 \mid (W,A,E)\right) > \mathcal{P}\left(S=0 \mid (W,A,E)\right) , \]
that is, we choose the class with the largest probability given what we observe (the \emph{maximum a posteriori} rule). The difficulty is that the data tell us the reverse conditioning: within each sex, we can observe how wages, ages and education levels are distributed. Bayes' theorem is the tool that reverses a conditioning.

\subsubsection{Using Bayes' Theorem}

Bayes' theorem states:

\begin{equation}\label{eq:usingBayesT}
\mathcal{P}\left(S\mid (W,A,E)\right) = \frac{\mathcal{P}\left(S \right) \times \mathcal{P}\left( (W,A,E) \mid S\right) }{\mathcal{P}\left(W,A,E\right) }
\end{equation}

Here:
\begin{itemize}
    \item \( \mathcal{P}\left(S\right) \) is the \emph{prior} probability of the sex being male or female, before looking at the wage, the age and the education.
    \item \( \mathcal{P}\left( (W,A,E) \mid S\right) \) is the \emph{likelihood} of observing the given wage, age, and education given the sex.
    \item \( \mathcal{P}\left(W,A,E\right) \) is the \emph{evidence}, the overall probability of observing the given wage, age, and education; by the law of total probability it equals $\mathcal{P}(S=1)\mathcal{P}((W,A,E)\mid S=1)+\mathcal{P}(S=0)\mathcal{P}((W,A,E)\mid S=0)$.
    \item \( \mathcal{P}\left(S\mid (W,A,E)\right) \) is the \emph{posterior} probability, the one we want.
\end{itemize}
In words, posterior $\propto$ prior $\times$ likelihood. From our sample of 289 men and 245 women, we know that:
\[ \mathcal{P}\left(S=1 \right) = \frac{289}{534} = 54\%, \qquad \mathcal{P}\left(S=0 \right) = 46\% . \]

\textbf{Nota bene:} In equation \ref{eq:usingBayesT}, we don't need to compute the denominator. It serves as a normalizing constant to ensure that the probabilities of being male or female sum to one, given that we consider only binary categories for individuals. To decide, it is enough to compare the two numerators, or to look at their ratio,
\begin{equation}\label{eq:posteriorodds}
\frac{\mathcal{P}\left(S=1\mid (W,A,E)\right)}{\mathcal{P}\left(S=0\mid (W,A,E)\right)} = \underbrace{\frac{\mathcal{P}\left(S=1\right)}{\mathcal{P}\left(S=0\right)}}_{\text{prior odds}} \times \underbrace{\frac{\mathcal{P}\left( (W,A,E) \mid S=1\right)}{\mathcal{P}\left( (W,A,E) \mid S=0\right)}}_{\text{likelihood ratio}} ,
\end{equation}
and to classify the observation as male when this ratio exceeds one.

\subsubsection{Dealing with the Curse of Dimensionality}

The likelihood is a joint distribution of three variables given the sex. If we discretize Wage, Age, and Education into \(x\) segments each, we then create a 3-dimensional array of \(x^3\) cells to fill: with \(x=10\), one thousand cells, from 289 men and 245 women. This expansion of the number of cells with the number of dimensions is often termed the 'curse of dimensionality'.

The conditional probability of observing \(W\), \(A\), and \(E\) given the sex can always be factorised, without any assumption (chain rule of conditional probabilities):
\[ \mathcal{P}\left( (W,A,E) \mid S\right) = \mathcal{P}\left( W\mid S\right) \times \mathcal{P}\left( A \mid (S,W)\right) \times \mathcal{P}\left( E \mid (S,W,A)\right) \]

However, to simplify calculations and reduce the complexity, we can make the naive assumption that wage, age, and education are conditionally independent given the sex. Therefore:
\begin{equation}\label{eq:naiveassumption}
\mathcal{P}\left( (W,A,E) \mid S\right) = \mathcal{P}\left( W\mid S\right) \times \mathcal{P}\left( A \mid S\right) \times \mathcal{P}\left( E \mid S\right)
\end{equation}

With this assumption, we are left to estimate only \(3 \times x\) values: three one-dimensional distributions per sex instead of one three-dimensional distribution. The assumption is naive because, within each sex, the wage and the education are positively correlated (a correlation of $0.47$ among women and $0.34$ among men in our sample) and the age and the education negatively; the classifier ignores these correlations, and the exercise at the end of this subsection shows how to take them into account.

\subsubsection{Naive Bayes Classifier - estimation}
We make the assumptions that Wage, Age, and Education are normally distributed within each sex, once we have the mean and the standard deviation, then the probability density can be simply derived using:
\begin{equation}\label{eq:gaussiandensityNB}
f\left( x\vert (\mu , \sigma )\right) = \frac{1}{\sqrt{2 \pi \sigma^2}} e^{-\frac{(x-\mu)^2}{2 \sigma^2}}
\end{equation}
Because the variables are continuous, the probabilities of equation \ref{eq:naiveassumption} are replaced by densities, and Bayes' theorem holds with densities in place of probabilities (see the primer). The whole classifier is therefore summarised by twelve numbers, a mean and a standard deviation per variable and per sex, estimated on the sample:

\begin{center}
\begin{tabular}{lrrrrrr}
\toprule
& \multicolumn{2}{c}{$W$ (log wage)} & \multicolumn{2}{c}{$A$ (age)} & \multicolumn{2}{c}{$E$ (education)} \\
& $\hat\mu$ & $\hat\sigma$ & $\hat\mu$ & $\hat\sigma$ & $\hat\mu$ & $\hat\sigma$ \\
\midrule
women ($S=0$) & 1.934 & 0.492 & 37.84 & 12.00 & 13.02 & 2.43 \\
men ($S=1$) & 2.165 & 0.534 & 35.98 & 11.44 & 13.01 & 2.77 \\
\bottomrule
\end{tabular}
\end{center}

The figure below shows the two estimated densities of the log wage against the histograms by sex: men earn more on average, but the two distributions overlap widely, which is why the wage alone, or the three variables together, cannot separate the sexes well.

\begin{center}
\includegraphics[width=0.6\textwidth]{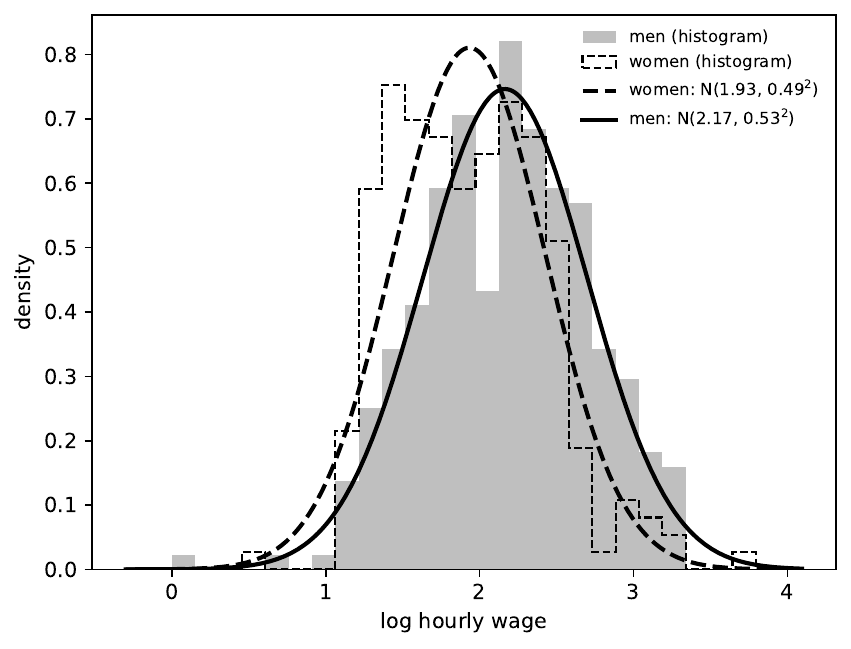}
\end{center}

For observation 6, $(W,A,E)=(1.493, 43, 10)$, an hourly wage of $4.45$ dollars. Plugging each value into the Gaussian density of its sex gives
\begin{align*}
S=0:\quad & f(W\mid S=0)\,f(A\mid S=0)\,f(E\mid S=0) = 0.542 \times 0.0303 \times 0.0757 = 1.24\times 10^{-3}, \\
S=1:\quad & f(W\mid S=1)\,f(A\mid S=1)\,f(E\mid S=1) = 0.338 \times 0.0289 \times 0.0797 = 7.79\times 10^{-4},
\end{align*}
so that the posterior odds of equation \ref{eq:posteriorodds} are $\frac{0.541}{0.459}\times\frac{7.79\times10^{-4}}{1.24\times10^{-3}} = 1.18 \times 0.63 = 0.74 < 1$: the posterior probability of being male is $\mathcal{P}(S=1\mid (W,A,E)) = 0.74/(1+0.74)=42\%$, and we classify observation 6 as a woman. The prior favours men, but a low wage for a worker aged 43 is more typical of the women of the sample, and the likelihood ratio wins. It turns out that observation 6 is a man: the classifier is wrong on this observation, which is why a classifier is never judged on one case but on its confusion matrix.

Applying the same rule to every observation of the sample, we obtain the following confusion matrix (actual sex in rows, predicted sex in columns):

\begin{tabular}{lrr}
&   Predicted 0 &  Predicted  1 \\
Actual 0 &  126 &  119 \\
Actual 1 &   94 &  195 \\
\end{tabular}

With the men as the positive class, the precision is $195/(195+119)=0.62$, the recall is $195/(195+94)=0.67$ and the accuracy is $(126+195)/534=0.60$. Which yields a F1 score of $0.65$, we tend to accept this rather simple model. Two remarks. First, the trivial rule ``everybody is a man'' has an accuracy of $0.54$ (the share of men) but an F1 score of $0.70$, higher than that of our classifier: the F1 score ignores the true negatives and rewards a recall of one when the positive class is the majority, so it must always be read next to a baseline. The gain of the classifier is visible in the accuracy ($0.60$ against $0.54$) and in the balance of its errors. Second, these numbers are computed on the sample used to estimate the twelve parameters; on the test sets of section \ref{sec:MLwhy} they would be slightly lower (five-fold cross-validation gives an accuracy of $0.59$).

\subsubsection{Exercise: relaxing the naive assumption}
The naive assumption \eqref{eq:naiveassumption} replaces the joint density of $(W,A,E)$ given the sex by a product of three one-dimensional densities. The multivariate normal distribution keeps the Gaussian shape but models the joint density directly, correlations included: with $x=(W,A,E)^\top\in\mathbb{R}^3$, a mean vector $\mu_s\in\mathbb{R}^3$ and a $3\times3$ variance--covariance matrix $\Sigma_s$ for each sex $s$,
\begin{equation}\label{eq:mvnormalNB}
f\left(x\mid S=s\right) = \frac{1}{\sqrt{(2\pi)^3 \det(\Sigma_s)}} \exp\left(-\frac{1}{2}\,(x-\mu_s)^\top \Sigma_s^{-1} (x-\mu_s)\right) ,
\end{equation}
where the exponent is a quadratic form in the deviations from the class mean (see the primer for the matrix notation). If $\Sigma_s$ is diagonal, the density factorises into the product of three one-dimensional Gaussian densities and we are back to equation \ref{eq:naiveassumption}: the naive Bayes classifier is the special case of \eqref{eq:mvnormalNB} in which the within-class covariances are set to zero. Using one full covariance matrix per class is known as quadratic discriminant analysis (QDA); imposing the same matrix to both classes is linear discriminant analysis (LDA), the method of section \ref{subsec:LDA}, whose decision boundary is linear in $x$.

After the main code of ML\_wage\_naive\_bayes.py, implement the following steps (a skeleton is provided in the last cell of the script):
\begin{enumerate}
\item Estimate $\mu_s$ and $\Sigma_s$ for each sex with the pandas methods \texttt{mean()} and \texttt{cov()} applied to the observations of the class. Count the parameters: three means and six distinct covariance entries per class, against three means and three standard deviations for the naive classifier.
\item Write a function computing the density \eqref{eq:mvnormalNB} for every row of the data set, with \texttt{numpy.linalg.inv} and \texttt{numpy.linalg.det}.
\item Combine the densities with the priors as in equation \ref{eq:usingBayesT}, classify every observation with the maximum a posteriori rule and compute the confusion matrix and the F1 score with the functions of the script.
\item Compare with the naive classifier and with the class \texttt{QuadraticDiscriminantAnalysis} of \texttt{sklearn.discriminant\_analysis}; then impose a common covariance matrix (the weighted average of $\hat\Sigma_0$ and $\hat\Sigma_1$) and compare with \texttt{LinearDiscriminantAnalysis}.
\item A model with more parameters always fits the estimation sample at least as well: check whether the gain survives a train/test split or a cross-validation (section \ref{sec:MLwhy}).
\end{enumerate}
You should find that the posterior probability of observation 6 being a man falls to $36\%$ and that the number of men classified as women falls from $94$ to $64$.

\subsection{K-nearest neighbors}
KNN is a "lazy" learner, meaning that it doesn't learn a discriminative function but rather memorize the training data set.

The KNN algorithm follows the following logist and steps:
\begin{enumerate}
\item Choose $k$ and a distance metric
\item Find the $k$ nearest neighbors of the point to classify
\item Assign by majority vote
\end{enumerate} 

\begin{center}
\includegraphics[scale=0.5]{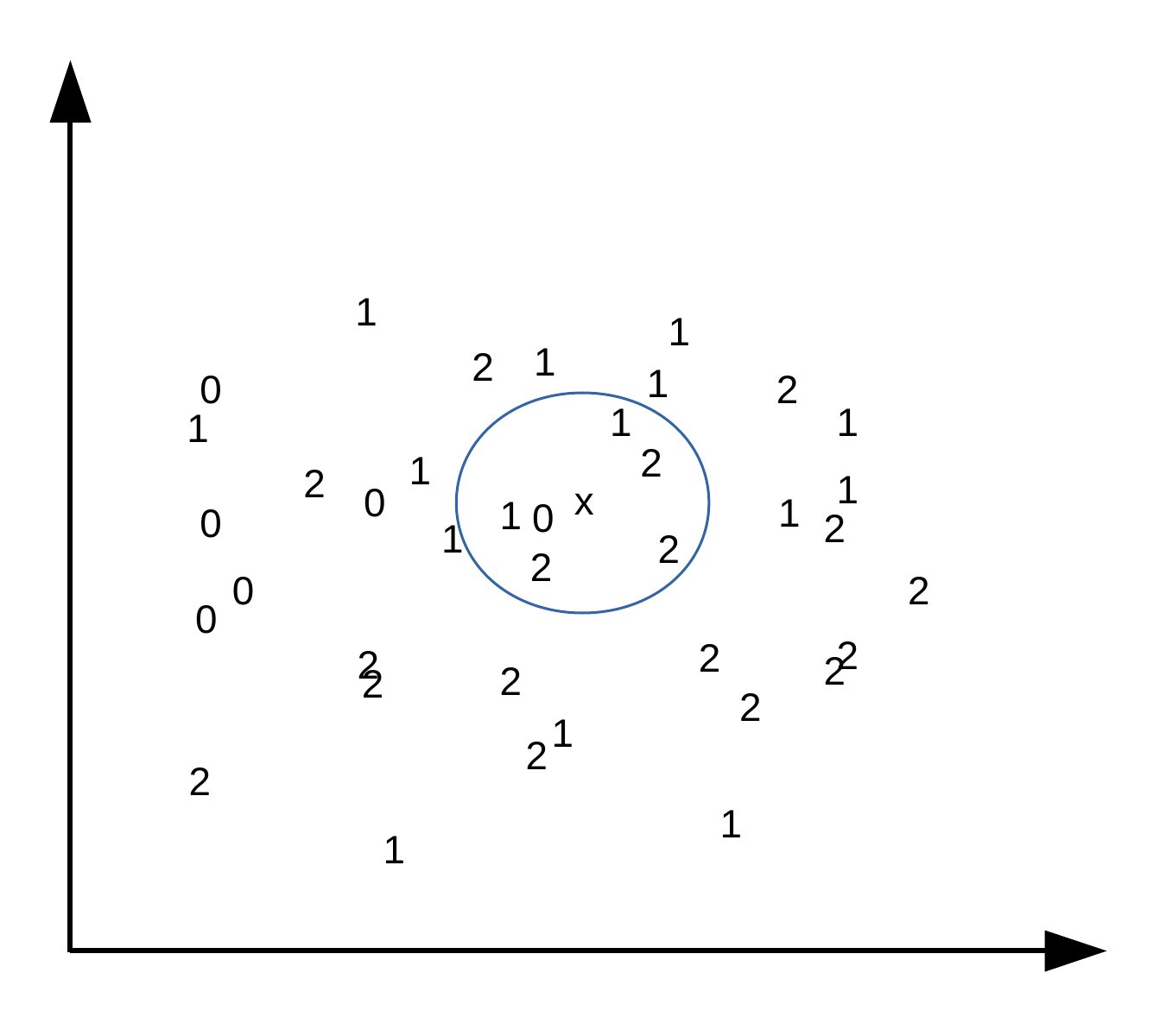}
\end{center}

In this illustrative example, $k=6$ and we use the Euclidean distance (assuming a two dimension problem). The value of $x$ will be $2$, by majority.

\subsubsection{K-nearest neighbors - illustration}
If we\footnote{ML\_wage\_KNN.py} use again Mincer's 1974 work. 

We take a random sample of $21$ elements from the data set. We change the last observation SEX value to $99$ and we want to predict its value ($0$ or $1$) using a 5-nearest neighbors method.

\begin{center}
\includegraphics[scale=0.5]{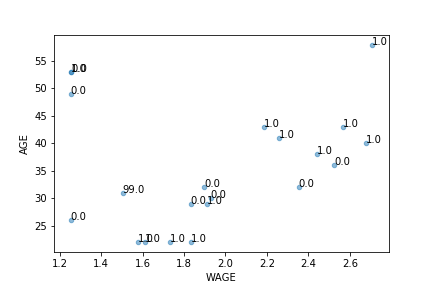}
\end{center}

We then compute the Euclidean distance between this point ($99$) and the rest of that sample. We then count the SEX of the $5$ nearest neighbors and decide on the SEX of our point based on majority vote.

In a two dimensions example the Euclidean distance is simply: $\sqrt{(x_2-x_1)^2 + (y_2-y_1)^2}$.

\subsection{Decision Trees}

If we\footnote{ML\_wage\_DTC.py} use again Mincer's 1974 work.

The obtained Decision Tree classifier is:
\begin{center}
\includegraphics[scale=0.5]{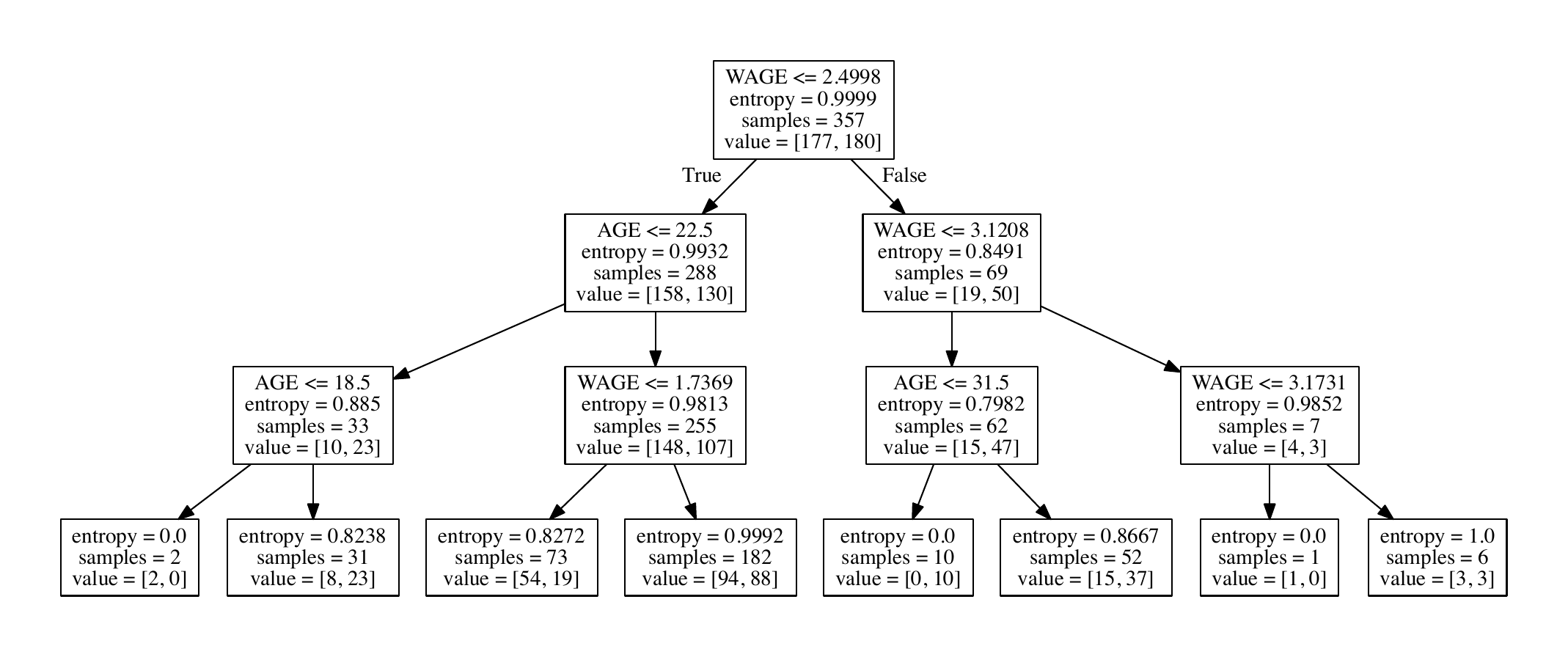}
\end{center}

In our case, we have limited the depth of our tree to $3$. This divide our 2-Dimension space as follow:
\begin{center}
\includegraphics[scale=0.4]{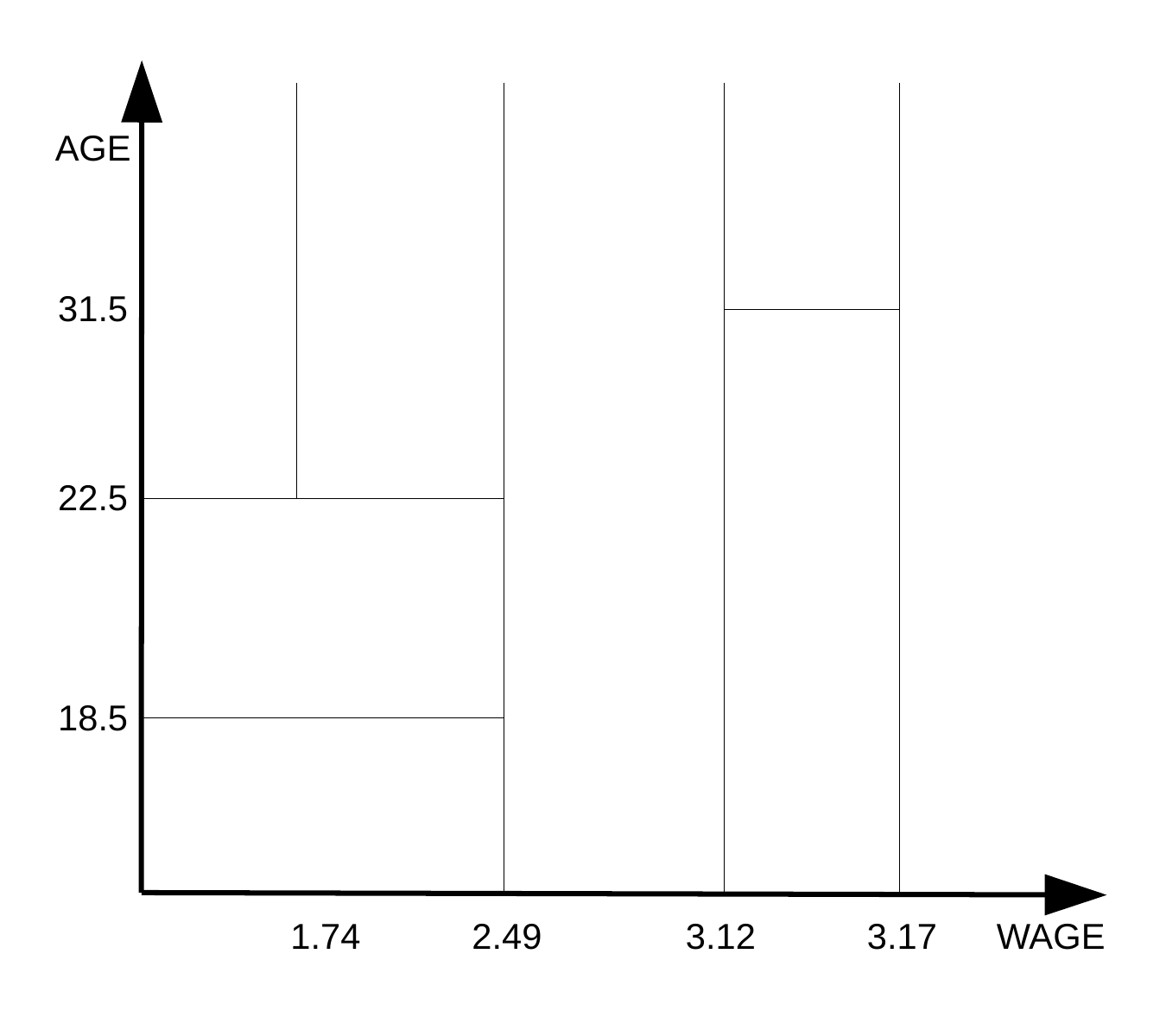}
\end{center}

In practice, we are using a binary tree, where each parent has two children. We can in this case simply define the entropy as:
$$Ent= - p(l) \log_2 p(l) - p(r) \log_2 p(r)$$

where $p(l)$ is the proportion of the parent node that belong to the left node, $p(r)$ to the right. The algorithm maximize this entropy, which in our case can be seen as information gain.

\subsubsection{Random Forest}
Random Forest is an ensemble of decision trees which algorithm (as described in \citep{raschka}:
\begin{enumerate}
\item Draw a random bootstrap sample of size $n$ (randomly choose $n$ samples from the training set with replacement).
\item Grow a decision tree from the bootstrap sample. At each node:
\begin{enumerate}
\item Randomly select $d$ features without replacement.
\item Split the node using the feature that provides the best split according to the objective function, for instance, by maximizing the information gain.
\end{enumerate}
\item Repeat the steps 1 to 2 $k$ times.
\item Aggregate the prediction by each tree to assign the class label by majority vote.
\end{enumerate}

\subsection{Before neural network, the "traditional" Ordinary Least Squares for linear regressions}

If we take the usual linear regression with no constant, we are searching for the weights $W = {w_i^1}, i \in [0,2]$. The variables $X^1 = {X_i^1}, i \in [0,2]$ are called features. We have observations of $X_0^2$ and we want to estimate them from the features $X_i^1, i \in [0,2]$, that is finding $\hat{X}_0^2 = W'X^1$. 

\begin{center}
\includegraphics[scale=0.5]{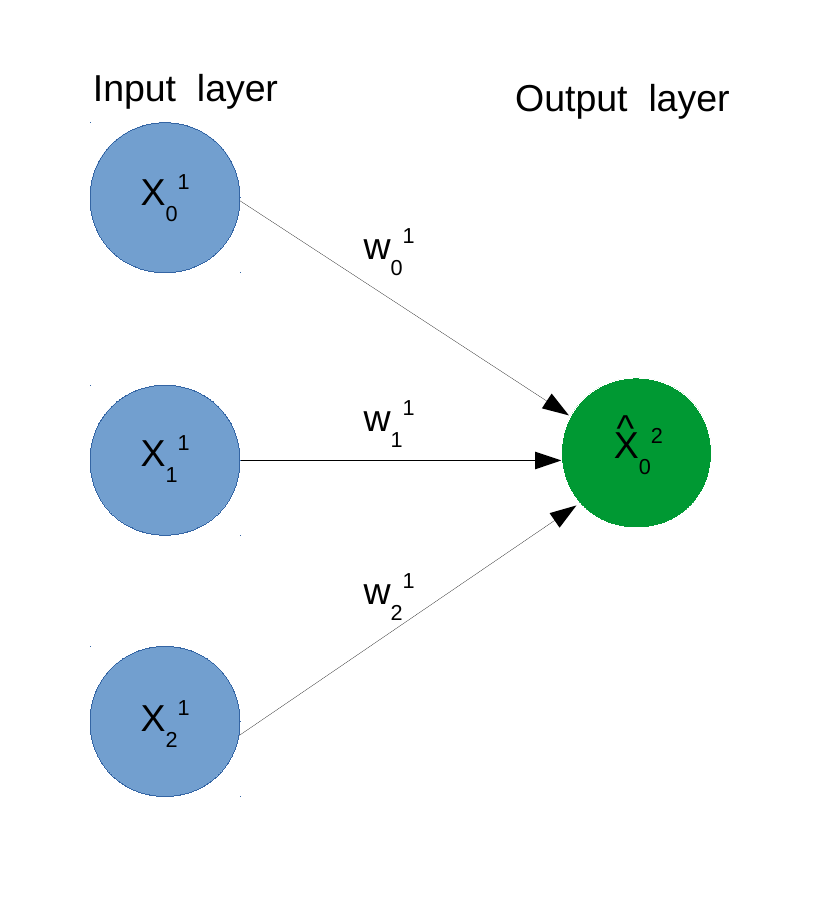}
\end{center}

To estimate the weights, that is finding $\hat{w}_i^1, i \in [0,2]$, the Ordinary Least Squares method. Our programme is:

\begin{equation}
\min_{w_i^1, i \in [0,2]} \left( \hat{X}_0^2 - X_0^2   \right)^2
\end{equation}

With first order conditions, we can find the estimate of $w_i^1, i \in [0,2]$ and we could prove that it is linear, unbiased, and efficient. For simplicity, if there were only one feature, the result would be:
$$\hat{w}_0^1= \frac{(X_0^1)' X_0^2}{(X_0^1)'X_0
^1}$$
which is echoing equation \ref{eq:betahat}, we could also add a constant in our model.

\subsubsection{Adding an activation function - perceptron algorithm}

In his work in \citep{perceptron}, Frank Rosenblatt proposed in 1957 a perceptron algorithm where for a binary classification task, an activation function $\sigma ( )$ is added in series.

\begin{center}
\includegraphics[scale=0.5]{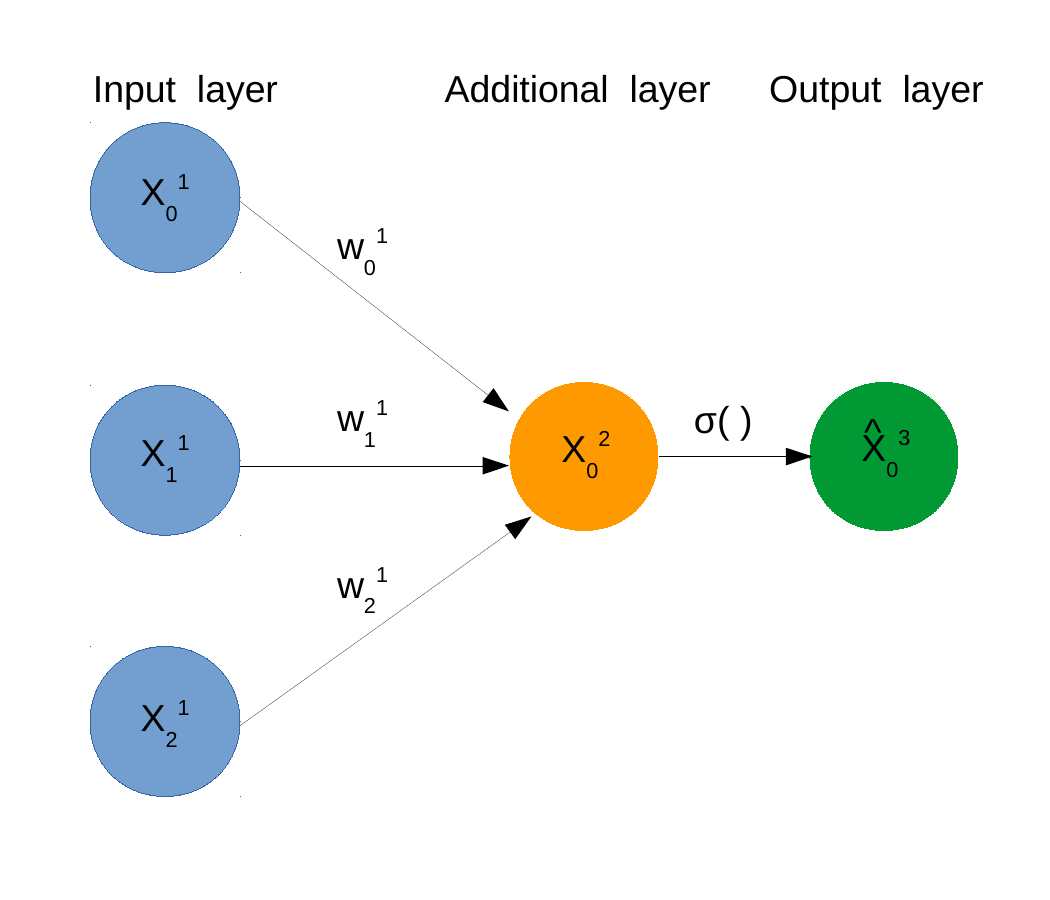}
\end{center}

Frank Rosenblatt defined a threshold $\theta$ for which:
\begin{equation}
    \sigma (W'X)=
    \begin{cases}
1,& \text{if} \quad W'X \geq \theta \\        -1, & \text{otherwise}
    \end{cases}
\end{equation}

This linear classification might not be satisfactory for scattered data sets and we might need to either use more complex activation function or add some layers. In which case we move to Artificial Neural Network.

\quad

We\footnote{perceptron\_example.py} propose a simple illustration of the perceptron algorithm. We generated two features, $X_1$ and $X_2$ and we define the output $Y$ as:

\begin{equation}
    Y=
    \begin{cases}
1,& \text{if} \quad \alpha_1 X_1 + \alpha_2 X_2 - \theta > 0 \\        -1, & \text{otherwise}
    \end{cases}
\end{equation}

That is indeed splitting the "observation" as follow:
\begin{center}
\includegraphics[scale=0.5]{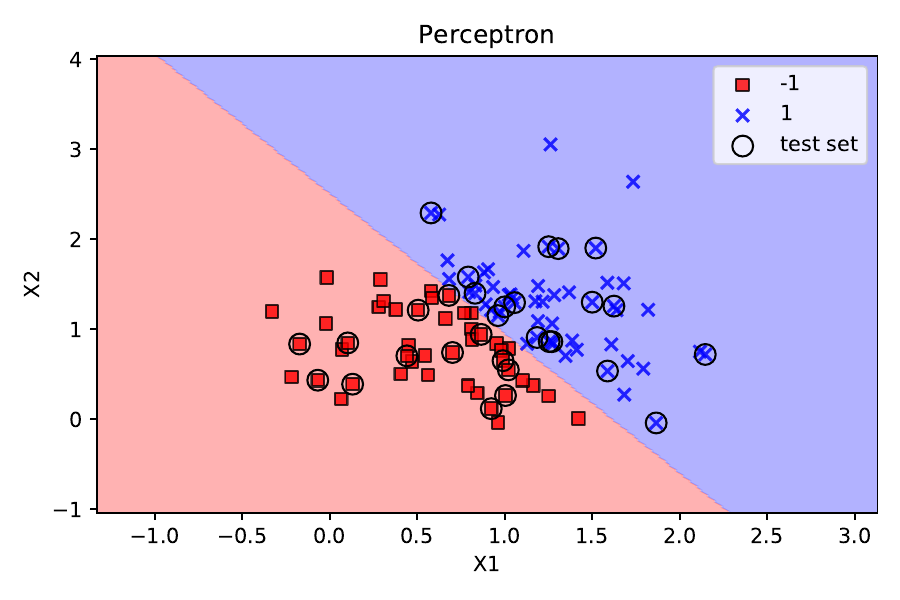}
\end{center}

With our perceptron rule, we want to find the estimates $\hat{\alpha}_1$, $\hat{\alpha}_2$, and $\hat{\theta}$ (pretending we did not know their values).

\quad
We follow the steps:
\begin{enumerate}
\item we initialize with vectors of $0$s or random small values of the estimates
\item we loop until $\hat{Y} - Y < \epsilon$ or until we reach the maximum number of allowed iteration (our algorithm might never converge):
\begin{itemize}
\item for each observation $i$ in our data set:
\end{itemize}
\begin{itemize}
\item compute the output value $\hat{Y}$
\item compute the estimation error for observation $i$
\item update the weights according to the learning rate $\eta$
$$
\begin{cases}
 w_j = w_j + \eta (\hat{Y}(i) - Y(i)) X_j, \quad j={1,2}\\
 \theta = \theta + \eta (\hat{Y}(i) - Y(i))
\end{cases}
$$
\end{itemize}
\end{enumerate}

We propose as an illustrative exercise to implement this algorithm with: $\epsilon = 10^{-4}$ and $\eta = 10^{-2}$ and limit our iteration to $10^4$. In \citep{raschka} you can find this algorithm in a more pythonic way, defining classes which can be found \href{https://github.com/rasbt/python-machine-learning-book-3rd-edition/blob/master/ch02/ch02.ipynb}{here} and illustrated:
\begin{center}
\includegraphics[scale=0.5]{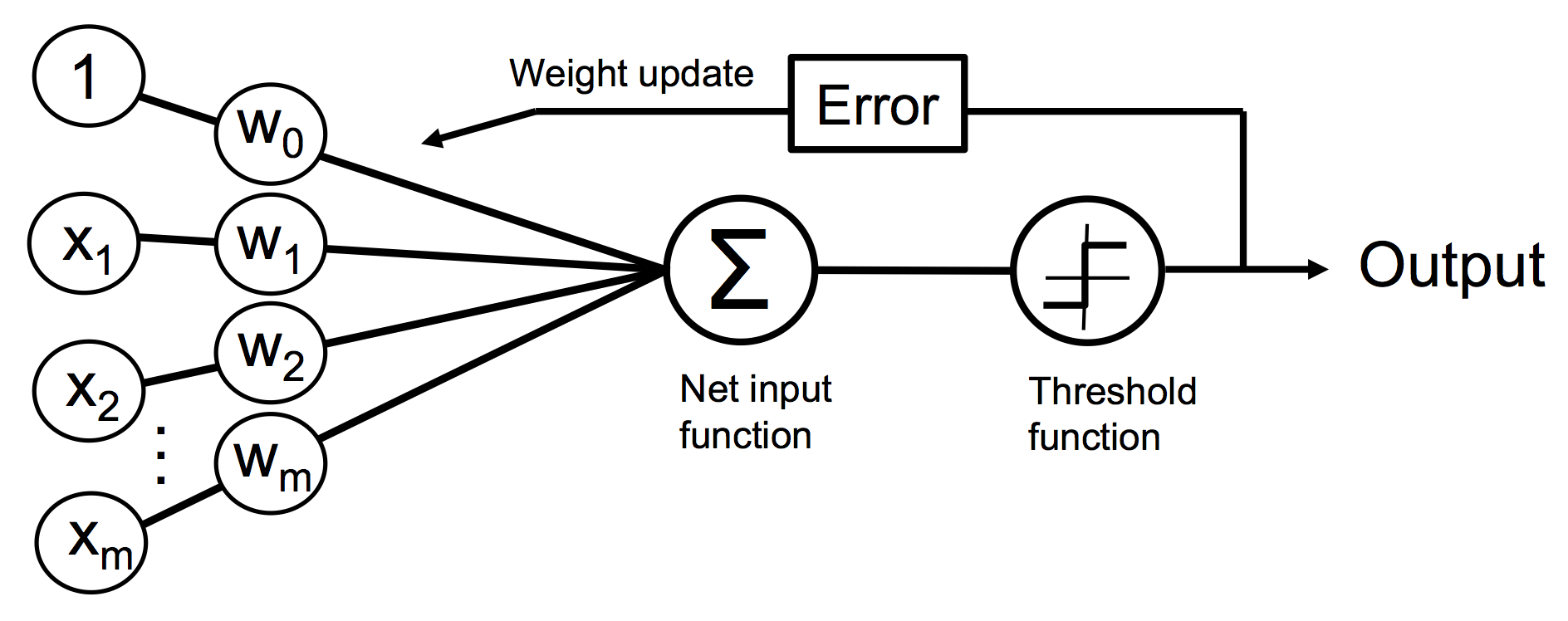}
\end{center}

\quad 

This separates the data in two categories as follow:
\begin{center}
\includegraphics[scale=0.5]{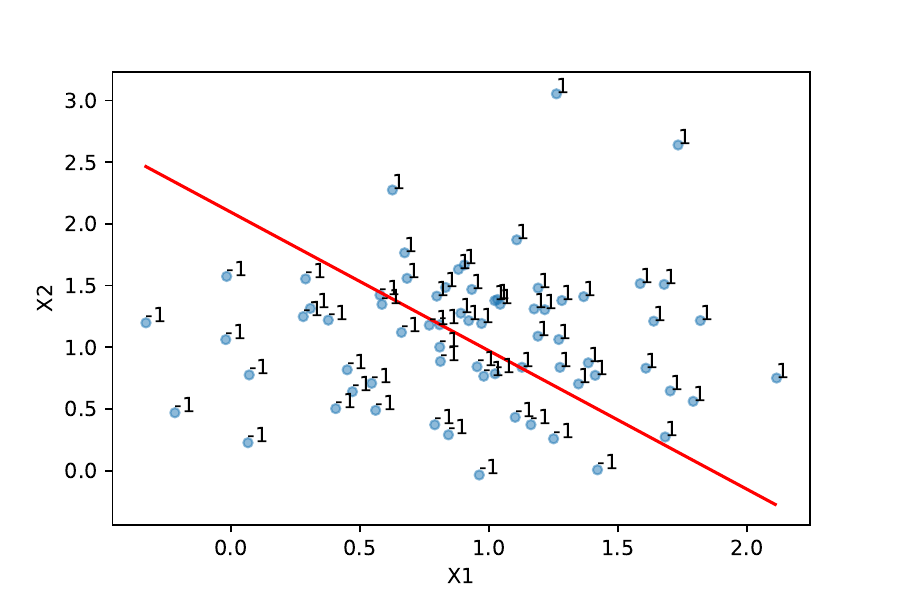}
\end{center}

We seperate our data set into a train and test one and compare our prediction accuracy with the scikit-learn Perceptron algorithm one.

\stepcounter{question}
\begin{itemize}
\item \textbf{Question \arabic{question}}
\begin{itemize}
\item Applyl a non-linear rule for the categorization, try to find a better algorithm to fit a non linear rule.
\end{itemize}
\end{itemize}

\subsubsection{Adaline algorithm}

With the perceptron algorithm, we updated observation by observation for each iteration. The threshold activation function we used is not differentiable. The ADAptive LInear NEuron algorithm  add a cost function:
\begin{equation}
J(w) = \frac{1}{2}\sum_i \left( \hat{Y}(i) - Y(i)  \right)^2 
\end{equation}

as this cost function is convex, we can use the gradient descent algorithm introduced section \ref{sec:gradientdescent}, we derive the cost function with respect to each weight, we find an update:
\begin{equation}
w = w + \eta X' \left( \hat{Y}(i) - Y(i)  \right)
\end{equation}

You should find this algorithm\footnote{code: adaline\_example.py} much faster than the simple perceptron.
\begin{center}
\includegraphics[scale=0.5]{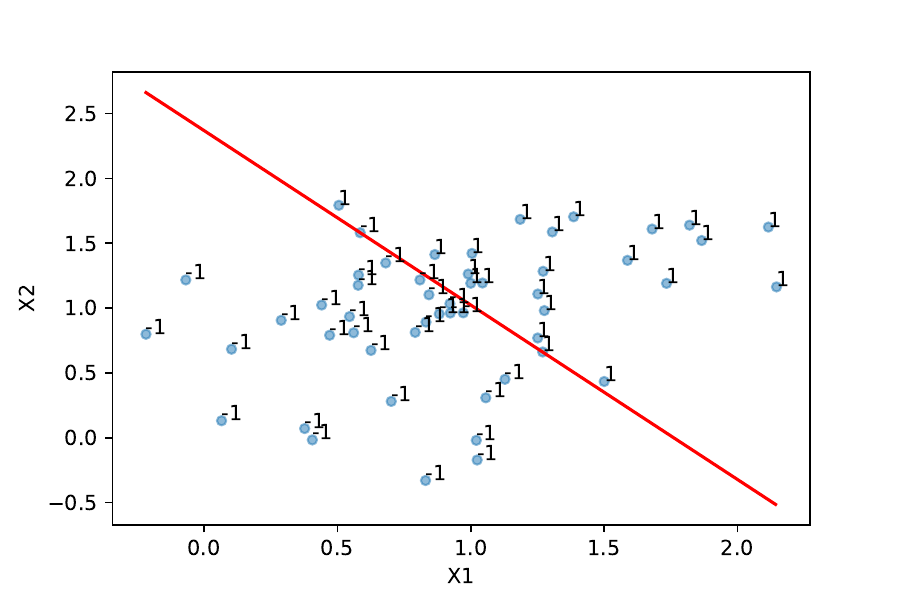}
\end{center}

\subsection{Artificial Neural Network characteristics}

\begin{itemize}
\item Layer: set of unconnected similar neurons
\item Neuron: processing unit
\item Parameters: edges weights ($\alpha_i$ and $\beta_j$)
\item Input layer: $X$
\item Hidden layer: $Z_m = \sigma(\alpha_{0m} + \alpha_m^T X)$, with $\sigma()$ an activation function
\item Output layer: $T_k=\beta_{0k}+\beta_k^T Z$ and $Y_k = g_k(T) = f_k(X)$, with $g_k()$ an output function
\end{itemize}

\subsubsection{Network diagram}

\begin{center}
\includegraphics[scale=0.5]{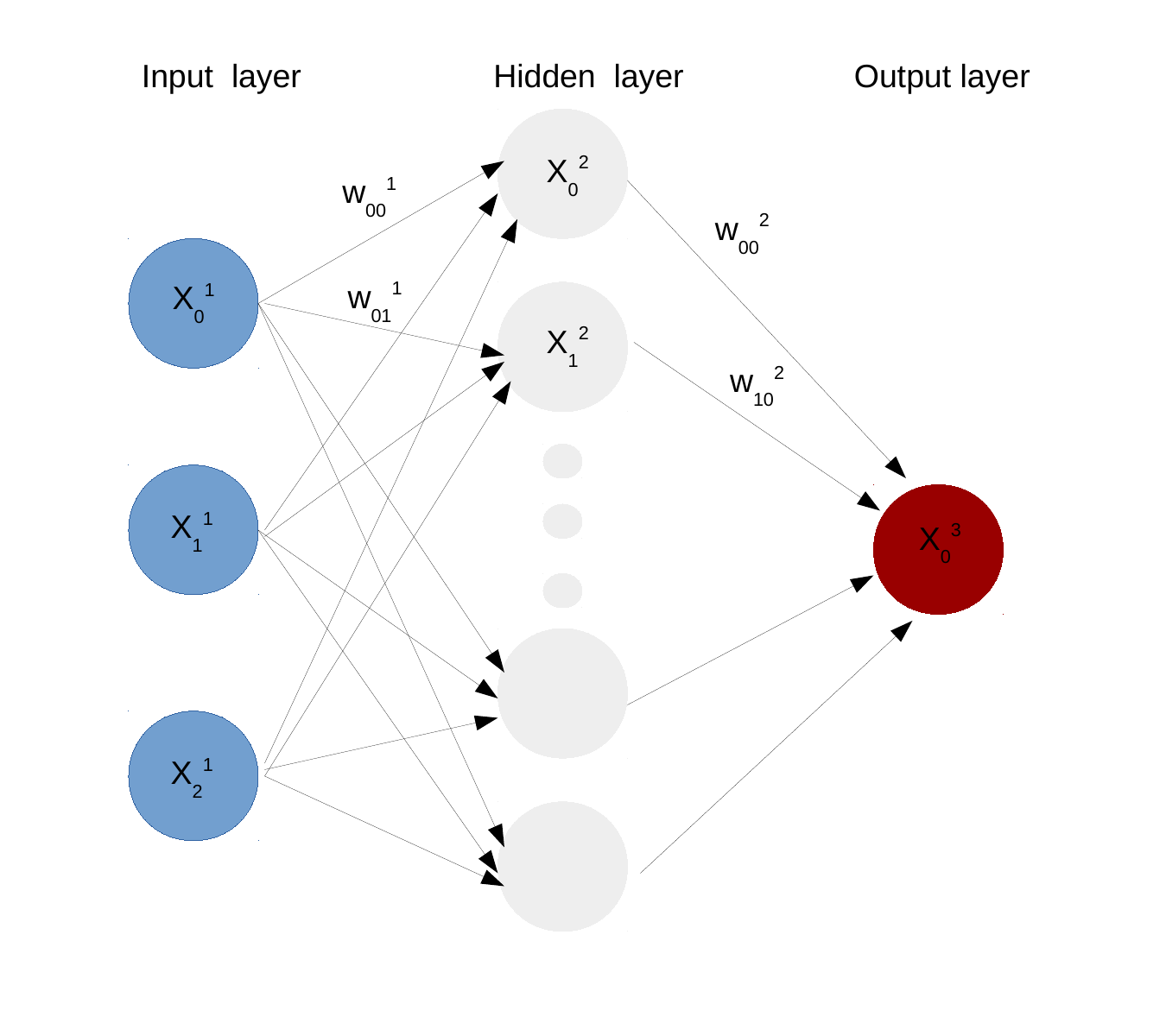}
\end{center}

In a simple example, we could have $\sigma(x) = \frac{1}{1 +  e^{-x}}$ as we use in a logistic regression for the activation function. We would then have:
$$X_0^2 = \sigma \left( w_{00}^1 X_0^1 + w_{10}^1 X_1^1  + w_{20}^1 X_2^1    \right) $$

The same philosophy goes for the next layer:
 
 $$X_0^3 = Y_k \left( w_{00}^2 X_0^2 + w_{10}^2 X_1^2  + \ldots  \right) $$

\subsubsection{Deep learning}

We call it \emph{deep} when there are multiple hidden layers, in order to estimate those models, we use \emph{deep learning} techniques:
\begin{center}
\includegraphics[scale=0.5]{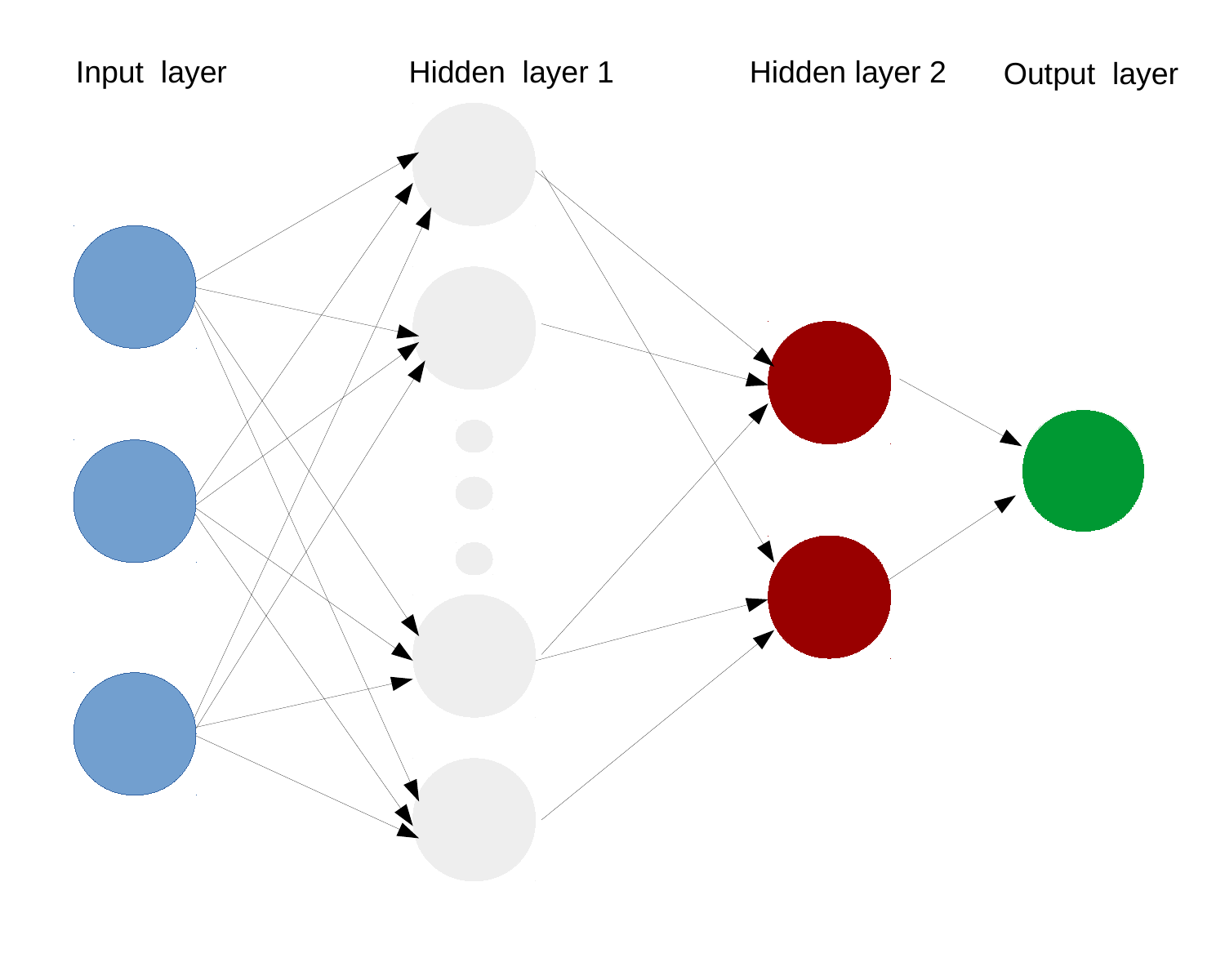}
\end{center}

\subsection{Support Vector Machine}

A support vector machine (SVM) constructs a hyper-plane or set of hyper-planes in a high or infinite dimensional space, which can be used for classification, regression or other tasks. Intuitively, a good separation is achieved by the hyper-plane that has the largest distance to the nearest training data points of any class (so-called functional margin), since in general the larger the margin the lower the generalization error of the classifier. We suggest to follow the \href{https://scikit-learn.org/stable/modules/svm.html}{scikit-learn section on SVM to go further}.

We illustrate\footnote{code SVM\_illustration.py} an implementation of a SVM on the same data set as what we used for the perceptron example and the logistic regression using \citep{raschka} typology:

\begin{center}
\includegraphics[scale=0.5]{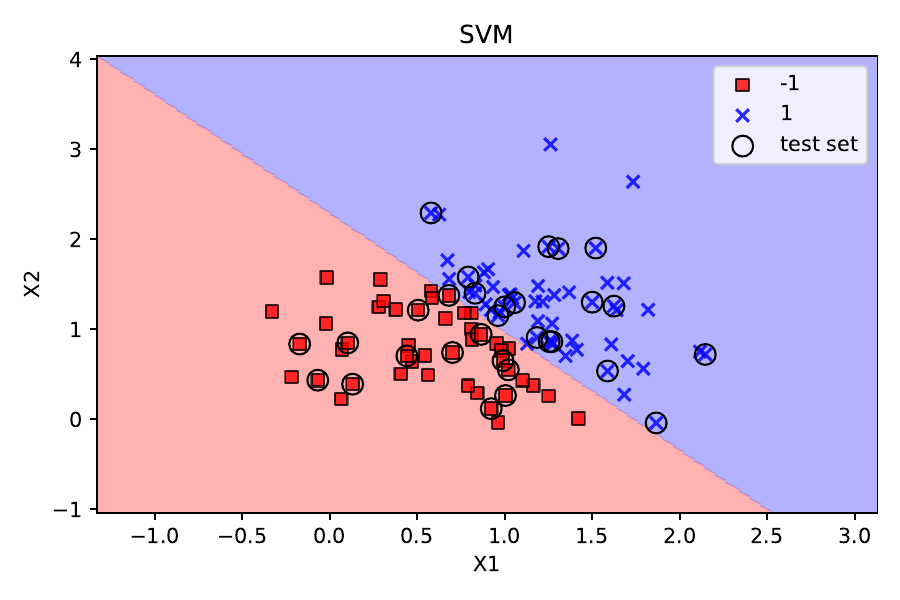}
\includegraphics[scale=0.5]{fig/perceptronillustration.pdf}
\includegraphics[scale=0.5]{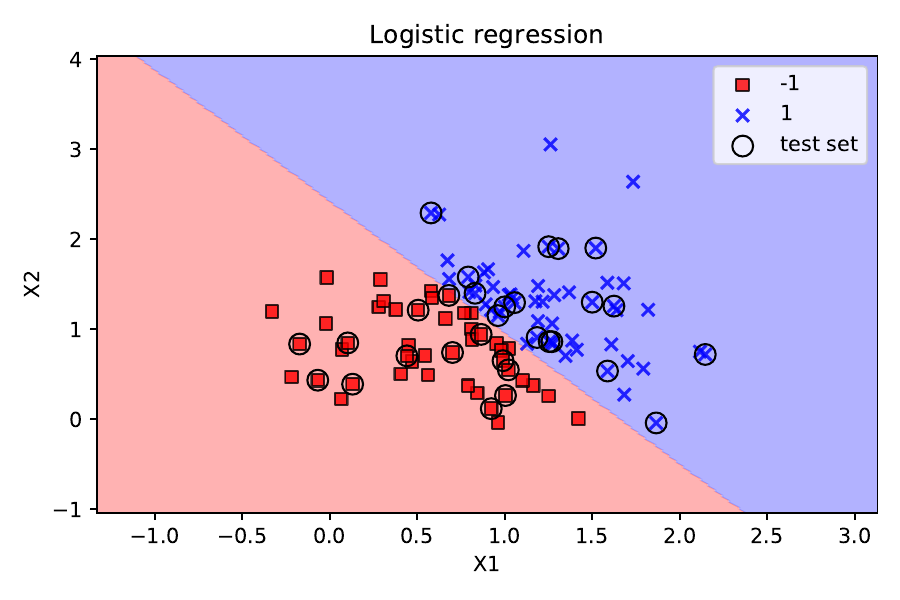}
\includegraphics[scale=0.5]{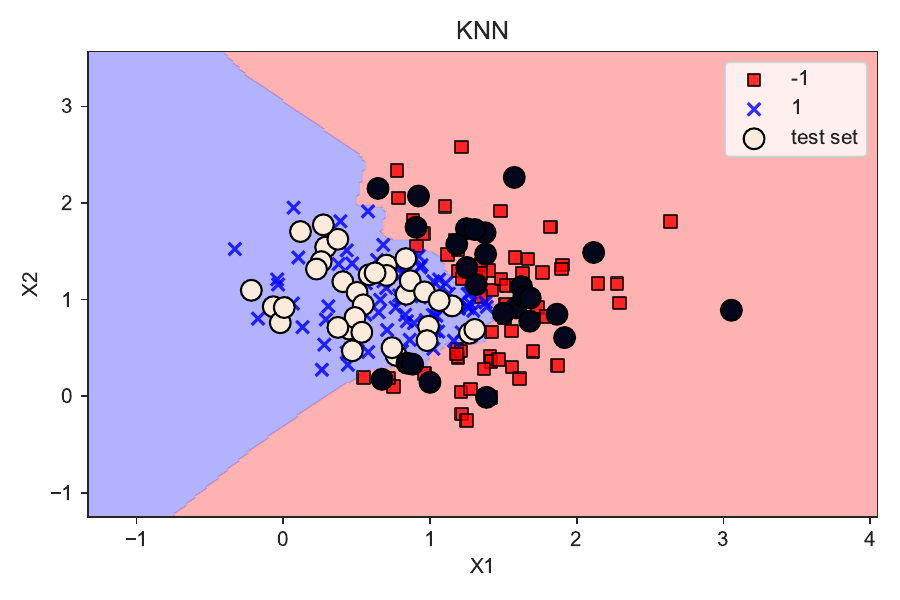}
\end{center}

\subsubsection{SVM kernel for non-linearly separable data}

We generate non-linearly separable data:
$$ \alpha_1\cos(X_1-.1) + \alpha_2 \sin\left[(1+X_1)X_2 \right] X_2 $$ 
We then use the radial basis function (rbf) kernel that transform the data set before apply the SVM:
\begin{center}
\includegraphics[scale=0.5]{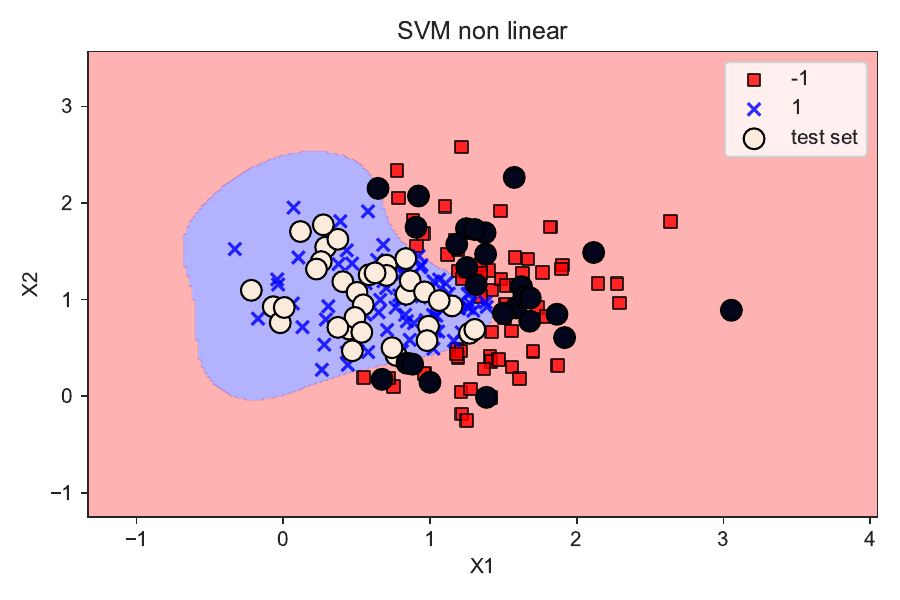}
\includegraphics[scale=0.5]{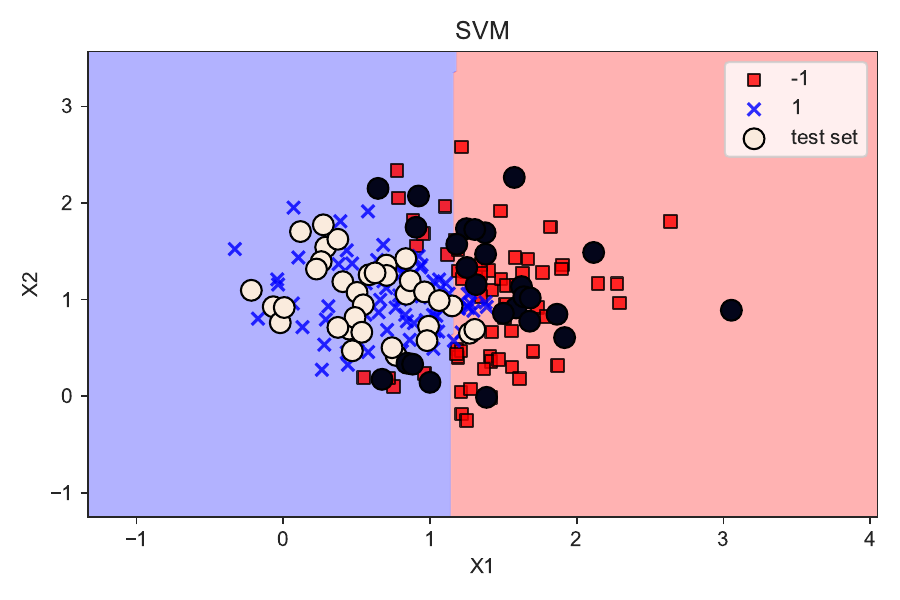}
\includegraphics[scale=0.5]{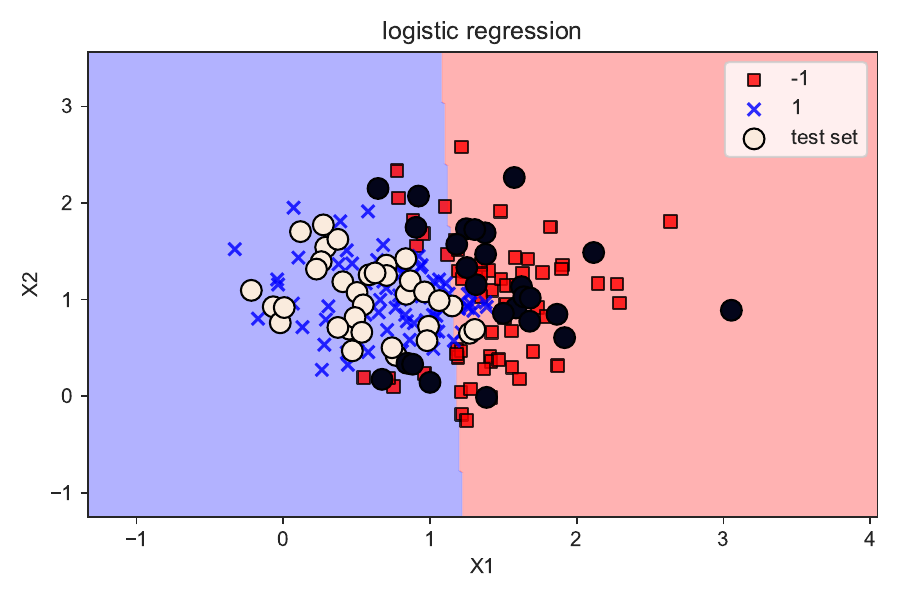}
\includegraphics[scale=0.5]{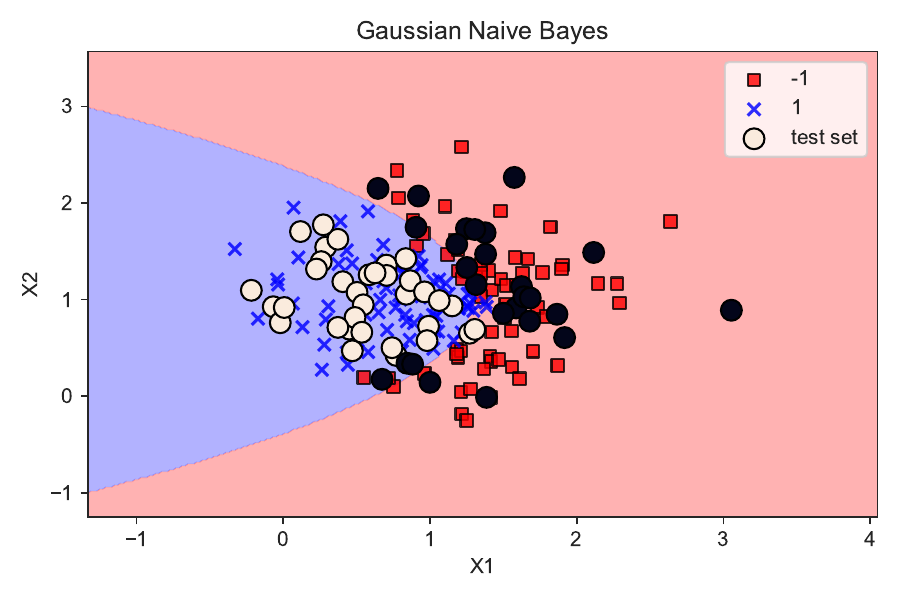}
\end{center}

We find the following accuracy scores:

\begin{tabular}{ll}
logistic regression & $85\%$\\
linear SVM&$85\%$\\
Gaussian Naive Bayes&$91.6\%$\\
Decision tree&$95\%$\\
KNN&$96.6\%$ \\
SVM with a rbf kernel&$100\%$\\
\end{tabular}

\subsection{Principal Component Analysis}\label{subsec:PCA}

We\footnote{vansteenberghe\_PCA\_example.py} can resort to a principal component analysis to reduce the dimensions of a data set to a minimum set of common factors. In the case where one factor would explain more than 70\% of the variance, we would keep only one factor. The linear algebra used below (matrix product and transpose, covariance matrix of a data matrix, quadratic forms, eigenvalues and eigenvectors, spectral theorem, Lagrange multipliers) is collected in the companion primer \citep{vansteenberghe2026mathematics}, whose notation we follow.

For this exercise, we focus on financial data of the biotech sector listed on Euronext. We use daily closing prices downloaded from Euronext, one file per company, for 60 biotech and pharmaceutical companies (listed in Paris for most of them, in Brussels or Amsterdam for a few). The companies were listed at different dates, so we keep the $N=28$ stocks with a complete price history over the window common to all the files, from 17 January 2013 to 6 September 2017, that is $T=1{,}174$ daily returns:
AB SCIENCE, ABLYNX, ADOCIA, BOIRON, CELLECTIS, DBV TECHNOLOGIES, DIAXONHIT, GALAPAGOS GENOMICS, GENFIT, GENOWAY, GUERBET, HYBRIGENICS, INNATE PHARMA, INTEGRAGEN, MDXH, NANOBIOTIX, NEOVACS, NICOX REGROUPEMENT, ONXEO, PCAS, PHARMING GROUP, THERADIAG, THROMBOGENICS, TIGENIX, TRANSGENE, VALNEVA SE, VETOQUINOL, VIRBAC.

\subsubsection{Method}

We standardize our returns (centre-reduit):
\begin{equation*}
R_{i,t}^{cr} = \frac{R_{i,t}-\bar{R_i}}{\sigma_i}
\end{equation*}

where, for each stock $i$, $\bar{R_i}$ is the average and $\sigma_i$ its standard deviation. Standardizing matters: without it, the most volatile stocks would dominate the analysis, since the principal components maximize a variance.

\quad

We collect the standardized returns in the data matrix $X\in\mathbb{R}^{T\times N}$, one row per date and one column per stock, and write $x_t\in\mathbb{R}^{N}$ for the column vector of the $N$ standardized returns of date $t$ (the $t$-th row of $X$, transposed). Since every column has mean zero and standard deviation one, the sample variance--covariance matrix
\begin{equation}\label{eq:PCAcov}
\hat\Sigma = \frac{1}{T-1}\,X^\top X \in \mathbb{R}^{N\times N}
\end{equation}
is the correlation matrix of the returns: it is symmetric and positive semidefinite, with ones on the diagonal, so that its trace, the total variance of the standardized returns, is $\operatorname{tr}(\hat\Sigma)=N$. In our sample the average pairwise correlation is $0.11$ (the largest is $0.32$): biotech stocks move together, but weakly.

\quad

A principal component (or factor) is a linear combination of the standardized returns, $F_{t} = w^\top x_t = \sum_{i=1}^{N} w_i R^{cr}_{i,t}$, with a vector of weights $w\in\mathbb{R}^N$, and its variance is the quadratic form $\Var(F_t) = w^\top \hat\Sigma\, w$ (variance of a linear combination). The first principal component is $F_{1,t} = w_1^\top x_t$, where $w_1$ solves the problem
\begin{equation}\label{eq:PCAproblem}
\max_{w} \; w^\top \hat\Sigma\, w \qquad \text{subject to the normalization} \qquad w^\top w = 1 .
\end{equation}
Without the normalization the variance could be made as large as we like by scaling $w$. With a Lagrange multiplier $\lambda$, the first-order condition of the Lagrangian $w^\top \hat\Sigma\, w-\lambda\,(w^\top w-1)$ is $2\hat\Sigma w - 2\lambda w = 0$, that is
\begin{equation}\label{eq:PCAeigen}
\hat\Sigma\, w = \lambda\, w :
\end{equation}
$w$ must be an eigenvector of $\hat\Sigma$ and $\lambda$ the associated eigenvalue. At such a $w$ the objective equals $w^\top \hat\Sigma\, w = \lambda\, w^\top w = \lambda$, so the maximum is reached for $w_1$, the eigenvector associated with the largest eigenvalue $\lambda_1$ of $\hat\Sigma$, and the eigenvalue $\lambda_1$ equals $\Var(F_{1,t}) = w_1^\top \hat\Sigma\, w_1$. The $j$th principal component solves the same problem, but under the additional restriction that $w_i^\top w_j =0$ for all $i< j$: it must be uncorrelated with the previous components. The solution is the eigenvector associated with the $j$th largest eigenvalue $\lambda_j$ (which equals $\Var(F_{j,t}) = w_j^\top \hat\Sigma\, w_j$).

\quad

The spectral theorem gives all the components at once. Since $\hat\Sigma$ is symmetric, it has $N$ real eigenvalues $\lambda_1\ge\lambda_2\ge\dots\ge\lambda_N\ge 0$ and an orthonormal basis of eigenvectors $w_1,\dots,w_N$; collecting them in the orthogonal matrix $W=(w_1\;\cdots\;w_N)$ and in $\Lambda=\operatorname{diag}(\lambda_1,\dots,\lambda_N)$,
\begin{equation}\label{eq:PCAspectral}
\hat\Sigma = W\Lambda W^\top, \qquad W^\top W = I_N, \qquad \sum_{j=1}^{N}\lambda_j = \operatorname{tr}(\hat\Sigma) = N .
\end{equation}
The matrix of the components, or scores, is $F = X W\in\mathbb{R}^{T\times N}$ (its $j$th column is $F_{j,t}=w_j^\top x_t$), and its covariance matrix is $F^\top F/(T-1) = W^\top\hat\Sigma W = \Lambda$: the components are uncorrelated and their variances are the eigenvalues. The share of the total variance explained by the $j$th component is $\lambda_j/N$, and the columns of $W$, the weights, are also called the loadings. Finally $X = F W^\top = \sum_{j=1}^{N} F_j w_j^\top$ writes each standardized return as a sum of contributions of the components; keeping only the first $k$ terms is the best approximation of $X$ by $k$ factors, in the sense that it explains the largest share $\sum_{j\le k}\lambda_j/N$ of the variance. The sign of an eigenvector is arbitrary ($-w_j$ solves the same problem as $w_j$), a convention has to be chosen to compare two implementations.

\quad

The code computes $\hat\Sigma$, diagonalizes it with numpy.linalg.eigh and checks every identity above numerically, then obtains the same eigenvalues, eigenvectors and scores with scikit-learn (PCA). We find $\lambda_1 = 4.20$, that is $15.0\%$ of the total variance, then $\lambda_2 = 1.26$ ($4.5\%$) and a slowly decreasing sequence of eigenvalues close to one: the first two components explain $19.5\%$ of the variance, and $17$ components would be needed to reach $70\%$ (left panel of the first figure below, the scree plot). Two criteria are commonly used to decide how many components to keep: the share of variance explained (the 70\% rule of thumb above), and the Kaiser criterion, which keeps the components whose eigenvalue exceeds one, that is the components which carry more variance than a single standardized stock; here it keeps nine. In practice the scree plot is read for an ``elbow'', and there is a clear one after the first component. The weights of the first component are all positive (between $0.09$ and $0.27$, second figure): $F_{1,t}$ is essentially the average standardized return of the sector (its correlation with the equal-weighted average is $0.99$), the sector factor. The second component contrasts the larger companies of the sample (positive weights for BOIRON, VIRBAC, GUERBET, ABLYNX, GALAPAGOS) with the small French biotechs (negative weights for INTEGRAGEN, NEOVACS, HYBRIGENICS, THERADIAG, ONXEO), a contrast that reads as a size factor.

\begin{center}
\includegraphics[width=0.9\textwidth]{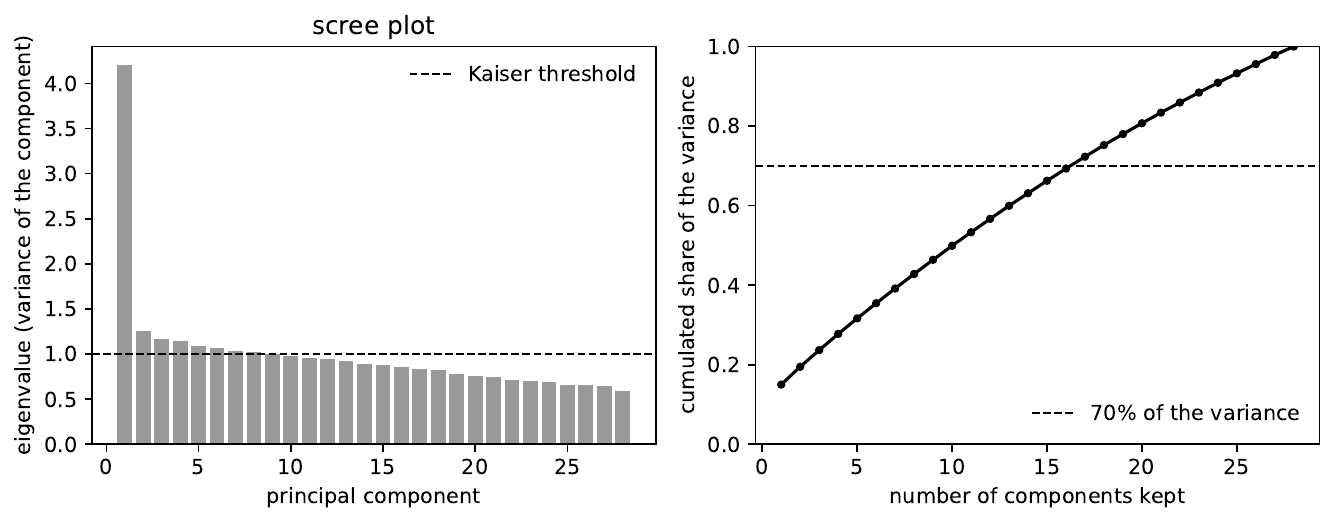}

\includegraphics[width=0.62\textwidth]{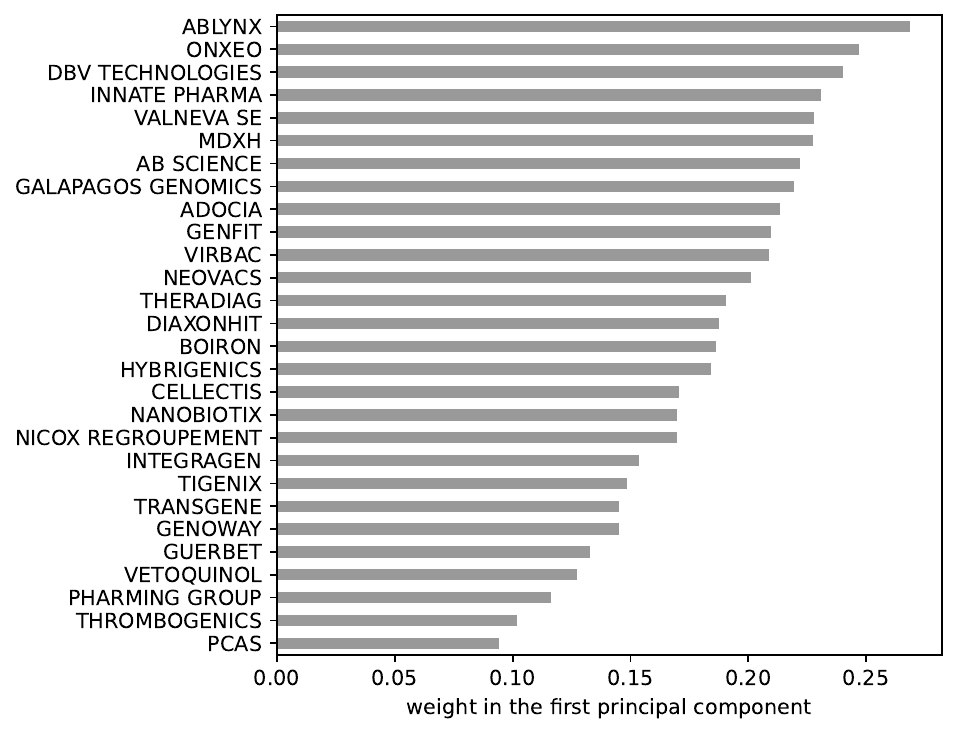}
\end{center}

We extract factor 1 and 2 ($F_{1,t}$ and $F_{2,t}$) with the principal component analysis. Then for each stock returns $i$, we use the following linear regression:
\begin{equation}\label{eq:PCAregression}
R_{i,t}^{cr} = \alpha_i F_{1,t} + \beta_i F_{2,t} + \epsilon_{i,t}
\end{equation}
Because the components are orthogonal, the OLS coefficients are exactly the loadings, $\hat\alpha_i = w_{i1}$ and $\hat\beta_i = w_{i2}$ (the $i$th entries of $w_1$ and $w_2$), a fact the code verifies, and the $R^2$ of the regression of stock $i$ is $\lambda_1 w_{i1}^2 + \lambda_2 w_{i2}^2$, whose average over the stocks is the share of variance explained by the two components, $19.5\%$. The residual $\epsilon_{i,t}$ is the idiosyncratic part of the return, which for these stocks is the largest part.

We assume that the factors and the residuals are following \textbf{normal distribution}\footnote{This is an area of improvement as during stress event the distribution is not normal, cf. Gourieroux - Dubecq}, with standard deviations $\sigma_j^F = \sqrt{\lambda_j}$ for the factors and $\sigma_i^\epsilon$, the standard deviation of the residuals of equation \eqref{eq:PCAregression}, for the residuals:
\begin{equation}
   \left \{
   \begin{array}{r c l}
\epsilon_i & \sim & \mathcal{N}(0,(\sigma_i^\epsilon)^2)\\
F_j & \sim & \mathcal{N}(0,(\sigma_j^F)^2) = \mathcal{N}(0,\lambda_j)
   \end{array}
   \right .
\end{equation}

\subsubsection{Shock generations - normal distribution assumption}

We chose a number $S$ of simulations we want to run and follow those 4 steps to generate a shock on the portfolio:
\begin{enumerate}
\item generate simulation for $F^s_{1,t+1}$, $F^s_{2,t+1}$ and $\epsilon^s_{i,t+1}$
\item estimate the standardize returns for each stock: $\hat{R}_{i,t+1}^{cr,s}=\hat{\alpha}_i F^s_{1,t+1} +\hat{\beta}_i F^s_{2,t+1} +\epsilon^s_{i,t+1}$
\item compute the actual returns $\hat{R}^s_{i,t+1}=\hat{R}_{i,t+1}^{cr,s} \sigma_i + \bar{R_i}$
\item compute the volume of the shock for the portfolio $\sum_i \hat{R}^s_{i,t+1} A_{i,t}$, with $A_{i,t}$ the volume of equity in the portfolio for the stock $i$ at time $t$, before the shock
\end{enumerate}
The code runs these steps for a portfolio holding one million euros of each of the 28 stocks, with $S=20$ simulations, then with $S=10{,}000$: the distribution of the one-day gain or loss has a standard deviation of about $310$ thousand euros and its $1\%$ quantile is a loss of about $855$ thousand euros, $3.1\%$ of the portfolio, the Monte Carlo counterpart of the Value-at-Risk of section \ref{sec:distribtuions}. Note that the model generates the shocks from two common factors and independent residuals: it reproduces the average co-movement of the stocks, not the joint extreme movements observed during stress events, hence the footnote above.

\subsubsection{Exercise: PCA on EBA stress test scenario in 2016}

Work on the \href{https://www.eba.europa.eu/documents/10180/1383302/2016+EU-wide+stress+test-Adverse+macro-financial+scenario.pdf}{EBA stress test scenario that were used in 2016}. You have the GDP, inflation and unemployment information stored in \verb|unemployment_EBA_2016.csv|, \verb|GDP_EBA_2016.csv|, and \verb|inflation_EBA_2016.csv|

\quad

Find the most important factor behind this scenarios of 2016. Once available, compare them to the 2018 scenarios. The pipeline is the one of the code: build the data matrix (one row per date, one column per country or variable), standardize, compute the covariance matrix and diagonalize it.

\subsection{Building an Indicator with PCA: The St.~Louis Fed Financial Stress Index}\label{subsec:PCAstress}

Financial stress is not directly observed. Each market indicator, an interest rate, a yield spread or an implied volatility, reflects it only partially and adds its own noise, but in periods of stress these indicators move together. \citet{kliesen2010measuring} exploit this comovement to build the St.~Louis Fed Financial Stress Index (STLFSI) from $N=18$ weekly US series starting on 31~December 1993: seven interest rates (among which the effective federal funds rate, the 10-year Treasury yield and the Baa corporate bond yield), six yield spreads (among which the 10-year minus 3-month Treasury spread and the Baa minus 10-year Treasury spread) and five other indicators (among which the VIX).\footnote{The VIX measures annualized 30-day implied volatility of the S\&P 500. The methodology is detailed \href{https://cdn.cboe.com/resources/indices/Cboe_Volatility_Index_Mathematics_Methodology.pdf}{here}.} The identifying assumption is stated in the appendix that documents the construction: ``We assume that financial stress is the primary factor influencing this comovement''.\footnote{Federal Reserve Bank of St.~Louis, \emph{National Economic Trends}, January 2010, Appendix, \url{https://files.stlouisfed.org/files/htdocs/publications/net/NETJan2010Appendix.pdf}.} Stress is then estimated by the first principal component of the panel.

\quad

Each series $y_{i,t}$ is standardized, $x_{i,t}=(y_{i,t}-\bar y_i)/\hat\sigma_i$, and $w_1$ is the eigenvector of the correlation matrix $\hat\Sigma$ of equation~\eqref{eq:PCAcov} associated with its largest eigenvalue $\lambda_1$. The index is the first principal component rescaled to unit variance,
\begin{equation}\label{eq:STLFSI}
S_t=\frac{F_{1,t}}{\sqrt{\lambda_1}}=\sum_{i=1}^{N}\frac{w_{i1}}{\sqrt{\lambda_1}}\,\frac{y_{i,t}-\bar y_i}{\hat\sigma_i},
\qquad \frac{1}{T}\sum_{t=1}^{T}S_t=0,
\qquad \Var(S_t)=\frac{w_1^\top\hat\Sigma\,w_1}{\lambda_1}=1 ,
\end{equation}
so that zero is the average level of stress over the sample, the ``normal'' state, and $S_t$ measures the deviation from it in standard deviations. The published ``adjusted coefficients'' are the weights $w_{i1}/\sqrt{\lambda_1}$, and the sign of $w_1$, which is arbitrary, is chosen so that a higher index means more stress. On the original sample, from 31~December 1993 to 11~December 2009, the first component explains $44.3\%$ of the variance of the 18 standardized series; the federal funds rate ($-0.083$) and the Treasury yields ($-0.092$ for the 10-year) enter negatively, as monetary easing and flights to quality lower them in stress episodes, while the Baa--Treasury spread ($0.125$) and the VIX ($0.110$) enter positively.

\quad

The label ``stress'' comes from the assumption, not from the algorithm: PCA returns the direction of largest common variance of the standardized panel, which measures stress only if stress dominates the comovement of the selected series. The choice and the transformation of the series are therefore part of the identification. Trends are a case in point: since interest rates trended down and stock prices up over the sample, version 2.0 of the index (March 2020) replaced their levels by their changes.\footnote{\url{https://www.stlouisfed.org/on-the-economy/2020/march/financial-stress-index-version-2}. Version 3 replaced LIBOR by SOFR in the computation of some spreads, and version 4 by a forward-looking SOFR rate; the index currently published on FRED is the STLFSI4, \url{https://fred.stlouisfed.org/series/STLFSI4}.} Finally, an index built on full-sample estimates of $\bar y_i$, $\hat\sigma_i$ and $w_1$ is revised whenever the sample is extended, and its past values use information that was not available at the time: it is not a real-time measure.

\subsubsection{Exercise: a financial stress index from five FRED series}

Download from FRED five ingredients of the STLFSI, the effective federal funds rate (\texttt{FEDFUNDS}), the 10-year Treasury yield (\texttt{DGS10}), Moody's Baa corporate bond yield (\texttt{BAA}), the 10-year minus 3-month Treasury spread (\texttt{T10Y3M}) and the VIX (\texttt{VIXCLS}), together with the published index (\texttt{STLFSI4}). Each series can be read with pandas directly from its FRED address, e.g.\ \url{https://fred.stlouisfed.org/graph/fredgraph.csv?id=STLFSI4}, without an API key.
\begin{enumerate}
\item \texttt{FEDFUNDS} and \texttt{BAA} are monthly averages, \texttt{DGS10}, \texttt{T10Y3M} and \texttt{VIXCLS} are daily and \texttt{STLFSI4} is weekly: convert every series to monthly averages and keep the months common to all of them, starting in January 1994.
\item Standardize the five series, compute $\hat\Sigma$ and diagonalize it. Report the share of the variance explained by each component and the loadings of the first two components, choosing the sign of each eigenvector so that its loading on the VIX is positive.
\item Build the index $S_t$ of equation~\eqref{eq:STLFSI} and compare it with the STLFSI4: plot the two series and compute their correlation. Compute also the correlation of the STLFSI4 with each of the five components, and show that the sum of their squares equals the $R^2$ of the OLS regression of the STLFSI4 on a constant and the five standardized series.
\item Is your first component a financial stress index? Interpret its loadings, find the component that tracks the STLFSI4, and relate your answer to the identifying assumption of \citet{kliesen2010measuring}.
\item Redo the analysis in the spirit of the current STLFSI: the three interest rates (\texttt{FEDFUNDS}, \texttt{DGS10}, \texttt{BAA}) in first differences, the two spreads (\texttt{T10Y3M} and the credit spread \texttt{BAA}$-$\texttt{DGS10}) and the VIX in levels. Compare the signs of the new loadings with the coefficients of the original STLFSI reported above, and the correlation of the new index with the STLFSI4 with that of the VIX alone. What does the composite index add?
\end{enumerate}

\subsection{Boosting}
Boosting\footnote{Text from \citep{MLBoE}} describes a set of methods which transform, or boost, a weak learner into a strong one (model). This means that one iteratively applies models to the data which individually have weak generalisation properties, but the final ensemble of models generalises well. Two popular boosting techniques are gradient boosting and adaptive boosting, or adaBoost. In gradient boosting, one sequentially updates the estimator in the direction of change of the loss function, i.e. a form of error correction. For adaBoost, one gives larger weights to misclassified or badly fitted observations at each consecutive step until one reaches a certain stopping criterion. Stopping criteria can be linked to target test or a flattening out-of-bag performance. AdaBoost requires a potentially large amount of data to converge and is susceptible to noise in the data. Boosting is often applied to tree models.

\subsection{Train and test sets}

We can randomly separate our data set into two sets: a Train and a Test data set. The Train data set will have two third of the data, and the Test set the rest. The practice is then to fit a model on the Train data set, then to use prediction on the test data set and see if the predictions are "close" to our empirical observations.

\subsection{Tuning hyper-parameters of an estimator}

We refer directly to sklearn tutorial on \href{https://scikit-learn.org/stable/modules/grid_search.html#}{Tuning the hyper-parameters of an estimator} and provide a detailed practical example section \ref{sec:NNtune}

\subsection{Metrics with our data sets and methods}
We sort our results by F1 score.

When lending, we can consider that false negatives (credit worthy clients get rejected) are more acceptable than false positives (non-credit worthy clients get a loan from us).

\resizebox{\textwidth}{!}{
\begin{tabular}{lrrrrrrr}
\toprule
{} &    FPR &  Classification Accuracy &  Sensitivity &    AUC &  precision &  recall &  F1score \\
\midrule
GaussianNB                 &  0.025 &                    0.804 &        0.221 &  0.597 &      0.667 &   0.225 &    0.337 \\
DecisionTreeClassifier     &  0.147 &                    0.721 &        0.221 &  0.607 &      0.377 &   0.403 &    0.390 \\
KNeighborsClassifier       &  0.057 &                    0.793 &        0.221 &  0.625 &      0.556 &   0.322 &    0.408 \\
MLPClassifier              &  0.050 &                    0.803 &        0.221 &  0.636 &      0.598 &   0.337 &    0.431 \\
GradientBoostingClassifier &  0.038 &                    0.819 &        0.221 &  0.652 &      0.673 &   0.353 &    0.463 \\
LogisticRegression         &  0.035 &                    0.821 &        0.221 &  0.653 &      0.687 &   0.351 &    0.465 \\
RandomForestClassifier     &  0.044 &                    0.816 &        0.221 &  0.656 &      0.649 &   0.368 &    0.470 \\
LinearDiscriminantAnalysis &  0.040 &                    0.822 &        0.221 &  0.661 &      0.675 &   0.372 &    0.480 \\
\bottomrule
\end{tabular}
}


\subsubsection{With optimized neural network}\label{sec:NNtune}

We optimize\footnote{code: ML\_credit\_card\_NN.py} the estimator for the neural network which requires computing power. We have to chose:
\begin{itemize}
\item hidden layer sizes
\item activation function for the hidden layer (logistic, tanh, etc.)
\item the learning rate
\item $\alpha$ the L2 penalty (regularization term) parameter
\end{itemize}

We find:
\begin{itemize}
\item CA: 0.82
\item S: 0.22
\item AUC: 0.77
\item F1 score: 0.47
\end{itemize}


\section{Python: computing the tide's heights at a given time}

Let's imagine you spend some time at Dunkerque end 2023. You want to sail and need to know\footnote{dunkerque\_tide.py} the tide's height at any time of the day. You have only access to the time of the low and high tide and both heights.

Hypothesize that the tide height follows:
\begin{equation}
a \sin\left( \omega t + \phi \right) + b
\end{equation}

where $a$ is half the tidal range, $b$ is the mean sea level, $\omega$ is angular frequency, $\phi$ fixes phase, and $t$ the time (in hours) since the low tide. With low tide at $t=0$ and period $T$, you should set $\omega = \tfrac{2\pi}{T}$ and $\phi = -\tfrac{\pi}{2}$ so that $h(0) = b-a$ matches the low tide. Between one low tide and the next high tide, the elapsed time is about 6h12. That corresponds to half of a full tidal period (since full period is 12h25). So $\omega = 2\pi / T$ with $T\approx 12.42$ h. You only have the following information:

\begin{table}[h!]
\centering
\begin{tabular}{|c|c|c|c|}
\hline
\textbf{Date} & \textbf{Heure} & \textbf{Hauteur} & \textbf{Coeff.} \\
\hline
Jeu. 28 & 01h21 & 5,68m & 78 \\
        & 08h12 & 0,87m &    \\
        & 13h38 & 5,80m &    \\
        & 20h32 & 1,06m & 78 \\
\hline
Ven. 29 & 02h00 & 5,68m & 78 \\
        & 08h51 & 0,85m &    \\
        & 14h19 & 5,76m &    \\
        & 21h09 & 1,12m & 77 \\
\hline
Sam. 30 & 02h37 & 5,64m & 75 \\
        & 09h28 & 0,89m &    \\
        & 14h56 & 5,67m &    \\
        & 21h44 & 1,22m & 73 \\
\hline
Dim. 31 & 03h11 & 5,55m & 71 \\
        & 10h04 & 0,98m &    \\
        & 15h33 & 5,54m &    \\
        & 22h18 & 1,33m & 68 \\
\hline
Lun. 01 & 03h46 & 5,43m & 65 \\
        & 10h39 & 1,09m &    \\
        & 16h10 & 5,40m &    \\
        & 22h53 & 1,44m & 61 \\
\hline
Mar. 02 & 04h24 & 5,30m & 58 \\
        & 11h15 & 1,20m &    \\
        & 16h50 & 5,24m &    \\
        & 23h30 & 1,55m & 54 \\
\hline
Mer. 03 & 05h06 & 5,15m & 51 \\
        & 11h54 & 1,32m &    \\
        & 17h35 & 5,08m &    \\
        &       &      & 47 \\
\hline
\end{tabular}
\caption{Dunkerque tide table}
\label{table:dktide}
\end{table}

\subsection{The tide's coefficient and the tide's height}

Note that the tide's coefficient $c$ will impact the tide's height. We import the data set for different tide coefficients and try to see if there is a systematic linear relationship between $a$ and $c$ on one side and $b$ and $c$ on the other.

\stepcounter{question}
\begin{itemize}
\item \textbf{Question \arabic{question}}
\begin{itemize}
\item Choose the 31th in the morning, the coefficient was 71, can you find the expected tide's heights?
\begin{itemize}
\item Nota bene: as the Sun and the Moon positions relative to the Earth evolve, it is unlikely to be the case.
\end{itemize}
\item On that day, what was the tide's height at noon?
\end{itemize}
\end{itemize}

\section{Python: Entertaining application - Perudo game}

To apply the skills developed so far, we suggest to play Liar's Dice and play the \href{http://www.perudo.com/perudo-rules.html}{Perudo}\footnote{code: vansteenberghe\_perudo.py} version and take the rules from perudo.com, but with some twists to make it simple:
\begin{quote}
 The object of perudo is to be the last player with a die or more. 

Perudo is played in rounds. Each player receives a cup and five dice. Each round begins by all players rolling their dice around in the cup. After shaking the dice, players turn the cups over on a table top, so that the dice are rolled and under the cups. Each player may peek in his own cup. 

Players bid, guessing at the number of rolls. When a player believes that another player has over-estimated, they say Dudo, which means "I doubt" in Spanish. 

The first player announces a number, and then the next player has the choice of doubting it, by saying Dudo or raising the bid, either by the number of dice or by the value of the dice (or by doing both). For example, if player one bid three twos, then player two could bid three threes, four twos, four fours, or even ten sixes. 
\end{quote}
The next player can also announce "calza" which means that he believes that the announcement of the player just before him was the correct guess. In fact, any player can call "calza" at any time even on his own announcement. If the player calling "calza" is correct, then we give him another dice, but if he is wrong then he loses one of his dice.

\begin{quote}
After player 2 bids, play goes on to the left. 

If a player calls Dudo, and is correct, then the player must show the dice, and each player must show their dice to verify whether the number was indeed too high. 

If there are enough dice of that number, then the player who called Dudo must place a die in the discard pile. If there are not, then the player who made the last bid must place a die in the middle. In either case, a new round begins. 

The player who lost a die in the last round is the first player in the new round. If the player lost his last die, then the player to his left plays first instead.

[\ldots]

The last player left with at least one die wins the game. ("Perudo" is a registered trademark of University Games Corporation, Burlingame, CA.) 
\end{quote}

\subsection{Should you call Calza, raise the bet or call Dudo?}
\subsubsection{If you don't use the information provided by the players}

Suppose there are 6 players in a game, which means there are initially 30 dice in play (let's denote this number by \( n_{\text{dice}} \)). The goal is to determine the most probable number of dice that show the same face value. This can be achieved through a Monte Carlo simulation where numerous random draws (``tirages'') are performed. The outcome of a single dice roll can range from 1 to 6. However, it's crucial to remember that an ace (represented by the value 1) is special and can stand for any value between 2 and 6, making it analogous to a joker.

From the simulations, we confirm that the average count of similar dice is 10. But what about the probability associated with making a ``Dudo'' call?

A ``Dudo'' is correctly called when the actual count of a particular dice face is \textbf{strictly} less than a player's announced count. We can accumulate all such counts that validate a ``Dudo'' call. For this particular game setup, if your predecessor announces a count of 13, calling a ``Dudo'' gives you an over 80\% probability of being correct.

However, it's essential to recognize that prior players might have provided hints about their dice counts (even though they might be bluffing). This additional information can change the odds.

\subsubsection{If you use the information provided by the players and your dices}

Now before you speak, several players did before you, but be careful, they might be bluffing.
Let's imagine that you have three dices of $2$, one ace and so far two players spoke and told the audience $3$ twos and then $12$ twos. If you say $13$ twos, are the odds still at $80\%$ against you?

\stepcounter{question}
\begin{itemize}
\item \textbf{Question \arabic{question}}
\begin{itemize}
\item Now consider that you are a player and so you know the values of your 5 dices, this changes the odds, you randomize the other players dices only.
\item Also consider the 5 other players with 5 dices. Some of those other players spoke befor you have to speak. Assign for each player a probability to "bluff". 
\begin{itemize}
\item As an illustration, if a player calls "there are 13 twos". If he is not bluffing, then he believes that there are round(25 dices / 3) so 8 twos in the 25 other dices and you can deduct that this player has 13-8 = 5 twos in his hidden hand (or some twos and some aces to be correct). This is highly unlikely, but this player can take into account what two players just before him said, maybe the players before him called "11 twos" then "12 twos" so it can be believed that those two players had 5 twos in their 10 dices and they both believed that the others (except them both) had round(20 dices / 3) = 6 twos... Well, the point is that guesses are iterative.
\item You have to be conservative and believe that sometimes players are too optimistic. So maybe a player that announces "there are 13 twos" actually is bluffing and only has 3 twos in his hidden hand... To take this into account, you can use a random variable taking value between $[0,1]$ that will diminish your estimation of dices based on players bet.
\end{itemize}
\end{itemize}
\end{itemize}

\section{Python : Pobability, population, samples and absolutely continuous distributions and maximum likelihood estimation}

In the following we will study the probability laws governing random variables (rv), their distributions. We are interested in their cumulative distribution function $P(X \leq x)$. 

If you really want to understand probability, you'd need to dig to measure theory \citet{tao2011} to be able to understand the following definition:
\begin{quote}
A probability space is a measure space of total measure $1$.
\end{quote}
In this book, you'll need to get to the following notions:
\begin{itemize}
\item interval, length
\item box, volume
\item elementary set
\item elementary measure
\item Jordan measurable set
\item Riemann integral (Darboux integral)
\item Lebesgue measurability
\item unsigned Lebesgue measurable function, integral
\item Boolean algebra
\item $\sigma$-algebra, measurable space
\item Borel $\sigma$-algebra, Borel measurable set
\item additive measure
\item measurable functions
\item integration on a measure space
\item probability space and probability measure
\end{itemize}

For an introduction to probability, rv, see \citet{hansen2022}.

The probability that a discrete rv $X$ equals the value $x$, $P(X=x)$, is called the probability mass function.
The expectation of $X$ is the  probability-weighted average. For a discrete rv, taking the value $x_i$ such that $P(X=x_i)=p_i$: $E(X)=\sum_{i=1}^\infty p_i x_i$.
For a continuous rv, if it's cumulative distribution function (cdf) $P(X \leq x)$ is differentiable, their probability density function (pdf) $P(X=x)=\frac{d P(X \leq x)}{dx}$.
Notation: we write the pdf $f_X(x)$ and the cdf $F_X(x)$.

\subsection{Frequency and Statistics or Probability?}
We follow \citet{kaplan1981quantitative} to define frequency versus probability.
\textbf{Frequency} is an objective, measurable quantity defined as the outcome of repeated (actual or conceptual) experiments; it belongs to the empirical domain of data and statistics. 
\textbf{Probability}, by contrast, is a numerical measure of uncertainty or degree of belief, representing a state of knowledge rather than a physical property of the world.
\textbf{Uncertainty} is assumed to be comparable across propositions, which allows it to be quantified on a calibrated scale. Frequencies provide the reference standard for this calibration but probability, once defined, applies more broadly, including to unique or non-repeatable events where no frequency data exist.
Although \textbf{probability is subjective} in that it reflects information rather than reality, it is objective conditional on information: given the same background knowledge, rational agents must assign the same probabilities. Accordingly, statistics is the science of handling data (frequencies), whereas probability is the science of reasoning under incomplete information. Far from being unusable with limited data, probability is precisely the appropriate tool in such settings.

\subsection{Ellsberg paradox}

\citet{einhorn1985} writes:

\begin{quotation}
\citet{ellsberg1961} used the following example to show that the uncertainty people experience contains several aspects, one of which is not captured by probabilities: Imagine two urns, each containing red and black balls. In Urn 1, there are 100 balls, but the proportions of red and black are unknown; Urn 2 contains 50 red and 50 black balls. Now consider a gamble such that, if you bet on red and it is drawn from the urn you get USD 100; similarly for black. However, if you bet on the wrong color, the payoff is USD 0. Imagine having to decide which color to bet on if a ball is to be drawn from Urn 1; that is, the choices are red ($R_1$), black ($B_1$), or indifference ($I$). What about the same choices in Urn 2: ($R_2$), ($B_2$), or (I)? Most people are indifferent in both cases, suggesting that the subjective probability of red in Urn 1 is the same as the known proportion in Urn 2 -- namely, .5. However, would you be indifferent to betting on red if Urn 1 were to be used versus betting on red using Urn 2 ($R_1$, vs. $R_2$)? Similarly, what about $B_1$ versus $B_2$? Many people find that they prefer $R_2$ over $R_1$, even though their indifference judgments within both urns imply that $p(R_1) = p(R_2) = .5$. Furthermore, the same person who prefers $R_2$ over $R_1$ may also prefer $B_2$ over $B_1$ . This pattern of response is inconsistent with the idea that even a rank order of probabilities can be inferred from choices. Thus, if $R_2$ is preferred over $R_1$, this implies that $p(R_2) > p(R_1)$. Moreover, because red and black are complementary events, this means that $p(B_2) < p(B_1)$. However, if $B_2$ is preferred over $B_1$ then $p(B_2) > p(B_1)$, which contradicts the preceding inequality.

$[\ldots]$

From our perspective, the importance of Ellsberg's paradox lies in the difference in the nature of the uncertainty between Urns 1 and 2. In Urn 1, whereas one's best estimate of the probability may be .5, confidence in that estimate is low. In Urn 2, on the other hand, one is at least certain about the uncertainty in the urn. Although it may seem strange, and even awkward, to speak of uncertainty as being more or less certain itself, such a concept captures an important aspect of how people make inferences from unknown, or only partially known, \textbf{generating processes}.

$[\ldots]$ one could argue that the cost of urn models, coin-flipping analogies, and the like, is that they obscure the fact that most real-world generating processes are not precisely known.
\end{quotation}

\citet{einhorn1985} defines $0\leq \theta \leq 1$ the amount of ambiguity and $\beta$ an individual's attitude toward ambiguity.  The judged probability of an initial guessed probability $p_A$ is expressed as:
\begin{equation}
S(p_A) = p_A + \theta \left( 1  - p_A - p_A^\beta\right)
\end{equation}
To explain the Ellsberg's paradow, they suggest taking $\theta>0$ and $\beta<1$

\subsection{Concepts of Data-Generating Process and iid}\label{sec:DGP}

For any data set, as an empiricist, we are making assumptions. The strongest and least credible assumptions we make are on the data-generating process (DGP)\footnote{\citet{Davidson1978aa} consider that "one could characterise 'econometric modelling' as an attempt to match the hypothetical data generation process postulated by economic theory with the main properties of the observed data."} where we assume that there is an underlying mechanism that is generating the observations under study. This is discusses in \citet{breiman2001statistical} and we discuss further such approaches with machine learning section \ref{sec:MLwhy}.

In our example, this means that we assume that the CAC 40 daily returns are generated by a process, we will introduce possible distributions in the following sections. We further initially assume that the return at date $t$ is drawn from the DGP independently with respect to the outcome at date $t-1$, but for time series we will need later to tackle the fact that this hypothesis doesn't hold.

Further, when applying an ordinary least square method to estimate a linear regression, we will assume (and test) that the residuals have mean $0$, were generated by the same underlying distribution, and are independent of each other. For the last two conditions, we say that the residuals are \textbf{independent and identically distributed} (iid). This is often violated: residuals are serially correlated or their variance differ for some observations (heteroskedasticity).

\subsubsection{In a data rich world, should we drop our a priori?}

\citep{Armatte2017aa} review the work of Edmond Malinvaud and his position on "\emph{A Priorism versus Empiricism}":
\begin{quote}
Is economic science inductive? Does it begin from observation? Is it not simply a theoretical structure whose coincidence with reality is doubtful? He answers them in the following manner: yes, economic science has become inductive; the knowledge of economic phenomena is developed in the same way as that of physical phenomena, that is, through a \textbf{continual back and forth [va-et-vient] between observation and theory}; given the particular conditions in which it operates, economics employs an inductive approach which is, in part, original, and thanks to which it progressively ensures a better correspondence with reality.
\end{quote}
he stresses the \begin{quote}
three steps which were preconditions to the progress of an inductive economic science: 
\begin{enumerate}
\item the accumulation of data, 
\item the elaboration of descriptive syntheses such as series of indices and national accounting, 
\item the diffusion of a method use to establish laws on the basis of observed data, namely, econometrics. 
\end{enumerate}
[\ldots] What naivety, is there not, in supposing, as I have seen it done repeatedly, that the abundance of data, certainly a necessary condition, is also sufficient for understanding phenomena and for good forecasting.
\end{quote}

And he gives this recommendation: \begin{quote}
In order to make the best use of the observations at his disposal, the economist must first draw all the consequence of what he already knows, that is, construct a model expressing his initial knowledge, in terms of which he can circumscribe properly the supplementary information to be gleaned from the data. Effective research demands this type of approach. To advance on the path of knowledge, the economist can and must first proceed with a deductive analysis of the potential causes of phenomena. The model is precisely the result of this deductive analysis. [\ldots] To gain knowledge of reality, the economist must continually rationalize and refine his theoretical constructions as he deciphers the information contained in the increasingly abundant data at his disposal.
\end{quote}

\subsection{Distribution function, mean, variance}

We follow section \ref{sec:euronext} and use the history of CAC 40 index including dividends (net returns)\footnote{vansteenberghe\_samples.py}. Now if as an investor I believe that the history or daily returns are representative of the likely upcoming daily returns, I might want to know to likelihood of a future daily return of $0.045$ (i.e. $4.5\%$). I can approach this question two ways:
\begin{enumerate}
\item I build an empirical histogram
\item I fit a parametric\footnote{usually, the density function is not known, here we select an arbitrary function and determine the appropriate parameter for this function to describe the distribution of the sample} distribution (normal, student, etc.) on the empirical data
\end{enumerate}

In formal terms, I consider the daily returns to be a continuous rv $X$ and believe that I can find a pdf and cdf that is stable over time and can help me predict the realizations.

\subsubsection{Empirical histogram}\label{sec:empiricalhisto}

If we build an empirical histogram, we have to chose the granularity of the bins. If doing it manually, we can start by limiting our representation to 2 bins: $]-\infty,0]$ and $]0,+\infty[$. In our representation, we have:

\begin{tabular}{lrr}
&   neg &   pos \\
count &  3762 &  4072 \\
\end{tabular}

We can normalize this so that the sum of the bins is equal to $1$ ($100\%$):

\begin{tabular}{lrr}
&       neg &       pos \\
count &  0.48 &  0.52 \\
\end{tabular}
In other words, empirically, I have $52\%$ of probability to have a positive daily return\footnote{before concluding, we would need to perform statistical test to see if this is indeed statistically different than $50\%$}. $P(X > 0)=0.52$.

\quad

We can use directly the python command and chose both the number of bins ($20$ in this example) and whether we normalize the histogram or not:
\begin{center}
\includegraphics[scale=0.5]{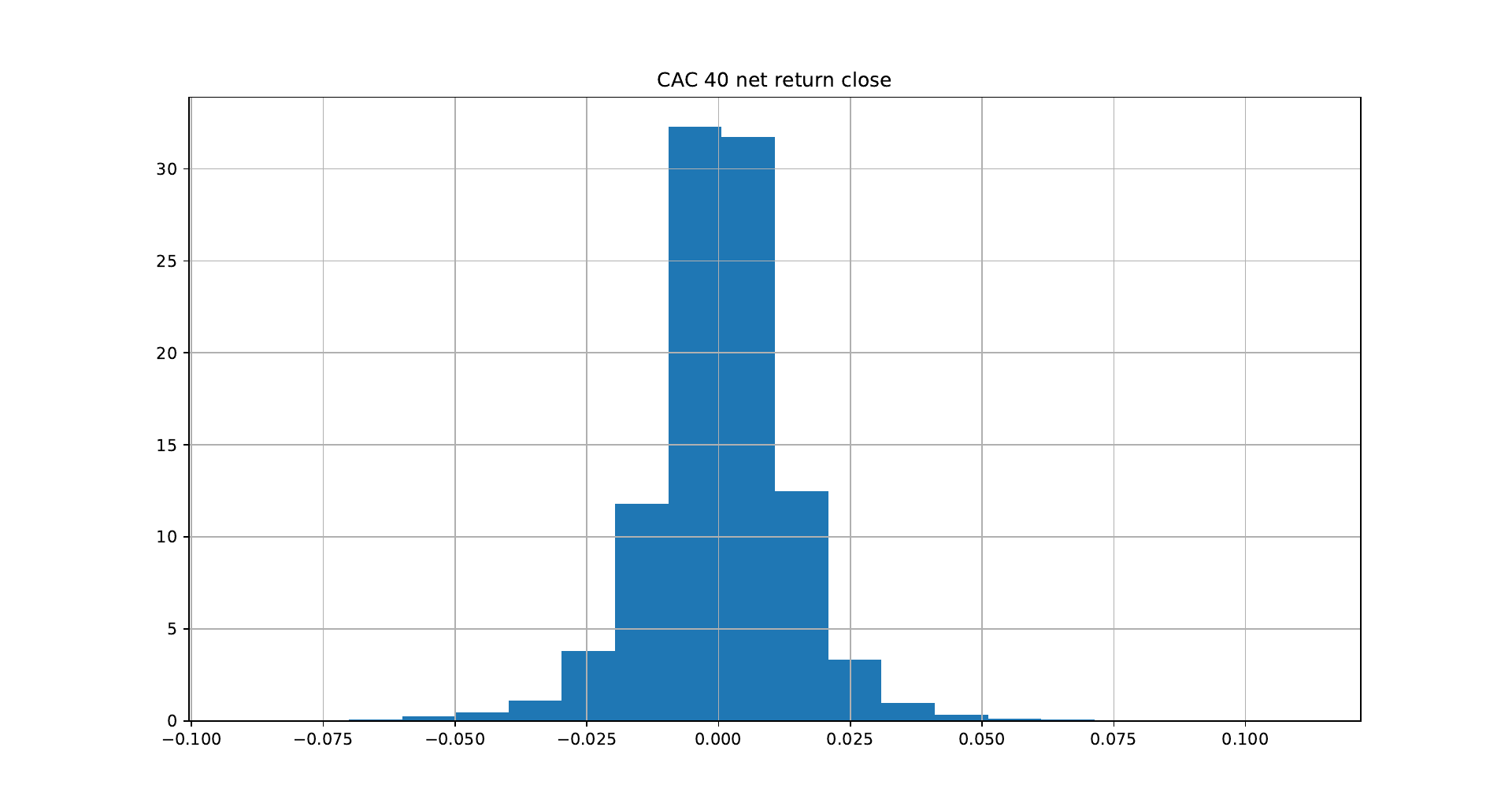}
\end{center}

So to answer our initial question on the likelihood to have a daily return of $4.5\%$, we can zoom in:
\begin{center}
\includegraphics[scale=0.5]{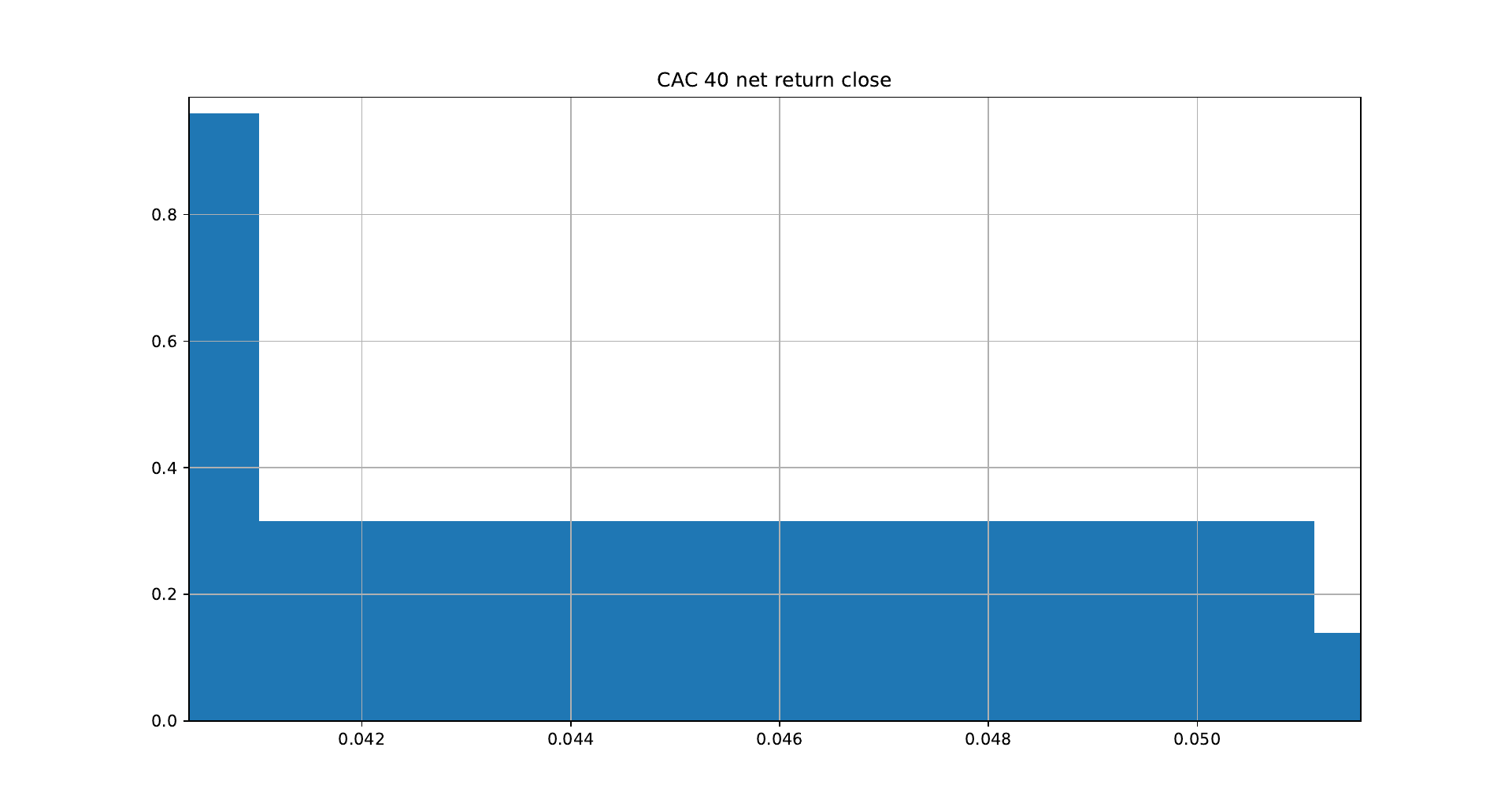}
\end{center}

For a continuous rv $X$, the pdf $P(X=0.045)=0$, \citep{hansen2022}, we are only able to quantify the probability for $X$ to fall in an interval. We visually observe a probability around $0.3\%$ (so on the y-axis between $0.2$ and $0.4$) which we confirm by counting how many daily return fell inside the bucket $[0.04,0.05]$ of size $0.01$ over the total count of daily return: $0.38\%$. You can look at equation (1.6) in \citet{li2023nonparametric}.

\subsection{Parametric density estimation}

We will present various parametric distributions in the coming sections\footnote{all of those are unimodal (that is, have a single local maximum), meaning that if you work with a multimodal distributed sample you will need to use other technics e.g. KDE we cover in section \ref{sec:KDE}}, we need to believe that it is possible to fit a "function" $P(X=x)$ that gives us the probability that the daily return is $x$. Note that the goodness of fit of that function will be relevant if we want to trust the answer on the probability, for this there exist various tests, one of which is presented section \ref{sec:KStest}.

\subsubsection{Mean}\label{sec:meandistrib}

We define the expectation as a weighed average: $$E(X) = \int_{-\infty}^\infty x P(X=x) dx$$ or in unified notation $E(X) = \int_{-\infty}^\infty x dF_X(x)$

in our case $E\left(X\right)=0.038\%$.

\subsubsection{Standard Deviation}
We define the standard deviation as:
$$s(X) = \sqrt{\int_{-\infty}^\infty \left[x-E(X)\right]^2 P(X=x) dx}$$
in our case we find $0.013$

\subsubsection{Moments}

In general, we consider the following moments first to fourth:
\begin{enumerate}
\item $\mu = E\left( X \right)$
\item variance $ = \sigma^2 = E\left[\left( X - \mu \right)^2\right]$
\item skewness $ = E\left[\left( \frac{X - \mu }{\sigma}\right)^3\right]$
\item kurtosis $ = E\left[\left( \frac{X - \mu }{\sigma}\right)^4\right]$
\end{enumerate}

\subsection{Normal distribution}\label{sec:nomaldist}

We introduce the normal distribution $\mathcal{N}(\mu,\,\sigma^2)$. Its probability density function, $P(X=x)$, is 
$$f(x;\mu,\sigma) = \frac{1}{\sqrt{2\pi \sigma^2}} e^{-\frac{(x-\mu)^2}{2\sigma^2}}$$ in simple words this function gives us the probability that a random variable $X$ takes the value $x$ if the random variable follow the normal distribution, we typically write $X \sim \mathcal{N}(\mu,\,\sigma^2)$. For illustration, we take the parameter $\mu$ and $\sigma$ as the empirical mean and standard deviation observed on the CAC 40 daily returns:
\begin{center}
\includegraphics[scale=0.5]{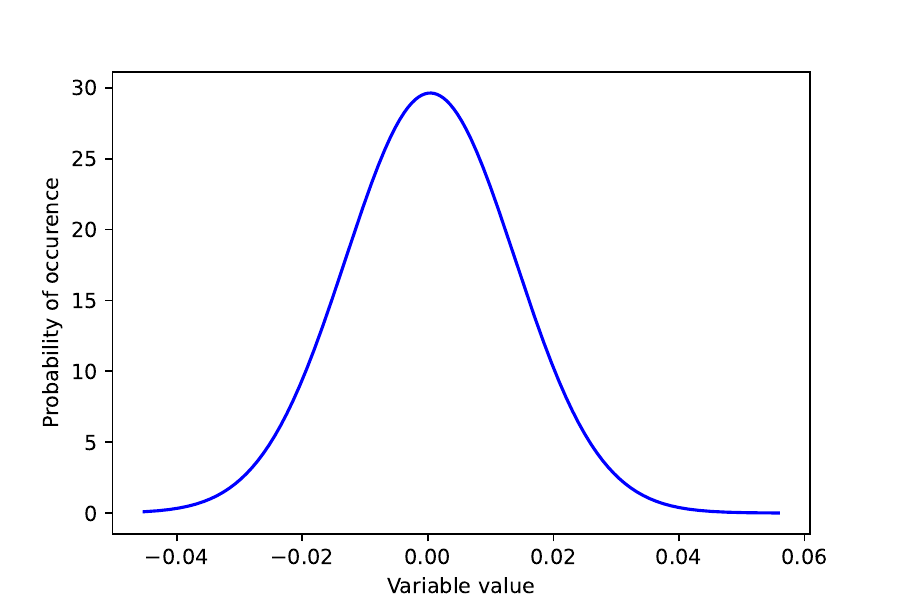}
\end{center}

\stepcounter{question}
\begin{itemize}
\item \textbf{Question \arabic{question}}
\begin{itemize}
\item Recall that $\frac{\partial e^{g(x)}}{\partial x} = g'(x) e^{g(x)}$ for any function $g$. Show that the (unique) maximum of the normal pdf is $\mu$.
\end{itemize}
\end{itemize}

The mean or first moment of the distribution can be computed following Section~\ref{sec:meandistrib}.  
We start from
\begin{equation}
\mu^1 = \int_{-\infty}^{+\infty} x f(x;\mu,\sigma)\, dx,
\end{equation}
with $f(x;\mu,\sigma)$ the $\mathcal{N}(\mu,\sigma^2)$ density.  

Make the change of variable
\[
z = \frac{x-\mu}{\sigma}, 
\qquad x = \mu + \sigma z, 
\qquad dx = \sigma dz.
\]

The integral becomes
\[
\mu^1 = \int_{-\infty}^{+\infty} (\mu + \sigma z)\, 
\frac{1}{\sqrt{2\pi}} e^{-z^2/2}\, dz.
\]

Split into two terms:
\[
\mu^1 = \mu \underbrace{\int_{-\infty}^{+\infty} \frac{1}{\sqrt{2\pi}} e^{-z^2/2}\, dz}_{=1}
+ \sigma \underbrace{\int_{-\infty}^{+\infty} z \frac{1}{\sqrt{2\pi}} e^{-z^2/2}\, dz}_{=0}.
\]

Hence
\[
\mu^1 = \mu.
\]

\subsubsection{Random draws}

We can draw randomly from a normal distribution (let's say $10^4$ draws) and then plot the histogram against the probability density:

\begin{center}
\includegraphics[scale=0.5]{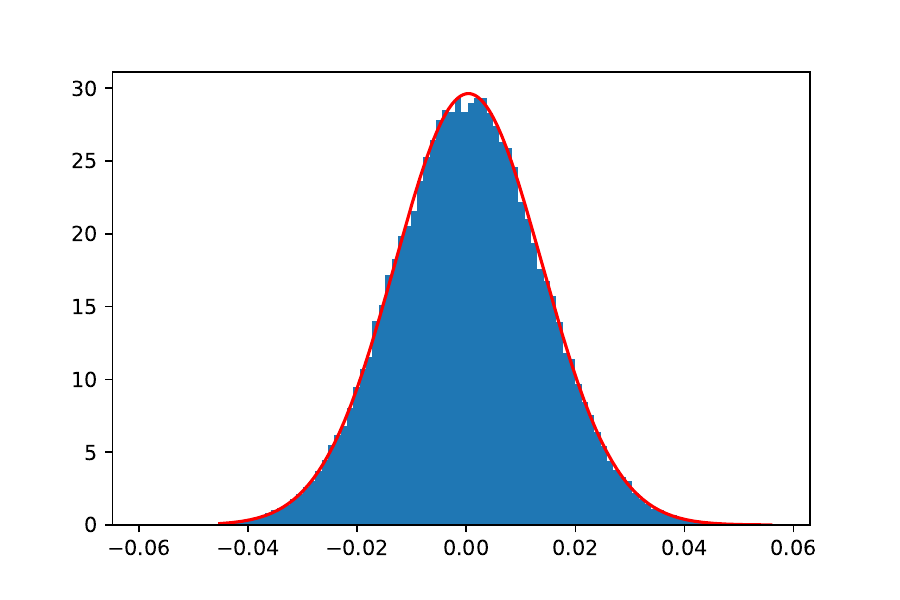}
\end{center}

The empirical data doesn't fit the theoretical normal distribution, visually you can observe fatter tails:
\begin{center}
\includegraphics[scale=0.5]{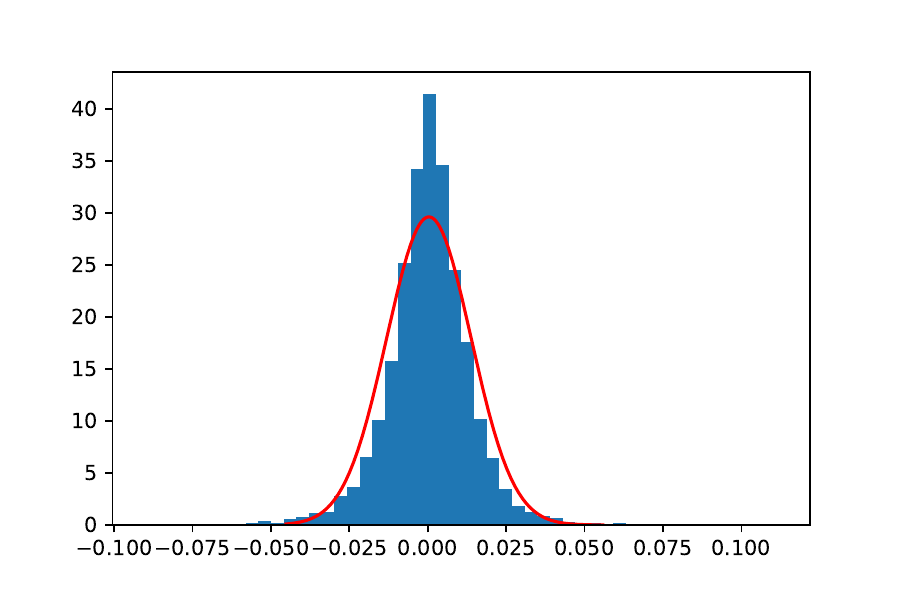}
\end{center}

\subsection{Introduction to Maximum Likelihood Estimation}\label{sec:MLestimation}

Let's imagine that you have a sample of size $n$, $[x_1, \ldots, x_n ]$, taken from a population governed by a normal law. We want to estimate two parameters of the parametric law, $\mu$ and $\sigma$. This means you are searching for the parameters that are making a normal law \textbf{most likely} to have generated the sample you observe. We call this method the Maximum Likelihood Estimation. Said differently:
\begin{quote}
The maximum likelihood estimator is the value of the parameter vector that maximizes the likelihood function, that is, the value of the parameter vector under which the observed data were most likely to have been observed.
\end{quote}

The probability density function for a sample of $n$ independent identically distributed normal random variables is the likelihood we want to maximize:
\begin{equation}\label{eq:MLE1}
L(\mu,\sigma) = f(x_1, \dots, x_n;\mu,\sigma)
\end{equation}
as the random variables are independent and identically distributed, equation \ref{eq:MLE1} can be written:
\begin{equation}
L(\mu,\sigma) = \prod_{i=1}^n f(x_i; \mu,\sigma)
\end{equation}

To say it differently, the probability that the random variable $X$ took the value $x_i$, for any $i$ ($\forall i$) is by definition the probability density function and here we use a parametric density function with two parameters that have to be estimated:
\begin{equation}
Pr\left( X = x_i \right) = f(x_i; \mu,\sigma)
\end{equation}

We want to choose $\mu$ and $\sigma$ that maximize $L(\mu,\sigma)$:
\begin{equation}
\max_{\mu, \sigma} L(\mu,\sigma)
\end{equation}

As the logarithm function is continuous strictly increasing and it is easier to deal with the log-likelihood, we maximize:
$$\log\left( L(\mu,\sigma)  \right) = -\frac{n}{2} \log\left( 2 \pi \sigma^2 \right) - \frac{1}{2 \sigma^2}\sum_{i=1}^n(x_i-\mu)^2$$

In a maximization problem, we can apply first order conditions, they yields the following parameters that will maximize our probabilities:
$$\hat{\mu} = \sum_{i=1}^n \frac{x_i}{n}$$
$$\hat{\sigma}^2 = \frac{1}{n}\sum_{i=1}^n (x_i-\hat{\mu})^2$$

It can be demonstrated that the estimator of the variance is biased:
\begin{equation}
E\left(\hat{\sigma}^2\right) = \frac{n-1}{n} \sigma^2.
\end{equation}

\noindent
\textbf{Intuition.}  
Why is there a bias? The key point is that in the sample variance formula we subtract the \emph{sample mean} $\hat{\mu}$, rather than the \emph{true mean} $\mu$.  
Because $\hat{\mu}$ itself is estimated from the data, it tends to be closer to the sample points than the true $\mu$.  
This makes the deviations $(x_i-\hat{\mu})^2$ systematically smaller than the deviations $(x_i-\mu)^2$, and therefore the variance estimator underestimates the true variance on average.  

\medskip

\textbf{Mathematical steps.} You can look for "Bessel's correction".

Let \(X_1,\ldots,X_n\) be independent and identically distributed, with
\[
\mathbb{E}[X_i]=\mu
\qquad \text{and} \qquad
\operatorname{Var}(X_i)=\sigma^2.
\]

Write
\[
\bar{X}=\frac{1}{n}\sum_{i=1}^n X_i.
\]

By definition,
\[
\operatorname{Var}(X)
=
\mathbb{E}\left[\left(X-\mathbb{E}[X]\right)^2\right]
=
\mathbb{E}\left[X^2\right]
-
\left(\mathbb{E}[X]\right)^2.
\]

The maximum-likelihood estimator of the variance is
\[
\widehat{\sigma}^2
=
\frac{1}{n}
\sum_{i=1}^n
\left(X_i-\bar{X}\right)^2.
\]

The true variance is
\[
\sigma^2
=
\mathbb{E}\left[\left(X_i-\mu\right)^2\right].
\]

Moreover,
\[
\mathbb{E}[\bar{X}]
=
\frac{1}{n}\sum_{i=1}^n \mathbb{E}[X_i]
=
\mu.
\]

Now observe that
\[
X_i-\mu
=
\left(X_i-\bar{X}\right)
+
\left(\bar{X}-\mu\right).
\]

Therefore,
\[
\begin{aligned}
\left(X_i-\mu\right)^2
&=
\left(X_i-\bar{X}\right)^2
+
\left(\bar{X}-\mu\right)^2 \\
&\quad
+
2\left(X_i-\bar{X}\right)
\left(\bar{X}-\mu\right).
\end{aligned}
\]

Summing over \(i\), we obtain
\[
\begin{aligned}
\sum_{i=1}^n \left(X_i-\mu\right)^2
&=
\sum_{i=1}^n \left(X_i-\bar{X}\right)^2
+
n\left(\bar{X}-\mu\right)^2 \\
&\quad
+
2\left(\bar{X}-\mu\right)
\sum_{i=1}^n \left(X_i-\bar{X}\right).
\end{aligned}
\]

Since
\[
\sum_{i=1}^n \left(X_i-\bar{X}\right)=0,
\]
it follows that
\[
\sum_{i=1}^n \left(X_i-\bar{X}\right)^2
=
\sum_{i=1}^n \left(X_i-\mu\right)^2
-
n\left(\bar{X}-\mu\right)^2.
\]

Hence,
\[
\begin{aligned}
\mathbb{E}\left[\widehat{\sigma}^2\right]
&=
\frac{1}{n}
\mathbb{E}\left[
\sum_{i=1}^n \left(X_i-\bar{X}\right)^2
\right] \\
&=
\frac{1}{n}
\left[
\mathbb{E}\left[
\sum_{i=1}^n \left(X_i-\mu\right)^2
\right]
-
n\mathbb{E}\left[
\left(\bar{X}-\mu\right)^2
\right]
\right].
\end{aligned}
\]

Finally,
\[
\operatorname{Var}(\bar{X})
=
\frac{\sigma^2}{n},
\]
and, because \(\mathbb{E}[\bar{X}]=\mu\),
\[
\mathbb{E}\left[\left(\bar{X}-\mu\right)^2\right]
=
\operatorname{Var}(\bar{X})
=
\frac{\sigma^2}{n}.
\]

Therefore,
\[
\mathbb{E}\left[\widehat{\sigma}^2\right]
=
\frac{1}{n}
\left[
n\sigma^2
-
n\frac{\sigma^2}{n}
\right]
=
\frac{n-1}{n}\sigma^2.
\]

\medskip

If we divide by $n$ as in the maximum likelihood formula, we get this downward bias of factor $\tfrac{n-1}{n}$.  
That is why the unbiased sample variance uses $n-1$ in the denominator:
\begin{equation}
s^2 = \frac{1}{n-1} \sum_{i=1}^n (x_i-\hat{\mu})^2,
\end{equation}
so that
\[
E[s^2] = \sigma^2.
\]

\medskip

\noindent
\textbf{Interpretation.}  
The difference between dividing by $n$ and $n-1$ becomes negligible for large $n$ (since $\tfrac{n-1}{n} \to 1$ as $n\to\infty$).  
But for small samples, the correction is important: the unbiased variance estimator avoids the systematic underestimation that comes from using $\hat{\mu}$ in place of $\mu$.

\subsubsection{Confidence interval of MLE estimators}\label{sec:ciMLEestimators}

We follow \citet{ruppert2015}.

The matrix of second partial derivatives of a function is called its Hessian matrix, so the Fisher information matrix is the expectation of the Hessian of the negative log-likelihood:
\begin{equation}
F_{ij}(\theta) = - E\left[ \frac{\partial^2}{\partial \theta_i \partial \theta_j}  \log \left( L(\theta) \right)\right]
\end{equation}

The standard errors are the square roots of the diagonal entries of the inverse of the Fisher information matrix:
\begin{equation}
s_{\hat{\theta_i}} = \sqrt{\left( F\left( \hat{\theta} \right)^{-1} \right)_{ii}}
\end{equation}

And finally, to be able to estimate a confidence interval:
\begin{quote}
Under \textit{suitable assumptions}, for large enough sample sizes, the maximum likelihood estimator is approximately normally distributed with mean equal to the true parameter vector and with covariance matrix equal to the inverse of the Fisher information matrix.
\end{quote}

For a normal law, we have:
$$\frac{\partial^2 \log L}{\partial \mu^2} = -\frac{\sum_i 1}{\sigma^2} = -\frac{n}{\sigma^2}$$
hence $F(\hat{\mu}) = \frac{n}{\sigma^2}$ and $s_{\hat{\mu}} = \frac{\sigma}{\sqrt{n}}$.
Thus, $\hat{\mu} \sim \mathcal{N}\left(\mu, \frac{\sigma^2}{n} \right)$ or:
\begin{equation}
\frac{\hat{\mu} - \mu}{\frac{\sigma}{\sqrt{n}}} \sim \mathcal{N}(0,1)
\end{equation}
and 
$$ z_{.05/2} \leq \frac{\hat{\mu} - \mu}{\frac{\sigma}{\sqrt{n}}} \leq z_{1-.05/2} $$
finally
\begin{equation}
\mu + z_{.05/2} \frac{\sigma}{\sqrt{n}} \leq \hat{\mu}\leq \mu + z_{1-.05/2} \frac{\sigma}{\sqrt{n}}
\end{equation}
and remember that $z_{.05/2} = - z_{1-.05/2}$.

\subsection{Estimate the probability of occurrence of a given return}
With our DGP in mind (a normal law for the returns) and our empirical observations, we can answer the previous question for the probability to see a daily return of $4.5\%$, or more specifically, the probability that our CAC 40 daily return will fall between $4$ and $5\%$:

\begin{itemize}
\item from empirical distribution, the probability is $0.38\%$;
\item from a parametric normal distribution, the theoretical probability is $0.12\%$.
\end{itemize}

\quad

With a normal distribution we tend to underestimate the probability of having absolute large daily returns.

\subsection{Cumulative distribution function}

The cumulative distribution function (cdf) answers the question: "What is the probability that my returns are below or equal to x?"

$$P(X\leq x)=\int_{-\infty}^x P(X) dX$$ 

We can apply a normal law cdf to the CAC 40 daily return:
\begin{center}
\includegraphics[scale=0.5]{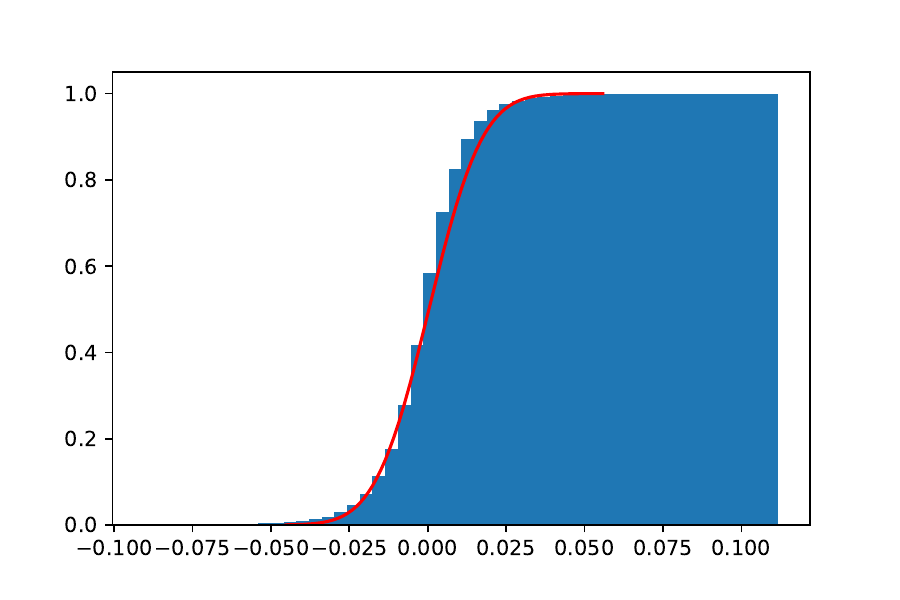}
\end{center}

With the empirical distribution function, for the cumulative up to $4.5\%$ daily return, we obtain $99.6\%$, but for the normal distribution we find $99.95\%$ where again with the normal distribution we underestimate the probability of having daily returns exceeding $4.5\%$: in practice we had $0.4\%$ of the daily returns above $4.5\%$ but the normal distribution function estimates that probability at $0.05\%$.

\stepcounter{question}
\begin{itemize}
\item \textbf{Question \arabic{question}}
\begin{itemize}
\item Find the return $x_i$ so that at this return, the probability to have daily return below is $0.4\%$.
\begin{itemize}
\item We'll see section \ref{seq:VaR}, that you are then searching for a Value at Risk (VaR).
\end{itemize}
\end{itemize}
\end{itemize}

\subsection{What is the mean of the absolute of a normally distributed random variable?}

Now let's imagine that we have a random variable $X$ that follows $\mathcal{N}(0,\,\sigma^2)$, we might wonder the mean of the $\vert X \vert$ random variable, $E(\vert X \vert)$. Writing $x=\sigma z$ where conveniently $Z \sim \mathcal{N}(0,\,1)$, we have $E(\vert X \vert) = \sigma E(\vert Z \vert)$. And:
$$ E(Z) =  \int_{-\infty}^{+\infty} \frac{x}{\sqrt{2\pi}} e^{-\frac{x^2}{2}} dx$$
$$ E(\vert Z \vert) =  \frac{2}{\sqrt{2\pi}} \int_{0}^{+\infty} x e^{-\frac{x^2}{2}} dx$$
$$ E(\vert Z \vert) = \sqrt{\frac{2}{\pi}} \left[-e^{-\frac{t^2}{2}} \right]_O^\infty = \sqrt{\frac{2}{\pi}}$$
Finally, $E(\vert X \vert) = \sigma \sqrt{\frac{2}{\pi}}$

We can verify this by drawing a sample and computing the formula to compare.

\subsection{What is the largest observation you expect to have in a sample?}

Let's imagine that we have a random variable $X$ that follows $\mathcal{N}(\mu,\,\sigma^2)$. If you have a sample of size $N$, what is the largest value you expect to observe in this sample?

\stepcounter{question}
\begin{itemize}
\item \textbf{Question \arabic{question}}
\begin{itemize}
\item Show\footnote{inspired by Jon Keating's lectures} that this is close to the formula equation \ref{eq:JKeating}:
\begin{itemize}
\item Each student chose her own $\mu$ and $\sigma$, then plot the largest value expected as a function of $N$;
\item Generate $N \in [1, 10, 10^2, 10^3, 10^4, 10^5 ]$ observations from a $\mathcal{N}(\mu,\,\sigma^2)$, use the numpy random seed\footnote{Note that numpy Random Generator should be the future option} at arbitrary $7$ so you all have the same results (except for your choices of $\mu$ and $\sigma$);
\item Plot both the expected function and the "empirical" largest values. You might want to adapt the range of $N$.
\end{itemize}
\end{itemize}
\end{itemize}

\begin{equation}\label{eq:JKeating}
\mu + \sigma \sqrt{2 \log N} - .5 \sigma \frac{\log \log N}{\sqrt{2\log N}}
\end{equation}

\subsection{Exponential distribution}

Exponential distributions are a particular case of gamma distribution. We write the probability density function of the exponential distribution:
\begin{equation*}
    f(x;\alpha) =
    \begin{cases}
\alpha e^{-\alpha x},& \text{for} \quad x \geq 0 \\     
0 & \text{for} \quad x < 0
    \end{cases}
\end{equation*}

For illustration here the exponential probability density function:
\begin{center}
\includegraphics[scale=0.5]{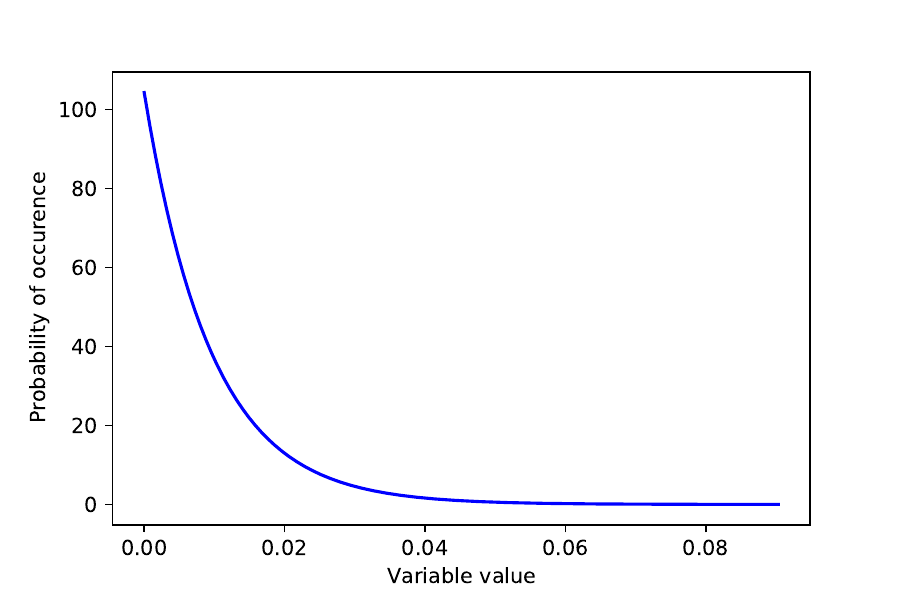}
\end{center}

The likelihood function on a sample of size $n$ is:
$$ L(x_1,\ldots,x_n\vert \alpha) = \alpha^n \exp\left(-\alpha \sum_{i=1}^n x_i \right) $$

Deriving this expression with respect to $\alpha$ to find an optimum $\alpha_o$:
$$n\alpha_o^{n-1} \exp\left(-\alpha_o \sum_{i=1}^n x_i \right) - \sum_{i=1}^n x_i \alpha_o^n \exp\left(-\alpha_o \sum_{i=1}^n x_i \right)=0$$ an optimum is thus find for the reciprocal of the sample mean:
$$ \alpha_o = \frac{n}{\sum_{i=1}^n x_i}$$

\quad

For the exponential law the first moment (the mean) is:
\begin{equation*}
\mu^1=\alpha \int_{0}^{+\infty} x e^{-\alpha x} dx = \alpha \left[ e^{-\alpha x} \left( \frac{-\alpha x - 1}{\alpha^2} \right) \right]_0^\infty = \frac{1}{\alpha}
\end{equation*}
and we can demonstrate that the variance of the exponential law is $\frac{1}{\alpha^2}$.

\quad

We can fit this to our negative CAC 40 return (we need to take the absolute values to be able to fit our exponential distribution):
\begin{center}
\includegraphics[scale=0.5]{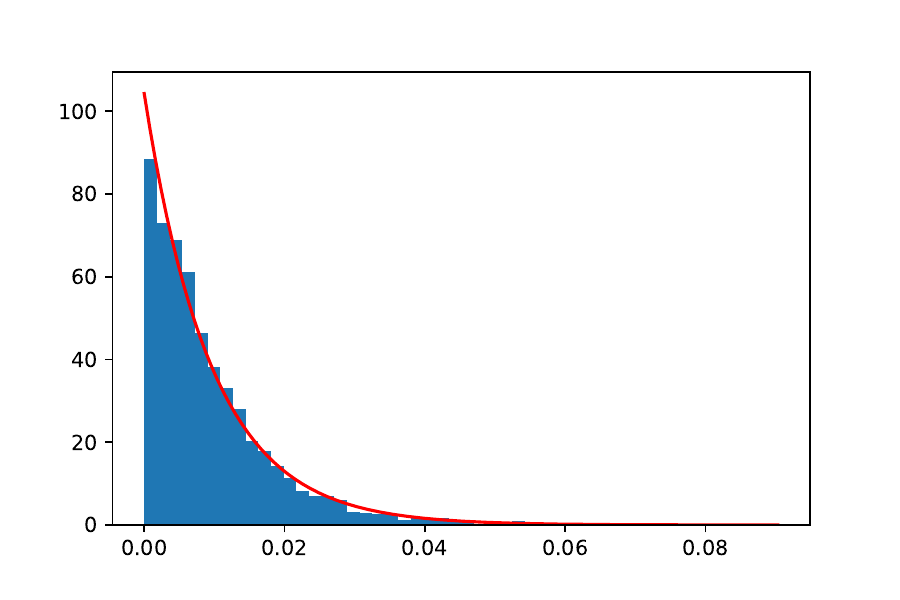}
\end{center}


\subsection{Gamma distribution}

We use the Gamma distribution for the probability density function $f_X(x) = \frac{r^s}{\Gamma(s)}x^{s-1}e^{-r x}$, $x>0$, $s>0$ the shape parameter, $r>0$ the rate parameter and $\Gamma(s) = \int_0^{+\infty}u^{s-1}e^{-u}du$, we have the property\footnote{some elements of demonstration \href{https://en.wikipedia.org/wiki/List_of_integrals_of_exponential_functions}{here}}
$\Gamma(s+1)=s \Gamma(s)$ and $\frac{\Gamma(s+1)}{\beta^{s+1}} = \int_0^{+\infty}u^{s}e^{-\beta u}du$.
Such distribution can be useful in insurance modelling (claims size distribution).
If $s>1$ then it is alike a log-normal distribution:
\begin{center}
\includegraphics[scale=0.5]{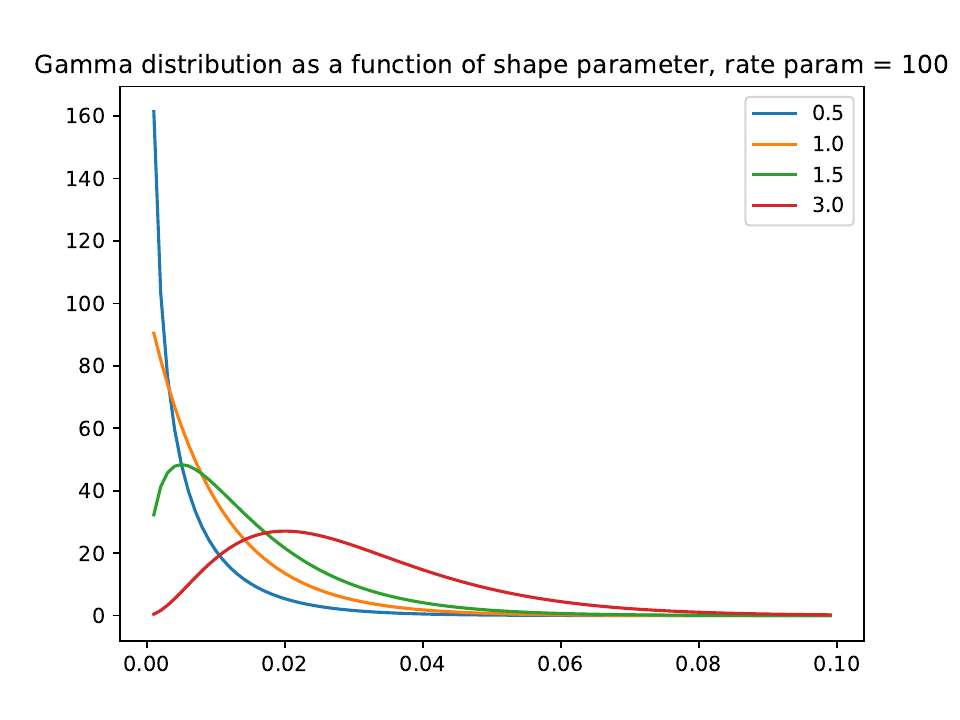}
\includegraphics[scale=0.5]{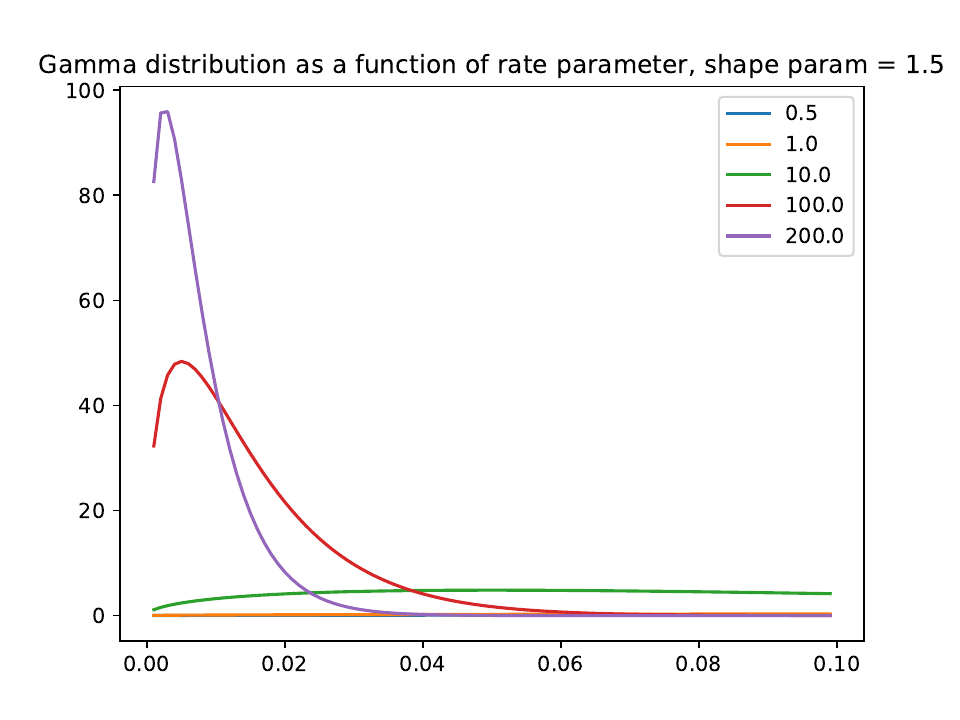}
\end{center}
For a random variable following a gamma distribution, we have:
$$E(X) = \int_0^{+\infty} \frac{r^s}{\Gamma(s)} u^{s}e^{-r u}du = \frac{r^s}{\Gamma(s)} \int_0^{+\infty}  u^{s} e^{-r u}du$$
$$E(X) = \frac{r^s}{\Gamma(s)} \frac{\Gamma(s+1)}{r^{s+1}} = \frac{s}{r} $$

For the variance, we have $Var(X) = E(X^2) - \left(E(X) \right)^2$ and $E(X^2) = \frac{s (s+1)}{r^2}$ so $Var(X) = \frac{s}{r^2}$.

\subsection{Generalized extreme value distribution}\label{sec:GEVintro}

We introduce the Generalized Extreme Value distribution which has a cumulative distribution function:
\begin{equation}
H_{\xi, \mu, \sigma}(x)= e^{-\left(1+\xi \frac{x-\mu}{\sigma} \right)^{-\frac{1}{\xi}} }
\end{equation}

for $1+\xi\left(\frac{x-\mu}{\sigma}\right)>0$, $\xi \neq 0$ the shape parameter, $\mu$ the location parameter and $\sigma>0$ the scale parameter. We can derive a probability density function
\begin{equation}\label{eq:gevpdf}
h_{\xi, \mu, \sigma}(x) =  \frac{1}{\sigma} \left(1+\xi \frac{x-\mu}{\sigma} \right)^{-\frac{1}{\xi}-1}e^{-\left(1+\xi \frac{x-\mu}{\sigma} \right)^{-\frac{1}{\xi}} }
\end{equation}

This distribution is sometimes called the Fisher-Tippett distribution. 

For $\xi = 0$, we are back to the Gumbel distribution introduced equation \ref{eq:gumbelfrechetweibull}:
\begin{equation}
G_{ \mu, \sigma}(x) =  \exp\left[ - exp\left(-\frac{x-\mu}{\sigma} \right) \right]
\end{equation}

We write the probability density function of the Gumbel distribution:
\begin{equation}
g_{ \mu, \sigma}(x) =\frac{1}{\sigma}  \exp\left[ -\frac{x-\mu}{\sigma}  - exp\left(-\frac{x-\mu}{\sigma} \right)  \right]
\end{equation}

For illustration here the standardized GEV probability density function for different values of $\xi$:
\begin{center}
\includegraphics[scale=0.3]{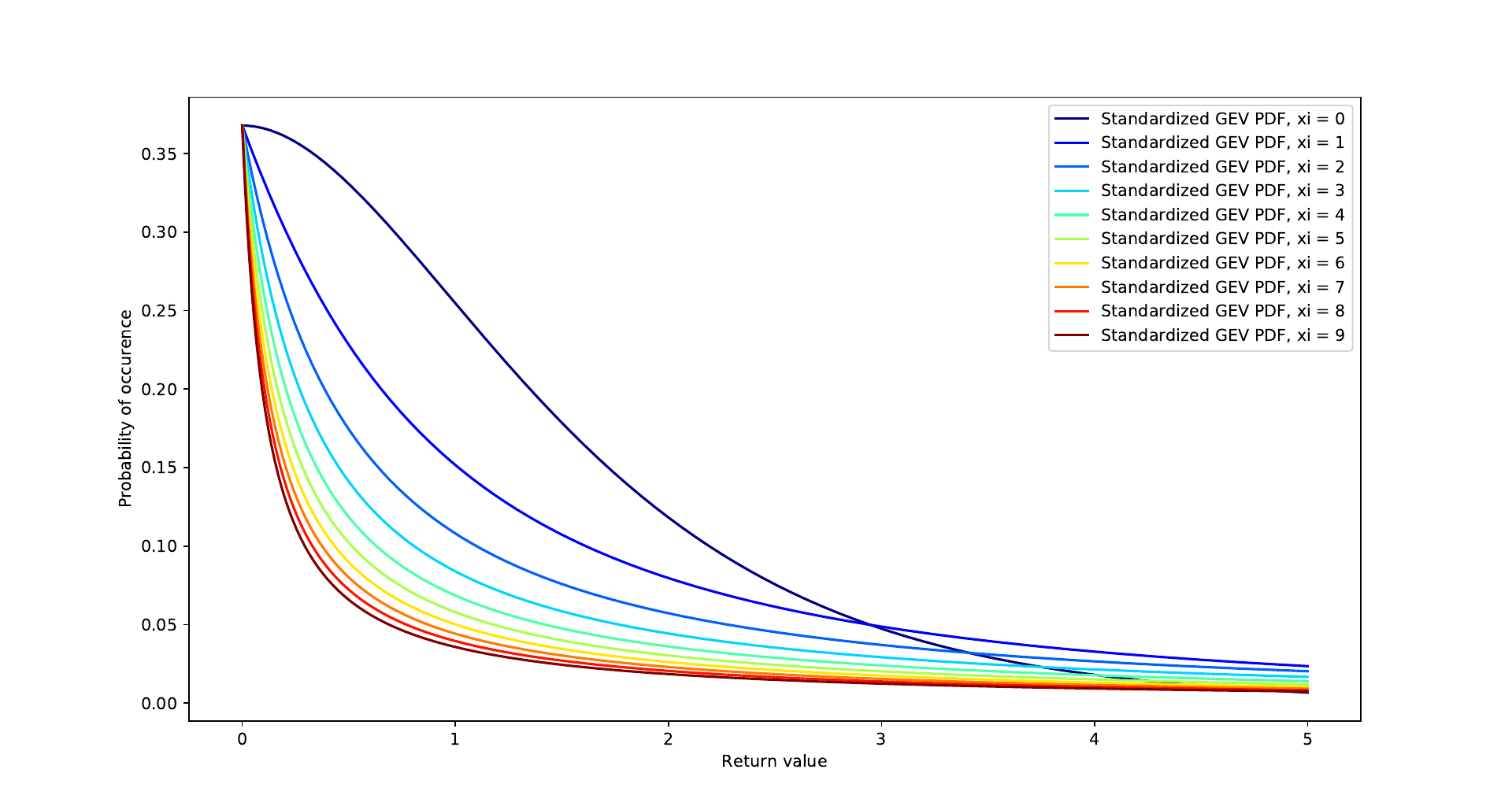}
\end{center}

For $\xi \geq 1$ this distribution has neither finite mean nor variance, illustration wih $\xi=1.5$:
\begin{center}
\includegraphics[scale=0.5]{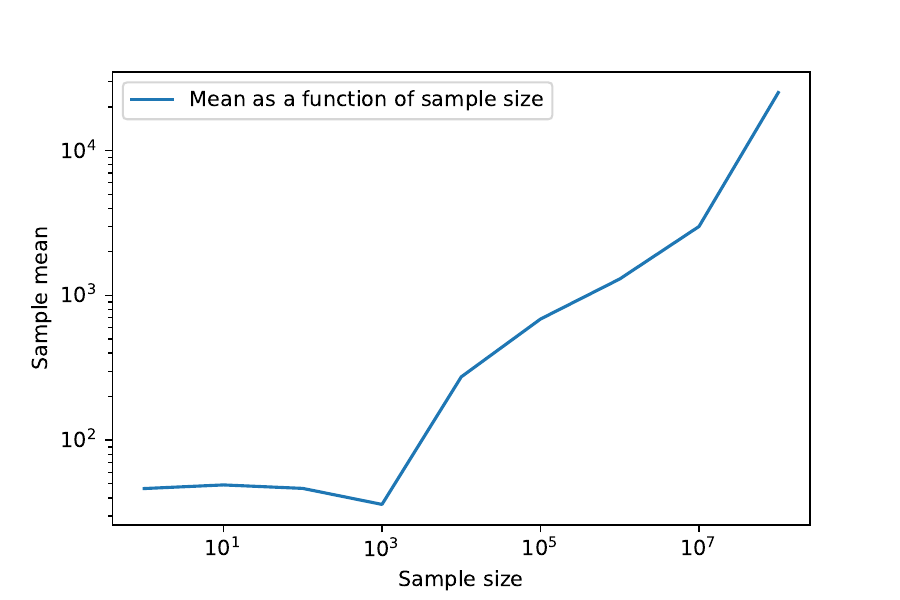}
\end{center}

To build confidence interval of the GEV estimators, you can follow section \ref{sec:ciMLEestimators} and the expressions of the Fischer information matrix are derived in \citet{Prescott:1983ur}.

\subsection{Compound Poisson-Exponential distributions}

Typically, in insurance claims modelling (non-life), we model the occurrence of claims as a Poisson\footnote{The number of claims up to time $t$, $N_t$, follows a Poisson distribution of parameter $\lambda t$, with probability mass function $P(N_t=k) = e^{-\lambda t} \frac{(\lambda t)^k}{k!}$, $k \in \{0,1,2,\dots\}$. In particular, $E[N_t] = Var[N_t] = \lambda t$.} process $(N_t)_{t \geq 0}$ of intensity $\lambda > 0$ and the magnitudes of claims as i.i.d.\ exponential random variables of rate $\alpha > 0$, with probability density function $\alpha e^{-\alpha c}$ for $c \geq 0$, independent of the claim arrivals. Writing that the claim epochs follow a Poisson process is equivalent to writing that the inter-occurrence times are i.i.d.\ exponential with rate parameter $\lambda$, with probability density function $\lambda e^{-\lambda x}$ for $x \geq 0$.

We model an insurance company facing claims arriving over time according to the Poisson process $N_t$ and whose claim sizes $C_i$ follow an exponential law. We have the compound process $X_t=\sum_{i=1}^{N_t} C_i$ (with $X_t = 0$ when $N_t = 0$) and its estimate $\hat{X}_t=\sum_{i=1}^{N_t} \hat{C}_i$.

Denote $\mu = E[C] = 1/\alpha$ and $\sigma^2 = Var[C] = 1/\alpha^2$, so that $E[C^2] = \sigma^2 + \mu^2 = 2/\alpha^2$. Conditioning on $N_t$ and using the independence between $N_t$ and the $C_i$, $E[X_t \mid N_t] = N_t \mu$ and $Var[X_t \mid N_t] = N_t \sigma^2$. Hence
\begin{equation}
E[X_t] = E\big[E[X_t \mid N_t]\big] = E[N_t]\, \mu = \frac{\lambda t}{\alpha},
\end{equation}
and, by the law of total variance,
\begin{equation}
\begin{split}
Var[X_t] &= E\big[Var[X_t \mid N_t]\big] + Var\big[E[X_t \mid N_t]\big] = E[N_t]\, \sigma^2 + Var[N_t]\, \mu^2 \\
&= \lambda t \left( \sigma^2 + \mu^2 \right) = E[N_t]\, E[C^2] = \frac{2 \lambda t}{\alpha^2}.
\end{split}
\end{equation}
The simplification $Var[X_t] = E[N_t]\, E[C^2]$ relies on the equidispersion of the Poisson distribution, $Var[N_t] = E[N_t]$. For $t = 1$, we recover $E[X_1] = \lambda/\alpha$ and $Var[X_1] = 2\lambda/\alpha^2$.

\subsection{Probability Integral Transform}\label{sec:PIT}

Let \(X\) be a continuous random variable with a continuous and strictly
increasing cDF \(F_X\). Define
\[
U = F_X(X).
\]
Then \(U\sim\mathcal U(0,1)\). In particular, its cDF and density are
\[
F_U(u)=u
\quad\text{and}\quad
f_U(u)=1,
\qquad u\in[0,1].
\]
This result is known as the probability integral transform
\citep{hansen2022}.

Indeed, \(U\) takes values in \([0,1]\) because \(F_X(x)=\Pr(X\leq x)\).
Moreover, for \(u\in(0,1)\),
\begin{equation}
\begin{aligned}
F_U(u)
&= \Pr(U\leq u) \\
&= \Pr\bigl(F_X(X)\leq u\bigr) \\
&= \Pr\bigl(X\leq F_X^{-1}(u)\bigr) \\
&= F_X\bigl(F_X^{-1}(u)\bigr) \\
&= u.
\end{aligned}
\end{equation}
The third equality follows because \(F_X\) is strictly increasing.
Differentiating \(F_U(u)=u\) yields \(f_U(u)=1\).

Conversely, let \(V\sim\mathcal U(0,1)\), and define
\[
\widetilde X = F_X^{-1}(V).
\]
Then \(\widetilde X\) has cDF \(F_X\). Indeed, for every \(x\),
\begin{equation}
\begin{aligned}
\Pr(\widetilde X\leq x)
&= \Pr\bigl(F_X^{-1}(V)\leq x\bigr) \\
&= \Pr\bigl(V\leq F_X(x)\bigr) \\
&= F_V\bigl(F_X(x)\bigr) \\
&= F_X(x).
\end{aligned}
\end{equation}
The last equality holds because \(V\sim\mathcal U(0,1)\), so that
\(F_V(v)=v\) for every \(v\in[0,1]\), and \(F_X(x)\in[0,1]\).
Thus, applying the inverse cDF to a uniform random variable generates a
random variable with cDF \(F_X\). This procedure is called inverse
transform sampling.

Let's imagine we want to generate a rv following the Pareto cdf:
\begin{equation}
F_X(x) = 1-\left(\frac{x_m}{x} \right)^\alpha
\end{equation}
with $x\geq x_m >0$ and $\alpha>0$. We take\footnote{vansteenberghe\_probability\_integral\_transformation.py} $Y \sim \mathcal U[0,1]$. We want $F_X(x)=y$, so we solve for $x$:
\begin{equation*}
y=1-\left(\frac{x_m}{x} \right)^\alpha
\;\Leftrightarrow\;
\left(\frac{x_m}{x}\right)^\alpha = 1-y
\;\Leftrightarrow\;
\frac{x_m}{x} = (1-y)^{\frac{1}{\alpha}}
\;\Leftrightarrow\;
x= x_m (1-y)^{-\frac{1}{\alpha}} = F_X^{-1}(y),
\end{equation*}
so we take $X= x_m (1-Y)^{-\frac{1}{\alpha}}$, which satisfies $X\geq x_m$ as $1-Y\in(0,1]$.

In the code, we start by generating a sample uniformly distributed, so we can check that as expected our sample mean is close to $0.5$ and our sample standard deviation close to $\sqrt{1/12}\approx 0.289$, remembering that the density is $f_Y(y)=1$ on $[0,1]$ (a density, not a probability: $P(Y=y)=0$ for a continuous rv):
\begin{align*}
E(Y) &= \int_0^1 y\, f_Y(y)\, dy = \int_0^1 y\, dy = \left[\frac{y^2}{2}\right]_0^1 = \frac{1}{2},\\
s(Y) &= \sqrt{\int_{0}^1 \left[y- \frac{1}{2} \right]^2 f_Y(y)\, dy} = \sqrt{\left[ \frac{\left( y-\frac{1}{2}\right)^3}{3} \right]_0^1} = \sqrt{\frac{1}{3}\left(\frac{1}{8}+\frac{1}{8}\right)} = \sqrt{\frac{1}{12}} \approx 0.289.
\end{align*}
We then transform each uniform draw $y_i$ into a Pareto draw $x_i= x_m (1-y_i)^{-\frac{1}{\alpha}}$.

\section{Python: Sample statistics, Law of Large Numbers and Central Limit Theorem}\label{sec:largen}

\subsection{Mean and Standard Deviation in theory}

If we have a sample of size $n$ from a population, we consider\footnote{vansteenberghe\_Central\_Limit\_Theorem.py} the random variable $X$ which have the outcome values $x$ (that would be $x \in [0,\infty [$ for a stock price), each outcome having the probability of occurring of $p(x)$. We can consider that this random variables are taken from a greater population with mean $\mu$ and standard deviation $\sigma$ (although a price process would never follow a normal law). In practice, we do not have access to the full population and have to work with finite samples.

Note that the outcome values $X$ are considered a random sample of independent and identically distributed random variables with the same probability density function. We will see in section \ref{sec:distribtuions} possible distributions function for financial returns.

In practical terms, if over 5 days we observe the following price values $x_t$ for a stock $X$, we have the probabilities:
\begin{center}

\begin{tabular}{|l|l|l|l|l|l|}
\hline
$t$&1&2&3&4&5\\
\hline
$x_t$&1.1&1.2&1.3&1.2&1.2\\
\hline
\end{tabular}

\quad

\begin{tabular}{lll}
$p(1.1) = \frac{1}{5}$&$p(1.2)=\frac{3}{5}$&$p(1.3)=\frac{1}{5}$\\
\end{tabular}

\quad

\end{center}

This is a very basic example and in practice you would need much more observations to work with plausible probability values.

The expected value of your random variable $X$ is called the sample mean, we can show that it is unbiased compared with the population true mean:
\begin{equation*}
E(\bar{X}_n)= E\left(\frac{1}{n} \sum_{i=1}^n X_i \right) = \frac{1}{n} \sum_{i=1}^n E(X_i)=\frac{1}{n} n \mu = \mu 
\end{equation*}

We can also write:
\begin{equation*}
E\left(\bar{X}_n\right) = \sum_x x p(x)
\end{equation*}

In our example, this would be: $E(X) = 1.1 \times \frac{1}{5} + 1.2 \times \frac{3}{5} + 1.3 \times \frac{1}{5} = \frac{6}{5} = 1.2$

We can state the weak law of large numbers:

\subsection{Weak Law of Large Numbers (WLLN)}\label{sec:WLLN}
\textbf{Statement:} Let $X_1, X_2, \ldots$ be a sequence of i.i.d. random variables with expected value $E(X_i) = \mu$ and variance $Var(X_i) = \sigma^2 < \infty$. Then, for any $\epsilon > 0$,
\[
\lim_{n \to \infty} P\left( \left| \frac{1}{n} \sum_{i=1}^{n} X_i - \mu \right| > \epsilon \right) = 0
\]
\textbf{Illustration:} Imagine flipping a fair coin, where Heads is $1$ and Tails is $0$. The expected value is $\mu = 0.5$. If you flip the coin many times and take the average of the results, this average will converge to 0.5 as the number of flips goes to infinity.

Another more mathematical way of writing the WLLN:
If we have $X_i, i\in[1,n]$ iid, with $E\left( \vert X \vert \right)<\infty$, then $\bar{X}_n = \frac{1}{n} \sum_{i=1}^n X_i \overset{p}{\to} E(X)$.

This second formulation only requires a finite mean, not a finite variance \citep[Theorem 7.4]{hansen2022}. The finite-variance version is the one we can prove in two lines, and the proof shows why the sample mean settles down.

\textbf{Why it holds:} the sample mean is unbiased, $E(\bar{X}_n)=\mu$ (see above), and its variance shrinks with the sample size:
\begin{equation*}
V(\bar{X}_n) = V \left( \frac{1}{n} \sum_{i=1}^n X_i \right) = \frac{1}{n^2} V \left(  \sum_{i=1}^n X_i \right) = \frac{1}{n^2}  n V \left( X \right) = \frac{\sigma^2}{n}
\end{equation*}
where the third equality uses independence (the variance of a sum of independent random variables is the sum of their variances) and the identical distribution. Chebyshev's inequality states that for any random variable $Z$ with finite variance and any $\epsilon > 0$, $P\left( \vert Z - E(Z) \vert > \epsilon \right) \leq V(Z) / \epsilon^2$. Applied to $Z = \bar{X}_n$:
\begin{equation*}
P\left( \left| \bar{X}_n - \mu \right| > \epsilon \right) \leq \frac{V(\bar{X}_n)}{\epsilon^2} = \frac{\sigma^2}{n \epsilon^2} \xrightarrow[n \to \infty]{} 0
\end{equation*}
which is the WLLN. The standard deviation of $\bar{X}_n$, $\sigma / \sqrt{n}$, is also the scale at which the central limit theorem below describes the error $\bar{X}_n - \mu$.

\subsection{Strong Law of Large Numbers (SLLN): which law do we use?}\label{sec:SLLN}
Under the same assumption $E\left( \vert X \vert \right)<\infty$, Kolmogorov's strong law of large numbers states that
\begin{equation*}
P\left( \lim_{n \to \infty} \bar{X}_n = \mu \right) = 1, \quad \mbox{written} \quad \bar{X}_n \overset{\text{a.s.}}{\to} \mu \quad \mbox{(almost surely, i.e.\ with probability one)}
\end{equation*}
The two laws differ by the position of the limit: the WLLN is the limit of a probability, the SLLN is the probability of a limit. Almost sure convergence implies convergence in probability, but not the converse, so the SLLN is the stronger statement; its proof needs heavier tools (Kolmogorov's inequality or the Borel--Cantelli lemma) and we refer to \citet[section 7.13]{hansen2022}.

In the coin example, the WLLN says: fix a large number of flips $n$, then the probability that the proportion of Heads differs from $0.5$ by more than $\epsilon$ is small. The SLLN says: keep flipping forever, then with probability one the running proportion of Heads converges to $0.5$. The first statement is about a sample of a given size, the second about one realised sequence of draws. This is what decides which law we use:
\begin{itemize}
\item consistency of an estimator, such as OLS in section \ref{sec:weightheightregression}, is defined as convergence in probability \citep{verbeek}, and the results we combine the law of large numbers with (the central limit theorem, the delta method) are stated in probability or in distribution. Econometrics therefore works with the WLLN \citep[section 7.13]{hansen2022}, which is usually called simply ``the law of large numbers'' and is the version used throughout these notes;
\item the SLLN is the natural statement when we watch a single sequence unfold: a Monte Carlo simulation run with $N$ increasing (section \ref{sec:MCintegration}, where we wrote $\widehat{I}_N \overset{\text{a.s.}}{\to} I$), or the time average of one realised time series, which is all we ever observe in time series; its extension to stationary dependent observations, the ergodic theorem, is what we invoke for the empirical moments of a time series in section \ref{sec:stationarity}.
\end{itemize}
Both laws require a finite mean. If the $X_i$ are Cauchy distributed, $\bar{X}_n$ has exactly the same distribution as $X_1$ for every $n$ and the sample mean never settles down, as illustrated in section \ref{sec:medianmean}: a warning to keep in mind with the heavy-tailed distributions of section \ref{sec:distribtuions}.

\subsection{Estimating the variance from the sample}\label{sec:samplevariance}
If we further assume $E(X)=\mu$ and $\sigma(X) = \sigma$, the plug-in estimator of the variance, $\frac{1}{n} \sum_{i=1}^n \big[X_i-\bar{X}_n\big]^2$, is biased downward because the deviations are measured from $\bar{X}_n$ rather than from $\mu$: writing $X_i - \bar{X}_n = (X_i - \mu) - (\bar{X}_n - \mu)$, expanding the square and using $V(\bar{X}_n)=\sigma^2/n$,
\begin{equation*}
E \left( \sum_{i=1}^n \big[X_i-\bar{X}_n\big]^2 \right) = E \left( \sum_{i=1}^n \big[X_i-\mu\big]^2 \right) - n E \left( \big[\bar{X}_n-\mu\big]^2 \right) = n \sigma^2 - n \frac{\sigma^2}{n} = (n-1) \sigma^2
\end{equation*}
If we want an unbiased variance\footnote{this can be measured using the standard deviation function with a degree of freedom of 1 ($\mbox{ddof} = 1$)} estimator from our sample, we therefore divide by $n-1$:
\begin{equation*}
S_n^2 = \frac{1}{n-1} \sum_{i=1}^n \big[X_i-\bar{X}_n\big]^2, \qquad E\left(S_n^2\right) = \sigma^2
\end{equation*}
Both estimators converge in probability to $\sigma^2$ by the law of large numbers applied to $X_i$ and $X_i^2$, the bias $-\sigma^2/n$ of the plug-in version vanishing as $n \to \infty$.

The variance of the population is:
\begin{equation*}
V(X) = E \bigg( \big[X-E(X)\big]^2 \bigg) = \sum_x p(x) \big[x-E(X)\big]^2
\end{equation*}
In our example, treating the five observed prices as the population ($\mbox{ddof} = 0$), this would be: $V(X) = (1.1-1.2)^2 \times \frac{1}{5} + (1.2-1.2)^2 \times \frac{3}{5} + (1.3-1.2)^2 \times \frac{1}{5} = 0.004$, while the unbiased estimator gives $S_n^2 = \frac{0.02}{4} = 0.005$.

\subsection{Central Limit Theorem (CLT)}\label{sec:CLT}

\textbf{Statement:} Let $X_1, X_2, \ldots$ be i.i.d. random variables with $E(X_i) = \mu$ and $Var(X_i) = \sigma^2 < \infty$. Then, as $n \to \infty$, the random variable
\[
Z_n = \frac{\sum_{i=1}^{n} X_i - n\mu}{\sigma \sqrt{n}}
\]
converges in distribution to the standard normal distribution, i.e.,
\[
Z_n \xrightarrow{d} Z \sim N(0,1)
\]

\stepcounter{question}
\begin{itemize}
\item \textbf{Question \arabic{question}}
\begin{itemize}
\item Consider rolling a six-sided die. The expected value is $ \mu = 3.5 $ and the variance is $ \sigma^2 = \frac{35}{12} $ (cf. 
\emph{Discrete uniform distribution}). If you roll the die many times, sum the results, subtract $n \times 3.5$, and then divide by the standard deviation times the square root of $n$, illustrate visually that the result will converge in distribution to the standard normal distribution as $n$ goes to infinity.
\end{itemize}
\end{itemize}

A more compact way to write the CLT:
If we have $X_i, i\in[1,n]$ iid, with $E\left( \vert X \vert \right)<\infty$, with $E(X) = \mu$ and $\sigma(X) = \sigma$, then if we define $\bar{Z}_n = \frac{\bar{X}_n - \mu}{\sqrt{\frac{\sigma^2}{n}}}$, then $\bar{Z}_n$ converges in distribution, $n \to \infty$ to the standard normal distribution $\mathcal{N}(0,\,1)$.

\quad

We suggest to illustrate this with samples taken from a population that has a uniform distribution function, with its probability density function:

\begin{equation*}
    f(x) =
    \begin{cases}
\frac{1}{b-a},& \text{for} \quad a \leq x \leq b \\     
0 & \text{for} \quad x < a \quad \text{or} \quad x>b
    \end{cases}
\end{equation*}

we take $a=0$ and $b=1$.

\stepcounter{question}
\begin{itemize}
\item \textbf{Question \arabic{question}}
\begin{itemize}
\item Use a standardized generalized extreme value distribution with $\xi \geq 1$, as this distribution has no finite mean, is the central limit theorem still applicable?
\end{itemize}
\end{itemize}

\subsubsection{Gosset (``Student'') and the Student's \texorpdfstring{$t$}{t} distribution}

Let \(X_1,\ldots,X_n\) be i.i.d.\ random variables with mean \(\mu\) and
variance \(\sigma^2<\infty\). The Central Limit Theorem states that
\[
\frac{\bar X_n-\mu}{\sigma/\sqrt{n}}
\xrightarrow{d} \mathcal{N}(0,1)
\qquad \text{as } n\to\infty.
\]

If, in addition, \(X_i\sim\mathcal{N}(\mu,\sigma^2)\), this result holds
exactly for every \(n\):
\[
\frac{\bar X_n-\mu}{\sigma/\sqrt{n}}
\sim \mathcal{N}(0,1).
\]

Define the unbiased sample variance as
\[
S_n^2=\frac{1}{n-1}\sum_{i=1}^n (X_i-\bar X_n)^2.
\]
For a normal sample, Gosset (writing under the pseudonym ``Student'')
showed that
\[
\frac{\bar X_n-\mu}{S_n/\sqrt{n}}
\sim t_{n-1},
\]
where \(t_{n-1}\) is the Student's \(t\) distribution with \(n-1\)
degrees of freedom.

Finally,
\[
t_\nu \xrightarrow{d} \mathcal{N}(0,1)
\qquad \text{as } \nu\to\infty.
\]

\begin{center}
\includegraphics[scale=0.5]{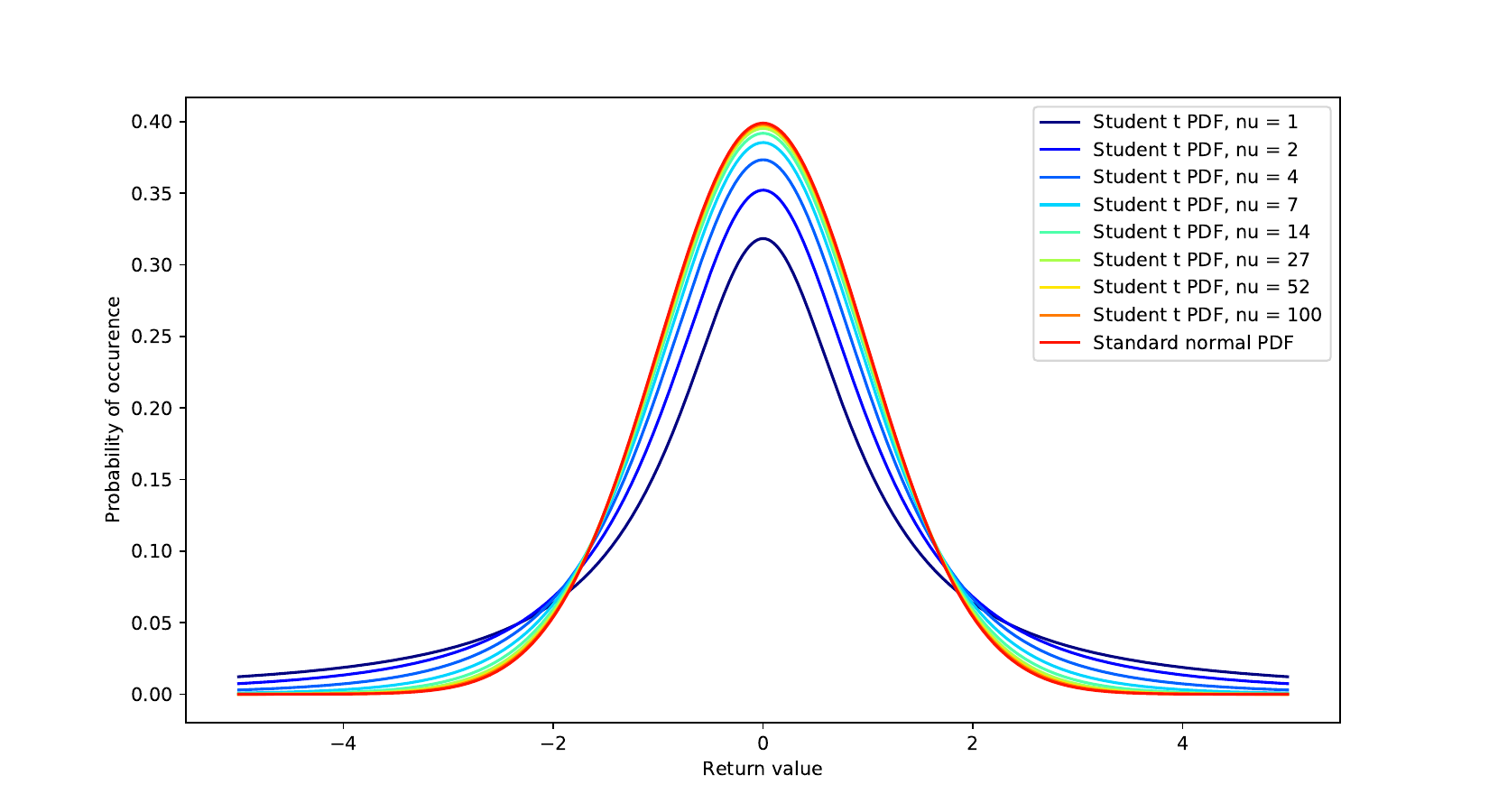}
\end{center}

\subsubsection{Skewness, Kurtosis and covariance in theory}

The standard deviation of our sample is defined as:
\begin{equation*}
\sigma_X = \sqrt{V(X)}
\end{equation*}
its skewness is defined as:
\begin{equation*}
S(X) = E\bigg[\left( \frac{X-E(X)}{\sigma_X} \right)^3 \bigg]
\end{equation*}
its kurtosis is defined as:
\begin{equation*}
K(X)=E\bigg[\left( \frac{X-E(X)}{\sigma_X} \right)^4 \bigg]
\end{equation*}

The covariance between two random variables $X$ and $Y$ indicates how likely they are to occur together:
\begin{equation*}
Cov(X,Y)= E\left[ \left(X-E(X)\right) \left(Y - E(Y)\right) \right] = E (XY) - E(X)E(Y)
\end{equation*}

\subsubsection{Mean, Standard Deviation and Correlation in practice}\label{sec:correlation}
If we look at an observation vector $X=[x_0,x_1,\ldots,x_{T-1}]$ (like the close price of a company's stock), we can call $E(X)$ the expected price, which in practice will be the average over the sample: $$E(X)=\frac{1}{T}\sum_{i=0}^{T-1} x_i$$

The standard deviation is a measure of risk expressed in $\%$: $$\sigma(X)=\sqrt{\frac{1}{T}\sum_{i=0}^{T-1} \left(x_i-E(X) \right)^2}$$

The covariance of $X$ with another vector of observations $Y=[y_0,y_1,\ldots,y_{T-1}]$ can be computed as: $$cov(X,Y)=\frac{1}{T}\sum_{i=0}^{T-1} \left[x_i-E(X) \right] \left[y_i-E(Y) \right]$$

The Pearson coefficient of correlation between $X$ and $Y$ is then:
$$corr(X,Y)=\frac{cov(X,Y)}{\sigma(X) \sigma(Y)}$$

\subsubsection{Correlation Measures techniques}\label{sec:corronreturns}

We can detail the correlation measures:
\begin{itemize}
\item Pearson correlation coefficient $\in [-1,1]$:	
$$r = \frac{\sum (x_i- \bar{x})(y_i- \bar{y})}{\sqrt{\sum (x_i- \bar{x})^2\sum (y_i- \bar{y})^2} }$$
\item Kendall rank correlation coefficient:
$$\tau=\frac{n_c-n_d}{\frac{1}{2}n(n-1)}$$
\item Spearman rank correlation coefficient:
$$\rho=1-\frac{6\sum d_i^2}{n(n^2-1)}$$
\end{itemize}

with:
\begin{itemize}
\item $n$: number of value in each data set
\item $n_c$: number of concordant, a pair $(x_i,y_i)$ is said to be concordant with the pair $(x_j,y_j)$ if either both $x_i < x_j$ and $y_i < y_j$ or $x_i > x_j$ and $y_i > y_j$
\item $n_d$: number of discordant, a pair $(x_i,y_i)$ is said to be discordant with the pair $(x_j,y_j)$ if either $x_i < x_j$ and $y_i > y_j$ or $x_i > x_j$ and $y_i < y_j$
\item $d_i$: the difference between the ranks\footnote{think of ranking from first, second to last, with possibly similar ranking when the values are equal, then you need to adjust for ties (take the mean rank not the upper nor lower rank of the tie)} of corresponding values $x_i$ and $y_i$
\end{itemize}

One might wonder: should correlation be applied to prices or returns? You can find detailed advice to apply correlation on asset returns \href{https://quantdare.com/correlation-prices-returns/}{here}, \href{http://t.www.na-businesspress.com/JBD/HaberJ_Web12_3_.pdf}{here} and \href{https://www.portfolioprobe.com/2011/01/12/the-number-1-novice-quant-mistake/}{here}.

\section{Modeling and generating random vectors}

I follow here \citet{johnson1987multivariate}.

\subsection{Inverse distribution function method}

We have seen how to generate random vectors from a uniform distribution. Now we want to generate a variable $X$ whose cumulative distribution function is specified by $F$.

The inverse distribution function method has two steps:
\begin{itemize}
\item Generate a variable $U$ that is uniform;
\item Let $X=\mbox{Sup}_x \left[F(x)\leq Y \right]$.
\end{itemize}
In the event that $F$ is strictly increasing, then $F^{-1}$ is well defined and step 2 becomes: $X=F^{-1}(U)$.

I test\footnote{numpy\_exercise\_part4.py} the following two:
\begin{itemize}
\item exponential: $F^{-1}(u) = -\ln(1-u)$
\item Pareto: $F^{-1}(u) = (1-u)^{-\frac{1}{\alpha}}$
\end{itemize}

When no closed form of $F^{-1}$ is available, the following recursion\footnote{Cf the Newton's method equation \ref{eq:newtonmethod}.} frequently converges to a fixed point $x^*$:
$$x_{n+1}=x_n - \frac{F(x_n)-U}{f(x_n)}$$

\subsection{Normal generator}

A standard normal law can be generated with two independent uniform vectors $U_1$ and $U_2$:
$$X = \sqrt{-2 \ln(U_1)}\cos(2\pi U_2)$$
or 
$$X = \sqrt{-2 \ln(U_1)}\sin(2\pi U_2)$$

Once a standard normal variable $X$ is generated, a normal variable $Y\sim \mathcal{N}(\mu,\sigma^2)$ can be generated with $Y = \sigma X + \mu$

Finally, a log-normal variable can be generated from a normal variable taking its exponential (Box–Muller transform).

\subsection{Multivariate normal distribution}

A $p$-dimensional random vector $X=\left(X_1,\dots,X_p \right)'$ is defined to have a multivariate normal distribtuion if and only if every nontrivial linear combination of the $p$-components of $X$ has a univariate normal distribution. The distribution of $X$ is denoted $\mathcal{N}_p\left(\mu, \Sigma \right)$, with $\mu_i=E(X_i)$ and $\Sigma_{i,j}=cov(X_i,X_j)$. If $\Sigma$ has full rank $p$, then the density function of $X$ exists. Let $L$ be the Choleski factorization which is then unique lower triangular such that $LL'=\Sigma$. Take $p$ univariate independent random vector $Z=\left(Z_1,\dots,Z_p \right)'$  with $Z\sim \mathcal{N}_p(0,I)$ if I define $X=LZ+\mu$, then $X\sim \mathcal{N}_p\left(\mu, \Sigma \right)$.

\subsection{Multivariate distributions: an introduction to NORTA}
\label{sec:NORTA}

Suppose that we want to simulate a random vector
\[
\mathbf{X}=(X_1,\ldots,X_d)^\top
\]
with prescribed marginal cdfs \(F_{X_1},\ldots,F_{X_d}\), while also reproducing a desired dependence structure, for example a target covariance matrix \(\Sigma_X\).

NORTA stands for \emph{NORmal To Anything}. The idea is to start from independent standard Gaussian shocks, introduce Gaussian dependence, and then transform each Gaussian variable into a variable with the desired marginal distribution:
\[
\mathbf{W}
\quad \longrightarrow \quad
\mathbf{Z}
\quad \longrightarrow \quad
\mathbf{U}
\quad \longrightarrow \quad
\mathbf{X}.
\]

More precisely:

\begin{enumerate}
    \item Generate independent standard Gaussian variables:
    \[
    \mathbf{W}=(W_1,\ldots,W_d)^\top
    \sim \mathcal{N}_d(\mathbf{0},\mathbf{I}_d).
    \]
    Thus, the components of \(\mathbf{W}\) are independent and each satisfies
    \(W_i\sim\mathcal{N}(0,1)\).

    \item Choose a Gaussian correlation matrix \(\Sigma_Z\), and compute its
    Cholesky decomposition:
    \[
    \Sigma_Z=LL^\top,
    \]
    where \(L\) is lower triangular.

    \item Create a correlated Gaussian vector:
    \[
    \mathbf{Z}=L\mathbf{W}.
    \]
    Since \(\mathbf{W}\sim\mathcal{N}_d(\mathbf{0},\mathbf{I}_d)\), we have
    \[
    \mathbf{Z}\sim\mathcal{N}_d(\mathbf{0},\Sigma_Z).
    \]
    In particular, each \(Z_i\) is standard normal, but the components
    \(Z_i\) and \(Z_j\) may be correlated.

    \item Transform each component of \(\mathbf{Z}\) into a uniform random
    variable:
    \[
    U_i=\Phi(Z_i),
    \]
    where \(\Phi\) is the cdf of the standard normal distribution. Since
    \(Z_i\sim\mathcal{N}(0,1)\), the probability integral transformation
    implies that
    \[
    U_i\sim\mathcal{U}[0,1].
    \]

    \item Transform each uniform variable into the desired marginal
    distribution:
    \[
    X_i=F_{X_i}^{-1}(U_i)
    =F_{X_i}^{-1}\!\left[\Phi(Z_i)\right],
    \qquad i=1,\ldots,d.
    \]
    Therefore, \(X_i\) has cdf \(F_{X_i}\).
\end{enumerate}

The construction can be summarized as

\resizebox{\textwidth}{!}{
$
\underbrace{\mathbf{W}\sim\mathcal{N}_d(\mathbf{0},\mathbf{I}_d)}
_{\text{independent Gaussian shocks}}
\quad \xrightarrow{\ \mathbf{Z}=L\mathbf{W}\ } \quad
\underbrace{\mathbf{Z}\sim\mathcal{N}_d(\mathbf{0},\Sigma_Z)}
_{\text{correlated Gaussian vector}}
\quad \xrightarrow{\ U_i=\Phi(Z_i)\ } \quad
\underbrace{U_i\sim\mathcal{U}[0,1]}
_{\text{uniform variables}}
\quad \xrightarrow{\ X_i=F_{X_i}^{-1}(U_i)\ } \quad
\underbrace{\mathbf{X}}
_{\text{desired marginals}}.
$
}

The difficult part is selecting \(\Sigma_Z\). Although \(\Sigma_Z\) governs
the dependence between the latent Gaussian variables \(\mathbf{Z}\), the
transformation \(F_{X_i}^{-1}\!\circ\Phi\) generally changes both the
magnitude and the nature of dependence. For \(i\neq j\),
\[
\operatorname{Cov}(X_i,X_j)
=
\operatorname{Cov}
\left(
F_{X_i}^{-1}\!\left[\Phi(Z_i)\right],
F_{X_j}^{-1}\!\left[\Phi(Z_j)\right]
\right).
\]

Hence, if \(\rho_{Z,ij}=\operatorname{Corr}(Z_i,Z_j)\), then
\[
[\Sigma_X]_{ij}
=
g_{ij}(\rho_{Z,ij}),
\]
where the function \(g_{ij}\) depends on the two marginal distributions
\(F_{X_i}\) and \(F_{X_j}\). Except for special choices of marginals,
\(g_{ij}\) has no closed-form inverse. Therefore, the Gaussian correlation
matrix \(\Sigma_Z\) is usually obtained numerically; see
\citet{cario1997modeling} and \citet{henderson2000generating}.

\subsection{Two Pareto variables with a given variance-covariance matrix}\label{sec:NORTApareto}

Can we generate two Pareto variables $X_1$ and $X_2$ whose variance-covariance matrix is a given
\begin{equation*}
C=\begin{pmatrix} c_{11} & c_{12}\\ c_{12} & c_{22} \end{pmatrix}?
\end{equation*}
This is the central problem of Section~\ref{sec:NORTA} in its simplest non-Gaussian form, and with Pareto margins it can be solved completely. We write $X_i\sim\mathcal{P}(\alpha_i,x_{m,i})$ when $F_i(x)=1-\left(x_{m,i}/x\right)^{\alpha_i}$ for $x\geq x_{m,i}$, so that $F_i^{-1}(u)=x_{m,i}(1-u)^{-1/\alpha_i}$ (Sections~\ref{sec:PIT} and~\ref{sec:paretodist}).

\paragraph{The variances must be finite.} Since $\E[X_i^k]=\alpha_i x_{m,i}^k/(\alpha_i-k)$ for $k<\alpha_i$ and $\E[X_i^k]=\infty$ otherwise,
\begin{equation}\label{eq:NORTAparetomoments}
\E[X_i]=\frac{\alpha_i\, x_{m,i}}{\alpha_i-1},\qquad
\Var(X_i)=\frac{\alpha_i\, x_{m,i}^2}{(\alpha_i-1)^2(\alpha_i-2)},\qquad
\mathrm{CV}_i\equiv\frac{\sqrt{\Var(X_i)}}{\E[X_i]}=\frac{1}{\sqrt{\alpha_i(\alpha_i-2)}},
\end{equation}
where the variance, hence the coefficient of variation $\mathrm{CV}_i$, exists only if $\alpha_i>2$. Imposing $C$ therefore requires $\alpha_1>2$ and $\alpha_2>2$; the covariance is then finite by the Cauchy--Schwarz inequality. Both cases occur with insurance losses: of the tail indices quoted in Section~\ref{sec:paretodist}, $1.5$ for fire losses leaves no covariance matrix to impose, while $2.5$ for motor liability does. This is also why the fat-tailed example of Section~\ref{sec:kousky} replaces $\alpha=2$ by $\alpha=2.01$ before introducing micro-correlation.

\paragraph{Why the Cholesky construction fails.} The multivariate normal recipe $X=\mu+LZ$ delivers the target variance-covariance matrix in one line, so it is tempting to feed it Pareto variables instead of Gaussian ones. Let us follow the recipe step by step and see where it breaks.

\textit{(a) What the recipe guarantees, and what it does not.} Let $W$ be any random vector with $\E[W]=0$ and $\Cov(W)=I$, Gaussian or not, and let $X=\mu+LW$ with $LL'=C$. Then $\E[X]=\mu$ and
\begin{equation}\label{eq:NORTAcholcov}
\Cov(X)=\E\left[(LW)(LW)'\right]=L\,\E[WW']\,L'=LL'=C .
\end{equation}
This identity is pure linear algebra: it uses the first two moments of $W$ and nothing else, so it says nothing about the distribution of each $X_i$. In the Gaussian case that distribution comes for free: $X_i=\mu_i+\sum_j l_{ij}Z_j$ is a linear combination of independent normal variables, hence normal. The recipe thus rests on two pillars: the identity \eqref{eq:NORTAcholcov}, valid for any $W$, and the closure of the Gaussian family under linear combinations of independent variables, which is special to it, the Gaussian being the only stable distribution with a finite variance \citep{embrechts1997}.

\textit{(b) The Pareto version.} Draw $P_1$ and $P_2$ independently with the target margins, $P_i\sim\mathcal{P}(\alpha_i,x_{m,i})$, the scales being set so that $\Var(P_i)=c_{ii}$ (Step 1 below), and standardize them: $\tilde P_i=(P_i-\mu_i)/\sigma_i$ with $\mu_i=\E[P_i]$ and $\sigma_i=\sqrt{c_{ii}}$, so that $\E[\tilde P]=0$ and $\Cov(\tilde P)=I$. In dimension two the Cholesky factor is explicit (compute $LL'$ to check it):
\begin{equation}\label{eq:NORTAchol2}
L=\begin{pmatrix}\sigma_1 & 0\\ \rho\,\sigma_2 & \sqrt{1-\rho^2}\,\sigma_2\end{pmatrix},\qquad \rho=\frac{c_{12}}{\sigma_1\sigma_2},
\end{equation}
so that $X=\mu+L\tilde P$ reads
\begin{equation}\label{eq:NORTAcholX}
X_1=\mu_1+\sigma_1\tilde P_1=P_1,\qquad X_2=\mu_2+\sigma_2\left(\rho\,\tilde P_1+\sqrt{1-\rho^2}\,\tilde P_2\right).
\end{equation}
By (a), $\E[X]=\mu$ and $\Cov(X)=C$ exactly, and $X_1=P_1$ is exactly Pareto. The whole question is the distribution of $X_2$, a weighted sum of two independent Pareto variables. Three symptoms show that it is not Pareto.

\textit{(c) First symptom: the lower bound.} A $\mathcal{P}(\alpha_2,x_{m,2})$ variable never falls below $x_{m,2}$. Since $\mu_i-x_{m,i}=\mu_i/\alpha_i$, each standardized variable is unbounded above and bounded below, $\tilde P_i\geq-b_i$ with $b_i=1/(\alpha_i\mathrm{CV}_i)=\sqrt{(\alpha_i-2)/\alpha_i}$. If $c_{12}<0$, then $\rho<0$ and the term $\rho\sigma_2\tilde P_1$ is unbounded below: $X_2$ takes arbitrarily large negative values, a negative claim if $X_2$ is an insurance loss. If $c_{12}>0$, $X_2$ is bounded below by
\begin{equation*}
\ell_2=\mu_2-\sigma_2\left(\rho\, b_1+\sqrt{1-\rho^2}\, b_2\right),
\end{equation*}
whereas the target distribution starts at $x_{m,2}=\mu_2-\sigma_2b_2$, and the two differ in general. In the example below, $(\alpha_1,\alpha_2)=(3,4)$ and $\rho=0.354$ give $\ell_2=2.78<x_{m,2}=3$: $10.6\%$ of the draws of $X_2$ fall in $[2.78,3)$, an interval to which the target distribution gives probability zero.

\textit{(d) Second symptom: the shape at the bottom of the support.} For particular values of $\rho$ the two lower bounds could coincide, so the support alone does not settle the question; the shape of the distribution near its lower bound does. A Pareto variable piles up mass just above $x_{m,2}$, where its density is at its maximum:
\begin{equation*}
P\left(X\leq x_{m,2}+\varepsilon\right)=1-\left(\frac{x_{m,2}}{x_{m,2}+\varepsilon}\right)^{\alpha_2}=\frac{\alpha_2}{x_{m,2}}\,\varepsilon+o(\varepsilon).
\end{equation*}
By contrast, when $c_{12}>0$, $X_2$ is within $\varepsilon$ of its lower bound $\ell_2$ only if both terms of \eqref{eq:NORTAcholX} are within $\varepsilon$ of their own lower bounds. The two terms are independent and each has a bounded density, so each event has a probability of order $\varepsilon$, and
\begin{equation*}
P\left(X_2\leq\ell_2+\varepsilon\right)\leq P\left(\tilde P_1\leq-b_1+\varepsilon/(\rho\sigma_2)\right)P\left(\tilde P_2\leq-b_2+\varepsilon/(\sqrt{1-\rho^2}\,\sigma_2)\right)=O(\varepsilon^2).
\end{equation*}
The density of $X_2$ therefore starts from zero at $\ell_2$ and rises, whereas a Pareto density jumps from zero to its maximum $\alpha_2/x_{m,2}$: $X_2$ is not Pareto, whatever the parameters. The intuition is that of two dice: each die shows a one with probability $1/6$, but their sum equals two with probability $1/36$ only, because both must be small at the same time. The figure below displays the densities in the example: the density of the Cholesky draws of $X_2$ starts from zero at $\ell_2=2.78$, while the target Pareto density jumps to $4/3$ at $x_{m,2}=3$, a jump that the draws of Step 3 below reproduce.
\begin{center}
\includegraphics[width=0.75\textwidth]{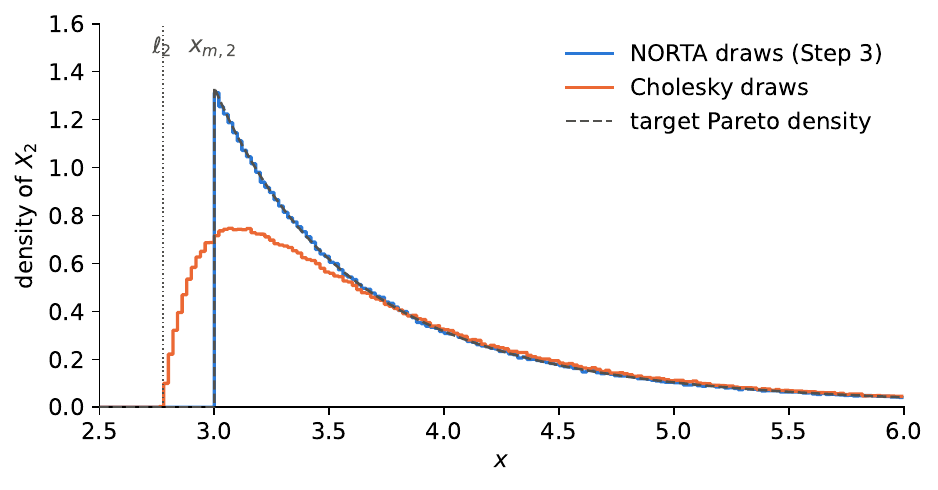}
\end{center}

\textit{(e) Third symptom: the right tail.} Far in the tail, a sum of independent heavy-tailed variables is large essentially because one of its terms is large: $P(A+B>x)\sim P(A>x)+P(B>x)$ as $x\to\infty$ for independent $A$ and $B$ with regularly varying tails \citep{embrechts1997}. Applied to \eqref{eq:NORTAcholX} with $c_{12}>0$ and $P(P_i>y)=(x_{m,i}/y)^{\alpha_i}$, this gives
\begin{equation}\label{eq:NORTAcholtail}
P\left(X_2>x\right)\sim\left(\frac{\rho\,\sigma_2\,x_{m,1}}{\sigma_1\,x}\right)^{\alpha_1}+\left(\frac{\sqrt{1-\rho^2}\,x_{m,2}}{x}\right)^{\alpha_2},\qquad x\to\infty,
\end{equation}
to be compared with the target $(x_{m,2}/x)^{\alpha_2}$. If $\alpha_1<\alpha_2$, the first term dominates and $X_2$ inherits the heavier tail of $X_1$: in the example, $P(X_2>x)$ decays like $x^{-3}$ instead of $x^{-4}$, so that $\E[X_2^3]=\infty$ although the target has a finite third moment. If $\alpha_1=\alpha_2=\alpha$, the scales are proportional to the standard deviations by \eqref{eq:NORTAparetoscale} below, and $P(X_2>x)\sim\left[\rho^\alpha+(1-\rho^2)^{\alpha/2}\right](x_{m,2}/x)^\alpha$, where the factor is smaller than $\rho^2+(1-\rho^2)=1$ because $\alpha>2$: large values are too rare, by $23\%$ for $\alpha=3$ and $\rho=0.5$. If $\alpha_1>\alpha_2$, the second term dominates and the tail is too thin by the factor $(1-\rho^2)^{\alpha_2/2}$. The distortion thus sits in the tail, where insurance applications look (capital requirements, reinsurance layers), and it can hide from a simulation check: in the example, the exceedance probabilities of $X_2$ and of the target differ by $5\%$ at $x=10$, but by a factor of $3$ at $x=10^3$.

\textit{(f) The lesson.} The Cholesky recipe controls exactly what it sees, the means and the covariances, and nothing else. It preserves the margins only when linear combinations of independent draws stay in the family, which, among distributions with a finite variance, is the case of the Gaussian alone. With Pareto margins, the dependence must be introduced without mixing the variables: the joint distribution has to be built from the margins and a copula, which is what we do now.

\paragraph{Margins first, dependence second.} By Sklar's theorem (Section~\ref{sec:Rcopula}), a joint distribution consists of margins tied together by a copula. The diagonal of $C$ involves the margins only, while the covariance involves both, as Hoeffding's formula makes explicit: for finite variances,
\begin{equation}\label{eq:hoeffdingcov}
\Cov(X_1,X_2)=\int_{\mathbb{R}^2}\left[F(x_1,x_2)-F_1(x_1)F_2(x_2)\right]dx_1\,dx_2 ,
\end{equation}
where $F$ is the joint cdf \citep{alexander2015}. The construction follows this split in three steps.

\paragraph{Step 1: the scales.} For given tail indices, the diagonal of $C$ fixes the scales through \eqref{eq:NORTAparetomoments}:
\begin{equation}\label{eq:NORTAparetoscale}
x_{m,i}=(\alpha_i-1)\sqrt{\frac{(\alpha_i-2)\,c_{ii}}{\alpha_i}} .
\end{equation}
We treat $\alpha_i$ as given because it is the parameter estimated from the data (Hill estimator, Section~\ref{sec:paretodist}). If the means $m_i$ are targeted as well, \eqref{eq:NORTAparetomoments} pins down both parameters, $\alpha_i=1+\sqrt{1+m_i^2/c_{ii}}>2$ and $x_{m,i}=m_i(\alpha_i-1)/\alpha_i$: the tail index is then dictated by the coefficient of variation.

\paragraph{Step 2: feasibility.} Pearson's correlation is scale invariant, so whether $\rho=c_{12}/\sqrt{c_{11}c_{22}}$ can be attained depends on $(\alpha_1,\alpha_2)$ only. The Fr\'echet--Hoeffding bounds of Section~\ref{sec:Rcopula}, $\max(F_1+F_2-1,0)\leq F\leq\min(F_1,F_2)$, bound the integrand of \eqref{eq:hoeffdingcov}, hence $\rho_{\min}\leq\rho\leq\rho_{\max}$, where $\rho_{\max}$ is attained by the comonotone pair $\left(F_1^{-1}(U),F_2^{-1}(U)\right)$ and $\rho_{\min}$ by the countermonotone pair $\left(F_1^{-1}(U),F_2^{-1}(1-U)\right)$, with $U\sim\mathcal{U}[0,1]$ \citep{embrechts2002correlation,alexander2015}. With Pareto margins, the scales cancel from $\rho$ and, for $x_{m,1}=x_{m,2}=1$, the two cross moments are elementary integrals:
\begin{align*}
\text{comonotone:}\quad & \E[X_1X_2]=\int_0^1(1-u)^{-\frac{1}{\alpha_1}-\frac{1}{\alpha_2}}\,du=\frac{\alpha_1\alpha_2}{\alpha_1\alpha_2-\alpha_1-\alpha_2},\\
\text{countermonotone:}\quad & \E[X_1X_2]=\int_0^1(1-u)^{-\frac{1}{\alpha_1}}u^{-\frac{1}{\alpha_2}}\,du=B\left(1-\tfrac{1}{\alpha_1},1-\tfrac{1}{\alpha_2}\right),
\end{align*}
with $B$ the Beta function. Writing $\rho=\kappa\,(\E[X_1X_2]/(\E[X_1]\E[X_2])-1)$, where $\kappa\equiv1/(\mathrm{CV}_1\mathrm{CV}_2)=\sqrt{\alpha_1\alpha_2(\alpha_1-2)(\alpha_2-2)}$, and using $B(a,b)=\Gamma(a)\Gamma(b)/\Gamma(a+b)$ and $\Gamma(z+1)=z\Gamma(z)$,
\begin{equation}\label{eq:NORTAparetobounds}
\rho_{\max}=\frac{\kappa}{\alpha_1\alpha_2-\alpha_1-\alpha_2},
\qquad
\rho_{\min}=\kappa\left[\frac{\Gamma\left(2-\frac{1}{\alpha_1}\right)\Gamma\left(2-\frac{1}{\alpha_2}\right)}{\Gamma\left(2-\frac{1}{\alpha_1}-\frac{1}{\alpha_2}\right)}-1\right].
\end{equation}
Three lessons follow (left panel of the figure below). (i) With a common tail index, $\rho_{\max}=1$, but $\rho_{\min}$ falls from $0$ as $\alpha\to2$ to $1-\pi^2/6\approx-0.645$ as $\alpha\to\infty$, the value for two exponential variables, the limit of a standardized Pareto: for $\alpha=3$, no correlation below $-0.262$ can be attained. (ii) With different tail indices, $\rho_{\max}<1$, and $\rho_{\max}\to0$ when the smaller tail index tends to $2$: two comonotone Pareto variables, one an increasing function of the other, with $(\alpha_1,\alpha_2)=(2.01,4)$ have a correlation of at most $0.20$. (iii) A positive definite $C$ is not enough: $C=\left(\begin{smallmatrix} 1 & -0.5\\ -0.5 & 2\end{smallmatrix}\right)$ is positive definite, but its correlation, $-0.354$, lies below $\rho_{\min}=-0.314$ for $(\alpha_1,\alpha_2)=(3,4)$, so no pair of Pareto variables with these tail indices has this variance-covariance matrix. \citet{embrechts2002correlation} count the belief that, for given margins, every correlation in $[-1,1]$ can be attained among the classical correlation fallacies of risk management.

\paragraph{Step 3: the dependence (NORTA).} Let $(Z_1,Z_2)$ be a standard normal pair with correlation $r$ and
\begin{equation}\label{eq:NORTApareto}
X_i=F_i^{-1}\left(\Phi(Z_i)\right)=x_{m,i}\left[1-\Phi(Z_i)\right]^{-1/\alpha_i},\qquad i=1,2 .
\end{equation}
Since $\Phi(Z_i)\sim\mathcal{U}[0,1]$ (Section~\ref{sec:PIT}), $X_i\sim\mathcal{P}(\alpha_i,x_{m,i})$ for every $r$: the margins are never distorted, and $r$ only moves the dependence. The correlation of $X$ depends on $r$ and on the tail indices only,
\begin{equation}\label{eq:NORTAparetog}
g(r)\equiv\Corr(X_1,X_2)=\kappa\left(\frac{(\alpha_1-1)(\alpha_2-1)}{\alpha_1\alpha_2}\,\E_r\left[\left(1-\Phi(Z_1)\right)^{-\frac{1}{\alpha_1}}\left(1-\Phi(Z_2)\right)^{-\frac{1}{\alpha_2}}\right]-1\right),
\end{equation}
a double integral against the bivariate normal density with no closed form, which we evaluate numerically. Three properties of $g$ settle the problem. (i) $g(0)=0$. (ii) $g(1)=\rho_{\max}$ and $g(-1)=\rho_{\min}$: $r=\pm1$ means $Z_2=\pm Z_1$, that is $\Phi(Z_2)=\Phi(Z_1)$ or $\Phi(Z_2)=1-\Phi(Z_1)$, the comonotone and countermonotone pairs of Step 2. (iii) $g$ is continuous and strictly increasing on $[-1,1]$. \citet{cario1997modeling} establish continuity and monotonicity of the NORTA correlation map under mild conditions, and a direct argument is available here: the joint cdf of $X$ is $\Phi_2\left(\Phi^{-1}(F_1(x_1)),\Phi^{-1}(F_2(x_2));r\right)$, with $\Phi_2(\cdot,\cdot;r)$ the bivariate normal cdf, and $\partial\Phi_2(z_1,z_2;r)/\partial r=\phi_2(z_1,z_2;r)>0$ (Plackett's identity), so the joint cdf increases pointwise with $r$; by \eqref{eq:hoeffdingcov}, so does the covariance, and dominated convergence gives continuity.

Hence every feasible $\rho$ is reached by exactly one $r^*=g^{-1}(\rho)$: in dimension two, NORTA matches every feasible covariance matrix \citep{ghosh2003behavior}. Moreover, $\left|g(r)\right|\leq|r|$:\footnote{Let $h_i(z)=F_i^{-1}\left(\Phi(z)\right)$, square integrable under the standard normal density when $\alpha_i>2$, and expand it on the normalized Hermite polynomials $H_k$, which are orthonormal under that density: $h_i=\sum_{k\geq0}a_{ik}H_k$. Mehler's formula gives $\E_r\left[H_j(Z_1)H_k(Z_2)\right]=r^k\delta_{jk}$, so $\Cov(X_1,X_2)=\sum_{k\geq1}a_{1k}a_{2k}r^k$ and, by the Cauchy--Schwarz inequality, $\left|\Cov(X_1,X_2)\right|\leq|r|\left(\sum_{k\geq1}a_{1k}^2\right)^{1/2}\left(\sum_{k\geq1}a_{2k}^2\right)^{1/2}=|r|\sqrt{\Var(X_1)\Var(X_2)}$.} the transformation attenuates correlation, so the Gaussian correlation must be inflated, $|r^*|\geq|\rho|$, and the more so the heavier the tails (right panel of the figure below). The algorithm\footnote{numpy\_exercise\_part4.py} is:
\begin{enumerate}
\item check that $\alpha_1>2$, $\alpha_2>2$ and that $C$ is positive definite;
\item compute the scales $x_{m,1}$ and $x_{m,2}$ with \eqref{eq:NORTAparetoscale};
\item check that $\rho=c_{12}/\sqrt{c_{11}c_{22}}\in\left[\rho_{\min},\rho_{\max}\right]$ with \eqref{eq:NORTAparetobounds};
\item solve $g(r^*)=\rho$ on $[-1,1]$ by Brent's method, evaluating \eqref{eq:NORTAparetog} by numerical integration;
\item draw $W_1,W_2$ iid $\mathcal{N}(0,1)$, set $Z_1=W_1$ and $Z_2=r^*W_1+\sqrt{1-r^{*2}}\,W_2$ (the Cholesky factor of the $2\times2$ correlation matrix), and apply \eqref{eq:NORTApareto}.
\end{enumerate}

With $(\alpha_1,\alpha_2)=(3,4)$ and $C=\left(\begin{smallmatrix} 1 & 0.5\\ 0.5 & 2\end{smallmatrix}\right)$, Step 1 gives $x_{m,1}=2/\sqrt{3}\approx1.155$ and $x_{m,2}=3$, the target correlation $0.354$ lies well inside $\left[\rho_{\min},\rho_{\max}\right]=[-0.314,0.980]$, and Step 3 gives $r^*=0.504$. With $10^6$ draws, both margins pass a Kolmogorov--Smirnov test against their Pareto target and the sample variance-covariance matrix is $\left(\begin{smallmatrix} 0.981 & 0.506\\ 0.506 & 2.026\end{smallmatrix}\right)$. The Cholesky construction delivers a similar sample matrix, but $10.6\%$ of its draws of $X_2$ fall below $x_{m,2}=3$, outside the support of the target distribution.

In the figure below, the left panel plots the bounds \eqref{eq:NORTAparetobounds} against the tail index, for a common tail index and for $\alpha_2=4$; the right panel plots the calibration curve $g$, with the example as a dot. With $\alpha_1=\alpha_2=2.2$, a Gaussian correlation of $0.5$ produces a Pareto correlation of only $0.14$.
\begin{center}
\includegraphics[width=0.9\textwidth]{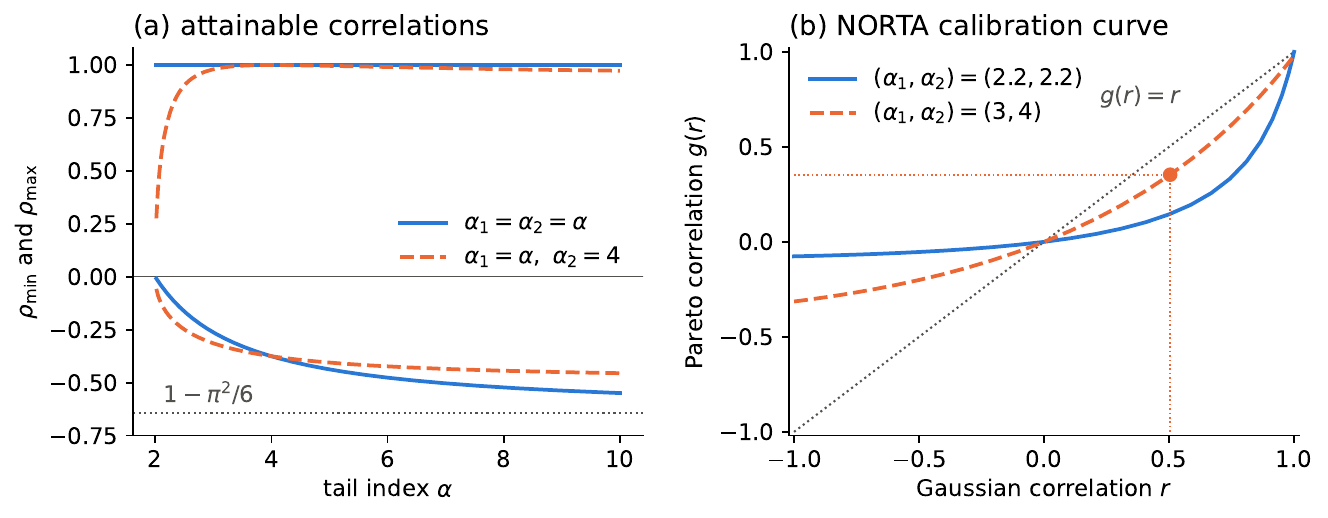}
\end{center}

\paragraph{Checking a heavy-tailed simulation.} For $\alpha_i\leq4$, $\E[X_i^4]=\infty$: the sample variance has an infinite variance, the usual $\sqrt{n}$ central limit theorem does not apply, and its sampling distribution is skewed, most samples underestimating $c_{ii}$ and a few overshooting it by far. Over $200$ samples of size $10^5$ from the example, the ratio of the sample to the target $c_{11}$ has a median of $0.97$ and ranges from $0.87$ to $1.15$ (5th and 95th percentiles); with $(\alpha_1,\alpha_2)=(6,8)$ and the same $C$, it ranges from $0.97$ to $1.04$. A sample covariance matrix is a weak check of a heavy-tailed simulation, and a weak estimator of dependence in heavy-tailed data.

\paragraph{Rank correlations.} Kendall's $\tau$ and Spearman's $\rho_S$ (Section~\ref{sec:corronreturns}) are invariant to increasing transformations: they depend on the copula only, and every value in $[-1,1]$ is attainable whatever the margins. Under the Gaussian copula, $\tau=\frac{2}{\pi}\arcsin r$ and $\rho_S=\frac{6}{\pi}\arcsin\frac{r}{2}$ \citep{alexander2015}, so a dependence target stated as a rank correlation is calibrated in closed form, e.g.\ $r=\sin\left(\pi\tau/2\right)$, without numerical integration.

\paragraph{Non-uniqueness and joint extremes.} $C$ does not pin down the joint distribution. Because \eqref{eq:hoeffdingcov} is linear in $F$, drawing $U_2=U_1$ with probability $\theta=\rho/\rho_{\max}$ and an independent $U_2$ otherwise also matches $C$ exactly when $\rho\geq0$, in closed form but with an unappealing law that puts a mass $\theta$ of the draws on a curve. The Gaussian copula has no tail dependence for $r<1$: joint extremes are rare. When joint large losses matter, as for natural catastrophes, a copula with upper tail dependence such as the Gumbel copula (Section~\ref{sec:Rcopula}) is more appropriate; Steps 1 and 2 are unchanged, and Step 3 calibrates the copula parameter instead of $r$.

\paragraph{Higher dimension.} For $d>2$, each $r_{ij}=g_{ij}^{-1}(\rho_{ij})$ is calibrated pair by pair, but the resulting matrix $\Sigma_Z$ may fail to be positive semi-definite even when $C$ is feasible, and this happens for an increasing proportion of correlation matrices as $d$ grows \citep{ghosh2003behavior}.

\section{Financial econometrics: what you must know}
\label{sec:fecore}

\subsection{The one idea: returns are unpredictable in level, predictable in scale}
\label{sec:fecore:oneidea}

Write the return of an asset between $t$ and $t+1$, conditionally on the information set
$\Fset_t$ available at $t$, as
\begin{equation}
r_{t+1} \;=\; \mu_{t} \;+\; \sigma_{t}\, z_{t+1},
\qquad
\E[z_{t+1}\mid \Fset_t]=0,
\qquad
\Var(z_{t+1}\mid \Fset_t)=1 ,
\label{eq:fecore:decomp}
\end{equation}
where $\mu_t = \E[r_{t+1}\mid\Fset_t]$ and $\sigma_t^2 = \Var(r_{t+1}\mid\Fset_t)$ are both
$\Fset_t$-measurable. Equation \eqref{eq:fecore:decomp} is not a model; it is a definition,
and it is the map of the entire field:

\begin{itemize}
\item \textbf{$\mu_t$ is small and hard to estimate.} On daily data the conditional mean of
      an equity index is of the order of $5\times 10^{-4}$ against a conditional standard
      deviation of $10^{-2}$: the signal-to-noise ratio is about $1/20$ per day. The econometrics of $\mu_t$ is
      therefore the econometrics of a very weak signal, and it is dominated by the problem
      of not being fooled.
\item \textbf{$\sigma_t$ is large, moves slowly, and is estimable.} Volatility is
      strongly autocorrelated: today's is an excellent predictor of tomorrow's. The
      econometrics of $\sigma_t$ is a genuine forecasting problem with a genuine payoff ---
      \S\ref{sec:fecore:garch}.
\item \textbf{$z_{t+1}$ is not Gaussian.} Its tails are heavier than normal even after
      conditioning on $\sigma_t$, which is why risk measures cannot be read off a normal
      table --- \S\ref{sec:fecore:risk}.
\item \textbf{Across assets, $\E[r^i]$ differs systematically.} Explaining the
      cross-sectional dispersion of unconditional means by exposure to a small number of
      risk factors is the asset-pricing half of the field --- \S\ref{sec:fecore:cross}.
\end{itemize}

Any question a student will meet in a master thesis on financial data is a question about
$\mu_t$, $\sigma_t$, the tail of $z$, or the cross-section of $\E[r^i]$.

\subsection{Returns: definitions, aggregation, and the pitfalls}
\label{sec:fecore:returns}

Let $P_t$ be the price of an asset at date $t$ and $D_t$ the cash flow paid over
$(t-1,t]$. The \emph{simple} net return and the \emph{log} (continuously compounded)
return are
\begin{equation}
R_t \;=\; \frac{P_t + D_t}{P_{t-1}} - 1 ,
\qquad\qquad
r_t \;=\; \ln\!\left(\frac{P_t+D_t}{P_{t-1}}\right) \;=\; \ln(1+R_t).
\label{eq:fecore:returns}
\end{equation}

The choice between them is not cosmetic; it is determined by the dimension along which
one aggregates.

\paragraph{Time aggregation is additive in logs.} Over $k$ periods,
$1+R_t(k)=\prod_{j=0}^{k-1}(1+R_{t-j})$, hence
\begin{equation}
r_t(k) \;=\; \sum_{j=0}^{k-1} r_{t-j}.
\label{eq:fecore:timeagg}
\end{equation}
A sum of weakly dependent terms is exactly the object to which the central limit theorem
applies, which is why log returns are the natural unit for time-series modelling, and why
monthly returns look far more Gaussian than daily ones.

\paragraph{Cross-sectional aggregation is additive in simple returns.} A portfolio with
weights $w_i$ summing to one has simple return $R_t^{p}=\sum_i w_i R_t^i$; the analogous
statement is false for log returns, since $\ln(\sum_i w_i e^{r^i}) \neq \sum_i w_i r^i$ by
Jensen's inequality. Portfolio and asset-pricing work --- \S\ref{sec:fecore:cross} --- is
therefore done in simple returns.

\paragraph{Excess returns.} Asset pricing is about compensation for risk, so the object of
interest is almost always the excess return $R^{e,i}_t = R^i_t - R^f_t$ over a
(conditionally) riskless rate. Mixing levels and excess returns across a paper is a
classic and avoidable error.

\paragraph{Three data pitfalls to state out loud.} First, \emph{dividends and splits}: a
raw price series is not a return series, and a series that ignores dividends
systematically understates the equity premium. Second, \emph{survivorship}: a
cross-section built from firms alive today mechanically excludes the failures and
inflates average returns. Third, \emph{stale and non-synchronous prices}: illiquid assets
do not trade at the closing instant, which induces spurious positive autocorrelation in
index returns and spurious low betas for small stocks. All three are discussed in
\citet[ch.~1]{tsay2010} and \citet[ch.~1]{linton2019}; all three have destroyed published
results.

\subsection{The stylized facts, and what each one forces on the econometrics}
\label{sec:fecore:facts}

Before any model: what do returns actually look like? \citet{Cont2001aa} organised the
empirical regularities that every model of \eqref{eq:fecore:decomp} must be consistent
with.

\begin{table}[H]
\centering
\small
\renewcommand{\arraystretch}{1.25}
\begin{tabular}{@{}p{4.6cm}p{5.1cm}p{4.9cm}@{}}
\toprule
\textbf{Stylized fact} & \textbf{Econometric consequence} & \textbf{Method it forces} \\
\midrule
Heavy tails: excess kurtosis of daily returns typically $3$--$10$ &
Gaussian inference on tails is wrong; fourth moments may not exist &
Student-$t$ innovations, EVT (\S\ref{sec:fecore:risk}) \\
$\Corr(r_t,r_{t-j})\approx 0$ for $j\geq 1$ &
The conditional mean is nearly unforecastable &
Efficiency tests (\S\ref{sec:fecore:efficiency}) \\
$\Corr(|r_t|,|r_{t-j}|)>0$, decaying slowly &
Returns are \emph{not} i.i.d.\ despite zero autocorrelation &
ARCH/GARCH (\S\ref{sec:fecore:garch}) \\
Volatility clustering &
Conditional heteroskedasticity is the rule, not a nuisance &
$\sigma_t$ modelled, not corrected for \\
Leverage effect: $\Corr(r_t,\sigma_{t+1})<0$ &
Symmetric volatility models are misspecified &
GJR, EGARCH (\S\ref{sec:garch}) \\
Volume--volatility correlation; intraday seasonality &
Sampling frequency is a modelling choice &
Realized (intraday) volatility measures \\
\bottomrule
\end{tabular}
\caption{The stylized facts of \citet{Cont2001aa} and the technique each one forces.}
\label{tab:fecore:facts}
\end{table}

\subsection{Market efficiency as an econometric hypothesis}
\label{sec:fecore:efficiency}

Markets are efficient if asset prices reflect all available information.

\subsubsection{What exactly is being tested}

``Markets are efficient'' is not testable as stated. What is testable is a restriction on
the conditional moments of prices. \citet[\S2.1]{campbell1997} organise the possible
restrictions in increasing order of strength; the taxonomy is worth putting on the board
because it disciplines the whole literature.

\begin{itemize}
\item \textbf{RW1}: $r_t$ i.i.d.\ with mean $\mu$. Increments independent \emph{and}
      identically distributed. Rejected by volatility clustering alone.
\item \textbf{RW2}: $r_t$ independent but not identically distributed. Allows
      unconditional heteroskedasticity; still rules out any nonlinear dependence.
\item \textbf{RW3}: $\Cov(r_t, r_{t-j})=0$ for all $j\geq 1$ --- increments merely
      \emph{uncorrelated}. This is the empirically relevant null, and it is compatible
      with GARCH.
\end{itemize}

The martingale-difference version of RW3, $\E[r_{t+1}-\mu \mid \Fset_t]=0$, is the
statement that no function of past information forecasts the return. Note the nesting:
RW1 $\Rightarrow$ RW2 $\Rightarrow$ RW3. Almost all classical ``tests of market
efficiency'' are tests of RW3.

\paragraph{The joint-hypothesis problem.} \citet{fama1970} makes the point that no test of
efficiency is a test of efficiency alone: any statement about whether an expected return
is ``abnormal'' presupposes a model of what the normal expected return is. A rejection is
therefore ambiguous between market inefficiency and a misspecified pricing model.

\subsection{The conditional variance: ARCH and GARCH}
\label{sec:fecore:garch}

\subsubsection{Testing for ARCH effects}
\label{sec:fecore:garch:test}

Start from the residuals $\hat\xi_t$ of a conditional-mean model (for daily returns, a
constant or an AR$(1)$ is enough). \citet{Engle:1982aa} proposes a Lagrange multiplier
test: regress $\hat\xi_t^2$ on a constant and $q$ of its own lags,
\begin{equation}
\hat\xi_t^2 \;=\; \gamma_0 + \sum_{j=1}^{q}\gamma_j \hat\xi_{t-j}^2 + w_t ,
\qquad
LM = T\,R^2 \;\overset{d}{\longrightarrow}\; \chi^2(q) \quad\text{under } H_0:\gamma_1=\dots=\gamma_q=0 .
\label{eq:fecore:archlm}
\end{equation}
On any daily equity series this rejects with a $p$-value indistinguishable from zero. That
is the empirical fact which the rest of the block models.

\subsubsection{From ARCH to GARCH}

\citet{Engle:1982aa} lets the conditional variance depend on past squared innovations,
$\sigma_t^2=\omega+\sum_{j=1}^{q}\alpha_j \xi_{t-j}^2$. In practice $q$ must be large to
capture the slow decay of $\Corr(|r_t|,|r_{t-j}|)$, and the parameters are poorly
estimated. \citet{bollerslev1986generalized} solves this with a parsimonious recursion, adding lagged
conditional variances. The workhorse of the field is the GARCH$(1,1)$:
\begin{equation}
\xi_t=\sigma_t z_t,\qquad
\sigma_t^2 \;=\; \omega + \alpha\,\xi_{t-1}^2 + \beta\,\sigma_{t-1}^2 ,
\qquad \omega>0,\ \alpha\geq 0,\ \beta\geq 0 .
\label{eq:fecore:garch11}
\end{equation}
Estimation by (quasi-)maximum likelihood, the asymmetric extensions that accommodate the
leverage effect, the forecasting recursion and the diagnostics are developed in
\S\ref{sec:garch}.

\subsection{From volatility to risk: Value at Risk, Expected Shortfall, backtesting}
\label{sec:fecore:risk}

Fix a coverage probability $p$ (small: $0.01$ or $0.025$). The conditional Value at Risk
and Expected Shortfall at horizon one are
\begin{equation}
\Pr\big(r_{t+1} < -\VaR^{p}_{t+1} \,\big|\, \Fset_t\big) = p,
\qquad
\ES^{p}_{t+1} = -\,\E\big[r_{t+1}\,\big|\, r_{t+1} < -\VaR^{p}_{t+1},\ \Fset_t\big] .
\label{eq:fecore:varesdef}
\end{equation}
Both are reported as positive loss numbers. $\VaR$ is a quantile of the predictive
distribution; $\ES$ is the mean of the tail beyond that quantile, so $\ES\geq\VaR$ always.

\paragraph{Why the regulator moved from VaR to ES.} \citet{artzner1999} define a risk
measure $\varrho$ as \emph{coherent} if it is monotone, translation-equivariant, positively
homogeneous and \emph{subadditive}:
$\varrho(X+Y)\leq\varrho(X)+\varrho(Y)$. Subadditivity is the formal statement that
diversification cannot increase risk. $\ES$ is coherent; $\VaR$ is not. The standard
counterexample takes two independent defaultable bonds each defaulting with probability
$3\%$: at $p=5\%$ neither individual $\VaR$ registers the default, but the portfolio's
does, so $\VaR$ of the sum exceeds the sum of the $\VaR$s. A risk measure that penalises
diversification is a poor basis for capital regulation, which is the argument behind the
Basel Committee's replacement of $\VaR$ at $99\%$ by $\ES$ at $97.5\%$ in the market-risk
framework. State this once: it is the clearest example in the course of an econometric
object being chosen for an economic property rather than a statistical one.

\subsubsection{Three estimators, in increasing order of credibility}

\paragraph{(i) Historical simulation.} $\widehat\VaR^p_{t+1}$ is minus the empirical
$p$-quantile of the last $m$ returns. It is assumption-free, which is its appeal, and
\emph{unconditional}, which is its flaw: after a volatility spike it reacts only with a
lag of $m$ days, and it produces the long runs of consecutive exceptions that the
independence test of \S\ref{sec:varbacktest} is designed to catch.

\paragraph{(ii) Parametric, from the GARCH fit.} With
$r_{t+1}=\mu_{t+1|t}+\sigma_{t+1|t}z_{t+1}$ and $F_z$ the (standardized) innovation
distribution,
\begin{equation}
\widehat\VaR^{p}_{t+1} = -\Big(\mu_{t+1|t}+\sigma_{t+1|t}\,F_z^{-1}(p)\Big),
\qquad
\widehat\ES^{p}_{t+1} = -\Big(\mu_{t+1|t}+\sigma_{t+1|t}\,\E\big[z\mid z\le F_z^{-1}(p)\big]\Big).
\label{eq:fecore:varparam}
\end{equation}
For $z\sim\mathcal{N}(0,1)$ the tail expectation has the closed form
$\E[z\mid z\le \Phi^{-1}(p)]=-\phi\big(\Phi^{-1}(p)\big)/p$, so that
$\widehat\ES^p_{t+1}=-\mu_{t+1|t}+\sigma_{t+1|t}\,\phi(\Phi^{-1}(p))/p$. For a
standardized Student-$t$ with $\nu>2$ degrees of freedom, whose $p$-quantile is
$t_\nu^{-1}(p)\sqrt{(\nu-2)/\nu}$, the corresponding expression is available in closed
form as well and is what \texttt{arch} reports. The entire content of
\eqref{eq:fecore:varparam} is that \emph{the dynamics come from $\sigma_{t+1|t}$ and the
tail shape comes from $F_z$}; these are two separate modelling decisions and students
should be able to say which one they are getting wrong when a backtest fails.

\paragraph{(iii) Filtered historical simulation and conditional EVT.} Keep the GARCH
dynamics but stop assuming a parametric $F_z$. Filtered historical simulation replaces
$F_z^{-1}(p)$ in \eqref{eq:fecore:varparam} by the empirical $p$-quantile of the
standardized residuals $\hat z_t$. Conditional EVT \citep{McNeil:2000aa} goes one step
further and fits a generalized Pareto distribution to the exceedances of $-\hat z_t$ over
a high threshold $u$: for $N_u$ exceedances out of $n$ residuals,
\begin{equation}
\widehat{F}_z^{-1}(p)
\;=\; -\left\{ u + \frac{\hat\beta}{\hat\xi}
\left[\left(\frac{n}{N_u}\,p\right)^{-\hat\xi}-1\right]\right\},
\qquad
\E\big[{-z}\mid -z>x_q\big]=\frac{x_q}{1-\hat\xi}+\frac{\hat\beta-\hat\xi u}{1-\hat\xi} ,
\label{eq:fecore:evt}
\end{equation}
where $\hat\xi<1$ is the tail index, $\hat\beta$ the scale, and $x_q$ denotes the brace in
the first expression, that is the upper $q=1-p$ quantile of $-z$; the $\ES$ then follows
from \eqref{eq:fecore:varparam} through $\E[z\mid z\le F_z^{-1}(p)]=-\E[-z\mid -z>x_q]$. This two-step structure ---
\emph{filter the dynamics, then model the tail of what is left} --- is the single most
useful idea in quantitative risk management, and it is the same logic used to detect
insurers exposed to climate-driven tail losses in \citet{vansteenberghe2025insurance}.

\subsection{The cross-section: testing an asset pricing model}
\label{sec:fecore:cross}

\subsubsection{One equation behind every factor model}

A stochastic discount factor $m_{t+1}$ prices every excess return:
\begin{equation}
\E\big[m_{t+1} R^{e}_{t+1}\big]=0 .
\label{eq:fecore:sdf}
\end{equation}
Suppose $m_{t+1}=a-b'f_{t+1}$ is linear in $K$ factors. Expanding
\eqref{eq:fecore:sdf} as $\E[m]\E[R^e]+\Cov(m,R^e)=0$ and substituting gives the
\emph{beta representation}
\begin{equation}
\E\big[R^{e,i}\big] \;=\; \beta_i' \lambda,
\qquad
\beta_i=\Sigma_f^{-1}\Cov(f,R^{e,i}),
\qquad
\lambda=\frac{\Sigma_f\, b}{\E[m]} .
\label{eq:fecore:beta}
\end{equation}
Equation \eqref{eq:fecore:beta} is the whole of empirical asset pricing on one line:
average excess returns are linear in factor exposures, with a common price of risk
$\lambda$ \citep[ch.~6]{cochrane}. The CAPM of \citet{sharpe1964capital} is the case
$K=1$ with $f$ the market excess return; \citet{fama1993} add size and value;
\citet{fama2015} add profitability and investment. \emph{The econometric question is always
the same: are the pricing errors zero?}

\section{Probability distributions from option prices or surveys}

We aim to recover the subjective probability distribution of average inflation over $n$ years. Let
\[
\pi(n)\quad\text{be the $n$-year average annual CPI inflation,}\qquad
f^{\mathbb{Q}^{n}}_{\pi(n)}(\cdot)\quad\text{its pdf under the $n$-forward measure}.
\]
There are two complementary routes. First, use zero-coupon inflation caps/floors across strikes to extract the \emph{risk-neutral} (forward-measure) distribution by differentiating option prices with respect to strike (Breeden--Litzenberger) and discounting with the $n$-year zero rate \citep{breeden1978prices,ross1976options,kitsul2013economics}. Second, use survey data to obtain a \emph{physical} or stated-belief distribution and compare it to the options-implied one.

To extract a probability distribution of future inflation from option prices, remember three key distinctions:

\begin{enumerate}
  \item \textbf{What we study.} $\pi(n)$ is the average inflation over $n$ years. When $n=1$, this reduces to one-year-ahead inflation, $\pi(1)$.
  \item \textbf{How probabilities are measured.} Option prices reflect probabilities under the \emph{risk-neutral} (or \emph{forward}) measure $\mathbb{Q}^{n}$, not real-world beliefs. Under $\mathbb{Q}^{n}$, expected returns are adjusted for risk so that all assets earn the risk-free rate. Hence, the pdf we recover from prices is a market-implied, not a subjective, distribution.
  \item \textbf{What enters the estimation.} We use \emph{pricing inputs} such as the term structure (for discounting) and option quotes across strikes, then apply \emph{estimation choices} to smooth the data and enforce no-arbitrage conditions (for example, monotonicity and convexity) using tools like local polynomial regression.
\end{enumerate}

\quad

\begin{enumerate}
  \item \textbf{Observed:} Inflation caps/floors $\Rightarrow$ option prices $P_t(k,n)$.
  \item \textbf{Differentiate:} $\displaystyle f^{\mathbb{Q}^n}_{\pi(n)}(k) \propto \frac{\partial^2 P_t(k,n)}{\partial k^2}$ (Breeden--Litzenberger).
  \item \textbf{Interpret:} $f^{\mathbb{Q}^n}$ = market-implied (risk-neutral) pdf.
  \item \textbf{Compare:} With survey-based (physical) pdf to infer \emph{pricing kernel}.
\end{enumerate}

\subsection{Options-implied inflation probability
density functions}

\citet{kitsul2013economics} define a zero-coupon inflation cap as a contract entered into at time $t$. The seller of the cap promises to pay a fraction
\[
\max\!\left( (1+\pi(n))^n - (1+k)^n,\, 0 \right)
\]
of a notional underlying principal as a single payment in $n$ years' time, 
where $\pi(n)$ denotes the average annual total consumer price index (CPI) inflation rate from $t$ to $t+n$ 
(not seasonally adjusted) and $k$ denotes the strike of the cap. 
Without loss of generality, the notional underlying principal is normalized to $1$. In exchange for this, the buyer makes an up-front payment of $P_t(k,n)$. The contract is effectively a call option on inflation. If $\pi(n)<k$, then the option is out-of-the-money.

Consider a one-year zero-coupon inflation cap with strike $k=2\%$.  
If realized inflation is $\pi(1)=3\%$, the payoff is
\[
(1+\pi(1))-(1+k)=1.03-1.02=0.01,
\]
that is, one cent per dollar of notional.  
If both parties are certain that $\pi(1)=3\%$ (i.e.\ $\mathbb{P}[\pi(1)=3\%]=1$) and there is no risk premium, then the fair price equals the discounted expected payoff:
\[
P_t(0.02,1)=e^{-y(1)}\times 0.01.
\]

\subsection*{Discounting: why the factor $e^{-n y(n)}$ appears}

\textbf{Principle.} A price equals the expected \emph{present value} of its future payoff under a pricing measure.  
If $y(n)$ is the continuously compounded $n$-year zero rate observed at $t=0$ (so here we take a bond with known return at time $t$ up to maturity and not the risk-free rate, which is unknown at time $t$ up to $t+n$), the price of one unit delivered at $t=n$ is
\[
B(0,n)=e^{-n y(n)}.
\]
This follows from continuous compounding: investing $e^{-n y(n)}$ today ($t=0$) at rate $y(n)$ for $n$ years grows deterministically to $1$ at $t=n$.

Suppose I receive $m$ interest payments at a rate of $\frac{r}{m}$ per annum. After $t$ years, I will receive: 
$$ \left( 1 + \frac{r}{m} \right)^{mt}$$ 
which can be rewritten as $\exp\left(m t \log\left[ 1 + \frac{r}{m} \right] \right) \simeq \exp\left( rt \right)$, when $ \frac{r}{m}$ is in the vicinity of 0.

We can get the same approximation with a Taylor expansion. If you have the amount $M(t)$ and invest it at interest rate $r$, then 
$$M(t+dt) = M(t) + \frac{d M(t)}{dt} \frac{t + dt - t}{1!} + \ldots$$ hence $M(t+dt) - M(t) = M(t) r dt \simeq \frac{d M(t)}{dt} dt$, and $$\frac{d M(t)}{dt} = r M(t)$$ which has the solution $M(0)e^{rt}$.

\textbf{Applied to the cap.} The zero-coupon inflation cap pays only at $t+n$. Therefore
\begin{equation}
P_t(k,n)=e^{-n y(n)}\,\mathbb{E}^{\mathbb{Q}^{n}}\!\big[\text{payoff at }t+n\big].
\label{eq:disc_general}
\end{equation}
For $n=1$, the payoff is $(\pi(1)-k)_{+}$, so
\begin{equation}
P_t(k,1)=e^{-y(1)}\,\mathbb{E}^{\mathbb{Q}^{1}}\!\big[(\pi(1)-k)_{+}\big].
\label{eq:disc_cap_n1}
\end{equation}

\textbf{Integral form.} Let $f^{\mathbb{Q}^{1}}_{\pi(1)}$ be the pdf of $\pi(1)$ under $\mathbb{Q}^{1}$. Then
\begin{equation}
P_t(k,1)=e^{-y(1)}\int_{k}^{\infty}(\pi-k)\,f^{\mathbb{Q}^{1}}_{\pi(1)}(\pi)\,\mathrm{d}\pi.
\label{eq:disc_integral}
\end{equation}

\paragraph{Meaning of $\mathbb{Q}^{1}\!\big(\pi(1)>k\big)$.}
$\mathbb{Q}^{1}$ denotes the $1$-forward (or risk-neutral) probability measure. 
Thus $\mathbb{Q}^{1}\!\big(\pi(1)>k\big)=1-F^{\mathbb{Q}^{1}}_{\pi(1)}(k)$ is the tail probability of $\pi(1)$ under $\mathbb{Q}^{1}$.

\paragraph{Differentiation with a $k$-dependent lower limit.}
From the integral form
\[
P_t(k,1)=e^{-y(1)}\int_{k}^{\infty}(\pi-k)\,f^{\mathbb{Q}^{1}}_{\pi(1)}(\pi)\,\mathrm{d}\pi,
\]
set 
\[
g(k):=\int_{k}^{\infty}(\pi-k)\,f^{\mathbb{Q}^{1}}_{\pi(1)}(\pi)\,\mathrm{d}\pi.
\]
Apply Leibniz’s rule with $h(\pi,k):=(\pi-k)f^{\mathbb{Q}^{1}}_{\pi(1)}(\pi)$, $a(k)=k$, and $b(k)=\infty$:
\[
\frac{\mathrm{d}}{\mathrm{d}k}\int_{a(k)}^{b(k)} h(\pi,k)\,\mathrm{d}\pi
= h\big(b(k),k\big)b'(k)-h\big(a(k),k\big)a'(k)+\int_{a(k)}^{b(k)}\frac{\partial h}{\partial k}(\pi,k)\,\mathrm{d}\pi.
\]
Here $b'(k)=0$, $a'(k)=1$, and $h\big(a(k),k\big)=(k-k)\,f^{\mathbb{Q}^{1}}_{\pi(1)}(k)=0$, while 
\[
\frac{\partial h}{\partial k}(\pi,k)=-\,f^{\mathbb{Q}^{1}}_{\pi(1)}(\pi).
\]

Here, we have
  \[
  \frac{\partial P_t(k,1)}{\partial k}
  =-\,e^{-y(1)}\,\mathbb{Q}^{1}\!\big(\pi(1)>k\big),
  \qquad
  \frac{\partial^{2} P_t(k,1)}{\partial k^{2}}
  =e^{-y(1)}\,f^{\mathbb{Q}^{1}}_{\pi(1)}(k).
  \]

\[
\qquad\Rightarrow\qquad
\boxed{
f_{\pi(1)}^{\mathbb{Q}^1}(k)
  = e^{y(1)} \frac{\partial^2 P_t(k,1)}{\partial k^2}
  }.
\]

\subsection*{Recovering implied pdfs from option prices}

\citet{kitsul2013economics} write:

\begin{quote}
We wish to use quotes on inflation caps at different strike prices to
find the pdf for $n$-year inflation $\pi(n)$ that solves the equation
\begin{equation}
P_t(k,n) \;=\; \exp\!\big(-ny(n)\big) \,
\mathbb{E}^{\mathbb{Q}^n}\!\left[ \max\!\left( (1+\pi(n))^n - (1+k)^n , 0 \right) \right],
\label{eq:capprice}
\end{equation}
where $P_t(k,n)$ denotes the price of an $n$-year inflation cap
with a strike of $k$ and $y(n)$ is the continuously compounded
$n$-year zero-coupon bond yield at time $t$.

\quad

$\ldots$

\quad

The idea for forming implied pdfs is based on the result
that the second derivative of the price of a call option with
respect to the strike price represents the risk-neutral pdf 
\citep{breeden1978prices, ross1976options}. 

The specific method used for solving for the implied pdf follows 
\citet{li2009nonparametric}, but without conditioning information, and it is based on the local polynomial regression of \citet{ait2003nonparametric}. 

We first filter the data, replacing the prices of inflation floors 
and caps by the closest prices that satisfy slope and convexity 
constraints. Consider a set of $L$ inflation caps on a given day 
at a given maturity. Let $y_i$ denote the price of the $i$th cap, 
with a strike of $k_i$. We seek to approximate the price of an 
inflation cap at a strike price $k'$ in a neighborhood around $k$ 
by a locally linear function
\[
\beta_0(k) + \beta_1(k)\,(k'-k).
\]

We estimate the coefficients as
\[
\hat\beta_0(k), \; \hat\beta_1(k) 
= \arg\min_{\beta_0(k), \beta_1(k)} 
\sum_{i=1}^L \Big( y_i - \beta_0(k) - \beta_1(k)(k_i-k) \Big)^2 
\, K\!\left( \frac{k_i-k}{h} \right) / h,
\]
where $K(\cdot)$ is a kernel function and $h$ is a bandwidth. 
The bandwidth is optimal for the minimization of integrated 
mean squared error, as described in \citet{ait2003nonparametric}. 

The second derivative of the cap price can then be obtained as 
$\hat\beta'_1(k)$. Using the nominal Treasury term-structure data 
set of \citet{gurkaynak2007us} to obtain the risk-free $n$-year yield, 
we then have the pdf of inflation under the forward measure. 
This method constructs the risk-neutral pdf on a single day at a 
single maturity from options at different strike prices.
\end{quote}

\subsection*{From option quotes to implied pdfs: what each paper contributes}

\paragraph{Step 0: Notation.}
Let $\pi(n)$ be the average annual CPI inflation from $t$ to $t+n$.
For a zero-coupon $n$-year inflation cap with strike $k$, define
\[
X \equiv (1+\pi(n))^{n}, \qquad K \equiv (1+k)^{n}.
\]
The cap payoff at $t+n$ is $(X-K)_{+}$.

\paragraph{Step 1: Pricing under the forward measure (foundation in \citep{ross1976options}).}
Let $y(n)$ be the continuously compounded $n$-year zero yield and $\mathbb{Q}^{n}$ the $n$-forward measure.
The time-t price equals discounted forward-measure expectation:
\begin{equation}
P_t(k,n) \;=\; e^{-n y(n)}\,\mathbb{E}^{\mathbb{Q}^{n}}\!\big[(X-K)_{+}\big].
\label{eq:pricing_forward}
\end{equation}

\paragraph{Step 2: Breeden--Litzenberger identity on strikes \citep{breeden1978prices,ross1976options}.}
View $C(K)\equiv e^{-n y(n)}\,\mathbb{E}^{\mathbb{Q}^{n}}[(X-K)_{+}]$ as a call on $X$ with strike $K$. Then
\begin{equation}
\frac{\partial C}{\partial K} \;=\; -\,e^{-n y(n)}\,\mathbb{Q}^{n}(X>K),
\qquad
\frac{\partial^{2} C}{\partial K^{2}} \;=\; e^{-n y(n)}\,f^{\mathbb{Q}^{n}}_{X}(K),
\label{eq:BL_in_K}
\end{equation}
where $f^{\mathbb{Q}^{n}}_{X}$ is the pdf of $X$ under $\mathbb{Q}^{n}$.

\paragraph{Step 3: Chain rule to work in inflation strike $k$.}
Since $K=(1+k)^{n}$, define $P(k)\equiv P_t(k,n)=C(K(k))$. By the chain rule,
\begin{align}
\frac{\partial P}{\partial k}
&= \frac{\partial C}{\partial K}\,\frac{\partial K}{\partial k}
= -\,e^{-n y(n)}\,\mathbb{Q}^{n}(X>K)\, \big[n(1+k)^{n-1}\big], \label{eq:first_deriv_k} \\
\frac{\partial^{2} P}{\partial k^{2}}
&= \frac{\partial^{2} C}{\partial K^{2}} \Big(\frac{\partial K}{\partial k}\Big)^{2}
+ \frac{\partial C}{\partial K}\, \frac{\partial^{2} K}{\partial k^{2}} \nonumber\\
&= e^{-n y(n)}\,f^{\mathbb{Q}^{n}}_{X}(K)\, \big[n(1+k)^{n-1}\big]^{2}
\;-\; e^{-n y(n)}\,\mathbb{Q}^{n}(X>K)\, \big[n(n-1)(1+k)^{n-2}\big].
\label{eq:second_deriv_k_general}
\end{align}

\paragraph{Step 4: Nonparametric recovery method \citep{li2009nonparametric}.}
Given option quotes $\{(k_i,y_i)\}_{i=1}^{L}$ at one maturity and date, recover a smooth function
$k \mapsto P_t(k,n)$ whose second derivative yields the pdf via \ref{eq:second_deriv_k_general}. The Li--Zhao framework estimates $P$ under slope/convexity constraints implied by \ref{eq:first_deriv_k}--\ref{eq:second_deriv_k_general}. Higher convexity of $P(k)$ at a given strike means markets assign more probability mass near that strike. Left-skewed pdf $\Rightarrow$ greater deflation concern.

\paragraph{Step 5: Local polynomial regression to estimate derivatives \citep{ait2003nonparametric}.}
Estimate $P(k)$ and its derivatives by local polynomial regression. Around a target $k$, fit a quadratic locally:
\begin{equation}
(\hat\beta_0(k),\hat\beta_1(k),\hat\beta_2(k))
= \arg\min_{\beta_0,\beta_1,\beta_2} 
\sum_{i=1}^{L} \Big( y_i - \beta_0 - \beta_1(k_i-k) \Big)^2
\, K\!\left(\frac{k_i-k}{h}\right),
\label{eq:LPR}
\end{equation}
with kernel $K$ and bandwidth $h$. Then
\[
\widehat{P}(k)=\hat\beta_0(k), \qquad
\widehat{P'}(k)=\hat\beta_1(k), \qquad
\widehat{P''}(k)=\hat\beta'_1(k),
\]

\paragraph{Step 6: Putting it together with discounting \citep{gurkaynak2007us}.}
Obtain $y(n)$ from the Treasury zero-coupon term structure. Then, for $n=1$,
\begin{equation}
\boxed{\, f^{\mathbb{Q}^{1}}_{\pi(1)}(k) \;=\; e^{\,y(1)}\, \widehat{P_t''}(k,1) \,}
\label{eq:boxed_n1}
\end{equation} For $n>1$, combine \ref{eq:second_deriv_k_general} with the change of variables
\[
f^{\mathbb{Q}^{n}}_{\pi(n)}(\pi) 
= n(1+\pi)^{n-1}\, f^{\mathbb{Q}^{n}}_{X}\!\big((1+\pi)^{n}\big).
\]

\paragraph{No-arbitrage filters used in practice.}
Impose $P'(k)\le 0$ (monotonicity in strike) and $P''(k)\ge 0$ (convexity) before derivative extraction. Replace observed quotes by nearby values that satisfy these constraints, then re-estimate via \ref{eq:LPR}.

\paragraph{Main takeaway for students.}
Recover the pdf from option quotes in three moves:
(i) price as discounted forward-measure expectation \ref{eq:pricing_forward};
(ii) differentiate in strike (Breeden--Litzenberger) \ref{eq:BL_in_K} and map to inflation strike via the chain rule \ref{eq:second_deriv_k_general};
(iii) estimate the required derivatives nonparametrically with local polynomials \ref{eq:LPR}, using $y(n)$ from the term structure.

\subsection{Preparing the implementation}

\begin{align*}
X &= (1+\pi(n))^{n}, \qquad 
K = (1+k)^{n}, \qquad 
B(0,n) = e^{-n y(n)}, \\
C(K) &= B(0,n)\,\mathbb{E}^{\mathbb{Q}^n}\!\big[(X-K)_+\big], \\[0.5ex]
\frac{\partial^2 C}{\partial K^2}(K) &= B(0,n)\,f_X^{\mathbb{Q}^n}(K), \qquad
f_{\pi(n)}^{\mathbb{Q}^n}(\pi) = n(1+\pi)^{n-1}\,f_X^{\mathbb{Q}^n}\!\big((1+\pi)^n\big), \\[0.5ex]
\text{Cap}(K) - \text{Floor}(K) &= B(0,n)\,\big(\mathbb{E}^{\mathbb{Q}^n}[X] - K\big), \\[0.5ex]
n &= 1:\quad \frac{\partial^{2} P_t(k,1)}{\partial k^{2}} 
= e^{-y(1)}\,f_{\pi(1)}^{\mathbb{Q}^1}(k).
\end{align*}

\section{Python: Weight and height distributions}\label{sec:weightheightdistrib}

Let's consider $H$ (height) and $W$ (weight) two continuous rv, the joint distribution function of $(H,W)$ is $F(h,w)=P\left(H \leq h, W \leq w \right)$. The rv $H$ and $W$ are statistically independent if 
\begin{equation}
\forall h,w \quad F(h,w) = F_H(h) F_W(w)
\end{equation}

If we believe that height and weight are statistically independent, then we should\footnote{code: vansteenberghe\_height\_normal\_distrib.py} just study both rv separately. We test $H_0:\rho=0$, no linear association between $H$ and $W$, against $H_1:\rho\neq0$ with Pearson's correlation test and reject $H_0$: $r=0.412$ with $n=15\,494$, and the $p$-value is numerically zero. Since independence implies $\rho=0$, this also rejects the independence of weight and height. The converse fails: nonlinear dependence may exist even when $\rho=0$, so a non-rejection would not establish independence.

Before estimating a relationship between height and weight, we study their individual distribution.

\quad

\citep{Hermanussen2001aa} found that \begin{quote}
Body weight is not normally distributed, but skewed to the right. Also power transformation was inadequate to sufficiently describe the shape of this distribution. The right tail of weight distributions declines exponentially, beyond a cut-off of $+0.5$ standard deviations. 
\end{quote} 

We further explore the concept of DGP introduced secion \ref{sec:DGP}. Now, the weight and height of an individual is likely to be related to his parent's, but appart from those relations, we are closer to an i.i.d. process than with a time series.

In the following, we use Indian 2005-6 Demographic and Health Surveys and would like to thank Professor Dean Spears for his help on collecting the data:

\quad

\begin{tabular}{lrrr}
\toprule
{} &     age  &     weight in .1 kg  &     height in mm\\
\midrule
count &  15494 &  15494 &  15494 \\
mean  &     32.2 &    567.5 &   1573.6 \\
std   &      9.4 &    121.1 &     67.8 \\
min   &     15.0 &    234.0 &   1267.0 \\
25\%   &     24.0 &    486.0 &   1530.0 \\
50\%   &     32.0 &    546.0 &   1573.0 \\
75\%   &     40.0 &    625.0 &   1617.0 \\
max   &     49.0 &   1599.0 &   1967.0 \\
\bottomrule
\end{tabular}

\quad

Our sample body weights are not normally distributed, but skewed to the right:
\begin{center}
\includegraphics[scale=0.5]{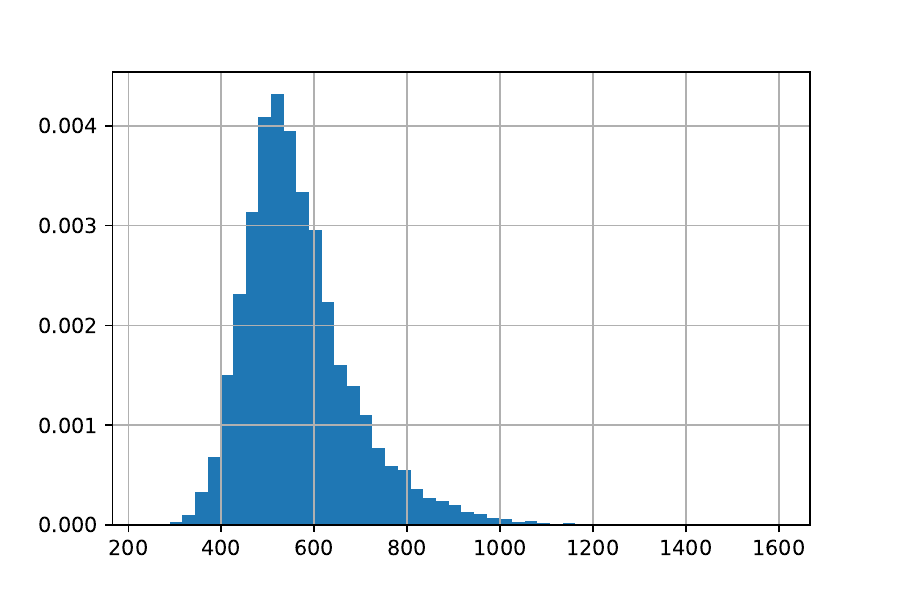}
\end{center}

\subsection{One-Sample $t$-Test}

We apply a two-sided one-sample $t$-test and do not reject
$H_0:\mu=567.5$, where $\mu$ denotes the population mean.
This does not establish that $\mu=567.5$; the data provide
insufficient evidence against this value.

For $n$ i.i.d.\ observations with sample mean $\bar X$ and sample
standard deviation $S$, we test $H_0:\mu=\mu_0$ against
$H_1:\mu\neq\mu_0$ using
\[
T=\frac{\bar X-\mu_0}{S/\sqrt{n}}.
\]
Under $H_0$, $T\sim t_{n-1}$ exactly for normal observations.
Without normality, provided the variance is finite and positive,
the Central Limit Theorem and Slutsky's theorem imply
$T\xrightarrow{d}\mathcal N(0,1)$, justifying the test asymptotically.

\paragraph{Example: bottle filling.}
A machine targets an average fill of $\mu_0=500$~mL.
Assuming independent normal fill volumes, suppose a sample of
$n=25$ bottles gives $\bar x=498$~mL and $s=4$~mL. Then
\[
t=\frac{498-500}{4/\sqrt{25}}=-2.5.
\]
With $24$ degrees of freedom, the two-sided $p$-value is approximately
$0.020$. We reject $H_0$ at the $5\%$ level, providing evidence
of average underfilling and motivating a check of the machine.

\subsection{Test for Normality}\label{sec:normalitytests}

We might want to ask ourselves the question: \textbf{Are the weights normally distributed?} For the normal law, this is so common that several tests have been developed. They share the null hypothesis
\begin{equation}\label{eq:H0normal}
H_0: X_1, \ldots, X_n \text{ are i.i.d. } \mathcal{N}(\mu,\sigma^2) \text{ for some unknown } \mu \text{ and } \sigma^2,
\end{equation}
and differ in what they compare with the normal law:
\begin{itemize}
\item the whole distribution function: Kolmogorov--Smirnov, Lilliefors and Anderson--Darling tests;
\item the skewness and the kurtosis: Jarque--Bera test;
\item the frequencies in bins: chi-square goodness-of-fit test.
\end{itemize}
Q-Q plots (section \ref{sec:qqplot}) are their visual counterpart. We first look at the data, then present the tests, and finally apply them to the weights, the log weights and the heights.

\subsubsection{The Jarque--Bera Test}

A normal distribution has a skewness of $0$ and a kurtosis of $3$. The Jarque--Bera test checks both, with the sample skewness $S=\hat m_3/\hat m_2^{3/2}$ and kurtosis $K=\hat m_4/\hat m_2^{2}$, where $\hat m_k=\frac{1}{n}\sum_{i=1}^n (X_i-\bar X)^k$. Its statistic follows asymptotically a $\chi_2^2$ under $H_0$:
\begin{equation}
JB = \frac{n}{6} S^2 + \frac{n}{24} (K - 3)^2.
\end{equation}
We reject $H_0$ when $JB$ is above the critical $\chi_2^2$ threshold ($5.99$ at $5\%$). Note that the \texttt{kurtosis} function of \texttt{scipy.stats} returns the \emph{excess} kurtosis $K-3$ by default. Relying on two moments only, the test cannot detect departures from normality with $S=0$ and $K=3$.

\subsubsection{Visual Inspection}\label{sec:logweights}

Our sample body weights are skewed to the right, with a skewness of $1.35$ (see the histogram above). Taking the log reduces the skewness to $0.46$, and the log weights look closer to a normal distribution:
\begin{center}
\includegraphics[scale=0.5]{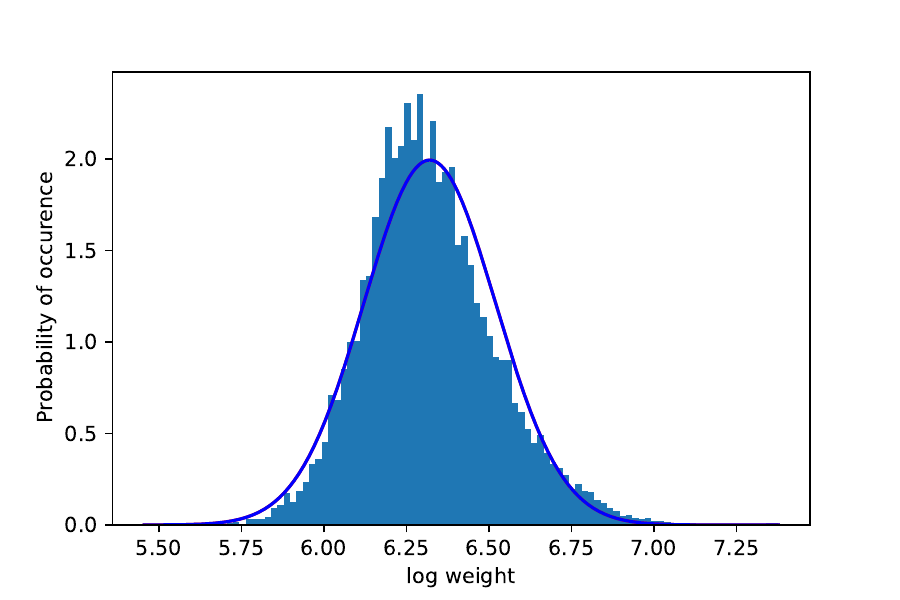}
\end{center}
Our sample body height is almost normally distributed, with a skewness of $0.11$:
\begin{center}
\includegraphics[scale=0.5]{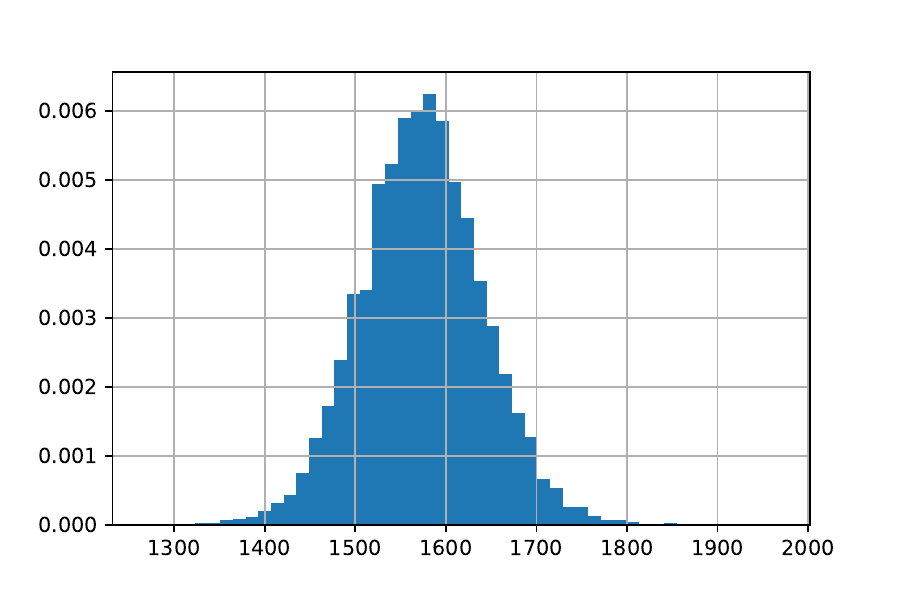}
\end{center}
A visual impression is not a test: we need a decision rule that controls the probability of wrongly rejecting normality.

\subsubsection{The Kolmogorov--Smirnov Test}\label{sec:KStest}

The Kolmogorov--Smirnov (KS) test compares the empirical cumulative distribution function (ECDF) $F_n(x)=\frac{1}{n}\sum_{i=1}^n\mathbf{1}\{X_i\leq x\}$ with a \emph{fully specified} continuous CDF $F_0$. Under $H_0$: $X_1,\ldots,X_n$ i.i.d.\ with CDF $F_0$, the distribution of
\begin{equation}\label{eq:KSstat}
D_n=\sup_{x\in\mathbb{R}}\big|F_n(x)-F_0(x)\big|
\end{equation}
does not depend on $F_0$, so that one table of critical values (for large $n$, the Kolmogorov distribution of $\sqrt{n}D_n$) serves all continuous $F_0$. We reject $H_0$ for large $D_n$.

For normality, $F_0$ must therefore be chosen \emph{before} looking at the data. Calling \texttt{kstest} with the standard normal CDF tests $\mathcal{N}(0,1)$, which is not our $H_0$ \eqref{eq:H0normal}. Plugging in the sample mean and variance instead, $F_0$ is fitted to the data and lies closer to $F_n$ than the true CDF would: $D_n$ tends to be smaller than the KS table assumes, and the test rejects too rarely. The Lilliefors test corrects this.

\paragraph{Two-sample version.} The KS test also compares the distributions of two observed samples, e.g.\ two return series, under $H_0$: both samples were drawn from the same distribution. The statistic is the maximum absolute (vertical) difference between their ECDFs $\hat{F}_1$ and $\hat{F}_2$:
\begin{equation}
KS=\max_x \lvert \hat{F}_1(x) - \hat{F}_2(x) \rvert.
\end{equation}
If the KS statistic is small (but be cautious in how you define small, as we talk about differences in distribution functions) or the $p$-value is high, we cannot reject $H_0$. Comparing our weights with a sample of the same size drawn from the fitted normal distribution (in red), we reject $H_0$:
\begin{center}
\includegraphics[scale=0.5]{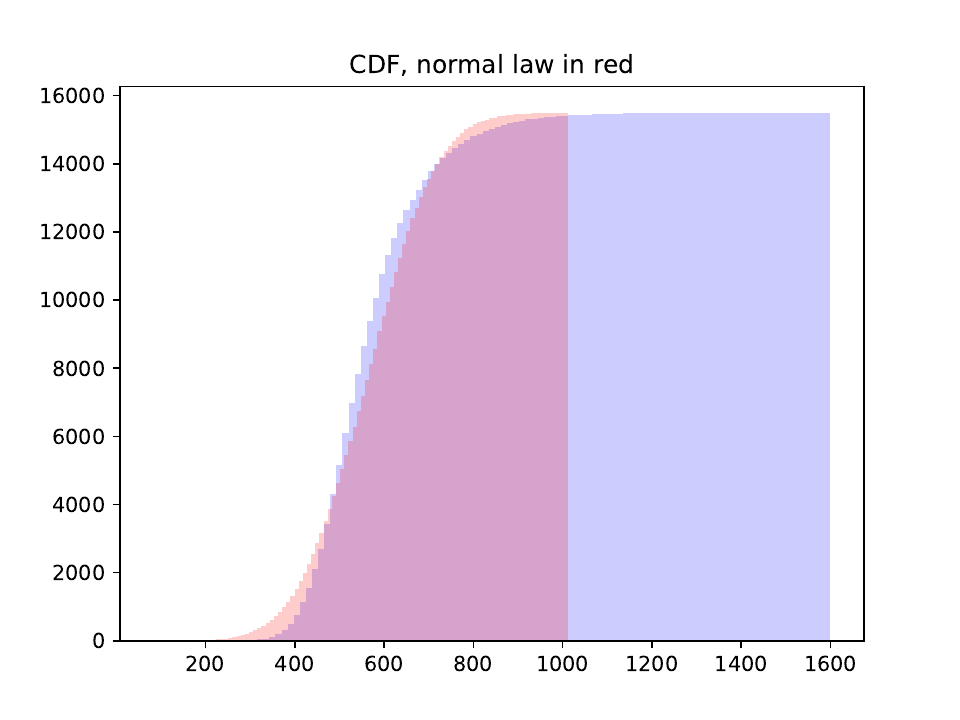}
\end{center}
The outcome depends on a random draw: for the heights, about half of the draws lead to a rejection at the $1\%$ level, and the other half do not. The one-sample tests below avoid this extra noise.

\subsubsection{The Lilliefors Test}

The Lilliefors test is a modification of the KS test, designed to test whether a sample comes from a normal distribution when the mean and variance are \emph{unknown} and must be estimated from the data.

\paragraph{Step 1. Null hypothesis.}
We test $H_0$ \eqref{eq:H0normal}, with unknown parameters $\mu$ and $\sigma^2$.

\paragraph{Step 2. Standardization.}
From the data we compute the sample mean and variance:
\[
\hat{\mu} = \frac{1}{n}\sum_{i=1}^n X_i, 
\qquad 
\hat{\sigma}^2 = \frac{1}{n-1}\sum_{i=1}^n (X_i - \hat{\mu})^2.
\]
We then standardize:
\[
Z_i = \frac{X_i - \hat{\mu}}{\hat{\sigma}}.
\]

\paragraph{Step 3. Empirical distribution.}
Order the standardized data:
\[
Z_{(1)} \leq Z_{(2)} \leq \cdots \leq Z_{(n)}.
\]
Define the empirical cumulative distribution function (ECDF):
\[
F_n(z) = \frac{1}{n} \sum_{i=1}^n \mathbf{1}\{Z_i \leq z\}.
\]

\paragraph{Step 4. Compare with fitted normal CDF.}
Let $\Phi(z)$ denote the CDF of the standard normal $\mathcal{N}(0,1)$.  
Compute the Kolmogorov–Smirnov type statistic:
\[
D_n = \sup_{z \in \mathbb{R}} \, \big| F_n(z) - \Phi(z) \big|.
\]

\paragraph{Step 5. Distribution under $H_0$.}
If $\mu$ and $\sigma$ were known, $D_n$ would follow the classical KS distribution.
But here they are estimated from the sample, which makes the critical values different, and smaller.
Lilliefors (1967) computed these critical values by Monte Carlo simulation, as we do in Table \ref{tab:normalitytests}: with $n=15\,494$, the $5\%$ critical value is $0.0074$, against $0.0109$ in the KS table.

\paragraph{Step 6. Decision rule.}
Reject $H_0$ at significance level $\alpha$ if
\[
D_n > c_\alpha(n),
\]
where $c_\alpha(n)$ is the Lilliefors critical value.

\subsubsection{The Anderson--Darling Test}\label{sec:ADtest}

Even with the Lilliefors critical values, the KS test is less powerful against most departures from normality than the Anderson--Darling and Shapiro--Wilk tests \citep{stephens1974edf}. If $\mu$ and $\sigma$ were known, $nF_n(z)$ would be binomial with parameters $n$ and $\Phi(z)$, so that $\Var\left(F_n(z)\right)=\Phi(z)\left(1-\Phi(z)\right)/n$ vanishes in the tails: the unweighted supremum $D_n$ mostly reacts to discrepancies near the centre of the distribution. The Anderson--Darling statistic divides the squared discrepancy by this variance and integrates it:
\begin{equation}\label{eq:ADstat}
\begin{aligned}
A^2 &= n\int_{-\infty}^{+\infty}\frac{\left(F_n(z)-\Phi(z)\right)^2}{\Phi(z)\left(1-\Phi(z)\right)}\,d\Phi(z)\\
&= -n-\frac{1}{n}\sum_{i=1}^{n}(2i-1)\left[\ln\Phi\left(Z_{(i)}\right)+\ln\left(1-\Phi\left(Z_{(n+1-i)}\right)\right)\right],
\end{aligned}
\end{equation}
so that discrepancies in the tails, where financial returns typically depart from normality, weigh more. We reject $H_0$ for large $A^2$, with critical values that, as in Step 5, account for the estimation of $\mu$ and $\sigma$.

\subsubsection{The Chi-Square Goodness-of-Fit Test}

The chi-square test tests if a sample of data came from a population with a specific distribution. This test is "better" suited for categorical or frequency data, but we introduce it here for completeness. The data are divided into $k$ bins, and the test statistic compares the observed frequency $O_i$ of bin $i$ with the frequency $E_i$ expected under $H_0$:
\begin{equation}
\chi^2 = \sum_{i=1}^k \frac{(O_i - E_i)^2}{E_i}.
\end{equation}
Under $H_0$, $\chi^2$ follows approximately a $\chi^2_{k-1-m}$, where $m$ is the number of parameters estimated to compute the $E_i$. The approximation requires expected frequencies that are not too small; a common rule is $E_i\geq 5$.

\paragraph{Example: is a coin fair?} With the $k=2$ bins heads and tails, we test $H_0: p_H = p_T = 0.5$, and no parameter is estimated. After $n$ flips, the expected counts are $E_H = E_T = n/2$, and
\[
\chi^2 = \frac{(O_H - E_H)^2}{E_H} + \frac{(O_T - E_T)^2}{E_T}
\]
follows a $\chi^2_1$ under $H_0$. With $n=24$ flips\footnote{code: in-class\_true\_false\_positives.py}, $14$ heads give $\chi^2=0.67$ ($p$-value $0.41$) and we do not reject that the coin is fair, whereas $20$ heads give $\chi^2=10.67$ ($p$-value $0.001$) and we reject it.

\paragraph{Application to normality.} We use $k=50$ bins that are equiprobable under the fitted normal distribution, so that $E_i=n/50\approx 310$ in every bin, and $m=2$ estimated parameters ($\mu$ and $\sigma$), hence $47$ degrees of freedom. Compared with the flat expected line, the weights have too few observations in the left tail, and too many around the mode and in the extreme right tail; we reject $H_0$:
\begin{center}
\includegraphics[scale=0.5]{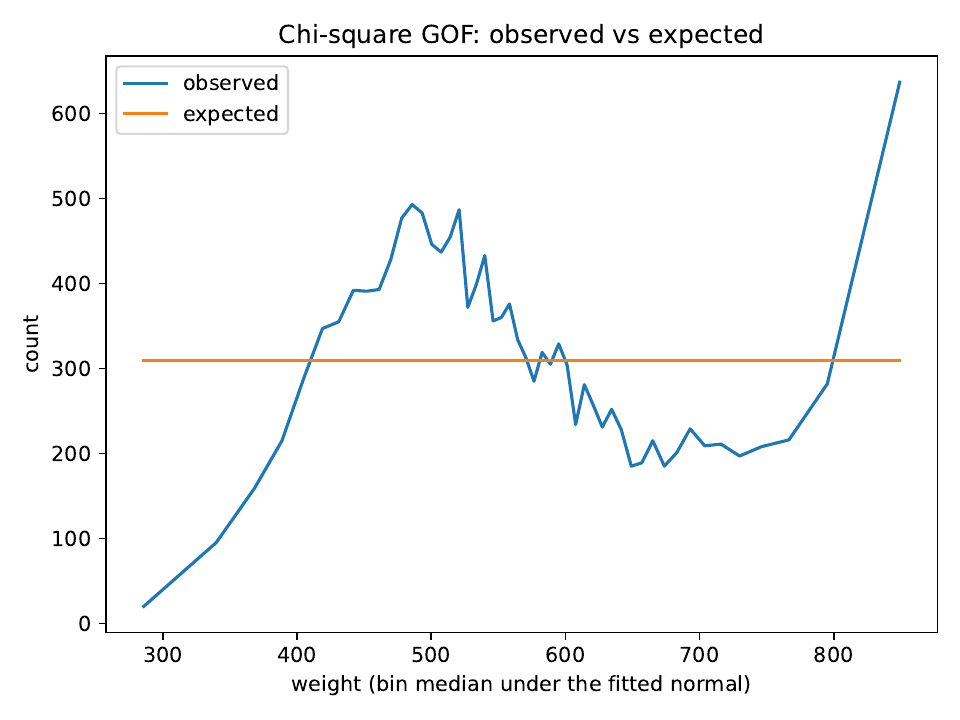}
\end{center}

\subsubsection{Results}

Table \ref{tab:normalitytests} applies the tests to our three samples of $n=15\,494$ observations. As Lilliefors did, we simulate the $5\%$ critical values: under $H_0$, the statistics do not depend on $\mu$ and $\sigma$, so we compute them on $1\,000$ samples of size $n$ drawn from $\mathcal{N}(0,1)$ and take their $95\%$ quantiles.

\begin{table}[H]
\centering
\caption{Normality tests on the weights, log weights and heights ($n=15\,494$)}\label{tab:normalitytests}
\begin{tabular}{lrrrr}
\toprule
 & Weight & Log weight & Height & $5\%$ critical value \\
\midrule
Skewness $S$ & $1.35$ & $0.46$ & $0.11$ & \\
Kurtosis $K$ & $7.04$ & $3.77$ & $3.67$ & \\
$D_n$ (Kolmogorov--Smirnov, Lilliefors) & $0.083$ & $0.041$ & $0.018$ & $0.0074$ \\
$A^2$ (Anderson--Darling) & $236.2$ & $50.5$ & $4.6$ & $0.79$ \\
$JB$ (Jarque--Bera) & $15\,210$ & $936$ & $320$ & $6.10$ \\
$\chi^2$ (goodness of fit, $k=50$) & $2\,178$ & $598$ & $380$ & $63.8$ \\
\bottomrule
\end{tabular}
\end{table}

All tests reject normality for the three samples. Taking the log shrinks every statistic, but the log weights still fail all tests. Even the heights, which look normal, are rejected: with $n=15\,494$ observations, the tests detect departures from normality too small to be seen on a histogram. Whether such departures matter depends on what we use the normal law for; Q-Q plots show where the data depart from it.

For $D_n$, the simulated critical value is well below the $0.0109$ of the KS table for a fully specified $F_0$: this is why the naive KS test rejects too rarely. For $JB$ and $\chi^2$, the simulated critical values are close to the asymptotic $\chi^2_2$ and $\chi^2_{47}$ ones, $5.99$ and $64.0$.

\subsection{Q-Q plots and probability plots}\label{sec:qqplot}

\subsubsection{Quantile-Quantile Plot (Q-Q Plot)}

\textbf{Definition:}
A Q-Q plot is a graphical tool used to compare the quantiles of two distributions. Typically, one of the distributions is empirical (from the data), and the other is theoretical. By plotting the quantiles of the empirical distribution against the quantiles of the theoretical distribution, we can visually assess how well the data conforms to the theoretical distribution. If the data lies along a straight line (usually a 45-degree line), it suggests that the data is well-represented by the theoretical distribution.

\textbf{Illustration:}
Imagine we have a dataset and want to check if it follows a parametric distribution. We can:

\begin{enumerate}
    \item Rank the data in ascending order.
    \item Calculate the quantiles of the empirical data.
    \item Calculate the corresponding quantiles for this parametric distribution (using the inverse of the cumulative distribution function).
    \item Plot the empirical quantiles against the theoretical quantiles.
\end{enumerate}

If the points in the Q-Q plot lie roughly along a straight line, it suggests that the data generating process could be the parametric distribution.

Quantile-Quantile plots\footnote{vansteenberghe\_qqplots.py} are an important tool to select a theoretical distribution to apply on a given sample. It has both the advantage and the drawback to be a visual tool.

In our approach, we assume that we are observing realisations $x_t$ from a random variable $X$ with theoretical probability density function $F$, a quantile of order $p$ is defined as\footnote{when the random variable $X$ is continuous, then $Pr(X<x)=Pr(X \leq x)$}
$$Q(p) = F^{-1}(p) = \min {x \in \mathbb{R} : F(x) \geq p} = \min {x \in \mathbb{R} : Pr(X<x) \geq p}$$
If we have $n$ observations, we can write:

\quad

\begin{tabular}{rllll}
Value&$x_1$&$x_2$&$\ldots$&$x_n$\\
Probability&$\frac{1}{n}$&$\frac{1}{n}$&$\ldots$&$\frac{1}{n}$
\end{tabular}

\quad

If there are $i,j$ such that $x_i = x_j$, then we regroup them into $y_1,\ldots,y_k$ for $k\leq n$, we now have, with $n_j$ the number of $x$'s regrouped in $y_j$:

\quad

\begin{tabular}{rllll}
Value&$y_1$&$y_2$&$\ldots$&$y_k$\\
Probability&$\frac{n_1}{n}$&$\frac{n_2}{n}$&$\ldots$&$\frac{n_k}{n}$
\end{tabular}

\quad

We define the empirical distribution function (EDF) as
\[
\hat{F}_n(x) \;=\; \frac{1}{n}\sum_{i=1}^n \mathbf{1}_{\{X_i \leq x\}},
\]
where $\mathbf{1}_{\{X_i \leq x\}}$ denotes the indicator function.

According to the law of large number\footnote{cf. section \ref{sec:largen}}, this empirical function converges to the theoretical repartition function in probability.

We can now define the empirical quantile of order $p$:
$$\hat{Q}_n(p) =\min \{x \in \mathbb{R} : \hat{F}_n(x) \geq p\}$$

$\hat{Q}_n(p)$ converges in probability to $Q(p)$.

Given an ordered sample of independent observations:
\begin{equation}
x_{(1)} \leq x_{(2)} \leq \ldots \leq x_{(n)}
\end{equation}
A quantile-quantile plot consists of the points:
\begin{equation}
\left( Q \left(\frac{i}{n+1}\right) , x_{(i)} \right), \quad i \in [1,n]
\end{equation}
or if we want to limit the number of plotted points to $k$
\begin{equation}
\left( Q \left(\frac{i}{k+1}\right) , \hat{Q}_k\left(\frac{i}{k+1}\right)  \right), \quad i \in [1,k]
\end{equation}

\subsubsection{Q-Q plot from a standard normal population}
If we take $n$ observations from a standard normal population, we can order those observations (we should also regroup the observations of same values). Then for each ordered observation $i$, we can compute the quantile of the theoretical standard normal distribution of probability $\frac{i}{n}$.

\begin{center}
\includegraphics[scale=0.5]{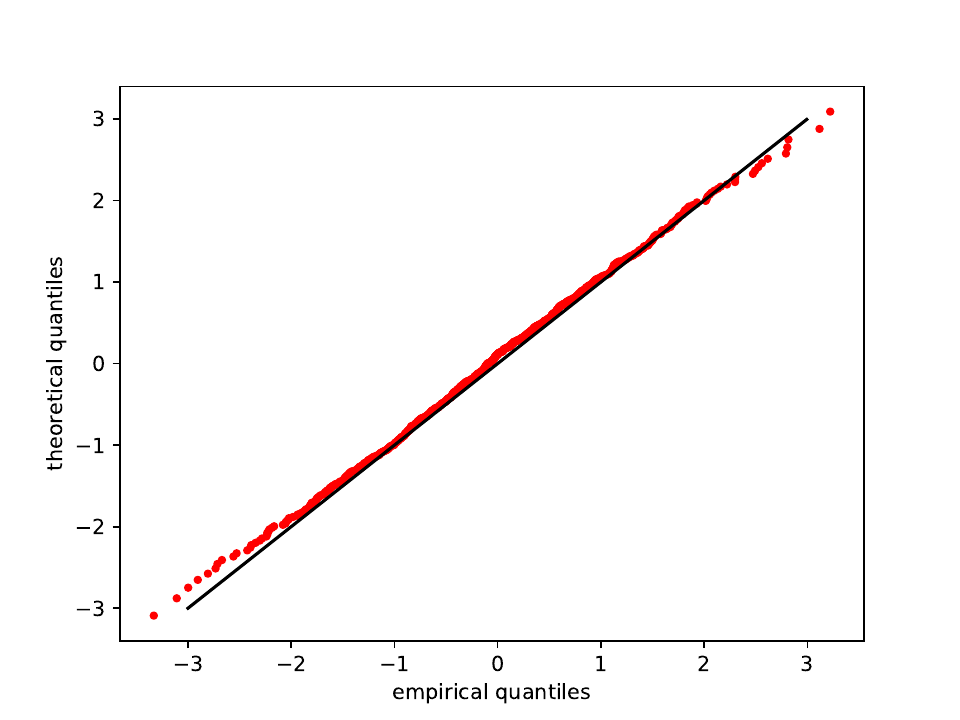}
\end{center}

\subsection{Searching for a more suitable distribution}

We run\footnote{code: vansteenberghe\_height\_normal\_distrib.py} a loop searching for alternative distributions and based on the p-values of Kolmogorov-Smirnov test, this can be very specific to the time frame we look at and might not be generalized, we select:
\begin{center}
\includegraphics[scale=0.5]{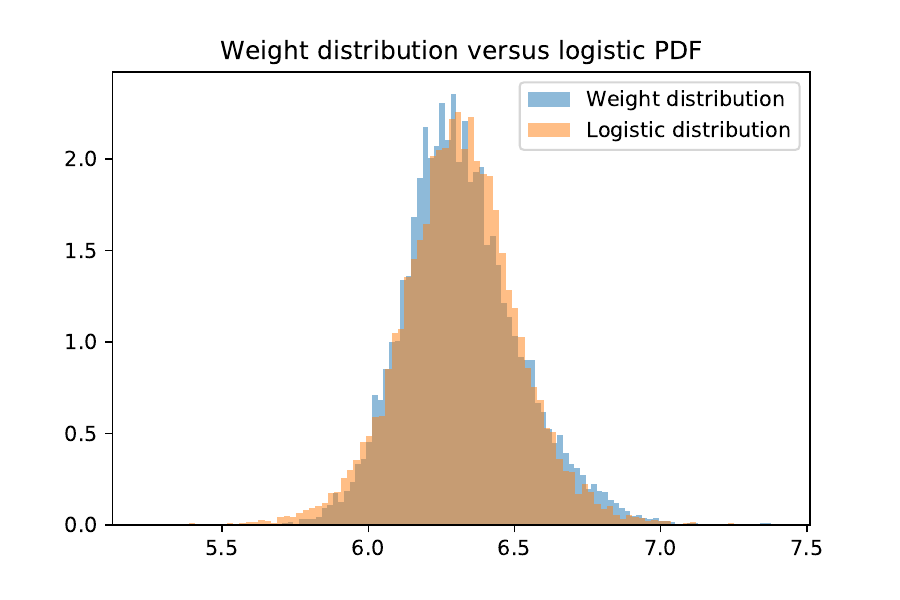}
\end{center}

\subsection{Pearson's Chi-Squared Test for Independence}

Pearson's chi-squared test also assesses the independence between two categorical variables, by comparing the observed frequencies of a contingency table with those expected under independence. It is useful to evaluate a binary classifier, for instance in medical diagnostics, where true and false positives matter greatly. The counts of true positives, false negatives, false positives and true negatives form a $2 \times 2$ contingency table:

\begin{center}
\begin{tabular}{|c|c|c|}
\hline
\textbf{} & \textbf{Predicted Positive} & \textbf{Predicted Negative} \\
\hline
\textbf{Actual Positive} & True Positives (TP) & False Negatives (FN) \\
\hline
\textbf{Actual Negative} & False Positives (FP) & True Negatives (TN) \\
\hline
\end{tabular}
\end{center}

The null hypothesis is that the predicted classification is independent of the actual classification, as for a classifier that guesses at random:
\[
H_0: \text{Predicted classification is independent of actual classification.}
\]
Under $H_0$, the expected frequency of cell $(i,j)$ is
\[
E_{ij} = \frac{(\text{Row Total}_i)(\text{Column Total}_j)}{\text{Total Observations}},
\]
and, with $O_{ij}$ the observed frequency,
\[
\chi^2 = \sum_{i=1}^{2} \sum_{j=1}^{2} \frac{(O_{ij} - E_{ij})^2}{E_{ij}}
\]
follows asymptotically a $\chi^2_1$ (a $\chi^2_{(r-1)(c-1)}$ for an $r \times c$ table). If the statistic exceeds the critical value, we reject $H_0$ and conclude that there is a significant relationship between predicted and actual classifications.

\textit{Both chi-square tests compare observed and expected frequencies: the goodness-of-fit test checks whether data follow a specified distribution, the test of independence whether two categorical variables are independent.}

\section{Econometrics: Main theory before the practice}

\subsection{Ordinary Least Squares}

Consider the linear regression model
\[
y=X\beta+\varepsilon,
\]
where \(y\) is a \(T\times 1\) vector and \(X\) is a
\(T\times k\) matrix. The standard assumptions are:
\begin{itemize}
    \item[H1.] \(\mathbb{E}(\varepsilon\mid X)=0\).
    \item[H2.] \(X\) has full column rank:
    \[
    \operatorname{rank}(X)=k.
    \]
    \item[H3.] The disturbances are conditionally homoskedastic
    and uncorrelated:
    \[
    \operatorname{Var}(\varepsilon\mid X)=\sigma^2 I_T,
    \qquad 0<\sigma^2<\infty.
    \]
\end{itemize}

Under H2, the Ordinary Least Squares estimator is
\[
\widehat{\beta}_{\mathrm{OLS}}
=
\underset{b\in\mathbb{R}^k}{\operatorname{arg\,min}}
(y-Xb)'(y-Xb)
=
(X'X)^{-1}X'y.
\]
Under H1 and H2,
\[
\mathbb{E}(\widehat{\beta}_{\mathrm{OLS}}\mid X)=\beta.
\]
Under H1--H3,
\[
\operatorname{Var}(\widehat{\beta}_{\mathrm{OLS}}\mid X)
=
\sigma^2(X'X)^{-1}.
\]

To interpret \(X'X/T\), recall that
\[
\operatorname{Var}(y)
=\mathbb{E}\left[
(y-\mathbb{E}(y))(y-\mathbb{E}(y))'
\right].
\]
Similarly, the sample variance-covariance matrix of the
columns of \(X\) is
\[
S_X=\frac{1}{T-1}
(X-\mathbf{1}_T\bar{x}')'(X-\mathbf{1}_T\bar{x}'),
\qquad
\bar{x}=\frac{1}{T}X'\mathbf{1}_T,
\]
where \(\mathbf{1}_T\) is a vector of ones.
Thus,
\[
\frac{X'X}{T}
=\frac{T-1}{T}S_X+\bar{x}\bar{x}'.
\]
Therefore, \(X'X/T\) is a second-moment matrix:
it includes both the sample covariances and the
products of the sample means.
\paragraph{Gauss--Markov theorem.} The Gauss--Markov theorem states that, under H1--H3,
\(\widehat{\beta}_{\mathrm{OLS}}\) is BLUE: the Best Linear
Unbiased Estimator. Here, linearity refers to linearity in \(y\),
and efficiency is relative to estimators that are conditionally
unbiased given \(X\).

\subsection{Consistency and Asymptotic Normality}

Throughout the asymptotic analysis, the number of regressors
\(k\) is fixed as \(T\to\infty\). For consistency, assume:
\begin{itemize}
    \item[H4.]
    \[
    \frac{X'X}{T}\xrightarrow{p}V_X,
    \]
    where \(V_X\) is finite and positive definite.
    \item[H5.]
    \[
    \frac{X'\varepsilon}{T}\xrightarrow{p}0.
    \]
\end{itemize}
Since
\[
\widehat{\beta}_{\mathrm{OLS}}-\beta
=
\left(\frac{X'X}{T}\right)^{-1}
\frac{X'\varepsilon}{T},
\]
H4 and H5 imply
\[
\widehat{\beta}_{\mathrm{OLS}}\xrightarrow{p}\beta.
\]
These consistency conditions do not require H3.

For asymptotic normality, retain H1--H3 and assume H4 together
with conditions ensuring the central limit theorem
\[
\frac{X'\varepsilon}{\sqrt{T}}
\xrightarrow{d}
\mathcal{N}(0,\sigma^2 V_X).
\]

\paragraph{Hint: applying the central limit theorem.}
For a simple example, suppose that the observations
\((x_t,\varepsilon_t)\) are i.i.d., where \(x_t'\) is
row \(t\) of \(X\), and
\(\mathbb{E}\|x_t\|^2<\infty\). Matrix multiplication gives
\[
\frac{X'\varepsilon}{\sqrt{T}}
=\frac{1}{\sqrt{T}}\sum_{t=1}^T x_t\varepsilon_t.
\]
Under H1, this vector has mean zero. Its covariance
matrix can therefore be calculated as
\[
\begin{aligned}
\operatorname{Var}\left(\frac{X'\varepsilon}{\sqrt{T}}\right)
&=\frac{1}{T}\mathbb{E}\left[X'\varepsilon\varepsilon'X\right]\\
&=\frac{1}{T}\mathbb{E}\left[
X'\mathbb{E}(\varepsilon\varepsilon'\mid X)X
\right]\\
&=\frac{\sigma^2}{T}\mathbb{E}(X'X),
\end{aligned}
\]
where H1 and H3 give
\(\mathbb{E}(\varepsilon\varepsilon'\mid X)=\sigma^2 I_T\).

Now observe that in this i.i.d.\ setting,
\[
\mathbb{E}\left(\frac{X'X}{T}\right)=V_X.
\]
Thus,
\[
\operatorname{Var}\left(\frac{X'\varepsilon}{\sqrt{T}}\right)
=\sigma^2 V_X.
\]

Now, write the scaled estimation error as
\[
\sqrt{T}\left(\widehat{\beta}_{\mathrm{OLS}}-\beta\right)
=
\underbrace{\left(\frac{X'X}{T}\right)^{-1}}_
{\xrightarrow{p}\,V_X^{-1}}
\underbrace{\frac{X'\varepsilon}{\sqrt{T}}}_
{\xrightarrow{d}\,\mathcal{N}(0,\sigma^2 V_X)}.
\]
By Slutsky's theorem, the limiting distribution is that
of \(V_X^{-1}Z\), where
\(Z\sim\mathcal{N}(0,\sigma^2 V_X)\).
Recall that multiplying a random vector by a constant
matrix \(B\) transforms its covariance into
\(B\operatorname{Var}(Z)B'\). Hence,
\[
\begin{aligned}
\operatorname{Var}(V_X^{-1}Z)
&=V_X^{-1}\operatorname{Var}(Z)(V_X^{-1})'\\
&=V_X^{-1}(\sigma^2 V_X)V_X^{-1}\\
&=\sigma^2 V_X^{-1},
\end{aligned}
\]
since \(V_X\) is symmetric. Therefore,
\[
\sqrt{T}\left(\widehat{\beta}_{\mathrm{OLS}}-\beta\right)
\xrightarrow{d}
\mathcal{N}\left(0,\sigma^2 V_X^{-1}\right).
\]

To see how sample size affects the conditional variance
of the unscaled estimator, write
\[
X'X=T\left(\frac{X'X}{T}\right),
\qquad
(X'X)^{-1}
=\frac{1}{T}\left(\frac{X'X}{T}\right)^{-1}.
\]
Thus, under H1--H4,
\[
\operatorname{Var}(\widehat{\beta}_{\mathrm{OLS}}\mid X)
=
\frac{\sigma^2}{T}
\underbrace{\left(\frac{X'X}{T}\right)^{-1}}_
{\xrightarrow{p}\,V_X^{-1}}
\xrightarrow{p}0.
\]
The matrix factor converges to a finite matrix, while
\(1/T\) tends to zero: the conditional variance decreases
at rate \(1/T\).

\subsection{Generalized Least Squares}

Retain H1 and H2, but replace H3 by
\[
\operatorname{Var}(\varepsilon\mid X)=\sigma^2\Omega,
\]
where \(\Omega\) is symmetric, positive definite, and known
for GLS estimation. It may be a function of \(X\).

Since \(\Omega\) is symmetric, its eigenvectors
\(p_1,\ldots,p_T\) can be chosen to have unit length and
to be perpendicular to one another:
\[
p_i'p_j=
\begin{cases}
1 & \text{if }i=j,\\
0 & \text{if }i\neq j.
\end{cases}
\]
Collect these eigenvectors as columns of
\(P=[p_1,\ldots,p_T]\). Matrix multiplication then gives
\[
P'P=
\begin{pmatrix}
p_1'p_1 & \cdots & p_1'p_T\\
\vdots & \ddots & \vdots\\
p_T'p_1 & \cdots & p_T'p_T
\end{pmatrix}
=I_T.
\]
Thus, \(P\) is an orthogonal matrix. Since it is square,
\(P'=P^{-1}\), and consequently \(PP'=I_T\) as well.

Let \(d_i\) be the eigenvalue associated with \(p_i\).
The eigenvector equations \(\Omega p_i=d_i p_i\) give
\[
\Omega P=PD,
\qquad D=\operatorname{diag}(d_1,\ldots,d_T).
\]
Multiplying on the right by \(P'\) gives the spectral
decomposition:
\[
\Omega PP'=PDP'
\quad\Longrightarrow\quad
\Omega=PDP'.
\]
Because \(\Omega\) is positive definite, all \(d_i>0\).

This is the spectral decomposition,
\[
\Omega=PDP',
\]
where \(P\) is orthogonal and \(D\) is diagonal with strictly
positive diagonal elements \(d_1,\ldots,d_T\). Define
\[
\Omega^{-1/2}=PD^{-1/2}P',
\qquad
D^{-1/2}
=
\operatorname{diag}
\left(
\frac{1}{\sqrt{d_1}},\ldots,\frac{1}{\sqrt{d_T}}
\right).
\]
Premultiplying the model by \(\Omega^{-1/2}\) gives
\[
\Omega^{-1/2}y
=
\Omega^{-1/2}X\beta+\Omega^{-1/2}\varepsilon.
\]
The transformed disturbance satisfies
\[
\mathbb{E}(\Omega^{-1/2}\varepsilon\mid X)=0,
\qquad
\operatorname{Var}(\Omega^{-1/2}\varepsilon\mid X)
=
\sigma^2 I_T.
\]

Since \(\Omega^{-1/2}=PD^{-1/2}P'\), it is symmetric:
\[
(\Omega^{-1/2})'
=(PD^{-1/2}P')'
=PD^{-1/2}P'
=\Omega^{-1/2}.
\]
Under H1, the transformed disturbance has conditional
mean zero. Hence,
\[
\begin{aligned}
\operatorname{Var}(\Omega^{-1/2}\varepsilon\mid X)
&=\mathbb{E}\left[
(\Omega^{-1/2}\varepsilon)
(\Omega^{-1/2}\varepsilon)'\mid X
\right]\\
&=\mathbb{E}\left[
\Omega^{-1/2}\varepsilon\varepsilon'
(\Omega^{-1/2})'\mid X
\right]\\
&=\Omega^{-1/2}
\mathbb{E}(\varepsilon\varepsilon'\mid X)
\Omega^{-1/2}\\
&=\sigma^2\Omega^{-1/2}\Omega\Omega^{-1/2}\\
&=\sigma^2 P D^{-1/2}D D^{-1/2}P'\\
&=\sigma^2 PP'
=\sigma^2 I_T.
\end{aligned}
\]
The matrices \(\Omega^{-1/2}\) can be taken outside the
conditional expectation because they are known given \(X\).

Thus, applying OLS to the transformed model yields the
Generalized Least Squares estimator:
\[
\widehat{\beta}_{\mathrm{GLS}}
=
(X'\Omega^{-1}X)^{-1}X'\Omega^{-1}y.
\]
Its conditional variance is
\[
\operatorname{Var}(\widehat{\beta}_{\mathrm{GLS}}\mid X)
=
\sigma^2(X'\Omega^{-1}X)^{-1}.
\]

If \(\Omega=S(\alpha)\) is unknown and depends on a parameter
vector \(\alpha\), the Feasible GLS procedure is:
\begin{enumerate}
    \item Estimate the model by OLS and obtain residuals
    \(\widehat{\varepsilon}\).
    \item Use the residuals to estimate \(\widehat{\alpha}\).
    \item Compute
    \[
    \widehat{\Omega}=S(\widehat{\alpha}).
    \]
    \item Provided \(\widehat{\Omega}\) is positive definite,
    compute
    \[
    \widehat{\beta}_{\mathrm{FGLS}}
    =
    (X'\widehat{\Omega}^{-1}X)^{-1}
    X'\widehat{\Omega}^{-1}y.
    \]
\end{enumerate}

For GLS consistency, assume
\[
\frac{X'\Omega^{-1}X}{T}\xrightarrow{p}V_{\Omega X},
\]
where \(V_{\Omega X}\) is finite and positive definite, and
\[
\frac{X'\Omega^{-1}\varepsilon}{T}\xrightarrow{p}0.
\]
Then
\[
\widehat{\beta}_{\mathrm{GLS}}\xrightarrow{p}\beta.
\]
If, additionally, a central limit theorem gives
\[
\frac{X'\Omega^{-1}\varepsilon}{\sqrt{T}}
\xrightarrow{d}
\mathcal{N}(0,\sigma^2 V_{\Omega X}),
\]
then
\[
\sqrt{T}\left(\widehat{\beta}_{\mathrm{GLS}}-\beta\right)
\xrightarrow{d}
\mathcal{N}\left(0,\sigma^2V_{\Omega X}^{-1}\right).
\]
FGLS has the same limiting distribution if estimating
\(\Omega\) is asymptotically negligible, that is,
\[
\sqrt{T}\left(
\widehat{\beta}_{\mathrm{FGLS}}
-
\widehat{\beta}_{\mathrm{GLS}}
\right)\xrightarrow{p}0.
\]
This requires additional regularity conditions; it does not
follow automatically from the GLS assumptions above.

\subsection{Endogeneity and Instrumental Variables}

Consider again
\[
y=X\beta+\varepsilon,
\]
but suppose that
\[
\mathbb{E}(\varepsilon\mid X)\neq 0.
\]
Failure of H1 alone does not necessarily imply OLS
inconsistency: consistency depends on the limiting
regressor--disturbance sample moments.

Indeed,
\[
\widehat{\beta}_{\mathrm{OLS}}
=
\beta+(X'X)^{-1}X'\varepsilon
=
\beta+
\left(\frac{X'X}{T}\right)^{-1}
\frac{X'\varepsilon}{T}.
\]
If
\[
\frac{X'X}{T}\xrightarrow{p}V_X,
\]
where \(V_X\) is finite and positive definite, and
\[
\frac{X'\varepsilon}{T}
\xrightarrow{p}\Sigma_{X\varepsilon}\neq 0,
\]
then
\[
\widehat{\beta}_{\mathrm{OLS}}
\xrightarrow{p}
\beta+V_X^{-1}\Sigma_{X\varepsilon}
\neq\beta.
\]
OLS is therefore inconsistent in this case.

Let \(Z\) be a \(T\times q\) matrix of instruments, with
\(q\) fixed as \(T\to\infty\). For identification and consistency
of the IV estimator, assume:
\begin{enumerate}
    \item \(q\geq k\), so there are at least as many instruments
    as regressors.

    \item Instrument exogeneity and a suitable law of large
    numbers ensure
    \[
    \frac{Z'\varepsilon}{T}\xrightarrow{p}0.
    \]
    For example, writing \(z_t'\) for row \(t\) of \(Z\), the
    population orthogonality conditions
    \(\mathbb{E}(z_t\varepsilon_t)=0\), together with appropriate
    moment and dependence conditions, imply this convergence.
    Orthogonality alone is not sufficient.

    \item Instrument relevance:
    \[
    \frac{Z'X}{T}\xrightarrow{p}\Sigma_{ZX},
    \qquad
    \operatorname{rank}(\Sigma_{ZX})=k,
    \]
    where \(\Sigma_{ZX}\) is finite.

    \item
    \[
    \frac{Z'Z}{T}\xrightarrow{p}\Sigma_{ZZ},
    \]
    where \(\Sigma_{ZZ}\) is finite and positive definite.
\end{enumerate}

Provided \(Z'Z\) is invertible, define the projection matrix
\[
P_Z=Z(Z'Z)^{-1}Z'.
\]
Provided \(X'P_ZX\) is also invertible, the Two-Stage Least
Squares estimator is
\[
\widehat{\beta}_{\mathrm{2SLS}}
=
(X'P_ZX)^{-1}X'P_Zy.
\]
The limiting rank conditions above ensure that these
invertibility requirements hold with probability approaching
one.

Substituting \(y=X\beta+\varepsilon\) gives
\[
\widehat{\beta}_{\mathrm{2SLS}}
=
\beta+(X'P_ZX)^{-1}X'P_Z\varepsilon.
\]
Equivalently,
\[
\widehat{\beta}_{\mathrm{2SLS}}
=
\beta+
\left[
\frac{X'Z}{T}
\left(\frac{Z'Z}{T}\right)^{-1}
\frac{Z'X}{T}
\right]^{-1}
\frac{X'Z}{T}
\left(\frac{Z'Z}{T}\right)^{-1}
\frac{Z'\varepsilon}{T}.
\]
Since
\[
\frac{Z'\varepsilon}{T}\xrightarrow{p}0
\]
and the matrix inside square brackets converges to the
positive definite matrix
\[
\Sigma_{ZX}'\Sigma_{ZZ}^{-1}\Sigma_{ZX},
\]
it follows that
\[
\widehat{\beta}_{\mathrm{2SLS}}\xrightarrow{p}\beta.
\]

\subsection{Too Many Weak Instruments}
\label{sec:manyweakIV}

The consistency of the IV estimator derived above assumes
that the number of instruments is fixed as $T\to\infty$.
When the number of instruments is large relative to the
sample size, the first-stage regression may overfit the
endogenous variable, introducing substantial bias into
2SLS estimation \citep{anatolyev2019many}.

To illustrate, consider one endogenous variable $x_t$:
\[
y_t=\beta x_t+u_t,
\qquad
x_t=\underbrace{\sum_{c\in\mathcal C}
\pi_c z_{c,t}}_{z_t'\pi}+v_t,
\]
where $z_{c,t}$ denotes instrument $c$, $K=|\mathcal C|$,
$\mathbb E(z_tu_t)=0$, and
$\mathbb E(v_tu_t)=\sigma_{vu}\neq0$.
Let $Z$ be the $T\times K$ instrument matrix and define
\[
P=Z(Z'Z)^{-1}Z'.
\]
The 2SLS estimator satisfies
\[
\widehat\beta_{\mathrm{2SLS}}-\beta
=(x'Px)^{-1}x'Pu.
\]
Since $x=Z\pi+v$, the fitted values contain both the
relevant component $Z\pi$ and the projected noise $Pv$.
For this illustration, assume
\[
\mathbb E(u\mid Z)=\mathbb E(v\mid Z)=0,
\qquad
\mathbb E(vv'\mid Z)=\sigma_v^2 I_T,
\qquad
\mathbb E(vu'\mid Z)=\sigma_{vu} I_T.
\]
Then, increasing the number of instruments also increases
the amount of endogenous noise fitted in the first stage.
In the extreme case $K=T$ and $\operatorname{rank}(Z)=T$,
we have $P=I_T$: 2SLS coincides exactly with the
potentially inconsistent OLS estimator.

Instrument strength must therefore be evaluated jointly.
In the homoskedastic, constant-coefficient reference model,
the concentration parameter is
\[
\boxed{
\mu_T^2=\frac{\pi'Z'Z\pi}{\sigma_v^2},
}
\]
where $\sigma_v^2=\operatorname{Var}(v_t)$.
It measures the identifying information contained in the
entire instrument set relative to first-stage noise.
In particular,
\[
\mathbb E(x'Px\mid Z)
=\sigma_v^2(\mu_T^2+K),
\]
so that the ratio of expected numerator and denominator
of the 2SLS estimation error is
\[
\frac{\mathbb E(x'Pu\mid Z)}
     {\mathbb E(x'Px\mid Z)}
=
\frac{\sigma_{vu}}{\sigma_v^2}
\frac{K}{\mu_T^2+K}.
\]
This ratio illustrates the bias mechanism, although it is
not the exact finite-sample bias of 2SLS. A large number
of individually weak instruments may be informative
collectively, but their number may also increase faster
than their identifying information.

As discussed by \citet{anatolyev2019many}, conventional 2SLS
may become inconsistent under many-instrument asymptotics
when $K/T\to\alpha>0$. LIML and the Fuller estimator
address this problem under homoskedasticity. However,
conventional LIML may also become inconsistent in the
presence of heteroskedasticity and many instruments.

For this reason, \citet{hausman2012instrumental} propose the
heteroskedasticity-robust Fuller estimator (HFUL).
Let
\[
D=\operatorname{diag}(P_{11},\ldots,P_{TT}),
\qquad Q=P-D,
\qquad \bar X=[y,X],
\]
and let $\widetilde\alpha$ be the smallest generalized
eigenvalue of
\[
\bigl(\bar X'Q\bar X,\bar X'\bar X\bigr).
\]
The Fuller correction with $C=1$ is
\[
\widehat\alpha
=
\frac{\widetilde\alpha-(1-\widetilde\alpha)/T}
     {1-(1-\widetilde\alpha)/T}.
\]
The HFUL estimator is then
\[
\boxed{
\widehat\beta_{\mathrm{HFUL}}
=
\left[X'(P-D-\widehat\alpha I_T)X\right]^{-1}
X'(P-D-\widehat\alpha I_T)y.
}
\]
Removing the diagonal elements of $P$ eliminates the
own-observation terms responsible for many-instrument
bias under the paper's independence assumptions.
The Fuller correction improves the moment properties
of the estimator relative to LIML.

In practice, instrument relevance should be assessed
through the joint first-stage relationship rather than
separate significance tests for individual instruments.
For a fixed number of instruments,
\citet{olea2013robust} propose an effective
$F$-statistic robust to heteroskedasticity and serial
correlation, with critical values designed to assess
weak identification. Conventional $F>10$ rules are not
generally reliable when instruments are numerous.

Finally, HFUL does not resolve identification when the
instruments are collectively too weak. Moreover, its
standard many-instrument theory assumes independence
across observations and does not directly extend to
serially dependent time series or locally weighted
time-varying IV estimation.

\section{Linear regression on weight and height}\label{sec:weightheightregression}

We study here\footnote{vansteenberghe\_weight\_height\_OLS.py} linear regression estimated by Ordinary Least Squares (OLS), on the $n=15\,494$ women aged 15 to 49 of section \ref{sec:weightheightdistrib}, with weight $w_i$ in kilograms and height $h_i$ in centimetres.\footnote{The figure files keep their historical prefix \texttt{Indian\_}, but the sample pools 20 Demographic and Health Surveys conducted in India and in 17 sub-Saharan African countries; 1\,952 of the women are Indian (variable \texttt{v000}). The data record weight (\texttt{v437}) in tenths of a kilogram and height (\texttt{v438}) in millimetres, so that the slope of \texttt{v437} on \texttt{v438} is in kilograms per centimetre, while intercepts and residuals must be divided by 10 to be in kilograms.} Section \ref{sec:weightheightdistrib} showed that weight and height are positively correlated ($r=0.412$). A regression goes further: it describes how \emph{average} weight changes with height, in kilograms per centimetre, and it tells us how precisely the data pin this change down.

We follow the sequence of \citet[ch.~2]{verbeek} and \citet[ch.~5--6]{fox2015}. OLS is first an algebraic tool, which fits a line to any data set without any assumption. It becomes an estimator of a population parameter once we specify a statistical model, and its properties then depend on assumptions that we state one at a time. \citet[ch.~4--5]{stock2020introduction} cover the same ground at an introductory level, and \citet[ch.~2--4 and~7]{hansen2022econometrics} at a more advanced one.

\subsection{The regression question: a conditional mean}\label{sec:regCEF}

\begin{figure}[!htbp]
\centering
\includegraphics[width=\textwidth]{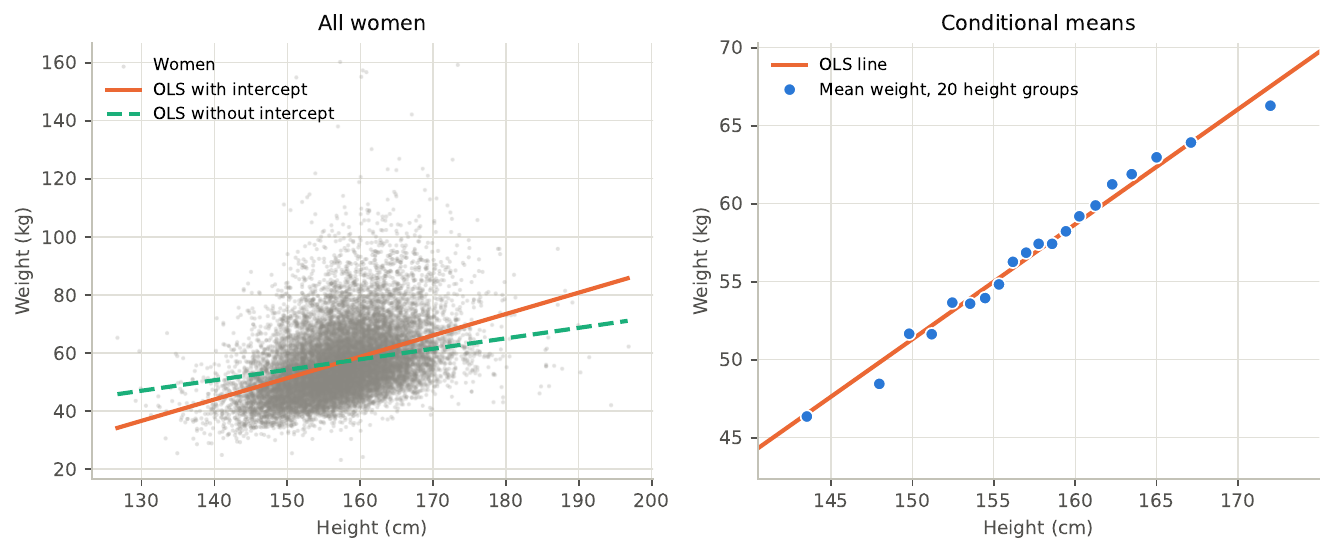}
\caption{Weight and height of $15\,494$ women. Left: the data, with the OLS lines estimated with and without an intercept. Right: the mean weight in 20 groups of roughly equal size sorted by height, and the OLS line.}\label{fig:binnedmeans}
\end{figure}

For each height $h$, the women of height $h$ have a distribution of weights. Regression analysis summarizes how this conditional distribution shifts with $h$, usually through its mean, the \emph{conditional expectation function} (CEF)
\begin{equation}\label{eq:CEF}
m(h)=\E(w\mid h).
\end{equation}
With $15\,494$ women, we can estimate $m(\cdot)$ without assuming a functional form, by averaging the weights of women of similar heights \citep[section 2.2]{fox2015}. The right panel of Figure \ref{fig:binnedmeans} sorts the women into 20 groups of roughly equal size by height and plots the mean weight of each group against its mean height: the 20 means lie within 1.4~kg of a straight line, the OLS line estimated below.

Two properties make the CEF the natural object of study. Define the CEF error $e=w-m(h)$. By construction, $\E(e\mid h)=0$; by the law of iterated expectations, $\E(e)=0$ and $\E[g(h)\,e]=0$ for any function $g$ such that the expectation exists. These properties hold by definition, not by assumption \citep[section 2.8]{hansen2022econometrics}. Consequently, if $w$ has a finite variance, for any predictor $g(h)$ with finite variance,
\begin{equation}\label{eq:CEFbestpredictor}
\E\left[(w-g(h))^2\right]=\E\left[(w-m(h))^2\right]+\E\left[(m(h)-g(h))^2\right]\geq \E\left[(w-m(h))^2\right],
\end{equation}
since the cross-product $2\,\E\left[e\,(m(h)-g(h))\right]$ is zero: the CEF is the best predictor of weight given height in mean squared error.

\paragraph{The best linear predictor.} Restricting the predictor to a line $b_0+b_1h$ and minimizing $\E[(w-b_0-b_1h)^2]$ gives the first-order conditions $\E(u)=0$ and $\E(h\,u)=0$, with $u=w-\beta_0-\beta_1h$. Their solution is the \emph{best linear predictor} (BLP)
\begin{equation}\label{eq:BLP}
\beta_1=\frac{\Cov(h,w)}{\Var(h)},\qquad \beta_0=\E(w)-\beta_1\,\E(h).
\end{equation}
It exists as soon as weight and height have finite variances and $\Var(h)>0$, whatever the shape of the CEF, and it is also the best linear approximation to the CEF \citep[sections 2.18 and 2.25]{hansen2022econometrics}. When the CEF is linear, the two coincide.

\paragraph{Three readings of a regression line.} Following the distinction of \citet[section 6.3]{fox2015} between empirical and structural relations:
\begin{itemize}
\item \emph{description}: $\beta_1$ is the slope of the BLP. It summarizes how mean weight differs across women of different heights and requires no assumption beyond finite variances. In a cross-section, ``one centimetre taller'' compares two women who differ by one centimetre; it does not describe a woman who grows by one centimetre \citep[section 5.1.1]{fox2015};
\item \emph{conditional mean}: if $\E(w\mid h)=\beta_0+\beta_1h$, then $\beta_1$ is the derivative of the CEF. This is the linear regression model below, under assumption (A4);
\item \emph{structure}: $\beta_1$ measures how weight would change if height changed, all other determinants of weight being held fixed. This requires that these other determinants, gathered in the error term, be unrelated to height (section \ref{sec:OVBweight}). For height, which cannot be manipulated, the counterfactual itself is delicate \citep[section 2.30]{hansen2022econometrics}.
\end{itemize}
For weight and height, the first two readings are the relevant ones.

\subsection{Linear regression in practice}

In \texttt{Python} or \texttt{R}, the formula notation is:
\begin{equation}
y \sim x
\end{equation}
which implicitly includes an intercept. 

If you apply it on our data set, you can focus on the following information:
\begin{itemize}
\item F-test p-value, see section \ref{sec:fishertestwage}: The F-statistic in statsmodels.formula.api (smf) OLS tests whether your overall regression model is statistically significant compared to a model with no predictors (where all slope coefficients except the intercept are zero);
\item $R^2$ and adjusted $R^2$;
\item $\beta_0$  and $\beta_1$ with their confidence interval, their t-stat and p-value;
\item the Durbin-Watson statistics, see section \ref{sec:durbinwatsontest};
\end{itemize}

\textbf{Definition.}
Linear regression models the relationship between a dependent variable and one or more explanatory variables. In the simple linear case, the model writes
\begin{equation}\label{eq:simpleLRM}
y_i = \beta_0 + \beta_1 x_i + \varepsilon_i, \qquad i=1,\dots,n,
\end{equation}
where:
\begin{itemize}
    \item $y_i$ is the dependent variable (e.g.\ weight $w_i$),
    \item $x_i$ is the explanatory variable (e.g.\ height $h_i$),
    \item $\beta_0$ is the intercept,
    \item $\beta_1$ is the slope coefficient,
    \item $\varepsilon_i$ is the error term, which gathers all the determinants of $y_i$ other than $x_i$, and measurement error \citep[section 6.1.1]{fox2015}.
\end{itemize}
The parameters $\beta_0$ and $\beta_1$ are fixed and unknown, and the errors $\varepsilon_i$ are not observed. ``Linear'' means linear in the parameters: transformations of the variables, such as the logarithm of weight in the log-linear regression below, are allowed.

In matrix form, with $\mathbf{y}=(y_1,\dots,y_n)'$, $\boldsymbol{\varepsilon}=(\varepsilon_1,\dots,\varepsilon_n)'$, $\boldsymbol{\beta}=(\beta_0,\beta_1)'$ and $\mathbf{X}$ the $n\times 2$ matrix whose $i$th row is $(1,x_i)$:
\[
\mathbf{y} = \mathbf{X}\boldsymbol{\beta} + \boldsymbol{\varepsilon}.
\]
This notation extends to several regressors (section \ref{sec:multreg}), where the OLS estimator of section \ref{sec:OLSalgebra} becomes $\hat{\boldsymbol\beta}=(\mathbf X'\mathbf X)^{-1}\mathbf X'\mathbf y$ \citep[section 2.1.4]{verbeek}.

Since $\mathbf X$ is generally rectangular, its inverse does not exist.
Moreover, $\mathbf y=\mathbf X\boldsymbol\beta$ generally has no exact
solution. OLS therefore minimizes the sum of squared residuals:
\[
\hat{\boldsymbol\beta}
=\arg\min_{\boldsymbol\beta}
(\mathbf y-\mathbf X\boldsymbol\beta)'
(\mathbf y-\mathbf X\boldsymbol\beta).
\]
The first-order condition yields the normal equations:
\[
\mathbf X'(\mathbf y-\mathbf X\hat{\boldsymbol\beta})=\mathbf 0
\quad\Longrightarrow\quad
\mathbf X'\mathbf X\hat{\boldsymbol\beta}=\mathbf X'\mathbf y.
\]
If $\mathbf X$ has full column rank, $\mathbf X'\mathbf X$ is invertible,
hence
\[
\boxed{\hat{\boldsymbol\beta}
=(\mathbf X'\mathbf X)^{-1}\mathbf X'\mathbf y.}
\]

\paragraph{The intercept.} Omitting the intercept imposes the restriction $\E(y\mid x=0)=0$: the line must go through the origin. If this restriction is false, the model is misspecified and the slope is biased, since it must both fit the slope and bring the line to the level of the data. For our women, the restriction is absurd, and the slope falls from 0.737 to 0.361~kg per cm (Figure \ref{fig:binnedmeans}, left, and Table \ref{tab:weightheightOLS}). It is generally unclear why an economist would estimate a regression without including an intercept. The exceptions are variables measured as deviations from their means, for which the OLS intercept is zero anyway, and a complete set of category indicators, which already spans the constant (section \ref{sec:multreg}). When a theory predicts a zero intercept, it is better to estimate the intercept and test it, as with the $\alpha$ of the CAPM (section \ref{sec:CAPM}). In academic papers, the intercept is sometimes omitted from the tables for brevity, but it was nevertheless included in the actual estimation. In such cases no bias arises, since the intercept is estimated but simply not reported because it is not relevant for the specific point being discussed. To estimate a regression without an intercept y $\sim$ x - 1.

\subsubsection{The conditional median and quantile regression}

Instead of the conditional mean, we may study the
\emph{conditional median function},
\[
q_{0.5}(h)=\inf\bigl\{a:\Pr(w\leq a\mid h)\geq 0.5\bigr\}.
\]
It describes the central weight at each height: at least half
of the conditional distribution lies on either side of a median.
It can be estimated nonparametrically by computing median weights
within groups of similar heights.

Whereas the conditional mean minimizes expected squared loss,
a conditional median minimizes expected absolute loss:
\[
m(h)=\operatorname*{arg\,min}_{a\in\mathbb R}
\E\bigl[(w-a)^2\mid h\bigr],
\qquad
q_{0.5}(h)\in\operatorname*{arg\,min}_{a\in\mathbb R}
\E\bigl[|w-a|\mid h\bigr],
\]
assuming the relevant moments exist.

If we specify a linear conditional median,
$q_{0.5}(h)=\alpha+\beta h$, the estimation problem is
\[
(\hat\alpha,\hat\beta)
\in\operatorname*{arg\,min}_{(\alpha,\beta)\in\mathbb R^2}
\sum_{i=1}^{n}|w_i-\alpha-\beta h_i|.
\]
This is called \emph{median regression}, or
\emph{least absolute deviations} (LAD) regression.
It is a special case of \emph{quantile regression}, with
quantile level $\tau=0.5$.

\subsection{Ordinary least squares}\label{sec:OLSalgebra}

Given the sample $\{(x_i,y_i)\}_{i=1}^n$, define the OLS estimators $(\hat\beta_0,\hat\beta_1)$ as the minimizers of the sum of squared residuals:
\[
(\hat\beta_0,\hat\beta_1)
=\arg\min_{\beta_0,\beta_1}\; S(\beta_0,\beta_1),
\qquad
S(\beta_0,\beta_1)=\sum_{i=1}^n \bigl(y_i-\beta_0-\beta_1 x_i\bigr)^2.
\]
Squaring prevents positive and negative deviations from cancelling out, and it penalizes large deviations more than proportionally; minimizing the sum of absolute deviations instead gives the median regression of section \ref{sec:quantilereg}.

\subsubsection*{Step 1: First-order conditions (FOC)}

Differentiate $S(\beta_0,\beta_1)$ with respect to $\beta_0$ and $\beta_1$.

\paragraph{FOC w.r.t.\ $\beta_0$.}
\[
\frac{\partial S}{\partial \beta_0}
=
\sum_{i=1}^n 2\bigl(y_i-\beta_0-\beta_1 x_i\bigr)(-1)
=
-2\sum_{i=1}^n \bigl(y_i-\beta_0-\beta_1 x_i\bigr)=0.
\]
Equivalently,
\[
\sum_{i=1}^n y_i - n\beta_0 - \beta_1\sum_{i=1}^n x_i=0
\quad\Longrightarrow\quad
\hat\beta_0=\bar y-\hat\beta_1\bar x,
\]
where $\bar y=\frac{1}{n}\sum_{i=1}^n y_i$ and $\bar x=\frac{1}{n}\sum_{i=1}^n x_i$.

\paragraph{FOC w.r.t.\ $\beta_1$.}
\[
\frac{\partial S}{\partial \beta_1}
=
\sum_{i=1}^n 2\bigl(y_i-\beta_0-\beta_1 x_i\bigr)(-x_i)
=
-2\sum_{i=1}^n x_i\bigl(y_i-\beta_0-\beta_1 x_i\bigr)=0.
\]

\subsubsection*{Step 2: Solve for $\hat\beta_1$}

Substitute $\hat\beta_0=\bar y-\hat\beta_1\bar x$ into the second FOC:
\[
\sum_{i=1}^n x_i\Bigl(y_i-(\bar y-\hat\beta_1\bar x)-\hat\beta_1 x_i\Bigr)=0.
\]
Use $\sum_{i=1}^n x_i(y_i-\bar y)=\sum_{i=1}^n (x_i-\bar x)(y_i-\bar y)$ (because $\sum_{i=1}^n (y_i-\bar y)=0$), and note that
\[
\sum_{i=1}^n x_i(x_i-\bar x)=\sum_{i=1}^n (x_i-\bar x)^2.
\]
Therefore,
\[
\sum_{i=1}^n (x_i-\bar x)(y_i-\bar y)
-
\hat\beta_1\sum_{i=1}^n (x_i-\bar x)^2=0,
\]
so that
\begin{equation}\label{eq:beta1OLS}
\hat\beta_1
=
\frac{\sum_{i=1}^n (x_i-\bar x)(y_i-\bar y)}
{\sum_{i=1}^n (x_i-\bar x)^2}
=
\frac{\sum_{i=1}^n x_i (y_i - \bar{y})}{\sum_{i=1}^n x_i ( x_i - \bar{x})}
=\frac{S_{xy}}{S_{xx}},
\qquad
\hat\beta_0=\bar y-\hat\beta_1\bar x,
\end{equation}
with $S_{xx}=\sum_{i=1}^n (x_i-\bar x)^2$ and $S_{xy}=\sum_{i=1}^n (x_i-\bar x)(y_i-\bar y)$.

The function $S(\beta_0,\beta_1)$ is a convex quadratic whose Hessian, $2\begin{pmatrix} n & \sum_i x_i\\ \sum_i x_i & \sum_i x_i^2\end{pmatrix}$, has determinant $4nS_{xx}$: the solution is the unique minimum if and only if $S_{xx}>0$, i.e.\ if $x$ varies in the sample. Dividing $S_{xy}$ and $S_{xx}$ by $n-1$, $\hat\beta_1$ is the sample covariance of $x$ and $y$ divided by the sample variance of $x$: OLS is the sample analogue of the BLP \eqref{eq:BLP} \citep[ch.~3]{hansen2022econometrics}.

\paragraph{Algebraic properties.} The two FOC state that the residuals $\hat\varepsilon_i=y_i-\hat\beta_0-\hat\beta_1x_i$ satisfy, in every sample,
\begin{equation}\label{eq:OLSorthogonality}
\sum_{i=1}^n\hat\varepsilon_i=0, \qquad \sum_{i=1}^n x_i\,\hat\varepsilon_i=0.
\end{equation}
Hence the residuals have a zero mean and a zero sample correlation with $x$, the line goes through the point of means $(\bar x,\bar y)$, and the fitted values $\hat y_i=\hat\beta_0+\hat\beta_1x_i$ are orthogonal to the residuals \citep[section 2.1]{verbeek}. These properties hold whatever the data-generating process. The residuals $\hat\varepsilon_i$ are not the errors $\varepsilon_i$, and a zero correlation between the residuals and $x$ tells us nothing about the errors \citep[section 2.2]{verbeek}.

\paragraph{Slope and correlation.} With $s_x$ and $s_y$ the sample standard deviations and $r$ the sample correlation,
\begin{equation}\label{eq:slopecorr}
\hat\beta_1=r\,\frac{s_y}{s_x}.
\end{equation}
Measured in standard deviations, the slope is the correlation: a woman one standard deviation taller is predicted to be $r=0.41$ standard deviations heavier. The slope has units, here kilograms per centimetre, whereas the correlation has none \citep[section 5.1.2]{fox2015}.

\paragraph{Regression through the origin.} Without an intercept, minimizing $\sum_{i}(y_i-\beta x_i)^2$ gives the single FOC $\sum_{i}x_i(y_i-\tilde\beta x_i)=0$, hence
\begin{equation}\label{eq:betahat}
\tilde{\beta} = \frac{\sum_i x_i y_i}{\sum_i x_i^2}.
\end{equation}
The residuals remain orthogonal to $x$ but no longer sum to zero.

\paragraph{Application.} On our sample, OLS gives (heteroskedasticity-robust standard errors in parentheses, section \ref{sec:Randtinterpret})
\begin{equation}\label{eq:weightheightfit}
\hat w_i=\underset{(2.18)}{-59.18}+\underset{(0.014)}{0.737}\,h_i,\qquad R^2=0.170,\qquad s=11.03\text{ kg},\qquad n=15\,494,
\end{equation}
where $s$ is the standard error of the regression (SER), the typical size of a residual $\hat\varepsilon_i=w_i-\hat w_i$, measured in kilograms:
\begin{equation}\label{eq:SER}
s=\sqrt{\frac{1}{n-2}\sum_{i=1}^n\hat\varepsilon_i^2},
\end{equation}
with $n-2$ rather than $n$ because two coefficients are estimated (section \ref{sec:OLSproperties}), and $R^2$ is the share of the sample variance of weight accounted for by the regression (section \ref{sec:Rsquared}). Two women who differ by 10~cm in height differ on average by 7.4~kg in weight. The intercept, $-59.2$~kg at a height of zero, is an extrapolation far outside the data, whose heights range from 127 to 197~cm, and it has no meaning. Centring height, $w_i=\beta_0^c+\beta_1(h_i-\bar h)+\varepsilon_i$, leaves the slope unchanged and turns the intercept into the mean weight at the mean height, $\hat\beta_0^c=\bar w=56.75$~kg (Table \ref{tab:weightheightOLS}).

\subsection{Goodness of fit}\label{sec:Rsquared}

Decompose each observation into a fitted value and a residual, $y_i-\bar y=(\hat y_i-\bar y)+\hat\varepsilon_i$. Squaring and summing,
\begin{equation}\label{eq:ANOVA}
\underbrace{\sum_{i=1}^n(y_i-\bar y)^2}_{\text{TSS}}=\underbrace{\sum_{i=1}^n(\hat y_i-\bar y)^2}_{\text{ESS}}+\underbrace{\sum_{i=1}^n\hat\varepsilon_i^2}_{\text{RSS}},
\end{equation}
because the cross-product $\sum_i(\hat y_i-\bar y)\hat\varepsilon_i=\hat\beta_1\sum_i(x_i-\bar x)\hat\varepsilon_i$ is zero by \eqref{eq:OLSorthogonality}. The decomposition is exact in every sample, provided the regression includes an intercept. The $R^2$ is the share of the sample variation of $y$ accounted for by the regression:
\begin{equation}\label{eq:R2}
R^2=\frac{\text{ESS}}{\text{TSS}}=1-\frac{\text{RSS}}{\text{TSS}}\in[0,1].
\end{equation}
In simple regression, $\hat y_i-\bar y=\hat\beta_1(x_i-\bar x)$, so that $\text{ESS}=\hat\beta_1^2S_{xx}=S_{xy}^2/S_{xx}$ and $R^2=S_{xy}^2/(S_{xx}S_{yy})=r^2$, with $S_{yy}=\text{TSS}$. An absolute measure of fit, in the units of $y$, is the SER \eqref{eq:SER}, $s=\sqrt{\text{RSS}/(n-2)}$, the typical size of a residual; \citet[section 5.1.2]{fox2015} argues that it deserves more attention than the $R^2$.

For our women, $R^2=0.170=0.412^2$, and $s=11.0$~kg against a standard deviation of weight of $12.1$~kg: knowing a woman's height reduces the typical error made in guessing her weight from 12.1 to 11.0~kg. A low $R^2$ is common with individual cross-sectional data, and it does not mean that the slope is poorly estimated: the 95\% confidence interval of the slope is $[0.709,\,0.764]$ (section \ref{sec:Randtinterpret}). The $R^2$ measures the quality of the linear approximation, not the quality of the model or of its causal interpretation \citep[section 2.4]{verbeek}. It depends on the dependent variable, so that the $R^2$ of regressions of $w$ and of $\log w$ are not comparable, and it never decreases when a regressor is added (section \ref{sec:multreg}).

Without an intercept, \eqref{eq:ANOVA} fails, and software reports the \emph{uncentred} $R^2_u=\sum_i\hat y_i^2/\sum_iy_i^2$, which compares the fit with the line $y=0$ instead of $y=\bar y$. If the series do not have a zero mean and we nonetheless force the intercept to be null, $R^2_u$ will seem larger than the explanatory power of the model: it is 0.962 for our regression through the origin, whose centred $R^2$ is only 0.126.

\subsection{Regression is not correlation: reverse regression and regression to the mean}\label{sec:regtomean}

The correlation is symmetric in $x$ and $y$; the regression is not. The slope of the reverse regression, of height on weight, is $\hat\delta_1=S_{xy}/S_{yy}=r\,s_x/s_y$, hence
\begin{equation}\label{eq:reverseregression}
\hat\beta_1\,\hat\delta_1=r^2\leq1,
\end{equation}
and the two lines coincide only if $|r|=1$. For our women, $\hat\delta_1=0.231$~cm per kg: inverting it gives $4.33$~kg per cm, not $0.737$. The two regressions answer two different questions, the mean weight given height and the mean height given weight \citep[section 2.27]{hansen2022econometrics}.

In standard units, \eqref{eq:slopecorr} reads $\hat z_y=r\,z_x$: the prediction is closer to the mean, in standard deviations, than the predictor. The word ``regression'' comes from this property: \citet{galton1886regression} found that the children of tall parents are, on average, taller than the mean but less so than their parents, a ``regression towards mediocrity''. In our data, a woman two standard deviations taller than the mean (170.9~cm) is predicted to be only $2r=0.82$ standard deviations heavier (66.7~kg), and a woman two standard deviations heavier (81.0~kg) is predicted to be only 0.82 standard deviations taller (163.0~cm). Regression to the mean thus runs in both directions: it follows from $|r|<1$ and is neither a causal nor a dynamic mechanism. Reading it as a force that compresses the dispersion over time is the \emph{regression fallacy}, which \citet[section 2.26]{hansen2022econometrics} illustrates with the analysis of firms' profits by Secrist (1933). In finance, the same arithmetic implies that, when performance is partly luck, the best-performing funds of a year are expected to be less exceptional the next year, even if the skill of their managers is unchanged.

\subsection{Classical linear regression assumptions}\label{sec:CLRMassumptions}

We state the standard assumptions underlying OLS in the cross-sectional case, as in \citet[section 2.3.1]{verbeek}, \citet[section 6.1.1]{fox2015} and \citet[Key Concept 4.3]{stock2020introduction}. Conditioning on $X=(x_1,\dots,x_n)$ allows us to treat the regressors as fixed.

\begin{itemize}
\item[(A1)] \textbf{Linearity in parameters}: $y_i=\beta_0+\beta_1x_i+\varepsilon_i$. Nonlinear transformations of $x_i$ or $y_i$ are allowed (e.g.\ $x_i^2$, $\log x_i$), provided the model remains linear in $\boldsymbol\beta$.
\item[(A2)] \textbf{Random sampling}: $\{(y_i,x_i)\}_{i=1}^n$ are independent draws from the same population. Surveys such as the DHS draw households in clusters, so that (A2) holds only approximately (section \ref{sec:Randtinterpret}).
\item[(A3)] \textbf{No perfect multicollinearity}: the regressors are not perfectly linearly dependent. In simple regression, this requires sample variation in $x$, $S_{xx}>0$.
\item[(A4)] \textbf{Zero conditional mean (exogeneity)}: $\E(\varepsilon_i\mid x_i)=0$, which under (A2) is equivalent to $\E(\varepsilon_i\mid X)=0$. With (A1), it makes the regression line the CEF, $\E(y_i\mid x_i)=\beta_0+\beta_1x_i$. This is the key assumption for unbiasedness. Its content depends on the reading of the error. If $\varepsilon_i$ is defined as $y_i-\beta_0-\beta_1x_i$, (A4) only says that the CEF is linear, which the data can check (section \ref{sec:weightheightdiagnostics}). If $\varepsilon_i$ gathers the other determinants of $y_i$, (A4) is the exogeneity assumption required for a causal reading of $\beta_1$. It is not tested by a $t$-test on residuals, which are uncorrelated with $x$ by construction \eqref{eq:OLSorthogonality}; rather, it is a structural assumption about the data-generating process, which must be defended by economic arguments \citep{angrist2009mostly}.
\item[(A5)] \textbf{Homoskedasticity}: $\Var(\varepsilon_i\mid x_i)=\sigma^2$, the same for all values of $x_i$ (tests in section \ref{sec:weightheightdiagnostics}).
\item[(A6)] \textbf{No autocorrelation}: $\Cov(\varepsilon_i,\varepsilon_j\mid X)=0$ for $i\neq j$. Random sampling (A2) implies it. It matters for ordered data such as time series, where it is assessed with Durbin--Watson or Breusch--Godfrey tests (section \ref{sec:breusch}), and for clustered samples.
\item[(A7)] \textbf{Normality (for exact finite-sample inference)}: $\varepsilon_i\mid X\sim\mathcal{N}(0,\sigma^2)$. Normality is not required for unbiasedness or consistency, but it delivers exact $t$ and $F$ distributions in small samples.
\end{itemize}

Taken together, (A1), (A4), (A5) and (A7) state that, at every value of $x$, $y$ is normally distributed around the regression line with the same variance $\sigma^2$, a strong restriction on the data \citep[section 6.1.1]{fox2015}. Most of these assumptions are not needed to justify OLS: they make its finite-sample properties easy to derive \citep[section 2.3.1]{verbeek}. Table \ref{tab:assumptionsproperties} summarizes what each property requires; the next subsections prove each line.

\begin{table}[H]
\centering
\caption{What each property of OLS requires}\label{tab:assumptionsproperties}
\begin{tabular}{@{}p{0.56\textwidth}p{0.40\textwidth}@{}}
\toprule
Property & Assumptions \\
\midrule
$\hat\beta_0,\hat\beta_1$ exist and are unique; algebraic properties \eqref{eq:OLSorthogonality}--\eqref{eq:ANOVA} & (A3) \\
Unbiasedness, $\E(\hat\beta_1\mid X)=\beta_1$ & (A1)--(A4) \\
$\Var(\hat\beta_1\mid X)=\sigma^2/S_{xx}$, unbiased $s^2$, Gauss--Markov theorem & (A1)--(A6) \\
Exact $t_{n-2}$ and $F$ distributions & (A1)--(A7) \\
Consistency & (A1)--(A3), $\Cov(x_i,\varepsilon_i)=0$, finite variances \\
Asymptotic normality, robust standard errors & as for consistency, with finite fourth moments \\
\bottomrule
\end{tabular}
\end{table}

\subsection{Properties of the OLS estimator}\label{sec:OLSproperties}

\subsubsection*{OLS is linear in $y$}

From \eqref{eq:beta1OLS}, $\hat\beta_1=\sum_{i=1}^n a_i\,y_i$, with weights $a_i=(x_i-\bar x)/S_{xx}$ that depend on $X$ only and satisfy $\sum_i a_i=0$, $\sum_i a_i x_i=1$ and $\sum_i a_i^2=1/S_{xx}$.

\subsubsection*{Express $\hat\beta_1$ in terms of $\beta_1$ and the errors}

Start from the model and its sample mean:
\[
y_i=\beta_0+\beta_1 x_i+\varepsilon_i,
\qquad
\bar y=\beta_0+\beta_1\bar x+\bar\varepsilon,
\qquad
\bar\varepsilon=\frac{1}{n}\sum_{i=1}^n \varepsilon_i.
\]
Subtract the mean equation from the individual equation:
\[
y_i-\bar y
=
\beta_1(x_i-\bar x)+(\varepsilon_i-\bar\varepsilon).
\]
Plug this into the expression of $\hat\beta_1$:
\begin{align*}
\hat\beta_1
&=
\frac{\sum_{i=1}^n (x_i-\bar x)\Bigl[\beta_1(x_i-\bar x)+(\varepsilon_i-\bar\varepsilon)\Bigr]}
{\sum_{i=1}^n (x_i-\bar x)^2}
=
\beta_1
+
\frac{\sum_{i=1}^n (x_i-\bar x)(\varepsilon_i-\bar\varepsilon)}
{\sum_{i=1}^n (x_i-\bar x)^2}.
\end{align*}
Since $\sum_{i=1}^n (x_i-\bar x)=0$, the term in $\bar\varepsilon$ vanishes, and the key decomposition is
\begin{equation}\label{eq:beta1decomp}
\boxed{
\hat\beta_1
=
\beta_1
+
\frac{\sum_{i=1}^n (x_i-\bar x)\varepsilon_i}
{\sum_{i=1}^n (x_i-\bar x)^2}
=\beta_1+\sum_{i=1}^n a_i\,\varepsilon_i
}.
\end{equation}

\subsubsection*{Unbiasedness of $\hat\beta_1$}

Take expectations conditional on $X=(x_1,\dots,x_n)$.
Under the exogeneity assumption $\E(\varepsilon_i\mid X)=0$, we obtain
\[
\E(\hat\beta_1\mid X)
=
\beta_1
+
\frac{\sum_{i=1}^n (x_i-\bar x)\,\E(\varepsilon_i\mid X)}
{\sum_{i=1}^n (x_i-\bar x)^2}
=
\beta_1.
\]
Hence, by iterated expectations,
\[
\E(\hat\beta_1)=\E\bigl(\E(\hat\beta_1\mid X)\bigr)=\beta_1,
\]
and similarly for $\hat\beta_0=\bar y-\hat\beta_1\bar x$. Neither (A5)--(A6) nor (A7) is used: heteroskedasticity does not bias OLS. If $\E(\varepsilon_i\mid X)\neq0$, the second term of \eqref{eq:beta1decomp} does not vanish in expectation, and OLS is biased.

\subsubsection*{Variance of $\hat\beta_1$}

Compute the variance conditional on $X$:
\[
\Var(\hat\beta_1\mid X)=\Var\!\left(\frac{1}{S_{xx}}\sum_{i=1}^n (x_i-\bar x)\varepsilon_i\;\middle|\;X\right)
=\frac{1}{S_{xx}^2}\Var\!\left(\sum_{i=1}^n (x_i-\bar x)\varepsilon_i\;\middle|\;X\right).
\]
Assume (i) \textbf{homoskedasticity} (A5) and (ii) \textbf{no conditional correlation} (A6):
\[
\Var(\varepsilon_i\mid X)=\sigma^2,
\qquad
\Cov(\varepsilon_i,\varepsilon_j\mid X)=0\quad (i\neq j).
\]
Then
\[
\Var\!\left(\sum_{i=1}^n (x_i-\bar x)\varepsilon_i\;\middle|\;X\right)
=
\sum_{i=1}^n (x_i-\bar x)^2\Var(\varepsilon_i\mid X)
=
\sigma^2 S_{xx},
\]
and therefore
\begin{equation}\label{eq:varbeta1}
\boxed{
\Var(\hat\beta_1\mid X)=\frac{\sigma^2}{S_{xx}}=\frac{\sigma^2}{n\,\hat\sigma_x^2}
},
\end{equation}
where $\hat\sigma_x^2=S_{xx}/n$ is the sample variance of $x$ with divisor $n$. Similarly, $\Var(\hat\beta_0\mid X)=\sigma^2\sum_ix_i^2/(nS_{xx})$ and $\Cov(\hat\beta_0,\hat\beta_1\mid X)=-\bar x\,\sigma^2/S_{xx}$, which is zero when $x$ is centred. The slope is estimated more precisely when the errors are small, the sample is large and the regressor is spread out \citep[section 2.3.2]{verbeek}: a wide range of heights provides a long lever to pin down the slope.

\subsubsection*{The Gauss--Markov theorem}

Under (A1)--(A6), OLS is the Best Linear Unbiased Estimator (BLUE): any other estimator that is linear in $y$ and conditionally unbiased, $\check\beta_1=\sum_ic_iy_i$ with weights $c_i$ depending on $X$ only, satisfies $\Var(\check\beta_1\mid X)\geq\Var(\hat\beta_1\mid X)$.

\emph{Proof.} Since $\E(\check\beta_1\mid X)=\beta_0\sum_ic_i+\beta_1\sum_ic_ix_i$, unbiasedness for all $(\beta_0,\beta_1)$ requires $\sum_ic_i=0$ and $\sum_ic_ix_i=1$. Under (A5)--(A6), $\Var(\check\beta_1\mid X)=\sigma^2\sum_ic_i^2$. Write $c_i=a_i+d_i$, where the $a_i$ are the OLS weights. Then $\sum_ia_id_i=\sum_ia_ic_i-\sum_ia_i^2=\bigl(\sum_ic_ix_i-\bar x\sum_ic_i\bigr)/S_{xx}-1/S_{xx}=0$, so that
\begin{equation}\label{eq:GaussMarkov}
\Var(\check\beta_1\mid X)=\sigma^2\sum_ia_i^2+\sigma^2\sum_id_i^2\geq\sigma^2\sum_ia_i^2=\Var(\hat\beta_1\mid X),
\end{equation}
with equality only if $\check\beta_1=\hat\beta_1$. \hfill$\square$

The theorem relies on homoskedasticity: with heteroskedastic errors, weighted least squares is more efficient, provided the form of the heteroskedasticity is known \citep[section 12.2]{fox2015}. Under normality (A7), OLS is moreover the most efficient of all unbiased estimators, linear or not \citep[section 6.1.2]{fox2015}.

\subsubsection*{Estimating $\sigma^2$}

The variance \eqref{eq:varbeta1} involves the unknown error variance $\sigma^2$. Its estimator is the square of the SER \eqref{eq:SER}, which divides the residual sum of squares by $n-2$:
\begin{equation}\label{eq:s2}
s^2=\frac{1}{n-2}\sum_{i=1}^n\hat\varepsilon_i^2,\qquad \E(s^2\mid X)=\sigma^2\ \text{under (A1)--(A6)}.
\end{equation}
Indeed, from the model and \eqref{eq:beta1OLS}, $\hat\varepsilon_i=(\varepsilon_i-\bar\varepsilon)-(\hat\beta_1-\beta_1)(x_i-\bar x)$, and $\sum_i(x_i-\bar x)(\varepsilon_i-\bar\varepsilon)=(\hat\beta_1-\beta_1)S_{xx}$ by \eqref{eq:beta1decomp}, so that
\[
\sum_{i=1}^n\hat\varepsilon_i^2=\sum_{i=1}^n(\varepsilon_i-\bar\varepsilon)^2-(\hat\beta_1-\beta_1)^2S_{xx},
\]
whose conditional expectation is $(n-1)\sigma^2-\sigma^2=(n-2)\sigma^2$. The two estimated coefficients impose the two restrictions \eqref{eq:OLSorthogonality} on the residuals, which are therefore less dispersed than the errors: dividing by $n$ would underestimate $\sigma^2$ \citep[section 2.3.2]{verbeek}.

\subsubsection*{Consistency}

An estimator is consistent if it converges in probability to the true value: for every $\delta>0$, $\lim_{n\to\infty}P(|\hat\beta_1-\beta_1|>\delta)=0$, which we write $\hat\beta_1\xrightarrow{p}\beta_1$ (section \ref{sec:largen}).

\paragraph{The variance goes to zero.} Under (A1)--(A6), Chebyshev's inequality applied conditionally on $X$ gives
\begin{equation}\label{eq:chebyshevOLS}
P\left(|\hat\beta_1-\beta_1|>\delta\;\middle|\;X\right)\leq\frac{\Var(\hat\beta_1\mid X)}{\delta^2}=\frac{\sigma^2}{\delta^2\,n\,\hat\sigma_x^2}.
\end{equation}
Under random sampling with $\Var(x_i)=\sigma_x^2\in(0,\infty)$, the law of large numbers gives $\hat\sigma_x^2\xrightarrow{p}\sigma_x^2$, so that the bound goes to zero in probability. A probability is bounded by one, and taking expectations (bounded convergence) yields $P(|\hat\beta_1-\beta_1|>\delta)\to0$ \citep[section 2.6.1]{verbeek}. As the sample grows, $S_{xx}$ increases proportionally to $n$ (provided $x$ has a nonzero variance): the variance of $\hat\beta_1$ shrinks at rate $1/n$ and its standard deviation at rate $1/\sqrt n$, so that halving a standard error requires four times as many observations.

\paragraph{Consistency under weaker assumptions.} Dividing the numerator and the denominator of \eqref{eq:beta1decomp} by $n$,
\[
\hat\beta_1-\beta_1=\frac{\frac1n\sum_{i=1}^n(x_i-\bar x)\,\varepsilon_i}{\frac1n\sum_{i=1}^n(x_i-\bar x)^2}.
\]
By the law of large numbers, the numerator converges in probability to $\Cov(x_i,\varepsilon_i)$ and the denominator to $\Var(x_i)>0$; by the continuous mapping theorem,
\begin{equation}\label{eq:plimOLS}
\operatorname{plim}\hat\beta_1=\beta_1+\frac{\Cov(x_i,\varepsilon_i)}{\Var(x_i)}.
\end{equation}
OLS is therefore consistent as soon as the error is uncorrelated with the regressor, which is weaker than (A4) and requires neither (A5), (A6) nor (A7) \citep[section 2.6.1]{verbeek}. In the descriptive reading, $\varepsilon_i$ is the BLP error $u$ of \eqref{eq:BLP}, uncorrelated with $x_i$ by construction: under random sampling with finite variances, OLS estimates the BLP slope consistently, whatever the shape of the CEF \citep[section 7.2]{hansen2022econometrics}. In a structural reading, \eqref{eq:plimOLS} gives the bias that remains however large the sample, when the error is correlated with the regressor (section \ref{sec:OVBweight}).

\subsubsection*{Asymptotic normality}

Up to a term that vanishes, the numerator above is the sample mean of the independent terms $(x_i-\mu_x)\varepsilon_i$, with $\mu_x=\E(x_i)$, to which the central limit theorem applies (section \ref{sec:CLT}) when the variables have finite fourth moments, i.e.\ when large outliers are unlikely \citep[Key Concept 4.3]{stock2020introduction}. Hence, when $\Cov(x_i,\varepsilon_i)=0$,
\begin{equation}\label{eq:asymptnormalOLS}
\sqrt n\left(\hat\beta_1-\beta_1\right)\xrightarrow{d}\mathcal N\left(0,\;\frac{\E\left[(x_i-\mu_x)^2\varepsilon_i^2\right]}{\Var(x_i)^2}\right),
\end{equation}
where the variance reduces to $\sigma^2/\Var(x_i)$ under (A4)--(A5), since then $\E(\varepsilon_i^2\mid x_i)=\sigma^2$ \citep[section 4.5]{stock2020introduction}. Normal errors are not needed: in large samples, $\hat\beta_1$ is approximately normal whatever the distribution of the errors \citep[section 2.6.2]{verbeek}. Under (A7), $\hat\beta_1$ is exactly normal in every sample, $\hat\beta_1\mid X\sim\mathcal N(\beta_1,\sigma^2/S_{xx})$, being a linear combination of normal errors.

\begin{figure}[!htbp]
\centering
\includegraphics[width=\textwidth]{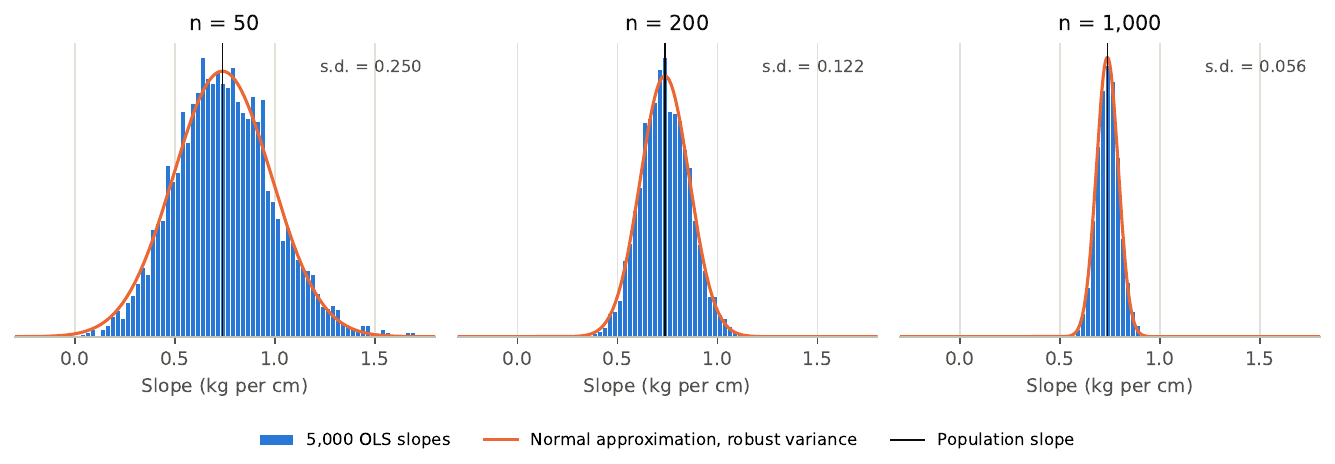}
\caption{Sampling distribution of the OLS slope: $5\,000$ samples of $n$ women drawn with replacement from our $15\,494$ women. The vertical line marks the population slope, $0.737$~kg per cm; the curve is the normal approximation \eqref{eq:asymptnormalOLS}.}\label{fig:OLSsampling}
\end{figure}

\paragraph{A Monte Carlo illustration.} We treat our $15\,494$ women as the population, whose slope is $0.737$~kg per cm, and draw from it, with replacement, $5\,000$ samples of $n=50$, $200$ and $1\,000$ women (section \ref{sec:MCmethod}). Figure \ref{fig:OLSsampling} shows the $5\,000$ OLS slopes obtained for each sample size. First, they are centred on the population slope, with means of 0.746, 0.739 and 0.737 (Monte Carlo standard errors 0.004, 0.002 and 0.001). The small bias at $n=50$ arises because the population CEF is not exactly linear (section \ref{sec:weightheightdiagnostics}), so that (A4) holds only approximately; it is of order $1/n$ and vanishes as $n$ grows. Second, they concentrate at the rate $1/\sqrt n$: their standard deviations, 0.250, 0.122 and 0.056, match the formula \eqref{eq:asymptnormalOLS}, 0.245, 0.123 and 0.055, whereas the homoskedastic formula $\sigma/(\sigma_x\sqrt n)$ gives 0.230, 0.115 and 0.051, too small because the errors are heteroskedastic: the women of extreme heights, who weigh most in the variance of the slope, also have the largest errors (section \ref{sec:weightheightdiagnostics}). Third, they are close to normal, although the errors are right-skewed.

\subsection{Standard errors, $t$-tests and confidence intervals}\label{sec:Randtinterpret}

Replacing $\sigma^2$ by $s^2$ in the variances above gives the classical (homoskedasticity-only) estimated variances, valid under (A1)--(A6):
\begin{equation}\label{eq:classicalSE}
\widehat{\Var}(\hat\beta_1)=\frac{s^2}{S_{xx}}=s^2\,\frac{n}{n\sum_ix_i^2-\left(\sum_ix_i\right)^2},\qquad
\widehat{\Var}(\hat\beta_0)=s^2\,\frac{\sum_ix_i^2}{n\sum_ix_i^2-\left(\sum_ix_i\right)^2},
\end{equation}
since $n\sum_ix_i^2-(\sum_ix_i)^2=nS_{xx}$. The standard error $\mathrm{SE}(\hat\beta_j)$ is the square root of the estimated variance. In the regression through the origin \eqref{eq:betahat}, $\widehat{\Var}(\tilde\beta)=\tilde s^2/\sum_ix_i^2$, where $\tilde s^2$ divides the residual sum of squares by $n-1$: the number of observations minus the number of estimated parameters.

\paragraph{The $t$-test.} To test $H_0:\beta_1=\beta_1^0$ against $H_1:\beta_1\neq\beta_1^0$, we use
\begin{equation}\label{eq:tstat}
t=\frac{\hat\beta_1-\beta_1^0}{\mathrm{SE}(\hat\beta_1)}.
\end{equation}
Under (A1)--(A7) and $H_0$, $t$ computed with the classical standard error \eqref{eq:classicalSE} follows exactly a Student distribution with $n-2$ degrees of freedom, because $\hat\beta_1$ is normal and $(n-2)s^2/\sigma^2\sim\chi^2_{n-2}$ is independent of it \citep[section 2.5]{verbeek}; with $K$ estimated coefficients, the degrees of freedom are $n-K$. Without normality, $t$ is asymptotically $\mathcal N(0,1)$ by \eqref{eq:asymptnormalOLS}, provided the standard error is consistent: the classical one under (A5), the robust one below in general. At the 5\% level, we reject $H_0$ if $|t|>t^*$, where $t^*$ solves $P(|T|>t^*)=0.05$ for $T\sim t_{n-2}$: $t^*$ is the 97.5\% quantile of the Student distribution, close to $1.96$ for large $n$. The $p$-value is the probability, under $H_0$, of a statistic at least as large in absolute value as the one observed. Regression outputs report the $t$-ratio $\hat\beta_1/\mathrm{SE}(\hat\beta_1)$, which tests $H_0:\beta_1=0$. A 95\% confidence interval,
\[
\left[\hat\beta_1-t^*\,\mathrm{SE}(\hat\beta_1),\;\hat\beta_1+t^*\,\mathrm{SE}(\hat\beta_1)\right],
\]
gathers all the values $\beta_1^0$ that the test does not reject; in repeated samples, 95\% of such intervals contain $\beta_1$.

\paragraph{Statistical and economic significance.} If $|t|>t^*$, the slope is statistically different from zero, provided our econometric methods are correct. This says nothing about its economic importance, nor about a causal effect of $x$ on $y$. In large samples, small effects are significant; in small samples, large effects may not be. We should therefore report and interpret the magnitude of the coefficient together with its confidence interval \citep[section 2.5.1]{verbeek}.

\paragraph{Heteroskedasticity-robust standard errors.} Without (A5), the derivation of \eqref{eq:varbeta1} gives $\Var(\hat\beta_1\mid X)=\sum_i(x_i-\bar x)^2\sigma_i^2/S_{xx}^2$, with $\sigma_i^2=\Var(\varepsilon_i\mid X)$, and the classical formula is wrong, even in large samples. \citet{white1980heteroskedasticity} shows that replacing $\sigma_i^2$ by $\hat\varepsilon_i^2$ gives an estimator that, multiplied by $n$, is consistent for the asymptotic variance in \eqref{eq:asymptnormalOLS}, whatever the form of the heteroskedasticity:
\begin{equation}\label{eq:HCSE}
\widehat{\Var}_{\mathrm{HC0}}(\hat\beta_1)=\frac{\sum_{i=1}^n(x_i-\bar x)^2\,\hat\varepsilon_i^2}{S_{xx}^2},\qquad
\widehat{\Var}_{\mathrm{HC1}}(\hat\beta_1)=\frac{n}{n-2}\,\widehat{\Var}_{\mathrm{HC0}}(\hat\beta_1).
\end{equation}
HC1 is the formula of \citet[equation (5.4)]{stock2020introduction} and of the \texttt{robust} option of Stata; the variants HC2 and HC3 also correct for leverage (section \ref{sec:weightheightdiagnostics}) and behave better in small samples \citep[section 4.14]{hansen2022econometrics}. Under homoskedasticity, the ratio of the robust to the classical standard error converges to one, so that in large samples robust standard errors can be used by default \citep[section 5.4]{stock2020introduction}; their use has become standard practice \citep[section 4.3.4]{verbeek}. With \texttt{statsmodels}: \verb|smf.ols('v437 ~ v438', data=df).fit(cov_type='HC1')|.

\paragraph{Application.} Table \ref{tab:weightheightOLS} reports the regressions with an intercept, with centred height and through the origin. The robust standard error of the slope, 0.0139, exceeds the classical one, 0.0131, by about 6.5\%, a first sign of heteroskedasticity (section \ref{sec:weightheightdiagnostics}); HC0 to HC3 coincide to four decimals with $n=15\,494$. The robust $t$-ratio, 52.9, rejects $\beta_1=0$ decisively, which is not very informative. The relevant information is that two women 10~cm apart in height differ on average by between 7.1 and 7.6~kg in weight (95\% confidence interval).\footnote{The DHS draws households in clusters (primary sampling units). If the errors of women of the same cluster are correlated, (A2) and (A6) fail and the standard errors must be clustered: with the 3\,573 clusters of our sample, the cluster-robust standard error of the slope is 0.0158, and the 95\% confidence interval widens to $[0.706,\,0.768]$.}

\begin{table}[H]
\centering
\caption{OLS regressions of weight (kg) on height (cm), $n=15\,494$}\label{tab:weightheightOLS}
\begin{tabular}{lccc}
\toprule
 & (1) Intercept & (2) Centred height & (3) No intercept \\
\midrule
Slope (kg per cm) & 0.737 & 0.737 & 0.361 \\
\quad classical SE & (0.0131) & (0.0131) & (0.0006) \\
\quad robust SE (HC1) & [0.0139] & [0.0139] & [0.0006] \\
Intercept (kg) & $-59.18$ & 56.75 & -- \\
\quad robust SE (HC1) & [2.18] & [0.09] & -- \\
$R^2$ & 0.170 & 0.170 & 0.126 \\
$R^2$ reported by \texttt{statsmodels} & 0.170 & 0.170 & 0.962 \\
SER $s$ (kg) & 11.03 & 11.03 & 11.32 \\
Mean residual (kg) & 0.00 & 0.00 & $-0.11$ \\
\bottomrule
\end{tabular}

\smallskip
\begin{minipage}{0.9\textwidth}\footnotesize Classical standard errors in parentheses, heteroskedasticity-robust (HC1) standard errors in brackets. $R^2$ is the centred $1-\text{RSS}/\text{TSS}$; for the regression without intercept, \texttt{statsmodels} reports the uncentred $R^2_u$.\end{minipage}
\end{table}

\paragraph{Predicting a mean, predicting a woman.} The regression line estimates the mean weight of the women of a given height precisely, but it predicts the weight of one woman poorly \citep[section 2.10]{verbeek}. At $h_0=160$~cm, $\hat w_0=\hat\beta_0+\hat\beta_1h_0=58.69$~kg. The estimated mean has variance $\Var(\hat w_0\mid X)=\sigma^2\left[1/n+(h_0-\bar h)^2/S_{xx}\right]$, which gives a 95\% confidence interval of $[58.50,\,58.88]$~kg. The prediction error for a new woman of this height, $w_0-\hat w_0=\varepsilon_0-(\hat w_0-\beta_0-\beta_1h_0)$, has variance $\sigma^2\left[1+1/n+(h_0-\bar h)^2/S_{xx}\right]$, and the 95\% prediction interval is $[37.1,\,80.3]$~kg.\footnote{Both intervals use the classical variance and Student quantiles. The prediction interval also assumes normal errors: given the right skew of the residuals (section \ref{sec:weightheightdiagnostics}), it is only indicative.} A precise slope and a low $R^2$ are compatible: the first is about the mean, the second about individuals.

\stepcounter{question}
\begin{itemize}
\item \textbf{Question \arabic{question}}
\begin{itemize}
\item Compute the variance of $\hat\beta_1$, the $R^2$ and the $t$ statistic with the formulas of this section, for the regressions with and without constant term (intercept), and compare them with the \texttt{statsmodels} output of Table \ref{tab:weightheightOLS}.
\end{itemize}
\end{itemize}

\subsection{From association to causation: omitted variables}\label{sec:OVBweight}

From the literature on the subject, we know we are missing explanatory variables: parents' height and weight at a minimum, but we do not have that information. Whether this matters depends on the reading of the regression \citep[section 6.3]{fox2015}. In the descriptive reading, nothing is missing: $\beta_1$ is the slope of the BLP of weight on height in the population sampled, and OLS estimates it consistently. In a structural reading, suppose that weight is determined by
\[
w_i=\beta_0+\beta_1h_i+\beta_2z_i+v_i,\qquad \Cov(h_i,v_i)=\Cov(z_i,v_i)=0,
\]
where $z_i$ is, for instance, the nutrition received during childhood, which raises adult height and, at a given height, adult weight ($\beta_2>0$). Omitting $z_i$ puts $\beta_2z_i$ in the error of the simple regression, and \eqref{eq:plimOLS} gives the omitted variable bias
\begin{equation}\label{eq:OVB}
\operatorname{plim}\hat\beta_1=\beta_1+\beta_2\,\frac{\Cov(h_i,z_i)}{\Var(h_i)}.
\end{equation}
The bias requires both $\beta_2\neq0$ and $\Cov(h_i,z_i)\neq0$, and its sign is that of $\beta_2\Cov(h_i,z_i)$, positive in this example. Its interpretation depends on the causal structure \citep[section 6.3]{fox2015}. If $z$ is a common cause of $h$ and $w$, a ``lurking variable'', the bias is a spurious component of the association, and $z$ must be controlled for. If $z$ lies between $h$ and $w$ on the causal path, a mediator, the ``bias'' is part of the effect of $h$, and controlling for $z$ would remove it. Parents' height and weight, through genes and family resources, are common causes of their daughters' height and weight.

Our pooled sample has an obvious lurking variable, the country, since mean height and mean weight differ across the countries surveyed. Controlling for it requires a multiple regression (section \ref{sec:multreg}). The next section and section \ref{sec:mpendog} discuss identification strategies that do not require observing every confounder.

\stepcounter{question}
\begin{itemize}
\item \textbf{Question \arabic{question}}
\begin{itemize}
\item Re-estimate the slope on the Indian women only (\texttt{v000} equal to \texttt{IA5}), then on the whole sample with one indicator per country. Use \eqref{eq:OVB} to discuss the direction of the change in the slope.
\end{itemize}
\end{itemize}

\subsection{Diagnostics of the weight--height regression}\label{sec:weightheightdiagnostics}

Some assumptions can be confronted with the data \citep[ch.~11--12]{fox2015}: the linearity of the CEF, homoskedasticity, the influence of single observations and the normality of the errors. Exogeneity (A4), in the structural sense, cannot.

\paragraph{Functional form.} The binned means of Figure \ref{fig:binnedmeans} lie within 1.4~kg of the OLS line. The RESET test of \citet{ramsey1969tests} adds the squared fitted value to the regression and tests its coefficient; in simple regression, this is equivalent to adding $h_i^2$. It rejects linearity ($F=22.5$, $p<0.001$), yet the quadratic term raises the $R^2$ only from 0.1700 to 0.1712: the CEF departs slightly from the line, a departure that is statistically significant with $n=15\,494$ and economically negligible. As with the normality tests of section \ref{sec:normalitytests}, a large sample detects departures that are too small to matter.

\paragraph{Heteroskedasticity.} Table \ref{tab:residsdquintile} sorts the women into five groups by height: the standard deviation of the residuals rises from 9.5~kg among the shortest women to 12.6~kg among the tallest. The Breusch--Pagan test \citep{breusch1979simple} regresses the squared residuals on the regressor,
\begin{equation}\label{eq:BPtest}
\hat{\varepsilon}_i^2 = \alpha + \gamma h_i + \nu_i,
\end{equation}
and tests $H_0:\gamma=0$ with the statistic $LM=nR^2$ of this auxiliary regression, asymptotically $\chi^2_1$ under homoskedasticity \citep[section 4.4.2]{verbeek}; this $nR^2$ form, the studentized version of \citet{koenker1981note}, is the default of \texttt{statsmodels}. The White test adds $h_i^2$ to the auxiliary regression and has two degrees of freedom \citep{white1980heteroskedasticity}. Both reject homoskedasticity: $LM=90.8$ and $101.0$, against 5\% critical values of 3.84 and 5.99. Heteroskedasticity does not bias OLS, but it invalidates the classical standard errors and the Gauss--Markov theorem, hence our use of robust standard errors (section \ref{sec:Randtinterpret}). Across height groups, it is mild: the standard deviation of the errors varies by a factor of 1.3, below the factor of about 2 from which \citet[section 12.2.4]{fox2015} recommends to worry. The robust standard error is nevertheless 6.5\% above the classical one, because the 91 women more than three standard deviations from the mean height, who weigh most in $\sum_i(x_i-\bar x)^2\hat\varepsilon_i^2$, have a residual standard deviation of 15.4~kg, against 11.0~kg for the others.

\begin{table}[H]
\centering
\caption{Standard deviation of the OLS residuals by height quintile}\label{tab:residsdquintile}
\begin{tabular}{lccccc}
\toprule
Height quintile & 1 & 2 & 3 & 4 & 5 \\
\midrule
Mean height (cm) & 148.0 & 153.9 & 157.4 & 160.8 & 166.9 \\
Weight on height (kg) & 9.51 & 10.22 & 11.08 & 11.53 & 12.59 \\
Log weight on height & 0.178 & 0.175 & 0.180 & 0.177 & 0.184 \\
\bottomrule
\end{tabular}
\end{table}

\stepcounter{question}
\begin{itemize}
\item \textbf{Question \arabic{question}}
\begin{itemize}
\item Sort the data set by ascending heights. Do two regressions on the first and second half of the sorted data set. Compare the residual distribution for each regression, this should give you a hint about homoscedasticity. The ratio of the two residual variances is the statistic of the \citet{goldfeld1965some} test, $F$-distributed with $(n_2-2,\,n_1-2)$ degrees of freedom under homoskedasticity and normal errors; with the heavy tails of our residuals, this $F$ distribution is not reliable.
\end{itemize}
\end{itemize}

\paragraph{Influential observations.} OLS minimizes squared residuals, so that a single observation far from the others can move the line \citep[section 2.9.1]{verbeek}. Its potential to do so, its leverage, is measured in simple regression by $\ell_i=1/n+(x_i-\bar x)^2/S_{xx}$, the $i$th diagonal element of the hat matrix, whose average is $2/n$ \citep[section 11.2]{fox2015}. Deleting observation $i$ changes the slope by
\begin{equation}\label{eq:dfbeta}
\hat\beta_1-\hat\beta_{1(-i)}=\frac{(x_i-\bar x)\,\hat\varepsilon_i}{S_{xx}\,(1-\ell_i)},
\end{equation}
the product of an unusual regressor and a large residual: this is the ``influence = leverage $\times$ discrepancy'' heuristic of \citet[section 11.1]{fox2015}, made exact \citep[section 3.21]{hansen2022econometrics}. In our sample, the largest leverage, 0.0022 for the tallest woman (196.7~cm), is 17 times the average, but no single woman moves the slope by more than 0.0022~kg per cm (0.3\%): with $15\,494$ observations, no genuine observation matters. Data errors do. The DHS codes a missing measure as 9999, and the script filters out such codes and impossible heights. Without the height filter, 57 more women enter the sample, 32 of them with a height coded 9999 (i.e.\ 999.9~cm), and the slope collapses from 0.737 to 0.024~kg per cm. Before running a regression, plot the data \citep[section 4.4]{stock2020introduction}.

\paragraph{Normality of the errors.} The residuals have a skewness of 1.53 and a kurtosis of 8.2 (Jarque--Bera statistic: 23\,319). Like weights (section \ref{sec:normalitytests}), the errors are right-skewed and heavy-tailed, and (A7) fails. Exact $t_{n-2}$ inference is therefore not available, but it is not needed: with $n=15\,494$, the central limit theorem makes the $t$-statistic approximately normal, and Figure \ref{fig:OLSsampling} shows that the slope is already nearly normal with $n=50$.

\paragraph{Taking the log of weight.} Heteroskedasticity and right-skewed errors both point to the same transformation, the logarithm of weight, with which our data set better fits the classical assumptions:
\begin{center}
\includegraphics[scale=0.5]{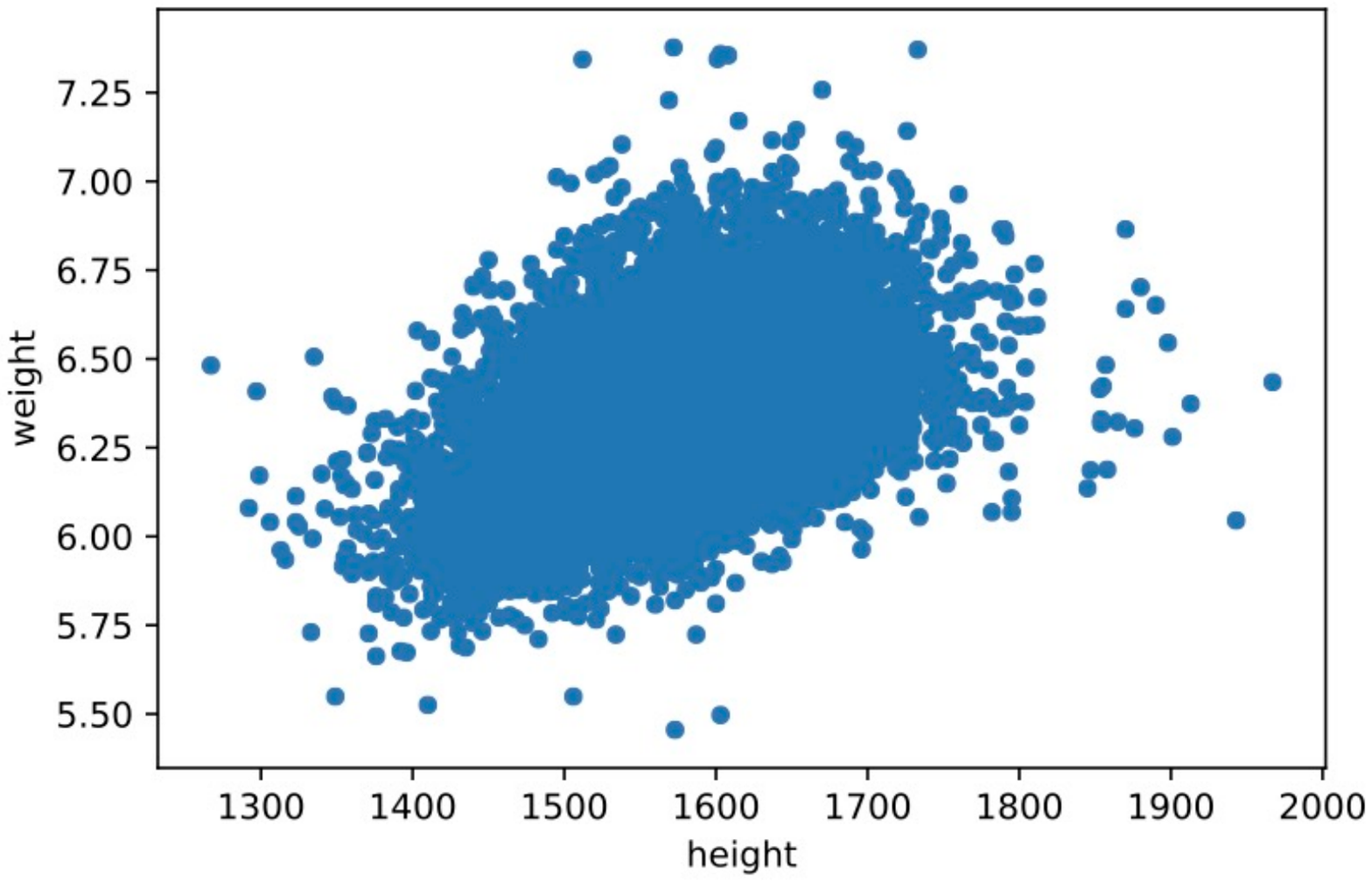}
\end{center}
The spread of the residuals is now nearly flat across height groups (Table \ref{tab:residsdquintile}), and the Breusch--Pagan statistic falls from 90.8 to 4.75 ($p=0.03$). We assume that $\log(w)$ and $h$ follow given data-generating processes and might want to fit a linear regression:
\begin{equation}
\log\left(w_i\right) = \alpha + \beta h_i + \epsilon_i,
\end{equation}
whose interpretation is the object of the next subsection.

\subsection{Log-linear Regression}

Linear regressions are powerful tools, but they often stumble when the relationship between variables isn't strictly linear or when the data is highly skewed. One of the strategies to mitigate such issues is the log-linear regression. 

The logarithm is a monotonic and concave function. This concavity property aids in reducing the skewness in datasets (for further details, refer to Section \ref{sec:logweights}). By transforming the dependent variable with a logarithm, we can often achieve a more linear relationship with the predictor and ensure that the assumptions of linear regression are better met. This improved linearity often translates into more efficient and robust regression estimates.

For our study of the relationship between weight and height, the log-linear regression is particularly useful. The relationship can be expressed as:
\begin{equation}\label{eq:loglineareq}
\log(y) = \alpha + \beta x + \epsilon
\end{equation}
Where:
\begin{itemize}
    \item $\log(y)$ is the natural logarithm of the dependent variable.
    \item $x$ is the independent variable (in our case, height).
    \item $\alpha$ and $\beta$ are the intercept and slope coefficients, respectively.
    \item $\epsilon$ is the error term.
\end{itemize}

\textbf{Illustration:}
We hypothesize that the logarithm of an individual's weight, $\log(w_i)$, is linearly related to their height, $h_i$. Our model is:
\[
\log(w_i) = \alpha + \beta h_i + \epsilon_i
\]

This model suggests that the percentage change in weight is linearly related to changes in height.

\citet{cohn2022count} in their Takeaway 2, demonstrate that bias due to heteroskedastic model error can cause log-linear regression coefficients to have the wrong sign. For this, they suppose that $y=\exp\left( \beta x \right) \eta$  where $x$ is normally distributed and $\eta$ is log-normally distributed with a standard deviation depending on the realization of $x$ (this is the modeling of heteroskedasticity), $\sigma_\eta(x)=\exp\left( \delta x \right)$ with $\delta$ a constant and they show that if $\delta > 2 \beta$ then this bias can cause the sign flip.

\textbf{Note on Model Completeness:} 
From prior research, we understand that crucial explanatory variables, such as parents' height and weight, influence an individual's weight. However, in this model, we don't account for these variables due to a lack of data. Such omission may lead to omitted variable bias.

\begin{center}
\begin{tabular}{lclc}
\toprule
\textbf{Dep. Variable:}    &     logv437      & \textbf{  R-squared:         } &     0.197   \\
\textbf{Model:}            &       OLS        & \textbf{  Adj. R-squared:    } &     0.197   \\
\textbf{Method:}           &  Least Squares   & \textbf{  F-statistic:       } &     3801.   \\
\textbf{Date:}             & Wed, 25 Mar 2020 & \textbf{  Prob (F-statistic):} &     0.00    \\
\textbf{Time:}             &     16:16:21     & \textbf{  Log-Likelihood:    } &    4646.1   \\
\textbf{No. Observations:} &       15494      & \textbf{  AIC:               } &    -9288.   \\
\textbf{Df Residuals:}     &       15492      & \textbf{  BIC:               } &    -9273.   \\
\textbf{Df Model:}         &           1      & \textbf{                     } &             \\
\bottomrule
\end{tabular}
\begin{tabular}{lcccccc}
                   & \textbf{coef} & \textbf{std err} & \textbf{t} & \textbf{P$> |$t$|$} & \textbf{[0.025} & \textbf{0.975]}  \\
\midrule
\textbf{Intercept} &       4.2587  &        0.033     &   127.231  &         0.000        &        4.193    &        4.324     \\
\textbf{v438}      &       0.0013  &     2.13e-05     &    61.654  &         0.000        &        0.001    &        0.001     \\
\bottomrule
\end{tabular}
\begin{tabular}{lclc}
\textbf{Omnibus:}       & 1317.052 & \textbf{  Durbin-Watson:     } &    1.664  \\
\textbf{Prob(Omnibus):} &   0.000  & \textbf{  Jarque-Bera (JB):  } & 1972.605  \\
\textbf{Skew:}          &   0.669  & \textbf{  Prob(JB):          } &     0.00  \\
\textbf{Kurtosis:}      &   4.125  & \textbf{  Cond. No.          } & 3.66e+04  \\
\bottomrule
\end{tabular}
\end{center}

Given the regression output:

- \textbf{Intercept ($\alpha$):} 4.2587. This is the expected logarithm of weight for an individual with a height of 0 (which is not meaningful in real-world context but provides a baseline for the model).

- \textbf{Coefficient of $h_i$ ($\beta$):} $0.0013$.  
Because $\tfrac{d \log w_i}{d h_i} = \tfrac{1}{w_i}\tfrac{dw_i}{dh_i}$, $\beta$ measures the proportional change in $y$:  
a one-unit increase in $h_i$ increases $w_i$ by about $100\times \beta = 0.13\%$, holding other factors constant.

- \textbf{R-squared:} 0.197. This means only 19.7\% of the variability in the logarithm of weight is explained by height based on this model.

Given the statistically significant p-values for both intercept and slope, we infer that the relationship between height and the logarithm of weight is statistically significant. However, with an R-squared value of 0.197, height explains a relatively small fraction of the variation in the logarithm of weight.

\subsection{Log (one plus)-linear regression}

When a data set has many 0s (e.g. $69\%$ fo Compustat firms are granted zero patents in a given year). To take those 0s into account, one might want to do a log1plus regression, because then 0s remain 0s with the log transformation.
\begin{equation}
\log\left(1+y\right) = \alpha + \lambda x + \epsilon
\end{equation}

\citet{cohn2022count} demonstrates that the log1plus estimates is linked to the log-linear estimates by
$$\lambda = \frac{E\left[ y\vert x \right]}{1+E\left[ y\vert x \right]} \beta$$
When $E\left[ y\vert x \right]$ is large then the two estimates are almost equal, but the authors note that when $E\left[ y\vert x \right]$ is large then it is unlikely that many observations are 0s.

For illustration only, here we perform a log1plus regression and as $E\left[ y\vert x \right]$ is large, then the two specification, log1plus and log-linear are similar and log1plus here doesn't add any value as we do not have any real $0$ in our data set.

\subsection{Poisson Regression}

While linear and log-linear regressions are commonly used for continuous data, count data or non-negative integer values often demand different models. One such model is the Poisson regression, typically used for count data.

In this context, citing \citet{cohn2022count}, we introduce the Poisson regression mainly for illustrative purposes. It's worth noting that while our data of weight isn't typical 'count' data, the Poisson regression can still provide valuable insights. 

The specification of the Poisson regression model is:
\begin{equation}\label{eq:poissoneq}
\log\left( E[y|x] \right) = \alpha + \omega x + \nu
\end{equation}
Where:
\begin{itemize}
    \item $E[y|x]$ represents the expected value of $y$ given $x$.
    \item $\log\left( E[y|x] \right)$ is the natural logarithm of the expected value of the dependent variable given the predictor.
    \item $x$ is the independent variable.
    \item $\alpha$ and $\omega$ are the intercept and slope coefficients, respectively.
    \item $\nu$ is the error term.
\end{itemize}
It's crucial to interpret the coefficients of Poisson regression in terms of percentage change, given its multiplicative nature.

\citet{bellego2022dealing} provide some critics\footnote{(1) Existence of a solution
is not guaranteed; (2) They can exhibit convergence issues; (3) They can be very imprecise
if the log-scale error is heavy-tailed; (4) They can be difficult to estimate with many fixed-effects; (5) instrumental variables require stronger assumptions and may dramatically
increase computational complexity; and (6) they suffer from an incidental parameter problem in the standard errors when including two-way fixed effects.} of the Poisson approach and suggest an alternative iOLS that goes back to log (one plus)-linear regression. 

\subsection{Exponential regression}
Recent econometrics paper study the implications of misspecified exponential regressions \citep{silva2026misspecified}. This model is typically used for gravity equations for trade, migration, etc. It is also in use for health economics to study ``the impact of out-of-pocket price, income, health status or some other economic or health covariates on the expenditures or visits by users of health care or the impact on some other positive economic outcome'' \citep{manning2001estimating}.

\subsection{Random Forest regressor}

We can also use machine learning techniques, but these will be oriented toward answering the question: given a height $h_i$ what is the predicted weight $w_i$, but the model being nonparametric, we cannot easily derive a sensitivity.

As an illustration, we fit a Random Forest regression model on the sample, we might want to restrict our fit not to overfit:
\begin{center}
\includegraphics[scale=0.5]{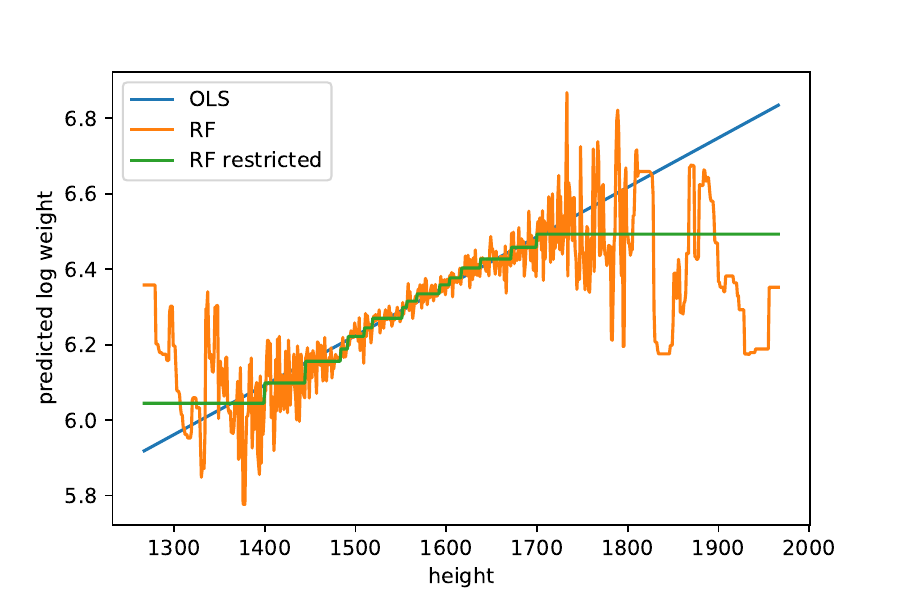}
\end{center}

knowing its limitation, the linear model allows an interpretation of the impact of height on weight.

\section{Correlation is not causation}

In one of the empirical exercises of this course, we merge official WHO data on alcohol consumption per adult aged 15+ with World Bank data on GDP per capita, and we produce a cross-country scatter plot. This is useful econometric work, but it is \emph{descriptive} work first. The purpose is to learn how to collect official data, harmonize country definitions, check data quality, separate more reliable observations from weaker ones, and visualize an empirical pattern in a transparent way.

What we do \emph{not} do at this stage is claim that the slope of the scatter plot has a causal interpretation. A positive cross-country association between alcohol consumption and GDP per capita should not be read as evidence that alcohol consumption improves productivity per person. On the contrary, such an interpretation is very unlikely. At the individual level, excessive alcohol consumption is more plausibly associated with lower productivity, worse health, and lower human capital accumulation. At the country level, the positive association is much more likely to reflect differences in development, institutions, demographics, regulation, tourism, religion, tax capacity, and measurement quality.

To see this formally, suppose that one estimates the cross-country regression
\begin{equation}
\log\left(\text{GDP per capita}_{i}\right)=\alpha+\beta\,\text{Alcohol consumption}_{i}+u_{i}.
\end{equation}
The OLS estimate of $\beta$ is causal only if alcohol consumption is orthogonal to the error term $u_i$, that is, only if countries that differ in alcohol consumption do not also differ systematically in other determinants of productivity once the controls are included. In practice, this assumption is rarely credible in a simple cross-section.

There are several reasons for this.
\begin{itemize}
\item \textbf{Omitted variables.} Richer countries differ from poorer countries in education, institutions, health systems, sectoral composition, urbanization, age structure, climate, conflict exposure, and many other determinants of GDP per capita. If these variables are correlated with alcohol consumption, OLS is biased.
\item \textbf{Reverse causality.} Higher income may increase the demand for restaurant services, leisure goods, and formally recorded alcohol. In that case, income affects alcohol consumption, not only the other way around.
\item \textbf{Measurement error.} Alcohol consumption is difficult to measure perfectly, especially where unrecorded production is important. This is precisely why data quality work matters. Measurement error can attenuate coefficients and also distort cross-country comparisons.
\item \textbf{Composition effects.} Recorded alcohol consumption may proxy for broader differences in market formalization, tourism intensity, or tax collection rather than for a biologically or economically meaningful ``effect of alcohol'' on productivity.
\end{itemize}

This is why a statistically significant correlation, even with a nice scatter plot and a significant slope coefficient, is not enough to make a causal statement. Significance tells us that the correlation is unlikely to be zero in the sample under the model assumptions; it does not tell us that the identifying assumptions are valid.

How should an econometrician handle this?
\begin{enumerate}
\item \textbf{Start with the right language.} Say that the figure documents a \emph{correlation}, a \emph{descriptive association}, or a \emph{stylized fact}. Do not write that alcohol consumption ``causes'' higher productivity from this exercise alone.
\item \textbf{Let theory guide the controls.} If you run a multivariate regression, add controls motivated by economics: education, demographics, institutions, health indicators, sectoral structure, and possibly region fixed effects. Do not simply throw variables into the regression without an identification argument.
\item \textbf{Use panel data when possible.} Country and year fixed effects can help absorb time-invariant country heterogeneity and common macro shocks. This is often better than a pure cross-section, but fixed effects alone do not solve all endogeneity problems.
\item \textbf{Exploit quasi-experimental variation.} If the question is causal, search for policy shocks or natural experiments: alcohol taxes, sales restrictions, legal reforms, or staggered policy changes. Then methods such as Difference-in-Differences or event studies become relevant, provided the identifying assumptions are carefully discussed and tested.
\item \textbf{Be demanding about instruments.} An instrumental-variables strategy is acceptable only if the exclusion restriction is credible. This is difficult here, because many candidate instruments are likely to affect productivity through channels other than alcohol consumption.
\item \textbf{Perform robustness checks.} Check sensitivity to the sample, the functional form, data quality thresholds, outliers, and alternative measures. If a result disappears after minimal robustness work, it was not a strong result.
\end{enumerate}

The pedagogical lesson is simple: descriptive analysis is not a second-rate activity. It is an essential first step. A careful scatter plot with transparent data cleaning can reveal whether a question is interesting, whether the data are credible, and what identification challenges will have to be solved. But the causal question is different from the descriptive one. The descriptive question is: \emph{do countries with higher recorded alcohol consumption also tend to have higher GDP per capita?} The causal question is: \emph{holding all other determinants constant, what exogenous change in alcohol consumption would alter productivity?} Econometrics begins precisely when we stop confusing these two questions.

\section{Stationarity, Unit Roots}

Stationarity, Dickey-Fuller test (original and augmented), auto-regressions\footnote{stationarity\_adf\_ar.py}.

Before introducing the theory of stationarity and the estimation of
AR, MA, ARMA, ARIMA, SARIMA, ARIMA--GARCH, and SARIMA--GARCH models,
we address a set of frequently asked questions.

\begin{itemize}
\item \textbf{What is the appropriate order of transformations and model fitting for a univariate time series?}

\begin{itemize}
\item \textbf{Primary recommendation.} Conduct a thorough literature review and replicate, as closely as possible, the transformations and specifications adopted in top academic journals for comparable time series. Statistical testing should be used to validate that similar treatments are appropriate for your data.

\item \textbf{In the absence of guidance from the literature.} This typically indicates a novel dataset. In that case, transformations and dynamic components should be specified \emph{jointly}, rather than sequentially.\footnote{This is analogous to the ARMA$(p,q)$--GARCH$(r,s)$ case (Section~\ref{sec:uARMAGARCH}), where one should not estimate an ARMA model first and a GARCH model on the residuals afterward, but rather estimate an ARMA--GARCH system simultaneously.}

\begin{itemize}
\item If the series exhibits seasonality but no clear deterministic trend, estimate a SARIMA model.\footnote{For pedagogical purposes, we often proceed step by step: (i) remove seasonality, (ii) test for unit roots and difference if needed, (iii) estimate an ARMA--GARCH model.} Recall that a SARIMA model for a non-stationary series is equivalent to an ARMA model applied to the appropriately differenced (stationary) series.

\item If a trend is present, first-order differencing is often sufficient. As a conservative approach, estimating a SARIMA model allows one to address both integration (I) and potential seasonality (S) simultaneously.

\item If conditional heteroskedasticity is observed, consider variance-stabilizing transformations such as the Box--Cox transformation, of which the logarithmic transformation is a special case.
\end{itemize}

\item The \href{https://en.wikipedia.org/wiki/Box–Jenkins_method}{Box--Jenkins methodology} provides a useful general framework, though it should not be applied mechanically.
\end{itemize}

\item \textbf{What if the time series is intended for use in a multivariate model?}

\begin{itemize}
\item If available, use the seasonally adjusted series.
\item If no adjusted series exists and the literature is silent, begin with a univariate analysis of the raw series to understand its properties, typically starting with a SARIMA--GARCH specification.
\end{itemize}
\end{itemize}

\vspace{0.5em}
\noindent\textbf{Formal decomposition of a SARIMA--GARCH model.}

Let $\{y_t\}$ denote a univariate time series. A SARIMA$(p,d,q)\times(P,D,Q)_s$--GARCH$(r,a)$ model can be written as
\[
\Phi_P(L^s)\,\phi_p(L)\,(1-L)^d(1-L^s)^D y_t
= \Theta_Q(L^s)\,\theta_q(L)\,\varepsilon_t,
\qquad \varepsilon_t = \sigma_t z_t,
\]
where $z_t \sim \text{i.i.d. }(0,1)$ and $\sigma_t^2$ follows a GARCH process.

\begin{itemize}
\item \textbf{Autoregressive (AR) component.}  
An AR$(p)$ process captures linear dependence on past realizations:
\[
\phi_p(L)y_t = \sum_{i=1}^p \phi_i y_{t-i}.
\]

\item \textbf{Moving-average (MA) component.}  
An MA$(q)$ process models dependence on past shocks:
\[
\theta_q(L)\varepsilon_t = \sum_{j=1}^q \theta_j \varepsilon_{t-j}.
\]

\item \textbf{Integrated (I) component.}  
Differencing operators $(1-L)^d$ and $(1-L^s)^D$ remove stochastic trends and seasonal unit roots, respectively.

\item \textbf{Seasonal (S) component.}  
Seasonal AR and MA polynomials $\Phi_P(L^s)$ and $\Theta_Q(L^s)$ capture dependence at multiples of the seasonal frequency $s$.

\item \textbf{GARCH component.}  
Conditional variance dynamics are modeled as
\[
\sigma_t^2 = \omega + \sum_{i=1}^r \alpha_i \varepsilon_{t-i}^2
+ \sum_{j=1}^a \beta_j \sigma_{t-j}^2,
\]
allowing for time-varying volatility and volatility clustering.
\end{itemize}

\paragraph{Definition (White noise).}
A stochastic process $\{\varepsilon_t\}_{t\in\mathbb{Z}}$ is said to be a
\emph{white-noise process} with mean zero and variance $\sigma_\varepsilon^2$,
denoted
\[
\varepsilon_t \sim \text{WN}(0,\sigma_\varepsilon^2),
\]
if it satisfies:
\begin{enumerate}
  \item \textbf{Zero mean:}
  \[
  \mathbb{E}[\varepsilon_t] = 0 \quad \text{for all } t;
  \]
  \item \textbf{Constant variance:}
  \[
  \mathrm{Var}(\varepsilon_t) = \sigma_\varepsilon^2 \quad \text{for all } t;
  \]
  \item \textbf{No serial correlation:}
  \[
  \mathrm{Cov}(\varepsilon_t,\varepsilon_{t-k}) = 0 
  \quad \text{for all } k \neq 0.
  \]
\end{enumerate}

\paragraph{Definition (Moving Average process).}
A time series $\{y_t\}_{t\in\mathbb{Z}}$ is said to follow a
\emph{moving-average process of order $q$}, denoted MA($q$), if it can be written as
\[
\boxed{\;
y_t = \mu + \varepsilon_t + \sum_{j=1}^{q} \theta_j\,\varepsilon_{t-j},
\qquad \varepsilon_t \sim \text{WN}(0,\sigma^2).
\;}
\]

\subsection{Concept of stationarity}\label{sec:stationarity}
Before regressing (or for AR), we need to go step by step on the integration order and the cointegration\footnote{For the cointegration test, the null hypothesis is no cointegration} of the variables.

A time series \(\{y_t\}\) is said to be \textit{weakly stationary} (or second-order stationary) if it satisfies the following conditions for all time points \(t\) and for all lags \(s\):
\begin{enumerate}
    \item The mean \(E(y_t)\) is constant over time, that is,
    \begin{equation}
    E(y_t) = \mu < \infty, \quad \forall t.
    \end{equation}
    \item The variance \(\text{Var}(y_t)\) is finite and constant over time, that is,
    \begin{equation}
    \text{Var}(y_t) = \sigma^2 < \infty, \quad \forall t.
    \end{equation}
    \item The covariance between \(y_t\) and \(y_{t+s}\) depends only on the lag \(s\) and is constant over time, that is,
    \begin{equation}
    \text{Cov}(y_t, y_{t+s}) = \gamma(s) < \infty, \quad \forall t.
    \end{equation}
\end{enumerate}
\noindent These conditions ensure that the time series has stable moments (mean and variance) across different time points and does not exhibit trends or seasonality in its first and second moments.

It can be shown\footnote{Using the law of large numbers and assuming that \(y_t\) is stationary.} that the following empirical moments are consistent estimators of the above moments when \(T \rightarrow \infty\):
\begin{equation}
\begin{cases}
    \hat{\mu} = \frac{1}{T}\sum_{t=1}^T y_t\\
    \hat{\gamma_0} = \frac{1}{T} \sum_{t=1}^T \left(y_t - \hat{\mu}\right)^2 \\
    \hat{\gamma_s} = \frac{1}{T-s} \sum_{t=1+s}^T \left(y_t - \hat{\mu}\right)\left(y_{t-s} - \hat{\mu}\right) & \text{with} \quad T \gg s
\end{cases}
\end{equation}
The most basic weakly stationary processes are white noises and moving averages.

A time series \(\{y_t\}\) is said to be \textit{strictly stationary}\footnote{In financial time series, strict stationarity is rare but can be taken as an assumption and then discussed with robustness checks.} if the joint distribution of \((y_t, y_{t+1}, \ldots, y_{t+k})\) is identical to \((y_{t+h}, y_{t+h+1}, \ldots, y_{t+h+k})\) for all time points \(t\) and for all lags \(h\) and \(k\), that is,
\begin{equation}
F(y_t, y_{t+1}, \ldots, y_{t+k}) = F(y_{t+h}, y_{t+h+1}, \ldots, y_{t+h+k}), \quad \forall t, h, k.
\end{equation}
\noindent In other words, the distribution of the series does not change over time when shifted along the time axis, making the series consistent in distributional properties at all time points.

\paragraph{Estimating the variance (home exercise)}
Let $\{y_t\}$ be weakly stationary, with mean $\mu$ and
autocovariance $\gamma(s)=\operatorname{Cov}(y_t,y_{t-s})$,
and let $T>1$.
Writing $\hat{\mu}=T^{-1}\sum_{t=1}^T y_t$, we have
\[
\sum_{t=1}^T(y_t-\hat{\mu})^2
=
\sum_{t=1}^T(y_t-\mu)^2-T(\hat{\mu}-\mu)^2.
\]
Since $\mathbb{E}(\hat{\mu})=\mu$, taking expectations gives
\[
\mathbb{E}\left[\sum_{t=1}^T(y_t-\hat{\mu})^2\right]
=
T\gamma(0)-T\operatorname{Var}(\hat{\mu}).
\]
Under weak stationarity,
\[
\operatorname{Var}(\hat{\mu})
=
\frac{1}{T^2}\sum_{t=1}^T\sum_{u=1}^T\gamma(t-u)
=
\frac{\gamma(0)}{T}
+\frac{2}{T^2}\sum_{s=1}^{T-1}(T-s)\gamma(s).
\]
Therefore, the estimator with denominator $T$ satisfies
\[
\hat{\gamma}_0
=
\frac{1}{T}\sum_{t=1}^T(y_t-\hat{\mu})^2,
\qquad
\mathbb{E}(\hat{\gamma}_0)
=
\gamma(0)-\operatorname{Var}(\hat{\mu}).
\]
Its bias is thus nonpositive, regardless of the signs of
the autocovariances.

If observations are serially uncorrelated, then
$\gamma(s)=0$ for $s\neq 0$, so
\[
\mathbb{E}(\hat{\gamma}_0)=\frac{T-1}{T}\gamma(0).
\]
In this case, Bessel's correction gives the unbiased estimator
\[
\tilde{\gamma}_0
=
\frac{1}{T-1}\sum_{t=1}^T(y_t-\hat{\mu})^2,
\qquad
\mathbb{E}(\tilde{\gamma}_0)=\gamma(0).
\]
With serial correlation, however,
\[
\mathbb{E}(\tilde{\gamma}_0)
=
\gamma(0)
-\frac{2}{T(T-1)}
\sum_{s=1}^{T-1}(T-s)\gamma(s),
\]
so dividing by $T-1$ generally does not eliminate the bias.

Both estimators are consistent under suitable additional
assumptions; weak stationarity alone is insufficient.
For example, if $\{y_t\}$ is strictly stationary and ergodic
with $\mathbb{E}(y_t^2)<\infty$, the ergodic theorem implies
\[
\hat{\mu}\xrightarrow{\mathrm{a.s.}}\mu,
\qquad
\frac{1}{T}\sum_{t=1}^T y_t^2
\xrightarrow{\mathrm{a.s.}}\mathbb{E}(y_t^2).
\]
Consequently,
\[
\hat{\gamma}_0
=
\frac{1}{T}\sum_{t=1}^T y_t^2-\hat{\mu}^2
\xrightarrow{\mathrm{a.s.}}
\mathbb{E}(y_t^2)-\mu^2
=
\gamma(0).
\]
Since $\tilde{\gamma}_0=\frac{T}{T-1}\hat{\gamma}_0$,
it has the same almost-sure limit.

\subsubsection{Concept of Integration of order $d$}

If \(\Delta\) is the difference operator defined as \(\Delta y_t = y_t - y_{t-1}\), \(\Delta^2 y_t = \Delta(\Delta y_t)\), etc., we say that \(y_t\) is \emph{integrated of order d} or \(I(d)\) if \(y_t\), \(\Delta y_t\), \(\ldots\), \(\Delta^{d-1} y_t\) are non-stationary and \(\Delta^{d} y_t\) is stationary\footnote{The concept is detailed in \citep{Engle:1987aa}. Definitions vary; for example, \citep{alexanderII} defines integrated of order \(d\) as \(\Delta^{d} y_t\) having a stationary ARMA representation. Keep in mind that integration orders greater than two are not very relevant for economic applications.}.

\subsection{Wold decomposition theorem}

The Wold Decomposition Theorem plays a pivotal role in the theory of stationary time series. This theorem essentially describes any stationary time series as a combination of two orthogonal components: a deterministic component and a stochastic component.

Let \(\{y_t\}\) be a zero-mean weakly stationary time series. According to the Wold Decomposition Theorem, \(\{y_t\}\) can be decomposed as
\begin{equation}
    y_t = \sum_{j=0}^{\infty} b_j \varepsilon_{t-j} + \eta_t,
\end{equation}
where
\begin{itemize}
    \item \(\{\varepsilon_t\}\) is a white noise process, implying that \(E(\varepsilon_t) = 0\), \(\text{Var}(\varepsilon_t) = \sigma^2 < \infty\), and \(\text{Cov}(\varepsilon_t, \varepsilon_s) = 0\) for \(t \neq s\),
    \item \(\{b_j\}\) are coefficients such that the series \(\sum_{j=0}^{\infty} b_j^2\) converges, assuring that the linear combination of white noise processes is well-defined,
    \item \(\{\eta_t\}\) is a deterministic component that could represent a polynomial trend or seasonality,
    \item and \(\{\varepsilon_t\}\) and \(\{\eta_t\}\) are uncorrelated processes, meaning that \(\text{Cov}(\varepsilon_t, \eta_s) = 0\) for all \(t\) and \(s\).
\end{itemize}

The first component \(\sum_{j=0}^{\infty} b_j \varepsilon_{t-j}\) is known as the \textit{innovations} or \textit{moving-average} representation of \(\{y_t\}\). The deterministic component \(\eta_t\) is typically of less interest and is often taken as zero, especially when the focus is on the stochastic properties of the time series.

More compact wording for part of this theorem:

$\epsilon \sim \mbox{WN}(0,\sigma^2)$ and $\sum_{i=0}^\infty \vert b_i\vert < \infty$.

\paragraph{What does it mean for an infinite sum to converge?}

An infinite sum
\[
\sum_{j=0}^{\infty} a_j
\]
is said to \emph{converge} if the sequence of partial sums
\[
S_N \equiv \sum_{j=0}^{N} a_j
\]
approaches a finite limit as $N \to \infty$. Intuitively, this means that adding more and more terms eventually has a negligible effect on the total sum.

\medskip
\noindent
\textbf{Deterministic intuition.}
Consider the geometric series
\[
\sum_{j=0}^{\infty} \rho^j,
\quad 0 < |\rho| < 1.
\]
Each additional term becomes smaller and smaller, and the total sum converges to $1/(1-\rho)$. In contrast, if $|\rho|\ge 1$, the terms do not decay fast enough and the sum diverges.

\subsubsection{Impulse Response Function}

Impulse Response Function aims at answering how much our process ${y_t}$ is affected at time $t$ by an innovation (a shock) that happened $h$ periods before $t$. 

The Impulse Response Function traces the effect of a one-unit shock to \(\varepsilon_t\) on the time series \(y_t\) at various time horizons. Mathematically, the \(h\)-step-ahead impulse response is given by
\[
IRF(h) = E\left( y_{t+h} \mid \varepsilon_t = 1, \varepsilon_{t-1} = 0, \varepsilon_{t-2} = 0, \ldots \right),
\]
where \(h\) represents the horizon of interest.

Graphically, the IRF represents the response of \(y_t\) to a one-unit impulse in \(\varepsilon_t\), holding all other past \(\varepsilon_{t-i}\) constant at zero, illustrating how an external shock or innovation permeates and affects the time series over subsequent periods. 

From the Wold theorem, for a stationary process the innovation at time $t-h$ is $\epsilon_{t-h}$ and the Impulse Response Function if $\frac{\partial y_t}{\partial \epsilon_{t-h}}$ which is thus $b_h$.

Consider a linear time series model (like an ARMA model):
\[
y_t = \phi(L) y_{t-1} + \theta(L) \varepsilon_t,
\]
where \(y_t\) is the time series, \(\phi(L)\) and \(\theta(L)\) represent polynomial lag operators, \(L\) is the lag operator, and \(\varepsilon_t\) represents white noise.

\subsection{AR(1) stationary process}\label{sec:AR1stationary}

We consider a stationary AR(1):
\begin{equation}\label{eq:AR1}
y_t = \theta y_{t-1} + \epsilon_t
\end{equation}

with $0<\vert \theta \vert < 1$ and $\epsilon \sim \mathcal{N}(0,1)$. We can rewrite:
$$y_t = \sum_{i=0}^{t-1}\theta^i \epsilon_{t-i} + \theta^t y_0$$ we have:
\begin{equation*}
\left\{
    \begin{array}{ll}
E\left(y_t\right) = \theta^t y_0 \rightarrow 0\\
Var\left(y_t\right) = \sum_{i=0}^{t-1} \theta^{2i} \sigma^2 \rightarrow \frac{1}{1-\theta^2} \sigma^2 \\
Cov\left(y_t, y_{t-s}\right) \sum_{i=0}^{t-s-1} \theta^{2i} \sigma^2 \rightarrow \frac{1}{1-\theta^2} \sigma^2 \\
\end{array}
\right.
\end{equation*}

We say that an AR(1) is asymptotically stationary, and usually just say stationary.

Another way to look at it, is that if this process is to be weakly stationary, then, we need:
\begin{equation*}
\left\{
    \begin{array}{ll}
\mu (1-\theta) = 0, \mbox{ as} \quad E\left(\epsilon_t\right)=0\\
\gamma_0  (1-\theta^2) = \sigma^2\\
Cov\left(y_t, y_{t-s}\right) = \gamma_s\\
\end{array}
\right.
\end{equation*}
which has solution if $0<\vert \theta \vert < 1$, as a variance is necessarily positive. We can rewrite:
\begin{equation*}
\left\{
    \begin{array}{ll}
\mu = 0 \\
\gamma_0  = \frac{\sigma^2}{1-\theta^2} \\
Cov\left(y_t, y_{t-s}\right) = \gamma_s\\
\end{array}
\right.
\end{equation*}

If $\theta = 1$ either we have no solution, either $\sigma = 0$, which would then mean that the error term is a constant $\forall t, \epsilon_t = 0$ and then $\forall t, y_t = y_0$ which is stationary but presents no interest for study.

From our initial exploration of the AR(1) model, we can derive certain insights, especially when applied to financial data like asset prices. First, to ensure that $E\left(y_t\right) \rightarrow 0$, we should refrain from using the price levels directly. If we were to use them, it might erroneously suggest a model where all prices converge to zero, implying a rather grim scenario where every company defaults. Instead of price levels, it's more practical and meaningful to work with price returns, thereby de-trending the data to respect the condition $E\left(y_t\right) \rightarrow 0$.

For the constant volatility assumption inherent in the AR(1) model, let's consider the CAC 40 index time series. Upon visual inspection, one can observe periods of 'volatility clustering', which implies that volatility isn't constant through time. This observation suggests that the basic AR(1) model might not be sufficient to capture the underlying dynamics of such financial time series. To address this, more advanced models that account for evolving volatility, such as the ARCH (Autoregressive Conditional Heteroskedasticity) model, become necessary.

\begin{center}
\includegraphics[scale=0.5]{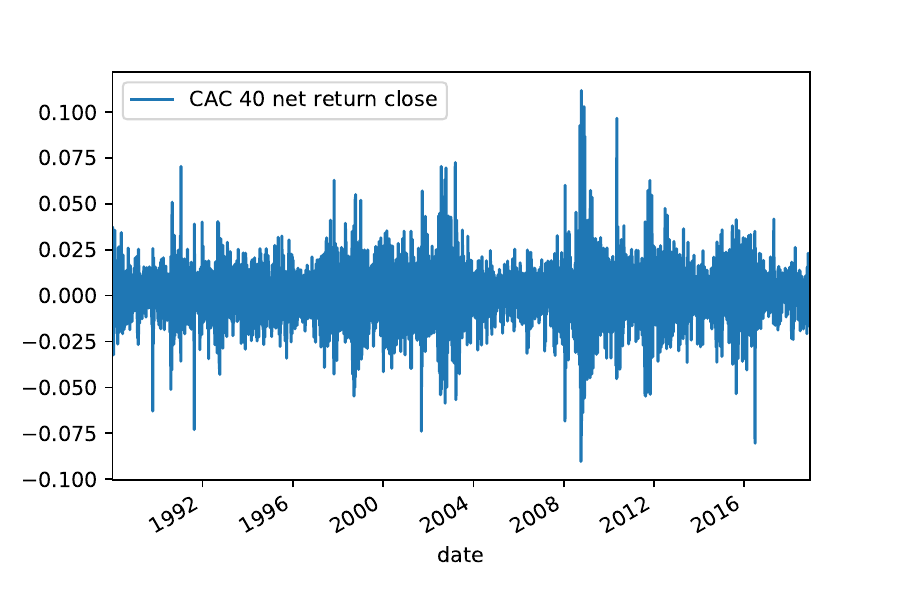}
\end{center}

\subsection{Finite-order lag polynomial}

Equation~\ref{eq:AR1} can be written using the lag operator $L$ as
\[
(1-\theta L)y_t = \varepsilon_t.
\]

For $\theta\neq 0$, the root of the autoregressive polynomial
$1-\theta z$ is $z=1/\theta$.
If $|\theta|<1$, this root lies outside the unit circle,
and the autoregressive lag polynomial admits a convergent
inverse.\footnote{The polynomial $\phi(L)=1-\theta L$
is called the autoregressive polynomial of $\{y_t\}$.}

In that case,
\[
(1-\theta L)^{-1}
=\sum_{i=0}^{\infty}\theta^i L^i,
\]
and equation~\ref{eq:AR1} admits the moving-average representation
\[
y_t=\sum_{i=0}^{\infty}\theta^i\varepsilon_{t-i}.
\]

Since $\sum_{i=0}^{\infty}|\theta|^i<\infty$,
the process $\{y_t\}$ is weakly stationary, provided that
$\{\varepsilon_t\}$ is white noise with finite variance.

This generalizes to an AR($p$) process written in
lag-polynomial form as
\[
\Phi(L)y_t=\varepsilon_t,
\qquad
\Phi(L)\equiv 1-\phi_1L-\cdots-\phi_pL^p.
\]

Let $\{\lambda_k^{-1}\}_{k=1}^p$ denote the roots of the
characteristic equation $\Phi(z)=0$.
If
\[
|\lambda_k^{-1}|>1
\quad\text{for all } k=1,\ldots,p,
\]
then $\Phi(L)$ can be factorized as
\[
\Phi(L)=\prod_{k=1}^p(1-\lambda_kL),
\]
and therefore
\[
\Phi(L)^{-1}
=\prod_{k=1}^p(1-\lambda_kL)^{-1}.
\]

Since $|\lambda_k^{-1}|>1$ implies $|\lambda_k|<1$,
each factor admits a geometric-series expansion,
\[
(1-\lambda_kL)^{-1}
=\sum_{j=0}^{\infty}\lambda_k^{j}L^j.
\]

Hence, the product defines a convergent power series in $L$,
\[
\Phi(L)^{-1}=\sum_{i=0}^{\infty}a_iL^i.
\]

Moreover, because all roots lie strictly outside the unit
circle, the coefficients $\{a_i\}$ decay geometrically.
In particular,
\[
\sum_{i=0}^{\infty}|a_i|<\infty.
\]

Thus, if $\{\varepsilon_t\}$ is white noise with finite
variance, the AR($p$) process admits the causal
MA($\infty$) representation
\[
y_t=\sum_{i=0}^{\infty}a_i\varepsilon_{t-i},
\]
which defines a weakly stationary process.

\subsection{Process with a unit root: a non-stationary process}

If we follow section \ref{sec:AR1stationary} but now set $\theta = 1$, $y_t = y_{t-1} + \epsilon_t$ which is a Random Walk and chose $y_0=100$. We can rewrite the process with $L$ the lag operator, $(1 - L) y_t = \epsilon_t$, the root of the equation $1 - x$ is 1, hence we say that this process has a unit root.

We can rewrite $y_t = \sum_{i=1}^{t-1}\epsilon_{t-i} + y_0$ and find that $Var\left(y_t\right) = t \sigma^2 \rightarrow \infty$ which is against our definition of weak stationarity.

\quad

\stepcounter{question}
\begin{itemize}
\item \textbf{Question \arabic{question}}
\begin{itemize}
\item create a stationary AR(1) ($\theta < 1$) and a unit root one, with $1000$ observation for each. Compute the variance of the series and compare them to the expected variance as described above.
\item for the stationary AR(1), introduce at time $500$ an additional shock (innovation) and evaluate the impulse response function at time $505$.
\end{itemize}
\end{itemize}

\quad

In simple terms, a time series has a unit root, or follows an integrated process, if shocks have permanent effects on the level, but not on the rate of change in the variable.
Indeed, if at a time $t_{\mbox{choc}}$ the time series experience a shock $\Delta = 50$, then the future level of $y_t$ will be impacted ($\forall t \geq t_{\mbox{choc}}$) but not the rate of change $\frac{y_{t+1} - y_t}{y_t}$.

For illustration, we apply a shock $\Delta$ at mid-period to our time series:

\begin{center}
\includegraphics[scale=0.5]{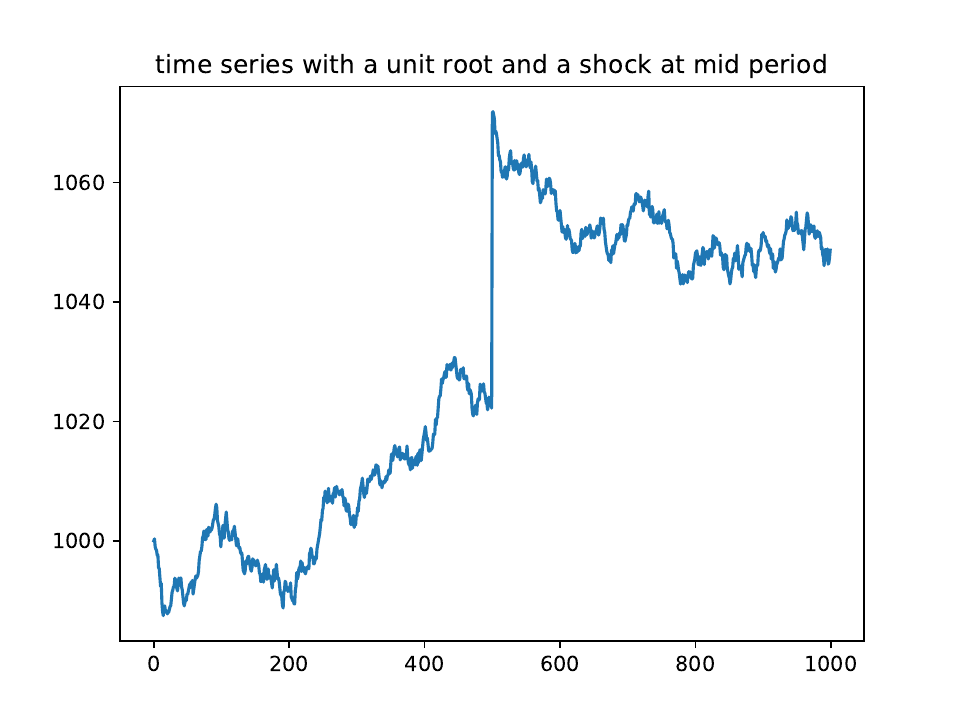}
\end{center}
\begin{center}
\includegraphics[scale=0.5]{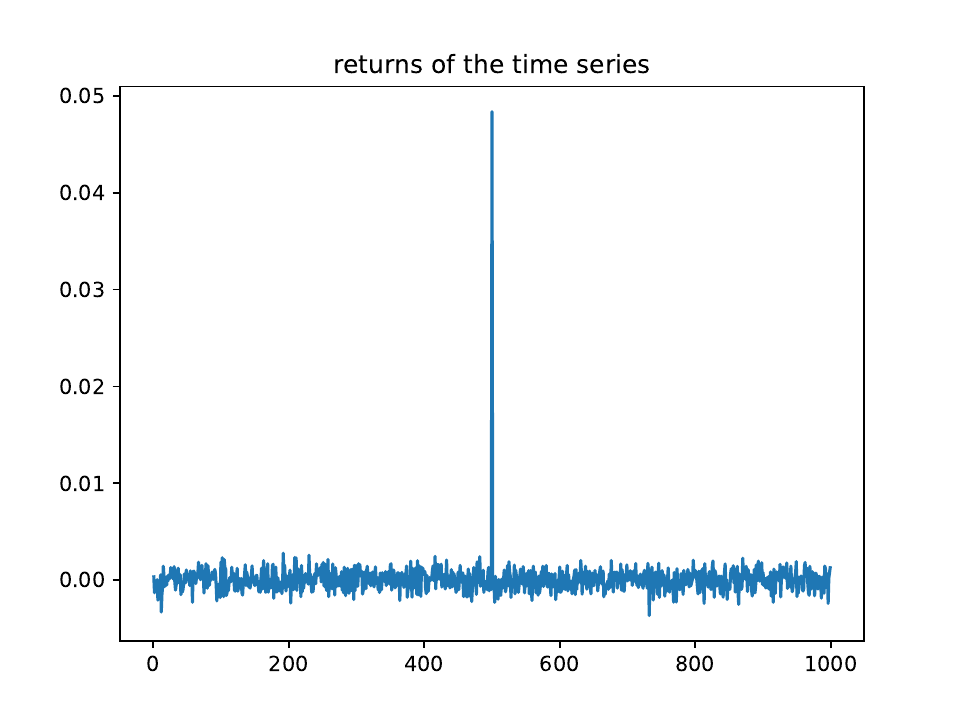}
\end{center}

\subsection{Dickey--Fuller test}\label{sec:DFtest}

The Dickey--Fuller (DF) test, introduced by Dickey and Fuller
(1979), tests whether a time series contains a unit root.

Consider the AR(1) process without a constant or deterministic trend:
\[
y_t=\rho y_{t-1}+\varepsilon_t,
\]
where $\varepsilon_t$ is i.i.d.\ white noise with zero mean
and constant variance. Subtracting $y_{t-1}$ gives
\begin{equation}
\Delta y_t=(\rho-1)y_{t-1}+\varepsilon_t.
\end{equation}
Defining $\gamma=\rho-1$, we obtain
\begin{equation}\label{eq:DFeq}
\Delta y_t=\gamma y_{t-1}+\varepsilon_t.
\end{equation}

The null hypothesis of a unit root is $H_0:\gamma=0$
($\rho=1$), against the alternative of stationarity,
$H_1:|\rho|<1$ (equivalently, $-2<\gamma<0$).

The test statistic is the conventional OLS $t$-ratio:
\[
\tau_{\mathrm{DF}}
=\frac{\hat{\gamma}}{\operatorname{se}(\hat{\gamma})}.
\]
However, under $H_0$, this statistic does not follow
Student's $t$-distribution, even asymptotically.
Instead, it follows a nonstandard Dickey--Fuller distribution
whose critical values depend on the deterministic terms
included in the regression \citep{MacKinnon2010}.

This test is not powerful enough to have full confidence as there is likely to be autoregressive component in time series: use an augmented test.

\subsection{Augmented Dickey-Fuller test}\label{sec:adf}

The Augmented Dickey-Fuller test can be used to  test whether a series is stationary, and take autoregressive effects into account. $H_0$: $\gamma = 0$.

The formula being used for the augmented Dickey-Fuller test is of the form:
\begin{equation}\label{eq:adf}
\Delta y_t = \beta_0 + \beta_1 t + \beta_2 t^2 + \gamma y_{t-1} + \sum_{j=1}^{p-1} \delta_j \Delta y_{t-j} + \epsilon_t
\end{equation}

\citep{MacKinnon2010} discuss the number of lags, selected via information criteria methods, and discuss the test statistics for the "no-constant" ('n'), "no-trend" ('c'), and "with-trend" ('ct') version of this test with reference to the table to be used.

\quad

With the python function adfuller in statsmodels, you can chose the parameter of the regression:
\begin{itemize}
\item 'c' : constant only (default), $\beta_1 = \beta_2 = 0$
\item 'ct' : constant and trend $\beta_2 = 0$
\item 'ctt' : constant, and linear and quadratic trend
\item 'n' : no constant, no trend $\beta_0 = \beta_1 = \beta_2 = 0$
\end{itemize}

\quad

The output of the test are:
\begin{itemize}
\item The statistics
\item \textbf{The p-value}
\item The number of lags used for the test
\item The number of observations used
\item The critical values
\end{itemize}

\quad
with the null hypothesis:

\textbf{$H_0$: the time series has a unit root.}

\quad

To apply this test, you need to have a "testing strategy" as suggested in \citep{Elder:2001aa} which "exploits known information about the growth character of the variable under investigation". So:
\begin{itemize}
\item if your original times series is evolving erratically around a non-zero mean like the unemployment rate, i.e. it doesn't grow in the long run ($\beta_0\neq 0$), then you do not assume any \textit{drift} in the first difference and on conduct an F test on the joint null that $\beta_0=0$ and $\gamma=0$;
\item if the original times series has a constant rate of growth ($\beta_1\neq 0$), we say that the series have a \textit{drift}, therefore conduct a F test on the joint null that $\beta_1=0$ and $\gamma=0$;
\item \citep{Elder:2001aa} discard the option $\beta_1 \neq 0$ as it implies an ever increasing (or decreasing) rate of change.
\end{itemize}

Additional data, you can refer to the book \citep{Lutkepohl2004} and the detailed answer posted \href{https://stats.stackexchange.com/questions/44647/which-dickey-fuller-test-for-a-time-series-modelled-with-an-intercept-drift-and}{here}.

\citep{Elder:2001aa}  stress the importance of the correct model as \begin{quote}
if the trend term is erroneously omitted, the tests
are biased toward finding a unit root. This is easy to explain. If there is a trend but no trend term included in the regression, the only way the regression can capture the trend is by estimating a unit root and using the intercept (drift) to reflect the trend. On the other hand, including a trend term in the regression when it is inappropriate reduces the power of the unit root tests, for the same reason that including irrelevant explanatory variables increases variance.
\end{quote}

\quad

The monthly variables of unemployment, population and GDP are integrated of order 1.

\subsubsection{Thorough augmented Dickey-Fuller test}

If no obvious conclusion can be taken from observations of the data set, then we can resort to a thorough procedure for the augmented Dickey-Fuller test, with three competing models:
\begin{equation}\label{eq:adf1}
\Delta y_t = \beta_0 + \beta_1 t+ \gamma y_{t-1} + \sum_{j=1}^{p-1} \delta_j \Delta y_{t-j} + \epsilon_t
\end{equation}

\begin{equation}\label{eq:adf2}
\Delta y_t = \beta_0 + \gamma y_{t-1} + \sum_{j=1}^{q-1} \delta_j \Delta y_{t-j} + \epsilon_t
\end{equation}

\begin{equation}\label{eq:adf3}
\Delta y_t = \gamma y_{t-1} + \sum_{j=1}^{r-1} \delta_j \Delta y_{t-j} + \epsilon_t
\end{equation}

For each model, the lag order ($p$, $q$ or $r$) can be selected by t-values of information criteria and one should check that with the chosen lag order the residuals have no serial correlation.

Then one starts following \citep{Dolado:1990aa} with the model \ref{eq:adf1} and tests for the null of $\beta_1$ if the null is rejected, the augmented Dickey-Fuller test can be performed with this model. If the null is not rejected, then one move on to model \ref{eq:adf2}. Note that for the null tests we must use the asymptotic distributions tabulated in \citep{Dickey:1981aa}.

With the model  \ref{eq:adf2}, one tests for the null of $\beta_0$, if the null is rejected, the aumented Dickey-Fuller test can be performed with this model. If the null is not rejected, then one move on to model \ref{eq:adf3} and perform the augmented Dickey-Fuller test with this model.

\subsection{Illustration of a spurious regression}
We can define two time series as in \citep{Granger:1974aa}:
$$y_t = y_{t-1} + \epsilon_t$$
$$x_t= x_{t-1} + \nu_t$$
with $\epsilon_t$ and $\nu_t$ two independent white noise from a normal law and we set $y_0=0$ and $x_0=0$. When regressing:
$$y_t = \alpha + \beta x_t + \gamma_t$$
one would expect the $\beta$ not to be significantly different than $0$ and the $R^2$ of the regression to be low. This is not the case as both series present a unit root.

The OLS estimator for \( \beta \) is:
\[ \hat{\beta}_{OLS} = (X'X)^{-1}X'y. \] Given the non-stationary nature of both series the quadratic mean grows with time (see also the formula section \ref{sec:AR1stationary} to get a sense of why this variance tends to infinity with a unit root):
\[
E(y_t^2) \propto T, \quad E(x_t^2) \propto T.
\]
Thus, as \( T \rightarrow \infty \), the variances of both series grow without bound.
The covariance between the two series will also grow with \( T \) due to their shared trending behavior, affecting the \( X'y \) term.
Given the nature of the OLS estimator:
\[ \hat{\beta}_{OLS} = (X'X)^{-1}X'y, \]
the matrices \( X'X \) and \( X'y \) grow to infinity due to the non-stationary nature of \( x_t \) and \( y_t \). This growth can lead to non-convergent behavior of \( \hat{\beta}_{OLS} \) even if the true \( \beta \) is zero (also get back to equation \ref{eq:beta1OLS} to get a sense of this result).

Moreover, the variance of the OLS estimator (given by the diagonal elements of the matrix \( \hat{\sigma}^2 (X'X)^{-1} \), where \( \hat{\sigma}^2 \) is the variance of the error term) can also be impacted by the non-stationarity of the series.

For the detailed of how to compute both the estimator and its variance, cf section \ref{sec:Randtinterpret}, I implement it directly in the code to illustrate how the numerator and/or the denominator can 'explode'.

\citet{phillips1986understanding} demonstrates that he OLS estimator does not converge in probability as the sample size increases, the t- and F-test statistics do not have well-defined asymptotic distributions and the Durbin-Watson statistic converges to zero. The reason is that, with $y_t$ and $x_t$ being I(1) variables, the error term $\epsilon_t$ will also be a nonstationary I(1) variable.

\quad

We illustrate this with an example of two draws and their $R^2$, coefficients, t-values and Durbin-Watson statistics:
\begin{center}
\includegraphics[scale=0.5]{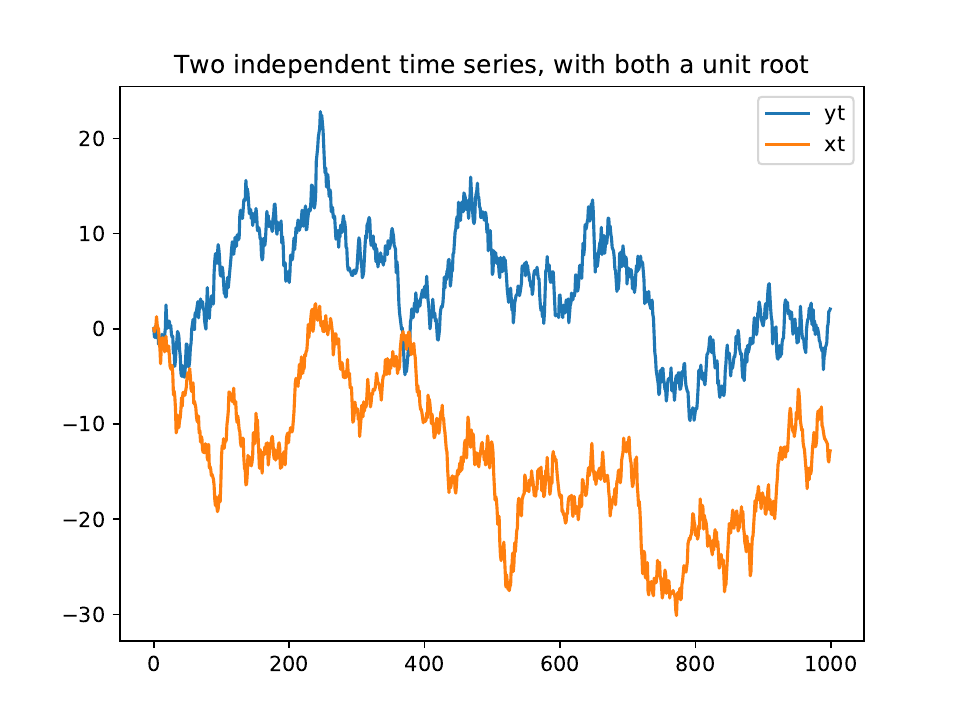}
\end{center}

\begin{tabular}{ll}
$R^2$&$19\%$\\
Intercept coefficient&9.1\\
Intercept coefficient t-value&25.3\\
$\beta$& 0.4\\
$\beta$ t-value&15.2\\
Durbin-Watson statistics&$0.038$\\
\end{tabular}

The $R^2$ seems "attractive"\footnote{and the time series have a correlation of $43\%$}, both the intercept and $\beta$ seems to have high t-value which means that they seems significantly different than $0$, but there is a low Durbin-Watson statistics which suggest that the residuals of the regressions are strongly autocorrelated. These are the symptoms of two I(1) series not cointegrated being regressed on one another, a spurious regression where two independent time series seems related. So in fact the true value of $\beta$ is zero but as demonstrated by Philips (1986) the estimate of $\beta$ doesn't converge in probability to zero (estimates of the regression coefficients are inefficient). Also the usual significance tests on the coefficients are invalid. This condition (high $R^2$ and low Durbin-Watson statistics) is not a necessary condition for a regression to be spurious as demonstrated in \citep{Granger:1974aa}.

\subsubsection{Durbin-Watson test and statistics}\label{sec:durbinwatsontest}

Autocorrelation, specifically serial correlation, refers to the correlation of a variable with itself through different time lags. This phenomenon violates the ordinary least squares (OLS) assumption that the error terms are uncorrelated, leading to inefficiency in estimators, and potentially distorting statistical inference.

The \textbf{Durbin-Watson (DW) test} is a formal statistical test of whether autocorrelation is present in the residuals of a regression analysis. It is commonly used in time-series data to detect the presence of serial correlation.

The \textbf{Durbin-Watson statistic} is calculated using the following formula:
\[
d = \frac{\sum_{t=2}^{T} (\hat{\epsilon}_t - \hat{\epsilon}_{t-1})^2}{\sum_{t=1}^{T} \hat{\epsilon}_t^2}
\]
where:
\begin{itemize}
    \item \(\hat{\epsilon}_t\) is the residual at time \(t\),
    \item \(T\) is the number of observations.
\end{itemize}

The test statistic \(d\) usually lies between 0 and 4, where:
\begin{itemize}
    \item \(d \approx 2\) indicates no autocorrelation,
    \item \(d < 2\) suggests positive autocorrelation,
    \item \(d > 2\) suggests negative autocorrelation.
\end{itemize}

Specific critical values for the DW statistic can be found in statistical tables, which can be used to more formally accept or reject the null hypothesis of no autocorrelation.

\textbf{Spurious regression} often occurs when working with time-series data. It refers to the scenario where two or more time series variables are correlated due to shared trends over time, rather than because there is a genuine relationship between them.

A classic symptom of spurious regression is a very high \(R^2\) value along with significant t-ratios, despite no theoretical or logical connection between the variables. Additionally, residuals from spurious regressions often exhibit patterns, violating the assumption of no autocorrelation and potentially leading to the rejection of the null hypothesis in the Durbin-Watson test even in the absence of a meaningful relationship between the variables.

Identifying and rectifying spurious regressions typically involves:
\begin{itemize}
    \item Investigating the theoretical foundation of the relationship.
    \item Employing de-trending methods or differencing the series to achieve stationarity.
    \item Utilizing cointegration tests when dealing with non-stationary series to ascertain if a long-run relationship exists between the variables.
\end{itemize}

\subsection{AR(2) process, stationarity, and unit roots}

Consider the AR(2)
\begin{equation}\label{eq:AR2}
y_t = \alpha + \beta_1 y_{t-1} + \beta_2 y_{t-2} + \varepsilon_t,
\end{equation}
where $\varepsilon_t$ is white noise (hence stationary).

\paragraph{Lag-polynomial and characteristic roots.}
Using the lag operator $L$, \ref{eq:AR2} becomes
\begin{equation}\label{eq:AR2_lag}
(1-\beta_1 L-\beta_2 L^2)y_t=\alpha+\varepsilon_t.
\end{equation}
Define the \emph{lag roots} $\lambda_1,\lambda_2$ as the solutions to
\begin{equation}\label{eq:lag_roots_poly}
1-\beta_1 \lambda-\beta_2 \lambda^2=0.
\end{equation}
Equivalently,
\[
-\beta_2 \lambda^2-\beta_1 \lambda+1=0
\quad\Longrightarrow\quad
\lambda_{1,2}=\frac{-\beta_1\pm\sqrt{\beta_1^2+4\beta_2}}{2\beta_2}\qquad (\beta_2\neq 0).
\]
Then the factorization is
\begin{equation}\label{eq:factor_lambda}
(1-\beta_1 L-\beta_2 L^2)=\left(1-\frac{L}{\lambda_1}\right)\left(1-\frac{L}{\lambda_2}\right).
\end{equation}

\paragraph{Stationarity condition (and the role of $|\lambda|$).}
A necessary and sufficient condition for (covariance) stationarity of the AR(2) part is
\begin{equation}\label{eq:stationarity_lambda}
|\lambda_1|>1 \quad\text{and}\quad |\lambda_2|>1,
\end{equation}
i.e. all \emph{lag roots} lie \emph{outside} the unit circle. Under \ref{eq:stationarity_lambda}, the AR admits a convergent MA($\infty$) representation (mean-square convergence).

\paragraph{Companion matrix / eigenvalues.}
Let $x_t=(y_t,y_{t-1})'$. Then
\[
x_t=
\begin{bmatrix}\alpha\\0\end{bmatrix}
+
A x_{t-1}
+
\begin{bmatrix}\varepsilon_t\\0\end{bmatrix},
\qquad
A=
\begin{bmatrix}
\beta_1 & \beta_2\\
1 & 0
\end{bmatrix}.
\]
The eigenvalues $r_1,r_2$ of $A$ solve the \emph{characteristic equation}
\begin{equation}\label{eq:char_r}
r^2-\beta_1 r-\beta_2=0.
\end{equation}
Stationarity is equivalently
\begin{equation}\label{eq:stationarity_r}
|r_1|<1 \quad\text{and}\quad |r_2|<1.
\end{equation}
The two parametrizations are consistent because $\lambda_i=1/r_i$: “$|\lambda_i|>1$” (lag roots) $\iff$ “$|r_i|<1$” (eigenvalues).

\paragraph{Unit root condition.}
The process has a unit root at $L=1$ if and only if
\begin{equation}\label{eq:unit_root_sum}
1-\beta_1-\beta_2=0
\quad\Longleftrightarrow\quad
\beta_1+\beta_2=1,
\end{equation}
since $L=1$ is a root of $1-\beta_1 L-\beta_2 L^2$ exactly under \ref{eq:unit_root_sum}. In that case the AR(2) is not stationary (it is $I(1)$ unless there is a second unit root).

\quad

\stepcounter{question}
\begin{itemize}
\item \textbf{Question \arabic{question}.}
Apply a one-time impulse $\varepsilon_0$ and take $\beta_1+\beta_2=1$ (e.g.\ $\beta_1=1.6$, $\beta_2=-0.6$). Verify that the level $y_t$ does \emph{not} revert to a constant mean (non-stationarity) because one root is exactly $\lambda=1$ (equivalently one eigenvalue is $r=1$).
\end{itemize}

\paragraph{Example $\beta_1=1.6$, $\beta_2=-0.6$.}
Here $\beta_1+\beta_2=1$, hence there is a unit root. Indeed the eigenvalues of $A$ solve
$r^2-1.6r+0.6=0$, giving $r_1=1$ and $r_2=0.6$, hence $\lambda_1=1$ and $\lambda_2=1/0.6\approx 1.67$.
Because $|\lambda_1|=1$, condition \ref{eq:stationarity_lambda} fails: the level $y_t$ is not stationary.

\paragraph{First difference under a single unit root.}
If \ref{eq:unit_root_sum} holds but the other root satisfies $|\lambda_2|>1$ (equivalently $|r_2|<1$), then
\[
(1-L)(1-r_2 L)y_t=\alpha+\varepsilon_t
\quad\Longrightarrow\quad
(1-r_2 L)\Delta y_t=\alpha+\varepsilon_t.
\]
Thus $\Delta y_t$ is AR(1) (with drift $\alpha$) and is stationary around a constant mean provided $|r_2|<1$.

\stepcounter{question}
\begin{itemize}
\item \textbf{Question \arabic{question}.}
\begin{itemize}
\item Take $\Delta y_t=y_t-y_{t-1}$ when $\beta_1+\beta_2=1$ and the second root satisfies $|r_2|<1$. Is $\Delta y_t$ stationary?
\item Choose $\beta_1+\beta_2\neq 1$ with $|\lambda_1|>1$ and $|\lambda_2|>1$ (equivalently $|r_1|<1$ and $|r_2|<1$) and verify that $y_t$ is stationary and can be inverted into an MA($\infty$).
\item Choose parameters such that at least one eigenvalue satisfies $|r_i|>1$ (equivalently $|\lambda_i|<1$). Show that the process is explosive.
\item Generate two independent random walks $y_t=y_{t-1}+\varepsilon_t$ and $z_t=z_{t-1}+\gamma_t$ with $\varepsilon_t,\gamma_t\sim\mathcal N(0,1)$, $t=0,\dots,100$. Compare the correlation of levels $(y_t,z_t)$ and of returns $(\Delta y_t,\Delta z_t)$.
\end{itemize}
\end{itemize}

\subsection{Logarithmic Transformations}

Log transformations are often used in finance to stabilize variance and interpret relationships in terms of percentage changes. Key points include:

\begin{itemize}
    \item \textbf{Stabilizing Variance:} Logs compress large values, reducing heteroscedasticity and making data more suitable for linear models.
    \item \textbf{Approximating Returns:} The log difference, \(\Delta \ln Y_t = \ln Y_t - \ln Y_{t-1}\), approximates the percentage change: 
    \[
    \Delta \ln Y_t \approx \frac{Y_t - Y_{t-1}}{Y_{t-1}}
    \]
    for small changes.
    \item \textbf{Interpreting Coefficients:} In log-linear models, coefficients represent elasticities, showing the percentage change in the dependent variable for a 1\% change in the independent variable.
\end{itemize}

Log transformations are especially useful for series with multiplicative relationships or exponential growth but should be avoided if the data contains zeros or negative values.

\section{AR(p) model for French Population Growth}\label{sec:ARpop}

If we believe\footnote{ARp\_French\_population.py} that that the time series of the French population is I(1) and there is no seasonality, we would want to fit an AR(p) process on the French population growth rate.

A first naive approach is to fit an AR(1), we de-trend the growth rate, $\tilde{y}_ t = y_t-\bar{y}$, with $\bar{y}$ the average growth rate observed over the period and then perform an ordinary least squares estimation of the linear regression:
$$\tilde{y}_t = \theta \tilde{y}_{t-1} + \epsilon_t$$

As we are working with a finite sample of size $T$ and exogeneity can be safely assumed for shocks on past population growth rates but the exogeneity assumption doesn't hold for future population growth rate, then the OLS estimator $\hat{\theta}$ of $\theta$ is biased, $E\left(\hat{\theta}\right) - \theta \neq 0$. The demonstration will not be provided here. To put it simply:
$$\hat{\theta} = \frac{\sum_{t=1}^T y_t y_{t-1}}{\sum_{t=1}^T y_{t-1}^2}$$ is unbiased if $E\left[ \epsilon_t\vert y_0, \ldots, y_t, \ldots, y_T\right] = 0$ which for time series is not the case.

But it can be shown that in case $\vert \theta \vert < 1$, then the series is strictly stationary and ergodic and the the OLS estimator is consistent, meaning that as the size of the sample increases (as $T \to \infty$) the estimate  tends to the quantity being estimated.

\textbf{Log-likelihood estimate of an AR(1) if we assume that errors are normally distributed:}

If we assume that $y_t-\theta y_{t-1} \sim \mathcal{N}\left( 0, \sigma^2 \right)$ and 

For the first observation $y_0$, we have a specific estimation of its distribution: $y_0 \sim \mathcal{N}\left(0, \frac{\sigma^2}{1-\theta^2} \right)$ (remember that $Var(y_t) = \theta^2 Var(y_{t-1}) + \sigma^2$).

as seen in section \ref{sec:MLestimation} on maximum-likelihood estimation, the log-likelihood is additive
$$\log\left(L(\theta)\right) = -\frac{T-1}{2}\log(2\pi\sigma^2)-\frac{1}{2\sigma^2}\sum_{t=1}^{T-1}\left(y_t - \theta y_{t-1} \right)^2 -\frac{1}{2}\log(2\pi) -\frac{1}{2}\log\left( \frac{\sigma^2}{1-\theta^2} \right) - \frac{1}{2}\frac{y_0^2}{\frac{\sigma^2}{1-\theta^2}}$$

We implement the log-likelihood estimate for this AR(1) on the French population and compare this estimate with the parameter from the OLS.

We might want here to try manually to fit a AR(3) and look at the significance of the regression coefficients:
$$\tilde{y}_t = \theta_1 \tilde{y}_{t-1} + \theta_2 \tilde{y}_{t-2} + \theta_3 \tilde{y}_{t-3} + \epsilon_t$$

We will see starting section \ref{sec:acf} how to select the order $p$ of the AR, also look at moving average (MA) component and how to fit the model to the data. Methods used are \href{https://math.unice.fr/~frapetti/CorsoP/Chapitre_4_IMEA_1.pdf}{maximum likelihood estimations and method of moments}.

In simple terms, the Partial Autocorrelation Function (PACF) of order $k$ can be estimated as $P_{k}$ in the following AR(k) estimations:
$$\tilde{y}_t = P_{1} \tilde{y}_{t-1}  + \epsilon_{1,t}$$
$$\tilde{y}_t = \theta_{1,2} \tilde{y}_{t-1} + P_{2} \tilde{y}_{t-2}  + \epsilon_{2,t}$$
$$\tilde{y}_t = \theta_{1,3} \tilde{y}_{t-1} + \theta_2 \tilde{y}_{t-2} + P_{3} \tilde{y}_{t-3} + \epsilon_{3,t}$$
$$ \ldots $$

The significance level at $95\%$ for a series with $T$ observations is\footnote{which can be approximated directly with $\pm \frac{2}{\sqrt{T}}$} $\pm \Phi^{-1}\left(\frac{1+0.95}{2}\right) \frac{1}{\sqrt{T}}$ with $\Phi^{-1}$ the inverse of the cumulative distribution function of the normal distribution.

To conclude on this AR models, to chose the order $p$, you can introduce and minimize the Akaike's Information Criterion (AIC) which you compute at each order $p$:
$$AIC_p = - \frac{2}{T} log\left(L(\theta,p)\right) + \frac{2}{T} p $$

or the Schwarz-Bayesian information criterion (BIC):
$$BIC_p =  \log\left(\hat{\sigma}^2(p)\right) + \frac{\log(T)}{T} p $$

with $\hat{\sigma}(p)$ the estimate of the innovation standard deviation from the maximum likelihood of the AR at order $p$. BIC tends to penalize more lag order $log(T)$ for large data set.

We usually strive for model parsimony (limiting the order $p$ in our example), the first reason is the Occam's razor principle where the least assumptions you make the more likely your explanation. The second is the risk of overfitting where if you have $T$ observations, you could reach $p \to T$ and we can mention in that sense \citep{Ledolter:1981aa} that argued that each unnecessary parameter increases the variance of the prediction error by a factor of $\frac{\sigma^2}{T}$.

One might wonder which criteria to use between AIC and BIC, but we are not aware of one that outperform the other in every applications. BIC might lead to model that are too parsimonious, this is why as a practitioner we compare both. We discuss further lag order selection in section \ref{sec:lagselect}.

We use the procedure to fit an ARMA (although restricting MA order $q$ to null for this exercise).

If our assumptions are correct, the residuals from the model should be a white noise. We apply a Ljung-Box test\footnote{see section \ref{sec:ljungbox} for more detail on this test} removing the order $p$ of the AR to the degree of freedom of the chi-square distribution.

\subsection{Are residuals independently distributed? Ljung-Box test}\label{sec:ljungbox}

To test for serial correlation with a maximum lag $h$ in the residual time series returns, we compute the Ljung-Box q-statistic against a $\chi_h^2$ distribution as in \citep{LJUNG:1978aa}:
\begin{equation}
Qstat = T (T+2) \sum_{k=1}^h \frac{\tau_k^2}{T-k}
\end{equation}

with $\tau_k$ the sample autocorrelation at lag $k$.

For residuals of an AR model, we remove the order $p$ of the AR to the degree of freedom of the chi-square distribution.

The $H_0$ of the Ljung-Box test is: no serial correlation at lag max.

With first a visual inspection of the AR(3) model residuals, they don't seem to be iid:
\begin{center}
\includegraphics[scale=0.5]{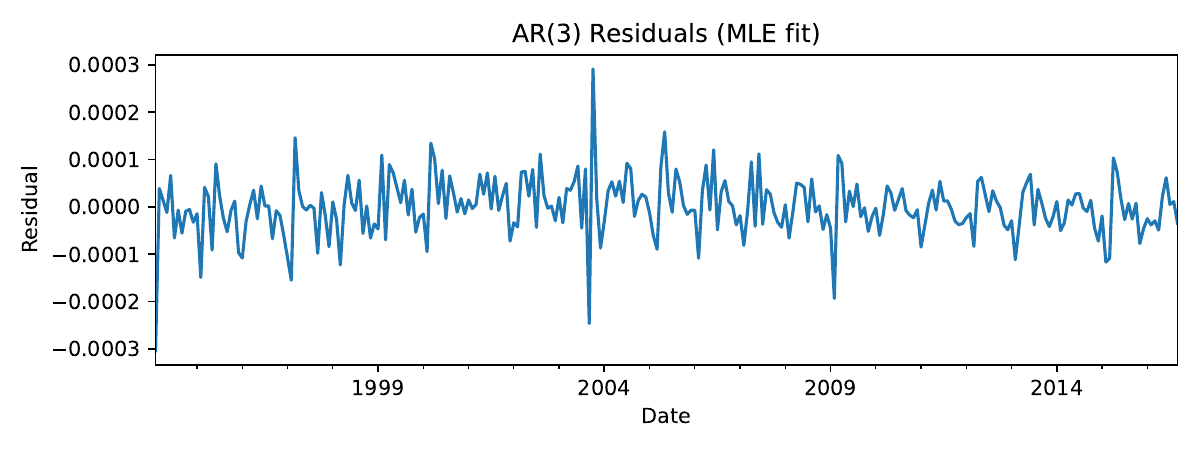}
\end{center}

With this test, we reject the hypothesis that residuals are independently distributed, but we conclude that residuals are rather serially correlated.

Our model is not satisfactory.

In fact, seasonality for birth rate in France is a well documented fact \href{https://www.ined.fr/fichier/s_rubrique/168/pop_soc474.fr.fr.pdf}{as INED showed}, as such the 23rd of September has $5\%$ additional births as a usual days, this day is exactly 256 days after the 31st of December.

\section{R: Treatment of seasonality in our time series}\label{sec:seasonality}

The main question is: how to treat seasonality (in French population, in a price index)?
And especially, should you first treat seasonality and then do the first difference to have a time series stationary. We know that a time series with seasonality cannot be stationary, so we would like to think that we first need to treat seasonality and then make the time series stationary.
But in fact, my take is that you cannot do this sequentially if you want to do it properly from an econometric point of view. This is that same as for the question of the ARMA(p,q)-GARCH(r,s) section \ref{sec:uARMAGARCH}. You have to fit with MLE p, q, r, and s at the same time and not do first an ARMA and then fit a GARCH on the residuals.
This is why in fact I would advise to learn time series until you master SARIMA and then treat a univariate time series that exhibit seasonality with an SARIMA.
My advice would then be: 
\begin{itemize}
\item if you work with only one time series (univariate) and nobody has treated the seasonality before you, then fit an SARIMA to fit the "optimal" seasonailty (S) and integration (I) at the same time and not sequentially. If a Central Bank or any other public institution has treated seasonality (most of the time with X-13), then most likely you won't beat them and you might just use their adjusted time series.
\item if you want to work with multivariate time series (like a VAR), then it is not common to work with seasonality and you want to work with "seasonnaly adjusted" time series. But this is already provided for by the main Central Banks and other public institutions so you don't have to treat seasonality yourself.
\end{itemize}

If the original time series data you work with has trend\footnote{to deal with trend, a first difference or computing the rate of change should do} and/or seasonality, then this series is unlikely to be stationary (or the power of your ADF test can be questioned). You will want to deal with this seasonality and trend before you try to fit a model. But even before this, if your time series display increasing variability, you need to address this.

In this section we follow\footnote{seasonality\_vansteenberghe.R}, the material from the books \textbf{Econometric models and economic forecasts} from Robert Pindyck and Daniel Rubinfeld published in 1998 and \textbf{Analyse des series temporelles} from Regis Bourbonnais and Michel Terraza published in 2010.

\subsection{Taking care of increasing variability}

Starting with the artificial data from \textbf{Analyse des series temporelles} from Regis Bourbonnais and Michel Terraza published in 2010, we observe three problems: increasing variability, trend and seasonality.

We first need to tackle the increasing variability issue, with a Box-Cox transformation, the most famous of which is the $\log()$.

Then we deal with the monthly seasonality with a difference with 12 lag, this also takes care of the trend.

\subsubsection{Box-Cox transformation of a time series}

If the time series present some long term non-linear behaviour (e.g. logarithm, exponential, power), we can use a Box-Cox transformation to revert to a linear process. The transformation can be written:

  \begin{equation}
    y_t=
    \begin{cases}
\frac{x_t^\lambda-1}{\lambda},& \text{if} \quad \lambda \neq 0 \\        ln(x_t), & \text{if} \quad \lambda=0
    \end{cases}
  \end{equation}

The transformation requires $x_t>0$ and nests the untransformed series ($\lambda=1$), the square root ($\lambda=1/2$) and the logarithm ($\lambda=0$). Box and Cox (1964) proposed it with three objectives: normalising the error distribution, stabilising the error variance and linearising the relationship between the dependent variable and its regressors \citep[section 12.5.1]{fox2015}. Conditional on $\lambda$, the model is linear and is estimated by OLS; $\lambda$ is then chosen on a grid by profiling. When the transformation applies to a regressor, this is a nonlinear least squares estimator that minimises the concentrated sum of squared residuals \citep[chapter 23]{hansen2022econometrics}. When it applies to the dependent variable, residual sums of squares are not comparable across values of $\lambda$, and the profile log-likelihood $-\frac{n}{2}\ln\hat{\sigma}^2_\varepsilon(\lambda)+(\lambda-1)\sum_i \ln Y_i$ includes the Jacobian of the transformation. In \citet{fox2015}'s regression of hourly wages on sex, age and education, $\hat{\lambda}=0.09$, with a 95\% confidence interval of $[0.04,0.13]$. Although $\lambda=0$ is rejected, the correlation between $\ln Y$ and $Y^{0.09}$ is 0.9996, and the logarithm is retained for interpretability. A statistically significant departure from $\lambda=0$ need not be economically meaningful.

\citet{proietti2013does} ask whether the transformation improves forecast accuracy. They estimate $\lambda$ nonparametrically as the minimiser of the one-step-ahead prediction error variance of the Jacobian-normalised transformed series, computed in the frequency domain as a bias-corrected geometric average of the periodogram, so no normality assumption is required. Across 530 seasonal monthly series of industrial production and retail turnover, they find four results. First, $\lambda=1$ is rejected for about 20\% of the series, and among these, $\lambda$ differs significantly from 0 in only about 5\% of cases, so the logarithm is the relevant transformation. Second, the transformation significantly improves one-step-ahead forecasts for about a fifth of the series, but the gain fades as the horizon increases. Third, the na\"ive predictor that simply inverts the transformation, which is the conditional median (e.g.\ $\exp(\hat{y}_{t+h|t})$ for $\lambda=0$), generally beats the conditional-mean predictor derived under normality (e.g.\ $\exp(\hat{y}_{t+h|t}+\sigma^2_h/2)$). Fourth, the in-sample interval estimate of $\lambda$ is a reliable guide to which series to transform.

\citet{hernandez1980large} show that the method serves mainly to normalise the distribution, and only approximately. Except in the log-normal case, a power transformation of a positive variable cannot be exactly normal, so the Box-Cox likelihood is misspecified. Its maximum likelihood estimator $\hat{\lambda}$ nevertheless converges almost surely to the value $\lambda^*$ that minimises the Kullback-Leibler divergence between the density of the transformed variable and the closest normal density. It is asymptotically normal, with a sandwich covariance matrix that reduces to the inverse information matrix only under log-normality. For gamma-distributed data, $\lambda^*\approx 1/3$, which recovers the Wilson-Hilferty cube-root transformation. Normality is therefore not guaranteed and must be checked on the transformed data, as in the question below.
  
\stepcounter{question}
\begin{itemize}
\item \textbf{Question \arabic{question}}
\begin{itemize}
\item Plot the distributions before and after the transformations, are we approaching normality?
\item use data(hare) to make some Box-Cox transformation tests on the data from the TSA package. Fit an ARIMA model. Use tsdiag to visualize the autocorrelation of the residuals. Perform a Shapiro test of normality of the residuals.
\end{itemize}
\end{itemize}

\subsection{Data set: coffee price in Brazil}
We focus on the coffee world prices with 240 observation for Brazil downloadable \href{http://www.economicswebinstitute.org/data/coffeeprices.zip}{here}.

We get the following Buys-Ballot table, it is not obvious to observe month or year effect:

\resizebox{\textwidth}{!}{
\begin{tabular}{rrrrrrrrrrrrr}
  \hline
 & Jan & Feb & Mar & Apr & May & Jun & Jul & Aug & Sep & Oct & Nov & Dec \\ 
  \hline
1982 & 56.26 & 56.72 & 57.44 & 60.23 & 60.00 & 63.28 & 58.88 & 57.27 & 55.13 & 55.92 & 57.94 & 62.46 \\ 
  1983 & 63.09 & 59.01 & 48.97 & 46.38 & 42.09 & 40.65 & 37.55 & 35.80 & 36.56 & 35.41 & 36.03 & 40.15 \\ 
  1984 & 41.17 & 40.49 & 41.12 & 42.48 & 43.51 & 49.41 & 53.22 & 52.55 & 54.67 & 55.59 & 57.10 & 60.33 \\ 
  1985 & 77.57 & 84.01 & 82.12 & 74.81 & 67.78 & 57.45 & 54.01 & 54.52 & 51.80 & 72.74 & 96.06 & 162.50 \\ 
  1986 & 190.75 & 172.90 & 171.23 & 151.56 & 152.11 & 140.93 & 137.73 & 152.28 & 156.67 & 141.40 & 140.31 & 160.99 \\ 
  1987 & 84.73 & 69.50 & 58.99 & 58.97 & 54.73 & 38.60 & 33.64 & 33.45 & 33.89 & 36.36 & 35.48 & 33.26 \\ 
  1988 & 36.35 & 43.80 & 44.07 & 43.65 & 39.88 & 48.53 & 47.50 & 46.49 & 50.15 & 50.32 & 53.82 & 70.18 \\ 
  1989 & 69.11 & 59.33 & 56.55 & 75.15 & 83.31 & 75.61 & 49.14 & 38.77 & 38.30 & 35.54 & 44.28 & 43.47 \\ 
  1990 & 50.53 & 60.05 & 66.03 & 62.57 & 58.95 & 54.64 & 49.14 & 59.09 & 54.78 & 46.88 & 39.76 & 40.49 \\ 
  1991 & 39.04 & 42.49 & 48.32 & 48.19 & 45.15 & 42.74 & 40.34 & 44.84 & 43.44 & 49.40 & 42.07 & 41.89 \\ 
  1992 & 47.61 & 42.92 & 43.12 & 40.47 & 35.38 & 35.34 & 39.19 & 32.86 & 35.45 & 61.01 & 66.50 & 62.54 \\ 
  1993 & 57.82 & 53.52 & 51.35 & 41.57 & 45.60 & 45.13 & 52.78 & 59.89 & 59.66 & 58.99 & 59.38 & 48.85 \\ 
  1994 & 59.93 & 63.78 & 67.31 & 74.43 & 102.85 & 112.34 & 165.95 & 156.90 & 158.67 & 148.94 & 139.68 & 124.06 \\ 
  1995 & 134.50 & 128.51 & 136.03 & 130.56 & 128.92 & 115.18 & 107.84 & 111.57 & 102.12 & 99.68 & 99.52 & 91.85 \\ 
  1996 & 102.93 & 118.56 & 114.21 & 105.28 & 112.58 & 97.04 & 78.20 & 79.25 & 86.91 & 86.47 & 86.84 & 87.88 \\ 
  1997 & 114.21 & 139.42 & 155.65 & 157.13 & 173.06 & 158.17 & 130.80 & 143.47 & 150.61 & 136.31 & 135.39 & 150.72 \\ 
  1998 & 162.11 & 159.49 & 133.33 & 125.89 & 105.37 & 88.36 & 81.05 & 84.71 & 77.42 & 76.65 & 79.15 & 89.47 \\ 
  1999 & 85.62 & 71.28 & 73.73 & 71.40 & 80.81 & 77.10 & 67.20 & 61.89 & 57.76 & 63.10 & 82.20 & 96.71 \\ 
  2000 & 89.44 & 80.15 & 80.57 & 74.68 & 72.52 & 63.80 & 60.91 & 56.36 & 55.26 & 56.34 & 53.32 & 48.02 \\ 
  2001 & 47.41 & 44.04 & 44.30 & 39.30 & 41.87 & 39.04 & 35.05 & 33.15 & 29.79 & 26.19 & 30.97 & 31.34 \\ 
\end{tabular}
}

\subsection{Influence test of factors}

We have a data set with:
\begin{itemize}
\item $N$ year
\item $p$ month (12)
\item $x_{ij}$ the price for a year $i$ at a month $j$
\end{itemize}

We can define the following "means":
$$x_{i.}=\frac{1}{p}\sum_{j=1}^p x_{ij}$$
$$x_{.j}=\frac{1}{N}\sum_{i=1}^N x_{ij}$$
$$x_{..}=\frac{1}{N p} \sum_{j=1}^p \sum_{i=1}^N x_{ij}$$

The following sum of squares:
$$S_p=N \sum_{j=1}^p (x_{.j}-x_{..})^2$$
$$S_A=p \sum_{i=1}^N (x_{i.}-x_{..})^2$$
$$S_r=N \sum_{j=1}^p (x_{ij}-x_{i.}-x_{.j}+x_{..})^2$$

The following variances:
$$V_p=\frac{S_p}{p-1}$$
$$V_A=\frac{S_A}{N-1}$$
$$V_r=\frac{S_r}{(p-1)(N-1)}$$

\quad

We cant compare the Fisher statistics:
if the empirical Fisher statistics $$F_c = \frac{V_p}{V_r}$$ is greater than the Fisher statistics $F_{(p-1) , (N-1)(p-1)}^{0.05}$ then we can reject $H_0$ and the series is seasonal.
if the empirical Fisher statistics $$F_c = \frac{V_A}{V_r}$$ is greater than the Fisher statistics $F_{(N-1) , (N-1)(p-1)}^{0.05}$ then we can reject $H_0$ and the series has a trend.

\subsection{Montyly dummy}

We have monthly data, for each observation we create a dummy, which will be equal to $1$ when the month of the observation is that given month, $0$ otherwise. Testing for seasonality is testing for the month significance. None can be find here.

\subsection{Data set: rainfall in Australia}
We use a data set from the \href{http://www.bom.gov.au/climate/change/index.shtml#tabs=Tracker&tracker=timeseries}{Autralian Governement - Bureau of Meteorology} which offers time series on monthly rainfall data in Australia:
\begin{center}
\includegraphics[scale=0.7]{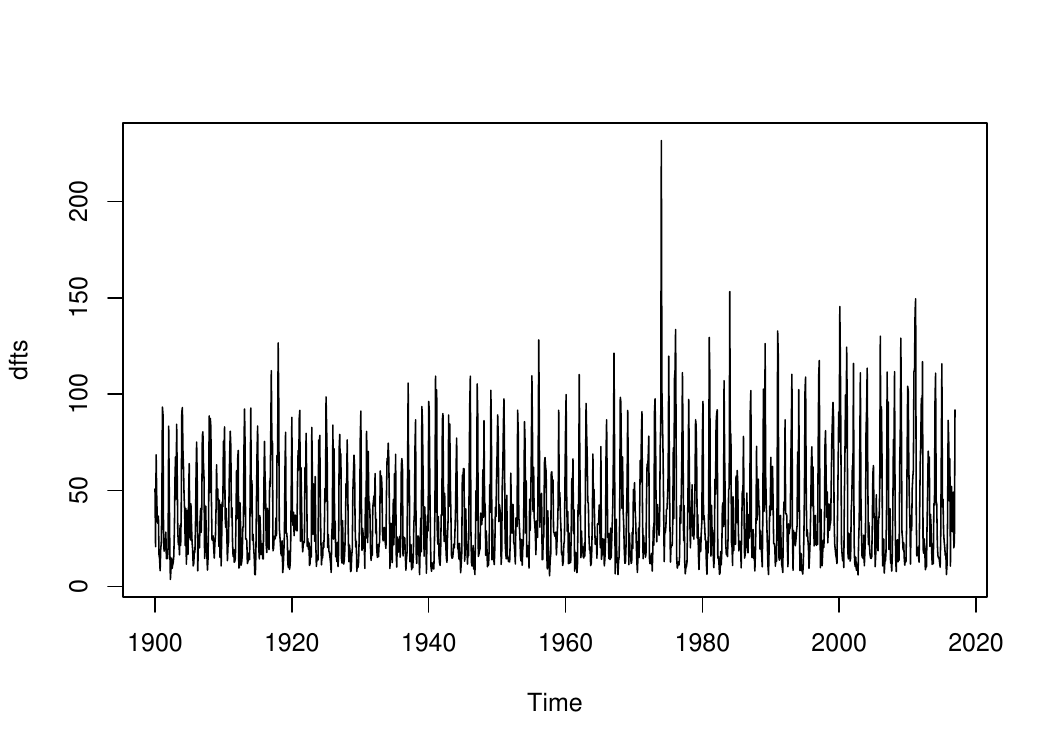}
\end{center}

\subsection{Decomposition principle, multiplicative}

When the seasonal component seems visually on the plot to depend on the level, we apply a multiplicative decomposition.

\textbf{Multiplicative seasonal adjustment} rely on the fact that each time series $y_t$ can be decomposed as:
\begin{equation}\label{eq:decomp}
y_t= L_t \times S_t \times C_t \times I_t
\end{equation}
with
\begin{itemize}
\item $L$ the long-term secular trend in series
\item $S$ the seasonal component
\item $C$ the long-term cyclical component
\item $I$ the irregular component
\end{itemize}

The aim is to decompose the time series, isolate and combine $L_t \times C_t$ (estimate it) then we can deduct an estimate of $S_t \times I_t$.

If we have monthly data, we estimate:
\begin{equation}
\hat{L}_t \times \hat{C}_t=\frac{1}{12} \left(y_{t+6}+\ldots+y_{t-5}\right)
\end{equation}

Then:
\begin{equation}
\hat{S}_t \times \hat{I}_t = \frac{y_t}{\hat{L}_t \times \hat{C}_t}
\end{equation}
Then $I$ has to be eliminated, we average each $\hat{S}_t \times \hat{I}_t $ element is average per month ($\tilde{z}_1$ for January, $\tilde{z}_2$ February, etc.). Those terms should sum to 12. If they do not, we define $\bar{z}_i=\tilde{z}_i \frac{12}{\sum_j \tilde{z}_j}$.

Finally, the seasonally adjusted series are:
\begin{itemize}
\item $y_1^a=\frac{y_1}{\bar{z}_1}$
\item $\ldots$
\item $y_{12}^a=\frac{y_{12}}{\bar{z}_{12}}$
\end{itemize}

\subsubsection{Decomposition of rain in Australia}

For rain, as the seasonal component should not vary with time nor the level of rain, we tend to use the simplest additive model $Y_t = T_t + S_t + \epsilon_t$ that we can rewrite as $Y_t= L_t + S_t + C_t + I_t$ where $T_t= L_t +  C_t$. To 'deseasonalize', we want to work with $Y_t - S_t$.

In order to separate the long-term secular trend from the long-term cyclical component, we can apply a Hodrick-Prescott filter on $T_t$, for monthly data we use $\lambda=129 600$.

\subsubsection{X-13-ARIMA}
R should allow the use of the United Sates Census Bureau seasonal adjustment program, \href{https://www.census.gov/srd/www/x13as/}{X-13ARIMA-SEATS}.

\section{R: Introduction to SARIMA}

We use\footnote{SARIMA\_vansteenberghe.R} the data from the astsa package.

The dataset, graciously provided by Dr. Roy Mendelssohn of the Pacific Environmental Fisheries Group (personal communication), is part of the \texttt{astsa} package and spans 453 months from 1950 to 1987. It includes:
\begin{itemize}
\item \texttt{soi}: Southern Oscillation Index representing changes in air pressure in the central Pacific Ocean.
\item \texttt{rec}: Recruitment, or the number of new fish.
\end{itemize}

The objective is to fit an AR(2) model that accounts for seasonality and compare its performance against models that disregard seasonality, like a simple ARMA(3,1):

\begin{center}
\includegraphics[scale=0.5]{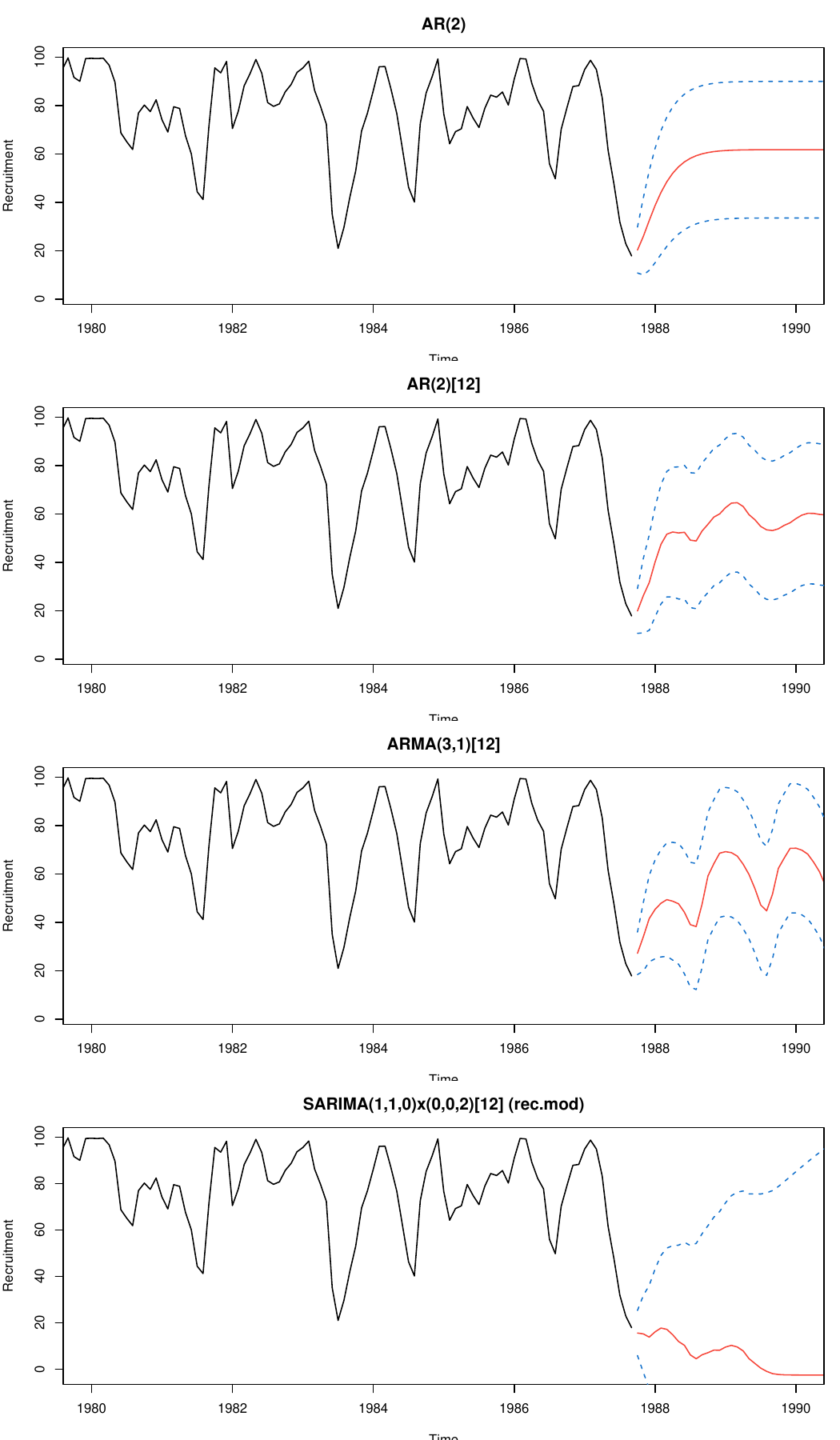}
\end{center}

\subsection{SARIMA}

The SARIMA model, a multiplicative extension of the traditional ARIMA, accounts for both non-seasonal and seasonal components in the data. Formally, a SARIMA model is denoted as ARIMA$(p,d,q) \times (P,D,Q)_s$, and its mathematical representation is:

\begin{equation*}
(1- \phi_1 B - \ldots - \phi_p B^p)(1- \Phi_1 B^s - \ldots - \Phi_P B^{sP}) y_t = (1+ \theta_1 B + \ldots + \theta_q B^q)(1+\Theta_1 B^s +\ldots + \Theta_Q B^{sQ}) \epsilon_t
\end{equation*}

To determine the optimal values for the parameters, it's common to inspect the autocorrelation function (ACF) and the partial autocorrelation function (PACF)\footnote{\href{https://www.otexts.org/fpp/8/9}{Rob J Hyndman and George Athanasopoulos}}. The seasonal components of an AR or MA model manifest as lags in the PACF and ACF:

\begin{itemize}
\item An ARIMA$(0,0,0)(0,0,1)_{12}$ model will show a spike at lag 12 in the ACF but no other significant spikes, with the PACF exhibiting exponential decay at seasonal lags: 12, 24, 36, and so on.
\item Conversely, an ARIMA$(0,0,0)(1,0,0)_{12}$ model will display exponential decay in the seasonal lags of the ACF, with a single significant spike at lag 12 in the PACF.
\end{itemize}

SARIMA models, given their ability to account for seasonal fluctuations, offer a robust approach to time series forecasting. By applying SARIMA to fisheries data, we aim to glean insights that are both economically significant and environmentally informative.

\stepcounter{question}
\begin{itemize}
\item \textbf{Question \arabic{question}}
\begin{itemize}
\item find the "optimal" values for p,d,q and P,D,Q and s of the SARIMA
\item use the AIC or loglikelyhood criterion
\end{itemize}
\end{itemize}

\subsection{Fractionally integrated ARMA}

Varves are melt of glacier deposit. With the package asta, we have varves data in New England for 634 years. We can observe from the original series that the thicker the deposited amount, the higher the volatility, a logarithmic transform can correct this nonstationarity in the variance.

\stepcounter{question}
\begin{itemize}
\item \textbf{Question \arabic{question}}
\begin{itemize}
\item suggest a test of the varve, the differenced varve and log(varve) time series for normality of their distribution.
\item Explain which transformation can be modelled with an MA(1), apply it.
\end{itemize}
\end{itemize}

\subsubsection{Finding $\theta$}

For this section, we use the detailed Gauss-Newton method presented in the book \citep{ngai}.

An MA(1) can be written as: $y_t = \epsilon_t - \theta \epsilon_{t-1}$. If $\theta < 1$, then we can write: $\epsilon_n = y_n + \theta y_{n-1} + \ldots + \theta^{n-1} y_1$. Note that in this case, we have $n$ observations of our process $y$. We can search $\theta$ that minimizes: $S(\theta) = \sum_{t=1}^n \epsilon_t^2$. 

We can compute this iteratively as we know that:
$$\epsilon_1 = y_1$$
$$\epsilon_2 = y_2 + \theta \epsilon_1$$
$$\ldots$$
$$\epsilon_n = y_n + \theta \epsilon_{n-1}$$

We find that our $\theta$ lies between $-0.8$ and $-0.6$:
\begin{center}
\includegraphics[scale=0.5]{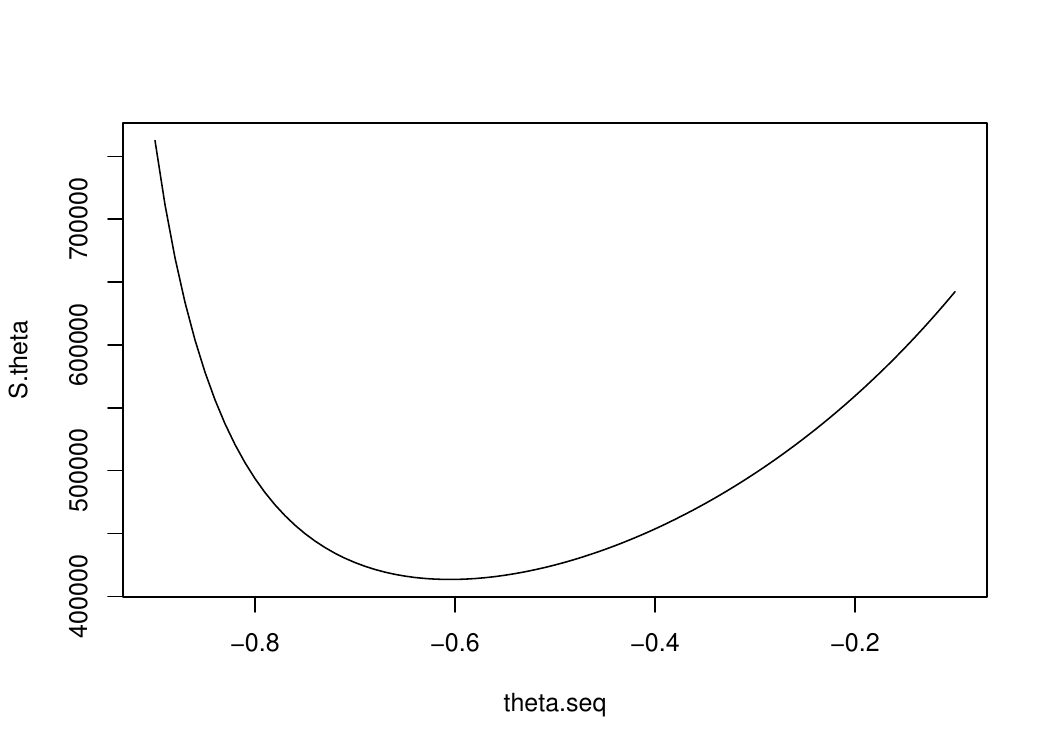}
\end{center}

\subsubsection{Residuals}

\stepcounter{question}
\begin{itemize}
\item \textbf{Question \arabic{question}}
\begin{itemize}
\item Do you find autocorrelation for the residuals of the MA(1)? What do you suggest?
\end{itemize}
\end{itemize}

\subsection{Long term memory}

As observed in the ACF of the varves time series, there is a long memory in the process. If we want to avoid first differencing our data, we can revert to a fractional differenced series: $-.5 < d < .5$ 

\begin{center}
\includegraphics[scale=0.5]{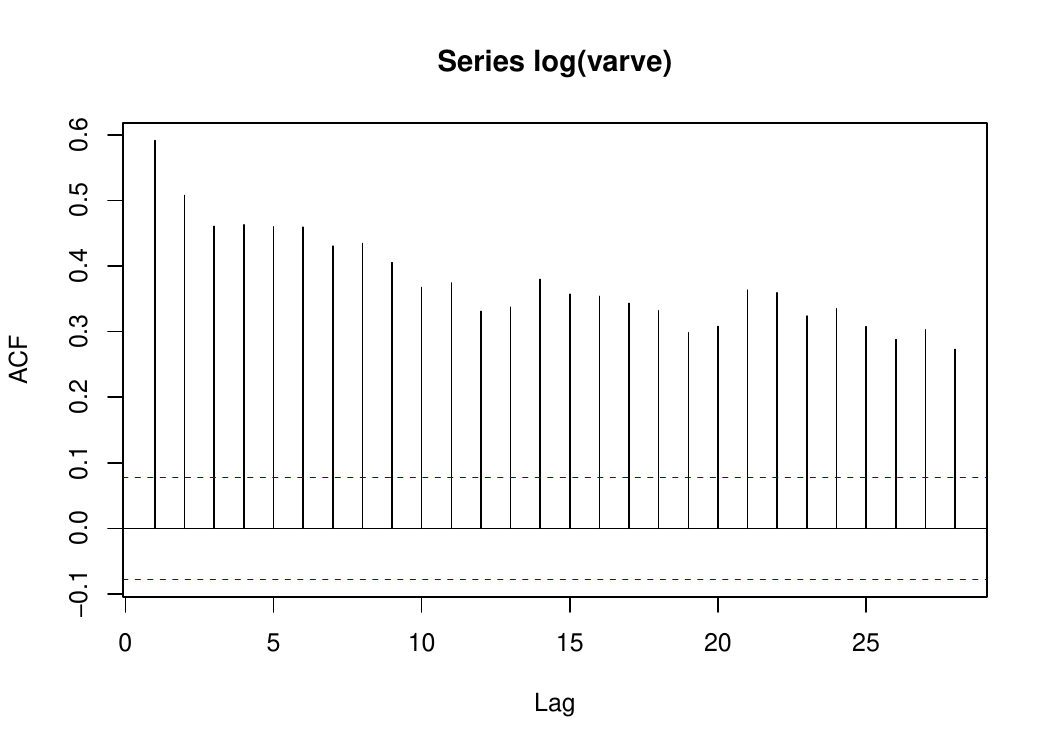}
\end{center}

Using the fracdiff function and fitting it on the varve data, we find a value of $d=0.38$

\stepcounter{question}
\begin{itemize}
\item \textbf{Question \arabic{question}}
\begin{itemize}
\item Forecast using both the ARIMA and the fractionally integrated ARMA.
\end{itemize}
\end{itemize}

\section{Modeling and Analysis of French Population Data Using SARIMA}

In this section\footnote{pandas\_exercise\_part\_2\_seasonality\_French\_population.py}, we focus on analyzing the monthly population data of France obtained from INSEE (Institut National de la Statistique et des Études Économiques). The dataset, identified as "Démographie - Population au début du mois - France métropolitaine" (ID: 000436387), provides a valuable insight into the demographic changes over time. Our objective is to investigate the presence of seasonality in this time series and apply appropriate econometric techniques to model this phenomenon.

\textbf{Preliminary Data Analysis:}
The initial step involves loading and preprocessing the data using Python's pandas library. We ensure that the time series is indexed by the date, facilitating easier analysis. A preliminary visual inspection is conducted to identify any noticeable trends or seasonal patterns. 

\textbf{Seasonality in Population Data:}
Given the nature of demographic data, it is reasonable to hypothesize the presence of seasonal effects. For instance, certain months might exhibit higher population growth rates due to factors like birth rates, migration patterns, etc. 

There are four main methods to deal with monthly seasonality in growth rates:
\begin{itemize}
\item \textbf{Regression with Monthly Dummy Variables}
\item \textbf{Seasonal Differencing}: subtract the value of a data point in a given month from the value in the same month in the previous year;
\item \textbf{Seasonal Decomposition of Time Series by Loess} (STL), which allows us to decompose the series into trend, seasonal, and residual components. Here you'll need to use the option\footnote{In STL decomposition, a seasonal window of 13 for monthly data is used instead of 12 to ensure symmetric LOESS smoothing around each data point. This odd-numbered window spans the full yearly cycle while providing robustness against noise and anomalies in the time series.}
 \textit{seasonal=13};
 \begin{itemize}
 \item if you have strong belief that the seasonality is stable, you can use the simpler additive decomposition: in the \texttt{seasonal\_decompose} model, the observed time series \( Y[t] \) at time \( t \) is decomposed into three components: the trend component \( T[t] \), the seasonal component \( S[t] \), and the residual component \( e[t] \), such that \( Y[t] = T[t] + S[t] + e[t] \).
 \end{itemize}
\item \textbf{SARIMA Models}: Seasonal AutoRegressive Integrated Moving Average.
\end{itemize}

\subsection{Regression Analysis with Dummy Variables}

To assess the presence of seasonal effects, we initially employ a regression model incorporating dummy variables for each month. This method allows us to isolate the impact of each month on the population growth rate.

\begin{itemize}
    \item \textbf{Creating Dummy Variables:}
    We transform the month information into dummy variables. This is achieved by encoding each month as a binary variable, where the presence of a specific month is indicated by 1, and absence by 0.

    \begin{verbatim}
    dx['month'] = dx.index.month
    dx_dummies = pd.get_dummies(dx['month'], drop_first=True, prefix='m')
    dx_with_dummies = pd.concat([dx, dx_dummies], axis=1)
    \end{verbatim}

    \item \textbf{Regression Model for September Growth Rate:}
    We then fit a linear regression model with the population growth rate as the dependent variable and the dummy variable for September as the independent variable.

    \begin{verbatim}
    model = smf.ols('pop ~ m_9', data=dx_with_dummies).fit()
    \end{verbatim}

    The coefficient of m\_9 represents the difference in growth rate for the month of September. The statistical significance of this coefficient is determined via a t-test, as indicated by the p-value.
\end{itemize}

To extend our analysis, we conduct a comparative regression analysis considering each month as a categorical variable. Here, September serves as the reference category.

\begin{itemize}
    \item \textbf{Regression with Categorical Months:}
    This model assesses if any month exhibits a statistically significant difference in growth rate compared to September.

    \begin{verbatim}
    model_categorical = smf.ols('pop ~ C(month, Treatment(reference=9))', 
                                data=dx_with_dummies).fit()
    \end{verbatim}

    In this model, each month is treated as a separate category, with September as the baseline. The coefficients and their respective p-values indicate whether the growth rates for other months are significantly different from September.
\end{itemize}

\subsubsection{Interpretation of Results}

Upon examining the results from the regression models, we observe that:

\begin{itemize}
    \item The initial model focusing solely on September provides insights into how this specific month differs from the average.
    \item The extended model with all months treated categorically reveals a more nuanced picture of how each month compares to September.
    \item The p-values associated with the coefficients for each month are crucial for understanding statistical significance. A low p-value (typically less than 0.05) suggests that the growth rate for that month is significantly different from September.
\end{itemize}

\subsection{SARIMA}
\textbf{Stationarity and Unit Root Testing:}
Before proceeding with the SARIMA model, it is crucial to ensure that the time series is stationary. We employ the Augmented Dickey-Fuller (ADF) test under different specifications (with constant, with constant and trend, and without both) to test for unit roots. The presence of unit roots would imply non-stationarity, necessitating differencing or transformation of the series.

\textbf{SARIMA Model Specification:}
The SARIMA model, which stands for Seasonal AutoRegressive Integrated Moving Average, is a natural choice for modeling time series data with seasonality. The model is specified as SARIMA(p, d, q)(P, D, Q, S), where:

\begin{itemize}
    \item \textbf{p, d, q} - Represent the autoregressive order, degree of differencing, and moving average order of the non-seasonal component.
    \item \textbf{P, D, Q} - Represent the autoregressive order, degree of differencing, and moving average order of the seasonal component.
    \item \textbf{S} - Is the length of the seasonal cycle (12 in the case of monthly data).
\end{itemize}

\textbf{Parameter Selection and Model Tuning:}
Selecting the optimal parameters for SARIMA is done through a systematic grid search approach. We iterate over a range of values for each parameter and select the combination that minimizes an information criterion, typically the Akaike Information Criterion (AIC). This process helps in identifying a model that balances goodness-of-fit with parsimony.

The optimal SARIMA model identified for the monthly population growth rate data is SARIMA(0, 1, 0)x(1, 0, 1, 12). The model can be expressed as follows:

\begin{equation}
    Y_t' = \alpha Y_{t-12}' + \epsilon_t + \theta \epsilon_{t-12}
\end{equation}

In this equation, \( Y_t' \) denotes the first-order differenced time series at time \( t \). The model includes a seasonal autoregressive term (\( \alpha Y_{t-12}' \)) and a seasonal moving average term (\( \theta \epsilon_{t-12} \)), both of which operate on a 12-month lag, reflecting the annual seasonal pattern in the monthly data. 
\begin{itemize}
    \item The coefficient \(\alpha\) for the seasonal AR term (denoted as \texttt{ar.S.L12} in the output) is 0.8452. This value represents the influence of the population growth rate from the same month in the previous year on the current month.
    \item The coefficient \(\theta\) for the seasonal MA term (represented as \texttt{ma.S.L12}) is -0.7361. It reflects how the previous year's error influences the current month's error term.
    \item Additionally, the model's residual variance, denoted as \(\sigma^2\) (sigma2 in the output), is approximately \(1.877 \times 10^{-9}\). This value provides an estimate of the variance in the prediction errors of the model. It's a measure of how much the observed values deviate from the values predicted by the model
\end{itemize}

\textbf{Diagnostic Tests for SARIMA Model Residuals}

\begin{itemize}
    \item \textbf{Visual Inspection of Residuals:}
    Plotting the residuals of the model provides a quick check of any obvious patterns. Ideally, residuals should appear as white noise, meaning they should be randomly scattered around zero without any discernible pattern.
    
    \item \textbf{Autocorrelation Function (ACF) and Partial Autocorrelation Function (PACF) Plots:}
    These plots are used to detect any autocorrelation in the residuals of the model. In the ACF and PACF plots, the absence of significant spikes suggests that there is no autocorrelation, indicating that the model has captured the time series dynamics effectively.
    
    \item \textbf{Ljung-Box Test:}
    This statistical test checks for the absence of autocorrelation in the residuals at multiple lag lengths. A high p-value (typically above 0.05) in this test suggests that the residuals are independently distributed, indicating a good fit of the model.
\end{itemize}

\subsubsection{Projection and comparison with the actual data}

We start by forecasting future population growth rates using our fitted SARIMA model. The model generates predictions for the growth rate from March to October 2020. This forecast is based on the underlying patterns and seasonal cycles identified in the historical data.

\begin{verbatim}
forecast_results = best_results.get_prediction(start=pd.to_datetime('2020-03-01'), 
                                               end=pd.to_datetime('2020-10-01'), 
                                               dynamic=True)
forecast = forecast_results.predicted_mean
\end{verbatim}

We then plot these forecasted growth rates alongside the actual growth rates observed during the same period to visually assess the model's accuracy.

To understand the impact of these growth rates on actual population numbers, we convert the predicted growth rates into estimated population levels. This is done by first adjusting the forecasted growth rates and then cumulatively multiplying these rates with the last known population level:

\begin{enumerate}
    \item \textbf{Adjusting the Forecasted Growth Rates:}
    The forecasted growth rates (\(f_t\)) are first adjusted to represent the proportional change. This is done by adding 1 to each forecasted growth rate. The adjusted forecast (\(f_t^{adj}\)) is given by the equation:
    \begin{equation}
        f_t^{adj} = 1 + f_t
    \end{equation}
    where \(f_t\) is the forecasted growth rate at time \(t\).

    \item \textbf{Cumulative Growth Factor:}
    We then compute the cumulative product of these adjusted forecasts to obtain the cumulative growth factor (\(G_t\)). This factor represents the compounded growth over time and is calculated as follows:
    \begin{equation}
        G_t = \prod_{i=1}^{t} f_i^{adj}
    \end{equation}
    where \(\prod\) denotes the cumulative product up to time \(t\).

    \item \textbf{Projected Population Level:}
    The projected population level (\(P_t\)) at any time \(t\) is then calculated by multiplying the last known actual population level (\(P_{last}\)) by the cumulative growth factor. The equation is:
    \begin{equation}
        P_t = P_{last} \times G_t
    \end{equation}
    where \(P_{last}\) is the last known population level before the forecast period begins.
\end{enumerate}

Similar adjustments are applied to the forecast's confidence intervals to estimate the range of possible future population levels.

Finally, we plot the projected population levels against the actual population levels observed from March to October 2020. This comparison provides a clear visual representation of the model's performance in predicting not just the rate of population growth but also the actual population size, or can help us understand the impact of the Covid-19 crisis.

\begin{center}
\includegraphics[scale=.5]{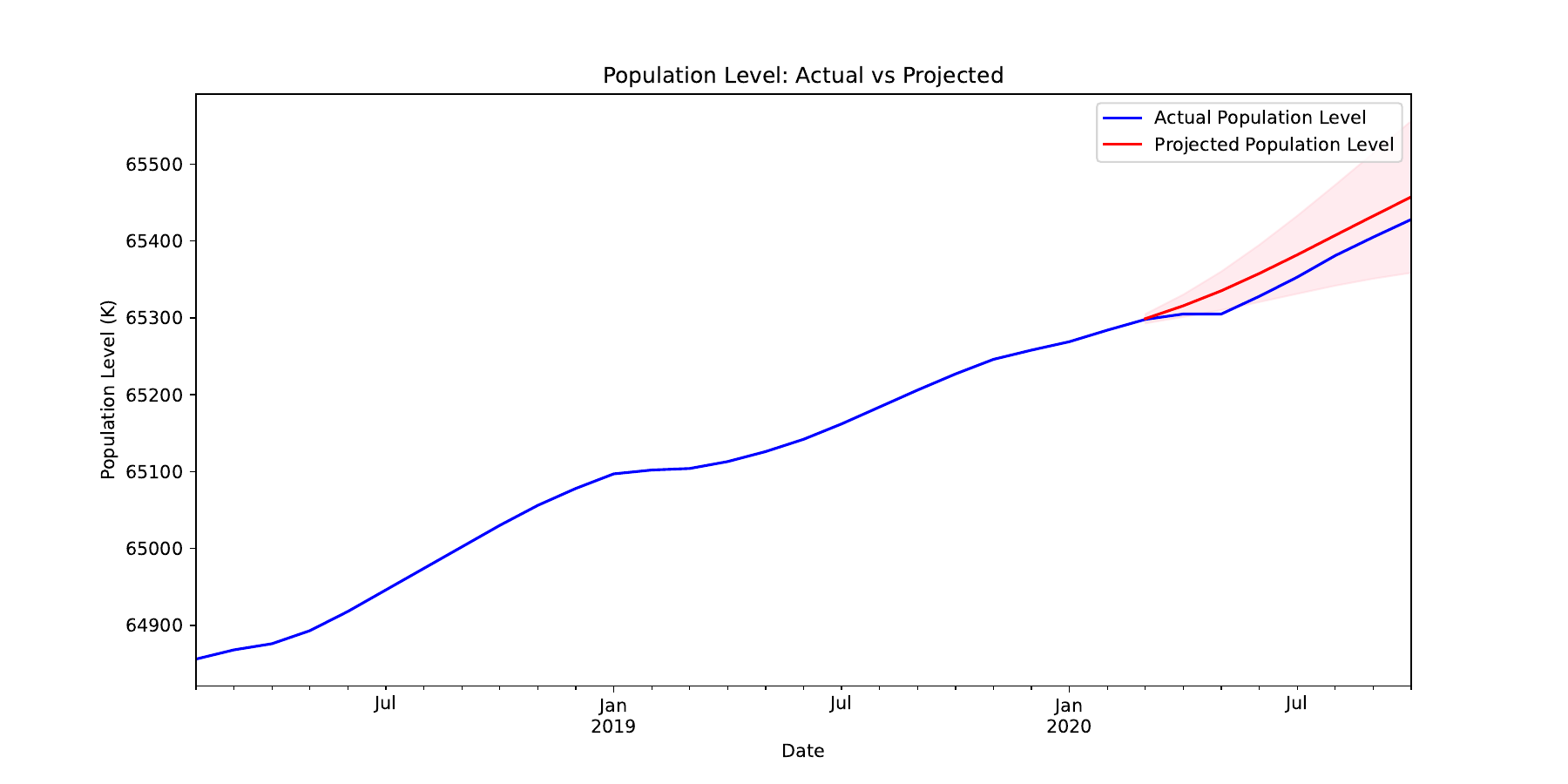}
\end{center}

\section{Inflation persistence exercise}

Inflation\footnote{\texttt{vansteenberghe\_inflation\_persistence.py}} is one of the key variables in monetary economics, both for policy design and for forecasting. A central question concerns its \emph{persistence}, that is, the long-run effect of a temporary shock to inflation: how strongly and for how long does a 1\% increase in inflation today affect future inflation rates, and how quickly does inflation return to its previous level, if at all?

In this exercise, we study inflation persistence using monthly data from the Eurostat \emph{Harmonised Index of Consumer Prices} (HICP), excluding tobacco. Following Chapter~8.8 of \citet{verbeek} and the empirical approach in \citet{pivetta2007persistence}, inflation dynamics are modeled using autoregressive processes.

The objective is threefold.
\begin{enumerate}
\item First, we assess the degree of persistence by testing for a unit root using the Augmented Dickey--Fuller (ADF) test, and challenge this conclusion with a KPSS test (see below for details):
\begin{itemize}
\item In \citet{verbeek}, inflation is treated as an $I(0)$ process. This assumption is not derived mechanically from unit root testing alone, but reflects a modelling choice consistent with earlier contributions such as \citet{fuhrer1995inflation}. Conditional on stationarity, persistence is then analysed through a systematic exploration of finite-order ARMA specifications rather than through differencing.
\item This choice contrasts with alternative views in the literature, most notably \citet{stock2007has}, who argue that inflation is better approximated by an $I(1)$ process.
\end{itemize}
\item Second, we estimate an AR$(p)$ models and compare them using information criteria (AIC and BIC) and residual diagnostics (Ljung--Box test):
\begin{itemize}
\item Check how \citet{verbeek} systematically explores several autoregressive specifications, none of which is entirely satisfactory, reflecting the tension between theoretical parsimony and the empirical properties of inflation data. Conditional on stationarity, the analysis starts from the ACF and PACF, leading to an AR(3) baseline model, motivated by the PACF becoming insignificant beyond lag~3. Residual diagnostics then motivate successive extensions—additional autoregressive lags, the inclusion of a moving-average term, and higher-order AR models. Model selection relies jointly on information criteria (AIC versus BIC) and Ljung--Box portmanteau tests, ultimately yielding two defensible specifications (AR(3) and a reduced AR(6)), which deliver similar conclusions regarding inflation persistence.
\end{itemize}
\item Third, we quantify persistence in economic terms by computing the sum of autoregressive coefficients
\[
\text{SARC} = \sum_{j=1}^{p} \theta_j,
\]
which governs the cumulative response of inflation to a one-off shock,
\[
\frac{1}{1-\text{SARC}},
\]
and the corresponding half-life of a shock,
\[
h = \frac{\log(0.5)}{\log(\text{SARC})}.
\]
\end{enumerate}

\begin{quote}\small
\textbf{Intuition and derivation.}
Inflation follows an AR($p$) process
\[
\pi_t = \mu + \theta_1 \pi_{t-1} + \dots + \theta_p \pi_{t-p} + \varepsilon_t .
\]
\emph{Cumulative response.} Let $\psi_h = \partial \pi_{t+h}/\partial \varepsilon_t$ be the impulse
response at horizon $h$ (with $\psi_0 = 1$). Summing the AR equation over all horizons gives
$\sum_h \psi_h = 1 + \text{SARC}\sum_h \psi_h$, so
\[
\sum_{h=0}^{\infty} \psi_h = \frac{1}{1-\text{SARC}} .
\]
A one-unit shock therefore raises inflation by a total of $1/(1-\text{SARC})$ units over time.
For example, $\text{SARC}=0.8$ gives a long-run multiplier of $5$. The same factor links the
intercept to the mean: $\mathbb{E}[\pi_t] = \mu/(1-\text{SARC})$.

\emph{Half-life.} In an AR(1) with $\rho = \text{SARC}$, the effect after $h$ periods is $\rho^h$.
The half-life solves $\rho^h = 0.5$:
\[
h \log \rho = \log 0.5 \;\Longrightarrow\; h = \frac{\log 0.5}{\log \rho}.
\]
For example, $\text{SARC} = 0.8 \Rightarrow h \approx 3.1$ periods, and
$\text{SARC} = 0.9 \Rightarrow h \approx 6.6$ periods. The base of the logarithm does not matter.
When $p>1$ the impulse response is not exactly geometric, so this $h$ is an AR(1)-equivalent
\emph{approximation}.

\emph{Valid range.} The formulas need $0 < \text{SARC} < 1$ (a stationary process).
As $\text{SARC} \to 1$, both $1/(1-\text{SARC})$ and $h$ go to infinity: a shock never dies
out (unit root).
\end{quote}

\paragraph{Data source.}
The inflation series used in this exercise comes from \emph{Eurostat} and corresponds to the Harmonised Index of Consumer Prices (HICP) for the European Union (EU aggregate).
We focus on the headline index (all items), expressed as an index with base year 2015.

\paragraph{Seasonal adjustment.}
Monthly inflation indices exhibit strong and recurrent seasonal patterns (e.g.\ sales periods, energy price seasonality).
To isolate medium- and long-run movements relevant for persistence analysis, we remove the seasonal component using a standard seasonal decomposition.

We rely on a \emph{multiplicative} decomposition rather than an additive one.
For price indices, seasonal fluctuations tend to scale with the level of the series: a given percentage seasonal effect corresponds to a larger absolute change when the index level is higher.
The multiplicative specification is therefore more appropriate and yields a seasonally adjusted series obtained by dividing the observed index by its estimated seasonal component.

\paragraph{About X-13ARIMA-SEATS.}
As a robustness check, one could use the \texttt{X-13ARIMA-SEATS} procedure, the reference seasonal adjustment method employed by statistical agencies.
X-13 combines ARIMA-based pre-adjustment, outlier detection, trading-day corrections, and signal extraction.
While more powerful and flexible, it requires an external installation and is unnecessarily complex for a first exposure to seasonal adjustment.
For pedagogical purposes, the simple decomposition approach is sufficient and fully transparent.

\paragraph{Unit root testing: ADF vs.\ KPSS.}
Following \citet{verbeek}, we rely primarily on the Augmented Dickey--Fuller (ADF) test,
whose null hypothesis is a unit root.
For inflation series, ADF results are often marginal and sensitive to the number of lags,
so sharp conclusions are rarely robust.
The KPSS test, which reverses the null (stationarity), typically leads to similarly
ambiguous conclusions due to its dependence on bandwidth choices.

\paragraph{KPSS test and interpretation.}
To circumvent the low power of standard unit root tests, \citet{kwiatkowski1992testing} propose a test
in which (trend-)stationarity is the null hypothesis and the existence of a unit root is the alternative.
This test, known as the KPSS test, decomposes the observed series as
\[
y_t = d_t + r_t + \varepsilon_t,
\]
where $d_t$ is a deterministic component (intercept or intercept and linear trend),
$r_t$ is a random walk component, and $\varepsilon_t$ is a stationary error term,
which is not required to be white noise.
Under the null hypothesis of (trend-)stationarity, the variance of the random walk innovation is zero,
so that $r_t$ is degenerate.

The KPSS test is a Lagrange multiplier test.
Its computation proceeds as follows.
First, an auxiliary regression of $y_t$ on $d_t$ is estimated by OLS and the residuals $\hat e_t$ are saved.
Next, partial sums $S_t = \sum_{s=1}^t \hat e_s$ are formed.
The test statistic is then given by
\[
\mathrm{KPSS} = \frac{1}{T^2} \sum_{t=1}^T \frac{S_t^2}{\hat\sigma^2},
\]
where $\hat\sigma^2$ is an estimator of the long-run variance
$\sigma^2 = \sum_{j=-\infty}^{\infty} \mathbb{E}(\varepsilon_t \varepsilon_{t-j})$.

Estimation of $\hat\sigma^2$ requires the choice of a kernel (weighting scheme) and a bandwidth.
Common choices include Bartlett weights, as in \citet{kwiatkowski1992testing}, and the quadratic spectral kernel
\citep{andrews1991heteroskedasticity}.
In practice, the KPSS test is highly sensitive to these choices, particularly in finite samples,
and tends to over-reject stationarity when persistence is high.

Because the ADF and KPSS tests place opposite hypotheses in the null and both suffer from low power
and tuning-parameter sensitivity in persistent macroeconomic series,
they are best interpreted as complementary diagnostic tools rather than decisive tests.
Accordingly, in this course, inflation persistence is quantified directly using autoregressive models,
through the sum of AR coefficients and the implied half-life of shocks.

\subsection{Univariate AR benchmarks for inflation - academic literature}

\paragraph{AR models in Stock and Watson (2007).}
In \citet{stock2007has}, the central univariate benchmark for inflation forecasting is a finite-order autoregressive model, denoted \textbf{AR(AIC)}.
\textbf{Inflation is modeled in first differences} to allow for a unit root in levels, and the lag order is selected recursively by the Akaike Information Criterion (AIC).
This AR(AIC) specification serves as the baseline against which all alternative models are evaluated.

\paragraph{Method and lag order.}
Operationally, the AR(AIC) model is estimated as a direct multi-step regression of future
average inflation on lagged changes in inflation, with the number of lags chosen by minimizing AIC
at each forecast origin.
This approach allows for flexible short-run dynamics while remaining fully univariate and
transparent.
In the postwar U.S.\ data, Stock and Watson show that low-order AR representations provided a good
approximation during the Great Inflation, but their forecasting performance deteriorated after the
mid-1980s.

\subsection{AR model of inflation}

\citet{verbeek} works on a "simple" AR model, citing the literature (older) \begin{quote}
To continue our analysis, we shall assume that inflation is I(0), as is done in \citet{fuhrer1995inflation}.
\end{quote}

\paragraph{AIC vs.\ BIC.}
Both the Akaike Information Criterion (AIC) and the Bayesian Information Criterion (BIC) trade off model fit against parsimony. AIC tends to select richer models, while BIC applies a stronger penalty that increases with sample size and favors more parsimonious specifications.

\paragraph{Practice in this exercise.}
Here, both AIC and BIC select an AR(6) specification, which strengthens confidence in the chosen lag order.  
When criteria diverge, AIC is usually favored for inflation dynamics, while BIC provides a useful robustness benchmark.

\subsection{Inflation persistence}

Following Chapter 8.8 in \citet{verbeek} and \citet{pivetta2007persistence}, measure the inflation persistence. Use Eurostat Harmonised Index of Consumer Prices (HICP) excluding tobacco.

\stepcounter{question}
\begin{itemize}
\item \textbf{Question \arabic{question}}
\begin{itemize}
\item Apply the ADF test.
\item As in \citet{verbeek}, consider AR(p) for $p \in [1,6]$, create a table with AIC, BIC and Ljung-Box test on residuals.
\item Compute the sum of the autoregressive coefficients (SARC) as \[
SARC = \sum_{j=1}^{p} \theta_j
\] the cumulative effect of a shock to inflation is measured by \[
\frac{1}{1-SARC}
\] and the half-life of a shock as:
\[
h = \frac{\log(0.5)}{\log\left( SARC \right)}
\]
\end{itemize}
\end{itemize}

\subsection{Forecast accuracy}

An $ARMA(p, q)$ model can be transformed into an $MA(\infty)$ model to facilitate the computation of forecast errors. The $ARMA(p, q)$ model is written as:

\[
y_t = \sum_{j=0}^{\infty} \alpha_j \varepsilon_{t-j} \quad \text{with} \quad \alpha_0 = 1
\]

where \( \alpha_j \) are the coefficients to be determined from the model. The \( h \)-period-ahead forecast, in terms of the errors \( \varepsilon_t \), is given by:

\[
y_{T+h\vert T} = E\left[y_{T+h}\vert y_T, y_{T-1}, \dots \right] = \sum_{j=h}^{\infty} \alpha_j \varepsilon_{T+h-j}
\]

The variance of the forecast error can be expressed as:

\[
E \left[(y_{T+h} - y_{T+h\vert T})^2\right] = \sigma^2 \sum_{j=0}^{h-1} \alpha_j^2
\]

This formula allows us to compute the variance of the forecast errors for any \( h \)-period-ahead forecast. The variance increases with the number of forecast steps, reflecting the increased uncertainty over time. Using the forecast error variance, we can construct 95\% confidence intervals for the forecasts. For a one-step-ahead forecast, the interval is given by:

\[
y_{T+1|T} \pm 1.96 \sqrt{c_1^2}
\]

where \( c_1^2 = \sigma^2 \sum_{j=0}^{0} \alpha_j^2 = \sigma^2 \). For \( h \)-period-ahead forecasts, the confidence interval is given by:

\[
y_{T+h|T} \pm 1.96 \sqrt{c_h^2}
\]

where

\[
c_h^2 = \sigma^2 \sum_{j=0}^{h-1} \alpha_j^2
\]

The wider the confidence interval, the greater the uncertainty in the forecast due to the accumulation of forecast errors as \( h \) increases.

\subsubsection{Example: AR(1) Model}

For an AR(1) model, we have \( \alpha_j = \theta^j \). The expected quadratic forecast errors are given by:

\[
c_1^2 = \sigma^2, \quad c_2^2 = \sigma^2 (1 + \theta^2), \quad c_3^2 = \sigma^2 (1 + \theta^2 + \theta^4), \dots
\]

For \( h \to \infty \), the forecast error variance converges to:

\[
c_{\infty}^2 = \frac{\sigma^2}{1 - \theta^2}
\]

This represents the unconditional variance of \( y_t \), showing how forecast uncertainty increases over time but stabilizes in the long run for stationary processes.

\subsection{A model for the French population}

\subsubsection{Plot some previsions}

As an illustration of in-sample "precision" of our model, we plot the normal distribution of the residuals and the shaded area equals $95\%$, we plot similarly the boundaries on the time series plot forecast:

\begin{center}
\includegraphics[scale=0.5]{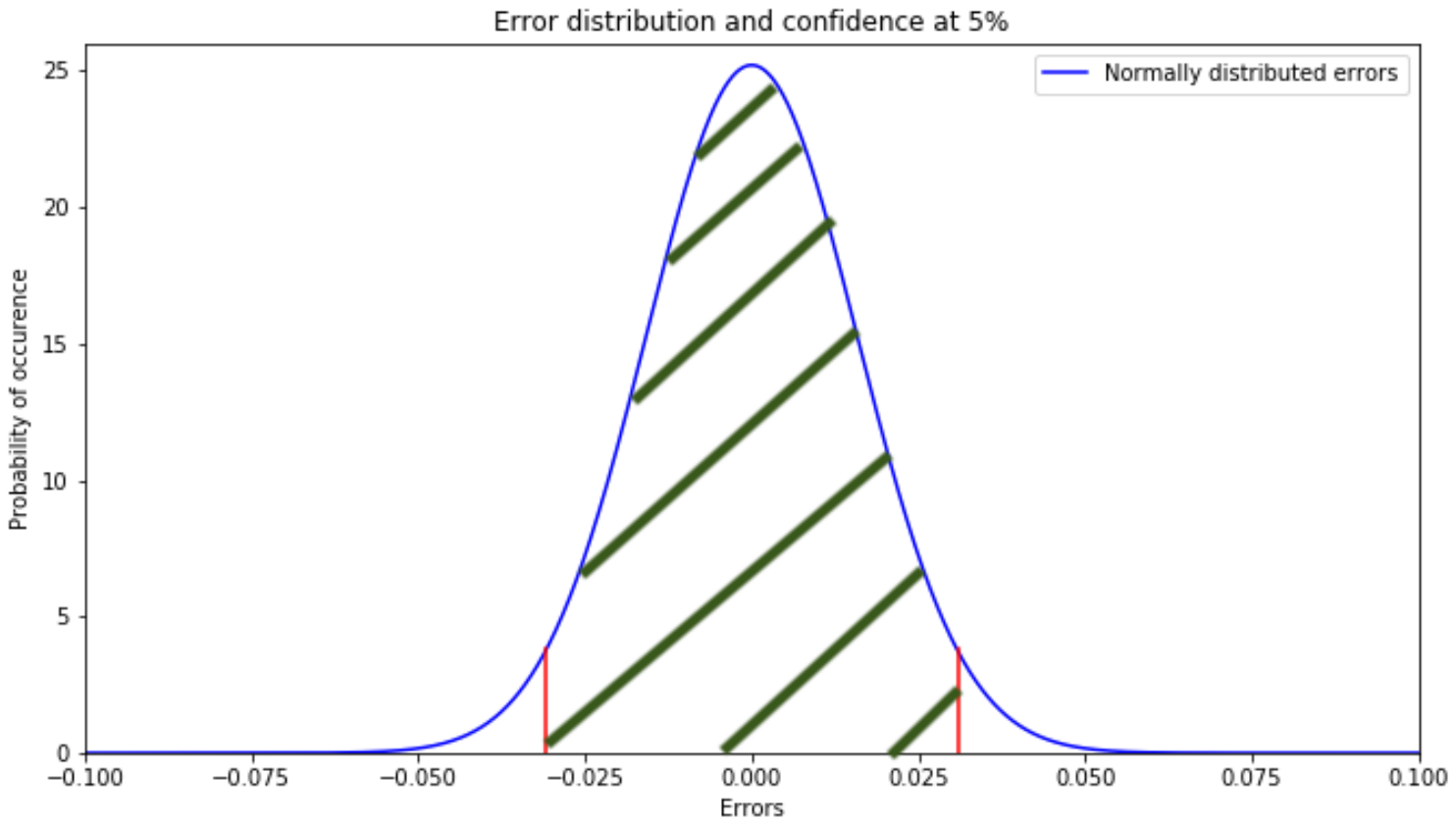}
\end{center}

For a $5\%$ two-sided confidence interval we take a t distribution with n degrees of freedom (which we approximate by a normal distribution as $n >> 120$): $Z_{\alpha/2}=1.96$. 

\stepcounter{question}
\begin{itemize}
\item \textbf{Question \arabic{question}}
\begin{itemize}
\item Work first on the monthly seasonality, then compute a monthly average population change and use it to project the population up to 2020.
\end{itemize}
\end{itemize}

\subsection{Barnard 1963}

Following \citet{barnard1963}, plot the data of airline passengers and follow the Box-Jenkins method and the adaptive forecasting method.

\section{Financial returns distributions and forecast}\label{sec:ARIMA}

\subsection{CAC 40 index data: mean and two-sided t-test, histogram}

We fetch\footnote{code: ARIMA\_vansteenberghe.py} CAC 40 index historical data from 1995-1-16 to 2017-1-16. We compute the daily return on business days only. 

\subsection{Modelling and forecasting the CAC 40 daily return}
One of the simplest question we could be interested in when it comes to financial time series is: "what is its value most likely to be tomorrow?"

If we assume (wrongly) that the daily returns are iid and follow a Gaussian DGP, then we only need to look at the histogram of the observed daily returns and can follow the logic presented section \ref{sec:empiricalhisto}, we model the CAC 40 return as a white noise section \ref{sec:whitenoise}.

But if we call $y_t = \frac{P_t}{P_{t-1}}-1$ the daily returns, we can expect that the realization of $y_t$ will depend on the realization of $y_{t-k} , k\in\mathbb{N}$.

One would take the historical data available for this time series, in our example, we have 5 638 observations from 1995-1-16 to 2017-1-16. We could choose the parameters based on the first $60\%$ of our data set (train our model) and then verify with the remaining $40\%$ observation that our model performs well (test our model). We might also want to cross-validate it, \href{http://scikit-learn.org/stable/modules/cross_validation.html}{more on this topic can be found here}.

\subsection{White Noise: model and tests}\label{sec:whitenoise}

A white noise is a is a sequence of independent and identically distributed random variables $y_t$ with finite mean and variance. Most of the time we take the assumption that $y_t$ is normally distribution

We use the simple model:
\begin{equation*}
y_t=\mu + \epsilon_t
\end{equation*}

with $\epsilon_t \sim \mathcal{N}(0,\sigma^2)$ and as we are working on modelling and forecasting the CAC 40 returns: $\mu$ is the mean of the daily returns.


As defined in section \ref{sec:stationarity}, this series is by definition stationary and as expected, when we perform an augmented Dickey-Fuller test: the series comes out as stationary.

\subsubsection{Trend stationarity}

We modify our model with a trend:
\begin{equation*}
y_t=\mu + \delta t + \epsilon_t
\end{equation*}
with $\epsilon_t \sim \mathcal{N}(0,\sigma^2)$.

We take $\delta=0.1\%$.

We perform an augmented Dickey-Fuller test taking the trend into account: the series is stationary.

If we are interested in the significance of the trend component of this test, we might have to do it by hand. Let's consider:

\begin{equation*}
\Delta y_t = \beta_0 + \beta_1 t + \gamma y_{t-1} +  \delta \Delta y_{t-1} + \epsilon_t
\end{equation*}

We find that in this case, the trend component is significant.

\quad

We might want to remove the trend. We can start with a first difference.

\quad

We also study quadratic trend model.


\subsection{Random Walk}

If we think that the CAC 40 returns are following a random walk, then:
\begin{equation*}
y_t=y_{t-1}+\epsilon_t
\end{equation*}

s.t.
$$ \epsilon \sim \mathcal{N}(0,\sigma^{2})$$

\quad

Here it is just that we use the previous business day's return to predict today's return: $E(y_t) = y_{t-1}$

This brings us to some of the measurements of forecasting errors:

\subsection{Forecast error measures}

\begin{itemize}
\item Mean Forecast Error: average the differences between the forecast and the actual data
$$ \mbox{Mean Forecast Error} = \frac{1}{N} \sum_{i=1}^{N} (\hat{y_t} - y_t)$$
\item Mean Absolute Forecast Error
$$ \mbox{Mean Absolute Forecast Error} = \frac{1}{N} \sum_{i=1}^{N} \mid \hat{y_t} - y_t \mid$$
\item Root Mean Square Forecast Error
$$ \mbox{Root Mean Square Forecast Error} = \sqrt{ \frac{1}{N} \sum_{i=1}^{N} ( \hat{y_t} - y_t )^2 }$$
\end{itemize}

\subsection{Auto Regressive model}\label{sec:ARmodel}
We follow section \ref{sec:AR1stationary} that introduced AR(1) models.

Before starting, intuitively, an AR(1) model would mean that $y_t$ and $y_{t-1}$ related, therefore we can test for serial correlation of our daily returns, e.g. with a Ljung-Box test introduced section \ref{sec:ljungbox}.

There are two reasons to ditch the CAC 40 returns for an AR(1) model (no serial correlation, our series looks more like a withe noise) and the AR(1) parameters are not significant. For a further study of ARCH effect in CAC 40 returns, see section \ref{sec:ARCHtest}.

We introduce a stationary AR(1) on time series of better returns candidates and for stationary, we check that: $0<\vert \theta \vert < 1$:
\begin{equation*}
y_t= c + \theta y_{t-1} + \epsilon_t
\end{equation*}
with $\epsilon_t \sim \mathcal{N}(0,\sigma^2)$, thus by definition (or assumption in our model if we apply it to a real time series) we assume that $y_t$ and $\epsilon_t$ are uncorrelated.

We can search for the optimal value of $\phi$, once we re-ordered the AR(1), detrending the $y_t$ time series to get rid of the intercept in the AR(1):
\begin{equation*}
y_t - \mu = \phi (y_{t-1} - \mu) +\epsilon_t
\end{equation*}

If $\phi<1$ we can iterate backward :
\begin{equation*}
y_t - \mu = \sum_{j=0}^\infty \phi^j \epsilon_{t-j}
\end{equation*}

If we note $x_t=y_t-\mu$, $E_t(x_t) = \phi x_{t-1}$. That is why AR(1) is usually a benchmark model against which one assess the relevance of a more complicated model. If a complex model cannot beat an AR(1) forecast, then we might just stick to an AR(1) model to base our predictions on.


\begin{equation*}
\gamma(h) = cov(x_{t+h},x_t) = \frac{\sigma^2 \phi^h}{1-\phi^2}
\end{equation*}

Hence the Autocorrelation Function (ACF) of an AR(1) is: $\rho(h)=\frac{\gamma(h)}{\gamma(0)}=\phi^h$, $h\geq 0$

If $\vert \phi \vert > 1$ the AR(1) process is explosive.

\stepcounter{question}
\begin{itemize}
\item \textbf{Question \arabic{question}}
\begin{itemize}
\item as defined in section \ref{sec:stationarity}, show numerically that this process is weakly stationary for a certain range of values for $\phi$.
\end{itemize}
\end{itemize}

\subsubsection{AR(p)}

The definition of an AR(p) model is simply:
\begin{equation*}
y_t= \phi_1 y_{t-1} + \ldots +  \phi_p y_{t-p} + \epsilon_t
\end{equation*}

\subsection{Moving Average model}\label{sec:MAmodel}
MA(1) model:
\begin{equation*}
y_t =  \epsilon_t + \theta \epsilon_{t-1}
\end{equation*}

with
$$E(y_t)=\mu$$
$$ \epsilon \sim \mathcal{N}(\mu,\sigma^{2})$$

If we note: $\gamma(h) = cov(y_{t+h},y_t)$
	
\begin{equation}
    \gamma(h)=
    \begin{cases}
(1+\theta^2)\sigma^2,& \text{if} \quad h=0 \\        \theta \sigma^2, & \text{if} \quad h=1 \\ 0, & \text{if} \quad h>1
    \end{cases}
\end{equation}
Hence the ACF of an MA(1) is:
\begin{equation}
    \begin{cases}
\frac{\theta}{1+\theta^2},& \text{if} \quad h=1 \\       0, & \text{if} \quad h>1 \\ 
    \end{cases}
\end{equation}

We can demonstrate that there is a problem of non-uniqueness of a MA(1) model (take $\theta$ or $\frac{1}{\theta}$ and an adjusted value of $\sigma$ for example).

\subsubsection{MA(q)}

The definition of an MA model of order $q$ is simply:
\begin{equation*}
y_t =  \epsilon_t + \theta_1 \epsilon_{t-1} + \ldots + \theta_q \epsilon_{t-q}
\end{equation*}

\subsection{ACF and PACF}\label{sec:acf}

The autocorrelation function (ACF) of a stationary time series is:
$$\rho_k = corr(y_t,y_{t+k})$$

We call it a partial autocorrelation function (PACF) of a stationary process when we filter out the effect of the random variables $y_{t+1}, \ldots , y_{t+k-1}$. The PACF is used to identify the order ($p$) of an autoregressive model, i.e. AR($p$).

\begin{equation}
\Phi_{k}=corr(y_{t+k+1}-\hat{y}_{t+k+1},y_{t+1}-\hat{y}_{t+1})
\end{equation}
and
\begin{equation}
\Phi_{1}=\rho_1
\end{equation}
with $ \hat{y}_{t+k+1}$ (respectively $ \hat{y}_{t+1}$) being the orthogonal projection of $y_{t+k+1}$ (respectively $y_{t+1}$) on $y_{t+1}, \ldots , y_{t+k}$.

We can use the ACF and the PACF as visual tools to identify respectively the order of the Moving Average and the Auto Regressive one:
\begin{itemize}
\item if for a time series the auto-correlation are non significant $\forall h> q$, we choose $q$ as the order of the MA component of our model.
\item if for a time series the partial auto-correlation are non significant $\forall h> p$, we choose $p$ as the order of the AR component of our model.
\end{itemize}

\subsection{ARMA model}
A process $y_t$ is an ARMA(p,q) if it is stationary and:
\begin{equation*}
y_t=\phi_1 y_{t-1}+\ldots+\phi_p y_{t-p}+\epsilon_t+\theta_1 \epsilon_{t-1}+\ldots+\theta_q \epsilon_{t-q}
\end{equation*}
with $\phi_p\neq 0$ and $\theta_q \neq 0$

We can follow the Box-Jenkins methodology:
\begin{enumerate}
\item identify the order of the model $p$ and $q$, this can be done with the help of ACF and PACF following section \ref{sec:acf}. We should also investigate for seasonality, see section \ref{sec:seasonality}.
\item estimate the model with the maximum likelihood method\footnote{by default, the ARIMA funcion maximizes the conditional sum of squares likelihood}.
\item model residuals diagnosis, e.g. with Breusch-Godfrey test described in section \ref{sec:breusch}.
\end{enumerate}

\quad

For our time series, we reject $H_0$ for the Breusch-Godfrey test and residuals seems to be autocorrelated, our model is not acceptable and we would need to search for a better model.

\quad

Topics to further explore: parameters redundancy, causality, invertibility

\subsubsection{Selection of the orders $p$ and $q$}

\citet{cairns2000discussion}:

\begin{quote}
the simplicity postulate (often known also as the \textbf{principle of parsimony}). It is well known that people prefer simple models or explanations to complex ones. Jeffreys states "that the simplest law is chosen because it is most likely to give correct predictions". Here the simplest law refers to the striking of a balance between consistency of a model with the observed data and the prior degree of belief in a model. Jeffreys' view was that the prior degree of belief in a model should be inversely related to the number of parameters.
\end{quote}

For the orders selection, section \ref{sec:lagselect} also provide some guidelines (in the VAR context). 

\subsection{ARIMA}

Most time series can be decomposed in a nonstationary trend component and a zero-mean stationary composant:
$$y_t=\mu_t +x_t$$
such that
$$\mu_t=\beta_0 +\beta_1 t$$

$x_t$ is stationary thus $\Delta y_t = \beta_1 + \Delta x_t$ is stationary.

\quad

A process $x_t$ is said to be ARIMA(p,d,q) if $\Delta^d x_t$ is ARMA(p,q).

\subsection{R: Box-Jenkins methodology for the ARIMA}

We follow\footnote{20190625\_Box\_Jenkins\_methodo.R} the Box-Jenkins methodology to fit an ARIMA model as introdued in \citep{box1970}.

\subsubsection{Identification}

\begin{enumerate}
\item To identify the order of the ARIMA(p,d,q), we plot the ACF and PACF.
\item We difference the time series until the ACF looks like one of a stationary process.
\item We use the inverse autocorrelation function to check for over-differencing.
\begin{itemize}
\item Following the \href{https://dms.umontreal.ca/~duchesne/chap7.pdf}{details here}, we compute the dual model with the Yule-Walker method. The ACF of this model is called the inverse autocorrelation function, if this latter decays slowly it is likely we over-differenced the original time series.
\end{itemize}
\end{enumerate}

\subsubsection{Parameters estimation}

Once we identified the orders p,d and q of the ARIMA, we estimate the parameters, e.g. using AIC criterion and we can simulate our model. Again, here we are not looking for a perfect reproduction of the historical time series, rather we want to check that our model can replicate the behavior of the original time series.

\subsubsection{Checking the model}

Once estimated, we check the error terms using the Ljung-Box\footnote{see section \ref{sec:ljungbox} for more detail on this test} test to test the null hypothesis: $H_0$: the error terms are independent.

\stepcounter{question}
\begin{itemize}
\item \textbf{Question \arabic{question}}
\begin{itemize}
\item Us the long history of U.S. GDP available \href{https://www.measuringworth.com/datasets/usgdp/}{here}, apply the tests and then an ARIMA model. Discuss the performance of your model.
\end{itemize}
\end{itemize}

\subsection{Main take away and the need for ARCH}

We want to forecast the the returns and for this explore a DGP that would be 'behind' the observations. If the observations were iid, then $E\left(y_t\vert y_{t-1} \right) = E\left(y_t \right) = \bar{y}$. Here this is not the case and a simple AR(1) model beat a forecast only based on the mean. We next need to explore the residuals of this AR(1) model and check if they are iid with a 'well behaved' DGP.

If we observe the residuals $\epsilon_t$ of an ARMA(1,1) model of the daily returns, we observe as in \citet{mandelbrot1963}  writes it: \begin{quote}
large changes tend to be followed by large changes --of either sign-- and small changes by small changes
\end{quote} and following \citet{Black1976aa} volatility tends to rise with "bad" news and fall with "good" news:
\begin{center}
\includegraphics[scale=0.5]{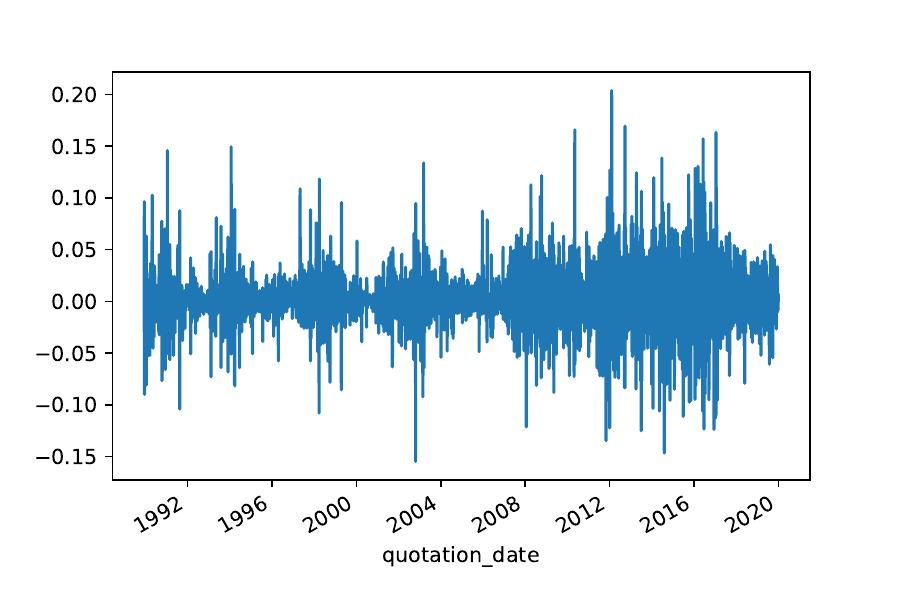}
\end{center}

To account for those properties in our residuals, we follow \citet{Nelson:1991aa} for the model notations.

\subsection{Testing for ARCH effects (conditional heteroskedasticity)}
\label{sec:ARCHEngle}

\subsubsection{Setup: mean and variance equations}
Let $y_t$ denote (demeaned) asset returns. A standard decomposition writes
\begin{equation}
y_t = \mu_t + \varepsilon_t, 
\qquad \varepsilon_t = \sigma_t z_t,
\qquad \mathbb{E}[z_t]=0,\ \mathbb{V}[z_t]=1,
\end{equation}
where $\mu_t$ captures predictable dynamics in the conditional mean (often an ARMA$(p,q)$), while $\sigma_t^2$ is the conditional variance (conditional on past/selected information: $\sigma_t^2=\sigma^2\left(\mathcal{F}_{t-1}\right)$).

\subsubsection{ARCH($m$) model}
The ARCH$(m)$ model \citep{Engle:1982aa} specifies the conditional variance as a function of past squared innovations\footnote{See \texttt{ARCH\_vansteenberghe.py} for the implementation.}
:
\begin{equation}
\sigma_t^2 = \alpha_0 + \sum_{i=1}^{m} \alpha_i \varepsilon_{t-i}^2,
\qquad \alpha_0>0,\ \alpha_i \ge 0.
\label{eq:archm}
\end{equation}
The ARCH$(1)$ case is the special case
\begin{equation}
\sigma_t^2 = \alpha_0 + \alpha_1 \varepsilon_{t-1}^2.
\label{eq:arch1}
\end{equation}

\paragraph{Unconditional variance (ARCH(1)).}
Using $\mathbb{E}[z_t^2]=1$ and $\varepsilon_t=\sigma_t z_t$,
\begin{equation}
\mathbb{E}[\varepsilon_t^2] 
= \mathbb{E}[\sigma_t^2]
= \alpha_0 + \alpha_1 \mathbb{E}[\varepsilon_{t-1}^2].
\end{equation}
If $0 \le \alpha_1 < 1$, the stationary solution yields
\begin{equation}
\mathbb{V}(\varepsilon_t)=\mathbb{E}[\varepsilon_t^2]=\frac{\alpha_0}{1-\alpha_1},
\end{equation}
which is constant over time even though $\sigma_t^2=\mathbb{V}(\varepsilon_t\mid\mathcal{F}_{t-1})$ varies with past shocks.

\subsubsection{How to test for ARCH in practice}
\label{sec:arch_lm}
ARCH effects are tested on the residuals from the model. Let $\hat{\varepsilon}_t$ denote those residuals. Under the null of no ARCH, the conditional variance is constant:
\[
H_0:\ \alpha_1=\cdots=\alpha_m=0
\quad\Longleftrightarrow\quad
\mathbb{E}[\hat{\varepsilon}_t^2\mid\mathcal{F}_{t-1}] = \alpha_0.
\]
Two common diagnostics are:

\paragraph{(i) Portmanteau test on squared residuals.}
If $\{\hat{\varepsilon}_t\}$ is i.i.d., then $\{\hat{\varepsilon}_t^2\}$ should be serially uncorrelated. One can therefore apply Ljung--Box tests to $\hat{\varepsilon}_t^2$ (Section~\ref{sec:ljungbox}).

\paragraph{(ii) Engle's ARCH LM test (recommended).}
Run the auxiliary regression
\begin{equation}
\hat{\varepsilon}_t^2 = \alpha_0 + \sum_{i=1}^{m} \alpha_i \hat{\varepsilon}_{t-i}^2 + u_t,
\label{eq:arch_aux}
\end{equation}
and test $H_0:\alpha_1=\cdots=\alpha_m=0$.

Let $R^2$ be the coefficient of determination from \ref{eq:arch_aux} and let $T$ be the sample size used in the regression. The LM statistic is
\begin{equation}
\mathrm{LM} = T \cdot R^2 \ \ \overset{H_0}{\approx}\ \ \chi^2(m).
\end{equation}
A small $p$-value indicates rejection of $H_0$ and supports the presence of ARCH effects, motivating an ARCH/GARCH-type variance specification.

\medskip
\noindent
\textbf{Choice of the order $m$.}
In practice, $m$ should be large enough to capture potential short-run volatility clustering, but not so large as to lose power. Several pragmatic approaches are common:
\begin{itemize}
    \item \emph{Rule of thumb:} try a grid of small values (e.g.\ $m=1,2,5,10,20$) and assess robustness of the rejection.
    \item \emph{Data-driven upper bound:} set $m$ proportional to the sample size, e.g.\ $m \approx \lfloor T^{1/3} \rfloor$ or $\lfloor \log T \rfloor$, as in other portmanteau-type tests.
    \item \emph{Graphical guidance:} inspect the sample autocorrelation function of $\hat{\varepsilon}_t^2$ and choose $m$ to cover the lags where autocorrelations appear economically or statistically significant.
\end{itemize}

In empirical work, it is good practice to report results for several values of $m$ and verify that conclusions about the presence of ARCH effects are not driven by a particular lag choice.

\paragraph{(iii) Li--McLeod test (robust portmanteau for heavy-tailed samples).}

In finite samples with heavy-tailed innovations, the Ljung--Box statistic applied to 
$\hat{\varepsilon}_t^2$ may exhibit size distortions. 
\citet{Li:1981aa} propose a modified portmanteau statistic that remains valid under weaker moment conditions and performs better in small samples.

Let $\hat{\rho}_i$ denote the sample cross-correlation matrix at lag $i$ of the (possibly multivariate) residual process, and define
\[
b_i = \mathrm{vec}(\hat{\rho}_i').
\]
Let $\hat{\rho}_0$ be the sample contemporaneous correlation matrix and $\otimes$ denote the Kronecker product. The Li--McLeod statistic is
\begin{equation}
\mathrm{LMc} 
= T^2 \sum_{i=1}^m 
\frac{1}{T-i}
\, b_i'
\left( \hat{\rho}_0^{-1} \otimes \hat{\rho}_0^{-1} \right)
b_i,
\end{equation}
which under $H_0$ (no serial dependence up to lag $m$) satisfies
\[
\mathrm{LMc} \ \overset{H_0}{\approx}\ \chi^2(k^2 m),
\]
where $k$ is the dimension of the residual vector.

\medskip
\noindent
\textbf{Univariate case.}
When $k=1$, the statistic simplifies substantially. Let $r_i$ denote the sample autocorrelation of $\hat{\varepsilon}_t^2$ at lag $i$. Then
\[
\mathrm{LMc} 
= T^2 \sum_{i=1}^m \frac{r_i^2}{T-i}
\ \ \overset{H_0}{\approx}\ \chi^2(m).
\]

\medskip
This test can be viewed as a heavy-tail–robust portmanteau test on squared residuals and is particularly useful in financial applications where excess kurtosis is common.

\subsection{GARCH models}
\label{sec:garch}

We use the notation in \citet{Nelson:1991aa}. Let $\xi_t$ denote the one-step-ahead innovation (prediction error) from the conditional mean equation, and let $\sigma_t^2$ denote its conditional variance given the information set $\mathcal{F}_{t-1}$:
\[
\mathbb{E}[\xi_t \mid \mathcal{F}_{t-1}] = 0,
\qquad 
\mathbb{V}(\xi_t \mid \mathcal{F}_{t-1}) = \sigma_t^2.
\]
Assume the standardized innovation is $z_t$:
\begin{equation}
\xi_t = \sigma_t z_t,
\qquad z_t \ \text{i.i.d. with } \ \mathbb{E}[z_t]=0,\ \mathbb{V}[z_t]=1,
\end{equation}
where $z_t$ is often taken Gaussian, Student-$t$, or GED (Generalized Error Distribution) to allow for heavy tails.

\subsubsection{GARCH($p,q$)}
The GARCH$(p,q)$ model generalizes ARCH by letting the conditional variance depend on both past squared innovations and past conditional variances:
\begin{equation}
\sigma_t^2
=
\omega
+
\sum_{i=1}^{q} \beta_i \sigma_{t-i}^2
+
\sum_{j=1}^{p} \alpha_j \xi_{t-j}^2,
\label{eq:garch_pq}
\end{equation}
with parameter restrictions
\[
\omega>0,\qquad \alpha_j\ge 0,\qquad \beta_i\ge 0.
\]
(Equivalently, since $\xi_{t-j}^2=\sigma_{t-j}^2 z_{t-j}^2$, the ARCH term can be written as
$\sum_{j=1}^{p} \alpha_j z_{t-j}^2\sigma_{t-j}^2$.)

\paragraph{Interpretation.}
The $\alpha_j$ terms capture the short-run impact of past shocks on volatility (news effect), while the $\beta_i$ terms govern volatility persistence through variance feedback. The most common specification is GARCH$(1,1)$:
\begin{equation}
\sigma_t^2 = \omega + \alpha_1 \xi_{t-1}^2 + \beta_1 \sigma_{t-1}^2.
\label{eq:garch_11}
\end{equation}

\paragraph{Second-order stationarity (GARCH(1,1)).}
If $\alpha_1+\beta_1<1$, then $\{\xi_t\}$ has finite unconditional variance and
\begin{equation}
\mathbb{E}[\sigma_t^2]=\mathbb{V}(\xi_t)=\frac{\omega}{1-\alpha_1-\beta_1}.
\label{eq:garch_uncond_var}
\end{equation}
When $\alpha_1+\beta_1$ is close to one, volatility is highly persistent, producing long clusters of high or low variance.

\paragraph{Remark on distributions.}
Assuming a non-Gaussian $z_t$ (Student-$t$, GED) typically affects likelihood-based inference and tail risk measures, but not the conditional variance recursion \ref{eq:garch_pq}.

\subsubsection{GARCH-M model}

For the GARCH-M model, we add the following equation
$$R_t=a + b \sigma_t^2+\xi_t$$

\subsection{AR(1)-ARCH(1)}

\citet{Nelson:1991aa} details how to use maximum likelihood methods to fit an ARMA(p,q) (and potentially exponential) ARCH model and discuss the regularity conditions to be checked, but as in this paper, we will not check this as this stage ans assume that the ML estimator is consistent and asymptotically normal.

We fit an AR(1)-ARCH(1) $\sigma_t^2 = \omega +  \alpha_1 z_{t-1}^2\sigma_{t-1}^2$ and an AR(1)-GARCH(1,1) $\sigma_t^2 = \omega + \beta_i \sigma_{t-1}^2 + \alpha_1 z_{t-1}^2\sigma_{t-1}^2$ model on the financial index daily returns.

\subsection{AR(1)--EGARCH}
\label{sec:egarch}

EGARCH models were introduced by \citet{Nelson:1991aa} to accommodate three empirical features often observed in returns:
(i) \emph{leverage/asymmetry} (negative shocks increase future volatility more than positive shocks),
(ii) fewer parameter restrictions than standard GARCH (the variance is positive by construction),
and (iii) a more transparent separation between \emph{size} and \emph{sign} effects of shocks.

\subsubsection{Specification}
Let $R_t$ denote returns. Consider an AR(1) conditional mean and an EGARCH$(p,q)$ conditional variance:
\begin{equation}
\begin{cases}
R_t = a + b R_{t-1} + \xi_t,\\[3pt]
\xi_t = \sigma_t z_t,\qquad z_t\ \text{i.i.d. with } \mathbb{E}[z_t]=0,\ \mathbb{V}[z_t]=1,\\[6pt]
\log(\sigma_t^2)
=
\omega
+\displaystyle\sum_{i=1}^{p}\alpha_i\Big(|z_{t-i}|-\mathbb{E}|z_t|\Big)
+\displaystyle\sum_{i=1}^{p}\gamma_i z_{t-i}
+\displaystyle\sum_{k=1}^{q}\beta_k \log(\sigma_{t-k}^2).
\end{cases}
\label{eq:egarch}
\end{equation}

\paragraph{Normal benchmark.}
If $z_t\sim\mathcal{N}(0,1)$, then $\mathbb{E}|z_t|=\sqrt{2/\pi}$, hence the centering term
$\big(|z_{t-i}|-\sqrt{2/\pi}\big)$ ensures that the intercept $\omega$ controls the unconditional level of log-variance.

\subsubsection{Interpretation}
The EGARCH decomposition is convenient:
\begin{itemize}
\item The $\alpha_i$ terms capture the effect of the \emph{magnitude} of standardized shocks (volatility response to ``surprises'').
\item The $\gamma_i$ terms capture the \emph{sign} effect (asymmetry/leverage). Typically $\gamma_i<0$ implies that negative shocks raise volatility more than positive shocks of the same magnitude.
\item The $\beta_k$ terms govern volatility persistence in log-variance.
\end{itemize}
Because the model is written for $\log(\sigma_t^2)$, $\sigma_t^2=\exp(\log(\sigma_t^2))>0$ automatically, so nonnegativity constraints on $\alpha_i,\beta_k$ are not required (unlike standard GARCH).

\subsubsection{A parsimonious workhorse: EGARCH(1,1)}
The most common specification is EGARCH$(1,1)$:
\begin{equation}
\log(\sigma_t^2)
=
\omega
+\alpha\Big(|z_{t-1}|-\mathbb{E}|z_t|\Big)
+\gamma z_{t-1}
+\beta \log(\sigma_{t-1}^2).
\label{eq:egarch11}
\end{equation}

\subsubsection{Remark on $c\sigma_t^2$ in the mean}
Some formulations allow $R_t$ to depend on $\sigma_t^2$ (a risk--return effect). For lecture-note simplicity and to focus on volatility dynamics, we set $c=0$ and estimate an AR(1)--EGARCH model as in \ref{eq:egarch}. Model orders $(p,q)$ can be selected using information criteria (AIC/BIC) and checked via residual diagnostics on standardized innovations $\hat{z}_t=\hat{\xi}_t/\hat{\sigma}_t$.

\citet{Nelson:1991aa} suggest the EGARCH model:
\begin{equation}\label{eq:Nelson1991}
\begin{cases}
R_t = a + bR_{t-1} + c\sigma_t^2 + \xi_t\\
\xi_t = \sigma_t z_t\\
\ln\left( \sigma_t^2 \right) = \omega + \sum_{i=1}^{p} \alpha_i \left(\vert z_{t-i}\vert - \sqrt{\frac{2}{\pi}}\right)+\sum_{j=1}^o \gamma_j z_{t-j} + \sum_{k=1}^q \beta_k \ln\sigma_{t-k}^2
\end{cases}
\end{equation}
as in \citet{Nelson:1991aa}, we keep $c=0$ and we would need to select the order $p$, $j$, and $q$ with information criterion.

\stepcounter{question}
\begin{itemize}
\item \textbf{Question \arabic{question}}
\begin{itemize}
\item Are the standardized residuals $z_t$ of the AR(1)-GARCH or EGARCH model iid?
\end{itemize}
\end{itemize}

In equation \ref{eq:Nelson1991}, we have a linear form of "GARCH-in-mean" with the constant $c$. We could also include any type of function (possibly non-linear) $g()$ and update equation \ref{eq:Nelson1991} with $R_t = a + bR_{t-1} + g\left(\sigma_t\right) + \xi_t$. We will explore section \ref{sec:AREGARCH} how with the package rugarch we can use the options in "mean.model" with "archm Whether to include ARCH volatility in the mean regression" and "archpow Indicates whether to use st.deviation (1) or variance (2) in the ARCH in mean regression".

\subsection{GARCH model checking}

\subsubsection{Significance of the coefficients}

Beyond checking coefficients p-values, there is the Lagrange Multiplier Test (LM Test), for a parametric model with true parameters $\theta_0$, $H_0:R\theta_0=r$, where we can test that some coefficients in $\theta_0$ are null (imposing some $0$s in $R$ and $r$).

Following \citet{Francq:2009tv}, use the Wald, LM or Likelihood Ratio statistics.

\subsubsection{Test on the residuals and squared residuals}

Ljung-Box test on the residuals (auto-correlation), squared residuals (ARCH) and ARCH LM test (presence of ARCH).

\subsubsection{Test of second order stationarity}

To test the second order stationarity, you want to test $H_0: \sum_i \beta_i + \sum_j \alpha_j < 1$, with the parameters taken from equation \ref{eq:garch_pq}, against $H_1: \sum_i \beta_i + \sum_j \alpha_j \geq 1$.

\subsection{ARCH/GARCH Models and the Measurement of Uncertainty}

In many financial and macroeconomic time series, the conditional mean may be relatively stable, while the conditional variance exhibits substantial fluctuations over time. Periods of calm alternate with periods of turbulence, a phenomenon commonly referred to as \emph{volatility clustering}. Models with time-varying second-order moments provide a natural framework to formalize and measure such uncertainty. Uncertainty is modeled as the \emph{conditional variance} of shocks. In Autoregressive Conditional Heteroskedasticity-type models, uncertainty is not an exogenous index but an endogenous, model-implied object:
\begin{equation}
\text{Uncertainty}_t \equiv \sigma_t^2 
= \mathbb{E}(\varepsilon_t^2 \mid \mathcal{F}_{t-1}).
\end{equation} where $\mathcal{F}_{t-1}$ is the information set available at time $t-1$. 

Thus, ARCH/GARCH models deliver a real-time measure of perceived risk based solely on past realizations. This makes them particularly useful for:
\begin{itemize}
    \item financial volatility modeling (e.g.\ stock returns),
    \item macroeconomic uncertainty proxies (e.g.\ inflation or output shocks),
    \item risk management and Value-at-Risk calculations,
    \item testing whether uncertainty affects economic decisions (investment, credit supply, consumption).
\end{itemize}

\paragraph{Application to Inflation Uncertainty.}

While ARCH-type models were initially developed for financial returns, they are equally relevant for macroeconomic variables such as inflation. Inflation dynamics typically exhibit persistence in the mean and episodes of heightened volatility (e.g.\ oil shocks, regime shifts, pandemic episodes). ARCH-type models have been used to measure inflation uncertainty, as in \citet{grier2000effects}. A natural specification therefore combines:

\begin{enumerate}
    \item a dynamic \emph{mean equation} for inflation,
    \item a time-varying \emph{conditional variance} capturing inflation uncertainty.
\end{enumerate}

Let $\pi_t$ denote (annualized) inflation. A standard specification is:
\begin{align}
\pi_t &= \mu_t + \varepsilon_t, \\
\mu_t &= \phi_0 + \sum_{i=1}^p \phi_i \pi_{t-i}, \\
\varepsilon_t &= \sigma_t z_t, \qquad z_t \sim \mathcal{D}(0,1),
\end{align}
where $\mathcal{D}$ may be Gaussian or Student-$t$ to allow for fat tails.

Inflation uncertainty is then defined as:
\begin{equation}
\text{Inflation Uncertainty}_t \equiv \sigma_t^2
= \mathbb{E}(\varepsilon_t^2 \mid \mathcal{F}_{t-1}).
\end{equation}

A GARCH(1,1) specification,
\begin{equation}
\sigma_t^2 = \omega + \alpha \varepsilon_{t-1}^2 + \beta \sigma_{t-1}^2,
\end{equation}
captures both the immediate impact of inflation surprises ($\alpha$) and the persistence of inflation uncertainty ($\beta$).

\paragraph{Economic Interpretation.}

In this framework, inflation uncertainty is the conditional variance of unexpected inflation. It reflects how difficult inflation is to forecast given available information. Importantly:

\begin{itemize}
    \item High $\sigma_t^2$ does not necessarily imply high inflation; it implies high \emph{unpredictability}.
    \item Inflation may be close to target, yet uncertainty can be elevated.
    \item Episodes such as the Great Inflation, the Zero Lower Bound period, or the post-pandemic inflation surge are characterized by sharp movements in $\sigma_t^2$.
\end{itemize}

\paragraph{Mean–Variance Interactions.}

A central question in macroeconomics concerns the relationship between the level of inflation and its uncertainty. Two influential hypotheses are:

\begin{itemize}
    \item \textbf{Friedman--Ball hypothesis:} Higher inflation leads to higher inflation uncertainty (policy credibility channel). \citep{friedman1977nobel,ball1992does}.
    \item \textbf{Cukierman--Meltzer hypothesis:} Higher uncertainty may induce policymakers to generate surprise inflation. \citep{cukierman1986theory}.
\end{itemize}

Empirically, this can be tested by augmenting the mean equation with conditional variance:
\begin{equation}
\pi_t = \phi_0 + \sum_{i=1}^p \phi_i \pi_{t-i}
+ \gamma \sigma_t^2 + \varepsilon_t,
\end{equation}
where $\gamma$ captures the effect of uncertainty on inflation dynamics.

\paragraph{Distributional Considerations.}

Inflation shocks may exhibit fat tails during crisis periods. Replacing the normal distribution with a Student-$t$ distribution often improves model fit and produces more robust measures of uncertainty:
\begin{equation}
z_t \sim t_\nu(0,1),
\end{equation}
with degrees of freedom $\nu$ estimated jointly with the volatility process.

\subsection{From univariate GARCH to Vector GARCH and Factor-ARCH}
\label{subsec:vector-garch-factor-arch}

The univariate GARCH model describes the dynamics of the conditional
variance of a single return. In a multivariate setting, however, one is
interested in the entire conditional variance--covariance matrix. Let
\[
    \mathbf{r}_t
    =
    \begin{pmatrix}
        r_{1,t} & \cdots & r_{N,t}
    \end{pmatrix}^{\prime}
\]
denote a vector of returns, and let $\mathcal{F}_{t-1}$ be the information
available at time $t-1$. For simplicity, suppose that the conditional mean
has already been removed, so that
\[
    \mathbb{E}\!\left(\mathbf{r}_t \mid \mathcal{F}_{t-1}\right)
    = \mathbf{0}.
\]
The conditional covariance matrix is
\[
    \mathbf{H}_t
    =
    \operatorname{Var}\!\left(\mathbf{r}_t \mid \mathcal{F}_{t-1}\right)
    =
    \mathbb{E}\!\left(
        \mathbf{r}_t\mathbf{r}_t^{\prime}
        \mid \mathcal{F}_{t-1}
    \right),
\]
whose $(i,j)$th element is denoted by
\[
    \sigma_{ij,t}
    =
    \operatorname{Cov}\!\left(
        r_{i,t},r_{j,t}
        \mid \mathcal{F}_{t-1}
    \right).
\]

A simple nonparametric estimate of this covariance is the rolling-window
estimator
\[
    \widehat{\sigma}_{ij,t}
    =
    \frac{1}{L}
    \sum_{h=1}^{L}
    r_{i,t-h}r_{j,t-h},
\]
assuming that returns have zero conditional mean. This estimator assigns
the same weight to the last $L$ observations. A Vector GARCH model instead
specifies a parametric law of motion for $\sigma_{ij,t}$, giving different
weights to recent shocks and allowing past conditional covariances to
affect current covariances.

\subsubsection{A general Vector GARCH representation}

A direct multivariate extension of GARCH can be written, element by
element, as
\begin{align}
    \sigma_{ij,t}
    &=
    \omega_{ij}
    +
    \sum_{\ell=1}^{p}
    \sum_{k=1}^{N}
    \sum_{m=1}^{N}
    \alpha_{ij,km,\ell}\,
    r_{k,t-\ell}r_{m,t-\ell}
    \notag\\
    &\quad
    +
    \sum_{\ell=1}^{q}
    \sum_{k=1}^{N}
    \sum_{m=1}^{N}
    \beta_{ij,km,\ell}\,
    \sigma_{km,t-\ell}.
    \label{eq:general-vector-garch}
\end{align}
Because the covariance matrices are symmetric, one may equivalently
restrict the inner sums to $k\geq m$ and work only with the
$N(N+1)/2$ distinct elements.

Equation~\eqref{eq:general-vector-garch} has the same intuition as a
univariate GARCH model:

\begin{itemize}
    \item $\omega_{ij}$ is the constant component of the conditional
          covariance between assets $i$ and $j$;
    \item $r_{k,t-\ell}r_{m,t-\ell}$ is a lagged covariance shock;
    \item $\alpha_{ij,km,\ell}$ measures how this shock affects
          $\sigma_{ij,t}$;
    \item $\sigma_{km,t-\ell}$ is a lagged conditional covariance;
    \item $\beta_{ij,km,\ell}$ measures covariance persistence.
\end{itemize}

An asymmetric specification can additionally distinguish negative from
non-negative shocks. For example, define
\[
    d_{km,t-\ell}
    =
    \mathbbm{1}
    \left\{
        r_{k,t-\ell}<0,\;
        r_{m,t-\ell}<0
    \right\}.
\]
A Vector GJR-GARCH-type extension is then
\begin{align}
    \sigma_{ij,t}
    &=
    \omega_{ij}
    +
    \sum_{\ell=1}^{p}
    \sum_{k=1}^{N}
    \sum_{m=1}^{N}
    \left(
        \alpha_{ij,km,\ell}
        +
        \gamma_{ij,km,\ell}d_{km,t-\ell}
    \right)
    r_{k,t-\ell}r_{m,t-\ell}
    \notag\\
    &\quad
    +
    \sum_{\ell=1}^{q}
    \sum_{k=1}^{N}
    \sum_{m=1}^{N}
    \beta_{ij,km,\ell}\,
    \sigma_{km,t-\ell}.
    \label{eq:asymmetric-vector-garch}
\end{align}
The coefficient $\gamma_{ij,km,\ell}$ captures any additional effect of
jointly negative innovations on the conditional covariance.

The flexibility of this model is also its main weakness. There are
$N(N+1)/2$ distinct conditional variances and covariances, and each can
depend on all past cross-products and conditional covariances. Consequently,
the number of parameters increases rapidly with $N$. Moreover, unrestricted
element-by-element equations do not automatically ensure that
$\mathbf{H}_t$ is positive semidefinite. These considerations motivate
more parsimonious parameterizations.

\subsubsection{Factor-ARCH}

Factor-ARCH assumes that the time variation in the covariance matrix is
driven by a small number $K$ of common factors, with $K\ll N$. Consider
the factor representation
\[
    \mathbf{r}_t
    =
    \boldsymbol{\mu}_t
    +
    \boldsymbol{\Lambda}\mathbf{f}_t
    +
    \mathbf{u}_t,
\]
where $\mathbf{f}_t$ is a $K\times 1$ vector of common factors,
$\boldsymbol{\Lambda}$ is the $N\times K$ matrix of factor loadings, and
$\mathbf{u}_t$ contains idiosyncratic innovations. If the conditional
covariance matrix of the factors is
\[
    \mathbf{H}^{F}_t
    =
    \operatorname{Var}\!\left(
        \mathbf{f}_t\mid\mathcal{F}_{t-1}
    \right),
\]
and the idiosyncratic covariance matrix is the time-invariant matrix
$\boldsymbol{\Omega}$, then
\begin{equation}
    \mathbf{H}_t
    =
    \boldsymbol{\Omega}
    +
    \boldsymbol{\Lambda}
    \mathbf{H}^{F}_t
    \boldsymbol{\Lambda}^{\prime}.
    \label{eq:factor-arch-matrix}
\end{equation}
Element by element, this becomes
\begin{equation}
    \sigma_{ij,t}
    =
    \omega_{ij}
    +
    \sum_{f=1}^{K}
    \sum_{f'=1}^{K}
    \lambda_{if}\lambda_{jf'}\,
    \sigma^{F}_{ff',t},
    \label{eq:factor-arch-general}
\end{equation}
where $\lambda_{if}$ is the loading of asset $i$ on factor $f$, and
$\sigma^{F}_{ff',t}$ is the conditional covariance between factors $f$
and $f'$.

When the factors are conditionally uncorrelated,
$\mathbf{H}^{F}_t$ is diagonal. Equation~\eqref{eq:factor-arch-general}
then simplifies to
\begin{equation}
    \sigma_{ij,t}
    =
    \omega_{ij}
    +
    \sum_{f=1}^{K}
    \lambda_{if}\lambda_{jf}\,h_{f,t},
    \label{eq:factor-arch-diagonal}
\end{equation}
where
\[
    h_{f,t}
    =
    \operatorname{Var}\!\left(
        f_{f,t}\mid\mathcal{F}_{t-1}
    \right).
\]
Each $h_{f,t}$ may itself follow a univariate ARCH or GARCH process, for
example
\[
    h_{f,t}
    =
    \omega_f
    +
    \alpha_f f_{f,t-1}^{2}
    +
    \beta_f h_{f,t-1}.
\]

The specification proposed by \citet{EngleNgRothschild1990} can be
expressed as
\begin{equation}
    \mathbf{H}_t
    =
    \boldsymbol{\Omega}
    +
    \sum_{f=1}^{K}
    h_{f,t}\,
    \boldsymbol{\lambda}_f
    \boldsymbol{\lambda}_f^{\prime},
    \label{eq:engr-factor-arch}
\end{equation}
where $\boldsymbol{\lambda}_f$ is the $N\times 1$ loading vector associated
with factor $f$. This representation has two important advantages. First,
the number of dynamic variance processes is reduced from
$N(N+1)/2$ to $K$. Second, if $\boldsymbol{\Omega}$ is positive
semidefinite and $h_{f,t}\geq 0$, then $\mathbf{H}_t$ is automatically
positive semidefinite.

The model also has a portfolio interpretation. A
\emph{factor-representing portfolio} is a portfolio whose return isolates
the variation associated with one of the factors. Under suitable
restrictions, forecasting the conditional covariance matrix of all $N$
assets can therefore be reduced to forecasting the conditional variances
of only a small number of factor-representing portfolios. In the empirical
application of \citet{EngleNgRothschild1990}, one- and two-factor
specifications are used to model the conditional covariance structure of
Treasury-bill excess returns.

\paragraph{Interpretation.}
The difference between the unrestricted Vector GARCH model and
Factor-ARCH can be summarized as follows. Vector GARCH allows each
conditional covariance to react directly to a large set of past shocks
and past covariances. Factor-ARCH instead assumes that these
co-movements originate from a small number of common sources of risk.
The latter restriction sacrifices some flexibility, but substantially
reduces the number of parameters and facilitates the construction of a
valid conditional covariance matrix.

\paragraph{Remark on estimation.}
The parameters can be estimated by Gaussian quasi-maximum likelihood.
Normality is then used to construct the criterion function but need not
be the true conditional distribution. Under appropriate stationarity,
moment, identification, and regularity conditions, quasi-maximum
likelihood estimators of Vector GARCH-type models are consistent and
asymptotically normal; see \citet{LingMcAleer2003}.

\section{Spurious Regression, Cointegration, and Robust Inference}

\subsection{Before regressing or computing correlation, plot your data}

Before\footnote{We suggest the companion script \texttt{vansteenberghe\_cointegration\_ecm\_regression.py}.}
applying any regression model or computing correlations, it is essential to
\emph{visualize the data}. In time-series analysis, plotting the levels and
their transformations is the first diagnostic step: it allows us to detect
trends, breaks, seasonality, outliers, and potential non-stationarity.

In this section, we refer to the second variable as ``GDP'' for notational
simplicity. However, the series actually used in the code corresponds to:

\begin{quote}
\emph{Masse salariale versée -- Total branches -- Valeur aux prix courants -- Série CVS}
\end{quote}

published by \href{https://www.insee.fr/fr/statistiques/serie/001689203}{INSEE, series 001689203}.
This variable is \textbf{not equivalent to GDP}. It measures the aggregate wage
bill (in current prices, seasonally adjusted), whereas GDP captures total value
added in the economy. The two series are economically related but conceptually
distinct. 

We use this wage-bill series for \emph{educational purposes only}, in order to
illustrate the mechanics of regression, correlation, unit-root testing, and
cointegration in a macroeconomic context.

\vspace{0.3cm}

We perform the same data cleaning operations as for the population CSV file.
Since the wage-bill series is quarterly, we align frequencies by converting
the monthly population series to quarterly averages before merging the two
datasets.

We then concatenate the two time series using the date index (see the
\href{http://pandas.pydata.org/pandas-docs/stable/merging.html}{\texttt{pandas} merging documentation})
and plot them, using a secondary $y$-axis for population.

You should obtain:

\begin{center}
\includegraphics[scale=0.5]{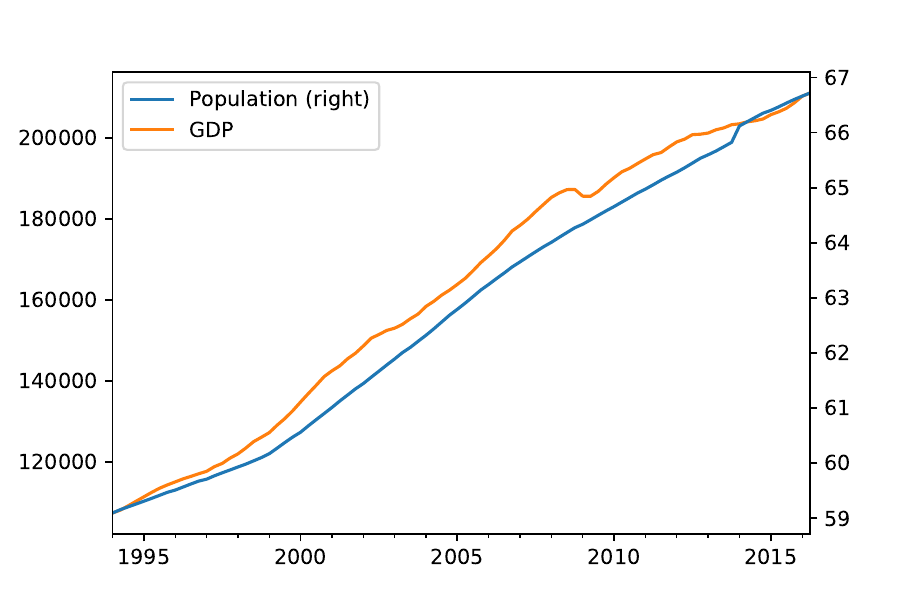}
\end{center}

\subsubsection*{Growth rates}

We work with quarterly growth rates:
\[
g_t = \frac{X_t - X_{t-1}}{X_{t-1}}
\qquad \text{or equivalently} \qquad
g_t \approx \Delta \log X_t .
\]

We compute the quarterly changes and expect a relatively low contemporaneous
correlation between wage-bill growth (labeled ``GDP'' in the code) and
population growth.

\begin{center}
\includegraphics[scale=0.5]{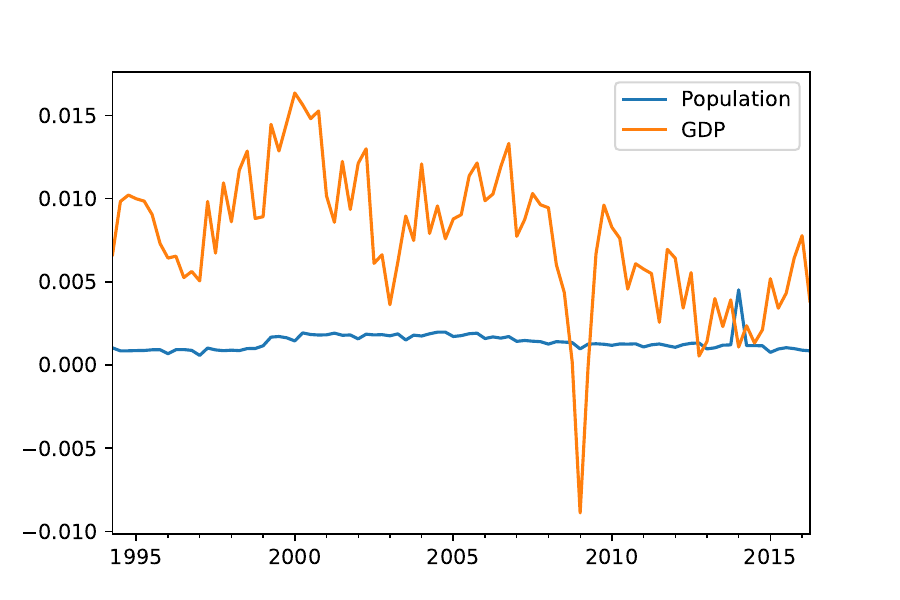}
\end{center}

We then inspect a scatter plot of wage-bill growth against population growth:

\begin{center}
\includegraphics[scale=0.5]{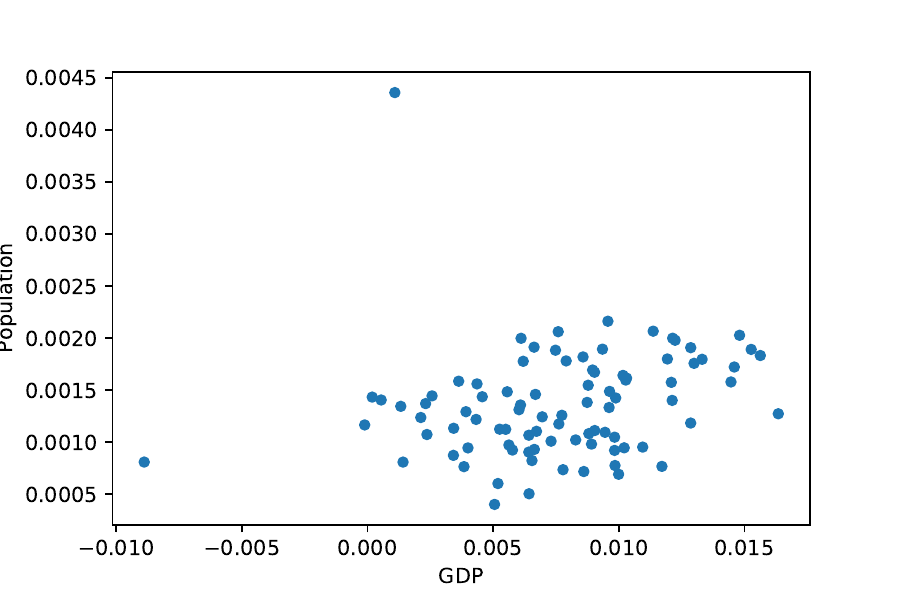}
\end{center}

Two observations clearly appear as outliers: one associated with a strongly
negative wage-bill growth rate (macro shock) and another with an unusually
high population growth rate. We discuss formal treatment of outliers in
Section~\ref{sec:outlier}.

\subsubsection{Correlation}

Following the methodology detailed in Section~\ref{sec:correlation}, we compute
the sample correlation between the two growth rates. We obtain a relatively
low correlation of approximately $18\%$.

This modest correlation illustrates an important empirical lesson:
even variables that are economically related in levels (e.g.\ activity and
demographics) may display weak short-run comovement in growth rates.

\subsection{Unit Root test - Dickey-Fuller test}

We test "manually" the GDP for a unit root following the test described in section \ref{sec:DFtest}.

We test that both the French population and GDP time series are I(1).

\newpage

\subsection{Cointegration, ECM and linear regression}
\subsubsection{Cointegration test}\label{sec:cointegration}

After we differenced the variables, we cannot check for their \textbf{long term} relation in level, when differenced the variables are only compared in short term models.

With two I(1) series\footnote{we define by I(n) integrated series of order n} $y_t$ and $x_t$, level comparison via linear regression would be difficult because subject to spurious regression. But two I(1) series are said to be cointegrated, if there exist a linear combination\footnote{called long-run equilibrium relationship} of them (e.g. $\beta_0 + \beta_y y_t + \beta_x x_t$) that is I(0). This means that $y_t$ and $x_t$ share a common stochastic trend and the two time series, after having applied some coefficients, cannot drift too far apart from one another in the long-run\footnote{unless the "forces" that kept them together in the past do not hold any longer}. If there exist\footnote{$\beta_y$ and $\beta_z$ are not unique (one can add a constant and the relationship would still hold), also note that $\beta_0$ can be null and the following still applies} $\beta_y$ and $\beta_x$ such that $E(\beta_0 + \beta_y y_t + \beta_x x_t) = 0$, then there is a long term relationship between $y_t$ and $x_t$ and one can test the following model:
\begin{equation}
y_t = \alpha + \beta x_t + \epsilon_t
\end{equation}
and from our statements above, one can deduce that the cointegrating residuals $\epsilon_t$ have to be $I(0)$ if $y_t$ and $x_t$ are cointegrated\footnote{but as demonstrated in Phillips and Ouliaris (1990), one cannot use the standard Augmented Dickey-Fuller critical values to test the \emph{estimated} residuals $\hat{\epsilon}_t$ for a unit root: OLS picks $\hat{\alpha}$ and $\hat{\beta}$ precisely so as to make the residuals look as stationary as possible, which biases the test toward finding cointegration, so dedicated (more negative) critical values must be used; furthermore, the t-statistics for $\beta$ will not be asymptotically normal if $\epsilon_t$ are serially correlated, see section~\ref{sec:cointcoefinterpretation}}.

We can then test for cointegration between variables with a cointegration test: the null hypothesis is no cointegration. The second output gives the p-value of the test.

For our variables, we do not reject the null hypothesis and conclude that there is no cointegration.

\subsubsection{Interpreting the coefficient of a cointegrating regression}\label{sec:cointcoefinterpretation}

A word on vocabulary first. The \emph{spurious regression} problem \citep{Granger:1974aa,phillips1986understanding} concerns levels regressions between I(1) variables that are \emph{not} cointegrated: $\hat{\beta}$ then converges not to a true parameter but to a random limit, the $R^2$ does not vanish even asymptotically, the Durbin--Watson statistic collapses toward zero, and the conventional t-statistic diverges, so that one routinely ``finds'' highly significant relationships between independent random walks. When $y_t$ and $x_t$ \emph{are} cointegrated, the levels regression $y_t = \alpha + \beta x_t + u_t$ is no longer spurious: it estimates the cointegrating relationship. Even so, its coefficient must be interpreted with great care.

If $y_t$ and $x_t$ are cointegrated, then there exists a $\beta$ such that the equilibrium error
\begin{equation}
u_t = y_t - \alpha - \beta x_t
\end{equation}
is stationary, $u_t \sim I(0)$. In that case, $\beta$ can be interpreted as a \textbf{long-run equilibrium relationship} between $y_t$ and $x_t$: depending on whether the variables are in logs, levels, etc., $\beta$ is a long-run elasticity, slope, or equilibrium coefficient.

There are, however, important caveats.

First, $\beta$ is \textbf{not a short-run marginal effect}. It tells us how $y_t$ and $x_t$ move together in the long run, not how $y_t$ reacts immediately to a change in $x_t$. The immediate reaction is governed by the short-run dynamics of the error correction model introduced below (the impact coefficient on $\Delta x_t$ and the speed of adjustment on the lagged equilibrium error $u_{t-1}$), and there is no reason for these short-run coefficients to coincide with $\beta$.

Second, \textbf{standard OLS inference on $\hat{\beta}$ is problematic}. Under cointegration, the OLS estimator of the cointegrating coefficient is \emph{superconsistent}: it converges to $\beta$ at rate $T$ rather than the usual $\sqrt{T}$\footnote{Stock (1987).}. Superconsistency, however, does not validate the usual standard errors: the asymptotic distribution of $\hat{\beta}$ is generally non-standard (it involves functionals of Brownian motions and second-order bias terms), so the usual t-statistics and standard errors are unreliable, especially when the regressor is endogenous (innovations of $x_t$ correlated with $u_t$) and when $u_t$ is serially correlated; with macro-financial data, both are the rule rather than the exception. Despite superconsistency, $\hat{\beta}$ can also be noticeably biased in samples of the size used in applied work. Hypothesis tests on $\beta$ (e.g.\ $H_0: \beta = 1$) should therefore rely on estimators designed for cointegrating regressions, such as fully modified OLS or dynamic OLS, or on maximum likelihood estimation of a vector error correction model, which restore asymptotically (mixed) normal inference\footnote{Fully modified OLS: Phillips and Hansen (1990); dynamic OLS: Saikkonen (1991), Stock and Watson (1993); maximum likelihood in the VECM: Johansen (1991).}.

In short: with cointegrated I(1) variables, the levels regression consistently estimates a long-run equilibrium coefficient, but $\hat{\beta}$ should not be read as the immediate effect of $x_t$ on $y_t$, and the OLS t-statistics on $\hat{\beta}$ should not be trusted unless endogeneity and serial correlation have been dealt with.

\subsubsection{Two I(0) series: regression in levels versus in first differences}\label{sec:I0levelsvsdiff}

So far we have discussed I(1) variables. Consider now the symmetric question\footnote{We suggest the companion script \texttt{vansteenberghe\_levels\_vs\_differences\_I0.py}.}: suppose the two series are \emph{stationary}, for instance a nominal policy rate $i_t$ and inflation $\pi_t$, both I(0), and compare the regression in levels with the regression in first differences:
\begin{align}
i_t &= c + \phi_\pi \pi_t + u_t, \label{eq:I0levels}\\
\Delta i_t &= c' + \phi^\Delta_\pi \Delta \pi_t + v_t. \label{eq:I0diffs}
\end{align}
What can we say, and how does $\phi^\Delta_\pi$ relate to $\phi_\pi$?

\paragraph{Both regressions are legitimate, but they do not estimate the same parameter.}
With I(0) variables there is no spurious regression problem, and differencing a stationary series yields another stationary series\footnote{$\Delta i_t$ is in fact \emph{over-differenced}: its moving-average representation contains a unit root (it is non-invertible) and its spectral density is zero at frequency zero. This does not affect the consistency of OLS, but it matters for efficiency and inference.}. OLS therefore consistently estimates, in each equation, the corresponding linear projection coefficient. For jointly stationary series these estimands are
\begin{equation}\label{eq:I0estimands}
\phi_\pi = \frac{\gamma_{i\pi}(0)}{\gamma_\pi(0)},
\qquad
\phi^\Delta_\pi = \frac{2\gamma_{i\pi}(0) - \gamma_{i\pi}(1) - \gamma_{i\pi}(-1)}{2\left(\gamma_\pi(0) - \gamma_\pi(1)\right)},
\end{equation}
where $\gamma_\pi(h) = \mathrm{Cov}(\pi_t, \pi_{t-h})$ and $\gamma_{i\pi}(h) = \mathrm{Cov}(i_t, \pi_{t-h})$. Equation~\eqref{eq:I0estimands} is the general answer to how $\phi^\Delta_\pi$ relates to $\phi_\pi$: the two coefficients are different functions of the joint autocovariance structure, and there is no reason for them to be equal. Intuitively, differencing acts as a high-pass filter: $\phi^\Delta_\pi$ is driven by the high-frequency (short-run) comovement of the two series, whereas $\phi_\pi$ is dominated by their persistent (low-frequency) comovement\footnote{In the frequency domain, differencing reweights all frequencies by $|1-e^{-\mathrm{i}\omega}|^2 = 2(1-\cos\omega)$, which is zero at frequency $\omega=0$: the long-run comovement receives zero weight in $\phi^\Delta_\pi$.}. Note also that for jointly stationary series without deterministic trend, $E(\Delta i_t) = E(\Delta \pi_t) = 0$, so the population constant of \eqref{eq:I0diffs} is $c' = 0$, whatever the dynamics; the level information contained in $c$ cannot be recovered from the differenced regression, and a significantly nonzero $\hat{c}'$ indicates a drift, i.e.\ that the series were not I(0) around a constant mean in the first place.

\paragraph{When do the two coefficients coincide?}
If the true relationship is purely contemporaneous, $i_t = c + \phi_\pi \pi_t + u_t$ with $u_t$ uncorrelated with $\pi_s$ at all leads and lags, then $\gamma_{i\pi}(h) = \phi_\pi \gamma_\pi(h)$ and \eqref{eq:I0estimands} yields $\phi^\Delta_\pi = \phi_\pi$: both regressions estimate the same coefficient. Even in this favorable case differencing is not innocuous: the error $v_t = \Delta u_t$ is a non-invertible MA(1), so the regression in differences is less efficient than the regression in levels and requires HAC standard errors; the constant is lost; and any classical measurement error in $\pi_t$ produces a worse attenuation bias in differences than in levels (differencing shrinks the signal of a persistent regressor but not the noise).

\paragraph{With dynamics, levels lean towards the long run, differences towards the short run.}
Differences between $\phi_\pi$ and $\phi^\Delta_\pi$ appear as soon as the relationship is dynamic. Consider a Taylor-type rule with interest-rate smoothing,
\begin{equation}\label{eq:I0smoothing}
i_t = \rho\, i_{t-1} + (1-\rho)\left(c + \phi_\pi \pi_t\right) + \varepsilon_t,
\qquad
\pi_t = (1-a)\,\pi^* + a\, \pi_{t-1} + e_t,
\end{equation}
with $|\rho|<1$ and $|a|<1$, so that both series are I(0). Here $\phi_\pi$ is the \emph{long-run} response of the policy rate to inflation, while $(1-\rho)\phi_\pi$ is its \emph{impact} (short-run) response. Working out the autocovariances, the OLS slopes of the static regressions \eqref{eq:I0levels} and \eqref{eq:I0diffs} converge to
\begin{equation}\label{eq:I0plims}
\mathrm{plim}\; \hat{\phi}_\pi = \frac{(1-\rho)\,\phi_\pi}{1-\rho a},
\qquad
\mathrm{plim}\; \hat{\phi}^\Delta_\pi = \frac{(1-\rho)\,\phi_\pi\,\left(2-\rho-2a+\rho a^2\right)}{2\,(1-a)(1-\rho a)} .
\end{equation}
The levels coefficient lies between the impact and the long-run responses, and tends to the long-run $\phi_\pi$ as inflation becomes more persistent ($a \to 1$); the differences coefficient stays close to the impact response. For $\phi_\pi = 1.5$, $\rho = 0.8$, $a = 0.9$: long-run response $1.5$, impact response $0.3$, $\mathrm{plim}\, \hat{\phi}_\pi \approx 1.07$ and $\mathrm{plim}\, \hat{\phi}^\Delta_\pi \approx 0.26$. The same stationary data thus deliver two very different ``inflation coefficients'' depending on the chosen transformation, and \emph{neither} equals the long-run $\phi_\pi$: to recover it, estimate the dynamic (partial-adjustment) regression of $i_t$ on $i_{t-1}$ and $\pi_t$ and compute the long-run response $\hat{b}/(1-\hat{\rho})$. The companion script verifies \eqref{eq:I0plims} by simulation and produces:

\begin{center}
\includegraphics[scale=0.5]{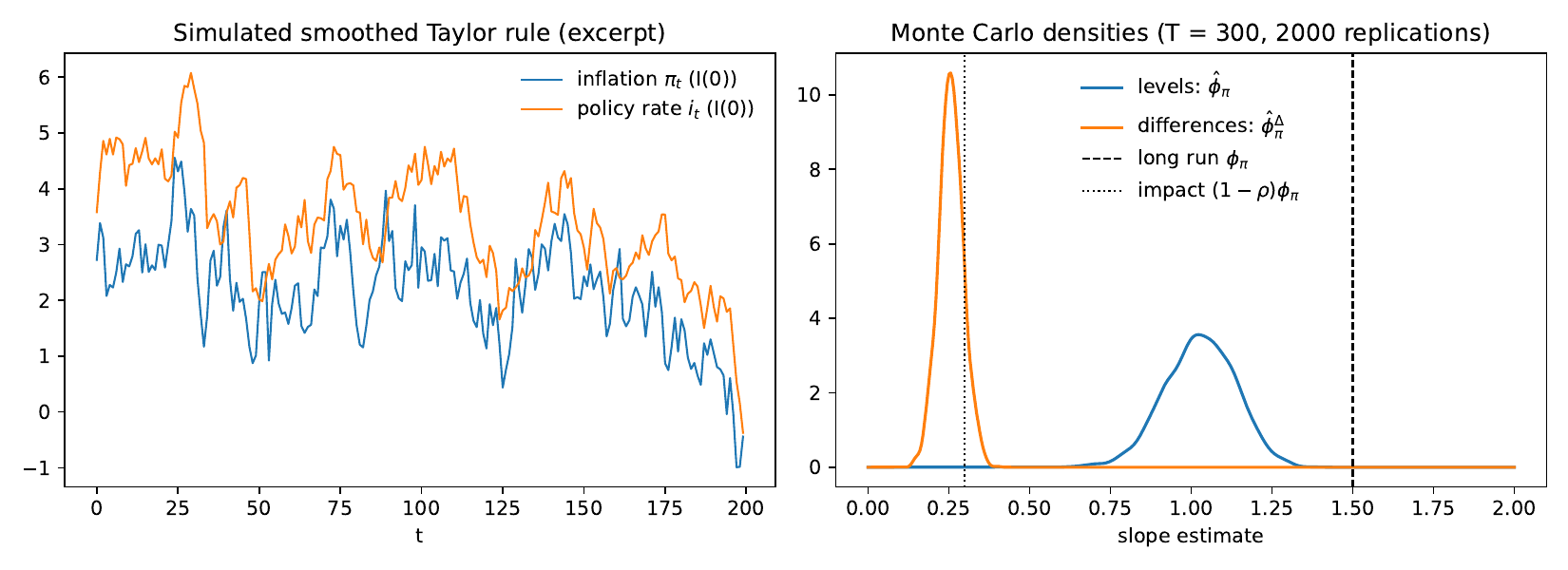}
\end{center}

The takeaway mirrors the I(1) discussion, with the logic reversed: differencing is the remedy for unit roots, not a harmless default. With stationary series, the choice between levels and differences is a choice of \emph{question} (long-run versus short-run comovement), not of statistical validity; and inference in either regression should rely on HAC standard errors when the series are persistent\footnote{Even with stationary variables, when both series are highly persistent the conventional t-statistic of the static levels regression can be severely oversized in finite samples; see Granger, Hyung and Jeon (2001), ``Spurious regressions with stationary series''.}.

\stepcounter{question}
\begin{itemize}
\item \textbf{Question \arabic{question}}
\begin{itemize}
\item Simulate model \eqref{eq:I0smoothing} and verify the plims in \eqref{eq:I0plims}. Set $\rho = 0$ and check that $\hat{\phi}^\Delta_\pi \approx \hat{\phi}_\pi$ and $\hat{c}' \approx 0$. Finally, estimate $i_t = c + \rho i_{t-1} + b \pi_t + \varepsilon_t$ and verify that $\hat{b}/(1-\hat{\rho})$ recovers the long-run $\phi_\pi$.
\end{itemize}
\end{itemize}

\subsubsection{Building two cointegrated time series}
We suggest to build two cointegrated time series with the system of equations:
\begin{equation}
\begin{cases}
y_t = \beta_y y_{t-1} + \beta_x x_{t} + \epsilon_t\\
x_t= x_{t-1} + \nu_t
\end{cases}
\end{equation}
with both $\epsilon_t$ and $\nu_t$ white noises.

We can show that both series are I(1) and cointegrated. Visually the link might not be obvious:
\begin{center}
\includegraphics[scale=0.5]{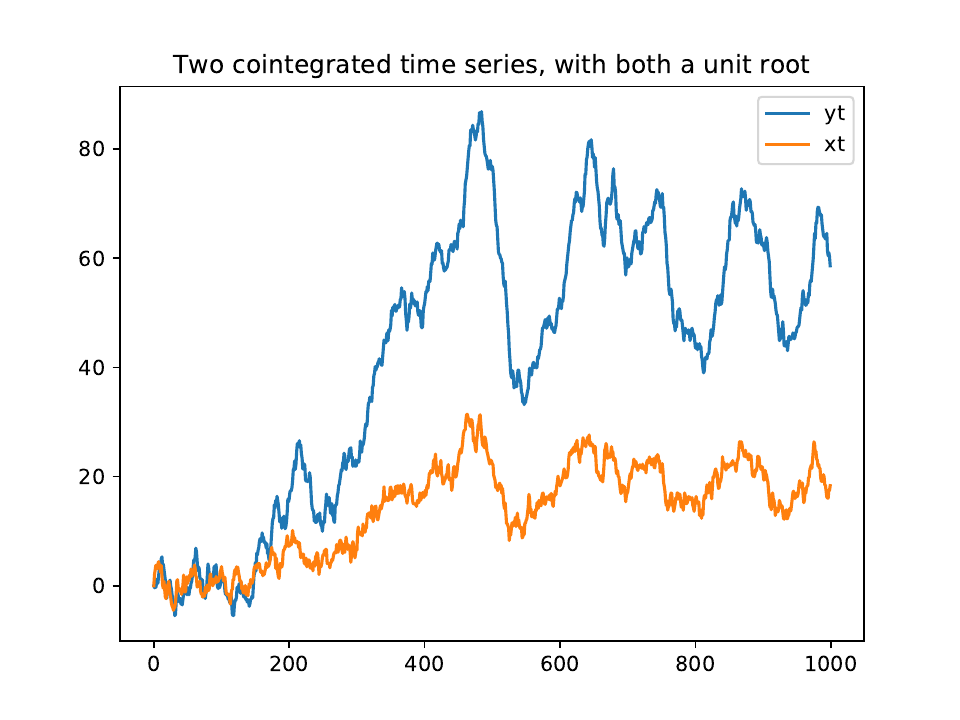}
\end{center}

Once we applied the cointegrating regression and found the coefficients, we can show the "forces de rappel" where there seem to be some forces that brings $y_t$ to "follow" $x_t$:
\begin{equation}
y_t = \gamma_0 +  \gamma_1 x_{t} + \chi_t\\
\end{equation}
\begin{center}
\includegraphics[scale=0.5]{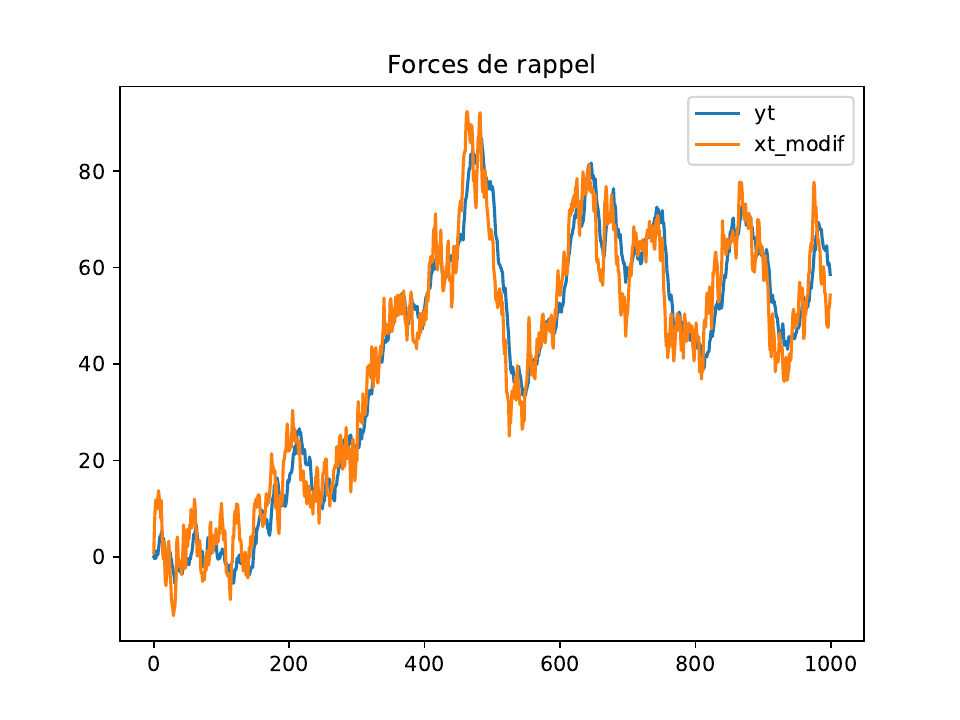}
\end{center}

We can then study the following Error Correction Model, with a grain of salt\footnote{e.g. "ECM system may be estimated by seemingly unrelated regressions (SUR) to increase efficiency if the number of lags in the two equations are different"} with regard to the consistency of the estimation by OLS as detailed in \citep{zivot2003}:
\begin{equation}
\begin{cases}
\Delta y_t = \alpha^y_1 +\alpha^y_2 \chi_{t-1} + \alpha^y_3 \Delta y_{t-1} + \alpha^y_4 \Delta x_{t-1}  + \epsilon^y_t\\
\Delta x_t = \alpha^x_1 +\alpha^x_2 \chi_{t-1} + \alpha^x_3 \Delta y_{t-1} + \alpha^x_4 \Delta x_{t-1}  + \epsilon^x_t\\
\end{cases}
\end{equation}

\subsubsection{Cointegrated variables - example}

We take the Euribor 1 year (FM.M.U2.EUR.RT.MM.EURIBOR1YD\_.HSTA) and the Euribor 3 months (FM.M.U2.EUR.RT.MM.EURIBOR3MD\_.HSTA) from the ECB statistical warehouse, check for cointegration between the two time series and build an ECM if relevant.

Noting that both time series are I(1) and cointegrated, there is no unit root in the error terms $\epsilon_t$ of the regression:
$$Y_t = \alpha + \beta X_t + \epsilon_t$$

Remember that if our model is correct, then $E(\epsilon_t) = 0$


\subsubsection{Error Correction Model - examples}

We first estimate the long-run relationship from the static levels regression (not spurious here, since the two series are cointegrated, but see section~\ref{sec:cointcoefinterpretation} for the caveats on interpreting and testing $\beta$):
$$Y_t = \alpha + \beta X_t + \epsilon_t$$

and we estimate the disequilibrium errors or the cointegrating residuals:
$$\hat{\epsilon}_t = Y_t - \hat{\alpha} - \hat{\beta} X_t$$

We then compute the error correction model were all variable are now stationary:
$$\Delta Y_t = \gamma_0 + \gamma_1 \Delta X_t + \gamma_2 \hat{\epsilon}_{t-1} + \nu_t$$

We can say that our model explains $78\%$ of the variance of the observed changes. And that \textbf{a one percentage point increase in the Euribor 3 months rate is associated, contemporaneously, with a $0.92$ percentage point increase in the Euribor 1 year rate} ($\gamma_1 = 0.92$ and is significantly different than $0$ with a t-value of $31$ well above $1.96$). Note that $\gamma_1$ is the \emph{short-run} (impact) effect; the \emph{long-run} equilibrium relationship between the two rates is captured by $\hat{\beta}$ from the cointegrating regression (section~\ref{sec:cointcoefinterpretation}). Since all the variables of the ECM are stationary, conventional OLS inference applies to the short-run coefficients\footnote{Using the estimated $\hat{\epsilon}_{t-1}$ instead of the true $\epsilon_{t-1}$ is asymptotically innocuous here because $\hat{\beta}$ is superconsistent.}.

\stepcounter{question}
\begin{itemize}
\item \textbf{Question \arabic{question}}
\begin{itemize}
\item In the ECM above, add the lagged variables of the expenditure and absorption, are the results still holding? $\Delta Y_t = \gamma_0 + \gamma_1 \Delta X_t + \gamma_2 \hat{\epsilon}_{t-1} + \gamma_3 \Delta Y_{t-1} + \gamma_4 \Delta X_{t-1} + \nu_t$
\end{itemize}
\end{itemize}

\subsubsection{Error Correction Model - consumption and income}

Noting that the permanent income model implies cointegration between consumption and income, we use some proxy data from the Penn World Table and focus on data for Italy:
\begin{itemize}
\item cda: Real domestic absorption, (real consumption plus investment), at current PPPs (in mil. 2011USD)
\item cgdpe: Expenditure-side real GDP at current PPPs (in mil. 2011USD)
\end{itemize}


\subsubsection{Before regressing two I(1) time series, test for cointegration}

We test the French population and GDP for cointegration. As they don't seem cointegrated we perform a linear regression.

\subsubsection{Linear regression}\label{sec:linreg}

Now we want to perform a simple linear\footnote{For non linear regression, the reader can start \href{https://scipy-cookbook.readthedocs.io/items/robust_regression.html}{with this reference}} regression of population change on French GDP change:
\begin{equation}\label{eq:popgdpreg}
\Delta Y_{\mbox{Pop},t} =\alpha + \beta \Delta X_{\mbox{GDP},t} + \epsilon_t
\end{equation}

We perform an Ordinary Least Square method.

\subsubsection{Visuals}
We can store the $\alpha$ and $\beta$ value of that regression for future use and display our regression line over a scatter plot of the original data:
\begin{center}
\includegraphics[scale=0.5]{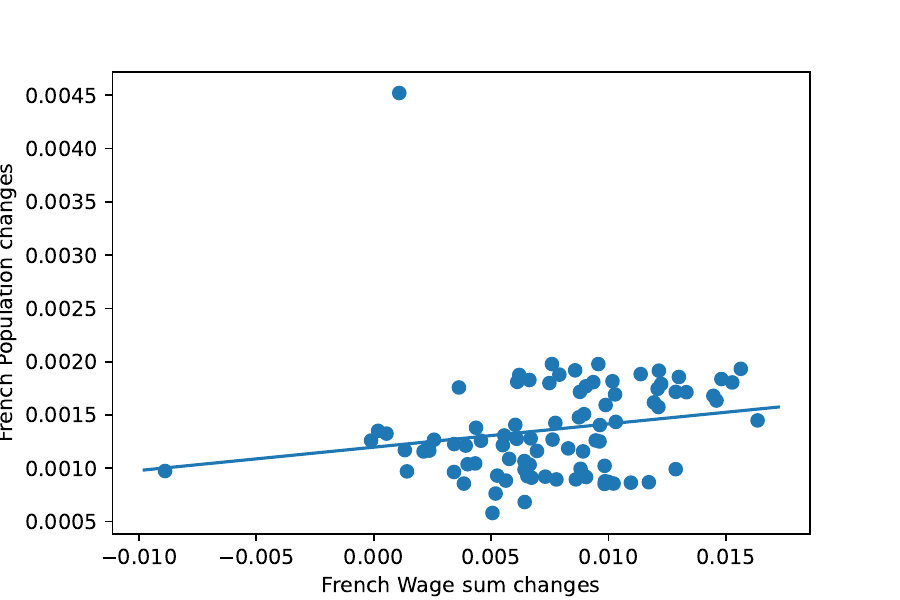}
\end{center}

In our visual of the regression, we can identify visually at least two outliers. We would need to check for the robustness of our model if we remove those two outliers as a first robustness check.

\subsubsection{Outlier detection and regression coefficient robustness}
We identified visually two potential outliers\footnote{To go further on the topic of outliers and regressions, refer to \href{http://polisci.msu.edu/jacoby/icpsr/regress3/lectures/week3/11.Outliers.pdf}{this lecture} and \citet{Krasker:1983aa}}. We want to determine if they influence the regression coefficients out of proportion. We identify one outlier with the minimum GDP growth rate and one with the maximum population growth rate.

We create a dummy variable $\mbox{Dummy}$ that has the value $0$ for all observation except for the outlier one where the value is $1$, we then do a new regression:

\begin{equation}
\Delta Y_{\mbox{Pop},t} = \alpha + \beta \Delta X_{\mbox{GDP},t} +  \gamma \mbox{Dummy}_t
\end{equation}

We test for both outliers if the $\gamma$ is significantly different to $0$. Instead of using the typical t-Student threshold $t_{1-\frac{\alpha}{2},n-p-1}$ we rather use the Bonferroni correction t-Student threshold: $t_{1-\frac{\alpha}{2n},n-p-1}$

We find that one of the outliers coefficient is significantly different to $0$ and we remove it from the observations before running a new linear regression and notice that the GDP growth rate sensitivity to the population growth rate is significantly different and the explanatory power of our model is improved.

\subsection{Q-Q plots of our linear regressions' normalized residuals against a normal law}

From the orginal regression, we can check whether the residuals are normally distributed:
\begin{center}
\includegraphics[scale=0.5]{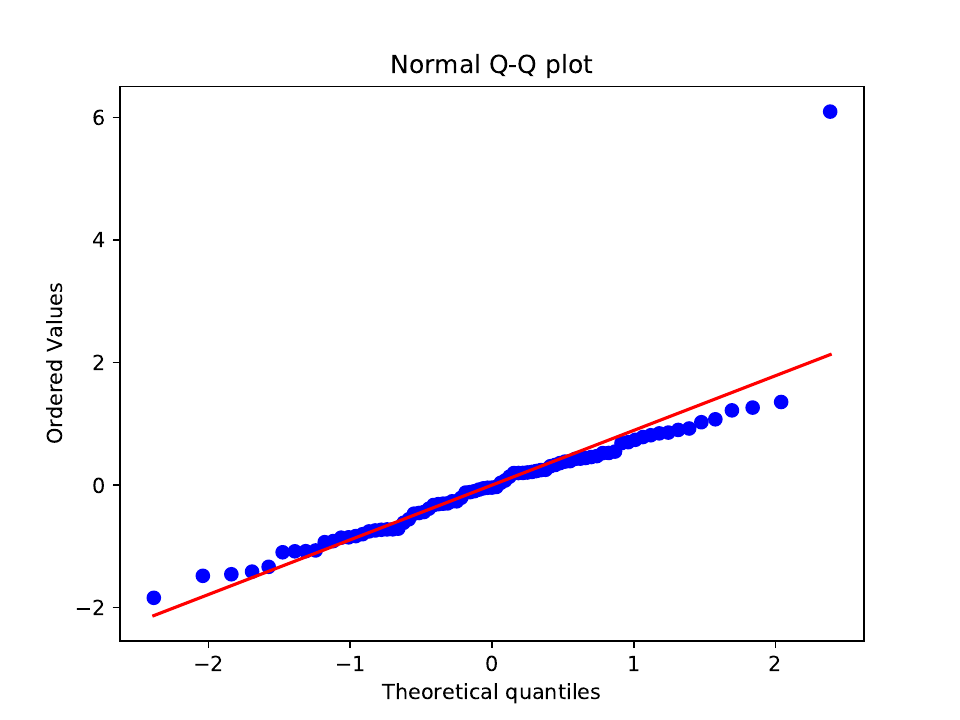}
\end{center}

We can see that if we remove the outlier number 2, then our residuals seems to be normally distributed and the model improved:
\begin{center}
\includegraphics[scale=0.5]{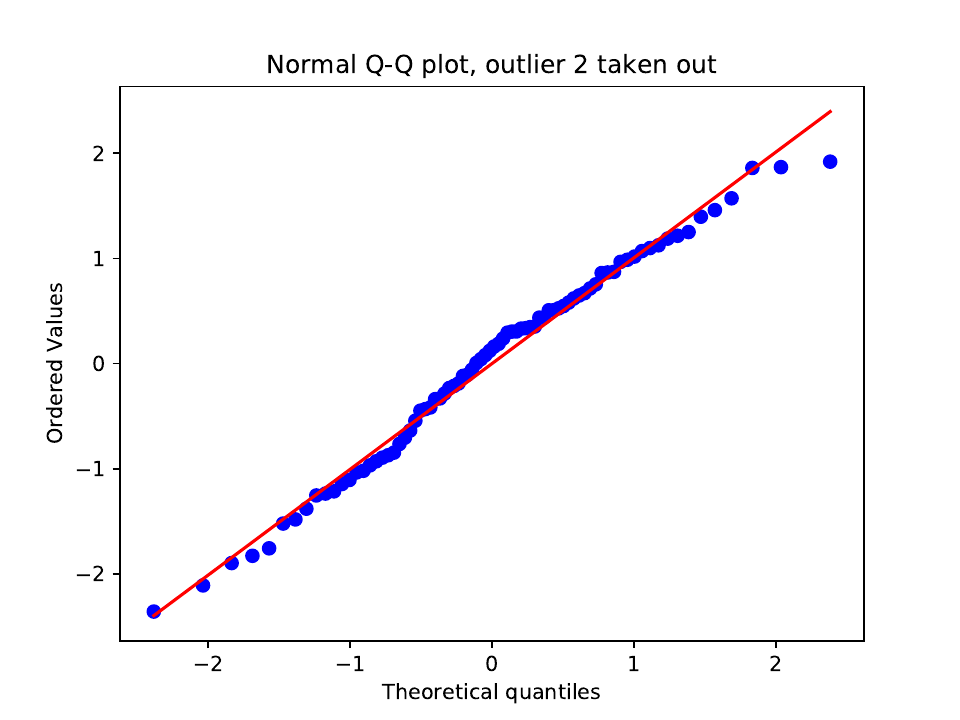}
\end{center}

\newpage

\subsection{Robust regression}\label{subsec:robust_regression}

We complement OLS with robust regression methods\footnote{Code: \texttt{vansteenberghe\_robust\_regressions.py}.}. Table~\ref{table:quantpop} reports OLS regressions in columns (1)--(2) and robust linear models (RLM) based on M-estimators in columns (3)--(4). In addition, we illustrate a \emph{consensus-based} approach (RANSAC) that explicitly separates inliers from outliers.

In our application, we include a dummy variable equal to $1$ for the atypical observation where the change in French population is unusually high (and $0$ otherwise). This allows us to diagnose whether results are driven by that observation, and to compare (i) \emph{explicit outlier control} (adding the dummy) to (ii) \emph{implicit robustness} (downweighting or excluding atypical points).

\begin{center}
\includegraphics[scale=0.5]{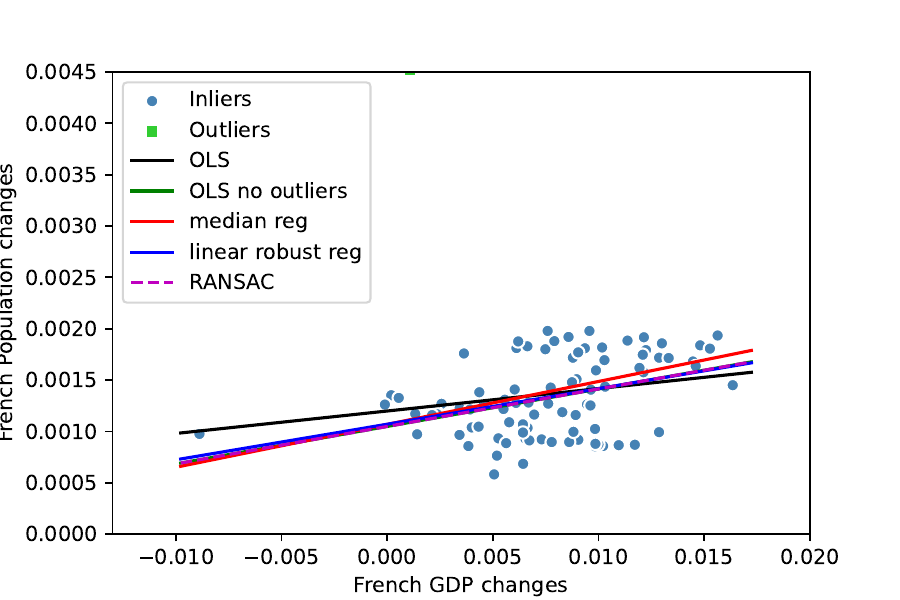}
\end{center}

\paragraph{Setup.}
Let $\{(x_i,y_i)\}_{i=1}^n$ denote observations with $x_i=\text{GDP}_i$ and $y_i=\text{Population}_i$.
We consider the linear model
\begin{equation}
y_i = \alpha + \beta x_i + \varepsilon_i, \qquad i=1,\dots,n,
\end{equation}
where $\varepsilon_i$ captures deviations from the linear relation. Let $r_i(\alpha,\beta)=y_i-\alpha-\beta x_i$ denote the residual.

\subsubsection{OLS: squared-loss and sensitivity to outliers}

In Section~\ref{sec:linreg}, we saw that OLS solves a least-squares problem:
\begin{equation}\label{eq:ols_obj}
(\hat{\alpha},\hat{\beta}) \in \arg\min_{\alpha,\beta}\sum_{i=1}^n r_i(\alpha,\beta)^2.
\end{equation}
Equivalently, OLS is an M-estimator with loss $\rho(u)=u^2$. The squared loss heavily penalizes large residuals: if one observation has a residual $10$ times larger than another, it receives $100$ times more weight in \ref{eq:ols_obj}. This is efficient under homoskedastic Gaussian errors, but it makes OLS \emph{non-robust}: a small number of aberrant observations can strongly affect $(\hat{\alpha},\hat{\beta})$.

\paragraph{Dummy for the outlier.}
A simple diagnostic is to include a dummy $d_i$ for the suspected outlier:
\begin{equation}\label{eq:dummy_model}
y_i = \alpha + \beta x_i + \gamma d_i + \varepsilon_i.
\end{equation}
If $d_i$ flags a single observation, then $\gamma$ absorbs a one-off level shift for that point, often stabilizing $\hat{\beta}$ relative to the baseline model.

\subsubsection{Median (quantile) regression: absolute-loss at the median}

Quantile regression generalizes least squares by targeting conditional quantiles rather than conditional means. For a quantile index $\tau\in(0,1)$, the quantile regression estimator solves
\begin{equation}\label{eq:qr_obj}
(\hat{\alpha}_\tau,\hat{\beta}_\tau)\in\arg\min_{\alpha,\beta}\sum_{i=1}^n \rho_\tau\!\left(r_i(\alpha,\beta)\right),
\qquad 
\rho_\tau(u)=u\left(\tau-\mathbf{1}\{u<0\}\right),
\end{equation}
where $\rho_\tau$ is the \emph{check (pinball) loss}. For $\tau=0.5$ (the median), this reduces to an absolute-loss criterion:
\begin{equation}\label{eq:median_obj}
(\hat{\alpha}_{0.5},\hat{\beta}_{0.5})\in\arg\min_{\alpha,\beta}\sum_{i=1}^n \lvert r_i(\alpha,\beta)\rvert.
\end{equation}
Because $\lvert r\rvert$ grows linearly rather than quadratically, median regression is less sensitive to large residuals than OLS. Intuitively, the fitted line is chosen so that about half the residuals are positive and half are negative, and extreme points do not dominate the criterion as they do under squared loss.

Columns (3)--(4) of Table~\ref{table:quantpop} compare the baseline median slope and the median slope net of the flagged outlier.

\subsubsection{Robust Linear Model (RLM): M-estimation and Huber loss}

Robust linear models in \texttt{statsmodels} are implemented as \emph{M-estimators}. The general M-estimation problem is:
\begin{equation}\label{eq:mest_obj}
(\hat{\alpha},\hat{\beta})\in\arg\min_{\alpha,\beta}\sum_{i=1}^n \rho\!\left(\frac{r_i(\alpha,\beta)}{s}\right),
\end{equation}
where:
\begin{itemize}
\item $\rho(\cdot)$ is a robust loss function (less aggressive than $u^2$ in the tails),
\item $s>0$ is a \emph{scale} parameter used to standardize residuals (so that the notion of ``large'' residual is unit-free).
\end{itemize}
A canonical choice is the \emph{Huber loss}:
\begin{equation}\label{eq:huber_rho}
\rho_k(u)=
\begin{cases}
\frac{1}{2}u^2, & \text{if } |u|\le k,\\[4pt]
k|u|-\frac{1}{2}k^2, & \text{if } |u|>k,
\end{cases}
\end{equation}
which behaves like least squares for small standardized residuals ($|u|\le k$), but switches to an absolute-loss regime for large residuals ($|u|>k$). Hence, outliers are \emph{downweighted} rather than allowed to dominate.

\paragraph{From minimization to iteratively reweighted least squares (IRLS).}
Huber M-estimation can be computed by IRLS. Define $\psi_k(u)=\rho_k'(u)$ and weights
\begin{equation}\label{eq:huber_weights}
w_k(u)=\frac{\psi_k(u)}{u}=
\begin{cases}
1, & |u|\le k,\\[4pt]
\frac{k}{|u|}, & |u|>k.
\end{cases}
\end{equation}
At each iteration, one (i) computes standardized residuals $u_i=r_i/s$, (ii) assigns weights $w_i=w_k(u_i)$, and (iii) runs a weighted least squares regression. Large residuals (large $|u_i|$) receive weights below $1$.

\paragraph{The tuning constant $k$ and the role of MAD.}
The constant $k$ controls robustness: smaller $k$ downweights more aggressively, larger $k$ is closer to OLS. A conventional value for Huber is $k=1.345$, which (under Gaussian errors) yields about 95\% efficiency relative to OLS.
Because residuals must be standardized, a robust scale estimate is needed. A common choice is the median absolute deviation (MAD),
\[
\mathrm{MAD} = \mathrm{median}_i\Bigl(\,|r_i-\mathrm{median}_j(r_j)|\,\Bigr),
\]
and, under normality, $\mathrm{MAD}/0.6745$ is a consistent estimator of the standard deviation. Therefore, one often sets
\begin{equation}\label{eq:scale_mad}
s \approx \frac{\mathrm{MAD}}{0.6745},
\end{equation}
so that $u=r/s$ is comparable across samples. In practice, \texttt{statsmodels} performs this scaling internally for \texttt{RLM}, and the Huber threshold $k=1.345$ is applied to these standardized residuals.

\paragraph{Implementation in Python (statsmodels).}
\texttt{statsmodels.RLM} expects a design matrix $X$; unlike formula-based OLS, we add an explicit intercept. This estimates the robust M-estimator \ref{eq:mest_obj} with $\rho=\rho_k$ in \ref{eq:huber_rho}. Adding the dummy corresponds to replacing $X=[\mathbf{1},x]$ with $X=[\mathbf{1},x,d]$. Comparing these two RLM estimates clarifies whether robustness alone is sufficient, or whether the flagged observation induces a distinct level shift better captured by \ref{eq:dummy_model}.

\subsubsection{RANSAC: consensus estimation by explicit inlier selection}

RANSAC (\emph{RANdom SAmple Consensus}) follows a different philosophy: rather than downweighting outliers smoothly, it attempts to \emph{identify} inliers and fit the model on a subset that is consistent with the dominant linear relation.

\paragraph{Algorithm (linear regression case).}
A simplified description is:
\begin{enumerate}
\item Randomly draw a minimal subset of points (for a line: a small subset sufficient to estimate $\alpha,\beta$).
\item Fit the model on this subset.
\item Compute residuals for all observations and classify as \emph{inliers} those with $|r_i|\le \tau$, where $\tau$ is a residual threshold.
\item Refit the model using all inliers; keep the model with the largest inlier set (or best inlier fit).
\end{enumerate}
The method is robust when the data contain a large fraction of gross outliers, provided there exists a majority inlier structure.

\paragraph{Implementation in Python (scikit-learn).}
In the code, \texttt{residual\_threshold} corresponds to the inlier cutoff $\tau$: observations with residuals below this threshold are treated as inliers. The printed slope and intercept come from the final refitted inlier model stored in \texttt{ransac.estimator\_}.

\paragraph{How to interpret differences across methods.}
\begin{itemize}
\item OLS estimates the \emph{conditional mean} slope but can be pulled by extreme observations.
\item Median regression estimates the \emph{conditional median} slope, reducing sensitivity to tail observations.
\item RLM (Huber) estimates a robust mean-like relationship by downweighting large standardized residuals via \ref{eq:huber_weights}.
\item RANSAC estimates the slope of the \emph{dominant inlier relationship} by selecting a consensus subset; results can change if the residual threshold is too tight/loose.
\end{itemize}
In practice, agreement between OLS, median regression, RLM, and RANSAC strengthens credibility; divergence indicates that atypical observations materially affect the fitted relation and should be investigated (data quality, structural breaks, or genuinely rare events).

\begin{table}[!htbp] \centering
\resizebox{\textwidth}{!}{ 
\begin{tabular}{@{\extracolsep{5pt}}lcccc}
\\[-1.8ex]\hline
\hline \\[-1.8ex]
& \multicolumn{4}{c}{\textit{Dependent variable: Population}} \
\cr \cline{2-5}
\\[-1.8ex] & (1) & (2) & (3) & (4) \\
\hline \\[-1.8ex]
 GDP & 0.022$^{*}$ & 0.037$^{***}$ & 0.042$^{***}$ & 0.042$^{***}$ \\
& (0.013) & (0.009) & (0.014) & (0.014) \\
 Intercept & 0.001$^{***}$ & 0.001$^{***}$ & 0.001$^{***}$ & 0.001$^{***}$ \\
& (0.000) & (0.000) & (0.000) & (0.000) \\
 dummy & & 0.003$^{***}$ & & 0.003$^{***}$ \\
& & (0.000) & & (0.001) \\
\hline \\[-1.8ex]
 Observations & 89 & 89 & 89 & 89 \\
 $R^2$ & 0.033 & 0.540 & nan & nan \\
 Adjusted $R^2$ & 0.022 & 0.529 & nan & nan \\
 Pseudo $R^2$ &  &  & 0.099 & 0.208 \\
 Residual Std. Error & 0.000 (df=87) & 0.000 (df=86) & 0.001 (df=87) & 0.000 (df=86) \\
 F Statistic & 2.955$^{*}$ (df=1; 87) & 50.403$^{***}$ (df=2; 86) & nan$^{***}$ (df=1; 87) & nan$^{***}$ (df=2; 86) \\
\hline
\hline \\[-1.8ex]
\textit{Notes:} & \multicolumn{4}{p{0.75\textwidth}}{\footnotesize 
Columns (1)--(2) report OLS estimates of population growth on GDP growth; column (2) adds a dummy equal to one for the largest population-growth observation. Columns (3)--(4) report median (quantile, $q=0.5$) regressions with and without the same dummy. OLS columns display $R^2$ and the F-statistic; quantile regressions report a pseudo-$R^2$. Robust standard errors in parentheses. $^{*}p<0.1$, $^{**}p<0.05$, $^{***}p<0.01$.} \\
\end{tabular}
}
\caption{OLS and median (quantile) regressions}\label{table:quantpop}
\end{table}

\paragraph{Economic interpretation and robustness.}

Across specifications, the estimated slope on GDP lies in a narrow interval between $0.03$ and $0.04$. Interpreting both variables in percentage changes, this coefficient can be read as an elasticity: a $1\%$ increase in French GDP is associated with an increase in French population growth of roughly $0.03$--$0.04\%$, ceteris paribus. 

Importantly, this magnitude is remarkably stable across (i) OLS on the trimmed sample, (ii) median (quantile) regression, and (iii) the Huber-based robust linear model. Such stability suggests that the positive association between GDP growth and population change is not mechanically driven by the single atypical observation. In other words, the relationship appears to reflect the central tendency of the data rather than the influence of extreme residuals.

For comparison, the RANdom SAmple Consensus (RANSAC) estimator delivers a slope close to $0.04$, which is economically and statistically comparable to the other robust estimates. However, unlike M-estimators or quantile regression, RANSAC relies on a hard classification between inliers and outliers determined by a residual threshold parameter. As a result, its estimates can vary non-negligibly when this threshold is modified. 

Overall, the convergence of OLS (after trimming), median regression, and Huber RLM reinforces the credibility of the estimated elasticity, while the RANSAC exercise illustrates both the usefulness and the sensitivity of consensus-based robust methods.

\section{Monetary Policy and Endogeneity}\label{sec:mpendog}

Before turning to multivariate systems, we pause on a single number that has occupied
empirical monetary economics for three decades: the strength with which a central bank
raises its policy rate when inflation rises. The previous section warned that a regression
between two persistent series can be spurious (Section~\ref{sec:cointegration}); this
section adds a second, logically distinct warning---\emph{endogeneity}---and shows that
even a perfectly stationary reaction function cannot be read off an ordinary least squares
(OLS) regression. The resolution, high-frequency ``surprises'' used as instruments, is
also the natural bridge to the structural VAR machinery of the next section and to the
external-instrument identification of Section~\ref{sec:wrapup_monetary}.\footnote{The
high-frequency and external-instrument identification discussed here is implemented in the
companion codes \texttt{monetary\_policy\_SVAR\_IV\_vansteenberghe.py} and
\texttt{macro\_model\_VAR\_vansteenberghe.py}.}

\subsection{The central question: the reaction to inflation, $\phi_\pi$}

Modern central banks are best described as setting a very short nominal interest rate in
response to the state of the economy. The canonical description is the \citet{taylor1993discretion}
rule: writing $i_t$ for the policy rate, $\pi_t$ for inflation, $\pi^\ast$ for the target,
$x_t$ for the output gap, $r^\ast$ for the natural real rate, and $\varepsilon_t$ for a
\emph{monetary-policy shock} (the part of the rate not explained by the rule),
\begin{equation}\label{eq:mptaylor}
i_t = r^\ast + \pi^\ast + \phi_\pi\,(\pi_t-\pi^\ast) + \phi_x\, x_t + \varepsilon_t .
\end{equation}
The coefficient of interest is $\phi_\pi$, the systematic response of the policy rate to
inflation. It carries a sharp theoretical meaning. Because \eqref{eq:mptaylor} is written
in nominal terms, the \emph{real} rate rises with inflation only if $\phi_\pi>1$; this is
the \citet{taylor1993discretion} principle, and in the New Keynesian model it is the condition for a
determinate, non-explosive equilibrium. Estimating whether $\phi_\pi$ is above or below one
is therefore not an academic curiosity: it is a test of whether policy is stabilising.

Two refinements matter for what follows. First, central banks are forward-looking:
\citet{clarida2000monetary} replace realised inflation by expected future inflation and add
interest-rate smoothing,
\begin{equation}\label{eq:mpcgg}
i_t = (1-\rho)\Big[\, i^\ast + \phi_\pi\big(\E_t \pi_{t+k}-\pi^\ast\big) + \phi_x\,\E_t x_{t}\,\Big] + \rho\, i_{t-1} + \varepsilon_t ,
\end{equation}
with $0\le\rho<1$. Second, the object in \eqref{eq:mptaylor}--\eqref{eq:mpcgg} is a
\emph{structural} reaction function: $\varepsilon_t$ is an exogenous policy disturbance,
orthogonal to the systematic response, and $\phi_\pi$ is the causal slope we would trace
out by exogenously varying inflation and watching how the central bank responds.

\subsection{Why $\phi_\pi$ is hard to estimate: simultaneity}

The temptation is to regress $i_t$ on $\pi_t$ (and $x_t$) by OLS and read off $\phi_\pi$.
This fails, and the reason is not measurement or functional form but \emph{simultaneity}:
inflation is not exogenous to the interest rate. The central bank moves rates in response
to inflation, but rates in turn move inflation through the transmission mechanism (the
IS/Phillips block). We are trying to recover one structural relationship from data that are
the equilibrium of two, exactly the identification problem met when correlation is
contrasted with causation and in the estimation of demand and supply. A cloud of
$(\pi_t,i_t)$ points is generated by \emph{both} curves at once, and OLS fits a single line
through it.

Formally, pair the reaction function with a stylised aggregate-demand/transmission
equation. Keep only inflation for transparency and write both relations in deviations from
means:
\begin{align}
\text{(policy / reaction)}\qquad & i_t = \phi_\pi\, \pi_t + \varepsilon_t, && \phi_\pi>0, \label{eq:mpreact}\\[2pt]
\text{(economy / transmission)}\qquad & \pi_t = -\gamma\, i_t + u_t, && \gamma>0. \label{eq:mpad}
\end{align}
Here $\gamma>0$ encodes that a tighter policy rate lowers inflation, and $u_t$ collects all
\emph{non-policy} disturbances (demand, cost-push, supply) to which the bank reacts. Assume
the two shocks are mutually and serially uncorrelated, with $\E[\varepsilon_t u_t]=0$,
$\Var(\varepsilon_t)=\sigma_\varepsilon^2$, $\Var(u_t)=\sigma_u^2$. Solving the two
equations simultaneously for inflation gives the reduced form
\begin{equation}\label{eq:mppireduced}
\pi_t = \frac{u_t - \gamma\,\varepsilon_t}{1+\gamma\phi_\pi}.
\end{equation}
Equation~\eqref{eq:mppireduced} is the crux. The regressor $\pi_t$ in the reaction
function \eqref{eq:mpreact} is itself a function of the very error $\varepsilon_t$ that the
regression treats as noise:
\begin{equation}\label{eq:mpcov}
\operatorname{Cov}(\pi_t,\varepsilon_t) = \frac{-\gamma\,\sigma_\varepsilon^2}{1+\gamma\phi_\pi} \;<\; 0 .
\end{equation}
The regressor is endogenous, and---this is the point the exercise turns on---the sign of
the correlation is \emph{negative}.

\subsection{The expected sign of the bias: attenuation, and beyond}

Because $\operatorname{Cov}(\pi_t,\varepsilon_t)\neq 0$, OLS is inconsistent. Using
\eqref{eq:mppireduced} and \eqref{eq:mpcov}, its probability limit is
\begin{equation}\label{eq:mpbias}
\operatorname{plim}\ \hat\phi_\pi^{\,\mathrm{OLS}}
= \phi_\pi + \frac{\operatorname{Cov}(\pi_t,\varepsilon_t)}{\Var(\pi_t)}
= \phi_\pi - \frac{\gamma\,\sigma_\varepsilon^2\,(1+\gamma\phi_\pi)}{\sigma_u^2 + \gamma^2\sigma_\varepsilon^2}
\;<\; \phi_\pi .
\end{equation}
The bias term is unambiguously negative: \textbf{the endogeneity of inflation biases the
estimated inflation response \emph{downward}}. The intuition is mechanical and worth
stating in words. A contractionary policy shock ($\varepsilon_t>0$) raises $i_t$ and, along
the transmission curve \eqref{eq:mpad}, simultaneously \emph{lowers} $\pi_t$. Such episodes
place high interest rates next to low inflation, which is exactly the co-movement of a
\emph{weak} anti-inflation response. OLS cannot tell this reverse-causation channel from
the central bank's genuine reaction, so it splits the difference and understates
$\phi_\pi$. A central bank that in truth respects the Taylor principle can look, through the
OLS lens, as if it violates it.

How large is the bias? Equation~\eqref{eq:mpbias} makes the answer transparent through two
limiting cases:
\begin{itemize}
\item \textbf{Non-policy shocks dominate} ($\sigma_\varepsilon^2/\sigma_u^2 \to 0$). The
bias term vanishes and $\operatorname{plim}\ \hat\phi_\pi^{\,\mathrm{OLS}} \to \phi_\pi$.
When almost all of the variation in inflation comes from disturbances the bank
\emph{reacts to} rather than \emph{causes}, the data trace out the reaction function and
OLS is (nearly) consistent.
\item \textbf{Policy shocks dominate} ($\sigma_u^2/\sigma_\varepsilon^2 \to 0$). Then
$\operatorname{plim}\ \hat\phi_\pi^{\,\mathrm{OLS}} \to -1/\gamma < 0$. When most of the
variation in inflation is \emph{caused by} policy, the data trace out the transmission
curve \eqref{eq:mpad}, and OLS recovers its negative slope---the wrong equation entirely.
\end{itemize}
The estimated slope is thus a variance-weighted blend of the reaction-function slope
$\phi_\pi$ and the transmission slope $-1/\gamma$, the weights being the shares of the two
shocks in the variance of inflation. This is precisely the modern reading of
\citet{carvalho2021taylor}: the asymptotic OLS bias in a Taylor rule is proportional to the
fraction of the regressors' variance that is due to monetary-policy shocks. Their empirical
verdict is reassuring---because policy shocks explain only a small share of the variance of
inflation and the output gap, the bias is quantitatively modest, and OLS and instrumental
estimates often coincide---but it is reassuring \emph{conditionally}: the argument rests on
the maintained assumption that policy shocks are a small part of macroeconomic volatility,
which need not hold in every sample or every episode (the 1970s, or a sharp policy
tightening, are natural exceptions).

Simultaneity is the deepest reason $\phi_\pi$ is hard to estimate, but not the only one, and
the others compound it. The bank responds to \emph{real-time} and \emph{forecast} data that
are heavily revised ex post \citep[the output gap is the notorious case;][]{coibion2012can};
the coefficients themselves \emph{drift} across regimes, so a fixed-coefficient regression
averages a moving target \citep{primiceri2005,cogley2005drifts};
and the regression can be spurious if the series are treated in non-cointegrated levels.
Endogeneity, however, is the one that survives even with perfect data and stable parameters.

\subsection{The inflation surprise as an instrument}

The textbook cure for an endogenous regressor is an instrument. We need a variable $z_t$
that is \emph{relevant}---correlated with inflation, $\operatorname{Cov}(z_t,\pi_t)\neq
0$---and \emph{valid}---uncorrelated with the monetary-policy shock,
$\operatorname{Cov}(z_t,\varepsilon_t)=0$. Economically, $z_t$ must shift inflation for
reasons \emph{other than} policy: it must load on the non-policy disturbance $u_t$ but be
orthogonal to $\varepsilon_t$. An instrument that shifts the economy's curve while leaving
the policy shock untouched traces out the reaction function, just as a supply shifter
identifies a demand curve.

The state-of-the-art construction of such an instrument is the \emph{inflation surprise}.
Define the surprise as the innovation in an inflation data release relative to the market's
prior expectation, typically the median professional forecast recorded just before the
release:
\begin{equation}\label{eq:mpsurprise}
s_t \;=\; \pi_t^{\text{release}} - \E\big[\pi_t^{\text{release}}\mid \mathcal{F}_{t^-}\big],
\end{equation}
where $\mathcal{F}_{t^-}$ is the information set an instant before the announcement. Under
efficient forecasts, $s_t$ is \emph{news}: by construction it is orthogonal to everything
dated before the release, and in particular to the systematic component of policy and to
the policy shock $\varepsilon_t$, which is realised at the policy decision, not at the
statistical office. Yet the surprise plainly moves inflation and inflation expectations, so
it is relevant. It is, therefore, a valid instrument for inflation in the reaction function.

This is the logic of \citet{hamiltonpruittborger2011}, who estimate the
\emph{market-perceived} monetary policy rule by relating macroeconomic news to the revision
in market-based forecasts of both the policy rate and economic conditions. Their
identification is an event study: in a narrow window around a data release, essentially the
only thing moving is the surprise, so the co-movement of the expected policy path with $s_t$
isolates the perceived response coefficient, free of the reverse-causation and slow-moving
confounders that plague a levels regression. Concretely, with a first-stage relevance
$\operatorname{Cov}(\Delta\pi_t, s_t)\neq0$, the instrumental-variables (Wald) estimand is
\begin{equation}\label{eq:mpiv}
\phi_\pi^{\,\mathrm{IV}} = \frac{\operatorname{Cov}(\Delta i_t,\, s_t)}{\operatorname{Cov}(\Delta \pi_t,\, s_t)},
\end{equation}
the ratio of the rate response to the inflation response, both measured over the release
window; this is exactly two-stage least squares with $s_t$ instrumenting inflation.

The inflation surprise is one member of a broader family of high-frequency and narrative
instruments developed to isolate exogenous variation around policy and data events:
\begin{itemize}
\item \textbf{Fed funds futures surprises} \citep{kuttner2001monetary}: the jump in the
futures-implied rate in a tight window around an FOMC decision isolates the
\emph{unanticipated} component of policy.
\item \textbf{Asset-price responses to announcements} \citep{gurkaynak2007us}: intraday
changes in a panel of yields around policy communications summarise the surprise; the
euro-area analogue is the EA-MPD dataset.
\item \textbf{Narrative shocks} \citep{romerromer2004}: intended rate changes are purged of
the Federal Reserve's own real-time (Greenbook) forecasts, leaving a residual plausibly
orthogonal to the state of the economy.
\item \textbf{Information-robust shocks} \citep{jarocinski2020deconstructing}: the
co-movement of interest rates and stock prices around announcements is used to separate a
pure policy component from a central-bank ``information'' component.
\end{itemize}

Two caveats keep this honest, and both are active research frontiers. First, the validity
of the surprise can fail through a \emph{central-bank information effect}: if an announcement
or a release reveals information about fundamentals, rates and the economy move together for
non-policy reasons, contaminating the exclusion restriction
\citep{nakamura2018high}. \citet{bauerswanson2023} show that once a pre-event drift in asset
prices is controlled for, much of the apparent information effect disappears, restoring
confidence in narrow-window surprises---but the debate is not closed. Second, even a valid
instrument need not \emph{point-identify} $\phi_\pi$: Taylor-rule coefficients are notoriously
\emph{weakly identified}, so that conventional standard errors are unreliable and the Taylor
principle cannot simply be estimated and tested \citep{cochrane2011determinacy,mavroeidis2010monetary}. Best
practice is therefore identification-robust inference---reporting Anderson--Rubin confidence
\emph{sets} (which may be wide, disjoint, or unbounded) rather than a point estimate with a
$t$-statistic. This is not a counsel of despair but of discipline: it states honestly what
the data can and cannot pin down about the central bank's response to inflation.

\subsection{Oil supply shocks and the demand/supply problem}\label{sec:mpoil}

A natural candidate for the instrument $z_t$ of the previous subsection is an \emph{oil price
shock}. Energy is a direct input to the price level, so a jump in the price of crude is a
textbook cost-push shifter of inflation---it moves the economy's curve \eqref{eq:mpad}
through the non-policy disturbance $u_t$---and, for an economy that is small in the world oil
market, it looks exogenous to the domestic monetary-policy shock $\varepsilon_t$. This is
also the oldest identified macro shock: \citet{hamilton1983} showed that oil price increases
preceded almost every U.S. recession since 1945, and \citet{hamilton2003} refined the object
to a nonlinear ``net'' oil-price increase. The temptation is therefore to instrument
inflation with the change in the price of oil.

\paragraph{Why the raw oil price is not a valid instrument.} The temptation must be resisted,
because the price of oil is itself an \emph{endogenous} equilibrium object. A large part of
its variation reflects \emph{global aggregate demand}: when world activity booms, the demand
for crude rises and so does its price. That same global boom raises domestic activity and
inflation and is exactly the kind of common disturbance the central bank reacts to, so the
demand component of the oil price is correlated with $u_t$ \emph{and} with the state that
drives $\varepsilon_t$, violating the exclusion restriction. Decompose the (log) real oil
price change into structurally distinct, mutually orthogonal components,
\begin{equation}\label{eq:mpoildecomp}
\Delta o_t = \underbrace{\theta_s\, w^{s}_t}_{\text{supply}}
           + \underbrace{\theta_d\, w^{d}_t}_{\text{aggregate demand}}
           + \underbrace{\theta_o\, w^{o}_t}_{\text{oil-specific demand}} .
\end{equation}
Only the \emph{supply} shock $w^{s}_t$ is a clean cost-push instrument, plausibly orthogonal
to $\varepsilon_t$; the aggregate-demand shock $w^{d}_t$ is precisely what makes $\Delta o_t$
invalid. Using the raw oil price re-injects the endogeneity of the reaction function: one
must first \emph{isolate the supply component}.

\paragraph{The main paper: \citet{kilian2009}.} The reference that makes
\eqref{eq:mpoildecomp} operational is \citet{kilian2009}, ``Not All Oil Price Shocks Are
Alike.'' He estimates a monthly structural VAR ($24$ lags, $1973$--$2007$) of the
\emph{global crude-oil market} in three variables---the percentage change in world oil
production, a global real-activity index built from dry-cargo ocean shipping freight rates,
and the real price of oil---and identifies it recursively, exactly as in the Cholesky SVAR
of the next section. Writing $e_t$ for the reduced-form innovations to (oil production, global
activity, real oil price) and $w_t=(w^s_t,w^d_t,w^o_t)'$ for the structural (supply,
aggregate-demand, oil-specific-demand) shocks, the impact matrix $S=A_0^{-1}$ is lower
triangular,
\begin{equation}\label{eq:mpoilsvar}
e_t = S\,w_t,
\qquad
S=\begin{bmatrix} s_{11} & 0 & 0 \\ s_{21} & s_{22} & 0 \\ s_{31} & s_{32} & s_{33} \end{bmatrix}.
\end{equation}
The zero restrictions are economic timing assumptions: crude supply does not respond
\emph{within the month} to oil demand (a vertical short-run supply curve, because producers
face adjustment costs and uncertainty about the nature of the shock), and global activity
responds to oil-\emph{specific} demand only with a lag. The recursive logic is identical to
that of the structural VAR introduced in the next section; here it is applied to the oil
market itself to separate its demand and supply drivers.

\paragraph{Results.} The three shocks move the oil price very differently: a supply
disruption produces a small, short-lived increase, whereas aggregate-demand and precautionary
(oil-specific) demand shocks produce large and persistent increases. Historically, the major
run-ups---$1979$, and above all $2003$--$2008$---were driven mainly by \emph{demand}, not by
supply disruptions. The macro consequences depend on the cause: a demand-driven oil-price
increase initially reflects a global boom (stimulative before it bites), whereas a supply- or
precautionary-driven increase is more purely stagflationary. The headline lesson is that the
oil price cannot be treated as exogenous, and ``an oil-price increase'' has no well-defined
effect on inflation or output until its \emph{cause} is identified---overturning the
exogenous-oil-price tradition that \eqref{eq:mpoildecomp} refines.

\paragraph{Does the central bank respond to oil? Identifying the systematic response.}
Applied to the reaction function, this decomposition reshaped a well-known debate.
\citet{bernankegertlerwatson1997} ran a VAR counterfactual---hold the policy rate on its
systematic path but shut down the response to oil---and concluded that much of the
post-oil-shock recession was caused by the Fed's \emph{endogenous tightening}, i.e.\ by its
response to oil-driven inflation rather than by the oil shock itself: a statement about the
systematic $\phi_\pi$ / $\phi_{\text{oil}}$ response. \citet{hamiltonherrera2004} showed the
result is fragile to lag length. Using the demand/supply decomposition,
\citet{kilianlewis2011} overturn the reading: there is no robust evidence of a systematic Fed
response to oil price shocks after $1987$, and even before, the Fed appears to have reacted to
oil prices \emph{directly and pre-emptively} rather than to oil-driven inflation, with little
effect on the funds rate or on output. Their methodological verdict echoes this whole section:
the single-equation reaction-function/counterfactual framework should give way to structural
models that treat the real oil price as endogenous and let the policy response depend on the
underlying cause of the shock.

\paragraph{The honest limitation.} Even a \emph{pure} oil supply shock is an imperfect
instrument for $\phi_\pi$ in isolation, because it does not raise inflation \emph{only}: it
also moves the output gap directly. If the central bank responds to the gap---as any Taylor
rule \eqref{eq:mptaylor} does---then the exclusion restriction ``the oil shock reaches $i_t$
only through $\pi_t$'' fails, and the instrument is contaminated by the very $\phi_x$ channel
it ignores. This is why modern work uses identified oil shocks to trace \emph{responses} and
to discipline structural/DSGE models rather than to claim a clean single-equation estimate of
$\phi_\pi$. The predeterminedness of oil prices with respect to U.S. macro news
\citep{kilianvega2011} supports treating the identified supply shock as exogenous, but not as
satisfying the exclusion restriction for $\phi_\pi$ on its own.

\subsection{From reaction functions to structural shocks}

The through-line of this section is that the endogeneity of macroeconomic variables makes
single-equation OLS uninformative about a structural coefficient, and that exogenous
``surprises'' restore identification. The same idea scales up. In the next section we build
the reduced-form VAR, whose innovations $e_t$ are themselves endogenous blends of structural
disturbances; recovering an interpretable \emph{monetary-policy shock} from them is once
again an identification problem, and one modern solution---the external-instrument or
``proxy'' SVAR of Section~\ref{sec:wrapup_monetary}---uses exactly the high-frequency
surprises introduced here \citep{stockwatson2018,gertlerkaradi2015,mertensravn2013}.
Endogeneity is thus not a nuisance to be dispatched in one section but the organising
problem that connects reaction functions, instrumental variables, and structural VARs.

\paragraph{Key takeaways.}
\begin{itemize}
\item The central object in empirical monetary policy is $\phi_\pi$, the systematic response
of the policy rate to inflation; $\phi_\pi>1$ (the Taylor principle) is the condition for a
stabilising, determinate equilibrium.
\item $\phi_\pi$ cannot be estimated by OLS because inflation is \emph{endogenous}: the bank
reacts to inflation while policy moves inflation, so the regressor is correlated with the
policy shock (simultaneity/reverse causation).
\item The bias has a definite sign. Because a contractionary policy shock raises the rate
and lowers inflation at the same time, $\operatorname{Cov}(\pi_t,\varepsilon_t)<0$ and OLS
is biased \emph{downward}; in the limit where policy shocks dominate inflation's variance,
OLS recovers the negative slope of the transmission curve, $-1/\gamma$.
\item The magnitude of the bias is proportional to the share of the regressors' variance
due to monetary-policy shocks \citep{carvalho2021taylor}; small in normal times, but not to be
assumed away.
\item An \emph{inflation surprise}---actual minus expected at release---is orthogonal to the
policy shock yet moves inflation, so it is a valid instrument; this is the perceived-rule
approach of \citet{hamiltonpruittborger2011} and the high-frequency identification toolkit
\citep{kuttner2001monetary,romerromer2004,gurkaynak2007us,jarocinski2020deconstructing}.
\item Even with instruments, $\phi_\pi$ is often only weakly identified
\citep{cochrane2011determinacy,mavroeidis2010monetary}; report identification-robust (Anderson--Rubin) sets,
and beware the central-bank information effect \citep{nakamura2018high,bauerswanson2023}.
\end{itemize}

\section{VAR and Structural VAR}

So far we applied univariate models or regressions, we will now introduce\footnote{we suggest the code vansteenberghe\_VAR.py} multivariate models such as Vector autoregressions (VARs) that were introduced into empirical economics by \citet{sims1980}.

We suggest the \href{https://python.quantecon.org/linear_models.html}{companion website} on linear state space dynamic system.

\subsection{VAR model}

Model equation \ref{eq:popgdpreg} links the the French population growth with the GDP growth. This models that immigration and birth rate simultaneously increase when the economy is performing well. If no cointegration exists between the population and GDP level times series\footnote{If they are cointegrated, then we would need to use a VECM.} and if we agree to discard contemporaneous relationships between variables, we are limited to model that the GDP growth is linked to \emph{lagged} immigration and \emph{lagged} birth rate, so.
Let:
\begin{itemize}
\item $p_t$ the population at quarter $t$
\item $g_t$ the GDP at quarter $t$
\item $e_{i,t}$ idd error terms $\sim \mathcal{N}(0,\sigma_i)$
\end{itemize}

\begin{equation}
\begin{cases}
\Delta p_t = b_1 + c_{11}\,\Delta p_{t-1} + c_{12}\,\Delta g_{t-1} + e_{p,t}, \\
\Delta g_t = b_2 + c_{21}\,\Delta p_{t-1} + c_{22}\,\Delta g_{t-1} + e_{g,t}.
\end{cases}
\end{equation}

which can be written in matrix form as:
\begin{equation}\label{eq:VARreduced}
\begin{bmatrix}
\Delta p_t \\
\Delta g_t
\end{bmatrix}
=
\begin{bmatrix}
b_1 \\
b_2
\end{bmatrix}
+
\begin{bmatrix}
c_{1,1} & c_{1,2}\\
c_{2,1} & c_{2,2}\\
\end{bmatrix}
\begin{bmatrix}
\Delta p_{t-1} \\
\Delta g_{t-1}
\end{bmatrix}
+
\begin{bmatrix}
e_{p,t} \\
e_{g,t}
\end{bmatrix}
\end{equation}

If we write:
$$\Phi = \begin{bmatrix}
c_{1,1} & c_{1,2} \\
c_{2,1} & c_{2,2} \\
\end{bmatrix}$$
we can rewrite equation \ref{eq:VARreduced}
\begin{equation}
y_t = b + \Phi y_{t-1} + e_t \qquad 
e_t \sim (0,\Omega),
\end{equation}
where $y_t \in \mathbb{R}^2$, $b \in \mathbb{R}^2$, and $\Phi \in \mathbb{R}^{2\times2}$.

\subsubsection{Stability, Stationarity, and Estimation of the VAR(1)}

\paragraph{Stability condition.}

Iterating forward,
\[
y_t = \sum_{j=0}^{k-1} \Phi^j b 
+ \Phi^k y_{t-k}
+ \sum_{j=0}^{k-1} \Phi^j e_{t-j}.
\]

The process is \emph{stable} if
\[
\Phi^k \longrightarrow 0 
\qquad \text{as } k \to \infty.
\]

This holds if and only if the spectral radius of $\Phi$ satisfies
\[
\rho(\Phi) = \max_i |\lambda_i(\Phi)| < 1,
\]
that is, all eigenvalues of $\Phi$ lie strictly inside the unit circle.

Equivalently, the roots of the characteristic polynomial
\[
\det(I - \Phi z) = 0
\]
must lie outside the unit circle. If one root lies on the unit circle, the process is non-stationary; if cointegrating relations exist, a VECM representation is required.

\paragraph{Wold representation and covariance stationarity.}

If $\rho(\Phi)<1$, then
\[
\sum_{j=0}^{\infty} \|\Phi^j\| < \infty,
\]
and the VAR admits the convergent moving-average representation
\[
y_t = \mu + \sum_{j=0}^{\infty} \Phi^j e_{t-j},
\qquad 
\mu = (I-\Phi)^{-1} b.
\]

Since $e_t$ is white noise with finite second moments and the coefficient matrices are absolutely summable, $y_t$ has:

\begin{itemize}
\item a constant mean $\mu$,
\item a finite, time-invariant covariance matrix,
\item autocovariances depending only on the lag.
\end{itemize}

Hence stability implies covariance stationarity.

\paragraph{Estimation.}

Under covariance stationarity and the exogeneity condition
\[
\mathbb{E}(e_t \mid y_{t-1}, y_{t-2}, \dots)=0,
\]
each equation of the reduced-form VAR can be consistently estimated by OLS. In fact, because the regressors are identical across equations, OLS coincides with equation-by-equation GLS.

If, in addition, $e_t \sim \mathcal{N}(0,\Omega)$, OLS is equivalent to maximum likelihood estimation for the reduced-form VAR.

Therefore, the eigenvalue condition $\rho(\Phi)<1$ is not merely technical: it guarantees existence of a finite second-moment solution, a convergent MA representation, and standard asymptotic properties of OLS estimators.

\subsubsection{From Reduced-Form Innovations to Structural Shocks}

In the reduced-form VAR \ref{eq:VARreduced}, the innovation vector satisfies
\[
e_t \sim \mathcal{N}(0,\Omega), 
\qquad 
\Omega = \mathbb{E}(e_t e_t') =
\begin{bmatrix}
\sigma_{pp} & \sigma_{pg} \\
\sigma_{pg} & \sigma_{gg}
\end{bmatrix}.
\]

In general, $\Omega$ is symmetric positive definite but not necessarily diagonal.  
If $\Omega$ is diagonal, the reduced-form shocks are contemporaneously uncorrelated.

If instead $\sigma_{pg}\neq 0$, the reduced-form residuals are correlated:
\[
\text{Cov}(e_{p,t},e_{g,t}) \neq 0,
\]
which implies that each $e_t$ combines multiple underlying economic disturbances. In that case, impulse responses computed directly from the reduced form do not correspond to economically meaningful structural shocks.

To recover orthogonal structural shocks, assume the existence of a structural representation
\begin{equation}\label{eq:impactmatrix}
A_0 y_t = A_1 y_{t-1} + u_t,
\qquad 
u_t \sim \mathcal{N}(0,I)
\end{equation}
where $u_t$ are structural shocks normalized to have identity covariance matrix. And $S=A_0^{-1}$ is called the impact matrix in equation \ref{eq:impactmatrix}. The reduced-form and structural innovations are linked by
\[
e_t = S u_t,
\qquad 
\Omega = S S'.
\]

Identification therefore requires imposing restrictions on $A_0$.  
Under recursive (triangular) identification, one uses the Cholesky factorization
\[
\Omega = PP',
\]
with $P$ lower triangular, and sets $S=A_0^{-1}=P$. Structural shocks are then obtained as
\[
u_t = S^{-1} e_t.
\]

The triangular structure implies that the ordering of variables determines the contemporaneous causal relations: variables ordered earlier are not contemporaneously affected by those ordered later. This highlights that structural identification is not purely statistical but embeds economic assumptions about contemporaneous interactions.

\subsubsection{Get the data}
As we did with the GDP series, we fetch the unemployment series from INSEE website and import the data.

\subsubsection{Preparing the VAR model}

We prepare the VAR model, first we import the corresponding package and then resample the population series in quarterly. Then we concatenate the time series into a DataFrame, take the differences related to the integration order of each variable and drop the rows where there are some missing values (NaN). 
VAR models are not good to deal with seasonality nor trend, we need to remove those effects before applying the model.
Finally we apply the VAR model and print the results summary.

\subsection{Select the lag order}\label{sec:lagselect}

For a VAR model of the type: $y_t = c +A_1 y_{t-1} + \dots + A_p y_{t-p} + \epsilon_t$, the order of the VAR is the value $p$ such that $A_p \neq 0$ and $A_k = 0, \forall k > p$.

We suggest two main references \citep{helmut2005} and \citep{Kilian:2017vd} to select the lag order, keeping in mind as Luetkepohl mentions: \begin{quote}
we know that the approximate Mean Square Error matrix of the 1-step ahead predictor will increase with the order $p$. Thus, choosing $p$ unnecessarily large will reduce the forecast precision of the corresponding estimated VAR($p$) model
\end{quote}

\stepcounter{question}
\begin{itemize}
\item \textbf{Question \arabic{question}}
\begin{itemize}
\item As in \citep{helmut2005}, we want to experiment fitting a VAR model of higher order than the data generating process:
\begin{enumerate}
\item consider the following VAR model:
$$ \begin{bmatrix}
Y_{1,t} \\
Y_{2,t} \\
\end{bmatrix}
=
\begin{bmatrix}
17.12 \\
-12.86 \\
\end{bmatrix}
+
\begin{bmatrix}
0 & -0.61\\
-0.17 & 0.29\\
\end{bmatrix}
\begin{bmatrix}
Y_{1,t-1} \\
Y_{2,t-1} \\
\end{bmatrix}
+
\begin{bmatrix}
\epsilon_{1,t} \\
\epsilon_{2,t} \\
\end{bmatrix}$$
\item consider $\epsilon_{1} \sim \mathcal{N}(0,0.1)$ and $\epsilon_{2} \sim \mathcal{N}(0,0.9)$
\item generate $200$ time series for $T=30$ and $T=100$
\item on the generated series, fit a VAR(1), a VAR(2) and a VAR(3), and for each model compute the average squared forecast errors at horizon $h=1$ and $h=3$.
\end{enumerate}
\end{itemize}
\end{itemize}

We create the VAR model and select the lag, search for up to lag order 15.

If the chosen lag order is unnecessarily large, the forecast precision of the corresponding VAR(p) model will be reduced.

The VAR order selected is 1, 
the statistics for four criterio are given:
\begin{enumerate}
\item Akaike's Information Criterion (AIC)
\item Bayesian information criterion (BIC) or Schwarz information criterion
\item Final Prediction Error (FPE)
\item Hannan-Quinn Criterion (HQIC)
\end{enumerate}

Following \citep{helmut2005}, there is no systematic rule to chose which coefficient to base your decision on: HQIC and BIC are consistent but \begin{quote}
In small samples, AIC and FPE may have better properties (choose the correct order more often) than HQ and SC [BIC]. Also, the former two criteria are designed for minimizing the forecast error variance. Thus, in small as well as large samples, models based on AIC and FPE may produce superior forecasts although they may not estimate the orders correctly. [...] AIC and FPE asymptotically overestimate the true order with positive probability and underestimate the true order with probability zero
\end{quote}

\begin{quote}
In a simulation study based on many other processes, Luetkepohl (1985) obtained similar results. In that study, for low order VAR processes, the most parsimonious SC criterion was found to do quite well in terms of choosing the correct VAR order and providing good forecasting models. Unfortunately, in practice we often don't even know whether the underlying data generation law is of finite order VAR type. Sometimes we may just approximate an infinite order VAR process by a finite order model. In that case, for moderate sample sizes, some less parsimonious criterion like AIC may give superior results in terms of forecast precision. Therefore, it may be a good strategy to compare the order estimates obtained with different criteria and possibly perform analyses with different VAR orders.
\end{quote}

\begin{center}
VAR Order Selection 
\end{center}
            
\begin{tabular}{lllll}
lag order &            aic      &    bic      &    fpe     &    hqic\\
0    &     14.58    &    14.65  &  2.138e+06   &     14.60\\
1     &   14.27*     &  14.48*   &1.572e+06*    &   14.35*\\
2      &   14.33      &  14.69    &1.683e+06     &   14.47\\
\end{tabular}                     
* Minimum

\quad

AIC and BIC are (Information Criteria) statistics that measure the distance between observations and the model. We wish to minimize that distance to fit the data generating process.
IC have two additive parts:
\begin{itemize}
\item a goodness-of-fit measure (e.g. minus the maximized likelihood);
\item a penalty that increases with the model's complexity.
\end{itemize}

A study of lag order selection for VAR Impulse Response Analysis is proposed by \citep{Ivanov2005}.

\citep{Hirano2017aa} suggest to add bootstrap aggregation (bagging) :
\begin{quote}
It is common practice to select the model based on pseudo out-of-sample fit from a sequence of recursive or rolling predictions. Parameters are then estimated over the whole sample period. The idea of using an out-of-sample criterion was discussed by Clark (2004) and West (2006), and has a long history, going back to Wilson (1934), Meese and Rogoff (1983), and Ashley, Granger, and Schmalensee (1980). Instead, one might select the model based on in-sample fit, but adjust for potential overfitting by using an information criterion, such as the Akaike information criterion (AIC) (Akaike (1974)), as advocated by Inoue and Kilian (2006).

We consider various methods of model selection and forecasting, including: using in-sample fit with the AIC information criterion; selecting the model based on recursive pseudo out-of-sample forecast accuracy and then using the whole data set for parameter estimation; and splitting the sample into two parts, using one part for model selection and the other for parameter estimation. We call this last method the split-sample approach.
\end{quote}

\subsection{Apply a VAR model and checking for impact}

We apply a VAR(1) to our data. We perform test on our model (all eigenvalues of $\Phi$ have modulus less than $1$, autocorrelations and cross-correlations of model residuals, normality tests of residuals, ARCH-LM test on residuals). 

The coefficient $c_{1,2}$ (from equation \ref{eq:VARreduced}) seems to be significantly different from $0$, so we expect that a shock on the GDP will have an impact on the population dynamics. That seems to be the only non diagonal element different from zero. We can apply a Granger causality test\footnote{For non-stationary data, the \citet{Toda:1995aa} test should be applied, \href{https://davegiles.blogspot.com/2011/04/testing-for-granger-causality.html}{cf. Dave Gile} and \href{Christoph Pfeiffer's work}{https://christophpfeiffer.org/2012/11/07/toda-yamamoto-implementation-in-r/}.} (more detail on this test can be found section \ref{sec:granger}) and also visually check that with an impulse response function (irf).

\subsection{(structural) VAR model}\label{sec:sVAR}

Let $y_t=(\Delta p_t,\Delta g_t)'$. A reduced-form VAR(1) writes
\begin{equation}
y_t = b + C y_{t-1} + u_t,
\qquad \mathbb{E}(u_t)=0,\quad \mathbb{E}(u_tu_t')=\Omega .
\end{equation}
This representation is statistically convenient, but the innovations $u_t$ generally combine multiple
economic disturbances: when $\Omega$ is not diagonal, the reduced-form innovations are contemporaneously correlated and cannot be interpreted as independent structural shocks. As a result, impulse responses based on $u_t$ do not have a causal interpretation without further identifying assumptions.

To recover economically meaningful shocks, we introduce a structural VAR:
\begin{equation}\label{eq:svar}
A y_t = a + B y_{t-1} + \varepsilon_t,
\qquad \mathbb{E}(\varepsilon_t\varepsilon_t')=\Sigma_\varepsilon,
\end{equation}
where $A$ captures contemporaneous interactions and $\varepsilon_t$ are \emph{structural} shocks
(normalized so that $\Sigma_\varepsilon$ is diagonal, typically the identity).
Premultiplying~\ref{eq:svar} by $A^{-1}$ yields the reduced form with
\begin{equation}
u_t = A^{-1}\varepsilon_t
\quad\Rightarrow\quad
\Omega = A^{-1}\Sigma_\varepsilon (A^{-1})'.
\end{equation}
Identification consists in imposing economically motivated restrictions on $A$ (and/or on
$\Sigma_\varepsilon$) so that $A$ can be recovered from the estimated $\Omega$.

Let's take a step back, we would need to identify the following set of simultaneous equations:

\begin{equation}
\begin{cases}
\Delta p_t = a_1 + \beta_0 \,\Delta g_t 
+ d_{11}\,\Delta p_{t-1} 
+ d_{12}\,\Delta g_{t-1} 
+ \varepsilon_{p,t}, \\[6pt]
\Delta g_t = a_2 + \beta_1 \,\Delta p_t 
+ d_{21}\,\Delta p_{t-1} 
+ d_{22}\,\Delta g_{t-1} 
+ \varepsilon_{g,t}.
\end{cases}
\end{equation}

If both relationships hold, then the model is not identified. To identify the following equations with $n$ variables, on top of assuming that $\epsilon^1$ and $\epsilon^2$ are uncorrelated we need $\frac{n(n-1)}{2}=1$ restrictions\footnote{those assumptions cannot be tested}, we impose $\beta^0=0$: we believe that people cannot "nowcast" GDP. We believe that the current population does contribute to the current GDP, $\beta^1 \neq 0$. We will show section \ref{sec:Chholesky} that this is equivalent with a Cholesky decomposition (see section \ref{sec:choleskydecomposition}) with a careful ordering of variables.

We estimate the following structural\footnote{structural because it is assumed to be derived by some underlying economic theory} VAR:

\begin{equation}
\begin{bmatrix}
1& -\beta^0\\
-\beta^1&1
\end{bmatrix}
\begin{bmatrix}
\Delta p_t \\
\Delta g_t
\end{bmatrix}
=
\begin{bmatrix}
a_1 \\
a_2
\end{bmatrix}
+
\begin{bmatrix}
d_{1,1} & d_{1,2}\\
d_{2,1} & d_{2,2}\\
\end{bmatrix}
\begin{bmatrix}
\Delta p_{t-1} \\
\Delta g_{t-1}
\end{bmatrix}
+
\begin{bmatrix}
\epsilon_{p,t} \\
\epsilon_{g,t}
\end{bmatrix}
\end{equation}

We assume that $\epsilon^0$ and $\epsilon^1$ are iid, orthogonal and exogenous. If this orthogonality condition is not respected, then changes in $\epsilon^0$ affects $\epsilon^1$ and then changes in $g_t$ will be due to changes in $p_t$ and in $\epsilon^1$, it is then not possible to identify (or isolate) the change in $g_t$ caused by $p_t$. 

We write: $A=\begin{bmatrix}
1& -\beta^0\\
-\beta^1&1
\end{bmatrix}$

As long as $A^{-1}$ exists\footnote{so at least $\prod_i \beta^i\neq 1 $}, the system of equations can be written in \textbf{reduced form} exactly as in equation \ref{eq:VARreduced}. But now regressors are correlated with $e_t = \begin{bmatrix}
e_{p,t} \\
e_{g,t}
\end{bmatrix}$ is the forecast error and both forecast errors are affected by both shocks $\epsilon_{p}$ and $\epsilon_{g}$. Indeed, $e_t = A^{-1} \epsilon_t$.

If our assumptions are correct that $A^{-1}$ exists and the innovations follow:
$$\epsilon_t = \begin{bmatrix}
\epsilon_{p,t} \\
\epsilon_{g,t}
\end{bmatrix} \sim \mathcal{N}(0,D)$$
with $D$ a diagonal matrix.

The main issue we have with the reduced  VAR form is that we do not identify the $\beta^i$, we only have correlation of variables with forecast and cannot properly apply orthogonal shock $\epsilon_{i}$, that is why we implemented the \textbf{structural} VAR, in our example, this is equivalent with a Cholesky decomposition (provided we have the right ordering of variables). Our main reference text book is \citep{Kilian:2017vd}.

\subsubsection{Getting the intuition about the necessity for parameter restrictions}

Let's imagine we want to work with the simple model:

\begin{equation}
\begin{cases}
\Delta p_t &= a_1 + \beta^0 \Delta g_t + \epsilon^1_t \\
\Delta g_t &= a_2 + \beta^1 \Delta p_t + \epsilon^2_t
\end{cases}
\end{equation}

If $\Delta p_t= a_1 + \beta^0 \Delta g_t + \epsilon^1_t$, then
$ \Delta g_t = - \frac{a_1}{\beta^0} +\frac{1}{\beta^0}\Delta p_t   -\frac{1}{\beta^0} \epsilon^1_t$, so you do not have the liberty to identify $\beta^0$ and $\beta^1$ separately.

\subsubsection{Cholesky or LDL decomposition, aka short-run restrictions}\label{sec:Chholesky}

We have made the assumptions:
$$\epsilon_t = \begin{bmatrix}
\epsilon_{p,t} \\
\epsilon_{g,t}
\end{bmatrix} \sim \mathcal{N}(0,D)$$
with $D$ a diagonal matrix and 
$e_t \sim \mathcal{N}(0,\Omega)$

In order to identify the model, we can make the assumption that the population get knowledge of the GDP growth only with a lag, thus the immigration or birth rate is only "reacting" to the GDP growth rate with a lag and we can assume a lower triangular matrix, from this structure it is as if we assume that:
$$\Omega = A^{-1} D (A^{-1})'$$ which enables us to identify $A$, the inverse of a lower (resp. upper) non-singular triangular matrix is a lower (resp. upper) triangular matrix.

The Cholesky decomposition delivers a unique triangular matrix (with strictly positive diagonal elements), which allows us to recover $A^{-1}$ and therefore $A$.

\medskip

\noindent Recall that:
\begin{itemize}
    \item The inverse of a non-singular triangular matrix is triangular of the same type (lower remains lower, upper remains upper).
    \item Identification relies entirely on the zero restrictions imposed on the contemporaneous matrix.
    \item The ordering of variables is crucial: changing the ordering modifies the implied contemporaneous structure and therefore the structural impulse responses.
\end{itemize}

It is then possible to do a moving average or Wold representation of the VAR to determine the impact of shocks on the variables.

\subsubsection{Impulse response function}

We can also plot the impulse responses: these are the responses to a unit impulse to a variable that our VAR model predicts. In our case, we can believe that it is possible to have an isolated shock on the population that is not contemporaneous with a shock on the GDP, hence it makes economic sense to study such an isolated shock.

As we expected from our Granger causality test, a shock on the GDP has an impact on the population significantly different from $0$ (but marginally):

\begin{center}
\includegraphics[scale=0.5]{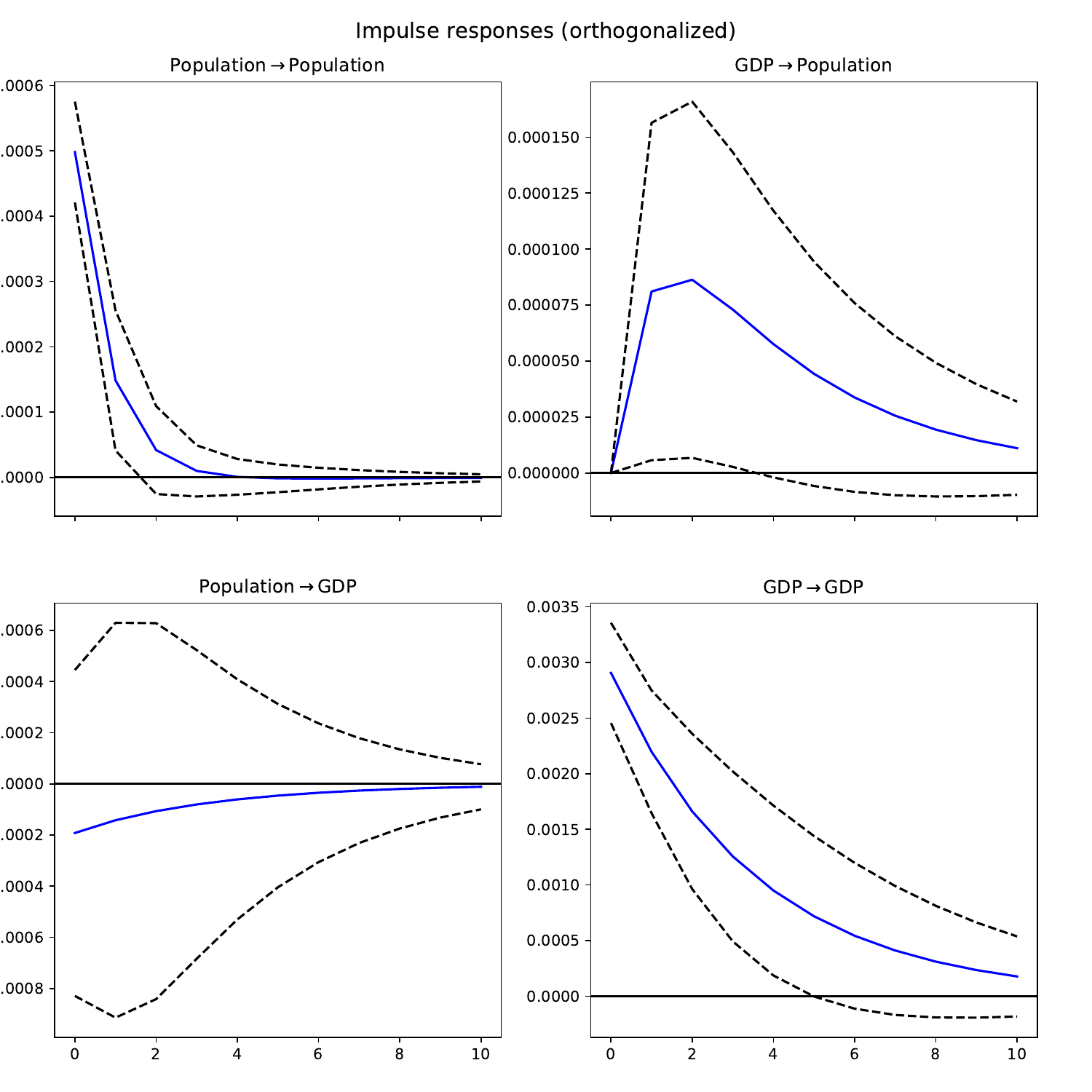}
\end{center}

\stepcounter{question}
\begin{itemize}
\item \textbf{Question \arabic{question}}
\begin{itemize}
\item An economist can also imagine that the unemployment rate both is cause and will cause effect on the population growth of France. Add the unemployment, do a three-variable SVAR (which restrictions do you suggest?):
\end{itemize}
\end{itemize}

\begin{equation}
\begin{bmatrix}
\Delta p_t \\
\Delta g_t \\
\Delta u_t
\end{bmatrix}
=
\begin{bmatrix}
b_1 \\
b_2 \\
b_3
\end{bmatrix}
+
\begin{bmatrix}
c_{1,1} & c_{1,2}&c_{1,3}  \\
c_{2,1} & c_{2,2} &c_{2,3}\\
c_{3,1}&c_{3,2}&c_{3,3}\\ 
\end{bmatrix}
\begin{bmatrix}
\Delta p_{t-1} \\
\Delta g_{t-1} \\
\Delta u_{t-1}
\end{bmatrix}
+
\begin{bmatrix}
\epsilon_{p,t} \\
\epsilon_{g,t} \\
\epsilon_{u,t}
\end{bmatrix}
\end{equation}

\stepcounter{question}
\begin{itemize}
\item \textbf{Question \arabic{question}}
\begin{itemize}
\item Perform a Granger causality test to test if the GDP is Granger causing the unemployment.
\item We have built a VAR with $3$ variables. Compare the forecast precision of such a complex model with the simpler AR(1) model we try to manually design section \ref{sec:ARpop}. After you covered the sections on ARMA models, try to conclude on the model you would chose as an economist to predict French population "next year".
\end{itemize}
\end{itemize}

\subsubsection{SVAR vs VAR (orthogonalized) IRFs — are they the same?}

\paragraph{Setup.}
Let $y_t\in\mathbb{R}^n$ follow a reduced-form VAR($p$):
\begin{equation}\label{eq:var}
y_t = c + \sum_{i=1}^p B_i y_{t-i} + u_t,
\qquad
\mathbb{E}[u_t]=0,\quad \mathbb{E}[u_tu_t']=\Sigma_u.
\end{equation}
Let $\Psi_h$ denote the MA coefficients of the reduced-form VAR:
\begin{equation}\label{eq:ma}
y_{t} = \mu + \sum_{h=0}^\infty \Psi_h u_{t-h},\qquad \Psi_0=I_n.
\end{equation}

\paragraph{Structural form.}
A (static) SVAR representation writes
\begin{equation}\label{eq:svarstruct}
A_0 y_t = a + \sum_{i=1}^p A_i y_{t-i} + \varepsilon_t,
\qquad
\mathbb{E}[\varepsilon_t\varepsilon_t']=I_n,
\end{equation}
so that the mapping between reduced-form and structural shocks is
\begin{equation}\label{eq:map}
u_t = A_0^{-1}\varepsilon_t \equiv S\,\varepsilon_t,
\qquad
\Sigma_u = \mathbb{E}[u_tu_t'] = SS'.
\end{equation}

\paragraph{IRFs.}

\textbf{Structural (non-orthogonalized) IRF in an SVAR:}
the response of $y_{t+h}$ to a unit structural shock $\varepsilon_{j,t}$ is
\begin{equation}\label{eq:irf_struct}
\mathrm{IRF}^{\text{SVAR}}_h(:,j)
=
\frac{\partial y_{t+h}}{\partial \varepsilon_{j,t}}
=
\Psi_h S e_j
=
\Psi_h A_0^{-1} e_j,
\end{equation}
where $e_j$ is the $j$-th canonical basis vector.

\textbf{Orthogonalized IRF in a reduced-form VAR (Cholesky):}
choose a factor $P$ such that
\begin{equation}\label{eq:chol}
\Sigma_u = PP', \qquad P \text{ lower triangular (Cholesky).}
\end{equation}
Define orthogonal innovations $\eta_t$ by $u_t=P\eta_t$ with $\mathbb{E}[\eta_t\eta_t']=I_n$.
Then
\begin{equation}\label{eq:irf_orth}
\mathrm{IRF}^{\text{VAR,orth}}_h(:,j)
=
\frac{\partial y_{t+h}}{\partial \eta_{j,t}}
=
\Psi_h P e_j.
\end{equation}

\paragraph{Equivalence conditions.}
Comparing \ref{eq:irf_struct} and \ref{eq:irf_orth}:
\begin{equation}\label{eq:equiv}
\mathrm{IRF}^{\text{SVAR}}_h(:,j) = \mathrm{IRF}^{\text{VAR,orth}}_h(:,j)
\quad \forall h,j
\quad \Longleftrightarrow \quad
S = A_0^{-1} = P.
\end{equation}

\subsection{"The Demise of Granger Causality Tests in Macroeconomics"}

As \citep{Kilian:2017vd} put it:
\begin{quote}
Initially, economists thought that bivariate Granger causality tests established that money leads income, but that result weakened once more variables were included in the VAR model. It also proved highly sensitive to different forms of detrending and to changes in the model specification. Moreover, it was shown that apparent Granger noncausality from money to income may simply reflect the omission of a third variable, whereas a finding of bivariate Granger causality may likewise reflect the omission of a third variable, calling into question even further the usefulness of Granger causality tests [\ldots].
 Granger causality tests were replaced by the new idea that we are not interested in the contemporaneous correlation between money and income growth or the lead-lag pattern, but in the question of how income \textbf{responds to unanticipated changes} in money growth (also known as innovations or shocks). The hope was that these shocks would be exogenous, even if the underlying money growth time series was not.
\end{quote}

\subsection{Short-run and long-run restrictions}\label{sec:shortlongrun}

In sections~\ref{sec:sVAR}--\ref{sec:Chholesky} we identified structural shocks by restricting the \emph{contemporaneous} impact matrix~$A$ (or, equivalently, by choosing a Cholesky ordering).
These are called \textbf{short-run restrictions} because they constrain the \emph{impact} (horizon~$h=0$) response of one variable to a shock in another.
A conceptually different strategy restricts the \emph{cumulative long-run} effect of a shock, i.e.\ the response as $h\to\infty$.
These \textbf{long-run restrictions}, introduced by \cite{blanchard1989}, are the topic of this section.

\subsubsection{Recap: short-run (Cholesky) restrictions}

Consider a bivariate VAR(1) in GDP growth~$\Delta g_t$ and inflation~$\pi_t$:
\begin{equation}\label{eq:var_sr}
y_t \;=\; M\, y_{t-1} + u_t,
\qquad y_t = \begin{pmatrix}\Delta g_t\\ \pi_t\end{pmatrix},
\qquad \mathbb{E}[u_t u_t'] = \Omega.
\end{equation}

We seek a matrix $S$ such that $u_t = S\,\varepsilon_t$ with $\mathbb{E}[\varepsilon_t\varepsilon_t'] = I$, which requires $\Omega = SS'$.
This factorisation is \emph{not} unique: for any orthogonal matrix~$Q$, $\tilde S = SQ$ also satisfies $\Omega = \tilde S\tilde S'$.
To pin down $S$ we need $n(n-1)/2 = 1$ additional restriction (for $n=2$).

\paragraph{Cholesky identification.}
Assume that a demand shock $\varepsilon_t^D$ has \emph{no contemporaneous effect on GDP growth}:
the entry $(1,2)$ of $S$ is zero, i.e.\ $S$ is lower-triangular.
Then $S$ is the unique lower-triangular Cholesky factor of $\Omega$ (with positive diagonal).
In practice, this means the ordering of variables matters: the variable ordered first is not affected contemporaneously by the shock to the second variable.

\paragraph{Implementation.}  With the \texttt{statsmodels} SVAR interface we write\footnote{See \texttt{SVAR\_vansteenberghe.py}.}
\begin{equation}
A =
\begin{pmatrix}
1 & 0\\ ? & 1
\end{pmatrix}
\qquad\Longleftrightarrow\qquad
S = A_0^{-1} =
\begin{pmatrix}
1 & 0\\ -a_{21} & 1
\end{pmatrix},
\end{equation}
where $?$ (denoted \texttt{'E'} in the code) is estimated.
The restriction $a_{12}=0$ means that inflation shocks do not move GDP on impact.

\paragraph{Ex-post economic checks.}
If we label the first structural shock as ``supply'' and the second as ``demand'', basic theory predicts:
\begin{itemize}
\item A positive \textbf{supply shock} should \emph{increase} GDP and \emph{decrease} prices.
\item A positive \textbf{demand shock} should \emph{increase} both GDP and prices.
\end{itemize}
We verify these sign predictions by inspecting the estimated impulse responses. If the signs are violated, the Cholesky identification (or the ordering) may not be adequate and an alternative scheme should be considered.

\subsubsection{Long-run (Blanchard--Quah) restrictions}\label{sec:BQ}

\paragraph{Motivation.}
Short-run restrictions are sometimes hard to justify theoretically: why should a demand shock have \emph{zero} contemporaneous effect?
An alternative, proposed by \cite{blanchard1989}, restricts the \emph{cumulative long-run} effect of a shock.
The classic application uses GDP growth and the unemployment rate and assumes:
\begin{quote}
A demand shock has no \emph{long-run} effect on the level of GDP (but it may affect GDP in the short run).
\end{quote}
This is motivated by the classical dichotomy / long-run neutrality: demand-side disturbances (monetary policy, fiscal spending, etc.) shift the economy along a vertical long-run aggregate supply curve and therefore cannot permanently alter real output.
Supply shocks (technology, productivity), on the other hand, \emph{can} shift the long-run level of output.

\paragraph{Formal derivation.}
Consider the reduced-form VAR($p$):
$$
y_t = M_1\,y_{t-1} + \cdots + M_p\,y_{t-p} + u_t, \qquad y_t = \begin{pmatrix}\Delta g_t\\ u_t^{\text{unemp}}\end{pmatrix},
$$
where $\Delta g_t$ is the GDP \emph{growth rate} (first difference of log GDP) and $u_t^{\text{unemp}}$ is the unemployment rate.
The moving-average (Wold) representation is:
\begin{equation}\label{eq:wold}
y_t = \sum_{h=0}^{\infty} \Phi_h\, u_{t-h}, \qquad \Phi_0 = I_n,
\end{equation}
where the $\Phi_h$ are the MA coefficient matrices computed recursively from the VAR coefficients (for a VAR(1), simply $\Phi_h = M^h$).

Using the structural decomposition $u_t = S\,\varepsilon_t$, the long-run cumulative impact on $y_t$ is:
\begin{equation}\label{eq:longrun}
\underbrace{\sum_{h=0}^{\infty} \Phi_h\, S}_{\displaystyle \equiv\; F}
= \underbrace{(I - M_1 - \cdots - M_p)^{-1}}_{\displaystyle \equiv\;\Theta}\, S,
\end{equation}
where $\Theta = \left(\sum_{h=0}^\infty \Phi_h\right)$ is the long-run reduced-form multiplier (the sum converges when the VAR is stable, i.e.\ all companion-form eigenvalues lie inside the unit circle).

\paragraph{Blanchard--Quah restriction.}
Since $y_t$ contains the GDP \emph{growth rate}, the cumulative response of the growth rate gives the \emph{permanent} (long-run) effect on the \emph{level} of GDP.
The restriction ``demand shocks have no long-run effect on GDP'' translates to:
\begin{equation}\label{eq:BQrestriction}
\boxed{F_{12} = 0,}
\end{equation}
i.e.\ the $(1,2)$ entry of the long-run matrix $F = \Theta\, S$ is zero.
This means $F$ is lower-triangular.

\paragraph{Algorithm.}
\begin{enumerate}
\item Estimate the reduced-form VAR($p$) and obtain $\hat M_1,\ldots,\hat M_p$ and $\hat\Omega$.
\item Compute the long-run reduced-form multiplier $\hat\Theta = (I - \hat M_1 - \cdots - \hat M_p)^{-1}$.
\item Form $\hat\Theta\,\hat\Omega\,\hat\Theta' = \hat F\,\hat F'$ (since $\Omega = SS'$ and $F = \Theta S$).
\item Apply a Cholesky decomposition to $\hat\Theta\,\hat\Omega\,\hat\Theta'$: its unique lower-triangular Cholesky factor is $\hat F$.
\item Recover $\hat S = \hat\Theta^{-1}\,\hat F = (I- \hat M_1 - \cdots - \hat M_p)\,\hat F$.
\end{enumerate}

Note that unlike Cholesky short-run restrictions, the matrix $S$ recovered from the Blanchard--Quah procedure is in general \emph{not} lower-triangular: the short-run impact $S$ is unrestricted (demand shocks \emph{can} affect GDP on impact).

\paragraph{Economic interpretation.}
Once $\hat S$ is recovered, structural impulse responses are computed as $\Phi_h\,\hat S$.  In the Blanchard--Quah application:
\begin{itemize}
\item A \textbf{supply shock} ($\varepsilon_t^S$, first column of $S$): permanent effect on GDP, can also affect unemployment in the long run.
\item A \textbf{demand shock} ($\varepsilon_t^D$, second column of $S$): only temporary effect on GDP (by construction), but may have a permanent effect on unemployment.
\end{itemize}

\paragraph{Implementation.}
The R package \texttt{vars} provides the function \texttt{BQ()} which directly applies the Blanchard--Quah decomposition.\footnote{See \texttt{SVAR\_long\_run\_restrictions\_vansteenberghe.R}.}
In our companion code we also replicate the algorithm step by step (extracting $\hat M$, computing $\hat\Theta$, Cholesky-decomposing $\hat\Theta\hat\Omega\hat\Theta'$, and recovering $\hat S$) and verify that the manual result matches \texttt{BQ()} to machine precision.

\begin{figure}[htbp]
\centering
\includegraphics[width=0.75\textwidth]{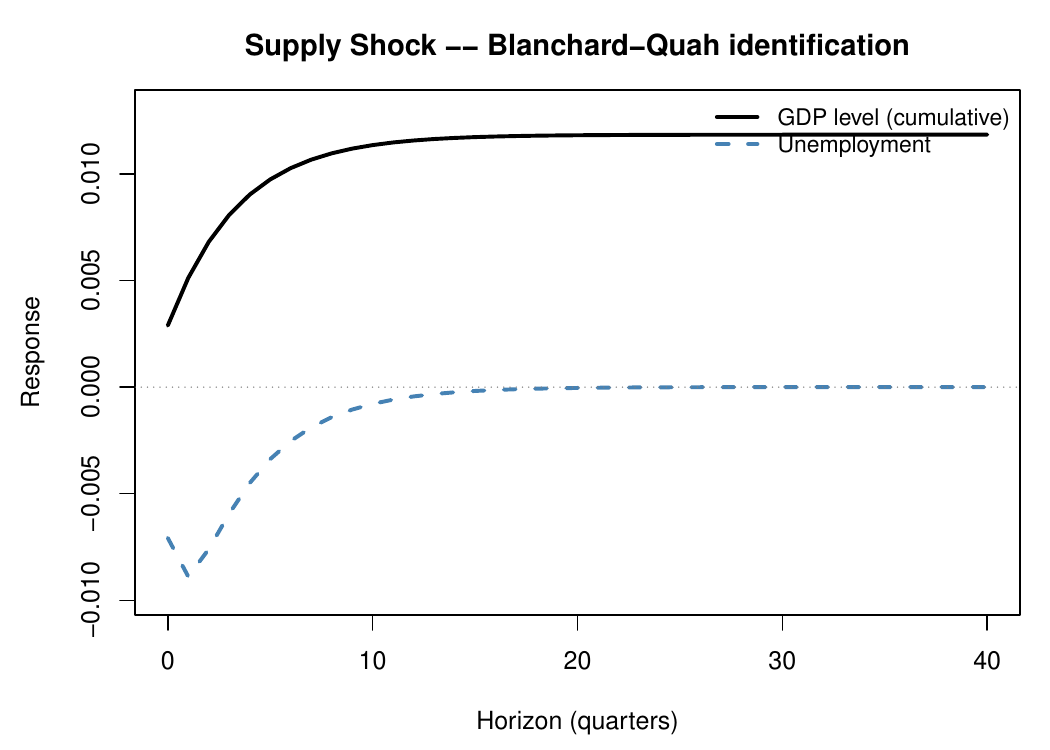}
\caption{Blanchard--Quah supply shock: cumulative GDP level response (black) and unemployment response (blue).  The supply shock has a \emph{permanent} positive effect on the GDP level.}
\label{fig:BQ_supply}
\end{figure}

\begin{figure}[htbp]
\centering
\includegraphics[width=0.75\textwidth]{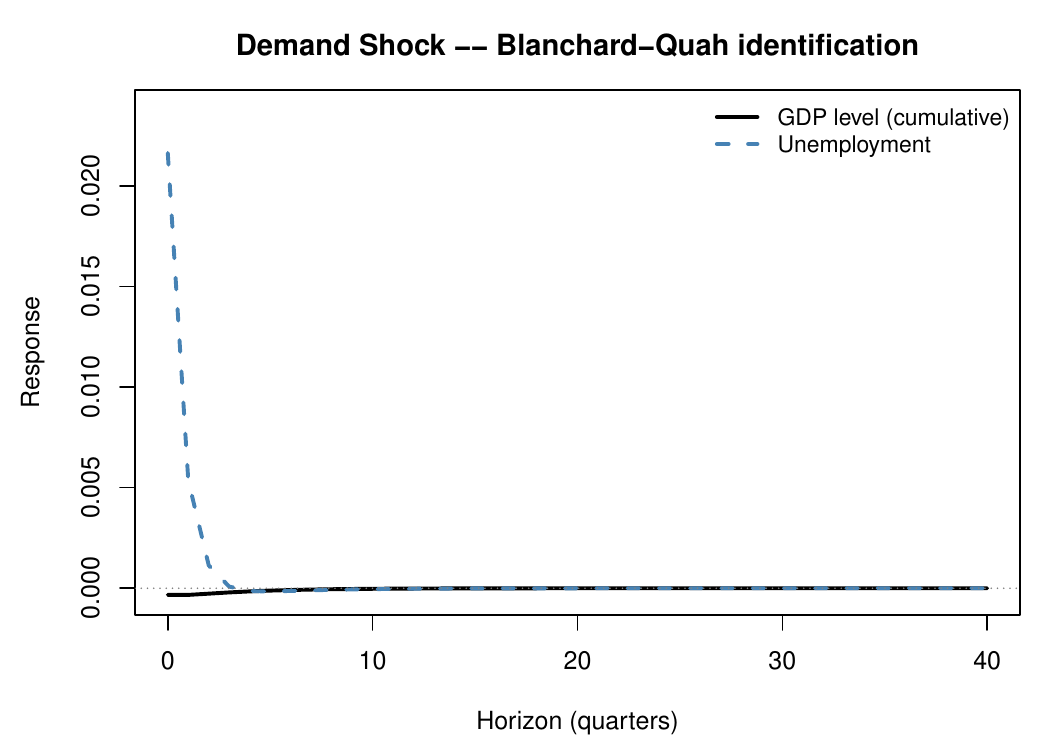}
\caption{Blanchard--Quah demand shock: cumulative GDP level response (black) and unemployment response (blue).  By the long-run restriction, the cumulative effect on GDP converges to zero.}
\label{fig:BQ_demand}
\end{figure}

The cumulative impulse response functions are shown in Figures~\ref{fig:BQ_supply} and~\ref{fig:BQ_demand}.  As expected, the supply shock has a permanent positive effect on GDP while the demand shock's cumulative effect on GDP converges to zero by construction.

\subsubsection{Comparison and practical guidance}

\begin{center}
\begin{tabular}{lll}
\toprule
& \textbf{Short-run (Cholesky)} & \textbf{Long-run (Blanchard--Quah)} \\
\midrule
Restriction on & impact matrix $S$ & long-run cumulative matrix $F$ \\
Typical zero & $S_{ij}=0$ & $F_{ij}=0$ \\
Amounts to & recursive ordering & neutrality / no permanent effect \\
Pros & simple, exact Cholesky & economically motivated \\
Cons & ordering-sensitive & imprecise in small samples \\
Classic reference & Sims (1980) & Blanchard \& Quah (1989) \\
\bottomrule
\end{tabular}
\end{center}

\medskip
In both cases, a $n$-variable SVAR requires $n(n-1)/2$ restrictions for exact identification.
Sign restrictions (e.g.\ ``a supply shock raises GDP'') provide an alternative that yields a \emph{set} of admissible models rather than a point estimate; see \cite{Kilian:2017vd} for a thorough treatment.

Economic theory posits that money is neutral in the long run: the level of money has no long-run impact on the level of GDP. A full VAR model of the macroeconomy would include: GDP, consumption, investment, employment, nominal wages, money, prices and nominal interest rates.

\subsection{A macroeconomic model}

As a basic macro-economist, assume the GDP growth $g_t$, the nominal interest rate $i_t$ and the inflation rate $\pi_t$  follow:
\begin{equation}\label{eq:macromodelsimple}
\begin{cases}
g_t &= \alpha_g +\beta_{i,g} i_{t-1} + \beta_{\pi,g} \pi_t + u_{g,t}\\
i_t &= \alpha_i +\beta_{i,i} i_{t-1} + \beta_{\pi,i} \pi_t + u_{i,t} \mbox{ Taylor rule}\\
\pi_t &= \alpha_\pi + \beta_{\pi,\pi} \pi_{t-1} + \beta_{g,\pi} g_{t-1} + u_{\pi, t} \mbox{ Phillips curve}
\end{cases}
\end{equation}

where the structural shocks $u_t$ are uncorrelated.

\stepcounter{question}
\begin{itemize}
\item \textbf{Question \arabic{question}}
\begin{itemize}
\item Write the model in reduced form $Y_t = \Phi Y_{t-1} + \epsilon_t$ show that $\epsilon_t = B u_t$ and describe the restrictions on $B$ you impose based on your macroeconomic model \ref{eq:macromodelsimple}
\item get data from a country of your choice and estimate the model. What is the impact of a positive (orthogonal) monetary policy shock $u_{i,t}$
\end{itemize}
\end{itemize}

\noindent\textbf{Remark (The Price Puzzle).}
When estimating this model on U.S.\ data, you will likely find a counterintuitive result: a contractionary monetary policy shock (a positive surprise increase in $i_t$) is followed by a \emph{rise} in inflation over several quarters, before eventually declining. This is the so-called \textbf{price puzzle}, first documented by \citet{sims1992interpreting}.

The leading explanation is an \emph{omitted variable} problem: the central bank observes forward-looking signals of future inflation (e.g.\ commodity prices) that are not included in the VAR.  When the Fed raises rates in response to anticipated inflation, the VAR attributes the subsequent rise in prices to the rate increase rather than to the underlying inflationary pressure the Fed was reacting to.

Three fixes are proposed in the literature:
\begin{enumerate}
\item \textbf{Add a broad commodity price index in log-levels} \citep{sims1992interpreting}. Use a broad index (e.g.\ \texttt{PPIACO} from FRED, PPI All Commodities) entered in \emph{log-levels}---not growth rates. Growth rates strip out the persistent level information that proxies for cumulative inflationary pressure. The variable should be ordered \emph{first} in the Cholesky ordering because commodity prices, being fast-moving asset prices, react within the quarter before domestic macro variables and monetary policy. Note that using oil prices \emph{alone} in growth rates is typically insufficient: oil is too narrow a proxy, and \citet{sims1992interpreting} originally used a broad commodity index, not oil.
\item \textbf{Restrict the sample to the post-Volcker era} (1984--2019). The puzzle is concentrated in the pre-Volcker period (1960--1979) when monetary policy was passive---the Fed did not raise rates aggressively enough in response to inflation, violating the Taylor principle. Under active monetary policy (post-1983), the puzzle is much smaller or vanishes entirely \citep{castelnuovo2010monetary}.
\item \textbf{Use the output gap} instead of output in levels \citep{giordani2004alternative}. The output gap better captures the demand pressure the Fed reacts to.
\end{enumerate}

\stepcounter{question}
\begin{itemize}
\item \textbf{Question \arabic{question} (Price Puzzle Resolution)}
\begin{itemize}
\item Import the PPI All Commodities index from FRED (series \texttt{PPIACO}), take its \emph{log-level} $(\log P^{com}_t \times 100)$, and augment your VAR to form a 4-variable system $[P^{com}_t, \pi_t, g_t, i_t]$.
\item Estimate the VAR on the full sample (1960--2019). Does adding $P^{com}_t$ reduce the price puzzle compared to the 3-variable baseline?
\item Now restrict the sample to 1984--2019 (post-Volcker). Compare the response of $\pi_t$ to a monetary policy shock across all four specifications (3-var / 4-var $\times$ full / post-Volcker). What do you conclude?
\end{itemize}
\end{itemize}

\subsection{A macroeconomic model with expectations}

A standard benchmark in modern macroeconomics is the canonical New Keynesian model, which combines an aggregate demand equation, a Phillips curve, and a monetary policy rule. Let $x_t$ denote the output gap, $\pi_t$ inflation, and $i_t$ the nominal short-term interest rate. The model is:
\begin{equation}\label{eq:NKmodel}
\begin{cases}
x_t &= \mathbb{E}_t x_{t+1} - \sigma\left(i_t - \mathbb{E}_t \pi_{t+1} - r_t^n\right) + u_{x,t} 
\qquad \mbox{(dynamic IS curve)}\\[0.3em]
\pi_t &= \beta \mathbb{E}_t \pi_{t+1} + \kappa x_t + u_{\pi,t}
\qquad \mbox{(New Keynesian Phillips curve)}\\[0.3em]
i_t &= \rho_i i_{t-1} + (1-\rho_i)\left(\phi_\pi \pi_t + \phi_x x_t\right) + u_{i,t}
\qquad \mbox{(Taylor rule)}
\end{cases}
\end{equation}

where:
\begin{itemize}
\item $x_t$ is the output gap,
\item $\pi_t$ is inflation,
\item $i_t$ is the nominal policy rate,
\item $\mathbb{E}_t x_{t+1}$ and $\mathbb{E}_t \pi_{t+1}$ are expectations formed at time $t$,
\item $r_t^n$ is the natural real interest rate,
\item $\sigma > 0$ measures the sensitivity of demand to the real interest rate,
\item $\beta \in (0,1)$ is the discount factor,
\item $\kappa > 0$ is the slope of the Phillips curve,
\item $\rho_i \in [0,1)$ captures interest-rate smoothing,
\item $\phi_\pi > 1$ and $\phi_x \geq 0$ are policy reaction coefficients,
\item $u_{x,t}$, $u_{\pi,t}$, and $u_{i,t}$ are structural shocks, assumed mutually uncorrelated.
\end{itemize}

\noindent
This system is structural and forward-looking. It cannot be directly estimated as a backward-looking VAR without additional assumptions. In empirical work, one common approach is to estimate a reduced-form VAR in observable variables and then recover structural shocks using identification restrictions motivated by \eqref{eq:NKmodel}.

\stepcounter{question}
\begin{itemize}
\item \textbf{Question \arabic{question}}
\begin{itemize}
\item Explain why the model in \eqref{eq:NKmodel} is structural and forward-looking, and why it does not directly take the reduced-form VAR representation
\[
Y_t = \Phi_1 Y_{t-1} + \cdots + \Phi_p Y_{t-p} + \varepsilon_t.
\]
\item Consider instead the empirical VAR
\[
Y_t = \begin{pmatrix}
\pi_t\\ x_t\\ i_t
\end{pmatrix},
\qquad
Y_t = \Phi_1 Y_{t-1} + \cdots + \Phi_p Y_{t-p} + \varepsilon_t,
\]
with $\varepsilon_t = B u_t$, where
\[
u_t = \begin{pmatrix}
u_{\pi,t}\\ u_{x,t}\\ u_{i,t}
\end{pmatrix}
\]
is the vector of orthogonal structural shocks.
Write down the restrictions on $B$ implied by a recursive identification consistent with the macroeconomic structure in \eqref{eq:NKmodel}.
\item Get data from a country of your choice and estimate the VAR. Identify a positive orthogonal monetary policy shock $u_{i,t}$ and discuss its effects on inflation, the output gap, and the interest rate.
\end{itemize}
\end{itemize}

\noindent\textbf{Remark (Identification and Timing).}
A standard recursive identification orders variables as
\[
Y_t = (\pi_t,\; x_t,\; i_t)'.
\]
This ordering assumes that within the period, monetary policy can react contemporaneously to inflation and real activity, while inflation and activity do not respond contemporaneously to the monetary policy shock. Under this assumption, the impact matrix $B$ is lower triangular:
\[
B=
\begin{pmatrix}
b_{11} & 0 & 0\\
b_{21} & b_{22} & 0\\
b_{31} & b_{32} & b_{33}
\end{pmatrix}.
\]
Hence, the monetary policy shock $u_{i,t}$ affects $i_t$ contemporaneously, but affects $\pi_t$ and $x_t$ only with a lag.

\noindent\textbf{Remark (The Price Puzzle).}
When estimating this model on U.S.\ data, you may find a counterintuitive result: a contractionary monetary policy shock (a positive surprise increase in $i_t$) is followed by a \emph{rise} in inflation over several quarters before eventually declining. This is the so-called \textbf{price puzzle}, first documented by \citet{sims1992interpreting}.

The leading explanation is an omitted-variable problem: the central bank observes forward-looking signals of future inflation that are not included in the VAR. When the Fed raises rates in response to anticipated inflation, the VAR may attribute the subsequent rise in inflation to the policy shock itself rather than to the inflationary pressures the central bank was reacting to.

Three standard fixes are proposed in the literature:
\begin{enumerate}
\item \textbf{Add a broad commodity price index in log-levels} \citep{sims1992interpreting}. Use a broad index such as \texttt{PPIACO} from FRED, entered in log-levels. Commodity prices are often ordered first because they react quickly to information.
\item \textbf{Restrict the sample to the post-Volcker era} (1984--2019), when monetary policy was more systematic and consistent with the Taylor principle \citep{castelnuovo2010monetary}.
\item \textbf{Use the output gap} rather than output in levels \citep{giordani2004alternative}, since the central bank reacts to cyclical pressure rather than to the level of output per se.
\end{enumerate}

\stepcounter{question}
\begin{itemize}
\item \textbf{Question \arabic{question} (Price Puzzle Resolution)}
\begin{itemize}
\item Import the PPI All Commodities index from FRED (series \texttt{PPIACO}), take its log-level $(100 \times \log P_t^{com})$, and estimate the augmented 4-variable VAR
\[
Y_t = \begin{pmatrix}
P_t^{com}\\ \pi_t\\ x_t\\ i_t
\end{pmatrix}.
\]
\item Estimate the VAR on the full sample. Does adding $P_t^{com}$ reduce the price puzzle relative to the 3-variable baseline?
\item Restrict the sample to 1984--2019 (post-Volcker). Compare the impulse response of inflation to a monetary policy shock across the four specifications (3-variable vs.\ 4-variable; full sample vs.\ post-Volcker). What do you conclude?
\end{itemize}
\end{itemize}

\subsection{How oil price shocks affect U.S. real GDP and inflation?}

\stepcounter{question}
\begin{itemize}
\item \textbf{Question \arabic{question}}
\begin{itemize}
\item Following \citep{Kilian:2017vd}, import the data on the \href{https://sites.google.com/site/lkilian2019/textbook/code}{author's website} and import "Responses of the U.S. economy to an unexpected increase in the real price of oil (Cholesky decomposition)" data:
\begin{itemize}
\item WTI price of crude oil;
\item the U.S. GDP deflator inflation rate;
\item U.S. real GDP growth.
\end{itemize}
\item estimate a structural VAR, how oil price shocks affect U.S. real GDP and inflation?
\end{itemize}
\end{itemize}

\subsection{A VAR model from a textbook}

It is not easy to find $I(0)$ data and to model a VAR that makes sense.

We use the data in Bourbonnais' Econometrie, Dunod. You can download the file online \href{http://www.regisbourbonnais.dauphine.fr/fr/publications/econometrie-manuel-et-exercices-corriges.html}{here}.

Demand $Y_1$ and prices $Y_2$ of some commodity are given.

\subsubsection{Import the data}
We import the data and clean it.

\subsubsection{Verify that the series are stationary}
We verify that both given series are stationary. We find indeed that both series are $I(0)$. We define a function that print the result of the test and that allow to chose for a threshold for the p-value.

\subsubsection{Estimate the VAR coefficients}
Now that we selected the lag order 1, we can estimate the VAR coefficients. This time we find all the coefficients significative except one:

\begin{equation*}
\begin{bmatrix}
Y_{1,t} \\
Y_{2,t} \\
\end{bmatrix}
=
\begin{bmatrix}
17.12 \\
-12.86 \\
\end{bmatrix}
+
\begin{bmatrix}
0 & -0.61\\
-0.17 & 0.29\\
\end{bmatrix}
\begin{bmatrix}
Y_{1,t-1} \\
Y_{2,t-1} \\
\end{bmatrix}
+
\begin{bmatrix}
\epsilon_{1,t} \\
\epsilon_{2,t} \\
\end{bmatrix}
\end{equation*}

The impact of $Y_2$ on $Y_1$ seems to be significantly different than $0$. We again check this with an impulse response function.

\subsubsection{A rolling regression}
If we want to perform a rolling regression of $Y_2$ on $Y_1$ with a window of $10$ quaters, we can simply use the function or define the rolling regression ourselves.

And apply our function. Note that we need to add some zeros at the beginning of our beta list with the size of our window.

\subsection{Rice and wheat prices related to world supplies of rice and wheat}

We use data from \href{http://faostat3.fao.org/download/P/PI/E}{FOOD AND AGRICULTURE ORGANIZATION OF THE UNITED NATIONS}:

\quad

\stepcounter{question}
\begin{itemize}
\item \textbf{Question \arabic{question}}
\begin{itemize}
\item Based on the exercises done so far, perform:
\begin{enumerate}
\item apply the same approach above to rice and wheat prices and world supplies.
\item add lagged variables.
\item once the section \ref{sec:granger} has been covered, apply Granger causality tests to this data set.
\end{enumerate}
\end{itemize}
\end{itemize}

\subsection{Pseudo maximum likelihood, GMM, Bayesian estimation of VARs}

Let $\left( Y_t\right)$ assume stationary solution of
\begin{equation}
\Phi_0 Y_t = \Phi_1 Y_{t-1} + \epsilon_t
\end{equation}
with $E\left( \epsilon_t\right)=0$ and $V\left(\epsilon_t\right)=\Sigma$, then in reduced form
\begin{equation}
Y_t = \Phi_0^{-1}\Phi_1 Y_{t-1} + \Phi_0^{-1}\epsilon_t
\end{equation}
but in fact there are two main issues:
\begin{enumerate}
\item it is impossible to identify $\Phi_0$, $\Phi_1$, and $\Sigma$ separately;
\item an OLS systematically provides estimators such that the eigenvalues of $\Phi_0^{-1}\Phi_1$ have their modulus $<1$. This is why you want to consider pseudo maximum likelihood, GMM or Bayesian inference of the VAR \citep{anttonen2024statistically}.
\end{enumerate}

Impulse response function (IRF) are studied considering:
\begin{equation}
Y_t = \Phi Y_{t-1}+C \eta_t
\end{equation}
where$V\left(\eta_t\right)= I_d$, assuming $C$ is lower triangular, $CC'$ is the variance-covariance matric of $Y_t$. The you study
$\delta Y_{T+h} = \hat \Phi \delta Y_{T+h-1} + \hat \delta \eta_{T+h}$ where $\delta Y_{T-1}=0$ and $\delta \eta_{T}=[\delta \eta_{1,T} 0]'$ and $\delta \eta_{T+h}=0, h>0$.

You'll find that you don't get the same results if you order $\pi_t$, $g_t$, $i_t$ or $\pi_t$, $i_t$, $g_t$ and whether you are using norminal or real variable ($g_t-\pi_t$ and $i_t-\pi_t$). Writing $\Delta$ a shock on $C \eta_t$ at time $T$, then $\delta Y_{T+h}=\hat \Phi^h \Delta$, $h>0$. And we are back at the question of the eigenvalues of $\hat \Phi$


\subsection{Why there is debate about VAR-based measures of policy shocks}
\label{subsubsec:var_policy_shocks_debate}

\paragraph{Why this matters.}
Since \citet{sims1980}, VAR models have been the workhorse for studying the macroeconomic effects of monetary policy.
The basic idea is simple: estimate a VAR on a handful of macro variables (output, prices, an interest rate, perhaps reserves or credit), impose identifying restrictions to isolate the ``monetary policy shock,’’ and trace its dynamic effects through impulse responses.
This approach has produced some of the most influential results in empirical macroeconomics---including estimates of how output and prices respond to an unexpected tightening.
But the entire exercise hinges on one question: \emph{is the extracted shock really an exogenous policy surprise?}

\subsubsection{The identification problem: a brief recap}

We gave the full derivation in Section~\ref{sec:sVAR}; here we only recall the key result.
A reduced-form VAR($p$) produces forecast errors $e_t$ with covariance $\Omega = \mathbb{E}(e_t e_t’)$, which are unpredictable given past $y$ but contemporaneously correlated ($\Omega$ is not diagonal).
Each $e_{j,t}$ mixes multiple structural disturbances---policy, demand, supply, information---so $e_t$ has no direct causal interpretation.

A structural decomposition writes $e_t = S\,\varepsilon_t$ with $\mathbb{E}(\varepsilon_t \varepsilon_t’) = I_n$, so $\Omega = SS’$.
But for any orthogonal $Q$, the rotation $\tilde S = SQ$ also satisfies $\Omega = \tilde S \tilde S’$.
We need $n(n-1)/2$ restrictions to pin down $S$, and the debate is about whether the standard choices deliver a credible ``policy shock.’’

\subsubsection{The recursive (Cholesky) approach}

The most common scheme orders the interest rate $i_t$ after ``slow-moving’’ variables (output, prices) and before ``fast-moving’’ ones (reserves, exchange rates).
This makes $S$ lower-triangular and implies:
\begin{itemize}
\item Output and prices do \emph{not} respond to a policy shock within the period (they are ``predetermined’’).
\item The policy rate \emph{can} respond contemporaneously to output and price innovations.
\end{itemize}
The orthogonalized innovation to the interest-rate equation is then labelled the ``monetary policy shock’’ $\varepsilon_{m,t}$, and the impulse response at horizon $h$ is $\Psi_h\, S\, e_m$, where $\Psi_h$ are the MA coefficients of the reduced-form VAR.

This is the approach used in the influential work of \citet{christiano2005nominal} and the earlier survey by \citet{christiano1999monetary}, and it typically finds that a contractionary monetary policy shock reduces output with a hump-shaped pattern peaking around 4--6 quarters.

\subsubsection{Three problems with VAR-based policy shocks}

\paragraph{Problem 1: the reaction-function critique \citep{rudebusch1998measures}.}
A key insight of \citet{rudebusch1998measures} is that the interest-rate equation in a VAR is not just any equation: since the central bank controls the short rate, the VAR equation for $i_t$ is implicitly a \emph{reaction function}, and its residual is implicitly a \emph{policy shock}.
This invites structural scrutiny that other VAR equations do not face.

Formally, let the \emph{true} policy rule be
\begin{equation}
i_t = \phi’\, x_t + \eta_t,
\label{eq:true_rule}
\end{equation}
where $x_t$ is the central bank’s full information set and $\eta_t$ is the true exogenous shock.
The VAR conditions on a much smaller set $\mathcal{I}^{\text{VAR}}_t = \{y_{t-1},\ldots,y_{t-p}\}$.
If important variables are omitted---real-time forecasts, financial conditions, forward guidance signals---the VAR residual absorbs them:
\begin{equation}
\boxed{u_t^{\text{VAR}}
= \underbrace{\eta_t}_{\substack{\text{true}\\\text{policy shock}}}
\;+\;
\underbrace{\phi’\, v_t}_{\substack{\text{systematic policy}\\\text{response to omitted}\\\text{information}}}}
\label{eq:contamination}
\end{equation}
where $v_t$ is the part of $x_t$ not spanned by $\mathcal{I}^{\text{VAR}}_t$.
The Cholesky orthogonalisation does \emph{not} remove $\phi’ v_t$: endogenous policy responses to information missing from the VAR are recorded as ``shocks.’’

\paragraph{Problem 2: the price puzzle.}
A long-standing difficulty in recursive VAR identification is the \emph{price puzzle} \citep{sims1992interpreting}: a contractionary monetary policy shock (a surprise increase in the policy rate) is followed by a temporary \emph{rise} in inflation, or in the price level, before prices eventually decline. This pattern is difficult to reconcile with standard monetary transmission mechanisms.

The leading interpretation, due to \citet{sims1992interpreting}, is that the VAR omits forward-looking information used by the central bank. If the Fed raises rates because it anticipates future inflationary pressures not captured by the VAR, the econometrician may misclassify this endogenous policy response as an exogenous monetary policy shock. The subsequent increase in inflation then appears to be caused by the shock, whereas it is in fact the development the central bank was reacting to. Adding a broad commodity price index to the VAR often mitigates the puzzle, because such a variable helps proxy for omitted information about future inflationary pressures; this provides support for the omitted-information interpretation.

We illustrate a related identification problem in the simulation exercise of Section~\ref{subsec:wrap_wrong}: using the ordering $(i_t, g_t, \pi_t)$ instead of the more standard recursive ordering $(g_t, \pi_t, i_t)$ can mechanically generate a puzzling positive response of prices to a contractionary policy shock (see Figure~\ref{fig:wrapup_wrong}). This simulation is best understood as a pedagogical illustration of how incorrect identification can produce misleading impulse responses, rather than as the main explanation of the empirical price puzzle itself.

\paragraph{Problem 3: instability and low correlation with market surprises.}
\citet{rudebusch1998measures} provides two striking empirical findings:
\begin{enumerate}
\item \textbf{The implied reaction function is unstable.}
The VAR imposes a time-invariant, linear rule with a small information set and long distributed lags.
The implied predictability of future rates is inconsistent with term-structure evidence from the yield curve.
\item \textbf{VAR shocks do not match market-based surprises.}
Using federal funds futures to construct unanticipated policy moves (\`a la \citealt{kuttner2001monetary}), Rudebusch finds that VAR-based policy residuals have \emph{low correlation} with futures-based surprises, and moreover that policy-shock series differ substantially across VAR specifications (low cross-VAR correlation).
\end{enumerate}

\paragraph{An alternative: sign restrictions \citep{uhlig2005effects}.}
The problems above all stem from the \emph{zero restrictions} imposed by the Cholesky ordering (certain variables cannot respond on impact).
\citet{uhlig2005effects} proposes replacing these with weaker \emph{sign restrictions}: for instance, requiring that a contractionary monetary shock raises the interest rate and does not raise output or prices for some horizons.
Sign restrictions are agnostic about the precise magnitude of responses and yield a \emph{set} of admissible impulse responses rather than a point estimate---providing a range of plausible effects rather than a single (potentially misleading) answer.
The price one pays is wider identified sets, but the benefit is that the identifying assumptions are more transparent and easier to justify economically.

\subsubsection{The errors-in-variables perspective}

These problems have a clean econometric interpretation.
Suppose the VAR ``shock’’ is $\hat\varepsilon_t = \varepsilon_t + \nu_t$, where $\varepsilon_t$ is the true structural shock and $\nu_t$ is measurement error (from omitted information, misspecification, etc.).
Then impulse responses estimated by projecting outcomes on $\hat\varepsilon_t$ suffer from classical \emph{attenuation bias}: the estimated response is biased toward zero.
Worse, if $\nu_t$ is correlated with other shocks (as in equation~\ref{eq:contamination}), the bias can go in any direction, producing puzzles like the price puzzle.

This is the deeper reason why the debate matters: it is not merely about which Cholesky ordering to choose, but about whether the VAR interest-rate equation is a credible approximation to the central bank’s reaction function and information set.
This realisation motivates the search for \emph{external} sources of identification, which is the subject of the next section.

\subsubsection{Where does the literature stand?}

The debate has evolved considerably since the original exchange between Sims and Rudebusch.

\paragraph{Defenders of recursive VARs.}
\citet{christiano2005nominal} and \citet{christiano1999monetary} argue that carefully specified recursive VARs---with enough variables to approximate the central bank’s information set---produce reliable estimates of monetary transmission.
Augmenting the VAR with commodity prices, credit spreads, or expectations-based indicators can mitigate the price puzzle and stabilise the identified shocks.

\paragraph{The external-instruments camp.}
\citet{rudebusch1998measures} and the proxy-VAR literature \citep{stockwatson2018, gertlerkaradi2015} argue that no reasonable number of VAR variables can fully capture the central bank’s information advantage, and that credible identification requires ``going outside the VAR’’ via external instruments (see Section~\ref{subsec:exogenous_shocks_VAR_IV}).

\paragraph{A recent synthesis.}
\citet{mirandaagrippino2021} show that the price puzzle largely disappears when the high-frequency monetary surprise is purged of the central bank’s ``information advantage’’---the fact that the Fed’s own forecasts are correlated with the surprise.
Their ``informationally robust’’ instrument produces clean impulse responses that are consistent across recursive VAR, LP-IV, and SVAR-IV, suggesting that much of the earlier disagreement was driven by a common information confound rather than by fundamental model differences.

\paragraph{The information-effect debate.}
\citet{nakamura2018high} documented that FOMC surprises move not only interest rates but also private-sector expectations about output---the so-called ``Fed information effect.’’
If the central bank reveals private information when it acts, then high-frequency surprises are no longer purely exogenous policy shocks.
However, \citet{bauerswanson2023} show that once the pre-FOMC drift in asset prices is properly accounted for, the apparent information effect is much smaller than originally estimated, restoring confidence in the exogeneity of narrow-window monetary surprises.

\paragraph{Where the consensus is heading.}
The debate has shifted from ``should we use VARs at all?’’ to ``what is the right instrument and how should it be constructed?’’
\citet{ramey2016} provides a useful survey that organises the broader field around the question of what constitutes a credible macroeconomic shock.
In current practice, the gold standard is to combine (i)~a reduced-form VAR for dynamic propagation with (ii)~an external instrument for structural identification, and to report both LP-IV and SVAR-IV as a robustness check.

\medskip
\noindent\fbox{\parbox{0.97\textwidth}{
\textbf{Key takeaway for students.}\quad
A VAR is a powerful tool for summarizing dynamics, but extracting \emph{causal} effects of policy requires identification assumptions that go beyond choosing a Cholesky ordering.
The VAR residual is a policy shock only if the VAR equation is the true reaction function---an assumption that is often violated.
This motivates external instruments, narrative identification, and high-frequency approaches.
The companion simulation in Section~\ref{sec:wrapup_monetary} demonstrates these issues concretely.
}}

\subsection{Estimating Dynamic Effects of ``Exogenous'' Shocks with External Instruments and LP-IV}
\label{subsec:exogenous_shocks_VAR_IV}

\paragraph{The big question.}
A central object in applied macroeconometrics is the \emph{dynamic causal effect} of a shock: how does an unexpected monetary tightening at date~$t$ propagate to output, prices, and credit conditions at horizons $h=0,1,2,\dots$?
As discussed in Section~\ref{subsubsec:var_policy_shocks_debate}, recovering such effects from a VAR requires identifying assumptions, and standard recursive schemes can be controversial.
A major modern alternative is to use an \emph{external instrument}: a variable that is correlated with the structural shock of interest, but orthogonal to other shocks.

\paragraph{From internal restrictions to external instruments.}
Traditional SVAR identification imposes restrictions \emph{inside} the model: Cholesky orderings, long-run restrictions, or sign restrictions.
External-instrument methods instead bring in variation from \emph{outside} the VAR.
Typical examples are high-frequency monetary surprises around policy announcements \citep{kuttner2001monetary}, narrative measures constructed from historical records \citep{romerromer2004}, and information-robust monetary shocks \citep{jarocinski2020deconstructing}.
A key point emphasised by \citet{stockwatson2018} is that such series are often best viewed not as the structural shock itself, but as a \emph{proxy} for it.
They may be noisy or partial measures, but IV methods remain valid as long as the proxy is relevant and exogenous.

\subsubsection{Recap: reduced-form VAR and structural shocks}

Let $Y_t\in\mathbb{R}^n$ denote a vector of macroeconomic variables.
A reduced-form VAR($p$) is
\begin{equation}
Y_t = A_1 Y_{t-1} + \cdots + A_p Y_{t-p} + u_t,
\qquad
\mathbb{E}(u_t)=0,\quad \mathbb{E}(u_tu_t')=\Sigma_u,
\label{eq:var_reduced}
\end{equation}
where $u_t$ is a vector of one-step-ahead forecast errors.
Under standard stability conditions, the VAR admits a moving-average representation
\begin{equation}
Y_t = \sum_{h=0}^{\infty}\Psi_h u_{t-h},
\qquad \Psi_0=I_n.
\label{eq:var_ma}
\end{equation}
The matrices $\Psi_h$ describe the dynamic propagation of reduced-form innovations, but these innovations are mixtures of structural shocks.
A structural decomposition writes
\begin{equation}
u_t = B\varepsilon_t,
\qquad
\mathbb{E}(\varepsilon_t\varepsilon_t')=I_n,
\qquad\Longrightarrow\qquad
\Sigma_u = BB'.
\label{eq:svar_impact}
\end{equation}
The structural impulse response to shock $j$ at horizon $h$ is the $j$-th column of
\begin{equation}
\Theta_h = \Psi_h B.
\label{eq:svar_irf}
\end{equation}
The identification problem is therefore to recover the relevant column of $B$.
External instruments provide one way of doing so without relying exclusively on internal restrictions.

\subsubsection{Where do external instruments come from?}

\paragraph{High-frequency monetary surprises.}
A leading example comes from financial-market movements in very narrow windows around monetary policy announcements.
In \citet{kuttner2001monetary}, changes in Fed funds futures around FOMC decisions isolate the \emph{surprise} component of policy.
Because the window is extremely short, one argues that most other macroeconomic shocks are approximately constant within that interval.
The surprise therefore satisfies the instrument conditions (correlated with the policy shock, orthogonal to others) ``by design.''

\paragraph{Narrative measures.}
Another influential strategy is narrative identification.
\citet{romerromer2004} construct a measure of exogenous monetary policy changes by purging intended interest-rate movements of the component explained by the Federal Reserve's own real-time forecasts.
The residual is then interpreted as the unexpected policy innovation.

\paragraph{Information effects and purer policy shocks.}
A complication is that financial-market reactions to policy announcements may combine true policy shocks with central-bank information about the economy.
\citet{nakamura2018high} stress this information effect, while \citet{jarocinski2020deconstructing} propose ways to separate a pure policy component from an information component using the co-movement of interest rates and stock prices around FOMC announcements.
This illustrates an important practical point: external instruments are rarely perfect, and much empirical work consists in improving their exogeneity.

\subsubsection{Local projections with external instruments (LP-IV)}

Local projections, introduced by \citet{jorda2005estimation}, provide a direct way to estimate impulse responses horizon by horizon.
Rather than fitting a full dynamic system and then iterating it forward, one estimates the response at each horizon separately.

\paragraph{The basic regression.}
For variable $i$ and horizon $h$, consider
\begin{equation}
Y_{i,t+h} = \alpha_{h,i} + \beta_{h,i} X_t + \gamma_{h,i}'W_t + e_{i,t+h},
\label{eq:lp_reg}
\end{equation}
where $X_t$ is the policy variable or treatment of interest, and $W_t$ contains controls, typically lags of the endogenous variables and possibly deterministic terms.
The coefficient $\beta_{h,i}$ is interpreted as the impulse response of variable $i$ at horizon $h$.

\paragraph{Why IV is needed.}
In macroeconomic applications, $X_t$ is generally endogenous.
For example, the policy rate responds systematically to inflation and output, so OLS applied to \eqref{eq:lp_reg} does not recover a causal effect.
To isolate the exogenous component of policy, one instruments $X_t$ with an external proxy $Z_t$.

\paragraph{Instrument conditions.}
For LP-IV to identify a causal impulse response, the instrument must satisfy three conditions:
\begin{enumerate}
\item \textbf{Relevance:} $Z_t$ must be correlated with the structural shock of interest.
\item \textbf{Contemporaneous exogeneity:} $Z_t$ must be orthogonal to other shocks hitting the system at date $t$.
\item \textbf{Dynamic exogeneity:} $Z_t$ must not load on past or future structural shocks that also affect $Y_{i,t+h}$.
\end{enumerate}
The third condition is especially important in macroeconomics.
Because outcomes at horizon $h$ depend on the full sequence of shocks between $t$ and $t+h$, the instrument must be dynamically clean, not merely contemporaneously valid.
Including lags of macro variables, and sometimes lags of the instrument itself, helps make this assumption more plausible \citep{stockwatson2018}.

\paragraph{Estimation.}
For each horizon $h$, equation \eqref{eq:lp_reg} is estimated by 2SLS, using $Z_t$ as an instrument for $X_t$.
Since the regression error $e_{i,t+h}$ is serially correlated across overlapping horizons, inference typically relies on HAC standard errors such as Newey--West.

\paragraph{Why LPs are attractive.}
LP-IV has several advantages.
First, it is transparent: each horizon corresponds to a separate regression.
Second, it is flexible: nonlinearities, state dependence, interactions, and asymmetries are easy to incorporate.
Third, it is relatively robust to dynamic misspecification because it does not require specifying and iterating a full VAR.
Its main drawback is lower efficiency: since each horizon is estimated separately, confidence intervals tend to become wide at medium and long horizons.

\subsubsection{Weak instruments and diagnostics}

As in any IV setting, the strength of the first stage is crucial.
If the instrument is only weakly correlated with the structural shock, LP-IV estimates can be badly biased and conventional confidence intervals can be misleading.

\paragraph{First-stage strength.}
The basic diagnostic is the first-stage F-statistic from the regression of the endogenous regressor on the instrument and controls.
A standard rule of thumb is that values below 10 should trigger caution.
In macro time series, however, heteroskedasticity and serial correlation complicate this benchmark.

\paragraph{Robust inference.}
When the instrument is weak, it is preferable to use weak-IV-robust procedures.
A standard option is Anderson--Rubin inference.
In addition, the effective F-statistic of Montiel Olea and Pflueger (2013) can be more informative than the conventional first-stage F-statistic in the presence of heteroskedasticity or autocorrelation.
In applied work, reporting first-stage diagnostics is not optional: it is part of the identification argument.

\subsubsection{How LP-IV relates to VAR methods}

LP-IV is often presented as an alternative to VAR-based estimation of impulse responses.
However, \citet{plagborg2021local} prove a remarkable equivalence result: under correct specification, LP and VAR estimate the \emph{same} population impulse responses.
The practical comparison is therefore not one of fundamentally different estimands, but of implementation trade-offs.

LP-IV remains attractive because it is simple, transparent, and robust to some forms of VAR misspecification.
VAR-based methods may be more statistically efficient when the dynamic model is correctly specified, because they impose cross-equation and cross-horizon restrictions.
But the key conceptual point is that, in population and under correct specification, both approaches target the same dynamic causal object.

\noindent\emph{Remark.} An alternative strategy, known as SVAR-IV or ``proxy VAR,'' uses the external instrument to identify the impact column of $B$ \emph{within} the VAR rather than via separate LP regressions; see \citet{stockwatson2018} and \citet{mertensravn2013} for details, and the companion simulation in Section~\ref{sec:wrapup_monetary} for an implementation.

\subsubsection{Practical checklist}

For an empirical project using external instruments and local projections, a useful workflow is:
\begin{enumerate}
\item Choose a stationary specification for the macro variables of interest and define the controls $W_t$.
\item Construct or import a plausible external instrument: high-frequency surprise, narrative measure, or another shock proxy.
\item Argue carefully for relevance and exogeneity.
\item Report first-stage diagnostics and, if needed, weak-IV-robust inference.
\item Estimate LP-IV horizon by horizon and inspect the impulse responses together with their confidence bands.
\item Check robustness to lag length, sample period, alternative instruments, and alternative control sets.
\end{enumerate}

\medskip
\noindent\fbox{\parbox{0.97\textwidth}{
\textbf{Key takeaway for students.}\quad
External-instrument methods identify structural shocks using variation that comes from outside the VAR.
Local projections then trace out the dynamic causal response horizon by horizon using IV.
Their appeal is transparency and flexibility.
Moreover, as shown by \citet{plagborg2021local}, LPs and VARs target the same population impulse responses under correct specification, so the choice between them is mainly about robustness, efficiency, and ease of implementation rather than about estimating different objects.
}}

\section{Wrap-up exercise: monetary policy shocks---from stationarity to causal inference}\label{sec:wrapup_monetary}

This section\footnote{Companion code: \texttt{monetary\_policy\_SVAR\_IV\_vansteenberghe.py}.} ties together all the time-series tools covered in this course---unit-root tests, cointegration, VAR, Granger causality, SVAR, forecast error variance decomposition, and external instruments---inside a single, self-contained Monte Carlo exercise grounded in standard macroeconomic theory.
We simulate data from a textbook \textbf{three-equation New-Keynesian model}, estimate the causal effect of monetary policy shocks using three different strategies, and compare them to the known truth.

\medskip
\noindent\fbox{\parbox{0.97\textwidth}{
\textbf{Why a Monte Carlo exercise?}\quad
With real-world data, we never observe the true structural shocks.
By simulating from a model with \emph{known} parameters, we can evaluate whether each identification strategy recovers the correct impulse responses.
This is the gold standard for understanding econometric methods before applying them to actual data.
}}

\subsection{The data-generating process}

The economy is described by three equations:

\begin{align}
g_t &= \rho_g \,g_{t-1} - \sigma_r\,(i_{t-1}-\pi_{t-1}) + \varepsilon_t^d
  &&\text{(IS curve: aggregate demand)}\label{eq:wrap_IS}\\[4pt]
\pi_t &= \rho_\pi\,\pi_{t-1} + \kappa\, g_{t-1} + \varepsilon_t^s
  &&\text{(Phillips curve: aggregate supply)}\label{eq:wrap_PC}\\[4pt]
i_t &= \phi_\pi\,\pi_t + \phi_g\, g_t + \varepsilon_t^m
  &&\text{(Taylor rule: monetary policy)}\label{eq:wrap_TR}
\end{align}

where:
\begin{itemize}
\item $g_t$ = output gap (deviation of GDP from potential);
\item $\pi_t$ = inflation rate;
\item $i_t$ = nominal interest rate (central-bank policy rate);
\item $r_t = i_t - \pi_t$ = ex-post real interest rate.
\end{itemize}
The three structural shocks are mutually independent:
$$
\varepsilon_t^d \sim \mathcal{N}(0,\sigma_d^2),\qquad
\varepsilon_t^s \sim \mathcal{N}(0,\sigma_s^2),\qquad
\varepsilon_t^m \sim \mathcal{N}(0,\sigma_m^2).
$$

\paragraph{Economic intuition and interpretation.}
The first equation describes aggregate demand through a simple IS curve. The output gap is persistent ($\rho_g$) and decreases with the real interest rate $r_t = i_t - \pi_t$, since higher borrowing costs discourage consumption and investment. Consequently, tighter monetary policy contracts demand while monetary easing stimulates economic activity. The second equation is a backward-looking Phillips curve linking inflation to past economic slack. Inflation is persistent ($\rho_\pi$) and increases with the lagged output gap ($\kappa$): when the economy operates above potential ($g_t>0$), stronger demand and tighter labor markets put upward pressure on prices, whereas negative output gaps generate disinflationary forces. The parameter $\kappa$ governs the inflation-output trade-off. The third equation represents the central bank's policy rule. Following a Taylor-type rule, the monetary authority raises the nominal interest rate when inflation or the output gap are above target ($\phi_\pi>0$, $\phi_g>0$) and lowers it when economic activity weakens. Choosing $\phi_\pi>1$ is standard in monetary economics because it implies an aggressive response to inflation and is associated with stabilizing policy in canonical New-Keynesian environments. The disturbance $\varepsilon_t^m$ captures unexpected deviations from the systematic component of monetary policy and therefore represents monetary policy shocks.

\paragraph{Recursive structure and identification.}
Notice that $g_t$ depends on \emph{lagged} variables only (not on contemporaneous $\pi_t$ or $i_t$), and $\pi_t$ depends on \emph{lagged} $g_{t-1}$ only.
The interest rate $i_t$, however, responds to \emph{contemporaneous} $g_t$ and $\pi_t$ through the Taylor rule.
This means the contemporaneous impact matrix\footnote{cf. $A_0^{-1}$ in equation \ref{eq:impactmatrix}.} is lower-triangular:
\begin{equation}\label{eq:wrap_S}
S =
\begin{pmatrix}
\sigma_d & 0 & 0 \\
0 & \sigma_s & 0 \\
\phi_g\,\sigma_d & \phi_\pi\,\sigma_s & \sigma_m
\end{pmatrix},
\end{equation}
i.e.\ the model is \emph{recursive by construction}, and the Cholesky ordering $(g_t,\,\pi_t,\,i_t)$ exactly recovers the true structural shocks.
This is the ordering used by \citet{christiano2005nominal}.

\paragraph{Why this DGP?}
This three-equation system is a deliberately simplified, backward-looking New-Keynesian data-generating process.
It captures the core economic channels emphasized in monetary macroeconomics: the central bank sets the policy rate, real activity reacts to the real interest rate, and inflation responds to economic slack.
Its main advantage for teaching is that the structural shocks and the contemporaneous timing are transparent, so that econometric identification can be evaluated against a known truth.

\subsection{Stationarity tests (Section~\ref{sec:adf} recap)}

The three macro variables $(g_t,\pi_t,i_t)$ are stationary ($I(0)$) by construction, and we verify this with the Augmented Dickey--Fuller test.

\subsection{Reduced-form VAR (Section~\ref{sec:sVAR} recap)}

We estimate a reduced-form VAR($p$) on the three stationary variables $(g_t,\pi_t,i_t)$, select the lag order via AIC, and verify diagnostics (stability, residual normality, whiteness).
The estimated companion matrix should be close to the true reduced-form matrix derived from the DGP.

\subsection{Granger causality tests (Section~\ref{sec:granger} recap)}

Before moving to structural identification, we examine the \emph{predictive} linkages between variables using the Granger causality test.
Recall: ``$X$ Granger-causes $Y$'' means that past values of $X$ help predict $Y$ beyond $Y$'s own past.
The null hypothesis is that the coefficients on lagged $X$ in the $Y$ equation are jointly zero.

In our DGP, we expect the following Granger-causal links:
\begin{itemize}
\item $g_t \to \pi_t$ (Phillips curve: the output gap enters the inflation equation with a lag).
\item $g_t \to i_t$ and $\pi_t \to i_t$ (Taylor rule: both enter the interest-rate equation).
\item $i_t \to g_t$ (IS curve: the lagged interest rate affects the output gap through the real rate).
\end{itemize}

\medskip\noindent\textit{Practical lesson:} Granger causality captures \emph{predictive} content, not \emph{structural} causality.
Finding that $i_t$ Granger-causes $g_t$ does \emph{not} identify a monetary policy shock---it merely confirms that the interest rate is a useful predictor of future output.
The interest rate could be predictive because the central bank systematically responds to economic conditions, not because policy shocks drive output.
This is precisely why we need \emph{structural} identification (SVAR or external instruments).

\subsection{Structural VAR with Cholesky identification}

Following Section~\ref{sec:Chholesky}, we impose a lower-triangular $A$ matrix:
\begin{equation}
A =
\begin{pmatrix}
1 & 0 & 0 \\
E & 1 & 0 \\
E & E & 1
\end{pmatrix}
\end{equation}
where \texttt{'E'} denotes parameters to be estimated.
Because the DGP is recursive, the Cholesky ordering $(g_t,\pi_t,i_t)$ is correctly specified in population.
Accordingly, in sufficiently large samples, the Cholesky-SVAR impulse responses should converge to the true structural responses.
Finite-sample deviations may nevertheless arise from estimation error; we quantify this uncertainty using a \textbf{residual bootstrap} (500 replications) to construct 90\% confidence bands around the Cholesky IRFs.

The third structural shock (innovation to the interest-rate equation, orthogonal to output and inflation innovations) is the \textbf{monetary policy shock}.
Its estimated impulse response should show:
\begin{itemize}
\item A \emph{decline} in the output gap (contractionary policy reduces demand).
\item A \emph{decline} in inflation (via the Phillips curve, with a lag).
\item A \emph{rise and decay} in the interest rate (the shock itself plus endogenous adjustment).
\end{itemize}

\subsection{Forecast error variance decomposition (FEVD)}
\label{subsec:wrap_fevd}

The impulse response function (IRF) shows the \emph{direction and magnitude} of the effect of each shock.
The \textbf{forecast error variance decomposition} complements it by answering a different question: \emph{how important is each shock in explaining the variability of each variable?}

At horizon $h$, the fraction of the forecast error variance of variable $j$ due to structural shock $k$ is:
\begin{equation}\label{eq:wrap_fevd}
\mathrm{FEVD}_{j,k}(h) \;=\;
\frac{\displaystyle\sum_{\ell=0}^{h} (\Psi_\ell S)_{j,k}^{\,2}}
     {\displaystyle\sum_{\ell=0}^{h}\sum_{m=1}^{n} (\Psi_\ell S)_{j,m}^{\,2}},
\end{equation}
where $\Psi_\ell$ is the reduced-form MA coefficient at lag $\ell$ and $S$ is the structural impact matrix.

\begin{figure}[htbp]
\centering
\includegraphics[width=0.95\textwidth]{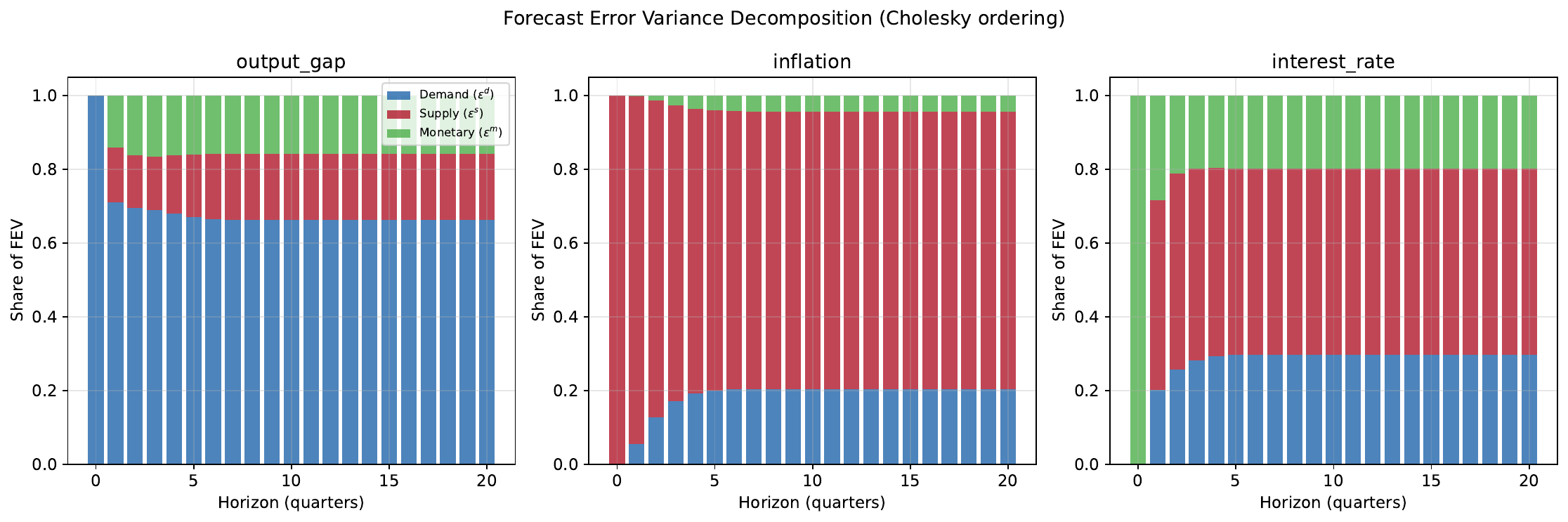}
\caption{Forecast error variance decomposition under the correct Cholesky ordering $(g_t,\pi_t,i_t)$.}
\label{fig:wrapup_fevd}
\end{figure}

\medskip\noindent\textit{Key results from the simulation} (Figure~\ref{fig:wrapup_fevd}):
\begin{itemize}
\item At $h=0$, the output gap is driven entirely by demand shocks (by the recursive structure).
At longer horizons, monetary policy shocks account for a significant share (about 16\%) of output gap variability.
\item Supply shocks dominate inflation at all horizons (about 75\%), with demand shocks contributing about 20\% via the Phillips curve.
\item The interest rate variance is split across all three shocks, reflecting the Taylor rule: the central bank reacts to both demand and supply conditions.
\end{itemize}

\subsection{External instrument: a noisy proxy for the policy shock}
\label{subsec:wrap_instrument}

Following Section~\ref{subsec:exogenous_shocks_VAR_IV}, we now suppose the econometrician does \emph{not} trust the Cholesky ordering and instead has access to an external instrument:
\begin{equation}\label{eq:wrap_instrument}
z_t = \varepsilon_t^m + \eta_t,\qquad \eta_t \sim \mathcal{N}(0,\sigma_\eta^2),
\end{equation}
where $\eta_t$ is measurement noise independent of all structural shocks.
Think of $z_t$ as a \emph{high-frequency monetary surprise} (\`a la \citealt{kuttner2001monetary}): it is correlated with the true policy shock $\varepsilon_t^m$ (relevance) but orthogonal to $\varepsilon_t^d$ and $\varepsilon_t^s$ (exogeneity), because the announcement window is too short for demand or supply shocks to contaminate the futures-price change.

The theoretical first-stage $R^2$ is $\sigma_m^2/(\sigma_m^2+\sigma_\eta^2)$, and we verify instrument strength with the first-stage F-statistic (rule of thumb: $F>10$).

\subsection{LP-IV and SVAR-IV estimation}

We estimate the impulse response of the monetary policy shock using two external-instrument strategies:

\paragraph{LP-IV (Section~\ref{subsec:exogenous_shocks_VAR_IV}).}
For each horizon $h$, we estimate the local-projection regression
\begin{equation}\label{eq:wrap_lp_reg}
Y_{j,t+h} = \beta_{h,j}\, i_t + \gamma_{h,j}'\, W_t + e_{j,t+h},
\end{equation}
by \textbf{two-stage least squares} (2SLS), using $z_t$ as an instrument for $i_t$.
The control vector $W_t$ includes lagged values of the endogenous variables.
In practice, the two stages are:

\medskip
\noindent\textit{First stage.}\; Regress the endogenous regressor $i_t$ on the instrument $z_t$ and the controls $W_t$:
$$
i_t \;=\; \delta\, z_t \;+\; \phi'\, W_t \;+\; v_t.
$$
Compute the fitted value $\hat\imath_t = \hat\delta\, z_t + \hat\phi'\, W_t$, which captures only the variation in $i_t$ that is driven by the exogenous instrument $z_t$ (i.e.\ the monetary-policy shock component), \emph{purged} of the endogenous Taylor-rule response.
The first-stage F-statistic on $\hat\delta$ tests instrument relevance (rule of thumb: $F > 10$).

\medskip
\noindent\textit{Second stage.}\; Replace $i_t$ by $\hat\imath_t$ in \ref{eq:wrap_lp_reg} and run OLS:
$$
Y_{j,t+h} \;=\; \beta_{h,j}\, \hat\imath_t \;+\; \gamma_{h,j}'\, W_t \;+\; \tilde e_{j,t+h}.
$$
The coefficient $\hat\beta_{h,j}$ is the LP-IV estimate of the impulse response of variable $j$ at horizon~$h$.
Because $\hat\imath_t$ varies only with the exogenous instrument, $\hat\beta_{h,j}$ captures the \emph{causal} effect of a monetary-policy shock, free of the simultaneity bias that would afflict a simple OLS regression of $Y_{j,t+h}$ on $i_t$.

\medskip
We use Newey--West (HAC) standard errors because the LP residuals are serially correlated by construction at horizons $h > 0$.
LP-IV requires no VAR model and is robust to VAR misspecification, but it is noisier at long horizons because each horizon is estimated independently.

\paragraph{SVAR-IV.}
We regress the VAR interest-rate residual $\hat u_{3,t}$ on $z_t$ (first stage), then regress the other residuals $\hat u_{1,t}$, $\hat u_{2,t}$ on the fitted value $\hat u_{3,t}^{\textit{IV}}$ (second stage).
This identifies the impact column $B_{\cdot 3}$ up to scale:
$$
\hat b_{j,3} = \frac{\operatorname{Cov}(\hat u_{3,t}^{\textit{IV}},\;\hat u_{j,t})}
                     {\operatorname{Var}(\hat u_{3,t}^{\textit{IV}})},
\qquad j = 1, 2.
$$
The full impulse response is then propagated through the VAR dynamics: $\Theta_h = \Psi_h\, B_{\cdot 3}$.
SVAR-IV is more efficient than LP-IV (because it reuses the VAR dynamics for all horizons) but inherits any VAR misspecification.

\subsection{Comparison of the four methods}

\begin{figure}[htbp]
\centering
\includegraphics[width=0.95\textwidth]{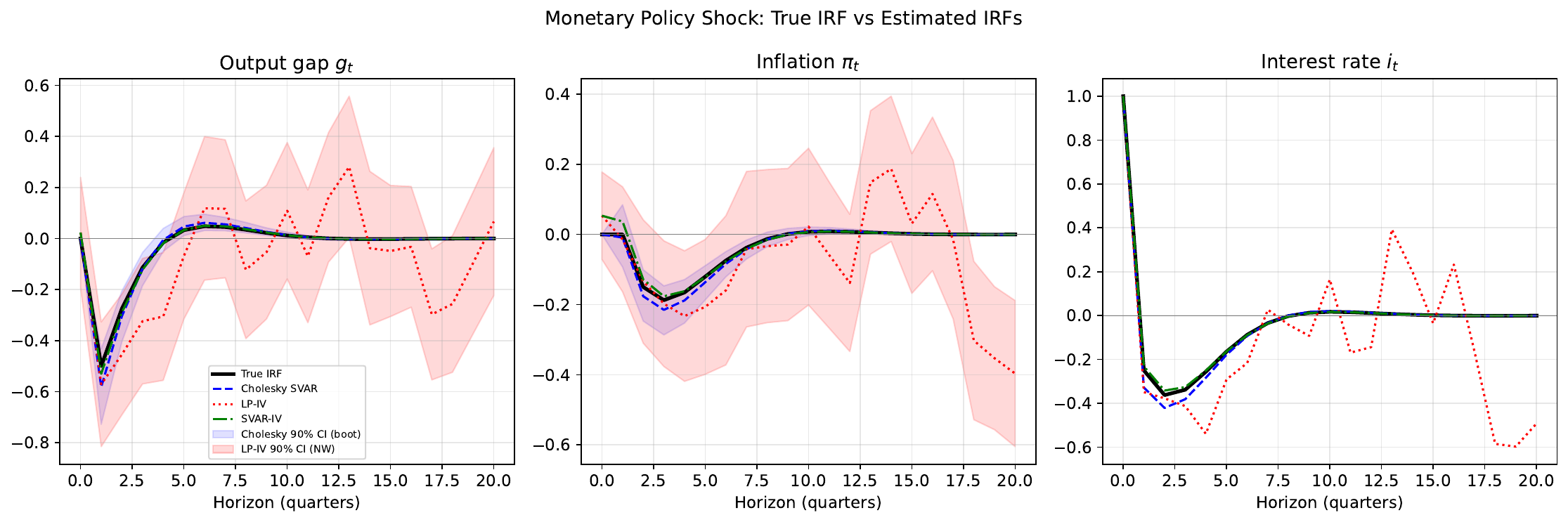}
\caption{Monetary policy shock: comparison of the true IRF (black solid) with Cholesky SVAR (blue dashed), LP-IV (red dotted), and SVAR-IV (green dash-dot). The blue shaded band is a 90\% bootstrap confidence interval for the Cholesky SVAR; the red shaded band is a 90\% Newey--West confidence interval for LP-IV.}
\label{fig:wrapup_irf}
\end{figure}

Because we know the true DGP, we can compare four impulse-response estimates (Figure~\ref{fig:wrapup_irf}):

\begin{center}
\begin{tabular}{lp{9cm}}
\toprule
\textbf{Method} & \textbf{What it estimates and when it works} \\
\midrule
True IRF & Computed analytically from the structural parameters. Benchmark. \\[3pt]
Cholesky SVAR & Exploits the recursive ordering $(g,\pi,i)$. Works \emph{here} because the DGP is recursive. Would fail if the ordering is wrong (cf.\ Section~\ref{subsec:wrap_wrong}). \\[3pt]
LP-IV & Uses the external instrument. Robust to VAR misspecification but noisier. HAC standard errors recommended. \\[3pt]
SVAR-IV & Uses the instrument to identify the VAR impact column. More efficient than LP-IV but inherits VAR specification. \\
\bottomrule
\end{tabular}
\end{center}

In this simulation, all three estimated methods should be close to the truth.
With \emph{real data}, the Cholesky ordering may be incorrect (the central bank may react to information not in the VAR), while LP-IV and SVAR-IV remain valid as long as the instrument satisfies relevance and exogeneity.

\subsection{What goes wrong with the wrong Cholesky ordering}
\label{subsec:wrap_wrong}

To illustrate \emph{why identification matters}, we re-estimate the SVAR using the \emph{wrong} ordering: $(i_t,\, g_t,\, \pi_t)$ instead of the correct $(g_t,\, \pi_t,\, i_t)$.

\paragraph{What changes.}
Placing $i_t$ first in the Cholesky ordering imposes that the interest rate is ``predetermined''---it does not respond to contemporaneous output or inflation innovations.
The ``monetary policy shock'' in this wrong ordering is the raw interest-rate innovation, \emph{not} orthogonalised against the Taylor-rule response to $g_t$ and $\pi_t$.
In other words, it confounds the exogenous policy shock $\varepsilon_t^m$ with the central bank's \emph{endogenous} reaction $\phi_\pi\,\pi_t + \phi_g\, g_t$.

\begin{figure}[htbp]
\centering
\includegraphics[width=0.95\textwidth]{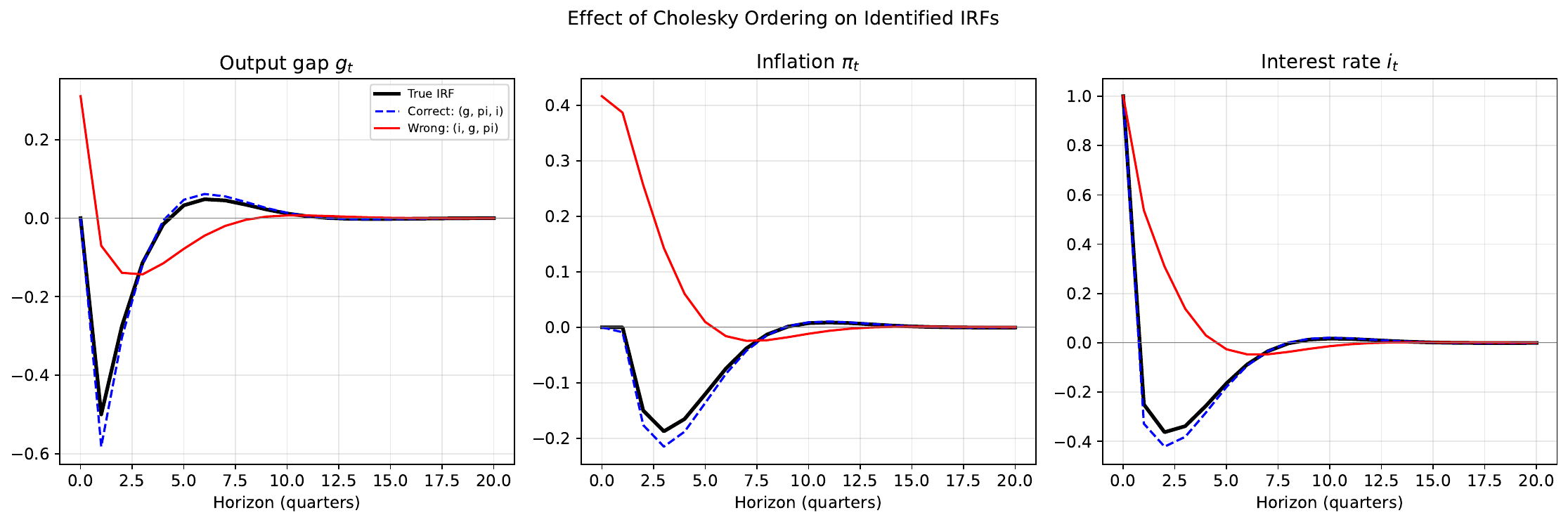}
\caption{Monetary policy shock: correct ordering $(g,\pi,i)$ vs wrong ordering $(i,g,\pi)$. The wrong ordering produces a ``price puzzle'' (contractionary shock appears to raise inflation) because the identified shock is contaminated by the Taylor-rule response.}
\label{fig:wrapup_wrong}
\end{figure}

\paragraph{The price puzzle in this exercise.}
In this simulation, the wrong Cholesky ordering generates a price puzzle mechanically: the identified ``policy shock'' is contaminated by the central bank's endogenous response to current macroeconomic conditions.
More generally, in empirical applications, price puzzles may also reflect omitted information, imperfect measurement, or an inadequate identification scheme.
Figure~\ref{fig:wrapup_wrong} shows the result: the wrongly identified ``tightening shock'' appears to \emph{raise} both output and inflation on impact.
This is the \textbf{price puzzle} phenomenon documented in the literature \citep{sims1992interpreting}---an apparent increase in inflation following a monetary tightening.
The puzzle disappears when using the correct ordering or an external instrument, confirming that the puzzle is a \emph{misidentification artefact}, not an economic anomaly.

\medskip\noindent\textit{Practical lesson:} the Cholesky identification is only valid when the assumed ordering matches the true contemporaneous structure of the economy.
Since we never know this ordering with certainty in real data, external instruments (LP-IV, SVAR-IV) provide a robust alternative: they identify the shock through an \emph{economically motivated exclusion restriction} rather than a \emph{mechanical ordering assumption}.

\subsection{Non-recursive DGP: when Cholesky fails but IV succeeds}
\label{subsec:wrap_nonrecursive}

The previous section showed that the \emph{wrong} ordering biases the Cholesky SVAR.
But what if the true contemporaneous structure is not recursive \emph{at all}?
To illustrate this, we modify the Phillips curve so that inflation responds to the \emph{contemporaneous} interest rate:
\begin{equation}\label{eq:wrap_PC_nr}
\pi_t = \rho_\pi\,\pi_{t-1} + \kappa\, g_{t-1} + \delta_i\, i_t + \varepsilon_t^s,
\end{equation}
where $\delta_i > 0$ captures, for instance, an exchange-rate pass-through channel (a rate hike appreciates the currency and lowers import prices) or a forward-looking expectations channel.

\paragraph{Why Cholesky fails.}
Because $\pi_t$ now depends on $i_t$ and $i_t$ depends on $\pi_t$ (via the Taylor rule), no ordering of $(g_t,\pi_t,i_t)$ makes the impact matrix $S$ lower-triangular.
The structural impact matrix becomes:
$$
S_{\text{NR}} =
\begin{pmatrix}
\sigma_d & 0 & 0 \\
\frac{\delta_i\,\phi_g\,\sigma_d}{1 - \delta_i\,\phi_\pi} & \frac{\sigma_s}{1 - \delta_i\,\phi_\pi} & \frac{\delta_i\,\sigma_m}{1 - \delta_i\,\phi_\pi} \\[4pt]
\frac{\phi_g\,\sigma_d}{1 - \delta_i\,\phi_\pi} & \frac{\phi_\pi\,\sigma_s}{1 - \delta_i\,\phi_\pi} & \frac{\sigma_m}{1 - \delta_i\,\phi_\pi}
\end{pmatrix},
$$
which has a non-zero $(2,3)$ entry: monetary policy shocks move inflation \emph{on impact}.
Cholesky---even with the ``correct'' ordering---forces this entry to zero and therefore misidentifies the shock.

\begin{figure}[htbp]
\centering
\includegraphics[width=0.95\textwidth]{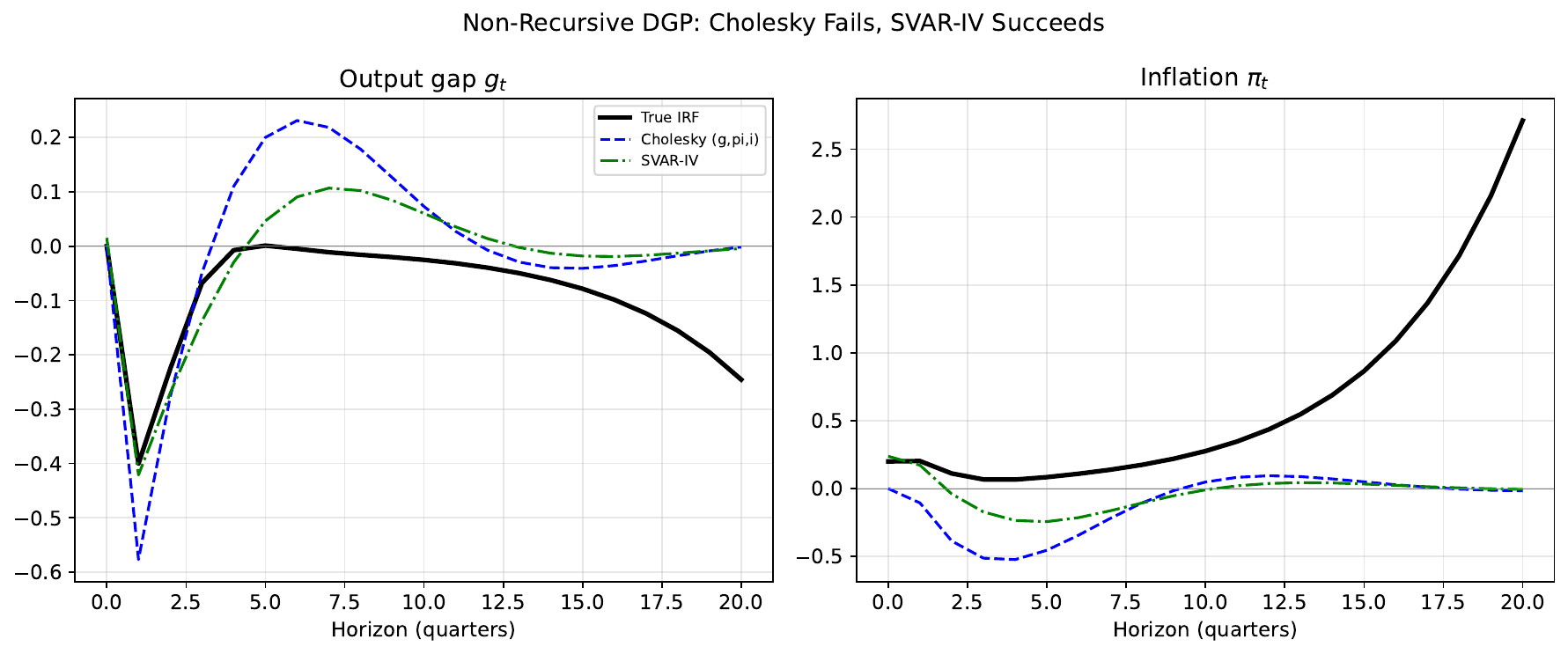}
\caption{Non-recursive DGP ($\delta_i = 0.2$): Cholesky with the ``correct'' ordering $(g,\pi,i)$ is misspecified and produces biased IRFs, while SVAR-IV recovers the true response.}
\label{fig:wrapup_nonrecursive}
\end{figure}

\paragraph{IV methods still work.}
SVAR-IV identifies the impact column through the external instrument, not through a triangular decomposition.
As long as $z_t$ satisfies relevance and exogeneity, the SVAR-IV estimate remains consistent regardless of whether the DGP is recursive.
Figure~\ref{fig:wrapup_nonrecursive} confirms this: the Cholesky IRF diverges from the truth, while SVAR-IV tracks it closely.

\medskip\noindent\textit{Practical lesson:} external instruments do not require knowledge of the contemporaneous causal structure. This is their fundamental advantage over Cholesky identification.

\subsection{Contaminated instrument: when exogeneity fails}
\label{subsec:wrap_contaminated}

The previous sections assumed that the instrument $z_t = \varepsilon_t^m + \eta_t$ is \emph{exogenous}: orthogonal to $\varepsilon_t^d$ and $\varepsilon_t^s$.
In practice, this assumption can fail.
A prominent concern is the \textbf{information effect} (\citealt{nakamura2018high}): if the central bank's announcement reveals information about the state of the economy, the high-frequency surprise $z_t$ may partly reflect macro news (demand shocks), not just the policy shock.

To illustrate, we construct a \emph{contaminated} instrument:
\begin{equation}\label{eq:wrap_contam}
z_t^c = \varepsilon_t^m + \eta_t + \lambda\,\varepsilon_t^d, \qquad \lambda > 0,
\end{equation}
where $\lambda$ controls the degree of contamination.
This instrument still satisfies \emph{relevance} ($\operatorname{Cov}(z_t^c, \varepsilon_t^m) \neq 0$) but violates \emph{exogeneity} ($\operatorname{Cov}(z_t^c, \varepsilon_t^d) = \lambda\,\sigma_d^2 \neq 0$).

\begin{figure}[htbp]
\centering
\includegraphics[width=0.75\textwidth]{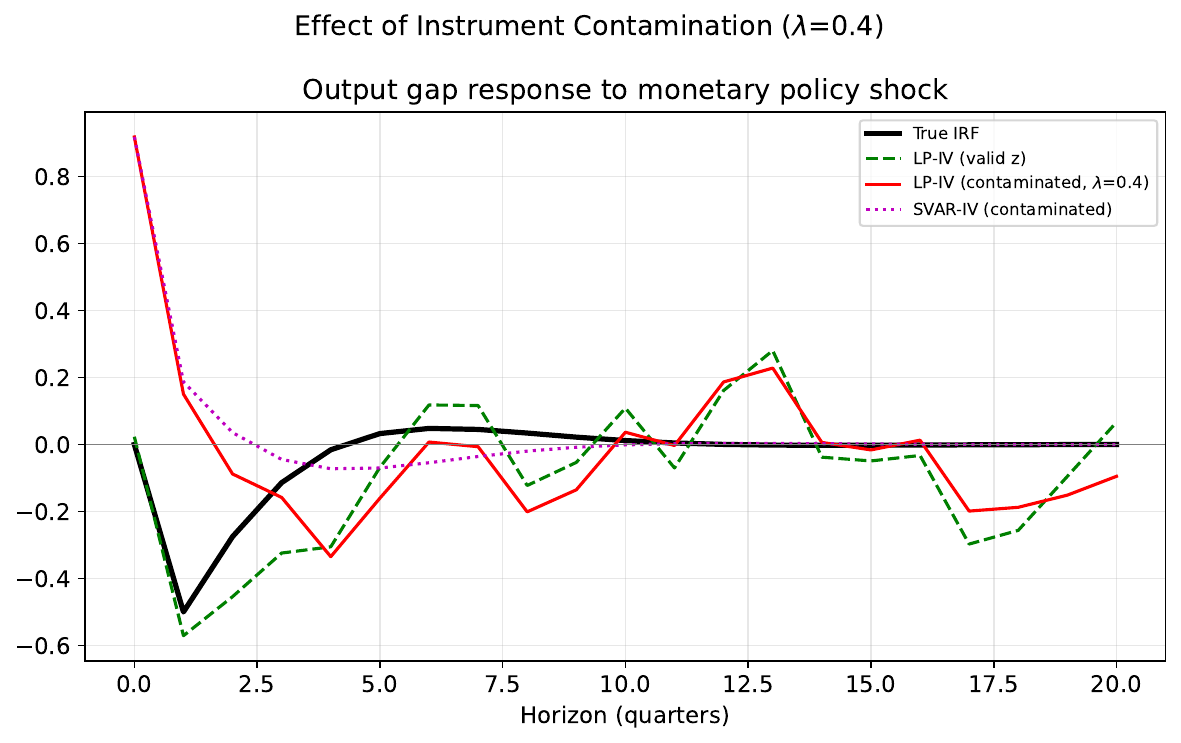}
\caption{Effect of instrument contamination ($\lambda = 0.4$). Both LP-IV and SVAR-IV using the contaminated instrument (red, magenta) are biased relative to the true IRF (black) and the valid-instrument LP-IV (green).}
\label{fig:wrapup_contaminated}
\end{figure}

Figure~\ref{fig:wrapup_contaminated} shows the result: both LP-IV and SVAR-IV produce \emph{biased} estimates when using $z_t^c$.
The bias is towards less contractionary effects, because the contaminated instrument partly captures expansionary demand shocks alongside the contractionary monetary shock.

\medskip\noindent\textit{Practical lesson:} instrument exogeneity is \emph{critical} and \emph{untestable}---it must be justified by economic reasoning about the identification strategy (e.g., the width of the surprise window, whether Fed information effects are present).

\subsection{Monte Carlo repetitions: sampling distributions}
\label{subsec:wrap_montecarlo}

A single simulation draw illustrates qualitative patterns but cannot distinguish estimation noise from systematic bias.
To confirm the properties of each identification strategy, we repeat the full DGP-estimation pipeline 500 times, each time drawing fresh structural shocks and estimating the output-gap IRF at horizon $h=4$.

\begin{figure}[htbp]
\centering
\includegraphics[width=0.95\textwidth]{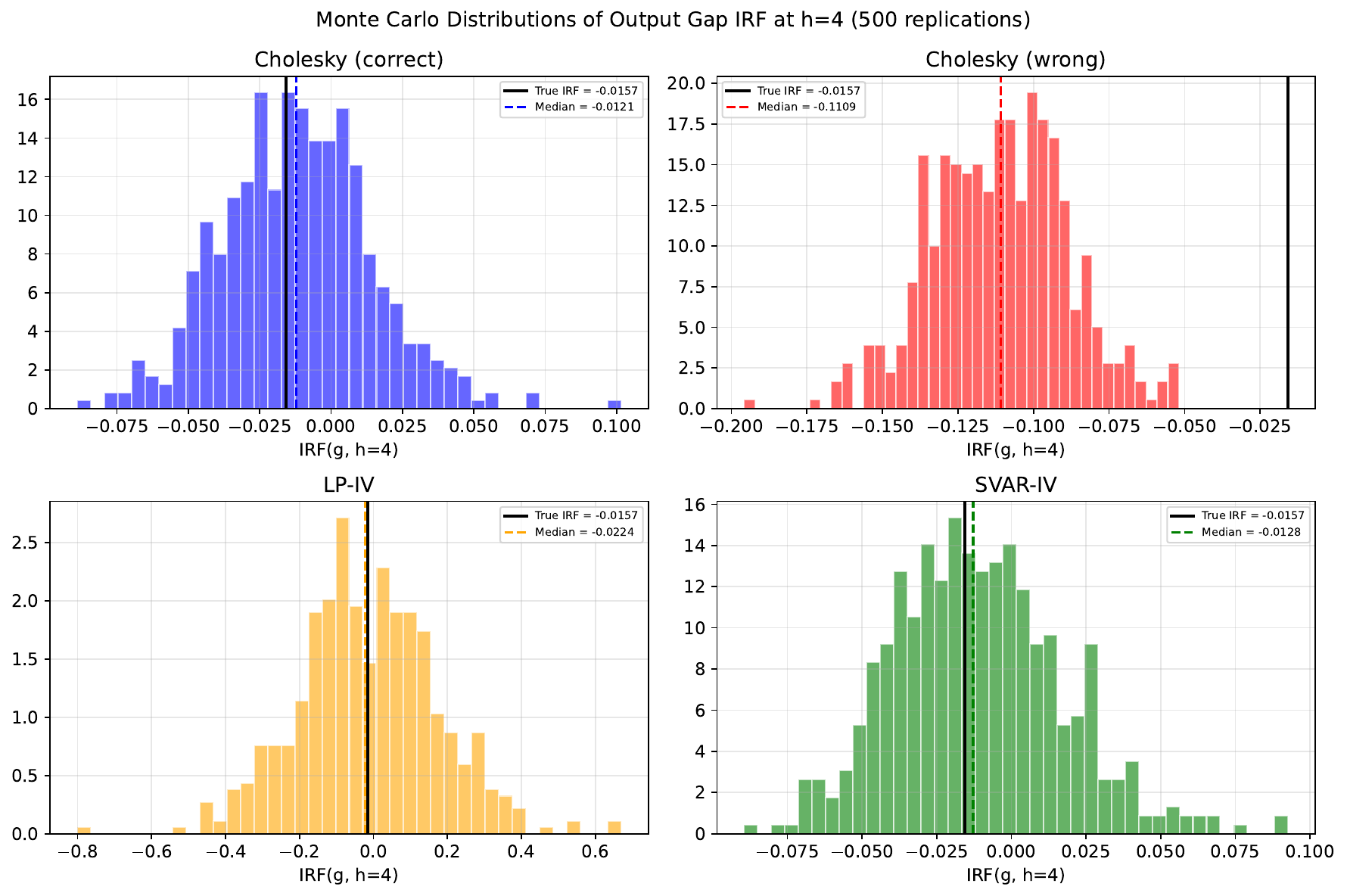}
\caption{Monte Carlo distributions of the output-gap IRF at $h=4$ (500 replications). The vertical black line is the true IRF; the dashed line is the median across replications. Cholesky (correct) and SVAR-IV are centred on the truth with small variance; Cholesky (wrong) is biased; LP-IV is unbiased but noisy.}
\label{fig:wrapup_montecarlo}
\end{figure}

Figure~\ref{fig:wrapup_montecarlo} shows the resulting sampling distributions:
\begin{itemize}
\item \textbf{Cholesky (correct ordering)} is centred on the true IRF with small dispersion---it is \emph{unbiased} and \emph{efficient} when the recursive structure is correct.
\item \textbf{Cholesky (wrong ordering)} is shifted away from the truth---the distribution is centred on a \emph{biased} value, confirming that ordering errors produce inconsistent estimates.
\item \textbf{LP-IV} is centred on the truth (\emph{unbiased}) but has much larger dispersion, reflecting the efficiency cost of model-free, horizon-by-horizon estimation.
\item \textbf{SVAR-IV} is centred on the truth with dispersion comparable to the Cholesky---it is both \emph{unbiased} and \emph{efficient}, gaining power by reusing the VAR dynamics.
\end{itemize}

\medskip\noindent\textit{Practical lesson:} this Monte Carlo confirms the \textbf{bias--variance trade-off} at the heart of causal inference in macroeconomics.
The ``best'' method depends on what assumptions you are willing to maintain: if the recursive ordering is known, Cholesky is hard to beat; if not, SVAR-IV with a valid instrument offers the best combination of consistency and efficiency; LP-IV provides robustness insurance at the cost of precision.

\subsection{Summary and key takeaways}

\noindent\fbox{\parbox{0.97\textwidth}{
\textbf{What this exercise teaches.}
\begin{enumerate}
\item \textbf{Stationarity first.} Always test for unit roots before fitting a VAR. Working with $I(1)$ series in a VAR produces spurious results.
\item \textbf{Granger causality $\neq$ structural causality.} Finding that $i_t$ Granger-causes $g_t$ does not identify a monetary policy shock; it only confirms predictive content.
\item \textbf{Identification is the hard part.} The same reduced-form VAR can be combined with different identification strategies (Cholesky, external instruments), and the choice depends on \emph{economic reasoning}---not statistical criteria.
\item \textbf{Ordering matters for Cholesky.} The wrong Cholesky ordering produces biased IRFs and can generate puzzles like the price puzzle.
\item \textbf{Non-recursive DGPs break Cholesky entirely.} When the true contemporaneous structure is not triangular, \emph{no} Cholesky ordering recovers the true shock---only IV methods remain consistent.
\item \textbf{External instruments are robust to ordering---but require exogeneity.} LP-IV and SVAR-IV do not require a correct ordering; they only require a valid instrument. When exogeneity fails (contaminated instrument), IV methods become biased.
\item \textbf{FEVD complements IRFs.} The IRF shows the direction of effects; the FEVD shows how important each shock is for driving variability.
\item \textbf{LP-IV vs SVAR-IV: a bias-variance trade-off.} LP-IV is more robust (no VAR model needed) but noisier. SVAR-IV is more efficient but depends on the VAR being correctly specified. Monte Carlo simulations confirm this trade-off quantitatively.
\end{enumerate}
}}

\section{Time-varying coefficients: rolling windows, ridge regressions, and modern approaches}
\label{sec:TVP}

So far, every regression in this course has carried a hidden assumption: the coefficients are the \emph{same} at the beginning and at the end of the sample\footnote{Companion code: \texttt{vansteenberghe\_time\_varying\_parameters.py}.}.
Yet much of what makes macroeconomics and finance interesting is precisely that relationships evolve.
The sensitivity of inflation to economic slack---the slope of the Phillips curve---is widely believed to have declined over recent decades \citep{stock2007has,delnegro2020phillips}; the transmission of monetary policy has changed with financial innovation and policy regimes \citep{primiceri2005,cogley2005drifts}; a stock's CAPM beta (Section~\ref{sec:CAPM}) moves as the firm's leverage and business mix change; and the predictive power of valuation ratios for returns comes and goes \citep{dangl2012}.
There is also a deep theoretical reason to \emph{expect} instability: the \citet{lucas1976} critique states that the coefficients of reduced-form relationships are functions of agents' expectations and of the policy regime, so that when the regime shifts, the coefficients shift with it.
A constant-parameter regression estimated across regimes is then, at best, an average of different economies.

This section confronts the problem head-on.
We first explain why the seemingly natural move---simply indexing the coefficients by time---produces a model that is \emph{not identified}, and why every usable method must take a stand on \emph{how} coefficients are allowed to move.
We then study the three workhorses in increasing order of sophistication: rolling-window regressions, the state-space (Kalman filter) approach, and the ridge-regression reformulation of \citet{gouletcoulombe2025ridge}, which turns out to unify the previous two and to connect time-varying parameters to the machine-learning toolbox of Section~\ref{sec:MLwhy}.
Finally, we map the modern landscape---Bayesian shrinkage, score-driven models, and the ``macroeconomy as a random forest'' of \citet{gouletcoulombe2024forest}---and we test everything inside the New-Keynesian laboratory of Section~\ref{sec:wrapup_monetary}, where the true coefficient path is known by construction.

\subsection{Why we cannot ``simply'' let the coefficients move}
\label{subsec:TVPident}

Consider the univariate regression we have used many times, but now with time-indexed coefficients:
\begin{equation}
\label{eq:TVPnaive}
y_t \;=\; \alpha_t \;+\; \beta_t\, x_t \;+\; \varepsilon_t,
\qquad t = 1, \dots, T.
\end{equation}
Writing equation~\eqref{eq:TVPnaive} costs nothing---and that is exactly the problem.
As written, it is not a model but an accounting identity waiting to happen: it has $2T$ free coefficients $(\alpha_1,\beta_1),\dots,(\alpha_T,\beta_T)$, plus the residual variance, for only $T$ observations.
Least squares (or maximum likelihood) applied to \eqref{eq:TVPnaive} without further restrictions does not produce estimates; it produces an embarrassment of riches.
To see this, consider the two polar ``solutions'':

\paragraph{Pole 1: attribute everything to the residuals.}
Impose $\alpha_t = \alpha$ and $\beta_t = \beta$ for all $t$ and run OLS.
The model is identified again, but any genuine movement in the coefficients has been swept into the error term.
The misspecification does not disappear---it resurfaces as symptoms we have already learned to detect: serially correlated residuals, unstable subsample estimates, apparent structural breaks, and poor out-of-sample performance.
Formally, if the true slope drifts, the OLS estimand is (approximately) a variance-weighted average of the path $\{\beta_t\}$, a single number summarizing a moving target.

\paragraph{Pole 2: attribute everything to the coefficients.}
Go to the other extreme and let every period have its own coefficients.
For each $t$, equation~\eqref{eq:TVPnaive} is \emph{one} equation with \emph{two} unknowns $(\alpha_t, \beta_t)$: we can set $\varepsilon_t = 0$ and pick any pair on the line $\alpha + \beta x_t = y_t$---for instance $\alpha_t = 0$, $\beta_t = y_t / x_t$.
The fit is perfect in every period: $\mathrm{SSR} = 0$, $R^2 = 1$.
But the ``estimates'' are pure noise: the path $\{\beta_t\}$ absorbs every shock $\varepsilon_t$ into the coefficients, the sampling variance of each $\beta_t$ is unbounded, and the model has no forecasting content whatsoever---which $\beta$ should we even use for $T+1$?
The last estimate $\beta_T$ reflects nothing but the last observation's noise.

\paragraph{Everything in between is also feasible.}
The two poles are only the endpoints of a continuum.
Pick \emph{any} candidate residual series $\{e_t\}$ you like; then choose the coefficients so as to fit $y_t - e_t$ exactly in every period.
The data alone cannot tell ``coefficient variation'' apart from ``noise'': every decomposition of $y_t$ into signal and residual is observationally equivalent.
In likelihood language, the likelihood is flat (indeed unbounded) along directions that trade off the variance of the coefficient path against the variance of the residuals.
This is an \emph{identification} failure, not an estimation difficulty: no amount of data fixes it, because each new observation brings a new pair of free parameters.

\medskip
\noindent\fbox{\parbox{0.97\textwidth}{
\textbf{The identification problem in one sentence.}\quad
Without an a priori restriction on \emph{how} $(\alpha_t, \beta_t)$ can move, the split of $y_t$ between ``signal through the coefficients'' and ``residual noise'' is arbitrary: we can attribute everything to $\varepsilon_t$ (constant-parameter OLS) or nothing (perfect fit with $R^2=1$), and anything in between.
Every practical method in this section is, at bottom, a different way of rationing the coefficients' freedom to move---a choice of \emph{effective} degrees of freedom strictly between $K$ (constant coefficients) and $T$ (interpolation).
}}

\paragraph{Effective degrees of freedom.}
A useful way to organize everything that follows: all the estimators of this section are \emph{linear smoothers}, meaning that the fitted values can be written $\hat{y} = S_{\lambda}\, y$ for some $T \times T$ matrix $S_{\lambda}$ that depends on the data $x$ and on a tuning choice $\lambda$ (a window length, a bandwidth, a signal-to-noise ratio, a penalty).
The trace $\mathrm{df}(\lambda) = \operatorname{tr}(S_{\lambda})$ counts the \emph{effective number of parameters} \citep{hastie2009elements}.
OLS uses $\mathrm{df} = K$; the interpolation pole uses $\mathrm{df} = T$; every credible time-varying-coefficient estimator lives strictly in between and must justify \emph{where} in between.
The three classical routes for doing so are:
\begin{itemize}
\item \textbf{Deterministic restrictions}: the path is constant except for a small number of discrete breaks---at known dates \citep{chow1960}, at estimated dates \citep{baiperron1998}---or switches between a small number of recurring regimes \citep[][and the Markov-switching models of the nonlinear time-series section]{Hamilton1989aa}.
\item \textbf{Local estimation}: the path is smooth enough to be treated as locally constant, so that nearby observations are approximately draws from the same model---rolling windows, and their kernel-weighted refinements \citep{giraitis2014}.
\item \textbf{Stochastic-process priors}: the path follows a parsimonious stochastic process, typically a random walk with small innovations---the state-space route \citep{cooley1976,harvey1989}, which, as \citet{gouletcoulombe2025ridge} makes explicit, is \emph{the same thing} as penalizing coefficient changes in a ridge regression.
\end{itemize}

\subsection{Rolling-window least squares}
\label{subsec:TVProlling}

The rolling window is the oldest and most transparent answer.
We already met it briefly in the VAR section (Section~\ref{sec:sVAR}); here we treat it as an estimator of a coefficient \emph{path}.
For a window of length $W$, the one-sided (``real-time'') rolling estimator at date $t$ is
\begin{equation}
\label{eq:TVProll}
\hat{\beta}_t^{(W)} \;=\; \operatorname*{arg\,min}_{\beta} \sum_{s = t-W+1}^{t} \left( y_s - x_s'\beta \right)^2 ,
\end{equation}
i.e.\ OLS on the most recent $W$ observations only.
A \emph{centered} variant uses the window $[t - W/2,\; t + W/2]$ and is preferable for historical description (but infeasible in real time, since it peeks at the future).
The \emph{recursive} (expanding) scheme sets the left endpoint at $s=1$ and lets the window grow---it converges to full-sample OLS and is best seen as a compromise between the two poles that drifts toward Pole~1 as the sample accumulates.

\paragraph{The window is the identification assumption.}
Note what \eqref{eq:TVProll} implicitly assumes: within the window, the coefficients are \emph{constant}; across windows, they may be anything.
The rolling window thus resolves the identification problem of Section~\ref{subsec:TVPident} by \emph{local constancy}.
The two poles reappear as the endpoints of the window choice: $W = T$ is exactly full-sample OLS (Pole~1), while $W = K$ makes each window an exactly determined system with a perfect fit (Pole~2).
Choosing $W$ \emph{is} choosing a position on the continuum---there is no escaping the choice, only making it well or badly.

\paragraph{Bias--variance trade-off.}
The mechanics are the familiar ones from kernel density estimation (Section~\ref{sec:KDE}).
A short window tracks genuine movements quickly but is noisy: the variance of $\hat{\beta}_t^{(W)}$ is of order $1/W$.
A long window averages away noise but smears genuine movements: if the true path drifts, the estimator is biased toward the window average of the path, and the one-sided version is additionally \emph{late}---it is centered, informationally, about $W/2$ periods in the past.
Figure~\ref{fig:TVProlling} displays this trade-off in our laboratory (described in Section~\ref{subsec:TVPlab} below): with $W=40$ the path is erratic; with $W=160$ it is smooth but arrives at the truth with a long delay; $W=80$ is the least bad, and the path-RMSE ranking across windows traces the usual U-shape.

\begin{figure}[htbp]
\centering
\includegraphics[width=0.9\textwidth]{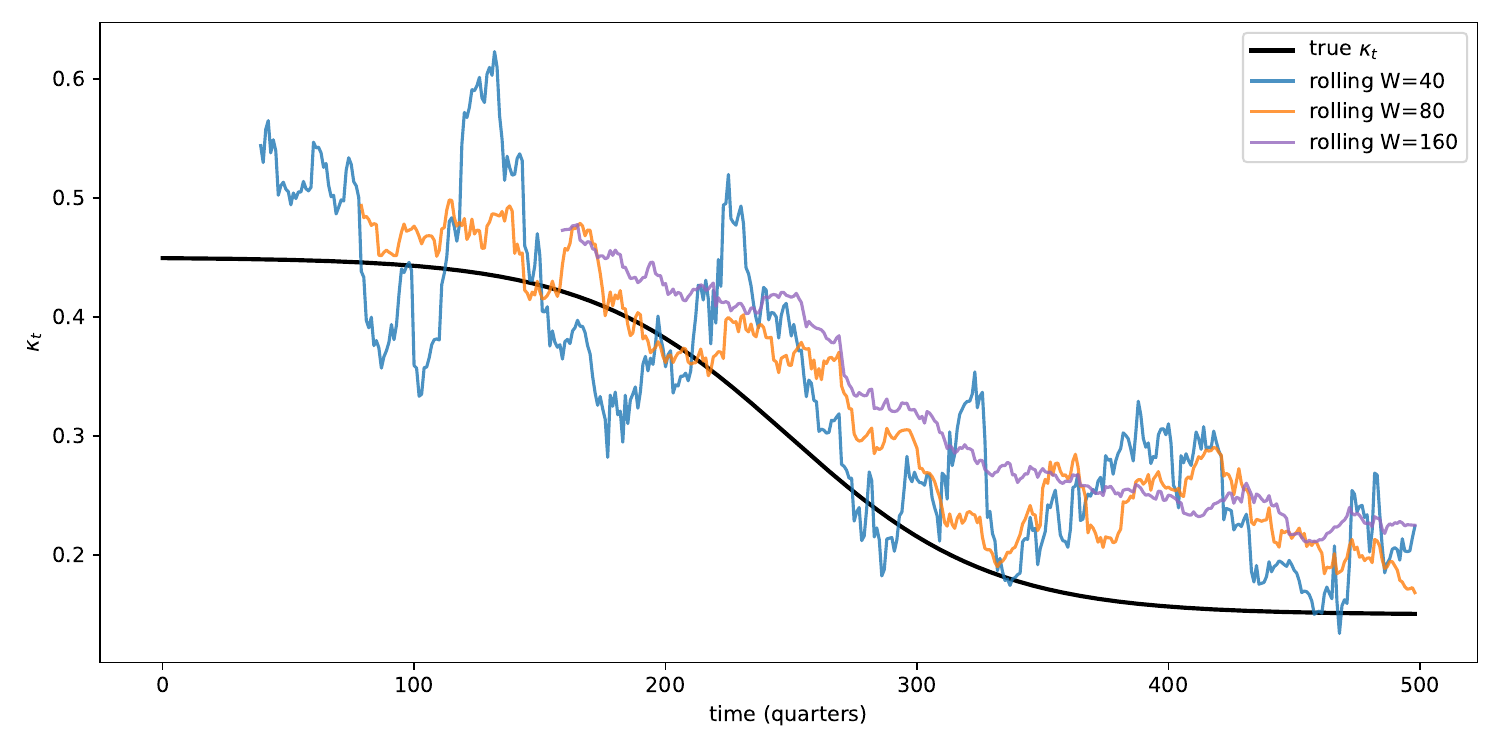}
\caption{Rolling-window estimates of the Phillips-curve slope $\kappa_t$ in the New-Keynesian laboratory, for three window lengths (one-sided windows). Short windows are noisy; long windows are smooth but biased and late. No window dominates: the choice of $W$ is a bias--variance choice.}
\label{fig:TVProlling}
\end{figure}

\paragraph{Refinements: from hard windows to smooth weights.}
The boxcar window treats observation $t-W+1$ as fully informative and observation $t-W$ as worthless---a discontinuity that produces jagged paths, since each estimate jumps whenever an influential observation enters or leaves the window.
Two classical refinements smooth the weights instead:
\begin{itemize}
\item \textbf{Exponentially weighted least squares} (EWLS, or \emph{discounted least squares}): minimize $\sum_{s \le t} \delta^{\,t-s} (y_s - x_s'\beta)^2$ with a forgetting factor $\delta \in (0,1)$; the effective window is $\approx 1/(1-\delta)$. This is the regression analogue of the EWMA volatility estimator and is popular in real-time finance applications.
\item \textbf{Kernel-weighted regression}: replace $\delta^{\,t-s}$ by a kernel $K\!\left(\frac{t-s}{H}\right)$ with bandwidth $H$. \citet{giraitis2014} show that such nonparametric estimators consistently recover coefficient paths that are themselves stochastic (e.g.\ bounded random walks), providing formal inference for what practitioners had long done informally.
\end{itemize}
Both refinements keep the defining feature (and the defining limitation) of the local approach: the econometrician picks the weighting scheme and its width \emph{a priori}, the same width applies to every coefficient, and nothing in the procedure measures \emph{how much} time variation the data actually support.

\subsection{The state-space route: drifting coefficients and the Kalman filter}
\label{subsec:TVPss}

The classical econometric answer replaces the deterministic window with a probabilistic law of motion for the coefficients \citep{cooley1976}.
The canonical specification makes the coefficient vector $\beta_t \in \mathbb{R}^K$ a multivariate random walk:
\begin{align}
y_t &= X_t'\beta_t + \varepsilon_t, & \varepsilon_t &\sim \mathcal{N}(0, \sigma_\varepsilon^2), \label{eq:TVPmeas}\\
\beta_t &= \beta_{t-1} + u_t, & u_t &\sim \mathcal{N}(0, \sigma_u^2 I_K), \label{eq:TVPstate}
\end{align}
with $\varepsilon_t$ and $u_t$ mutually independent.
The random walk is the workhorse for three reasons.
It is maximally parsimonious: instead of $KT$ free coefficients, the path is governed by \emph{one} new parameter, $\sigma_u^2$ (or a $K \times K$ matrix $\Omega_u$ in the general case).
It nests the constant-coefficient model: $\sigma_u^2 = 0$ shuts the time variation down, so ``is there instability?'' becomes a hypothesis about a variance.
And it is agnostic about direction: a martingale implies no pull toward a long-run value, encoding the humility that we do not know where the coefficients are heading.
(A mean-reverting AR(1) state is the natural variant when theory suggests a stable anchor.)

Identification is restored because the \emph{signal-to-noise ratio}
\begin{equation*}
q \;=\; \frac{\sigma_u^2}{\sigma_\varepsilon^2}
\end{equation*}
now disciplines the entire path: given $q$, the smoothness of $\{\beta_t\}$ is no longer a free choice, and the likelihood of the data---evaluated recursively by the Kalman filter \citep{kalman1960,harvey1989}---is informative about $q$ itself.
The filter delivers $\E[\beta_t \mid y_1,\dots,y_t]$ (the real-time, one-sided estimate); the associated \emph{smoother} delivers $\E[\beta_t \mid y_1,\dots,y_T]$ (the two-sided historical estimate).
This machinery powered the classic empirical macro literature on drifting coefficients: \citet{cogley2005drifts} and \citet{primiceri2005} on U.S. monetary policy (time-varying VARs with stochastic volatility), and \citet{stock2007has} on inflation---whose unobserved-components model with stochastic volatility is exactly a time-varying-coefficients model, and which we already met when discussing inflation persistence.
\citet{pivetta2007persistence}, also encountered there, track the drifting persistence of U.S. inflation with both rolling windows and Bayesian state-space methods---a nice illustration that the two routes answer the same question.

\paragraph{Why this is not the end of the story.}
Three practical frictions motivate what comes next.
First, maximum likelihood estimation of $\sigma_u^2$ suffers from the \emph{pile-up problem}: when true time variation is modest, the MLE of $\sigma_u^2$ has a point mass at exactly zero, understating instability.
Second, the multivariate versions used in practice (drifting VAR coefficients \emph{and} stochastic volatility, as in \citealp{primiceri2005}) require Bayesian MCMC machinery that is computationally heavy and delicate to tune.
Third, the state dimension is $K T$: with many regressors the filter-based approach becomes unwieldy exactly when modern, data-rich applications need it most.

\subsection{The Goulet Coulombe equivalence: TVP models \emph{are} ridge regressions}
\label{subsec:TVPridge}

\citet{gouletcoulombe2025ridge} starts from an observation that is easy to state and surprisingly consequential: the random-walk TVP model \eqref{eq:TVPmeas}--\eqref{eq:TVPstate} is a \emph{ridge regression} in disguise.
This single reformulation (i) makes the two poles of Section~\ref{subsec:TVPident} visible as the two ends of a regularization path, (ii) replaces filtering machinery with two lines of linear algebra, and (iii) hands the tuning problem to cross-validation, the standard tool of the machine-learning toolbox of Section~\ref{sec:MLwhy}.

\subsubsection{A detour: ridge regression in a nutshell}

Since these lecture notes have not yet used ridge regression, here is the two-paragraph version.
OLS minimizes the sum of squared residuals; \emph{ridge} \citep{hoerl1970} adds a penalty on the size of the coefficients:
\begin{equation*}
\hat{\theta}^{\text{ridge}} \;=\; \operatorname*{arg\,min}_{\theta}\; \|y - Z\theta\|^2 + \lambda\, \|\theta\|^2
\;=\; (Z'Z + \lambda I)^{-1} Z' y .
\end{equation*}
The penalty parameter $\lambda \ge 0$ trades bias for variance: $\lambda = 0$ gives OLS (unbiased, possibly wildly variable when regressors are many or collinear), $\lambda \to \infty$ shrinks every coefficient to zero (maximal bias, zero variance).
Crucially, ridge remains well-defined even when $Z$ has \emph{more columns than rows}---a case where OLS breaks down entirely because infinitely many coefficient vectors fit the data perfectly.
Ridge resolves that indeterminacy by picking, among all solutions, a shrunken one; $\lambda$ is chosen by cross-validation (CV), i.e.\ by out-of-sample fit rather than in-sample fit.
Readers will recognize the ``L2 penalty'' $\alpha$ of scikit-learn estimators from Section~\ref{sec:MLwhy}; a full treatment is in \citet{hastie2009elements}.
Keep the italicized clause in mind: \emph{more columns than rows} is exactly the situation of the unrestricted TVP model, with its $KT$ coefficients for $T$ observations.

\subsubsection{Time variation as a penalty}

Estimating the path in \eqref{eq:TVPmeas}--\eqref{eq:TVPstate} by penalized least squares means solving
\begin{equation}
\label{eq:TVPpls}
\min_{\beta_1, \dots, \beta_T}\;
\sum_{t=1}^{T} \left( y_t - X_t'\beta_t \right)^2
\;+\; \lambda \sum_{t=2}^{T} \left\| \beta_t - \beta_{t-1} \right\|^2 .
\end{equation}
The first term wants the path to fit the data; the second charges a price $\lambda$ per unit of squared coefficient \emph{change}.
This is not an ad hoc device: problem \eqref{eq:TVPpls} delivers the posterior mode (equal to the posterior mean, in this Gaussian setting) of the state-space model, provided the penalty equals the inverse signal-to-noise ratio,
\begin{equation}
\label{eq:TVPlambda}
\lambda \;=\; \frac{\sigma_\varepsilon^2}{\sigma_u^2} \;=\; \frac{1}{q} ,
\end{equation}
up to a normalization\footnote{With a single $\lambda$ shared by $K$ coefficients, \citet{gouletcoulombe2025ridge} writes $\lambda = \frac{\sigma_\varepsilon^2}{\sigma_u^2}\frac{1}{K}$; the constant is immaterial once $\lambda$ is cross-validated.}.
A large $\lambda$ (high $\sigma_\varepsilon^2$, low $\sigma_u^2$) says ``most of what you see is noise---keep the path stiff''; a small $\lambda$ says ``the residuals are quiet---let the coefficients do the moving.''
Equation \eqref{eq:TVPlambda} is the bridge between the statistical and the machine-learning views of the same object: \emph{a signal-to-noise ratio is a regularization strength}.

\subsubsection{From penalty to ridge: the basis expansion}

Problem \eqref{eq:TVPpls} penalizes \emph{differences} of coefficients, not coefficients themselves---so it is not yet literally a ridge regression.
The fix is a change of variables that any student of unit roots will find familiar: parametrize the path by its \emph{innovations}, $\theta_t \equiv \beta_t - \beta_{t-1}$, so that the level is the cumulative sum $\beta_t = \sum_{s \le t} \theta_s$.
Stacking, $\boldsymbol{\beta} = C \boldsymbol{\theta}$ where (for one coefficient and $T=4$)
\begin{equation*}
C \;=\;
\begin{bmatrix}
1 & 0 & 0 & 0\\
1 & 1 & 0 & 0\\
1 & 1 & 1 & 0\\
1 & 1 & 1 & 1
\end{bmatrix},
\end{equation*}
the lower-triangular ``random-walk basis'' (in general, $I_K \otimes C$).
Substituting into \eqref{eq:TVPpls}, the fitted value becomes $X_t'\beta_t = Z_t'\boldsymbol{\theta}$ with an \emph{expanded design matrix}
\begin{equation*}
Z \;=\; \big[\, \operatorname{diag}(x_1)\, C \;\big|\; \cdots \;\big|\; \operatorname{diag}(x_K)\, C \,\big]
\qquad (T \times KT),
\end{equation*}
whose column $(k, s)$ equals $x_{k,t}\,\mathbf{1}\{t \ge s\}$: regressor $k$ ``switched on'' from date $s$ onward.
The penalized problem \eqref{eq:TVPpls} is then \emph{exactly}
\begin{equation}
\label{eq:TVPridge}
\min_{\boldsymbol{\theta}}\; \|\tilde{y} - Z\boldsymbol{\theta}\|^2 + \lambda \|\boldsymbol{\theta}\|^2 ,
\qquad
\hat{\boldsymbol{\beta}} = C\,\hat{\boldsymbol{\theta}} = C\,(Z'Z + \lambda I_{KT})^{-1} Z'\tilde{y} :
\end{equation}
a completely standard ridge regression, with $KT$ ``regressors'' for $T$ observations\footnote{Two practical details. First, the initial level $\beta_1$ plays a different role from the subsequent increments: it should be left (almost) unpenalized, otherwise the path is artificially pinned at zero---or, in the convenient centered implementation we use in the companion code, at the full-sample OLS value, which makes $\lambda \to \infty$ recover OLS \emph{exactly}. \citet{gouletcoulombe2025ridge} treats this ``constant part'' of the coefficients separately for the same reason. Second, $\tilde{y}$ denotes the dependent variable after this centering.}.
Nothing about time series remains visible: any ridge solver---indeed, any penalized-regression routine from the machine-learning sections---can now estimate a time-varying-parameter model.

\begin{figure}[htbp]
\centering
\includegraphics[width=0.9\textwidth]{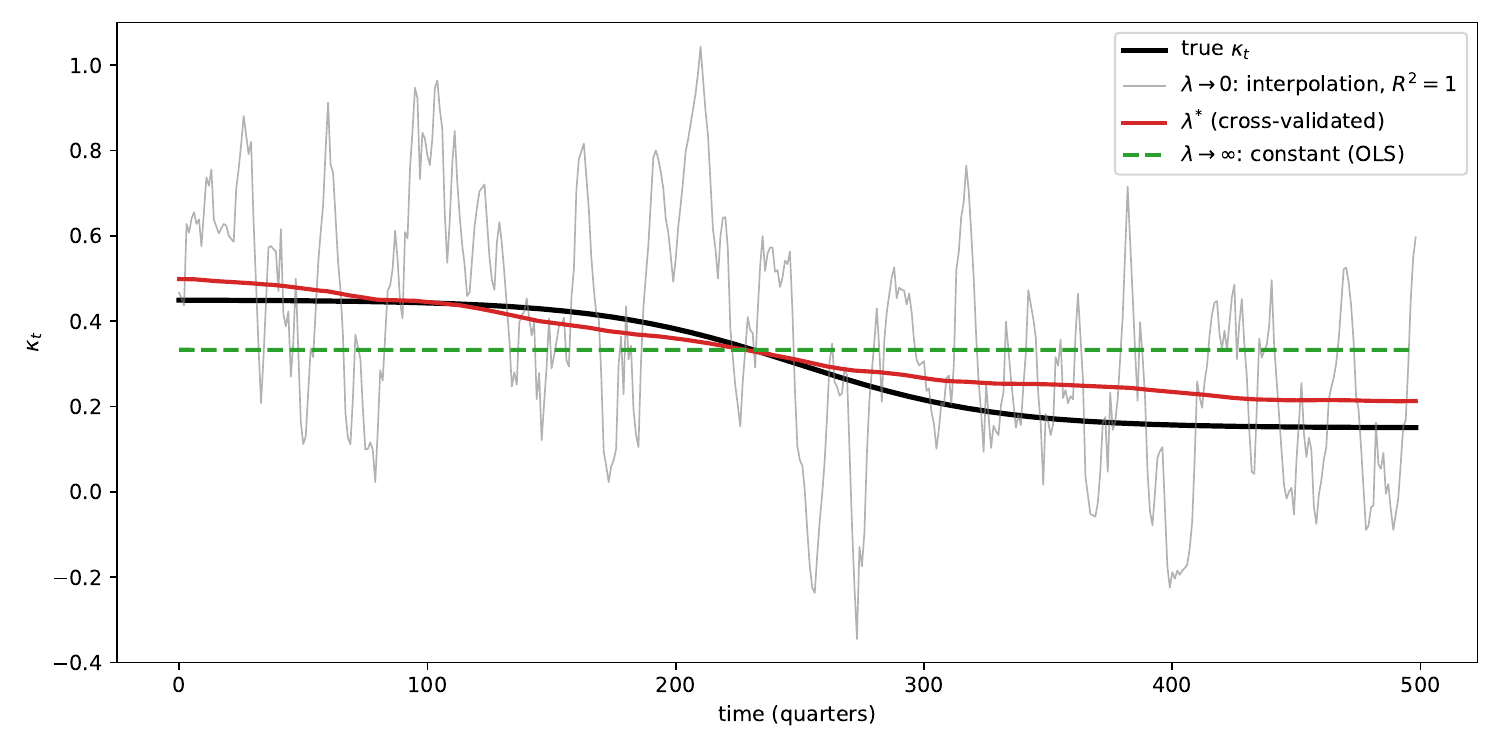}
\caption{The identification problem as a regularization path. Ridge-TVP estimates of $\kappa_t$ for three penalties: $\lambda \to 0$ reproduces Pole 2 (interpolation: the coefficient path absorbs the noise, in-sample $R^2 = 0.995$), $\lambda \to \infty$ reproduces Pole 1 (constant coefficients, i.e.\ full-sample OLS), and the cross-validated $\lambda^{*}$ recovers the true path (thick black line) as well as the noise permits.}
\label{fig:TVPlambda}
\end{figure}

\paragraph{The two poles, once more.}
The regularization path of \eqref{eq:TVPridge} \emph{is} the continuum of Section~\ref{subsec:TVPident}.
As $\lambda \to \infty$, all increments are shrunk to zero and we recover constant-coefficient OLS: Pole 1, everything attributed to the residuals.
As $\lambda \to 0$, we approach the unpenalized problem with $KT > T$ free parameters, which fits the sample exactly: Pole 2, $R^2 = 1$, everything attributed to the coefficients.
Figure~\ref{fig:TVPlambda} shows both poles and the cross-validated solution in our laboratory: with $\lambda \approx 0$ the estimated slope path oscillates wildly between $-0.3$ and $+1.0$ and the in-sample $R^2$ reaches $0.995$; with $\lambda \to \infty$ the path is the flat OLS line ($R^2 = 0.669$); the cross-validated $\lambda^*$ ($R^2 = 0.696$, about $14$ effective parameters out of a possible $499$) tracks the true path.
The lesson deserves emphasis: \textbf{writing $y_t = \alpha_t + \beta_t x_t + \varepsilon_t$ was never the obstacle---the equation only becomes an estimator once $\lambda$ (equivalently the signal-to-noise ratio $q$, equivalently a window width) is chosen, and the entire art of time-varying-parameter econometrics is the disciplined choice of that one number.}

\subsubsection{Computation: the dual trick}

At first sight \eqref{eq:TVPridge} looks expensive: inverting a $KT \times KT$ matrix is $O(K^3 T^3)$ and quickly prohibitive (for $T=500$, $K=10$: a $5000 \times 5000$ system \emph{per} value of $\lambda$).
The rescue is the \emph{dual} form of ridge---the same trick that powers kernel methods in machine learning.
The solution of \eqref{eq:TVPridge} necessarily lies in the row space of $Z$, so it can be written $\hat{\boldsymbol{\theta}} = Z'\hat{\alpha}$ with $\hat{\alpha} \in \mathbb{R}^T$, and
\begin{equation}
\label{eq:TVPdual}
\hat{\boldsymbol{\beta}} \;=\; C\,Z'\,(Z Z' + \lambda I_T)^{-1}\, \tilde{y} .
\end{equation}
Only a $T \times T$ system remains, whatever $K$ is.
\citet{gouletcoulombe2025ridge} reports that his two-stage procedure runs in about one second where a Bayesian TVP-VAR with stochastic volatility needs minutes, and demonstrates scalability by estimating roughly $4{,}600$ time-varying parameters in large local projections on Canadian data---orders of magnitude beyond what filter-based MCMC can comfortably handle.

\subsubsection{Tuning by cross-validation}

In the state-space world, $q$ is estimated by maximum likelihood or given a prior; in the ridge world, $\lambda$ is cross-validated, i.e.\ chosen to \emph{predict held-out data}.
This is a genuinely different---and refreshingly agnostic---answer to ``how much time variation is there?'': as much as improves out-of-sample fit, no more.
Two remarks are in order for time-series data.
First, the folds should respect time: we hold out \emph{contiguous blocks} rather than random observations, and predict a held-out block from the path estimated on the remaining data (the random-walk basis interpolates across the gap).
Second, one may worry that CV is invalid under serial dependence; \citet{bergmeir2018} show that k-fold CV remains valid for model selection provided the model's residuals are uncorrelated---in practice, provided enough lags of the dependent variable are included, as in our application below.
Note finally the humbling arithmetic of regularization in large models: with many regressors, predictive tuning pushes $\lambda$ up, and, as Goulet Coulombe puts it, there is little hope of finding a lot of time variation in a large model \emph{that is tuned to predict well}---a useful antidote to overenthusiastic claims of instability everywhere.

\subsubsection{The two-stage ridge regression (2SRR)}

One $\lambda$ for all coefficients embodies a strong homogeneity assumption: every coefficient drifts at the same speed, and the residual variance is constant.
Both assumptions are relaxed in a second stage, giving the \emph{two-stage ridge regression} (2SRR) of \citet{gouletcoulombe2025ridge}:
\begin{enumerate}
\item \textbf{Stage 1.} Estimate \eqref{eq:TVPdual} with a single cross-validated $\lambda$; recover the path $\hat{\beta}_t$, the coefficient innovations $\hat{u}_{k,t} = \hat{\beta}_{k,t} - \hat{\beta}_{k,t-1}$, and the residuals $\hat{\varepsilon}_t$.
\item \textbf{Stage 2a (who moves?).} Estimate each coefficient's innovation variance, $\hat{\sigma}_{u,k}^2 = \frac{1}{T}\sum_t \hat{u}_{k,t}^2$, and give coefficient $k$ its own penalty $\lambda_k \propto 1/\hat{\sigma}_{u,k}^2$: coefficients that stage 1 found stable are frozen harder; coefficients that moved get more freedom.
\item \textbf{Stage 2b (when is the data informative?).} Fit a simple volatility model (an EWMA, or a GARCH from Section~\ref{sec:uARMAGARCH}) to $\hat{\varepsilon}_t$ and reweight observations by $1/\hat{\sigma}_{\varepsilon,t}$---a GLS step. This matters more than it may seem: episodes of high residual volatility (crises!) otherwise masquerade as coefficient movement, generating spurious ``instability'' precisely when the data are least informative.
\item \textbf{Stage 2c.} Re-estimate the weighted, coefficient-specific ridge, cross-validating the global scale of the penalties again.
\end{enumerate}
In our laboratory below, stage 2a estimates relative innovation scales of $1.15$ (intercept), $0.61$ (lagged inflation), and $1.43$ (output-gap slope): without being told, the procedure discovers that the Phillips-curve \emph{slope} is the coefficient doing the moving, and that the persistence coefficient should be held closer to constant.

\subsubsection{Same object, two algorithms: ridge path = Kalman smoother}

It bears repeating that nothing statistical was changed by the reformulation: with the variances held at values consistent with $\lambda$, the ridge path \eqref{eq:TVPdual} and the Kalman \emph{smoother} mean of model \eqref{eq:TVPmeas}--\eqref{eq:TVPstate} are the \emph{same object}.
The companion code verifies this to machine precision: the maximum absolute difference between the two paths in our application is $4 \times 10^{-14}$.
What the ridge view adds is not a new estimator but (i) a tuning principle (CV in place of a fragile variance MLE), (ii) scalability (the dual trick), and (iii) a conceptual bridge: time-varying-parameter econometrics is revealed to be regularized regression, so the entire machine-learning apparatus---cross-validation, penalties, even random forests, as we will see---applies natively.
Two things are genuinely lost relative to the full Bayesian treatment: the posterior uncertainty about the variances themselves, and, with it, finite-sample credible bands that account for it (though frequentist bands are available, since the estimator is linear in $y$).

\subsection{Every method is a weighting scheme}
\label{subsec:TVPweights}

\begin{figure}[htbp]
\centering
\includegraphics[width=0.9\textwidth]{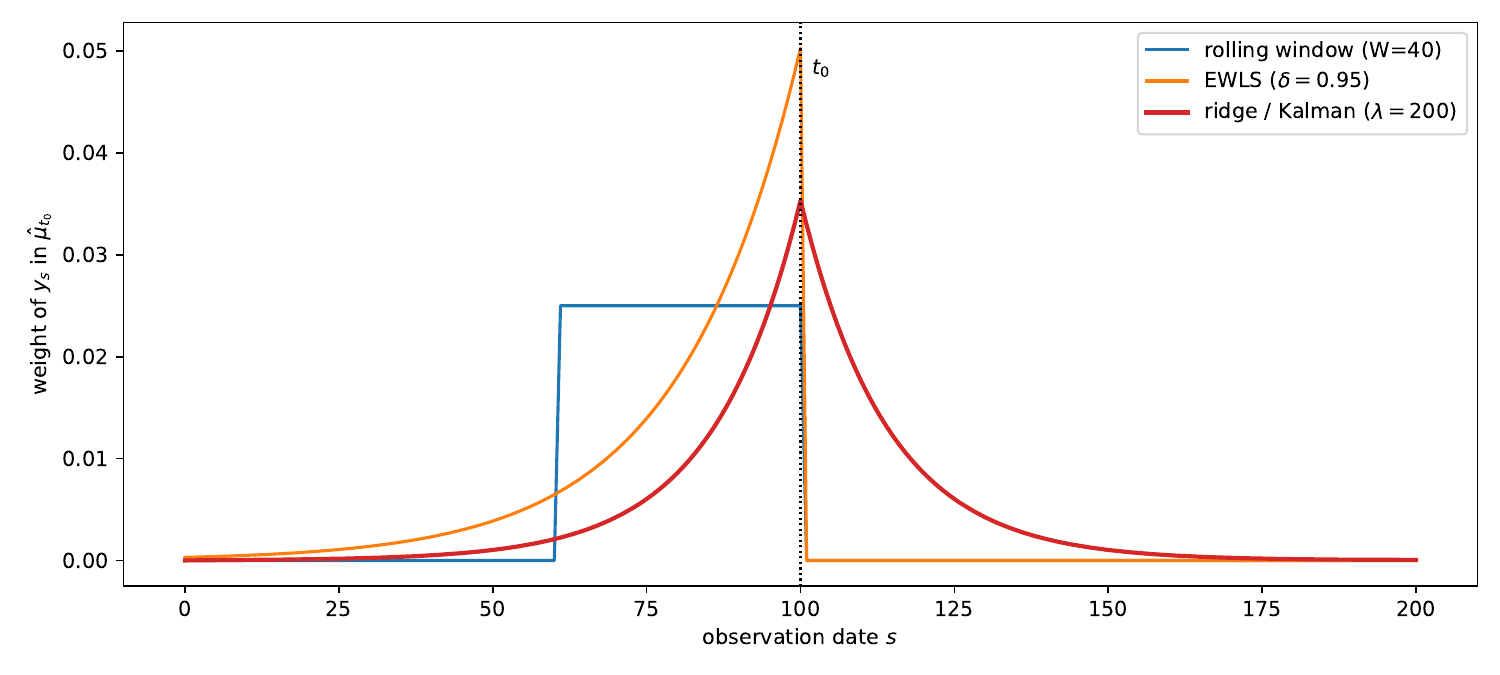}
\caption{Implicit weights on observation $s$ in the estimate of the local level at $t_0 = 100$, for three methods. Rolling window: one-sided boxcar. Exponentially weighted least squares: one-sided geometric decay. Ridge/Kalman smoother: two-sided kernel whose bandwidth is governed by $\lambda$. Time-varying-parameter estimators differ only in their weighting schemes.}
\label{fig:TVPweights}
\end{figure}

A final unification before the laboratory.
All the estimators of this section are linear in $y$, so each estimate $\hat{\beta}_{t_0}$ is a weighted average of the observations, and the methods differ \emph{only} in the weights they place on observation $s$ when estimating the coefficient at date $t_0$.
Figure~\ref{fig:TVPweights} extracts these weights in the simplest possible setting (a local-level model, i.e.\ a regression on a constant): the rolling window is a one-sided boxcar---flat weights that end abruptly; EWLS is a one-sided decaying exponential; and the ridge/Kalman smoother is a smooth, \emph{two-sided} kernel centered at $t_0$, whose width is set by $\lambda$---not chosen by hand, but cross-validated.
Seen this way, the ``sophisticated'' state-space estimator is simply a rolling window with better-shaped, data-tuned weights (and the kernel estimators of \citealp{giraitis2014} choose the shape directly).
This also exposes a subtlety worth remembering: \emph{smoothed} (two-sided) estimates use future information and are the right object for historical description, while \emph{filtered} (one-sided) estimates are the right object for real-time analysis and forecasting---compare the discussion of look-ahead bias in the event-study and backtesting sections.

\subsection{The laboratory: a drifting Phillips curve}
\label{subsec:TVPlab}

We now test every method where it cannot cheat: inside the three-equation New-Keynesian laboratory of Section~\ref{sec:wrapup_monetary}, where the truth is known by construction.
The data-generating process is identical---IS curve, Phillips curve, Taylor rule, with the same parameters---except for one change: the slope of the Phillips curve is no longer constant but flattens smoothly over the sample,
\begin{equation}
\label{eq:TVPkappa}
\pi_t = \rho_\pi\,\pi_{t-1} + \kappa_t\, g_{t-1} + \varepsilon_t^s,
\qquad
\kappa_t = 0.15 + \frac{0.30}{1 + \exp\!\big(\frac{t - T/2}{40}\big)} ,
\end{equation}
so that $\kappa_t$ declines from $0.45$ to $0.15$ with the transition centered mid-sample ($T = 500$ quarters).
This is the stylized ``flattening Phillips curve'' that has preoccupied central banks for two decades \citep{delnegro2020phillips}: the same output gap buys less disinflation than it used to.
The econometrician observes $(\pi_t, g_t)$ and estimates
\begin{equation*}
\pi_t = \alpha_t + \rho_t\, \pi_{t-1} + \kappa_t\, g_{t-1} + \varepsilon_t ,
\end{equation*}
allowing \emph{all three} coefficients to move (she does not know that only $\kappa_t$ does).

\paragraph{Pole 1 in action.}
Full-sample OLS returns $\hat{\alpha} = -0.01$, $\hat{\rho} = 0.67$, $\hat{\kappa} = 0.33$ with $R^2 = 0.67$: a single slope, close to the average of the true path, blind to the flattening---exactly the behaviour predicted in Section~\ref{subsec:TVPident}.
An analyst relying on it at the end of the sample would overstate the output--inflation trade-off by a factor of two ($0.33$ against a true $0.15$).

\begin{figure}[htbp]
\centering
\includegraphics[width=0.9\textwidth]{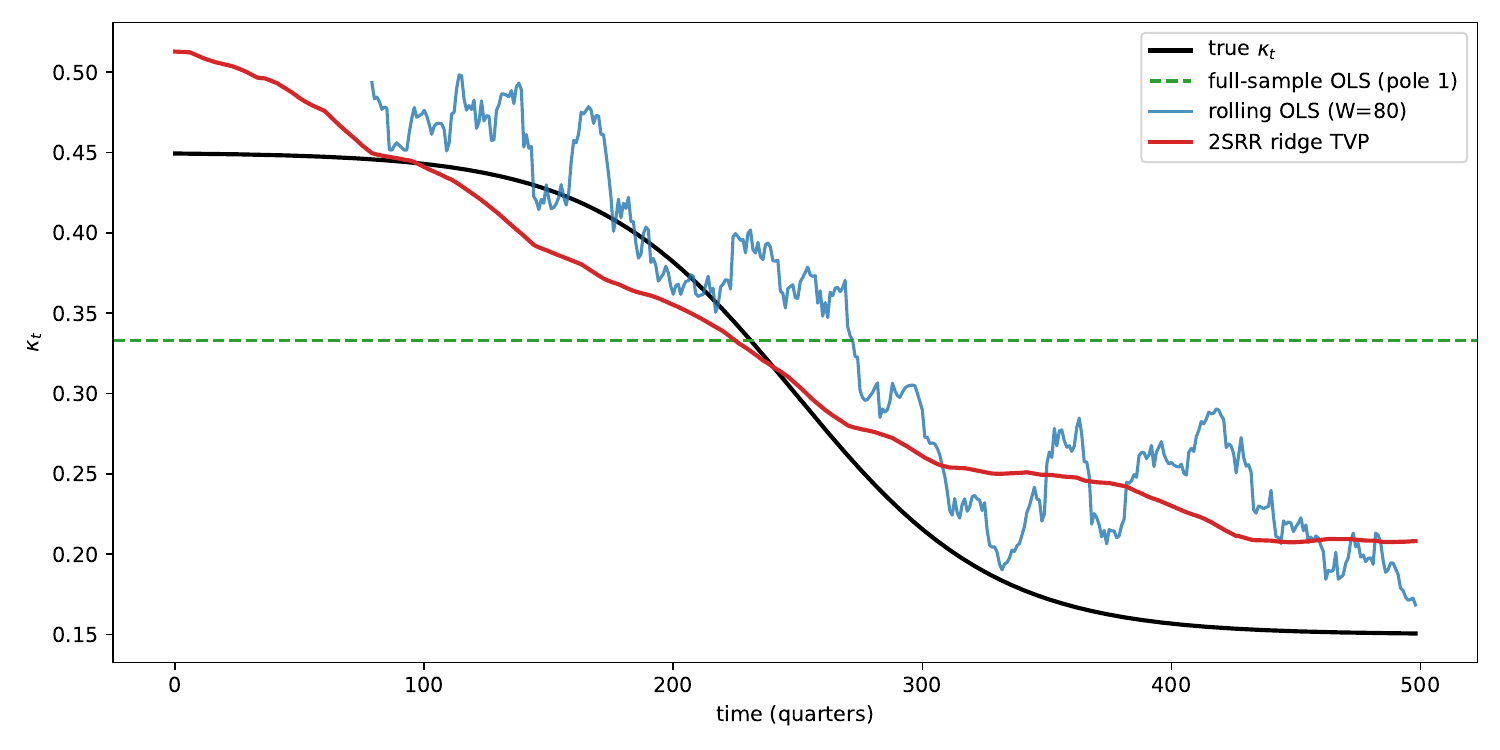}
\caption{Estimating a flattening Phillips-curve slope in the New-Keynesian laboratory. Full-sample OLS (Pole 1) reports a single average slope. The rolling window tracks the decline but is jagged and late. The two-stage ridge regression (2SRR) follows the true transition closely, at the cost of mild smoothing.}
\label{fig:TVPkappa}
\end{figure}

\paragraph{The horse race.}
Figure~\ref{fig:TVPkappa} plots the true path against full-sample OLS, the rolling window ($W=80$), and 2SRR; Table~\ref{tab:TVPrmse} reports the root-mean-squared error of each estimated $\kappa_t$ path against the truth, together with a pseudo-out-of-sample one-step-ahead forecast comparison for $\pi_t$ over the last $200$ quarters (each method re-estimated in real time, using one-sided information only).

\begin{table}[htbp]
\centering
\begin{tabular}{lcc}
\toprule
 & RMSE of $\hat{\kappa}_t$ path & OOS 1-step RMSE for $\pi_t$ \\
\midrule
Full-sample OLS          & 0.128 & --    \\
Recursive (expanding) OLS& --    & 0.310 \\
Rolling OLS, $W=40$      & 0.081 & --    \\
Rolling OLS, $W=80$      & 0.062 & 0.299 \\
Rolling OLS, $W=160$     & 0.085 & --    \\
Ridge TVP (stage 1, $\lambda^{*}$ by CV) & 0.049 & 0.297 \\
2SRR (stage 2)           & 0.048 & --    \\
\bottomrule
\end{tabular}
\caption{Performance in the laboratory (simulated flattening Phillips curve, $T=500$). Left column: RMSE of the estimated slope path against the true $\kappa_t$ (smoothed/two-sided estimates, over dates where each estimator is defined). Right column: pseudo-out-of-sample one-step-ahead forecast RMSE for inflation over the last 200 quarters, using one-sided (real-time) estimates. Numbers produced by the companion script; they vary slightly across simulation seeds.}
\label{tab:TVPrmse}
\end{table}

Three lessons stand out.
First, the rolling window displays the predicted U-shape in $W$ ($0.081 \to 0.062 \to 0.085$): both too little and too much smoothing hurt, and nothing internal to the method says which $W$ to pick.
Second, the cross-validated ridge cuts the best rolling window's path error by roughly a quarter further, and 2SRR adds a small refinement on top---mostly by holding $\hat{\rho}_t$ and $\hat{\alpha}_t$ closer to constant, which frees precision for $\hat{\kappa}_t$.
Third, and importantly for perspective: the \emph{forecasting} gains are modest (0.310 vs 0.299 vs 0.297), because one-step-ahead errors are dominated by the irreducible shock $\varepsilon^s_t$; the large gains are in \emph{measurement} of the structural path itself.
This asymmetry is a general finding in the literature \citep{gouletcoulombe2022ml}: time variation improves forecasts modestly and unevenly, while its main value is often the economic reading of the path---which is precisely what is unavailable from a constant-coefficient regression.

\subsection{When smoothness is the wrong prior: breaks and regimes}
\label{subsec:TVPbreaks}

\begin{figure}[htbp]
\centering
\includegraphics[width=0.9\textwidth]{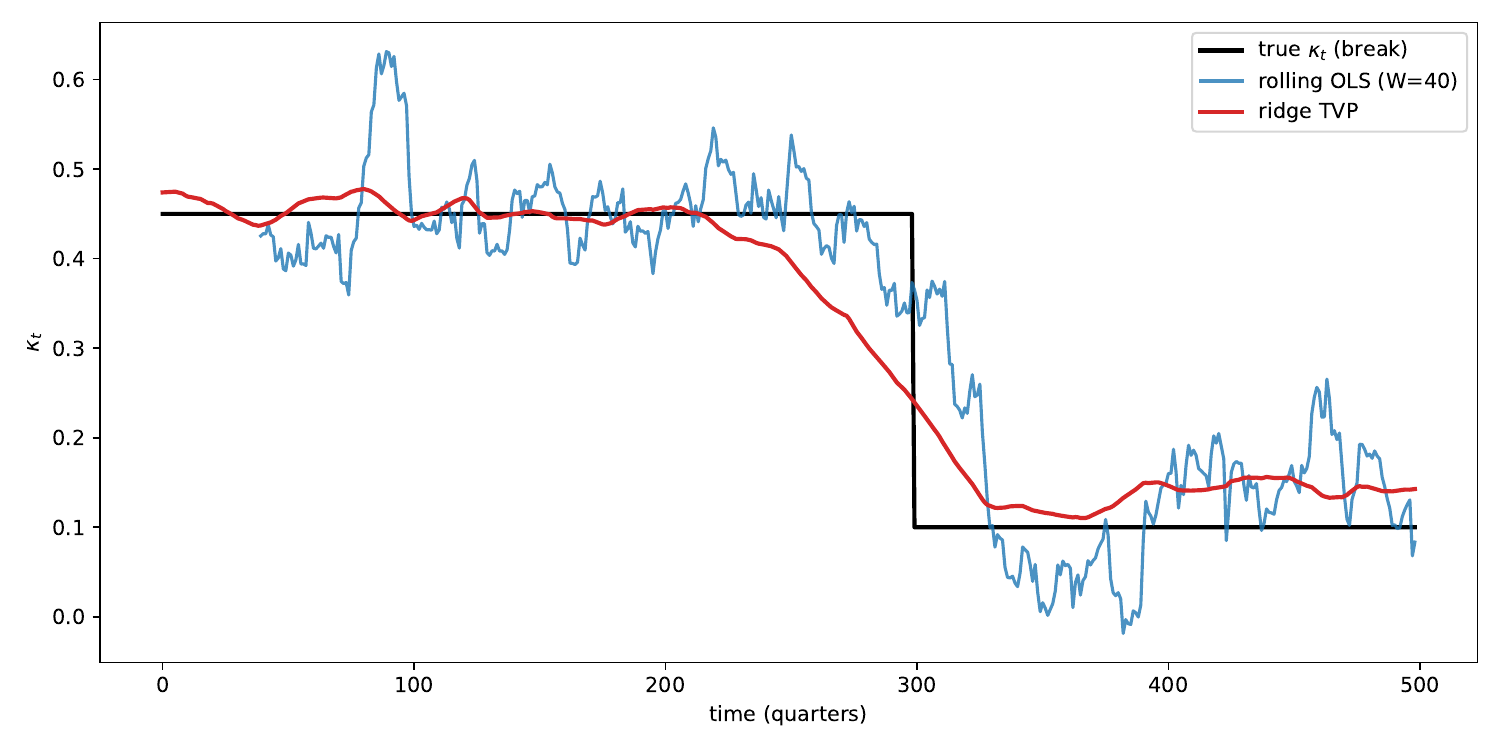}
\caption{An abrupt break in $\kappa_t$ (from $0.45$ to $0.10$ at $t=300$). The ridge/Kalman smoother---built for frequent small changes---smears the jump and begins adjusting before the break (it uses future observations); the short rolling window adapts within a window length but noisily. Discrete changes call for break tests \citep{chow1960,baiperron1998} or regime-switching models \citep{Hamilton1989aa}.}
\label{fig:TVPbreak}
\end{figure}

The random-walk penalty encodes a specific belief: coefficients move often, by small amounts.
Its mirror image---coefficients move rarely, by large amounts---calls for different tools, and confusing the two produces characteristic artifacts.
Figure~\ref{fig:TVPbreak} repeats the laboratory with a single abrupt break ($\kappa_t$ drops from $0.45$ to $0.10$ at $t=300$): the ridge/Kalman smoother spreads the jump over roughly a decade on each side---and, being two-sided, starts adjusting \emph{before} the break even occurs, an anticipation that is purely an artifact of smoothing with future data.
The short rolling window, in contrast, reaches the new level within a window length, at the price of its usual noise.
When the underlying change is plausibly discrete---a policy reform, a currency changeover, a regulatory regime---the honest tools are structural-break methods: the \citet{chow1960} test when the candidate date is known, the \citet{baiperron1998} framework to estimate multiple unknown break dates, and CUSUM-type diagnostics \citep{brown1975} to detect instability without specifying its form.
If instead the economy \emph{recurrently} visits a small number of states (recessions and expansions, high- and low-volatility regimes), the Markov-switching models of \citet{Hamilton1989aa}---introduced in the nonlinear time-series section, including their time-varying-transition-probability extension \citep{filardo1994}---are the natural framework: there, the coefficient path is itself a stochastic process, but a discrete-valued one.
Rolling windows, penalized smooth paths, discrete breaks, and switching regimes are not competitors so much as different priors about the \emph{shape} of change; choosing among them is an economic judgment before it is a statistical one.

\subsection{The modern landscape: beyond the random walk}
\label{subsec:TVPmodern}

The ridge equivalence opened time-varying parameters to the full machine-learning toolbox, and the last decade has produced a rich ecosystem.
We sketch the main branches---students should be able to place any paper they encounter on this map.

\paragraph{(a) Full Bayesian treatment with stochastic volatility.}
The direct descendants of \citet{primiceri2005} estimate drifting coefficients \emph{jointly} with stochastic volatility by MCMC, delivering full posterior uncertainty about paths, variances, and everything else.
The joint treatment is not a luxury: assuming constant residual volatility when volatility actually moves generates \emph{spurious} coefficient variation, as calm and turbulent periods are forced through the same error variance (2SRR's stage 2b is the frequentist shortcut to the same concern).
The price remains computational and increases steeply with dimension.

\paragraph{(b) Shrinkage: not everything moves.}
In a model with many coefficients, letting all of them drift is both implausible and statistically wasteful.
Hierarchical and sparsity-inducing priors let the data decide \emph{which} coefficients are time-varying and which are effectively constant \citep{belmonte2014,huber2021}, while \emph{dynamic} shrinkage processes \citep{kowal2019} allow the shrinkage itself to vary over time---strong smoothing in quiet periods, room to move around events.
This is the Bayesian counterpart of 2SRR's coefficient-specific penalties, taken further.

\paragraph{(c) Nonparametric and observation-driven paths.}
The kernel approach of \citet{giraitis2014} dispenses with a stochastic model for the path altogether and brings the guarantees of nonparametric regression.
A different philosophy underlies \emph{score-driven} (generalized autoregressive score, GAS) models \citep{creal2013}: the parameter is updated each period in the direction of the scaled score of the conditional likelihood---a filter specified directly, with no latent state to integrate out.
Students have already met the most famous member of this family without knowing it: a GARCH equation (Section~\ref{sec:uARMAGARCH}) is exactly a score-driven update for a time-varying \emph{variance} parameter.

\paragraph{(d) Generalized TVPs: the macroeconomy as a random forest.}
All methods so far let coefficients move \emph{with time}---an agnostic choice that describes change but cannot explain it.
\citet{gouletcoulombe2024forest} takes the next step: keep the linear equation $y_t = X_t'\beta_t + \varepsilon_t$ for interpretability, but let the coefficients be a \emph{function of observable state variables},
\begin{equation*}
\beta_t = F(S_t),
\end{equation*}
where $S_t$ can contain many candidate drivers (lags, credit conditions, uncertainty proxies, oil prices, a time trend\ldots) and $F$ is learned by a random forest---the algorithm introduced in Section~\ref{sec:MLwhy}---suitably regularized (trees are grown on random subsamples; within each leaf a ridge regression is fitted, with a random-walk penalty ensuring the implied paths remain smooth).
The output, \emph{generalized time-varying parameters}, nests our random-walk TVP (put only $t$ in $S_t$) but can also answer \emph{why} a coefficient moves: variable-importance measures reveal, for instance, whether the Phillips-curve slope tracks trend inflation, openness, or slack itself.
The cost is the usual one for flexible machine learning: more data hunger, more tuning, and inference that is less crisp than for the linear methods.
For forecasting, the evidence in \citet{gouletcoulombe2022ml} suggests that nonlinearities of this kind help most at longer horizons and in turbulent periods.

\paragraph{A map.}
Table~\ref{tab:TVPmap} organizes the approaches by the restriction each one places on the coefficient path---the ``identification currency'' every method must pay.

\begin{table}[htbp]
\centering
\small
\begin{tabular}{p{3.6cm}p{3.6cm}p{2.4cm}p{5.2cm}}
\toprule
Method & Implicit law of motion for $\beta_t$ & Tuning object & Comparative advantage / caveat \\
\midrule
Full-sample OLS & constant & --- & benchmark; blind to change \\
Chow test / known break \citep{chow1960} & one jump at a known date & break date & sharp test; date must be known \\
Bai--Perron \citep{baiperron1998} & few jumps at unknown dates & number of breaks & policy reforms; discrete changes only \\
Markov switching \citep{Hamilton1989aa} & recurrent discrete regimes & number of states & business cycles; regimes must recur \\
Rolling / recursive OLS & locally constant & window $W$ & transparent; jagged, late, $W$ ad hoc \\
EWLS / kernel \citep{giraitis2014} & locally smooth & $\delta$ or bandwidth & smooth weights; still one speed for all \\
Kalman MLE / Bayes \citep{primiceri2005} & random walk, small steps & $q$ (SNR), priors & full uncertainty; heavy, pile-up risk \\
Ridge TVP / 2SRR \citep{gouletcoulombe2025ridge} & random walk via penalty & $\lambda$ by CV & fast, scalable, ML-native; smooths breaks \\
Sparse / dynamic shrinkage \citep{huber2021,kowal2019} & most coefficients constant & priors & selects \emph{which} move; MCMC cost \\
Score-driven \citep{creal2013} & deterministic filter of past data & score scaling & no latent state; misspecification risk \\
Random forest GTVP \citep{gouletcoulombe2024forest} & function of observables $S_t$ & forest hyperpars. & explains \emph{why}; data-hungry \\
\bottomrule
\end{tabular}
\caption{A map of time-varying-coefficient methods. Each row is a different answer to the identification problem of Section~\ref{subsec:TVPident}: a restriction on how the path may move, purchased with a tuning choice.}
\label{tab:TVPmap}
\end{table}

\paragraph{Choosing in practice.}
A reasonable default workflow for applied work: start from constant coefficients and \emph{test} for instability (CUSUM, Chow, Bai--Perron); if instability is diffuse rather than discrete, report a rolling-window path for transparency alongside a 2SRR (or Kalman) path for efficiency, with the volatility correction if the sample spans calm and crisis periods; reserve regime-switching for genuinely recurrent states, and reach for generalized TVPs when the question is \emph{why} the coefficients move, not just whether.
And always be explicit about the one-sided/two-sided distinction: smoothed paths for history, filtered paths for real time.

\medskip
\noindent\fbox{\parbox{0.97\textwidth}{
\textbf{What this section teaches.}
\begin{enumerate}
\item \textbf{The unrestricted TVP model is not identified.} $y_t = \alpha_t + \beta_t x_t + \varepsilon_t$ has more parameters than observations: one can attribute everything to residuals (constant-coefficient OLS) or nothing (perfect fit, $R^2 = 1$, coefficients absorbing the noise), and the data cannot arbitrate. Structure on the path is not optional.
\item \textbf{Every method rations coefficient movement.} Breaks, regimes, windows, penalties, and priors are different currencies for the same purchase: effective degrees of freedom strictly between $K$ and $T$.
\item \textbf{Rolling windows assume local constancy.} The window length is a bias--variance dial with the two degenerate poles at its ends ($W=T$ is OLS; $W=K$ is interpolation); the one-sided version is also systematically late.
\item \textbf{The state-space model prices movement probabilistically.} A random-walk path with signal-to-noise ratio $q$ restores identification with a single parameter; the Kalman filter/smoother deliver real-time and historical paths.
\item \textbf{TVP models are ridge regressions} \citep{gouletcoulombe2025ridge}. With the random-walk basis $Z = W C$, penalized path estimation is literally ridge; $\lambda = \sigma^2_\varepsilon / \sigma^2_u$; the dual form needs only a $T \times T$ inversion; $\lambda$ is cross-validated on contiguous blocks; 2SRR adds coefficient-specific penalties and a volatility correction. The ridge path and the Kalman smoother coincide to machine precision.
\item \textbf{Smoothness is a prior, not a fact.} Penalized paths smear discrete breaks (and two-sided smoothers ``react'' before them); use break tests or switching models when change is plausibly abrupt or recurrent.
\item \textbf{Path measurement $\neq$ forecasting.} In the laboratory, ridge TVP halves the path-measurement error of the best rolling window, but one-step forecast gains are small---the shock variance dominates. The economic value of TVP methods is usually in reading the path.
\item \textbf{Modern extensions decide what moves and why.} Shrinkage selects \emph{which} coefficients vary \citep{huber2021,kowal2019}; the macroeconomic random forest makes coefficients functions of observables and identifies the drivers \citep{gouletcoulombe2024forest}.
\end{enumerate}
}}

\subsection{Exercises}

\stepcounter{question}
\begin{itemize}
\item \textbf{Question \arabic{question}}
\begin{itemize}
\item Prove the two poles formally. (i) Show that for any data $\{(y_t, x_t)\}_{t=1}^T$ with $x_t \neq 0$, the model \eqref{eq:TVPnaive} admits parameter paths achieving $\mathrm{SSR} = 0$, and exhibit two distinct such paths (hence non-uniqueness). (ii) Show that the solution of the penalized problem \eqref{eq:TVPpls} converges to the full-sample OLS coefficients as $\lambda \to \infty$.
\end{itemize}
\end{itemize}

\stepcounter{question}
\begin{itemize}
\item \textbf{Question \arabic{question}}
\begin{itemize}
\item Run the companion script and extend the rolling-window grid to $W \in \{20, 40, 60, \dots, 240\}$. Plot the path RMSE of $\hat{\kappa}_t$ against $W$ and locate the minimum. Explain the shape of the curve using the bias--variance decomposition, and explain why the minimizing $W$ would change if (i) $\sigma_s$ doubled, (ii) the transition in \eqref{eq:TVPkappa} were twice as fast.
\end{itemize}
\end{itemize}

\stepcounter{question}
\begin{itemize}
\item \textbf{Question \arabic{question}}
\begin{itemize}
\item In the companion code, the Kalman smoother reproduces the ridge path when the state innovation variance is set to $\sigma_u^2 = \sigma_\varepsilon^2 / \lambda^*$. Verify the equivalence for two other values of $\lambda$, and use the mapping \eqref{eq:TVPlambda} to translate the cross-validated $\lambda^{*}$ into an implied signal-to-noise ratio $q$. Comment: does CV ``believe'' in a lot of time variation in this sample?
\end{itemize}
\end{itemize}

\stepcounter{question}
\begin{itemize}
\item \textbf{Question \arabic{question}}
\begin{itemize}
\item Modify the DGP so that $\kappa$ is constant but the \emph{persistence} $\rho_{\pi,t}$ drifts from $0.9$ to $0.5$. Re-run 2SRR: do the stage-2 coefficient-specific scales correctly reallocate the freedom to move from $\hat{\kappa}_t$ to $\hat{\rho}_t$? Relate your finding to the discussion of drifting inflation persistence in \citet{pivetta2007persistence} and \citet{cogley2005drifts}.
\end{itemize}
\end{itemize}

\stepcounter{question}
\begin{itemize}
\item \textbf{Question \arabic{question}}
\begin{itemize}
\item Real data. Using the euro-area HICP inflation series constructed in the inflation sections and a measure of slack of your choice (e.g.\ the unemployment gap from an HP-filtered unemployment rate, or the OECD output gap), estimate the backward-looking Phillips curve $\pi_t = \alpha_t + \rho_t \pi_{t-1} + \kappa_t\, \text{slack}_{t-1} + \varepsilon_t$ by (i) rolling OLS with two windows and (ii) 2SRR. Plot the two $\hat{\kappa}_t$ paths with their timing conventions made explicit (one-sided vs two-sided). Has the euro-area Phillips curve flattened? How confident can you be, given the lessons of this section?
\end{itemize}
\end{itemize}

\stepcounter{question}
\begin{itemize}
\item \textbf{Question \arabic{question}}
\begin{itemize}
\item Challenge. A time-varying CAPM: using the data of Section~\ref{sec:CAPM}, estimate a drifting market beta for one stock by rolling OLS ($W = 250$ trading days) and by ridge TVP. Compare the paths around a major market event. Then read the abstract of \citet{gouletcoulombe2024forest} and propose a set of state variables $S_t$ that could \emph{explain} the movements of your $\hat{\beta}_t$; describe (without necessarily implementing it) how you would evaluate whether the generalized-TVP explanation beats the pure time-drift one out of sample.
\end{itemize}
\end{itemize}

\section{An Inflation Panel for Empirical Macroeconomics}

This section introduces a simple yet instructive panel of monthly inflation data, combining harmonised euro-area inflation series with national inflation data for Ukraine. The objective is to illustrate how economists construct comparable macroeconomic time series from heterogeneous official sources and to clarify the conceptual meaning of inflation measures used in empirical work. 

All data construction steps, transformations, and replication codes used in this section are fully documented and publicly available in the accompanying GitHub repository: 
\href{https://github.com/skimeur/pioneer-detection-method}{\texttt{https://github.com/skimeur/pioneer-detection-method}}. 
The repository ensures transparency and reproducibility, and allows readers to replicate the figures and extend the analysis to alternative countries, inflation measures, or sample periods.

\subsection{What is inflation?}

Inflation measures the rate at which the general level of consumer prices increases over time, thereby eroding the purchasing power of money. Empirically, inflation is derived from a \emph{consumer price index} (CPI), which aggregates prices of a representative basket of goods and services consumed by households.

Throughout this lecture, inflation is expressed as a \emph{year-on-year} (y/y) rate:
\[
\pi_t = \frac{P_t - P_{t-12}}{P_{t-12}} \times 100,
\]
where $P_t$ denotes the price index in month $t$. This transformation mitigates seasonal effects and is standard in macroeconomic analysis and monetary policy.

\subsection{Euro-area inflation: HICP from the ECB}

For euro-area countries, inflation data are obtained from the European Central Bank via the ECB Data Portal. The underlying series is the \emph{Harmonised Index of Consumer Prices} (HICP).

Key features of HICP data:
\begin{itemize}
    \item Monthly frequency.
    \item Headline (all-items) index.
    \item Reported directly as annual rates of change.
    \item Methodology harmonised across EU countries under Eurostat regulations.
\end{itemize}

The HICP is specifically designed to ensure cross-country comparability within the European Union and serves as the reference inflation measure for the conduct of euro-area monetary policy.

\subsection{Ukraine inflation: CPI from the State Statistics Service of Ukraine}

For Ukraine, inflation data are sourced from the State Statistics Service of Ukraine (SSSU) via its SDMX API. The available series is a consumer price index normalized to \emph{previous month = 100}.

To obtain a measure comparable to HICP inflation, the following steps are applied:
\begin{enumerate}
    \item Monthly inflation factors are computed as $f_t = \text{CPI}_t / 100$.
    \item Year-on-year inflation is reconstructed by chaining twelve consecutive monthly factors:
    \[
    \pi_t^{y/y} = \left( \prod_{k=0}^{11} f_{t-k} - 1 \right) \times 100.
    \]
\end{enumerate}

This procedure yields a year-on-year inflation rate expressed in percentage points, consistent with the transformation applied to HICP data.

\subsection{Are ECB and Ukrainian inflation measures comparable?}

The two inflation measures are \emph{conceptually comparable but not fully harmonised}.

\paragraph{Comparable aspects.}
\begin{itemize}
    \item Both are consumer price indices covering household consumption.
    \item Both are expressed as year-on-year percentage changes.
    \item Both are produced by official statistical institutions.
\end{itemize}

\paragraph{Key differences.}
\begin{itemize}
    \item HICP follows a strictly harmonised EU methodology, while the Ukrainian CPI follows national statistical standards.
    \item Consumption baskets, weights, and price-collection practices differ.
    \item Territorial coverage and price dynamics in Ukraine are affected by geopolitical conditions.
\end{itemize}

As a result, inflation \emph{levels and volatility} should not be interpreted as directly comparable in a welfare or policy sense. However, for time-series purposes—such as studying persistence, co-movement, or predictive relationships—the series are sufficiently comparable.

\subsection{Purpose of the panel}

The inflation panel constructed in this lecture is intended as a pedagogical tool. It serves to illustrate:
\begin{itemize}
    \item how to collect macroeconomic time series from official APIs,
    \item how to transform raw price indices into inflation rates,
    \item how to combine heterogeneous national data into a unified panel,
    \item and how such data can be used for standard econometric exercises (unit-root tests, Granger causality, VAR models).
\end{itemize}

The emphasis is methodological rather than structural: the panel is used to demonstrate empirical techniques commonly employed in applied macroeconomics and finance.

\subsection{Course Exercise: Extending the Pioneer Detection Method via GitHub}

In this exercise, students are asked to work directly on a research-grade codebase and to contribute an original extension using standard GitHub collaboration tools.

\paragraph{Objective.}
The goal is to extend the existing script
\begin{center}
\texttt{ecb\_hicp\_panel\_var\_granger\_be.py}
\end{center}
from the repository
\begin{center}
\texttt{https://github.com/skimeur/pioneer-detection-method}
\end{center}
by adding an implementation of the \emph{Pioneer Detection Method}(s) from \citet{vansteenberghe2025insurance}. The empirical question is whether inflation dynamics in Ukraine exhibit a pioneer (leading) role relative to other countries in the sample.

\paragraph{Student workflow.}
Each student (or group) must follow the standard open-source contribution workflow:
\begin{enumerate}
    \item \textbf{Fork} the original GitHub repository to their own GitHub account.
    \item \textbf{Clone} their fork locally and work on the code using VS Code.
    \item \textbf{Implement} the Pioneer Detection extension (preferably in a clearly identified section or auxiliary script).
    \item \textbf{Commit} changes with clear and informative commit messages.
    \item \textbf{Push} the changes to their fork.
    \item \textbf{Open a Pull Request} (PR) proposing their contribution to the original repository.
\end{enumerate}

\paragraph{Expected contribution.}
The proposed extension should:
\begin{itemize}
    \item be functional and reproducible,
    \item be clearly documented in the code,
    \item produce interpretable outputs (tables or figures),
    \item include a brief economic interpretation of the results.
\end{itemize}

\paragraph{Evaluation and integration.}
All submitted pull requests will be reviewed by the instructor. Evaluation will be based on:
\begin{itemize}
    \item correctness and robustness of the implementation,
    \item code clarity and structure,
    \item relevance of the economic interpretation.
\end{itemize}
The best contributions may be merged into the main repository, giving students direct experience with academic code review and integration.

\paragraph{Pedagogical rationale.}
This exercise is designed to expose students to real research practices: reading existing code, extending a method, and contributing through GitHub. It mirrors professional workflows used in collaborative empirical research and open-source scientific projects.

\section{Inflation and Economic Growth}

We work on \citet{barro1995inflation}, which documents two stylized facts that motivate the empirical analysis of inflation and growth: (i) the cross-country distribution of inflation rates is highly skewed, with a small number of high-inflation episodes, and (ii) inflation volatility increases systematically with the level of inflation. 

The empirical exercises in this section are based on a fully reproducible replication of Barro’s original analysis. All data construction steps, figures, and regression codes are available in the accompanying GitHub repository, which also serves as the starting point for the practical exercises:
\href{https://github.com/skimeur/barro1995-inflation-growth-replication}{\texttt{https://github.com/skimeur/barro1995-inflation-growth-replication}}.

\subsection{Replication of Figures 1 and 2}

We replicate Figures~1 and~2 of \citet{barro1995inflation} using a balanced panel of countries observed annually over the period 1960--2019. Inflation data are taken from the Global Macro Data (GMD) database. To focus on economically meaningful variation, we trim extreme observations corresponding to severe deflation episodes and hyperinflation regimes, which would otherwise dominate distributional moments.

\begin{figure}[h!]
\centering
\includegraphics[width=0.7\textwidth]{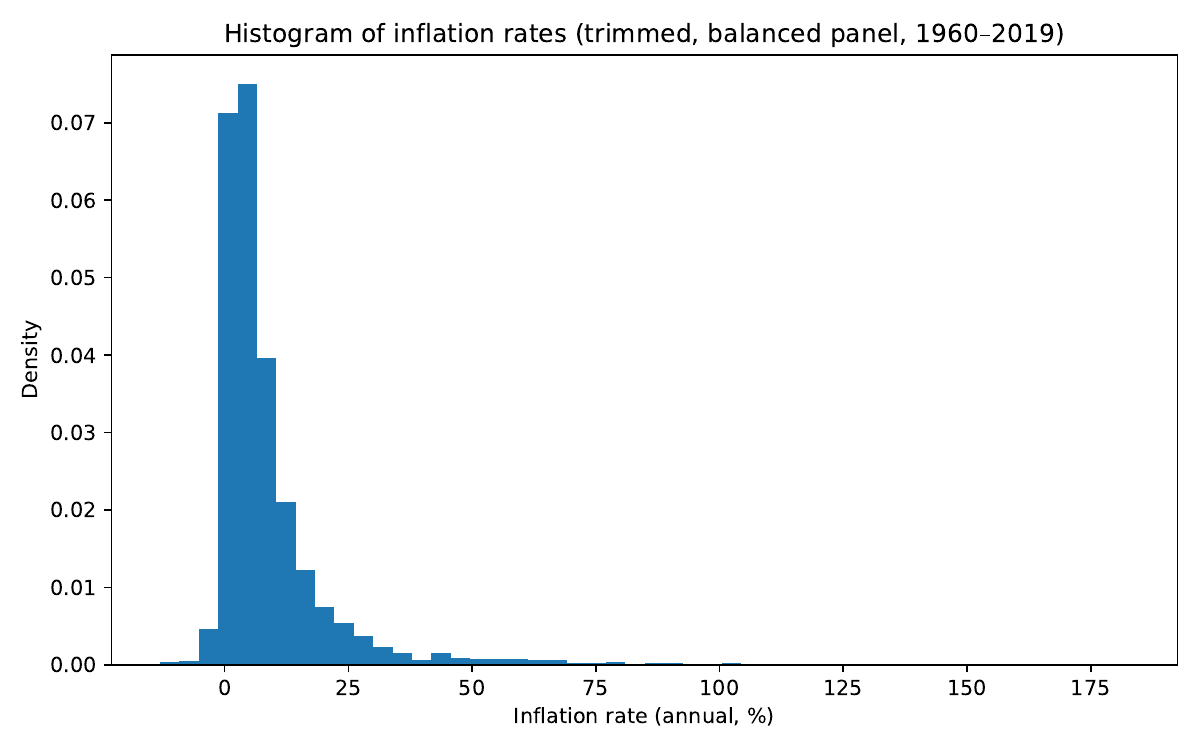}
\caption{Distribution of annual inflation rates (balanced panel, 1960--2019).}
\label{fig:barro1995_fig1_replication}
\end{figure}

\paragraph{Figure 1: Distribution of Inflation Rates.}
Figure~\ref{fig:barro1995_fig1_replication} displays the histogram of annual inflation rates pooled across countries and years. Inflation is measured at the annual frequency, and the sample is restricted to countries observed in all years of the panel. The figure illustrates the strong right skewness of inflation outcomes emphasized by \citet{barro1995inflation}: most observations correspond to low and moderate inflation, while a small fraction of high-inflation episodes generate a long right tail.

\begin{figure}[h!]
\centering
\includegraphics[width=0.7\textwidth]{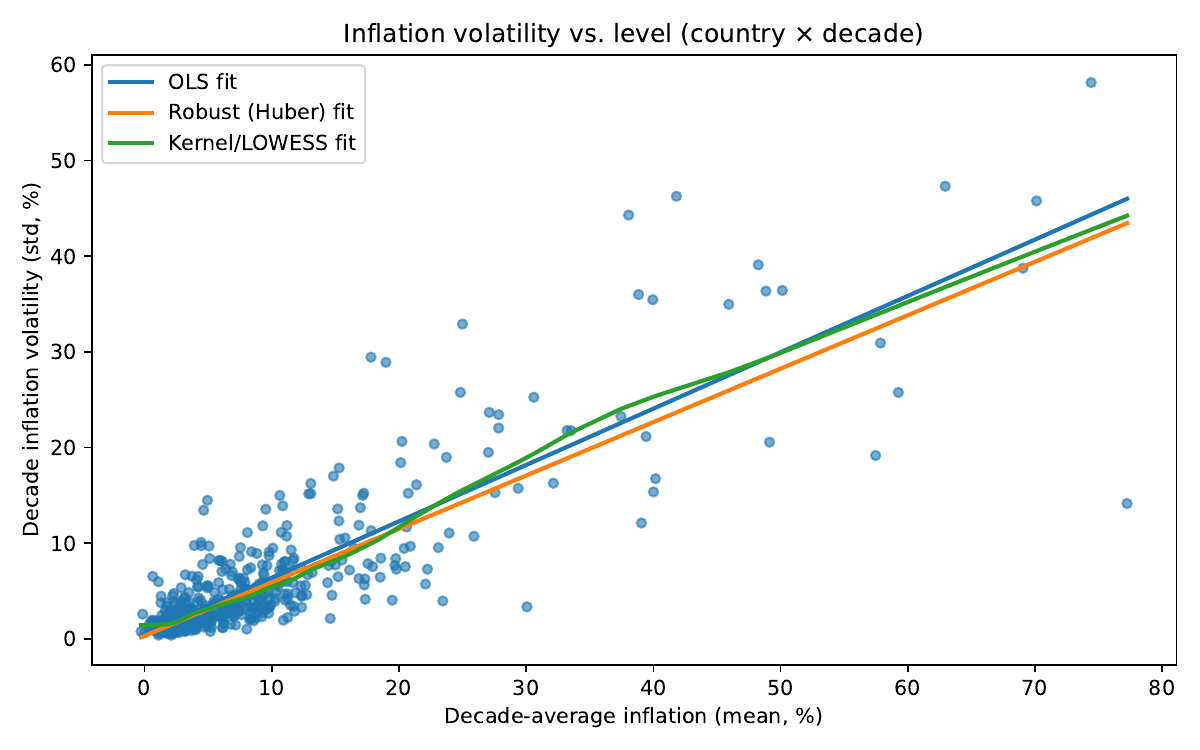}
\caption{Inflation volatility versus inflation level (country--decade data).}
\label{fig:barro1995_fig2_replication}
\end{figure}

\paragraph{Figure 2: Inflation Level and Volatility.}
Figure~\ref{fig:barro1995_fig2_replication} reproduces the relationship between inflation volatility and the level of inflation. For each country and decade, we compute the average inflation rate and the corresponding standard deviation of annual inflation within the decade. The scatter plot relates these two moments across country--decade observations. To characterize the relationship, we overlay three fitted curves: an ordinary least squares regression, a robust Huber M-estimator that downweights outliers, and a nonparametric kernel (LOWESS) fit.

Consistent with \citet{barro1995inflation}, the figure shows a strong positive association between mean inflation and inflation volatility. This pattern highlights that high-inflation environments are not only characterized by higher average inflation but also by greater macroeconomic instability.

Together, these two figures summarize the empirical motivation for treating inflation as both a level and a risk variable in growth regressions, a central insight of \citet{barro1995inflation}.

\subsection{Barro--Lee variables used to replicate Barro (1995), Table 2}

To replicate Table~2 of \citet{barro1995inflation}, we use the Barro--Lee panel variables listed below. The dataset is organized in five-year subperiod averages indexed by $x\in\{1,2,3,4,5,6\}$. The growth block $\texttt{grwb}x$ measures growth over the interval \emph{ending} at year $1960+5x$, so $x=1$ covers 1960--65, $x=2$ covers 1965--70, \dots, and $x=6$ covers 1985--90.\footnote{It is tempting, but wrong, to read block $x$ as sitting at the calendar year $1960+5x$ and then to average over calendar decades: that misalignment places every growth rate and flow one period too early. \citet{barro1995inflation} estimates on the periods 1965--75, 1975--85 and 1985--90, which are assembled from these blocks as the averages of blocks $\{2,3\}$, $\{4,5\}$ and $\{6\}$ respectively; initial conditions (income, schooling, life expectancy) are taken at the start of each period, i.e.\ in 1965, 1975 and 1985. Level variables are reported quinquennially with suffix $\texttt{XX}\in\{60,65,70,75,80,85,90\}$, and 1960 values are kept to construct lags.}

\paragraph{Outcome.}
Per-capita GDP growth over subperiod $x$:
\[
g_{i,x} \equiv \texttt{grwbx}_{i,x},
\]
the growth rate of real GDP per capita $\texttt{GDPWBXX}$.

\paragraph{Initial income.}
Real GDP per capita in constant (1980) local-currency prices:
\[
y_{i,XX} \equiv \texttt{gdpwbXX}_{i,XX}, \qquad XX\in\{60,65,70,75,80,85,90\},
\]
used to form initial income (and, when required, its lagged value) at the beginning of each subperiod.

\paragraph{Human capital and demographics.}
Average years of schooling in the population aged 25+ (male and female):
\[
h^M_{i,XX}\equiv \texttt{HUMANMXX}_{i,XX},\qquad 
h^F_{i,XX}\equiv \texttt{HUMANFXX}_{i,XX},
\]
and life expectancy at birth and fertility (subperiod averages):
\[
\ell_{i,x}\equiv \texttt{LIFEE0x}_{i,x},\qquad
f_{i,x}\equiv \texttt{FERTx}_{i,x}.
\]

\paragraph{Policy and macro controls.}
Government consumption ratio and investment ratio (subperiod averages):
\[
g^{G}_{i,x}\equiv \texttt{GOVSH5x}_{i,x},\qquad
inv_{i,x}\equiv \texttt{INVSH5x}_{i,x},
\]
total education spending ratio (subperiod average),
\[
edu_{i,x}\equiv \texttt{GEETOTx}_{i,x},
\]
black-market premium (subperiod average),
\[
bmp_{i,x}\equiv \texttt{BMPx}_{i,x},
\]
and terms-of-trade shock (subperiod average),
\[
tot_{i,x}\equiv \texttt{TOTx}_{i,x}.
\]

\paragraph{Political rights (time-constant in our replication).}
The political-rights index is reported with a different time coding in the Barro--Lee files (e.g., $x\in\{3,4,5,6,B\}$, with $B$ spanning 1972--89). For replication consistency, we collapse it to a single country-level constant:
\[
pr_i \equiv \frac{1}{|\mathcal{X}_i|}\sum_{x\in\mathcal{X}_i}\texttt{POLRIGHTx}_{i,x},
\]
where $\mathcal{X}_i$ is the set of available political-rights observations for country $i$.

\subsection{Inflation uncertainty and Normalized Inflation Uncertainty (NIU)}

In addition to replicating the core growth regressions of \citet{barro1995inflation}, we extend the analysis by incorporating a variance-stabilized measure of inflation uncertainty inspired by \citet{vansteenberghe2024uncertain}. The objective is to distinguish pure dispersion in inflation outcomes from mechanical heteroskedasticity induced by the inflation level itself.

\paragraph{Inflation moments.}
For each country $i$ and period $p\in\{1965\text{--}75,\,1975\text{--}85,\,1985\text{--}90\}$ we compute inflation moments from annual CPI inflation, measured as the log price change $\pi_{i,t}=100\,\ln(1+\text{CPI inflation}/100)$ and left untrimmed, as in \citet{barro1995inflation}:
\[
\bar{\pi}_{i,p} \equiv \frac{1}{T_p}\sum_{t\in p}\pi_{i,t},
\qquad
\sigma_{\pi,i,p} \equiv \sqrt{\frac{1}{T_p-1}\sum_{t\in p}(\pi_{i,t}-\bar{\pi}_{i,p})^2}.
\]
\citet{barro1995inflation} is explicit that the negative inflation--growth relation ``still shows up only when the high-inflation observations are included,'' so we do not trim the extreme episodes.

\paragraph{Normalized Inflation Uncertainty (NIU).}
Following the logic developed in \citet{vansteenberghe2024uncertain}, raw inflation dispersion $\sigma_{\pi}$ is normalized by a function of the inflation level to account for mean--variance dependence:
\[
\text{NIU}_{i,p}
\;\equiv\;
\frac{\sigma_{\pi,i,p}}{\sqrt{1+\lvert \bar{\pi}_{i,p}\rvert}},
\qquad \bar\pi \text{ in percentage points.}
\]
The normalization uses \emph{no} inflation target: none of these countries in these periods had an announced target, so there is no reference level for the deflator to sit at. It is a variance-stabilizing device that isolates residual uncertainty conditional on the inflation level, rather than conflating uncertainty with high-inflation regimes. The deflator is not scale-free, so $\bar\pi$ is measured in percentage points, the units in which the underlying variance law is fitted.

\paragraph{Growth regressions.}
Table~\ref{tab:barlee_growth_decade_specs} presents growth regressions on \citet{barro1995inflation}'s three periods, built from the Barro--Lee covariates and period-level inflation moments. All five columns include the full control set with period fixed effects and country-clustered standard errors, and differ only in how the inflation experience enters. Column~(1) uses mean inflation alone: it is negative and significant, the classic \citet{barro1995inflation} result. Column~(2) uses the raw within-period standard deviation of inflation alone, also negative and significant; column~(3) uses the normalized measure NIU alone, which is negative and the most significant of the three ($t\approx-2.8$).

Columns~(4) and~(5) put the level and a volatility measure on the right-hand side together. Mean inflation and its raw standard deviation are strongly collinear---high-inflation periods are also the volatile ones---so in column~(4) neither survives: each absorbs the other's explanatory content and both become insignificant. This is the identification problem \citet{barro1995inflation} confronts, and it is exactly what the normalization is meant to relieve.

Column~(5) replaces the raw standard deviation with NIU, which strips out the part of dispersion that the inflation level mechanically predicts. NIU then carries a coefficient several times larger than the raw term and remains negative and significant at the ten-percent level ($-0.118$, $t\approx-1.7$), while mean inflation does not: the cleaned measure retains information about growth that the level alone does not. Its incremental significance beside the level is marginal in this teaching specification and, as the companion research of \citet{vansteenberghe2024uncertain} documents, leans on the high-inflation observations; the fuller specification there, which adds \citet{barro1995inflation}'s log(GDP)$\times$human-capital interaction, sharpens it to conventional significance. The lesson stands either way: a regressor cleaned of level-driven heteroskedasticity can be more informative about growth than the raw dispersion it replaces.

\begin{table}[!htbp]\centering
\caption{Growth regressions on \citet{barro1995inflation}'s periods, raw and normalized inflation uncertainty}
\label{tab:barlee_growth_decade_specs}
\resizebox{\textwidth}{!}{
\begin{tabular}{l c c c c c}
\toprule
 & (1) & (2) & (3) & (4) & (5) \\
\midrule
Mean inflation & $-0.0118^{**}$ &  &  & $-0.0070$ & $-0.0059$ \\
 & $(0.0052)$ &  &  & $(0.0101)$ & $(0.0072)$ \\
Std.\ dev.\ inflation $\sigma_\pi$ &  & $-0.0139^{**}$ &  & $-0.0071$ &  \\
 &  & $(0.0055)$ &  & $(0.0092)$ &  \\
NIU $=\sigma_\pi/\sqrt{1+|\bar\pi|}$ &  &  & $-0.1549^{***}$ &  & $-0.1178^{*}$ \\
 &  &  & $(0.0556)$ &  & $(0.0699)$ \\
$\log$ Real GDP per capita (int'l prices) & $-0.0223^{***}$ & $-0.0222^{***}$ & $-0.0225^{***}$ & $-0.0222^{***}$ & $-0.0223^{***}$ \\
 & $(0.0039)$ & $(0.0039)$ & $(0.0039)$ & $(0.0039)$ & $(0.0039)$ \\
Male schooling years (25+) & $0.0046^{**}$ & $0.0050^{***}$ & $0.0052^{***}$ & $0.0047^{**}$ & $0.0049^{**}$ \\
 & $(0.0020)$ & $(0.0019)$ & $(0.0019)$ & $(0.0020)$ & $(0.0019)$ \\
Female schooling years (25+) & $-0.0048^{**}$ & $-0.0052^{***}$ & $-0.0053^{***}$ & $-0.0049^{**}$ & $-0.0050^{***}$ \\
 & $(0.0019)$ & $(0.0019)$ & $(0.0019)$ & $(0.0019)$ & $(0.0019)$ \\
$\log$ Life expectancy at birth & $0.0569^{***}$ & $0.0552^{***}$ & $0.0568^{***}$ & $0.0559^{***}$ & $0.0563^{***}$ \\
 & $(0.0204)$ & $(0.0205)$ & $(0.0204)$ & $(0.0209)$ & $(0.0205)$ \\
$\log$ Fertility rate & $-0.0250^{***}$ & $-0.0246^{***}$ & $-0.0240^{***}$ & $-0.0248^{***}$ & $-0.0243^{***}$ \\
 & $(0.0068)$ & $(0.0067)$ & $(0.0067)$ & $(0.0068)$ & $(0.0069)$ \\
Government consumption / GDP & $-0.0501^{*}$ & $-0.0502^{*}$ & $-0.0495^{*}$ & $-0.0502^{*}$ & $-0.0498^{*}$ \\
 & $(0.0292)$ & $(0.0293)$ & $(0.0296)$ & $(0.0293)$ & $(0.0296)$ \\
Gov. education spending / GDP & $0.1294$ & $0.1389$ & $0.1445$ & $0.1325$ & $0.1366$ \\
 & $(0.1052)$ & $(0.1043)$ & $(0.1015)$ & $(0.1042)$ & $(0.1018)$ \\
Black-market premium & $-0.0045^{***}$ & $-0.0049^{***}$ & $-0.0050^{***}$ & $-0.0046^{***}$ & $-0.0047^{***}$ \\
 & $(0.0015)$ & $(0.0015)$ & $(0.0014)$ & $(0.0015)$ & $(0.0015)$ \\
Terms-of-trade shock & $0.1561^{***}$ & $0.1567^{***}$ & $0.1585^{***}$ & $0.1560^{***}$ & $0.1569^{***}$ \\
 & $(0.0505)$ & $(0.0508)$ & $(0.0508)$ & $(0.0509)$ & $(0.0510)$ \\
Investment / GDP & $0.0869^{**}$ & $0.0866^{**}$ & $0.0842^{**}$ & $0.0861^{**}$ & $0.0833^{**}$ \\
 & $(0.0350)$ & $(0.0343)$ & $(0.0335)$ & $(0.0347)$ & $(0.0341)$ \\
Political rights index & $0.0156^{***}$ & $0.0151^{***}$ & $0.0150^{***}$ & $0.0156^{***}$ & $0.0157^{***}$ \\
 & $(0.0047)$ & $(0.0045)$ & $(0.0046)$ & $(0.0047)$ & $(0.0047)$ \\
Political rights$^2$ & $-0.0022^{***}$ & $-0.0022^{***}$ & $-0.0021^{***}$ & $-0.0022^{***}$ & $-0.0022^{***}$ \\
 & $(0.0006)$ & $(0.0006)$ & $(0.0006)$ & $(0.0006)$ & $(0.0006)$ \\
Constant & $-0.0278$ & $-0.0227$ & $-0.0256$ & $-0.0249$ & $-0.0254$ \\
 & $(0.0729)$ & $(0.0728)$ & $(0.0726)$ & $(0.0741)$ & $(0.0731)$ \\
\midrule
Periods & \multicolumn{5}{c}{1965--75, 1975--85, 1985--90} \\
Period fixed effects & \multicolumn{5}{c}{Yes} \\
Country-clustered SE & \multicolumn{5}{c}{Yes} \\
Observations & 233 & 233 & 233 & 233 & 233 \\
Countries & 88 & 88 & 88 & 88 & 88 \\
$R^2$ & 0.501 & 0.500 & 0.504 & 0.501 & 0.505 \\
\bottomrule
\end{tabular}
}
\begin{minipage}{0.95\linewidth}\vspace{0.2em}\footnotesize
\emph{Notes:} Dependent variable: average annual growth of real GDP per capita over each of \citet{barro1995inflation}'s periods (1965--75, 1975--85, 1985--90). Initial income, schooling and life expectancy are measured at the start of each period; the remaining controls are period averages. Every column includes the full control set with period fixed effects; standard errors are clustered by country. Columns differ only in the inflation summary entered: (1) mean inflation; (2) the raw within-period standard deviation $\sigma_\pi$; (3) the normalized measure $\text{NIU}=\sigma_\pi/\sqrt{1+|\bar\pi|}$ ($\bar\pi$ in percentage points, no inflation target); (4) mean inflation with $\sigma_\pi$; (5) mean inflation with NIU. Inflation moments use the annual log change in the consumer price index (Global Macro Data, GMD), untrimmed. This teaching specification omits \citet{barro1995inflation}'s log(GDP)$\times$human-capital interaction, so it reproduces the shape of his results rather than every coefficient. Significance: $^{***}p<0.01$, $^{**}p<0.05$, $^{*}p<0.1$.
\end{minipage}
\end{table}

\section{Inflation, Expectations, and Uncertainty}

This section studies the joint dynamics of realized inflation and inflation expectations, with a particular emphasis on the role of inflation uncertainty in shaping expectation formation. We use two monthly French time series obtained from Banque de France Webstat. Realized inflation is measured by the year-on-year Harmonised Index of Consumer Prices (ICP, series \texttt{ICP.M.FR.N.000000.4.ANR}). Inflation expectations are proxied by a press-based indicator (PAI, series \texttt{PAI.M.FR.N.PR.\_Z.INPR03.TX}), which captures media-implied inflation anticipations. Both series are downloaded through the Webstat Explore API, cleaned, transformed into numeric format, and indexed by date at monthly frequency.

The two time series are then merged into a single dataset and plotted on a common axis. The figure displays the co-movement between observed inflation and press-based expectations over time. Periods of heightened inflation volatility are visually associated with stronger deviations between the two series, motivating the econometric analysis that follows. In particular, we will examine (i) whether expectations are unbiased predictors of inflation, (ii) whether a long-run equilibrium relationship exists between the two series, and (iii) whether inflation uncertainty—measured through conditional volatility—affects the dynamics of expectation formation.

\subsection{Mincer--Zarnowitz regression and underreaction}\label{subsec:mz_underreaction}

A standard benchmark to evaluate the informational content of inflation expectations is the
\citet{mincer1969evaluation} regression. Let $\pi_t$ denote realized inflation (year-on-year)
and $\pi_t^{e}$ the expectation proxy observed at (or formed by) date $t$.\footnote{In our application,
$\pi_t$ is the ICP year-on-year inflation series and $\pi_t^{e}$ is the press-based expectations indicator (PAI),
both at monthly frequency.} We estimate:
\begin{equation}\label{eq:mz}
\pi_t = \alpha + \beta\,\pi_t^{e} + \varepsilon_t ,
\end{equation}
where $\varepsilon_t$ is a forecast error. Under the joint hypothesis of \emph{unbiased expectations}
(and no systematic forecast error), we should have:
\begin{equation}\label{eq:mz_H0}
H_0:\ \alpha = 0 \quad \text{and} \quad \beta = 1.
\end{equation}
Equation~\ref{eq:mz} is estimated by OLS. Because monthly macroeconomic forecast errors are typically
serially correlated (and may exhibit heteroskedasticity), inference is conducted using
HAC (Newey--West) standard errors with $L=12$ lags:
\begin{equation}\label{eq:hac}
\widehat{\mathrm{Var}}(\hat{\theta}) = \left(X'X\right)^{-1}\!\left(\sum_{h=-L}^{L} w_h\, \widehat{\Gamma}_h\right)\!\left(X'X\right)^{-1},
\qquad \theta=(\alpha,\beta)',
\end{equation}
where $w_h$ are Bartlett weights and $\widehat{\Gamma}_h$ denotes the sample autocovariance of $X_t\hat{\varepsilon}_t$ at lag $h$.

\paragraph{Testing underreaction.}
A common deviation from rational-unbiased expectations is \emph{underreaction}:
expectations move in the correct direction but by too little. In the Mincer--Zarnowitz regression,
this corresponds to a slope coefficient below unity:
\begin{equation}\label{eq:underreaction}
H_0:\ \beta = 1
\qquad \text{against} \qquad
H_1:\ \beta < 1.
\end{equation}
Intuitively, if $\beta<1$, realized inflation varies more than expectations: expectations fail to fully
incorporate available information and respond sluggishly to shocks.

\paragraph{Joint unbiasedness test.}
Beyond underreaction, we test the joint restriction~\ref{eq:mz_H0} using a Wald test:
\begin{equation}\label{eq:wald_joint}
W = \left(R\hat{\theta}-r\right)'\left[ R\,\widehat{\mathrm{Var}}(\hat{\theta})\,R' \right]^{-1}\left(R\hat{\theta}-r\right),
\qquad
R=
\begin{pmatrix}
1 & 0\\
0 & 1
\end{pmatrix},\ 
r=
\begin{pmatrix}
0\\
1
\end{pmatrix}.
\end{equation}
Under $H_0$ and standard regularity conditions, $W$ is asymptotically $\chi^2(2)$.

\paragraph{Implementation details (data alignment).}
The regression is run on the balanced sample obtained by merging the two series by date and dropping
observations with missing values:
\begin{equation}\label{eq:balanced_sample}
\mathcal{T}=\{t:\ \pi_t\ \text{and}\ \pi_t^e\ \text{are both observed}\}.
\end{equation}
In Python, this corresponds to \verb|df_mz = df_plot.dropna()| and OLS estimation of~\ref{eq:mz}
with HAC standard errors (\verb|cov_type="HAC"|, \verb|maxlags=12|).

\medskip
\noindent\textbf{Notes.}
(i) If expectations are measured for a horizon $h>0$ (e.g., one-year-ahead), replace $\pi_t$ by $\pi_{t+h}$
to match the forecasting target.
(ii) The test in~\ref{eq:underreaction} is one-sided; many software routines report the two-sided test
by default, in which case the one-sided $p$-value is obtained by halving the two-sided $p$-value
when $\hat{\beta}<1$.

\subsection{Engle--Granger Cointegration and Johansen VECM}

The Mincer--Zarnowitz regression tests short-run unbiasedness. However, even if expectations
are biased in the short run, the two series may still share a common long-run stochastic trend.
Cointegration analysis is therefore appropriate when both inflation and inflation expectations
are integrated of order one, $I(1)$.

\paragraph{Precondition: Integration order.}
The Engle--Granger approach is meaningful only if:
\[
\pi_t \sim I(1), 
\qquad 
\pi_t^{e} \sim I(1).
\]
If both series are stationary in levels ($I(0)$), standard VAR methods should be preferred.

\subsubsection*{1. Engle--Granger Two-Step Procedure}

We first estimate the long-run equilibrium relationship:
\begin{equation}\label{eq:eg_longrun}
\pi_t = \alpha + \beta \pi_t^{e} + u_t .
\end{equation}

The residuals $\hat u_t$ represent deviations from the potential long-run equilibrium.

\paragraph{Null hypothesis.}
\[
H_0: u_t \text{ has a unit root} \quad (\text{no cointegration})
\]
\[
H_1: u_t \sim I(0) \quad (\text{cointegration})
\]

\paragraph{Test.}
An Augmented Dickey--Fuller test is applied to the estimated residuals:
\begin{equation}\label{eq:eg_adf}
\Delta \hat u_t = \rho \hat u_{t-1} + \sum_{j=1}^{p} \gamma_j \Delta \hat u_{t-j} + \varepsilon_t.
\end{equation}

The null hypothesis corresponds to:
\[
H_0: \rho = 0,
\]
which implies a unit root in $\hat u_t$.

Because $\hat u_t$ is generated regressor residuals, the test statistic follows
non-standard critical values (MacKinnon critical values must be used).

\paragraph{Interpretation.}
If $H_0$ is rejected, inflation and expectations are cointegrated:
they share a stable long-run relationship despite short-run deviations.
This is often interpreted as a form of \emph{long-run anchoring}.

\subsubsection*{2. Johansen System Approach}

The Engle--Granger test is limited to a single cointegrating relationship
and treats one variable as dependent. The Johansen procedure instead estimates
a full vector autoregression in levels:

\begin{equation}
Y_t =
\begin{pmatrix}
\pi_t \\
\pi_t^{e}
\end{pmatrix},
\qquad
Y_t = A_1 Y_{t-1} + \cdots + A_k Y_{t-k} + \varepsilon_t.
\end{equation}

Rewriting in Vector Error Correction (VECM) form:

\begin{equation}\label{eq:vecm}
\Delta Y_t = \Pi Y_{t-1}
+ \sum_{i=1}^{k-1} \Gamma_i \Delta Y_{t-i}
+ \varepsilon_t,
\end{equation}

where
\[
\Pi = \alpha \beta'.
\]

\paragraph{Cointegration rank.}
The rank of $\Pi$ determines the number of cointegrating relationships:

\begin{itemize}
\item $\mathrm{rank}(\Pi)=0$ : no cointegration.
\item $\mathrm{rank}(\Pi)=1$ : one long-run equilibrium relation.
\item $\mathrm{rank}(\Pi)=2$ : both variables stationary.
\end{itemize}

\paragraph{Johansen Trace Test.}

\[
H_0: \mathrm{rank}(\Pi) \le r
\quad \text{vs} \quad
H_1: \mathrm{rank}(\Pi) > r.
\]

The trace statistic is:
\begin{equation}
\mathrm{Trace}(r)
=
- T \sum_{i=r+1}^{n} \ln(1 - \hat{\lambda}_i),
\end{equation}
where $\hat{\lambda}_i$ are the ordered eigenvalues of $\Pi$.

Critical values are non-standard and provided by Johansen tables.

\subsubsection*{3. Economic Interpretation via VECM}

If one cointegration relationship exists, the VECM for inflation reads:

\begin{equation}\label{eq:infl_vecm}
\Delta \pi_t =
\alpha_{\pi}
\left(
\pi_{t-1} - \beta \pi_{t-1}^{e}
\right)
+ \Gamma \Delta X_t
+ u_t,
\end{equation}

where:
\begin{itemize}
\item $\pi_{t-1} - \beta \pi_{t-1}^{e}$ is the long-run deviation,
\item $\alpha_{\pi}$ measures the speed of adjustment.
\end{itemize}

If $\alpha_{\pi} \neq 0$, inflation adjusts to restore equilibrium.
If instead the adjustment coefficient in the expectations equation is significant,
expectations absorb the disequilibrium.

\paragraph{Link with anchoring.}
Cointegration between inflation and expectations implies that
expectations do not drift permanently away from realized inflation.
Failure to find cointegration may signal de-anchoring.

\medskip

In what follows, we will extend this framework by incorporating
time-varying inflation uncertainty and studying whether volatility
modifies either the long-run equilibrium or the short-run adjustment dynamics.

\subsection{Information Rigidities: \citet{coibion2012can}}

Even if expectations are biased or cointegration fails, they may still adjust
gradually to new information. \citet{coibion2012can} propose a simple and
powerful test for information rigidities based on forecast errors.

Let inflation be denoted by $\pi_t$ and expectations by $\pi_t^{e}$.
Define the forecast error:
\begin{equation}
FE_t = \pi_t - \pi_t^{e}.
\end{equation}

Under full-information rational expectations, forecast errors should not be
predictable using past information. In particular, they should not be serially correlated.

Coibion and Gorodnichenko test this by estimating:
\begin{equation}\label{eq:cg}
FE_t = \lambda FE_{t-1} + \varepsilon_t.
\end{equation}

\paragraph{Null hypothesis (Full Information).}
\[
H_0: \lambda = 0.
\]

Under $H_0$, forecast errors are not persistent, and expectations adjust immediately
to new information.

\paragraph{Alternative hypothesis (Information rigidity).}
\[
H_1: \lambda > 0.
\]

If $\lambda > 0$, forecast errors are positively autocorrelated, implying that
expectations adjust sluggishly. Large shocks are incorporated gradually,
consistent with sticky-information or noisy-information models.

\paragraph{Estimation.}
Equation~\ref{eq:cg} is estimated by OLS using HAC (Newey--West) standard
errors with 12 lags to account for serial correlation and heteroskedasticity.

\paragraph{Interpretation.}
\begin{itemize}
\item $\hat{\lambda} \approx 0$ : fast adjustment, close to full-information benchmark.
\item $\hat{\lambda} > 0$ : sluggish adjustment, informational rigidity.
\item Larger $\hat{\lambda}$ : stronger persistence in forecast errors.
\end{itemize}

This framework provides a direct and model-light measure of expectation rigidity,
which is particularly useful before introducing volatility or uncertainty effects
into the expectation formation process.

\subsection{Does inflation volatility affect inflation expectations?}\label{subsec:vol_affects_exp}

This subsection studies whether \emph{inflation uncertainty}, proxied by time-varying inflation volatility, influences inflation expectations. The motivating idea is classical: inflation and inflation uncertainty interact through multiple channels. First, higher volatility may raise perceived inflation risk and induce more cautious (or more dispersed) expectations \citealp{friedman1977nobel,ball1992does}. Second, under imperfect information, agents may rationally downweight noisy signals, implying slower or biased expectation updates \citealp{coibion2012can}. Third, monetary policy credibility and stabilization motives may generate non-trivial feedback between inflation and uncertainty \citealp{cukierman1986theory}. Our empirical goal is not to adjudicate among all structural mechanisms, but to provide a transparent testing framework that links a measurable volatility object to observed expectations.

\subsubsection*{1. Measuring inflation uncertainty}

Let $\pi_t$ denote observed inflation (monthly, year-on-year). Inflation uncertainty is measured as a conditional variance (or conditional standard deviation) extracted from an ARCH/GARCH model \citealp{engle1982autoregressive,bollerslev1986generalized}. A baseline specification is:
\begin{align}
\pi_t &= \mu + \varepsilon_t, \qquad \varepsilon_t = \sigma_t z_t, \qquad z_t \sim \mathcal{N}(0,1), \\
\sigma_t^2 &= \omega + \alpha \varepsilon_{t-1}^2 + \beta \sigma_{t-1}^2. \label{eq:garch11}
\end{align}
We denote the resulting volatility proxy by $\widehat{\sigma}_t$ (or $\widehat{\sigma}_t^2$). Extensions (often useful for inflation) include asymmetric models that allow volatility to react differently to positive and negative inflation surprises (e.g., GJR/EGARCH variants).

\subsubsection*{2. Baseline expectations equation}

Let $\pi_t^e$ denote inflation expectations (here: press-based expectations). A simple reduced-form test is:
\begin{equation}\label{eq:exp_vol_baseline}
\pi_t^e = c + \beta \pi_t + \gamma \widehat{\sigma}_t + u_t.
\end{equation}

\paragraph{Null hypothesis (no volatility effect).}
\[
H_0: \gamma = 0.
\]
\paragraph{Alternative.}
\[
H_1: \gamma \neq 0 \quad (\text{or } \gamma>0 \text{ if uncertainty raises expectations}).
\]

Because $\pi_t$ and $\pi_t^e$ are persistent and sampled monthly, inference should use HAC (Newey--West) standard errors. A pedagogically useful extension adds dynamics:
\begin{equation}\label{eq:exp_vol_dyn}
\pi_t^e
= c + \sum_{j=1}^{p}\phi_j \pi_{t-j}^e
+ \sum_{j=0}^{p}\beta_j \pi_{t-j}
+ \sum_{j=0}^{p}\gamma_j \widehat{\sigma}_{t-j}
+ u_t,
\end{equation}
so that short-run adjustment and delayed volatility effects can be tested jointly:
\[
H_0:\ \gamma_0=\gamma_1=\cdots=\gamma_p=0.
\]

\subsubsection*{3. Volatility and information rigidity (interaction with forecast errors)}

Volatility may not only shift the level of expectations; it may also affect \emph{how quickly} expectations adjust. Building on the Coibion--Gorodnichenko forecast-error persistence regression, define:
\[
FE_t = \pi_t - \pi_t^e.
\]
The baseline rigidity test is:
\[
FE_t = \lambda FE_{t-1} + \varepsilon_t,
\]
with $H_0:\lambda=0$ vs $H_1:\lambda>0$ (\citealp{coibion2012can}). To test whether volatility amplifies rigidity, estimate:
\begin{equation}\label{eq:cg_vol_interaction}
FE_t = a + \lambda FE_{t-1} + \delta \widehat{\sigma}_t
+ \kappa \left(FE_{t-1}\times \widehat{\sigma}_t\right) + \eta_t.
\end{equation}
\paragraph{Key hypothesis.}
\[
H_0:\ \kappa=0 \qquad \text{vs} \qquad H_1:\ \kappa>0.
\]
A positive $\kappa$ means that forecast errors become more persistent when inflation volatility is high, consistent with slower updating in uncertain environments.

\subsubsection*{4. System evidence: VAR with volatility and Granger causality}

A complementary approach treats expectations and volatility jointly. Consider:
\begin{equation}\label{eq:var_vol}
Z_t =
\begin{pmatrix}
\pi_t^e \\
\pi_t \\
\widehat{\sigma}_t
\end{pmatrix},
\qquad
Z_t = A_1 Z_{t-1} + \cdots + A_p Z_{t-p} + e_t.
\end{equation}
One can test whether volatility helps predict expectations:
\[
H_0:\ \text{lags of } \widehat{\sigma}_t \text{ do not enter the } \pi_t^e \text{ equation}
\quad (\text{no Granger causality from volatility to expectations}).
\]
This does not establish structural causality, but it provides an informative diagnostic on predictive content.

\subsubsection*{5. Long-run anchoring with short-run volatility effects (VECM with exogenous volatility)}

If $\pi_t$ and $\pi_t^e$ were cointegrated (rank $=1$), volatility could enter the short-run dynamics while preserving the long-run relationship:
\begin{equation}\label{eq:vecm_vol}
\Delta Y_t
=
\alpha\beta' Y_{t-1}
+
\sum_{i=1}^{p-1}\Gamma_i \Delta Y_{t-i}
+
\Psi \widehat{\sigma}_t
+
\varepsilon_t,
\qquad
Y_t=
\begin{pmatrix}
\pi_t \\
\pi_t^e
\end{pmatrix}.
\end{equation}
The key test is $H_0:\Psi=0$. In our current application, the Johansen trace test suggests rank $=2$ (both series behave as stationary in levels), so a VECM is not required; the volatility effect can be studied in levels or in a stationary VAR.

\section{Interpolation of time series}

When working with multiple time series, e.g. to estimate a VAR, it might be that all time series are at a monthly frequency except one that is quarterly. If your research question imposes that you work at monthly frequency (e.g. you have to choose a monetary policy and cannot wait for next quarter figures), you might want to do interpolation of time series. There exists several methods. We will start to present in detail the \citep{chow1971best} method, when you have a related time series with a higher frequency. But again, as detailed in \citep{salazar1999monthly}, 
``the monthly model has little to add to a quarterly model when projecting one quarter ahead. However, the monthly model has an important role to play in building up a picture of the current quarter once one or two months' hard data becomes available.''

\subsection{\citep{chow1971best} method}

They find the best linear unbiased estimates of a monthly series by regression of a related series, starting from a quarterly series.

Assuming monthly observations of a series during $3 n$ months is satisfying a multiple regression relationship with $p$ related series:
\begin{equation}
    y = X\beta + u \label{eq:basic_regression}
\end{equation}
where \(y\) is \(3n \times 1\), \(X\) is \(3n \times p\), and \(u\) is a random vector with mean 0 and covariance matrix \(V\).

\textbf{Conversion from Monthly to Quarterly Observations}
Let \(C\) be the \(n \times 3n\) matrix converting \(3n\) monthly observations into \(n\) quarterly observations.
The matrix \(C_I\) selects one monthly observation per quarter, typically the first month. It is thus for interpolation and is defined as:
\begin{equation}
    C_I = \begin{bmatrix}
    1 & 0 & 0 & 0 & \cdots & 0 \\
    0 & 0 & 0 & 1 & \cdots & 0 \\
    0 & 0 & 0 & 0 & \cdots & 1 \\
    \vdots & \vdots & \vdots & \vdots & \ddots & \vdots \\
    0 & 0 & 0 & 0 & \cdots & 0
    \end{bmatrix} \label{eq:CI}
\end{equation}

This matrix takes the first, fourth, seventh, etc., months of each quarter to represent the quarterly values. Now quarterly variables, subscripted by a dot, will satisfy Equation \ref{eq:quarterly_conversion}.

\begin{equation}
    y. = Cy = CX\beta + Cu = X. \beta + u. \label{eq:quarterly_conversion}
\end{equation}
where \(E(u.u.') = V. = CVC'\).

\textbf{Estimation of Monthly Observations}
The problem is to estimate a vector  \(z\) of \(m\) observations on the dependent variables, where  \(z\) would be identical with  \(y\) in the cases of interpolation. 
A linear unbiased estimator \(\hat{z}\) of \(z\) satisfies for some \(m \times n\) matrix \(A\):
\begin{equation}
    \hat{z} = A y. = A (X. \beta + u.) \label{eq:condition_5}
\end{equation}
and
\begin{equation}
    E\left( \hat{z} - z \right) =  E\left( A (X. \beta + u.) - A (X \beta + u) \right) \label{eq:condition_6}
\end{equation}
Conditions \ref{eq:condition_5} and \ref{eq:condition_6} imply 
\begin{equation}\label{eq:condition_7}
A X. - X = 0
\end{equation}
and  \(\hat{z} - z = Au.-u\).

\textbf{Minimization of Covariance Matrix}
To find the best linear unbiased estimator $\hat{z}$, we minimize the covariance matrix of \((\hat{z} - z)\) subject to equation \ref{eq:condition_7}. We find a solution for $A$ and plug in back into equation \ref{eq:condition_5}.
For interpolation, they show in the paper that the resulting estimator is:
\begin{equation}\label{eq:resulting_estimator}
\hat{z} = X\hat{\beta} + V C' \left( CVC' \right)^{-1} \hat{u}
\end{equation}

\textbf{Estimation of the Covariance Matrix}
In practice $V$ is unknown, and has to be estimated by assuming some structure to the residuals in the monthly regression \ref{eq:basic_regression}. In the Chow-Lin approach, the covariance matrix of the residuals is assumed to follow an autoregressive process of order 1 (AR1). Specifically, the residuals \( u_t \) are modeled as:
\begin{equation}
    u_t = \rho u_{t-1} + \epsilon_t, \quad \text{where } \epsilon_t \text{ is white noise (WN)}(0, \sigma^2_{\epsilon})
\end{equation}
with \( |\rho| < 1 \). The covariance matrix \( \Sigma(\rho) \) for the AR1 process takes the form:
\begin{equation}
    \Sigma(\rho) = \frac{\sigma^2_{\epsilon}}{1-\rho^2} \begin{bmatrix}
    1 & \rho & \cdots & \rho^{n-1} \\
    \rho & 1 & \cdots & \rho^{n-2} \\
    \vdots & \vdots & \ddots & \vdots \\
    \rho^{n-1} & \rho^{n-2} & \cdots & 1
    \end{bmatrix}
\end{equation}
This structure implies that each entry \(\Sigma_{ij}\) is \(\rho^{|i-j|}\), reflecting the decreasing influence of autocorrelations as the lag increases.
The parameter \( \rho \) is estimated using an iterative procedure based on the observed autocorrelations of the low frequency residuals:
\begin{enumerate}
    \item Compute the autocorrelation of the observed low frequency residuals.
    \item Use this computed autocorrelation to estimate \( \rho \).
    \item Re-estimate the residuals using the new \( \rho \) and repeat until convergence.
\end{enumerate}

\subsection{Example}
One of the main influential macroeconomic paper, \citet{sims1980} applies a similar approach: \begin{quote}
Real GNP: [$\ldots$] preparation of this series required interpolation to obtain quarterly from published semi-annual data over much of the sample
period. The interpolation was carried out by a regression of observed semi-annual data on monthly values of industrial production for the current and three preceding months.
\end{quote}

\subsection{Open question}
Isn't it strange to assume that Equation \ref{eq:basic_regression} holds and then create a VAR model by stacking $y_l$ and $X_l$?
Idea: create a monthly VAR with 2 variables. Modify one variable to become yearly. Apply the Chow-Lin method to this variable assuming it is related to the second variable. Then calibrate a VAR on the two recreated monthly variables, how well does this perform?

\subsubsection*{Creating a Monthly VAR} We start with a two-variable monthly VAR model: 
\begin{equation} \begin{bmatrix} y_t \\ x_t \end{bmatrix} = A \begin{bmatrix} y_{t-1} \\ x_{t-1} \end{bmatrix} + \begin{bmatrix} \epsilon_{y,t} \\ \epsilon_{x,t} \end{bmatrix} 
\end{equation} 
where \( A \) is the matrix of autoregressive coefficients and \( \epsilon_t \) is the noise vector. 
We follow the work of \citep{helmut2005}and generate $200$ time series for $T=30$ and $T=100$. $\epsilon_{y,t} \sim \mathcal{N}(0,\,0.1)$ and $\epsilon_{x,t} \sim \mathcal{N}(0,\,0.9)$, and $y_0=x_0=0$.

\subsubsection*{Modifying Frequency} 

We modify \( y_t \) to create \( y_l \), an annual series derived from \( y_t \), and keep \( x_t \) as a monthly series. 
\subsubsection*{Applying Chow-Lin} 
Apply the Chow-Lin method to estimate the missing monthly values of \( y_l \) using \( x_t \) as the related series: 
\begin{equation} 
\hat{y}_t = \text{Chow-Lin}(y_l, x_t) 
\end{equation} 
\subsubsection*{Calibrating a New VAR} 
We then calibrate a new VAR model using the recreated monthly series \( \hat{y}_t \) and the original \( x_t \): 
\begin{equation} \begin{bmatrix} \hat{y}_t \\ x_t \end{bmatrix} = B \begin{bmatrix} \hat{y}_{t-1} \\ x_{t-1} \end{bmatrix} + \mu_t
\end{equation} 
 where \( B \) represents the new estimated coefficients. 

\section{First steps with R - functions, loops, imports and exports}

We suggest in the code \verb|helloworld_vansteenberghe.R| to get familiar with some R basics features essential before starting a research economics project:
\begin{itemize}
\item use the \textbf{help} of R to use existing functions
\item set your working directory to access your data sets and save figures and outputs
\item install and load \textbf{libraries}
\item assign values to variables, work with strings
\item perform mathematical operations
\item \textbf{plot} your data
\item learn to define your own \textbf{functions}
\item \textbf{if- else} conditions
\item use of sapply(), eval(), and fsolve()
\item learn to implement \textbf{for} and \textbf{while} loops
\item work with arrays, matrices, \textbf{data frames}, lists
\begin{itemize}
\item R data frames are tightly coupled collections of variables which share many of the properties of matrices and of lists, used as the fundamental data structure by most of R's modelling software; a matrix-like structure whose columns may be of differing types (numeric, logical, factor and character and so on)
\end{itemize}
\item handle errors (and bypass them while going through data)
\item use of random data generation according to distribution (e.g. normal, uniform, binomial)
\item export and import data from external csv files
\item perform basic linear regressions
\end{itemize}

\subsection{Some frequent questions on R}
\subsubsection{How to indicate the path to files to be imported}
With RStudio, there is a simple way to indicate the path to files that have to be imported:
\begin{enumerate}
\item File > Import Dataset > From Excel...
\item use Browses and select the a file in the folder where all the data sets are located
\item clikc on "Import", the path to your file is displayed and can be used from now on to change the working directory with setwd()
\end{enumerate}

\subsubsection{How to convert dates data to Date in R}
To convert dates data to R date format, one can follow \href{https://www.statmethods.net/input/dates.html}{the steps here}. In simple terms, one need to "indicate" R the format the dates have, e.g. if the date is of the format \verb|31/12/1959|, then in the format is in fact \verb|day/month/yearwith4digits| which in R is written \verb|%d/%m/%Y|. And the column "Date" of the dataframe df can be converted with the command: 

\verb|df$Date <- as.Date(df$Date, format = "%d/%m/%Y")|

\section{R: Multivariate regressions and tests (wages)}\label{sec:multreg}

\subsection{Introduction with the 2023 Claudia Goldin Nobel laureate}

Claudia Goldin, a distinguished Harvard professor, was awarded the Nobel Memorial Prize in Economic Sciences for her extensive research concerning women's evolution and involvement in the work force.

\begin{itemize}
    \item Investigated the \textbf{causes} of the gender wage gap and the \textbf{progression} of women’s participation in the labor market over two centuries.
    \item Revealed historical shifts in female employment with decreases in the 1800s due to industrialization and increases in the 1900s related to the expansion of the service sector.
    \item Identified the 1970s as a pivotal period when women began marrying later, advancing in higher education, and making significant strides in the labor market, partially influenced by the increased availability of birth control pills.
    \item Demonstrated that the contemporary gender wage gap is largely due to disparities between men and women \textit{within} the same occupation, notably becoming apparent following the birth of a woman's first child.
\end{itemize}

\citet{goldin2014grand} states that women without children generally have higher earnings than women with children and the former's earnings are almost equal to those of comparable men. It emphasizes the crucial role of labor market transformations, specifically in job structure and compensation, to advance gender convergence and potentially eliminate the gender pay gap, without necessarily requiring government intervention or altering men's domestic roles. Highlighting a notable disparity across sectors, it suggests that promoting temporal flexibility\footnote{Workplace flexibility incorporates the number of hours to be worked and also the particular hours worked, being 'on call', providing 'face time', being around for clients, group meetings, and the like.} and dismantling the disproportionate rewards for long and specific working hours are pivotal for achieving gender equality in the labor market.

Here is a weird computation one could do to advise Elon Musk and Donald Trump, in the vein of \citet{swift1995modest}: 
\begin{quote}
The societal benefit of an additional child can be estimated based on their GDP contribution over a lifetime. Using the U.S. GDP per capita of \$80K and assuming 40 working years, the total lifetime contribution is approximately: \$3.2 million. On the individual side, women face a significant cost due to the "child earnings penalty." After childbirth, women experience a reduction in earnings of about 33\%. For a median annual income of \$50,000, this translates into an annual loss of \$16,500. Over a 20-year period, the total earnings loss is \$330 K.
\end{quote}

It uses the data from the US Census and the American Community Survey (ACS) for the years from 1970 to 2010. And find that the ratio of (mean) annual earnings between male and female workers (full-time, full-year, 25 to 69 years) was 0.72 in 2010 and that of the medians was 0.77.

It mentions that \begin{quote}
The wage is also a summary statistic for an individual's education, training, prior labor force experience, and expected future participation. 
\end{quote}
The wage discrimination is the difference in earnings between observationally identical males and females.

\citet{goldin2014grand} uses the following model for the study (we omit the year fixed effect):
\begin{align}
\log(\text{Wage}_i) = \, & \beta_0 + \beta_1 \text{female}_i + \beta_2 \text{age}_i^2 + \beta_3 \text{race}_i + \beta_4 \text{education}_i \nonumber \\
& + \beta_5 \text{occupation}_i + \beta_6 \log(\text{hours}_i) + \beta_7 \log(\text{weeks}_i) \nonumber \\
& + \beta_8 \text{female}_i \times \text{occupation}_i + \epsilon_i \label{eq:goldin2014}
\end{align}

with:
\begin{itemize}
\item education: categorical variable;
\item occupation: categorical variable;
\item hours: hours worked;
\item weeks: weeks worked.
\end{itemize}

\subsection{Our study for this lecture}

We focus on the a salary equation\footnote{with the code multreg\_vansteenberghe.R} as in Schooling, Experience and Earnings \citep{mincer1974schooling}. We find a sample with 534 observation \href{http://www.economicswebinstitute.org/data/wagesmicrodata.xls}{here}.

"Return to education" has been widely studied in social science and as detailed by this St Louis Fed \href{https://www.stlouisfed.org/publications/regional-economist/january-2010/the-return-to-education-isnt-calculated-easily}{article}, it is not trivial to isolate the effect of the level of education on earnings, but studies usually estimate that  the return to one year of schooling is around 9 \citep{psacharopoulos2018returns} or 10 percent \citep{Card1999aa}, we find coherent results in the following sections.

\subsection{Descriptive statistics}

From our data set, we are interesting in the descriptive statistics, looking into mean, standard deviation, counts, percentiles, histograms to identify the dispersion, shape and potential outliers.

\resizebox{\textwidth}{!}{
\begin{tabular}{@{\extracolsep{5pt}} ccccccccccccc} 
\\[-1.8ex]\hline 
\hline \\[-1.8ex] 
 & n & mean & sd & median & trimmed & mad & min & max & range & skew & kurtosis & se \\ 
\hline \\[-1.8ex] 
WAGE & $534$ & $9.02$ & $5.14$ & $7.78$ & $8.28$ & $4.12$ & $1$ & $44.50$ & $43.50$ & $1.69$ & $4.90$ & $0.22$ \\ 
OCCUPATION & $534$ & $4.15$ & $1.66$ & $4$ & $4.31$ & $1.48$ & $1$ & $6$ & $5$ & $$-$0.49$ & $$-$0.94$ & $0.07$ \\ 
SECTOR & $534$ & $0.28$ & $0.54$ & $0$ & $0.16$ & $0$ & $0$ & $2$ & $2$ & $1.83$ & $2.39$ & $0.02$ \\ 
UNION & $534$ & $0.18$ & $0.38$ & $0$ & $0.10$ & $0$ & $0$ & $1$ & $1$ & $1.66$ & $0.77$ & $0.02$ \\ 
EDUCATION & $534$ & $13.02$ & $2.62$ & $12$ & $13.05$ & $1.48$ & $2$ & $18$ & $16$ & $$-$0.20$ & $0.81$ & $0.11$ \\ 
EXPERIENCE & $534$ & $17.82$ & $12.38$ & $15$ & $16.75$ & $11.86$ & $0$ & $55$ & $55$ & $0.68$ & $$-$0.40$ & $0.54$ \\ 
AGE & $534$ & $36.83$ & $11.73$ & $35$ & $36.07$ & $11.86$ & $18$ & $64$ & $46$ & $0.55$ & $$-$0.60$ & $0.51$ \\ 
FEMALE & $534$ & $0.46$ & $0.50$ & $0$ & $0.45$ & $0$ & $0$ & $1$ & $1$ & $0.16$ & $$-$1.98$ & $0.02$ \\ 
MARR & $534$ & $0.66$ & $0.48$ & $1$ & $0.69$ & $0$ & $0$ & $1$ & $1$ & $$-$0.65$ & $$-$1.58$ & $0.02$ \\ 
RACE & $534$ & $2.70$ & $0.68$ & $3$ & $2.87$ & $0$ & $1$ & $3$ & $2$ & $$-$1.94$ & $1.97$ & $0.03$ \\ 
SOUTH & $534$ & $0.29$ & $0.46$ & $0$ & $0.24$ & $0$ & $0$ & $1$ & $1$ & $0.91$ & $$-$1.17$ & $0.02$ \\ 
EXPERIENCE2 & $534$ & $470.60$ & $570.23$ & $225$ & $359.80$ & $296.52$ & $0$ & $3,025$ & $3,025$ & $1.61$ & $2.05$ & $24.68$ \\ 
\hline \\[-1.8ex] 
\end{tabular} 
}

\subsubsection{Scatter plots}
As expected, if we scatter plot experience versus age we can see a trend, which doesn't appear for education versus age. This means (roughly speaking) that we will not want to have both age and experience in the left hand side of our regression (LHS) as they might be "too" correlated, but having age and education both in the LHS should not be an issue.

\subsection{Gender}

The average dummy "FEMALE" is close to $\frac{1}{2}$ so our data set is rather balanced between men and women.

\begin{tabular}{lll}
\textbf{Variable}&\textbf{Mean}&\textbf{Standard Deviation}\\
log Wage (men)&2.17&0.53\\
log Wage(women)&1.93&0.49\\
\end{tabular}

In our data set, the average log wage for male is $2.17$ while for women it is $1.93$. One question we want to answer is whether this difference is significant if we control for the other variables (if men have in our sample more experience or education than women, then gender might not be the main explanation for this wage difference). We test the importance of being a man. We can also use boxplots.

Nota Bene: not all data sets would allow testing "ceteris paribus", you need enough "heterogeneity" in both sub-samples female and male. For example, if all males in your sample have PhDs and all females in your sample have no degree, it would be close to impossible to conclude whether there is a "gender wage gap".

\subsubsection{Boxplots}

A \textbf{boxplot}, also known as a box-and-whisker plot, is a graphical representation of a dataset that provides a visualization of its central tendency, spread, and shape. The major components of a boxplot include the median, quartiles, and potential outliers, which help to effectively summarize the distribution of a dataset. 

\begin{itemize}
    \item \textbf{Quartiles:} Quartiles divide a rank-ordered dataset into four equal parts. The three points that separate these parts are called the first quartile \((Q_1)\), second quartile \((Q_2)\) or median, and third quartile \((Q_3)\). Specifically,
    \begin{itemize}
        \item \(Q_1\) is the median of the lower half not including the overall median,
        \item \(Q_2\) (Median) is the central value that separates the lower half from the upper half of the data sample,
        \item \(Q_3\) is the median of the upper half not including the overall median.
    \end{itemize}
    Mathematically, if we have a dataset \(X\) and \(F(Q)\) is the cumulative distribution function of \(X\), then \(Q_1\), \(Q_2\), and \(Q_3\) are defined as the values for which 
    \[
    F(Q_1) = 0.25, \quad F(Q_2) = 0.50, \quad \text{and} \quad F(Q_3) = 0.75,
    \]
    respectively.

    \item \textbf{Outliers:} Outliers are data points that are significantly different from the others in the dataset. In a boxplot, outliers are typically represented as individual points outside the whiskers, which extend to \(1.5 \times \text{IQR}\) (Interquartile Range) above \(Q_3\) and below \(Q_1\). \(\text{IQR} = Q_3 - Q_1\) 

    \item \textbf{Box and Whiskers:} The "box" part of a boxplot is created using \(Q_1\), \(Q_2\), and \(Q_3\). The "whiskers" are lines extending from the box to the smallest and largest observations within the \(1.5 \times \text{IQR}\) boundary.
\end{itemize}

\begin{center}
\includegraphics[scale=0.5]{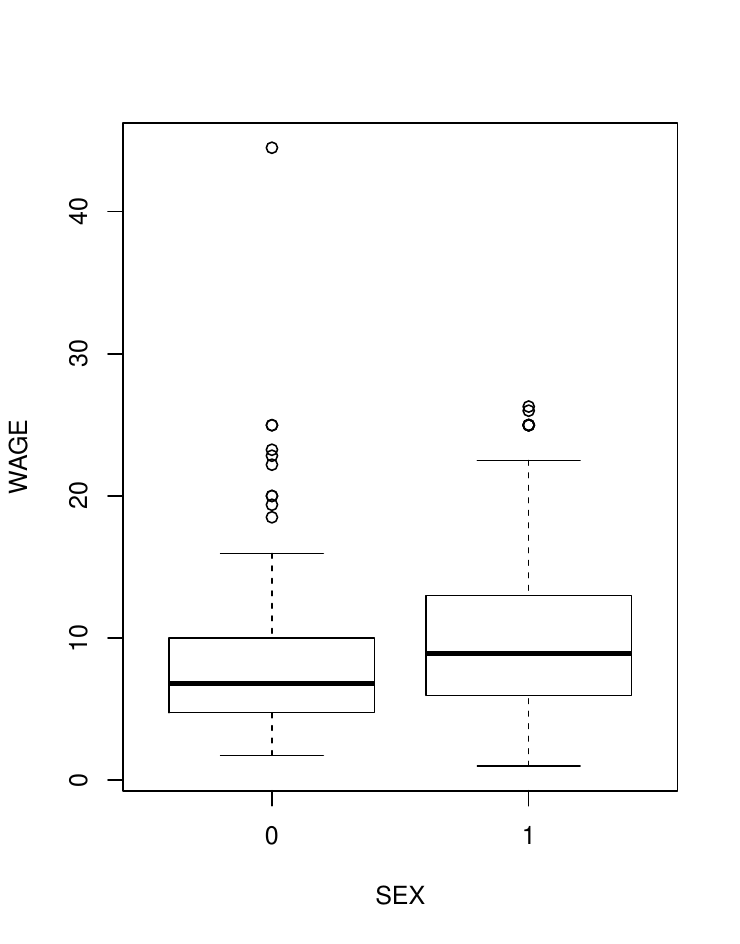}
\includegraphics[scale=0.5]{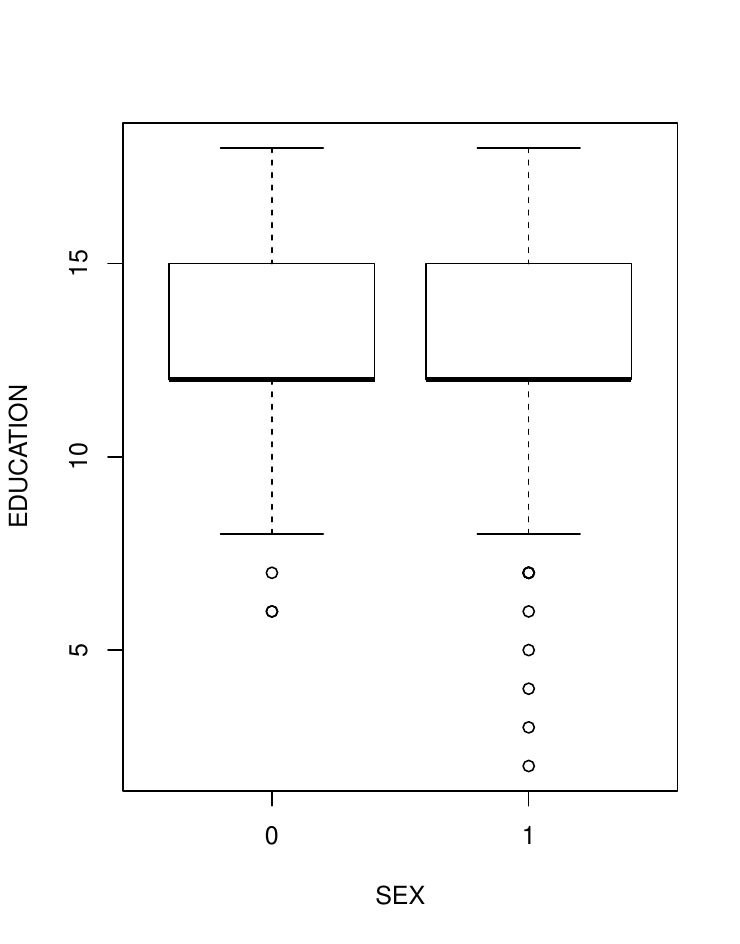}
\end{center}

While our sample indicates a wide spread of education levels among both genders, a noticeable wage gap persists, with women earning less. Men, tend to secure higher wages than women, yet no clear visual conclusion arises concerning educational disparities within this sample. This is "encouraging" for our purpose (testing gender wage gap hypothesis). To scrutinize these wage and education disparities further, we will employ various statistical tests and visualization tools, such as boxplots, in subsequent sections.

\subsection{Test on the importance of being a man}

We might do a simple model with a log-level specification, here we ignore the other features:
\begin{equation}\label{eq:wagemale}
\log(\mbox{Wage}_i)=\beta_0  + \beta_1 \mbox{male}_i+\epsilon_i
\end{equation}
As expected, we find: $\beta_0 = 1.93$ and $\beta_0  + \beta_1 = 2.16$ and we also see in our test that $\beta_1$ is significantly different to $0$, as we have for $H_0$: $\beta_1 = 0$ a p-value\footnote{remember that a p-value is the probability to wrongly reject $H_0$} of $4.31e-07$. We can say that the wage difference between women and men is significantly different than zero in our sample.

\subsection{Wage's level versus log}

Wage are typically log-normally distributed, which is a key assumption we want to have in our models. Applying the log will also attenuate the impact of outliers on the model and the coefficients can then be interpreted as percentage changes of the outcome variable.

We can decide to work with wages' log versus their level. If we work with a model in levels, we face homoskedasticity issues for the model's residuals, the higher the level of the fitted values, the higher the residuals standard deviation, thus we violate the hypothesis that the standard deviation of the residuals is the same for all observations:

\begin{equation}\label{eq:wagemalelevel}
\mbox{Wage}_i=\beta^l_0  + \beta^l_1 \mbox{male}_i+\epsilon^l_i
\end{equation}


If we take the log of the wage, this problem is less relevant by using the log\footnote{remember that $\log(a \times b) = \log(a) + \log(b)$, this means that by taking the log it is as if we move from an additive model to a multiplicative model and now the error terms in a multiplicative model have the same volatility irrespective of observation} of the wage in our model \ref{eq:wagemale}


\subsection{Determining Sample Size for Estimating Gender Wage Gap}

\citet{kahneman2011thinking} about \citet{cohen1962statistical}:

\begin{quote}
For a research psychologist, sampling variation is not a curiosity; it is a nuisance and a costly obstacle, which turns the undertaking of every research project into a gamble. Suppose that you wish to confirm the hypothesis that the vocabulary of the average six-year-old girl is larger than the vocabulary of an average boy of the same age. The hypothesis is true in the population; the average vocabulary of girls is indeed larger. Girls and boys vary a great deal, however, and by the luck of the draw you could select a sample in which the difference is inconclusive, or even one in which boys actually score higher. If you are the researcher, this outcome is costly to you because you have wasted time and effort, and failed to confirm a hypothesis that was in fact true. Using a sufficiently large sample is the only way to reduce the risk. Researchers who pick too small a sample leave themselves at the mercy of sampling luck.

The risk of error can be estimated for any given sample size by a fairly simple procedure. Traditionally, however, psychologists do not use calculations to decide on a sample size. They use their judgment, which is commonly flawed. An article I had read shortly before the debate with Amos demonstrated the mistake that researchers made (they still do) by a dramatic observation. The author pointed out that psychologists commonly chose samples so small that they exposed themselves to a 50\% risk of failing to confirm their true hypotheses! No researcher in his right mind would accept such a risk. A plausible explanation was that psychologists' decisions about sample size reflected prevalent intuitive misconceptions of the extent of sampling variation.
\end{quote}

In studies investigating the gender wage gap, determining an adequate sample size\footnote{sample\_size\_and\_test.py} is crucial for reliable inference. We have a sample of $n = 534$ observations, where the log wages for men and women are assumed to be normally distributed. The sample statistics are as follows:
\begin{itemize}
    \item Men: $\mu_1 = 2.0592$, $\sigma_1 = 0.5277$.
    \item Women: $\mu_2 = 1.9340$, $\sigma_2 = 0.4921$.
\end{itemize}

To estimate the difference between the population means, $\mu_1 - \mu_2$, with a specified precision $d$, we assume equal variances and sample sizes for the two populations. The distribution of the sample mean difference $\bar{x}_1 - \bar{x}_2$ under these assumptions is given by:
\[
\bar{x}_1 - \bar{x}_2 \sim N\left(\mu_1 - \mu_2, \frac{2\sigma^2}{n}\right),
\]
where $\sigma^2$ is the pooled variance estimate.

The required sample size $n$ to achieve a maximum difference $d$ between the true and estimated mean difference at a $(1-\alpha)100\%$ confidence level is determined by the formula:
\[
n = \frac{2Z_{\alpha/2}^2\sigma^2}{d^2}.
\]

Here:
\begin{itemize}
    \item $Z_{\alpha/2}$ is the critical value of the standard normal distribution corresponding to the desired confidence level.
    \item $\sigma^2 = \frac{\sigma_1^2 + \sigma_2^2}{2}$ is the pooled variance.
\end{itemize}

Using the provided data:
\[
\sigma^2 = \frac{(0.5277)^2 + (0.4921)^2}{2} \approx 0.2591.
\]

For a confidence level of $95\%$ ($Z_{\alpha/2} \approx 1.96$) and a desired precision $d$, the sample size from each population is:
\[
n = \frac{2(1.96)^2(0.2591)}{d^2}.
\]

Table \ref{tab:sample_size} illustrates the required sample size for varying levels of $d$. We would want a precision smaller than the observed sample mean difference of $ 0.1252$. With approximately 500 observations half/half between men and women, we can expect a precision of $0.08$.

\begin{table}[h!]
    \centering
    \resizebox{\textwidth}{!}{
    \begin{tabular}{ccc}
        \toprule
        \textbf{Precision ($d$)} & \textbf{Minimum Detectable Difference} ($5\%$ level and $90\%$ power) & \textbf{Sample Size ($n$)} \\
        \midrule
        0.045 & 0.07 & 1000\\
        0.08 & 0.15 & 250 \\
        0.14 & 0.23 & 100 \\
        0.2 & 0.33 & 50 \\
        0.28 & 0.47 & 25 \\
        \bottomrule
    \end{tabular}
    }
    \caption{Required Sample Size for Estimating Gender Wage Gap}
    \label{tab:sample_size}
\end{table}

If the population variances are unknown, the value of $Z_{\alpha/2}$ in the formula can be replaced with the $t_{\alpha/2,\nu}$ value based on degrees of freedom $\nu = n - 1$. This approach ensures robust estimation through an iterative process, particularly when sample sizes are small.

\subsection{Hypothesis Testing for Difference Between Two Means}

To determine the sample size \(n\) required for a two-sample \(t\)-test to detect a true difference \(d\) between population means at a significance level of \(\alpha = 0.05\) with a power of \(1 - \beta = 0.90\), we use the following formula:

\[
n = \frac{2\sigma^2 \cdot \left(Z_{\alpha} + Z_{\beta}\right)^2}{d^2}
\]

Where:
\begin{itemize}
    \item \(Z_{\alpha}\) is the critical value of the standard normal distribution for the significance level (two-tailed, \(\alpha = 0.05\)).
    \item \(Z_{\beta}\) is the critical value of the standard normal distribution corresponding to the desired power (\(1 - \beta = 0.90\)). The power of a test is the probability in rejecting a null hypothesis when alternative
hypothesis is true
    \item \(\sigma\) is the standard deviation of the population.
    \item \(d\) is the smallest true difference between the population means to be detected.
\end{itemize}

For this example, we assume:
\begin{itemize}
    \item \(\alpha = 0.05\), \(Z_{\alpha} = 1.96\).
    \item \(1 - \beta = 0.90\), \(Z_{\beta} = 1.28\).
    \item \(\sigma = \sqrt{0.2591}\).
\end{itemize}

With a sample of size 534, we can expect to be able to detect a difference in mean of approximately $0.15$ at the $0.05$ significance level with a $90\%$ chance of detecting the true difference.

\subsection{Wage determinants estimation}

We want to determine if the wages of male and female are different even when they have the same age, level of education and experience (that is typically said \textit{ceteris paribus}), now if $\beta_1$ is significantly different than $0$, then ceteris paribus men and women don't earn the same wages.

\subsection{Variables correlations and some models exploration}

We might first want to compute the variables correlations and explore some possible models:

\quad

\resizebox{\textwidth}{!}{
\begin{tabular}{@{\extracolsep{5pt}}lD{.}{.}{-3} D{.}{.}{-3} D{.}{.}{-3} D{.}{.}{-3} } 
\\[-1.8ex]\hline 
\hline \\[-1.8ex] 
 & \multicolumn{4}{c}{\textit{Dependent variable:}} \\ 
\cline{2-5} 
\\[-1.8ex] & \multicolumn{4}{c}{lWAGE} \\ 
\\[-1.8ex] & \multicolumn{1}{c}{(1)} & \multicolumn{1}{c}{(2)} & \multicolumn{1}{c}{(3)} & \multicolumn{1}{c}{(4)}\\ 
\hline \\[-1.8ex] 
 MALE &  &  & 0.256^{***} & 0.257^{***} \\ 
  &  &  & (0.039) & (0.040) \\ 
  & & & & \\ 
 AGE &  &  &  & -0.080 \\ 
  &  &  &  & (0.114) \\ 
  & & & & \\ 
 EDUCATION & 0.076^{***} & 0.096^{***} & 0.098^{***} & 0.177 \\ 
  & (0.008) & (0.008) & (0.008) & (0.114) \\ 
  & & & & \\ 
 EXPERIENCE &  & 0.012^{***} & 0.013^{***} & 0.092 \\ 
  &  & (0.002) & (0.002) & (0.114) \\ 
  & & & & \\ 
 Constant & 1.064^{***} & 0.598^{***} & 0.421^{***} & 0.898 \\ 
  & (0.108) & (0.125) & (0.123) & (0.693) \\ 
  & & & & \\ 
\hline \\[-1.8ex] 
Observations & \multicolumn{1}{c}{533} & \multicolumn{1}{c}{533} & \multicolumn{1}{c}{533} & \multicolumn{1}{c}{533} \\ 
R$^{2}$ & \multicolumn{1}{c}{0.143} & \multicolumn{1}{c}{0.209} & \multicolumn{1}{c}{0.267} & \multicolumn{1}{c}{0.268} \\ 
Adjusted R$^{2}$ & \multicolumn{1}{c}{0.142} & \multicolumn{1}{c}{0.206} & \multicolumn{1}{c}{0.263} & \multicolumn{1}{c}{0.263} \\ 
Residual Std. Error & \multicolumn{1}{c}{0.489 (df = 531)} & \multicolumn{1}{c}{0.470 (df = 530)} & \multicolumn{1}{c}{0.453 (df = 529)} & \multicolumn{1}{c}{0.453 (df = 528)} \\ 
F Statistic & \multicolumn{1}{c}{88.905$^{***}$ (df = 1; 531)} & \multicolumn{1}{c}{70.175$^{***}$ (df = 2; 530)} & \multicolumn{1}{c}{64.367$^{***}$ (df = 3; 529)} & \multicolumn{1}{c}{48.351$^{***}$ (df = 4; 528)} \\ 
\hline 
\hline \\[-1.8ex] 
\textit{Note:}  & \multicolumn{4}{r}{$^{*}$p$<$0.1; $^{**}$p$<$0.05; $^{***}$p$<$0.01} \\ 
\end{tabular} 
}

We might want to add in the 5th column the Claudia Goldin's closest form of the model \ref{eq:goldin2014}.
Here, column 5 we do:
\begin{align}
\log(\text{Wage}_i) = \, & \beta_0 + \beta_1 \text{male}_i + \beta_2 \text{age}_i^2 + \beta_3 \text{race}_i + \beta_4 \text{education}_i \nonumber \\
& + \beta_5 \text{sector}_i + \beta_8 \text{male}_i \times \text{sector}_i + \epsilon_i
\label{eq:goldinlinke}
\end{align}

This model is very demanding with "only" 533 observations, so we might unnecessarily loose significance for some of our parameters.

\resizebox{\textwidth}{!}{
\begin{tabular}{@{\extracolsep{5pt}}lD{.}{.}{-3} D{.}{.}{-3} D{.}{.}{-3} D{.}{.}{-3} D{.}{.}{-3} } 
\\[-1.8ex]\hline 
\hline \\[-1.8ex] 
 & \multicolumn{5}{c}{\textit{Dependent variable:}} \\ 
\cline{2-6} 
\\[-1.8ex] & \multicolumn{5}{c}{lWAGE} \\ 
\\[-1.8ex] & \multicolumn{1}{c}{(1)} & \multicolumn{1}{c}{(2)} & \multicolumn{1}{c}{(3)} & \multicolumn{1}{c}{(4)} & \multicolumn{1}{c}{(5)}\\ 
\hline \\[-1.8ex] 
 MALE &  & 0.256^{***} & 0.257^{***} & 0.257^{***} & 0.212^{***} \\ 
  &  & (0.039) & (0.039) & (0.040) & (0.045) \\ 
  & & & & & \\ 
 AGE &  &  &  & -0.080 &  \\ 
  &  &  &  & (0.114) &  \\ 
  & & & & & \\ 
 I(AGE$\hat{\mkern6mu}$2) &  &  &  &  & 0.0001^{***} \\ 
  &  &  &  &  & (0.00002) \\ 
  & & & & & \\ 
 factor(RACE)2 &  &  &  &  & 0.008 \\ 
  &  &  &  &  & (0.104) \\ 
  & & & & & \\ 
 factor(RACE)3 &  &  &  &  & 0.097 \\ 
  &  &  &  &  & (0.060) \\ 
  & & & & & \\ 
 EDUCATION & 0.076^{***} & 0.098^{***} & 0.091^{***} & 0.177 & 0.088^{***} \\ 
  & (0.008) & (0.008) & (0.008) & (0.114) & (0.008) \\ 
  & & & & & \\ 
 EXPERIENCE &  & 0.013^{***} & 0.036^{***} & 0.092 &  \\ 
  &  & (0.002) & (0.005) & (0.114) &  \\ 
  & & & & & \\ 
 I(EXPERIENCE$\hat{\mkern6mu}$2) &  &  & -0.001^{***} &  &  \\ 
  &  &  & (0.0001) &  &  \\ 
  & & & & & \\ 
 factor(SECTOR)1 &  &  &  &  & 0.063 \\ 
  &  &  &  &  & (0.080) \\ 
  & & & & & \\ 
 factor(SECTOR)2 &  &  &  &  & -0.100 \\ 
  &  &  &  &  & (0.325) \\ 
  & & & & & \\ 
 MALE:factor(SECTOR)1 &  &  &  &  & 0.126 \\ 
  &  &  &  &  & (0.104) \\ 
  & & & & & \\ 
 MALE:factor(SECTOR)2 &  &  &  &  & 0.260 \\ 
  &  &  &  &  & (0.340) \\ 
  & & & & & \\ 
 Constant & 1.064^{***} & 0.421^{***} & 0.345^{***} & 0.898 & 0.484^{***} \\ 
  & (0.108) & (0.123) & (0.122) & (0.693) & (0.128) \\ 
  & & & & & \\ 
\hline \\[-1.8ex] 
Observations & \multicolumn{1}{c}{533} & \multicolumn{1}{c}{533} & \multicolumn{1}{c}{533} & \multicolumn{1}{c}{533} & \multicolumn{1}{c}{533} \\ 
R$^{2}$ & \multicolumn{1}{c}{0.143} & \multicolumn{1}{c}{0.267} & \multicolumn{1}{c}{0.295} & \multicolumn{1}{c}{0.268} & \multicolumn{1}{c}{0.270} \\ 
Adjusted R$^{2}$ & \multicolumn{1}{c}{0.142} & \multicolumn{1}{c}{0.263} & \multicolumn{1}{c}{0.289} & \multicolumn{1}{c}{0.263} & \multicolumn{1}{c}{0.258} \\ 
Residual Std. Error & \multicolumn{1}{c}{0.489 (df = 531)} & \multicolumn{1}{c}{0.453 (df = 529)} & \multicolumn{1}{c}{0.445 (df = 528)} & \multicolumn{1}{c}{0.453 (df = 528)} & \multicolumn{1}{c}{0.454 (df = 523)} \\ 
F Statistic & \multicolumn{1}{c}{88.905$^{***}$ (df = 1; 531)} & \multicolumn{1}{c}{64.367$^{***}$ (df = 3; 529)} & \multicolumn{1}{c}{55.143$^{***}$ (df = 4; 528)} & \multicolumn{1}{c}{48.351$^{***}$ (df = 4; 528)} & \multicolumn{1}{c}{21.539$^{***}$ (df = 9; 523)} \\ 
\hline 
\hline \\[-1.8ex] 
\textit{Note:}  & \multicolumn{5}{r}{$^{*}$p$<$0.1; $^{**}$p$<$0.05; $^{***}$p$<$0.01} \\ 
\end{tabular} 
} 

\subsection{The issue with Multicollinearity}

Multicollinearity arises when predictor variables in a regression model are closely correlated, thereby complicating the isolation of individual predictors' effects on the dependent variable.  
We seek to test the following Wage Determination Model:
\begin{equation}
\log(\mbox{Wage}_i)=\beta_0+\alpha \mbox{age}_i + \beta_1 \mbox{male}_i +\beta_2 \mbox{education}_i +\beta_3 \mbox{experience}_i+\epsilon_i
\end{equation}

Expressed in matrix form, this model can be articulated as: \(Y = X \beta + U\) and, assuming \(X'X\) is not singular, the coefficients are estimated as: \(\hat{\beta} = (X'X)^{-1}X'Y\).

Testing this model, the coefficients for age, education, and experience seem non-significant. Notably, a high correlation is observed between age and experience level. In matrix form, severe multicollinearity means that either \(X\) is not of full rank\footnote{No column of \(X\) should be a linear combination of other \(X\) columns.} or \(\vert X'X\vert\) is close to \(0\), rendering the estimate \(\hat{\beta}\) unstable\footnote{Practically, one can check if the determinant of the correlation matrix of the explanatory variables is close to \(0\).}. Therefore, unless the sample size is substantially increased, the impact of age and education cannot be isolated. An alternative approach is to exclude one variable, such as age, from our example.

\subsubsection{Variance Inflation Factor (VIF)}

The \textbf{Variance Inflation Factor} (VIF) serves as a metric for diagnosing multicollinearity, quantifying the extent to which the variance of the estimated regression coefficients is increased due to multicollinearity.

\textbf{VIF definition}

Given the linear regression model:
\begin{equation}
Y = \beta_0 + \beta_1X_1 + \beta_2X_2 + \ldots + \beta_pX_p + \epsilon
\end{equation}

The VIF for each predictor \(X_j\) is calculated as:
\begin{equation}
\text{VIF}_j = \frac{1}{1 - R_j^2}
\end{equation}
where \(R_j^2\) is the coefficient of determination obtained by regressing \(X_j\) on all other predictors. Essentially, \(R_j^2\) represents the proportion of the variance in \(X_j\) explained by the other predictors. A general guideline is that if \(\mbox{VIF}_j > 5\), multicollinearity might be problematic, potentially requiring further investigation or adjustments to the model.

\subsubsection{Farrar and Glauber test}

The Farrar and Glauber test is used to identify multicollinearity within regression models. It involves three primary tests: a test for overall multicollinearity, a test for specific variables, and a test to identify which variable(s) are responsible for causing multicollinearity.

\textbf{Overall Multicollinearity Test}

The overall multicollinearity test uses the eigenvalues \(\lambda_i\) of the \(X'X\) matrix, where larger eigenvalues indicate stronger multicollinearity. 

\textbf{Individual Multicollinearity Test}

The individual multicollinearity test focuses on the variable causing multicollinearity by analyzing the condition index, defined as:
\begin{equation}
\text{Condition Index} = \sqrt{\frac{\lambda_{\max}}{\lambda_j}}
\end{equation}
where \(\lambda_{\max}\) is the largest eigenvalue, and \(\lambda_j\) is the eigenvalue associated with variable \(j\).

To identify which variables are causing multicollinearity, we examine the variance decomposition proportions, which allocate the variance inflation caused by multicollinearity among the variables.

In application to our wage determination model, we find that the coefficients for age, education, and experience are non-significant, hinting at potential multicollinearity. As an initial measure, VIF is applied to check the severity of multicollinearity among these predictors. Subsequent to VIF analysis, the Farrar and Glauber test would be applied to perform an in-depth diagnosis and understand the roots of the multicollinearity observed.

\subsection{Multicollinearity Diagnostics: Insight into the Klein Test}

Multicollinearity in a regression model arises when two or more explanatory variables are highly correlated, making it difficult to isolate the individual impact of each variable on the dependent variable. Detecting and addressing multicollinearity is pivotal in model building to ensure reliable and interpretable coefficient estimates. One approach to inspect multicollinearity is the Klein test, which furnishes a straightforward diagnostic technique without necessitating additional model estimations.

\textbf{Theoretical Backdrop of the Klein Test:}
The Klein test is fundamentally premised on examining the relationships amongst the independent variables. Specifically, it sets a criterion to scrutinize:
\[
R^2 < | \text{corr}(x_i, x_j) |
\]
Where:
\( R^2 \) is the coefficient of determination from the regression model.
\(\text{corr}(x_i, x_j) \) denotes the correlation coefficient between two explanatory variables \( x_i \) and \( x_j \).

The rationale behind the test is that if the absolute correlation between any pair of independent variables is greater than the \( R^2 \) of the model, the model may suffer from harmful multicollinearity. This is underpinned by the intuition that two predictors sharing a strong linear relationship can potentially disturb the stability and interpretability of the coefficient estimates.

\textbf{An Applied Example:}
In the context of the example provided, the correlation matrix is:
\begin{center}
\begin{tabular}{lccc}
          & MALE  & EDUCATION & EXPERIENCE\\
\hline
MALE      &  1.000 & -0.002 &  -0.075\\
EDUCATION & -0.002  & 1.000 &  -0.353\\
EXPERIENCE& -0.075 & -0.353  &  1.000\\
\end{tabular}
\end{center}

Observe that the correlation coefficient between EDUCATION and EXPERIENCE is -0.353, which might prompt scrutiny when compared against the model's \( R^2 \). If -0.353 surpasses the \( R^2 \), it could be indicative of multicollinearity between these variables, although, as noted, a level of 35\% might not be immediately alarming depending on the field and context of the study.

\stepcounter{question}
\begin{itemize}
\item \textbf{Question \arabic{question}}
\begin{itemize}
\item Develop a function in R that:
    \begin{itemize}
    \item Calculates the correlation matrix for all possible pairs of independent variables in a dataset.
    \item Compares each non-diagonal element of the matrix to the \( R^2 \) of the model, checking if \( | \text{corr}(x_i, x_j) | > R^2 \).
    \item Returns pairs of variables that violate the above condition, signaling potential multicollinearity.
    \end{itemize}
\end{itemize}
\end{itemize}

\subsubsection{A model with less multicollinearity}

\begin{equation}\label{eq:wage}
\log(\mbox{Wage}_i)=\beta_0+\beta_1 \mbox{male}_i +\beta_2 \mbox{education}_i +\beta_3 \mbox{experience}_i+\epsilon_i
\end{equation}

We apply an ordinary least square regression and obtain the following results:

\begin{tabular}{@{\extracolsep{5pt}}lc} 
\\[-1.8ex]\hline 
\hline \\[-1.8ex] 
 & \multicolumn{1}{c}{\textit{Dependent variable:}} \\ 
\cline{2-2} 
\\[-1.8ex] & lWAGE \\ 
\hline \\[-1.8ex] 
 MALE & 0.256$^{***}$ \\ 
  & (0.039) \\ 
  & \\ 
 EDUCATION & 0.098$^{***}$ \\ 
  & (0.008) \\ 
  & \\ 
 EXPERIENCE & 0.013$^{***}$ \\ 
  & (0.002) \\ 
  & \\ 
 Constant & 0.421$^{***}$ \\ 
  & (0.123) \\ 
  & \\ 
\hline \\[-1.8ex] 
Observations & 533 \\ 
R$^{2}$ & 0.267 \\ 
Adjusted R$^{2}$ & 0.263 \\ 
Residual Std. Error & 0.453 (df = 529) \\ 
F Statistic & 64.367$^{***}$ (df = 3; 529) \\ 
\hline 
\hline \\[-1.8ex] 
\textit{Note:}  & \multicolumn{1}{r}{$^{*}$p$<$0.1; $^{**}$p$<$0.05; $^{***}$p$<$0.01} \\ 
\end{tabular} 

And an adjusted $R^2$ close to $26\%$. That is to say that $26\%$ of the variance of the log of wages can be explained by gender, education and experience.

The more explanatory variables are added to a model\footnote{with $k$ the number of variables, $+1$ intercept, and $N$ the number of observations}, the greater the $R^2$, it is better to penalize for the addition of new variables, hence the use of an adjusted $R^2$ defined as:
$$\mbox{adj} R^2 = 1 - (1-R^2)\frac{N-1}{N-k-1}$$

\subsection{Multicollinearity and sample size to identify coefficients}

To determine the minimum sample size required to detect statistically significant coefficients in a regression model with highly collinear predictors (correlation $\rho = 0.9779$), we\footnote{collinearity\_sample\_size.py} use Monte Carlo simulations. The true model is $y = X_1 + X_2 + \varepsilon$, where $\varepsilon \sim N(0, 1)$ and both $X_1$ and $X_2$ have unit variance and specified correlation. For each sample size $n$, we simulate the proportion of runs where the p-values for both coefficients are below $\alpha = 0.05$. The required $n$ is the smallest sample size where at least 95\% of simulations yield significant p-values, ensuring that $P(p\text{-value} < \alpha) \geq 0.95$.
Next, we would need to update the code so that $X_1$ and $X_2$ have the same distribution as age and experience in our study.

\subsection{The Fisher Test in Regression Analysis}\label{sec:fishertestwage}

The Fisher Test, also referred to as Fisher's F-test, plays a pivotal role in hypothesis testing within the realm of regression analysis. Named after Ronald A. Fisher, the test facilitates the comparison of models in the context of their fit to the data, serving as an invaluable tool in model selection.

Given a regression model
\begin{equation}
Y = \beta_0 + \beta_1 X_1 + \beta_2 X_2 + \ldots + \beta_p X_p + \epsilon
\end{equation}
one common application of the Fisher Test involves testing the null hypothesis that a set of regression coefficients are equal to zero, essentially evaluating the joint significance of the predictors in the model. Formally, this can be expressed as:
\begin{equation}
H_0: \beta_{j_1} = \beta_{j_2} = \ldots = \beta_{j_k} = 0
\end{equation}
versus the alternative hypothesis that at least one of the coefficients is non-zero:
\begin{equation}
H_1: \text{At least one } \beta_{j_i} \neq 0
\end{equation}

The test statistic for the Fisher Test is given by
\begin{equation}
F = \frac{(RSS_0 - RSS_1)/(p_1 - p_0)}{RSS_1/(n - p_1)}
\end{equation}
where \(RSS_0\) and \(RSS_1\) represent the residual sum of squares for the reduced and full model respectively, \(p_0=1\) and \(p_1=4\) denote the number of predictors in each model, and \(n\) is the sample size. The test statistic follows an \(F\)-distribution with \((p_1 - p_0, n - p_1)\) degrees of freedom under the null hypothesis.

The Fisher Test assesses whether the additional predictors in the full model significantly enhance the fit of the model. A significant F-test suggests that the full model represents a statistically better fit to the data compared to the reduced model, while a non-significant F-test may indicate that the additional predictors do not significantly improve the model.

In the context of our model \ref{eq:wage}, we posit the null hypothesis that the coefficients for \textit{male}, \textit{education}, and \textit{experience} are all zero, formally:
\begin{equation}
H_0: \beta_1 = \beta_2 = \beta_3 = 0 
\end{equation}

If we consult an F-distribution table for an \(\alpha\) level of 5\%, with $3$ degrees of freedom in the numerator (due to testing 3 coefficients) and $N-4$ in the denominator (where \(N\) is the total sample size), we find a critical value around 2.6. If our calculated F-statistic is higher than this critical value, we reject \(H_0\) at the 5\% significance level, providing evidence that our model has some explanatory power relative to a model with no predictors.

\subsubsection{Fisher test on education and experience}
We want to perform a Fisher test on $H_0$: $\beta_2 =\beta_3=0$ for the model \ref{eq:wage}, in other words, were we correct in adding education and experience as explanatory variables to the model. We test that $J$ variables of the $K$ explanatory variables ($J<K$) are not jointly null.

We obtain the following statistics, F-statistic:  
$$\frac{\frac{SSR_r - SSR_u}{J}}{\frac{SSR_u}{N-J}}$$
or equivalently:                   
$$\frac{\frac{R^2 - R_{\mbox{restricted}}^2}{J}}{\frac{1-R^2}{N-J}}$$
where $R_{\mbox{restricted}}^2$ is the R square from the model \ref{eq:wagemale}. We compare our statistic to the threshold $F_{N-J,\alpha}^J$ and find that we can reject $H_0$.

\subsection{t-Test in Regression Analysis}

In the context of regression analysis, the t-test is utilized to assess the statistical significance of each coefficient in the model. For each coefficient, the null hypothesis states that it equals zero (\(H_0: \beta_j = 0\)), implying no effect. A t-test decides whether to reject this null hypothesis.

Given the regression model:
\begin{equation}
\log(\mbox{WAGE}_i) = \beta_0 + \beta_1 \mbox{MALE}_i + \beta_2 \mbox{EDUCATION}_i + \beta_3 \mbox{EXPERIENCE}_i + \epsilon_i
\end{equation}

The t-statistic for each coefficient is computed as:
\begin{equation}
t_j = \frac{\hat{\beta}_j - 0}{SE(\hat{\beta}_j)}
\end{equation}
where \(\hat{\beta}_j\) is the estimated coefficient and \(SE(\hat{\beta}_j)\) is its standard error.

For our regression, all the computed t-statistics exceed the critical value of \(1.96\), which is the critical value for a two-tailed test with infinite degrees of freedom and a significance level of \(\alpha=5\%\).

Therefore, we can infer that gender, education, and experience have a statistically significant impact on observed wages.

\subsection{How much do men earn more than women according to our model}

According to our model, if a woman earn a wage $w$, a man will earn (ceteris paribus):
$\exp\left(\log(w)+\beta_1\right)=w \exp(0.255)=1.29 \times w$

According to our model, a man earns around $29\%$ more than a woman, ceteris paribus.

In fact, the difference between the means of our sample for men and women was $0.229$ and controlling for education and experience this difference doesn't vanish and is $0.257$.

\subsection{Non-linear effect of experience}\label{sec:experiencesquare}

We might want to test the non-linear effect of experience adding its square in the model:
\begin{equation}\label{eq:wageEXP}
\log(\mbox{Wage}_i)=\beta_0+\beta_1 \mbox{male}_i +\beta_2 \mbox{education}_i +\beta_3 \mbox{experience}_i+\beta_4 \mbox{experience}^2_i+\epsilon_i
\end{equation}

\begin{tabular}{@{\extracolsep{5pt}}lD{.}{.}{-3} D{.}{.}{-3} } 
\\[-1.8ex]\hline 
\hline \\[-1.8ex] 
 & \multicolumn{2}{c}{\textit{Dependent variable:}} \\ 
\cline{2-3} 
\\[-1.8ex] & \multicolumn{2}{c}{lWAGE} \\ 
\\[-1.8ex] & \multicolumn{1}{c}{(1)} & \multicolumn{1}{c}{(2)}\\ 
\hline \\[-1.8ex] 
 MALE & 0.256^{***} & 0.257^{***} \\ 
  & (0.039) & (0.039) \\ 
  & & \\ 
 EDUCATION & 0.098^{***} & 0.091^{***} \\ 
  & (0.008) & (0.008) \\ 
  & & \\ 
 EXPERIENCE & 0.013^{***} & 0.036^{***} \\ 
  & (0.002) & (0.005) \\ 
  & & \\ 
 EXPERIENCE2 &  & -0.001^{***} \\ 
  &  & (0.0001) \\ 
  & & \\ 
 Constant & 0.421^{***} & 0.345^{***} \\ 
  & (0.123) & (0.122) \\ 
  & & \\ 
\hline \\[-1.8ex] 
Observations & \multicolumn{1}{c}{533} & \multicolumn{1}{c}{533} \\ 
R$^{2}$ & \multicolumn{1}{c}{0.267} & \multicolumn{1}{c}{0.295} \\ 
Adjusted R$^{2}$ & \multicolumn{1}{c}{0.263} & \multicolumn{1}{c}{0.289} \\ 
Residual Std. Error & \multicolumn{1}{c}{0.453 (df = 529)} & \multicolumn{1}{c}{0.445 (df = 528)} \\ 
F Statistic & \multicolumn{1}{c}{64.367$^{***}$ (df = 3; 529)} & \multicolumn{1}{c}{55.143$^{***}$ (df = 4; 528)} \\ 
\hline 
\hline \\[-1.8ex] 
\textit{Note:}  & \multicolumn{2}{r}{$^{*}$p$<$0.1; $^{**}$p$<$0.05; $^{***}$p$<$0.01} \\ 
\end{tabular} 

\quad

Our t-student test statistics confirm that $\beta_3$ and $\beta_4$ are significant. $\beta_4$ being negative means that the higher the level of a worker's experience, additional experience compare to his peers will have less impact.

We can plot:
\begin{equation}
\log \mbox{Wage difference}=\beta_3 \mbox{experience}+\beta_4 \mbox{experience}^2
\end{equation}

So we plot the wage difference according to the number of years of experience:
\begin{center}
\includegraphics[scale=0.5]{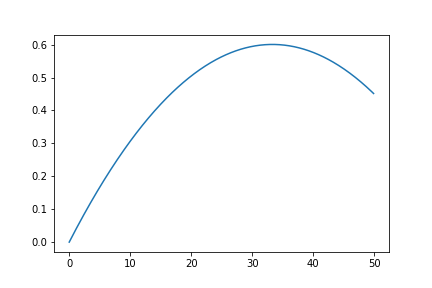}
\end{center}

Experience and its square are highly correlated (96\%), hence it will inflate the estimated coefficients standard errors. We can reduce the correlation by using the square of the demeaned experience (but then the coefficients have to be interpreted differently).

\subsubsection{Does it matter, to add experience squared?}\label{sec:educationsquared}

One might wonder if it matters to have experience squared in the model. If we go back to our regression in its simplest form:
$$y_i =  \beta x_i + u_i$$
If in fact the real data generation process for $y$ was:
$$y_i =  \beta x_i + \gamma x_i^2 + \epsilon_i$$
with $\epsilon \sim \mathcal{WN}(0,\sigma^2)$, then our estimated residuals will be: $u_i = \gamma x_i^2 + \epsilon_i$ and the conditional expectation of $y_i$ knowing $x_i$ is no longer $\beta x_i$.

\subsection{Validating Model Specification - Utilizing the Ramsey RESET Test}

Ensuring a model is correctly specified is pivotal to making reliable inferences. The Ramsey RESET (Regression Equation Specification Error Test) test offers a robust method to validate the functional form of the model by examining the statistical significance of higher-order terms of the fitted values.

\begin{enumerate}
    \item Compute the initial regression model and obtain the fitted values:
    \[
    y_i = \beta_0 + \beta_1 x_{i1} + \beta_2 x_{i2} + \ldots + \beta_k x_{ik} + \epsilon_i
    \]
    Acquire \(\hat{y}_i\) from this model.
    
    \item Construct an auxiliary regression, introducing powers of \(\hat{y}_i\) to the original model:
    \[
    y_i = \alpha_0 + \alpha_1 x_{i1} + \alpha_2 x_{i2} + \ldots + \alpha_k x_{ik} + \delta_2 \hat{y}_i^2 + \ldots + \delta_Q \hat{y}_i^Q + \mu_i
    \]
    Typically, \(Q = 2\) or \(3\) is chosen to avoid overfitting, though consideration of sample size and model intricacy is warranted.
    
    \item Apply a hypothesis test, examining the statistical significance of \(\delta_2, \ldots, \delta_Q\). The null hypothesis stipulates that the model is well-specified, denoted as:
    \[
    H_0: \delta_2 = \ldots = \delta_Q = 0
    \]
    A rejection of the null hypothesis, via statistically significant \(F\)-test or \(t\)-statistics, indicates possible model mis-specification.
\end{enumerate}

In the context of the provided example, suppose we retain $H_0$:

\begin{enumerate}
\item The model $y_i=\beta x_i +\epsilon$ is initially tested, yielding \(\hat{y}_i\) and subsequently, \(\hat{y}_i^2\).
\item The auxiliary model $y_i=\alpha x_i + \delta \hat{y}_i^2 +\mu_i$ is assessed.
\item The coefficient \(\delta\) undergoes a t-test, with a well-specified model indicated by \(\hat{\delta}=0\).
\end{enumerate}

\textit{Note:} Passing the Ramsey RESET test doesn’t confirm correct model specification; rather, it signals no detection of the mis-specification types it's designed to identify. Thus, it should be complemented with additional diagnostic tests and substantiated with theoretical underpinning.

\subsection{One model we might be happy about and how to interpret it}

For instance, the partial derivative of the wage with respect to experience, denoted as its marginal effect, can be expressed as:
\begin{equation*}
\frac{\partial \log \mbox{WAGE}}{\partial \mbox{EDUCATION}} = \beta_3
\end{equation*}

In our model, \(\beta_2\) quantifies the percentage change in the wage for an additional year of education, holding other factors constant. A \(\beta_2\) of \(9.1\%\) suggests that each additional year of education is associated with a \(9.1\%\) increase in the hourly wage, ceteris paribus. This underscores the economic principle that investments in human capital, like education, are correlated with wage increments.

In a log-level model of the form 
\[
\log(y) = \beta_0 + \beta_1X + \epsilon
\]
where $\log(y)$ is the natural logarithm of the dependent variable, $X$ is the independent variable, $\beta_0$ and $\beta_1$ are parameters to be estimated, and $\epsilon$ is the error term, the coefficient $\beta_1$ can be interpreted as the percentage change in $y$ for a one-unit change in $X$, holding all other variables constant. Mathematically, this can be expressed as:
\[
\beta_1 = \frac{\% \Delta y}{\Delta X}
\]

\textbf{Justification:}

Consider a one-unit change in $X$ ($\Delta X = 1$):
\[
\log(y_2) - \log(y_1) = \beta_1(X_2 - X_1) + (\epsilon_2 - \epsilon_1)
\]

Given that the percentage change in $y$ is defined as
\[
\% \Delta y = \frac{y_2 - y_1}{y_1} \times 100\%
\]
and using the logarithmic identity $\log(a) - \log(b) = \log(a/b)$, we rewrite the expression as:
\[
\log\left(\frac{y_2}{y_1}\right) = \beta_1(X_2 - X_1) + (\epsilon_2 - \epsilon_1)
\]

Assuming that $\epsilon_2 - \epsilon_1$ is sufficiently small or averages to zero with large enough samples, and substituting the expression for the percentage change, we get:
\[
\log\left(1 + \frac{y_2 - y_1}{y_1}\right) \approx \beta_1 \Delta X
\]
\[
\log\left(1 + \frac{\% \Delta y}{100}\right) \approx \beta_1 \Delta X
\]

Since $\Delta X = 1$ and for small percentage changes, $\log(1 + \frac{\% \Delta y}{100}) \approx \frac{\% \Delta y}{100}$ (using the approximation that $\log(1+x) \approx x$ for small $x$), we have:
\[
\beta_1 \approx \frac{\% \Delta y}{100}
\]

Thus, $\beta_1$ approximately represents the percent change in $y$ for a one-unit change in $X$.

\subsection{Addressing Heteroskedasticity - Employing the White Test}

Heteroskedasticity, a circumstance where the variance of the errors exhibits a pattern or is not constant across all levels of the explanatory variables, poses substantial concerns in regression analysis. It compromises the efficiency of the Ordinary Least Squares (OLS) estimators and skews statistical tests, potentially leading to incorrect inferences. Therefore, testing and correcting for heteroskedasticity is imperative.

One broadly applied approach to test for heteroskedasticity is the White test, which is a general test not restricted to a specific form of heteroskedasticity, making it aptly suitable when the source of heteroskedasticity is unknown or not directly observable. 

The null hypothesis of the White test is 'homoskedasticity'.

Let's delve into the methodology:

\begin{enumerate}
    \item Begin with estimating the initial model using OLS:
    \[
    y_i = \beta_0 + \beta_1 x_{i1} + \beta_2 x_{i2} + \ldots + \beta_k x_{ik} + \epsilon_i
    \]
    and compute the squared residuals, \(\hat{\epsilon}_i^2\).
    
    \item Construct an auxiliary regression by regressing the squared residuals on the original explanatory variables, their squares, and their cross-products\footnote{With the lm() function, we can use : or *: adds a new column to the design matrix with the product of the other two columns. * will also include the individual columns that were multiplied together.}
    \[
    \hat{\epsilon}_i^2 = \alpha_0 + \alpha_1 x_{i1} + \alpha_2 x_{i2} + \ldots + \alpha_{2k} x_{i1}x_{i2} + \mu_i
    \]
    
    \item Assess the overall significance of the explanatory variables in the auxiliary regression using an \(F\)-test or by comparing the \(N \times R^2\) with a critical value from the \(\chi^2\) distribution with \(p\) degrees of freedom (where \(p\) represents the number of independent variables in the auxiliary regression). 
\end{enumerate}

In the provided example, a White test statistic of \(W = 13.12\) we do not reject the null hypothesis of homoskedasticity at the \(5\%\) significance level.

Addressing heteroskedasticity often involves adjusting the standard errors to ensure they are robust. This adjustment does not alter the OLS coefficients but does provide more reliable standard errors, thereby improving the validity of the statistical inferences. 

If we did reject the null hypothesis of the White test, we would need to apply a correction of White in order to have non-biased estimators. Because so far, the coefficients found are acceptable but not the computed standard deviations. We would need to use the corrected formula to compute the standard deviations. The White robust standard errors\footnote{Has explained in the manual \href{http://thomasleeper.com/Rcourse/Tutorials/olsrobustSEs.html}{here}} can be computed in R as follows:

\begin{lstlisting}
sqrt(diag(vcovHC(model)))
\end{lstlisting}

Resulting in the following robust standard errors alongside the estimated coefficients:

\begin{center}
\begin{tabular}{lcc}
& Coefficient & Robust Std. Error (White) \\
\hline
Intercept & 0.3437 & 0.124 \\
MALE & 0.2570 & 0.039 \\
EDUCATION & 0.0913 & 0.008 \\
EXPERIENCE & 0.0361 & 0.006 \\
EXPERIENCE\(^2\) & -0.0005 & 0.000 \\
\end{tabular}
\end{center}

\subsection{Should we add RACE in our regression model?}

If we try to add RACE as an explanatory variable, we find a t-student coefficient below the threshold of $1.96$ hence we do not reject the null hypothesis that the coefficient for RACE in our model is $0$. Hence we could say that this variable doesn't bring much to our model.

Make a similar exercise with the marital status.

\subsection{Omitted Variable Bias (OVB) Formula}

When we estimate a regression model and omit a relevant variable, the estimated coefficients can be biased. This is called \textbf{Omitted Variable Bias} (OVB).

\paragraph{OVB Direction.}
Suppose the true model is:
\begin{equation}
Y = \beta_0 + \beta_1 X + \beta_2 Z + \epsilon
\end{equation}
but we omit $Z$ and instead estimate the \textit{short} regression:
\begin{equation}
Y = \beta_0^{s} + \beta_1^{s} X + u
\end{equation}

Then the bias on $\hat{\beta}_1^{s}$ is given by:
\begin{equation}
\text{Bias}(\hat{\beta}_1^{s}) = \hat{\beta}_2 \cdot \hat{\delta}_1
\end{equation}
where $\hat{\delta}_1$ is the coefficient in the \textit{auxiliary} regression:
\begin{equation}
Z = \delta_0 + \delta_1 X + u
\end{equation}

The \textbf{sign of the bias} is determined by:
\[
\text{sign of bias} = \text{sign}(\hat{\beta}_2) \times \text{sign}(\hat{\delta}_1)
\]

\paragraph{Application: omitting experience.}
Suppose we only observe gender and education, but not experience. The true model is:
\[
\log(\mbox{Wage}_i) = \beta_0 + \beta_1 \mbox{female}_i + \beta_2 \mbox{education}_i + \beta_3 \mbox{experience}_i + \epsilon_i
\]
If we omit experience:
\begin{itemize}
    \item $\hat{\beta}_3 > 0$: experience has a positive effect on wages.
    \item $\hat{\delta}_1$: the sign depends on how experience correlates with gender and education, conditional on the other regressors.
\end{itemize}
The product $\hat{\beta}_3 \cdot \hat{\delta}_1$ tells us the direction and magnitude of the bias on the coefficients in the short regression. We can verify this numerically: the difference between the short and long regression coefficients should match the OVB formula.

\subsection{Variables interactions}

So far we identified that education matters and being a female matters too. Now, and in a similar vein as when we squared experience, we can interact education with "being a woman" to see if being a women impact significantly the return to schooling:
\begin{equation}
\log(\mbox{Wage}_i)=\beta_0+ \beta_1 \mbox{sex}_i +\beta_2 \mbox{education}_i +\beta_3 \mbox{experience}_i+\beta_4 \mbox{experience}^2_i+ \beta_5 \mbox{sex}_i \times \mbox{education}_i  +\epsilon_i
\end{equation}

\begin{tabular}{@{\extracolsep{5pt}}lD{.}{.}{-3} D{.}{.}{-3} } 
\\[-1.8ex]\hline 
\hline \\[-1.8ex] 
 & \multicolumn{2}{c}{\textit{Dependent variable:}} \\ 
\cline{2-3} 
\\[-1.8ex] & \multicolumn{2}{c}{WAGE} \\ 
\\[-1.8ex] & \multicolumn{1}{c}{(1)} & \multicolumn{1}{c}{(2)}\\ 
\hline \\[-1.8ex] 
 MALE & 0.257^{***} & 0.755^{***} \\ 
  & (0.039) & (0.199) \\ 
  & & \\ 
 EDUCATION & 0.091^{***} & 0.114^{***} \\ 
  & (0.008) & (0.012) \\ 
  & & \\ 
 EXPERIENCE & 0.036^{***} & 0.037^{***} \\ 
  & (0.005) & (0.005) \\ 
  & & \\ 
 EXPERIENCE2 & -0.001^{***} & -0.001^{***} \\ 
  & (0.0001) & (0.0001) \\ 
  & & \\ 
 MALE:EDUCATION &  & -0.038^{**} \\ 
  &  & (0.015) \\ 
  & & \\ 
 Constant & 0.345^{***} & 0.037 \\ 
  & (0.122) & (0.172) \\ 
  & & \\ 
\hline \\[-1.8ex] 
Observations & \multicolumn{1}{c}{533} & \multicolumn{1}{c}{533} \\ 
R$^{2}$ & \multicolumn{1}{c}{0.295} & \multicolumn{1}{c}{0.303} \\ 
Adjusted R$^{2}$ & \multicolumn{1}{c}{0.289} & \multicolumn{1}{c}{0.297} \\ 
Residual Std. Error & \multicolumn{1}{c}{0.445 (df = 528)} & \multicolumn{1}{c}{0.442 (df = 527)} \\ 
F Statistic & \multicolumn{1}{c}{55.143$^{***}$ (df = 4; 528)} & \multicolumn{1}{c}{45.871$^{***}$ (df = 5; 527)} \\ 
\hline 
\hline \\[-1.8ex] 
\textit{Note:}  & \multicolumn{2}{r}{$^{*}$p$<$0.1; $^{**}$p$<$0.05; $^{***}$p$<$0.01} \\ 
\end{tabular} 

$\beta_5$ seems to be significantly different from $0$, thus the slopes of returns to education for men is less than for women, making it "less interesting" for men to study longer than it is for women.

\subsection{Chow Test: Testing for Structural Breaks}

The Chow Test tests for the presence of structural breaks in dataset. Specifically, it helps to verify if the independent variables have different impacts on the dependent variable in two or more groups, or time periods. Mathematically, it tests the null hypothesis that the coefficients in two regression models are equal:

\[
H_0: \beta_{1j} = \beta_{2j} \quad \forall j
\]

Given our regression model:

\[
\mbox{WAGE} = \beta_0 + \beta_1 \times \mbox{MALE} + \beta_2 \times \mbox{EDUCATION} + \beta_3 \times \mbox{EXPERIENCE} + \epsilon
\]

Let's consider that we want to evaluate whether union membership moderates the effects of the independent variables on wages, i.e., whether being a union member changes the impact of sex, education, and experience on wages.

To do this, we run separate regressions for union and non-union members, obtaining separate residual sum of squares, denoted \(SCR_0\) and \(SCR_1\) respectively. The combined regression for all samples gives a residual sum of squares denoted \(SCR\). With \(k\) denoting the number of parameters in the model and \(n\) representing the total sample size, the Chow Test statistic is calculated as:

\begin{equation*}
C = \frac{\frac{SCR - (SCR_0 + SCR_1)}{k}}{\frac{SCR_0 + SCR_1}{n - 2k}}
\end{equation*}

A statistically significant Chow test statistic, compared against the critical value from the F-distribution with \(k\) and \(n-2k\) degrees of freedom, allows us to reject the null hypothesis, suggesting the regression functions for the two groups are distinct.

In our analysis, we compute a Chow statistic of \(4.73\). Given the critical value of \(2.23\) at a \(5\%\) significance level, we reject the null hypothesis, providing sufficient evidence to conclude that the effect of the covariates (sex, education, and experience) on wages is not homogeneous across union and non-union workers. Thus, it indicates that the relationship between the dependent and independent variables is structurally different between the two groups.

\subsection{Fixed Effect}

We can study the region's fixed effect, specifying regions as factor, captured here by $\alpha_j$ for each region $j$:
\begin{multline}
\log(\mbox{Wage}_{i,j})=\beta_0 + \alpha_j + \beta_1 \mbox{education}_i +\beta_2 \mbox{experience}_i+\beta_3 \mbox{experience}^2_i+\beta_4 \mbox{is black}_i\\+\beta_5 \mbox{is metropolitan}_i+\beta_6 \mbox{is southerner}_i+\epsilon_i
\end{multline}

\subsection{To cluster or not to cluster}
\citep{fox2015} writes \begin{quote}
Clustered data commonly arise in two contexts: hierarchical data, in which lower-level units, such as individual students, are nested within higher-level units, such as schools, and longitudinal data, in which individuals (or other multiple units of observation) are followed over time. In both cases, observations within a cluster -- lower-level units within higher-level units or different measurement occasions for the same individual -- cannot reasonably be treated as statistically independent. Mixed-effect models take account of dependencies in hierarchical, longitudinal, and other dependent data.
\end{quote}

In this example, as discussed \href{https://blogs.worldbank.org/impactevaluations/when-should-you-cluster-standard-errors-new-wisdom-econometrics-oracle}{here} as workers were randomly sampled (we did not choose sub-regions then workers within these sub-regions), there is no need to adjust the standard errors for clustering at region or other level.

\section{Addressing Endogeneity: The Role of Instrumental Variables and Hausman Test}

\citet{jiang2017have} details why the OLS estimate might be overestimated:
\begin{quote}
    1. \textbf{Affirmative endogeneity:} \( \text{SignBiasOLS} > 0 \) \\
    In this case, there is a convincing argument that an OLS estimate overestimates the population average treatment effect. The classic example of this is a regression of earnings on years of education. There are two problems with a reduced-form analysis. First, there is the unobserved variable of “ability,” which tends to affect both earnings and education in the same direction. Second, agents who anticipate better career prospects tend to go to school longer and attain a better education. Both the missing variables and the reverse causality lead the OLS estimate to overstate the true effect.
\end{quote}

And later

\begin{quote}
This issue is well recognized in labor economics. For example, although economists generally believe that the correlation between education and earnings overstates the true causal impact of education, published studies routinely present estimates from elaborate identification schemes that are even more dramatic than the simple uninstrumented estimates (see a review and analysis by \citet{Card1999aa}).

Consider the following classical example, featured in \citet{angrist2009mostly}, in which a researcher is analyzing the relationship between earnings and college attendance. The structural model is:
\[
\text{Earnings}_i = b_0 + b_{1,i} \times \text{College}_i + b_2 \times \text{Control}_i + \varepsilon_i \quad (1)
\]
If the conditions \(E(\varepsilon_i|\text{College}_i, \text{Control}_i) = 0\) and \(E(b_{1,i}|\text{College}_i) = E(b_{1,i})\) are satisfied, the OLS regression yields a population average treatment effect, that is, \(\text{plim} \ \hat{b}_{1,OLS} = E(b_{1,i})\). Most researchers would probably agree that, a priori, it is likely that \(E(\varepsilon_i \times \text{College}_i|\text{Control}_i) > 0\), in other words, people with better earnings prospects based on attributes unobservable to econometricians are more likely to choose to receive more education. Hence, in expectation, \(\hat{b}_{1,OLS} > \hat{b}_{1,IV}\) if the compliers are representative of the population.

Realizing that \(\text{College}_i\) is potentially endogenous (that is, \(E(\text{College}_i \times \varepsilon_i) \neq 0\)), the researcher introduces an instrument \(\text{Proximity}_i\), defined as the distance between person \(i\) and the closest college campus. Suppose we accept that \(\text{Proximity}_i\) is a valid instrument; that is, \(\text{Proximity}_i\) affects the decision to attend the college in a significant way but does not affect \(\text{Earnings}_i\), except indirectly through actual college attendance. For simplicity, let both \(\text{College}_i\) (whether one goes to college or not) and \(\text{Proximity}_i\) (whether there is a college campus within commuting distance or not) be discrete \(\{0, 1\}\) variables. Then \(\hat{b}_{1,IV}\) identifies the average treatment effect of a subpopulation of students who attend college if and only if there is a campus close by.

If a careful execution produces a result indicating that \(\hat{b}_{1,IV} > \hat{b}_{1,OLS}\), then we learn the following: the subgroups whose decisions are affected by the supply-side shock (i.e., campus proximity) were constrained by the marginal cost of schooling, rather than by the lack of either desire or ability to benefit from education. Hence, the return to education for this subgroup could be substantially higher than the subpopulation that has no interest in attending college, even if there is a campus on the next block. The return to education for these compliers may also be higher than the return to the subpopulation of people who go to college regardless of distance because they can afford it. As such, the local treatment effect on the margin for the IV-compliers could exceed that of the population average treatment effect by more than the force of affirmative endogeneity embedded in the OLS estimate.
\end{quote}

Instrumental Variable (IV) techniques re pivotal tools in econometric analyses, especially when grappling with endogeneity issues. It is called 'unobserved heterogeneity' as ability affects both left and right-hand-side variables, but we cannot observe it and add it as a right-hand-side variable. Consider a wage model with education as a potentially endogenous variable because of ability. Selecting an apt instrument – one that is correlated with education but uncorrelated with the error term in the wage equation – is crucial for unbiased and consistent coefficient estimates.

\textbf{Commonly Used Instruments in Wage Models:}
\begin{itemize}
    \item \textbf{Parental Education:} Postulates that parents' educational attainment influences an individual's education but not directly their wages.
    \item \textbf{Proximity to Educational Institutions:} Assumes ease of access to educational facilities might influence one's educational attainment without directly impacting their wages.
    \item \textbf{Educational Policy Shifts:} Exploits policy changes as exogenous shocks to determine their influence on educational attainment while presuming they do not directly affect wage determination.
\end{itemize}

The Hausman test, in the context of IV regression, aids in testing whether the differences in the coefficients obtained from OLS and IV estimations are statistically significant, which can be an indicator of endogeneity in the variables of interest.

\begin{equation}
H_0: \text{The OLS estimator is consistent and efficient.}
\end{equation}
\begin{equation}
H_1: \text{The OLS estimator is inconsistent and the IV estimator should be preferred.}
\end{equation}

Remember, the process of validating an instrument fundamentally hinges on two key properties: relevance and validity, ensuring that the instrument is not only correlated with the endogenous variable but also does not have a direct causal relationship with the dependent variable.

\subsection{Endogeneity and Instrumental Variables}\label{sec:endoVerbeek}

We follow\footnote{instrument\_variable\_vansteenberghe.R} the example suggest in the book \citep{verbeek} whose data sets and variable details are available \href{http://eu.wiley.com/legacy/wileychi/verbeek2ed/datasets.html}{here}:
\begin{quote}
data on 3010 men taken from the US National Longitudinal Survey of Young Men, also employed in Card (1995). In this panel survey, a group of individuals is followed since 1966 when they were aged 14-24, and interviewed in a number of consecutive years. The labour market information that we use covers 1976. In this year, the average years of schooling in this sample is somewhat more than 13 years, with a maximum of 18. Average experience in 1976, when this group of men was between 24 and 34 years old, is 8.86 years, while the average hourly raw wage is USD5.77.
\end{quote}

\quad

It can be the case that some of the explanatory variables $x_2$ are correlated with the model's error terms, i.e. $y=\beta x +\epsilon$ with $x_t = [x_{1,t},x_{2,t}]$ and we do not respect $E(\epsilon_t x_{2,t})=0$. This violates the hypotheses we have taken so far and we can have biais and inconsistency in our model estimates.

An example for such a correlation: if we regress wages on schooling duration, the error term $\epsilon$ encompasses all other factors that influence wages. But it can be that a person's ability, let's write it $u$ as in \citet{verbeek}, influences both $x_2$, the schooling duration, and $y$ the wage. If we cannot observe $u$, then it 'remains' in the error term of our regression and if we test the model $y = \alpha +  \phi \mbox{controls} + \beta x_2 + \epsilon$, then $\epsilon = \gamma u + \nu$ and we are left to guess the sign of $\gamma$.

Our estimates $\hat{\beta}$ is not the sensitivity of $y$ on $x$ any longer, it is biased $\hat{\beta} = \beta + \frac{d \epsilon}{d x_2}$ as:
$$\frac{d y}{d x_2} = \beta + \frac{d \epsilon}{d x_2}$$

\quad

The simplest example we can think of is given in the book \citep{verbeek} with the Keynesian Model where we can test this model for real consumption per capita regressed on real per capita income:
\begin{equation}
	C_t = \beta_1 + \beta_2 Y_t + \epsilon_t
\end{equation}

we know that there is another structural equation to this model, taking real per capita investmen, $I_t$:
\begin{equation}
Y_t=C_t+I_t
\end{equation}

$Y_t$ and $C_t$ are endogenous variables, which are simultaneously determined in this model. As $C_t$ influences $Y_t$, we cannot assume that $E(\epsilon_t Y_t)=0$ and our $\hat{\beta_{2}}$ will be biased (it can be demonstrated that it will be \textit{overestimated}).

\stepcounter{question}
\begin{itemize}
\item \textbf{Question \arabic{question}}
\begin{itemize}
\item In the Mincer model and data set, is "being married" playing a role in the wage (variable MARR in the data set)? If yes, then  you should consider that this is due to unobservable characteristics of married and unmarried people (suggest some of those characteristics).
\end{itemize}
\end{itemize}

\subsubsection{Instrumental Variable}

An instrumental variable $z_2$ is a variable that can be assumed to be uncorrelated with the models error $\epsilon$ but correlated with the endogenous explanatory variable $x_2$. It must be that $z_2$ is not a regressor candidate for $y$. In our wage / schooling example, the proximity with a school $z_2$ can be expected to influence "directly" $x_2$ the education level while not necessarily $y$ (nor the other factors as "ability"), and is thus a good instrumental variable candidate suggested in \citep{Card:1993aa}. This is the case if $z_2$ is not correlated with $\epsilon$ and if $z_2$ is strongly correlated with $x_2$, hypotheses that will have to be confirmed.

The original model can then be estimated and, under certain conditions detailed in \citep{verbeek}, $\hat{\beta}_{IV}$ will be consistent and asymptotically normal:
\begin{equation}\label{eq:IVformula}
\hat{\beta}_{IV} = (z' x)^{-1} z' y
\end{equation}
where $z = [x_1, z_2]$

If we replace $z_2$ in this expression by $x_2$ we find the OLS estimator (which should not be used here as $E\left( x_2 \epsilon\right)\neq 0$). A necessary condition for the instrumental variables estimator in equation \ref{eq:IVformula} to be consistent and asymptotically normal is that the instrument variable $z_2$ is correlated with the endogenous variable $x_2$ and not a linear combination of elements in $x_1$. If the correlation is weak, then we will observe an "inflation" of the estimator variance.

In practice, using R command ivreg(), you should include endogenous variables ($x_2$) only before the vertical line of R's ivreg() command; include instruments $z_2$ only after the vertical line; finally include exogenous variables that are not instruments ($x_1$) both before and after the vertical line. The ivreg() command is implementing a regression by two-stages least squares. With this method, you can have more (but never less) instrument variables than endogenous variables. First it runs:
\begin{equation}
x_2 = z_2 \pi + \nu_2 
\end{equation} 
and then uses the estimated $\hat{x}_2$ to estimate:
\begin{equation}
\hat{\beta}_{IV} = (\hat{x}' \hat{x})^{-1} \hat{x}' y
\end{equation}

\subsection{Interpretation of the different stages}

\subsubsection{First Stage}
In the first stage of IV, we regress the endogenous variable on the instrument(s) and any other exogenous regressors. This stage helps us to determine whether the instrument is relevant, i.e., it actually correlates with the endogenous regressor.

\begin{align*}
X &= \pi_0 + \pi_1 Z + \pi_2 W + \nu
\end{align*}

Where:
\begin{itemize}
\item $X$ is the endogenous explanatory variable.
\item $Z$ is the instrumental variable.
\item $W$ represents other exogenous regressors.
\item $\nu$ is the error term.
\end{itemize}

We should expect to see a statistically significant coefficient on $Z$ to confirm its relevance.

\subsubsection{Reduced Form}
Before proceeding to the second stage, we estimate the reduced form of the equation, where the dependent variable is regressed on the instrumental variable(s) and other exogenous variables. The reduced form equation helps us understand how the instruments are related to the dependent variable.

\begin{align*}
Y &= \gamma_0 + \gamma_1 Z + \gamma_2 W + \eta
\end{align*}

Where:
\begin{itemize}
\item $Y$ is the dependent variable.
\item $Z$ is the instrumental variable.
\item $W$ represents other exogenous regressors.
\item $\eta$ is the error term.
\end{itemize}

The significance of $\gamma_1$ indicates the instrument's explanatory power for variations in $Y$ and is crucial for assessing the strength of the instrument.

\citet{verbeek}:
\begin{quote}
The usual rule of thumb is that an instrumental variable should have an F-statistic in the reduced form larger than 10, corresponding to a t-ratio exceeding 3.16.
\end{quote}

\subsubsection{Durbin–Wu–Hausman Test}

A computationally attractive version of the Hausman test for endogeneity is the Durbin–Wu–Hausman test, which uses a simple auxiliary regression. First, use the residuals from the first stage \( \hat{v} \) and include these residuals in the structural equation:

\begin{align*}
Y &= \gamma_0 + \gamma_1 X + \gamma_2 W + \hat{v}_i \gamma + e
\end{align*}

If \( \gamma = 0 \), then \( X \) is exogenous. This can be tested using a simple t-test on \( \gamma = 0 \). Note that this requires the instrument \( Z \) to be valid and uncorrelated with the error term \( e \).

\subsubsection{Second Stage}
In the second stage, we take the predicted values from the first stage, $\hat{X}$, and use them to estimate the response variable $Y$.

\begin{align*}
Y &= \alpha_0 + \alpha_1 \hat{X} + \alpha_2 W + \epsilon
\end{align*}

The coefficient $\alpha_1$ on the predicted value $\hat{X}$ is the IV estimate of the effect of $X$ on $Y$. It is important to test for overidentifying restrictions to ensure the exogeneity of the instrument.

\subsubsection{Comparing OLS with IV}
When comparing OLS with IV estimates, we are interested in the consistency and unbiasedness of the estimators. If the OLS estimates are biased due to endogeneity, the IV estimates will differ. The magnitude and direction of the bias depend on the nature of the endogeneity.

Consider the classical example of estimating the returns to education on earnings. The OLS regression may suffer from endogeneity if there are unobserved factors that affect both education and earnings.

\begin{align*}
\text{Earnings} &= \beta_0 + \beta_1 \text{Education} + \beta_2 \text{Experience} + \beta_3 \text{Experience}^2 + u
\end{align*}

Assume that the ability is unobserved and affects both earnings and education. An instrument for education might be the proximity to a college when the individual was 18, under the assumption that it affects education but not earnings directly, except through education.

Below is an example of what the regression table may look like:

\begin{center}
\begin{tabular}{l D{.}{.}{-3} D{.}{.}{-3} }
\toprule
 & \multicolumn{1}{c}{OLS} & \multicolumn{1}{c}{IV} \\
\midrule
Intercept & 4.734^{***} & 4.066^{***} \\
 & (0.068) & (0.608) \\
Education (ed76) & 0.074^{***} & 0.133^{***} \\
 & (0.004) & (0.051) \\
Experience (exp76) & 0.084^{***} & 0.056^{**} \\
 & (0.007) & (0.026) \\
Experience Squared (exp762) & -0.002^{***} & -0.001 \\
 & (0.0003) & (0.001) \\
\bottomrule
\end{tabular}
\end{center}

\textit{Note:} $^{*}$p$<$0.1; $^{**}$p$<$0.05; $^{***}$p$<$0.01

\emph{* p<0.1; ** p<0.05; *** p<0.01}

The IV coefficient on education is larger than the OLS estimate.
First, keep in mind the low correlation between the endogenous variable and the instrument; our estimates is therefore expected to be unprecise.
 \citet{jiang2017have} details that this can also occur because IV identifies the return to education for individuals whose college attendance is influenced by proximity to a campus. These 'compliers' tend to have higher returns, reflecting constraints on access rather than ability or motivation, unlike those unaffected by proximity. Consequently, the local average treatment effect (LATE) identified by IV can exceed the population average treatment effect (ATE) captured by OLS.

\subsection{Estimating the Returns to Schooling with an IV}

People with higher education show higher wages, but is this due to their level of education or unobserved characteristics (more efficient people select longer education, but would have got higher wages anyway because they are more efficient)?

\quad

We import the data from \verb|schooling.dta| using the package foreign. As described in \citep{verbeek}, it contains data taken from the National Longitudinal Survey of Young Men (NLSYM) concerning the United States. The analysis focuses on 1976 but uses some variables that date back to earlier years.

\quad

We test the following model (the use of 3 dummy variables were chosen by the book's author after iterations to find variable that significantly affect the education and experience coefficients):
\begin{multline}
\log(\mbox{Wage}_i)=\beta_0+\beta_1 \mbox{education}_i +\beta_2 \mbox{experience}_i+\beta_3 \mbox{experience}^2_i+\beta_4 \mbox{is black}_i\\+\beta_5 \mbox{is metropolitan}_i+\beta_6 \mbox{is southerner}_i+\epsilon_i
\end{multline}

The return to schooling, ceteris paribus and according to our model, would be: $\hat{\beta}_1 \simeq 7\%$.

\quad

We suspect schooling to be endogenous, as age is not an exogenous choice we can assume experience and experience squared to also be endogenous as they are directly linke to age and age squared (the last two are then potential candidates). \citep{verbeek} suggests to instrument schooling variables with a dummy variable "lived near college" (we take "grew up near 4-yr college" from the data set). The latter, we could argue, impact mostly the schooling choice, but not so much the earning (but one can still argue that there is still endogeneity a family that chose to live near a college might stimulate more its children, etc.). We first test that "lived near college" is indeed related to the schooling choice:
\begin{multline}
\mbox{education}_i = \alpha_0+\alpha_1 \mbox{lived near college}_i + \alpha_2 \mbox{age}_i+\alpha_3 \mbox{age}^2_i\\+\alpha_4 \mbox{is black}_i+\alpha_5 \mbox{is metropolitan}_i+\alpha_6 \mbox{is southerner}_i+\epsilon_i
\end{multline}

We confirm that people that lived near a college attended on average $\alpha_1 \simeq 0.35$ more years at school.

\quad
We want, as in \citep{verbeek} use "lived near a college", "age" and "age squared" as instrumental variables to instrument education, experience and experience squared and remember that 
\begin{quote}
it is not the case that instruments have a one-to-one correspondence with the endogenous regressors. Implicitly, all instruments are jointly used for all variables.
\end{quote}

Using instrumental variables in our regression, we find an estimator for a returns to schooling $\hat{\beta}_{1,IV} \simeq 13 \%$.

\begin{tabular}{@{\extracolsep{5pt}}lD{.}{.}{-3} D{.}{.}{-3} D{.}{.}{-3} D{.}{.}{-3} } 
\\[-1.8ex]\hline 
\hline \\[-1.8ex] 
 & \multicolumn{4}{c}{\textit{Dependent variable:}} \\ 
\cline{2-5} 
\\[-1.8ex] & \multicolumn{1}{c}{lwage76} & \multicolumn{1}{c}{ed76} & \multicolumn{2}{c}{lwage76} \\ 
 & \multicolumn{1}{c}{OLS} & \multicolumn{1}{c}{First Stage} & \multicolumn{1}{c}{Reduced Form} & \multicolumn{1}{c}{Second Stage} \\ 
\\[-1.8ex] & \multicolumn{1}{c}{(1)} & \multicolumn{1}{c}{(2)} & \multicolumn{1}{c}{(3)} & \multicolumn{1}{c}{(4)}\\ 
\hline \\[-1.8ex] 
 ed76 & 0.074^{***} &  &  & 0.133^{***} \\ 
  & (0.004) &  &  & (0.051) \\ 
  & & & & \\ 
 exp76 & 0.084^{***} &  &  & 0.056^{**} \\ 
  & (0.007) &  &  & (0.026) \\ 
  & & & & \\ 
 exp762 & -0.002^{***} &  &  & -0.001 \\ 
  & (0.0003) &  &  & (0.001) \\ 
  & & & & \\ 
 nearc4 &  & 0.347^{***} & 0.032^{**} &  \\ 
  &  & (0.107) & (0.016) &  \\ 
  & & & & \\ 
 age76 &  & 1.061^{***} & 0.182^{***} &  \\ 
  &  & (0.301) & (0.046) &  \\ 
  & & & & \\ 
 age762 &  & -0.019^{***} & -0.002^{***} &  \\ 
  &  & (0.005) & (0.001) &  \\ 
  & & & & \\ 
 black & -0.190^{***} & -1.468^{***} & -0.239^{***} & -0.103 \\ 
  & (0.018) & (0.115) & (0.018) & (0.077) \\ 
  & & & & \\ 
 smsa76 & 0.161^{***} & 0.835^{***} & 0.186^{***} & 0.108^{**} \\ 
  & (0.016) & (0.109) & (0.017) & (0.050) \\ 
  & & & & \\ 
 south76 & -0.125^{***} & -0.460^{***} & -0.142^{***} & -0.098^{***} \\ 
  & (0.015) & (0.102) & (0.016) & (0.029) \\ 
  & & & & \\ 
 Constant & 4.734^{***} & -1.870 & 3.101^{***} & 4.066^{***} \\ 
  & (0.068) & (4.298) & (0.660) & (0.608) \\ 
  & & & & \\ 
\hline \\[-1.8ex] 
Observations & \multicolumn{1}{c}{3,010} & \multicolumn{1}{c}{3,010} & \multicolumn{1}{c}{3,010} & \multicolumn{1}{c}{3,010} \\ 
R$^{2}$ & \multicolumn{1}{c}{0.291} & \multicolumn{1}{c}{0.119} & \multicolumn{1}{c}{0.244} & \multicolumn{1}{c}{0.176} \\ 
Adjusted R$^{2}$ & \multicolumn{1}{c}{0.289} & \multicolumn{1}{c}{0.117} & \multicolumn{1}{c}{0.242} & \multicolumn{1}{c}{0.175} \\ 
Residual Std. Error (df = 3003) & \multicolumn{1}{c}{0.374} & \multicolumn{1}{c}{2.516} & \multicolumn{1}{c}{0.386} & \multicolumn{1}{c}{0.403} \\ 
F Statistic (df = 6; 3003) & \multicolumn{1}{c}{204.932$^{***}$} & \multicolumn{1}{c}{67.294$^{***}$} & \multicolumn{1}{c}{161.226$^{***}$} &  \\ 
\hline 
\hline \\[-1.8ex] 
\textit{Note:}  & \multicolumn{4}{r}{$^{*}$p$<$0.1; $^{**}$p$<$0.05; $^{***}$p$<$0.01} \\ 
\end{tabular} 

\quad

There is not a perfect correlation between instrumental variables and the endogenous variable, hence we are "inflating" the variance of the estimators. Also note that in the case of the instrumental variable model, the $R^2$ has no interpretation as we have endogeneity (correlation between explanatory variables and error terms).

\section{Python and R: Kernel Density Estimation, a non parametric method}\label{sec:KDE}

\epigraph{La th\'{e}orie des erreurs et la moyenne arithm\'{e}tique entre autres celle qui s'appuie sur l'affirmation que la probabilit\'{e} des \'{e}carts est proportionnelle aux \'{e}carts. Tout le monde y croit cependant, me disait un jour M. Lippmann, car les exp\'{e}rimentateurs s'imaginent que c'est un th\'{e}or\`{e}me de math\'{e}matiques, et les math\'{e}maticiens que c'est un fait exp\'{e}rimental.}{Calcul des probabilit\'{e}s \\ \textit{Henri Poincar\'{e}}}

Nonparametric methods try to attempt an estimation of the density directly from the data, without making any parametric assumptions about the underlying distribution. We present\footnote{code: vansteenberghe\_KDE.py} kernel density estimation (KDE) as a mature tool for data analysis and follow the work of \citep{Gramacki2018}.

\subsection{Definition}
If we observe realizations $X_1, X_2, \ldots, X_n$ independent and all drawn from the same distribution with unknown density $f$, we want to estimate the shape of this density with the kernel estimator:
\begin{equation}
\hat{f}_h(x) = \frac{1}{nh}\sum_{i=1}^n K\left(\frac{x-X_i}{h} \right)
\end{equation}
where $K$ is the kernel and $h$ > 0 is a smoothing parameter called the bandwidth. 

\subsubsection{Kernel requirements}

The kernel $K$ to be used for a KDE must satisfy the following requirements:
\begin{equation}
    \begin{cases}
\int K(x) dx = 1 \\
 \int x K(x) dx = 0 \\
  \int x^2 K(x) dx < \infty \\
  K(x) \geq 0, \quad \forall x \\
  K(-x) = K(x)
    \end{cases}
\end{equation}

\subsection{Example: univariate normal kernel}

The univariate normal\footnote{this is not to say that we need to have a parametric kernel, one could use Epanechnikov, Uniform, Triangular, etc. kernels for the estimation} kernel is described by:
\begin{equation}
K(u) = \frac{1}{\sqrt{2\pi}} \exp\left(- \frac{u^2}{2} \right)
\end{equation}

The univariate \emph{normal} KDE for a random sample $X_1, X_2, \ldots, X_n$ is given by:
\begin{equation}\label{eq:normalKDE}
\hat{f}(x,h) = 
\begin{cases}
\frac{1}{n} \sum_{i=1}^n \frac{1}{\sqrt{2\pi} h}\exp\left[ - \frac{(x-X_i)^2}{2h^2} \right] \\
\frac{1}{n} \sum_{i=1}^n \mathcal{N}(x; X_i,h^2)
\end{cases}
\end{equation}

We recognize in equation \ref{eq:normalKDE} the normal density we introduced section \ref{sec:nomaldist}. KDE can be understood as a weighted sum of density 'bumps' that are centred at each data point $X_i$. 

\subsection{KDE illustration from a known density}

Let's imagine that as in \citep{Gramacki2018} we have the following true density, a mixture of two Gaussians:
\begin{equation}
f(x) = \frac{7}{10}\mathcal{N}(x; 4, 0.8) + \frac{3}{10} \mathcal{N}(x;6,0.5)
\end{equation}

if we draw $1000$ observations from this density:
\begin{center}
\includegraphics[scale=0.4]{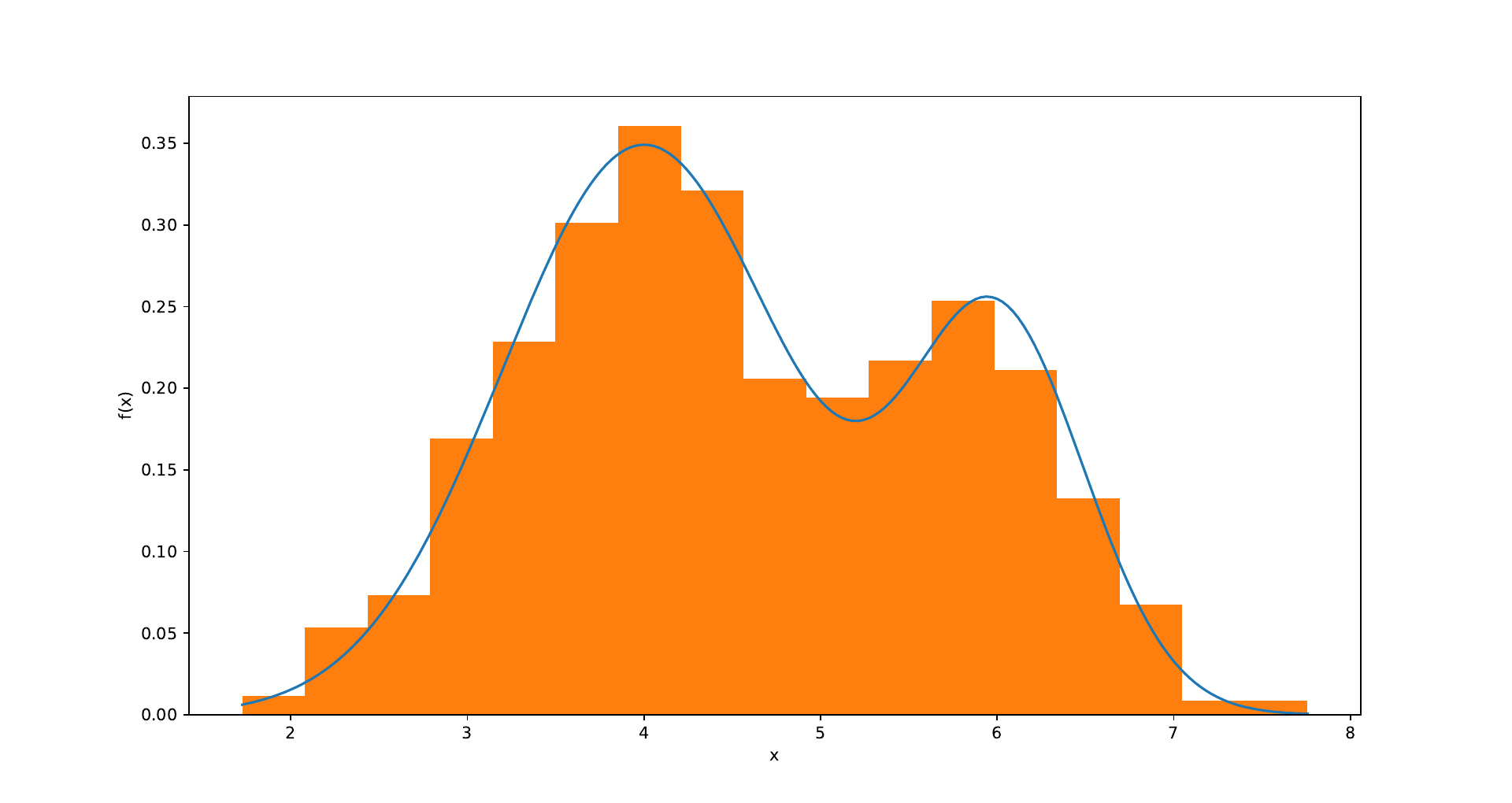}
\end{center}

then if we draw only $10$ "empirical" observations from this mixture of two Gaussians and plot a normal density above each observation, the KDE is the average of those density 'bumps'. An important choice for this is the bandwidth $h$, here we took $h=0.75$ as found using the method prescribed in \citep{Gramacki2018} and available with the R package ks\footnote{code: vansteenberghe\_ks\_hpi.R}:
\begin{center}
\includegraphics[scale=0.4]{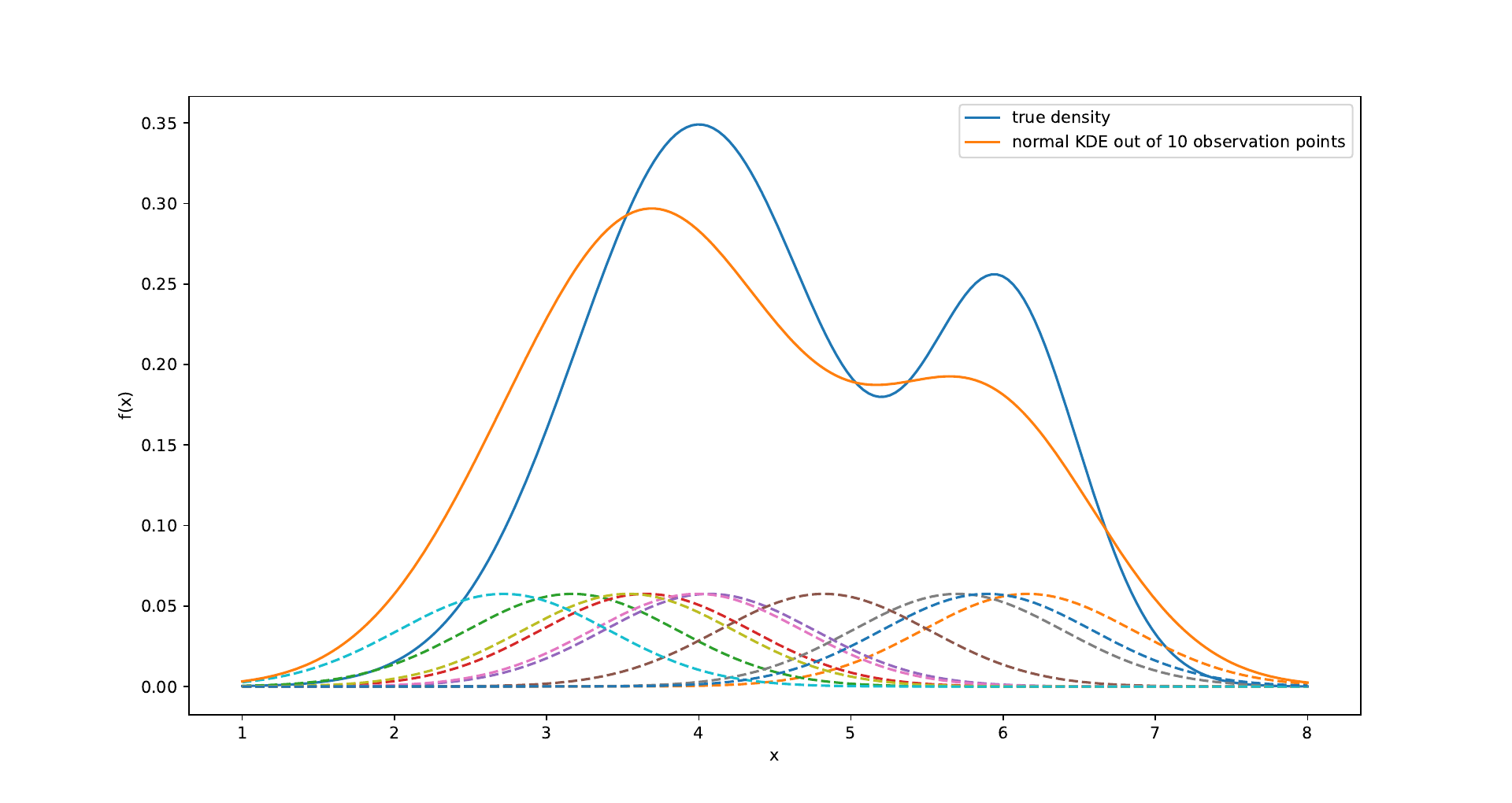}
\end{center}

For the $1000$ observation draw, we found an optimal bandwith of $h=0.22$ using the Plug-in bandwidth selector suggested in \citep{Gramacki2018}. Note that usually, it is not possible to provide a single best optimal bandwidth. The bandwith is important, as an illustration:
\begin{center}
\includegraphics[scale=0.2]{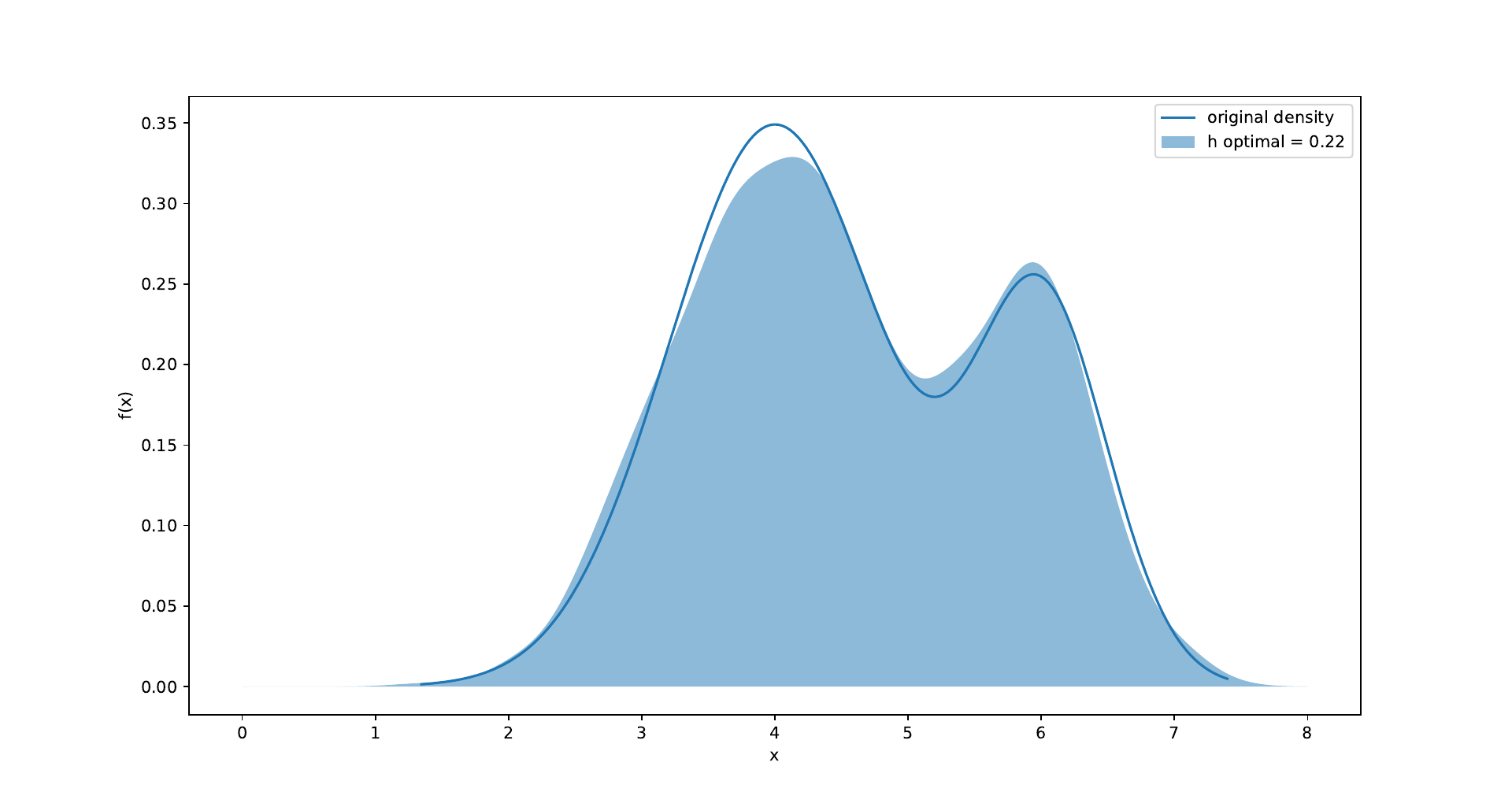}
\includegraphics[scale=0.2]{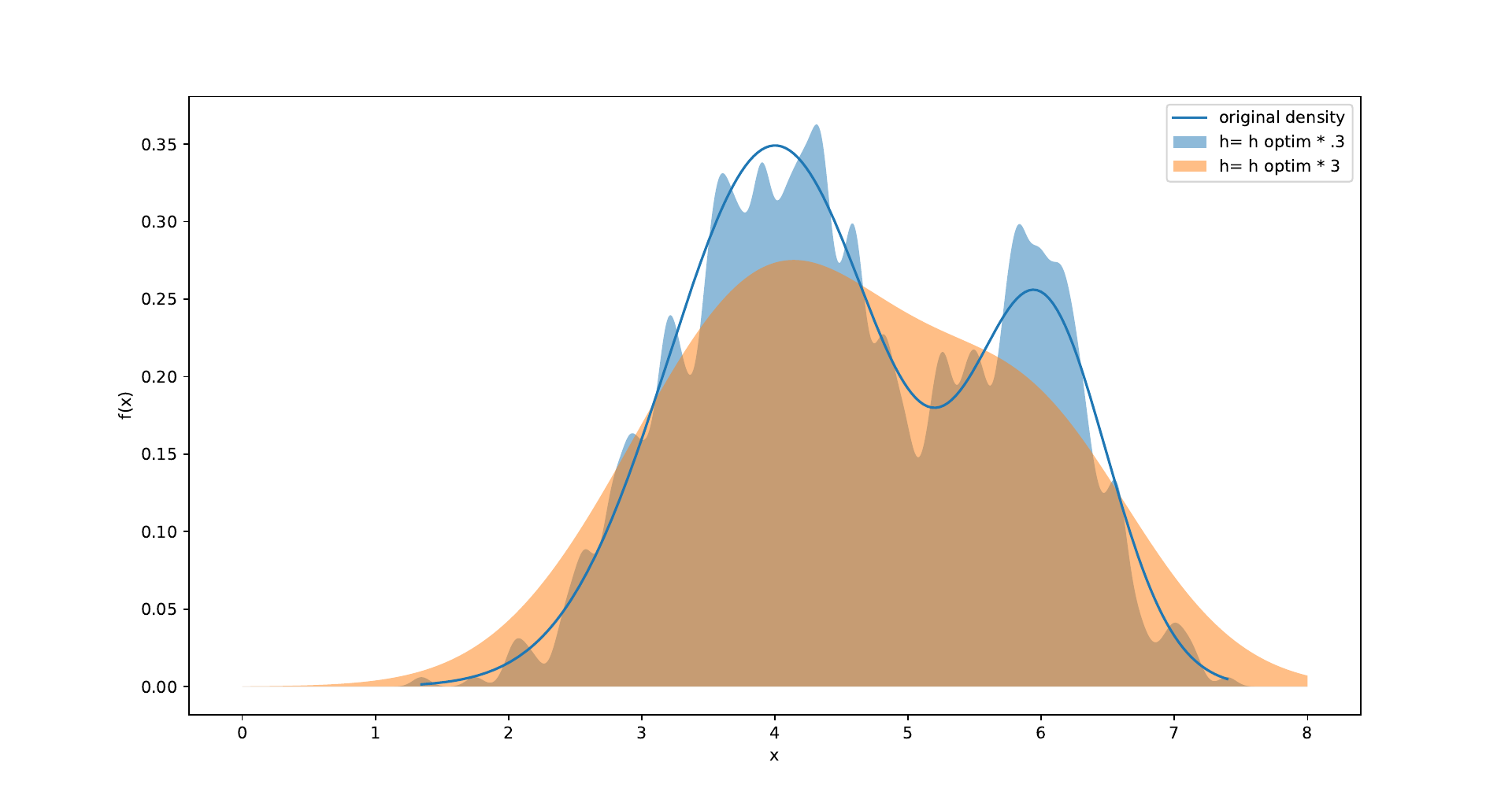}
\end{center}
If we take a much greater or smaller bandwidth than the optimal one, we get a bumpier or a smoother estimation.

\subsection{Bandwidth estimation}

\subsubsection{Univariate rule-of-thumb selectors}

Following \citep{Gramacki2018}, one can demonstrate that if one assume that the unknown density follows a normal distribution, it can be shown that an acceptable bandwidth (rule-of-thumb or Silverman's) is:
\begin{equation}
h_{\mbox{ROT}} \simeq 1.06 \hat{\sigma} n^{-\frac{1}{5}} 
\end{equation}
where $\hat{\sigma}$ is the standard deviation of the sample. If the data set has outliers, then one can use the interquartile range:
\begin{equation}
h_{\mbox{ROT bis}} \simeq 1.06 \min \left(\hat{\sigma}, \frac{\mbox{IQR}}{1.34} \right) n^{-\frac{1}{5}} 
\end{equation}

Sometimes, the factor $1.06$ is replaced with $0.9$.

\subsubsection{Univariate plug-in selector}

We use from the R package fs the hpi() function following \citep{wand1994multivariate}  that selects the optimal bandwidth, the procedure is also detailed in \citep{Gramacki2018}.

\subsection{Kernel selection}
So far we used the gaussian kernel, the Epanechnikov kernel is the most efficient one but \begin{quote}
the normal kernel generates the smoothest density curves and this is the reason why this type of kernel is most often used in practice.
\end{quote}

\begin{center}
\includegraphics[scale=0.2]{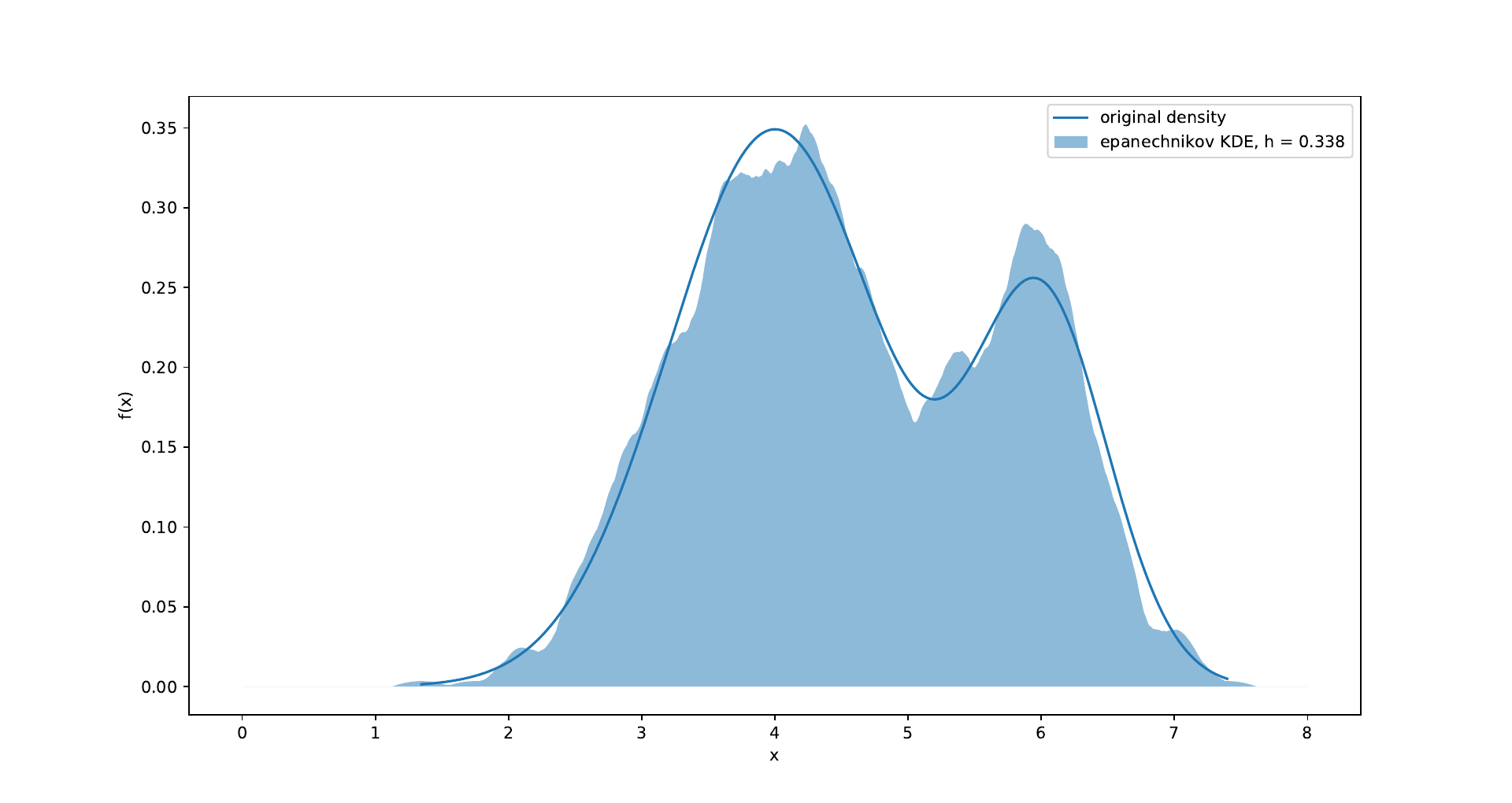}
\end{center}

\subsection{Kernel Efficiency}

Kernel efficiency is a measure of how well a given kernel performs in estimating the underlying density. The efficiency of a kernel can be quantified by comparing the performance of different kernels based on criteria such as bias, variance, and the mean integrated squared error. Kernel efficiency is particularly important because it helps in choosing a kernel that minimizes estimation error and produces a smoother density estimate.

\subsubsection{Mean Integrated Squared Error (MISE)}

The MISE is commonly used to evaluate the performance of a kernel density estimator. It measures the overall discrepancy between the true density $f(x)$ and the estimated density $\hat{f}_h(x)$, and is defined as:

\begin{equation} \text{MISE}(h) = \mathbb{E}\left[ \int \left( \hat{f}_h(x) - f(x) \right)^2 dx \right] \end{equation}

The MISE can be decomposed into two terms: bias and variance.

\subsubsection{Bias and Variance Trade-off}

The kernel estimator introduces both bias and variance. The bandwidth $h$ plays a key role in controlling this trade-off.

Bias: Bias refers to the systematic error introduced by the kernel. It is given by the difference between the expected value of the estimator and the true density: \begin{equation} \text{Bias}(\hat{f}_h(x)) = \mathbb{E}[\hat{f}_h(x)] - f(x) \end{equation}
Variance: Variance refers to the variability of the estimator around its expected value: \begin{equation} \text{Var}(\hat{f}_h(x)) = \mathbb{E}[\left( \hat{f}_h(x) - \mathbb{E}[\hat{f}_h(x)] \right)^2] \end{equation}
The total MISE is a function of both bias and variance: \begin{equation} \text{MISE}(h) \approx \int \left( \text{Bias}^2(\hat{f}_h(x)) + \text{Var}(\hat{f}_h(x)) \right) dx \end{equation}

For small bandwidths $h$, the estimator exhibits low bias but high variance. For large $h$, the estimator has low variance but high bias. The optimal bandwidth $h$ balances these two competing forces to minimize the MISE.

\subsubsection{Relative Efficiency of Kernels}

Different kernels have different levels of efficiency. A measure of relative efficiency is given by the ratio of MISE for different kernels to the MISE of the optimal kernel. It is known that the Epanechnikov kernel is the most efficient in the sense that it minimizes the MISE for a fixed bandwidth. However, other kernels, such as the Gaussian kernel, are often preferred because they produce smoother density estimates despite their slightly lower efficiency.

The relative efficiency $e(K)$ of a kernel $K$ is defined as: \begin{equation} e(K) = \frac{\text{MISE}(K)}{\text{MISE}(\text{Epanechnikov})} \end{equation}

The relative efficiencies of common kernels are as follows: \begin{itemize} \item Epanechnikov kernel: 1.00 (most efficient) \item Biweight kernel: 1.01 \item Triweight kernel: 1.03 \item Gaussian kernel: 1.06 \end{itemize}

\subsubsection{Choosing the Optimal Kernel}

The choice of kernel often depends on the smoothness of the resulting density estimates. While the Epanechnikov kernel is theoretically optimal in terms of MISE, the Gaussian kernel is often used in practice due to its smooth appearance and flexibility. In most cases, the choice of bandwidth $h$ has a greater impact on the quality of the density estimate than the choice of kernel.

\subsection{Kernel Nonparametric Regression}\label{sec:KNR}

As stated in \citet{Blundell1998aa}:
\begin{quote}
The obsession with linearity in empirical economic analysis clearly does not stem from any strong prior of economic theory. [...] Nonparametric regression analysis seems to provide a compelling alternative to linear regression, allowing the data to determine the "local" shape of the conditional mean relationship.
\end{quote}

In nonparametric regression, the assumption of a predetermined functional form is relaxed, allowing the data to "speak for itself." The goal is to estimate the unknown function \(m(x)\) in the model:
\begin{equation}
y = m(x) + \epsilon,
\end{equation}
where \(m(x)\) represents the conditional mean of \(y\) given \(x\), and \(\epsilon\) is the error term with \(\mathbb{E}[\epsilon|x] = 0\). Unlike parametric regression, nonparametric methods make no assumptions about the specific functional form of \(m(x)\), offering considerable flexibility.

\subsubsection{The Nadaraya-Watson Estimator}

One widely used technique in nonparametric regression is the Nadaraya-Watson kernel regression estimator. It estimates the conditional expectation \(m(x) = \mathbb{E}[y|x]\) by weighting the observed \(y_i\) values based on their proximity to the target point \(x\) using a kernel function \(K(\cdot)\). The Nadaraya-Watson estimator is given by:
\begin{equation}
\hat{m}_h(x) = \frac{\sum_{i=1}^n K\left(\frac{x-x_i}{h} \right)y_i}{\sum_{i=1}^n K\left(\frac{x-x_i}{h} \right)},
\end{equation}
where:
\begin{itemize}
    \item \(K(\cdot)\) is the kernel function, which assigns weights based on the distance between the observed data points \(x_i\) and the target point \(x\). Common choices for \(K(\cdot)\) include the Gaussian, Epanechnikov, and uniform kernels.
    \item \(h > 0\) is the bandwidth, a smoothing parameter controlling the kernel's width and the degree of smoothing.
\end{itemize}

The Nadaraya-Watson estimator is a weighted average of the observed \(y_i\) values, with weights determined by the distance between \(x_i\) and \(x\). Observations closer to \(x\) receive higher weights \(K\left(\frac{x-x_i}{h}\right)\), giving them a stronger influence on the estimate \(\hat{m}_h(x)\).

The choice of bandwidth \(h\) is critical:
\begin{itemize}
    \item For small bandwidths (\(h\)), the estimator places most weight on points near \(x\), resulting in less smoothing but potentially noisy estimates.
    \item For large bandwidths, the weights become more uniform, leading to more smoothing but potentially introducing bias if the true relationship is not smooth.
\end{itemize}

The optimal bandwidth minimizes the mean squared error (MSE), balancing bias and variance.

\subsubsection{Bandwidth Selection Methods}

Several methods exist for selecting the bandwidth:
\begin{itemize}
    \item \textbf{Cross-validation:} Leave-one-out cross-validation minimizes the prediction error by iteratively using each data point as a test set.
    \item \textbf{Plug-in Methods:} These methods estimate the optimal bandwidth by assuming specific smoothness properties of the underlying function.
    \item \textbf{Rule-of-Thumb:} Approximations such as Silverman’s rule-of-thumb, adapted from kernel density estimation, provide quick and straightforward estimates.
\end{itemize}

Selecting the appropriate bandwidth is essential for achieving a balance between overfitting (high variance) and oversmoothing (high bias).

\subsubsection{Kernel Choice}

The kernel function $K(\cdot)$ determines the weighting of nearby points in the regression. Common kernels include: \begin{itemize} \item Gaussian Kernel: Produces smooth estimates but may result in over-smoothing in some cases. \item Epanechnikov Kernel: The most efficient kernel in terms of minimizing the mean integrated squared error (MISE). \item Uniform Kernel: Applies equal weights to points within a fixed distance of $x$. \end{itemize}

While the choice of kernel affects the estimate, the bandwidth $h$ generally has a much larger impact on the final result. In practice, most kernels perform similarly, and the focus is usually on selecting an appropriate bandwidth.

\subsection{Introducing the \texttt{np} Library in R}

The \texttt{np} library in R provides tools for kernel nonparametric regression (KNR), allowing flexible estimation without assuming a specific functional form. Using\footnote{code: vansteenberghe\_KNR.R} the \texttt{cps71} dataset, we estimate the relationship between \texttt{log(wage)} and \texttt{age}.

First, we fit a parametric model (\texttt{lm}) as a benchmark. Then, we implement KNR using \texttt{npreg}, which employs local-linear regression (\texttt{regtype = "ll"}) and selects the bandwidth via cross-validation minimizing AIC (\texttt{bwmethod = "cv.aic"}). We enable gradient estimation (\texttt{gradients = TRUE}) to examine the slope of the fitted curve. Finally, we assess the significance of the nonparametric model using \texttt{npsigtest}, which compares the fit against a null model. 

Question: \textbf{are the two approaches leading to very distinct results?}

\subsection{\citet{Blundell1998aa} Replication}

We replicate\footnote{code: vansteenberghew\_KNR\_Blundell1998.R} the analysis in \citet{Blundell1998aa}, using their kernel density estimation approach to study the distribution of log total expenditure. The original dataset is derived from the British Family Expenditure Survey (FES). Figure 1 in \citet{Blundell1998aa} compares the kernel density estimate of log total expenditure with a fitted normal density curve. The kernel density estimate uses a Gaussian kernel, with the bandwidth chosen according to Silverman's rule of thumb. This figure illustrates how the nonparametric estimate captures subtle features in the data that a parametric normal density may overlook, although the two curves appear relatively similar for this dataset. In our replication, we use the same dataset but do not control for the number of children. Below is the replication of Figure 1:

\begin{center}
\includegraphics[scale=0.5]{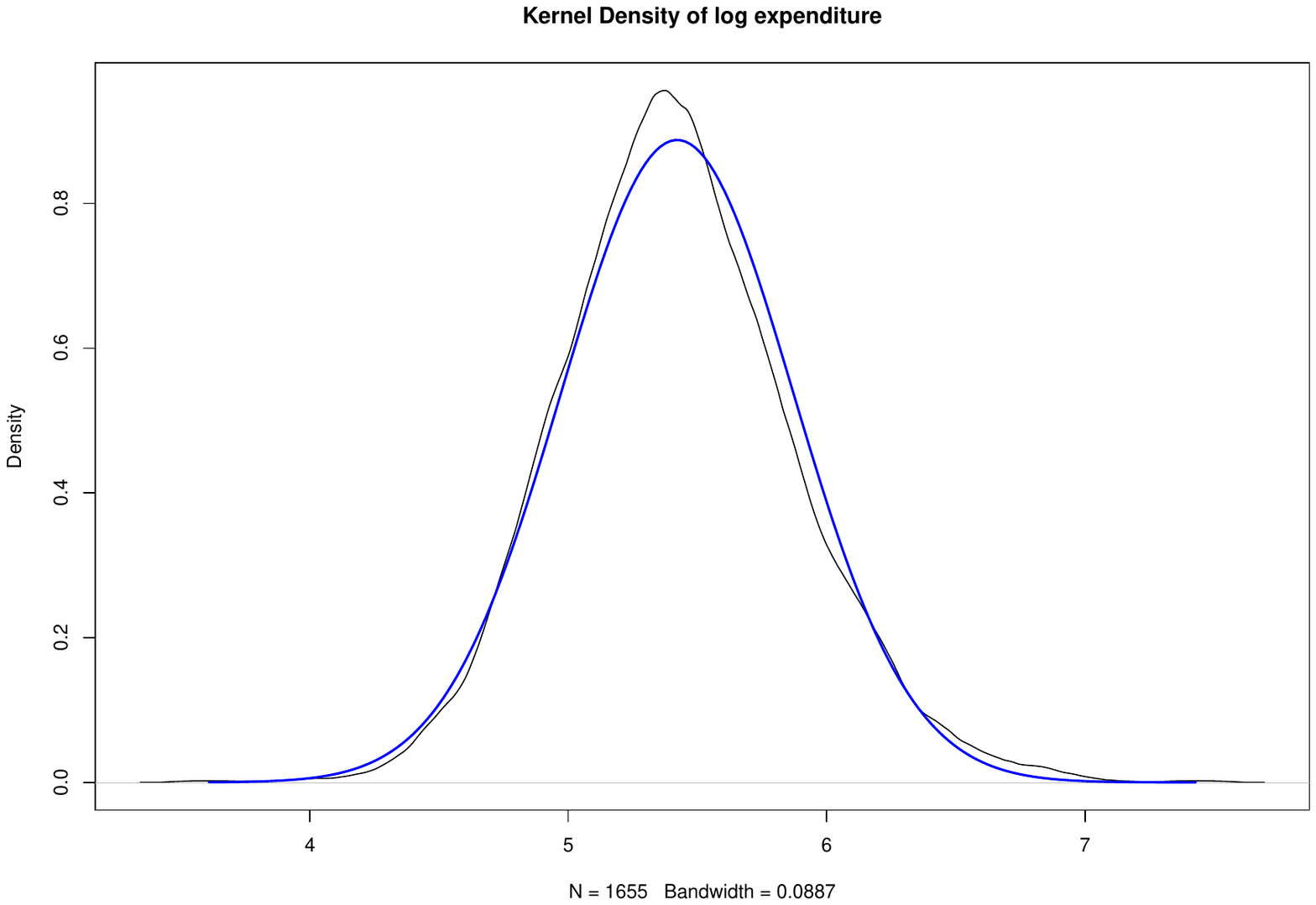}
\end{center}

\noindent The black curve represents the kernel density estimate, while the blue curve is the normal density fitted to the data. The alignment of the two curves suggests that, for this dataset, a normal distribution provides a reasonable approximation to the observed distribution of log total expenditure.

Figure 2 illustrates the bivariate relationship between the logarithm of total expenditure (\texttt{logexp}) and the share of food expenditure (\texttt{food}) for households in the dataset. The scatterplot captures the negative correlation between these two variables, indicating that as total expenditure increases, the share allocated to food decreases. The Pearson correlation coefficient is calculated as \texttt{cor(logexp, food)} to quantify this negative relationship.

We estimate the Engel curve for the share of food expenditure (\texttt{food}) as a function of the logarithm of total expenditure (\texttt{logexp}) using kernel nonparametric regression (KNR).\footnote{just because one writes $x1 + x2$ in the command in no way means or is meant to imply that the model will be linear and additive (why use fully nonparametric methods to estimate such models in the first place?). It simply means that there are, say, two covariates in the model, the first being $x1$ and the second $x2$, we are passing them to a routine with the formula interface, and nothing more is presumed nor implied} To provide a comparison, we also include a parametric linear regression fit. The KNR bandwidth is selected through cross-validation.

The plot below compares the KNR fit (red points with a flexible curve in black) with the linear regression fit (blue line). The KNR method allows for the data to determine the relationship's shape, capturing potential nonlinearities in the Engel curve.

\begin{center}
\includegraphics[scale=0.5]{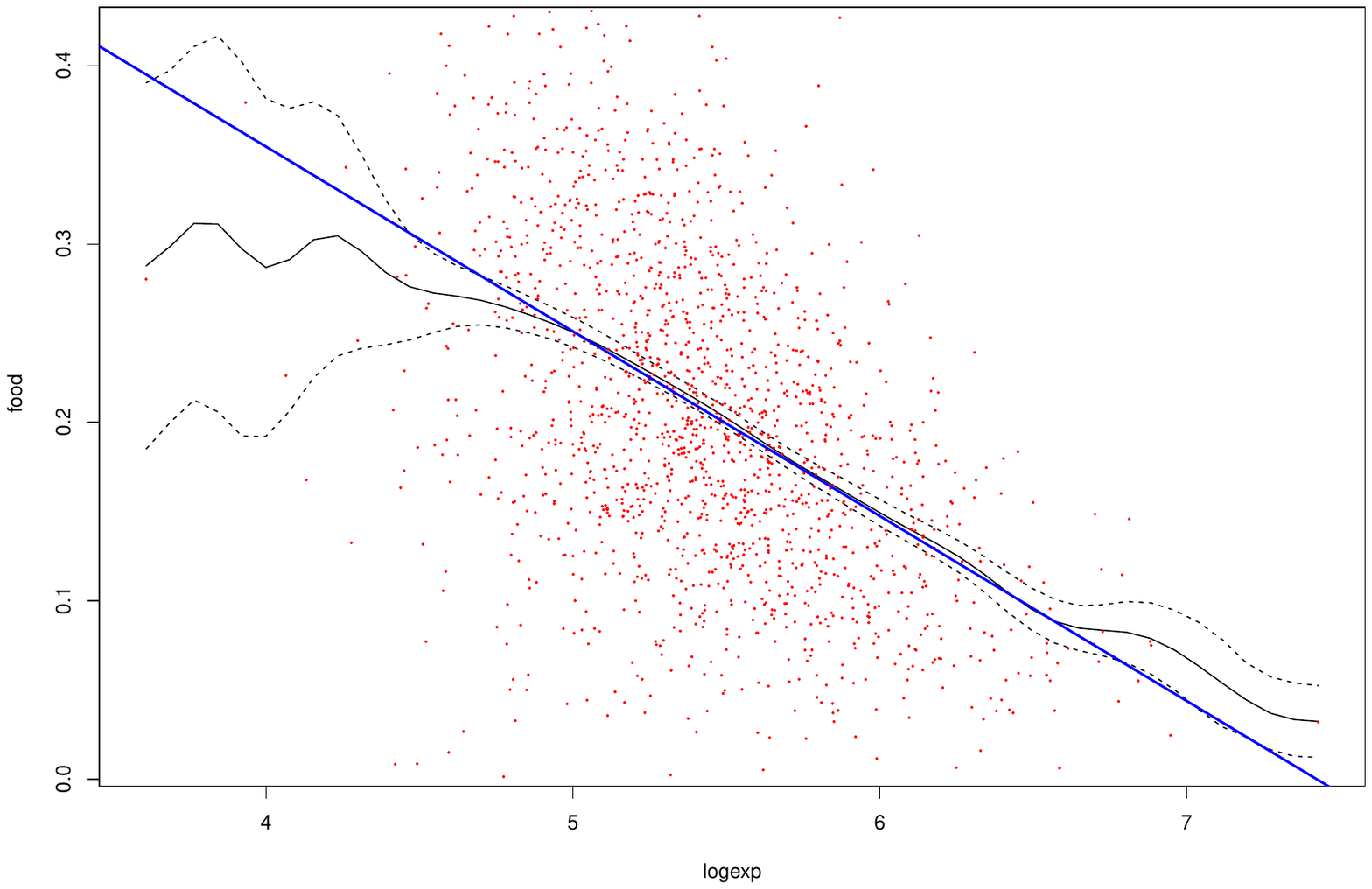}
\end{center}
The KNR results highlight how the relationship deviates from linearity in regions where there is sufficient data density, while the parametric linear regression assumes a fixed slope across all values of \texttt{logexp}.

We compare the slope of the kernel nonparametric regression (KNR) with the slope of the ordinary least squares (OLS) linear regression. While the KNR provides a flexible estimate of the derivative (slope), we observe that it aligns closely with the OLS slope across most of the range of \texttt{logexp}. However, at the extremes (high or low values of \texttt{logexp}), the KNR slope deviates, though the sparsity of observations in these regions limits the statistical confidence of these differences.

The figure below plots the derivative (slope) of the KNR along with the OLS slope (blue horizontal line). The KNR slope captures local variations in the data, while the OLS assumes a constant slope across the entire range.

\begin{center}
\includegraphics[scale=0.5]{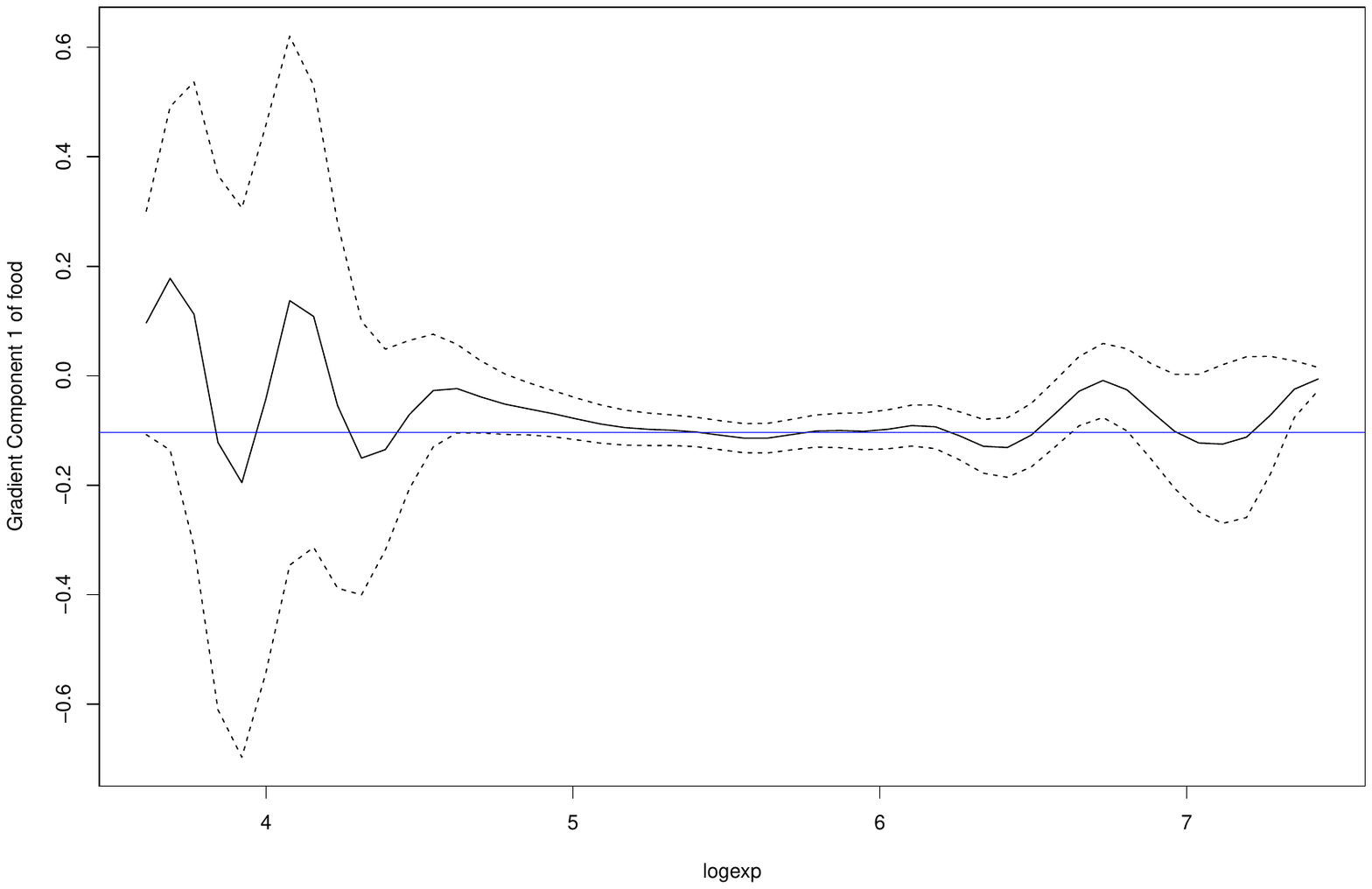}
\end{center}

In this case, the nonparametric regression does not offer clear advantages over the parametric (linear) specification, as the flexibility of KNR does not uncover meaningful deviations from linearity in most of the data range.

\subsubsection{KNR illustration}
For illustration purpose, we suppose that a phenomenon "x" is drawn from a $\mathcal{N}(\mu_x, \sigma_x)$ and the phenomenon "y" observation $i$ is related to "x" observation $i$:
\begin{equation}
y_i = 
\begin{cases}
\alpha_1 + \beta_1 x^\gamma_i + \epsilon_i & \mbox{if } x_i \leq \mu_x \\
\alpha_2 + \beta_2 x^\gamma_i + \epsilon_i & \mbox{if } x_i > \mu_x
\end{cases}
\end{equation}

If we fit a linear specification $y = \alpha + \beta x + \nu$ displayed in blue, this is likely to perform poorly. We fit a KNR displayed in black with bootstrap error in dashed lines:
\begin{center}
\includegraphics[scale=0.5]{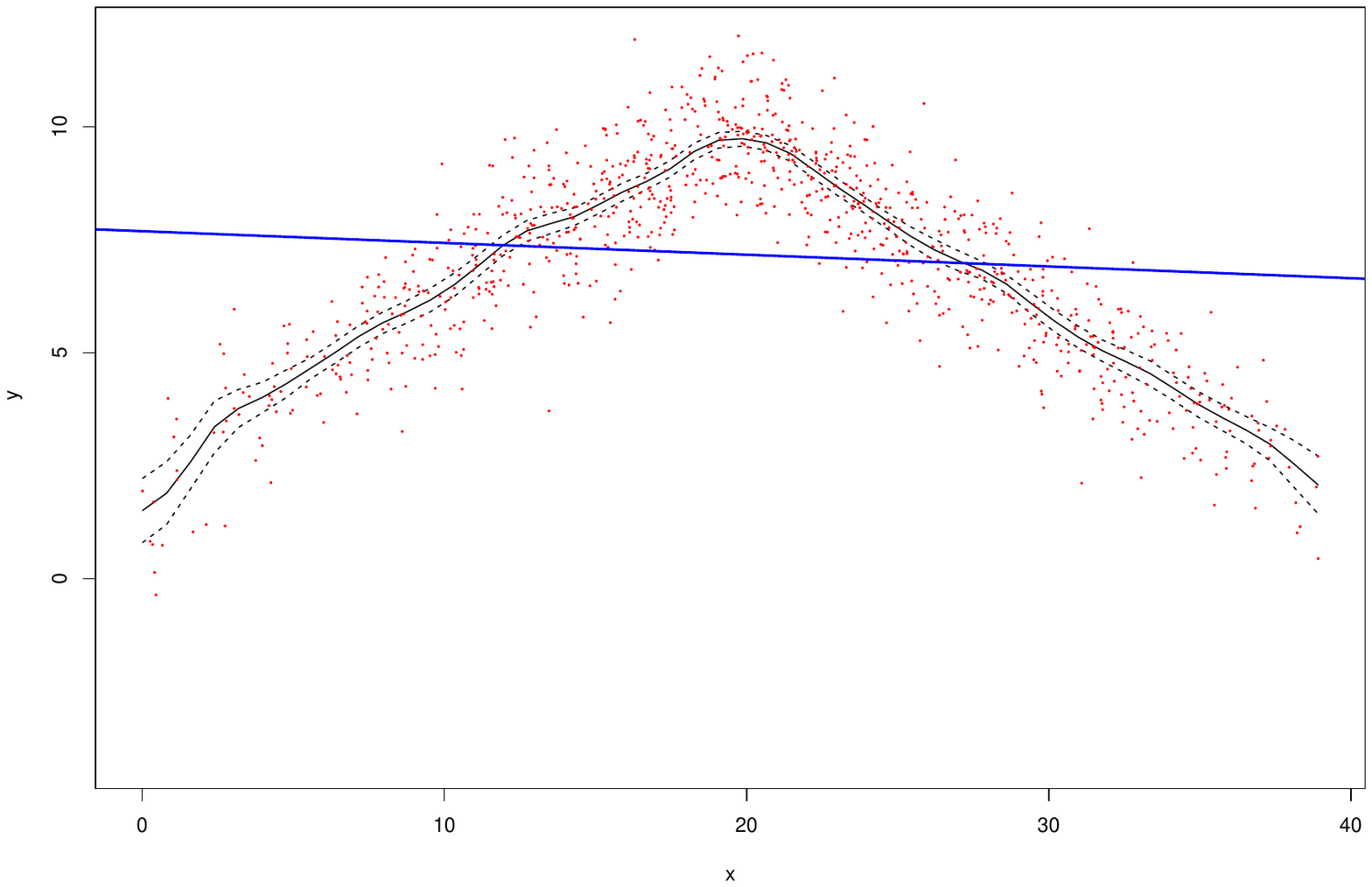}
\end{center}

and indeed, we find that the slope of the KNR (in black) is significantly different than the OLS estimator (in blue):
\begin{center}
\includegraphics[scale=0.5]{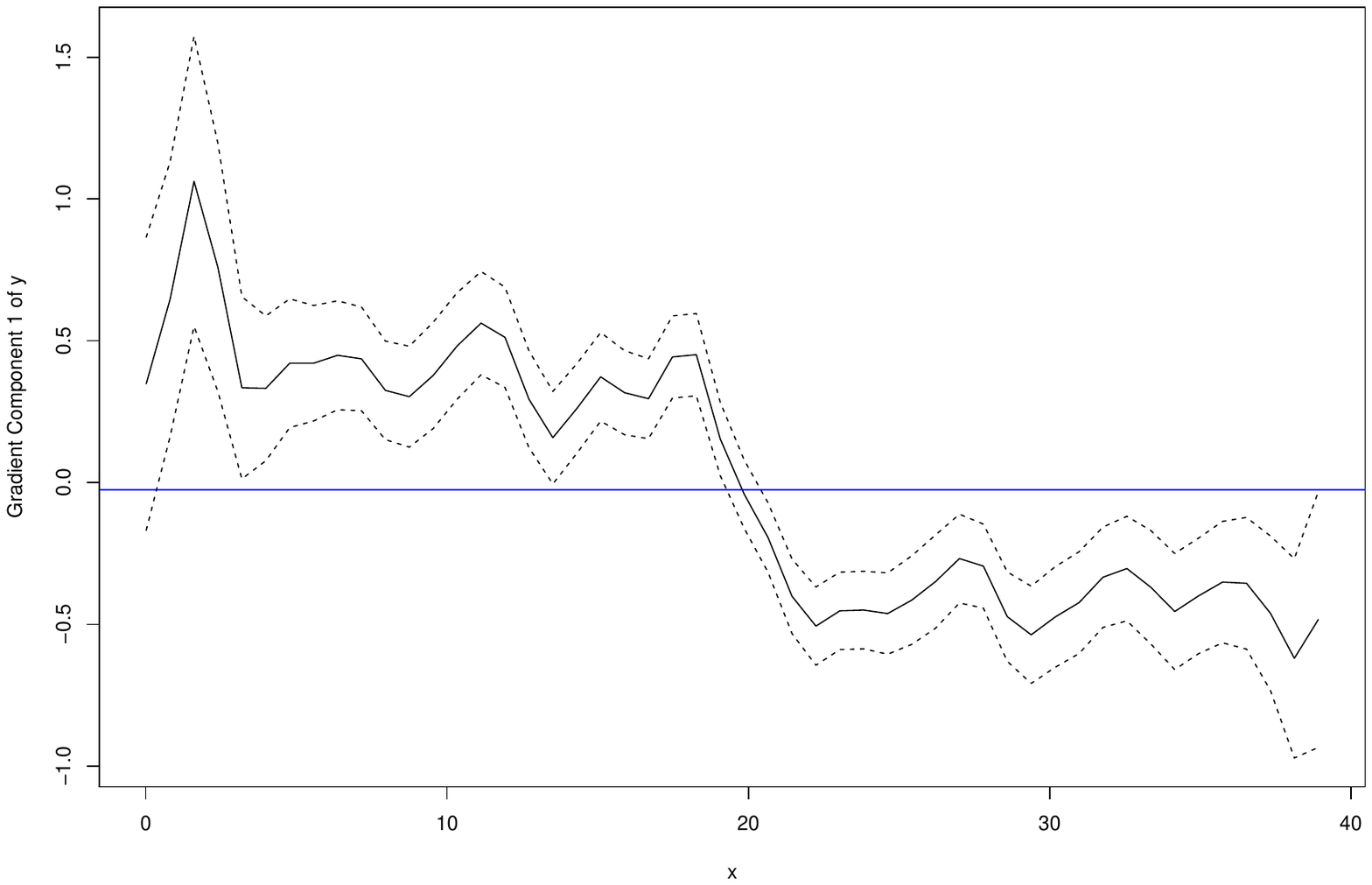}
\end{center}

\section{R and python: multivariate regressions of wages}
\subsection{Nonparametric model}

We detail KNR section \ref{sec:KNRwage} based on \citep{verbeek} data set which is richer. We also want to fit\footnote{multreg\_ML\_vansteenberghe.R} a KNR (nonparametric model $m$) on the Mincer 1974 data set:
\begin{equation}
\log(\mbox{Wage}_i)=m \left(\mbox{male}_i,\mbox{education}_i, \mbox{experience}_i, \mbox{experience}^2_i\right)+\epsilon_i
\end{equation}

The nonparametric model $R^2$ of $32\%$ is better than the parametric model $R^2$ of $30\%$. 

\stepcounter{question}
\begin{itemize}
\item \textbf{Question \arabic{question}}
\begin{itemize}
\item Following \citep{hayfield2007np}, discuss the parametric versus nonparametric models.
\item What is the parametric model conclusion of the effect of education on wage?
\item What is the nonparametric model conclusion of the effect of education on wage?
\end{itemize}
\end{itemize}

\subsection{Kernel nonparametric regression (KNR)}\label{sec:KNRwage}

Following section \ref{sec:KNR} and using the data set as in section \ref{sec:endoVerbeek}, we apply first apply a simple parametric specification $\log(\mbox{wage}) = \alpha + \beta \mbox{educ} +  \epsilon$:

\begin{tabular}{@{\extracolsep{5pt}}lD{.}{.}{-3} } 
\\[-1.8ex]\hline 
\hline \\[-1.8ex] 
 & \multicolumn{1}{c}{\textit{Dependent variable:}} \\ 
\cline{2-2} 
\\[-1.8ex] & \multicolumn{1}{c}{lwage76} \\ 
\hline \\[-1.8ex] 
 ed76 & 0.052^{***} \\ 
  & (0.003) \\ 
  & \\ 
 Constant & 5.571^{***} \\ 
  & (0.039) \\ 
  & \\ 
\hline \\[-1.8ex] 
Observations & \multicolumn{1}{c}{3,010} \\ 
R$^{2}$ & \multicolumn{1}{c}{0.099} \\ 
Adjusted R$^{2}$ & \multicolumn{1}{c}{0.098} \\ 
Residual Std. Error & \multicolumn{1}{c}{0.421 (df = 3008)} \\ 
F Statistic & \multicolumn{1}{c}{329.537$^{***}$ (df = 1; 3008)} \\ 
\hline 
\hline \\[-1.8ex] 
\textit{Note:}  & \multicolumn{1}{r}{$^{*}$p$<$0.1; $^{**}$p$<$0.05; $^{***}$p$<$0.01} \\ 
\end{tabular} 

\quad

And compare to a nonparametric specification, a KNR $\log(\mbox{wage}) = m( \mbox{educ}) +  \epsilon$:
the $R^2$ is then $11.6\%$, which is marginally better than the parametric one ($10\%$ and the variable educ is significant.

We compare the OLS (blue) and KNR (black) regressions and the differences are visually not compelling:

\begin{center}
\includegraphics[scale=0.4]{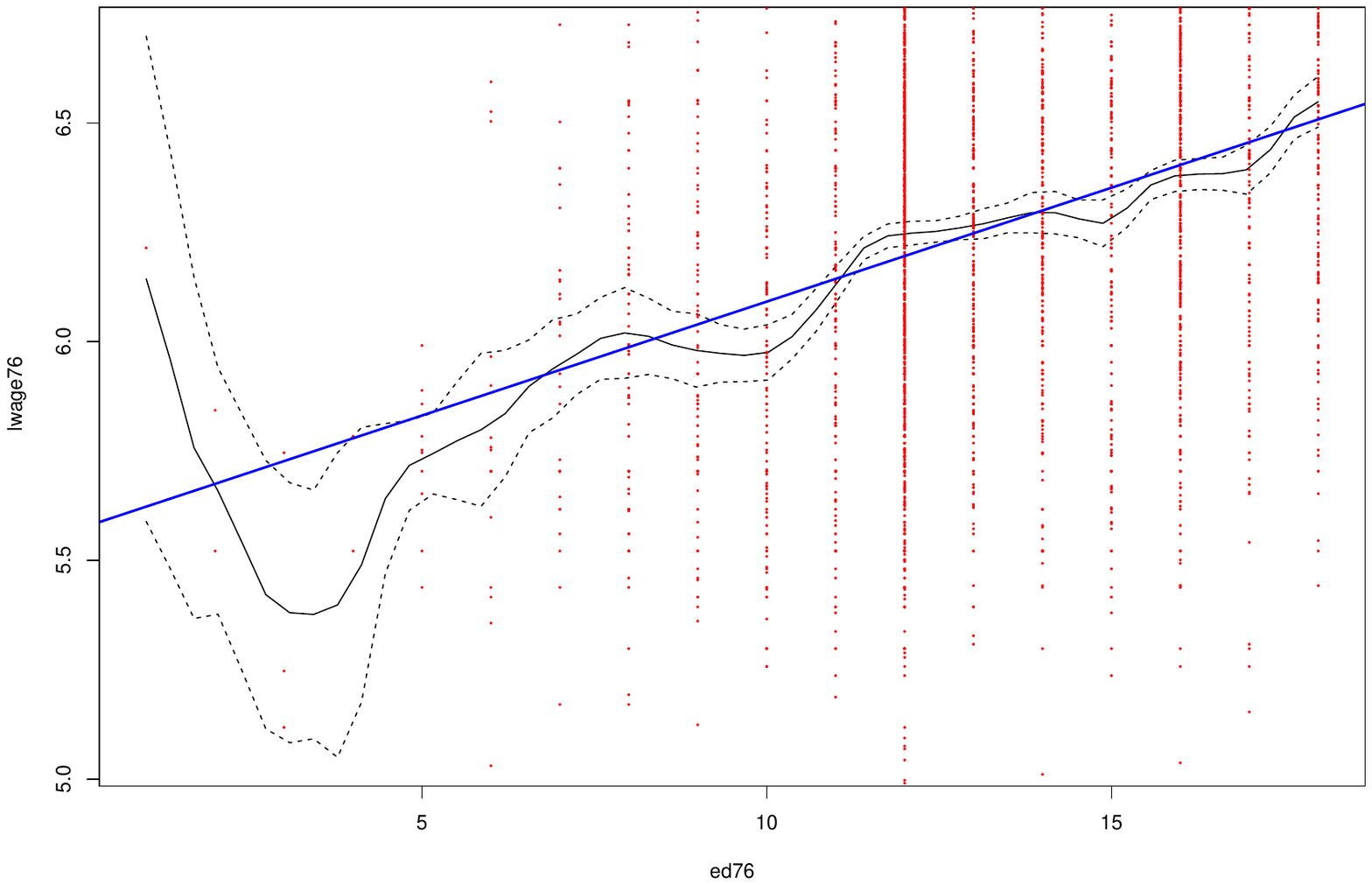}
\end{center}

We would conclude that the local slope of the KNR (in black) is not significantly different form the slope of a simple OLS (in blue):
\begin{center}
\includegraphics[scale=0.4]{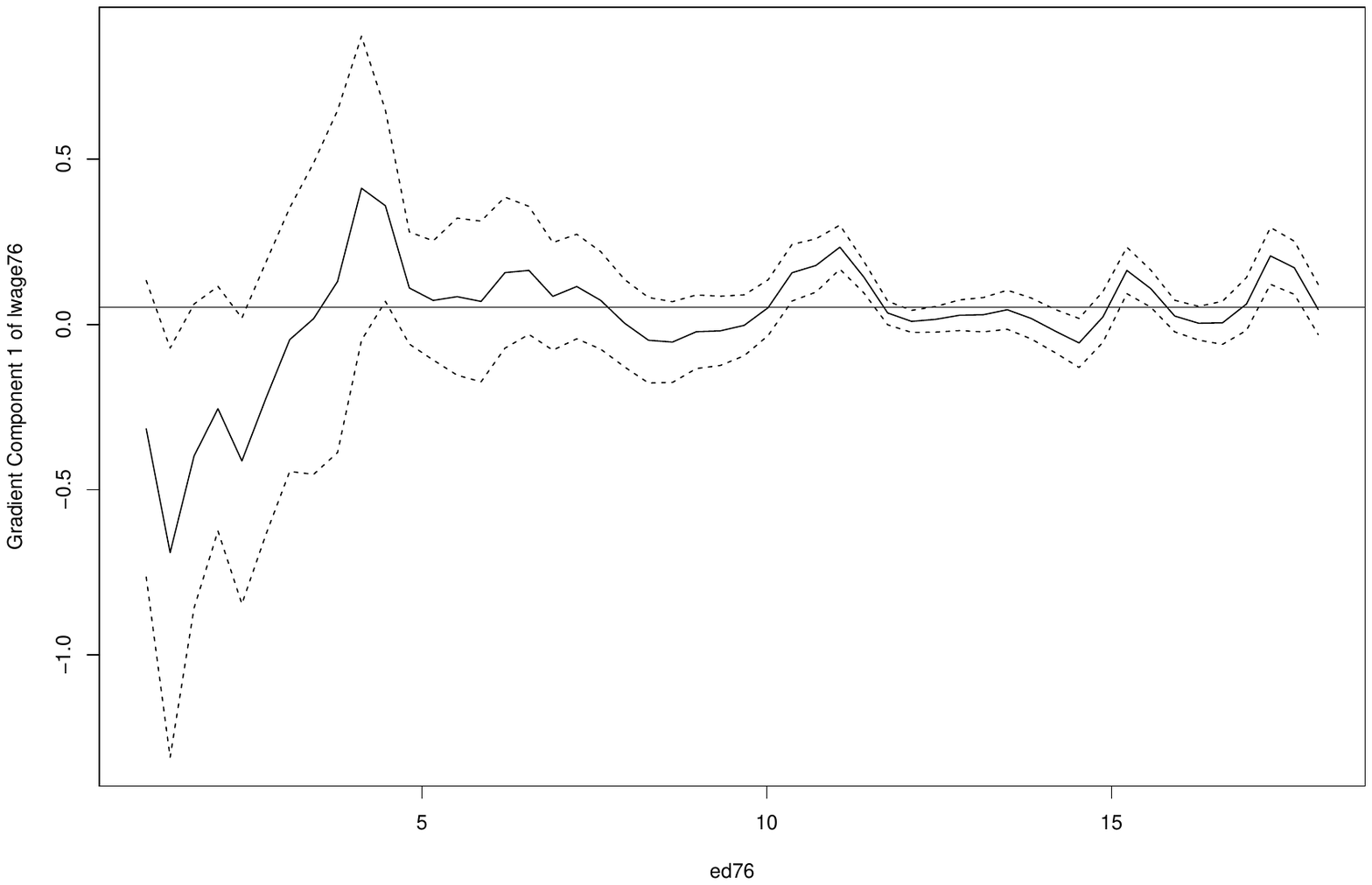}
\end{center}

Although bumpier, we would have the same conclusion for a regression of log wages on age (a proxy for experience).

Here using nonparametric regression analysis would slightly improve our precision but at the expense of then making it harder to make policy conclusion. \citep{hayfield2007np} detail why they believe for our analysis a nonparametric model is more appropriate and as described in \citep{Blundell1998aa}, with KNR we can compute a \textbf{local} slope estimator, but this requires more computing power than our traditional parametric specification.

\subsection{Horse race with machine learning estimators}

Section \ref{sec:MLwhy} we introduce some machine learning estimators and present how to tune their hyper-parameters and fit them to a data set.

I provide a brief illustration of linear and RF regression\footnote{ML\_regressor\_example.py} and classification\footnote{ML\_classification\_expample.py}.

We apply\footnote{code: vansteenberghe\_ML\_wage.py} machine learning estimators and tune their hyper-parameters. We then compare\footnote{using as a score the negative root mean squared error} the performance of these estimators with a very simple linear regression on the data set using three features: SEX, AGE and EDUCATION

\begin{tabular}{lr}
\toprule
{} &  score \\
\midrule
BaggingRegressor          & -0.499 \\
RandomForestRegressor     & -0.498 \\
GradientBoostingRegressor & -0.448 \\
LinearRegression          & -0.443 \\
KNeighborsRegressor       & -0.443 \\
SVR                       & -0.439 \\
MLPRegressor              & -0.439 \\
\bottomrule
\end{tabular}

We obtain different results than \citep{charpentier2018} where they do not tune the hyper-parameters before comparing the models. Nevertheless, our point and the point of \citep{charpentier2018} is to suggest that the linear regression estimates are a good "competitor" plus they present the advantage that elasticity can be interpreted.

\subsection{eXplainable Artificial Intelligence (XAI)}

I strongly recommend reading \citet{arrieta2020explainable}, a literature review on XAI. It tackles the relatively recent question of explainability in ML models.

One can also look into PiML, a Python toolbox for interpretable machine learning model development and validation.

\section{Mutivariate regression on wage: concluding comments}

\subsection{Suggested model}

Based on the Mincer 70's data, we suggest working with the model \ref{eq:wageEXP} that explains $29\%$ of the wage variations.

It could face endogeneity bias, but we could not find an instrument variable to perform an Hausman test in our main data set.

It faces some mutli-collinearity issues but we can accept this, considering the level of collinearity.

Based on the literature, we are also inclined to follow Claudia Goldin's model and use the model \ref{eq:goldinlinke}.

However, we know that the is some endogeneity and with this data set we cannot treat it, so we have a bias in our estimates of the return to schooling.

To go further on the topic of models selection (which variables to include or not), the reader can follow \href{http://statweb.stanford.edu/~jtaylo/courses/stats203/notes/selection.pdf}{this lecture} and \href{https://lagunita.stanford.edu/c4x/HumanitiesSciences/StatLearning/asset/ch6.html}{this page}.

\stepcounter{question}
\begin{itemize}
\item \textbf{Question \arabic{question}}
\begin{itemize}
\item Create a code to test all possible models (with any interaction two by two, so squares included), apply some tests to each and show the results in a LaTeX table. Comment below the table your results (you might need to break your results in several tables). Key words you should at a minimum include: parsimony, multicollinearity, Fisher test, Ramsey, endogeneity.
\end{itemize}
\end{itemize}

\subsection{Recent study around returns to schooling}

\href{https://www.nber.org/papers/w26719?utm_campaign=ntwh&utm_medium=email&utm_source=ntwg16}{THE WAGE PENALTY OF REGIONAL ACCENTS}

Jeffrey Grogger, Andreas Steinmayr and Joachim Winter test:

$$ w_i = \alpha D_i + X_i \beta + \epsilon_i$$

with:

\begin{itemize}
\item $w_i$ the log of hourly wage
\item $D_i$ a dummy for regional accent
\item $X_i$: education, gender, experience, experience squared, and other control variables
\end{itemize}

Even using the omitted variables bias correction (Oster 2019), they find an effect ranging from $-0.108$ to $-0.238$.

As their data follow individuals over time, they cluster at individual level.

\subsection{Discussion: how is gender pay gap still a thing?}

Some extracts from the Financial Times article of 6th August 2020: "How to close the gender pay gap - from the women trying to do it"

\begin{quote}
Across the world, the average woman's annual income is 11,500 USD, compared to 21,500 USD for a man, the WEF estimates. The latest \href{https://data.oecd.org/earnwage/gender-wage-gap.htm}{OECD data} put the UK gender pay gap at 16 per cent, while the US figure stands at 18.5 per cent. What is more, the economic consequences of coronavirus threaten to derail the slow progress that has been made.

[\ldots]

Sheryl Sandberg, chief operating officer of Facebook and author of Lean In, is unequivocal about the issue. “If you fix the pay gap, you would lift three million women out of poverty in the US and you would cut the child poverty rate in half,” she says. When you take into account differences in hours spent at work, experience and occupation, she argues there is still 38 per cent of the pay gap you can’t explain. “It is bias... it is gender,” she says. “There's no other explanation.”

Others call for changes to structural inequalities that have seen women earning less than men for decades - whether by giving companies tax incentives if they can document equal pay, publishing more transparent data or ­changing parental leave laws to make it easier for both men and women to balance careers with family responsibilities.

[\ldots]

Research carried out by LeanIn.Org, the organisation she co-founded, estimates that 60 per cent of male managers are now afraid to engage in common workplace activities with a woman, including one-on-one meetings and mentoring.

[\ldots]

Japan's 25 per cent gender pay gap is the largest in the G7 and the second-largest in the OECD. Kathy Matsui, the Goldman Sachs strategist who coined the term “womenomics” and inspired Prime Minister Shinzo Abe to put female economic empowerment at the top of his agenda, believes the high number of women taking up part-time work is crucial in explaining this.

[\ldots]

Women on the first rung of the corporate ladder are forced by a number of the larger Japanese companies to choose between "career track" (sogo-shoku) and "non-career" (ippan-shoku) positions when they join a company after university.

The former typically offer higher salaries, as well as training and development, but 82 per cent of non-career-track roles are accounted for by women starting their careers, she says. By contrast, the career-track figure for women is just over 20 per cent. This "rigid dual-track system" results in significant gaps in promotion opportunities and wages for women in the long term.

[\ldots]

Chou, a software developer who has worked at Pinterest and Quora, has since become an advocate for fixing the structural bias facing women in tech. When it comes to the gender pay gap in the sector, she suggests it should be illegal for employers to ask job applicants for their current or prior salaries - a common negotiation tactic that companies deploy with new hires.

"Let's say a woman of colour starts off in the industry and has a pretty low salary because of systemic bias. Every future employer who asks for her previous salary will just add on a little bit more on top of that, then the differences with men are compounded over time," she says, during an interview in February in London.

[\ldots]

"So, the people who get a lot of equity will be the founders, investors and early employees, and sometimes executives, who often tend to be male."

She points out that these men frequently use their wealth to start new companies, become investors themselves and take on a lot of risk, "so the ecosystem of inequality perpetuates itself".

[\ldots]

The solution, she says, is to bring greater diversity into the ranks of investors, hiring them into venture capital firms at first.

[\ldots]

Amsellem also wants France to follow the example of Sweden and replace maternity leave with parental leave. A Swedish study published in 2010 found that each month a father stays on parental leave increases a mother's earnings, which in turn led researchers to conclude that a lack of involvement by fathers in childcare and parental leave “could be one factor” behind the gender pay gap.

In France, says Amsellem, data show the gap between men and women starts to widen between the ages of 30 and 35, when many women tend to have their first child. If parental leave was equal, "women would be less disadvantaged in the labour market and employers would no longer have to consider the fact that a potential employee was at a child-bearing age", she says. "Women would also no longer be prevented from taking on higher responsibilities when they consider having a child."
\end{quote}

\section{Bootstrapping methods}\label{sec:boostrap}

We follow\footnote{code: bootstrap\_vansteenberghe.py} the work in section \ref{sec:multreg} based on the data on returns to schooling in Mincer 1974 and introduce boostrapping methods.

\subsection{Bootstrapped estimation of the population}

The data is a sample of the population. We have seen in section \ref{sec:largen} the weak law of large numbers and the Central Limit Theorem. We have a limited sample of size $n=534$ to estimate the "true" mean $\mu$ and standard deviation $\sigma$ of the population's log wage.
The mean of our sample is unbiased, while the standard deviation of the mean of the sample is biased. If we assume that the log wages of the population are normally distributed and if we define as in section \ref{seq:VaR} $\Phi()$ as the inverse of the c.d.f. of the normal law, then we can estimate the confidence interval at $\alpha \%$ of the population mean:
\begin{equation}
\mu = E\left[ \log\left( \mbox{wages} \right) \right] \pm \Phi(1-\alpha) \frac{\sigma}{\sqrt{n}}
\end{equation}

As detailed in section \ref{sec:largen}, as we use an estimate $S_n$ of the population standard deviation $\sigma$, then we rather use a t-Student distribution with $n-1$ degree of freedom:
\begin{equation}
\mu = E\left[ \log\left( \mbox{wages} \right) \right] \pm t_{n-1,\frac{1-\alpha}{2}} \frac{S_n}{\sqrt{n}}
\end{equation}

We find: $2.01 \leq \mu \leq 2.1$ but we can doubt the normality of the log wages distribution of the population.

As stated in \citep{fox2015} \begin{quote}
Bootstrapping uses the sample data to estimate relevant characteristics of the population. The sampling distribution of a statistic is then constructed empirically by resampling from the sample. The resampling procedure is designed to parallel the process by which sample observations were drawn from the population. For example, if the data represent an independent random sample of size $n$ (or a simple random sample of size $n$ from a much larger population), then each bootstrap sample selects $n$ observations with replacement from the original sample. The key bootstrap analogy is the following: \textbf{The population is to the sample as the sample is to the bootstrap samples}.
\end{quote}

We follow their notation and draw $r$ bootstrap replications with replacement ($Y^*_{b1},\ldots,Y^*_{bn}$ with $b\in [1,r]$) from the original sample and use $1000$ as advised. We have the bootstrap sample mean:
\begin{equation}
\bar{Y}^*_{b} = \frac{\sum_{i=1}^n Y^*_{bi}}{n}
\end{equation} and the estimate of the standard deviation of the bootstrap means:
\begin{equation}
\mbox{SE}^*(\bar{Y}^*) = \sqrt{\frac{\sum_{b=1}^r \left(\bar{Y}^*_{b} - \bar{\bar{Y}}^*  \right)^2}{r-1}}
\end{equation}
where $\bar{\bar{Y}}^* = \frac{\sum_{b=1}^r \bar{Y}^*_{b}}{r}$

From section \ref{sec:largen}, we can now estimate the confidence interval of the mean without any assumption on the distribution of the population:
\begin{equation}
\mu = \bar{\bar{Y}}^* \pm \Phi(1-\alpha) \mbox{SE}^*(\bar{Y}^*) \sqrt{\frac{n}{n-1}}
\end{equation}

With this method the confidence interval is marginally improved for this specific example, the "original sin" of having a sample with few observations cannot be overcome.

\subsection{Bootstrapping Regression Models}
\subsubsection{Fixed-X Resampling}

If we assume that our Mincer model, equation \ref{eq:wageEXP}, is correct and that the residuals $\epsilon_t$ are iid, then we can estimate the regression coefficients' standard error and confidence interval using bootstrapping by following these steps:

\begin{enumerate}
    \item Estimate the regression coefficients $\hat{\beta}$ and the fitted values of log wages $\hat{Y}_i$, where:
    \[
    \hat{Y}_i = X_i \hat{\beta}.
    \]
    \item Bootstrap $r$ replications of the residuals $\epsilon_{b,i}$, $b \in [1, r], i \in [1, n]$, and compute:
    \[
    Y_{b,i} = \hat{Y}_i + \epsilon_{b,i}.
    \]
    \item Regress $Y_b$ on the fixed $X$-values from the original sample:
    \[
    Y_b = X \beta_b + \nu_b,
    \]
    and extract the bootstrap regression coefficients $\beta_b$ for each replication.
    \item Use the bootstrap regression coefficients $\{\beta_b\}_{b=1}^r$ to compute:
    \begin{itemize}
        \item The standard error of the regression coefficients:
        \[
        \text{SE}(\hat{\beta}) = \sqrt{\frac{1}{r-1} \sum_{b=1}^r \left( \beta_b - \bar{\beta} \right)^2},
        \]
        where $\bar{\beta} = \frac{1}{r} \sum_{b=1}^r \beta_b$.
        \item The $(1-\alpha)100\%$ confidence interval:
        \[
        \left[ \beta^{\text{lower}}, \beta^{\text{upper}} \right] = \left[ \text{Quantile}_\alpha(\beta_b), \text{Quantile}_{1-\alpha}(\beta_b) \right].
        \]
    \end{itemize}
\end{enumerate}

e.g. at the $95\%$ confidence interval:

\begin{tabular}{lrrr}
 &     lower bound &  regression coefficient &     upper bound \\
Intercept  &  0.1163 &                  0.3437 &  0.5743 \\
SEX        &  0.1844 &                  0.2570 &  0.3352 \\
EDUCATION  &  0.0756 &                  0.0913 &  0.1068 \\
EXPERIENCE &  0.0254 &                  0.0361 &  0.0464 \\
EXP2       & -0.0008 &                 -0.0005 & -0.0003 \\
\end{tabular}

\subsubsection{Random-X resampling}
 
An alternative is to use random-X resampling, which is simply done by drawing bootstrap replication from the observations. We follow \citep{fox2015} example based on Duncan's regression of occupational prestige.

\stepcounter{question}
\begin{itemize}
\item \textbf{Question \arabic{question}}
\begin{itemize}
\item Apply random-X resampling to Mincer's model.
\end{itemize}
\end{itemize}

\subsection{Paper replication: bootstrap and Kolmogorov-Smirnov test}\label{sec:abadie}

We further illustrate\footnote{vansteenberghe\_Boostrap\_Abadie.py and the function kstats.py} boostrap techniques with a replication of the work of \citep{Abadie:2002aa} and sincerely thank Professor Alberto Abadie that provided us with the material. We focus on the empirical application of the paper using a sample of 11,637 white men, born in 1950-1953 and use their annual labor earnings, Vietnam veteran status and an indicator of draft-eligibility based on the Vietnam draft lottery outcome. Empirical evidence suggests that the earnings distribution for veterans exhibits higher lower quantiles and lower upper quantiles compared to non-veterans. Although the mean difference is small (with veterans earning, on average, about USD264 less), these distributional differences hint that military service may reduce the probability of extreme earnings outcomes. However, because veteran status is not randomly assigned (there is a selection process into military service), a direct comparison of observed earnings could be misleading. Here, draft eligibility serves as a valid instrument that permits us to recover the potential earnings distributions for the subpopulation of individuals whose treatment status is affected by the instrument.

\subsubsection*{Methodological Framework}
To capture the entire distributional effect, we estimate the cumulative distribution functions (CDFs) for both groups:
\begin{itemize}
    \item Let $F_1(y)$ denote the CDF of annual earnings for veterans.
    \item Let $F_0(y)$ denote the CDF of annual earnings for non-veterans.
\end{itemize}

We then test the following hypotheses:
\begin{enumerate}
    \item \textbf{Equality of Distributions:}
    \[
    H_0: F_0(y) = F_1(y) \quad \forall\, y \in \mathbb{R}.
    \]
    \item \textbf{First-Order Stochastic Dominance:} Non-veterans are said to dominate veterans if
    \[
    H_0: F_0(y) \le F_1(y) \quad \forall\, y \in \mathbb{R}.
    \]
    \item \textbf{Second-Order Stochastic Dominance:} This requires that the integrated CDF of non-veterans does not exceed that of veterans:
    \[
    H_0: \int_{-\infty}^{y} F_0(t)\,dt \le \int_{-\infty}^{y} F_1(t)\,dt \quad \forall\, y \in \mathbb{R}.
    \]
\end{enumerate}

For testing first-order stochastic dominance we use a Kolmogorov–Smirnov type statistic defined as
\[
T_{\text{FSD}} = \sqrt{\frac{n_0 n_1}{n}}\, \sup_{y \in \mathbb{R}} \Bigl( F_0(y) - F_1(y) \Bigr),
\]
where $n_0$ and $n_1$ are the number of non-veterans and veterans, respectively, and $n = n_0+n_1$.\footnote{the Kolmogorov-Smirnov test is further detailed in section \ref{sec:KStest}}

\subsubsection*{Bootstrap Procedure}
Because the asymptotic distribution of the test statistic under the null hypothesis depends on the unknown underlying distribution, we use a bootstrap procedure to approximate the p-values:
\begin{enumerate}
    \item \textbf{Compute the test statistic.} Calculate the observed statistic (e.g., $T_{\text{FSD}}$) from the original data.
    \item \textbf{Resample.} Draw a bootstrap sample by resampling with replacement from the pooled data. Partition this bootstrap sample into two groups of sizes $n_0$ and $n_1$ corresponding to non-veterans and veterans.
    \item \textbf{Recompute.} Calculate the test statistic for the bootstrap sample.
    \item \textbf{Repeat.} Repeat steps 2 and 3 a large number of times (e.g., 2000 replications) to obtain the bootstrap distribution of the test statistic.
    \item \textbf{Determine the p-value.} Estimate the p-value as the proportion of bootstrap replications for which the test statistic exceeds the observed statistic.
\end{enumerate}

A similar bootstrap strategy is used for the tests of equality of distributions and second-order stochastic dominance by modifying the KS statistic accordingly.

\subsubsection*{Empirical Findings}
With 2000 bootstrap replications we obtain:
\begin{itemize}
    \item \textbf{Test of Equality of Distributions:} p-value $\approx 12\%$.
    \item \textbf{Test for First-Order Stochastic Dominance (non-veterans dominating veterans):} p-value $\approx 63\%$.
    \item \textbf{Test for Second-Order Stochastic Dominance (non-veterans dominating veterans):} p-value $\approx 75\%$.
\end{itemize}

These results indicate that we cannot reject the null hypothesis of equality of distributions at conventional significance levels, and there is even less evidence against the dominance hypotheses.

\begin{figure}[h!]
\centering
\includegraphics[scale=0.5]{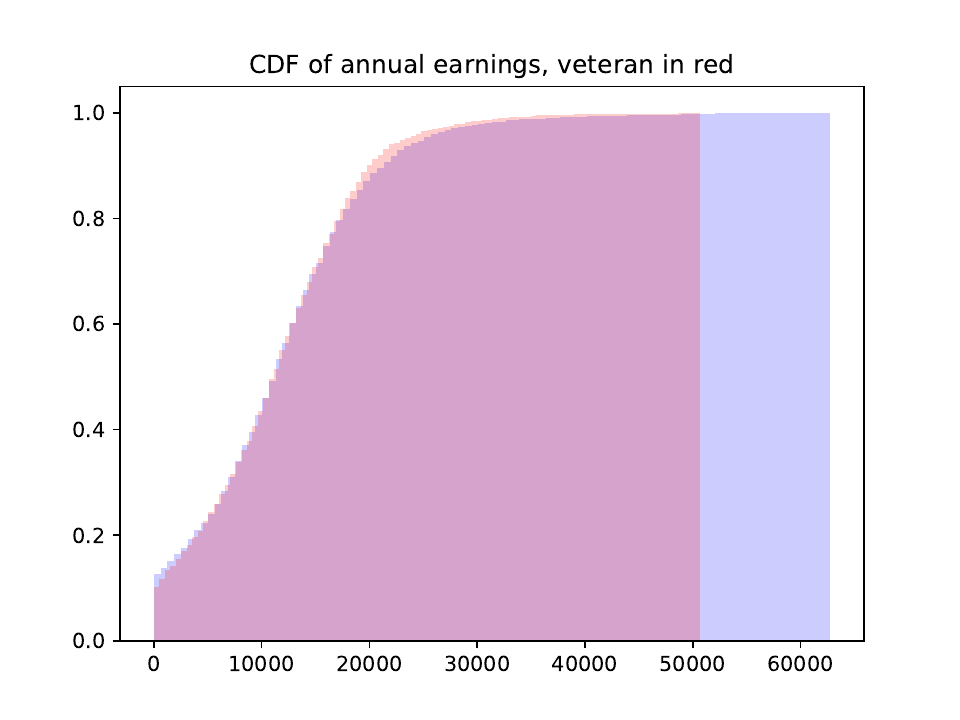}
\caption{Empirical cumulative distribution functions (cdfs) of realized annual earnings for veterans (red) and non-veterans (blue).}
\label{fig:earnings}
\end{figure}

\begin{figure}[h!]
\centering
\includegraphics[scale=0.5]{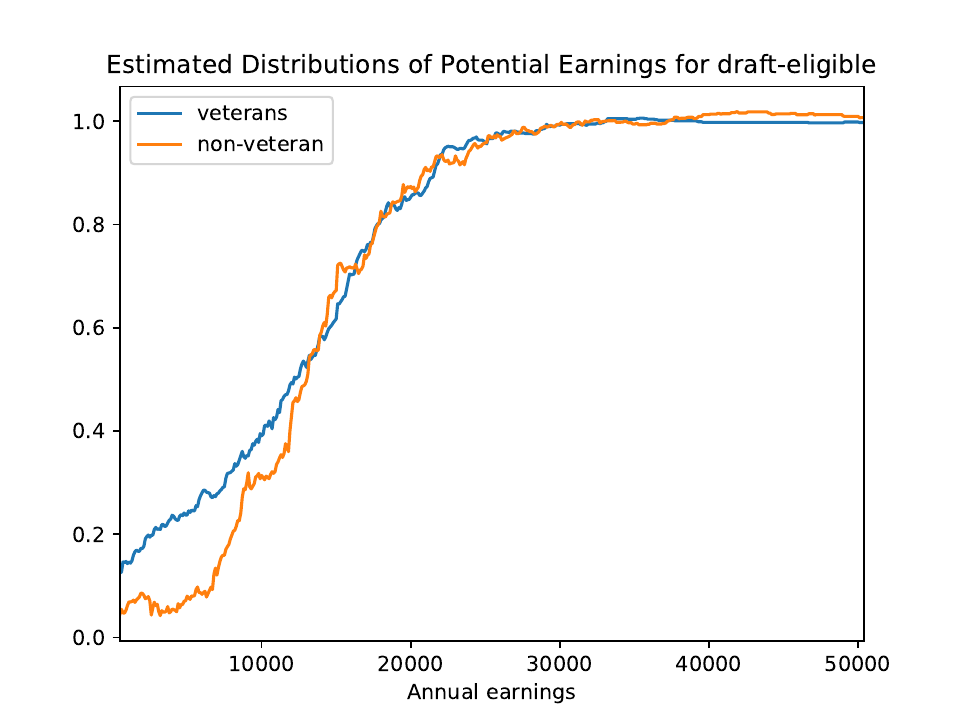}
\caption{Estimated cumulative distribution functions (cdfs) of potential earnings for compliers, obtained via instrumental variable methods (using equations (5) and (6)). The horizontal axis represents annual earnings thresholds, while the vertical axis shows the cumulative probability that a complier's earnings are below a given level. The two curves correspond to the estimated distributions if all compliers were treated (veterans) versus untreated (non-veterans). The gap in the lower tail indicates that military service mainly reduces earnings among low earners, even though the average difference may be small.}
\label{fig:potentialEarnings}
\end{figure}

\section{Python: Median versus mean and outlier detection}\label{sec:medianmean}

The mean of a sample is vulnerable to outlying observation, we talk about breakdown point as in \citep{Leys:2013aa} which is \begin{quote}
the maximum proportion of observations that can be contaminated (i.e., set to infinity) without forcing the estimator to result in a false value (infinite or null in the case of an estimator of scale). For example, when a single observation has an infinite value, the mean of all observations becomes infinite; hence the mean's breakdown point is 0. By contrast, the median value remains unchanged. The median becomes absurd only when more than 50\% of the observations are infinite. With a breakdown point of 0.5, the median is the location estimator that has the highest breakdown point.
\end{quote} We take\footnote{code: vansteenberghe\_median\_mean.py} for example a Cauchy distribution of location $\mu = 2$ as we know it is symmetric and interestingly it has no defined mean nor variance. We draw $10$ random points to form a sample where we observe a clear outlier at $-10$:
\begin{center}
\includegraphics[scale=0.5]{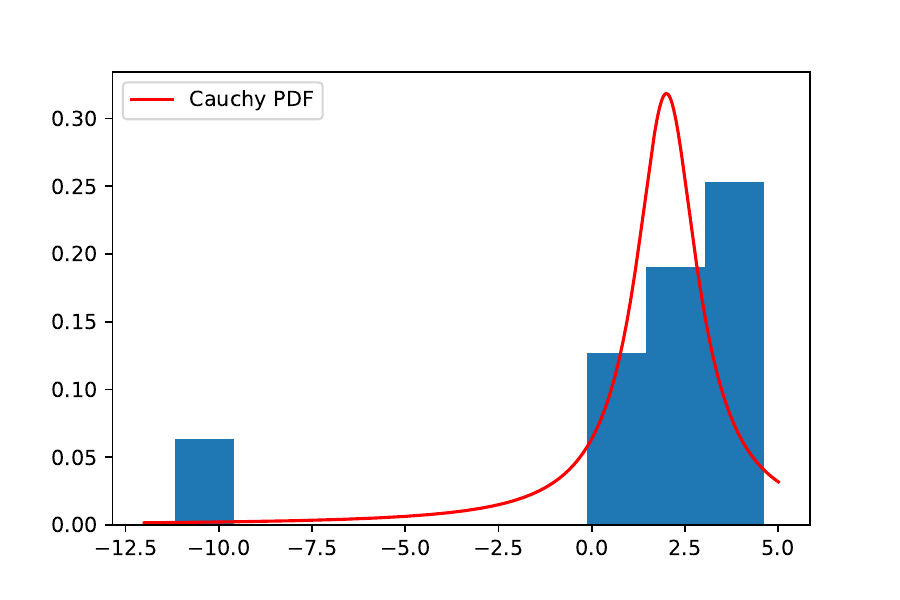}
\end{center}

If we observe the evolution of the mean and median, from a sample ranging from $2$ to $200$ draws, we observe that the median is robust around the location while the mean, depending on the random draw will be "polluted" by outliers and in fact increasing the sample size will not enable us to has a mean that converges:
\begin{center}
\includegraphics[scale=0.5]{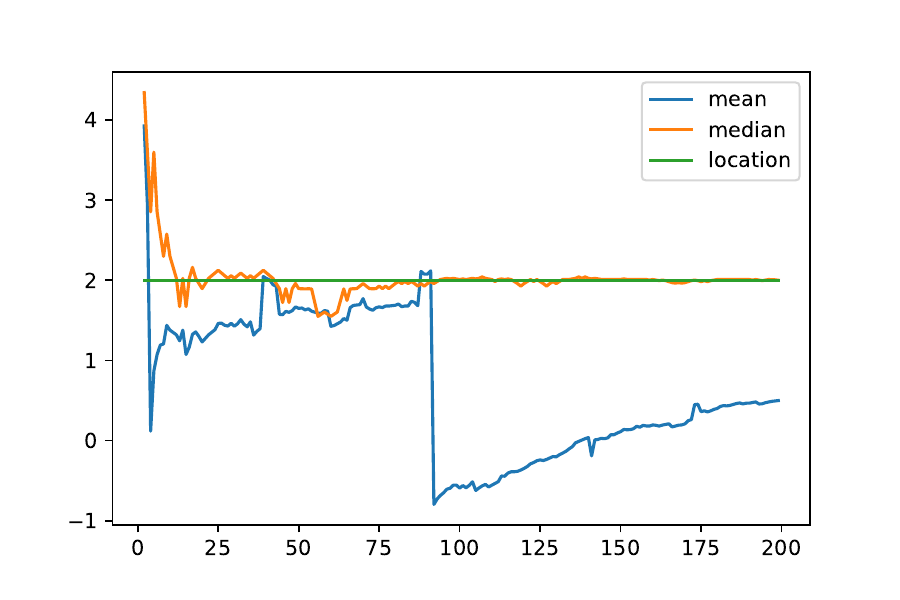}
\end{center}

So at this stage we might wonder why using the mean (which is at the core of the OLS we will see in the following sections) and not the median. We find an answer in the efficiency of the two measures. According to the central limit theorem, the mean of a sample with variance $\sigma^2$ has a variance of $\frac{\sigma^2}{n}$. Let's demonstrate the the median of a sample with variance $\sigma^2$ has a variance of $\frac{\pi \sigma^2}{2 n}$. For this, we follow Dr David A. Stephens demonstration:


Suppose $X_1 , \ldots, X_n$ are i.i.d. continuous random variables from distribution with cumulative distribution function $F_X$ and we define:
$$Y_n(x) = \frac{1}{n}\sum_{i=1}^n Z_i(x)$$ with
$$Z_i(x) = \begin{cases} 1 \quad \mbox{if} \quad X \leq 0 \\ 0 \quad \mbox{otherwise}\end{cases}$$

Then $Z_i(x)$ follows a Bernoulli distribution, hence:
$$E\left[Z_i(x)\right] = F_X(x)$$
$$Var\left[Z_i(x)\right] = F_X(x)\left(1 - F_X(x) \right)$$

Using the central limit theorem:
$$\sqrt{n} \left[Y_n(x)-F_X(x) \right] \overset{d}{\to} \mathcal{N}\left( 0, F_X(x)\left(1 - F_X(x) \right)  \right)$$
Applying the Delta method:
$$\sqrt{n} \left[F^{-1}_X Y_n(x) - x \right]  \overset{d}{\to}  \mathcal{N}\left( 0,\frac{F_X(x)\left(1 - F_X(x) \right) }{\left[f_X(x)\right]^2}  \right)$$

Writing $p= F_X(x)$ we have the random variable $F^{-1}_X Y_n(x)$ that is lying between the $(p-1)^{\mbox{th}}$ and $p^{\mbox{th}}$ sample quantile. So if we apply this to the median $p=\frac{1}{2}$ of a symmetric distribution with a defined standard deviation and a true median $\theta$, then the median $\tilde{X}_n$ of the sample follows:
$$\sqrt{n} \left[\tilde{X}_n - \theta \right]  \overset{d}{\to}  \mathcal{N}\left( 0,\frac{1}{4\left[f_X(\theta)\right]^2}  \right)$$
and if we assume that the population has a normal distribution of mean $\mu$, then $\theta=\mu$ and $f(\mu)^2=\frac{1}{2\pi \sigma^2}$:
$$\sqrt{n} \left[\tilde{X}_n - \theta \right]  \overset{d}{\to}  \mathcal{N}\left( 0,\frac{\pi \sigma^2}{2 n}  \right)$$

We suggest an empirical application, growing $100$ samples from a normal distribution of mean $2$ and standard deviation $3$, we also plot the Median Absolute Deviation we introduce section \ref{sec:MAD} as a benchmark of a measure which is known to be robust and efficient or as \citep{huber1981} put it the "single most useful ancillary estimate of scale":
\begin{center}
\includegraphics[scale=1]{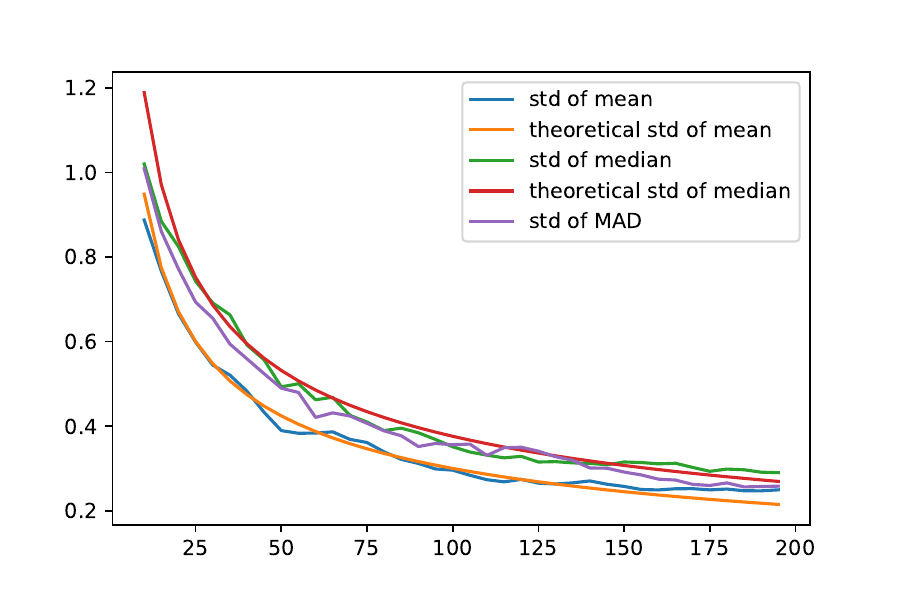}
\end{center}

We indeed find in this well-behaved example that the mean of the sample with a theoretical variance of $\frac{\sigma^2}{n}$ is more efficient than the median of the sample with a theoretical variance of $\frac{\pi \sigma^2}{2 n}$. The lack of efficiency the median is what leads to consider M estimators. But before, let's consider methods to detect outliers.

\subsection{Outlier detection}\label{sec:outlier}

Let's imagine\footnote{code: vansteenberghe\_outlier\_detection.py} that we have at our disposal a small sample of size $N$ from a population that is almost normally distributed with mean $\mu=0$, but with standard deviation $\sigma$ that changes randomly based on a latent variable (sigchoice) following a combination of discrete uniform distributions that determine whether the standard deviation is $1$ or $100$ for any given observation and we make it so that the low standard deviation is more likely than the high one. If the sample size is large enough, high and low observations from the population will compensate and the mean should be unbiased.

Now if we take a small sample from this population: $-173.29$,  $-1.03$,  $-14.63$,    $0.3$,    $0.4$,    $0.65$,    $1.88$, $-0.2$,   $ 0.46$,   $-1.79$
then the mean is strongly biased at $-18.7$ and less so the median at $0.05$.

One option to detect an outlier is to remove the extreme $1\%$, with our limited sample, we trim our sample and reject from it the extreme $10\%$ which is simply in our previous example limited to $10$ observations, the observation with the highest absolute value: $-173.29$.

Another approach is to discard outliers which are defined as observations which are further than plus or minus two standard deviation of the sample mean. With our example, we fail to reject the negative outlier which is in line with \citep{Leys:2013aa} \begin{quote}
the mean and standard deviation are strongly impacted by outliers. [...] this method (the mean plus or minus three standard deviations) is very unlikely to detect outliers in small samples.
\end{quote}

\subsubsection{Median absolute deviation}\label{sec:MAD}

In \citep{Leys:2013aa}, they state that: 
\begin{quotation}
A survey revealed that researchers still seem to encounter difficulties to cope with outliers. Detecting outliers by determining an interval spanning over the mean plus/minus three standard deviations remains a common practice. However, since both the mean and the standard deviation are particularly sensitive to outliers, this method is problematic. We highlight the disadvantages of this method and present \textbf{the median absolute deviation}, an alternative and more robust measure of dispersion that is easy to implement.
\end{quotation}

As we experienced just above, in outliers detections, we have three main issues:
firstly, distributions might not be normal. Secondly, the mean and standard deviation are strongly impacted by outliers. Thirdly, this method is very unlikely to detect outliers in small samples.

\quad

The median absolute deviation (MAD) is defined as the median of the absolute deviations from the sample median and is a robust alternative to the standard deviation measure of a sample:

\begin{equation}
\mbox{MAD} =b \quad \mbox{median} \left(\vert X_{i}-\mbox{median}(X) \vert \right)
\end{equation}
where $b$ is linked to the distribution assumption of the population and is taken to be the inverse of the 0.75 quantile of that underlying distribution, which for a normal is $b = 1.4826$.

We finally detect outliers by determining an interval spanning over the median plus/minus three MAD. The trimmed sample presents "well behaved" mean and median:
\begin{itemize}
\item MAD trimed sample mean $0.08$
\item MAD trimed sample median $0.35$
\end{itemize}

\subsubsection{Alternatives to the Median Absolute Deviation (MAD)}

While the MAD has been an effective tool in statistics for handling outliers and robustly estimating the variability, it relies on a central location estimate, which can be less meaningful in skewed or asymmetric distributions. In addition, MAD may not fully utilize the information available in the sample because it only considers the deviation with respect to the median.

\textbf{The \( S_n \) Method} introduced by \citet{rousseeuw1993alternatives}. An alternative approach, which we denote as \( S_n \), has been proposed to overcome these limitations. Unlike the MAD, \( S_n \) does not require any location estimate of the data. It is defined as a measure of scale based on the pairwise differences of the sample observations, representing a typical distance between observations. Formally, \( S_n \) can be expressed as:
\[
S_n = c \cdot \text{median}(\text{median}(|X_i - X_j|)),
\]
for \( i \neq j \), where \( c \) is a constant that makes \( S_n \) consistent at the normal distribution.

The method \( S_n \) has several advantages over the MAD:
\begin{itemize}
  \item It is a robust measure of scale that is not influenced by extreme values or asymmetries in the data distribution.
  \item No central value (such as the median) is needed, making \( S_n \) particularly useful for asymmetric distributions.
  \item \( S_n \) utilizes the distances between all pairs of observations, potentially providing a more comprehensive measure of variability.
\end{itemize}

\textbf{Calibration of \( c \) for Non-Central t-Distributions}

To ensure that the robust scale estimator \( S_n \) is consistent when applied to non-central t-distributions, the scale constant \( c \) must be calibrated accordingly. The non-central t-distribution, characterized by its degrees of freedom and non-centrality parameter, poses additional challenges due to its asymmetry and heavier tails compared to the normal distribution.

The calibration of \( c \) typically involves a simulation approach where the following steps are taken:

\begin{enumerate}
  \item Generate a large number of samples from a non-central t-distribution with known parameters.
  \item For each sample, compute \( S_n \) without the scale constant \( c \).
  \item Determine the constant \( c \) such that the median of these \( S_n \) values across all samples approximates the known scale of the non-central t-distribution.
\end{enumerate}

The simulation process can be described in pseudocode as follows\footnote{vansteenberghe\_Sn\_c\_calibration.py}

\begin{verbatim}
1. Define the non-centrality parameter delta and degrees of freedom nu
2. Choose a large number N of samples to generate
3. For each sample i from 1 to N do:
   a. Generate a sample of size n from the non-central t-distribution T(nu, delta)
   b. Compute S_n_i (without the scale constant c)
4. Compute c as the factor that scales the median of S_n_i to the known scale
\end{verbatim}

\subsubsection{Isolation Forest}

\citep{isolationforest} detail the use of isolation Forest (IF), a method rather suited toward "big data". We nevertheless implement it here to compare with our previous results.

\subsubsection{Outlier isolation comparison}

Finally, we generate $100$ samples of size $100$ and compare the different methods:

\begin{tabular}{lrr}
\toprule
 & mean & std \\
\midrule
mean & 0.204000 & 3.417000 \\
median & -0.010000 & 0.137000 \\
trimmed mean & 0.227000 & 3.471000 \\
trimmed median & -0.010000 & 0.137000 \\
std dev mean & -0.039000 & 1.721000 \\
std dev median & -0.009000 & 0.139000 \\
MAD mean & -0.015000 & 0.104000 \\
MAD median & -0.012000 & 0.137000 \\
Sn mean & -0.015000 & 0.103000 \\
Sn median & -0.013000 & 0.137000 \\
IF mean & -0.012000 & 0.145000 \\
IF median & -0.015000 & 0.138000 \\
\bottomrule
\end{tabular}

\begin{center}
\includegraphics[scale=0.5]{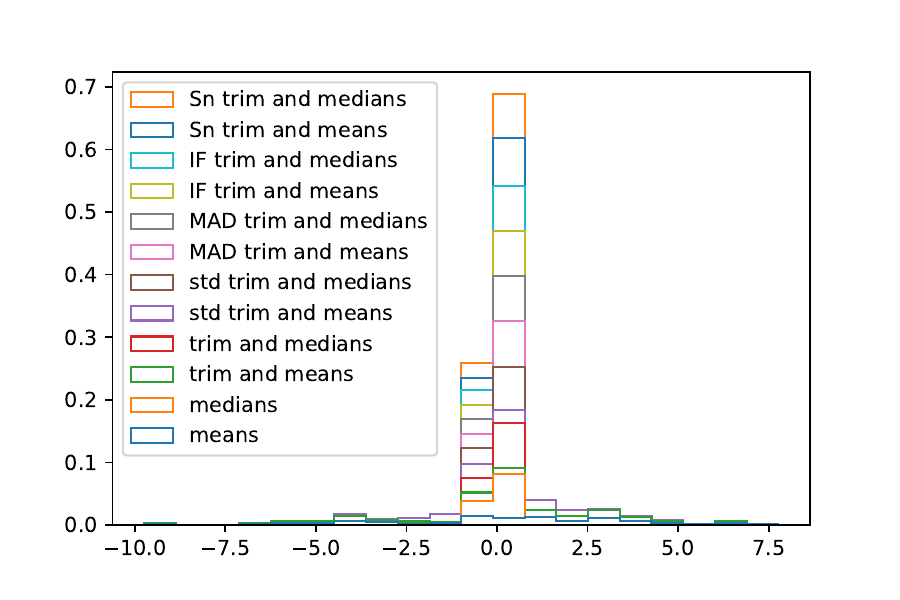}
\end{center}

To go further about robust statistics, we recommend \citep{maronna2019} who recommend the bisquare M-estimator for location with MAD  for scale. And for machine learning application, isolation forest has been extended in \citep{Hariri2019aa}.

\stepcounter{question}
\begin{itemize}
\item \textbf{Question \arabic{question}}
\begin{itemize}
\item increase progressively the size of the generated samples and comment on the convergence of the measures and/or their efficiency.
\end{itemize}
\end{itemize}

\section{Python: Monte Carlo method and econometric tests}\label{sec:MCmethod}

We suggest\footnote{pandas\_exercise\_Monte\_Carlo.py} Monte Carlo methods to help understanding standard econometrics hypothesis testing.

In section \ref{sec:linreg} we performed ordinary least square to estimate linear regressions of the form:
$$ y_t =\alpha + \beta x_t + \epsilon_t $$
our estimate $\hat{\beta}$ of the true parameter $\beta$ will depend on the size of the samples and on the validity of assumptions we made about underlying data generation processes for $\epsilon$ and $x$.

We assumed that $\epsilon$ was iid and $\epsilon \sim \mathcal{N}\left(0,\sigma_\epsilon\right)$ and $x \sim \mathcal{N}\left(\mu_x,\sigma_x\right)$. Even if those assumptions are true, we have a finite and relatively small number of observations so the precision of our estimate will be limited.

The t statistic for a coefficient is defined as:
$$t = \frac{\hat{\beta}}{\hat{\sigma}_{\hat{\beta}}}$$

This statistics follow a Student law with $T-k$ degree of freedom ($T$ being the number of observations and $k$ the number of estimated parameters).

The critical threshold can be computed at a given confidence level with the inverse of the cumulative Student t distribution, the percent point function (ppf) which in the $5\%$ case is $t^*$ such that: $ \mbox{Prob}(X \leq t^*) = 0.025$

The null hypothesis $H_0$ is : $\beta = 0$, if $\vert t \vert > t^*$, then we can reject $H_0$, again, if our econometric methods were correct, $x$ has a significant influence over $y$.

\subsection{Monte Carlo method for coefficient significance of a simple OLS}
We know that the t statistic for $\beta$ follows a Student law with $T-k$ degree of freedom. We can generate a "large" number of simulations (e.g. MCrun = $1000$), drawing the error and observable from the assumed distributions: $\epsilon \sim \mathcal{N}\left(0,\sigma_\epsilon\right)$ and $x \sim \mathcal{N}\left(\mu_x,\sigma_x\right)$ and extract for each simulation the t-value of the $\beta$. So step by step:

\begin{enumerate}
\item we estimate $ y_t =\alpha + \beta x_t + \epsilon_t $ and extract: $\hat{\alpha}$, $\hat{\beta}$, $\hat{\sigma}_\epsilon$, $\hat{\mu}_x$, and $\hat{\sigma}_x$;
\item we generate MCrun simulations generating $\tilde{\epsilon}_t$ and $\tilde{x}_t$ sample of length $T$, the original series' length;
\item we compute each simulation $i$'s $\tilde{y}$ series: $ \tilde{y}^i_t = \hat{\alpha} + 0 \times \tilde{x}^i_t + \tilde{\epsilon}^i_t $, so under the null hypothesis, $\tilde{x}^i_t $ has no "influence" over $\tilde{y}^i_t$
\item from each simulation, we estimate the regression: $ \tilde{y}^i_t = \delta + \gamma \tilde{x}^i_t + \nu_t $, from each regression we store in a list the t-statistic of the estimated coefficient $\hat{\gamma}$;
\item we are "sure" that $\hat{\gamma}$ should be $0$ rather than $\hat{\beta}$, hence $100\%$ of our t-statistics from the previous step should be in the $H_0$ acceptance region and are distributed as a t-Student with $T-2$ degree of freedom;
\begin{itemize}
\item at this point, we can compute the critical threshold in a similar manner as a parametric VaR for a t-Student distribution as we develop in section \ref{seq:VaR};
\end{itemize}
\item we extract the $\beta$ t-statistic from the original estimation $ y_t =\alpha + \beta x_t + \epsilon_t $ and count how many t-statistics we generated are above this value, because our test is two-sided, this is half the probability to wrongly reject $H_0$. In other words, if we are almost outside of the empirical distribution of t-values we generated we can safely reject $H_0$ and (as long as our assumptions are correct) $\beta$ is significantly different from $0$, but if we are almost in the middle of the empirical distribution, we cannot reject $H_0$ and it is likely that $\beta = 0$.
\end{enumerate}

Below is the distribution of the $\beta$ t-values under the null hypothesis. We draw left and right the threshold so that the green area in between is $95\%$. So for our test, we reject the null hypothesis at the $5\%$ confidence level is our t-value is outside of this zone:
\begin{center}
\includegraphics[scale=0.5]{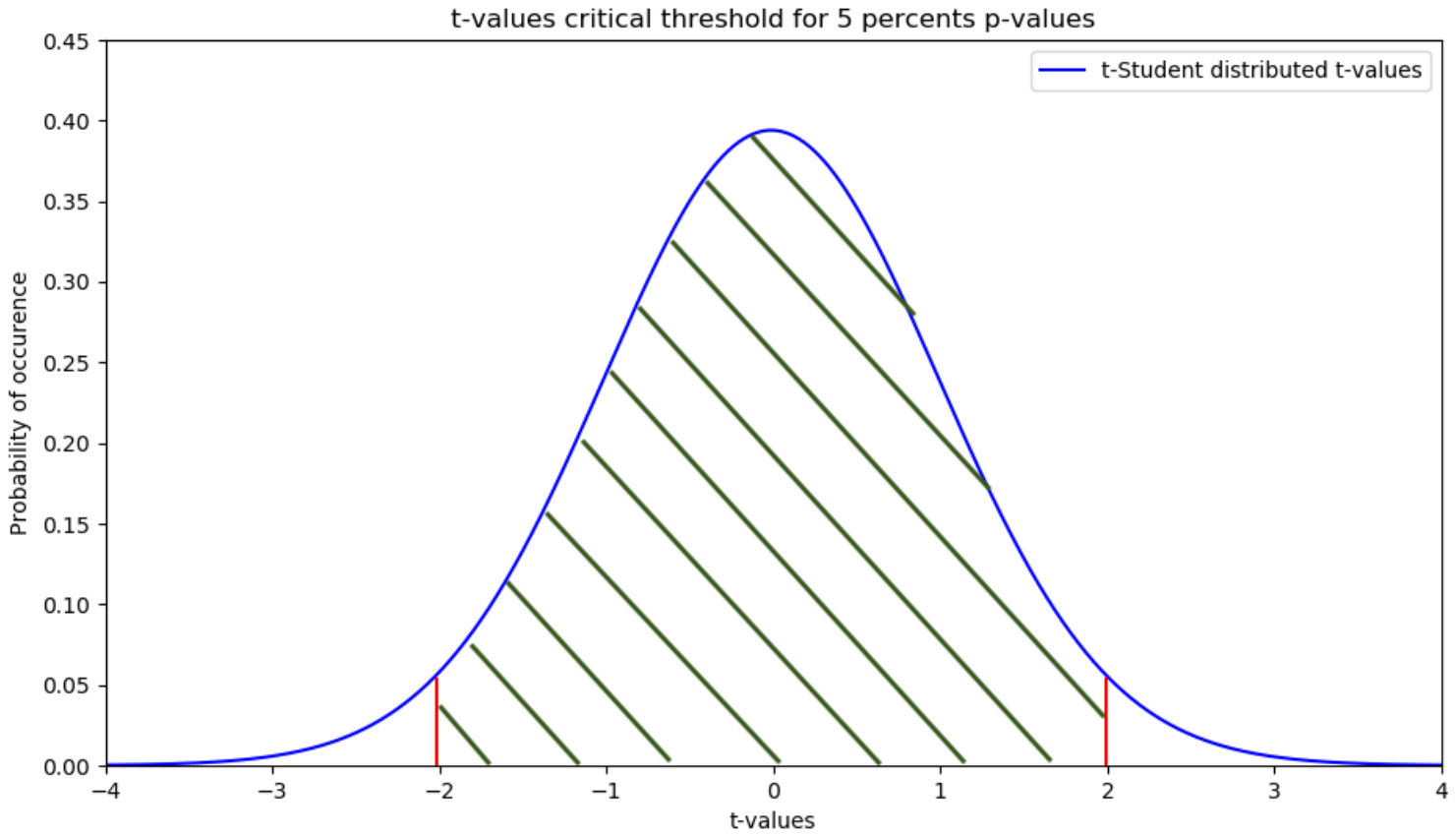}
\end{center}

\subsection{Monte Carlo method for augmented Dickey-Fuller test p-value}

We are now interested in the augmented Dickey-Fuller test we perform as in section \ref{sec:adf}. We apply this with the no constant (option 'nc') on the demeaned GDP growth rate and the comparison is simple as the test select no lags, hence equation \ref{eq:adf} simplifies into:

$$\Delta y_t = \gamma y_{t-1} + \epsilon_t$$

We want to test the null hypothesis $H_0$ under which $\gamma=0$.

For this, as in previous section, we will generate a "large" number of series that respect this null hypothesis, and compare our inital regression t-value of $\gamma$ with the distribution of t-value of $\gamma$ obtained, with the underlying assumption that $\epsilon \sim \mathcal{N}\left(0,\sigma_\epsilon\right)$:

\begin{enumerate}
\item we estimate $  \Delta y_t = \gamma y_{t-1} + \epsilon_t $ and extract: $\hat{\sigma}_\epsilon$;
\item we generate MCrun simulations generating $\tilde{\epsilon}_t$ sample of length $T$, the original series' length;
\item we compute each simulation $i$'s $\tilde{y}$ series: $ \tilde{y}^i_t = 1 \times \tilde{y}^i_{t-1} + \tilde{\epsilon}^i_t $, so under the null hypothesis, $\tilde{y}$ has a unit root
\item from each simulation, we estimate the regression: $ \Delta \tilde{y}_t = \gamma \tilde{y}_{t-1} + \nu_t $, from each regression we store in a list the t-statistic of the estimated coefficient $\hat{\gamma}$;
\item we are "sure" that $\hat{\gamma}$ should be $0$ (our series has a unit root), hence $100\%$ of our t-statistics from the previous step should be in the $H_0$ acceptance region;
\begin{itemize}
\item be careful: as demonstrated in \citep{MacKinnon2010}, this t-values are not distributed as a traditional t-Student distribution, we can use the tables in \citep{MacKinnon2010} and compute the critical value following as in the paper $ \beta_{\infty} + \beta_1 / T + \beta_2 / T^2 + \beta_3 / T^3$ or in our case we use a Monte Carlo method which doesn't need to take any assumption on the distribution of the t-values of $\hat{\gamma}$;
\end{itemize}
\item we extract the $\gamma$ t-statistic from the original estimation $ \Delta y_t = \gamma y_{t-1} + \epsilon_t $ and count how many t-statistics we generated are below this value and this is the probability to wrongly reject $H_0$. In other words, if we are almost outside of the empirical distribution of t-values we generated we can safely reject $H_0$ and (as long as our assumptions are correct) $\gamma$ is significantly different from $0$, but if we are almost in the middle of the empirical distribution, we cannot reject $H_0$ and it is likely that $\gamma = 0$ and our series has a unit root (as long as our assumptions on the constant, drift, trend, and quadratic terms were correct!).
\end{enumerate}

Note that the empirical distribution of the $\hat{\gamma}$ t-values is not "well-behaved", in the sense that for example even with 20 000 runs the distribution is not smooth and this can introduce some noise in our computed p-value:
\begin{center}
\includegraphics[scale=0.5]{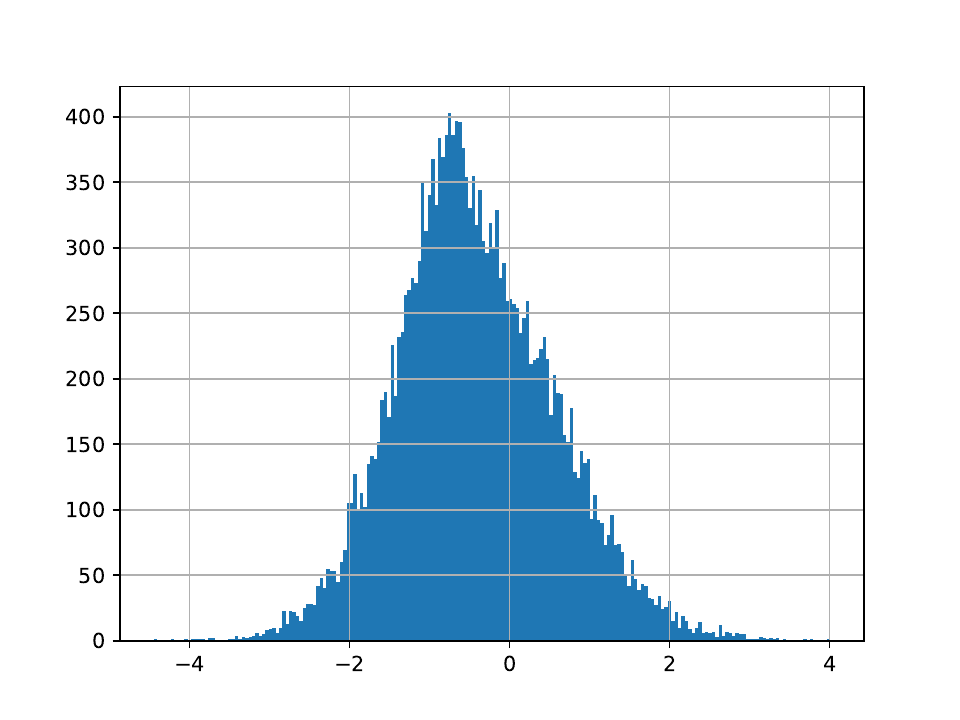}
\end{center}

Hence to get a powerful test we would need to perform "many" Monte Carlo simulations, here we have an estimated p-value of $0.06\%$ meaning that we expect to have $60$ t-values below our t-statistics out of 100 000 simulations (so in that respect, even 100 000 seems not enough to be confident over the results). In \citep{MacKinnon2010} they ran 200 000 simulations.

\section{Granger causality test: academic paper replication}\label{sec:granger}

In this section\footnote{broker\_dealer\_leverage\_vansteenberghe.py}, with data taken from the Federal reserve website \href{https://www.federalreserve.gov/releases/z1/20160310/data.htm}{here}, that provides historical data of financial accounts of the United States, namely for the broker-dealers, we suggest to replicate a Granger causality test done in the paper \citep{Serletis:2018aa}. Keep in mind that a cause cannot come after the effect, information on one variable helps to predict another one but this do not imply any sort of economic causality (e.g. a banks changes its business model, its stock price moves).

\citep{Serletis:2018aa} suggested to study the causality between broker-dealer leverage and the stock market in the United States, using quarterly data since 1967. 
Leverage defined as (total assets) / (total assets - total liabilities) is a measure of how much debt an investor assumes in making an investment.
The main idea is that a higher leverage means a higher amount of cash in circulation that should drive the asset prices up.
%

Note that as a first difference in this exercise, we do not divide the S\&P 500 by the CRB BLS spot index\footnote{we tried to divide with publicly available data, like the US GDP or the consumer price, but it did not help reach the same conclusion for the lag order selection}

From the naked eye, it would be very difficult to observe which returns seems to cause the other:
\begin{center}
\includegraphics[scale=0.7]{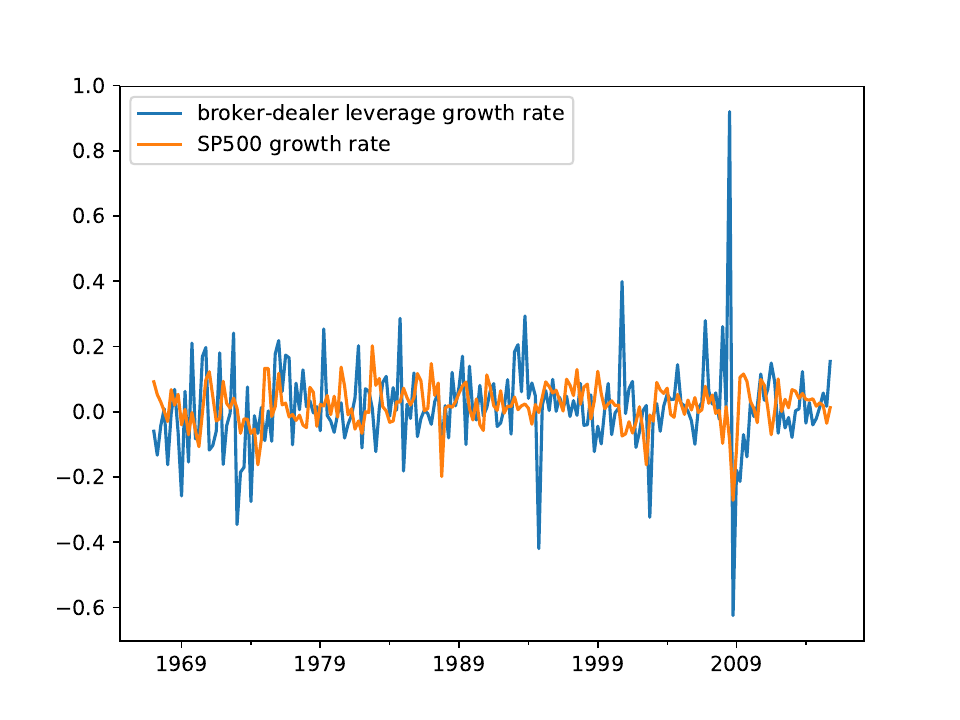}
\end{center}

\subsection{ADF and cointegration test}
As in the paper, we perform ADF and cointegration test. We find that the variables are cointegrated. Hence we can build an ECM to study the two time series.

\subsection{Granger causality definition and test}

We can say that a variable $x_t$ for which we have a stationary time series of length $T$ does not Granger cause another stationary time series $y_t$ of length $T$ if the information we have on that variable does not improve our ability to predict the other variable. In fact the Granger causality test is just a test whether certain autoregressive coefficients are zero (the null hypothesis is $H_0$: $\forall i, \beta^2_i = 0$ and $\gamma = 0$). We do a restricted ($r$) and unrestricted ($u$) regression presented in this order, adding the lagged estimated residuals $\hat{\epsilon}_{t-1}$from the cointegration regression as the two time series are cointegrated:

\begin{equation}
    \begin{cases}
y_t = \sum_{i=1}^m \alpha^1_i y_{t-i} + c^1 +\nu^1_t  \\       
y_t = \sum_{i=1}^m \alpha^2_i y_{t-i} + \sum_{i=1}^n \beta^2_i x_{t-i}  + \gamma \hat{\epsilon}_{t-1} +  c^2 +\nu^2_t \\ 
    \end{cases}
\end{equation}

The order of lag, $m$ and $n$, are determined by minimizing the Schwarz information criterion from the various regressions.

\quad

We compute an F statistics with the sums of squared residuals as in the Wald test\footnote{for further explanation, the reader can refer to \citep{verbeek} sections "2.5.4 A joint test of significance of regression coefficients" and "3.2.2 selecting regressors"}
$$ F : \frac{\frac{SSR_r -SSR_u}{n+1}}{\frac{SSR_u}{T-m-n-1-1}}$$

We need to compute the critical value\footnote{F statistic with a numerator degree of freedom of $n$ and a denominator degree of freedom of $T-m-n-1-1$} and then determine:
\begin{itemize}
\item our statistic $<$ critical value: fail to reject the null hypothesis
\item our statistic $\geq$ critical value: reject the null hypothesis
\end{itemize}

In our case, we look in a F-statistic table $F_{4,192}$ for both variables and our statistic is above the critical value, we reject the null hypothesis. Hence we assume the broker-dealer leverage Granger cause the stock price and vice-versa. Note that our lag selection with the BIC leave us with a different model for the tests on the stock market. As noted in the paper, this is a \textbf{linear} test and in times of crises, the relationship might present some non-linearities.

\quad

Also note that in their 2006 paper, Diks and Panchenko indicates that their test might under-reject their causality for sample with less than $500$ observations which is our case here.

\section{Recap on econometric models}

\citet{Intriligator:1983aa}:

A model is a simplified representation of an actual phenomenon, such as an actual system or process. The actual phenomenon is represented by the model in order to explain it, to predict it, and to control it, goals corresponding to the three purposes of econometrics, namely structural analysis, forecasting, and policy evaluation.

Let's consider a system of $g$ independent and consistent (i.e. mutually compatible) equations in the $g$ endogenous variables, $y_1, y_2, \ldots ,y_g$, the $k$ exogenous (or lagged endogenous) variables, $ x_1, x_2, \ldots,x_k$, and the $m$ parameters, $\delta_1, \delta_2, \ldots, \delta_m$. The model can be written in vector notation:
\begin{equation}\label{eq:eqsystem}
f\left(y, x, \delta \right) = 0
\end{equation}
 where $f$ is a column vector of $g$ functions.
 
 Assuming the functions are differentiable and that the Jacobian matrix of first-order partial derivatives is non-singular at a particular point:
\begin{equation}
\left\lvert \frac{\partial f}{\partial y}\right\rvert \neq 0 \mbox{ at } (y ,x)
\end{equation}

 the implicit function theorem implies that at this point it is possible to solve the system of equations \ref{eq:eqsystem} for the endogenous variables as differentiable functions of the exogenous variables and parameters: $ y = \Phi(x, \delta)$ where $\Phi$ a column vector of g functions.
 
\emph{Econometric models} are generally algebraic models that are stochastic in including random variables (as opposed to deterministic models which do not include random variables). The random variables that are included, typically as additive stochastic disturbance terms, account in part for the omission of relevant variables, incorrect specification of the model, errors in measuring variables, etc. The general econometric model with additive stochastic disturbance terms can be written as the non -linear structural form system of $g$ equations: $f\left(y, x, \delta \right) = \epsilon$  where $\epsilon$ is a vector of stochastic disturbance terms. If the conditions of the implicit function theorem are met these equations can be solved for the endogenous variables as differentiable functions of the exogenous variables and parameters, with the stochastic disturbance terms included as additive error terms. The resulting non-linear reduced form is the system of $g$ equations: $y = \Phi(x, \delta) + u$, where $u$ is the vector of the stochastic disturbance terms in the reduced form. The econometric model uniquely specifies not the endogenous variables but rather the probability distribution of each of the endogenous variables, given the values taken by all exogenous variables and given the values of all parameters of the model. Each equation of the model, other than definitions, equilibrium conditions, and identities, is generally assumed to contain an additive stochastic disturbance term, which is an unobservable random variable with certain assumed properties, e.g. mean, variance, and covariance. The values taken by that variable are not known with certainty; rather, they can be considered random drawings from a probability distribution with certain assumed moments. The inclusion of such stochastic disturbance terms in the econometric model is basic to the use of tools of statistical inference to estimate parameters of the model.

The basic econometric model  is a linear stochastic structural form:
\begin{equation}\label{eq:basicmodel}
y\Gamma + xB = \epsilon
\end{equation}

There is a trivial indeterminacy in the structural equations in that multiplying all terms in any one of these equations by a non-zero constant does not change the equation. This indeterminacy is eliminated by choosing a normalization rule, which is a rule for selecting a particular numerical value for one of the non-zero coefficients in each question. A convenient normalization rule is that which sets all elements along the principal diagonal of the $\Gamma$ matrix of coefficients of endogenous variables at $-1$, $\gamma_{h,h}=-1$. 

Certain stochastic assumptions are typically made concerning the $g$ stochastic disturbance vectors:
\begin{itemize}
\item zero mean, $E(\epsilon_i)=0$
\item same covariance matrix of $\epsilon_i$ at each observation, $cov(\epsilon_i)=\Sigma, \forall i$
\item uncorrelated over the sample, $E\left(\epsilon_i' \epsilon_j \right) = 0, \forall i\neq j$
\end{itemize} 

These conditions are satisfied if $\epsilon_i$ are iid\footnote{Sometimes the further assumption of normality is made: $\epsilon_i \sim \mathcal{N}(0,\Sigma)$}. Under these general assumptions, while the stochastic disturbance terms are uncorrelated over the sample, they can be correlated between equations. This latter phenomenon of correlation between stochastic disturbance terms in different equations (due to the fact that there is usually more than one endogenous variable in each equation) is an essential feature of the simultaneous-equation system econometric model and the principal reason why it must be estimated using simultaneous-equation\footnote{\citet{Hausman:1983aa}} (or VAR models) rather than single-equation techniques.

It is usually assumed that $\Gamma$ is non-singular, se we can write the model \ref{eq:basicmodel} as the reduced form:
\begin{equation}
y = x\Pi + u
\end{equation}
where $\Pi = B\Gamma^{-1}$ and $u = \epsilon \Gamma^{-1}$. We can write the assumptions:
\begin{itemize}
\item zero mean, $E(u_i)=0$
\item same covariance matrix of $u_i$ at each observation, $cov(u_i)=\Omega, \forall i$
\item uncorrelated over the sample, $E\left(u_i' u_j \right) = 0, \forall i\neq j$
\end{itemize} 

These assumptions summarize the stochastic specification of the reduced-form equations. Under these assumptions the conditions of both the Gauss-Markov Theorem and the Least Squares Consistency Theorem are satisfied for the reduced-form equations, so the least squares estimators: $\hat{\Pi}=\left(X'X \right)^{-1}X'Y$.

The problem of identification is that of using estimates of reduced-form parameters $\Pi$ and $\Omega$ to obtain estimates of structural-form parameters $\Gamma$, $B$, and $\Sigma$. Once estimated, we have $\hat{B}=\hat{\Pi}\hat{\Gamma}$ and $\hat{\Sigma}=\hat{\Gamma}'\hat{\Omega}\hat{\Gamma}$. But we can imagine a "bogus" system multiplied by a matrix $R$: $y\Gamma R + xB R = \epsilon R$, then for this system, the estimates of $\Pi$ and $\Omega$ would be the same as the one form model \ref{eq:basicmodel}. We say that the system is identified in that all structural parameters can be determined from the reduced-form parameters. In this case, we need a priori information, which are restrictions on the structural parameters imposed prior to the estimation of the reduced form.

\section{Python: Data and model of French hospitals deaths with Covid-19}

\subsection{Data and limitations}
We use\footnote{code: 20200807\_COVID\_19\_FR.py} data from \href{https://www.data.gouv.fr/fr/datasets/donnees-hospitalieres-relatives-a-lepidemie-de-covid-19/
}{Sante publique France} published each day at 19h00. We focus on the number of recorded deaths in French hospitals as we would not rely on confirmed cases as the test campaign is not systematic.

Nota bene: 
\begin{itemize}
\item under-reporting:
\begin{itemize}
\item as some Covid-19 people are transferred via trains to other regions, this is creating a downward bias in our data, so for the Parisian and East of France data death toll would be higher
\item there are limited capacities in hospitals, meaning that in highly infected areas, hospitals might not be able to treat all infected people which are then either transferred to other regions' hospital or left to struggle outside of hospitals' monitoring (at home, in the street...)
\end{itemize}
\item we talk about "death with covid-19" which seems more precise than "death due to covid-19", we have no information if this would make a difference in our analysis
\end{itemize}

\subsection{Overview of death with covid-19 in France}
We first have a look at the total recorded death in French hospitals with covid-19 since the 18th of March 2020:
\begin{center}
\includegraphics[scale=0.5]{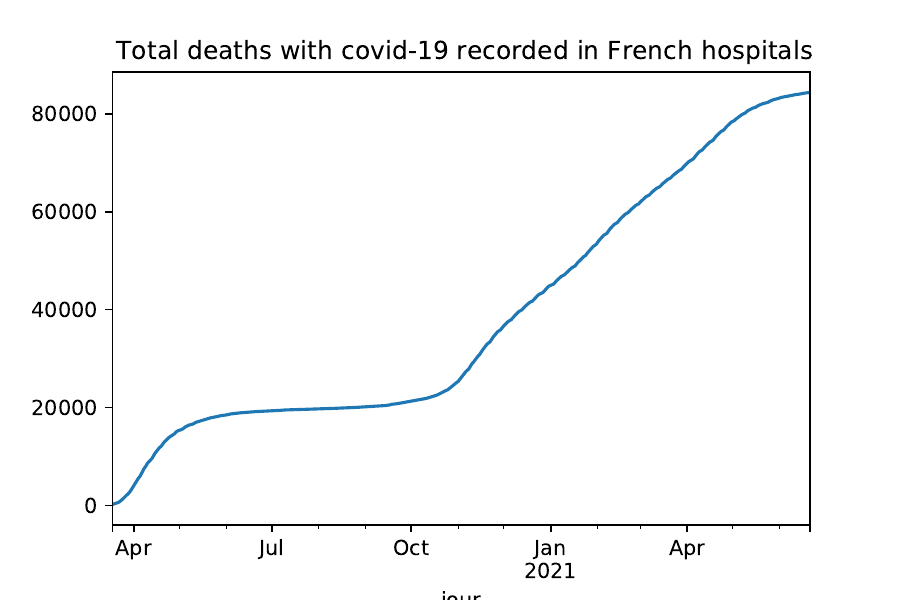}
\end{center}
\begin{center}
\includegraphics[scale=0.5]{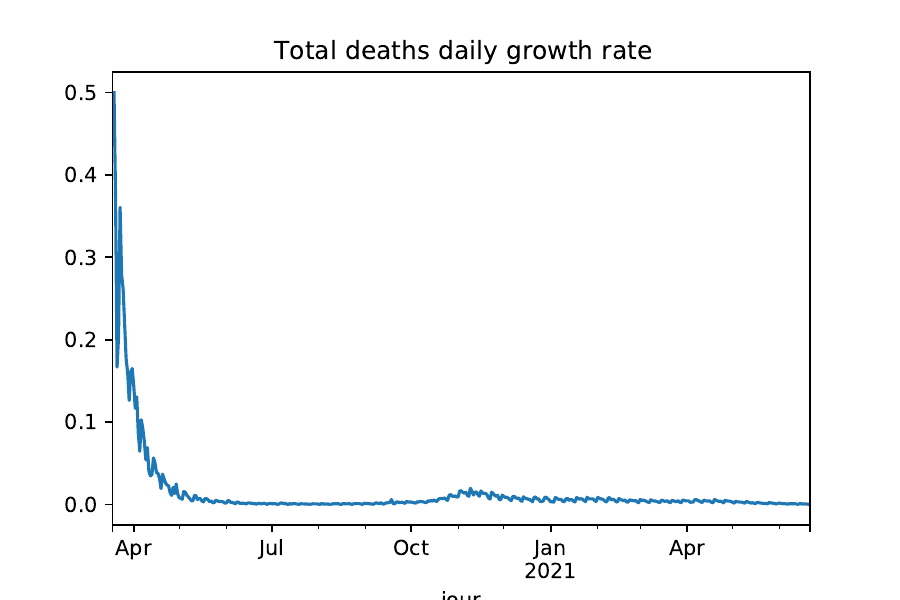}
\end{center}

\subsection{The SIR model}

Since the work of \citep{Kermack1927aa}, many models have been introduced to model the spread of epidemics. We will limit our analysis here to this SIR model. We write:
\begin{equation}
d_c(t) = \alpha I(t)
\end{equation}
with $I(t)$ the infectious compartment at time $t$, $d_c(t)$ the cumulative deaths at time $t$ and $\alpha$ the mortality rate of the epidemics, assumed constant.
We do not develop the model here, but then it can be approximated in the early phase of the epidemics:
\begin{equation}
\frac{\partial d_c(t)}{\partial t} = \alpha \mu (R_0 - 1) \exp \left[\mu (R_0-1)t \right]
\end{equation}
which is to say that at an early stage of the epidemics, the death dynamics in log should fit a linear regression:
\begin{equation}\label{eq:SIR}
\log\left( d_c(t) \right) = \beta_0 + \beta_1 t + \epsilon_t
\end{equation}

If the equation \ref{eq:SIR} is correct over the full sample, then simply taking the log of total death would be a linear function over time:
\begin{center}
\includegraphics[scale=0.5]{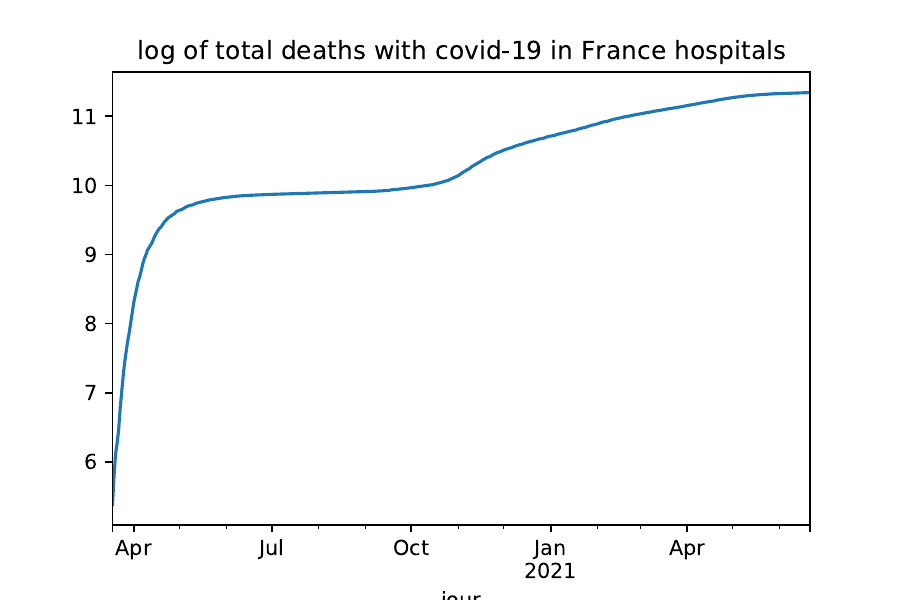}
\end{center}

If it seems linear in the early phase, then maybe the lock-down measures starting mid-March transformed the dynamics.

We will see that in very early days of the epidemics, an exponential model was worryingly fitting:
\begin{center}
\includegraphics[scale=0.5]{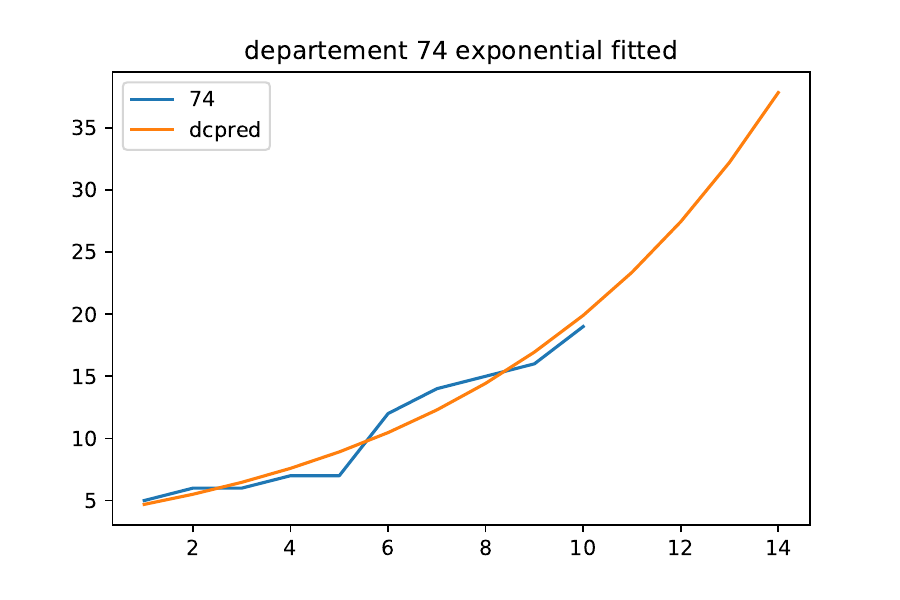}
\end{center}

\subsubsection{Fit SIR early epidemics on total French data}

We fit our data to the model \ref{eq:SIR}:

\begin{center}
\begin{tabular}{lclc}
\toprule
\textbf{Dep. Variable:}    &      logdc       & \textbf{  R-squared:         } &     0.522   \\
\textbf{Model:}            &       OLS        & \textbf{  Adj. R-squared:    } &     0.518   \\
\textbf{Method:}           &  Least Squares   & \textbf{  F-statistic:       } &     136.7   \\
\textbf{Date:}             & Fri, 07 Aug 2020 & \textbf{  Prob (F-statistic):} &  8.68e-22   \\
\textbf{Time:}             &     09:36:42     & \textbf{  Log-Likelihood:    } &   -125.45   \\
\textbf{No. Observations:} &         127      & \textbf{  AIC:               } &     254.9   \\
\textbf{Df Residuals:}     &         125      & \textbf{  BIC:               } &     260.6   \\
\textbf{Df Model:}         &           1      & \textbf{                     } &             \\
\bottomrule
\end{tabular}
\begin{tabular}{lcccccc}
                   & \textbf{coef} & \textbf{std err} & \textbf{t} & \textbf{P$> |$t$|$} & \textbf{[0.025} & \textbf{0.975]}  \\
\midrule
\textbf{Intercept} &       8.1724  &        0.117     &    69.895  &         0.000        &        7.941    &        8.404     \\
\textbf{day}       &       0.0185  &        0.002     &    11.690  &         0.000        &        0.015    &        0.022     \\
\bottomrule
\end{tabular}
\begin{tabular}{lclc}
\textbf{Omnibus:}       & 57.150 & \textbf{  Durbin-Watson:     } &    0.012  \\
\textbf{Prob(Omnibus):} &  0.000 & \textbf{  Jarque-Bera (JB):  } &  154.861  \\
\textbf{Skew:}          & -1.785 & \textbf{  Prob(JB):          } & 2.36e-34  \\
\textbf{Kurtosis:}      &  7.064 & \textbf{  Cond. No.          } &     148.  \\
\bottomrule
\end{tabular}
\end{center}

Warnings: \newline
 [1] Standard Errors assume that the covariance matrix of the errors is correctly specified.

We would not want to put too much meaning on the estimates of $\beta_0$ nor $\beta_1$, what we do in the following is to fit an OLS on the first $90\%$ of the points and then see if the following $10\%$ follows this fitted exponential or deviates from it:
\begin{center}
\includegraphics[scale=0.5]{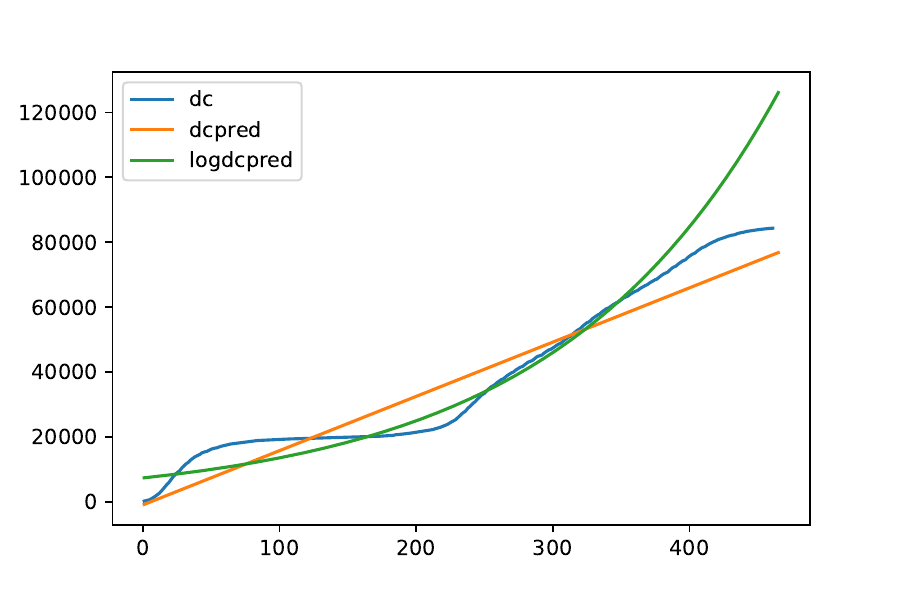}
\end{center}

In both cases, we see a very poor fit our the \ref{eq:SIR} or simple OLS model.

\subsection{Model by parts}
As the dynamic of the COVID-19 pandemic evolved over time in France, we might want to decompose the period:
\begin{enumerate}
\item an exponential phase as model \ref{eq:SIR} from 2020-03-18 to 2020-03-28
\item a linear phase from 2020-03-28 to 2020-04-18
\item a almost-logarithm model
\item then a linear part
\item etc.
\end{enumerate}

\subsection{Fit SIR early epidemics on some French departments}

We next focus on some of the most impacted regions (departement) in France:
\begin{itemize}
\item 13 Bouches-du-Rhone
\item 57 Moselle
\item 59 Nord
\item 67 Bas-Rhin
\item 68 Haut-Rhin
\item 74 Haute-Savoie
\item 75 Seine (Paris)
\item 76 Seine-Maritime
\item 93 Seine-Saint-Denis
\item 94 Val-de-Marne
\end{itemize}


\subsection{Conclusion on modelling the COVID-19 death evolution in French hospitals}

While it might be early to conclude, one can say while the death toll with covid-19 rises in France, the lock-down measures could be part of the explanation while the dynamics doesn't fit the expected exponential trend in the early phase and seems to follow a linear from day $20$ after the start of the lock-down period, then even a decreasing trend.

\citep{atkeson2020} challenge the efficacy of lock-down measure, finding that over several locations, \begin{quote}
the growth rates of daily deaths from COVID-19 fell from a wide range of initially high levels to levels close to zero within 20-30 days after each region experienced 25 cumulative deaths
\end{quote} and this could be attributed to voluntary social distancing, the network structure of human interactions, and the nature of the disease itself.

\section{python: Nonlinear time series model: an introduction}

The Covid-19 brought both behavioral changes (citizens changed their habits when it came to working, shopping, etc.) and structural changes (in places where a lock-down was imposed, some business were imposed to shut down). Such changes can be sudden and nonlinear. We introduce methods to test the nonlinearities and some models which allow a switch from different regimes.

\subsection{BDS test for nonlinearity: non iid time series}\label{sec:BDStest}

\citet{Broock1996aa} developed a test for nonlinearity with the null hypothesis $H_0$: independent and identical distribution of a random variable. If $H_0$ can be rejected on residuals from a fitted linear time series model, then the linear model is likely to be mis-specified.

For the quarterly US GNP growth rates\footnote{vansteenberghe\_MSAR.py}, we reject $H_0$, hence they are not iid.

\subsection{Markov-switching model of growth rates}

\citet{Hamilton1989aa} explored discrete shifts in regime where the economy might be either in a fast or slow growth phase. Taking $y_t$ as the quarterly growth rate of the real US GNP in percent:
\begin{equation}\label{eq:Hamilton1989}
    \begin{cases}
y_t = \alpha_1 S_t + \alpha_0 + z_t \\        
z_t = \sum_{i=1}^r \phi_i z_{t-i} + \epsilon_t \\
\epsilon_t \sim \mathcal{N}\left(0, \sigma^2 \right)\\
P\left(S_t=s_t\vert S_{t-1}=s_{t-1}\right) = \begin{bmatrix}
q & 1-p  \\
1-q & p 
\end{bmatrix}
    \end{cases}
\end{equation}

The original data set was taken from the \textit{Business Conditions Digest} we compare it with the modern Federal Reserve Bank  of St. Louis data, we observe some discrepancies while the observations are consistent both in terms of sign and relative magnitude:
\begin{center}
\includegraphics[scale=0.5]{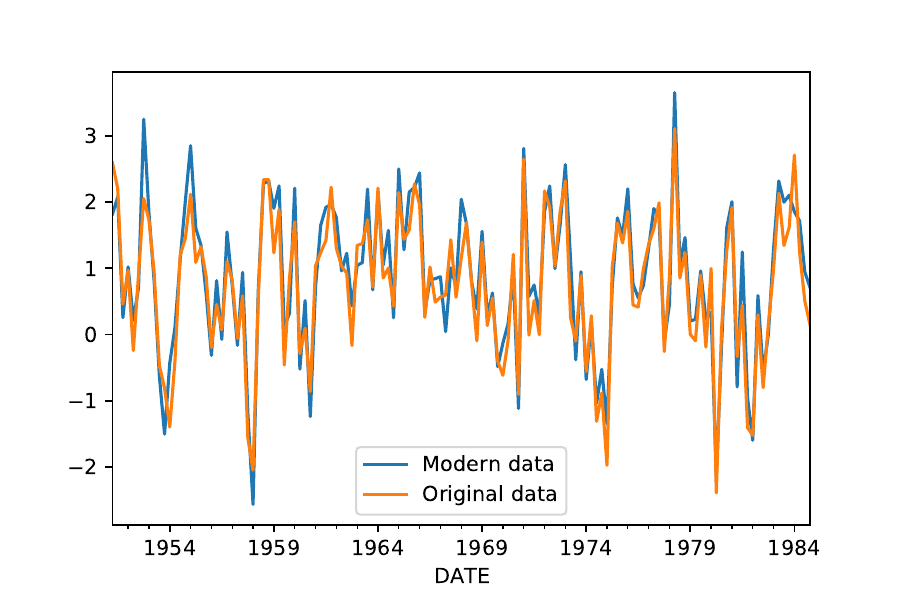}
\end{center}

With the original data set, we are able to replicate the model calibration and the inferred probability that the economy was in the falling GNP state at date $t$. But we find that this model is very sensitive to the data, with the updated data set, we do not find a model that is as precise to inferred probability was in the falling GNP state:
\begin{center}
\includegraphics[scale=0.5]{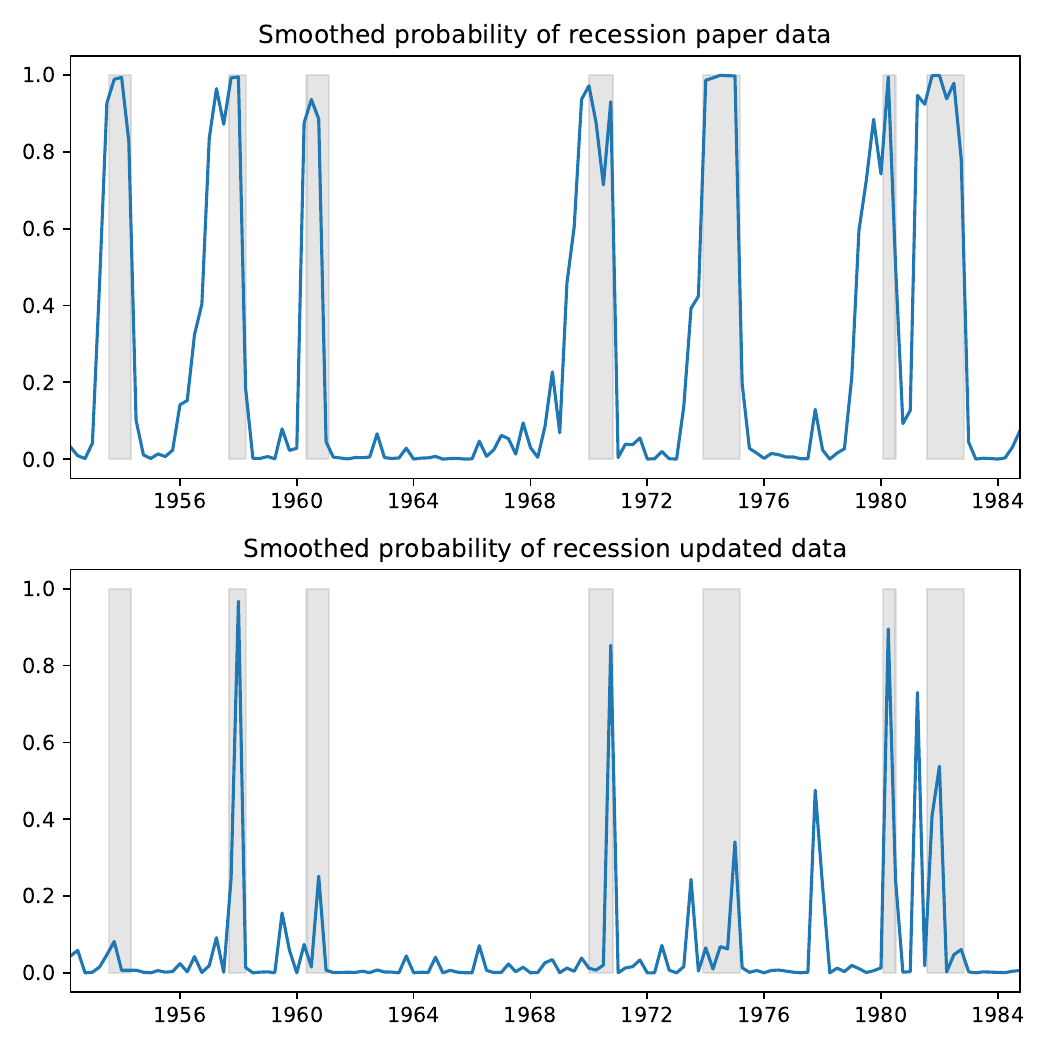}
\end{center}

If we now take the full period up to end of 2019, fitting an AR(3) as suggest by the AIC, we are able to infer correctly some states (1950, 1970, 2008) but with some false positives. If now we include the year 2020 in our data set to calibrate our model, 2020 was such an outlier that we cannot infer any other dates for the falling state:
\begin{center}
\includegraphics[scale=0.5]{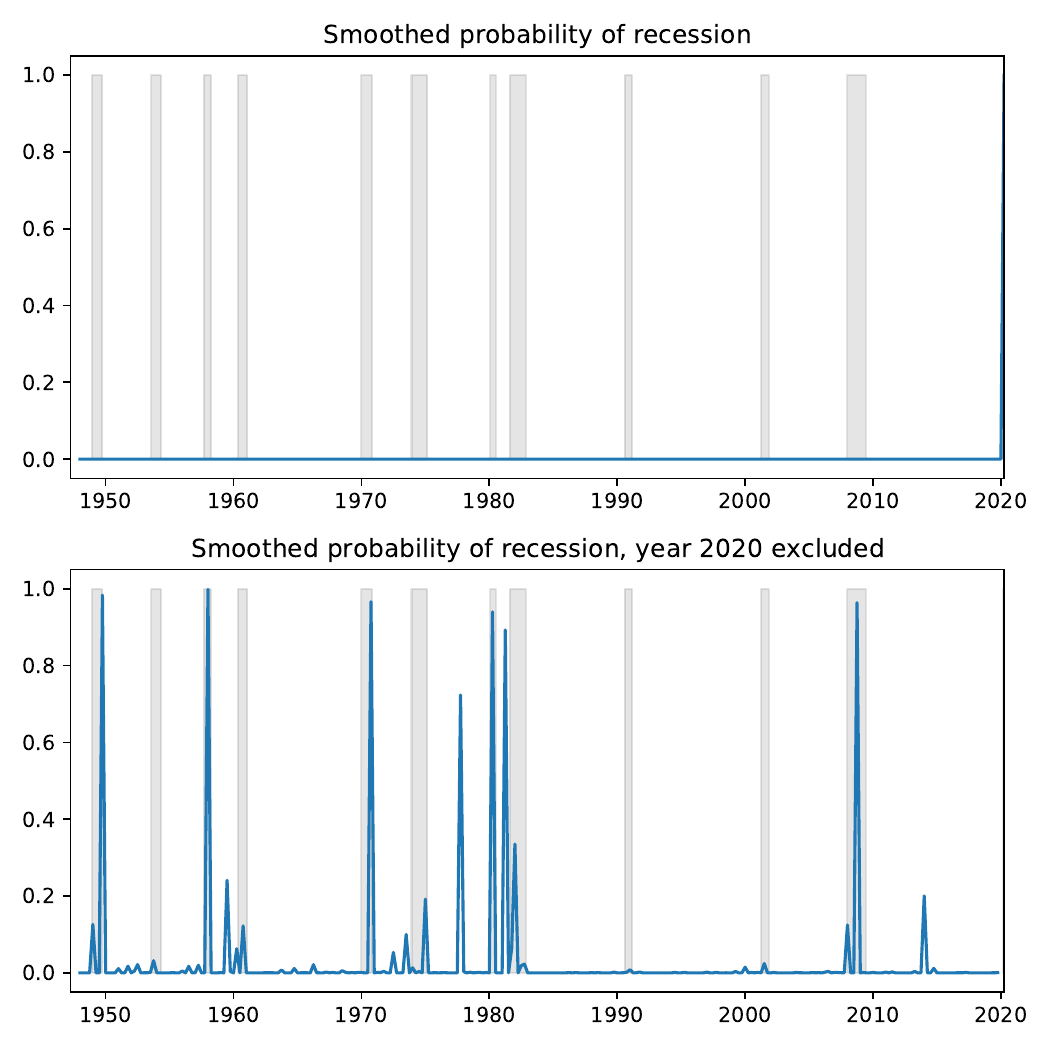}
\end{center}

This paper was discussed with some suggestions, namely by \citet{Hansen:1992aa}.

\subsection{Three state Markov-switching process}

\citet{Kim1998aa} estimates conditional distributions of parameters as well as the unobserved state they consider as random variables. Following \citet{Hamilton1994aa}, they model the demeaned monthly stock market returns $y_t$ as a three state Markov-switching process following:
\begin{equation}
    \begin{cases}
y_t \sim \mathcal{N}\left(0, \sigma_t^2 \right) \\        
\sigma_t^2 = \sum_{i=1}^3 \sigma_i^2 S_{it} \\
S_{kt}=1 \text{ if } S_t=k, \text{ and } S_{kt}=0 \text{ otherwise; } k=1,2,3\\
Pr\left[ S_t = j \vert S_t=i \right] = p_{ij} \text{  } i,j=1,2,3
    \end{cases}
\end{equation}

Applying\footnote{vansteenberghe\_variance\_switching.py} a BDS test on $y_t$ we reject that the returns are iid.

We find the same parameters as in the paper with the following smoothed probability of the variance to be in each of the three states:
\begin{center}
\includegraphics[scale=0.5]{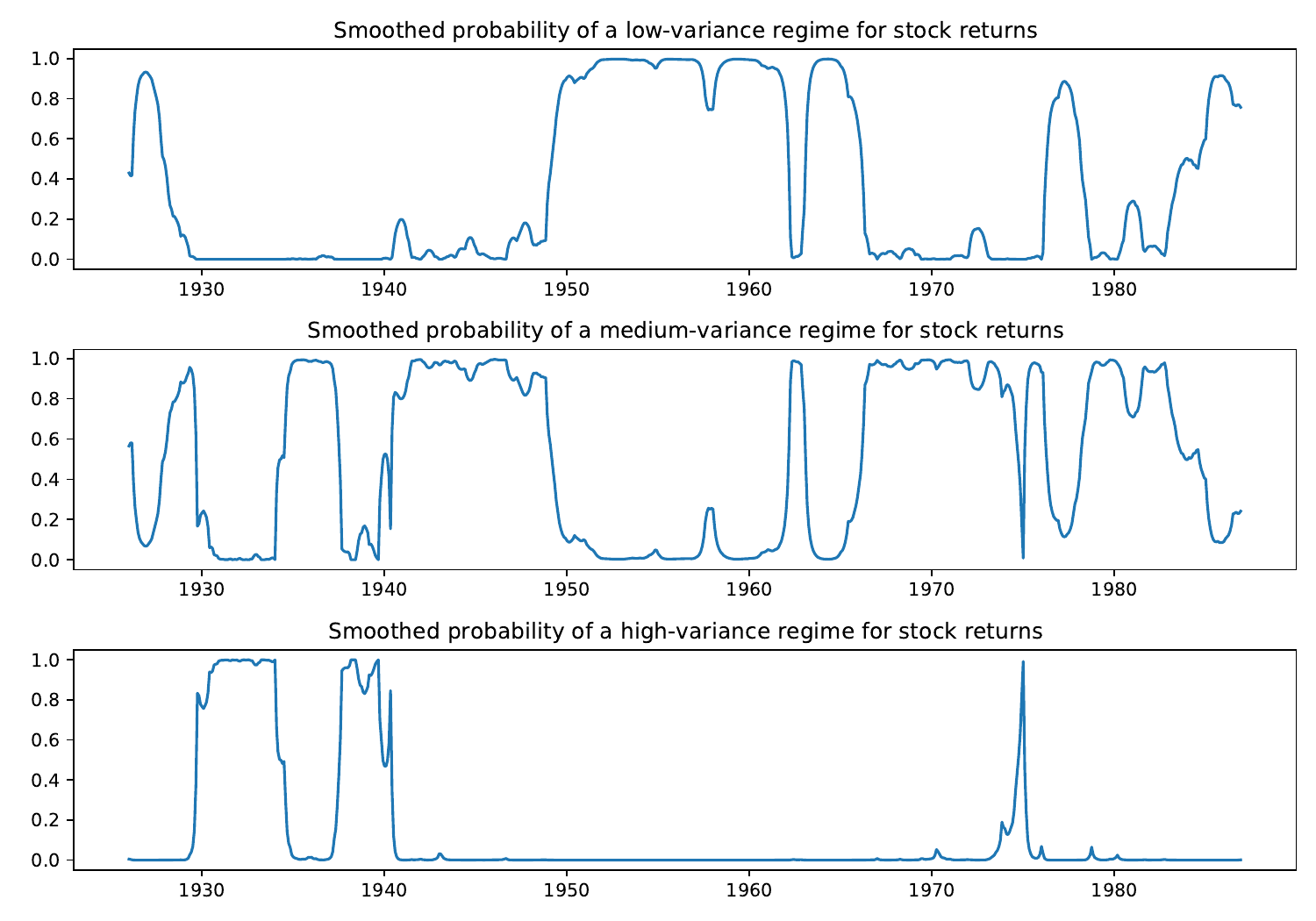}
\end{center}
with the estimated variance:
\begin{center}
\includegraphics[scale=0.5]{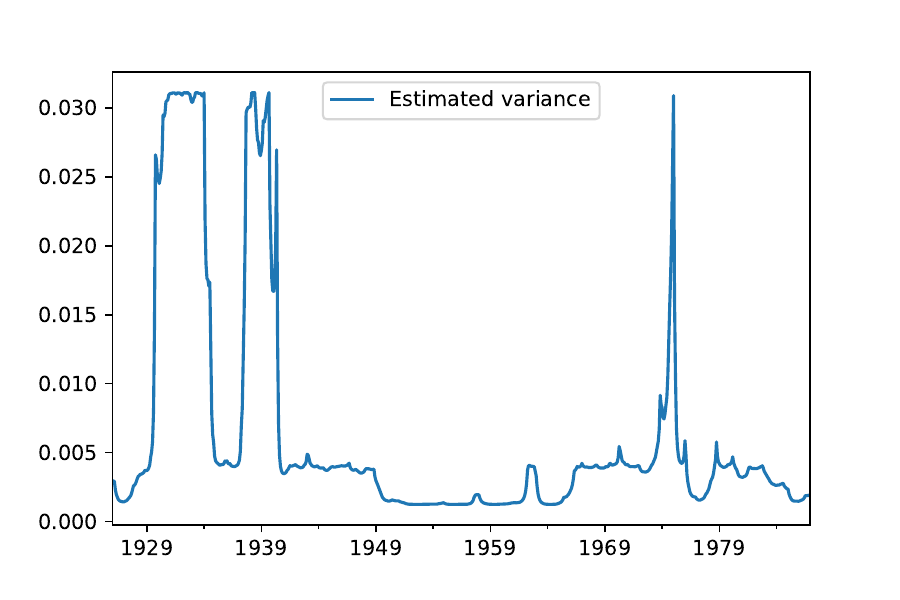}
\end{center}
the standardized returns $\frac{y_t}{\sigma_t}$ now looks more iid and a BDS test doesn't reject this assumption:
\begin{center}
\includegraphics[scale=0.5]{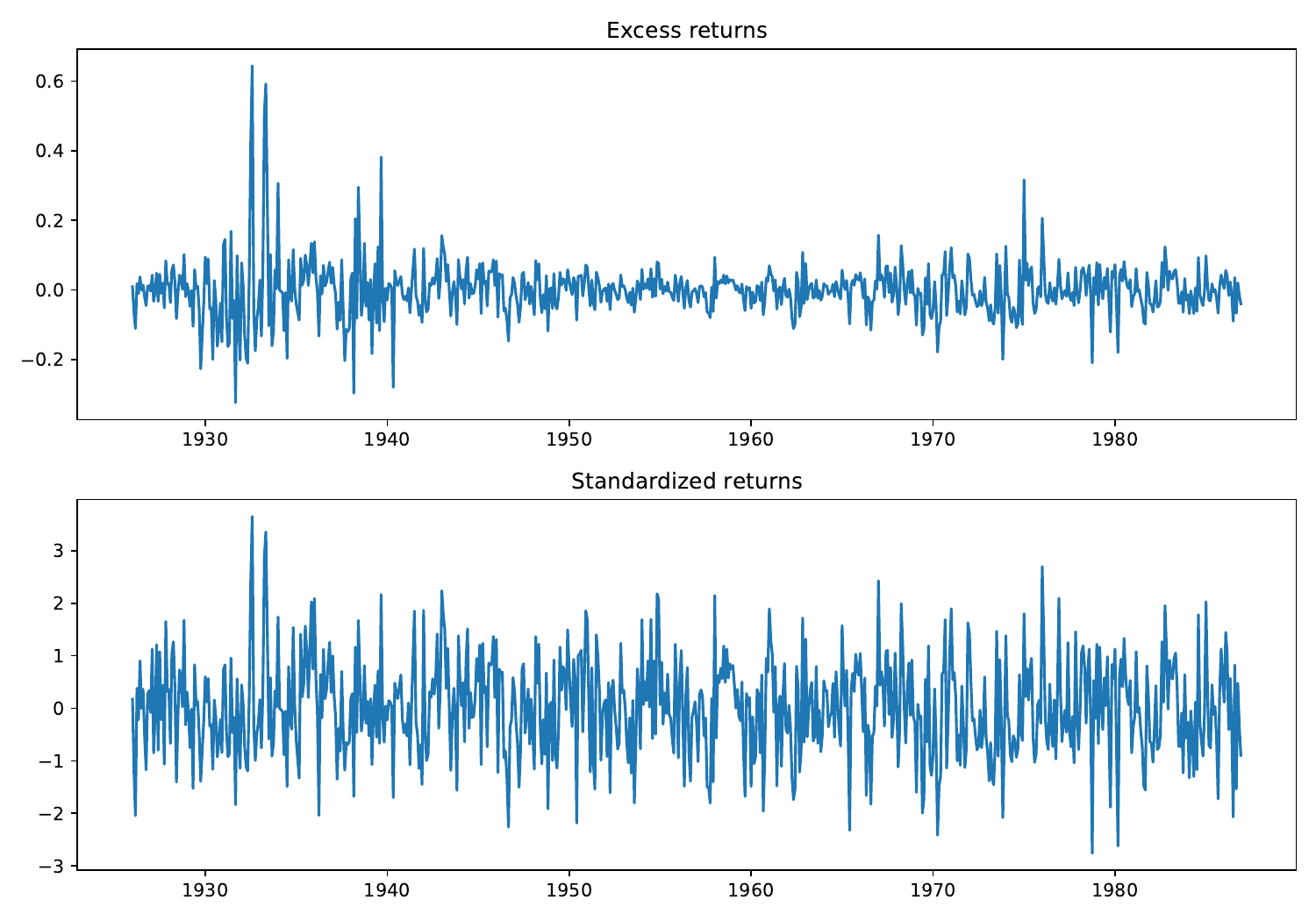}
\end{center}
and the histograms:
\begin{center}
\includegraphics[scale=0.5]{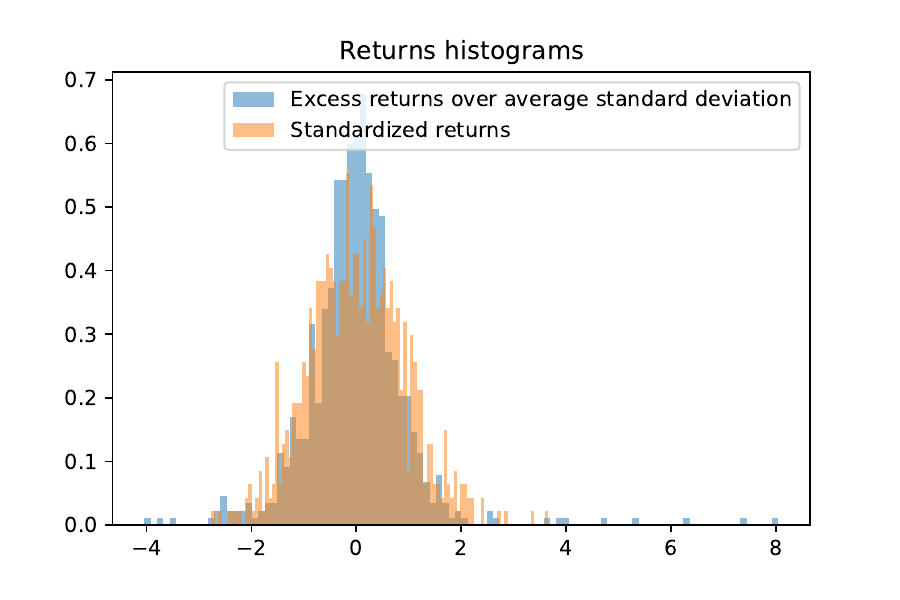}
\end{center}

\subsection{Time-Varying Transition Probabilities}

\citet{filardo1994} also assumes that the state of the economy cannot be known with certainty and allow for a time-varying transition probability between states. He modifies equation \ref{eq:Hamilton1989} with:
\begin{equation}
P\left(S_t=s_t\vert S_{t-1}=s_{t-1}\right) = \begin{bmatrix}
q\left(z_t\right) & 1-p\left(z_t\right)  \\
1-q\left(z_t\right) & p\left(z_t\right) 
\end{bmatrix}
\end{equation}
where $z_t$ is an economic-indicator variable and applies it to the seasonally adjusted US industrial production (NBER recessions in grey):
\begin{center}
\includegraphics[scale=0.5]{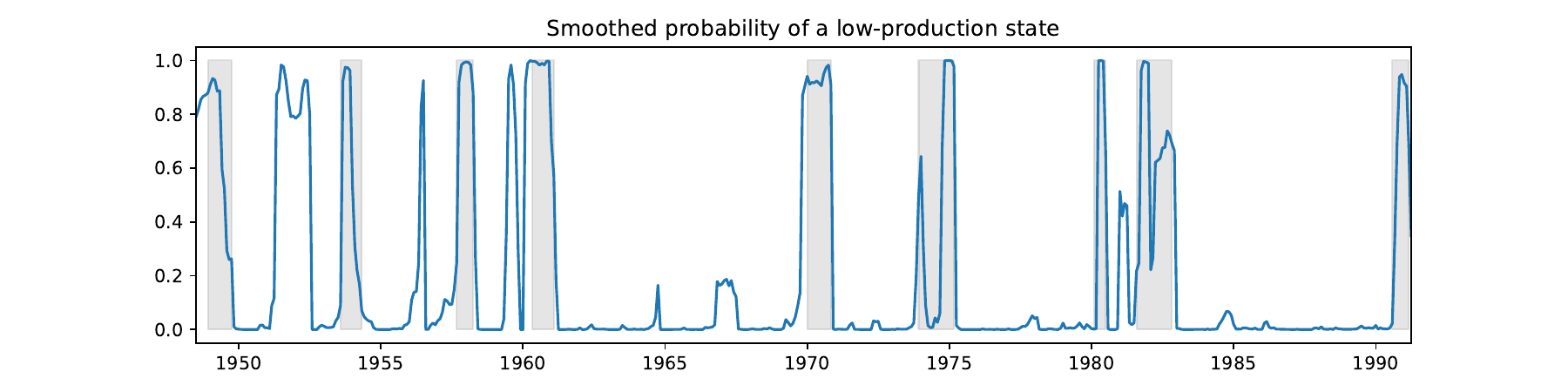}
\end{center}

We apply this model\footnote{20201104\_TVTP.py} to more recent data, seasonally adjusted US industrial production and use the S\&P 500 as an economic-indicator variable (as one configuration in \citet{filardo1994}:
\begin{center}
\includegraphics[scale=0.5]{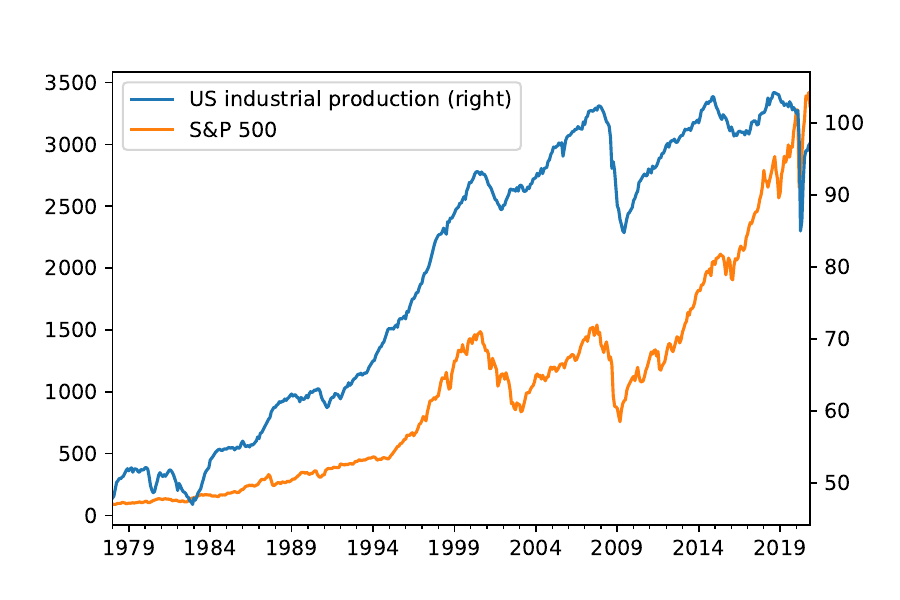}
\end{center}

If we leave year 2020 out of our sample, the outcome can be interpreted although it is difficult to avoid false positives:
\begin{center}
\includegraphics[scale=0.5]{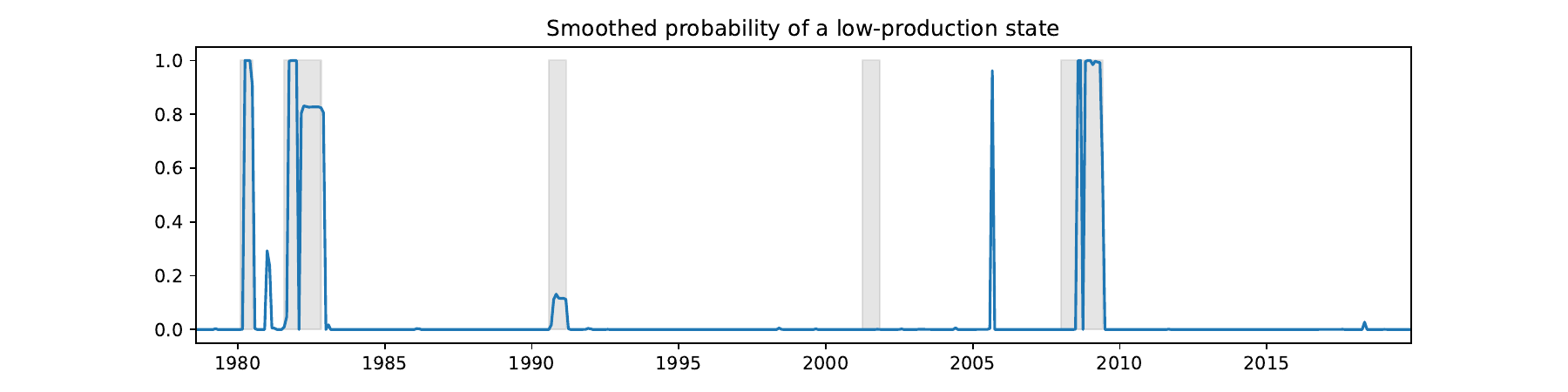}
\end{center}
But if we include year 2020 in our sample, then we cannot really interpret the model as it predicts low production state for most of the observation period. In 2020, the production dropped severely while the financial markets faced low turbulence, hence our model become more (too?) sensitive.

\subsection{Mean and variance switching states}

\citet{mcconnell2000} combined mean and variance state switching to model US output fluctuations and identify periods or recessions and expansions:
\begin{equation}\label{eq:McConnell2000}
    \begin{cases}
y_t = \mu^{S_t} + \Phi \left( y_{t-1} - \mu_{S_{t-1},V_{t-1}} \right) + \epsilon_t \\        
\epsilon_t \sim \mathcal{N}\left(0, \sigma_{V_t}^2 \right)\\
P\left(S_t=s_t\vert S_{t-1}=s_{t-1}\right) = \begin{bmatrix}
q & 1-p  \\
1-q & p 
\end{bmatrix}\\
P\left(V_t=v_t\vert V_{t-1}=v_{t-1}\right) = \begin{bmatrix}
u & 1-v  \\
1-u & v 
\end{bmatrix}
    \end{cases}
\end{equation}
 In our application\footnote{20201105\_McConnell.py}, we merge $S_t$ and $V_t$ into one latent variable and we find that low-production state are associated with higher variance. We find that the Hamiltonian model is not fit to identify low-production state\footnote{in the Hamiltonian model, equation \ref{eq:McConnell2000} is modified with $V_t$ removed}. Our implemented version of the \citet{mcconnell2000} model allows to identify low-production states probability in line with observed recessions, with still some false positives and false negatives:
 \begin{center}
 \includegraphics[scale=0.5]{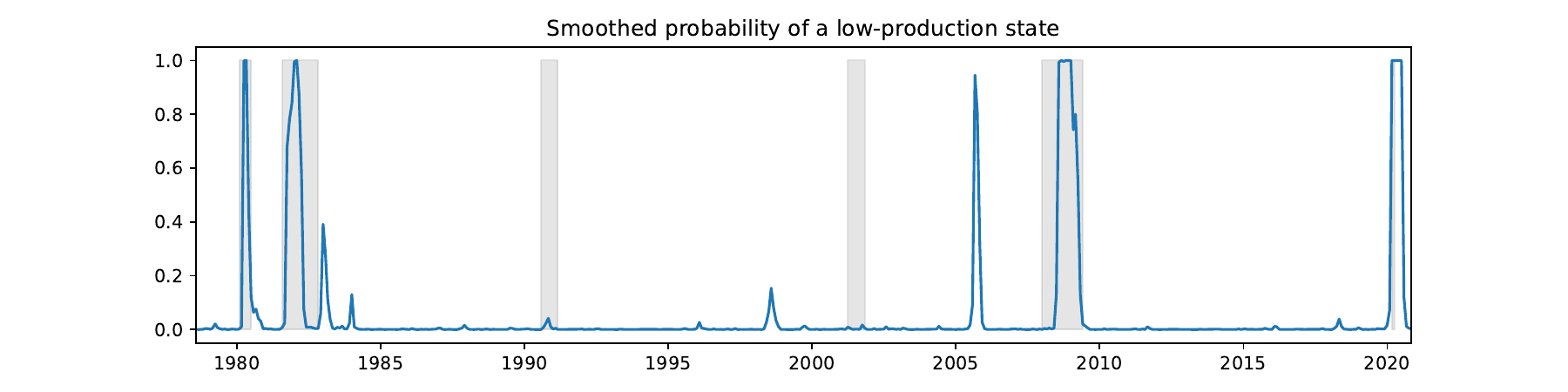}
 \end{center}
 
 We might want to use another economic-indicator variable, such as the National Financial Conditions Index (NFCI), the conclusions are similar:
 \begin{center}
 \includegraphics[scale=0.5]{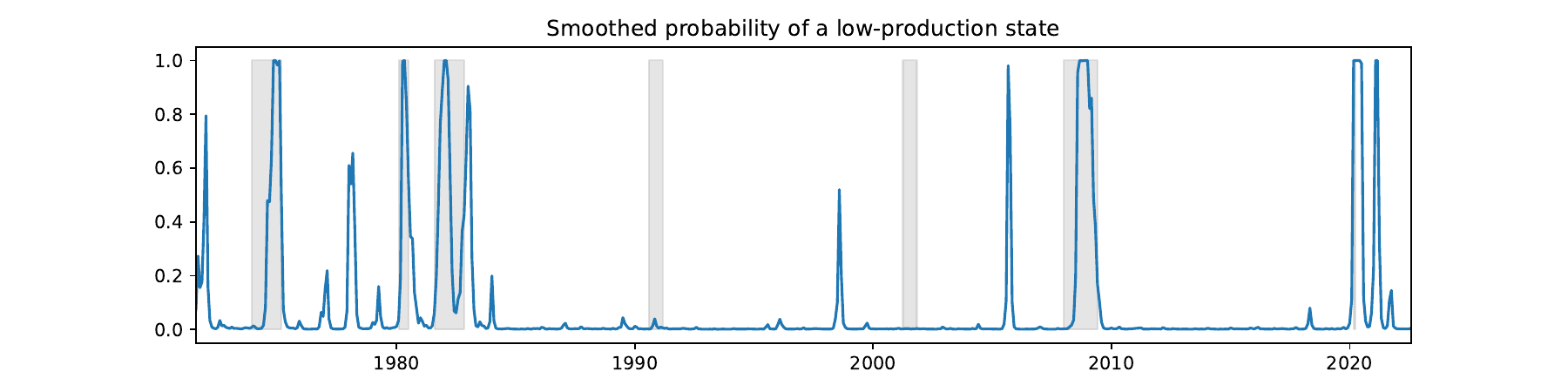}
 \end{center}
 
 Following these works, \citet{adrian2019} argue 
 \begin{quote}
 that the entire distribution of GDP growth evolves over time, with the left tail of the distribution positively correlated with slack in financial conditions.
 \end{quote}
 
\section{Python: Event studies}

There is an abundant academic literature on event studies. The main application is to measure the effects of an economic event on the value of a firm. Methods are described in \citep{MacKinlay:1997aa}.

\subsection{Event study on fines impact on banks} \label{sec:fines}

As in the 2016 paper on regulatory sanctions impact in the UK \citep{armour}, we suggest to work on the impact of US fines on banks.

\subsubsection{Methodology}
We recap the main methodology for an event study.

The abnormal return for a given bank $i$ at time t is:
\begin{equation}
AR_{i,t}=R_{i,t}-\alpha_{i,t}-\beta_{i} R_{m,t}
\end{equation}
With $R_{i,t}$ and $R_{m,t}$ the daily returns of the bank $i$ and of the market $m$. In Europe, the market can be proxied using Eurostoxx index.

The $\alpha$ and $\beta$ are obtained with a simple OLS regression on a one-year window, so from $t-261$ to $t-2$.

This measure can help decide if on a specific date for a specific bank $i$ there was a significant event for the market.

In order to assess whether all the $N$ banks were impacted, we can compute the average abnormal return:
\begin{equation}
AR_t=\sum_{i=i}^N \frac{AR_{i,t}}{N}
\end{equation}

Finally the cumulative average abnormal return for the period $[t_1,t_2]$ is:
\begin{equation}
CAR(t_1,t_2)=\sum_{t=t_1}^{t_2} AR_t
\end{equation}

As in the paper \citep{armour}, we encourage to winsorize the abnormal returns to 90\% percentile of data.

Finally, it is possible to compute a reputational loss as :

\begin{center}
Reputational Loss = $\Delta V_t$ - Fine
\end{center}

With $\Delta V_t$  the change in market capitalisation in the event window around the announcement.

\subsubsection{Data}

The journal \emph{Le Monde} has released a list of bank fines \href{https://docs.google.com/spreadsheets/d/1yR182ikRzJsb1l5a5-1cLnXsWb4IVJTjsZBOcRxIfv4/edit#gid=0}{here}.

The data range from 2007 to 2016.

We suggest to focus on banks impacted by descending order of fines.

So let say: Bank of America, JPMorgan, BNP, Deutsche Bank.

You will have to define the exact day of the announcement of the fine.

Start with daily returns, then weekly or monthly and identify if the announcement had an impact and generated an abnormal return on the bank, and what about other banks on the same day?

\subsection{Event study on stress test results impact on banks}

For an event study on the effects of the publication of stress test results on bank stock prices, we suggest to follow\footnote{code: market\_reaction\_ST\_vansteenberghe.py} the same method presented section \ref{sec:fines}.

We focus on 53 quoted banks over a 10 year horizon. The window for the regression is one year, so 365 calendar days.

The red dotted line is the day of announcement of the Brexit, 24th of June 2016 and the green dotted line indicates the day of the announcement of the EBA stress test results.

We can illustrate the data with the stock prices and returns for Deutsche Bank AG.

We "zoom" on Deutsche Bank stock prices around the dates of the Brexit and EBA stress test announcements:

Note that the code can take some time to run as we apply 200,000 regressions and computations on returns.

Then we can apply a threshold, for example $4\%$ and observe the selected dates related news that could explain such an abnormal return across the board:
\begin{itemize}
\item 18 Sep 2008 Les banques centrales des grands pays industrialises se sont concertees jeudi et ont a nouveau injecte massivement des liquidites sur les marches financiers
\item 18 Jan 2012 World Bank fears Europe's crisis could set off deeper global slump than Lehman collapse
\end{itemize}

Now if we apply a negative threshold of $-4\%$, the day of the announcement of the Brexit clearly appears.

The trading day after the EBA stress test results publication, on the first of August 2016, the observed average abnormal return is around $-1\%$. The day after, on the second of August 2016, the observed average abnormal return is around $-3.4\%$. Either the market needed some time to assess the results or those effects would rather be "dominant" compared with the results publication:
\begin{itemize}
\item 2nd August 2016: Deutsche Bank and Credit Suisse booted from top stock market index
\item 2nd August 2016: Monte dei Paschi rescue deal over its 10 billion euro worth of non-performing loans was "agreed" but "investors are starting to doubt just how viable the rescue plan" according to the Business Insider and Commerzbank results were published and disappointing
\end{itemize}

We could also observe bank by bank to identify which day saw significant abnormal returns and see if the trading day following the publication of the stress test results, so the first of August 2016 saw significant abnormal return or see the applicable threshold and how many trading days saw similar or above abnormal return over the last 9 years.

Finally, we plot the cumulative average abnormal returns over a window of one year (365 days) and "zoom" on early 2016, placing the date of the EBA stress test results announcements end of July 2016 or the day of the Brexit announcement on 24th of June 2016.

 \begin{center}
 \includegraphics[scale=0.5]{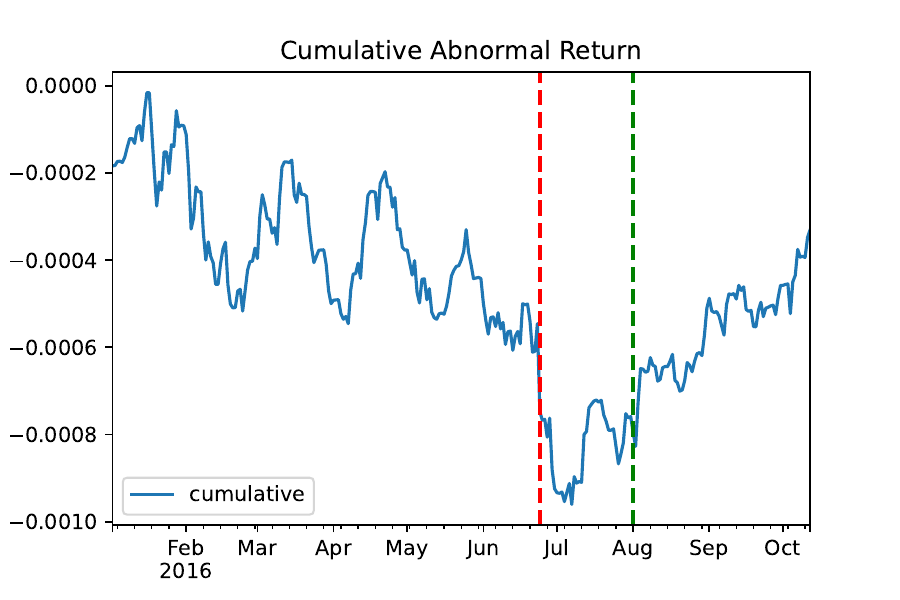}
 \end{center}

\section{Python: Manipulating financial data and some investment strategies}
\subsection{Share price history}

We wrap up what we've learn so far in simple examples\footnote{financial\_data\_vansteenberghe}:
\begin{itemize}
\item get data from the internet or from a file
\item \textit{clean} the dataset
\item \href{http://pandas.pydata.org/pandas-docs/stable/indexing.html}{indexing} and \href{http://pandas.pydata.org/pandas-docs/stable/merging.html}{joining, merging} a \href{http://www.python-simple.com/python-pandas/panda-intro.php}{DataFrame}
\item plot and save the data
\item create a function (\href{http://anh.cs.luc.edu/python/hands-on/3.1/handsonHtml/functions.html}{help on this} or \href{http://www.tutorialspoint.com/python/python_functions.htm}{here}) to improve the harvest of data
\item select and search through data via for and/or while loops (\href{https://www.tutorialspoint.com/python/python_for_loop.htm}{example})
\item observe descriptive statistics
\item simulate investment decisions, assess their performance
\end{itemize}

\subsubsection{Data set cleaning}

We inspect our data set and clean for missing data or "suspicious" data points (this can be peaks that cannot be explained). In our first approach, sort each stock price time series per maximum deviation factor to its mean, the deviation factor is computed as:
$$\max \frac{P_{i,t} - \bar{P}}{\sigma}$$
with $\bar{P}$ the mean of the price over the observed period and $\sigma$ the standard deviation of the price over the observed period. If we find some anomalies or outliers\footnote{you can find section \ref{sec:outlier} some further approaches to detect outliers}, we need to first remove those anomalies before proceeding to our analyses. Also, we can check for prices or returns stuck at $0$, another indicator of missing data.

Nota bene: if after some period the price is stuck at a given level, this could be a good indicator of a default of a company. In this exercise we are likely to remove such asset from the sample which give a positive bias to any trading strategy as we could have picked a company that would have defaulted and lost our position.

\subsubsection{Daily returns}

We can now compute the daily returns for the stock $i$:
\begin{equation*}
\mbox{return}_{i,t}=R_{i,t}=\frac{P_{i,t}}{P_{i,t-1}}-1
\end{equation*}

Applying a Taylor expansion of $\ln(x)$ when $x$ is close to $1$:
$$\ln(x) = \ln(1) + \frac{x-1}{1} - \frac{(x-1)^2}{2}+ \ldots$$

If the price is not too volatile over time, that is if $\frac{P_{i,t}}{P_{i,t-1}} \simeq 1$, then we could also compute\footnote{$\ln \left(\frac{x_{i,t}}{x_{i,t-1}}\right) \simeq \frac{x_{i,t}}{x_{i,t-1}} - 1$} the daily returns for the stock $i$:
\begin{equation}
R_{i,t}=\ln \left(\frac{P_{i,t}}{P_{i,t-1}}\right)
\end{equation}

We can then plot the evolution of the share price and its daily return, we clearly observe the emission scandal effect on Volkswagen share price and returns.


We provide 10 years of price data for stock which would be eligible for a French "PEA": a selection of 495 shares in \verb|pea_price.csv| and \verb|pea_price_2.csv|, with time series from 2006 to 2016.

\subsubsection{Share's return in practice}

In practice, as an investor, you are not only interested in the growth of the share price, but also in the dividend $D_i$ the company will pay\footnote{we saw in section \ref{sec:euronext} the importance of dividends for the total return for an investor in the stock market}
\begin{equation*}
\mbox{return}_{i,t}=R_{i,t}=\frac{P_{i,t}- P_{i,t-1}+  D_{i,t}}{P_{i,t-1}}
\end{equation*}

There are also some split decision where share can be split or repurchase. This is why we work with \textbf{Adjusted Closed }values which account for both the dividends and the splits.

\subsubsection{A simple regression, but a useless one}


We might want to work on two french banks' stock price evolution, BNP and Societe Generale (GLE). Just as illustrations, we suggest a simple regression (Ordinary Least Square) of BNP daily log prices on Societe General daily log prices, with a constant:
\begin{equation}
y_{\mbox{BNP},t}=\alpha+\beta x_{\mbox{Societe Generale},t} +  \epsilon_t
\end{equation}
we find a $\beta$ of $0.45$, but in fact this should not be interpreted as a sensitivity between the two time series.

We might want to separate and execute two regressions, one where the Societe Generale stock price ies below 50 and the rest, we obtain a plot that looks better.

It looks as if we had something, but it is a catastrophe in terms of econometrics. You can see in section \ref{sec:spurious} why (think about stationarity, cointegration, etc.). In other terms, we might be looking at a spurious regression.

\begin{itemize}
\item In fact, we perform the following
\begin{enumerate}
\item ADF test on the series in level and in first differences.
\item Both series are I(1)
\item We do a cointegration test over the full sample and the time series are not cointegrated. Doing a cointegration test for the dates after 2008 the time series seems to be cointegrated.
\end{enumerate}
\end{itemize}

\stepcounter{question}
\begin{itemize}
\item \textbf{Question \arabic{question}}
\begin{itemize}
\item Perform an ECM for the dates after 2008.
\item measure correlations of prices and returns of both times series. What can you say with respect to section \ref{sec:corronreturns}
\end{itemize}
\end{itemize}

\subsubsection{Comparing correlation measures}
On our samples, we are comparing our correlation measures on returns\footnote{we also suggest reading \href{https://www.r-bloggers.com/cointegration-correlation-and-log-returns/}{these elements}} Pearson, Kendall, and Spearman, following the notions introduced in \ref{sec:largen}.

We sort our DataFrame according to the Kendall coefficient and check whether this is also monotonous for the Pearson and Spearman.

There is some "noise" in the Pearson plot, meaning that our sorting based on Kendall is not the same as it would be for Pearson.

As this is intensive in terms of computing power, we suggest to run this via a command line\footnote{from more tips on how to run a python script via the command prompt, see \href{https://www.pythoncentral.io/execute-python-script-file-shell/}{this web page}}. With MacOS, you can run a Terminal and try 
\begin{verbatim}
python '/pathtofile/filename.py'
\end{verbatim}

With WindowsOS, you can run cmd.exe, and give a command line:
\begin{verbatim}
C:\pathtoAnaconda3\python.exe C:\pathtofile\filename.py
\end{verbatim}

\subsubsection{Manipulating those concepts on our sample}

As a first simple approach of risk deversification, we can look through our 496 stocks which pairs are historically the most positively or negatively correlated. For this, we first compute the correlation matrix.
We ignore the diagonal component, unstack this matrix to have a list of all pairs in this correlation matrix.

We are then in a position to sort this list from minimum to maximum and we can, as an illustration, select the first and last elements of this list to plot the two most negatively and positively correlated stocks of our 495 stocks sample.

In our sample, the two most correlated stock prices are the ones of the companies: Essilor International and Dassault Systemes.

\quad

With a similar approach we can detect the least and most volatile stock.

Is there a correlation between stock capitalisation and volatility in our sample?
\begin{enumerate}
\item We compute the average capitalisation of each firm over the period.
\item Then compute the standard deviation of each stock over the period.
\item Finally we regress the average capitalisation over the standard deviation and we reject the model.
\end{enumerate}

\subsubsection{Computing the return of a portfolio - an introduction}

If we are working with Adjusted Close values of stocks, we can focus on the returns of the holdings (assuming that the dividends, splits, and other effects are taken into account in the prices).

\quad

We make the following assumptions for an unsophisticated investor:
\begin{itemize}
\item the initial cash to invest is $100,000$ \euro
\item each transaction cost (buy or sell) is $7$ \euro
\item we assume that based on a given criteria, our investor purchase the $10$ best performers
\item our investor dispatch uniformly its investment
\item all the position are sold at the end of the investment period and the return is computed
\end{itemize}

That is, if we have a sample of $N = 10$ best performers, a transaction cost $\mbox{TC}$ the number of shares our investor will want to purchase for a performer $i$ is:
\begin{equation*}
\frac{\mbox{Cash}_{t_0} -  \mbox{TC} \times N}{N} \times \frac{1}{P_{i,t_0}}
\end{equation*}

We have to keep in mind that our cash is limited, hence we have to round this to the inferior unit and keep whatever is left in the form of cash.

\subsubsection{Defining a scoring to allow basic selection}
Historical correlation and volatility can be very basic selection criteria for a portfolio, the capitalisation as well. With the suggested sample, we can plot the biggest and smallest capitalisation (in K EUR) and observe different behaviours.

Based on what we have done so far, we can compute for each stock its daily returns. Then we compute the average return over the period, its volatility, and the average capitalisation of the company.

An investor could want to invest in stocks that historically showed positive return, low volatility, and high capitalisation (the stock is generally more liquid). We can define a naive scoring for a stock $i$ out of our $N$ sample:
\begin{equation}
\mbox{score}_i= \alpha \frac{E(R_i)}{\frac{1}{N}\sum_{j=0}^{N-1} E(R_j)}  + \beta \frac{E(Cap_i)}{\frac{1}{N}\sum_{j=0}^{N-1} E(Cap_j)} + \gamma \frac{\sigma (R_i)}{\frac{1}{N}\sum_{j=0}^{N-1} \sigma (R_j)}
\end{equation}

\stepcounter{question}
\begin{itemize}
\item \textbf{Question \arabic{question}}
\begin{itemize}
\item suggest some values for $\alpha$, $\beta$ and $\gamma$ based on investor's risk aversion. \begin{enumerate}
\item Start with simple values of $\alpha$, $\beta$ and $\gamma$, sort your scores and compare with the returns one could expect with this selection.
\item Separate the sample into a training and a test sample.
\item Calibrate the weights on the training sample and compute the portfolio return (of the 10 best stocks selected) on the test sample. Is it convincing? Remember that the financial crisis happened during the training sample.
\end{enumerate}
\end{itemize}
\end{itemize}

\stepcounter{question}
\begin{itemize}
\item \textbf{Question \arabic{question}}
\begin{itemize}
\item Introduce the skewness and kurtosis into the selection criteria.
\end{itemize}
\end{itemize}

To follow up on this questions, you might want to go to section \ref{sec:CAPM}.

\subsubsection{Testing a strategy}
\stepcounter{question}
\begin{itemize}
\item \textbf{Question \arabic{question}}
\begin{itemize}
\item Define a strategy over time to trigger buying and selling orders.
\begin{enumerate}
\item Using CAC 40 data (investment in CAC4 0 ETF is meant) compare the performance of your portfolio against the return of investing in a CAC 40 index.
\item Simulate x times (100 or more) a random selection and create a list of returns. Compare this to the performance of a portfolio selected with your method. The distribution looks something around $10\%$.
\item Try to vary the dates you buy and sell based on some criteria to improve the performance of your portfolio.
\end{enumerate}
\end{itemize}
\end{itemize}

\subsection{Share price history: focus on French biotech}

As a practical exercise\footnote{biotech\_price\_histories}, we suggest to focus on the French and Benelux biotech sector over 5 year since September 2012 with a capitalization over EUR 100 million. Use data directly from Euronext  and perform the following exercises:

\stepcounter{question}
\begin{itemize}
\item \textbf{Question \arabic{question}}
\begin{itemize}
\item Now you want to keep the information of each day's lowest (Valeur Basse) and highest values (Valeur Haute) taken by the stock
\item You start with a initial cash value of EUR 100,000, you buy or sell by amount of 2,000 maximum per stock per day, each transaction costs EUR 7 (buying or selling)
\item Each day, you are allow to choose to place an limit order, buy or sell, on all stocks based on your strategy, for the following day
\item If your buy order is above the lowest value of the following day, or your sell order below the highest value of the following day, your order is executed
\item Compute your return over the 5 years for your strategy and deliver an history day per day of how much of each stock you hold over the 5 years (this is for a future work on "\textbf{herding}")
\item Construct a portfolio selecting the 5 lowest volatility in the last 20 trading days, compare its performance with the performance above
\end{itemize}
\end{itemize}

\stepcounter{question}
\begin{itemize}
\item \textbf{Question \arabic{question}}
\begin{itemize}
\item Identify days and stock with "extreme variation", chose a criteria until you are left with a minimum of 10 event
\item On those dates and for the companies involved, search for relevant news that could explain the variations
\end{itemize}
\end{itemize}

\subsection{Efficient Markets Theory: an introduction}
If we go back to Eugene Fama in 1964 and confront his statement mentioned in \citep{exuberance} that \begin{quote}
the average correlation coefficient between successive day' log price changes over the thirty Dow Jones Industrial Average stocks between 1957 and 1962 was only 0.03
\end{quote}

\subsubsection{The debate: contrarian versus momentum strategies and the efficient market hypothesis}
When investing in the stock market, should one buy low and sell high or buy high and sell even higher?
Momentum and contrarian strategies are still debated in the press \href{http://www.ft.com/cms/s/0/58852d80-d910-11e3-837f-00144feabdc0.html#axzz4KjiGyRMG}{like in this FT article}.
In this section, we won't pretend to lead to an answer, but we will explore some aspects with the data we have at hand: a selection of 495 shares in \verb|pea_price.csv| and \verb|pea_price_2.csv|, with time series from 2006 to 2016.

De Bondt and Thaler in their 1985 paper \citep{BONDT:1985aa} show that over 3-to 5-year holding periods stocks that performed poorly over the previous 3 to 5 years achieve higher returns than stocks that performed well over the same period.

Jegadeesh and Titman 1993 paper \citep{JEGADEESH:1993aa} \begin{quotation}investigates the efficiency of the stock market \end{quotation} and provides an analysis of relative strength trading strategies, meaning buying stocks that showed high return over 3- to 12-month horizons. They find that \begin{quote}
 the portfolio formed on the basis of returns realized in the past 6 months generates an average cumulative return of 9.5\% over the next 12 months but loses more than half of this return in the following 24 months.
\end{quote}

\stepcounter{question}
\begin{itemize}
\item \textbf{Question \arabic{question}}
\begin{itemize}
\item Compute the average correlation coefficient between successive daily returns.
\item Perform a quick check of the De Bondt and Thaler theory in \citep{BONDT:1985aa} on another data set.
\item Describe and replicate the 16 strategies in Jegadeesh and Titman \citep{JEGADEESH:1993aa} (only Panel A and up to section II of the paper, leave aside the t-statistics for now).
\item What is your most successful zero-cost strategy? Compare it with the paper \citep{JEGADEESH:1993aa} findings.
\item Consider buying the losers versus the winner decile portfolios.
\end{itemize}
\end{itemize}

\section{Python: CAPM}\label{sec:CAPM}

Let $i$ be the index for an asset (e.g. a share), $f$ the risk-free asset and $M$ the market portfolio. We have the Capital Asset Pricing Model (CAPM) equation \ref{eq:CAPM} which derives from the work of \citet{Markowitz:1952aa} and \citet{sharpe1964capital}:
\begin{equation}\label{eq:CAPM}
E\left( R_i \right) =R_f+\beta_i \left[ E\left( R_M \right) -R_f\right]
\end{equation}

We will see step by step how we reconcile this very simple equation with some basic time series econometrics.

 \subsection{Minimum Variance Portfolio}

We consider\footnote{vansteenberghe\_capm.py} two assets, labeled $1$ and $2$, simplifying as a first step by assuming that each asset's worth is unity. The investor's choice concerning these two assets is recorded in a vector of weights, $w$:
\begin{equation*}
w=\begin{pmatrix}
  w_1\\
  w_2
\end{pmatrix}
\end{equation*}

To understand the risk associated with this portfolio, we need to consider the variance and covariance of the asset returns. Let $R_1$ and $R_2$ be the random variables representing the returns of asset $1$ and asset $2$, respectively. The standard deviation of asset $i$ is denoted by $\sigma_i$, and $\rho$ is the correlation coefficient between $R_1$ and $R_2$. 

The variance-covariance matrix $\Sigma$ is constructed from these statistics:
\begin{equation*}
\Sigma = \begin{pmatrix}
  \sigma_1^2 & \rho \sigma_1 \sigma_2\\
  \rho \sigma_1 \sigma_2 & \sigma_2^2
\end{pmatrix}
\end{equation*}
This matrix $\Sigma$ is computed from the historical returns of the assets. The diagonal elements $\sigma_1^2$ and $\sigma_2^2$ are the variances of the returns of assets $1$ and $2$, respectively, while the off-diagonal elements represent the covariance between the returns, calculated as $\rho \sigma_1 \sigma_2$.

Short-selling of assets is not allowed (i.e., $\forall i, w_i \geq 0$), and the asset weights are normalized ($w_1 + w_2 = 1$).

We seek the portfolio $w^*$ that minimizes the total variance of the portfolio, given by the quadratic form $w^T \Sigma w$. This is expressed in the following optimization program:
\begin{equation}\label{eq:programme}
\left\{
    \begin{array}{lll}
        w^* = \mbox{arg} \quad \mbox{min}  \quad w^T \Sigma w\\
        1^T w = 1
    \end{array}
\right.
\end{equation}

The expression $w^T \Sigma x$ represents the total variance of the portfolio. Here, $w^T$ is the transpose of the weight vector, and the multiplication $w^T \Sigma x$ results in a scalar that quantifies the expected variance of the portfolio's return, taking into account both the variances and covariances of the individual asset returns.

Nota bene: $\mbox{Var}\left(w_1 R_1\right) = w_1^2 \mbox{Var}\left(R_1\right)$ and $\mbox{Cov}\left(w_1 R_1, w_2 R_2 \right) = w_1w_2 \mbox{Cov}\left(R_1, R_2 \right)$.

We have:
$$w^T \Sigma w = w_1^2 \sigma_1^2 + w_2^2 \sigma_2^2 + 2 w_1 w_2 \rho \sigma_1 \sigma_2$$
and with the first constraint of equation \ref{eq:programme} $w_1+w_2=1$:
\begin{equation}\label{eq:expression}
w^T \Sigma w = w_2^2 \left(\sigma_1^2-2\rho \sigma_1 \sigma_2+\sigma_2^2\right)+2 w_2 \left(\rho \sigma_1 \sigma_2 -\sigma_1^2\right)+\sigma_1^2
\end{equation}

We compute the optimum of the expression given in equation \ref{eq:expression}:
\begin{equation}\label{eq:optimum}
\frac{\partial (w^T \Sigma w)}{\partial w_2} = 0 \iff w_2^* = 1 - w_1^* = \frac{\sigma_1^2 - \rho \sigma_1 \sigma_2}{\sigma_1^2 - 2 \rho \sigma_1 \sigma_2 + \sigma_2^2}
\end{equation}

Now, let's demonstrate that the denominator of equation \ref{eq:optimum}, $\sigma_1^2 - 2 \rho \sigma_1 \sigma_2 + \sigma_2^2$, is greater than zero. Observe that this expression can be rewritten as $(\sigma_1 - \rho \sigma_2)^2 + (1 - \rho^2) \sigma_2^2$. Since both terms are squares, they are non-negative. Furthermore, since the correlation coefficient $\rho$ satisfies $-1 \leq \rho \leq 1$, the term $(1 - \rho^2) \sigma_2^2$ is non-negative and is strictly positive unless $\rho = \pm 1$. Thus, the entire expression is always non-negative and is strictly positive unless $\rho = \pm 1$ and $\sigma_2 = 0$.

\stepcounter{question}
\begin{itemize}
\item \textbf{Question \arabic{question}}
\begin{itemize}
\item Will the term $w^T \Sigma w$ be increasing or decreasing on each side of $w_2^*$?
\end{itemize}
\end{itemize}

We notice that:
\begin{equation*}
\left\{
    \begin{array}{lll}
w_2^* < 0 \iff \sigma_1 < \rho \sigma_2\\
w	_2^* > 1 \iff \sigma_2 < \rho \sigma_1
\end{array}
\right.
\end{equation*}

On the contrary, if $\rho \leq \frac{\sigma_i}{\sigma_j}$ then it will be recommended to invest jointly in both assets in order to hold the minimum variance portfolio.

\subsubsection{Multiple assets}

We can write\footnote{cf. \href{https://bookdown.org/compfinezbook/introcompfinr/Determining-the-Global.html}{this bookdown}} the minimization problem in matrix form for a portfolio of $n$ assets with weights $w$. The weights should sum to $1$, and we write the variance-covariance matrix as:
\begin{equation}
\Sigma =
\begin{bmatrix}
\sigma^2_1 & \sigma_{1,2} & \ldots & \sigma_{1,n} \\
\sigma_{2,1} & \sigma^2_2 & \ldots & \sigma_{2,n} \\
\vdots & \vdots & \ddots & \vdots \\
\sigma_{n,1} & \sigma_{n,2} & \ldots & \sigma^2_n
\end{bmatrix}
\end{equation}

We are searching for the portfolio weights that answer the following programme:
\begin{equation}
\left\{
    \begin{array}{lll}
\min_w w^T \Sigma w\\
\mbox{s.t.} \; w^T 1 = 1
\end{array}
\right.
\end{equation}

We write the Lagrangian:
$$L(w_1,w_2,\ldots,w_n,\lambda) = \sum_{i,j} w_i^2 \sigma^2_i + w_j^2 \sigma^2_j + \dots w_n^2 \sigma^2_n + 2 w_i w_j \sigma_{i,j} + \lambda (w_1 + \ldots + w_n - 1)$$
And the First Order Conditions:
$$2w_i \sigma^2_i + 2 w_j \sigma_{i,j} + \ldots + 2 w_n \sigma_{i,n} + \lambda = 0$$
$$w_1 + w_2 + \ldots + w_n - 1 =0$$

In matrix form, we write the Lagrangian:
$$L(w,\lambda) = w^T \Sigma w + \lambda (1^T w - 1)$$
And the First Order Conditions (FOC):
$$\nabla_w L = 2 \Sigma w + \lambda 1 = 0$$
$$1^T w = 1$$
Thus, in matrix form:
$$ w = -\frac{1}{2} \lambda \Sigma^{-1} 1$$
$$1 = -\frac{1}{2} \lambda 1^T \Sigma^{-1} 1$$
And finally:
$$ w = \frac{\Sigma^{-1} 1}{1^T \Sigma^{-1} 1} $$

\subsubsection{Application}

Let's consider two stocks. We consider to simplify that the value of each share is unity. We invest in our portfolio so that $w_1+w_2=1$ and consider no short selling. In order to plot the frontiere of possible portfiolio we have to vary: $w_1 \in [0,1]$.

Identify the minimum variance portfolio on this frontiere, find the weights of this minimum variance portfolio and this variance.

\subsection{Maximum Sharpe ratio portfolio - two assets case}

The Sharpe ratio is defined as $S_R=\frac{E(R_p)-R_f}{\sigma(R_p)}$. With two assets, the expected return and variance of the portfolio is:

\begin{equation*}
    E(R_p) = w_1 E(R_1) + w_2 E(R_2), \quad V(R_p) = w_1^2 \sigma_{11} + w_2^2 \sigma_{22} + 2 w_1 w_2 \sigma_{12}
\end{equation*}

For maximizing the Sharpe Ratio, the derivative of \( S_R \) with respect to \( w_1 \) is set to zero. This results in a quadratic equation that can be solved to find the optimal weights for maximizing the Sharpe Ratio.

Our program is:
$$\frac{d S_R}{d w_1} =0 $$
If we develop, we find a second order equation:
$$w_1^2 \left((3 E(R_2)+E(R_1))(2\sigma_{12}-\sigma_{11}-\sigma_{22}))\right) +2 \mu_2 w_1 (\sigma_{22}-\sigma_{11})+ \sigma_{22} (E(R_1)+E(R_2)) -2 E(R_2) \sigma_{12} = 0$$

Solving this quadratic equation, we obtain:
\begin{equation*}
    w_1 = \frac{-b \pm \sqrt{b^2 - 4ac}}{2a}
\end{equation*}
where \( a, b, \) and \( c \) are the coefficients of the quadratic equation.

After algebraic manipulation, we find the optimal weight \( w^* \) as:
\begin{equation*}
    w^* = \frac{\left(E(R_2)-R_f\right) Cov(R_1,R_2) - \left(E(R_1)-R_f\right) V(R_2)}{\left(E(R_1)-R_f\right) \left( Cov(R_1,R_2)-V(R_2)\right) + \left(E(R_2)-R_f\right) \left( Cov(R_1,R_2)-V(R_1)\right)}
\end{equation*}

\subsubsection{Efficient frontier with multiple assets}

Given a set of assets with expected returns vector $\mu$ and variance-covariance matrix $\Sigma$, the mean-variance optimization problem for a portfolio with weights w can be formulated as:

\begin{equation}
\begin{aligned}
& \underset{w}{\text{minimize}}
& & w^T \Sigma w \\
& \text{subject to}
& & w^T \mu = \mu_p, \\
&&& w^T 1 = 1,
\end{aligned}
\end{equation}

where \(\mu_p\) is the desired portfolio return, and 1 is a vector of ones.

The efficient frontier can be generated by solving this optimization problem for a range of desired portfolio returns \(\mu_p\).

The efficient frontier can be represented as a set of portfolios that minimize portfolio variance for given levels of expected return. For each desired return \(\mu_p\), we find the weights w that solve the above optimization problem. The frontier is the plot of the minimum variance (or standard deviation) for each level of return.

\subsubsection{Determining Mean-Variance Efficient Portfolios Using Matrix Algebra}

The investment opportunity set is the set of portfolio expected return, $\mu_p$, and portfolio standard deviation, $\sigma_p$, values for all possible portfolios whose weights sum to one. As in the two risky asset case, this set can be described in a graph with $\mu_p$ on the vertical axis and $\sigma_p$ on the horizontal axis. With two assets, the investment opportunity set in $(\mu_p, \sigma_p)$-space lies on a curve (one side of a hyperbola). With three or more assets, the investment opportunity set in $(\mu_p, \sigma_p)$-space is described by a set of values whose general shape is complicated and depends crucially on the covariance terms $\sigma_{ij}$. However, we do not have to fully characterize the entire investment opportunity set. If we assume that investors choose portfolios to maximize expected return subject to a target level of risk, or, equivalently, to minimize risk subject to a target expected return, then we can simplify the asset allocation problem by only concentrating on the set of efficient portfolios. These portfolios lie on the boundary of the investment opportunity set above the global minimum variance portfolio. This is the framework originally developed by Harry Markowitz, the father of portfolio theory and winner of the Nobel Prize in economics.

Following Markowitz, we assume that investors wish to find portfolios that have the best expected return-risk trade-off. We can characterize these efficient portfolios in two equivalent ways. In the first way, investors seek to find portfolios that maximize portfolio expected return for a given level of risk as measured by portfolio variance. Let $\sigma^2_{p,0}$ denote a target level of risk. Then the constrained maximization problem to find an efficient portfolio is:

\begin{equation}
\max_x \quad \mu_p = x' \mu \quad \text{s.t.} \quad \sigma^2_p = x' \Sigma x = \sigma^2_{p,0} \quad \text{and} \quad x'1 = 1.
\end{equation}

The investor’s problem of maximizing portfolio expected return subject to a target level of risk has an equivalent dual representation in which the investor minimizes the risk of the portfolio (as measured by portfolio variance) subject to a target expected return level. Let $\mu_{p,0}$ denote a target expected return level. Then the dual problem is the constrained minimization problem:

\begin{equation}\label{eq:dualproblem}
\min_x \quad \sigma^2_{p,x} = x' \Sigma x \quad \text{s.t.} \quad \mu_p = x' \mu = \mu_{p,0}, \quad \text{and} \quad x'1 = 1.
\end{equation}

To find efficient portfolios of risky assets in practice, the dual problem \ref{eq:dualproblem} is most often solved. This is partially due to computational conveniences and partly due to investors being more willing to specify target expected returns rather than target risk levels. The efficient portfolio frontier is a graph of $\mu_p$ versus $\sigma_p$ values for the set of efficient portfolios generated by solving \ref{eq:dualproblem} for all possible target expected return levels $\mu_{p,0}$ above the expected return on the global minimum variance portfolio. Just as in the two asset case, the resulting efficient frontier will resemble one side of a hyperbola and is often called the “Markowitz bullet”. This frontier is illustrated in Figure below as the boundary of the set generated by random portfolios above the global minimum variance portfolio.

To solve the constrained minimization problem \ref{eq:dualproblem}, first form the Lagrangian function:

\begin{equation}
L(x, \lambda_1, \lambda_2) = x' \Sigma x + \lambda_1 (x' \mu - \mu_{p,0}) + \lambda_2 (x' 1 - 1).
\end{equation}

Because there are two constraints ($x' \mu = \mu_{p,0}$ and $x' 1 = 1$), there are two Lagrange multipliers $\lambda_1$ and $\lambda_2$. The FOCs for a minimum are the linear equations:

\begin{align}
\frac{\partial L(x, \lambda_1, \lambda_2)}{\partial x} &= 2 \Sigma x + \lambda_1 \mu + \lambda_2 1 = 0, \\
\frac{\partial L(x, \lambda_1, \lambda_2)}{\partial \lambda_1} &= x' \mu - \mu_{p,0} = 0, \\
\frac{\partial L(x, \lambda_1, \lambda_2)}{\partial \lambda_2} &= x' 1 - 1 = 0.
\end{align}

These FOCs consist of $N + 2$ linear equations in $N + 2$ unknowns ($x, \lambda_1, \lambda_2$). We can represent the system of linear equations using matrix algebra as:

\begin{equation}
\begin{bmatrix}
2 \Sigma & \mu & 1 \\
\mu' & 0 & 0 \\
1' & 0 & 0
\end{bmatrix}
\begin{bmatrix}
x \\
\lambda_1 \\
\lambda_2
\end{bmatrix}
=
\begin{bmatrix}
0 \\
\mu_{p,0} \\
1
\end{bmatrix},
\end{equation}

which is of the form $A z_x = b_0$, where

\begin{equation*}
A = 
\begin{bmatrix}
2 \Sigma & \mu & 1 \\
\mu' & 0 & 0 \\
1' & 0 & 0
\end{bmatrix},
\quad
z_x = 
\begin{bmatrix}
x \\
\lambda_1 \\
\lambda_2
\end{bmatrix},
\quad
b_0 = 
\begin{bmatrix}
0 \\
\mu_{p,0} \\
1
\end{bmatrix}.
\end{equation*}

The solution for $z_x$ is then:

\begin{equation}
z_x = A^{-1} b_0.
\end{equation}

The first $N$ elements of $z_x$ are the portfolio weights $x$ for the minimum variance portfolio with expected return $\mu_{p,x} = \mu_{p,0}$. If $\mu_{p,0}$ is greater than or equal to the expected return on the global minimum variance portfolio, then $x$ is an efficient (frontier) portfolio. Otherwise, it is an inefficient (frontier) portfolio.

\begin{center}
\includegraphics[scale=0.5]{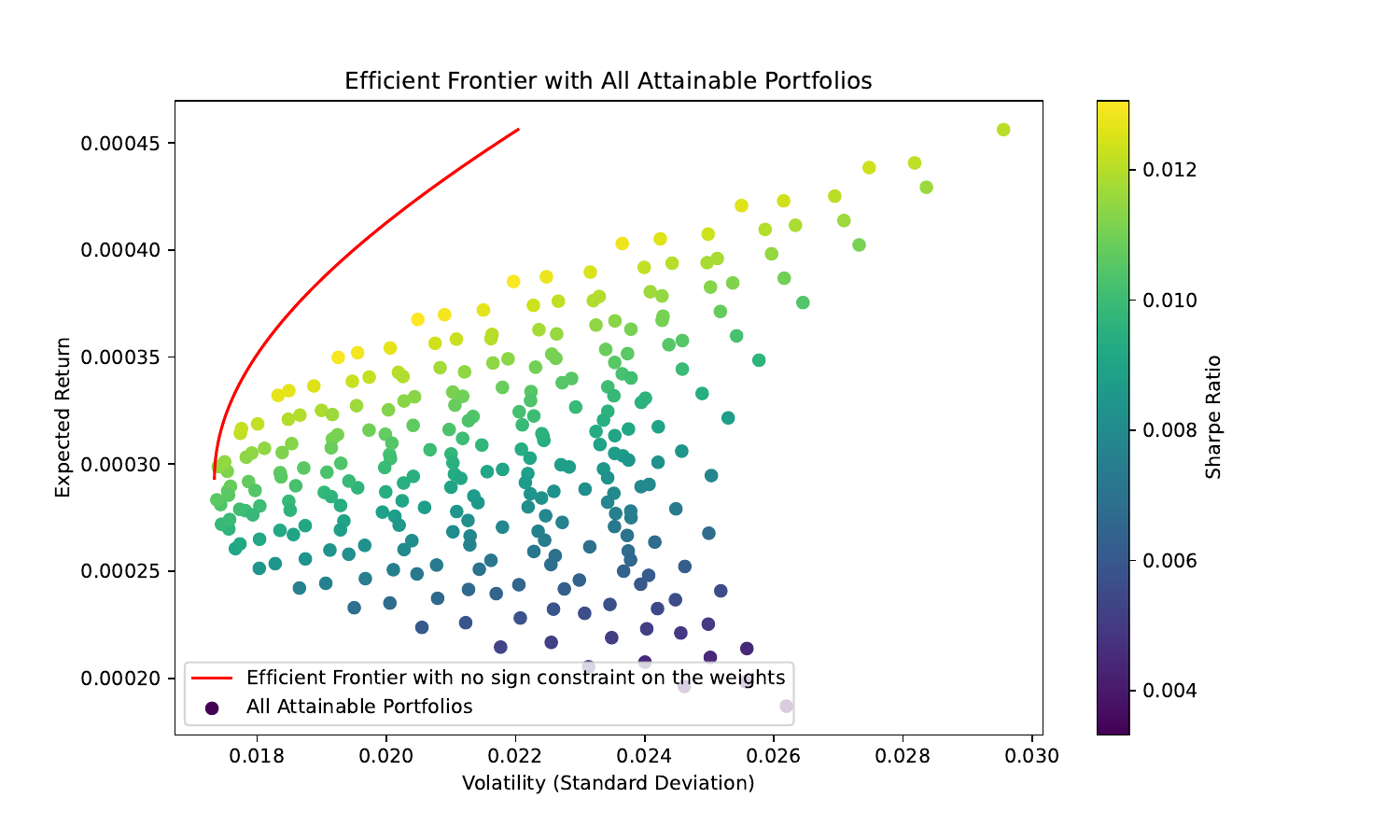}
\end{center}

\subsubsection{Numerical Portfolio Optimization with Positivity Constraint on the Weights}

In practical portfolio optimization, especially for retail investors, it is often required that the portfolio weights be non-negative, i.e., no short selling is allowed. Moreover, the weights should sum up to one, ensuring a fully invested portfolio. This section describes the numerical approach to finding the mean-variance efficient portfolios under these constraints.

The optimization problem can be formulated as follows:

\begin{equation}\label{eq:constrained_optimization}
\begin{aligned}
& \underset{w}{\text{minimize}}
& & w^T \Sigma w \\
& \text{subject to}
& & w^T \mu = \mu_{p,target}, \\
&&& w^T 1 = 1, \\
&&& 0 \leq w_i \leq 1 \quad \forall i,
\end{aligned}
\end{equation}
where $w$ represents the portfolio weights, $\Sigma$ is the covariance matrix of asset returns, $\mu$ is the vector of expected asset returns, and $\mu_{p,target}$ is the target portfolio return.

This optimization problem is a Quadratic Programming (QP) problem with linear equality and inequality constraints. Such problems are typically not solvable using analytical methods but can be efficiently solved using numerical optimization techniques.

In Python, the `scipy.optimize.minimize` function with the Sequential Least Squares Programming (SLSQP) method is used to solve this problem. The objective function to minimize is the portfolio variance $w^T \Sigma w$, and the constraints are implemented as follows:

\begin{itemize}
    \item The equality constraint $w^T \mu = \mu_{p,target}$ ensures the portfolio has the desired expected return.
    \item The equality constraint $w^T 1 = 1$ ensures the weights sum up to one, representing a fully invested portfolio.
    \item The inequality constraints $0 \leq w_i \leq 1$ ensure no short selling and that the weights do not exceed 100\% of the portfolio.
\end{itemize}

In python, implement this optimization taking the asset returns DataFrame, a list of assets, and the target return as inputs and returns the optimal portfolio weights under the specified constraints.

To visualize the efficient frontier under these constraints, generate a range of target returns and solve the constrained optimization problem for each target return. The target returns range from the minimum variance portfolio return with positive constraint on the weights and on the maximum return of the unique assets in the portfolio.

\begin{center}
\includegraphics[scale=0.5]{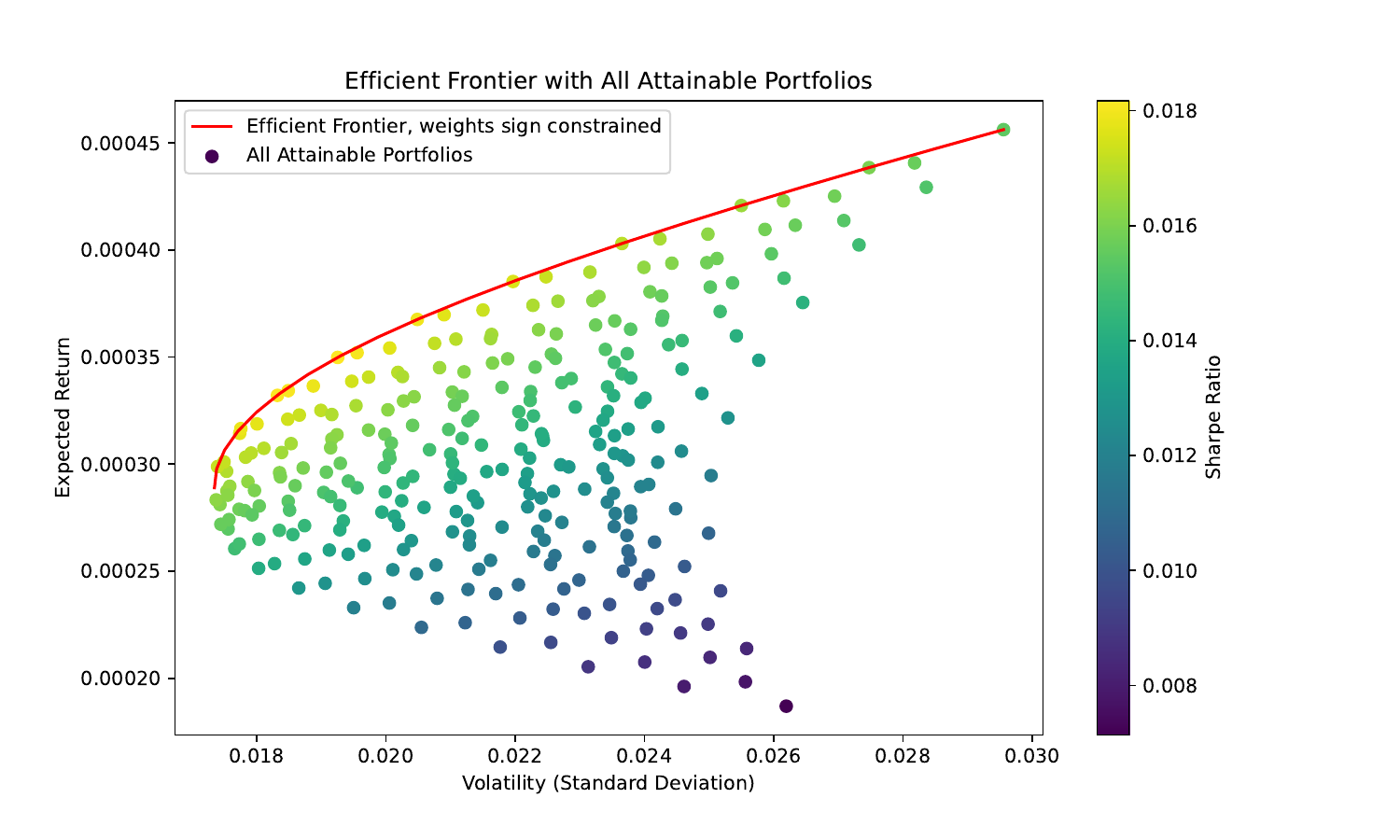}
\end{center}

\subsection{Investor utility and CAPM}

Consider a rational investor who allocates their investment between a risk-free asset and a market portfolio (the portfolio with the maximum attainable Sharpe ratio). The investor's portfolio consists of a fraction \( w \) invested in the market portfolio, with an expected return \( E(R_M) \) and volatility \( \sigma^2(R_M) \), and a fraction \( (1-w) \) in a risk-free asset with return \( R_f \). The investor's utility is assumed to increase with the portfolio's return and decrease with the portfolio's risk (volatility). We take the simplified approach of \citet{ross1976}:
$R_M$ is composed of iid assets such that $E(R_i)=\rho$ and $E(R_i^2)=\sigma^2$. Then we can model the investor utility as a function of its wealth. The expected return of the portfolio is a weighted average of the returns on the risky market portfolio and the risk-free asset:
\begin{equation*}
    E(R_p) = (1-w) R_f + w E(R_M)
\end{equation*} The utility function is a simple exponential here, for tractability:
\begin{equation}
U(x) = -\exp\left(-A z\right)
\end{equation}

The investor will maximize its expected utility:
\begin{equation}\label{eq:utilityoptimization1}
\begin{aligned}
& \underset{w}{\text{maximize}}
& & -E\left( \exp\left[-A (1-w)R_f\right] \right)E\left( \exp \left[ -A xR_M \right] \right)
\end{aligned}
\end{equation}

We use two tricks: a Taylor expansion around $R_M$:
$$\exp \left[ -A xR_M \right] = 1-AwR_M+\frac{1}{2}A w R_M^2$$
$$E\left( \exp \left[ -A xR_M \right] \right) = 1-Aw E\left(R_M\right)+\frac{1}{2}A w\left[\sigma^2 + E\left(R_M\right)^2\right]$$
and then we take the exponential of the log and consider for simplicity here that $E(R_M^2)=\sigma^2$, then
$$E\left( \exp \left[ -A xR_M \right] \right) = \exp \left[ -A w E\left(R_M\right)+\frac{1}{2}A w\sigma^2 \right]$$

The program then becomes
\begin{equation}\label{eq:utilityoptimization2}
\begin{aligned}
& \underset{w}{\text{maximize}}
& & -\exp\left[-A (1-w)R_f\right]  \exp \left[ -A w E\left(R_M\right)+\frac{1}{2}A w\sigma^2 \right] 
\end{aligned}
\end{equation}

And the first order condition yields:
\begin{equation*}
    w^* = \frac{E(R_M) - R_f}{A \sigma^2}
\end{equation*}

An alternative, simpler way, to model the investor's utility is:
\begin{equation*}
    U(w) = E(R_p) - \frac{1}{2} A \sigma^2(R_p)
\end{equation*}
where \( U(w) \) is the utility, \( E(R_p) \) is the expected return of the portfolio, \( \sigma^2(R_p) \) is the portfolio's variance, and \( A \) is a risk-aversion coefficient.

The portfolio's risk, represented by its variance, is assumed to be solely due to the risk of the market portfolio (as the risk-free asset has zero variance):
\begin{equation*}
    \sigma^2(R_p) = w^2 \sigma^2(R_M)
\end{equation*}

To find the optimal fraction \( w \) to invest in the market portfolio, the investor maximizes their utility. This is achieved by setting the first derivative of the utility function with respect to \( w \) to zero:
\begin{equation*}
    \frac{\partial U(w)}{\partial w} = 0
\end{equation*}

Solving this equation for \( w \) yields the optimal investment fraction in the risky asset:
\begin{equation*}
    w^* = \frac{E(R_M) - R_f}{A \sigma^2(R_M)}
\end{equation*}

The advice to give to the investor will depend on his risk-aversion.

\subsection{Efficient frontier and Capital Market Line (CML)}

The CAPM features the efficient frontier as a key concept, representing the set of portfolios that offer the highest expected return for a given level of risk or the lowest risk for a given level of expected return. This frontier is derived from analyzing historical returns data to determine the boundary where return and risk are optimally balanced.

A central element of the CAPM is the market portfolio, denoted as \( M \). This theoretical portfolio comprises all investable assets in the market, each weighted according to its market value. Representing the entirety of risky investment opportunities, the market portfolio is assumed to be fully diversified. In the idealized CAPM framework, the market portfolio is posited to lie on the efficient frontier with the highest Sharpe ratio, indicating that it offers the best possible risk-return trade-off under perfect market conditions. However, it is important to note that in practical scenarios, due to various market imperfections such as taxes, transaction costs, and information asymmetry, the actual market portfolio might not perfectly align with the efficient frontier as theorized in CAPM.

In CAPM, a typical portfolio \( p \) consists of a mix of a risky asset \( i \), a risk-free asset \( f \), and the market portfolio \( M \). Investors can modify their portfolio's composition, including through short-selling the risk-free asset to invest more in the risky asset. The expected return of such a portfolio is given by:
\begin{equation*}
    E(R_p) = w_M E(R_M) + w_i E(R_i) - w_i R_f
\end{equation*}
Here, \( w_M \) and \( w_i \) indicate the portfolio's proportions invested in the market and the risky asset, respectively, while \( R_f \) is the return on the risk-free asset.

The tangency portfolio, located at the point where the efficient frontier is tangent to the line from the risk-free rate, is characterized by \( w_M = 1 \) and \( w_i = 0 \), this is the Capital Market Line (CML). This portfolio represents the most efficient risk-return balance achievable in the market under the CAPM assumptions. It is significant for having the highest Sharpe ratio, indicating the most favorable risk-return balance available in the market.

\begin{center}
\includegraphics[scale=.5]{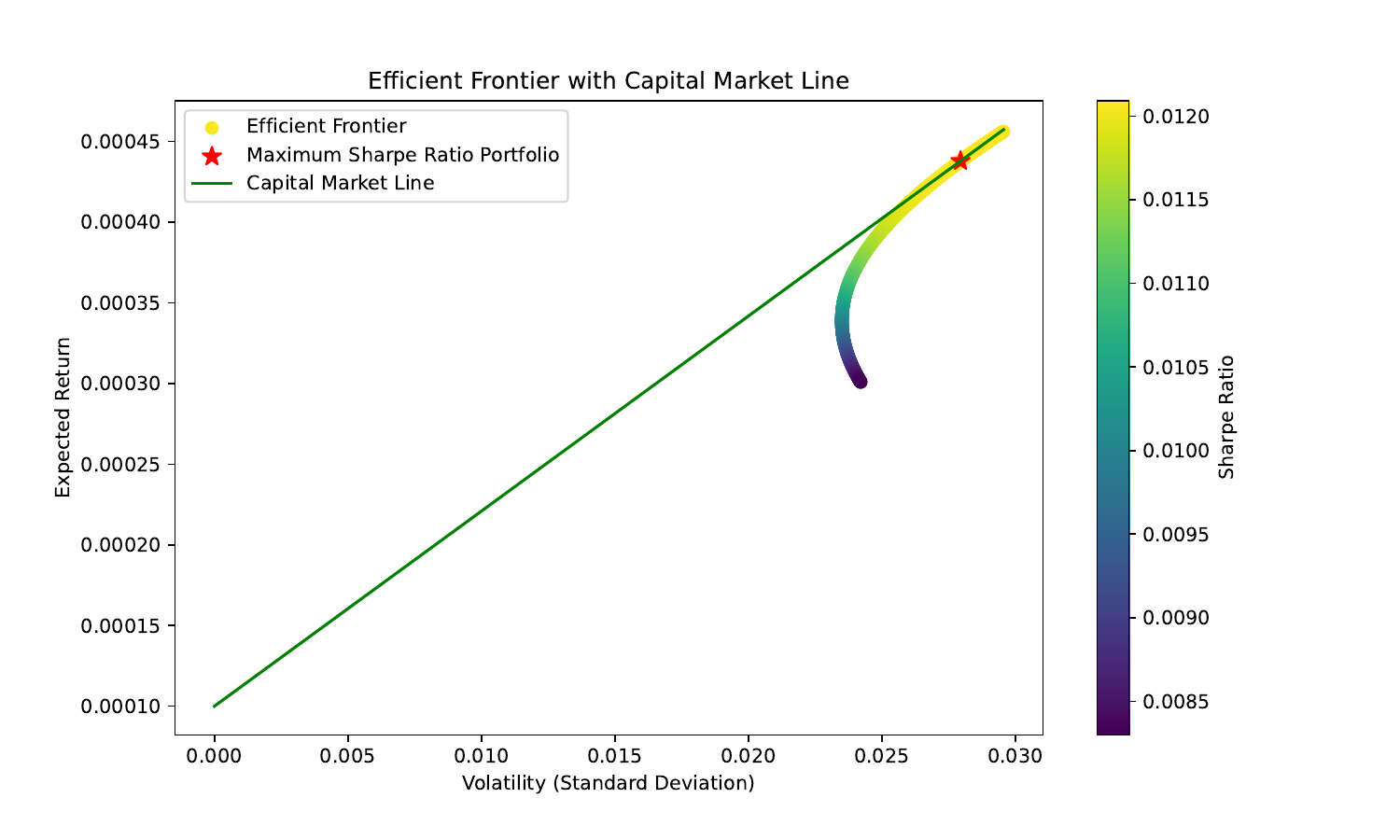}
\end{center}

The efficient frontier thus symbolizes the frontier of optimal investment choices for a given level of risk. Portfolios along this boundary provide the best possible returns for a specified risk level, guiding investors in their decision-making process.

The risk of portfolio \( p \) is quantified by its variance, which can be described as:
\begin{equation*}
    \sigma_p^2 = w_M^2 \sigma_M^2 + w_i^2 \sigma_i^2 + 2 w_M w_i \sigma_{i,M}
\end{equation*}

To determine the optimal weight \( w_i \) for the risky asset \( i \), the following derivative calculations are performed:
\begin{align*}
    \frac{\partial E(R_p)}{\partial w_i} &= E(R_i) - R_f \\
    \frac{\partial \sigma_p^2}{\partial w_i} &= 2 w_i \sigma_i^2 + 2 w_M \sigma_{i,M}
\end{align*}

At the tangency point of the efficient frontier:
\begin{equation*}
    \frac{\partial \sigma_p^2}{\partial w_i} = 2 \sigma_{i,M}
\end{equation*}

Therefore, at this crucial tangency point (\(w_M = 1\) and \(w_i = 0\)):
\begin{equation*}
    \frac{\frac{\partial E(R_p)}{\partial w_i}}{\frac{\partial \sigma_p^2}{\partial w_i}} = \frac{E(R_i) - R_f}{2 \sigma_{i,M}} = \frac{E(R_M) - R_f}{2 \sigma_M^2}
\end{equation*}

In CAPM, the beta \( \beta_i \) of an asset \( i \) is defined as:
\begin{equation}
    \beta_i = \frac{\text{cov}(R_i, R_M)}{\sigma_M^2}
\end{equation}
Beta serves as a measure of the relative risk of asset \( i \) compared to the market. An asset with a beta greater than 1 is considered riskier than the market, while a beta less than 1 indicates lower risk than the market.

For two distinct assets \( i \) and \( j \), their respective betas are differentiated as:
\begin{equation*}
    \beta_i = \frac{E(R_i) - R_f}{E(R_M) - R_f}
\end{equation*}

Illustrations of the risk premium \( (R_M - R_f) \) and beta comparisons between two assets can be visualized as follows:
\begin{center}
    \includegraphics[scale=0.5]{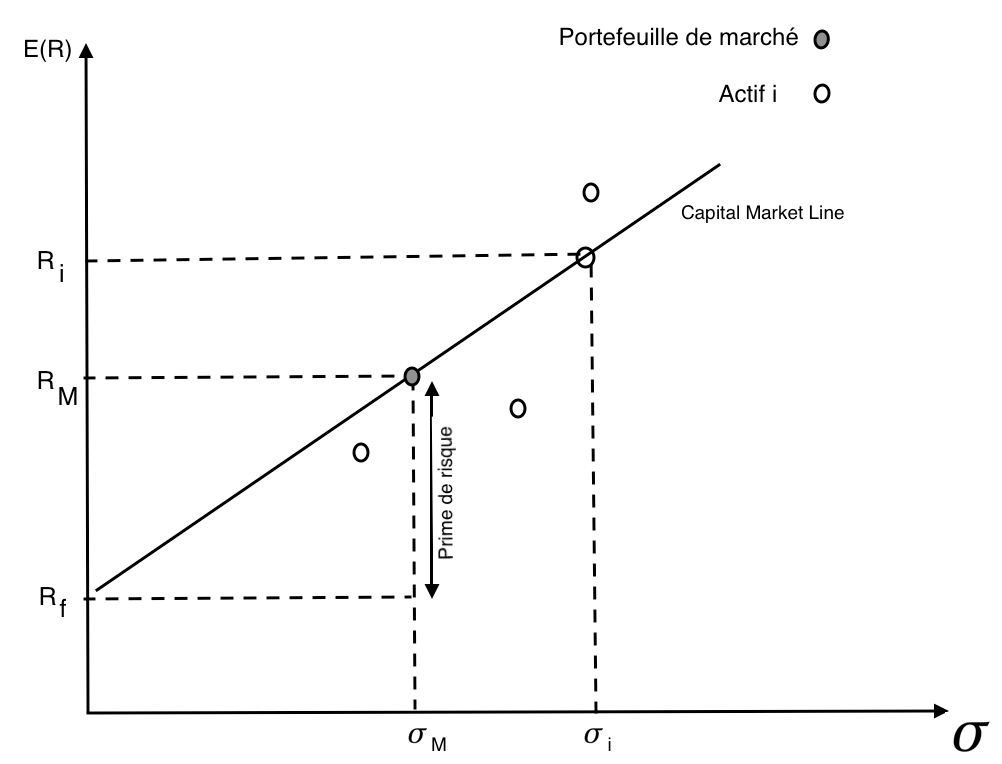}
    \includegraphics[scale=0.5]{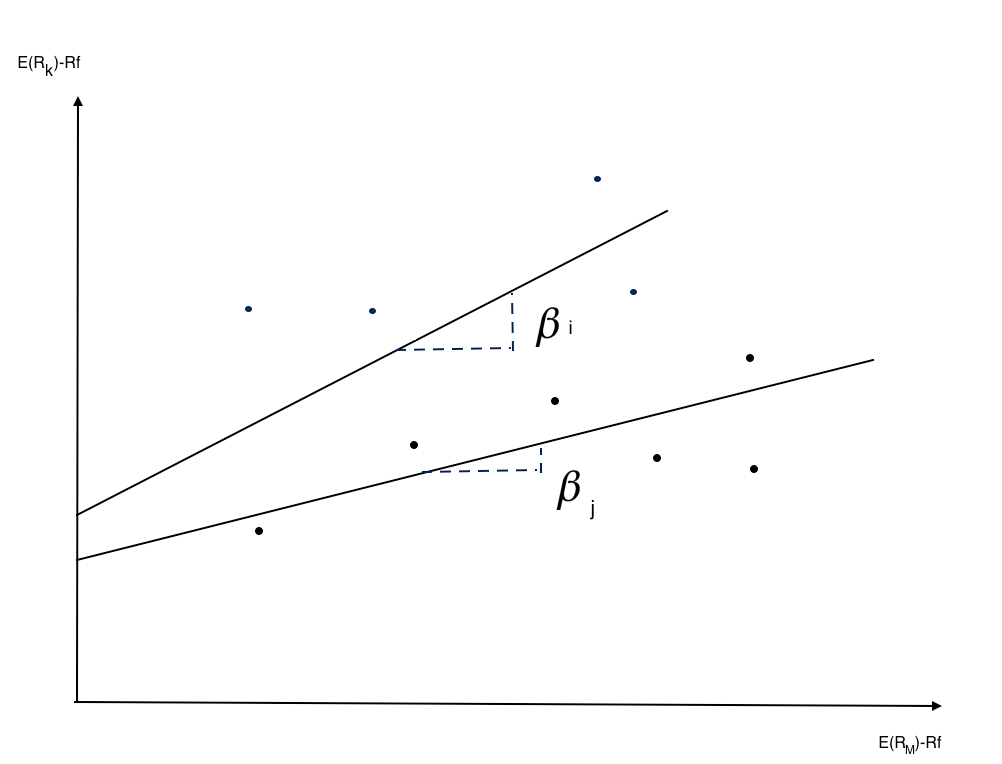}
\end{center}

\subsection{Linking CAPM Beta with Regression Beta}

In the CAPM context, beta (\( \beta_i \)) is a measure of how much the return of an asset \( i \) changes with respect to a change in the market return. This beta can also be interpreted using traditional econometric methods, particularly through a linear regression model.

Consider a linear regression of an asset's returns \( R_i \) against the market returns \( R_M \):
\begin{equation*}
    R_{i,t} = \alpha_i + \beta_i R_{M,t} + \varepsilon_{i,t}
\end{equation*}
where \( \alpha_i \) is the intercept term, \( \beta_i \) is the slope coefficient, and \( \varepsilon_i \) is the error term.

The slope coefficient \( \beta_i \) in this regression equation represents the sensitivity of the asset's returns to the market returns. It is calculated as:
\begin{equation*}
    \beta_i = \frac{\text{cov}(R_i, R_M)}{\text{var}(R_M)}
\end{equation*}
where \(\text{cov}(R_i, R_M)\) is the covariance between the asset's returns and the market's returns, and \(\text{var}(R_M)\) is the variance of the market's returns.

This definition of \( \beta_i \) in the regression context aligns with the definition of beta in the CAPM equation \ref{eq:CAPM}. Both represent the amount by which the asset's return is expected to change for a given change in the market return. Thus, we can see that the beta obtained from a regression analysis is the same as the beta used in the CAPM framework.

\subsection{CAPM Demonstration}

In the Capital Asset Pricing Model (CAPM), we assess portfolio efficiency based on expected return and risk. A portfolio \(A\) is more efficient than a portfolio \(B\) if it has a higher expected return for the same level of risk, or a lower risk for the same level of expected return:

\begin{equation*}
    \mu_A = \mu_B \implies \sigma_A < \sigma_B \quad \text{or} \quad \sigma_A = \sigma_B \implies \mu_A > \mu_B
\end{equation*}

In the presence of a risk-free asset \( R_f \), the market portfolio is found by maximizing the Sharpe ratio:

\begin{equation*}
    \text{Maximize} \quad \theta = \frac{E(R_p) - R_f}{\sigma(R_p)} \quad \text{subject to} \quad \sum_{i=1}^N w_i = 1
\end{equation*}

Where \( E(R_p) \) is the expected return and \( \sigma(R_p) \) is the standard deviation of the portfolio returns. Using matrix notation:

\begin{equation*}
    \theta = \frac{W^T(R - \overline{R_f})}{\sqrt{W^T \Omega W}}
\end{equation*}

Here, \( W \) is the weight vector, \( R \) is the vector of expected returns, and \( \Omega \) is the covariance matrix. The notation \( \overline{R_f} \) represents a vector with all elements equal to the risk-free rate \( R_f \).

To find the optimal weights, we take the derivative of \( \theta \) with respect to \( W \) and set it to zero:

\begin{equation*}
    \frac{\partial \theta}{\partial W} = \frac{(R - \overline{R_f})}{\sqrt{W^T \Omega W}} - \frac{1}{2} \frac{[W^T (R - \overline{R_f})] 2\Omega W}{(W^T \Omega W)^{\frac{3}{2}}} = 0
\end{equation*}

Rearranging terms, we get:

\begin{equation*}
    (R - \overline{R_f}) - \frac{[W^T (R - \overline{R_f})]}{W^T \Omega W} \Omega W = 0
\end{equation*}

We introduce an unknown function \( u(W) = \frac{[W^T (R - \overline{R_f})]}{W^T \Omega W} \) and reformulate the equation as:

\begin{equation*}
    (R - \overline{R_f}) -  u(W) \Omega W = 0
\end{equation*}

Introducing \( Z = u(W) W \) simplifies our optimization problem. The equation \( (R - \overline{R_f}) = \Omega Z \) can be solved to find the vector \( Z \), and the weights \( w_i \) are then obtained as:

\begin{equation*}
    w_i = \frac{z_i}{\sum_{i=1}^N z_i}
\end{equation*}

The value of \( u(W) \) can be found by enforcing the constraint, leading to a system of linear equations that can be solved to find the optimal weights. Based on our constraint, $\sum_{i=1}^N w_i u(W) = \sum_{i=1}^N z_i \implies u(W) = \sum_{i=1}^N z_i$, thus $w_i=\frac{z_i}{\sum_{i=1}^N z_i}$.

\begin{equation*}
\begin{cases}
R_1-R_f = z_1 \sigma_{11} + z_2 \sigma_{12} + \ldots + z_N \sigma_{1N}\\
\vdots\\
R_N-R_f = z_1 \sigma_{N1} + z_2 \sigma_{N2} + \ldots + z_{N} \sigma_{NN}
\end{cases}
\end{equation*}

Having found the weights for the market portfolio, we now find the weights needed for a targeted portfolio return \( E(R_p) \):

Minimize \( \frac{1}{2} W^T \Omega W \) subject to:

\begin{equation*}
\begin{cases}
W^T \overline{R} = E(R_p)\\
W^T \overline{1} = 1
\end{cases}
\end{equation*}

The Lagrangian is given by:

\begin{equation*}
\mathcal{L} = \frac{1}{2} W^T \Omega W + \lambda [E(R_p)-W^T \overline{R}]+\gamma [1-W^T \overline{1}]
\end{equation*}

Taking the derivatives, we get:

\begin{equation*}
\begin{cases}
\frac{\partial \mathcal{L}}{\partial W} = \Omega W - \lambda \overline{R} - \gamma \overline{1} = 0\\
\frac{\partial \mathcal{L}}{\partial \lambda} = E(R_p) - W^T \overline{R} = 0\\
\frac{\partial \mathcal{L}}{\partial \gamma} = 1 - W^T \overline{1} = 0
\end{cases}
\end{equation*}

From these equations, we find:

\begin{equation*}
\begin{cases}
W = \Omega^{-1}(\lambda  \overline{R} + \gamma \overline{1})\\
\overline{R}^T W = E(R_p)\\
\overline{1}^T W = 1
\end{cases}
\end{equation*}

We pre-multiply $W = \Omega^{-1}(\lambda \overline{R} + \gamma \overline{1})$ by $\overline{R}$ and $1^T$ and get
\begin{equation*}
\begin{cases}
 E(R_p) = A\lambda + B \gamma\\
1 = B\lambda + C \gamma
\end{cases}
\end{equation*}

We define the following constants:

\begin{align*}
A &= \overline{W}^T \Omega^{-1} W \\
B &= \overline{W}^T \Omega^{-1} \overline{1}\\
C &= \overline{1}^T \Omega^{-1} \overline{1}\\
D &= BC - A^2
\end{align*}

Solving for \( \lambda \) and \( \gamma \):

\begin{equation*}
\begin{cases}
\lambda = \frac{C E(R_p) - B}{D}\\
\gamma = \frac{A - B E(R_p)}{D}
\end{cases}
\end{equation*}

Finally, we express \( W \) as a linear combination of \( g \) and \( h \):

\begin{equation}\label{eq:frontierweights}
W = g + h E(R_p)
\end{equation}

with

\begin{equation*}
\begin{cases}
g = \frac{A \Omega^{-1} \overline{1} - B \Omega^{-1} E(R_p)}{D}\\
h = \frac{C \Omega^{-1} E(R_p) - B \Omega^{-1} \overline{1}}{D}
\end{cases}
\end{equation*}

and the variance is
\begin{equation*}
\sigma_w^2=g^T \Omega g + 2 E(R_p) g^T\Omega h + E(R_p)^2 h^T \Omega h
\end{equation*}
The efficient frontier is the upper boundary of the $(\mu_w,\sigma_w)$ hyperbola.

The weights in equation \ref{eq:frontierweights} are linear in $E(R_p)$. If we know two frontier portfolio weights, then we know all minimum-variance portfolios. This observation is often called the two mutual fund theorem.

\subsection{Naive implementation with 4 assets}

We suggest to do a naive implementation with 4 assets. The investor has a budget to spread between the 4 stocks, so we spread the weight from $0$ ot $1$ and the sum has to be $1$. We first need to compute the possible weights:

\quad

\begin{tabular}{lllll}
Stock &1&2&3&4\\
Weights&1&0&0&0\\
Weights&0.9&0.1&0&0\\
Weigths&0.8&0.1&0.1&0\\
Weights&etc.&&&\\
\end{tabular}

We would get the following for a 4 stock portfolio:
\begin{center}
\includegraphics[scale=0.5]{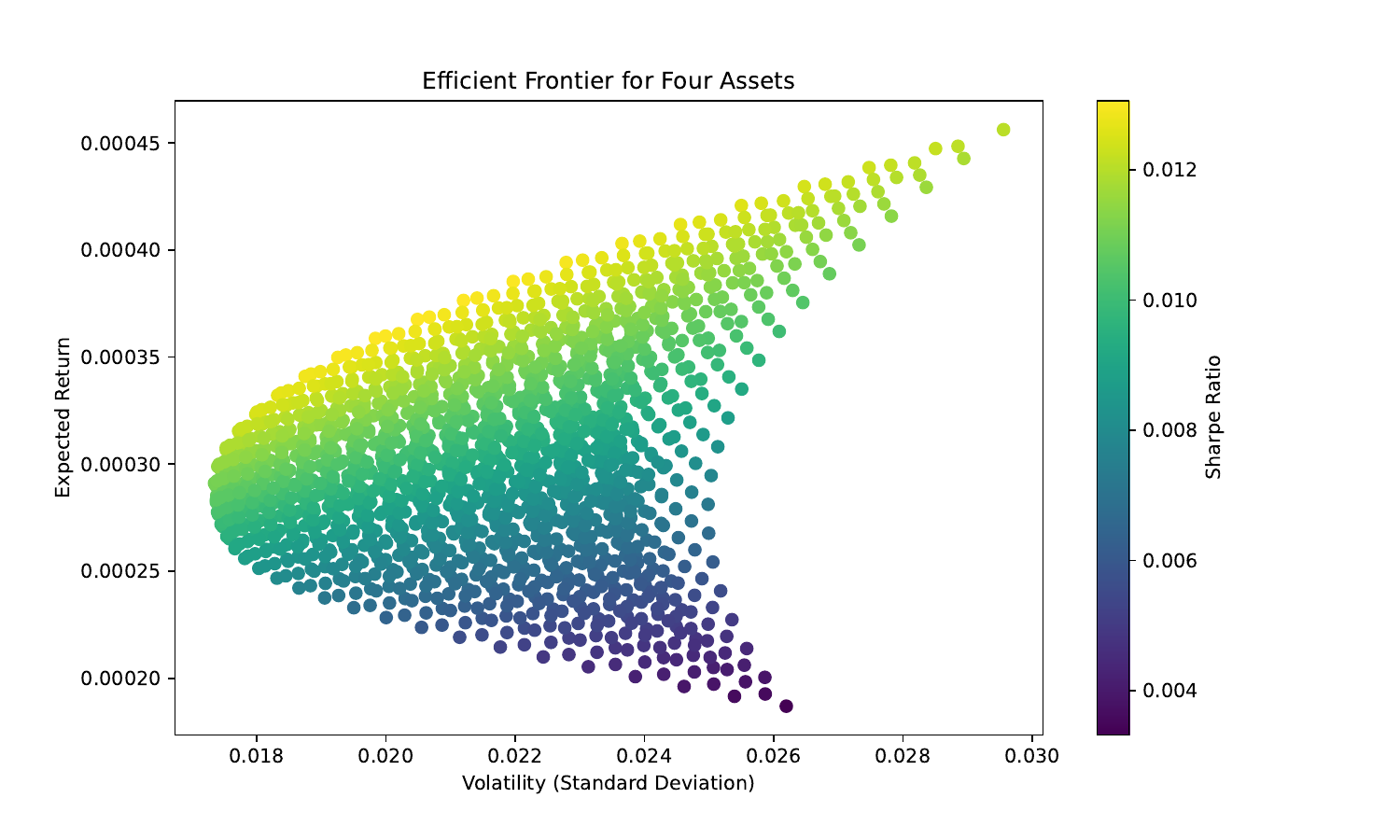}
\end{center}

\stepcounter{question}
\begin{itemize}
\item \textbf{Question \arabic{question}}
\begin{itemize}
\item Randomize the choice of the $4$ stocks, rank the maximum reachable Sharpe ratio, and make a proposition for an investor that would want to invest in $4$ stocks based on historical returns.
\item Implement a proper Markovitz frontier algorithm.
\end{itemize}
\end{itemize}

\subsection{Capital Market Line}
From the equation \ref{eq:CAPM}, we can define for each return the minimum associated risk (using the standard deviation as a proxy for risk). All those portfolio give us the \emph{efficient frontier}. The tangent of that efficient frontier that goes through the risk-free asset is call the \emph{Capital Market Line}. The portfolio at the intersection of the efficient frontier and this tangent is call the \emph{market portfolio}. This is the mentioned portfolio in equation \ref{eq:CAPM}.
\begin{center}
\includegraphics[scale=0.5]{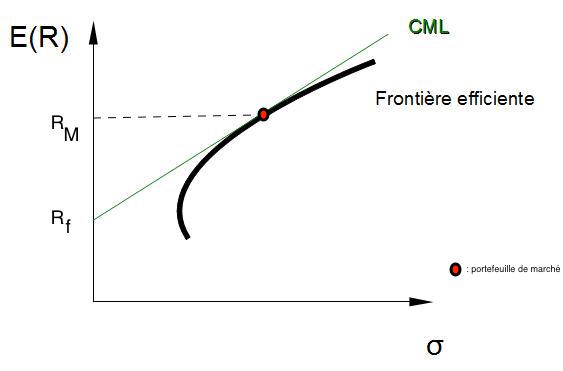}
\end{center}

In order to find the market portfolio, we have to search for the portfolio configuration that maximize the Sharpe ratio:
$$S_M=\frac{E(R_M)-R_f}{\sigma_M}$$

\stepcounter{question}
\begin{itemize}
\item \textbf{Question \arabic{question}}
\begin{itemize}
\item Demonstrate that the market portfolio $M$ is such that, when it is composed of $n$ assets:
\end{itemize}
\end{itemize}
$$\forall i,j \in [1,n], i\neq j, \frac{E(R_i)-R_f}{cov(R_i,R_M)}=\frac{E(R_j)-R_f}{cov(R_j,R_M)}$$

 \subsection{Security Market Line}

The \emph{Security Market Line} (SML) indicates the expected return of an asset, $R_i$, knowing its $\beta_i$:
\begin{equation}\label{eq:SML}
R_i=\alpha + \beta_i R_M
\end{equation}
Depending on the position of the asset relative to this SML, the asset is either undervalued or overvalued:
\begin{center}
\includegraphics[scale=0.5]{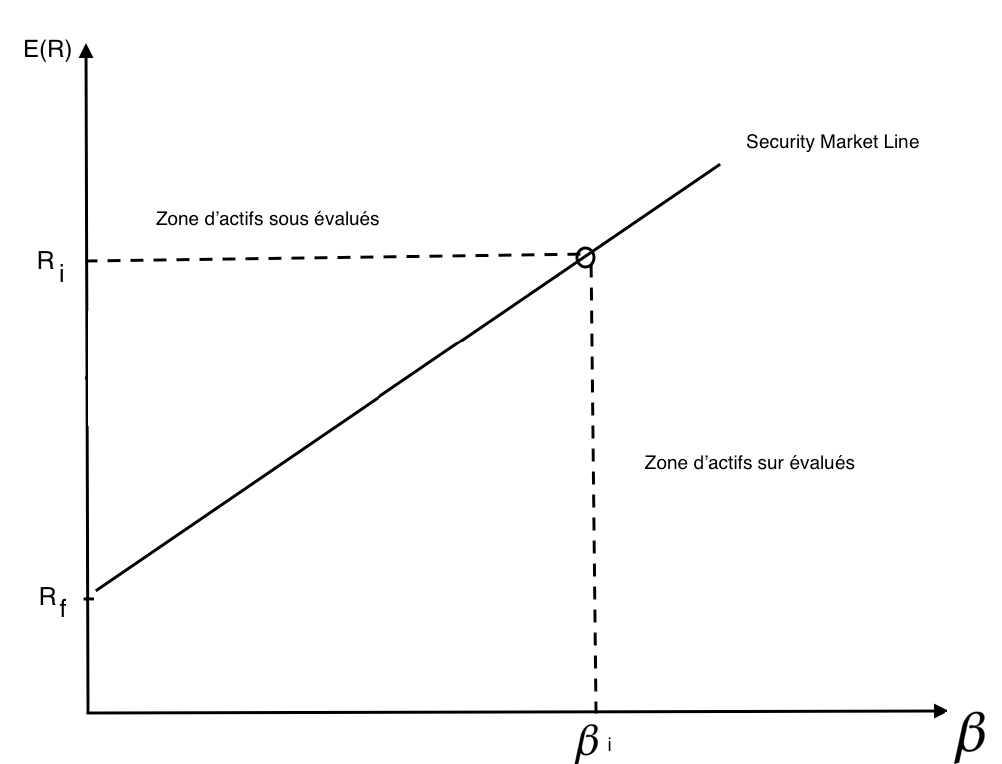}
\end{center}
A negative $\alpha$, means that the expected return of the asset is lower than the market after taking into account the systematic risk.

\begin{center}
\includegraphics[scale=0.5]{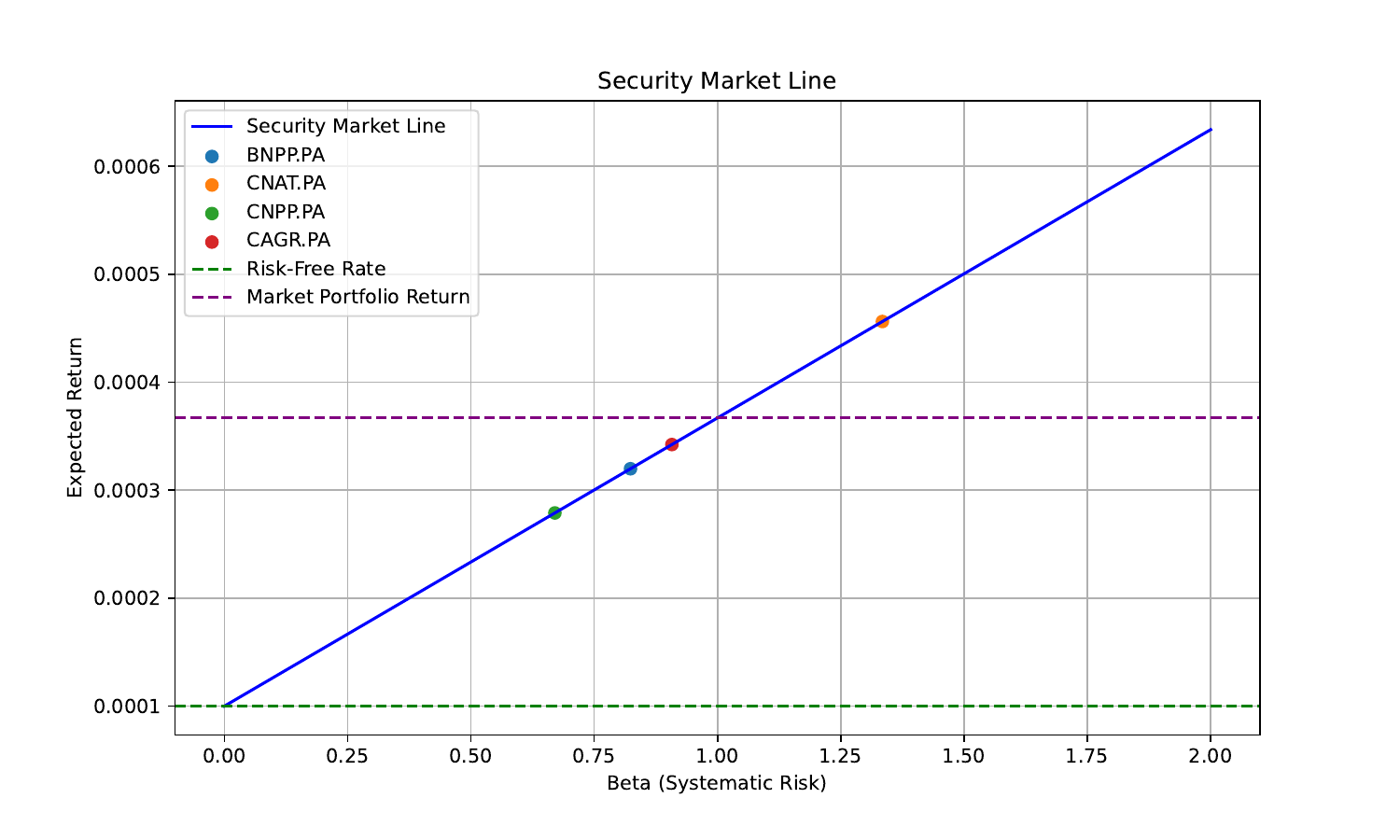}
\end{center}

\section{Arbitrage Pricing Theory (APT)}\label{sec:APT}

The model from section \ref{sec:CAPM} has to main limitations:
\begin{itemize}
\item
the model is based on the mean variance
\item
the returns of an assets are captured by a single risk factor: the risk associated with the market porftolio (with a factor $\beta_i$ that amplify or reduce this risk)
\end{itemize}

There are some empirical anomalies with this CAPM: when portfolios are formed on the basis of firm size, small firms systematically experienced average rates of return nearly 20\% per year greater than those of large firms, even after accounting for differences in estimated betas, \citep{reinganum1981arbitrage}.

The Arbitrage Princing Theory (APT) model was defined in 1976 by Stephen Ross, \citep{ross1976}. This model takes into account other factors to explain the performance of an asset. The asset is supposed to react to those factors when the latter show unexpected variation. While both APT and CAPM are used to determine expected returns, APT offers a more flexible framework by allowing for multiple factors. Unlike CAPM, which is a single-factor model (only considering market risk), APT can incorporate different types of systematic risk, providing a potentially more accurate and comprehensive model of risk and return.

The arbitrage pricing theory is essentially based on three assumptions. First, capital markets are perfectly competitive. Secondly, investors always prefer more wealth to less wealth with certainty. And, lastly, the stochastic process generating asset returns can be represented as a k-factor model of the form of equation\ref{eq:APT}. In equilibrium, the return on a zero-investment, zero-systematic-risk portfolio is zero (no arbitrage assumption), as long as the idiosyncratic effects vanish in a large portfolio. This economic reasoning, combined with theory from linear algebra, implies that the expected return on any asset i can be expressed as equation \ref{eq:APT2}.
\begin{equation}\label{eq:APT}
R_i =\alpha_i +\sum_{k}\beta_{i,k} f_k + \epsilon_i
\end{equation}
where $R_i$ is the return on asset $i$, $\alpha_i$ is a constant for asset $i$, $\beta_{ij}$ is the sensitivity of the $i$th asset to the $j$th factor, $F_j$ is the $j$th factor, $\epsilon_i$ is the idiosyncratic risk (firm-specific risk) of asset $i$. We assume that $E\left( \epsilon_i \right)=0$, $E\left(f_k \epsilon_i \right)=0$ and $E\left(\epsilon_i \epsilon_j \right)=0$.
\begin{equation}\label{eq:APT2}
E\left( R_i \right) = R_f +\sum_{k}\beta_{i,k} \mbox{RP}_k 
\end{equation}
where $RP_k$ is the risk premium associated with the $k$th factor, defined as $RP_k = E(F_k) - R_f$, and $R_f$ is the risk-free rate, that is the expected return on an asset with zero systematic risk.

Examples of possible factors $F_j$ :
\begin{itemize}
\item investor confidence
\item inflation
\item central bank interest rate
\item GDP growth rate
\item market index
\end{itemize}

\citet{onatski2010determining} designs a method to estimate the number of factors to be one, maximum two.

\subsection{APT and Diversification}

APT suggests that diversification can reduce but not eliminate systematic risk. By holding a well-diversified portfolio, investors can eliminate idiosyncratic risk. The remaining risk, which cannot be diversified away, is the systematic risk associated with the factors in the model.

\subsection{Limitations of APT}

APT, while more flexible than CAPM, has its own limitations:
\begin{itemize}
\item Identification and selection of appropriate factors are challenging and can be subjective.
\item Estimation of factor sensitivities requires extensive historical data and sophisticated statistical techniques.
\item The assumption that idiosyncratic risks are uncorrelated and have zero mean may not hold in practice.
\end{itemize}

\subsection{Empirical estimation of the APT}

\subsubsection{Factor analysis}

\href{https://online.stat.psu.edu/stat505/lesson/1}{Factor Analysis} (FA) is a statistical approach used to model the relationships between observed variables and a smaller number of latent variables, known as factors. These latent factors are hypothesized to represent underlying processes or constructs that influence the observed measurements. For instance, underlying social desires could explain variations in consumer behavior patterns. FA is particularly valuable for exploring complex phenomena in fields such as psychology, sociology, and economics, where direct measurement of these constructs is often impractical. FA is primarily exploratory, serving to identify underlying structures in datasets. It is subject to interpretative flexibility, leading to diverse opinions on its findings. Despite its resemblance to Principal Component Analysis (PCA), FA provides a more nuanced understanding by modeling observed variables as linear combinations of latent factors, in contrast to PCA's approach of transforming observed variables into principal components. Let $\mathbf{X}$ represent the vector of observed variables for an individual, where $\mathbf{X_i}$ denotes the $i^{th}$ variable. This vector is expressed as:
\begin{equation*}
\mathbf{X} = \left( \begin{array}{c}
X_1 \\
X_2 \\
\vdots \\
X_p
\end{array} \right),
\end{equation*}
with a corresponding population mean vector $\boldsymbol{\mu}$.

The factor model posits that each observed variable $X_i$ is a linear function of $m$ unobservable factors $f_j$ and a random error term $\epsilon_i$:
\begin{align*}
X_1 &= \mu_1 + l_{11}f_1 + l_{12}f_2 + \dots + l_{1m}f_m + \epsilon_1,\\
X_2 &= \mu_2 + l_{21}f_1 + l_{22}f_2 + \dots + l_{2m}f_m + \epsilon_2,\\
&\vdots \\
X_p &= \mu_p + l_{p1}f_1 + l_{p2}f_2 + \dots + l_{pm}f_m + \epsilon_p.
\end{align*}
The coefficients $l_{ij}$ are known as factor loadings, and the model is succinctly represented in matrix form as:
\begin{equation*}
\mathbf{X} = \boldsymbol{\mu} + \mathbf{Lf} + \boldsymbol{\epsilon}.
\end{equation*}
where $\boldsymbol{\epsilon}$ is a diagonal matrix with the variance of the random errors.

The model presupposes that both common factors and specific factors have zero mean and that factors are uncorrelated among themselves and with specific errors. These assumptions ensure the identifiability and parsimony of the model.  The variance-covariance matrix is expressed as shown below:
\(\Sigma = \mathbf{LL'} + \boldsymbol{\Psi}\)
The Principal Component Method approximates the factor model using the sample variance-covariance matrix \(\textbf{S} = \dfrac{1}{n-1}\sum\limits_{i=1}^{n}\mathbf{(X_i - \bar{x})(X_i - \bar{x})'}\), eigenvalues $\hat{\lambda}_i$, and eigenvectors $\hat{\mathbf{e}}_i$. Factor loadings and specific variances are estimated, facilitating the interpretation of underlying factors.

\begin{equation*}
\mathbf{L} = \begin{pmatrix}
\sqrt{\hat{\lambda}_1}\hat{\mathbf{e}}_1 & \sqrt{\hat{\lambda}_2}\hat{\mathbf{e}}_2 & \cdots & \sqrt{\hat{\lambda}_m}\hat{\mathbf{e}}_m
\end{pmatrix}
\end{equation*}
with specific variances computed as $\hat{\Psi}_i = s^2_i - \sum_{j=1}^{m}\lambda_j \hat{e}^2_{ji}$.

\subsubsection{PCA Analysis in APT}

Principal Component Analysis (PCA) is integrated into the Arbitrage Pricing Theory (APT) framework to identify key factors affecting asset returns, offering a systematic approach to dimensionality reduction and factor estimation.

\textbf{Data Preparation:} Historical return data for a range of assets is collected and standardized.

\textbf{Applying PCA:} PCA is applied to the asset returns to extract principal components (PCs), which represent potential factors in the APT model.

\textbf{Factor Identification:} Significant PCs that explain a substantial portion of the total variance are selected as factors for the APT model.

\textbf{Estimating Factor Loadings:} Factor loadings, $\beta_{ij}$, are estimated by projecting asset returns onto these significant PCs.

\textbf{Factor Model Construction:} The APT factor model for each asset is constructed using the identified factors and estimated loadings, along with an estimated expected return, $E(R_i)$.

\textbf{Key Benefits:}

\textit{Dimensionality Reduction:} Simplifies APT by identifying major factors from complex datasets.
\textit{Objective Factor Identification:} Offers a methodical approach to uncovering underlying factors.

\textbf{Limitations:}

\textit{Interpretability:} Factors derived may lack clear economic meaning.
\textit{Assumptions:} PCA assumes linearity and may miss non-linear relationships.
\textit{Historical Bias:} Relies on past data, potentially limiting future applicability.

\subsection{5-factor Model}

Import data from the Kennth French website: \href{http://mba.tuck.dartmouth.edu/pages/faculty/ken.french/data_library.html}{here}.

In this dataset, we have 5 factors enabling the estimation of a 5-factor model:
\begin{enumerate}
    \item 'Mkt-RF': Market risk premium
    \item 'SMB': Size factor (Small Minus Big)
    \item 'HML': Value factor (High Minus Low book-to-market)
    \item 'RMW': Profitability factor (Robust Minus Weak)
    \item 'CMA': Investment factor (Conservative Minus Aggressive)
\end{enumerate}

The first step involves performing a linear regression:
\begin{equation}
R_{i,t} - R_{f,t} = \alpha_i + \sum_{k=1}^5 \beta_{i,k} \mbox{RP}_{k,t} + u_{i,t}
\end{equation}

Under the no-arbitrage condition, we set $\alpha_i = 0$, leading to the following regression estimation:
\begin{equation}
R_{i,t} - R_{f,t} = \sum_{k=1}^5 \beta_{i,k} \mbox{RP}_{k,t} + u_{i,t}
\end{equation}

The steps are as follows:
\begin{enumerate}
    \item Loop through each asset and estimate the $\beta_{i,k}$.
    \item Store the estimated $\beta_{i,k}$ in a data frame, dfAPT.
\end{enumerate}

\subsubsection{Risk Premia under the APT}

Once the $\beta_{i,k}$ have been estimated, we can estimate the risk premia $\gamma_k$ as follows:
\begin{equation}
\overline{R_{i,t} - R_{f,t}} =  \sum_k \beta_{i,k} \gamma_k
\end{equation}

This regression is performed to provide the estimated risk premia.

\subsection{Empirical Testing of APT}

Empirical testing of APT involves selecting a set of factors and testing whether these factors can explain the cross-section of asset returns. Commonly used factors include market risk, size, value, momentum, and various macroeconomic indicators.

\citep{reinganum1981arbitrage} empirically rejects the APT model.

\section{python and R: Some portfolio performance measures}

\subsection{Downloading data and importing it}

Import data from the Kennth French website: \href{http://mba.tuck.dartmouth.edu/pages/faculty/ken.french/data_library.html}{here}.

\begin{enumerate}
\item We are using the the 30 Industry Portfolios in csv format, \href{http://mba.tuck.dartmouth.edu/pages/faculty/ken.french/ftp/30_Industry_Portfolios_daily_CSV.zip}{link here}.
\item It is already in the  data forlder under the name \verb|30_Industry_Portfolios_Daily.CSV|
\item Import the data as a data frame, for that to work, you might want to use the first column as index, those are the dates\footnote{index\_col = 0 if you use python}. These are already returns so you can directly work and regress them (after doing ADF test).
\item You might want to convert the dates into date format.
\end{enumerate}

\subsubsection{Plotting the data}

\begin{enumerate}
\item Plot one of the 30 industries daily returns.
\end{enumerate}

\subsection{Risk and performance measures}

Nota bene: in this section, for simplicity, we consider that $R_{f,t} = 0, \forall t$.

\subsubsection{Risk: variance and standard deviation of returns}

Compute the sample variance of each of the 30 industry returns as a risk measure. The historical variance of an asset $i$ over the time horizon $T$ is measured as:
\begin{equation}
\hat{\sigma}^2_i = \frac{1}{T-1} \sum_{t=1}^T  (R_{i,t} - \overline{R}_i)^2
\end{equation}
with $\overline{R}_i$ the average return over the entire period $T$.
Here you can use the standard deviation function directly to compute $\hat{\sigma}_i$.

\subsubsection{Sharpe ratio}

For an asset $i$, the Sharpe ratio is defined as:
\begin{equation}
SR(R_i) = \frac{\overline{R_i - R_f}}{\sigma(R_i - R_f)}
\end{equation}

You might want to decompose this in two steps:
\begin{enumerate}
\item Compute the average return for each column of the data set
\item Divide this average return per asset over its standard deviation computed before
\end{enumerate}

You should find the best Sharpe Ratio for the "Food" at $0.0467$.

\subsubsection{Semi-volatility}

As an investor, we are mostly concerned about negative returns. Therefore we might want to compute the semi-volatility, which only takes into account the returns below the mean return over the horizon:
\begin{equation}
sv(R_i) = \sqrt{ \frac{1}{T-1} \sum_{t=1}^T \max(0, \overline{R_i} - R_{i,t})^2 }
\end{equation}

For this, you might want to follow these steps:
\begin{enumerate}
\item Create a new data frame dfneg where for each column you remove its mean return
\item Take the minus of this data frame
\item Replace negative values by $0$
\item Take the squares of the values of this matrix
\item Take the mean for each column
\item Take the square roots of those values
\end{enumerate}

The worst semi-volatility you should find is for Servs at $1.66$

\subsection{Sortino ratio}
The Sortino ratio is a performance measure computed as:
\begin{equation}
SoR(R_i) = \frac{\overline{R_i - R_f}}{sv(R_i - R_f)}
\end{equation}

Compute it for each column. The best Sortino ratio is for Foods at $0.065$.

\subsection{Maximum Drawdown}

The concept of Maximum Drawdown (MDD) measures the largest single drop from peak to trough in the value of a portfolio, before a new peak is achieved. This metric is particularly relevant in the study of investor behavior and risk aversion, as it captures the potential loss an investor might face during the holding period. For a review of the MDD see \citet{geboers2023review}.

\paragraph{Definition}
Formally, the Maximum Drawdown at time $t$ over a historical window of $s$ time units (for example, $252$ trading days in a year) for a portfolio with liquidation value $P_t$ is defined as the largest percentage drop from any peak to any subsequent trough:

\begin{equation}
\mbox{MDD}_{t,s} = \max_{t-s\leq i < j\leq t} \left( \frac{P_i - P_j}{P_i} \right)
\end{equation}

Here, $P_i$ and $P_j$ represent the portfolio values at times $i$ and $j$, respectively, with $i$ and $j$ being any two times within the considered window where $i$ is a peak and $j$ is the trough following $i$.

\paragraph{Computational Approach}
To calculate the MDD, one needs to implement a scanning algorithm over the specified time period. The process can be outlined as follows:

\begin{enumerate}
    \item Initialize the maximum drawdown as $0$ and the peak value as the portfolio value at the start of the period.
    \item Iterate over each time point $t$ in the window:
    \begin{enumerate}
        \item Update the peak value if the current portfolio value exceeds it.
        \item Compute the drawdown at $t$ as the percentage drop from the peak to the current value.
        \item If this drawdown is larger than the maximum drawdown recorded so far, update the maximum drawdown.
    \end{enumerate}
    \item The maximum drawdown at the end of the iteration is the desired MDD.
\end{enumerate}

\subsection{Mathematical Framework for Computing Maximum Drawdown}

This section translates the computational logic of the Python functions, designed to calculate portfolio returns and Maximum Drawdown (MDD), into a formal mathematical framework. This elucidation facilitates a deeper understanding of the underlying principles guiding the MDD computation.

\paragraph{Portfolio Return Calculation}
The first step in analyzing portfolio performance and subsequently calculating MDD involves determining the average return of a portfolio over a given period. The portfolio return is computed as the weighted average of individual asset returns, mathematically expressed as:

\begin{equation}
R_{\text{portfolio}} = \frac{1}{N} \sum_{k=1}^{N} w_k \cdot R_k
\end{equation}

where:
\begin{itemize}
    \item $R_{\text{portfolio}}$ is the average return of the portfolio over the period.
    \item $N$ is the number of assets in the portfolio.
    \item $w_k$ represents the weight of the $k^{th}$ asset in the portfolio.
    \item $R_k$ is the return of the $k^{th}$ asset.
\end{itemize}

This equation serves as the basis for calculating the portfolio's performance, essential for the subsequent computation of MDD.

\paragraph{Maximum Drawdown Calculation}
The Maximum Drawdown (MDD) is determined through the analysis of cumulative returns over the investment period. The process involves identifying peak values in the cumulative return series and calculating the drawdown from these peaks. The mathematical representation of this process is as follows:

\begin{equation}
\text{MDD} = \min_{t \in [0, T]} \left( \frac{CR_t}{\max_{\tau \in [0, t]} CR_{\tau}} - 1 \right)
\end{equation}

where:
\begin{itemize}
    \item $CR_t$ is the cumulative return at time $t$, defined as $CR_t = \prod_{i=1}^{t} (1 + R_i)$, with $R_i$ being the return at time $i$.
    \item $T$ is the total duration of the period analyzed.
    \item The expression $\max_{\tau \in [0, t]} CR_{\tau}$ identifies the peak cumulative return up to time $t$.
    \item The result of this equation, $\text{MDD}$, represents the largest percentage drop observed in the portfolio's value from a peak to a subsequent trough within the analyzed period.
\end{itemize}

\begin{center}
\includegraphics[scale=.5]{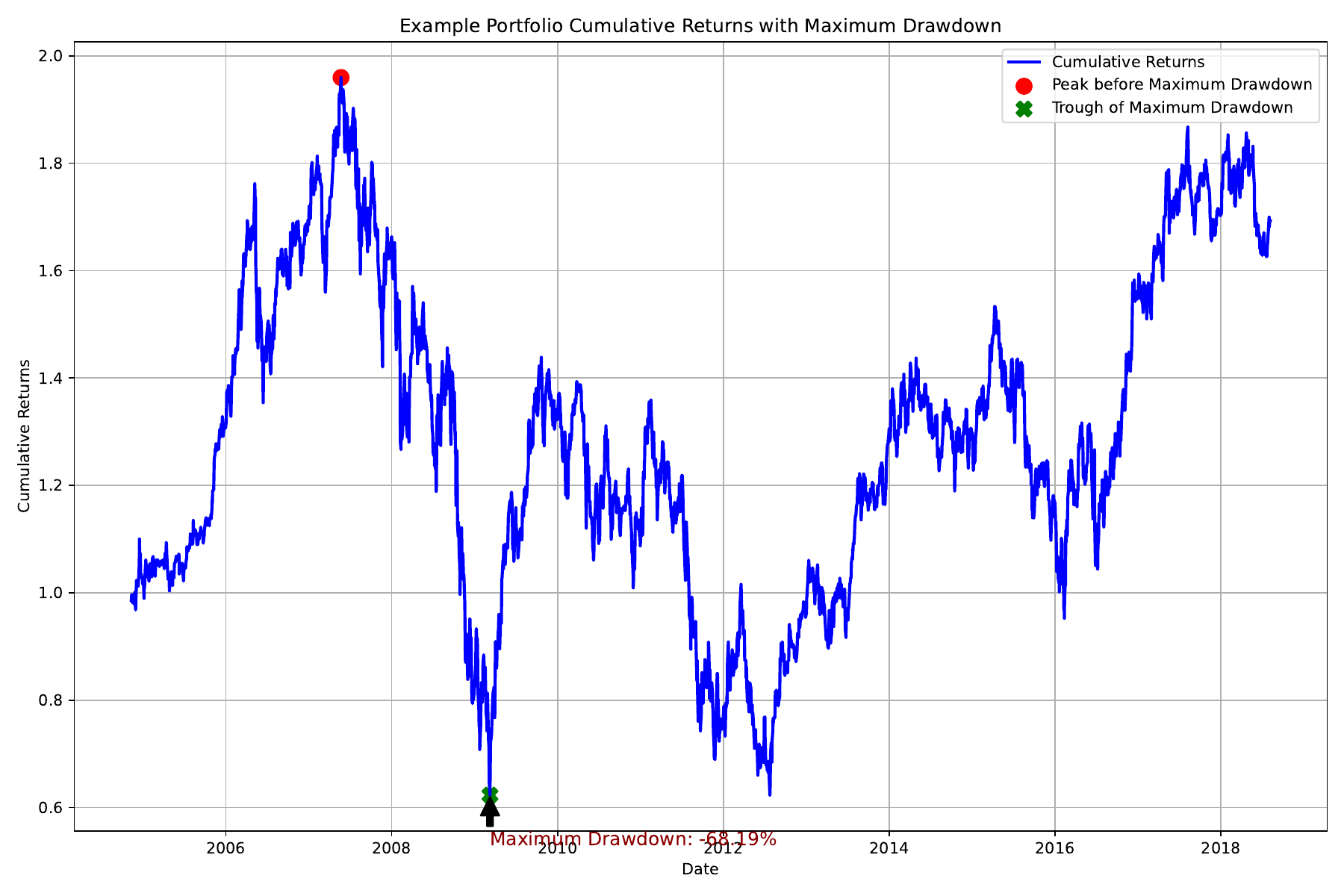}
\end{center}

\paragraph{Investor Utility and Risk Aversion}
The MDD is a critical metric from the perspective of investor utility and risk aversion. Unlike standard deviation, which considers both upward and downward volatility, MDD focuses solely on the downside risk, which is often a primary concern for investors. The concept of MDD aligns with the behavioral finance principles where investors are particularly sensitive to losses, as described in \citet{tversky1992advances} prospect theory.

\paragraph{Recent Academic Insights}
Recent studies have extended the understanding of MDD in portfolio optimization and risk management. For instance, \citet{chekhlov2005drawdown} discuss the application of MDD in the context of portfolio optimization. Moreover, \citet{goldfarb2003robust} provide insights into incorporating MDD constraints in mean-variance portfolio optimization models.

\subsubsection{Length}
The length of the Maximum Drawdown is the time between the peak to bottom.

\subsubsection{Recovery time}
Here we define the recovery time as the time it take, once the bottom of the Maximum Drawdown has been reach, to recover at least the portfolio value at the time of the peak of the Maximum Drawdown.

\subsection{List of assets}

From our first data frame, each column represent an asset, you can create a list of asset by converting these columns names into a list, call it assetlist\footnote{with python, you can use list(df.columns)}.

\subsection{Fama-French factors}

Nota bene: in this section, we now have $R_{f,t} \neq 0$.

\begin{itemize}
\item Import the daily times series of Fama-French factors, this is in the  data folder under the file \verb|F-F_Research_Data_5_Factors_2x3_daily.csv|
\item Import the data as a data frame you name df2, for that to work, you might want to use the first column as index, those are the dates\footnote{index\_col = 0 if you use python}.
\item You might want to convert the dates into date format.
\end{itemize}

\subsubsection{Merge both data set}

You now join both data frame df and df2 into a unique data frame keeping only days for which you have the full information (no NAs in the row)\footnote{with python you can use pd.concat and then dropna(), df here is the data frame with the industry returns}.

\subsubsection{Market returns and CAPM}

From your merged data frame, you can consider that the market returns are the column 'Mkt-RF'. In the following we call these returns $z_{m,t}$.

\subsubsection{CAPM for each asset}

For each asset, you can compute the following linear regression:
\begin{equation}
R_{i,t}-R_{f,t} = \alpha_i + \beta_i z_{m,t} + u_{i,t}
\end{equation}

You might want to proceed as follow:
\begin{enumerate}
\item Create a data frame dfCAPM with the assets as index and two columns, alpha and beta
\item Loop through all assets and compute $R_{i,t}-R_{f,t} = \alpha_i + \beta_i z_{m,t} + u_t$
\item Store the $\alpha_i$ and $\beta_i$ of each regression for each corresponding asset.
\end{enumerate}

\subsubsection{Treynor ratio}

We can now compute the Treynor ratio as:
\begin{equation}
TrR_i = \frac{\overline{R_i-R_f}}{\beta_i}
\end{equation}

The best Treynor ratio is for Beer at $0.079$.

\subsubsection{Jensen's alpha}

Jensen's alpha characterise the over- or under-performance of an asset return compared with the return expected with the CAPM:
\begin{equation}
\hat{\alpha}_i = \overline{R_i-R_f} - \hat{\beta_i} \overline{z_{m}}
\end{equation}

The highest Jensen's alpha is for Smoke at $0.029$.

\section{Pairs trading and statistical arbitrage}

Pairs trading, the ancestor of statistical arbitrage, has over 400 citations in the literature, for which a review is done by \citep{Krauss:2017aa}.

\subsection{Pairs trading strategy based on cointegration}

The following\footnote{pyton code: cointegration\_trading\_vansteenberghe} is inspired by the academic paper \citep{Rad:2016aa} and the work \href{https://www.quantconnect.com/tutorials/pairs-trading-strategy/}{here}. Cointegration test and some economic examples have been introduced in section \ref{sec:cointegration}.\citep{Gatev:2006aa} found that between 1962 and 2002 \begin{quote}
find[ing] two stocks whose prices have moved together historically. When the spread between them widens, short the winner and buy the loser. If history repeats itself, prices will converge and the arbitrageur will profit. [...] we show that our profits are not caused by simple mean reversion as documented in the previous literature.
\end{quote} Such a strategy yield up to 11\% average annualized excess returns for self-financing portfolios of top pairs. Noting that "If the U.S. equity market were efficient at all times, risk-adjusted returns from pairs trading should not be positive."

\quad

Two $I(1)$ time series $x_t$ and $y_t$ are said to be cointegrated if there exist a linear combination of them (e.g. $\alpha + y_t + \beta_x x_t$) that is I(0). If $E(\alpha + y_t + \beta_x x_t) = 0$, then there is a long term relationship between $y_t$ and $x_t$, with an equilibrium relationship: $y_t = -\alpha  - \beta_x x_t$.

We can then test for cointegration between variable with a cointegration test: 

\begin{center}
\textbf{the null hypothesis $H_0$ is no cointegration.} 
\end{center}

The second output of the python test gives the p-value of the test.

\quad

In simple terms, two cointegrated time series have a long term relationship: if they are departing from one another in level, one can expect the series multiplied by some coefficients to "meet again" (if past behaviours would be a good predictor for future behaviours). If you believe in this, then you might spot some investment opportunities (\href{https://www.quantconnect.com/tutorials/pairs-trading-strategy/}{inspired from here}).

\quad

In other terms, there is no unit root in the error terms of the regression:
$$Y_t = \alpha + \beta X_t + \epsilon_t$$

To get an intuition, you can think of a system:
\begin{equation}
\begin{cases}
\Delta Y_t = \beta_Y \epsilon_{t-1} + \nu_{Y,t}\\
\Delta X_t = \beta_X \epsilon_{t-1} + \nu_{X,t}
\end{cases}
\end{equation}

Then if we imagine that we have significant coefficients $\beta_Y < 0$ and $\beta_X > 0$, then in case the spread depart from its long-term mean (e.g. $\epsilon_{t-1} > E(\epsilon_{t-1})$) then the stock price $Y_t$ is expected to decrease and the stock price $X_t$ is expected to increase to correct and bring the spread to its long-term mean.

We can also think of an error-correction model by OLS:
$$\Delta Y_t = \gamma_0 + \gamma_1 \Delta X_t + \gamma_3 \epsilon_{t-1} + \nu_t$$
with
$$\epsilon_t = Y_t - \hat{\alpha} -  \hat{\beta} X_t$$

We search for cointegrated time series candidates:
\begin{enumerate}
\item we focus on the biggest capitalization to avoid liquidity issues when trading
\item we work with series in $\log$ to avoid level issues	
\item we keep only $I(1)$ time series
\item we rank the couples of time series by cointegration test p-values
\end{enumerate}

\quad

Here is an example with two candidates:
\begin{center}
\includegraphics[scale=0.5]{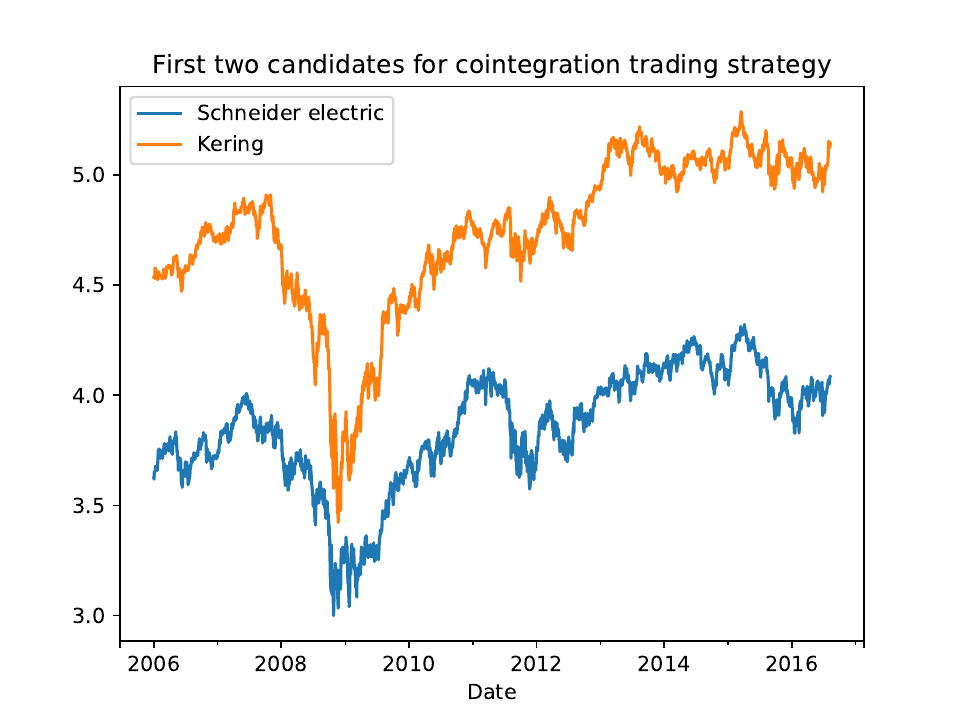}
\end{center}

\quad

Then at time $t$:

\begin{enumerate}
\item we estimate a coefficient $\beta$ of sensitivity between $y$ and $x$ based on all the history of the time series, the value at time $t$ not included.
\item we estimate the spread between both time series:
$$\mbox{Spread}_t = y_t - \hat{\alpha} - \hat{\beta} x_t$$
\item if the value of the Spread at time $t$ is away from its historical mean $\mu$ by more than 2 times its historical standard deviation $\sigma$, we are:
\begin{itemize}
\item short $y$ and long $x$ if $\mbox{Spread}_t > \mu + 2 \sigma$
\item long $y$ and short $x$ if $\mbox{Spread}_t < \mu - 2 \sigma$
\end{itemize}
\item we exit a trade once we are only only or less than $2$ standard deviations away from the mean.
\end{enumerate}

Nota Bene: as we assume both series to be cointegrated, we expect:
$$E(\mbox{Spread}_t) = \mu$$

hence if the Spread in absolute value is significantly different than $\mu$ we consider that there is a trading opportunity.

\begin{center}
\includegraphics[scale=0.5]{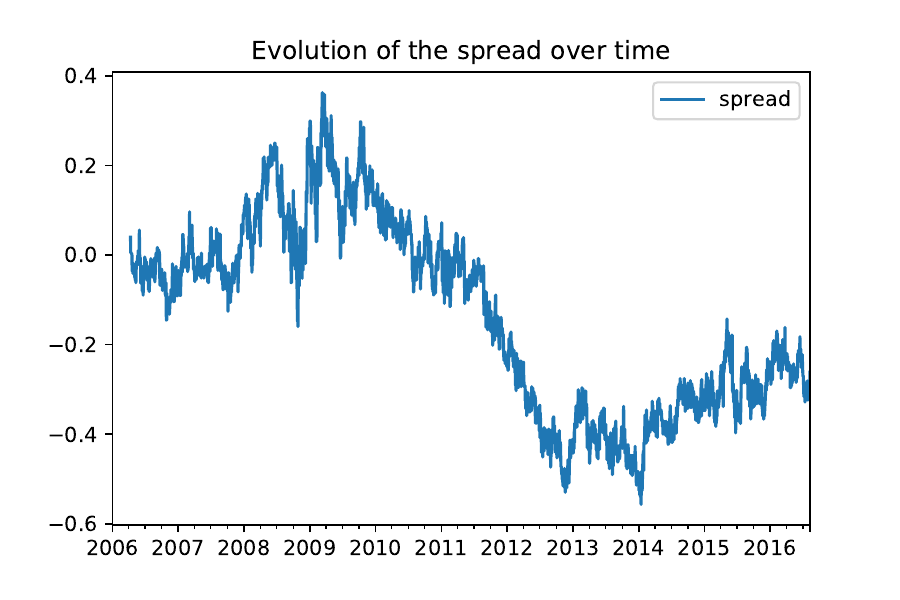}
\end{center}

\stepcounter{question}
\begin{itemize}
\item \textbf{Question \arabic{question}}
\begin{itemize}
\item Compute the return of this strategy. Apply this strategy on some other pair candidates, is the strategy robust?
\item In fact, to form your decision on whether the series are cointegrated or not, you can only work on past historical data, change the code to reflect this.
\item To reduce the number of short/long orders you might want to initiate order only when crossing the $2$ standard deviations distance from the spread mean.
\end{itemize}
\end{itemize}

\subsection{Introduction to copulas}

We follow\footnote{copula\_vansteenberghe} the illustration for the paper \citep{copulas}:

\quad

\begin{itemize}
\item First we rank the stock log return pairs per Kendall $\tau$
\item Use the Clayton copulas: $\theta = 2 \tau ( 1 - \tau )^{-\theta}$
\item Assume the pdf: $C(u,v) = (u^{-\theta} + v^{-\theta} -1 ) ^{-\frac{1}{\theta}}$ 
\item Identify if you are in the $95\%$ confidence interval to determine trading opportunities
\end{itemize}

\section{Python: introduction to power laws}\label{sec:powerlaw}

We have seen section \ref{sec:weightheightdistrib} distribution that peaks around a typical value (the median or the mean). But some distributions (e.g. city populations, individual wealth) can have bulk around small values and a small proportion can reach "very high" values, leading to fat right tails, up to a point were the variance becomes infinite and even in some cases the expectation becomes infinite too.

A quantity $x$ obeys a power law if it is drawn from a probability distribution
\begin{equation}\label{eq:powerlaw}
p(x) = C x^{-\gamma}
\end{equation}
where both $\gamma$ and $c$ are constant. Distribution of the form \ref{eq:powerlaw} were introduced by \citet{pareto1895} and are said to follow a power law with an exponent $\gamma$, typically between $2$ and $3$. $C$ is not interesting, as once $\gamma$ is calibrated, it is chosen so that $p(x)$ integrates to $1$, $C=(\gamma-1) x_{min}^{\gamma-1}$ provided $\gamma>1$. $x_{min}$, also written $x_m$ is the scale of a Pareto distribution.

Nota bene:
\begin{itemize}
\item typically, it is only the right tail (e.g. above $95\%$ quantile) of distributions that follow a power law;
\begin{itemize}
\item Hence it is not usually possible to collect enough data to accurately estimate the tail directly. The power law is a way of overcoming this problem.
\end{itemize}
\item  if we want to reconcile the probability density function of a power law presented equation \ref{eq:powerlaw} and the Pareto probability density function presented section \ref{sec:paretodist}, then we need to transfrom: $\gamma = \alpha + 1$, where $\alpha$ is the shape of the Pareto.
\end{itemize}

We assume in the following that the exponent is greater than $1$, we display here different pdf for different Pareto shapes ($\alpha$):
\begin{center}
\includegraphics[scale=0.5]{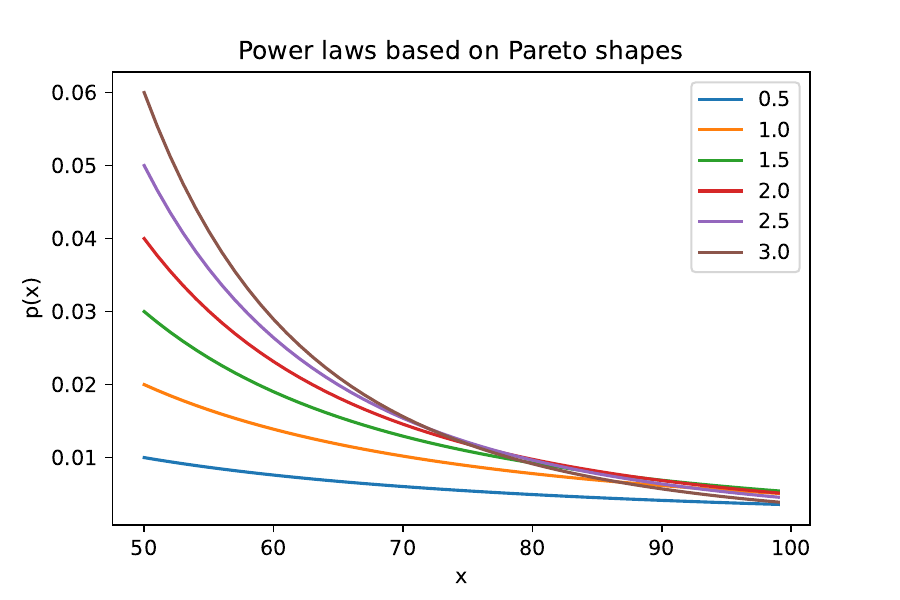}
\end{center}

Provided that the exponent is greater than $1$, i.e. the Pareto shape parameter is greater than $0$, when replotted with logarithmic horizontal and vertical axes, the histogram follows quite closely a straight line. Let $p(x) dx$ be the fraction of observations between $x$ and $x+dx$, then we have a relationship, reformulated from equation \ref{eq:powerlaw}:
\begin{equation*}
\ln p(x) = - \gamma \ln x + c
\end{equation*}

\begin{center}
\includegraphics[scale=0.5]{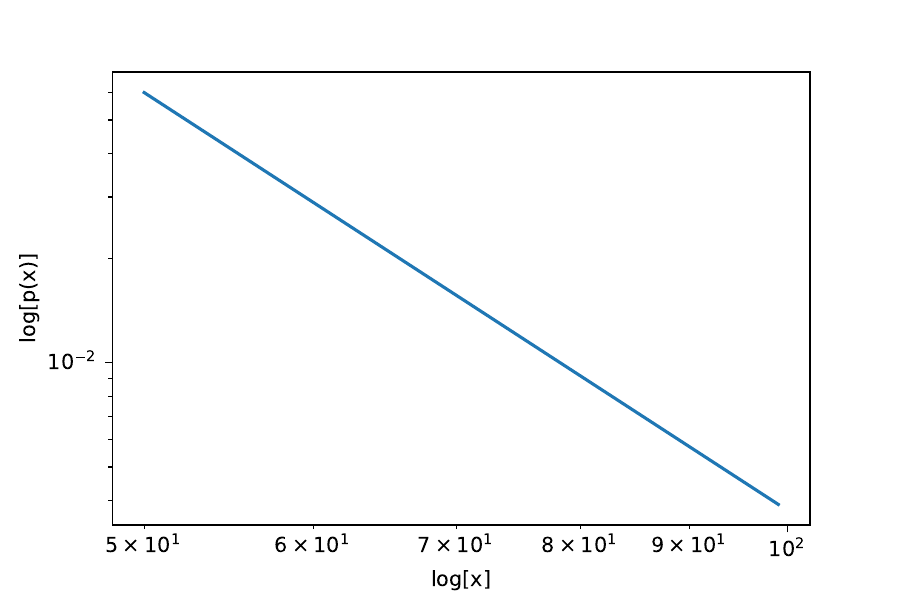}
\end{center}

Most of the time, only the tail of the distribution follows a power law, in other words the power law applies only for values greater than some minimum $x_{min}$.

\subsection{Power laws in nature}

\citep{Newman2005}, \citep{Pinto2012aa} and \citep{gabaix2016} suggested applications and data sets for power laws observable in nature. We follow their work and apply it to a data set we manually download next section.

\subsection{Application to French city sizes}
As in \citep{Newman2005}, we already studied the distribution of human heights section \ref{sec:weightheightdistrib}. Now as claimed, we check that the distribution of French city sizes\footnote{data taken from \href{https://www.insee.fr/fr/statistiques/2521169}{INSEE}} follow a power law\footnote{code: powerlaw\_introduction\_vansteenberghe.py}. \citep{Gabaix1999aa} argues that this is more applicable to agglomerations which is coherent with our data set. We find similar histogram as FIG. 2 of that paper, with a highly right-skewed histogram:
\begin{center}
\includegraphics[scale=0.5]{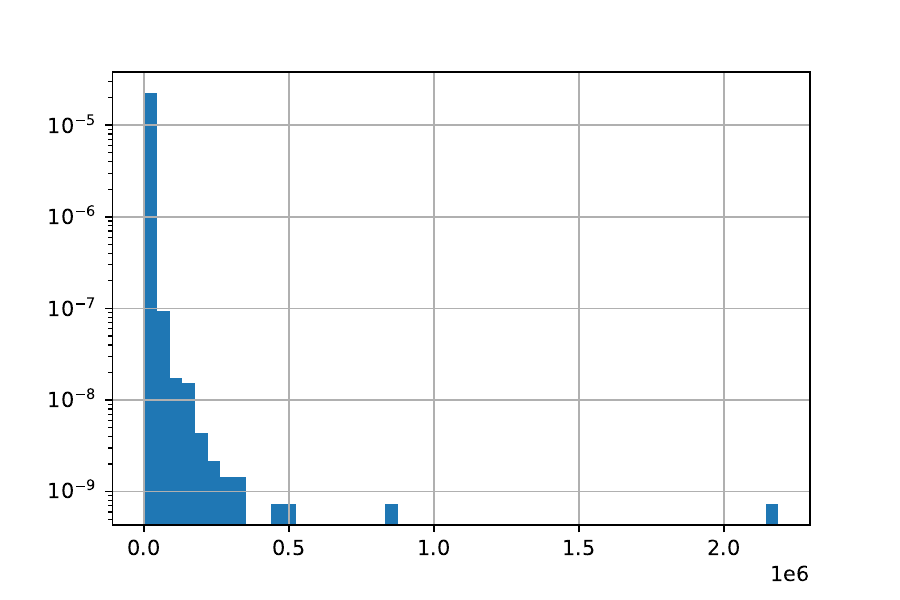}
\includegraphics[scale=0.5]{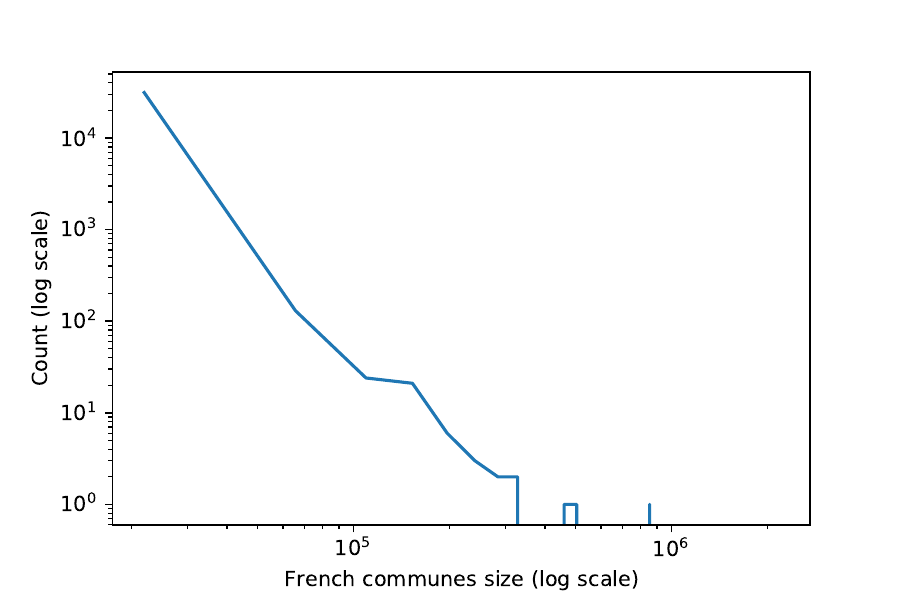}
\end{center}

To construct the log-log histogram, we first have to construct an empirical probability density function:
\begin{itemize}
\item we chose a number of bins for this histogram, nbins
\item we divide our observation segment from $x_{min}$ to $x_{max}$ into nbins segments of equal sizes
\item we count for each segment the number of observations that falls within
\item we report at each point in the middle of a segment how many observations where counted.
\end{itemize}

We can see that the right-hand end of the distribution is noisy because of sampling errors if we keep the bin size evenly distributed. We logarithmically spaced bins:
$$t = \left(\frac{x_{max}}{x_{min}} \right)^\frac{1}{nbins}$$
and for $i \in [1, nbins]$, the bin range is $[x_{min} t^{i-1};x_{min} t^{i}]$:

\begin{center}
\includegraphics[scale=0.5]{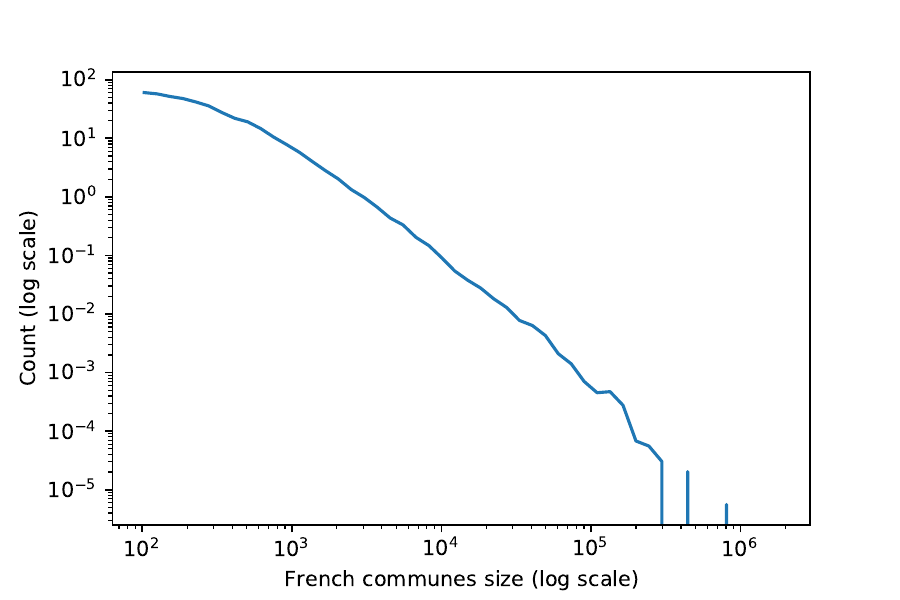}
\end{center}

We also see that we might have to increase the minimum threshold under which we censor the data set.

If we want to avoid modifying the bin width, we can use the complementary cumulative distribution function:
$P(X \geq x) = 1 - P(X \leq x) $ and we have
\begin{equation}\label{eq:cdfpowerlaw}
P(X \geq x) = C \int_x^\infty x^{-\gamma} dx = \frac{C}{\gamma - 1} x^{1-\gamma}
\end{equation}
thus $P(X \geq x)$  follows a power law but of exponent $\gamma-1$:
\begin{center}
\includegraphics[scale=0.5]{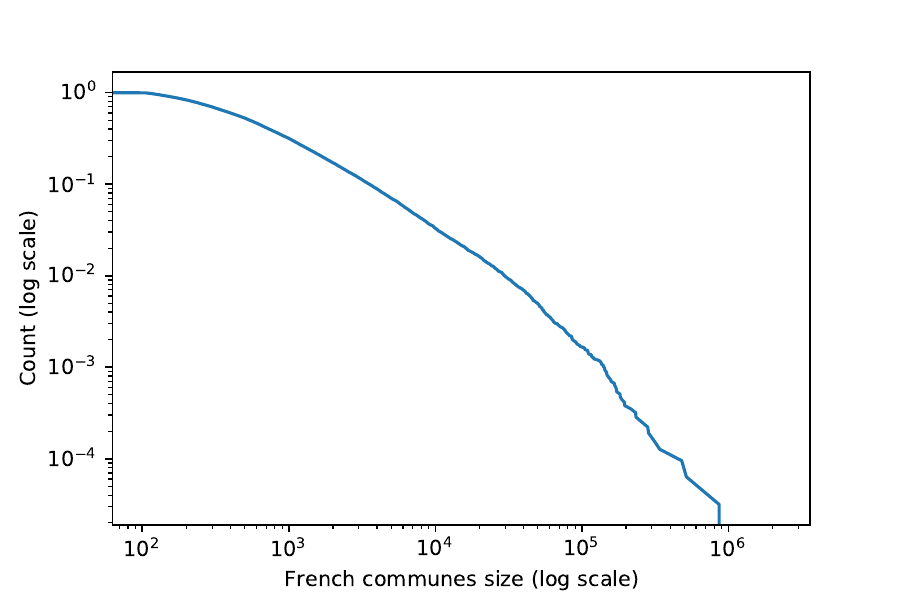}
\end{center}

\stepcounter{question}
\begin{itemize}
\item \textbf{Question \arabic{question}}
\begin{itemize}
\item As in \citep{Gabaix1999aa}, draw a log-rank against log-size. Do you get a straight line, what is the slope?
\end{itemize}
\end{itemize}

\stepcounter{question}
\begin{itemize}
\item \textbf{Question \arabic{question}}
\begin{itemize}
\item manually implement a fake data set from a power law distribution as suggest in \citep{Newman2005} footnote 3 to reproduce all four sub-figures of FIG. 3
\end{itemize}
\end{itemize}

If we try a very basic OLS fit with the pdf and cdf, method dating back to the 19\textsuperscript{th} century work of Pareto:

\quad

\begin{tabular}{@{\extracolsep{5pt}}lcc}
\\[-1.8ex]\hline
\hline \\[-1.8ex]
& \multicolumn{2}{c}{\textit{Dependent variable:}} \
\cr \cline{2-3}
\\[-1.8ex] & equation \ref{eq:powerlaw} & equation \ref{eq:cdfpowerlaw} \\
\hline \\[-1.8ex]
 Intercept & 37.443$^{***}$ & 3.609$^{***}$ \\
  & (3.765) & (0.007) \\
 x & -2.875$^{***}$ & -0.729$^{***}$ \\
  & (0.307) & (0.001) \\
\hline \\[-1.8ex]
 Observations & 11 & 34,077 \\
 $R^2$ & 0.907 & 0.934 \\
 Adjusted $R^2$ & 0.896 & 0.934 \\
 Residual Std. Error & 1.002(df = 9) & 0.257(df = 34075)  \\
 F Statistic & 87.445$^{***}$ (df = 1.0; 9.0) & 479684.588$^{***}$ (df = 1.0; 34075.0) \\
\hline
\hline \\[-1.8ex]
\textit{Note:} & \multicolumn{2}{r}{$^{*}$p$<$0.1; $^{**}$p$<$0.05; $^{***}$p$<$0.01} \\
\end{tabular}

\quad

We lose less observations with equation \ref{eq:cdfpowerlaw}  so we might be more precise, but we know that fitting an OLS is a poor method. Applying the step by step procedure introduced in \citep{Clauset2009aa}, we adapt the $x_{min}$ threshold of our data set and improve the exponent estimation.

\subsubsection{Step by step approximation}\label{sec:approxPL}

We first have to assume that $\gamma > 1$, otherwise we would need apply another method and our power law distribution would lose its interest.

Given empirical observations sorted ${x_1, \ldots, x_N}$ so that $x_1 \leq x_2 \leq \ldots \leq x_N$, we compute the Kolmogorov-Smirnov (KS, see section \ref{sec:KStest}) statistic $KS_i$ considering that $x_{min} = x_i$:
\begin{equation}
D_i = \max_{x\geq x_i} \vert S(x) - P(x) \vert
\end{equation}
where $S(x)$ is the empirical CDF and $P(x)$ is the CDF of the power law that best fit the data choosing $x_{min}=x_i$. We chose $x_{min}$ that minimizes $D_i$.

At each step, to fit $P(x)$, we use the Maximum Likelihood Estimation to estimate $\alpha$, method that we introduced section \ref{sec:MLestimation}.  The probability density function for a sample of $n$ independent identically distributed normal random variables is the likelihood we want to maximize:
$$L(\gamma) = f(x_1, \dots, x_n;\gamma) = \prod_{i=1}^n f(x_i; \gamma) = \prod_{i=1}^n (\gamma-1) x_{min}^{\gamma-1} x_i^{-\gamma}$$
As the logarithm function is continuous strictly increasing and it is easier to deal with the log-likelihood, we maximize:
$$\log\left( L(\gamma)  \right) = n \ln\left(\gamma-1\right) - n \ln x_{min}- \gamma \sum_{i=1}^n \ln \frac{x_i}{x_{min}}$$

The first order condition yields:
$$\hat{\gamma} = 1 + n \left[ \sum_{i=1}^n \ln \frac{x_i}{x_{min}} \right]^{-1}$$

Finally, \citep{Clauset2009aa} suggest using a goodness-of-fit test where the p-value is the fraction of synthetic distances (taken from the estimated power law distribution) that are greater than the observed empirical distance. They suggest to rule out the power law distribution if $p\leq 1$. This goodness-of-fit is not implemented in the powerlaw package and \citep{alstott2014} rather argue to compare the power law fit with competing distribution such as the exponential.

\subsubsection{Step by step approximation applied to French city sizes}

We now use the step by step procedure suggested in \citep{Clauset2009aa}, we find an exponent of 2.13, which is close to the US city sizes exponent of 2.3 reported in \citep{Newman2005} and 2.37 reported in \citep{Clauset2009aa}.

Using equation \ref{eq:cdfpowerlaw} and the fact that $\gamma \simeq 2$, then as in \citep{Gabaix1999aa} we can say that cities with population greater than $S$ is proportional to $\frac{1}{S}$.

\subsection{Application to word frequency in Marcel Proust's work}

We import Marcel Proust's \textit{A la recherche du temps perdu - Du c\^{o}t\'{e} de chez Swann} and find that the word frequencies follow a power law of exponent $2.1$:
\begin{center}
\includegraphics[scale=0.5]{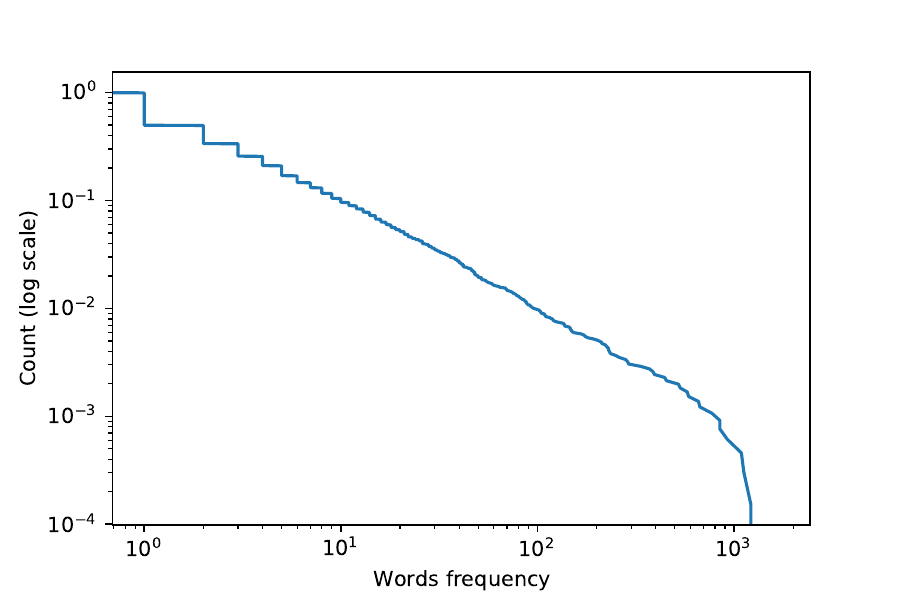}
\end{center}

\subsection{Why power-law emerges for city sizes, an explanation by Xavier Gabaix}

When $\gamma=2$, then equation \ref{eq:cdfpowerlaw} simplifies into Zipf's law:
\begin{equation}
P(X \geq x) \propto \frac{1}{x}
\end{equation}

\citep{Gabaix1999aa} suggest that random growth process can be an explanation for the emergence of Zipf's law, if the following hypotheses are met:
\begin{itemize}
\item Gibrat's law: the growth rate is independent on $x$
\item there are common mean and variance of growth rate process
\end{itemize}

\stepcounter{question}
\begin{itemize}
\item \textbf{Question \arabic{question}}
\begin{itemize}
\item Following \citep{Gabaix1999aa}:
\begin{itemize}
\item Take $N$ cities
\item Chose initial city sizes of $S_0^i=\frac{1}{N}$
\item Impose a growth rate of $2\% \pm .2\%$
\item Normalize the city sizes $\sum_{i=1}^N S_t^i=1$
\end{itemize}
\item Show that city sizes converges to a Zipf's law
\end{itemize}
\end{itemize}

\subsection{Fit power law to GDP or stock market returns}

\citep{Pagan1996aa} reviewed the econometric literature on financial markets up to 1996, suggesting:
\begin{itemize}
\item unit root in asset prices;
\item financial series are not independently distributed over time;
\item stock returns are not normally distributed;
\item stock returns density have fat tails, exponent $\gamma > 2$ both left and right and for daily or monthly returns.
\begin{itemize}
\item but as the returns are not i.i.d., the estimation of $\gamma$ suffers from lack of precision as demonstrated in \citep{kearns1997};
\item \citep{Jondeau2003aa} tested and found that left and right tails are not statistically different.
\item A theoretical foundation for fat tailed return densities was explored by \citep{lux2002}.
\end{itemize}
\end{itemize}

\citep{Cont2001aa} reviewed the literature on extreme stock returns with a power-law tail exponent: $2 < \gamma(T) \leq 5$ while warning on the significant digits that can be estimated. \citep{Plerou2001aa} found for stock returns of 5 minutes intervals:
\begin{equation}
    \gamma=
    \begin{cases}
3.10 \pm 0.03,& \text{positive tail}\\      2.84 \pm 0.12, & \text{negative tail}
    \end{cases}
\end{equation}

\stepcounter{question}
\begin{itemize}
\item \textbf{Question \arabic{question}}
\begin{itemize}
\item Try to fit power law to GDP or stock market returns, take the frequency (daily, quarterly, yearly) into account before comparing their exponent.
\item Which power law have higher exponent, what does it mean?
\end{itemize}
\end{itemize}

\subsection{Alternative distribution with fat tails}

\citep{Mantegna1995aa} found that the probability distribution of the Standard \& Poor's 500 can be characterized by a L\'{e}vy stable process for the central part of the distribution and an exponential in the tails.

Nonetheless,  \citep{Williams2017aa} stated that \begin{quote}
A consensus has developed in the literature that the distribution of GDP growth rates can be approximated by the Laplace distribution in the central part and power-law distributions in the tails.
\end{quote} they study a panel of GDP growth rates and find that the distribution of GDP growth rates can be fitted using the heavy-tailed Cauchy distribution for almost all countries. We can identify several shortcomings in the approach: there are a maximum of $61$ data points for each countries, which would make it difficult to fit power-law in the tails. They do within-country growth distribution. 
\begin{itemize}
\item What about between-country growth distribution per year?
\item What about you use all data to fit a generic law? To fit a power law in the tail?
\end{itemize}

\citep{Fagiolo2008aa} approximate output growth-rate distribution by symmetric exponential power (EP) densities whose functional form reads:
\begin{equation}\label{eq:exponentialpower}
f \left( x; b, a, m \right) = \frac{1}{2 a b^{\frac{1}{b}}\Gamma\left( 1 + \frac{1}{b} \right)} e^{-\frac{1}{b}\vert \frac{x-m}{a} \vert^b}
\end{equation}
they explain that \begin{quote}
The shape parameter is the crucial one for our analysis: the larger is $b$, the thinner are the tails. In fact, the EP density encompasses both the Laplace and the Gaussian distributions. If $b = 2$, the distribution reduces to a Gaussian. If $b < 2$, the distribution displays tails fatter than those of a Gaussian (henceforth 'super-Normal' tails). If $b = 1$, one recovers a Laplace. Finally, values of $b$ smaller than one indicate tails fatter than those of a Laplace ('super-Laplace' tails in what follows). The above property is the value-added of the EP density, as it allows one to precisely measure how far the empirical distribution is from the normal benchmark and how close it is instead to the Laplace one. Another important property of the EP density is that it is characterized by exponentially shaped tails, which are less thick than those of power-law distributions.
\end{quote}

Their data set is available \href{http://qed.econ.queensu.ca/jae/datasets/fagiolo001/}{here} and they used the package \href{http://cafim.sssup.it/~giulio/software/subbotools/}{subbotools}

\section{Parametric distributions with heavy tail}\label{sec:heavytail}

\citet{resnick2007heavy} explains that roughly speaking, a rv $X$ has a heavy (right) tail if there exist a positive parameter $\alpha$, called the tail index such that
\begin{equation}
\lim_{x\to\infty} \frac{P[X>x]}{x^{-\alpha}} = 1
\end{equation}

Prominent examples of heavy-tailed distributions are the lognormal, Pareto and Weibull distributions with shape parameter smaller than $1$. In practice\footnote{we use vansteenberghe\_heavy\_tail.py}, actuaries model motor insurance with lognormal and fire claims with Pareto distributions. \citet{mcneil1997estimating} review the work of reinsurance using Extreme Value Theory (EVT).

Writing $F$ the cumulative distribution function of a random variable $X$, it is said to have a heavy tail distribution if
\begin{equation}
\int_{-\infty}^{\infty} e^{tx}dF(x) = \infty , \quad \forall t>0
\end{equation}
we can write the survival function $\bar{F}(x)= 1-F(x) = Pr \left( X > x \right)$
and the above implies that
\begin{equation}
\lim_{x \to \infty} e^{tx} \bar{F}(x)  = \infty, \quad \forall t>0
\end{equation}

writing $-\log(\left(\bar{F}(x) \right)$ the hazard function, we also have that $X$ is heavy-tailed if 
\begin{equation}\label{eq:heavyT}
\lim \sup_{x \to \infty} \frac{ -\log \left(\bar{F}(x) \right) }{x} = 0
\end{equation}

\subsection{Regularly varying distribution functions and tail index}

\citet{cooke2014fat} define:
A distribution function $F$ is called regularly varying at infinity with \textbf{tail index} $\alpha \in (0,\infty)$ if
\begin{equation}
\lim_{x\to\infty} \frac{\bar{F}(tx) }{\bar{F}(x) } = t^{-\alpha}
\end{equation}

\subsection{Detection of heavy-tailed distributions}

We follow \citet{rolski1999}.

It goes without saying that the detection of dangerous claim size distributions is one of the main worries of the practicing actuary. Most practitioners have some personal concept of what they would call a large claim. However, a mathematically sound formulation is not always obvious. We need to introduce a bit of notation. Denote by ${U_i, 1\leq i \leq n}$ the successive claims in a portfolio. The total claim amount is then $X_n = U_l +U_2+\ldots+U_n$. We denote the sequence of ordered claims with
\begin{equation}
U_{(1)} = \min U_i \leq U_{(2)} \leq \ldots \leq U_{(n)} = \max U_i
\end{equation}
Often, a claim is called large when the total claim amount is predominantly determined by it. This rather vague formulation can be interpreted in a variety of ways. Let us give a number of possible examples.
\begin{itemize}
\item One sometimes hears that a claim within a portfolio is large if a value of that size is only experienced every so many years. It needs no explanation that this kind of description can hardly be forged into a workable definition.
\item Another interpretation could be that the ratio of $U_{(n)}$ and $X_n$, is "too large".  If there are no excessive claims then we expect $U_{(n)}$ to play an increasingly lesser role in the total $X_n$.
\item More generally, a claim is called large if it consumes more than a fair portion $p$ of the total claim amount. This means that we call $U_{(m)}$ large if $m \geq \min \{ k : U_{(k)} > p X_n \}$.
\item When the practitioner tries to estimate the mean and/or variance of the claim size distribution, he will use resampling techniques to obtain a reliable estimate. However it happens that the successive sample values are not averaging out to a limiting value. One possible and theoretically understandable reason is that the mean and/or the variance of the claim size distribution do not exist because there is too much mass in the tail. A possible parametrized distribution causing this type of phenomenon is any Pareto-type distribution with small exponent $\alpha$.
\end{itemize}

Writing the empirical distribution $F_n$ defined by
\begin{equation}
F_n(x) = \frac{1}{n} \max \{ i : U_{(i)} \leq x \}
\end{equation}
then, if equation \ref{eq:heavyT} hold:
\begin{equation}
\lim \sup_{n \to \infty} \frac{ -\log \left[1 - F_n\left(U_{(n-k)}\right) \right] }{U_{(n-k)}} = \lim \sup_{n \to \infty} \frac{ \log \left[\frac{n}{k} \right] }{U_{(n-k)}} = 0
\end{equation}
which is statistically unverifiable \citep{rolski1999}. In general, the heavy-tail property is hard to verify statistically. So we are using normal, lognormal, Pareto or Weibull quantile plots to decide on which distribution to use.

\subsection{Pareto distributions}\label{sec:paretodist}

\subsubsection{Strict Pareto distribution (type I)}

The Pareto distribution probability density function was introduced by \citet{pareto1895} and writes, for a given threshold $u>0$:
\begin{equation}\label{eq:ParetoI}
    f(x;\alpha, u) = \alpha \frac{u^\alpha}{x^{\alpha+1}} \quad , \quad x > u
\end{equation}
where $\alpha$ is called the \textbf{Pareto exponent} or the \textbf{shape parameter} and the cumulative distribution function:
\begin{equation*}
    F(x;\alpha, u) = 1-\left(\frac{u}{x}\right)^\alpha \quad , \quad x > u
\end{equation*}

We can simplify setting $u=1$, then we get the survival function $\bar{F}(x) = 1-F(x) = x^{-\alpha}$ and we can convince ourselves that for the Pareto type 1 distribution, the \textbf{tail index} is the Pareto parameter $\alpha$.

Per definition, the Pareto distribution is only to be fitted on observed values above a threshold $u$. The selection of the threshold $u$, also called the scale parameter, has to be estimated. For a reinsurance treaty, that would be the priority of the excess of loss treaty (meaning that the reinsurer covers only claims above this priority, or threshold). \citet{Mukhopadhyay:1987uu} suggest a sequential estimation for the scale parameter. In practice, the estimation of $\alpha$ is done on a limited number of observations, sometimes much less than a hundred, hence the variance of our estimate can be large.

\quad

If $\alpha >1$, then the mean value exists: $E(X) = \frac{\alpha}{\alpha - 1} u$. Indeed, we have:
$$E(X) = \int_{-\infty}^\infty \alpha \left(\frac{u}{x}\right)^\alpha dx = \alpha u^\alpha \int_{-\infty}^\infty x^{-\alpha} dx = \frac{\alpha}{1-\alpha} u^\alpha \left[ x^{1-\alpha} \right]_u^\infty$$

\quad

If $\alpha>2$ the variance exists and: $Var(X) = \frac{\alpha}{(\alpha-1)^2 (\alpha-2)} u^2$.

\quad

The mean excess function is defined as:
\begin{equation*}
e(x_m) = E\left( X-x_m \vert X>x_m \right) =\frac{x_m}{\alpha-1} \quad , \quad \alpha > 1 \quad , \quad x_m>u
\end{equation*}

We can use the \textbf{Maximum likelihood estimation} methods: \citet{Zajdenweber:1996ub} demonstrates (see also the demonstration section \ref{sec:approxPL}) that above a threshold $u$, the exponent $\alpha$ can be estimated with $N$ claims of value $x_i>u$:
\begin{equation}
\frac{1}{\hat{\alpha}} = \frac{1}{N} \sum_i \left( \log x_i - \log u \right)
\end{equation}
This is called the Hill estimator, \citep{hill1975}.

As $X_1, \ldots, X_n$ are independent and identically Pareto distributed, then $\ln\left(\frac{X_i}{u} \right)$ are independent and identically (exponentially) distributed\footnote{a thorough demonstration can be found in \citep{Rytgaard:1990aa}} with mean value $\frac{1}{\alpha}$ and variance $\frac{1}{\alpha^2}$. It follows from the Central Limit Theorem that the estimate $\hat{\alpha}^{-1}$ which is $\frac{1}{n}\sum_{i=1}^n \ln\left(\frac{X_i}{u} \right)$ is asymptotically normally distributed $\mathcal{N}\left( \frac{1}{\alpha}, \frac{1}{n \alpha^2} \right)$.

In other words (see also cited references in \citet{Charpentier2021}):
\begin{equation}
\sqrt{N} \left(\hat{\alpha} - \alpha \right) \to \mathcal{N}(0,\alpha^2)
\end{equation}

The confidence interval at $95\%$ can be approximated with\footnote{as a reminder, norm.ppf(0.025, loc=0, scale=1) $= -1.96 \simeq -2$}
\begin{equation}\label{eq:hillest}
\hat{\alpha} \left( 1 \pm \frac{z_{0.05/2}}{\sqrt{N}} \right)
\end{equation}

To illustrate this distribution, we chose $1 < \alpha = 1.95 < 2$ (for fire losses, \citep{Rytgaard:1990aa} suggests $\alpha = 1.5$ and for motor liability $2.5$) and we use for simplicity $x_m = \frac{\alpha - 1}{\alpha} * 1000$ (so that $E(X) =1000$).

For more literature on fire insurance losses, read \citet{resnick1997discussion}.

As $\alpha < 2$ this distribution has no finite variance so in the words of Gourieroux the "pseudo-true value of the parameter" doesn't exist.  The distribution has an infinite expectation when $\alpha < 1$.

As the population size increases (x axis), the mean of the distribution converges.

But as the population size increases (x axis), the variance of the distribution doesn't converge.

We estimate that $u$ is $\min_i (x_i)$ and assume that $x_1, \ldots,x_n$ are independent and identically distributed.

\stepcounter{question}
\begin{itemize}
\item \textbf{Question \arabic{question}}
\begin{itemize}
\item Chose\footnote{solutin: 20220206\_fat\_tail\_ex\_vansteenberghe.py} as above $\alpha = 1.95$ and $x_m = \frac{\alpha - 1}{\alpha} * 1000$;
\item Generate $N = 10^5 $ observations, use the numpy random seed at arbitrary $7$ so you all have the same results;
\item Use the equation \ref{eq:hillest} to estimate $\alpha$
\item Plot the empirical and theoretical histograms.
\end{itemize}
\end{itemize}

\subsubsection{Generalized Pareto Distribution (type II)}

We will use section \ref{sec:evtthresh} the Generalized Pareto Distribution, for $x\geq u$ and a \textbf{Pareto exponent} $\alpha \neq 0$:
\begin{equation}\label{eq:ParetoII}
F(x; \alpha, \sigma, u) = 1 - \left[ 1 +  \frac{1}{\alpha} \frac{x-u}{\sigma} \right]^{-\alpha}
\end{equation}

and the probability density function:
\begin{equation}
f(x; \alpha, \sigma, u) = \frac{1}{\sigma} \left(1 + \frac{1}{\alpha} \frac{x-u}{\sigma} \right)^{-\alpha - 1}
\end{equation}
Pareto I is a special case of the GPD when $u=\alpha \sigma$, indeed, replacing this in equation \ref{eq:ParetoII} we find equation \ref{eq:ParetoI}.

\quad

Some packages have implemented functions using a slightly different but equivalent notation:
\begin{equation}
F(x; \xi, \sigma, u) = 1 - \left[1+ \xi \frac{x-u}{\sigma} \right]^{-\frac{1}{\xi}}
\end{equation}
with:
\begin{itemize}
\item loc : $u$
\item scale : $\sigma$
\item shape : $\xi$. So to get back to equation \ref{eq:ParetoII}, we need to transform $\alpha = \frac{1}{\xi}$.
\end{itemize}

\subsubsection{Extended Pareto Distribution}

\citet{BEIRLANT2009} introduced an extended Pareto distribution:
\begin{equation}
F(x; u, \delta, \tau, \alpha) = 1 - \left[\frac{x}{u} \left( 1 +\delta - \delta \left( \frac{x}{u} \right)^\tau \right) \right]^{-\alpha} \quad , x \geq u
\end{equation}
where $\tau \leq 0$ and $\delta> \max\left(-1, \frac{1}{\tau}\right)$ and $\alpha \neq 0$ the \textbf{Pareto exponent}.

Pareto I is a special case of the EPD, when $\delta=0$ and the GDP is a special case of the EPD when $\tau=-1$ and $\delta=\frac{1}{\alpha \sigma} - 1$.

The package ReIns implement the EPD as follow:
\begin{equation}
F(x; \kappa, \tau, \gamma) = 1 - \left[x \left( 1 +\kappa - \kappa \left( x \right)^\tau \right) \right]^{-\frac{1}{\gamma}}
\end{equation}

\subsection{Norwegian fire data exercise}

We exploit as in \citet{rolski1999}. The reporting threshold for this data set was $500$ Kronen, we can take this as our scale parameter $u$, although this might not be the optimal threshold above which the claims follow a heavy-tailed distribution.

\subsection{Log-normal distribution}

Log-normal distributions are considered plausible models for motor insurance \citep{rolski1999}.
A random variable $X$ is said to follow a log-normal distribution if $\ln(X)$ follows a normal distribution. Another way is to consider that there exists a random variable $Z$ following a standard normal distribution ($\mathcal{N}(0,1)$) such that:
\begin{equation}
X = \exp\left( \mu + \sigma Z \right)
\end{equation}

It can then be demonstrated that the mean is $\exp\left(\mu + \frac{\sigma^2}{2} \right)$ and the median $\exp(\mu)$. 

\subsection{Exercise with a Pareto and a Normal law, and GEV}

\stepcounter{question}
\begin{itemize}
\item \textbf{Question \arabic{question}}
\begin{enumerate}
\item Generate a $10^5$ large sample $S_N$ taking the absolute values from a $\mathcal{N}(1, 1)$;
\item Generate a $10^5$ large sample $S_P$ from a Pareto type I distribution with $\alpha=1.5$ and a threshold $x_m = \max \{ S_N \}$;
\item Display each sample histograms, using logs for the Pareto one;
\item Combine both samples into one sample $S$, shuffle the data set (this will be useful when applying the block maxima method);
\item Display the quantile-quantile plots for $S$ against theoretical laws: normal, exponential, log-normal, weibull, Pareto, GEV as section \ref{sec:GEVintro}
\begin{itemize}
\item Before applying the GEV, you have to apply the block maxima method introduced section \ref{sec:blockmaxima} so the the theorem introduced section \ref{sec:gnedenkotheo} is verified. In our example, we use block of size $100$, this might not be the optimum, this is for illustration purpose only.
\item When you apply the Pareto type I and II on the full sample, is the "true" $\alpha$ inside the confidence interval at $95\%$?
\end{itemize}
\item Perform a Q-Q plot on the upper tail of $S$, defining the upper tail by a percentile you define;
\item Estimate the Pareto exponent $\alpha$ using both type I and type II method on the full sample or the upper tail;
\item Conclude on the threshold being important (or unimportant) when fitting a Pareto distribution.
\begin{itemize}
\item Complete the exercise by exploring different ways to chose a threshold $u$ for the Pareto distribution: ME plot introduced section \ref{sec:evtthresh}, test with a percentile, plot several Q-Q plots with different $u$, etc.
\end{itemize}
\end{enumerate}
\end{itemize}

\quad

scipy genextreme implement the GEV with a pdf:
\begin{equation}
\left(1-c x \right)^{\frac{1}{c}-1} \exp \left[ -\left(1 - c x \right)^{\frac{1}{c}} \right]
\end{equation}

where the shape is $c=-\xi = -\frac{1}{\alpha}$ if we want to link it to how we introduced the GEV equation \ref{eq:gevpdf}.

\citet{coles2001} summarizes a bridge between our GPD and GEV:
Let $X1,X2, \ldots, X_n$ be a sequence of independent random variables with common distribution funciton $F$ and let $M_n=\max{X_1,\ldots,X_n}$, then for large $n$,
\begin{equation}\label{eq:GEVbm}
Pr\left(M_n\leq z\right) \simeq G(z)=\exp\left(- \left[ 1+ \frac{1}{\alpha} \left( \frac{z-\mu}{\sigma}\right)^{-\alpha} \right] \right)
\end{equation}, then, for some large enough threshold $u$, the distribution of $(X-u)$, conditional on $X>u$ is approximately
\begin{equation}\label{eq:GPDalpha}
1 - \left( 1 + \frac{1}{\alpha} \frac{y}{\tilde{\sigma}} \right)^{-\alpha}
\end{equation}
where $\tilde{\sigma}=\sigma + \frac{1}{\alpha} (u-\mu)$.

And the associated GEV shape parameter $\alpha$ in equation \ref{eq:GEVbm} is equal to the corresponding GPD shape parameter $\alpha$ in equation \ref{eq:GPDalpha}

\section{Python: assets risk measures, VaR and ES}\label{sec:distribtuions}

Before investing in financial assets an investor might want to first study the risk it exposes itself to. If one believes that past performances are a good indication of future performances, one can use historical return data. A simple approach is to consider that returns of an assets followed a given distribution in the past and will follow a similar distribution in the future. In the following we will test the adequacy of fitting a normal, t Student or Levy stable law to a list of asset returns.

Once we consider that a distribution is suitably fitted to each asset, we can then either extract random draws from these distributions to assess the portfolio total risk, and/or compute the Value at Risk (VaR) of Expected Shortfall (ES) defined hereafter.

We suggest\footnote{VaR\_ES\_vansteenberghe.py} to use daily price data from 3 667 assets over the period 01/01/1998 to 06/08/2018. This is an interesting period as one captures the "dot com" and the "sub-prime" crises.


As defined in \citep{BCBS2009}, VaR is used by central bank regulator to determine the capital charge\footnote{for simplicity, a capital charge can be understood as the capital a financial institution has to "keep safely aside" to be able to perform some risky activities} required for certain activity \begin{quotation} the trading book capital charge for a bank using the internal models approach for market risk will be subject to a general market risk capital charge (and a specific risk capital charge to the extent that the bank has approval to model specific risk) measured using a 10-day value-at-risk at the 99 percent confidence level and a stressed value-at-risk. \end{quotation} and the financial institutions \begin{quotation} will also be required to use hypothetical backtesting at least for validation. \end{quotation}

\quad

Having in mind the historical or normal approach to compute a VaR, one might wonder if there is the need for a more complex approach. \citep{Danielsson::Canada} provide some answers to this question. "[E]stimate the worst case by taking the most negative outcome in the historical
sample" or "estimate the lower tail of the distribution by semi-parametric methods", they find "that either method is best, depending on how heavy the tails are and their specific shape. Generally, for the heaviest, the semiparametric approach is best, and as it thins, the historical minima eventually becomes better."

If we define the \textbf{rate of loss} of an asset $i$ over time by $R_{i,t} = - \frac{P_{i,t}-P_{i,t-1}}{P_{i,t-1}}$.

\subsection{Value at Risk}\label{seq:VaR}
For a portfolio composed of assets with iid returns, the Value-at-Risk (VaR) summarises the worst possible loss of a portfolio at a given confidence level $q\in (0,1)$ over a given time period $[t, t+1]$. $VaR_q$ is such that the rate of loss of the portfolio won't exceed that level over that period with a probability inferior or equal to $1-q$:
\begin{align*}
    VaR_q& =\inf\{x:\mathbb{P}(R>x)\leq 1-q\}  \text{ or} \\
    VaR_q& = F^{-1}(q)=\inf\{x:F(x)\geq q\} 
\end{align*}
where $F(x) = \mathbb{P}(R\leq x)$, $x\in\mathbb{R}$ is the distribution function of the rate of loss $R$ and $F^{-1}$ the generalized inverse of $F$.

For example, in a year with $252$ trading days we might want to know the worst return we are likely to face one day out of those $252$, that is VaR at $99.6\%$ over a year.

There are two categories of methods to compute VaR: those using historical rate of loss directly, called historical simulation or nonparametric methods, and those relying on statistical models, called parametric methods as they rely on parametric distribution.

Historical simulations rely on the assumption that the data set is representative  of the future behaviour of the asset. If we consider the sorted realizations of a random variable $(R_{n,1}\leq R_{n,2}\leq ... \leq R_{n,n})$ then the VaR can be estimated as:
\begin{equation*}
\widehat{VaR}_q = R_{n,s}, \text{  with  }  s=[nq]+1
\end{equation*}
where $[x]$ represents the integer part of $x$.

The main drawback of the historical computation is that it is not possible to make the VaR depend on the current state of the market, namely to update volatility.

Some parametric models explicitly include the volatility to predict the VaR, the exposure to risk is conditional on the period of observation. 

\subsubsection{VaR: parametric methods}

\begin{enumerate}
\item we suppose that the return are following a law
\item we compute the mean $\mu=E(X)$ and the standard deviation $\sigma$
\item if it is a Student t law, we need to choose a degree of freedom $\nu$
\item we use a given quantile of the law \footnote{for the normal law, it is 1.65 for 95\textsuperscript{th}, 2.33 for 99\% and 2.65 for 99.6\%}
\end{enumerate}

If we assume that $R$ follows a Gaussian distribution: $R\sim\mathcal{N}(\mu,\sigma^2)$, then its $VaR_q$ can be estimated as
 \begin{equation*}
 \mathbb{P}(R\leq VaR_q) = \mathbb{P}\Big(\frac{X-\mu}{\sigma}\leq \frac{VaR_q-\mu}{\sigma}\Big)=q.
 \end{equation*}
If $z_q$ is the q-quantile of a gaussian law $\mathcal{N}(0,1)$, then $VaR_q$ is defined as:
 \begin{equation*}
VaR_q = \mu + \sigma z_q.
\end{equation*}
We can estimate $VaR_q$ replacing $\mu$ and $\sigma^2$ by the estimations: $\bar{R}_n=\frac{1}{n} \overset{n}{\underset{i=1}{\sum}}R_i$ and $\bar{\sigma}^2_n=\frac{1}{n-1} \overset{n}{\underset{i=1}{\sum}}(R_i-\bar{R}_n)^2$, then $VaR_q$ can be estimated as:
\begin{equation*}
\widehat{VaR}_q = \bar{R}_n + \bar{\sigma}_n z_q.
\end{equation*}

\quad 

If we denote by $\Phi$ the percentage point function which is the inverse of the cumulative distribution function\footnote{in simple terms, $-\Phi(1-\alpha)$ find the return level $z$ on the left of which the area under the pdf is $\alpha$} is: $$\mbox{VaR}_\alpha = \mu - \Phi(1-\alpha) \sigma$$

\quad

For the generalized Student's-t distribution, we use $\mu'$ the location, $\sigma'$ the scale and $\nu$ the degree of freedom\footnote{For the details of the calculus, cf. \citep{alexanderVaR}}. For $\nu>2$, we have $\mu_{\mbox{sample}} = \mu'$ and $\sigma_{\mbox{sample}} =\sigma' \sqrt{\frac{\nu}{\nu-2}}$, hence with $t$ the percentage point function of the Student's-t:
$$\mbox{VaR}_q = \mu + t(1-q) \sigma \sqrt{\frac{\nu-2}{\nu}}$$

\begin{center}
\includegraphics[scale=0.5]{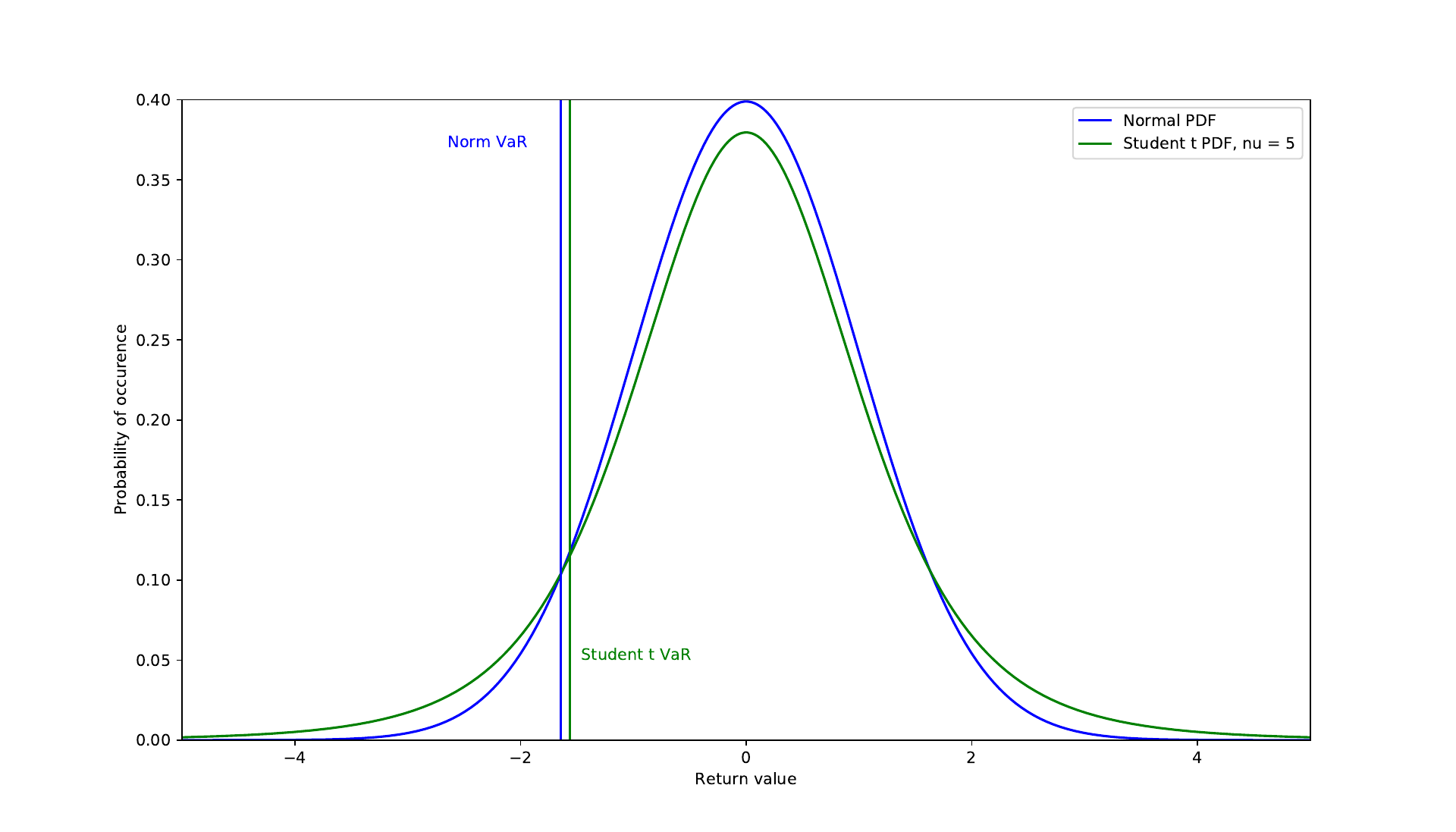}
\end{center}

\quad

Monte Carlo approach: If we are not certain about the inversion of the distribution function, we can generate a large number of returns from the distribution and then compute the $VaR$ and $ES$ for this sample.

Some asset managers declare their VaR policy in the funds brochure, e.g. from SEEYOND EQUITY VOLATILITY STRATEGIES:
\begin{quotation}
the Global Exposure Risk is managed using the absolute Value-at-Risk approach (VaR approach). The VaR approach measures the maximum potential loss at a given confidence level (probability level) over a specific time period, under normal market conditions. The absolute VaR for the Sub-fund cannot exceed 20\% of its net asset value in a confidence interval of 99\% for a 1-month holding period (20 available working days).
\end{quotation}

\subsubsection{historical VaR}

\begin{enumerate}
\item we rank the observed returns
\item we consider the 0.4\% worst for the VaR 99.6\% or the 5\% worst for the VaR 95\%
\end{enumerate}

As in \citep{guegan}, let $X_1, \ldots , X_n$ be a sequence of losses. The historical VaR at the p-quantile is:
$VaR_p = X_m$
where $m = np$ if $np $ is an integer and $m = [np]+1$ otherwise\footnote{The operator $[]$ is the largest integer inferior}.

\subsection{Expected Shortfall}
The \textbf{Expected Shortfall} is the mean of the returns once the VaR has been crossed, over a given period. It is also named C-VaR or Tail-Loss:

$$ES(\alpha)=E[r|r<VaR(\alpha)]$$
in practice, with $\psi$ the standard normal density function\footnote{for the Student's-t ES, one can look into the details \href{http://www.quantatrisk.com/2016/12/08/conditional-value-at-risk-normal-student-t-var-model-python/}{here}}
$$\mbox{ES}_\alpha = \mu - \alpha^{-1} \psi\left(\Phi(\alpha)\right) \sigma$$

\subsection{Other mean to compute the VaR and ES from a distribution}
If we are not certain about the inversion of the distribution function, we can generate a large number of returns from the distribution and then compute the VaR and ES for this sample.

\subsection{Data preparation}

Before exploiting our data set, we need to prepare the data.

First, we might want to limit our observations to trading days in order to make sure that we won't capture null returns on week-ends and overweight $0$ in our histogram. We filter our data frame rows on the trading days.

Second, we want to distinguish between real $0$'s and missing values. It can happen that instead of a NA the value $0$ is given in the data set. More problematic are negative price values. We investigate the minimum price value per asset. For those with negative and $0$ values we do a visual inspection. For example we can have one observation at $0$, in which case we migth want to keep the asset in the data set but remove the observation with a $0$ value (we might replace that $0$ with an average between the previous and the next observation:
\begin{center}
\includegraphics[scale=0.5]{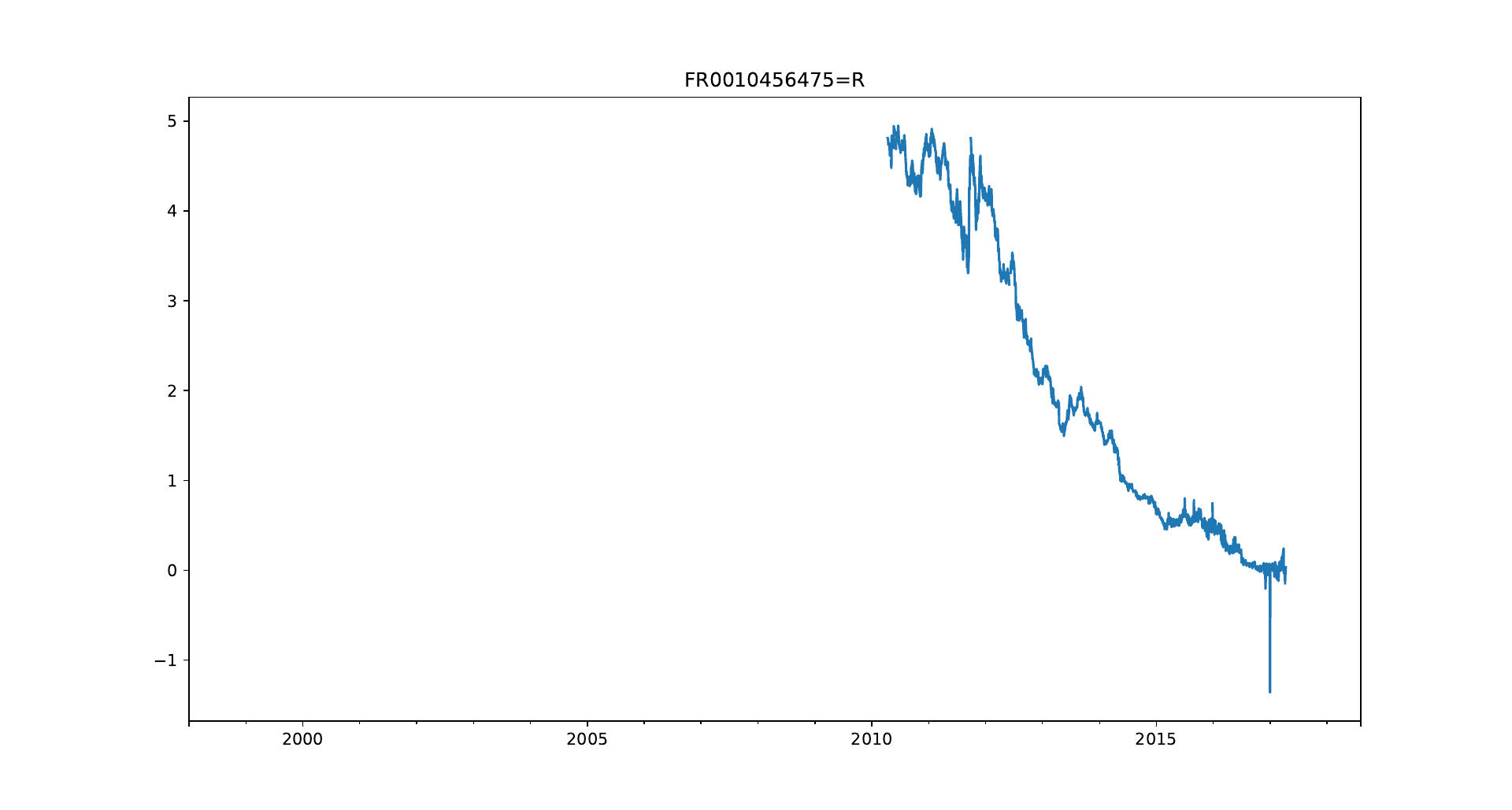}
\end{center}

We can also investigate price time series which are either stuck at $0$ by computing successive $0$'s in the time series or for returns stuck at $0$, then by visual inspection we can decide if we keep the asset or not.

If the data set is so large that we cannot commit to visual inspection we can be more brutal in dropping potential anomalous series.

We can drop series of which observation are below a given threshold compare to the rest of the data set.

We can also deal with anomalous return values by winsorizing the returns data set. Winsorizing is simply putting a positive and negative limit on the values the returns can take, for example if we winsorize returns between $-1$ and $1$: $R_{\mbox{winsorised},t} = \max\left[ \min\left(1, R_{\mbox{original},t}\right), -1\right]$. This approach is different than just getting rid of series where there exist a return for which $\vert R_{\mbox{original},t} \vert > 1$, if our data set is large enough, we might prefer to get rid of series with anomalous returns as winsorizing can introduce bias.

\subsection{Beyond normal law: t-Student, Levy stable}
Now we want to check whether those returns are normally distributed, meaning that the daily return. We do a first visual check by comparing the histogram of actual BNP's daily returns and draws from the normal distribution:
\begin{center}
\includegraphics[scale=0.5]{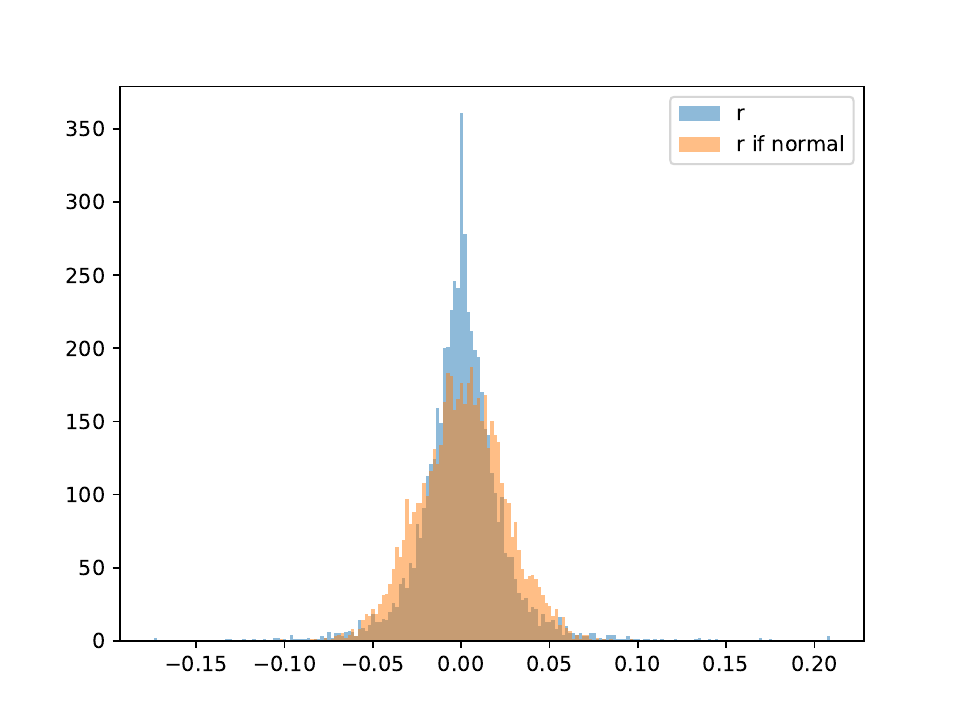}
\end{center}

Visually it seems that both distributions differ, we need to use a statistical test to determine whether both distributions can be consider similar or different: we use here the Kolmogorov-Smirnov\footnote{we suggest another normality test in section \ref{sec:GMMWald}} test where the null hypothesis $H_0$ is that both groups were sampled from populations with identical distributions. 

The two-sided test uses the maximum absolute difference between the cdfs of the distributions of the two data vectors. The test statistic is:
$$KS=\max_x \lvert \hat{F}_1(x) - \hat{F}_2(x)  \lvert $$

The p-value of the test is low enough that we can reject $H_0$ and the normal distribution is not adequate to model BNP's daily returns over the period.

\subsubsection{Student t distributions and stable laws}
The normality of financial asset returns distribution has been question since the work of Mandelbrot and Fama in the 60's. Other distributions have been suggested and we explore two:
\begin{itemize}
\item Student t distributions that present fat tails and variance;
\item Stable laws that present fat tails and skewness.
\end{itemize}

\begin{center}
\includegraphics[scale=0.5]{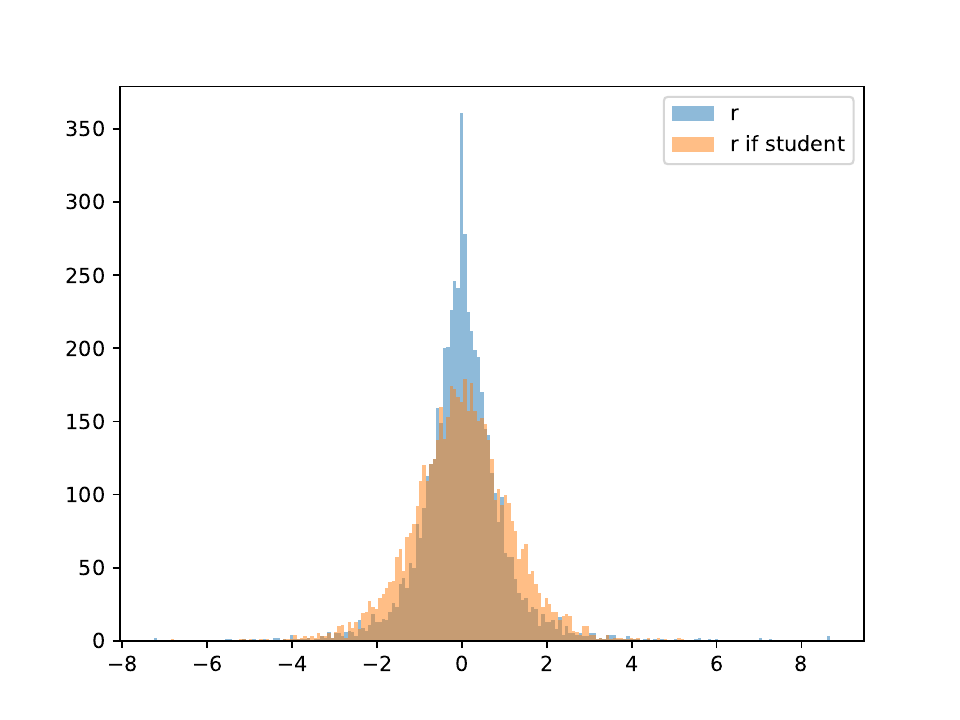}
\end{center}

\begin{center}
\includegraphics[scale=0.5]{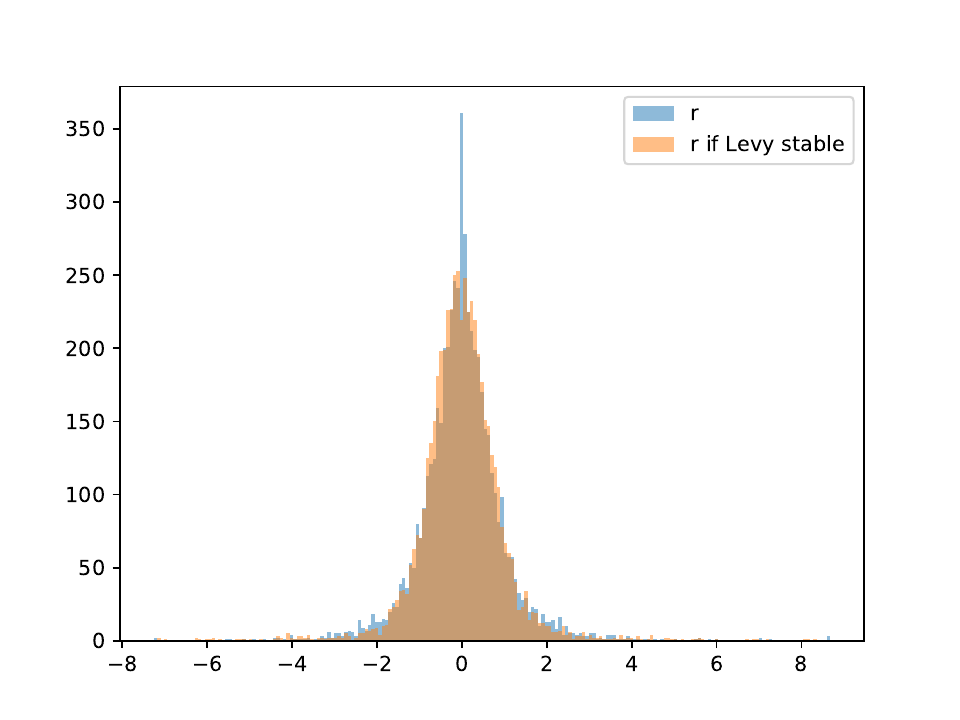}
\end{center}

We fit the normal, t Student and Levy stable distributions on the empirical returns for each of the 3 667 assets and compute the p-value of the two sided Kolmogorov-Smirnov test\footnote{a preliminary remark: we would need to generate several random samples from the distributions and compute each time the p-values in order to approach the real p-value of the test}.
We keep respectively 82, 71, and 419 candidates for the normal, t Student and Levy stable laws.

\subsection{Parametric VaR and ES when considering normal, Student's and Levy distributions}
For each of the 2 866 assets we compute the parametric and historical VaR and ES. We find that for a large share of the assets the parametric VaR and ES based on normal assumption are less conservative than the historical observation. This is likely due to a poor fit of the normal distribution, namely the existence of fat tails.

We find that the median VaR and ES are too optimistic in the normal and Student's t case while too extreme for the stable law:

\quad

\begin{tabular}{ll}
historical VaR         &  -0.028 \\
parametric VaR normal  &  -0.021 \\
parametric VaR student &  -0.024 \\
MC VaR levy    &  -0.049 \\
historical ES          &  -0.043 \\
parametric ES normal   &  -0.023 \\
parametric ES student  &  -0.031 \\
MC ES levy     & -0.347\\
\end{tabular}

We can observe that the Levy stable law can be too conservative while the normal and Student's t law are not conservative enough.

\stepcounter{question}
\begin{itemize}
\item \textbf{Question \arabic{question}}
\begin{itemize}
\item As established by \citet{Artzner2007}, show some numerical applications that the VaR does not verify the subadditivity property (incoherent risk measure).
\end{itemize}
\end{itemize}

\section{Focusing on the tail: Extreme Value Theory, an introduction}\label{sec:EVT}

A formal definition of heavy-tailed distributions is given section 2.5.1 in \citet{rolski1999}. As for fat-tail distribution, we will consider that the moment of order 3 and 4 are greater than either a normal or exponential distribution.

Extreme Value Theory (EVT) is used for estimating tail risk. Normal distributions tails can be considered too thin to capture outer limits returns. The field of EVT goes back to 1927, when \citep{Frechet1927} formulated the functional equation of stability for maxima, which later was solved with some restrictions by \citep{Fisher1928}, and finally by \citep{Gnedenko1943} and \citep{deHaan1970}. 

This introduction is greatly inspired from \citep{alexander2015}, \citet{bezat2000}, \citet{bensalah2000} and \citet{lezaud2013}.

Let us assume that we have access to a sample $(X_1, \ldots , X_n)$ of independent and identically distributed (iid), or even stationary and weakly dependent random variables (RV) from an underlying model $F$, defined as $F(x) = Pr (X \leq x)$. We can assume that we only look at losses (defined positively).
Writing $(X_{1:n} \leq \ldots \leq X_{n:n})$ the sample of associated ascending order statistics (OS).

\subsection{Gnedenko's extremal types theorem}\label{sec:gnedenkotheo}

We want to study\footnote{vansteenberghe\_EVT.py} the "behaviour" of extreme events so we focus on the RV:
$$M_n = \max(X_1,\ldots,X_n)$$
As the RV are iid, we have:
$$Pr(M_n \leq x) = Pr(X_1 \leq x, \ldots , X_n \leq x) = \left[ F(x) \right]^n$$

We do not know $F$, thus we study the asymptotic behavior of the RV $M_n$ to know the law toward which $M_n$ will converge, we will show this can be Gumbel, Frechet or Weibull $\implies$ all three can be generalized in a Generalized Extreme Value (GEV) distribution we introduce section \ref{sec:GEV}. You can think about this approach as we did for the Central Limit Theorem (CLT) section \ref{sec:CLT}, where now the GEV distribution plays the role of the normal distribution in the CLT.

Let us assume that it is possible to linearly normalize the sequence of maximum values,
$\{X_{n:n}\}_{n\geq1}$, so that we get a non-degenerate limit for the sequence $\frac{X_{n:n} - b_n}{ a_n}$, with $a_n > 0$ and $b_n \in\mathbb R$. 

\begin{equation}
\lim_{n\to\infty} Pr\left(\frac{X_{n:n} - b_n}{ a_n} \leq x \right) = G(x)
\end{equation}

Gnedenko's extremal types theorem (ETT) assures us that such a limiting RV has a cumulative distribution function (CDF) of the type of the extreme value distribution (EVD), given by

$$G_\xi(x) = 
    \begin{cases}
\exp\left( - (1+\xi x)^{-\frac{1}{\xi}} \right),& 1 + \xi x > 0 \quad \text{if} \quad \xi \neq 0 \\        \exp\left( - exp(-x) \right), &  x\in\mathbb R \quad \text{if} \quad  \xi = 0
    \end{cases}$$
we call $\xi$ the extreme value index (EVI), it can be understood as a weight of the right tail-function. The higher the value of EVI, the fatter the tail. The EVD can be rewritten on the three domains of attractions\footnote{we should define and demonstrate that the maximum of a sample converges in distribution to one of the following distribution}, with the cumulative distribution functions (you might want to normalize first, in python that would be using location and scale parameters which are omitted here for simplicity):
\begin{equation}\label{eq:gumbelfrechetweibull}
\begin{cases}
\text{Gumbel}: \exp\left( - exp(-x) \right), &  x\in\mathbb R \quad \text{if} \quad  \xi = 0 \\
\text{Frechet}:  \exp\left( - x^{-\frac{1}{\xi}} \right),& x \geq 0 \quad \text{if} \quad \xi > 0 \\
\text{Weibull}: 1 - \exp\left( - (-x)^{-\frac{1}{\xi}} \right),& x \leq 0 \quad \text{if} \quad \xi < 0 \\        
    \end{cases}
\end{equation}

If we define the \textbf{right endpoint} as $x_{F}= \sup\{x\in\mathbb{R}: 0<F(x)<1\}$, it is infinite for Frechet and Gumbel distribution, but finite for Weibull distributions.
    
\begin{center}
\includegraphics[scale=0.5]{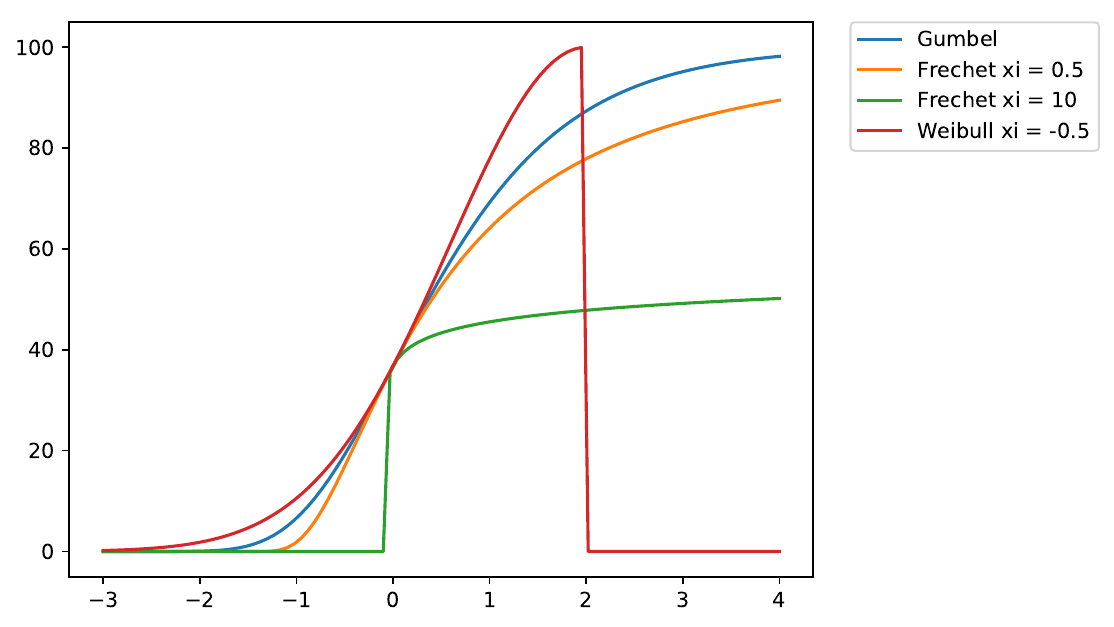}
\end{center}

The standard Gumbel distribution is defined by the following probability density function
$$
f(x) = e^{-x-e^{-x}},
$$
where $x\in\mathbb R$.

A Fr\'{e}chet type distribution is a distribution whose distribution function is 
$$
F(x) = e^{-\left(\frac{x-\mu}{\sigma}\right)^{-\frac{1}{\xi}}},
$$
for $x \geq \mu$ and $\xi > 0$. These distributions are the most studied distributions in EVT and they are of particular interest in financial applications because they are heavy-tailed distributions with infinite higher moments. \citet{Cont2001aa} compares the Fr\'{e}chet distribution shape parameter with the power-law tail parameter: $\alpha = \frac{1}{\xi} > 0$, cf section \ref{sec:powerlaw}. As long as $\alpha>1$, the smaller $\alpha$ (equivalently, the higher $\xi$), the fatter the tail of the distribution.

A Weibull type distribution is characterized by the following distribution function
$$
F(x) = 1- e^{-\left(\frac{x-\mu}{\sigma}\right)^{-\frac{1}{\xi}}},
$$
for $x \geq \mu$ and $\xi < 0$, the least interesting for financial applications since the distributions in this class all have finite right endpoints.

\subsection{Generalized extreme value distribution}\label{sec:GEV}

Usually, we apply the ETT in an approximate manner, we take $n$ "high enough", and $a>0$ and a $b$:
\begin{equation*}
Pr\left(\frac{M_n - b}{ a} \leq x \right) \simeq G(x)
\end{equation*}
or
\begin{equation*}
Pr\left(M_n  \leq x \right) \simeq G\left(\frac{x-b}{a}\right)
\end{equation*}
Thus we can cover the three types of distribution with a unique  \textbf{generalized extreme value distribution} (GEV) introduced section \ref{sec:GEVintro}.

\subsubsection{Block Maxima method, maximum likelihood}\label{sec:blockmaxima}

We assume that we have a sample of size $kn$, $(X_1, \ldots , X_{kn})$, we divide them in $k$ blocks of size $n$, notice the trade-off for choosing $k$ and $n$, and extract the maxima:
\begin{equation*}
M_{n,i} = \max_{(i-1)n<j\leq in} X_j
\end{equation*}

We want to estimate shape, location and scale parameters $\theta = [\mu, \sigma, \xi]$ that maximize the log-likelihood, summing over $i$:
$$l(M_{n,i} ; \theta) = -\ln(\sigma) - \left( \frac{1+\xi}{\xi} \right) \ln\left(1 + \xi \frac{M_{n,i}-\mu}{\sigma} \right) - \left[ 1 + \xi \frac{M_{n,i}-\mu}{\sigma}   \right]^{- \frac{1}{\xi}} $$

We are estimating:
$$\hat{\theta} = \arg\max_{\theta} \sum_{k=1}^N l( M_{n,k} ; \theta)$$
s.t.
\begin{itemize}
\item $\sigma >0$
\item $\forall i$, $1 + \xi \frac{M_{n,i}-\mu}{\sigma} > 0$
\end{itemize}

As detailed in \citet{lezaud2013}, regularity conditions for the ML approach are not met.
The assumption that the RV $M_{n,i}$ are iid can be accepted. Sampling over block of "correct" size can take care of clustering (ARCH effect removed), and a maxima in a block might not help predict a maxima in the succeding block.
This was formalized in \citet{leadbetter1974}.
Even if we can argue for stationarity of extremes, the problem with the block maxima method is that if in a block you had a lot of maxima, you only take one, so in fact you neglect the clustering of extreme events and loose information which can be problematic.

\subsubsection{VaR estimation}
Once the distribution of $M_{n,i}$ is estimated by a GEV, the VaR of condifence level $\alpha$ can be estimated as:
\begin{equation}
\widehat{VaR}_p = h_p(\hat{\theta}_n) = \hat{\mu}-\frac{\hat{\sigma}}{\hat{\xi}}\big[1-(-\log(p^n)^{-\hat{\xi}}\big].
\end{equation}

\subsubsection{Return level}

The return level of the event $\{M>u\}$ is given by: $\frac{1}{1-H_{\xi, \mu, \sigma}(u)}$, that is the average number of blocks until the event occurs.

\subsection{Estimation of the Generalized Extreme Value distribution parameters}

In our example, we are exploring the daily returns of the Copper prices and explore for extreme realisation of returns compare to a normal distribution.
%

\stepcounter{question}
\begin{itemize}
\item \textbf{Question \arabic{question}}
\begin{itemize}
\item Do you detect increasing variability, trend or seasonality? If yes, deal with it as in section \ref{sec:seasonality}.
\item Is it an I(1) time series?
\item Work on the standardized innovations of an "optimal" ARMA-GARCH fitted to the time series of the Cooper daily returns.
\item Are the standardized innovations normally distributed?
\item What VaR do you measure?
\end{itemize}
\end{itemize}

We apply the block maxima method.
\stepcounter{question}
\begin{itemize}
\item \textbf{Question \arabic{question}}
\begin{itemize}
\item Compute the VaR with this method and compare it to what you computed above.
\end{itemize}
\end{itemize}

\subsubsection{QQ plot}
We can define the quantile function $F^{-1}$ which can be considered as the inverse of the distribution function. We can define it:
$$F^{-1} (p) = \inf \{ u : F(u) \geq p \}$$

A quantile-quantile plot (Q-Q plot) is a plot of the quantile functions of two different distributions against each other. It can be seen as a non-parametric method to compare underlying distributions of two samples. The "theoretical Q-Q plot" of two distributions which have distribution functions $F_1$ and $F_2 $ is the set of all $(x_1 ,x_2) \in \mathbb R^2$ for
which there exists a $p$ such that $x_1 = F_1^{-1}(p)$ and $x_2 = F_2^{-1}(p)$.

If the Q-Q plot is a straight line, then the two distributions are either identical, or the underlying random variables differ only by a scale factor.

\quad

In our example, we use the empirical distribution of the copper daily returns and the normal distribution and find that the distribution is \textit{fat tailed}\footnote{extreme outcomes occur more frequently than predicted by normal distribution} as the graph shows curves at the top right end and at the
bottom left end:
\begin{center}
\includegraphics[scale=0.5]{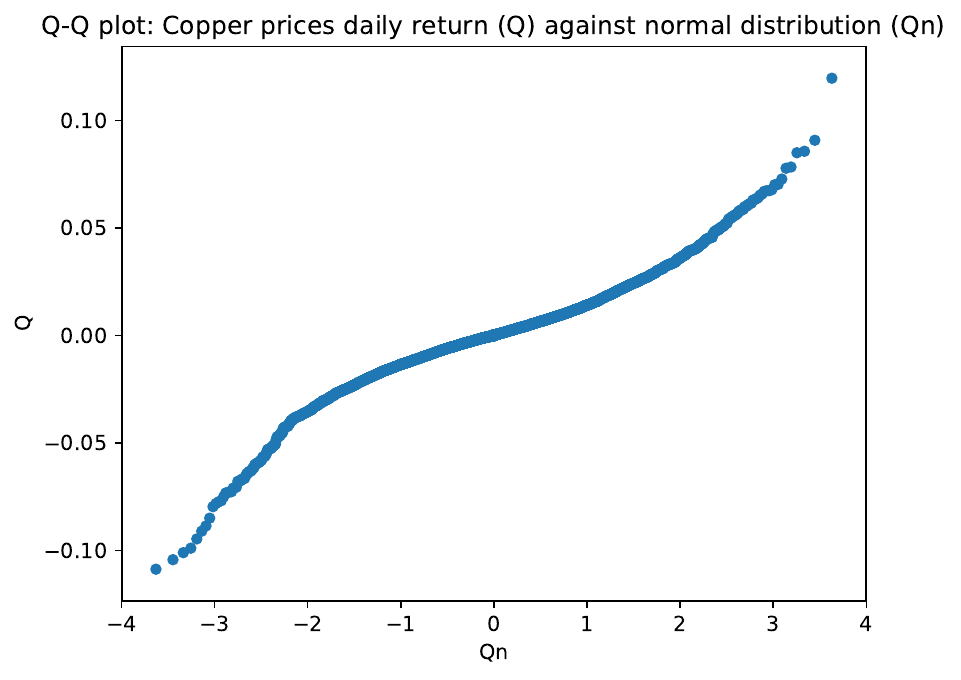}
\end{center}

\subsection{R: Application to S\&P 500 time series}

We suggest to follow \citep{alexander2015}\footnote{code vansteenberghe\_EVT\_SandP500.R} exemple 7.12 based on S\&P 500 index time series.

In their example, they limit the period from 31/12/1959 to 16/10/1987 just before the occurrence of the 19/10/1987 "black Monday". They compute the annual maximum fall and calibrate the GEV function:

\begin{tabular}{| c | c | c |}
\hline
$\hat{\sigma}$ & $\hat{\mu}$ & $\hat{\xi}$\\
 \hline
  0.72 & 2.04 & 0.27 \\
\hline
\end{tabular}

Finally, compute the VaR at $95\%$ level.

\stepcounter{question}
\begin{itemize}
\item \textbf{Question \arabic{question}}
\begin{itemize}
\item Do the same exercise with semesterly maxima.
\item Work on the standardized innovations of an "optimal" ARMA-GARCH fitted to the time series of the S\&P 500 index daily returns, what VaR do you measure?
\end{itemize}
\end{itemize}

\subsection{EVT: threshold exceedances method}\label{sec:evtthresh}

This is also called Peak-over-Threshold and is used in modeling cyber risks \citet{eling2019actual}. The Basel II Operational Risk measure was based on the paper \citet{de2006capital}.

We consider a series of loss rates observations $R_1,...,R_n$, taken from an unknown distribution function $F$. We focus on the number $N_u$ of extremes (strictly positive) that exceed a given threshold $u$, and on the excess over $u$, $E_i=R_i-u$, that forms the excess sample $E_1,..,E_n$, that we suppose to be $iid$. 

The principal distribution model for the excess over a threshold is the Generalize Pareto Distribution (GPD) that is defined by the following repartition function, with $\beta$ the scale and $\xi$ the shape parameter:
\begin{equation}
G_{\xi,\beta}(x)= \begin{cases}
1-(1+\xi x/\beta)^{-1/\xi} & \text{ if } \xi \neq 0, \\
 1-\exp(-x/\beta)& \text{ if } \xi = 0
\end{cases}
\end{equation}
where $\beta > 0$ and $x \geq 0$ when $\xi \geq 0 $ and where $0\leq x\leq -\beta/\xi$ when $\xi<0$. The $GPD$ regroups three types of distribution depending on the shape parameter: 
\begin{itemize}
\item if $\xi>0$, it is the usual Pareto distribution with $\alpha=1/\xi$ and $\kappa=\beta/\xi$;
\item if $\xi<0$ it is the Pareto distribution of type $II$ (short-tailed);
\item if $\xi=0$, it is the exponential function.
\end{itemize}
In terms of domains of attraction we have that $G_{\xi,\beta}\in MDA(H_\xi)$ for all $\xi\in \mathbb{R}$. Note that in the case of fat tailed distribution, $\xi\geq 1$, for the Pareto exponent it would be $\alpha\leq 1$, it can be demonstrated that the expectation of a $GPD$ is infinite i.e $\mathbb{E}(X^k)=\infty$ for $k\leq 1/\xi$. The expectation of the $GPD$ is defined as long as $\xi<1$ by
\begin{equation}\label{EGPD}
\mathbb{E}(X)=\frac{\beta}{1-\xi}.
\end{equation}

We define the distribution function of the excesses over a threshold $u$ by 
\begin{equation}\label{FDE}
F_u(x)=\mathbb{P}(R-u\leq x | R>u)=\mathbb{P}(E\leq x|R>u)=\frac{F(x+u)-F(u)}{1-F(u)},
\end{equation}
for $0\leq x <x_F-u$. $F_u(x)$ is then the probability that an extreme exceed the threshold $u$  by a quantity inferior of equal to $x$, knowing that the threshold is exceeded. We introduce the mean excess function:
\begin{equation} \label{MEF}
e(u)=\mathbb{E}(R-u | R>u).
\end{equation}
The mean excess function $e(u)$ express the expectation of the law $F_u$ as a function of $u$.
For the special case of the $GPD$, the excess law and the mean excess function can be computed. If $F$ is an exponential function, then $F_u(x)=F(u)$ for all $x$ (lack-of-memory property). More generally, if $R$ has the distribution function $F=G_{\xi,\beta}$, then following \ref{FDE}, we show that:
\begin{equation}\label{FDEGPD}
F_u(x)= G_{\xi,\beta(u)} \text{ with } \beta(u)=\beta+\xi u
\end{equation}
where $0\leq x<\infty$ if $\xi \geq 0$ and $0\leq x\leq-(\beta/\xi)-u$ if $\xi < 0$. i.e the distribution function of the excesses over $u$ is a $GPD$ with the same shape parameter $\xi$ and a scale parameter $\beta$ that increases linearly with $u$. 
The mean excess function of the $GPD$ can be computed from \ref{FDEGPD} and \ref{EGPD}:
\begin{equation}
e(u)= \frac{\beta(u)}{1-\xi}=\frac{\beta+\xi u}{1-\xi},
\end{equation} 
where $0\leq u<\infty$ if $0\leq\xi<1$ and $0\leq u\leq -\beta/\xi$ if $\xi<0$. Thus the mean excess function of a $GPD$ is linear with respect to $u$, this is a characterizing property of the $GPD$.  We can plot the 'Mean Excess plot' to determine the region where this linearity property is respected.

The Balkema-de Haan-Pickands theorem from 1974-1975 is essential to model the excesses. It shows that the $GPD$ is a natural limit for several excess distribution laws.

\begin{theorem}\label{BHP}
For a certain class of distributions\footnote{This class regroup all classical continuous distributions.}, the $GPD$ is the limit distribution for the excess distribution when the threshold tends to $x_F$. Formally, we can find a positive and measurable function $\beta(u)$ such that
\begin{equation}
\lim_{u\to x_F}\sup_{0\leq x < x_F-u}|F_u(x)-G_{\xi,\beta(u)}(x)|=0
\end{equation} 
if and only if, $F\in MDA(H_\xi)$.
\end{theorem}

Thus the distribution for which the maxima converge to a $GEV$ distribution constitute a set of distributions for which the excesses converge to a $GPD$ when the threshold increases. Moreover the shape parameter of the limit $GPD$ distribution for the excesses is the same as the parameter for the limit $GEV$ for the maxima.

\subsubsection{Modelling excess losses}
We model excess losses distribution following theorem \ref{BHP} and assume the excesses over some suitably chosen high threshold $u$ follow exactly a $GPD$ [i.e. $F \in MDA(H_\xi)$]. We make the following idealising assumption:

\begin{assumption}\label{hyp1}
Let $F$ be a distribution with right end point $x_F$, we assume that for some high threshold $u$ we have $F_u(x)=G_{\xi,\beta(u)}(x)$ for $0\leq x\leq x_F-u$ and some $\xi\in \mathbb{R}$ and $\beta>0$. 
\end{assumption}

\subsubsection{Threshold selection}
We have to select a threshold $u$ high enough so that assumption \ref{hyp1} is respected to avoid bias in the estimations. $u$ should also allow for the excess losses sample to be filled with enough data points, if $u$ is too high and we have too few data points, the standard errors of the estimators would be too important.

In practice, a tool to choose the threshold $u$ is the sample mean excess plot of the empirical function $e_n(u)$ ($ME$-plot) defined by $\{(u,e_u(u)): R_{n,1}<u<R_{n,n}\}$
where $R_{n,1}$ and $R_{n,n}$ are respectively the minimum and maximum of the excess loss sample. $e_n(u)$ is defined as the sum of excesses over a threshold $u$ divided by the number $N_u$ of observations exceeding $u$. The empirical mean excess function $e_n(u)$ is an empirical estimator of the mean excess function $e(u)$:
\begin{equation}
e_n(u)=\frac{\sum_{i=1}^n(R_i-u)^+}{\sum_{i=1}^n\mathbf{1}_{\{R_i>u\}}}=\frac{1}{N_u}\sum_{i=1}^n(R_i-u)^+.
\end{equation}

In order to interpret an $ME$-plot, accorind to hypothesis \ref{hyp1}, we have $F_v(x)=G_{\xi,\beta+\xi(v-u)}$ for any high threshold $v\geq u$, indeed:
\begin{align}\label{distv}
\overline{F}_v(x)&=1-F_v(x)\nonumber\\
&=1-\frac{F(x+v)-F(v)}{1-F(v)} \nonumber\\
&=\frac{\overline{F}(v+x)}{\overline{F}(v)}=\frac{\overline{F}(u+(x+v-u))}{\overline{F}(u)} \frac{\overline{F}(u)}{\overline{F}(u+(v-u))} \nonumber\\
&=\frac{\overline{F}_u(x+v-u)}{\overline{F}_u(v-u)}=\frac{\overline{G}_{\xi,\beta}(x+v-u)}{\overline{G}_{\xi,\beta}(v-u)} \nonumber\\
&=\overline{G}_{\xi,\beta+\xi(v-u)}(x).
\end{align} 

Thus the excess distribution over higher thresholds remains a $GPD$ with the same $\xi$ parameter but a scaling that grows linearly with the threshold $v$. Provided that $\xi<1$\footnote{The condition $\xi<1$ is generally respected for financial returns.}, the mean excess function is given by:
\begin{equation}\label{EMSeuilv}
e(v)= \frac{\beta+\xi(v-u)}{1-\xi}=\frac{\xi v}{1-\xi}+\frac{\beta-\xi u}{1-\xi}
\end{equation} 
where $u\leq v <\infty$ if $0\leq \xi<1$ and $u\leq v\leq u-\beta/\xi$ if $\xi<0$.

The linearity of the mean excess function in $v$ is commonly used as a diagnostic for data admitting a $GPD$ model for the excess distribution. It forms the basis for the following simple graphical method for choosing an appropriate threshold. Indeed, if the data over some threshold follow a $GPD$ we expect that the $ME$-plot becomes linear for any value over that threshold according to equation \ref{EMSeuilv}, an increasing trend indicates a $GPD$ with a positive shape parameter ($\xi>0$), a horizontal trend indicates a shape parameter converging to $0$ and a decreasing trend indicate a negative shape parameter ($\xi<0$).

In practice, reading a $ME$-plot is non trivial, we can choose the threshold $u$ on the right of which the $ME$-plot seems to become linear.

For memory, \citet{McNeil:2000aa} simply used the $10\%$ worst outcomes (100 out of 1000 in their paper) \begin{quotation}
We assume the tail of the underlying distribution begins at the threshold $u$. From our sample of $n$ [=1000] points a random number $N=N_u>0$ will exceed this threshold. If we assume that the $N$ excesses over the threshold are iid with exact GPD distribution, \citet{smith1987} has shown that maximum likelihood estimates $\hat{\xi} = \hat{\xi}_N$ and $\hat{\beta} = \hat{\beta}_N$ of the GPD parameters $\xi$ and $\beta$ are consistent and asymptotically normal as $N\to \infty$, provided $\xi>-\frac{1}{2}$. [\ldots]
In practice we will actually modify the procedure slightly and fix the number of data in the tail to be $N=k$ [=100].
\end{quotation}

Alternative method for the threshold selection was introduced by \citet{gonzalo2004}.

\subsubsection{Estimate of the $GPD$ model}\label{sec:gpdestim}

If we assume that the excess sample $E=(E_1,...,E_{N_u})$ is $iid$ following a $GPD$, $G_{\xi,\beta}$. We can estimate the parameters of the $GPD$, following the maximum likelihood method ($ML$) or the probability-weighted moments ($PWM$). We use the $ML$ method in the following.  \citep{Hosking:1987aa} showed that when $0\leq \xi\leq 0.4$ and for small sample size, the $PWM$ is more precise than the $ML$ method but this difference decrease as the sample size increases. Moreover, \citep{Rootzen:1997aa} showed that for $\xi\leq 0.5$, the $PWM$ is biased contrary to the $ML$ method. As demonstrated by \citep{Hosking:1987aa} for  $\xi\leq -0.5$, the regularity conditions for the $ML$ method are met and the $ML$ $(\hat{\xi}_{N_u}, \hat{\beta}_{N_u})$  estimated over the $N_u$ sample are asymptotically normal.

Writing $g_{\xi,\beta}$ for the density of the $G_{\xi,\beta}$ for $\xi\neq0$:
\begin{equation*}
g_{\xi,\beta}(x)=\frac{1}{\beta}\big(1+\xi\frac{x}{\beta}\big)^{-\frac{1+\xi}{\xi}}
\end{equation*}
The log-likelihood is given by:
\begin{equation*}
\ell(\xi,\beta;E)=-N_u \log(\beta)-\frac{1+\xi}{\xi}\sum_{i=1}^{N_u}\log\big(1+\xi\frac{E_i}{\beta}\big).
\end{equation*}
which must be maximized subject to the parameter constraints that $\beta>0$ and $1+\xi E_i/\beta>0$ for all $i=1,...,N_u$. Solving the maximization problem yields a $GPD$ model, $G_{\xi,\beta}$, for the excess distribution $F_u$.

\subsection{Machine learning and EVT}

Extreme events can be modeled using machine learning techniques. A recent research avenue is provided in \citet{farkas2021}.

\section{R: EVT paper replications}

In this chapter, we will focus on extreme value theory and its main application: the insurance and reinsurance businesses. We recommend reading \citet{Charpentier:2008vn} as an introduction.

The following R packages will be used: ReIns, POT, and extRemes.

\subsection{Zajdenweber 1996 - EVT for business interruptions}

We first replicate\footnote{Zajdenweber1996.R} the paper \citet{Zajdenweber:1996ub}. We use data set available from the CASdataset package. We then focus on the Pareto Distribution Function that we introduced section \ref{sec:paretodist}. 

\subsection{Charpentier 2007 - based on Zajdenweber 1996}
We next replicate\footnote{Charpentier2007.R} \citet{charpentier2007} that itself start by following \citet{Zajdenweber:1996ub} on the same although updated data set. 

\subsubsection{Hill estimator}
Using the fact that
\begin{equation}
Pr\left( X \leq x \vert X>u \right) \simeq 1 - \left( 1 + \xi \frac{x+u}{\sigma} \right)^{-\frac{1}{\xi}}
\end{equation}
for some $\sigma>0$, equivalently:
\begin{equation}
\log\left( 1 - \frac{F(x)}{1-F(u)} \right) \simeq -\frac{1}{\xi}\log x + \mbox{constant}
\end{equation}
Or in simple terms:
\begin{equation}
\log\left( \mbox{count} X > x \right) \simeq -\frac{1}{\xi}\log x + \mbox{constant}
\end{equation}

We can thus estimate $\xi$ with an OLS method, dating back to Pareto works.

\subsubsection{POT and Bloc-maxima estimations}

We next move to peak-over-threshold and bloc-maxima estimations.

\subsection{Charpentier and Flachaire 2021}

We follow\footnote{Charpentier\_Flachaire\_2021.R} \citet{Charpentier2021} to distinguish between, the distributions introduced section \ref{sec:paretodist}:
\begin{itemize}
\item strict Pareto distribution (type I)
\item Generalized Pareto distribution (type II)
\item Extended Pareto distribution
\end{itemize}

\section{Python: insurance, introduction}

For the link between the utility theory and insurance, we refer to the first chapter in \citet{kaas2008modern}. The first main results comes from \citet{arrow1963}, where optimal insurance is derived in a model, this is extended by \citet{raviv1992design}. \citet{ghossoub2017arrow} extends \citet{marshall1992optimum} under heterogeneous beliefs between the insurer and the insurance buyer. You can find an introduction in \citet{boland2007statistical} chapter 8, and more specifically section \textit{Utility and insurance}.

\section{Introduction to Bayesian statistics}\label{seq:BAM}

The easiest introduction to Bayesian statistics would be \citet{johnson2022bayes} \href{https://www.bayesrulesbook.com}{available online}.

I also recommend this \href{https://statswithr.github.io/book/}{Introduction to Bayesian Thinking}.

\subsection{Bayesian methods introduction}

\subsubsection{Schaum's example}

We use Schaum's introduction example \begin{quote}
A box contains two fair coins and a biased coin with probability for heads $P(H) = 0.2$. A coin is chosen at random from the box and tossed three times. If two heads and a tail are obtained, what is the probability of the event F, that the chosen coin is fair, and what is the probability of the event B, that the coin is biased?
\end{quote}

Bayes' theorem is stated as follow, if $D$ is an event (the realization of a sample, a data set) and $F$ is a characteristic (a fair coin is being tossed), then the theorem states:
\begin{equation}
P(F\vert D) = \frac{P(D\vert F) P(F)}{P(D)}
\end{equation}
and we can write: $P(D) =  P(D\vert F) P(F) + P(D\vert B) P(B)$

The we have:
\begin{equation}
P(F\vert D)= \frac{(0.5)^3\frac{2}{3}}{(0.5)^3\frac{2}{3} + (0.2)^2 0.8 \frac{1}{3}}\simeq 0.89
\end{equation}

\subsubsection{Tversky example}

Next, I take an illustration in \citet{tversky1974judgment}. \begin{quote}
Subjects were shown brief personality descriptions of several individuals, allegedly sampled at random from a group of 100 professionals - engineers ($A$) and lawyers ($B$). The subjects were asked to assess, for each description ($C$), the probability that it belonged to an engineer rather than to a lawyer. In one experimental condition, subjects were told that the group from which the descriptions had been drawn consisted of 70 engineers and 30 lawyers. In another condition, subjects were told that the group consisted of 30 engineers and 70 lawyers. The odds that any particular description belongs to an engineer rather than to a lawyer should be higher in the first condition, where there is a majority of engineers, than in the second condition, where there is a majority of lawyers. Specifically, it can be shown by applying Bayes' rule that the ratio of these odds should be $\left(\frac{.7}{.3}\right)^2$, or $5.44$, for each description.
\end{quote}
The descriptions were the same in both conditions; what was manipulated is the prior. Subjects who believed what they were told held $P(A)=.7$ in the first condition (let us write $P_1$ for that belief) and $P(A)=.3$ in the second (let us write $P_2$). Because the description $C$ is the same, the likelihood ratio $\frac{P(C\vert A)}{P(C\vert B)}$ is the same under both conditions and cancels. Writing Bayes' rule in odds form in each condition and taking the ratio of the two:
\begin{equation}\label{eq:tverskyodds}
\frac{\frac{P_1(A\vert C)}{P_1(B\vert C)}}{\frac{P_2(A\vert C)}{P_2(B\vert C)}}=\underbrace{\frac{\frac{P( C \vert A)}{P( C\vert B)}}{\frac{P( C \vert A)}{P( C\vert B)}}}_{=\,1}\frac{\frac{P_1(A)}{P_1(B)}}{\frac{P_2(A)}{P_2(B)}} = \frac{P_1(A)}{P_1(B)}\frac{P_2(B)}{P_2(A)} = \left(\frac{.7}{.3} \right)^2 \simeq 5.44
\end{equation}
so the posterior odds in favour of an engineer should be $5.44$ times larger in the first condition than in the second, for every description. Note that the whole effect comes from the prior odds: the evidence $C$ contributes the same factor in both conditions and therefore drops out of the comparison.

\begin{quote}
In a sharp violation of Bayes' rule, the subjects in the two conditions produced essentially the same probability judgments. Apparently, subjects evaluated the likelihood that a particular description belonged to an engineer rather than to a lawyer by the degree to which this description was representative of the two stereotypes, with little or no regard for the prior probabilities of the categories.
\end{quote}

Next, they mention
\begin{quote}
Imagine an urn filled with balls, of which 2/3 are of one color and 1/3 of another. One individual has drawn ($D_1$) 5 balls from the urn, and found that 4 were red and 1 was white. Another individual has drawn ($D_2$) 20 balls and found that 12 were red and 8 were white. Which of the two individuals should feel more confident that the urn contains 2/3 red balls and 1/3 white balls, rather than the opposite? What odds should each individual give?
\end{quote}
We assume equal prior probabilities, $P(R=2/3)=P(R=1/3)=0.5$, and we drop the binomial coefficients: as in the previous example they are common to both hypotheses and cancel.
\begin{equation}
P(R=2/3\vert D_1) = \frac{P(D_1\vert R=2/3)P(R=2/3)}{P(D_1)}=\frac{\left(\frac{2}{3}\right)^4\frac{1}{3} 0.5}{\left(\frac{2}{3}\right)^4\frac{1}{3} 0.5 + \left(\frac{1}{3}\right)^4\frac{2}{3} 0.5} = \frac{8}{9}
\end{equation}
so odds of 8 to 1, and
\begin{equation}
P(R=2/3\vert D_2) = \frac{P(D_2\vert R=2/3)P(R=2/3)}{P(D_2)}=\frac{\left(\frac{2}{3}\right)^{12}\left(\frac{1}{3}\right)^{8} 0.5}{\left(\frac{2}{3}\right)^{12}\left(\frac{1}{3}\right)^{8} 0.5 + \left(\frac{1}{3}\right)^{12}\left(\frac{2}{3}\right)^{8} 0.5} = \frac{16}{17}
\end{equation}
so odds of 16 to 1. The second individual should be much more confident. \begin{quote}
In this problem, the correct posterior odds are 8 to 1 for the 4 : 1 sample and 16 to 1 for the 12 : 8 sample, assuming equal prior probabilities. However, most people feel that the first sample provides much stronger evidence for the hypothesis that the urn is predominantly red, because the proportion of red balls is larger in the first than in the second sample. Here again, intuitive judgments are dominated by the sample proportion and are essentially unaffected by the size of the sample, which plays a crucial role in the determination of the actual posterior odds.
\end{quote}

\subsubsection{Monty Hall}

Let's imagine you play a game with three doors, $A$, $B$ and $C$. Behind one door there is a prize. You have to choose one door, without any information, you could choose any indifferently, let's imagine you choose $A$. Now the master tells you he will open one of the two remaining door where the prize is not, let's imagine he opens $B$. Should you change your mind?
You start with equal prior probability: $P(A)=P(B)=P(C)=1/3$. Now, the probability that the master opened $B$, knowing that $A$ was right is $1/2$, because the master could also have chosen $C$ indifferently: $P(B\vert A)=1/2$. But now if $C$ is the door with the prize, the master would have had only one choice, $B$: $P(B\vert C)=1$. Using Bayes formula:
$P(A\vert B) \propto P(B\vert A)P(A)$ and $P(C\vert B) \propto P(B\vert C) P(C)$ hence now the odd ratio is in favor of $C$ and you have better change your mind:
\begin{equation}
\frac{P(C\vert B)}{P(A\vert B)} = 2
\end{equation}

\subsection{Bayesian inference}
I follow the notations in \citet{ghosh2006introduction} and \citet{gelman1995bayesian}. Let's imagine a Bayesian expert believe a population $X$ follows a parametric distribution, with unkown parameter $\theta$. The conditional probability density (also called the posterior) of $\theta$ given $X=x$ follows the Bayes formula
\begin{equation}
\pi\left( \theta \vert x \right) = \frac{\pi(\theta)f\left(x \vert  \theta \right) }{\int_\Theta \pi(\theta')f\left(x \vert  \theta' \right) d\theta'}
\end{equation}
where $\pi(\theta)$ is the prior density function, if we chose a parametric density the prior distribution's parameters are referred to as hyperparameters, and $f\left(x \vert  \theta \right)$ the conditional density of $X$ given $\theta$ also called the sampling distribution. The numerator is the joint density of $\theta$ and $X$, $f(\theta, x)=\pi(\theta) f(x\vert \theta)$ and the denominator is the marginal density of $X$, $f(x)=\int_\Theta \pi(\theta')f\left(x \vert  \theta' \right) d\theta'$, also called the \emph{prior predictive distribution}: prior because it is not conditional on a previous observation of the process, and predictive because it is the distribution for a quantity that is observable. 
As probability density function have to be normalized so that when integrated from $-\infty$ to $\infty$ they sum to one, we can just focus on the numerator and work with the unnormalized posterior density:
\begin{equation}
\pi\left( \theta \vert x \right) \propto \pi(\theta)f\left(x \vert  \theta \right) 
\end{equation}

The posterior mean is
\begin{equation}
E\left( \theta \vert x \right) = \int_{-\infty}^\infty \theta \pi\left( \theta \vert x \right) d\theta
\end{equation}

Interestingly,
$$E\left[\mbox{var}\left(\theta\vert y \right)  \right] = E\left(E\left(\theta^2\vert y \right) -\left[E\left(\theta\vert y \right) \right]^2 \right)$$
$$\mbox{var}\left[E\left(\theta\vert y \right) \right] = E\left[\left(E\left(\theta\vert y \right) \right)^2 \right] - \left[E\left(E\left(\theta\vert y \right)\right) \right]^2$$
hence
\begin{equation}
E\left[\mbox{var}\left(\theta\vert y \right)  \right] + \mbox{var}\left[E\left(\theta\vert y \right) \right] = \mbox{var}\left( \theta \right)
\end{equation}

Therefore, the posterior variance is on average smaller than the prior variance, by an amount that depends on the variation in posterior means over the distribution of possible data. The greater the latter variation, the more the potential for reducing our uncertainty with regard to the unknown parameter $\theta$.

\subsubsection{Natural conjugate prior}
If $\mathcal{F}$ is a class of sampling distribution $p(y\vert\theta)$ and $\mathcal{P}$ is the set of all densities having the same functional form as the likelihood, then $\mathcal{P}$ is a natural conjugate for $\mathcal{F}$ if $p(\theta \vert y) \in \mathcal{P}$ for all $p(.\vert \theta)\in \mathcal{F}$ and $p(.)\in \mathcal{P}$. This is convenient for computation and interpretation. It has been shown that, in general, the exponential families are the only classes of distributions that have natural conjugate prior distributions. The class $\mathcal{F}$ is an exponential family if all its members have the form
\begin{equation}
p(y_i\vert\theta)=f(y_i)g(\theta)\exp\left[\phi(\theta)^T u(y_i) \right]
\end{equation}
if $y=(y_1,\dots,y_n)$ is a vector of iid observations, then
\begin{equation}
p(y\vert\theta)\propto g(\theta)^n \exp\left[\phi(\theta)^T \sum_{i=1}^n u(y_i) \right]
\end{equation}
if we specify the prior as 
$$p(\theta)= g(\theta)^\eta \exp\left[\phi(\theta)^T \nu \right]$$
then the posterior density is also of the exponential family:
\begin{equation}
p(\theta\vert y)\propto g(\theta)^{n+\eta} \exp\left[\phi(\theta)^T \left( \sum_{i=1}^n u(y_i) +\nu \right) \right]
\end{equation}

We provide an example where the Gamma distribution is the natural conjugate prior for the Pareto distribution section \ref{sec:bayesparetoclaims}.

\subsection{Bayesian inference - normal distribution with known variance}

Let's imagine we have $n$ independent and identically distributed observations $X_1,\ldots,X_n$ that follows a normal distribution with known variance $\sigma^2$ but unknown mean $\mu$. As a Bayesian expert, we might want to assume (have the prior) that $\mu \sim \mathcal{N}(\eta,\tau^2)$, then our posterior will be, according to Bayes formula
$$\pi\left(\mu \vert X_1, \ldots,X_n \right) \propto \frac{1}{\tau \sqrt{2\pi}}\exp\left(-\frac{1}{2}\left(\frac{\mu - \eta}{\tau} \right)^2 \right) \prod_{i=1}^n \frac{1}{\sigma \sqrt{2\pi}}\exp\left(-\frac{1}{2}\left(\frac{X_i - \mu}{\sigma} \right)^2 \right)$$
$$\pi\left(\mu \vert X_1, \ldots,X_n \right) \propto \exp\left(-\frac{1}{2} \left[ \left(\frac{\mu - \eta}{\tau} \right)^2 + \sum_{i=1}^n \left(\frac{X_i - \mu}{\sigma} \right)^2 \right] \right) $$
$$\pi\left(\mu \vert X_1, \ldots,X_n \right) \propto \exp\left(-\frac{1}{2} \left(\frac{\mu - \eta'}{\tau'} \right)^2 \right) $$

with
$$\eta' = \frac{\frac{\eta}{\tau^2}+\frac{n}{\sigma^2}\bar{X}}{\frac{1}{\tau^2}+\frac{n}{\sigma^2}}$$
and
$$\sigma' = \sqrt{\frac{\tau^2\sigma^2}{\sigma^2+ n \tau^2}}$$
We illustrate\footnote{bayesian\_inferance\_normal\_iid.py} this with samples from a normal distribution and display how the Bayesian expert update his prior and posterior distributions.

\subsection{Noninformative prior distributions}\label{sec:jeffreysprinciple}

When as a statistician you have little prior knowledge about the population (or the phenomenon) you want to study, you want the prior distribution to be almost "transparent" to carry little to no information. \citet{jeffreys1998theory} introduced an approach to defining noninformative prior distributions, his invariance principle where the prior density follows
\begin{equation}
p(\theta)\propto \left[J(\theta) \right]^{\frac{1}{2}}
\end{equation}
where $J(\theta)$ is the Fisher information for $\theta$:
\begin{equation}
J(\theta) = -E\left(\frac{d^2 \log p(y\vert\theta)}{d\theta^2} \middle\vert \theta \right)
\end{equation}

We provide an illustration for Pareto distribution section \ref{sec:paretoJeffrey}.

\subsection{Normal approximations to the posterior distribution}

If the posterior distribution $p(\theta \vert y)$ is unimodal and roughly symmetric, it can be convenient to approximate it by a normal distribution, as first demonstrated by \citet{le1953some}. We apply a Taylor expansion of $\log p(\theta \vert y)$ centered at the posterior mode\footnote{The mode is defined as $\max_\theta p(\theta \vert y)$, which is usually different from the posterior mean. At the mode, $\frac{d \log p(\theta \vert y)}{\theta}=0$ hence this term won't appear in the Taylor expansion.} $\hat{\theta}$:

\begin{equation}
\log p(\theta \vert y) = \log p(\hat{\theta} \vert y) + \frac{1}{2}\left( \theta - \hat{\theta}\right)' \left[\frac{d^2 \log p(\theta\vert y)}{d\theta^2} \right]_{\theta=\hat{\theta}}\left( \theta - \hat{\theta} \right) + \ldots
\end{equation}

Hence we can approximate that:
\begin{equation}
p(\theta \vert y)  \sim \mathcal{N}\left(\hat{\theta}, \left[I(\hat{\theta}) \right]^{-1} \right)
\end{equation}
where $I(\hat{\theta})$ is the observed information:
\begin{equation}
I(\theta) = - \frac{d^2 \log p(\theta \vert y)}{d\theta^2}
\end{equation}

We give an illustration with Pareto distributed claims section \ref{sec:posterirornormalapprox}

\subsection{Bayesian Model Averaging}\label{sec:BMA}

\citet{Morris:1974wk} formalized the use of experts in decision situation that would lead to the Bayesian Averaging Model literature. \citet{Clemen:1999vy} did a review of this branch. See section \ref{sec:bayespooling} for Bayesian methods for experts' opinion pooling.

While according to the classical rationale econometricians assume that true (unknown) model parameters are deterministic, the Bayesian approach considers them random variables. Bayes' rule concretizes the idea:
$$ h(\beta \vert y) = \frac{f(y\vert \beta)h(\beta)}{f(y)}$$

A density $h(\beta)$ reflects prior beliefs about the parameter before having subjected the model to data. The posterior parameter distribution $h(\beta\vert y)$ reflects the density of the parameter conditional on having seen the data. Further, $f(y)$ is the unconditional density of the data and $f(y\vert \beta)$ the density of the data conditional on the model (the parameter).

For seeing the rationale behind Bayesian model averaging, one can compartmentalize the model space (the parameter space) up to $I$ regions. Reformulating Bayes' rule gives an expression for the posterior density of a parameter as a function of the compartmentalized model space.
\begin{equation}\label{eq:BAM1}
h(\beta \vert y) = \sum_{i=1}^I P(M_i\vert y ) \frac{f(y\vert \beta)h(\beta \vert M_i)}{f(y \vert M_i)}
\end{equation}

The posterior distribution of $\beta$ is therefore a weighted average of the $I$ models' posterior parameter densities, with posterior model probabilities being used as weights.\footnote{The Bayesian averaging of classical estimates approach that is used for credit risk modelling application now envisages the use of diffuse priors which in case that the parameter space is bounded implies uniform weights for all models}

The weights $P(M_i\vert y)$ that are needed to obtain the posterior parameter densities in equation \ref{eq:BAM1} would be computed as follows:

$$P(M_i\vert y) = \frac{P(M_i) T^{-g_i/2} \mbox{SSE}_i^{-T/2}}{\sum_{j=1}^I P(M_j)T^{-g_i/2} \mbox{SSE}_i^{-T/2} }$$

where $g_j$ is the number of right-hand side variables included in model $j$. The final posterior mean of a parameter $\beta$ would be computed as:
$$E(\beta \vert y) = \sum_{i=1}^I P(M_i \vert y ) \hat{\beta}_i$$

with $\hat{\beta}_i= E(\beta \vert y,M_i)$ being the parameter's posterior mean estimate conditional on model $i$. The posterior variance is the following:

$$Var(\beta \vert y) = \sum_{i=1}^I P(M_i \vert y ) Var(\beta \vert y,M_i) + \sum_{i=1}^I P(M_i \vert y ) \left(\hat{\beta}_i - E(\beta \vert y) \right)^2$$

an additional object of interest is the probability for any particular predictor to be included in the model space, the \textit{posterior inclusion probability}. it is computed as the sum of the posterior model probabilities $P(M_i \vert y)$ which contain the particular predictor.

An Autoregressive Distributed Lag (ADL) model structure is the basis for defining the model space for all credit risk indicators. A dependent variable $Y_t$ is allowed to be a function of its own lags as well as contemporaneous and possibly further lags of a set of predictor variables $X_t^k$.

$$Y_t = \alpha + \rho_1 Y_{t-1} + \ldots + \rho_p Y_{t-p} + \sum_{k=1}^{k_i} \left( \beta_0^k X_t^k + \ldots + \beta_{q^k}^k X_{t-q^k}^k \right) + \epsilon_t$$

for each specific model $i$ with its predetermined set of predictor variables, the lag structures for autoregressive and distributed exogenous terms, $p$ and $q^k$, are chosen optimally by estimating all combinations of lag structures up to a limit $G$. The specification for which the Bayesian Information Criterion (BIC) is minimal is chosen. In the course of the additional search for the optimal lags, the lag structure for the autoregressive part and for lags of exogenous predictors are forced to be "closed" (without gaps).

\section{Python: Insurance claims, prior and posterior, \citet{meyers1996}}

We replicate\footnote{vansteenberghe\_bayes\_pareto\_distribution.py} the work of \citet{meyers1996} and also use the book \citet{klugman2012loss} example 13.1.

\begin{quote}
One of the more popular models used [to estimate the expected claim severity in an excess layer of insurance] is the single parameter Pareto.
\end{quote}

For an application, we follow equation \ref{eq:ParetoI}, with $u=1$ and $\alpha=2.5$, but where the $\alpha$ is unknown to the insurance expert.

The gamma distribution is typical for the prior distribution in this case (Bayesian estimation).

\subsection{Bayesian estimation - Pareto}\label{sec:bayesparetoclaims}

The prior distribution is a probability distribution over the space of possible parameter values. It is denoted $\pi(\alpha)$ and represents our opinion concerning the relative chances that various values of $\alpha$ are the true value of the parameter.

In our case, with a gamma distribution with shape of $s=2$ and rate of $r=1$:
\begin{equation}
\pi(\alpha) = \frac{1}{\Gamma(2)} \alpha e^{-\alpha}
\end{equation}

The model distribution is the probability distribution for the data as collected given a particular value for the parameter. The model pdf is denoted $f_{X \vert A} \left(x \vert \alpha \right)$, with $x$ a sample from $X$ and $\alpha$ our parameter from the parameter space $A$.

The posterior distribution is the conditional probability distribution of the parameters given the observed data. It is denoted $\pi_{A\vert X} \left(\alpha \vert x \right)$ and can be computer as:
\begin{equation}
\pi_{A\vert X} \left(\alpha \vert x \right) = \frac{f_{X \vert A} \left(x \vert \alpha \right) \pi(\alpha)}{\int f_{X \vert A} \left(x \vert \alpha \right) \pi(\alpha) d \alpha}
\end{equation}

\stepcounter{question}
\begin{itemize}
\item \textbf{Question \arabic{question}}
\begin{itemize}
\item Demonstrate that if we take a prior with a gamma distribution with shape of $s$ and rate of $r$, then after facing $n$ claims of intensity $x_i$ each,   the posterior distribution is a multiple of gamma function with a shape $s+n$ and a rate $r + \sum_{i=1}^n  \ln (x_i)$.
\end{itemize}
\end{itemize}

\begin{equation}
\pi_{A\vert X} \left(\alpha \vert x \right) = \frac{\alpha^{n+s-1} \exp\left[ - \alpha \left( r + \sum_i  \ln (x_i)\right) \right]}{(n+s-1)! \left( \frac{1}{ r + \sum_i  \ln (x_i)} \right)^{n+s}}
\end{equation}

A prior distribution is said to be a conjugate prior distribution for a given model if the resulting posterior distribution is from the same family as the prior (but perhaps with different parameters).

Nota bene: with the noninformative (and improper) prior $\pi(\alpha)=\alpha^{-1}$, \citet{meyers1996} demonstrate that the posterior is normally distributed (equation 4.7) with mean $\hat{\alpha}$ and variance $\frac{\hat{\alpha}^2}{n}$.

\stepcounter{question}
\begin{itemize}
\item \textbf{Question \arabic{question}}
\begin{itemize}
\item What is the shape of the posterior when the prior is normally distributed?
\end{itemize}
\end{itemize}

\subsubsection{Jeffreys' principle for an uninformative prior}\label{sec:paretoJeffrey}

If you have no information to form a prior distribution for the Pareto exponent $\alpha$, then you might want to use Jeffreys' principle introduced section \ref{sec:jeffreysprinciple}.

We assume that insurance claims $X$ follow a Pareto distribution with unknown $\alpha$ and a threshold $u=1$, then the likelihood writes
\begin{equation}
p(x\vert \alpha) = \frac{\alpha}{x^{\alpha+1}}
\end{equation}
therefore
\begin{equation}
\log p(x\vert \alpha) = \log \alpha - (\alpha+1) \log x
\end{equation}
If we derive twice:
\begin{equation}
\frac{d^2 \log p(x\vert \alpha)}{d\alpha^2} = -\frac{1}{\alpha^2}
\end{equation}
Hence the Fisher information for $\alpha$ writes $J(\alpha) = \alpha^{-2}$ and if we want to apply Jeffreys' principle, we have to choose a prior distribution:
\begin{equation}
p(\alpha) \propto \frac{1}{\alpha}
\end{equation}

\subsubsection{Normal approximations to the posterior distribution - Pareto exponent}\label{sec:posterirornormalapprox}

We have demonstrated that the posterior distribution of the Pareto exponent $\alpha$, when we use the Gamma distribution (natural conjugate), with shape $s$ and rate $r$:
\begin{equation}
p(\alpha \vert x) \propto \gamma\left( s(t), r(t) \right)
\end{equation}
with $s(t+1) = 1 + s(t)$ and $r(t+1)=r(t) + \ln x_t$.

Then 
\begin{equation}
\log p(\alpha \vert x) \propto s \log r - \log \Gamma(s) + (s-1)\log \alpha - r\alpha
\end{equation}
and when we differentiate twice
\begin{equation}
\frac{d^2 \log p(\alpha \vert x)}{d\alpha^2} = - \frac{s-1}{\alpha^2}
\end{equation}
Therefore
\begin{equation}
I(\hat{\alpha})=\frac{s-1}{\hat{\alpha}^2}
\end{equation}
Therefore, we can approximate that the posterior follows
\begin{equation}
p(\alpha \vert x)  \sim \mathcal{N}\left(\hat{\alpha}, \frac{\hat{\alpha}^2}{s-1} \right)
\end{equation}

As for a Gamma distribution, the mode is $(s-1)r^{-1}$ then this can be written as
\begin{equation}
p(\alpha \vert x)  \sim \mathcal{N}\left(\frac{s-1}{r}, \frac{s-1}{r^2} \right)
\end{equation}
If we study a process where at each period one claim $x_t$ is generated from a Pareto distribution with parameter $\alpha$ and a threshold of one, then for $t\geq2$

\begin{equation}\label{eq:posteriorapproxdist}
p(\alpha_t \vert x_1,\ldots,x_t)  \sim \mathcal{N}\left(\frac{t-1}{\sum_{i=1}^t \ln x_i}, \frac{t-1}{\left(\sum_{i=1}^t \ln x_i \right)^2} \right)
\end{equation}

\stepcounter{question}
\begin{itemize}
\item \textbf{Question \arabic{question}}
\begin{itemize}
\item Following \citet{lee2019exponentiated}, show that the rate parameter follows an inverted gamma distribution of shape $t$ and rate $\alpha$.
\item Using the approximation in equation \ref{eq:posteriorapproxdist}, express the evolution over time of the variance of the posterior, \href{https://stats.stackexchange.com/questions/133681/if-x-takes-on-a-gamma-distribution-how-can-i-find-x2-x3-etc}{hint here}.
\end{itemize}
\end{itemize}

\subsection{Bayes estimate of $X_{n+1}$ and B\"{u}hlmann credibility}

\citet{klugman2012loss} example 17.11, for claims exponentially distributed, the Bayes estimate of $X_{n+1}$, when the prior is a gamma distribution with shape $s$ and rate $r$ is:
\begin{equation}
\frac{n}{n+s-1} \bar{x} + \frac{s-1}{n+s-1}\frac{r}{s-1}
\end{equation}
in this case, this is also the B\"{u}hlmann estimate.

\stepcounter{question}
\begin{itemize}
\item \textbf{Question \arabic{question}}
\begin{itemize}
\item What is the Bayes estimate when the claims have a Pareto type I distribution?
\end{itemize}
\end{itemize}

\section{\citet{arrow1963} replication}

This section replicates\footnote{Arrow1963.py} the work of \citet{arrow1963}. 

\subsection{Optimality theorems}

\begin{enumerate}
\item "If a competitive equilibrium exists at all, and if all commodities relevant to costs or utilities are in fact priced in the market, then the equilibrium is necessarily optimal in the following precise sense (due to V. Pareto): There is no other allocation of resources to services which will make all participants in the market better off."
\item If there are no increasing returns in production, and if certain other minor conditions are satisfied, then every optimal state is a competitive equilibrium corresponding to some initial distribution of purchasing power.
\end{enumerate}

\begin{quote}
Operationally, the significance of this proposition is that if the conditions of the two optimality theorems are satisfied, and if the allocation mechanism in the real world satisfies the conditions for a competitive model, then social policy can confine itself to steps taken to alter the distribution of purchasing power [via taxes or subsidies].
\end{quote}

Major competitive preconditions:
\begin{enumerate}
\item existence of competitive equilibrium;
\item marketability of all goods and services relevant to costs and utilities;
\item nonincreasing returns.
\end{enumerate}

\subsection{Price discrimination}

We start by the \textbf{footnote 33}, where Arrow states that \begin{quotation}the pricing practices of the medical industry depart sharply form the competitive norm. [\ldots] Hypothetically, [\ldots] everyone would be better off if prices were made equal for all, and the rich compensated the poor for the changes in the relative positions. [\ldots] the discussion is simplified by considering only two income levels, rich and poor [\ldots] suppose the price of medical care to the rich is double that to the poor, the medical expenditures by the rich are 20 per cent of those by the poor, and the elasticity of demand for both classes is .5; then the net social gain due to the abolition of discrimination is slightly over 1 per cent of previous medical expenditures. [\ldots] Let us choose our quantity and monetary units so that the quantity of the medical services consumed by the poor and the price they pay are both 1. [\ldots] The supply of medical services is assumed fixed and therefore must equal 1.1.\end{quotation}

By definition, the price elasticity is $\frac{p}{q}\frac{\partial q}{\partial p}$, as for the poor $p=q=1$, then $\frac{\partial q}{\partial p} = -.5$, thus for the poor we have:
\begin{equation}
q_{poor}(p_{poor}) = c_{poor} p_{poor}^{-.5}
\end{equation}
as $q_{poor}=p_{poor}=1$, necessarily the constant $c_{poor}$ has to be one.

The rich have the same elasticity, hence:
\begin{equation}
q_{rich}(p_{rich}) = c_{rich} p_{rich}^{-.5}
\end{equation}
has we know that $p_{rich}=2$, then 
\begin{equation}
q_{rich} p_{rich} = .2 = c_{rich} 2^{.5}
\end{equation}
hence $c_{rich} = .2 \times 2^{-.5}$ and $q^*_{rich} = .1$.

The supply of medical services is assumed fixed, hence it is $q_{poor} + q_{rich} = 1.1$

Now if price discrimination between the rich and the poor is abolished, then the equilibrium price must satisfy:
\begin{equation}
q_{rich} \bar{p} + q_{poor} \bar{p} = 1.1
\end{equation}
hence $\bar{p} = 1.08$

\subsubsection{Consumers' surplus}

The consumer surplus is the monetary gain obtained by consumers because they are able to purchase a product for a price that is less than the highest price that they would be willing to pay. 

Let's focus on the rich class, they are willing to pay (inverse demand function):
\begin{equation}
p_{rich} = \left( \frac{c_{rich}}{q_{rich}} \right)^2
\end{equation}
intuitively, we see that for a quantity .1 if the price is fixed at $\bar{p} = 1.08$ they will be better of than the discriminating price they had before ($p_{rich}=2$). At this equilibrium, the rich we demand the quantity: $\bar{q}_{rich} = c_{rich} \bar{p}^{-.5}$

The supply curve is simply the constant $\bar{p}$, we illustrate the supply and demand curve and the consumer surplus (the red area):
\begin{center}
\includegraphics[scale=.5]{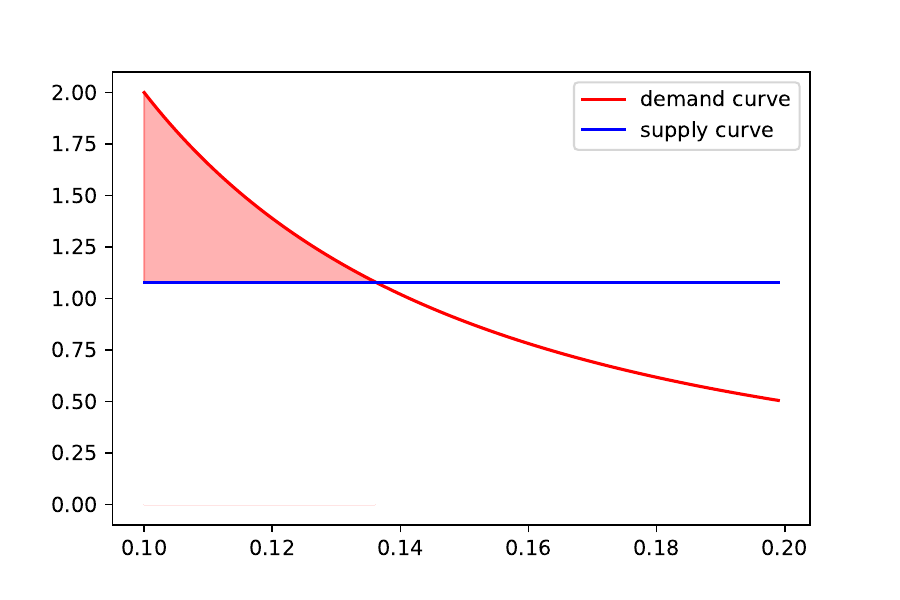}
\end{center}

The rich consumer surplus can be computed with:
\begin{equation}
\int_{q^*_{rich}}^{\bar{q}_{rich}} \left( \frac{c_{rich}}{q} \right)^2 dq - \bar{p}(\bar{q}_{rich}-q^*_{rich})
\end{equation}

\stepcounter{question}
\begin{itemize}
\item \textbf{Question \arabic{question}}
\begin{itemize}
\item Compute the poor surplus change and conclude on the overall consumers' surplus change.
\end{itemize}
\end{itemize}

\subsection{The theory of ideal insurance}

Each individual acts so as to maximize the expected value of utility function. 
\begin{itemize}
\item $W$: individual initial wealth
\item $X$: loss (random variable) with probability density function $\Phi()$
\item $I(X)$: insurance claims subject to $I(X)\geq 0, \forall X$
\item $P$: insurance premium
\end{itemize}
Individual final wealth: $Y(X)=W-P-W+I(X)$. Individual maximizes $E\left[ Y(X) \right]$
Illness enters the utility function as a separate variable.
Individuals are normally risk-averters, i.e. they have a diminishing marginal utility of income: $U'\left[ Y(X') \right] > U'\left[ Y(X) \right], \forall Y(X') < Y(X)$.

It follows from the assumption of risk aversion that if an individual is given a choice between a probability distribution of income with a given mean $m$, and the certainty of the income $m$, he would prefer the latter.

Consider a large insurance company and cost of medical care a random variable with mean $m$, the company charges the premium $m$ and agrees to indemnify the individual for all medical costs. There should be a social gain if medical risks on different individuals are basically independent. In the limit, the welfare loss, even assuming risk aversion on the part of the insurer would vanish (law of large number, etc.). In fact, there are interdependence among the risks and administrative costs, but then a premium, perhaps slightly above the actuarial level, would be sufficient to offset this welfare loss.

Suppose, to take a simple case, the insurance company is not willing to sell any insurance policy that a consumer wants but will charge a fixed-percentage loading above the actuarial value for its premium. Then it can be shown that the most preferred policy form the point of view of an individual is a coverage with a deductible amount; that is, the insurance policy provides 100 per cent coverage for all medical costs in excess of some fixed-dollar limit. If, however, the insurance company has some degree of risk aversion, its loading may also depend on the degree of uncertainty of the risk. In that case, the Pareto optimal policy will involve some element of coinsurance, i.e., the coverage for costs over the minimum limit will be some fraction less than 100 per cent.

\textbf{Proposition 1}: if an insurance company is willing to offer any insurance policy against loss desired by the buyer at a premium which depends only on the policy's actuarial value, then the policy chosen by a risk-averting buyer will take the form of 100 per cent coverage above a deductible minimum.

\textbf{Proof}: take an insurance policy $I_1(X)$ and $Y_1(X_1)>Y_2(X_2)$.
Choose $\delta$ sufficiently small so that
\begin{equation}\label{eq:arrow1}
I_1(X)>0 \quad \mbox{for} \quad X_1\leq X \leq X_1+\delta
\end{equation}
\begin{equation}\label{eq:arrow2}
Y_1(X')  < Y_1(X) \quad \mbox{for} \quad X_2\leq X' \leq X_2+\delta, \quad X_1\leq X \leq X_1+\delta
\end{equation}
Write $\pi_i$ the probability that $X\in[X_i, X_i+\delta ]$, from \ref{eq:arrow1} and \ref{eq:arrow2}, we can choose $\epsilon>0$ and sufficiently small so that
\begin{equation}\label{eq:arrow3}
I_1(X) -\pi_2 \epsilon \geq 0
\end{equation}
\begin{equation}\label{eq:arrow4}
Y_1(X') + \pi_1 \epsilon < Y_1(X) - \pi_2 \epsilon
\end{equation}

Define a new insurance policy:
\begin{equation}
    I_2(X)=
    \begin{cases}
I_1(X) - \pi_2 \epsilon,& \text{if} \quad X\in[X_1, X_1+\delta ] \\        I_1(X) + \pi_1 \epsilon, & \text{if} \quad X\in[X_2, X_2+\delta ]
    \end{cases}
\end{equation}
From equation \ref{eq:arrow3}, $I_2(X)\geq 0, \forall X$. We show that the two policies have the same actuarial value and, by assumption, the same premium $E[I_2(X)]=E[I_1(X)]$:

\begin{align}
E\left[I_2(X)-I_1(X)\right] &= \int_{X_1}^{X_1+\delta} \left[ I_2(X)-I_1(X) \right]\Phi(X) dX + \int_{X_2}^{X_2+\delta} \left[ I_2(X)-I_1(X) \right]\Phi(X) dX\\
&= -\pi_2 \epsilon \int_{X_1}^{X_1+\delta} \Phi(X) dX + \pi_1 \epsilon \int_{X_2}^{X_2+\delta} \Phi(X) dX\\
&= 0
\end{align}

We have $Y_2(X) - Y_1(X) = I_2(X) - I_1(X)$, then from equation \ref{eq:arrow4}:
\begin{equation}
Y_1(X') < Y_2(X')<Y_2(X) < Y_1(X)
\end{equation}
and
\begin{equation}
E\left[ U\left[Y_2(X) \right] - U\left[Y_1(X) \right] \right] = \int_{X_1}^{X_1+\delta} \left[ U\left[Y_2(X) \right] - U\left[Y_1(X) \right] \right]\Phi(X) dX + \int_{X_2}^{X_2+\delta} \left[ U\left[Y_2(X) \right] - U\left[Y_1(X) \right] \right]\Phi(X) dX
\end{equation}
By the Mean Value Theorem, $\forall X$
\begin{align}
U\left[Y_2(X) \right] - U\left[Y_1(X) \right] &= U'\left[Y(X)\right] \left[Y_2(X)  - Y_1(X) \right]\\
&= U'\left[Y(X)\right] \left[I_2(X)  - I_1(X) \right]
\end{align}
As $Y(X') < Y(X)$, as individuals are normally risk-averters, i.e. they have a diminishing marginal utility of income: $U'\left[ Y(X') \right] > U'\left[ Y(X) \right]$. Equivalently, for some $u$: $U'\left[Y(X')\right] >u$ and $U'\left[Y(X)\right] < u$ we get
\begin{align}
E\left[ U\left[Y_2(X) \right] - U\left[Y_1(X) \right] \right] &= -\pi_2 \epsilon \int_{X_1}^{X_1+\delta}  U'\left[Y(X) \right]\Phi(X) dX + \pi_1 \epsilon \int_{X_2}^{X_2+\delta}  U'\left[Y(X) \right]\Phi(X) dX\\
&> -\pi_2 \epsilon u \pi_1 + \pi_1 \epsilon u \pi_2 = 0
\end{align}
So the second policy is preferred. So, if for some $X$ and $X'$: $I(X)>0$ and $Y(X)>Y(X')$, then $I()$ cannot be optimal. Writing $Y_{min}$ the minimum value taken by $Y(X)$, then under the optimal policy, $I(X)=0$ if $Y(X)>Y_{min}$. Insurance would only be triggered if the final wealth would fall below this minimum to that the final wealth equals this minimum. This is a 100 per cent coverage for loss above a deductible.

\section{The Dismal Theorem}

\subsection{Jeffreys prior}

\citet{jeffreys1932theory} introduced a non-informative prior distribution for a parameter space with a density function proportional to $\frac{1}{x}$. 

\begin{quote}
Note that a prior proportional to $\frac{1}{x}$ does not integrate to a constant i.e. it is improper. As long as the pay off relevant range of a decision problem is finite the use of improper priors does not cause a problem from expected utility theory point of view, because it can be normalized on any finite interval.
\end{quote}

The Jeffreys prior (non-informative) can be extended to the square root of the determinant of the Fisher information matrix.

The Jeffreys' ignorance prior is not a proper distribution (i.e. it does not integrate to a constant). In contrast, the integral of the tails of the posterior distribution converges as long as $k > 1$.

\subsection{Michael Schwarz seminal work}

\citet{weitzman2009modeling} introduced the Dismal Theorem which is mainly based on the definitions, theorems and propositions of \citet{schwarz1999decision}. 

Let $X$ be a positive random variable and $I$ and information set consisting of values $x_1,x_2,\ldots,x_n$ of positive real valued variables $X_1,X_2,\ldots,X_n$. $X$ and $X_1,X_2,\ldots,X_n$ are iid, $\exists \rho(.)$, $\int_0^\infty \rho(x)dx=1$ such that the joint pdf of $(x,x_1,\ldots,x_n)$ is $\rho(x)\rho(x_1)\ldots\rho(x_n)$. 

The first important restriction are invariance principles. \begin{quote}
Invariance principles require that an agent's actions and beliefs be independent of the units of measurement that are being used; such a restriction may be interpreted as a basic rationality requirement. [$\ldots$] The invariant approach was introduced into statistics by \citet{jeffreys1932theory}. [$\ldots$]  Even when substantial data relevant for estimating a random variable is available, the data may be of limited value for estimating the tails of the distribution. Whenever a decision depends on an agent's beliefs about the behavior of the tails of the distribution, we will refer to such decision problems as \textbf{decision making under information scarcity}. In Savage's framework, decision making under information scarcity is characterized by the tails of the agent's prior. We show that invariance restrictions imply that even if some relevant data is available to an agent, her beliefs about the tails of the distribution (i.e. information scarcity region) are well approximated by the Jeffreys' ignorance prior. \citet{jeffreys1998theory} and the subsequent authors interpreted the Jeffreys' ignorance prior as the prior describing an agent's beliefs about a parameter in the absence of any relevant information (i.e., beliefs in an information vacuum).
\end{quote}

and interestingly

\begin{quote}
Note that in a Bayesian framework, if a variable is drawn from a distribution that belongs to some family of probability densities, the probability distribution that an agent assigns to this variable need not belong to the family or densities from which the variable is drawn. In fact, we show that an agent who has invariant beliefs will arrive to a posterior distribution of the random variable in question that has no finite moments. This is true even if the agent knows that the variable is drawn from a probability density that belongs to a family of densities such that all moments are finite for each density (this is because, the parameter of the distribution is unknown to the agent, consequently after integrating over distributions with thin tails one may obtain a distribution with fat tails). This point is essential for establishing applicability area of our results and applicability area of empirical studies of extreme values mentioned earlier. We investigate an agent's beliefs about the tails of the distributions and not the distribution of observed outcomes, consequently, the present theory can not be either supported or refuted based on statistical analysis of outliers of economic variables. We make statements about an agent's beliefs that can be tested by observing an agent's actions that are revealing of her beliefs about the relative likelihood of extreme values. For instance, the pricing of far out of the money options contains information about an agent's beliefs relating tails of the distribution of stock returns.
\end{quote}

This invariance based restrictions on agent's strategy give the Axiom 0: Consider decision problem $d= \left( X, p(x,\alpha),A,I \right)$ and $d'=\left(  X, p'(x,\alpha),A,I \right)$, action $\alpha \in R_{++}$ and $p'(x,\alpha)=p(x,h(\alpha))$, where $h(\alpha)$ is an invertible function. If an agent's strategy in the first decision problem is $\alpha^*$ then her strategy in the second decision problem is given by $\alpha'$ such that $h(\alpha')=\alpha^*$

Where:
\begin{itemize}
\item $p: R_+ \times R_+ \to R$ is a payoff function,
\item $A$ is an action space
\item $\alpha\in A$ is an action chosen before the value of $X$ is revealed
\end{itemize}

Let us define dimensionality compatible transformations $g$ as a function that leaves ratios unchanged with respect to changes in the units of measurement, i.e., for any $x_1, x_2,t>0$,
$$\frac{g(x_1)}{g(x_2)}=\frac{g(\lambda x_1)}{g(\lambda x_2)}$$ 
A function $g(x)$ belong to this set if and only if $g(x)=tx^k$. Invariance assumption can be written as $X$ and $g(X)$ are decision equivalent, meaning that the agent choses identical strategies in decision problem $\left( X, p(x,\alpha),A,I=I_{x_1\ldots x_n}^{X_1\ldots X_n} \right)$ and $\left( g(X), p(x,\alpha),A,I=I_{g(x_1)\ldots g(x_n)}^{g(X_1)\ldots g(X_n)} \right)$.

The invariance axiom on an agent's behavior requires the agent's beliefs to be data driven and the name of the unit of measurement is an uninformative label.

Define a group $G$: $\{x \to tx^k\}$, $t>0$ and $k\neq 0$. $A$ is a subset of $\{\rho\}$ with densities with finite mean and variance. Assume $A$ is closed with respect to $G$ i.e. $\forall g\in G$ and $\forall \rho \in A$, $g \circ \rho \in A$. $H$ is a subset of $A$ for which the expected value equals zero and variance equals one. Then all element of $A$ can be represented in the form $\rho(y)=h\left(\frac{y-\mu}{\sigma} \right) \frac{1}{\sigma}$, where $h\in H$.

Let's define the beliefs of a Bayesian agent as $B=\left(M,W(\theta)\right)$ with $M=\{\rho(x;\theta)\vert \theta\in\Theta\}$ and a prior $W(\theta)$.

Suppose that 
$$\int_\Theta \rho(x;\theta) \rho(x_1;\theta) \ldots \rho(x_n;\theta)W(\theta)d\theta < \infty$$
then the resulting posterior distribution after a standard Bayesian updating procedure is proportional to 
\begin{equation}
F_{X \vert X_1\ldots X_n ; B(x\vert x_1 \ldots x_n)} = \int_\Theta \rho(x;\theta) \rho(x_1;\theta) \ldots \rho(x_n;\theta)W(\theta)d\theta
\end{equation}

Beliefs are by definition equivalent if they correspond to the same probability distribution, that is $B$ is equivalent to $B'$ if $F_{X \vert X_1\ldots X_n ; B(x\vert x_1 \ldots x_n)} =\chi F_{X \vert X_1\ldots X_n ; B'(x\vert x_1 \ldots x_n)} $, with $\chi$ a constant.

Beliefs $B$ are invariant if $\forall t>0, k\neq 0$ these beliefs are equivalent to $B'=\left(\{ \frac{1}{kt^{\frac{1}{t}}} x^{\frac{1-k}{k}} \rho\left(\left(\frac{x}{t} \right)^\frac{1}{k};\theta \right)\vert \theta \in \Theta \}, W(\theta) \right)$.

\citet{schwarz1999decision} Theorem 20: Consider a set of beliefs $B^*=(M^*,W^*)$ where
\begin{equation}\label{eq:th20schwarz}
M^*=\{\frac{1}{x\sigma}h\left(\frac{\ln x - \mu }{\sigma};\nu \right)\vert
\mu\in R^1, \sigma \in R_+^1, \nu\in V \},\\
W^*(\mu,\sigma,\nu)=\frac{T(\nu)}{\sigma^s}
\end{equation}
and $\int_V T(\nu)d\nu < \infty$. Then the following is true:
\begin{enumerate}
\item beliefs $B^*$ are invariant
\item the asymptotic behavior of $F_{X \vert X_1\ldots X_n ; B^*(x\vert x_1 \ldots x_n)}$ as $x\to \infty$ is described by the formula $\frac{\mbox{constant}}{x\left(\ln x \right)^{N+s}}$ where $N$ is the number of observations and $s$ is the parameter in the expression for $W^*(\mu,\sigma,\nu)$
\end{enumerate}

Theorem 22: For any well behaved invariant beliefs there exist equivalent beliefs that are characterized by equation \ref{eq:th20schwarz}.

\subsection{The dismal theorem}

Let $C$ be reduced-form consumption that has been adjusted for welfare by subtracting out all damages from climate change. Suppose to begin with that the representative agent has a standard familiar utility function of CRRA (constant relative risk aversion) form
$$U(C)=\frac{C^{1-\eta}}{1-\eta}$$
With time-preference parameter $\beta\in(0;1]$, the 'stochastic discount factor' or 'pricing kernel' is
$$M(C)=\beta \frac{U'(C)}{U'(1)}$$ The amount of present consumption the agent would be willing to give up in the present period to obtain one extra sure unit of consumption in the future period is $E\left[ M \right]$.

Let $Y=\ln C$ be a random variable which "encapsulates the reduced-form uncertainty that is at the abstract core of an economic analysis of climate change". This RV has pdf $f(y)$ and
$$E\left[ M \right]=\beta \int_{-\infty}^\infty e^{-\eta Y} f(y) dy$$

The question is how to bound $E\left[ M \right]$? The model by placing an ad hoc positive lower bound on consumption, which is denoted D (for 'death'), so that always $C\geq D >0$. The lower bound $D$ is not completely arbitrary, however, because it can be related conceptually to a 'fear of ruin' or a 'value of statistical life' (VSL) parameter. The utility can be written
\begin{equation}
U(C;D) = \begin{cases}
\frac{C^{1-\eta}-D^{1-\eta}}{1-\eta}\quad \mbox{if } C\geq D >0\\
0 \quad \mbox{otherwise}
\end{cases}
\end{equation}
If we normalize $D$ to one and let $A(q)$ be the amount of extra consumption the individual requires within this period to exactly compensate for $P(C\leq D)=q$ within this period. In free translation, $q$ is the probability of death. Form EU theory, $A(q)$ statisfies the equation $(1-q)U(1+A(q);D)=U(1;D)$, which, when differentiated with respect to $q$ and evaluated at $q=0$ yields
\begin{equation}
D(\lambda)=\left[1+(\eta-1)\lambda \right]^{-\frac{1}{\eta-1}}
\end{equation}

The dismal theorem states that:
\begin{equation}
\lim_{\lambda \to \infty} E\left[ M \vert \lambda \right]  =+\infty
\end{equation}
Sketch of proof as in \citet{weitzman2009modeling} :

As detailed above there is an affine change of RV $Y=s Z + \mu$ where $\mu$ is a known constant and $s$ is an \textbf{unknown scale parameter}. The pdf of $Z$, $\phi(z)$ has weak regularity conditions: $\Phi(z)>0$ within some neighborhood of $z=0$ and $E\left(e^{-\alpha z} \right)<\infty, \forall \alpha >0$, furthermore $\int_{-\infty}^{\infty} z \Phi(x) dz = 0$ and $\int_{-\infty}^{\infty} z^2 \Phi(x) dz = 1$. The conditional pdf of $y$ is 
\begin{equation}
h(y\vert s)=\frac{1}{s}\Phi\left(\frac{y-\mu}{s} \right)
\end{equation}
Agents have a prior pdf of $S$ which is scale invariant (noninformative reference) where $k$ identifies the strength of prior knowledge
\begin{equation}
p_0(s) \propto s^{-k}
\end{equation}
$n$ observations are available $\tilde{y} = (y_1,\ldots,y_n)$, hence the posterior pdf of $S$ become
\begin{equation}
p_n(s\vert \tilde{y} ) \propto p_0(s) \prod_{j=1}^n h(y_j\vert s)
\end{equation}
the unconditional posterior-predictive pdf of $y$ is
\begin{equation}\label{eq:posteriorpdfy}
f(y) \propto \int_0^\infty \frac{1}{s^{n+k+1}}\Phi\left(\frac{y-\mu}{s} \right)\prod_{j=1}^n \Phi\left(\frac{y_j-\mu}{s} \right) ds
\end{equation}

\citet{schwarz1999decision} theorem 20 shows that, in the limit, the tails of $f(y)$ defined by equation \ref{eq:posteriorpdfy} are power-law of order $n+k$.

\section{Python: insurance, \citet{Kousky:2012wg} replication}\label{sec:kousky}

\citet{Kousky:2012wg} main question:
\begin{itemize}
\item Why do individuals fail to purchase disaster insurance?
\begin{itemize}
\item they show this is not due to lack of understanding, but can be rational even when individual homeowners have perfect information about the risk.
\end{itemize}
\end{itemize}

Catastrophic risks (see \citet{cooke2010limits} for definitions):
\begin{itemize}
\item fat tailed, see section \ref{sec:heavytail};
\item micro-correlations between risks;
\item tail dependence: extreme losses tend to occur together.
\end{itemize}

\citet{cooke2010limits}: \begin{quote}
Technically, upper tail dependence of variables $X$ and $Y$ is defined as the limit (if it exists) of the probability that $X$ exceeds its $100r$-percentile, $x_r$, given that $Y$ exceeds its $100r$-percentile, $y_r$, as $r$ goes to $1$
\end{quote}
\begin{equation}
\mbox{UTD}(X,Y) = \lim P\left(X > x_r | Y > y_r\right) 
\end{equation}
Note that the UTD does not depend on the distributions of $X$ and $Y$ (distributions called the marginals in copula theory).

\subsection{The insurance company model}

A solvency-constrained insurer manages aggregate risk to keep its probability of insolvency below some target level $\lambda$. It offered coverage of policies indexed by $i$, across lines of business indexed by $j$ and region $s$ at a price $q_{js}$. It faces aggregated claims $C_\alpha$ and has access to asset $A$ and reinsurance $K$ at a price $r$, then its expected profit is
$$\sum \sum \sum q_{js} \alpha_{ijs} - rK - E\left[ \mbox{Min} \left( C_\alpha,  \sum \sum \sum q_{js} \alpha_{ijs} + (1-r) K + A\right] \right] $$
and its constraint is
\begin{equation}
Pr\left(C_\alpha >  \sum \sum \sum q_{js} \alpha_{ijs} + (1-r) K + A \right) \leq \lambda
\end{equation}
and they define the required capital or surplus, $S_{\alpha, \lambda}$ as
\begin{equation}
F^{-1}(1-\lambda) = S_{\alpha, \lambda}
\end{equation}
When faced with the characteristic of catastrophic risks, the required surplus will increased as studied below in numerical simulations.

\subsection{Simulating required surplus and the impacts on pricing}

\subsubsection{Micro-correlation}

I simulate\footnote{vansteenberghe\_kousky\_cooke\_2012\_replication.py} from a lognormal distribution portfolios of $N=100$ policies following $\mbox{Lognormal} (\mu ,\,\sigma ^{2})$ with, once normalized, median $\exp(\mu)=1$, mean $\exp(\mu+\frac{\sigma^2}{2})=2$ and by definition a variance of $\exp\left( 2\mu + \sigma^2 \right) \left[ \exp\left( \sigma^2 \right) - 1 \right]$. As the normalized expected mean is $2$, it means that you'll need to divide any simulation results by this $2$ before multiplying by the true expected loss.

The first question that comes to mind: "Is a sum of lognormal rv following a closed form distribution?" \citet{asmussen2008asymptotics} shows that no, but it can be approximated as a lognormal distribution at the right tail. Hence for the following study, I cannot use a closed form approach and perform simulations as in the original paper.

\begin{itemize}
\item For random generation, I use default\_rng(0) so the outcomes are reproducible (seed).
\item As in \citet{Kousky:2012wg}, I generate $N=10^5$ \textbf{independent} flood insurance losses from a lognormal distribution with mean $2$ and median $1$ (consider it to be yearly, this information is only relevant when computing later the expected losses).
\begin{itemize}
\item Based on a lognormal distribution, I have to choose $\mbox{Lognormal} (\mu=0 ,\,\sigma\simeq 1.18)$ to respect the requirement, I will then have a rv with mean $2$ and a standard deviation of $\simeq 3.46$;
\item I can either directly generate the rvs from the lognormal or generate a normal rvs and take its exponential. I check that indeed this is equivalent.
\item In both cases, for $100$ rvs and $10^5$ Monte Carlo runs, we find a sum of rvs with a $99$th percentile of $302.7$ as in the paper. 
\end{itemize}
\item As in \citet{Kousky:2012wg}, I want to generate $N=10^5$ flood insurance losses from a lognormal distribution with mean $2$ and median $1$, but now with a positive correlation of $\rho=0.04$. 
\begin{itemize}
\item I am using numpy, as an illustration, I use the method multivariate\_normal, where as in the documentation  it draws N-dimensional samples, $X=[	x_1,x_2,\ldots]$. If $X \sim \mathcal{N}\left( \mu, \Sigma \right)$ then $Y_i=\exp(X_i)$ are multivariate normal with the mean $E\left[Y\right]_i=\exp\left[ \mu_i + \frac{1}{2}\Sigma_{ii} \right]$, with $\Sigma_{ii}=\sigma^2$. The covariance matrix element $C_{ij}$ is the covariance of $Y_i$ and $Y_j$. The element $C_{ii}$ is the variance of $Y_i$. Its expression is 
$$ C_{ij} = \exp\left( 2\mu + \sigma^2 \right) \left[ \exp\left( \gamma \sigma^2 \right) - 1 \right]$$
$$ C_{ii} = \exp\left( 2\mu + \sigma^2 \right) \left[ \exp\left( \sigma^2 \right) - 1 \right]$$
\item with $\gamma$ the correlation of the normal rvs. As introduced section \ref{sec:multchol}, I start with generating multivariate standard normally distributed rvs $Z$ with correlation $\gamma$, the question is how to calibrate this coefficient to have the log-normal multivariate variables to be correlated with $\rho=0.04$. We want to have for the log-normal rvs : $C_{ij} = \rho C_{ii}$, so:
$$ \gamma = \frac{1}{\sigma^2} \log \left( 1 + \rho \left[ \exp\left(\sigma^2\right)-1\right] \right)$$
\item I test another approach with the Cholesky decomposition. I want the covariance matrix for the lognormal rvs $C$ and apply an LDL decomposition so I have $LDL' = C$, I avoid the use of NORTA (cf section \ref{sec:NORTA}) and circumvent the issue by generating the rvs $Z$ such that I get $Var(Z) = D_{ii} = e^{\sigma^2}\left(e^{\sigma^2} - 1 \right)$, that is I need to find $\sigma$ knowing that $\mu=0$. Writing $H=e^{\sigma^2}$, I have $H^2-H-D_{ii}=0$, that is one root is $$\sigma = \sqrt{\log\left( \frac{1 + \sqrt{1 + 4 D_{ii}}}{2} \right)}$$
\end{itemize}
\end{itemize}
\begin{itemize}
\item For a 100 claim portfolio I find a sum of rvs with micro-correlation of 442 (close enough to the 434 in the paper).
\item For a 200 claim portfolio I find a sum of rvs for iid of 537 (534 in the paper) and with micro-correlation of 855 (close enough to the 840 in the paper).
\end{itemize}
I find similar figures as the paper Figure 1a and b:
\begin{center}
\includegraphics[scale=.5]{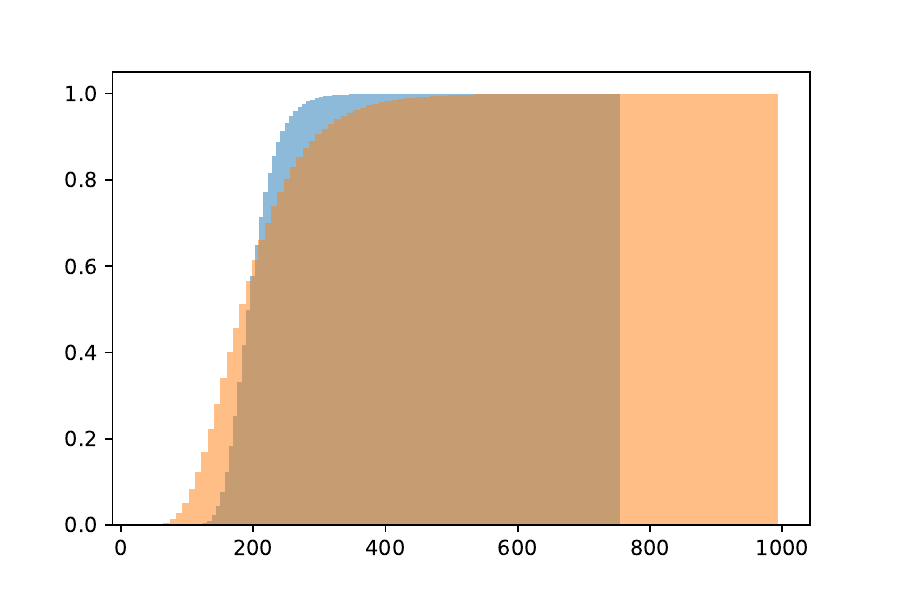}
\includegraphics[scale=.5]{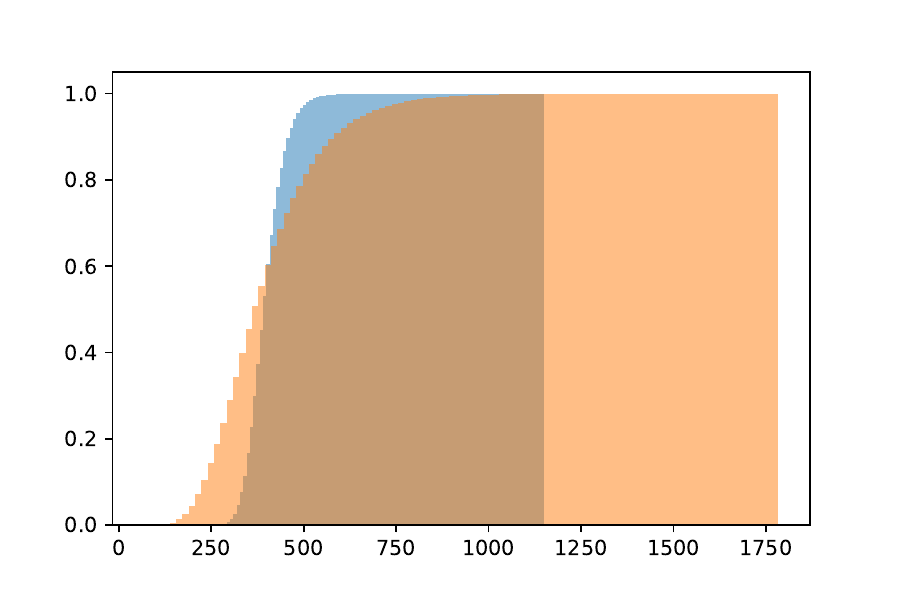}
\end{center}

\begin{itemize}
\item To follow the example of how quickly tiny, positive correlations policies can become pernicious, I use\footnote{vansteenberghe\_kousky\_cooke\_2012\_Binomial\_Section.py} a Bernoulli distribution for a rv, with $p=.01$. If I have $Z=[Z_1,Z_2,\ldots]$ with $Z_i \sim \mathcal{B}\left( p_i \right)$, then the variance-covariance matrix of $Z$ is a diagonal matrix $D$ with its diagonal element $D_{ii}=Var(Z)=p_i(1-p_i)$. I want to find a LDL decomposition so that $L D L^T = \Sigma$, with $\Sigma$ the desired variance-covariance matrix $\Sigma_{ii}=\sigma^2$ and $\Sigma_{ij}=\rho \sigma^2$. Once I generated this $Z$ rv, I can get my desired rv $X = L Z$. I extract from the LDL decomposition $D_{ii}=\sigma_i^2$, then I have a quadratic equation $p_i^2-p_i+\sigma_i^2=0$ and I chose the root $\frac{1\pm\sqrt{1-4\sigma_i^2}}{2}$ (I take + here). I take a simpler approach to ease the required computing power: a portfolio of 100 with $10^5$ runs
\end{itemize}
I find in the same spirit as Figure 2, expected loss multiplier as a function of micro-correlation:
\begin{center}
\includegraphics[scale=.5]{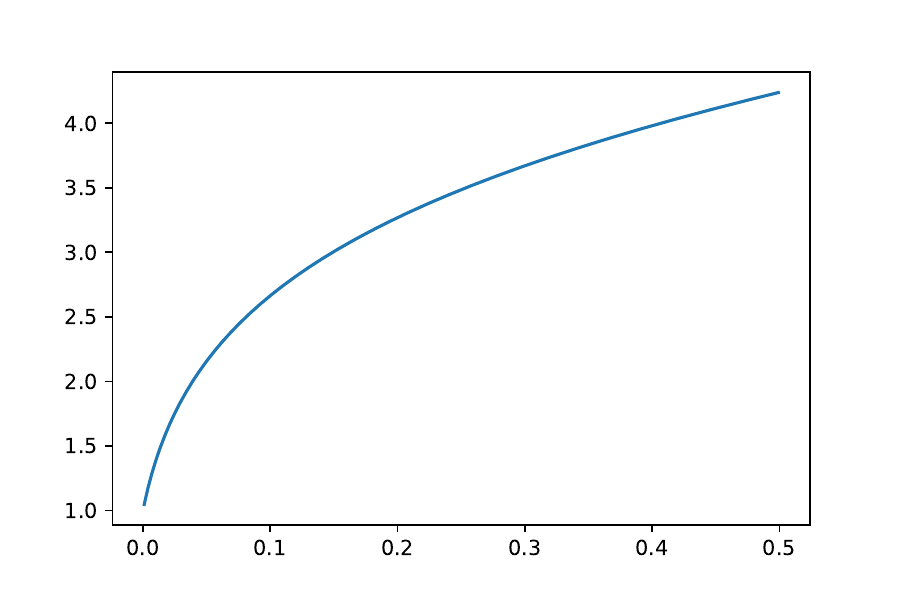}
\end{center}

\subsubsection{Tail dependence}

I introduce\footnote{vansteenberghe\_extreme\_copulas\_estimation.R} a Gumbel copula (for an introduction to copulas, cf section \ref{sec:Rcopula}) to study the tail dependence. Here the original paper cannot be exactly reproduced as the parameter of the Gumbel is not explicit so I take $\theta = 2$. The marginals are standard normal and I don't introduce the micro-correlation to really focus on the effect of the copula. I generate $10^5$ portfolios of size $100$ and $200$. Exactly as in the paper, I find that having a portfolio of greater size doesn't reduce the required price multiplier (no pooling benefit).

\subsubsection{Fat-tailed loss distribution}

Next we move\footnote{vansteenberghe\_kousky\_cooke\_2012\_replication.py} to the Pareto distribution, with a tail index $\alpha=2$ and a threshold set up so that the mean $\frac{\alpha x_m}{\alpha-1}=1$, so $x_m=.5$ and we write $P(x;\alpha=2, x_m=0.5)$. The fair price multiplier for independent policies of size $100$ and $200$ is respectively $1.52$ and $1.38$ (against respectively 1.77 and 1.49 in the paper).

To introduce micro-correlation, we need a variance-covariance matrix, which is not possible to explicit in the case $\alpha=2$ as then the variance of the rv is infinite. So I use an $\alpha=2.01$ and find that with the micro-correlation the benefit of aggregation still accrue.

I then introduce a Gumbel copula with $\theta=2$ (no micro-correlation) and for simplicity focus on a portfolio with one single policy versus a portfolio with 2 policies linked by a Gumbel copula. The multipliers difference is similar as in the paper, increasing with the portfolio size (as it doubles, the multiplier more than doubles).

\begin{center}
This is the key paper's result:\\ \textbf{with fat-tailed distribution, tail dependence makes aggregation worse for the insurer}.
\end{center}

\subsection{Impact of higher multiplier on demand}

\begin{itemize}
\item $i\in [1,N]$ policyholders
\item $L_i$ potential loss
\item $p$ loss probability
\item $q$ insurer coverage rate 
\item $x$ the fair price multiplier, such that $q = px$
\item $w_i$ individual wealth
\item $\alpha_i$ coverage purchased
\end{itemize}

Policyholders expected utility:
\begin{equation}
EU=(1-p)u(w_i-\alpha_i q) + p u(w_i-\alpha_i q -L_i+\alpha_i)
\end{equation}
the FOC yield
\begin{equation}
-q(1-p)u'(w_i-\alpha_i q)+p(1-q)u'(w_i-\alpha_i q -L_i+\alpha^*_i)=0
\end{equation}
For insurance priced actuarially fairly: $p=q$ and a risk-averse policyholder will fully insure: $\alpha^*_i=L_i$.

To get an estimate of the impact of the price of catastrophe insurance on  demand, assume policyholder's utility follows the common constant relative risk aversion (RRA) function, with $\eta$ the coefficient of RRA:
\begin{equation}
U = \begin{cases}
\frac{w^{1-\eta}}{1-\eta} \quad  \mbox{if} \quad \eta \neq 1\\
\ln(w) \quad \mbox{if} \eta=1
\end{cases}
\end{equation}

Policyholder faces two constraints:
\begin{itemize}
\item utility with insurance should be greater than without
\begin{equation}
U(w-Lq)\geq (1-p)U(w)+pU(w-L)
\end{equation}
\item budget constraint, letting $q=x^* p$, $w=1$ and $\eta=1$:
\begin{equation}
\ln(1-Lxp)\geq(1-p)\ln(1)+p\ln(1-L)
\end{equation}
\end{itemize}
which yields
\begin{itemize}
\item utility with insurance should be greater than without
\begin{equation}
x \leq \frac{1-L}{Lp}
\end{equation}
\item budget constraint, letting $q=x^* p$, $w=1$ and $\eta=1$:
\begin{equation}
x \leq \frac{1-\left(1-L \right)^p}{Lp}
\end{equation}
\end{itemize}

I replicate\footnote{vansteenberghe\_kousky\_replication\_demand.py} Figure 5
\begin{center}
\includegraphics[scale=.5]{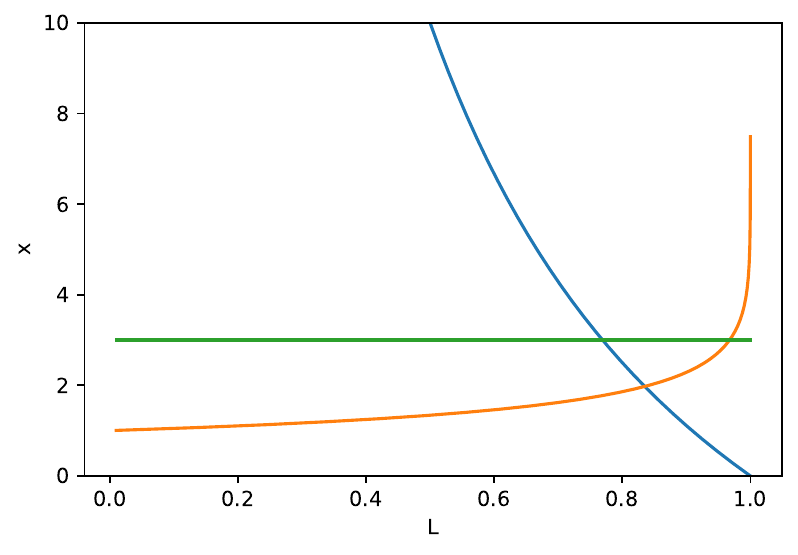}
\includegraphics[scale=.5]{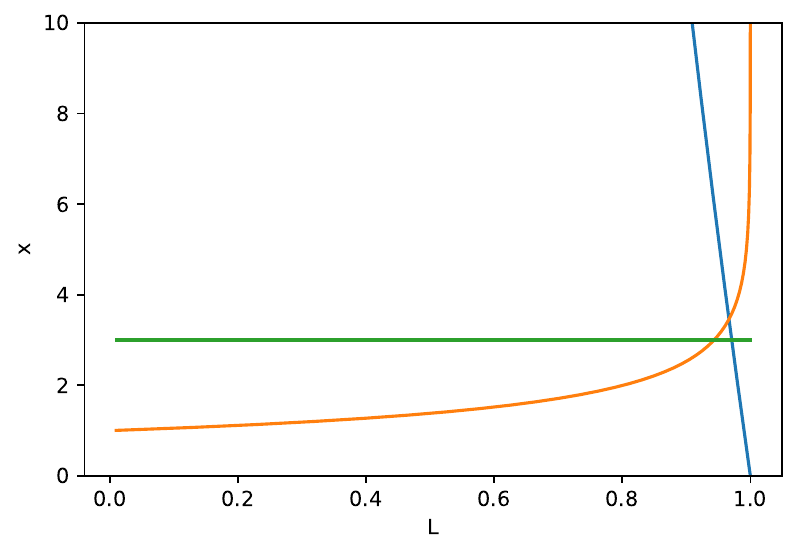}
\end{center}

I replicate Figure 7 showing the constraint as a function of $p$ and Figure 8 showing the optimal coverage $\alpha$ as a function of $x$ and $L$:
\begin{center}
\includegraphics[scale=.5]{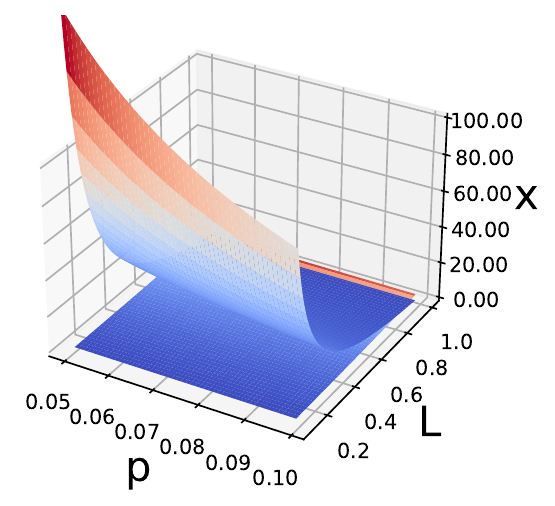}
\includegraphics[scale=.5]{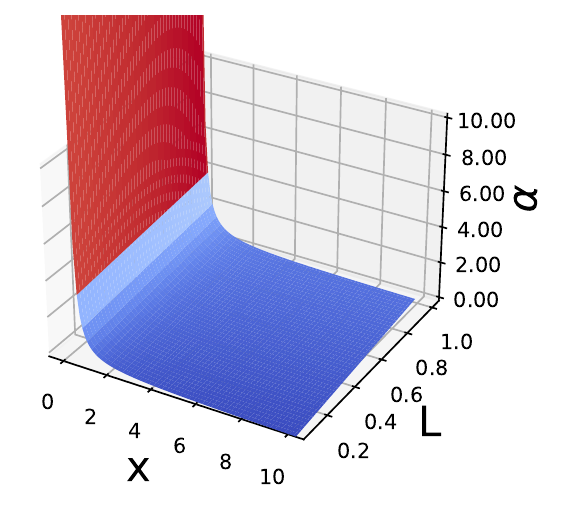}
\end{center}

\section{Quantile regression: an introduction}\label{sec:quantilereg}

So far we applied standard linear regressions of the type:
$$Y_t = \alpha + \beta X_t + \epsilon_t$$
to model conditional expectation of the dependent variable $Y_t$ given $X_t$, which writes $E\left(Y \vert X\right)$ for this to be identifiable (estimate $\beta$), we needed to make necessary assumptions for the consistency of the OLS estimator: $E\left( \epsilon_t X_t \right) = 0$.

If we assume that $X$ and $Y$ have a bivariate normal distribution, then a standard linear regression provides a model of the mean and variance of the dependent variable from which we can infer the whole conditional distribution of the dependent variable. If as in most cases $X$ and $Y$ have arbitrary joint distribution, a quantile regression can help complete the picture of the conditional distribution of the dependent variable. 

If as in \citep{Firpo2009aa} we want to estimate the direct effect on the log wage $Y$ of increasing the proportion of unionized workers, $p = Pr[X=1]$, where $X=1$ if the worker is unionized and $X=0$ otherwise. We can run the specification: $Y=\alpha + \beta X + \epsilon$ and $\beta$ is a measure of the impact of increasing the proportion of unionized workers on the mean wage $\frac{\partial \mu(p)}{\partial p}$ with $\mu(p)$ the mean wage which is a function of the proportion of unionized workers.

As in \citep{Firpo2009aa}, we use a large sample of 266,956 observations of U.S. males from the 1983-1985 Outgoing Rotation Group supplement of the Current Population Survey. The OLS estimator is the impact on the conditional mean $\beta = E(Y\vert X=1) - E(Y\vert X=0) = 0.2449$ which is the difference between $\mu(1)$ and $\mu(0)$ a visually shown below:
\begin{center}
\includegraphics[scale=0.5]{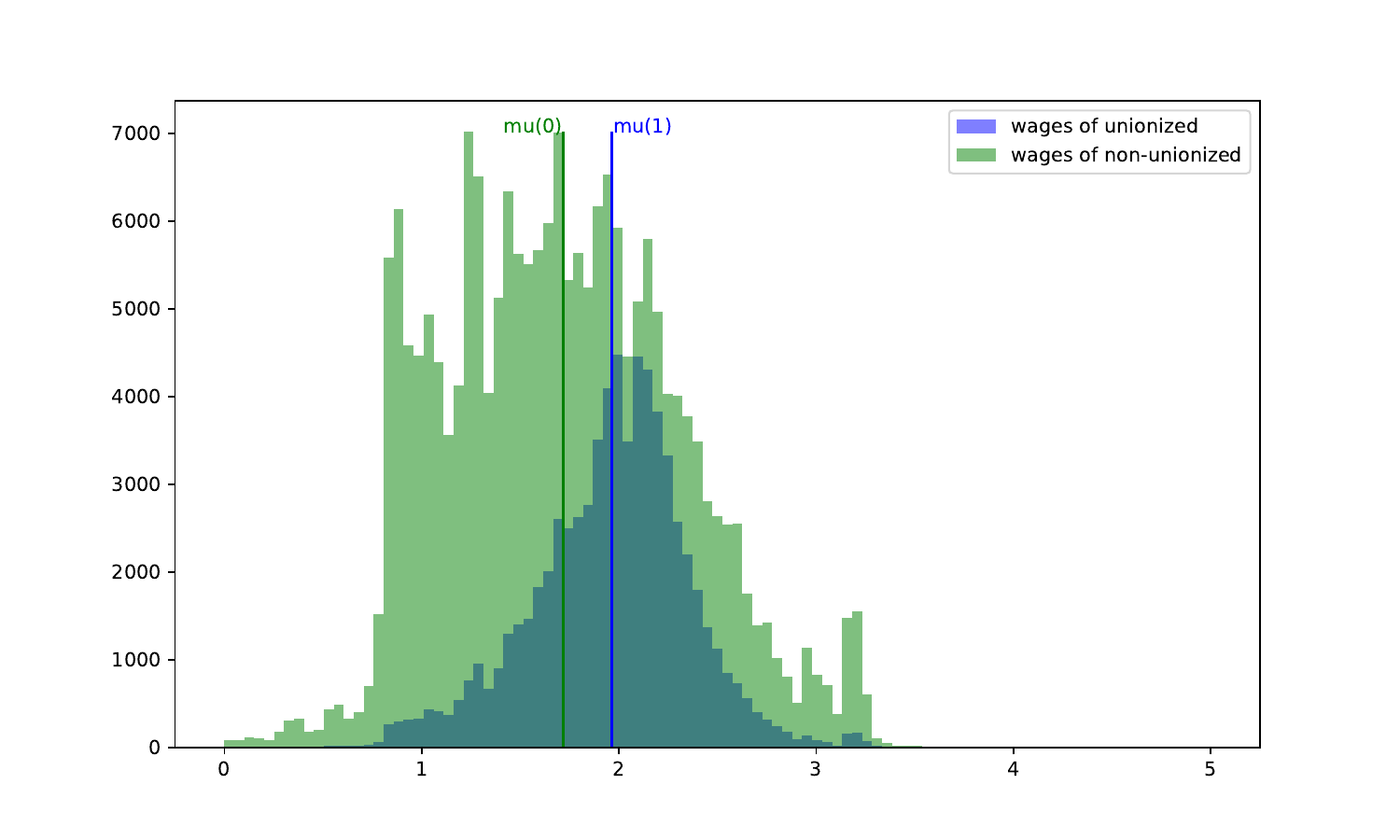}
\end{center}
We can have some scepticism as for the distribution of the log of wages are expected to be normally distributed, our sample seems to be well balanced for unionized workers (but clearly unbalanced with fewer unionized observations), but for non-unionized workers the sample of log wages is not distributed as expected, so we should treat the results with caution.

The unconditional mean $\mu$ is the weighed average of the conditional means: $ p E(Y\vert X=1) + (1-p) E(Y\vert X=0)$, deriving we have: $\frac{\partial \mu(p)}{\partial p} = E(Y\vert X=1) - E(Y\vert X=0) = \beta$

\subsection{Quantile regressions}

We might want to dig further in the relationship at different points in the conditional distribution of $Y$: the conditional quantile function $Q_q(Y\vert X)$. We have formally define a quantile in section \ref{sec:qqplot}, in words: a $q$ quantile is a value $x_q$ that cuts the distribution into two parts and the area under the density to the left of $x_q$ is $q$, $P(X\leq x_q)=q$, with $X$ a random variable.

We might want to estimate a line that rather than passing through the mean of the sample (OLS), divide the sample into two unequal parts (a given quantile).

If we want to regress two random variables $X$ and $Y$ that are not bivariate normal, quantile regression methods enable one to explore
potential effects on the shape of the distribution.

OLS minimizes $\sum_t \epsilon_t^2$ while (conditional) quantile regression minimizes a sum that gives asymmetric penalties $(1 - q)\vert \epsilon_t \vert$ for overprediction and $q \vert \epsilon_t \vert$ for underprediction. It can be demonstrated that the quantile regression estimator is asymptotically normally distributed.

Median regression is more robust to outliers than least squares regression, and is semiparametric as it avoids assumptions about the parametric distribution of the error process.

If we write as in \citep{alexanderII} the $q$ quantile of the errors as $F_\epsilon^{-1}(q)$, then a conditional quantile regression is the relation between the conditional $q$ quantile of the dependent variable which is found from the inverse of $F(Y\vert X)$:
\begin{equation}
F^{-1}(q \vert X) = \alpha + \beta(q) X + F_\epsilon^{-1}(q)
\end{equation}
an alternative notation: $Q_Y(q \vert X ) \simeq X' \beta(q)$ where $Q_Y(q \vert X )$ denotes the $q^{\mbox{th}}$ conditional quantile of $Y$ assumed to be linearly dependent on $X$, a vector containing the conditioning variables, including a constant. The dependence is given by $\beta(q)$ and we can consider it \textit{unconditional} if no exogenous control variables are added to the model, conditional otherwise (we will see later on how to consider the latter unconditional).

The coefficient estimate can be done via the programme over all observations $i$:
\begin{equation}\label{eq:programmequantreg}
\hat{\beta}(q) = \arg \min_\beta \sum_{i=1}^n \rho_q\left(Y_i - X_i' \beta(q)\right)
\end{equation}
The check function $\rho_q(u)$ is defined as:
\[
\rho_q(u) = 
\begin{cases} 
q \cdot u & \text{if } \vert u \vert \geq 0 \\
(1 - q) \cdot \vert u \vert & \text{if } u < 0
\end{cases}
\]

The conditional quantile regression estimator is unbiased and consistent only under \textit{zero conditional quantile assumption} as \citep{Wenz2019aa} puts it, and write: $Q_q\left( \epsilon_t \vert X_t\right) = 0$ which is the quantile $q$ of $\epsilon$ conditional on $X$.

Following \citet{koenker1999goodness} and \citet{geraci2007quantile}, a possible parametric link between the minimization of the sum of the absolute deviates in equation \ref{eq:programmequantreg} and maximum likelihood theory is given by the Asymmetric Laplace Distribution (ALD). We say that a random variable $Y$ is distributed as an ALD with parameters $\mu$, $\sigma$, and $\tau$, and we write it as $Y \sim \text{ALD}(\mu,\sigma,\tau)$, if the corresponding probability density function is:

\begin{equation}
f(y \vert \mu, \sigma, \tau) = \frac{\tau(1 - \tau)}{\sigma}  \exp \left(- \rho_\tau \left( \frac{y - \mu}{\sigma} \right) \right), \quad \text{for} \; 0 < \tau < 1, \; \sigma > 0, \; \mu \in \mathbb{R},
\end{equation}

where the check function $\rho_\tau(v)$ is given by:

\begin{equation}
\rho_\tau(v) = v\left( \tau - \mathbb{1}(v \leq 0) \right),
\end{equation}

and $\mathbb{1}( \cdot )$ is the indicator function. The parameter $\tau$ represents the skewness (with $0 < \tau < 1$), $\sigma$ is the scale parameter, and $\mu$ is the location parameter. The distribution assigns weight $\tau$ to observations greater than $\mu$, and weight $1 - \tau$ to those less than $\mu$, and the probability that $y$ is less than $\mu$ is given by $\Pr(y \leq \mu) = \tau$. Thus, the distribution divides into two parts: one with probability $\tau$ to the left of $\mu$ and one with probability $1 - \tau$ to the right.

This ALD framework provides a unifying bridge between the minimization problem in quantile regression and the maximum likelihood estimation. Specifically, if we assume that $y_i \sim \text{ALD}(\mu_i, \sigma, \tau)$ with $\mu_i = X_i^\top \beta$, the likelihood for $N$ independent observations is, up to a constant of proportionality:

\begin{equation}
L(\beta, \sigma; y, \tau) \propto \sigma^{-1} \exp \left( - \frac{1}{\sigma} \sum_{i=1}^{N} \rho_\tau \left( \frac{y_i - X_i^\top \beta}{\sigma} \right) \right).
\end{equation}

If we treat $\sigma$ as a nuisance parameter, then the maximization of this likelihood with respect to $\beta$ is equivalent to the minimization of the objective function in equation \ref{eq:programmequantreg} under the ALD assumption. Therefore, the ALD proves useful as a probabilistic foundation for quantile regression estimation. 

\citet{koenker1999goodness} introduced a goodness-of-fit test for quantile regression and related inference processes under this parametric assumption of an Asymmetric Laplace Distribution (ALD) for the error term. Moreover, they also considered the likelihood ratio statistic for quantile regression inference.

\subsection{How not to interpret a conditional quantile regression}
With a standard linear regression (OLS), we modelled:
$$E(Y \vert X=x) = \beta x$$ thus we can interpret the coefficient $\beta$ as $$\frac{\partial E(Y \vert X=x)}{\partial x} = \beta$$

We should not, however, use the same philosophy to interpret:
$$\frac{\partial Q_Y(\tau \vert X=x)}{\partial x}$$ as a particular subject who happens to fall at the $\tau^{\mbox{th}}$ quantile initially, and then receives an increment $\Delta x$, say, another year of education, will not necessarily fall again on the $\tau^{\mbox{th}}$ conditional quantile function following the increment.

\subsection{How not to envisage a quantile regression}\label{sec:sovbonds}

As noted in \citep{Koenker2001}, \begin{quote}
We have occasionally encountered the faulty notion that something like quantile regression could be achieved by segmenting the response variable into subsets according to its unconditional distribution and then doing least squares fitting on these subsets.
\end{quote} 

We illustrate\footnote{20200209\_not\_quantile\_regression.py} this with monthly sovereign bonds returns in Europe from January 1993 to November 2016:
\begin{center}
\includegraphics[scale=0.5]{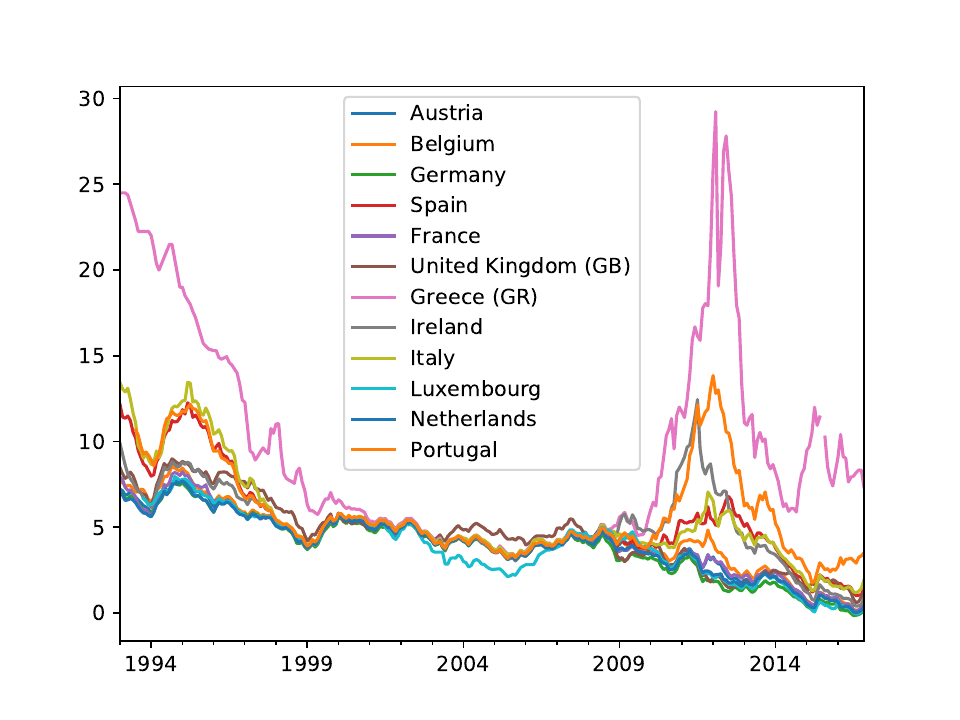}
\end{center}

We compute the monthly returns before applying a regression. An erroneous approach to quantile regression would be to truncate the return of the independent (or dependent) variable according to a quantile (here we chose $q=.1$):
\begin{center}
\includegraphics[scale=0.5]{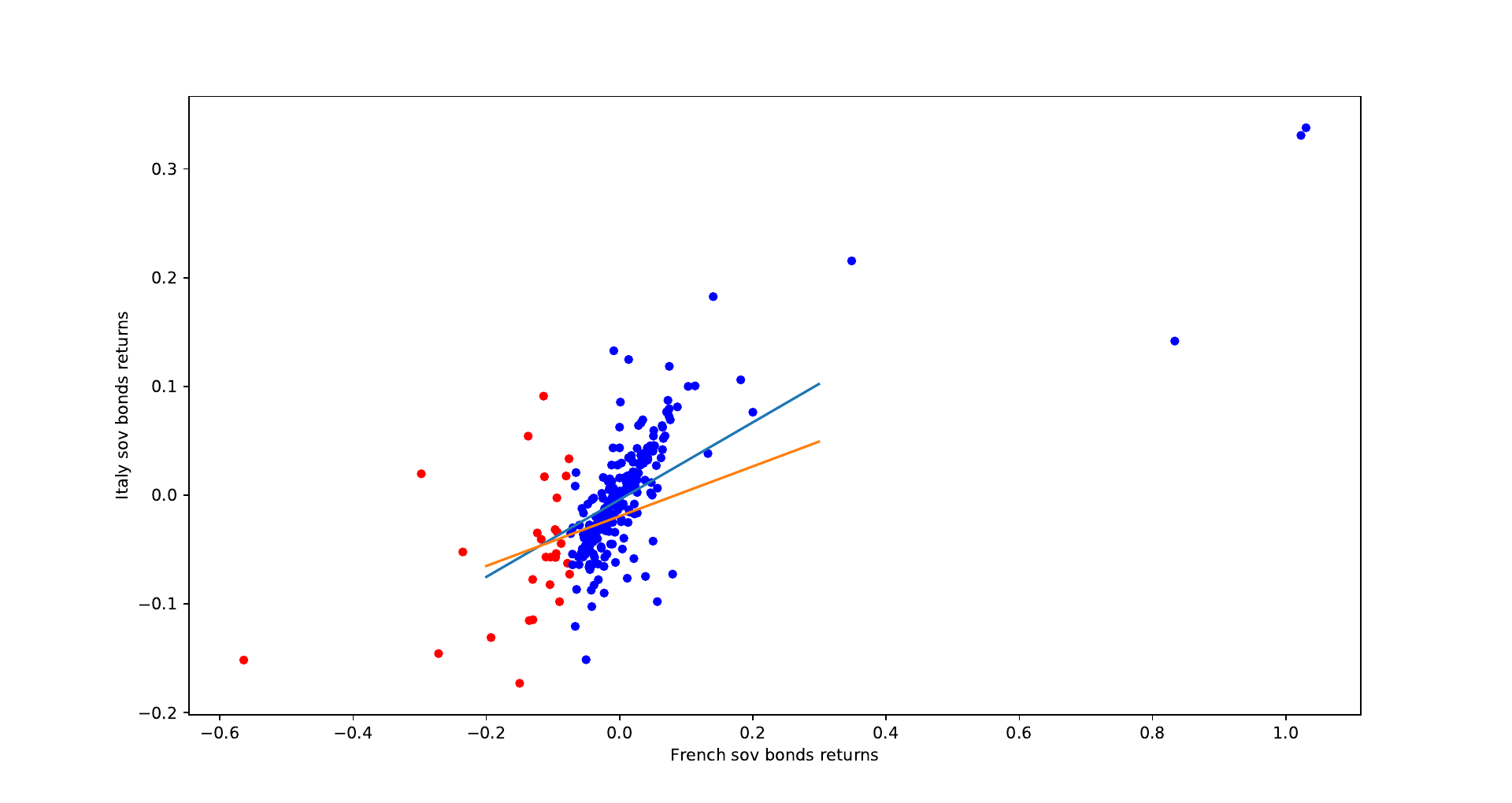}
\includegraphics[scale=0.5]{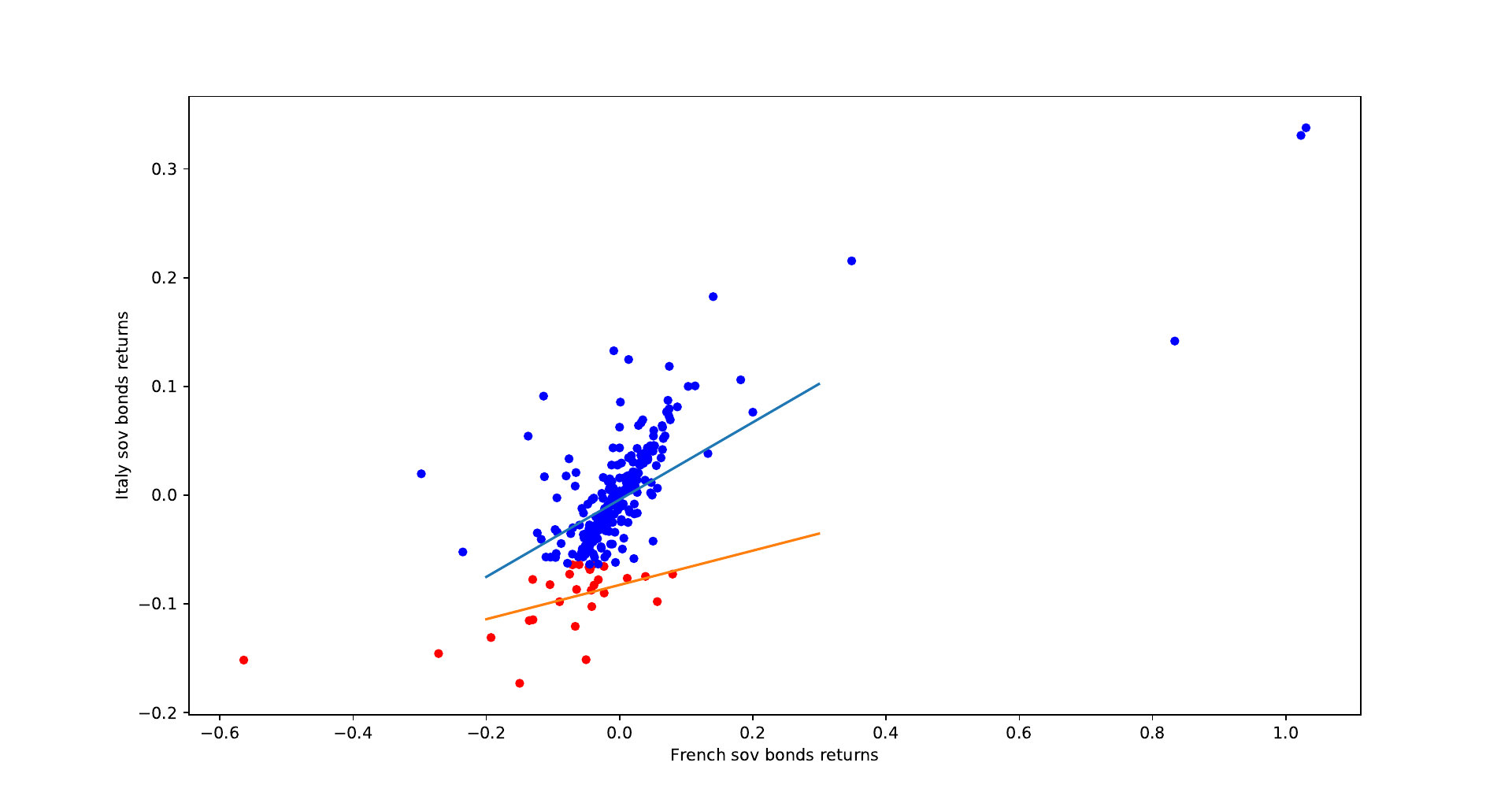}
\end{center}

We implement a proper quantile regression that gives asymmetric penalties but indeed uses the full set of observations:
\begin{center}
\includegraphics[scale=0.5]{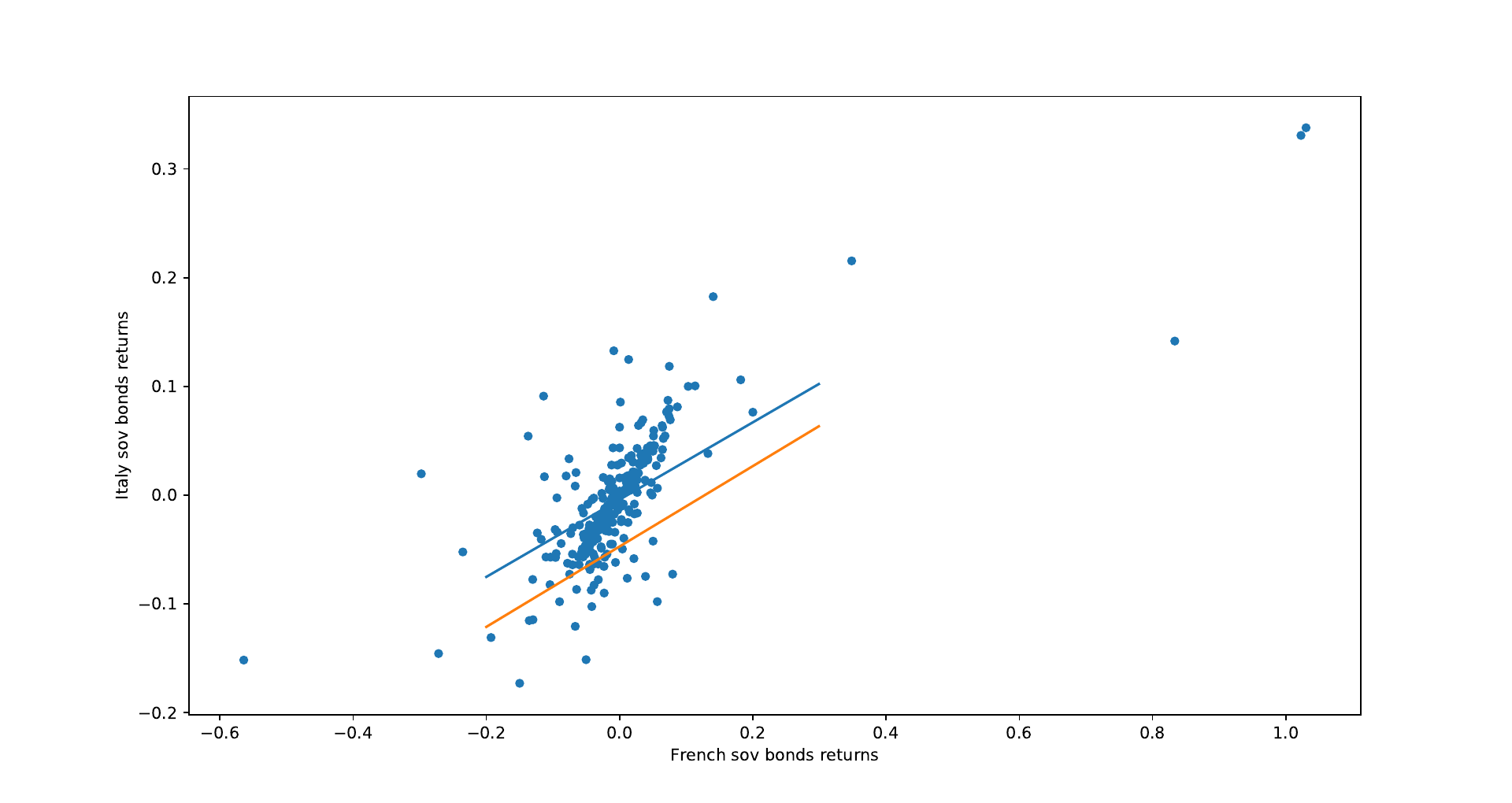}
\end{center}

\subsection{Conditional Quantile Regression: Relationship between household's expenditure and income}

We follow\footnote{20200208\_quantile\_regression\_Engel1857.py} Ernest Engel's (1857) analysis of the relationship between household food expenditure and household income for 235 European working-class households. The higher the income a family receives, the more dispersion we find in the food expenditure, we could implement an OLS and a Least Absolute Deviation (median regression) pictured dashed line:
\begin{center}
\includegraphics[scale=0.5]{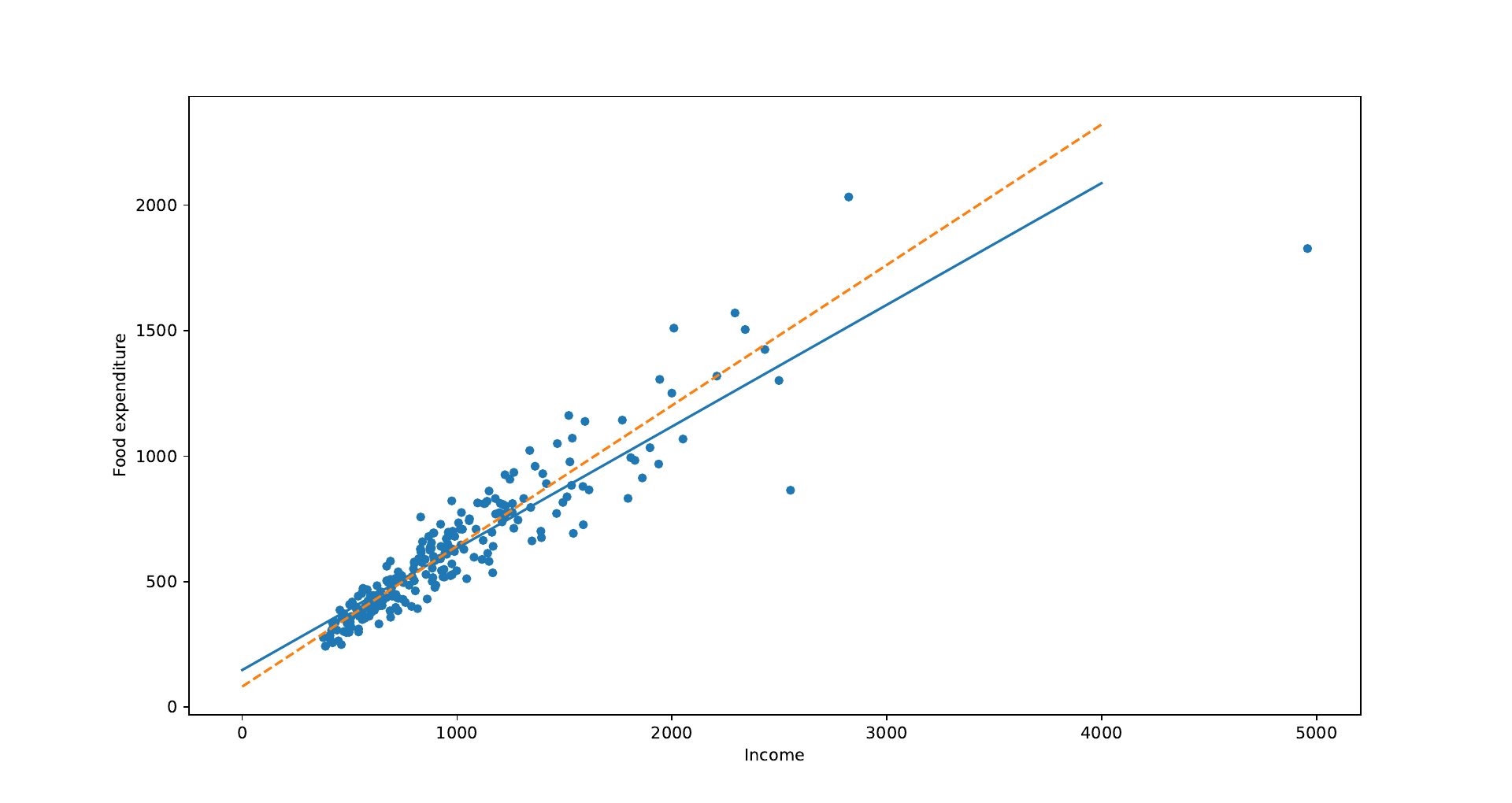}
\end{center}
as variance of expenditure seems to increase with income, we should test for heteroscedasticity and question the robustness of our estimators. 

We can note that the OLS is "polluted" by extreme values on the right hand side and hence loses precision for low income observations. The median is more robust to this issue.

We implement a quantile regression:

\begin{center}
\includegraphics[scale=0.5]{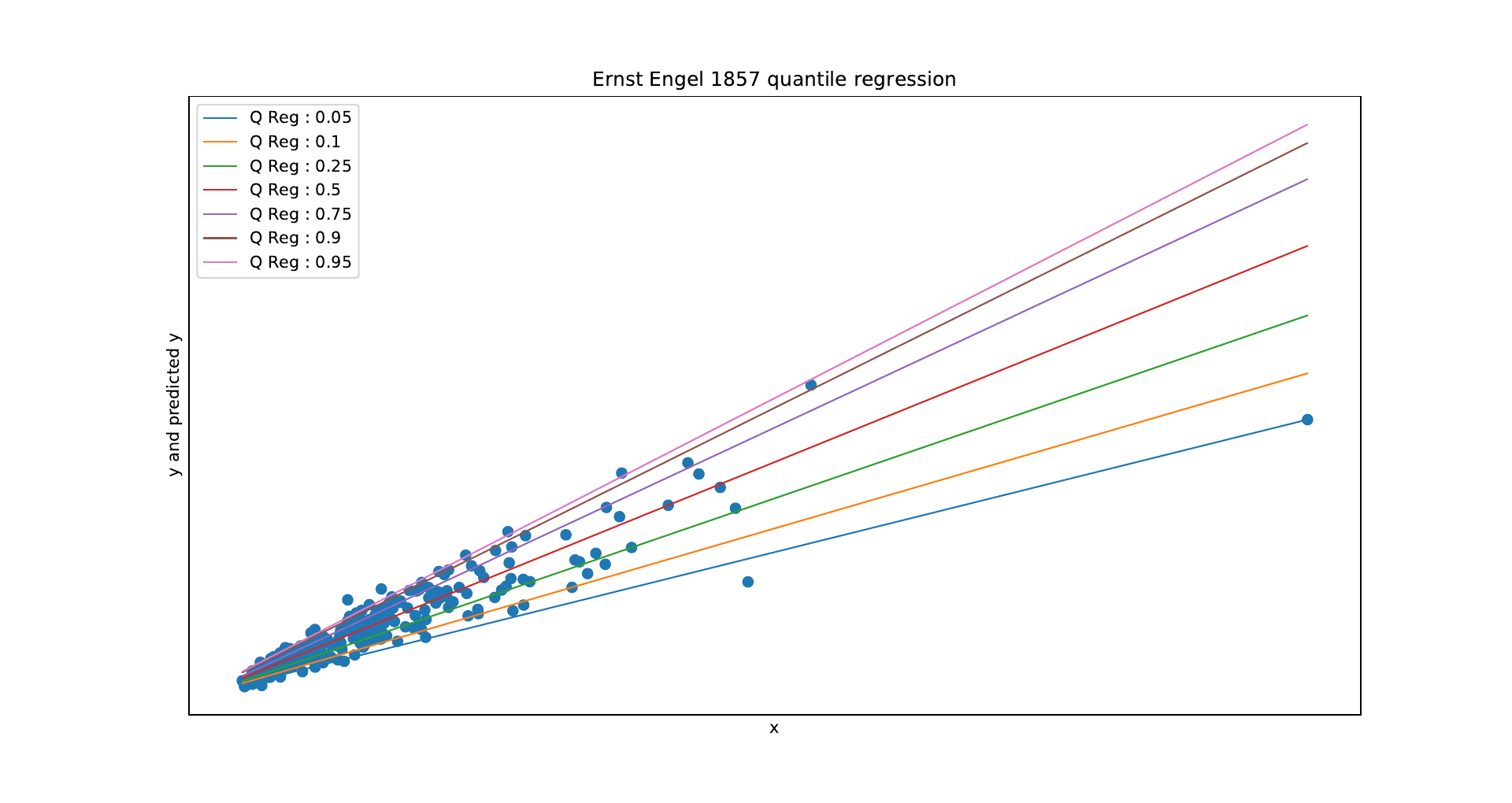}
\end{center}

as depicted in \citep{Koenker2001} \begin{quote}
The spacing of the quantile regression lines also reveals that the conditional distribution of food expenditure is skewed to the left: the narrower spacing of the upper quantiles indicating high density and a short upper tail and the wider spacing of the lower quantiles indicating a lower density and longer lower tail.
\end{quote}

We can plot the estimators for each quantile and most of the quantile slope estimates looks different from the least squares estimate\footnote{It might not be trivial to test for statistical difference, we would need a bootstrap method to test for significance (see the work of Domenico Giannone).}
\begin{center}
\includegraphics[scale=0.5]{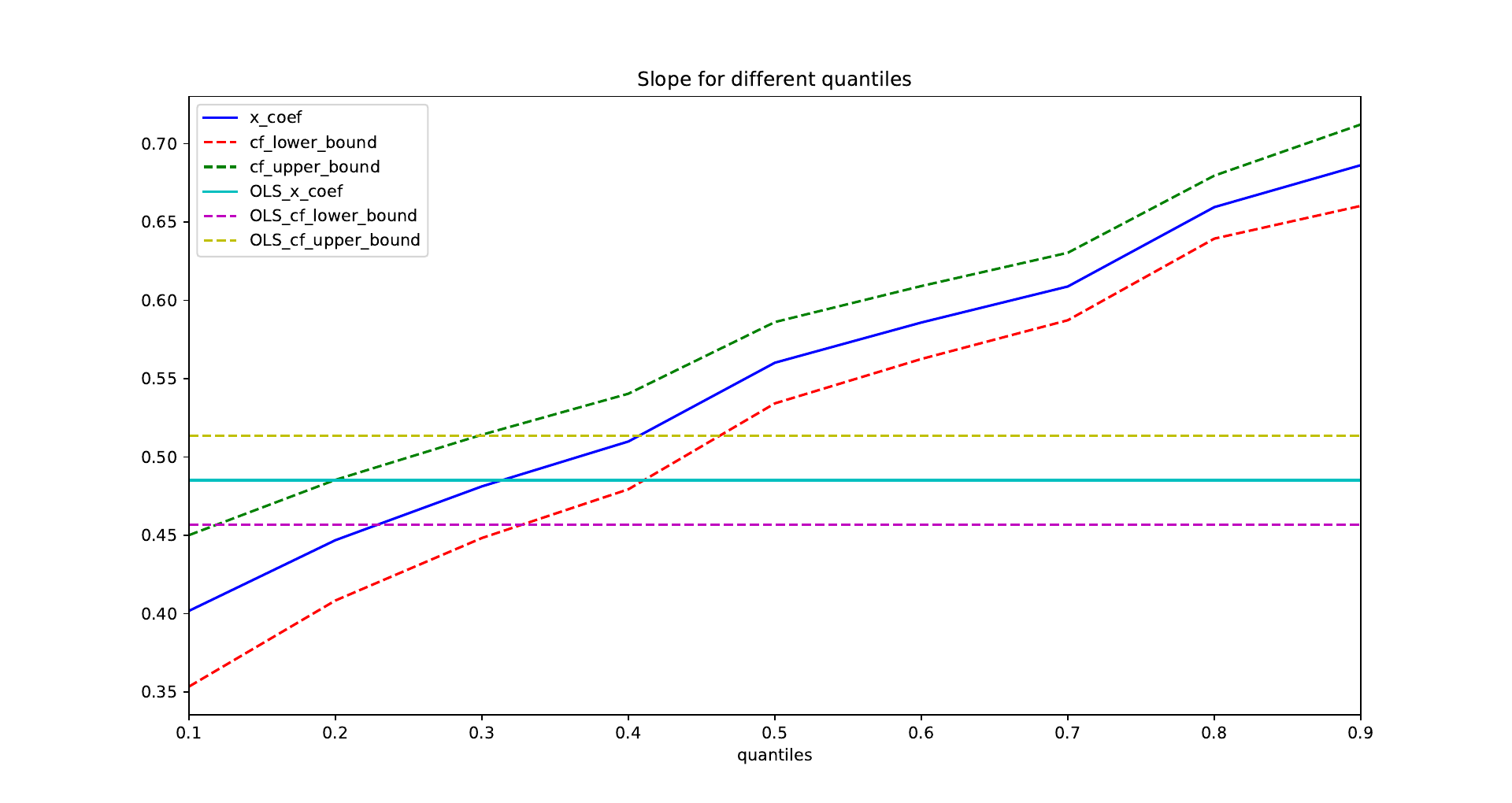}
\end{center}

\subsection{Conditional Quantile Regression: FTSE and VFTSE}

To  further our introduction to quantile regression\footnote{20200209\_quantile\_regression\_Alexander.py and 20190621\_copula\_quantile\_regression.R}, following the book \citep{alexanderII} section II.7.3.

We import daily VFTSE 30-day volatility index and daily FTSE 100 index:
\begin{center}
\includegraphics[scale=0.5]{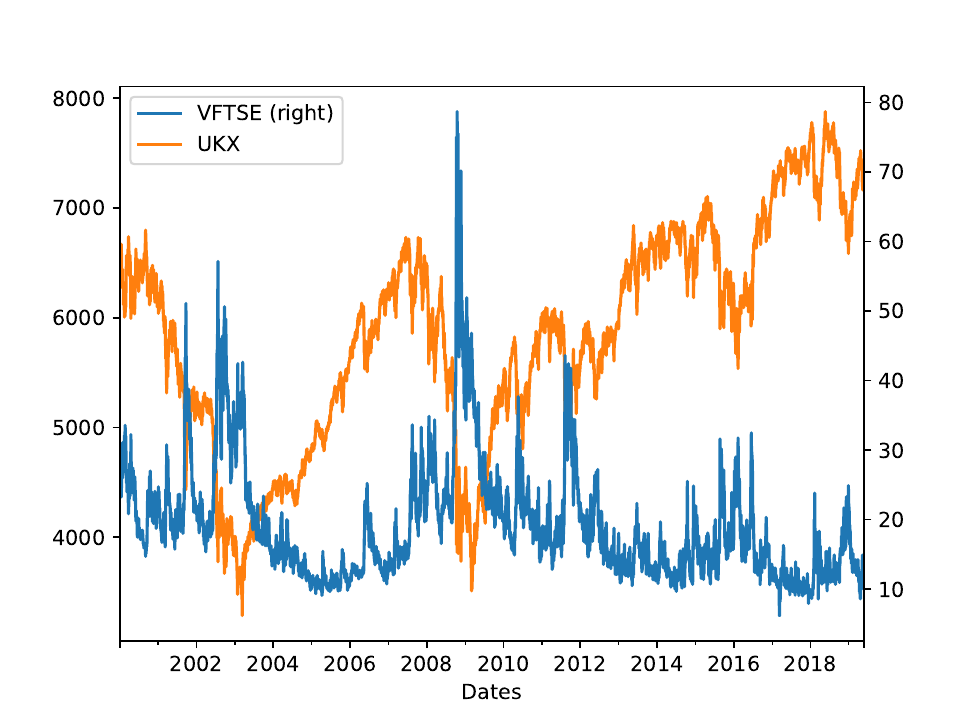}
\end{center}

We test for unit root and cointegration relationship before apply an OLS on the returns, limiting our sample as in \citep{alexanderII} to have similar results:
\begin{center}
\includegraphics[scale=0.5]{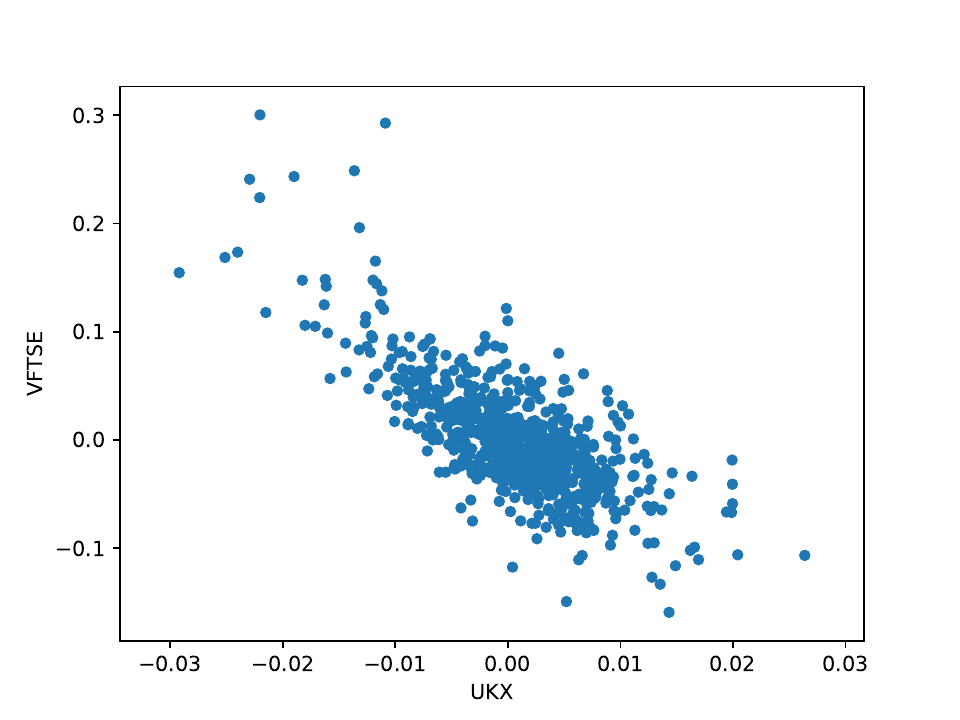}
\end{center}

Close to the book, we find:
$$R_{\mbox{VFTSE}} = 0.0038 - 5.9 R_{\mbox{FTSE}}$$ with a $R^2=55.4\%$. 

But a standard linear regression cannot be trusted as:
\begin{itemize}
\item Price and implied volatility returns certainly do not have a bivariate normal distribution
\item It cannot capture asymmetric dependence between price and implied volatility, i.e. that volatility increases much more following a large price fall than it decreases following a large price rise.
\item It cannot capture tail dependence, i.e. that volatility reacts strongly to extreme moves in price but changes very little when price moves are within the ordinary daily variations seen in the market.
\end{itemize}

And the quantile regression results are comparable:

\begin{tabular}{lllll}
quantile &       $\alpha$ & t-stat $\alpha$ &     $\beta$ & t-stat $\beta$ \\
0.1 &  -0.0370409 &     -21.3649 & -5.93893 &    -26.5641 \\
0.2 &  -0.0232976 &     -14.7147 & -5.71407 &    -28.9744 \\
0.3 &  -0.0132768 &     -9.12366 & -5.74898 &    -29.2566 \\
0.4 & -0.00562426 &     -4.05034 & -5.73885 &    -29.0187 \\
0.5 &  0.00119825 &     0.902543 & -5.64621 &    -28.3278 \\
0.6 &  0.00845477 &      6.15903 & -5.60836 &    -25.5528 \\
0.7 &   0.0166459 &      11.1161 & -5.54149 &    -21.5502 \\
0.8 &   0.0268208 &      14.6944 & -5.69384 &    -16.0327 \\
0.9 &   0.0464823 &      16.8802 & -5.66432 &    -9.63975 \\
\end{tabular}

Although we expected the returns to be affected with higher VFTSE dispersion when the FTSE return are lower, the visual is not striking:
\begin{center}
\includegraphics[scale=0.5]{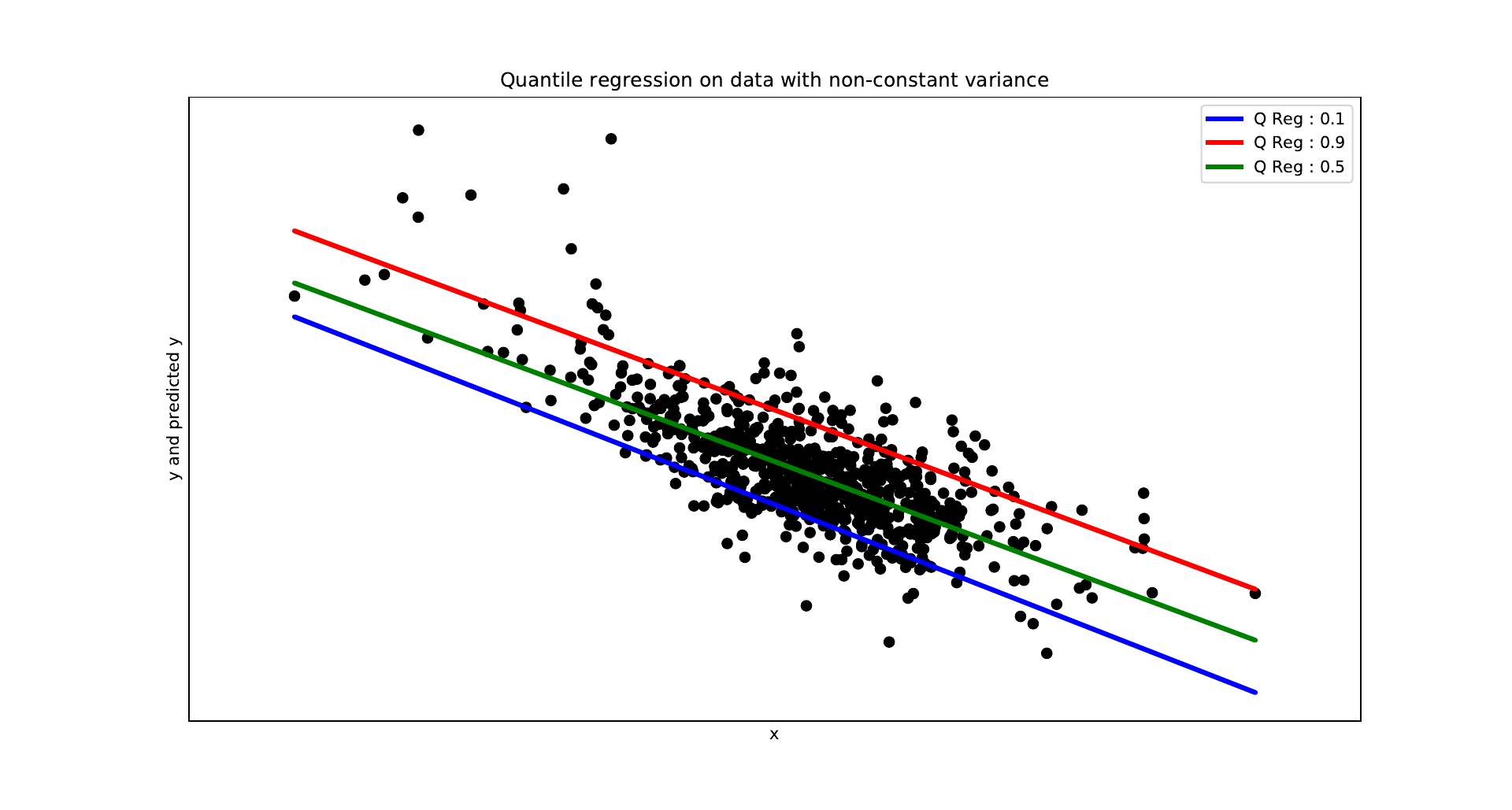}
\end{center}

\subsection{Conditional Quantile Regressions and copulas}

\citep{Baur2013aa} studies the structure and degree of dependence between pairs of random variables as empirical data sets show that dependence of assets or markets is non-linear and asymmetric.

Copulas (we introduce section \ref{sec:Rcopula}) facilitates the separation of the degree and structure of dependence. Empiricists have to chose copulas. \citep{Baur2013aa} proposes an alternative to the copula framework based on a quantile regression model, that has no prior on the structure of dependence: the structure of dependence is defined by the vector of conditional (or unconditional) quantile estimates $\hat{\beta} = \left[ \hat{\beta}(\tau=1) \ldots \hat{\beta}(\tau=99)\right]'$ and the degree of dependence is given by the average estimate over all quantile denoted by $\bar{\hat{\beta}}$.

For major equity indexes they can test for contagion in the degree and the structure of dependence and they conclude:
\begin{quote}
equity index returns during the global financial crisis in 2007 and 2008 relative to a pre-crisis period. [\ldots] The findings illustrate that contagion is a phenomenon that leads to changes both in the degree and the structure of dependence including changes in the lower and the upper tails of the distribution. The estimation results also show that a positive change in the degree of dependence can incorrectly indicate contagion if changes in the structure and in particular in the tails are neglected. It is found that many emerging markets display increased dependence and co-movement in the upper quantiles during the financial crisis. This would indicate contagion in a framework which neglects differences in the tails, for example if the contagion test is based on the correlation coefficient. The quantile regression results instead show that contagion occurred in "good" states and thus does not adversely affect the benefits of diversification
\end{quote}

\subsubsection{Copula quantile regressions}
Copulas are introduced section \ref{sec:Rcopula}. We move to copula quantile regression to capture non-linearity and tail dependence as in \citep{Bouye2009aa} that demonstrated:
\begin{equation}
\frac{\partial C(u,v)}{\partial u} = F_{Y\vert X} \left( F_Y^{-1}(v) \vert F_X^{-1}(u) \right)
\end{equation}
 and used the Frank Copula:
 \begin{equation}
 C(u,v,\delta) = -\delta^{-1} \log\left[ 1 + \frac{(e^{-\delta u} - 1)(e^{-\delta v}-1)}{e^{-\delta}-1} \right]
 \end{equation}
 
 That way we can model a quantile relationship:
 \begin{equation}
 Q_{Y\vert X} (\tau \vert x) =  -\delta^{-1} \log\left[ 1 + (e^{-\delta u} - 1)\left(1 - e^{-\delta F_X(x)}(\tau - 1)\right) \right]
 \end{equation}
\begin{center}
\includegraphics[scale=0.5]{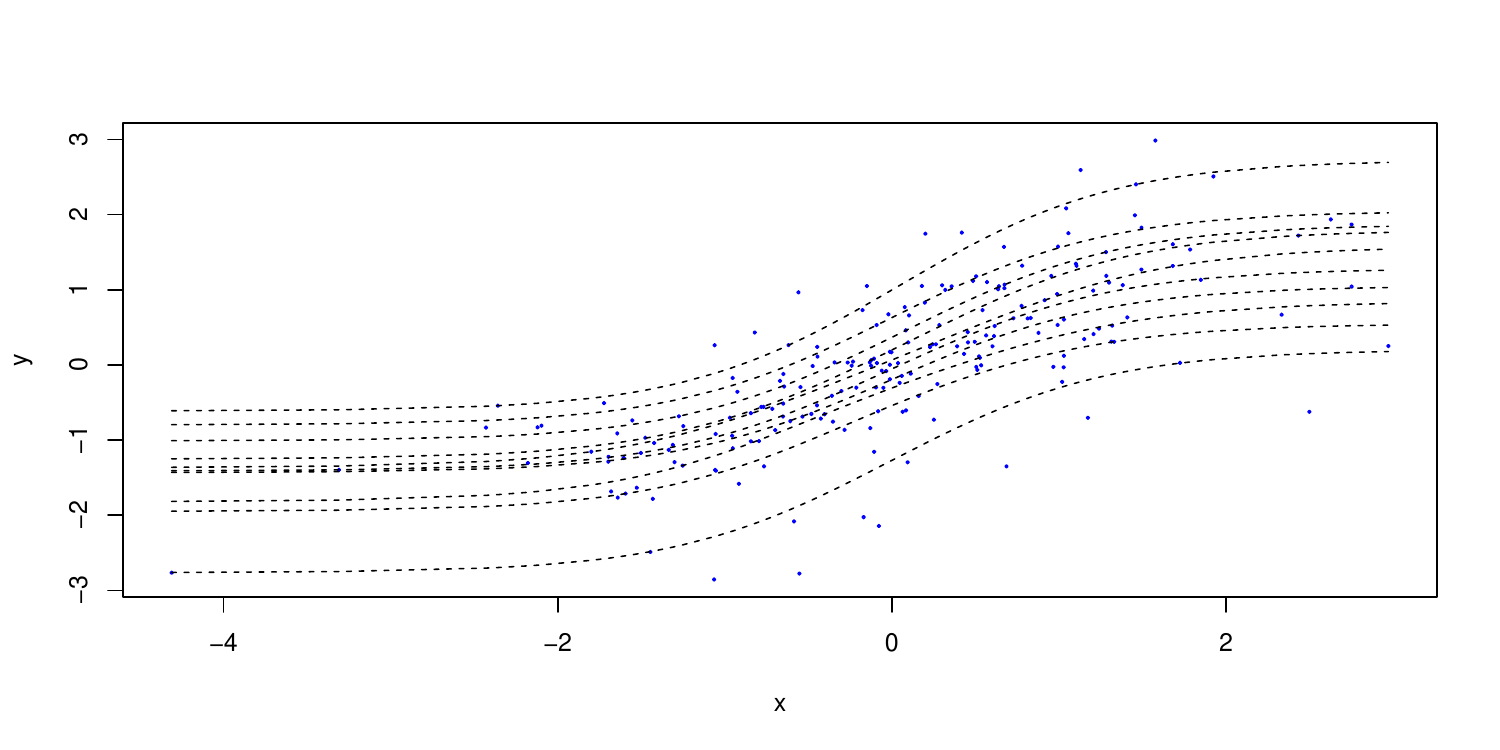}
\end{center}

\stepcounter{question}
\begin{itemize}
\item \textbf{Question \arabic{question}}
\begin{itemize}
\item As in \citep{Koenker2001}, download NCHS' Vital Statistics Natality Birth Data and perform quantile regression on Determinants of Infant Birthweight.
\end{itemize}
\end{itemize}

\subsection{Unconditional Quantile Regression (UQR)}

When we apply a (conditional) quantile regression, the interpretation of the results is in practice not very interesting for policy decision. If we write $q_\tau$ the $\tau^{\mbox{th}}$ quantile of the marginal (unconditional) distribution $F_Y(y)$. In the case of \citep{Firpo2009aa}, this would be the distribution of log wages. We would like to estimate $\frac{\partial q_\tau (p)}{\partial p}$, the direct effect of increasing the proportion of unionized workers on the $\tau^{\mbox{th}}$ quantile of the distribution of wages. In general, the coefficient $\beta_\tau$ of a conditional quantile regression is different from this direct effect we want to measure $\frac{\partial q_\tau (p)}{\partial p} = \frac{Pr[Y> q_\tau \vert X=1] - Pr[Y> q_\tau \vert X = 0]}{f_Y(q_\tau)}$.

To measure this effect, \citep{Firpo2009aa} suggest using the recentered influence function (RIF):
\begin{equation}
RIF(Y;q_\tau,F_Y) = q_\tau + \frac{\tau - \mathbb{1}_{Y\leq q_\tau }}{f_Y(q_\tau)}
\end{equation}
$\tau$ is a given quantile, $q_\tau$ is the value of $Y$ at the $\tau^{\mbox{th}}$ quantile and $f_Y(q_\tau)$ is the density of $Y$ at $q_\tau$ and $F_Y$ is the cumulative distribution function of $Y$. We
estimate the density $f_Y(q_\tau)$ using kernel methods\footnote{see section \ref{sec:KDE}} and then run an OLS regression of this new dependent variable on the covariates (they call it the RIF-OLS method). Or use a logistic regression of $\mathbb{1}_{Y\leq q_\tau }$ on the covariates to estimate the average marginal effect, which is then multiplied by $\frac{1}{f_Y(q_\tau)}$ (they call it the RIF-Logit method). Finally they also estimate a nonparametric estimator using polynomial series (they call it the RIF-NP method).

The authors show that \begin{quote}
the average derivative of the unconditional quantile regression corresponds to the marginal effect of the unconditional quantile of a small location shift in the distribution of covariates, holding everything else constant.
\end{quote} 

\subsubsection{UQR: Is the motherhood penalty large for low-wage women?}

We usually use quantile regression to generate multiple estimates of the motherhood penalty at various points in the wage distribution, but following the didactic paper of \citep{Killewald2014aa}

Using Conditional quantile regression (CQR), one cannot conclude that \begin{quote}
mothers who can least afford to lose wages are the ones incurring the largest proportionate penalties
\end{quote}

\begin{quote}
Conditional quantile regression (CQR) estimates the association between motherhood and wages at different points of the conditional wage distribution, thereby estimating variation in the motherhood penalty among women with different values of hourly wages but similar covariate values. Women at high quantiles of the conditional wage distribution have higher wages than we would expect, given their covariates, but are not necessarily high-wage in an absolute sense. CQR is therefore inappropriate for the question [which relates to variation in the motherhood penalty at different points of the unconditional wage distribution: high- versus low-wage workers], because conditional quantiles do not identify women along the unconditional wage distribution. The conditional distribution measures whether individuals have higher or lower wages than would be expected, given their other characteristics. [\ldots] results can be interpreted as indicating the motherhood penalty at different points of the wage distribution within each educational group. [\ldots] Using unconditional quantile regression (UQR), we show the motherhood penalty is larger for women in the middle of the wage distribution than for women below or above the median.
\end{quote}

\begin{center}
\includegraphics[scale=0.5]{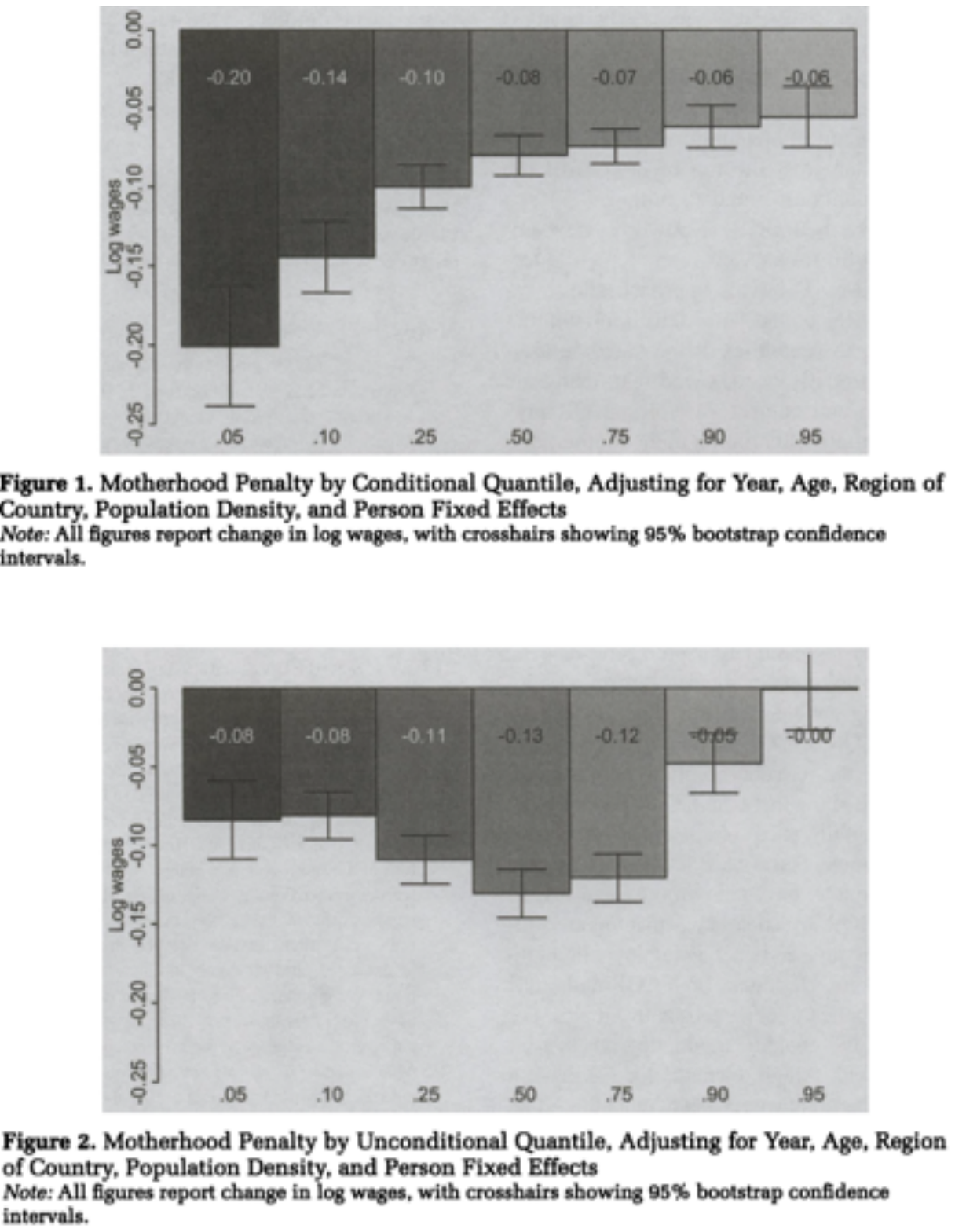}
\end{center}

Another simple and visual introduction of the limitation of conditional quantile regression is provided \href{https://stats.stackexchange.com/questions/80349/what-is-the-difference-between-conditional-and-unconditional-quantile-regression}{here}
If you wish to explore further quantile regressions, we recommend the work of \href{https://sites.google.com/site/blaisemelly/home/computer-programs/r-generic-inference-on-quantile-functions-for-discrete-outcomes}{Blaise Melly}.

\subsubsection{UQR: How does union status and wages varies across the wage distribution}

\citep{Firpo2009aa} write: 
\begin{quote}
Finding that the effect of unions (for short) estimated using conditional quantile regressions is smaller at the 90th than at the 10th quantile simply means that unions reduce within-group dispersion, where the 'group' consists of workers who share the same values of the covariates X (other than union status).
\end{quote}

\begin{center}
\includegraphics[scale=0.4]{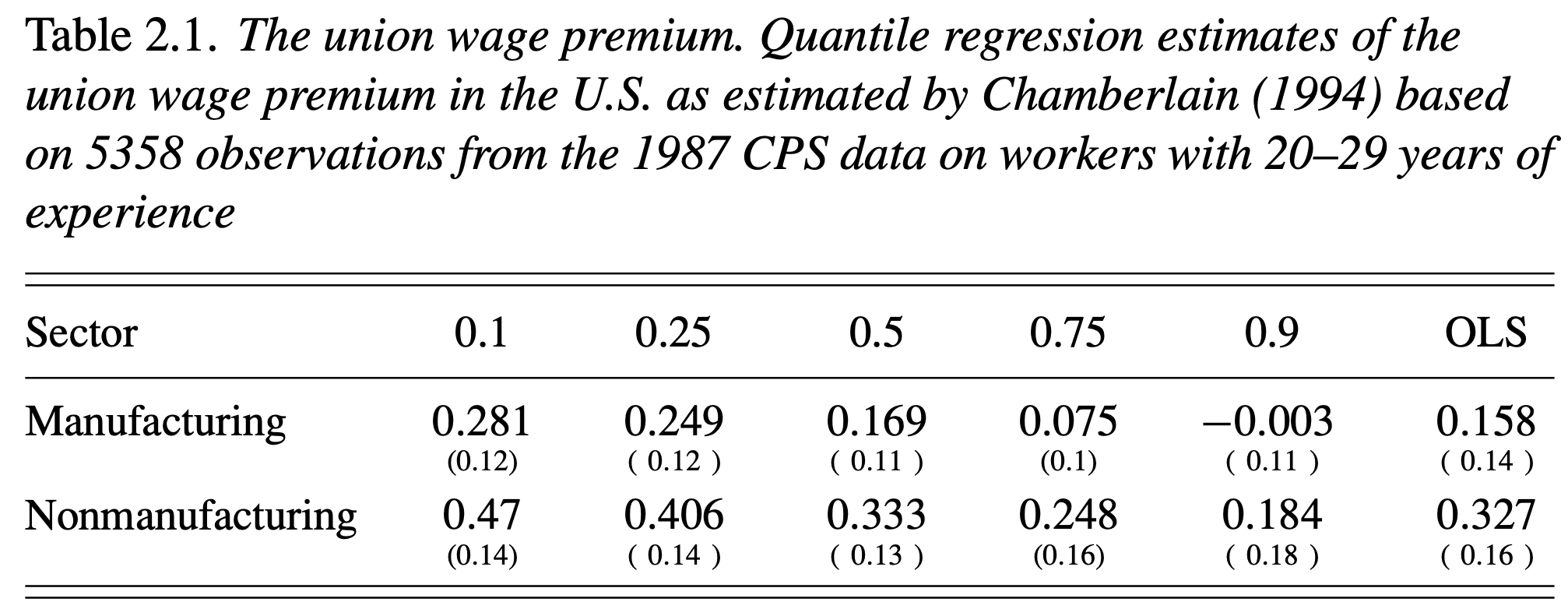}
\end{center}

\citep{Killewald2014aa} write:
\begin{quote}
By contrast, UQR addresses how the association between union status and wages (net of controls) varies across the wage distribution. \citep{Firpo2009aa} found that the association between unionization and wages rises over about the first one-third of the wage distribution, then declines, and is actually negative for workers in the top wage quantile. This is in contrast to the monotonic declines across conditional quantiles found by Chamberlain (1994). The two approaches thus answer different research questions and produce different point estimates
\end{quote}

We replicate\footnote{code: 20200316\_uqr\_vansteenberghe.py with the functions in regressionfunctions.py} figure 1 in \citep{Firpo2009aa} which illustrate the difference between the estimated effect of unions for conditional and
unconditional quantile regression, which plots both conditional and unconditional quantile regression estimates of union status at 19 different quantités (from the 5th to the 95th):

\begin{center}
\includegraphics[scale=0.3]{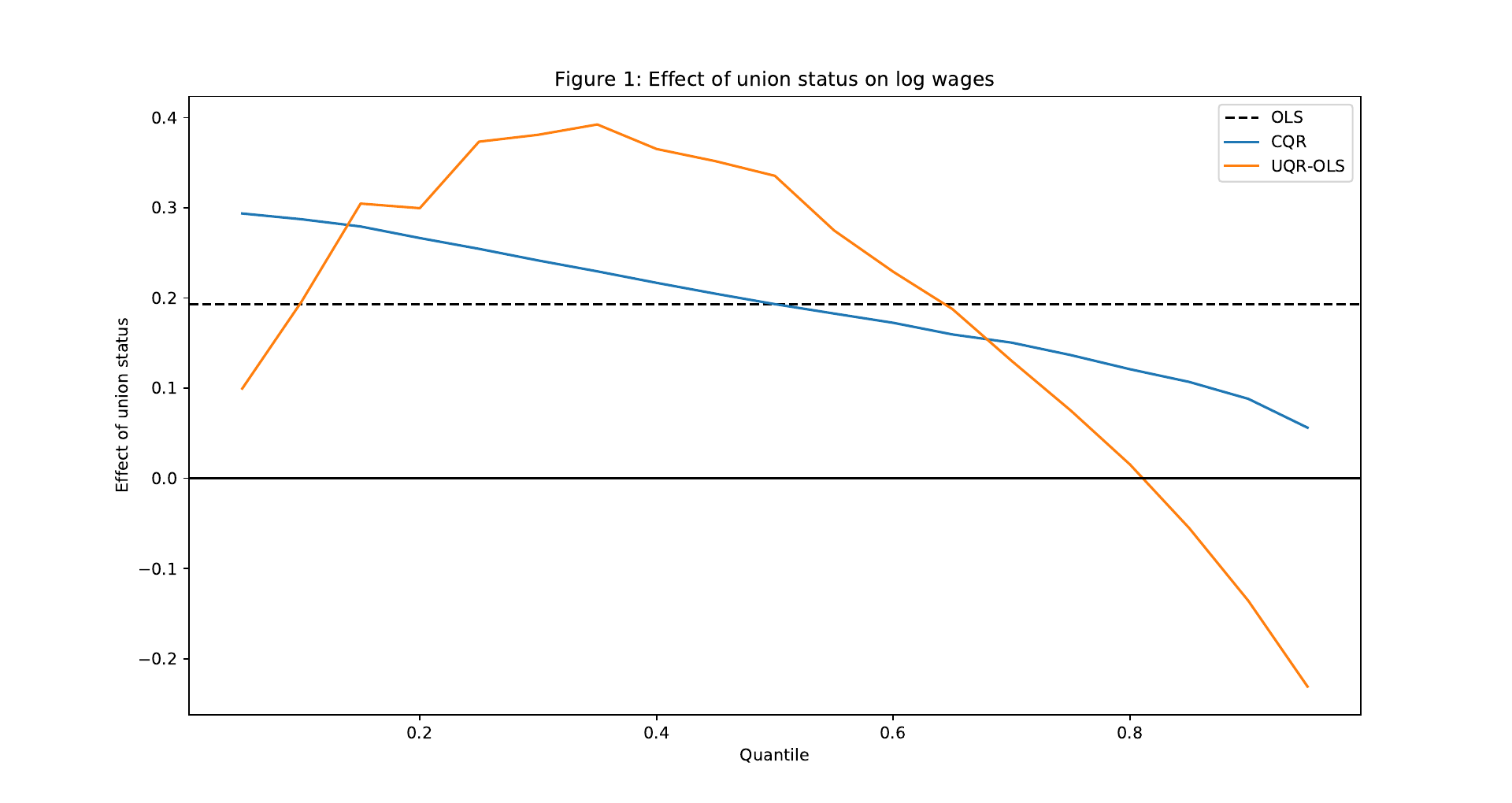}
\end{center}

Unconditional effect first increase before declining and eventually reaching a large negative effect for high quantiles. \begin{quote}
unionization progressively increases wages in the three lower quintiles of the distribution, peaking around the 35th quantile, and actually reduces wages in the top quintile of the distribution. As a result, the decline in unionization over the last three decades should have contributed to a reduction in wage inequality at the bottom end of the distribution and to an increase in wage inequality at the top end.
\end{quote}

\section{\citet{adrian2019} replication}
\citet{adrian2019} study\footnote{vansteenberghe\_vulnerable\_growth\_introduction.py and vansteenberghe\_vulnerable\_growth\_fit\_distibutions.py} the conditional distribution of GDP growth as a function of economic and financial conditions. They use real GDP growth between $t$ and $t+h$, $y_{t+h}$ and the National Financial Conditions Index (NFCI), $x_t$ the vector of conditioning variables, and find that ``extreme negative outcomes in GDP growth tend to coincide with extreme positive outcomes of the NFCI''.
\begin{center}
\includegraphics[scale=.5]{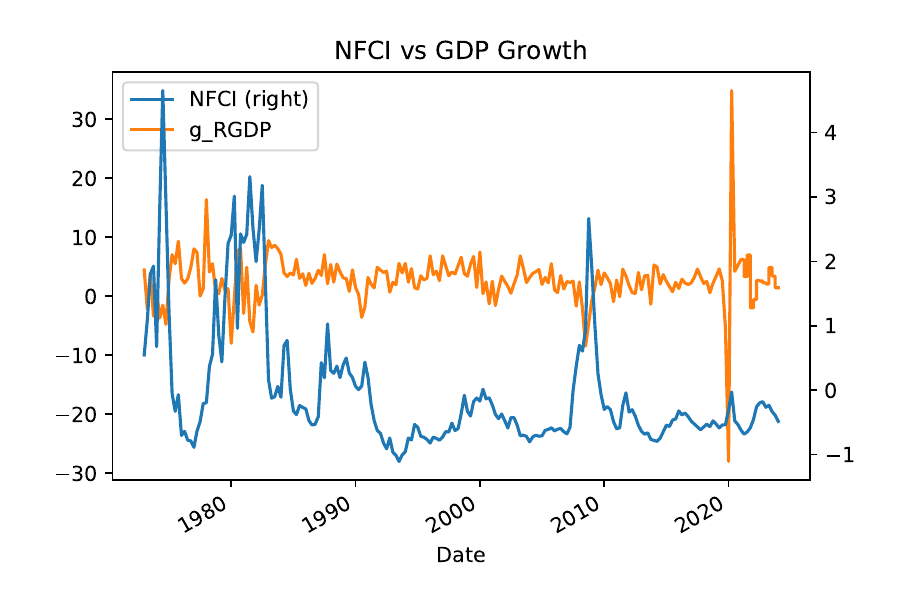}
\end{center}

Next, they fit an OLS and quantile regressions and find that ``at both the lower and the upper quantiles, the estimated slopes are significantly different, at the 10 percent level, from the OLS slope''.
\begin{center}
\includegraphics[scale=.5]{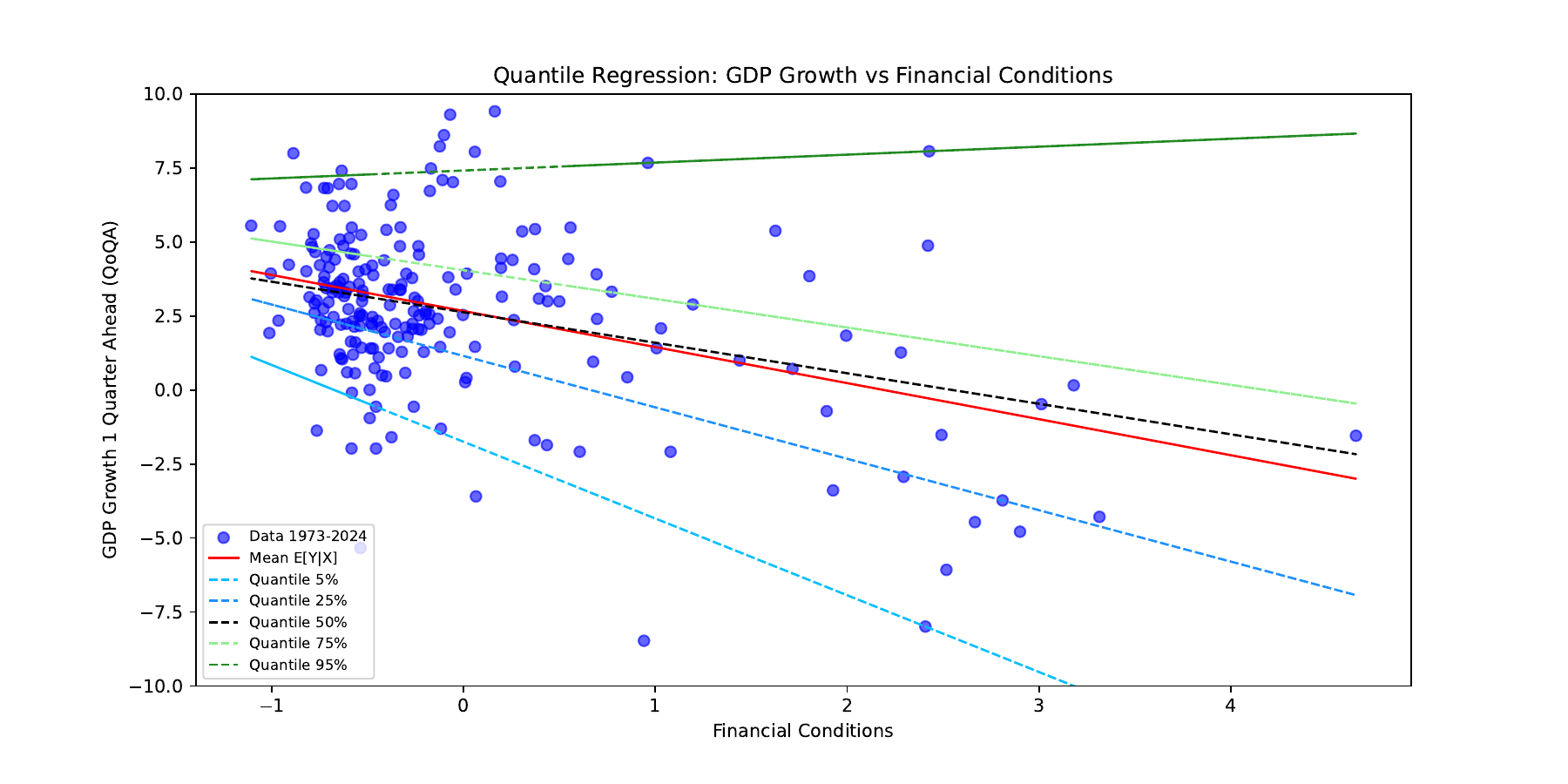}
\end{center}

And also, ``Importantly, the regression slopes for the NFCI do not change significantly when current GDP growth is also included in the regression, indicating that most of the explanatory power of future GDP vulnerability arises from the information content of financial conditions.''.

The next figure ``demonstrates the main result of the paper: the asymmetry between the upper and lower conditional quantiles. While the lower quantiles vary significantly over time, the upper quantiles are stable'' and state that ``changes in financial conditions have relatively little predictive information for the upper quantiles of future GDP growth'', here as in the paper, we will study whether lower quantiles have predictive information.
\begin{center}
\includegraphics[scale=.5]{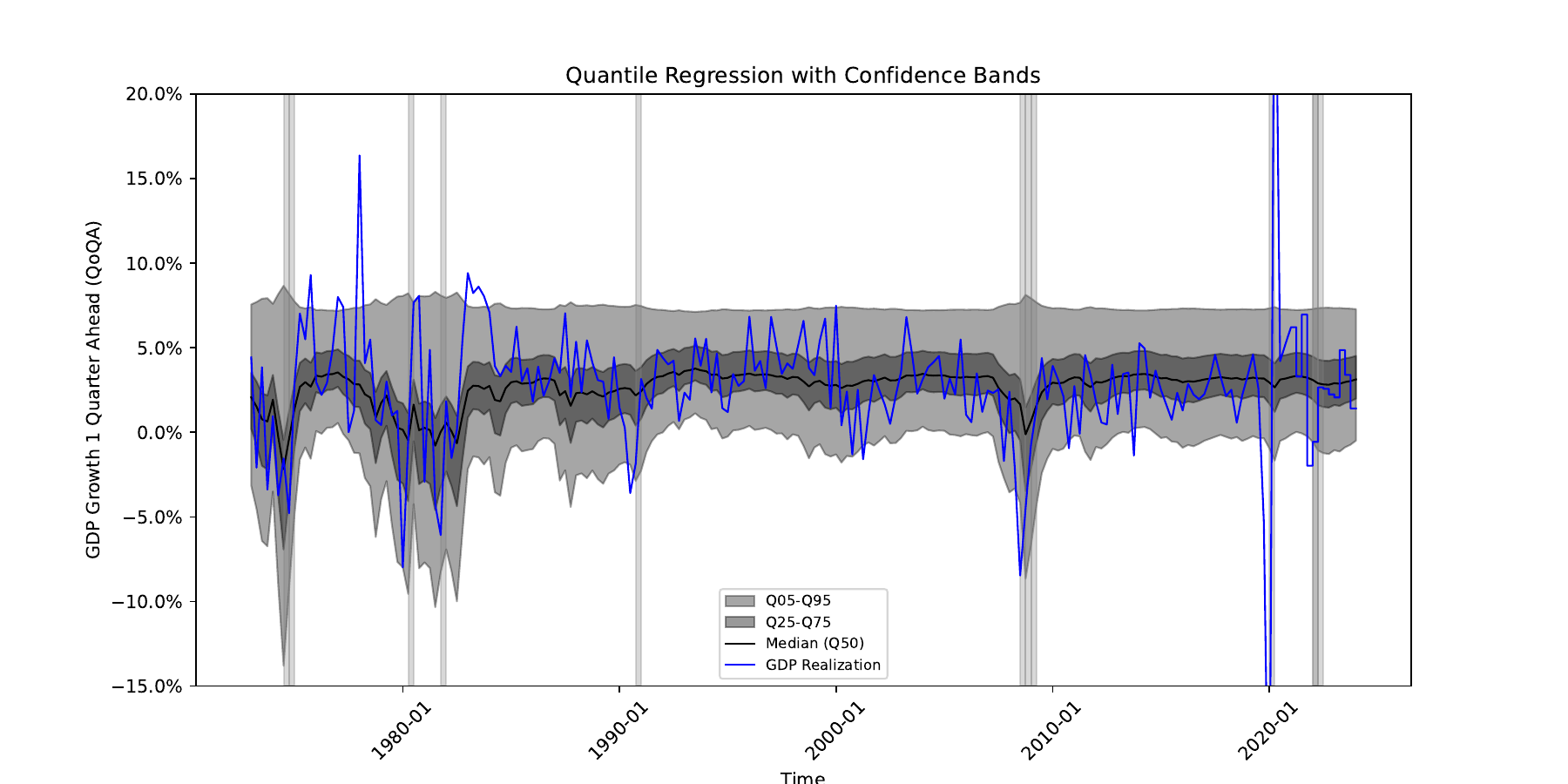}
\end{center}

\subsection{The quantile regression approach}
The main specification of the paper is a quantile regression of the real GDP growth between $t$ and $t+h$, $y_{t+h}$ with $h=1$, and a vector of conditioning variables $x_t$ with an intercept, the National Financial Conditions Index (NFCI), and the lagged real GDP growth. Equation \ref{eq:Adrian1} to get the quantile estimates of the predictive distribution Equation \ref{eq:Adrian2}. Once they have the quantile estimates, they fit a skewed t-distribution using 4 estimated quantiles (so 4 unknown and 4 data points, which is the bare minimum, the exactly identified approach) with the 5, 25, 75, and 95 quantiles Equation \ref{eq:Adrian3}, they could use any parametric distribution $F\left(.,\theta \right)$. They defined the shortfall and longrise, as downside and upside risk to the GDP growth as per Equation \ref{eq:Adrian4}. Lastly, they evaluate the out-of-sample performances of the methods. To assess the reliability of the predictive distribution, they measure the accuracy of a density forecast using the predictive score Equation \ref{eq:Adrian5}, computed as the predictive distribution generated by a model and evaluated at the realized value of the time series. To assess the calibration of the predictive distribution, they compute the empirical cumulative distribution function (CDF) of the Probability Integral Transform (PIT) values, where the PIT measures the probability that a realized value falls below a given quantile of the predictive distribution. The empirical CDF, $\hat{F}_{\text{PIT}}(\tau)$, represents the proportion of observations with PIT values less than or equal to $\tau$. In a well-calibrated model, $\hat{F}_{\text{PIT}}(\tau)$ closely follows the 45-degree line, indicating that the fraction of realizations below each quantile is approximately equal to $\tau$.
\begin{equation}\label{eq:Adrian1}
\hat{\beta}_\tau = \arg \min_{\beta_\tau \in \mathbb{R}^k} \sum_{t=1}^{T-h} \left( \tau \cdot 1(y_{t+h} \geq x_t \beta) \lvert y_{t+h} - x_t \beta_\tau \rvert + (1-\tau) \cdot 1(y_{t+h} < x_t \beta) \lvert y_{t+h} - x_t \beta_\tau \rvert \right)
\end{equation}
\begin{equation}\label{eq:Adrian2}
\hat{Q}_{y_{t+h} | x_t}(\tau | x_t) = x_t \hat{\beta}_\tau
\end{equation}
\begin{equation}\label{eq:Adrian3}
 \hat{\theta}_{t+h} = \arg \min_{\theta} \sum_{\tau} \left( \hat{Q}_{y_{t+h} | x_t}(\tau | x_t) - F^{-1}(\tau; \theta) \right)^2
\end{equation}

\begin{equation}\label{eq:Adrian4}
\text{SF}_{t+h} = \frac{1}{\pi} \int_0^{\pi} \hat{F}^{-1}_{y_{t+h} | x_t}(\tau | x_t) \, d\tau ; \quad \text{LR}_{t+h} = \frac{1}{\pi} \int_{1-\pi}^{1} \hat{F}^{-1}_{y_{t+h} | x_t}(\tau | x_t) \, d\tau
\end{equation}

\begin{equation}\label{eq:Adrian5}
\text{Predictive Score} = \log f(y_{t+h} | x_t; \hat{\theta})
\end{equation}

\stepcounter{question}
\begin{itemize}
\item \textbf{Question \arabic{question}}
\begin{itemize}
\item Why fitting a Skewed t-distribution? How would a Generalized Beta distribution perform?
\end{itemize}
\end{itemize}

\subsection{Forecast evaluation}

\citet{ferrara2022high} use four criteria to evaluate the forecast, cf details in \citet{jordan2017evaluating}:
\begin{enumerate}
\item average log score differentials (LS),
\item average continuous ranking probability score ratios (CRPS) \citep{gneiting2007strictly},
\item  average quantile-weighted probability score ratios (QWPS) \citep{gneiting2011comparing}, and
\item  average quantile score ratios (QS).
\end{enumerate}

We focus on the LS and CRPS as the last two criteria should be coherent. We get nuanced results with respect to \citet{adrian2019}, some elements are discussed in \citet{schick2024real}.
\begin{tabular}{lrr}
\toprule
Model & Average LS & Average CRPS \\
\midrule
Model 6: NFCI & 2.437314 & 1.661221 \\
Model 3: GDP + SPF & 2.787769 & 1.840065 \\
Model 1: GDP & 2.519937 & 2.005339 \\
Model 7: SPF & 2.813691 & 2.212663 \\
Model 2: GDP + NFCI & 2.639117 & 2.322547 \\
Model 5: NFCI + SPF & 3.167647 & 3.405092 \\
Model 4: GDP + NFCI + SPF & 3.160124 & 3.858428 \\
\bottomrule
\end{tabular}
 
\stepcounter{question}
\begin{itemize}
\item \textbf{Question \arabic{question}}
\begin{itemize}
\item Apply this to Euro Area data as in \citet{ferrara2022high}. Here focus on the ECB-SPF, does it bring any added value to the performance of the model?
\end{itemize}
\end{itemize}

\subsection{The ECB SPF}\label{sec:ecbspf}

The ECB Survey of Professional Forecasters (SPF) is a quarterly survey initiated in 1999, designed to gather expectations on key economic indicators from professional forecasters across a broad range of institutions. The survey serves as an important tool for monitoring the expectations of professionals on euro area inflation, GDP growth, and unemployment. The ECB SPF collects quantitative forecasts from around 90 forecasting institutions, including banks, research institutes, and other financial or economic bodies involved in analyzing the euro area economy. The institutions participating in the survey are relatively stable, allowing the ECB to track their individual forecasts over time. This provides a rich panel of data with significant continuity.

Forecasters participating in the SPF are asked to provide predictions for three key macroeconomic variables:
\begin{itemize}
    \item The year-on-year inflation rate in the euro area (measured by the Harmonized Index of Consumer Prices, or HICP)
    \item The year-on-year real GDP growth rate
    \item The euro area unemployment rate
\end{itemize}
Forecasts are provided for multiple time horizons, including one year and two years ahead, as well as a shorter-term horizon for the upcoming quarter. The forecasted values are matched with realized data (typically sourced from Eurostat) in order to assess their accuracy.

A notable feature of the SPF is the combination of two types of forecasting approaches:
\begin{enumerate}
    \item \textbf{Rolling Horizon Forecasts:} In this approach, forecasters are asked to update their projections for a fixed horizon, such as one or two years ahead. For example, in 2021Q2, a forecaster would be asked to predict the inflation rate for one year ahead (i.e., for 2022Q2). In the following quarter, the forecaster would provide an updated prediction for 2022Q3. This approach allows researchers to observe how forecasts for a particular time point evolve over time.
    \item \textbf{Calendar Horizon Forecasts:} In this type of forecast, forecasters are asked to provide their predictions for the end of the current and the next calendar year. For instance, in 2021Q2, forecasters would predict inflation rates for both the end of 2021 and the end of 2022. This method allows the ECB to gather consistent forecasts for fixed, recurring periods, irrespective of when the survey is conducted.
\end{enumerate}

One of the key advantages of the ECB SPF over other surveys, such as the U.S. SPF, is its larger sample size, with typically around 60 (anonymous) respondents each quarter. Moreover, because the survey tracks responses from the same institutions over time, it provides a stable panel of data that can be used to analyze individual forecaster behavior, revisions, and forecast errors. The main advantage of a survey with subjective probability distribution is the quantitative measure of the uncertainty surrounding the forecast \citep{garcia2003introduction} and the main variable of interest is the inflation expectations. For a Q2 questionnaire, the March HICP inflation is available to the forecasters. Therefore, the year-on-year inflation rate in March of the year+1 and year+2 are requested as the rolling horizon forecast. Forecasters are also asked to provide expectations five years ahead. The subjective probability bin width has to be considered as $0.5$ percentage point (hence we need to correct for the bin edges and $[-1,-0.6]$, $[-0.5,-0.1]$ in the survey becomes $[-1,-0.5]$, $[-0.5,0]$).

Special surveys led by the ECB \citep{ECB2018SpecialSurvey}, indicate that most forecasters use judgment or reduced-form models (ARIMA, other single equation model, VAR/VEC) to forecast inflation, and mostly a combination of models and judgment. A striking aspect is that the reported subjective probability distribution is mostly judgment based.

Inflation expectations are one of the key variables surveyed in the ECB SPF, specifically focusing on the Harmonised Index of Consumer Prices (HICP) inflation. The survey collects inflation forecasts over different time horizons, including short-term (one year ahead), medium-term (two years ahead), and long-term (five years ahead) horizons. Participants are asked to assign probabilities to various ranges of possible inflation outcomes, which allows the ECB to gauge both the central tendency and the uncertainty surrounding these expectations.
The probability intervals, or "bins," used in the questionnaire are carefully chosen to balance precision with practical limitations. Each bin typically spans 0.5 percentage points, such as \(0.0\%-0.4\%\), \(0.5\%-0.9\%\), \(1.0\%-1.4\%\), and so on, with the lower-end and upper-end intervals left open to capture extreme outcomes. For instance, for inflation expectations exceeding 3.5\%, participants assign cumulative probabilities without specifying how much higher the inflation rate could be. This structure ensures that forecasters provide a comprehensive probability distribution across all potential outcomes.

\section{Consensus, combination of forecasts and opinion pooling}

In this section, a decision maker $D$ obtain experts subjective probability distribution of an unknown value of some parameter and forms his own subjective distribution. We want to advise $D$ on how to "aggregate the statements of experts". A similar problem is to make as good a forecast as possible by combining forecasts. The experts' opinions aggregation problem can be formulated:

\begin{center}
We have $k$ experts forming subjective probability distributions $f_i$ of a parameter $\alpha$,\\ with information sets $z_i$ that can be independent,\\ overlapping or unique ($\forall i, z_i=z$).\\ How to increase $D$'s accuracy in estimating $\alpha$? (mathematical)\\ Or alternatively, how to forge a consensus? (behavioral)
\end{center}

\citet{stone1961} is credited for introducing the concept of "opinion pool". \citet{Bates:1969vx} introduce combinations of expert forecasts based on performance persistence over time. From the Bayesian perspective, $D$ is \textit{supra-Bayesian}. \citet{Roberts:1965wq} suggested a distribution which combines the distributions (or models) of two experts. The outcome, essentially a weighted averaged of posterior distributions of two models, is similar to Bayesian Averaging Model. \citet{Morris:1974wk} formalized the use of experts in decision situation that would lead to the Bayesian Averaging Model literature\footnote{for a literature review, see \citet{Clemen:1999vy}}. To apply a bayesian method, $D$ must construct his prior beliefs about what the experts are going to tell him, conditional on the true value of the parameter $\theta$: $f(k\vert \theta)$.


\subsection{Reaching a consensus (behavioral)}

An aspect of this literature is the question of how to reach a consensus (with possible interactions between experts) and whether it can be reached \citep{degroot1974reaching}. This is in the group decision making literature, which is a bit different than our main interest where one decision maker $D$ has to formulate his opinion. Writing $F_i$ the subjective probability distribution of individual $i$ and $p_{ij}$ the weight that individual $i$ assigns to distribution of individual $j$ when updating his subjective probability distribution, then this can be iterated $F^{(n)}=P^n F$ and a consensus is reach if there exist a positive integer $n$ such that every element in at least one column of the matrix $P^n$ is positive. Or analogously: if all the recurrent states of the Markov chain communicate with each other and are aperiodic, then a consensus is reached. \citet{degroot1974reaching} doesn't elicit how $P$ is built nor how opinion dynamics emerges, \citep{hegselmann2002opinion}.

\subsection{Combination of forecasts (mathematical)}

We follow the literature review of \citet{clemen1989combining} and \citet{diebold1996}.

A single forecast is optimal if
\begin{enumerate}
\item forecast  errors have a zero mean;
\begin{itemize}
\item if errors are iid a t-test is valid, otherwise a GMM can be applied.
\end{itemize}
\item 1-step-ahead errors are white noise;
\item k-step-ahead errors as at most MA$(k-1)$;
\item k-step-ahead error variance is non-decreasing in $k$.
\end{enumerate}

Optimality can also be assessed against an information set, errors should present unforecastability on the basis of information available, that is we can test $(\beta_0,\beta_1)=(0,1)$ in $y_{t+k}=\beta_0+\beta_1 \hat{y}_{t+k,t}+u_t$. In other words, the errors a forecaster makes at time $t$ when forecasting $t+k$ cannot be predicted.

Before combining, one need a tool to measure forecast accuracy.

\subsubsection{Comparison of opinion or forecast accuracy}

We need to choose a tool for comparing opinion accuracy if we known the target. 

\citet{brier1950verification} suggested verification method that satisfies a criterion: the verification scheme should influence the forecaster in no undesirable way. It supposes that $n$ occasions to forecast an event that can occur in $r$ possible way, the Brier score is
\begin{equation}
\frac{1}{n}\sum_{j=1}^r\sum_{i=1}^n\left(f_{ij} - E_{ij} \right)^2
\end{equation}
where $f_{ij}$ if a forecast probability and $E_{ij}$ takes the value 1 or 0 according to whether the event occured in class $j$ on occasion $i$ or not.

An alternative approach is the root mean square errors, alternatives are introduced in \citet{diebold1995}.

\subsubsection{Combination of forecasts}

\citet{Bates:1969vx} compared forecasts combinations (for forecasting a time series $x_t$) with the objective to make as good a forecast as possible. They suggested the following steps, given two forecasts $f_{1,t}$ and $f_{2,t}$ and their respective forecast erros $e_{1,T}$ and $e_{2,T}$:
\begin{enumerate}
\item check that the individual sets of forecasts are unbiased, and if biased to correct for the average percentage (or absolute) bias;
\item give greater weight to the set of forecasts which seemed to contain the lower (mean-square) errors to obtain the forecasts combination $C_T = k_T f_{1,T} + (1-k_T)f_{2,T}$ , for this:
\begin{enumerate}
\item start when $t=1$ with $k_1=0.5$;
\item $k_T=\frac{E_2}{E_1+E_2}$ with $E_1=\sum_{t=T-v}^{T-1} \left(e_{1,t} \right)^2$ with $v=6$ or less if less observations are available.
\end{enumerate}
\end{enumerate}

\citet{granger1984improved} considers three alternative approaches, with the best method to add a constant term and not to constraint the weights to add to unity: $C_T = \alpha_T f_{1,T} + \beta_T f_{2,T} + \gamma m$, where $m=E[x_T]$. They show that this method gives the smallest mean squared error and has an unbiased combined forecast, compared with two other methods, a first one where no constant is allowed and a second one where weights have to sum up to unity. The obvious way to proceed is to perform the least-squares regression, for this you'll need to have time series of past forecast from both forecasters and past realizations of $x$: $x_{t} = \alpha_1 f_{1,t} + \alpha_2 f_{2,t} + a + \epsilon_t$, where $a$ is a constant.

\citet{libby1978performance} report that the majority of the improvement in accuracy is achieved with the combination of the first two or three forecasts.

\citet{de2000review} suggested going beyond the RMSE point estimate (i.e. accuracy) and use:
\begin{enumerate}
\item error variance;
\item distribution asymmetry (skewness);
\item serial correlation.
\end{enumerate}
They focus on several combining methods:
\begin{itemize}
\item Simple average;
\item Outperformance \citep{bunn1975bayesian} where each individual weight is interpreted as the probability that its respective forecast will perform the best;
\item Optimal \citep{Bates:1969vx}: where linear weights are calculated to minimise the error variance of the combination;
\item Regression \citep{granger1984improved}.
\end{itemize}

\citet{raftery2005using} suggests using Bayesian Model Averaging (BMA), where the overall forecast PDF is a weighted average of forecast PDFs based on each of the individual forecasts; the weights are the estimated posterior model probabilities and reflect the models' forecast skill in the training period, relative to the other models. The weights can also provide a basis for selecting ensemble members: when they are small there is little to be lost by removing the corresponding ensemble member. This can be useful given the computational cost of running ensembles. For more details, see section \ref{sec:BMA}.

\citet{genre2013combining} find, working on the ECB Survey of Professional Forecasters, \begin{quote}
Overall, we conclude from this study that there exists only a very modest case for considering combinations other than equal weighting as a means of better summarising the information collected [\ldots].
\end{quote}

\subsubsection{Replications of \citet{bessler1981forecasting}, \citet{granger1984improved} and stocks watson}

I am using the data in the Appendix of \citet{bessler1981forecasting} for the hog prices from 1976-01 to 1979-02. I replicate Table 7, using $v=2$ for the "Minimum MSE adaptive". I also add the results for the methods when also including Expert (in the original paper, they discard the Expert forecast for the Minimum MSE adaptive and Minimum Variance methods). I do not find exactly the same MSE for the Minimum Variance nor the Minimum MSE adaptive as this can be due to approximation when inverting matrices or sum of squares. It is not clear how $v$ was chosen as increasing $v$ does improve the MSE for the adaptive method. The main conclusion that the simple average is the most successful combination forecast holds. I also implement method A, B and C from  \citet{granger1984improved}. Finally, as in \citet{stock2004combination}, I introduce the median and compute Meas Squared Errors relative to a benchmark, the Auto-Regressive model of order 1.

\begin{tabular}{lrllll}
 & MSE & Constant & Econometric & ARIMA & Expert \\
Expert & 1.09 &  &  &  &  \\
AR & 1.00 &  &  &  &  \\
Econometric & 0.98 &  &  &  &  \\
Minimum MSE adaptive, expert excluded & 0.59 &  &  &  &  \\
Two forecast Simple average & 0.55 &  &  &  &  \\
ARIMA & 0.51 &  &  &  &  \\
Minimum MSE adaptive & 0.51 &  &  &  &  \\
Minimum Variance, expert excluded & 0.50 &  &  &  &  \\
median & 0.49 &  &  &  &  \\
Minimum Variance & 0.43 &  &  &  &  \\
Three forecast Simple average & 0.42 &  &  &  &  \\
Method B & 0.41 & 0 & 0.42 & 0.30 & 0.28 \\
Method A & 0.41 & 0 & 0.41 & 0.30 & 0.28 \\
Method C & 0.41 & -2.84 & 0.49 & 0.28 & 0.28 \\
\end{tabular}

\subsection{Opinion pooling}

We follow the literature review of \citet{genest1986}. Each expert's opinion has been encoded as a \emph{subjective probability distribution}.

To decide on how to pool opinions, $D$ can define axioms the function has to satisfy (\emph{axiomatic method}).

\subsubsection{Intelligence of a crowd}

\citet{galton1907vox} writes an early description of the wisdom of a crowd (vox populi), discussed by \citet{wallis2014revisiting}. In an early laboratory experiment, addressing the question ``Is the judgment of a group of persons any better than the judgment of the average member of the group?", \citet{gordon1924} had 200 people ranking 10 items by weights. The true ranking was known to the designer and she computed the Spearman rank correlation coefficient. Individual guesses would have an average correlation of $41\%$ with the true ranking, while averaging each participant ranking would reach a $94\%$ correlation. \citet{kelley1925} demonstrated that the increase in reliability was given by the Spearman-Brown formula. \citet{Larrick:2006uz} demonstrated how those basic concept where misappreciated, hence not systematically implemented.

\stepcounter{question}
\begin{itemize}
\item \textbf{Question \arabic{question}}
\begin{itemize}
\item Use the original work of \citet{galton1907vox} and his percentiles to fit a gamma distribution \href{https://blogs.sas.com/content/iml/2018/03/07/fit-distribution-matching-quantile.html}{hints in this article}. What are the mode and the mean of your distribution?
\end{itemize}
\end{itemize}

\subsubsection{Linear opinion pool}

$D$ forms his own subjective probability distributions with a weighed-average method \citep{stone1961}:
\begin{equation}
f(\theta) = \sum_{i=1}^k \omega_i f_i(\theta \vert z_i)
\end{equation}

The linear pooling functions are the only independent and unanimity-preserving pooling functions (theorem 1 in \citet{dietrich2016probabilistic}). \citet{weerahandi1981} demonstrated that for simple normal priors, the weighed-average method is unsatisfactory when the estimated parameters are widely separated, which was the configuration of experts in early estimations of climate change. \citet{Madansky1964} demonstrated that this approach is not externally Bayesian. 

Subjective weights can be selected by $D$ \citep{winkler1968}. \citet{Cooke1991} suggested to choose weights based on an initial elicitation exercise to test experts performance in terms of calibration and information. This can also be interpreted as the trust $D$ assigned to each experts.

Instead of weights, $D$ can use ranking, if $r_i$ is the rank of each expert and we note the sequence of ordered experts $r_{(i)} = i$ then $w_{(i)} = \frac{(i)}{\sum_{r=1}^k r}$.

$D$ can also apply the median to have a opinion pool that is less sensitive to outliers.

\subsubsection{Logarithmic opinion pool}
Attributed to Peter Hammond \citep{genest1986}:
\begin{equation}
f(\theta) = c \prod_{i=1}^k f_i(\theta \vert z_i)^{\omega_i}
\end{equation}

with $c$ a normalizing constant $\int  \prod_{i=1}^k f_i(\theta \vert z_i)^{\omega_i} d \mu$ with $\mu$ a reference measure on the sample space.
This method satisfies the external Bayesianity axiom \citep{dietrich2016probabilistic} and \citep{Madansky1964}.

As the logarithmic opinion pool doesn't have the coherent marginalization, where the linear does, the properties might not be the decisive factors. The logarithmic pool would be preferred if $D$ can be certain that the data sets $z_.^t$ are distinct for some experts. 

\subsection{Bayesian methods}\label{sec:bayespooling}
If $f(\theta)$ represents a prior distribution and $z$ the information contained in a sample, then Bayes' Theorem can be written in the form:
\begin{equation}
f\left(\theta \vert z \right) = \frac{f(\theta)l(z\vert \theta)}{\int f(\theta)l(z\vert \theta) d\theta}
\end{equation}
where $l$ is the likelihood of the experts' opinions. Starting with a prior $f(\theta)$, the decision maker's posterior becomes
\begin{equation}
f\left( \theta \vert f_1,\ldots, f_n \right) \propto  \frac{f(\theta) l\left(f_1,\ldots, f_n\vert \theta \right)}{f(g_1,\ldots,g_n)}
\end{equation}

\citet{dietrich2010bayesian} presents  a Bayesian group belief method which incorporates all information spread across individuals without individuals having to explicitly communicate their information (that may be too complex or personal to describe, or not describable in principle in the language).

We follow \citet{raftery2005using}.

In the case of $y$ to be forecast on the basis of training data $y^T$ using $K$ statistical models $M_1,\ldots,M_k$, the law of total probability tells us that the forecast pdf is given by
\begin{equation}
p(y) = \sum_{k=1}^K p\left( y \vert M_k \right) p\left( M_k \vert y^T\right)
\end{equation}
where $p\left( y \vert M_k \right)$ is the forecast pdf based on model $M_k$ and $p\left( M_k \vert y^T\right)$ the posterior probability of model $M_k$ being correct given the training data.

The basic idea is that for any given forecast ensemble there is a "best" model, or member, but we do not know what it is, and our uncertainty about the best member is quantified by BMA. Writing $f_k$ a bias-corrected forecast and $g_k\left( y \vert f_k\right)$ the conditional pdf of $y$ conditional on $f_k$, given that $f_k$ is the best forecast in the ensemble. The BMA predictive model is
\begin{equation}
p\left( y\vert f_1,\ldots,f_k\right) = \sum_{k=1}^K w_k g_k\left( y \vert f_k\right)
\end{equation}
where $w_k$ is the posterior probability of forecast $k$ being the best one and is based on forecast $k$’s performance in the training period. They can be seen as weights summing to unity and can be estimated using maximum likelihood methods with the EM algorithm).
\subsection{Vincentization}

Vincentization consists of averaging the $\alpha$ per cent quantiles of the experts' distributions in order to construct the $\alpha$ per cent quantile of the consensus, $0<\alpha<1$. This is $q_i$ is the $\alpha$ per cent quantile of $f_i$, $F_i(q_i)=\alpha$, the consensual distribution $F$ would be defined by setting $F^{-1}(\alpha)=\sum_{i=1}^n \omega_i q_i$.

$D$ can follow a Vincent method, building a consensus cumulative distribution function construct $F$ such that if
$$F_j\left(\pi_{i,j}\right)=\epsilon, \quad \forall \epsilon \in [0,1]$$
then $$F^{-1}(\epsilon)=\sum_{j=1}^m h_j \pi_{t,j}$$, with arbitrary $h_j$ (e.g. $\frac{1}{m}$).
Nota bene: the quantile of a Gamma distributed random variable is the ratio of the inverse of the lower incomplete gamma function and the rate parameter of the Gamma distribution.

\subsection{Ranking methods}

In our set up, $n$ items have to be ranked by $m$ experts and the decision maker $D$ assign weights ($q$ vector of weights) to experts to form his final (subjective) ranking ($r$ is $D$'s ranking vector).

In matrix notation, if $\pi_{i,j}$ is the valuation of expert $j$ on item $i$, then the decision maker ranking $r$ is:
\begin{equation}
r = \pi q
\end{equation}

The \textbf{counting method} assigns scores proportional to the totals of the valuations:
$q = \frac{1}{m}$.

The \textbf{invariant method} can be called the eigenvalue methods, it chooses $r$ so that $r$ is an eigenvector of $\pi$, i.e. $r=\lambda \pi r$, for some positive $\lambda$.

It can be called Liebowitz-Palmer (LP) method. The Perron-Frobenius theorem states that if the matrix $\pi$ has positive elements ($\forall i,j \pi_{i,j}>0$) then such an eigenvector exists and is unique up to a scalar.

The hyperlink-induced topic search method (\textbf{HITS method}) is such that $ r = \lambda \pi \pi' r$, where$\pi'$ is the transpose of $\pi$.

\citet{demange2014} introduces a \textbf{handicap-based ranking} method which is the only method to be intensity invariant, homogeneous, and uniform. The handicap-based method is highly correlated with the invariant and HITS methods. 
$D$ assigns weight $w$ to experts and scores $h$ to the periods (handicap). Experts whose statements have a high correlation with the handicaps receive a lower weight than those whose statements have a low correlation. This methods amount to a standard matrix scaling problem, where the matrix of expert judgment is re-scaled by two vectors $h$ and $w$  through the iterative scaling algorithm RAS \citep{Bacharach:1965tx}, also known as the iterative proportional fitting for which \citet{Soules:1991vt} demonstrated convergence at a geometric rate:
 \begin{equation}\label{eq:bacharach1}
\forall t, \sum_j \left(\pi_{t,j} w_j \right) h_t = 1
\end{equation}
and
\begin{equation}\label{eq:bacharach2}
\forall j, \sum_t \left( \pi_{t,j} h_t \right) w_j = \frac{T}{m}
\end{equation}
Once the algorithm converged, the ranking can be defined following equation \ref{eq:bacharach1} as:
\begin{equation}
o_t = \frac{\lambda}{h_t}
\end{equation}
where $\lambda$ is chosen s.t. $\sum_t o_t =1$.

\citet{demange2014} introduces the handicap-based method where experts are being assigned weight $q$ and scores to the items $h$. When statements differ, the experts whose statements have a high correlation with the handicaps receive a lower weight than those whose statements have a low correlation. This methods amount to a standard matrix scaling problem, where the matrix $\pi$ is re-scaled by two vectors $h$ and $q$ such that:

\begin{equation}
\forall i, \sum_j \left(\pi_{i,j} q_j \right) h_i = 1
\end{equation}
where
\begin{equation}
\forall j, \sum_i \left( \pi_{i,j} h_i \right) q_j = \frac{n}{m}
\end{equation}

This method can be applied through the iterative scaling algorithm RAS \citep{Bacharach:1965tx}, also known as the iterative proportional fitting.

The starting points are:
\begin{equation}
h_i^0=\frac{m}{\sum_i \pi_{i,j}}, \quad q_j^0=\frac{1}{m}
\end{equation}

If $\sum_j h_i q_j \pi_{i,j}=1$ and $\sum_i h_i q_j \pi_{i,j}=\frac{n}{m}$, then the algorithm stops.
Otherwise, $h$ and $q$ are updated until the two conditions \ref{eq:bacharach1} and \ref{eq:bacharach2} are respected, indexing interation by $t$:
\begin{equation}
h_i^t= \frac{1}{\sum_j \pi_{i,j}q_j^{t-1}}
\end{equation}
and then
\begin{equation}
q_j^t=\frac{n}{m}\frac{1}{\sum_i h_i^t\pi_{i,j}}
\end{equation}

\citet{Soules:1991vt} demonstrated convergence at a geometric rate.
Once the algorithm converged, the ranking is defined as:
\begin{equation}
r_i = \frac{\lambda}{h_i}
\end{equation}
where $\lambda$ is chosen so that $\sum_i r_i =1$.

\subsection{An application: actuaries model aggregations}

Let's imagine an insurance supervisor $D$ want to understand the status of an insurance market segment. $D$ aggregates the estimate of $k$ insurance experts. Let's imagine that the underlying risk is a Pareto distribution of exponent $\alpha=1.5$ and mixed with a normal law. Two cases are presented: one where all experts have access to the same data set, another case where each expert has access to an independent data set. With a linear opinion pool, one can test if the independent case is better suited.

\section{R: Panel Data - Basics}

The following is greatly inspired by the work of \citep{hurlin}.

\subsection{Homogeneity tests}

We assume that we have $N$ individual processes $i$ that span over a time horizon $T$. We can first assume that we have the following unrestricted model:
\begin{equation}\label{eq:unconstrained}
\forall i, \forall t, \quad y_{it} = \alpha_i + \beta'_i x_{it} + \epsilon_{it}
\end{equation}

The reason to work with panel analysis on data, is to assume that the data generation processes for the individuals have some homogeneity. If there are completely heterogeneous, then panel data analysis has no sense. In other words, we will apply some homogeneity tests to our data, in order to check whether there is some homogeneity in $\alpha$ or $\beta$.

\subsubsection{Homogeneity test 1: $H_0^1$}
We test with a Fisher statistics:

$H_0^1$ : $\beta_i = \beta$  and $\alpha_i = \alpha, \forall i$

If we accept $H_0^1$, then we need to estimate the fully restricted model, the panel is homogeneous:
\begin{equation}\label{eq:constrained}
y_{it} = \alpha + \beta' x_{it} + \epsilon_{it}
\end{equation}

The test statistics is:
$$F_1=\frac{\frac{SCR_{1,c}-SCR_1}{(N-1)(K+1)}}{\frac{SCR_1}{NT-N(K+1)}}$$
With:
\begin{itemize}
\item $N$: the number of individuals
\item $K$: the dimension of $\beta$
\item $SCR_{1,c}$: the sum of squared residual of the equation of the constrained equation \ref{eq:constrained}
\item $SCR_1$: the sum of squared residual of the equation of the unconstrained equation \ref{eq:unconstrained}
\end{itemize}

\quad

If we accept the fully restricted model, this is quite powerful in terms of implications. For each process $i$ we have the same sensitivity and effects.

\quad

If we reject $H_0^1$, then we move on to the next test:

\subsubsection{Homogeneity test 2: $H_0^2$}

We test with a Fisher statistics:

$H_0^2$: $\beta_i = \beta, \forall i$

The test statistics is:
$$F_2=\frac{\frac{SCR_{2,c}-SCR_1}{(N-1)K}}{\frac{SCR_1}{NT-N(K+1)}}$$
With:
\begin{itemize}
\item $SCR_{2,c}$: the sum of squared residual of the equation of the constrained equation \ref{eq:constrained2}
\end{itemize}

If we reject $H_0^2$, then there is no real case for panel data analysis as we can only work with the unrestricted model.

If we accept $H_0^2$, then we move on to the next test:

\subsubsection{Homogeneity test 3: $H_0^3$}

We test with a Fisher statistics:
$H_0^3$: $\alpha_i = \alpha, \forall i$

The test statistics is:
$$F_3=\frac{\frac{SCR_{1,c}-SCR_{2,c}}{(N-1)}}{\frac{SCR_{2,c}}{N(T-1)-K}}$$

If we reject $H_0^3$, then we need to estimate the \textbf{fixed effects} model:
\begin{equation}\label{eq:constrained2}
y_{it} = \alpha_i + \beta' x_{it} + \epsilon_{it}
\end{equation}

The $\alpha_i$ capture the effects of each process $i$ which is constant over time.

In the case where we would reject $H_0^2$ and not reject $H_0^3$, then we do not have \emph{a} model for this panel but $N$ models.

\section{R: Panel Data - Strikes}
This section has the code: \verb|strikespanel_vansteenberghe.R|

We follow \citep{hurlin} and the data collection from Bruce Western\footnote{available \href{http://lib.stat.cmu.edu/datasets/}{here}} on $s_{it}$ the number of days where there is a strike for 1 000 workers for country $i$ at date $t$. We link this to the unemployment rate $u$ and the inflation $p$:
\begin{equation}
s_{it} = \alpha_i +  \beta_i u_{it} + \gamma_i p_{it} + \epsilon_{it}
\end{equation}

We find that the panel data is cylindrical for 17 countries over the period 1951-1985.

\subsection{Restricted model}
We estimate a first fully restricted model:
\begin{equation}
s_{it} = \alpha +  \beta u_{it} + \gamma p_{it} + \epsilon_{it}
\end{equation}

We obtain:

\begin{tabular}{rl}
\textbf{Variable}&\textbf{Estimate}\\
Constant&95.08\\
unemployment    &27.44 \\
inflation     &  18.61 \\
\end{tabular}

\subsection{Within estimator, Fixed effects}
We obtain the Within estimator with an ordinary least square (OLS) on equation \ref{eq:constrained2}

We can also estimate the model:
\begin{equation}
s_{it}-\bar{s}_{i} = \beta (u_{it}- \bar{u}_{i}) + \gamma (p_{it} - \bar{p}_{i}) + \epsilon_{it}
\end{equation}

This is also called Least-Square Dummy Variable (a dummy $\alpha_i$ per country). From this, the individual effect $\alpha_i$ can be estimated via:
\begin{equation}
\hat{\alpha}_i = \bar{s}_{i} - \hat{\beta}\bar{u}_{i} - \hat{\gamma} \bar{p}_{i}
\end{equation}

Within estimator, of Fixed Effect, assume that each individual $i$ (each country here) has a specific characteristic that do not vary over time. We remove this specific characteristic (or fixed effect) to assess the net effect on the independent variable on the dependent variable. For example, France might be more prone to strikes because of its labour law that we assume don't change over the time period of observation, we want to control for this effect and only focus on the effect of unemployment and inflation on strikes.

\quad

The fixed effect is also called within as it explains the difference within each individual $i$ between $s_{it}$ and $\bar{s}_{i}$, but it does not suggest any explanation on the difference between $\bar{s}_{i}$ and $\bar{s}_{j}$.

\quad

We obtain:

\begin{tabular}{rl}
\textbf{Variable}&\textbf{Estimate}\\
Constant& Not estimated \\
unemployment    &-21.60\\
inflation     &  16.27 \\
\end{tabular}

\quad

From the OLS to the Fixed Effect regression, we have results in Table \ref{table:olsFE} and the sensitivity to unemployment changes sign.

\begin{table}[!ht]
\caption{}
\label{} 
\begin{tabular}{ l D{.}{.}{2}D{.}{.}{2} } 
\hline 
  & \multicolumn{ 1 }{ c }{ OLS } & \multicolumn{ 1 }{ c }{ Fixed Effects } \\ \hline
(Intercept)       & 95.08 ^*      &              \\ 
                  & (43.14)       &              \\ 
unemployment      & 27.44 ^*      & -21.60 ^*    \\ 
                  & (7.54)        & (9.19)       \\ 
inflation         & 18.61 ^*      & 16.27 ^*     \\ 
                  & (4.99)        & (4.76)       \\ 
factor(country)B  &               & 355.49 ^*    \\ 
                  &               & (94.68)      \\ 
factor(country)BA &               & 1012.66 ^*   \\ 
                  &               & (108.41)     \\ 
factor(country)BB &               & 114.22       \\ 
                  &               & (90.18)      \\ 
factor(country)BC &               & 27.50        \\ 
                  &               & (93.26)      \\ 
factor(country)BD &               & 164.24       \\ 
                  &               & (92.72)      \\ 
factor(country)BE &               & 14.44        \\ 
                  &               & (90.72)      \\ 
factor(country)BF &               & 14.94        \\ 
                  &               & (91.68)      \\ 
factor(country)BG &               & -44.85       \\ 
                  &               & (86.87)      \\ 
factor(country)BH &               & 284.45 ^*    \\ 
                  &               & (95.24)      \\ 
factor(country)BI &               & 495.87 ^*    \\ 
                  &               & (99.80)      \\ 
factor(country)C  &               & -2.58        \\ 
                  &               & (90.71)      \\ 
factor(country)E  &               & 801.99 ^*    \\ 
                  &               & (102.27)     \\ 
factor(country)F  &               & 211.06 ^*    \\ 
                  &               & (102.28)     \\ 
factor(country)G  &               & 387.84 ^*    \\ 
                  &               & (93.68)      \\ 
factor(country)H  &               & 339.07 ^*    \\ 
                  &               & (94.37)      \\ 
factor(country)I  &               & 57.54        \\ 
                  &               & (90.56)      \\ 
factor(country)J  &               & 582.62 ^*    \\ 
                  &               & (113.77)      \\
 $N$               & 595           & 595          \\ 
$R^2$             & 0.05          & 0.41         \\ 
adj. $R^2$        & 0.05          & 0.39         \\ 
Resid. sd         & 557.26        & 505.11        \\ \hline
 \multicolumn{3}{l}{\footnotesize{Standard errors in parentheses}}\\
\multicolumn{3}{l}{\footnotesize{$^*$ indicates significance at $p< 0.05 $}} 
\end{tabular} 
\label{table:olsFE}
\end{table}
 
 \subsection{Between estimator}
\emph{Within} refers to the variation within one individual (over time), \emph{between} measures the variation between the individuals. The variance of the dependent variable may be decomposed into the sum of variance \emph{within} and the variance \emph{between}.
We obtain the Between estimator with an ordinary least square (OLS) on the following:
\begin{equation}
\bar{s}_{i} = \alpha +  \beta \bar{u}_{i} + \gamma \bar{p}_{i} + \epsilon_{i}
\end{equation}

With the bar operator the average over the time period for each country.

We obtain:

\begin{tabular}{rl}
\textbf{Variable}&\textbf{Estimate}\\
Constant&-341.55 \\
unemployment    &80.95\\
inflation     &  59.39 \\
\end{tabular}

\stepcounter{question}
\begin{itemize}
\item \textbf{Question \arabic{question}}
\begin{itemize}
\item Plot a visual for the between estimates, simplifying with one independent variable.
\end{itemize}
\end{itemize}

\subsection{Random effects}
The random effect estimator can be understood as an efficient combination of within and between dimencions\footnote{cf. for example \citep{verbeek}}.

\quad

In a random effect model, in which the variation across entities is assumed to be random and uncorrelated with the predictor or independent variables included in the model:
\begin{equation}
s_{it} = \alpha +  \beta u_{it} + \gamma p_{it} + \nu_{it} + \epsilon_{it}
\end{equation}
With:
\begin{itemize}
\item $\nu_{it}$ the between-entity error
\item $\epsilon_{it}$ the within-entity error
\end{itemize}

\subsection{Tests and model selection}

\subsubsection{Economic approach}
If individuals $i$ are randomly selected from a pool, we might want to apply random effect. If individuals $i$ are countries, which make each unique and particular, we might want to stick to fixed effect, as we can consider that each country has unobserved characteristics.

\subsubsection{Econometric test approach}
We can perform various econometric tests to help the decision making between several models.

Hausman test between the OLS and the fixed-effect model.
Hausman test between the fixed-effect and the random-effect model.
Hausman test to select or not time-fixed effect.

Based on the tests results, we estimate the fixed-effects model:
\begin{equation}
s_{it} = \alpha_i +  \beta u_{it} + \gamma p_{it} + \epsilon_{it}
\end{equation}

\subsection{Generalized Least Square}
\stepcounter{question}
\begin{itemize}
\item \textbf{Question \arabic{question}}
\begin{itemize}
\item Apply Generalize Least Square model, test for different model of autocorrelation
\end{itemize}
\end{itemize}

\section{R: Panel Data - Augmented Solow Model HOMEWORK}
This section has the code: \verb|productionpanel_vansteenberghe.py|
In the paper \citep{Arnold:2011aa}, the test an augmented Solow model:
\begin{equation}
Y(t)=K(t)^\alpha H(t)^\beta[A(t)L(t)]^{1-\alpha-\beta}
\end{equation}
with:
\begin{itemize}
\item $K$ the physical capital
\item $H$ the human capital
\item $L$ the labour
\item $A$ the level of technology
\end{itemize}

\subsection{Pooled mean group approach: PMG}
Following several transformation as detailed in the paper \citep{Arnold:2011aa}, they estimate the model:
\begin{equation}
\Delta \ln y_{it} = -\Phi_i [ \ln y_{i,t-1} - \theta_1 \ln s_{kit} - \theta_2 \ln h_{it} +\theta_3 n_{it} -a_{4i} t -\theta_{0i}]+ b_{1i} \Delta\ln s_{kit} - b_{2i} \Delta \ln h_{it} +b_{3i} \Delta n_{it} +\epsilon_{it}
\end{equation}

With some non-linear restrictions:

$$\frac{a_{ki}}{\Phi_i}=\theta_{k..}$$

\subsection{Data}
The data used are on 21 OECD countries, from 1971-2004.
\begin{itemize}
\item loggdp: log real GDP per head of population aged 15-64 years, expressed in national currency
\item loginv: logarithm of ratio of real non residential fixed capital formation to real GDP
\item logpop: log of working age population (15-64 years)
\item loghcap: log of of the average number of years of schooling of the population aged between 25 and 64 years
\end{itemize}

\subsection{Paper replication}
\stepcounter{question}
\begin{itemize}
\item \textbf{Question \arabic{question}}
\begin{itemize}
\item Use the OECD online database to reproduce and actualize the time series with newest data.
\item Reproduce the results and PMG approach from \citep{Arnold:2011aa}.
\end{itemize}
\end{itemize}

\section{R: Panel Data - Technology Diffusion HOMEWORK}
This section has the code: \verb|20190219_technodif_panel.R|

\subsection{The model}
This section is inspired by the paper \citep{Hultberg:1999aa}, where they suggest a dynamic model that extends the neoclassical growth model by including technology diffusion as \begin{quote}
The existence of a technology gap may therefore present an additional opportunity for growth through technology flows.
\end{quote} and possible inefficiency caused by institutional rigidities as \begin{quote}
nations differ in ability to adopt and absorb new knowledge.
\end{quote} And where only technology diffuses but physical capital and labour are immobile.

They start with the standard neoclassical model with a Cobb-Douglas production function:
\begin{equation}
Q_{it}=A_{it} K_{it}^{\beta_1} L_{it}^{\beta_2} H_{it}^{\beta_3}
\end{equation}

where output $Q$ depends on technology $A$, physical capital stock $K$, employment $L$, and human capital $H$ for country $i$ at time $t$.

They further assume: $A_{it} = A_{i0}e^{\gamma t}$ and $L_t = L_0 e^{nt}$, and the evolution of capital:
\begin{equation}
\dot{K}_{it} = s Q _{it} - \delta K_{it} +\xi_{it} K_{it}
\end{equation}

with $\xi_{it}=\rho_i \ln\left(\frac{Y^*_{i,t-1}}{Y_{i,t-1}} \right)$ the technology catch up term (the main novelty of this paper), $Y_{i,t}=\frac{Q_{i,t}}{L_{i,t}}$ the inverse ratio of labour productivity, and $Y^*$ the desired level of this ratio (or attainable level for each country). In fact this is determined by the country's level of efficiency $E_{it}$ as this efficiency is considered constant, this amounts to say:
$$ Y^*_{it} = \frac{Y_t^L}{E_i}$$ with $Y_t^L$ the leader's productivity (the U.S.A.'s).

\quad

By log-linearisation and first-differencing\footnote{note that the efficiency of the country, being constant is captured in $\phi'$}
\begin{equation}
y_{it} = \phi' + \beta_1 k_{it} + \beta_2 l_{it} + \beta_3 h_{it} +\rho_{i} \left[\ln Y^L_{i,t-1} - \ln Y_{i,t-1}  \right]
\end{equation}

\subsection{Data}

We focus on the Penn-World Tables version 9, available \href{http://www.rug.nl/ggdc/productivity/pwt/}{here} with more information in the paper \citep{pwt}, with information on relative levels of income, output, input and productivity, covering 182 countries between 1950 and 2014.

\subsection{Simplified application}
For this section, we download the latest Penn-World Tables.

From that paper, we suggest a simplistic application where we overlook some subtleties to focus on the panel data theory.

For the variables, we use for the:
\begin{itemize}
\item $y$, GDP per worker: rgdpna / emp
\item $k$, capital stock: ck
\item $l$, employment: emp
\item $h$, human capital: hc
\end{itemize}

With the following definitions:
\begin{itemize}
\item rgdpo: output-side real GDP at chained PPPs in million 2011 USD (to compare productivity capacity between countries, think "deflated by a producer price index")
\item rgdpna: real GDP using national-accounts growth rates, for studies comparing (output-based) growth rates across countries
\item pop: population (in million)
\item ck: capital stock at current PPPs in million 2011 USD
\item emp: number of persons engaged in millions
\item avh: average annual hours worked by persons engaged (we consider that there are 2 080 working hours in a year)
\item hc: human capital index, based on years of schooling and returns to education
\end{itemize}

\stepcounter{question}
\begin{itemize}
\item \textbf{Question \arabic{question}}
\begin{itemize}
\item Review the definition of the \citep{pwt} variables and discuss their correspondence for the model in \citep{Hultberg:1999aa}.
\item As in the paper, apply log linearization and differencing.
\end{itemize}
\end{itemize}

\subsubsection{OLS model}

We apply a simple OLS model. 
\begin{equation}
y_{it}  = \alpha + \beta_1 k_{it} + \beta_2 l_{it}+ \beta_3 h_{it} + \epsilon_{it}
\end{equation}

\subsubsection{Fixed effect model}

We demean our model (or use directly the fixed effect estimates):
\begin{equation}
y_{it} - \bar{y} = \beta_1 (k_{it} - \bar{k}) + \beta_2 (l_{it}-\bar{l}) + \beta_3 (h_{it}-\bar{h})
\end{equation}

Our results are different from the paper, note that the time periods differ and we made simplified assumptions on the variables:

\begin{tabular}{lrr}
Variable & Our coefficients & Original paper's\\
Constant & NA & $-0.26$\\
Capital stock & $0.04$ & $0.03$\\
Employment & $0.29$ & $0.32$\\
Human capital & $0.17$ & $0.05$\\
$R^2$&$78\%$ & $13\%$	\\
\end{tabular}

%
%
%

\subsection{Two-stage least squares - 2SLS}
In the paper, they apply a two-stage least squares instrumenting the technology gap variable with its lagged value.

\subsection{Generalized Method of Moments - GMM}

\citep{Arellano:1991aa} estimate a dynamic model from panel data by the generalized method of moments. In the section 5, "AN APPLICATION TO EMPLOYMENT EQUATIONS", they suggested a model based on GMM, where the employment of a firm $n_i$ is dependent on the wage $w_i$, the capital $k_i$ and the output $y_i$:
\begin{equation}
n_{i,t} = \alpha_1 n_{i,t-1}+\alpha_2 n_{i,t-2}+\beta'(L) x_{i,t,t-1,t-2}+\lambda_t+\eta_i+\nu_{i,t}
\end{equation}
with $x_{i,t,t-1,t-2} = [w_{i,t} w_{i,t-1} k_{i,t} k_{i,t-1} k_{i,t-2} y_{i,t} y_{i,t-1} y_{i,t-2}]$

We implement\footnote{an interesting introduction to the methods can be found in \emph{GMM Estimation of Empirical Growth Models} by Bond et al. 2001} \citep{Arellano:1991aa} Table 4 equation (a2) and find very similar results. Then we add some instruments (wage and capital and reproduce Table 4 equation (d).

\stepcounter{question}
\begin{itemize}
\item \textbf{Question \arabic{question}}
\begin{itemize}
\item Implement a GMM to replicate \citep{Hultberg:1999aa} Table 3.
\end{itemize}
\end{itemize}

%

\section{Networks study}

The theory applied to financial networks to study systemic risk historically stems from three domains: ecology, epidemic, and social network. That research contributed to help understanding ecosystem dynamics, optimizing drug posology in a network, maximizing information sharing, and controlling for the spread of rumors (or fake news). First, the ecology literature will help a regulator that is interested in managing diversity and/or interactions in a network if it judges that they either promotes or degrade the stability of the ecosystem. Second, the epidemic literature will help a regulator take preventive action to restrain the risk of shocks amplification and propagation in a network. Finally, the social network literature can help identify nodes that influence the others in a network while adapting the approach to the features of a financial network.

First, the bridges between ecology and financial network stability are detailed in \citep{Haldane:2011aa} where the authors derive the notion of systemic risk in banking ecosystems from earlier work of ecologists in 1960s such as Evelyn Hutchinson suggesting that \begin{quotation} oscillations observed in arctic and boreal fauna may be due in part to the communities not being sufficiently complex to damp out oscillations. \end{quotation} Also \begin{quotation} a closer examination of model ecosystems showed that a random assembly of $N$ species, each of which had feedback mechanisms that would ensure the population's stability were it alone, showed a sharp transition from overall stability to instability as the number and strength of interactions among species increased. \end{quotation} The author juded that \begin{quotation} Even less effort has been put into providing regulatory incentives to promote diversity of balance sheet structures, business models and risk management systems. In rebuilding and maintaining the financial system, this systemic diversity objective should probably be given much greater prominence by the regulatory community. 
Second, modularity within the financial system. The structure of many non-financial networks is explicitly and intentionally modular. This includes the design of personal computers and the world wide web and the management of forests and utility grids. Modular configurations prevent contagion infecting the whole network in the event of nodal failure. By limiting the potential for cascades, modularity protects the systemic resilience of both natural and constructed networks.
The same principles apply in banking. That is why there is an ongoing debate on the merits of splitting banks, either to limit their size (to curtail the strength of cascades following failure) or to limit their activities (to curtail the potential for cross-contamination within firms).\end{quotation} 

The authors point out that \begin{quotation}
In financial ecosystems, evolutionary forces have often been survival of the fattest rather than the fittest.
\end{quotation}

Second, to study and better understand epidemic, different measures have been suggested and new measures are still being developed. This field study how influential nodes can initiate and conduct the dissemination of diseases more efficiently than normal users. In \citep{Wen:2017aa}, the authors review the evolution of the suggested measures. Starting from works in the 1960s as in \citep{Sabidussi:1966aa} about the degree of node which count the number of edges going in or out of a node, but weighing every neighbor equally, the eigenvector centrality was designed in the 1970s as in \citep{Bonacich:1972aa} and weights the neighbors according to their importance. Another class of centrality emerged in the late 1970s as in \citep{Freeman:1977aa} with the betweenness which is defined as the fraction of shortest paths between node pairs that pass through the node of interest. The betweenness centrality measure has been modified and improved to better explain the observation of epidemic, with flow or random-walk betweenness, and more recently in \citep{Wen:2017aa} with the epidemic betweenness which is the expected number of nodes that are influenced directly or indirectly by a node after that node becomes influenced by epidemics, averaged over all possible incidents. The main issue when trying to directly apply those measures to financial networks, limitation that can also be found in the original DebtRank \citep{Battiston2012aa}, is that a node cannot continue an epidemic once it has already been influenced (in other words, one can only get infected once). Hence a node has only two states: susceptible or infected, and infected has two substates: contagious or dormant. With dormant being an absorbing state. In a financial network, shocks can spread in circles and hence be amplified passing several times through a given node.

Finally, spreading processes of information (rumor or knowledge for example) have been studied also with the aim of identifying influential nodes in a social network. In \citep{Xu:2017aa}, they point out the limitation of the closeness and betweenness measures which are based on shortest path. Another approach is to study the node position with respect to its peers, identified either by its coreness or k-core decomposition, as an indicator of epidemic spreading. Interestingly, \citep{Lu:2016aa} show that degree, H-index and coreness are the initial, intermediate and steady state of contagion process iterations in networks, but the low resolution of thos measures limit their applications. Another approach, matrix based, is derived from the eigenvector and eigenvalue computation as in PageRank, a measure that was formally introduced in \citep{Brin:1998}. This measure was modified with a random walk process to become LeaderRank \citep{Lu:2011aa} that has a faster convergence. \citep{Xu:2017aa} suggested a new algorithm called adaptive LeaderRank that is more robust to topological perturbations. The success of search engines based on PageRank-type measure is not directly applicable to financial network as the resilience of the nodes is not taken into account and the computation of the scores is iterative.

\subsection{Preliminary definitions}

We define a bilateral matrix $B$ with $B_{i,j}$ elements, with a number of $N$ financial groups. We record financial link between two nodes with $U$ (the adjacency matrix), that is $U_{i,j}=1$ if $B_{i,j} \neq 0$

\quad

$B_{i,j}$ represents in volume the contractual obligation that group $i$ owes to group $j$. This contract can be a loan, short-term funding, insurance or reinsurance contract. In any case, if a negative shock impacts group $i$, via its contract, group $j$ will also be impacted. In the case of a loan it would be a higher risk of default of $i$ and/or payment delays, in the case of a reinsurance contract it would be a shock affecting $i$'s capacity to provide reinsurance with due diligence. More generally, if $i$ becomes riskier, its risk spreads to $j$ via their connections.

\quad

We also define $\Pi$ the relative liabilities matrix 
\[
    \Pi_{i,j}= 
\begin{cases}
    \frac{B_{i,j}}{\sum_j B_{i,j}},& \text{if } \geq 0\\
    0,              & \text{otherwise}
\end{cases}
\]

By means of those matrices we will measure which group is important to the network or fragilized by the network.

\subsection{In-degree}

In-degree is the number of links that are going "into" a node $k_i^{in}$, divided by the maximum attainable degree. The formula for the in-degree of group $i$ is:

\begin{equation}
In\_deg_{i} = \frac{\sum_j U_{j,i}}{N-1} = \frac{k_i^{in}}{N-1}
\end{equation}

If the edges are lending, then $In\_deg$ represents the share of $i$'s borrowers over the total possible number of borrowers. It can be assimilated to the concentration of $i$'s lending. 

\subsection{Out-degree}

Out-degree is the number of links that are going "out of" a node $k_i^{out}$, divided by the maximum attainable degree.

The formula for the out-degree of group $i$ is:

\begin{equation}
Out\_deg_{i} = \frac{\sum_j U_{i,j}}{N-1} = \frac{k_i^{out}}{N-1}
\end{equation}

Unlike the $In\_deg$, the $Out\_deg$ estimate how $i$ manage its funding. 

\subsection{Closeness}

Closeness measures "how close" a node is from the rest of the nodes, it is proportional to the inverse of the average distance from one node to the rest of the network. Here the distance is measured by the minimum number of edges necessary to reach a given node, i.e. the shortest path. It can be demonstrated that the shortest path between group $i$ and group $j$ can be measured in a matrix form as:
\begin{equation}
\delta_{i,j} = \min_p \left[ U^p \right]_{i,j} > 0
\end{equation}

The closeness of group $i$ is then:
\begin{equation}
C_{i} = \frac{N-1}{\sum_j \delta_{i,j}}
\end{equation}

This measure estimates how quickly nodes can spread information or stress to the network.

\quad

NB: when faced with nodes that cannot be linked, we can decide to allocate the value $N$ to the shortest path between those two entities, to avoid division by 0.

\subsection{Betweenness Centrality}
The betweenness centrality is a measure of centrality that can be applied on directed and weighted networks, based on shortest paths between nodes.

We define by $\delta_{i,j}$ the minimum length of any path connecting $i$ and $j$, we denote by  $\sigma_{ij}$ the number of shortest paths from $i$ to $j$ and $\sigma_{ij}(k)$  the number of shortest paths from $i$ to $j$ where $k$ is on the path. The betweenness centrality of a node $k$ is defined as:

\begin{equation}
BC_{k} = \sum_{i \neq k \neq j} \frac{\sigma_{ij}(k)}{\sigma_{ij}}
\end{equation}

Betweenness Centrality measures node's importance as link in the shortest connection between two nodes. Regarding this measure in financial networks, we could assimilate an agent with high score to an agent which may have high probability to amplify shocks (concept of financial hub). We have to note that an agent could have a low betweenness centrality score and still be a threat to a network, one could think of a heavy and isolated node with a unique link to a central node.

We followed \citep{Brandes:2001aa} for the implementation of the algorithm.

Take the simple example of an exposure matrix:
\begin{center}
\includegraphics[scale=0.4]{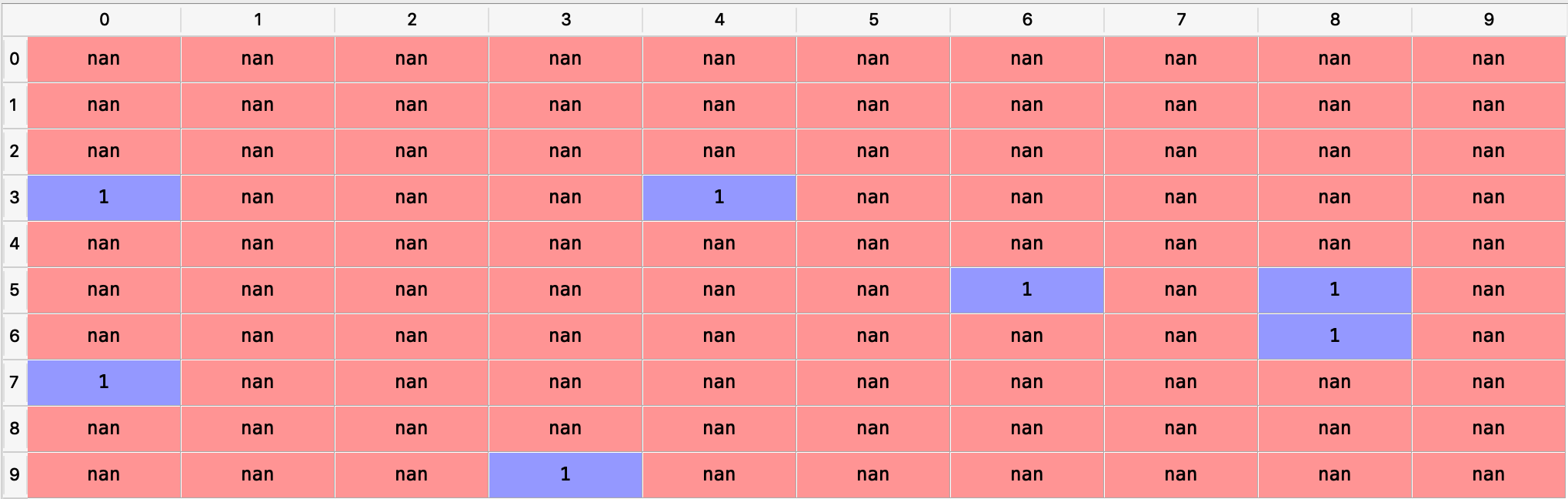}
\end{center}

We can identify the following paths:
\begin{itemize}
\item $3 \to 0$
\item $3 \to 4$
\item $5 \to 6$
\item $5 \to 6 \to 8$
\item $5 \to 8$
\item $6 \to 8$
\item $7 \to 0$
\item $9 \to 3$
\item $9 \to 3 \to 0$
\item $9 \to 3 \to 4$
\end{itemize}

So in fact, the "only" and thus "shortest" way to go from $9$ to $0$ and from $9$ to $4$ goes twice through $3$, hence the edge $3$ will have a betweenness centrality score of $2$, but if it is possible to go from $5$ to $8$ via $6$, it is in fact shorter to go from $5$ directly to $8$ because in this case all weights are 1.
Hence in this example all scores are $0$ except for the edge $3$ it is $2$.

Now we illustrate "competing" paths, with the code 20190809\_betweenness\_centrality.py, where we use the matrix:
\begin{center}
\includegraphics[scale=0.4]{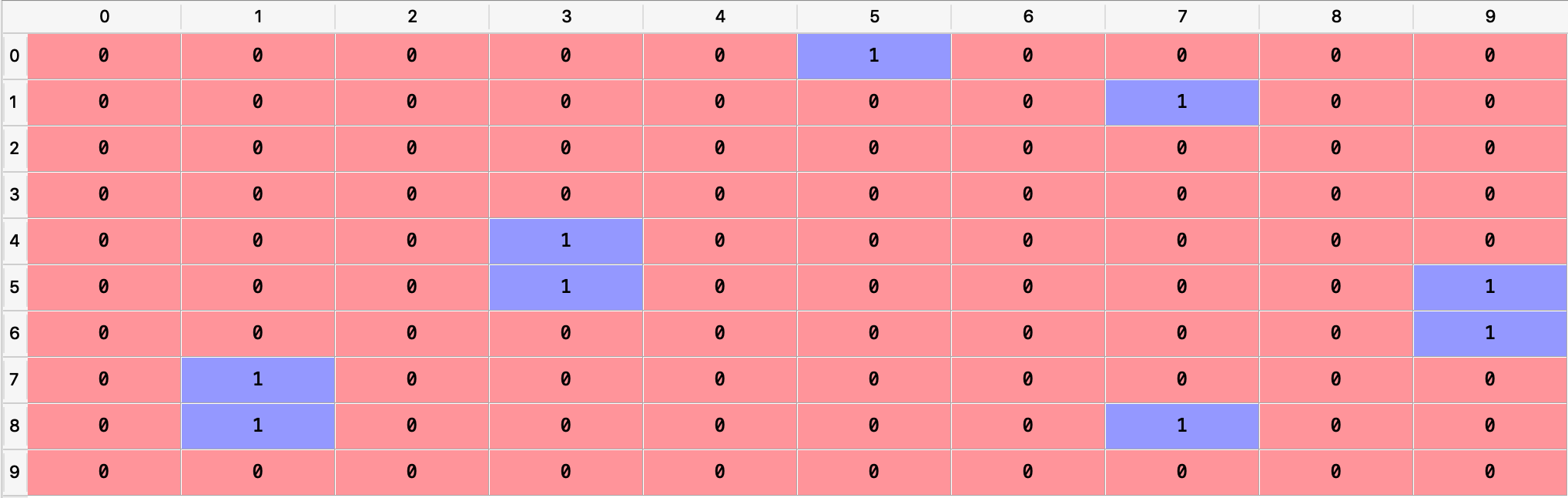}
\end{center}

We observe the following "competing" paths:
We can identify the following paths:
\begin{itemize}
\item $7 \to 0 \to 2$
\item $7 \to 9 \to 2$
\item $8 \to 2 \to 0$
\item $8 \to 7 \to 0$
\end{itemize}

As we are working with an adjacency matrix, weights are uniform and thus the vertices $0$, $9$, $2$ and $7$ only gain $0.5$ points with the four above mentioned paths, thus the final scores with $.5$

As an exercise, we implemented a naive algorithm that explore most paths, but we did not implement the check if paths have the same length in which case the score needs to be split in $.5$ and $.5$.

\subsubsection{Weighed Betweenness Centrality}

The Betweenness Centrality measure can also be weighed. The higher the weight of a path, the least likely it is to be included as a shortest path between two nodes. That is the distance is not only the sum of the number of edges between to vertices, but the sum of the weights of the edges between two vertices.

For our purpose, measuring the importance of a financial institution in a network, this measure is counter-intuitive and we are working with the inverse of the lending volumes.

With the example of this weighted vertices matrix given in code 20190812\_weighed\_betweennes\_centrality\_tests.py following the code modifications in \citep{Brandes:2008aa}:
\begin{center}
\includegraphics[scale=0.4]{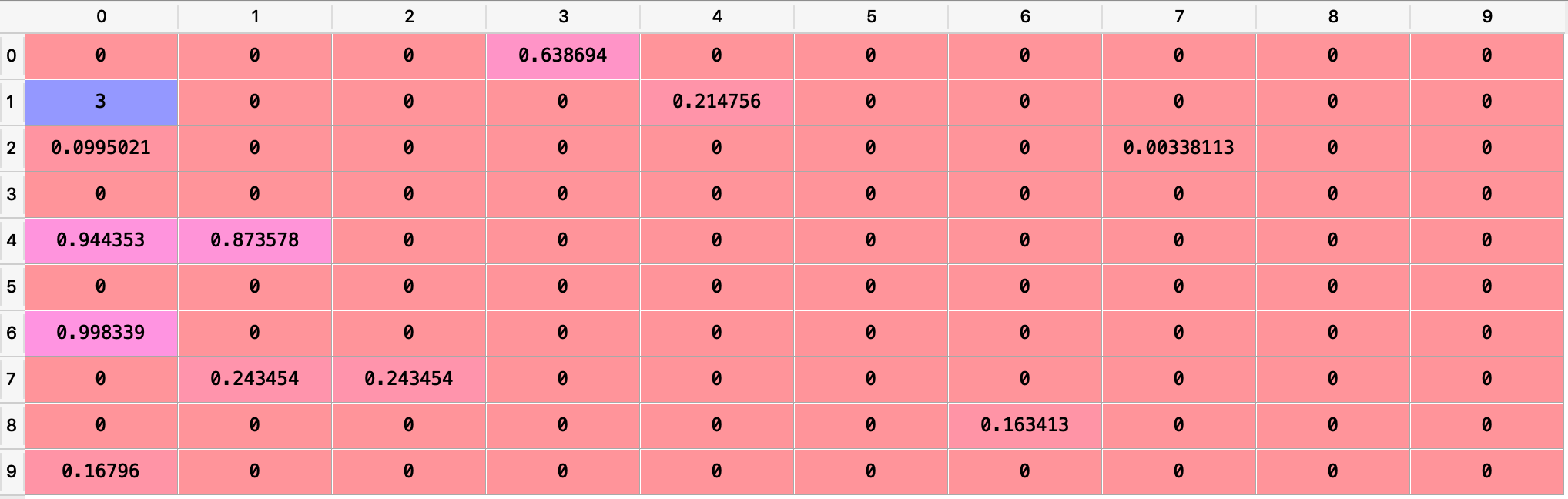}
\end{center}

We can identify the following \textbf{shortest} paths with at least two edges:
\begin{itemize}
\item $1 \to 4 \to 0$, weight $1.16$
\item $1 \to 4 \to 0 \to 3$
\item $2 \to 0 \to 3$
\item $2 \to 7 \to 1$
\item $2 \to 7 \to 1 \to 4$
\item $4 \to 0 \to 3$
\item $6 \to 0 \to 3$
\item $7 \to 1 \to 4$
\item $7 \to 2 \to 0$
\item $7 \to 2 \to 0 \to 3$
\item $8 \to 6 \to 0$
\item $8 \to 6 \to 0 \to 3$
\item $9 \to 0 \to 3$
\end{itemize}

Nota bene:
\begin{itemize}
\item $1 \to 4 \to 0$ with weight $1.16$ is shorter than the "direct" path $1 \to 0$ with weight $3$
\item similarly, $1 \to 4 \to 0 \to 3$ is shorter than $1 \to 0 \to 3$
\item $7 \to 1 \to 0$ with a weight of $3.24$ is not considered as the competing path $7 \to 2 \to 0$ with a weight of $0.33$ is shorter.
\end{itemize}

To have our weighted-betweenness-centrality score, we have to count how many time an edge lies on a shortest path:
\begin{itemize}
\item $0$, 7 times
\item $1$, 2 times
\item $2$, 2 times
\item $4$, 2 time
\item $6$, 2 times
\item $7$, 2 times
\end{itemize}

In fact, the current algorithm has we've implemented from \citep{Brandes:2008aa} seems to discriminate between $7 \to 1 \to 0$ and $7 \to 2 \to 0$, attributing a score to the vertice $2$ but not $1$, but fail to recognize "backward\footnote{by backward we mean that the algorithm fails to recognize that a weighed path of length $n+1$ is in fact shorter than a weighed path of length $n$}" that $1 \to 4 \to 0$ is shorter than $1 \to 0$ and thus fail to attribute scores to the vertice $4$.

The betweenness\_centrality function from networkx in its weighed version, weight="weight", provides the correct results.

One approach we suggest is to add before the algorithm a check of paths that have more edges but are lighter in sum of weight than shorter paths. We implement a function that checks if there are some to paths up to length $n$ that are shorter than the original paths in the matrix (of length $1$). If we identify such cases, we discard the cell in the original matrix before applying the weighed betweenness centrality algorithm.

\stepcounter{question}
\begin{itemize}
\item \textbf{Question \arabic{question}}
\begin{itemize}
\item Using the adjacency matrix up to power $n$ indicates whether there exist path of with $n$ edges between vertices $i$ and $j$ of the original matrix. Use this information to modify the algorithm in order to loop only on vertices where paths are expected. Time the execution, is the execution time improved compare to the approach above?
\end{itemize}
\end{itemize}

\subsection{H-index and coreness (k-core decomposition)}
The Hirsch index, or H-index, was originally used to measure the citation impact of a scholar, the H-index is the maximum value $h$ such that there exists at least $h$ papers, each with citation count $\geq h$. In \citep{Lu:2016aa}, they show that the H-index converge to the coreness. 

The coreness of a node is measures by k-core decomposition as in \citep{kcore}. K-core of graph measures the subgraph where each node has a minimum of $k$ connections with the subgraph (In or out-degree). It could be crudely linked to density of certain subgraph of the network. We can measure the importance of each node by using the k-core to estimate the maximum $k$ for each node. The higher $k$ value of node $i$ means that this node $i$ is connected to higher interlinked core on the network.

\subsection{Clustering}
This measures of unsupervised classification aims to estimate for a given node $i$ the connection of its neighbors between one another. It is the fraction of possible triangles through that node over the possible triangles:

\begin{equation}
C_{i} = \frac{2}{\sum_j U_{i,j} (\sum_j U_{i,j} - 1)} \times \left( \sum_{j,k \in N_i}  U_{j,k} \right)
\end{equation}

Where $N_{i}$ are the neighbors of the node $i$, if $N_{i} \leq 1$ then $C_i = 0$

For directed graphs, which is the use case for financial networks, we have to modify this measure to:
\begin{equation}
C_{i} = \frac{1}{\sum_j U_{i,j} (\sum_j U_{i,j} - 1)} \times \left( \sum_{j,k \in N_i}  [U_{j,k} + U_{k,j}] \right)
\end{equation}

\subsection{Assortativity}

Assortative mixing in networks, as in \citep{Peel:2018aa}, measure the tendency for nodes to connect to nodes that are like (or unlike) them with respect to some characteristics (we could consider the business model or the size of the financial institution). We do not consider this measure in this work as we could not establish a link to financial stability and assortativity.

\subsection{Eigenvector centrality}

This centrality is calculated by iteration, all nodes start with the same score and, by iteration, node scores are assigned relatively to the scores of their neighbors. 

\begin{equation}
EV_{i} = \frac{1}{\lambda} \times \sum_{j \neq i}B_{i,j} EV_{j}
\end{equation}

$\lambda$ being the largest eigenvalue of the matrix $B$.

\subsection{LeaderRank}

The LeaderRank algorithm presented in \citep{Lu:2011aa} aims to improve\label{LR} the way to measure nodes importance in networks. The authors introduced the idea of leaders and fans in the network. The score of each node is computed regarding how many fans (or followers) it has.

In the network of $N$ nodes and $M$ directed links, a ground node, which is connected to every nodes through bidirectional links, is added. This node is useful for the convergence of the recursive scores computation. The network is then formed by $N+1$ node and $M+2N$ links. If node i points to $j$ then $j$ is leader of $i$ (in finance that could be modeled by $i$ lends to $j$ so $i$ reacts regarding to the situation of $j$) 

The recursive formula used to compute scores is:
\begin{equation}
S_{i}(t+1)= \sum\limits_{j=1}^{N+1} \frac{U_{j,i}}{k^{out}_{j}} S_{j}(t)
\end{equation}

\begin{itemize}
\item $\textrm{S}_\textrm{i}$(0) = 1 for all nodes except the ground node where it is equal to 0
\item $\textrm{k}_\textrm{j}\textsuperscript{out}$ is the total of out-degree (in our simulation we use the sum of all outgoing loans)
\end{itemize}

We introduce a modification to this measure to tailor it for financial networks and use the exposition matrix calculated with loans divided by lenders equities.

\subsection{DebtRank}
Closeness and Betweenness Centrality take into account all nodes in a network by measuring "how far" nodes are or the fraction of the shortest paths running through each nodes. Measures based on shortest path may be difficult to justify in a financial contagion context. To quote \citep{Battiston2012aa}:

\begin{quotation}
In presence of a cycle in the network there is an infinite number of reverberations of the impact of a node to the others and back to itself, which leads to no simple and measurable economic interpretation. DebtRank overcomes this problem by only allowing for walks that do not visit the same edge twice.
\end{quotation}

The authors develop a measure inspired by the centrality measures. The method allows to measure impact not limited to those cases in which the capital buffer is exhausted, but where partial stress on solvency is spread through the system. 

The measure is captured by the individual cumulative relative equity loss in time:

$$h_i(t) = \frac{KP_i(0) - KP_i(t)}{KP_i(0)}$$

and the leverage of bank $i$ toward bank $j$ at time $t$:
$$l_{i,j} = \min\left( 1, \frac{B_{j,i}}{KP_i(t)} \right)$$

The dynamic is defined as:
$$h_i(t) = \min \left(  1, h_i(t-1) + \sum_j l_{i,j} h_j(t-1) \right)$$ where $j\vert s_j(t-1) = D$, meaning we sum over the distressed nodes:

\[s_i(t)
    = 
\begin{cases}
    D,& \text{if } h_i(t) > 0 ; s_i(t-1) \neq I,\\
    I,& \text{if } s_i(t-1) = D,\\
    s_i(t-1), & \text{otherwise}             
\end{cases}
\]

The distressed induced in the system is then weighed by the relative exposures:

$$ R = \sum_j h_j(T) \nu_j - \sum_j h_j(1) \nu_j$$

with $\nu_j = \sum_i \frac{B_{i,j}}{\sum_i\sum_j B_{i,j}}$

\quad

In the updated version of the algorithm \citep{Bardoscia2015aa}, the author show that when the modulus of the larger eigenvalue of the leverage matrix $L$ is smaller than one, the dynamics converges to the fixed point $\Delta h = 0$. On the contrary, if there exists an eigenvalue of the leverage matrix greater than one, the initial shock will be amplified and at least one bank will default.

\quad

In practice, in this paper, we implement three DebtRank algorithm:
\begin{enumerate}
\item DebtRank\_rec: an algorithm based on a simple matrix multiplication as detailed in equation \ref{eq:debtrankmat}
\item DebtRank\_dyn: the algorithm based on the original DebtRank paper of 2012
\item DebtRank\_new: the algorithm updated by the 2015 paper
\end{enumerate}

\subsubsection{Matrix notation, tentative}

If we discard the different cases (distress and idle: D and I) for which we question the economic rationale and the cap on the ratio of asset to equity that we question as well, in matrix notation the DebtRank measure would be:
$$H(t) = L(t-1) H(t-1) + H(t-1)$$

If we note: $L^{add}(t) = I + L(t)$:
$$H(rank) = L^{add}(rank-1) \times \ldots \times L^{add}(0) H(0)$$

If the shocks are considered small enough compare to the volumes of core equity, we can approximate that the variation of $L$ with time are negligible:
\begin{equation}\label{eq:debtrankmat}
H(rank) \simeq (L^{add})^{rank-1} H(0)
\end{equation}

\subsection{Google's PageRank (1998) applied to a bank network}

For this section we use the code \verb|PageRank_illustration_vansteenberghe.py|

\subsubsection{See a Page as a bank}

For illustration purpose, think instead of a web page as of a bank. Each bank is lending money to another bank, hence they are linked. A central and first question we want to answer as a financial supervisor is: "who are the 'central' banks of our network?"

Each bank points to another bank by lending, each web page points to another web page with hyperlinks.

\begin{center}
\includegraphics[scale=0.5]{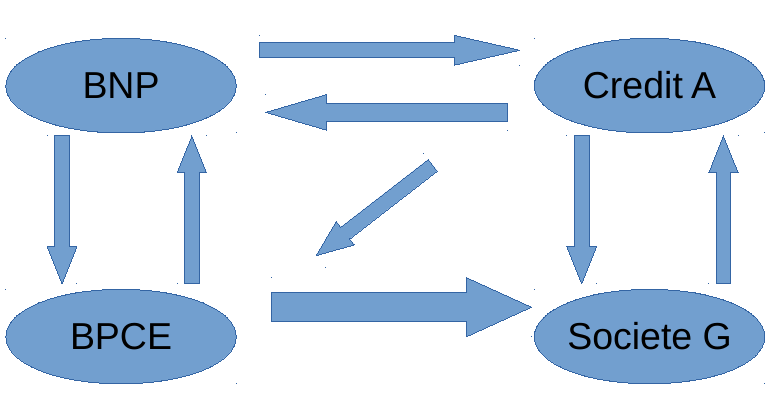}
\end{center}

In this visual example, the thickness of the arrow represent how much a bank lend to another. Here we can see that some bank do not lend to some others. We would like to be able to know from this set up which bank is the most "important" one to the network.

\subsubsection{Variables definitions}

\begin{itemize}
\item $N$ banks (or node)
\item $p_j$ the total assets of bank $j$
\item $p_{i,j}$ payment due from bank $i$ to bank $j$
\item $\Pi$ the relative liabilities matrix 
\[
    \Pi_{i,j}= 
\begin{cases}
    \frac{p_{i,j}}{p_j},& \text{if } \geq 0\\
    0,              & \text{otherwise}
\end{cases}
\]
\end{itemize}

\subsubsection{Naive implementation of PageRank}
In their first description of the PageRank algorithm, we can translate the PageRank for a bank $PR_i$ as:
\begin{equation}
PR_i = (1 - d) + d  \sum_{j\neq i} \frac{PR_j}{p_j}
\end{equation}
with $d \in [0,1]$ a damping factor.

In Matrix form:
\begin{equation}
PR = (1 - d) e_N + d  \Pi PR
\end{equation}
with $e_N$ a vector with $N$ elements which are $\frac{1}{N}$.

Per design, we set $\sum_j PR_j = 1$

We see that there is some sort of tautology in defining the $PR$'s. Indeed, a bank's $PR$ is computed from its neighbours' $PR$. Hence we need to proceed iteratively from a starting situation, assigning $PR$ to every bank and then converge to the "true" $PR$.

\begin{itemize}
\item set $d=0.85$, $\mbox{MaxIter}=100$, $N=4$, and $\epsilon=10^{-5}$
\item define $r_0$ your initial guess for the vector of $PR_j, j\in [1,N]$. $r_0=[\frac{1}{N},\ldots,\frac{1}{N}]$
\end{itemize}

Then we need to compute the relative liabilities matrix $\Pi$:

\begin{itemize}
\item Import the price data on banks from the file 2016\_10\_12\_bank\_prices.csv and select only the prices of the following 4 french banks, compute their daily returns:
\begin{itemize}
\item BNPPARIBAS
\item CREDITAGRICOLESA
\item BPCE
\item SOCIETEGENERALESA
\end{itemize}
\item Compute the correlation matrix but take the absolute value (negative lending would not make sense), compute (correlation matrix - identity), make sure the sum of each columns is 1 (you can divide the matrix by a vector of the sum of each column) and consider that this is your relative liabilities matrix $\Pi$ 
\item This is a very poor assumption, it is assuming that the market would react on one bank according to how much they lent to one another, but this will do for our exercise.
\end{itemize}

\begin{center}
\includegraphics[scale=0.5]{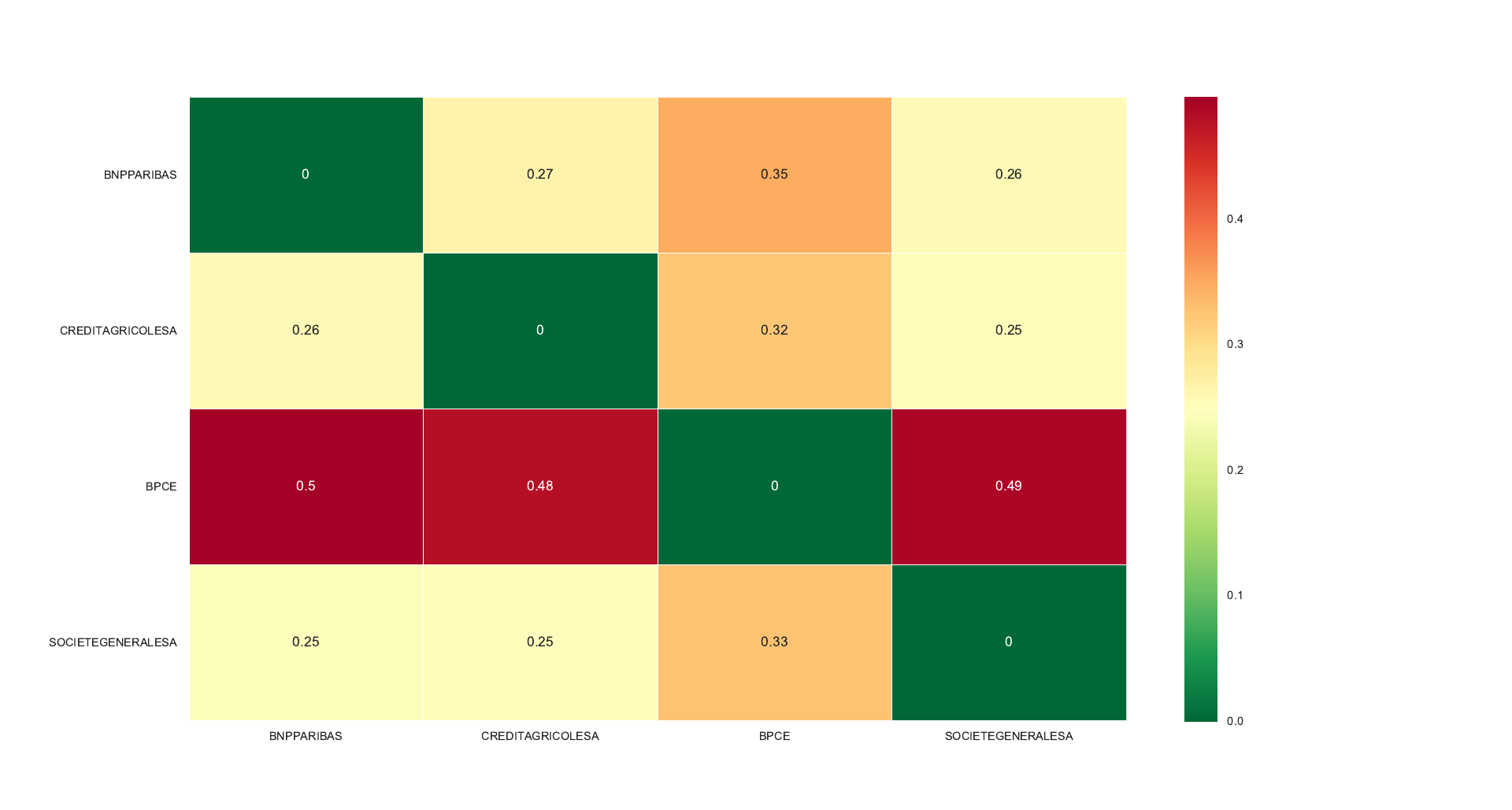}
\end{center}

\begin{itemize}
\item Then compute $r_1$ as: $r_1= (1-d)+ d \times \Pi \times r_0$
\item You then need to create a loop and stop your loop either once $(r_1-r_0)^2 < \epsilon$ or once we have reach the maximum number of iterations $\mbox{MaxIter}$.
\item List the bank and the PageRank score you find.
\end{itemize}

\quad

With our illustration, we obtain the following:

\begin{tabular}{lr}
 &     PageRank \\
BPCE              &  0.320 \\
BNPPARIBAS        &  0.232 \\
CREDITAGRICOLESA  &  0.224 \\
SOCIETEGENERALESA &  0.223 \\
\end{tabular}

Which would tend to say that BPCE is the most important bank in this constructed network of liabilities.

\subsubsection{With teleportation}
In order to treat dead-end nodes, we add a teleportation distribution $f$. In practice, it is a vector of length $N$ satisfying $\sum_i f_i =1$.

We then define compute the PageRank vector $r$ as:
\begin{equation}
PR_i = (1-d) f_i + d \left( \sum_{j\neq i} \frac{PR_j}{p_j} +  \sum_{j\neq i} f_j d_{out,j} PR_j \right)
\end{equation}

With:
\[
    d_{out,i}=
\begin{cases}
    1,& \text{if } \sum_j p_{j,i}=0\\
    0,              & \text{otherwise}
\end{cases}
\]

and in matrix form:
\begin{equation}
PR = (1-d) f + d \left( \Pi + f d_{out} \right) PR
\end{equation}

$d_{out}$ is a vector that identifies nodes that have no outgoing link to other nodes. If you have such nodes in your network, you need to have this "teleportation" factor $f$ so that you can jump randomly from one node to the other even if the node has no outgoing link. In other words, you don't get stuck in one node and the algorithm converges.

This can be rewritten as:
\begin{equation}
PR = \Delta PR
\end{equation}

it can be demonstrated that the matrix $\Delta$ is irreducible and aperiodic stochastic matrix and from the Perron-Frobenius theorem its largest eigenvalue is always equal to $1$ and its eigenvector as the sum of its elements summing to $1$. Here we just verify this with python.

\stepcounter{question}
\begin{itemize}
\item \textbf{Question \arabic{question}}
\begin{itemize}
\item Make a simple assumption about f: like $1\%$ uniform chance $f=0.01$ and implement an algorithm with the teleportation aspect\footnote{you should not observe any change in your result as all possible links are present in your network}.
\end{itemize}
\end{itemize}

\section{R: Before and after regressing financial time series - some steps}\label{sec:spurious}

\subsection{Example: exercise on the CAC40 and the Dow Jones}
We suggest an example\footnote{spurious\_vansteenberghe.R} inspired by the book \citep{bazen}.

We download the 'Close' value for the CAC 40 and the Dow Jones indexes from 1995-01-16 to 1999-02-26.

If we regress one price series on the other, we find a high value of $R^2$ and low value for the Durbin-Watson test statistics\footnote{In a Durbin-Watson test, the null hypothesis $H_0$ is that the autocorrelation of the disturbances (or error terms of our regression is 0). As the p-value in this case is low, we reject $H_0$ and suspect autocorrelation in the model error terms. Note that if there is a lagged dependent variable, we have to use a Breusch-Godfrey test instead as in section \ref{sec:breusch}.}. $R^2 > DW statistic$ is a rule of thumb to suspect that this regression is spurious, it indicates a non existing relationship between what looks like two random walks.

\subsection{Unit root tests}
Before regressing one variable onto another, we first test for unit root. With R we use the package \emph{urca} and perform an Augmented Dickey-Fuller test as detailed in \ref{sec:adf}.

If some variables follow a random walk, regressing one against the other might lead to spurious regression. We might thus need to first-difference to work with stationary series.

At the 99, 95 and 90\%, the critical values of the Augmented Dickey-Fuller test are: $-2.58$, $-1.96$, and $-1.64$

\quad

We download the 'Close' value for the CAC 40 and the Dow Jones indexes and we perform the augmented Dickey-Fuller test on the log of the index. For the CAC 40 index, we find a test statistic of $-0.2$ and for the Dow of $-1.34$ $\implies$ both series are non-stationary and regressing one onto the other could be spurious\footnote{our statistics are below the critical values and we fail to reject $H_0$: the time series has a unit root}.

\quad

We also apply the DF-GLS version of the test introduced by \citep{Elliott:1996aa} which has higher power than the augmented Dickey-Fuller test when the autoregressive root is large but less than one.

\subsection{Cointegration test}

We do a cointegration test on the two series and find that the two time series are not cointegrated.

\subsection{Are both series I(1)?}
We apply the first difference to the log of both index and we find that both series were I(1). 

\subsection{A first regression model}
Both time series have means different than $0$ hence we need to add an intercept in our regression model.

We use the model and find a $R^2$ of $13\%$:
\begin{equation}\label{eq:4}
\Delta \mbox{LCAC}_t=\alpha + \beta \Delta \mbox{LDow}_t + u_t
\end{equation}

As $\log(\mbox{CAC 40}_t)-\log(\mbox{CAC 40}_{t-1}) \simeq \mbox{Daily Returns}$, we can conclude that over the studied time period, the daily returns of the Dow explain $13\%$ of the daily returns of the CAC 40.

\subsection{Durbin-Watson test}
The errors of a regression model are supposed to be uncorrelated. This can break down with time series. The errors can be serially correlated. Hence an errors associated with one observation will then carry over to future observations. This is relevant namely with net present value model where future dividends impact the current stock price.

The serial correlation of the errors can be written as: $u_t=\rho u_{t-1} + \theta_t$ 

The Durbin-Watson test the null hypothesis $H_0$: $\rho=0$

Note that the Durbin-Watson test cannot be used if the regression equation contains a lagged dependent variable, in which case we use instead a Breusch-Godfrey test as in section \ref{sec:breusch}.

We apply the Durbin-Watson test to our equation \ref{eq:4}, with the above limitations in mind.

We find a value of $DW=2.23$ which could tend to show that there is no serial correlation ($\rho=0$) as $DW = 2 (1 - \hat{\rho})$, but in fact, looking at the Durbin-Watson table, we find that $d_u \simeq 1.78$ and $d_l \simeq 1.76$, we do not find that: $4 - d_l < DW < 4$ hence we do not reject the null hypothesis, and there is no strong evidence for negative serial correlation present in our initial model \ref{eq:4}.

With a lagged endogenous variable as in our final model \ref{eq:final}, we can use the statistics:
\begin{equation*}
h=\left(1-\frac{DW}{2}\right) \sqrt{\frac{T}{1-T Var (\hat{\alpha}_4)}}
\end{equation*}

With $Var (\hat{\alpha}_4)$ the square of the standard error observed in the summary of the regression \ref{eq:final}. h can be supposed to be normally distributed with unit variance and the normal distribution table an be used.

\quad

With lagged endogenous variable we can use the Breusch-Godfrey test and also choose for the lag order in our test :

\subsection{Breusch Godfrey test}\label{sec:breusch} 
We apply a Breusch Godfrey test on the series to check for second order autocorrelation, this test is robust even if the model includes lagged dependent variables.

We are working with the model:
\begin{equation}\label{eq:final}
\Delta \mbox{LCAC}_t=\alpha_1 + \alpha_2 \Delta \mbox{LDow}_t + \alpha_3 \Delta \mbox{LDow}_{t-1} + \alpha_4 \Delta \mbox{LCAC}_{t-1} + u_t
\end{equation}

We suspect that the residuals are following an AR(p):

\begin{equation}
u_t = \rho_1 u_{t-1} + \rho_2 u_{t-2} + \ldots + \rho_p u_{t-p} + \epsilon_t
\end{equation}

We test the hypothesis: $H_0$ : $\rho_i = 0, \forall i \in [1,p]$

In practice, the test follows the procedure:
\begin{enumerate}
\item regress the model \ref{eq:final}
\item use the residuals from the above regression and test: $\hat{u}_t=\alpha_1+\alpha_2 \hat{u}_{t-1}+ \ldots + \alpha_p \hat{u}_{t-p}+\beta_1 \Delta \mbox{LDow}_t + \beta_2 \Delta \mbox{LDow}_{t-1} + \beta_3 \Delta \mbox{LCAC}_{t-1} + v_t$
\item test the statistic $LM= n R^2$ against the threshold $\chi_{0.05}^2(p)$, at the order $2$, the threshold is $5.99$
\end{enumerate}

The regression \ref{eq:final} has a Breusch Godfrey test statistic below $5.99$, hence we accept both the hypothesise that there are neither first nor second order autocorrelation.

\subsection{Gauss-Markov hypotheses}\label{sec:GaussMarkov}

Underlying our linear regression, we assume the Gauss-Markov hypotheses for our Ordinary Least Square estimate. We follow \citep{verbeek}:

\subsubsection{$E(u_t) = 0$}
Assuming our error terms are normally distributed, we can apply a one-sample t-test that use the following statistics:
\begin{equation}
\frac{\overline{u_t} - \mu}{\frac{\sigma}{\sqrt{n}}}
\end{equation}
we want to test if the null hypothesis $H_0$: $\overline{u_t}=\mu$ with  $\overline{u_t}$ the mean of the residuals. In its two-sided form, the alternative hypothesis is: $H_1$: $\overline{u_t} \neq \mu$.

\subsubsection{Error terms and explanatory variables are independent}
${u_t}$ and ${X_t}$ are independent, where $X_t$ is the vector of explanatory variables.

\subsubsection{Homoskedasticity}
By homoskedasticity we mean that the conditional variance of the error term is constant over time and not correlated with the explanatory variable: $var(u_t) = \sigma^2$.
We apply the Breusch-Pagan test with $H_0$: homoskedasticity, which is based on the following regression, testing for the significance of the parameter $\beta$:
\begin{equation}
u_t^2 = \alpha + \beta X_t + \nu_t
\end{equation}

\subsubsection{$cov(u_i,u_j) = 0$ $\forall i \neq j$}
We check whether error terms are uncorrelated with a Durbin-Watson or Breusch-Godfrey test seen above.


\stepcounter{question}
\begin{itemize}
\item \textbf{Question \arabic{question}}
\begin{itemize}
\item check if those hypotheses are respected with our data set. If those hypotheses are respected, the OLS estimator is then BLUE (best linear unbiased estimator). Is the estimator biased if we have heteroskedasticity?
\item Manually compute the OLS estimator by solving the optimisation programme below:
\end{itemize}
\end{itemize}

$$\min_{\alpha,\beta} \sum_t \left( \Delta \mbox{LCAC}_t - \alpha - \beta \Delta \mbox{LDow}_t \right)^2$$

\quad

With no constant:
$$\hat{\beta} = \frac{\sum_t \Delta \mbox{LDow}_t \times \Delta \mbox{LCAC}_t}{\sum_t \Delta \mbox{LDow}_t^2}$$
With a constant $\alpha$:
$$\hat{\beta} =  \frac{T \sum_t \Delta \mbox{LDow}_t \times \Delta \mbox{LCAC}_t -  \sum_t \Delta \mbox{LDow}_t \sum_t \Delta \mbox{LCAC}_t}{T \sum_t \Delta \mbox{LDow}_t^2 - \left( \sum_t \Delta \mbox{LDow}_t\right)^2}$$

\subsection{Normality test: Shapiro-Wilk test}

We perform the Shapiro-Wilk test on both series returns to test for their normality. The null hypothesis $H_0$ is that the sample of returns came from a normally distributed population. The normality of the returns is rejected.

\subsubsection{Scaling the returns}
We can rescale the returns, for example with:
$$ \hat{R}_t = \frac{R_t - \min(R)}{\max(R)-\min(R)} $$

This would not change our conclusion on the normality of the distribution of return, although sometimes one call this "normalizing" the returns.

\subsection{Regression residuals - test the presence of ARCH}

We now follow \citep{bazen} and apply an ARCH(1) model on the errors terms of the model described by equation \ref{eq:final}. We start witht the ARCH(1) model introduced section \ref{sec:ARCHEngle}.

\subsubsection{Test for ARCH}\label{sec:ARCHtest}
First we use the Robert Engle method to test the presence of an ARCH process. From the regression \ref{eq:final} we did, we use the residuals $u_t$ and regress as in equation \ref{eq:residARCHtest}:
\begin{equation}\label{eq:residARCHtest}
\hat{u}_t^2 = \alpha + \beta\, \hat{u}_{t-1}^2 + \varepsilon_t .
\end{equation}

Then we use the $R^2$ of that regression and the length of the time series (here 1062 days) and compare it to the threshold 3.84, the $5\%$ critical value of a $\chi^2$ distribution with one degree of freedom (Engle's LM statistic is $T R^2$).

We find a $R^2=12\%$ and we have $T=1062$ the length of our time series. Hence our test statistic is above the threshold: $12.5> 3.84$, therefore the null hypothesis $H_0$: $\beta=0$ is rejected and we are facing an ARCH process.

\subsubsection{ARCH model of the residuals - PACF}

We use a PACF of the squared residual to determine the order $m$ of the ARCH process. Visually we decide to select an ARCH(4):
\begin{center}
\includegraphics[scale=0.5]{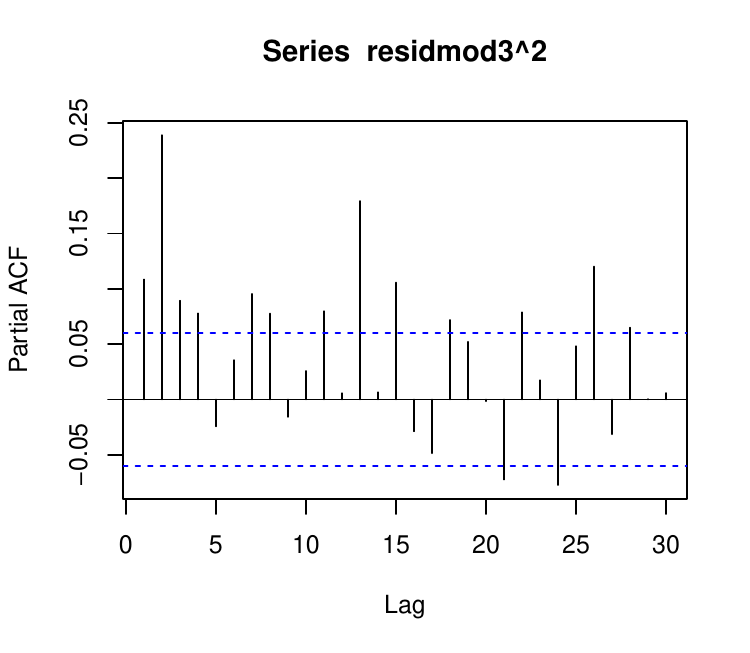}
\end{center}

\subsubsection{Estimation of an ARCH(m) and check relevence}

We estimate an ARCH(4) model on the residuals and then check with the Ljung-Box\footnote{see section \ref{sec:ljungbox} for more detail on this test} that we do not reject the null hypothesis $H_0$ of non serial correlation of the ARCH(4) residuals.
If we try to account for heavy tails using Student's-t distribution for the standardized residuals $\epsilon_t$, the estimation return $8$ as the degree of freedom (shape parameter in the summary), the QQ-plot are more satisfactory for the Student's-t distribution:
\begin{center}
\includegraphics[scale=0.5]{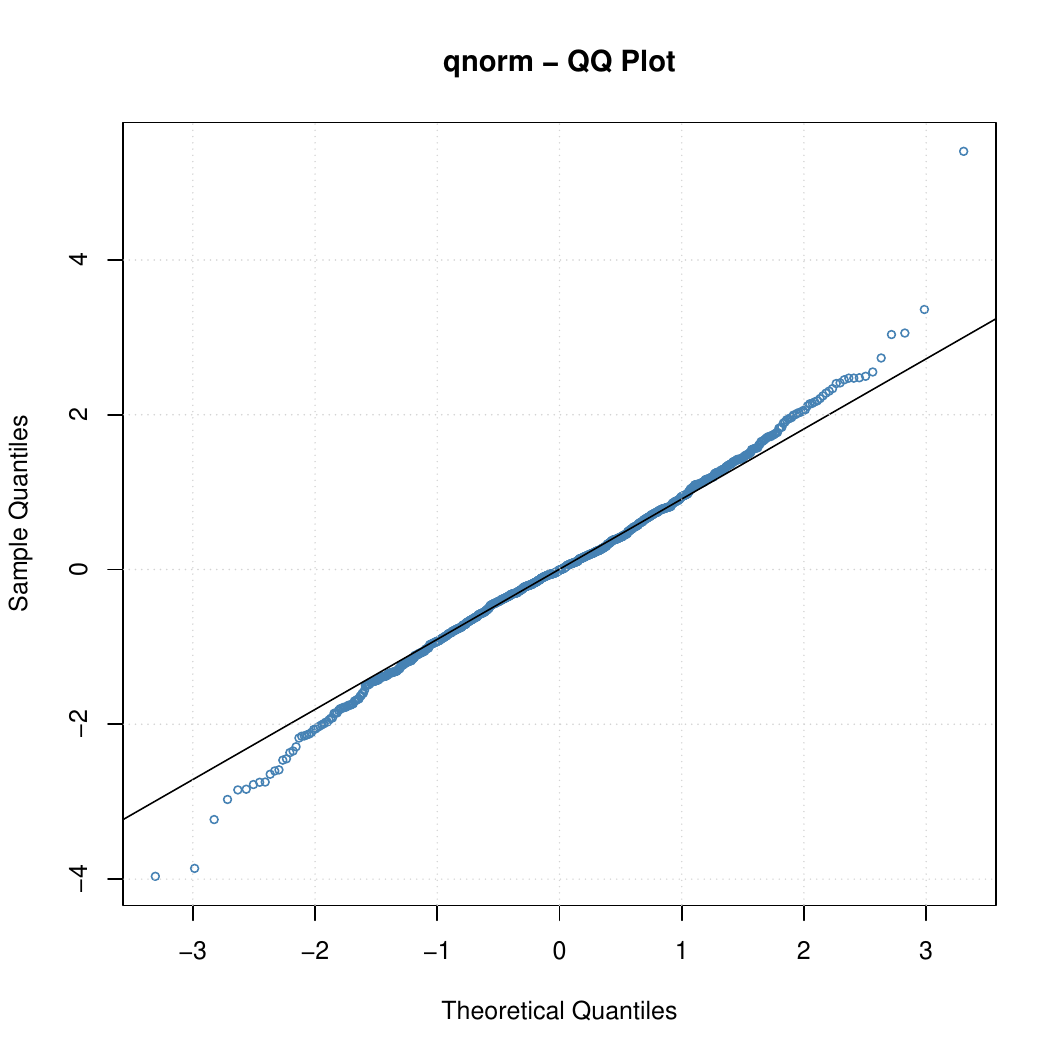}
\includegraphics[scale=0.5]{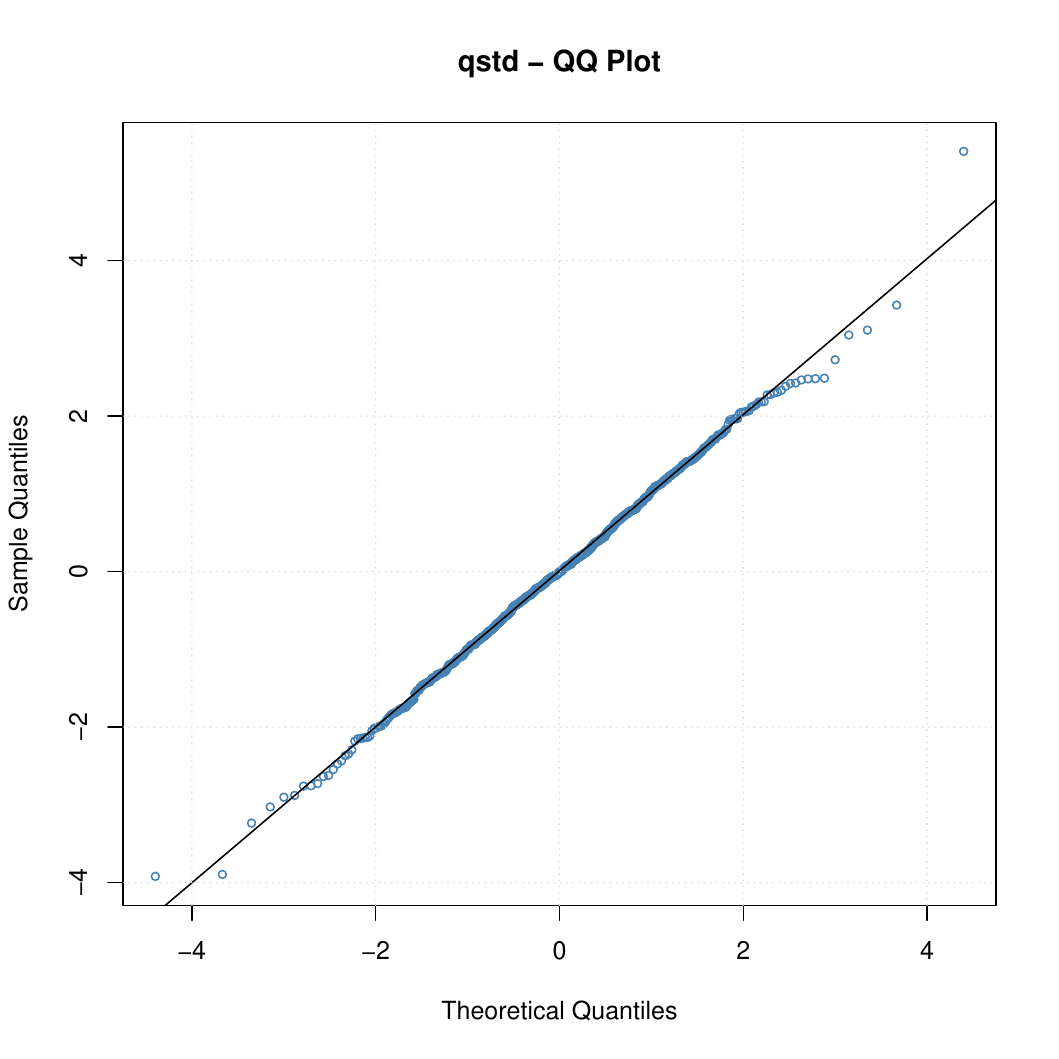}
\end{center}
The ARCH model standardized residuals should show less clustering issues than the original time series. This will be further explored in section \ref{sec:filter-GARCH}.

\section{Univariate ARMA-GARCH model}\label{sec:uARMAGARCH}

Following \citet{PERLIN2021}, we work\footnote{code: ARMA\_GARCH\_vansteenberghe} with the two returns series and identify volatility clusters, meaning that the volatility at date $t$ seems to depends of the volatility at date $t-h$:
\begin{center}
\includegraphics[scale=0.5]{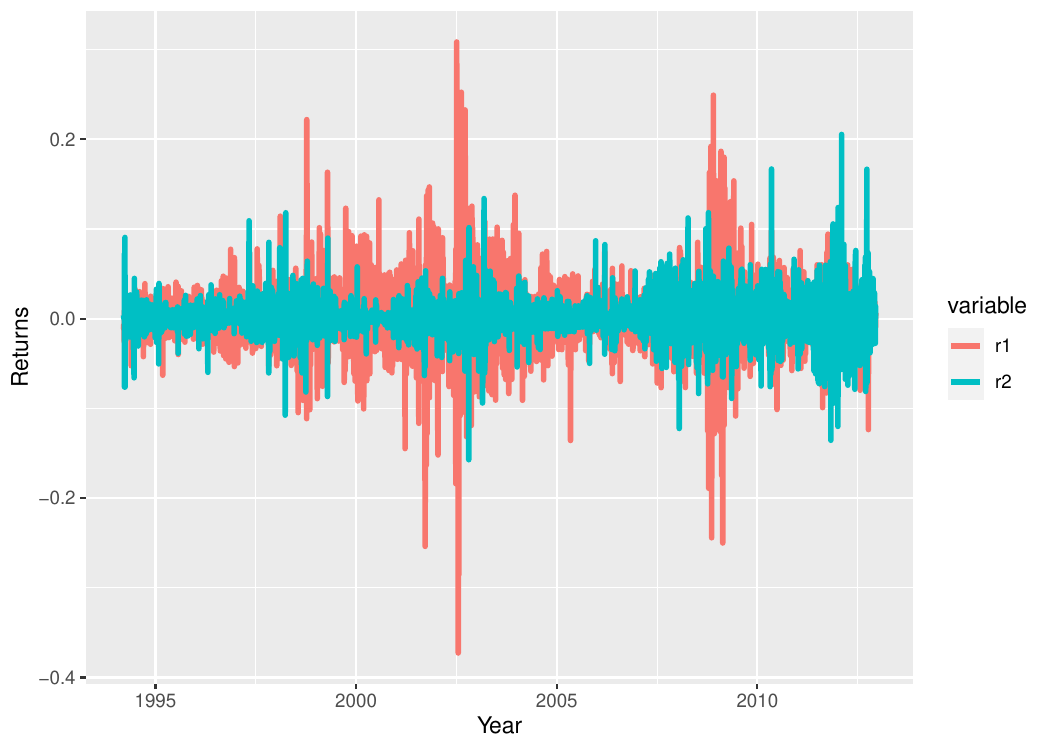}
\end{center}

We assume that we have autocorrelated financial returns (after a Ljung-Box test):
\begin{equation}\label{Mdyna}
R_t = \mu_t + \sigma_t Z_t = \mu_t + a_t
\end{equation}
where $\mu_t = E \left( R_t \vert \mathcal{F}_{t-1} \right)$ and $\sigma_t$ are $\mathcal{F}_{t-1}-$ measurable and $(\mathcal{F}_t)_t$ is the natural filtration\footnote{in simple terms, the filtration contains information up to time $t$ that is useful to determine $R_t$} of the process $(R_t)_t$. $(Z_t)_t$ is a time series of iid random variable of unknown distribution that we call the innovations process and has zero mean\footnote{and unit variance if we standardize them}. 

We now suppose that the process $(\mu_t)_t$ is an autoregressive model of order $p$, $p\in \mathbb{N}$, noted $AR(p)$  defined as
\begin{equation*}
\mu_t = \mu + \theta_{1} R_{t-1}+...+ \theta_{p} R_{t-p}.
\end{equation*}
In many cases in finance we simply use an $AR(1)$:
\begin{equation*}
\mu_t = \mu + \theta R_{t-1}.
\end{equation*}
We model the process $(\sigma^2_t)_t$ as an $GARCH(p,q)$, after testing for the ARCH effect, defined as:
\begin{equation}
\sigma^2_t = \alpha_0 +\sum_{i=1}^p \alpha_i(R_{t-i}-\mu_{t-i})^2+ \sum_{j=1}^q \beta_j\sigma^2_{t-j}.
\end{equation}
where\footnote{in R rugarch package, they defined "ar1" as $\theta$ and "omega" as $\alpha_0$} $\alpha_i \geq 0$, $\beta_i\geq0$, $\forall i$ and $\sum_{i=1}^p\alpha_i+\sum_{j=1}^q \beta_j<1$. It is the stationarity condition of second order, i.e. the non-conditional variance is constant and finite. This is a generalization of the $ARCH$ (Auto Regressive Conditionnal Heteroskedasticity) model introduced by \citep{Engle:1982aa}\footnote{once estimated, a Ljung-Box(see section \ref{sec:ljungbox} for more details) test can check the residuals independence, to verify that the heteroskedasticity has been correctly modelled. Models can be compared and chosen with AIC, BIC or likelihood ratios.}. In finance, the $GARCH(1,1)$ model is often used:
\begin{equation}
\sigma^2_t = \alpha_0 +\alpha_1(R_{t-1}-\mu_{t-1})^2+\beta\sigma^2_{t-1}.
\end{equation}
Thus the returns are modelled by the following equation system known as the  $AR(1)-GARCH(1,1)$:
\begin{equation}
\begin{cases}
R_t &= \mu_t + \sigma_t Z_t \\
\mu_t &= \mu + \theta R_{t-1} \\
\sigma^2_t &= \alpha_0 +\alpha_1(R_{t-1}-\mu_{t-1})^2+\beta\sigma^2_{t-1}
\end{cases}
\end{equation}

It is not advised to calibrate first an ARMA on the data and then a GARCH and this first model residuals. Indeed, when first estimating the ARMA model, you would assume conditional homoskedasticity.

We present a method of pseudo maximum liquelihood to estimate the parameters of the $AR(1)-GARCH(1,1)$ model. We assume that we have a data set of size $n+1$; $R_0,...,R_n$ and notice that the joint density of the random variable $R_0,...,R_n$ can be written as:
\begin{equation}\label{densite1}
f_{R_0,...,R_n}(r_0,...,r_n)=f_{R_0}(r_0)\prod_{t=1}^nf_{R_t|R_{t-1},...,R_0}(r_t|r_{t-1},...,r_0).
\end{equation} 
As the marginal density $f_{R_0}$ in the equation $\ref{densite1}$ is unknown, we compute the conditional density knowing $R_0$ and we have:
\begin{equation}\label{densite2}
f_{R_1,...,R_n|R_0}(r_1,...,r_n|r_0)=)\prod_{t=1}^nf_{R_t|R_{t-1},...,R_0}(r_t|r_{t-1},...,r_0).
\end{equation}
Furthermore, as $\sigma_t$ depends on $\sigma_{t-1}$, we need an initial value for $\sigma_0$ and we compute the conditional density knowing $R_0$ and $\sigma_0$ writing:
\begin{equation}\label{densite3}
f_{R_1,...,R_n|R_0,\sigma_0}(r_1,...,r_n|r_0,\sigma_0)=\prod_{t=1}^nf_{R_t|R_{t-1},...,R_0,\sigma_0}(r_t|r_{t-1},...,r_0,\sigma_0).
\end{equation}  
Thus the probability density $f_{R_t|R_{t-1},...,R_0,\sigma_0}$ only depend on the past via the value $\sigma_t$, that is obtained by a recursive approach from $\sigma_0, R_0,...,R_{t-1}$ using that $\sigma^2_t= \alpha_0 +\alpha_1(R_{t-1}-\mu_{t-1})^2+\beta\sigma^2_{t-1}$. We consider the conditional likelihood function defined as:
\begin{equation*}
L(R; \theta_1, \alpha_0,\alpha_1,\beta)=\prod_{t=1}^n\frac{1}{\sigma_t}g\Big(\frac{R_t-\mu - \theta_1 R_{t-1}}{\sigma_t}\Big).
\end{equation*} 
where $g$  is the unknown probability density of the iid random variables $(Z_t)_t$. In practice, it is often supposed that $g$ is the density of a standard normal law or a t-Student with an unknown degree of freedom $\nu$. In finance, the true law for $Z_t$  is seldom normal nor t-Student but \citep{Gourieroux:1984aa} demonstrate that under various regularity conditions, the maximum pseudo-likelihood estimators converge and are asymptotically normal.

\subsubsection{Fit an AR(1)-GARCH(1,1) to the Dow Jones and CAC 40 index returns}
We fit an AR(1)-GARCH(1,1) to the Dow Jones and CAC 40 index returns, and find for both $\alpha_0$ and $\theta$ not significantly different than $0$. The standardized innovations seems freed of clusters:
\begin{center}
\includegraphics[scale=0.5]{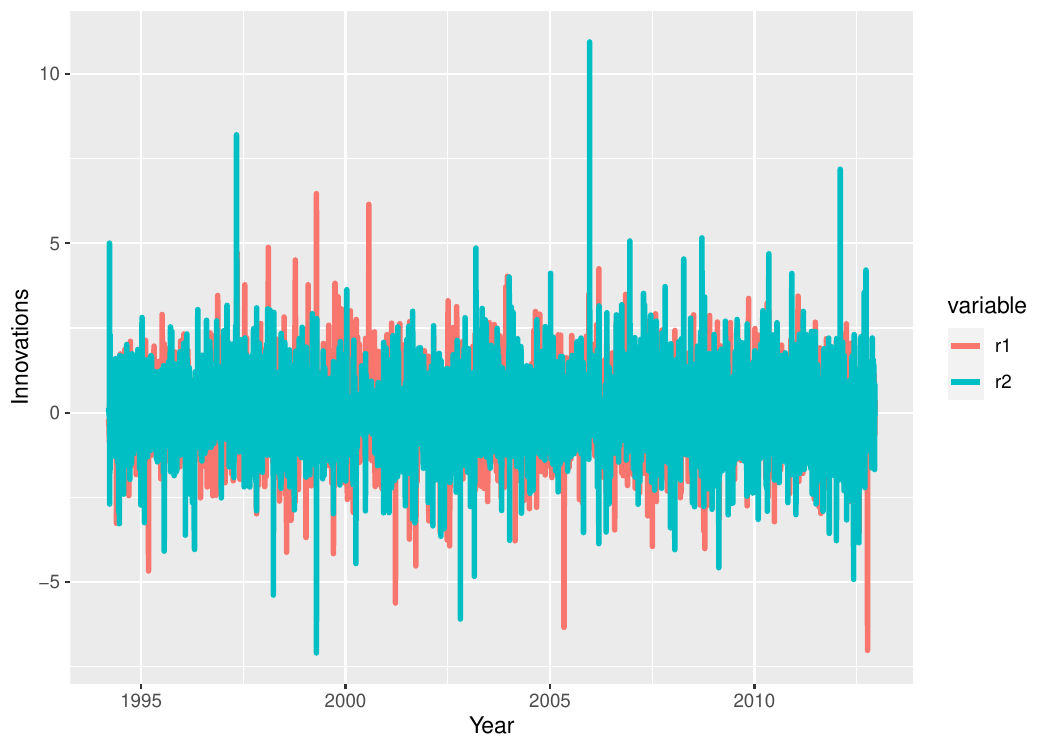}
\end{center}

We can test whether the standardized residuals (the innovations) are iid.

The above model can be applied to a univariate time series sample, in practice, we are dealing with multiple time series and we need to model the evolution of their correlation over time as empirically observed in section \ref{sec:correlationB} with a multivariate volatility model. We use the Dynamic conditional correlations model.

\subsection{Explore for the best ARMA-GARCH}

We also suggest an alternative package, fGarch, and explore for the best fit (in terms of AIC or BIC) for the orders $p$, $q$, $r$ and $s$: ARMA($p$,$q$)-GARCH($r$,$s$).

We fit an MA(1)-eGARCH(2,1) assuming the innovations to follow a Student's t distribution:
\begin{center}
\includegraphics[scale=0.5]{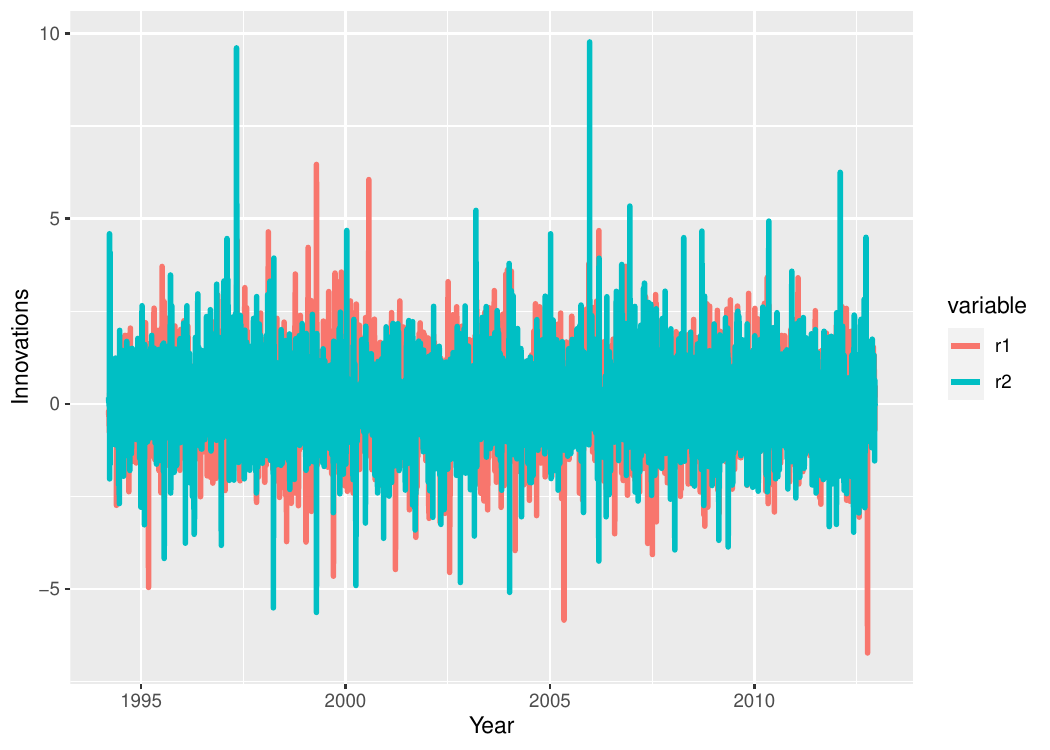}
\end{center}

\subsubsection{Dynamic conditional correlations model (DCC)}\label{sec:dcc}

\citep{Engle:2002aa} suggest dynamic conditional correlations (DCC) models that have the flexibility of univariate GARCH models coupled with parsimonious parametric models for the correlations\footnote{further details on multivariate GARCH models can be found \href{http://www.timberlake-consultancy.com/slaurent/G@RCH/default.htm}{here}}. \citep{Sampid:2018aa} follow a DCC-GARCH model before applying copulas to estimate VaR. We write the conditional correlation between two asset loss rates as:
\begin{equation*}
Cov(R_{t,i},R_{t,j})=\rho_{t,ij}
\end{equation*}
and the conditional covariance matrix $\Sigma_t = cov\left( a_t \vert \mathcal{F}_{t-1} \right)$:
\begin{equation*}
\begin{bmatrix}
\sigma_{t,11} & \sigma_{t,12} & \ldots & \sigma_{t,1N}  \\
 \sigma_{t,12} & \sigma_{t,22} & \ldots & \sigma_{t,2N} \\
\vdots & \vdots & \ldots & \vdots  \\
\sigma_{t,1N}  & \sigma_{t,2N} & \ldots & \sigma_{t,NN} \\
\end{bmatrix}
\end{equation*}

this model assumes that the conditional covariance matrix evolves as:
\begin{equation*}
\Sigma_t = D_t \rho_t D_t
\end{equation*}

with $D_t$ the diagonal matrix of loss rate volatilities and $\rho_t$ the conditional correlation matrix of the loss rates which can be expressed as a function of the conditional covariance matrix of the standardized innovations $Q_t$
\begin{equation*}
\rho_t = diag(\sqrt{Q}_t)^{-1} Q_t diag(\sqrt{Q}_t)^{-1}
\end{equation*}
where $diag(\sqrt{Q}_t)$ is the matrix with the square roots of the elements of $Q_t$.
The model assumes that the matrix $Q_t$ follows the dynamic:
\begin{equation}
Q_t = (1-\alpha - \beta)\bar{Q} + \alpha Z^{sd}_{t-1} Z^{sd'}_{t-1} + \beta Q_{t-1}
\end{equation}
where $\bar{Q}$ is the unconditional correlation matrix of the standardized innovations, $Z^{sd}$ the standardized innovations, $\alpha$ and $\beta$ are non-negative real numbers and $0 < \alpha+ \beta < 1$.

\stepcounter{question}
\begin{itemize}
\item \textbf{Question \arabic{question}}
\begin{itemize}
\item Following the instructions of the rmgarch R package for DCC model, fit various models on the two time series, suggest your model of choice.
\end{itemize}
\end{itemize}

Some starting elements below:

\quad

We apply a simple DCC-GARCH(1,1) model, for each asset returns we assume (we keep the same notation as R language output):
\begin{equation}
\begin{cases}
R_{t,i} &= \mu_i + \sigma_{t,i} Z_{t,i} \\
\sigma^2_{t,i} &= omega +alpha_1(R_{t-1,i}-\mu)^2+beta_1 \sigma^2_{t-1,i}
\end{cases}
\end{equation}

We plot the conditional correlations of the asset returns over time:
\begin{center}
\includegraphics[scale=0.45]{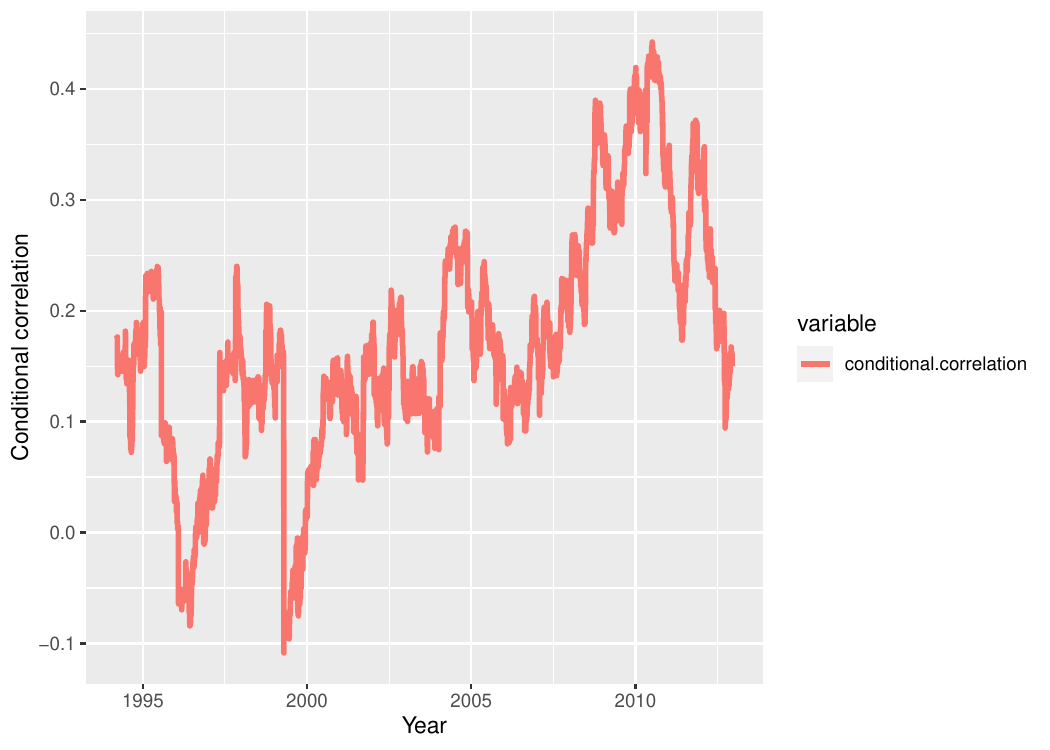}
\end{center}

We can extract the standardized residuals from the DCC-model, the innovations, we "filtered out" the volatility clusters:
\begin{center}
\includegraphics[scale=0.4]{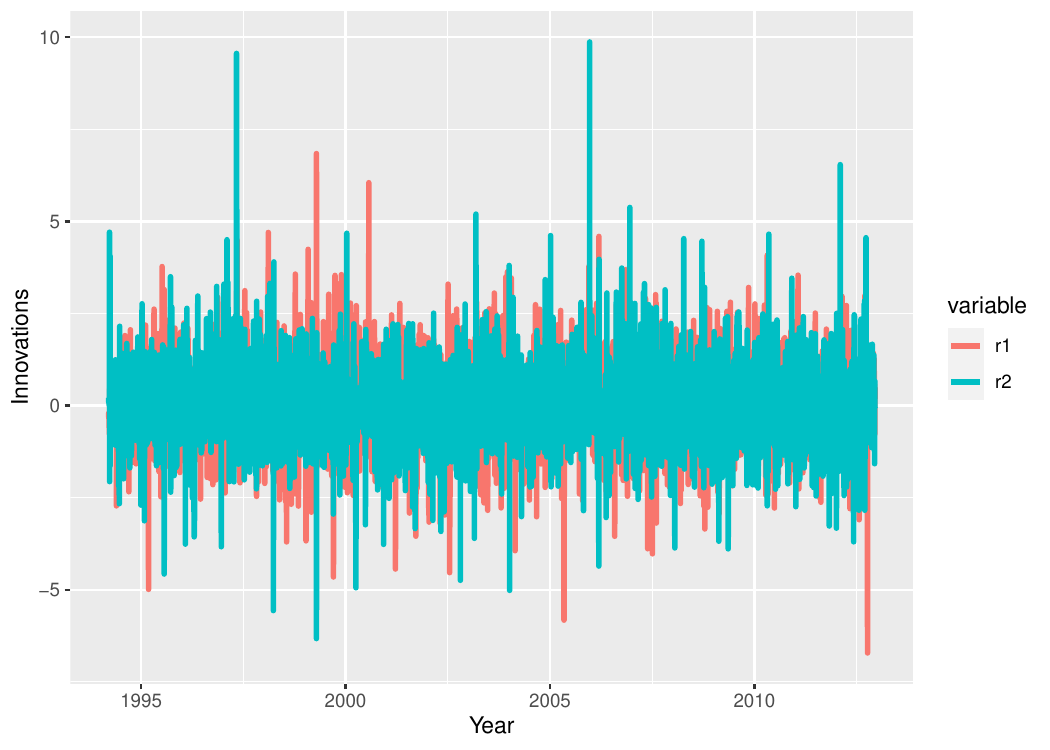}
\end{center}

 \section{Binary choice models}

Suppose as in \citep{verbeek} we want to explain\footnote{code: logit\_credit\_cards.py} whether a credit card customer default on its loan or not. Let the sole explanatory variable be the client income. We have data on $N$ agents ($i = 1, \ldots, N$), with observations on their income, $x_{i,2}$, and whether or not they default. This latter element is described by the binary variable $y_i$, defined as
\begin{itemize}
\item $y_i = 1$ if $i$ defaults
\item $y_i = 0$ if $i$ doesn't default
\end{itemize}

Suppose we were to use a regression model to explain $y_i$ from$x_{i2}$ and an intercept term ($x_{i,1} = 1$). This linear model would be given by
\begin{equation}
y_i =\beta_1 +\beta_2x_{i,2} + \epsilon_i
\end{equation}
It seems reasonable to make the standard assumption that $E\left(\epsilon_i\vert x_i \right)=0$ such that $E\left(y_i\vert x_i\right) = x_i'\beta $. This implies that
\begin{equation}
E\left(y_i\vert x_i\right) = 1 P\left(y_i = 1\vert x_i\right) + 0 P\left(y_i = 0\vert x_i\right) = P\left(y_i = 1 \vert x_i\right) = x_i' \beta
\end{equation}

Thus, the linear model implies that $x_i'\beta$ is a probability and should therefore lie between $0$ and $1$. This is only possible if the $x_i$ values are bounded and if certain restrictions on $\beta$ are satisfied. Usually this is hard to achieve in practice. In addition to this fundamental problem, the error term $\epsilon_i$ has a highly non-normal distribution and suffers from heteroskedasticity. Because $y_i$ has only two possible outcomes ($0$ or $1$), the error term, for a given value of $x_i$, has two possible outcomes as well. In particular, the distribution of $\epsilon_i$ can be summarized as
\begin{equation}
\begin{cases}
P\left(\epsilon_i = -x_i' \beta \vert x_i\right) = P\left(y_i = 0\vert x_i\right) = 1 - x_i' \beta\\
P\left(\epsilon_i = 1 -x_i' \beta \vert x_i\right)  = P\left(y_i = 1\vert x_i\right) = x_i' \beta
\end{cases}
\end{equation}
This implies that the variance of the error term is not constant but dependent upon the explanatory variables according to $V\left(\epsilon_i \vert x_i\right) = x_i'\beta (1 - x_i'\beta)$. Note that the error variance also depends upon the model parameters $\beta$.

\subsection{Probit, Logit models}
We introduce binary choice models as:
\begin{equation}
P\left(y_i=1 \vert x_i\right) = F\left( x_i' \beta \right)
\end{equation}
if $F()$ is
\begin{itemize}
\item the standard normal distribution function, then it is a probit model
\item the logistic function $F(w) = \frac{e^w}{1+e^w}$, then it is a logit model
\end{itemize}

\subsection{Credit card clients data set}\label{sec:defaultdata}
We use a data set proposed (\href{https://archive.ics.uci.edu/ml/datasets/default+of+credit+card+clients#}{available here}) by I--Cheng Yeh and Che-hui Lienon \footnote{This is inspired by Xavier Dupre's work and \href{http://www.xavierdupre.fr/app/ensae_teaching_cs/helpsphinx2/notebooks/solution_2016_credit_clement.html}{website}}  on default of credit card clients Data Set for their paper \citep{Yeh2009aa}. This research aimed at the case of customers' default payments in Taiwan and compares the predictive accuracy of probability of default among six data mining methods. From the perspective of risk management, the result of predictive accuracy of the estimated probability of default will be more valuable than the binary result of classification - credible or not credible clients. Because the real probability of default is unknown, this study presented the novel "Sorting Smoothing Method" to estimate the real probability of default. With the real probability of default as the response variable (Y), and the predictive probability of default as the independent variable
(X), the simple linear regression result (Y = A + BX) shows that the forecasting model produced by artificial neural network has the highest coefficient of determination; its regression intercept (A) is close to zero, and regression coefficient (B) to one. Therefore, among the six data mining techniques, artificial neural network is the only one that can accurately estimate the real probability of default.

\subsubsection{Presentation of the data set}
The data set has 30 000 observations with the following information:

This research employed a binary variable, $Y$ the default payment (Yes = 1, No = 0), as the response variable. This study reviewed the literature and used the following 23 variables as explanatory variables: 
\begin{itemize}
\item X1, LIMIT\_BAL: Amount of the given credit (NT dollar): it includes both the individual consumer credit and his/her family (supplementary) credit. 
\item X2, SEX: Gender (1 = male; 2 = female). 
\item X3, EDUCATION: Education (1 = graduate school; 2 = university; 3 = high school; 4 = others). 
\item X4, MARRIAGE: Marital status (1 = married; 2 = single; 3 = others). 
\item X5, AGE: Age (year). 
\item X6 - X11, PAY\_i: History of past payment. We tracked the past monthly payment records (from April to September, 2005) as follows: X6 = the repayment status in September, 2005; X7 = the repayment status in August, 2005; . . .;X11 = the repayment status in April, 2005. The measurement scale for the repayment status is: -1 = pay duly; 1 = payment delay for one month; 2 = payment delay for two months; . . .; 8 = payment delay for eight months; 9 = payment delay for nine months and above. 
\item X12-X17, BILL\_AMTi: Amount of bill statement (NT dollar). X12 = amount of bill statement in September, 2005; X13 = amount of bill statement in August, 2005; . . .; X17 = amount of bill statement in April, 2005. 
\item X18-X23, PAY\_AMTi: Amount of previous payment (NT dollar). X18 = amount paid in September, 2005; X19 = amount paid in August, 2005; . . .;X23 = amount paid in April, 2005. 
\end{itemize}

\begin{tabular}{lrrrrr}
&       LIMIT\_BAL &           AGE &     EDUCATION &      MARRIAGE &             Y \\
count &    30000.00 &  30000.00 &   30000.00 &  30000.00 &  30000.00 \\
mean  &   167484.32 &     35.49 &       1.85 &      1.55 &      0.22 \\
std   &   129747.66 &      9.22 &       0.79 &      0.52 &      0.42 \\
min   &    10000.00 &     21.00 &       0.00 &      0.00 &      0.00 \\
25\%   &    50000.00 &     28.00 &       1.00 &      1.00 &      0.00 \\
50\%   &   140000.00 &     34.00 &       2.00 &      2.00 &      0.00 \\
75\%   &   240000.00 &     41.00 &       2.00 &      2.00 &      0.00 \\
max   &  1000000.00 &     79.00 &       6.00 &      3.00 &      1.00 \\
\end{tabular}

In our data set, there is an empirically observed probability of default of $22\%$, with an average of $167$ K USD lent to people aged on average 35.

\subsection{Linear regression versus Probit and Logit models}

We use \ref{sec:multreg} and cheat a little bit (we use the outcome variable...) to create an artificial wage variable. We first fit a linear regression:
\begin{equation}
y_i = \alpha + \beta x_i + \epsilon_i
\end{equation}

\quad

\begin{center}
\begin{tabular}{lclc}
\toprule
\textbf{Dep. Variable:}    &        Y         & \textbf{  R-squared:         } &     0.455   \\
\textbf{Model:}            &       OLS        & \textbf{  Adj. R-squared:    } &     0.455   \\
\textbf{Method:}           &  Least Squares   & \textbf{  F-statistic:       } & 2.504e+04   \\
\textbf{Date:}             & Sun, 29 Mar 2020 & \textbf{  Prob (F-statistic):} &     0.00    \\
\textbf{Time:}             &     17:12:53     & \textbf{  Log-Likelihood:    } &   -7083.3   \\
\textbf{No. Observations:} &       30000      & \textbf{  AIC:               } & 1.417e+04   \\
\textbf{Df Residuals:}     &       29998      & \textbf{  BIC:               } & 1.419e+04   \\
\textbf{Df Model:}         &           1      & \textbf{                     } &             \\
\bottomrule
\end{tabular}
\begin{tabular}{lcccccc}
                   & \textbf{coef} & \textbf{std err} & \textbf{t} & \textbf{P$> |$t$|$} & \textbf{[0.025} & \textbf{0.975]}  \\
\midrule
\textbf{WAGE}      &      -0.6522  &        0.004     &  -158.253  &         0.000        &       -0.660    &       -0.644     \\
\textbf{intercept} &       1.2073  &        0.006     &   186.385  &         0.000        &        1.195    &        1.220     \\
\bottomrule
\end{tabular}
\begin{tabular}{lclc}
\textbf{Omnibus:}       & 2317.352 & \textbf{  Durbin-Watson:     } &    1.664  \\
\textbf{Prob(Omnibus):} &   0.000  & \textbf{  Jarque-Bera (JB):  } & 2903.263  \\
\textbf{Skew:}          &   0.762  & \textbf{  Prob(JB):          } &     0.00  \\
\textbf{Kurtosis:}      &   3.043  & \textbf{  Cond. No.          } &     7.96  \\
\bottomrule
\end{tabular}
\end{center}

 \quad


Then we fit a Logit model:

\quad

\begin{center}
\begin{tabular}{lclc}
\toprule
\textbf{Dep. Variable:} &        Y         & \textbf{  No. Observations:  } &    30000    \\
\textbf{Model:}         &      Logit       & \textbf{  Df Residuals:      } &    29998    \\
\textbf{Method:}        &       MLE        & \textbf{  Df Model:          } &        1    \\
\textbf{Date:}          & Sun, 29 Mar 2020 & \textbf{  Pseudo R-squ.:     } &   0.5662    \\
\textbf{Time:}          &     17:12:53     & \textbf{  Log-Likelihood:    } &   -6877.3   \\
\textbf{converged:}     &       True       & \textbf{  LL-Null:           } &   -15853.   \\
\bottomrule
\end{tabular}
\begin{tabular}{lcccccc}
                   & \textbf{coef} & \textbf{std err} & \textbf{z} & \textbf{P$> |$z$|$} & \textbf{[0.025} & \textbf{0.975]}  \\
\midrule
\textbf{WAGE}      &      -7.8812  &        0.104     &   -75.843  &         0.000        &       -8.085    &       -7.678     \\
\textbf{intercept} &       9.1620  &        0.134     &    68.382  &         0.000        &        8.899    &        9.425     \\
\bottomrule
\end{tabular}
\end{center}

\quad


We finally fit a Probit model:

\quad

\begin{center}
\begin{tabular}{lclc}
\toprule
\textbf{Dep. Variable:} &        Y         & \textbf{  No. Observations:  } &    30000    \\
\textbf{Model:}         &      Probit      & \textbf{  Df Residuals:      } &    29998    \\
\textbf{Method:}        &       MLE        & \textbf{  Df Model:          } &        1    \\
\textbf{Date:}          & Sun, 29 Mar 2020 & \textbf{  Pseudo R-squ.:     } &   0.5517    \\
\textbf{Time:}          &     17:12:53     & \textbf{  Log-Likelihood:    } &   -7106.7   \\
\textbf{converged:}     &       True       & \textbf{  LL-Null:           } &   -15853.   \\
\bottomrule
\end{tabular}
\begin{tabular}{lcccccc}
                   & \textbf{coef} & \textbf{std err} & \textbf{z} & \textbf{P$> |$z$|$} & \textbf{[0.025} & \textbf{0.975]}  \\
\midrule
\textbf{WAGE}      &      -4.0937  &        0.047     &   -86.751  &         0.000        &       -4.186    &       -4.001     \\
\textbf{intercept} &       4.7555  &        0.063     &    75.227  &         0.000        &        4.632    &        4.879     \\
\bottomrule
\end{tabular}
\end{center}

\quad


As expected in empirical work, Probit and Logit models yield similar results. As detailed in statsmodels documentation, the goodness of fit measure is based on McFadden's pseudo-R-squared which is $1 - log L_1 / log L_0$ with $log L_1$the maximum loglikelihood value of the model of interest and $log L_0$ the maximum value of the loglikelihood function when all parameters, except the intercept, are set to zero. In this application, we would tend to chose the Logit model to predict the default of agents.

\section{R: difference-in-differences}

For a full theoretical lecture on difference-in-differences (DiD), we refer to Professor De Chaisemartin material, available on his website as of early 2020. 

\subsection{Synthetic Control Method}

Difference-in-differences can be seen as a subset of Synthetic Control Method, for a full picture we recommend reading \citet{abadie2021using} equation 10 and his reference to \citet{bai2009panel}, the replication code is available from the authors and very explicit \citet{abadie2015comparative}. Hypothesis testing are done with a Monte Carlo experiment in \citet{firpo2018synthetic}.

\subsection{Fake data creation and logic}

I create\footnote{DiD\_lecture.py} a mock up data set for a clear picture on difference-in-differences and continuous treatment intensity.
There will be $2N$ entities (e.g. firms). $N$ will be non-treated (control) and $N$ will be treated.
There will be 24 periods (months). [-11 to 0] nothing happens, at $t=1$ the treatment starts, and [1, 12] treated units are being treated with various intensity we observe what happens to the outcome $y$.
For all units, the outcome is decomposed as:
\begin{equation}
y_{i,t} = y_t + T_{D_{i,t}} + X_i \left( 1 + \log\left[  1 + T_{D_{i,t}} \right] \right)
\end{equation}
Entities will have control characteristic X, for control $X_i \sim  \mathcal{N}(\mu_c,.1)$ and for treated X in $X_i \sim  \mathcal{N}(\mu_t,.1)$, with
$\mu_t = 1 < \mu_c = 1.5$.
The main variable of interest $y_t$ is taken from a standard normal law for control and treated and I add a time trend equal to $t \times 10^{-2}$.

The treatment intensity is uniformly distributed among treated entities, $T_{D_{i,t}}$ is taken from a uniform distribution between 0 and 1 (so an average treatment of .5), I also add a trend to the treatment effect so the effect grows over time with a time trend equal to $t \times 10^{-2}$.

We show the control (blue) and treated (red) and how the treated would have evolved without the treatment (red dashed).

\begin{center}
\includegraphics[scale=0.4]{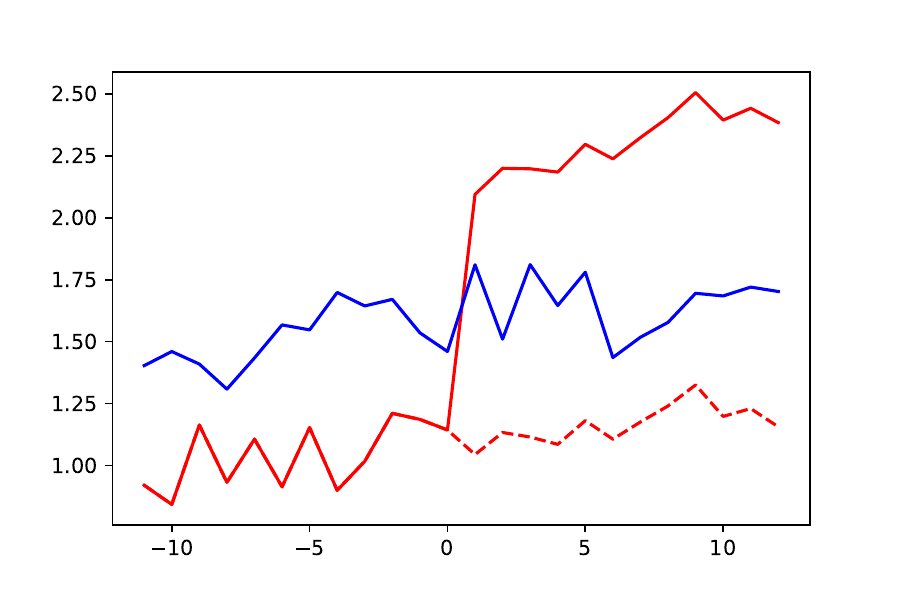}
\end{center}

\subsection{Main difference-in-differences specification}

The Difference-in-Differences can be measure as $\beta$ in specification \ref{eq:mainDiD} if assumptions are valid (see the three first assumptions in \citet{roth2023s}).
\begin{equation}\label{eq:mainDiD}
    Y_{i,t} = \alpha_i + \Phi_t + \beta D_i Post_t + \epsilon_{i,t}
\end{equation}

where $Post_t$ is a dummy equal to unity if the treatment has taken place (here $t \geq 1$) and $D_i$ is a dummy stable over time that indicates whether unit $i$ will be part of the treated group when treatment starts.

If the three assumptions are valid, then the results\footnote{DiD\_lecture.R} for our set is in Table \ref{tab:mainDiDresults}. We have a baseline value of y of 1.14 before the treatment, so (conditional on valid assumptions) we can interpret this coefficient as the treatment leads\footnote{$1.128 \times 100 / 1.14$} to a 98\% increase in y post-treatment. 

\begin{table}\label{tab:mainDiDresults}
\begingroup
\centering
\begin{tabular}{lc}
   \tabularnewline \midrule \midrule
   Dependent Variable:    & y\\  
   Model:                 & (1)\\  
   \midrule
   \emph{Variables}\\
   D $\times$ Post $=$ 1  & 1.128$^{***}$\\   
                          & (0.0780)\\   
   \midrule
   \emph{Fixed-effects}\\
   time                   & Yes\\  
   firm                   & Yes\\  
   \midrule
   \emph{Fit statistics}\\
   Observations           & 4,800\\  
   R$^2$                  & 0.23059\\  
   Within R$^2$           & 0.07477\\  
   \midrule \midrule
   \multicolumn{2}{l}{\emph{Clustered (firm) standard-errors in parentheses}}\\
   \multicolumn{2}{l}{\emph{Signif. Codes: ***: 0.01, **: 0.05, *: 0.1}}\\
\end{tabular}
\par\endgroup
\end{table}

We are typically worried about the parallel trend assumption that might be violated.

\subsection{Parallel trends assumption}

Let's define $R_{i,t}=t - G_i +1$ the time relative to the treatment (here treatment occurs at $1$, so there is no need to adjust our time labels [-11,12]), $D_{i,t}$ is a dummy whether unit $i$ received treatment in period $t$ (treatment is an absorbing state for a unit) and $G_i=\min\left( t:D_{i,t}=1 \right)$. Then we will plot $Y_{i,t}$ averaged for treated and control over time according to $R_{i,t}$, treatment occur at $1$ and is a capturing state. In dashed red line, we show how the treated would have evolved if no treatment had been applied to them. 

\begin{figure}[H]
\caption{Differences-in-Differences visual}\label{fig:DiD}
\centering
\begin{tikzpicture} \begin{axis}[
    xlabel={Time steps},
    xtick={-0,1,2},
    ylabel={$Y$},
    legend pos=north west, 
    ]
    \addplot+[blue] coordinates {
        (0,2) (1,3) (2,4) 
    };
    \addplot+[red] coordinates {
        (0,1) (1,2) (2,6)
    };
        \addplot+[red, dashed] coordinates {
        (0,1) (1,2) (2,3)
    };
    \legend{$c$,$i$}
    \plot [<->, black, thick] coordinates {(2.1,3.1)  (2.1,5.9)};
    
    \node[] at (axis cs: 2.15,4.2) {$\beta$};
\end{axis}
\end{tikzpicture}
\begin{center}
\begin{minipage}{.7\linewidth}
\footnotesize
\emph{Notes:} $i$ represents the average $y$ for the treated and $c$ the average $y$ for the control.
\end{minipage}
\end{center}
\end{figure}

Using the notation in \citet{roth2023s}, we test for pre-existing trends and if this trends were parallel for the control and treated: did outcomes for the treated and comparison groups (possibly conditional on covariates) move in parallel prior to the time of treatment? 
We replicate the same idea as equation (16) from \citet{roth2023s} in specification \ref{eq:roth16} where $\alpha_i$ an individual fixed effect and $\Phi_t$ a time fixed effect. 
\begin{equation}\label{eq:roth16}
    Y_{i,t} = \alpha_i + \Phi_t + \sum_{r\neq 0} 1\left[ R_{i,t}=r \right] \beta_r + \epsilon_{i,t}
\end{equation}

The difference-in-differences estimates are of the form of equation~\ref{eq:diddynamic} where $\bar{y}_{rd}$ is the sample mean of the outcome for units with $D_i=d$ in period $t=r\neq0$. $\hat{\beta}_r$ compares the change in the mean outcome between period $0$ and period $r$ for the treated and control units.
\begin{equation}\label{eq:diddynamic}
    \hat{\beta}_r = \left( \bar{y}_{r1} - \bar{y}_{r0}\right) - \left( \bar{y}_{01} - \bar{y}_{00} \right)
\end{equation}

\begin{figure}\label{fig:twfeplot}
\begin{center}
\includegraphics[scale=1]{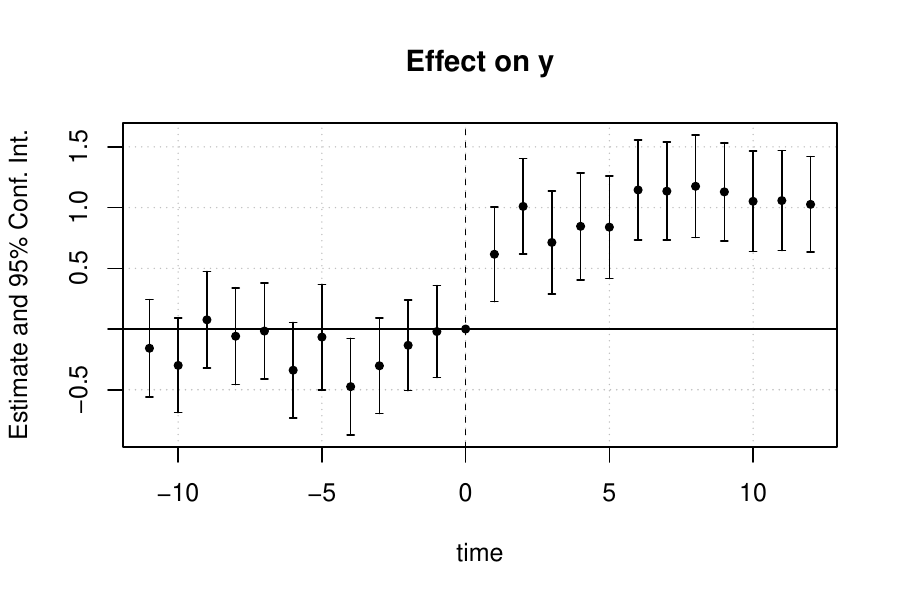}
\end{center}
\end{figure}

We find Table~\ref{tab:pretrendcoefs} and also in the visual \ref{fig:twfeplot} that all of the pre-treatment coefficients (i.e., $\hat{\beta}-r$ for $r < 0$) are insignificant, interpreted as a sign in favor of the validity of the design, since we cannot reject the null that parallel trends was satisfied in the pre-treatment period. We only display an extract in the table for visual comfort.

\begin{table}
\begingroup
\centering
\begin{tabular}{lc}
   \tabularnewline \midrule \midrule
   Dependent Variable:      & y\\  
   Model:                   & (1)\\  
   \midrule
   \emph{Variables}\\
   D $\times$ time $=$ -11  & -0.1582\\   
                            & (0.2041)\\ 
   $\ldots$  & -$\ldots$\\   
   D $\times$ time $=$ -1   & -0.0206\\   
                            & (0.1932)\\   
   D $\times$ time $=$ 1    & 0.6157$^{***}$\\   
                            & (0.1992)\\   
   D $\times$ time $=$ 2    & 1.010$^{***}$\\   
                            & (0.2007)\\ 
   $\ldots$  & -$\ldots$\\    
   \midrule
   \emph{Fixed-effects}\\
   time                     & Yes\\  
   firm                     & Yes\\  
   \midrule
   \emph{Fit statistics}\\
   Observations             & 4,800\\  
   R$^2$                    & 0.23611\\  
   Within R$^2$             & 0.08141\\  
   \midrule \midrule
   \multicolumn{2}{l}{\emph{Clustered (firm) standard-errors in parentheses}}\\
   \multicolumn{2}{l}{\emph{Signif. Codes: ***: 0.01, **: 0.05, *: 0.1}}\\
\end{tabular}
\par\endgroup
    \caption{Testing for pre-existing trends}
    \label{tab:pretrendcoefs}
\end{table}

\subsection{Violation of the parallel trend assumption}

Here, except one, no coefficient pre-treatment are significantly different from null, Figure \ref{fig:twfeplot}. But nevertheless we might be worried that the parallel trend assumption is violated, we will use \citet{rambachan2023more} to study the scope of this potential problem. As in \citet{roth2023s}, let's write the violation as the pre-treatment differences in trends between the treated and control.
\begin{equation}
\delta_r = \mathbb{E}\left[  y_{i,0}-y_{i,r} \vert D_i=1\right] - \mathbb{E}\left[  y_{i,0}-y_{i,r} \vert D_i=0\right]
\end{equation}

We typically want to check the level $M$ for which
\begin{equation}
\mbox{mean}_{r>0} (\vert \delta_r\vert ) \leq M \max_{r<0} \vert \delta_r  \vert
\end{equation}

In our case, the maximum M is between .1 and .2, hence this indicates that to invalidate the conclusion of a positive effect, we would need to allow for a post-treatment violation of parallel trends $10\%$ larger than the maximal pre-treatment violation.

\begin{center}
\includegraphics[scale=.5]{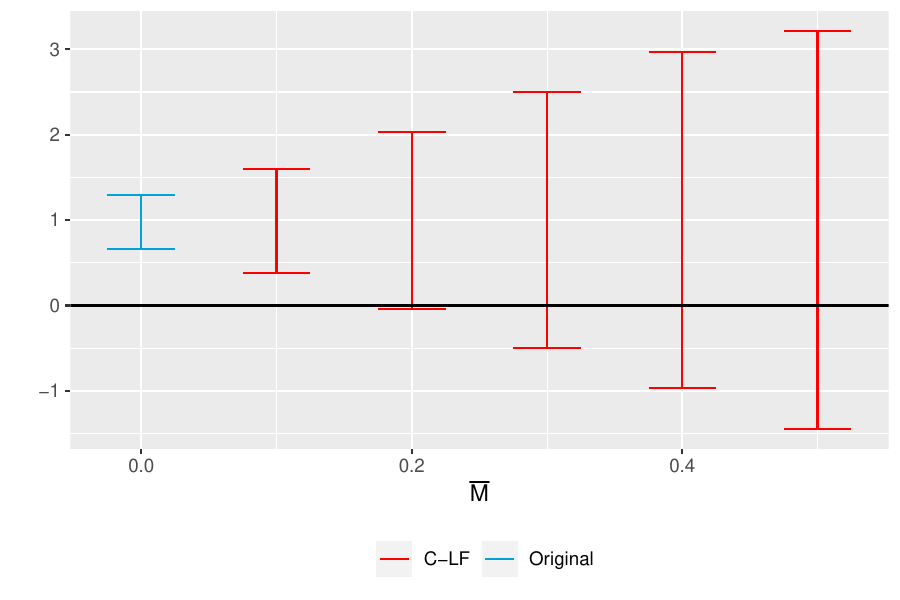}
\end{center}

The picture is much different if we focus only on the first post-treatment period (which is the default configuraiton of the HonestDiD package). It is unclear to me why such gap exists between a single period conclusion and the global conclusion.

\subsection{Continuous treatment}

Now let's consider that $D_i$ in Equation \ref{eq:mainDiD} is a continuous treatment, representing the intensity of the treatment. We use \citet{callaway2021difference} to explain what the $\beta$ (the two-way fixed effects regression estimator) estimates in the baseline case.

If we believe that our set up respects assumption 1, 2, 3 and 5 from \citet{callaway2021difference}, then \begin{quote}
In general, when the distribution of the dose is symmetric and closer to normal, TWFE weights can be close or even identical to weighting $ACR(d)$ parameters (Average Causal Response on the treated for the units who receive dose $d$) by the distribution of $D$. But when the distribution of the dose is skewed, TWFE weights $ACR(d)$ parameters close to the mean dose more than their population weights.
\end{quote}
but for uniformly distributed dose, the TWFE puts more weight on parameters in the middle of the distribution and less weights on the ends.

Especially, we assume that there is no selection bias into the doses, then we still have that $\hat{\beta}$ does not equal an average of ACR parameters weighted by the population distribution of the dose ($D$).

If we now change our estimates with the continuous treatment, we get an update of Table \ref{tab:mainDiDresults} with Table \ref{tab:mainCTresults}, but now the estimate of $\beta$ is not interpretable as the distribution of the treatment is uniform.

\begin{table}\label{tab:mainCTresults}
\begingroup
\centering
\begin{tabular}{lc}
   \tabularnewline \midrule \midrule
   Dependent Variable:            & y\\  
   Model:                         & (1)\\  
   \midrule
   \emph{Variables}\\
   treatment $\times$ Post $=$ 1  & 1.734$^{***}$\\   
                                  & (0.0770)\\   
   \midrule
   \emph{Fixed-effects}\\
   time                           & Yes\\  
   firm                           & Yes\\  
   \midrule
   \emph{Fit statistics}\\
   Observations                   & 4,800\\  
   R$^2$                          & 0.25565\\  
   Within R$^2$                   & 0.10491\\  
   \midrule \midrule
   \multicolumn{2}{l}{\emph{Clustered (firm) standard-errors in parentheses}}\\
   \multicolumn{2}{l}{\emph{Signif. Codes: ***: 0.01, **: 0.05, *: 0.1}}\\
\end{tabular}
\par\endgroup
    \caption{TWFE with unifrom continuous treatment}
\end{table}

\subsection{Gaussian continuous treatment}

Now I update with a Gaussian continuous treatment\footnote{DiD\_lecture\_gaussian\_treatment.R}, with the same average treatment of .5 and 1 as standard deviation.

We show the average output for control (blue) and treated (red) and how the average treated output would have evolved without the treatment (red dashed).

\begin{center}
\includegraphics[scale=0.4]{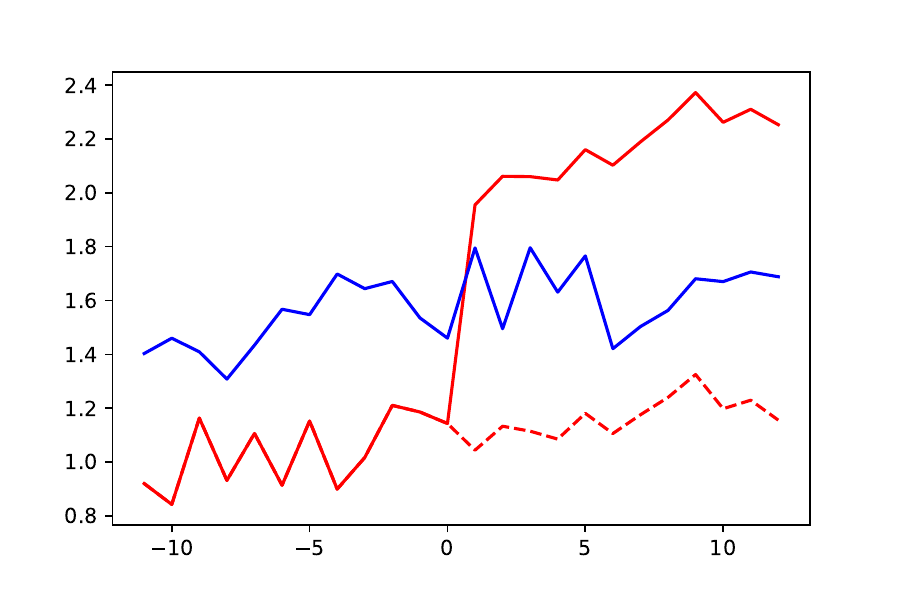}
\end{center}

The event plot becomes:
\begin{center}
\includegraphics[scale=1]{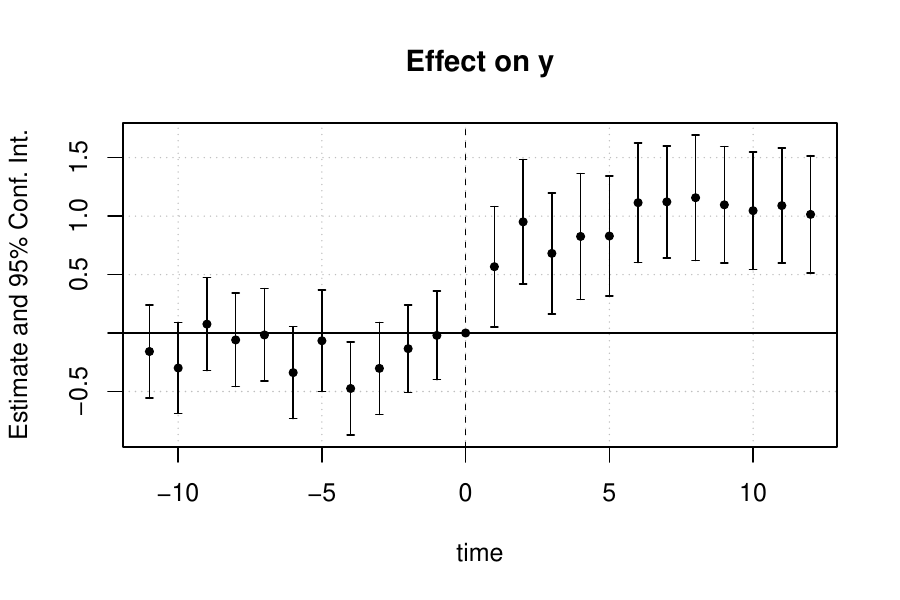}
\end{center}

The honest DiD plot:
\begin{center}
\includegraphics[scale=.5]{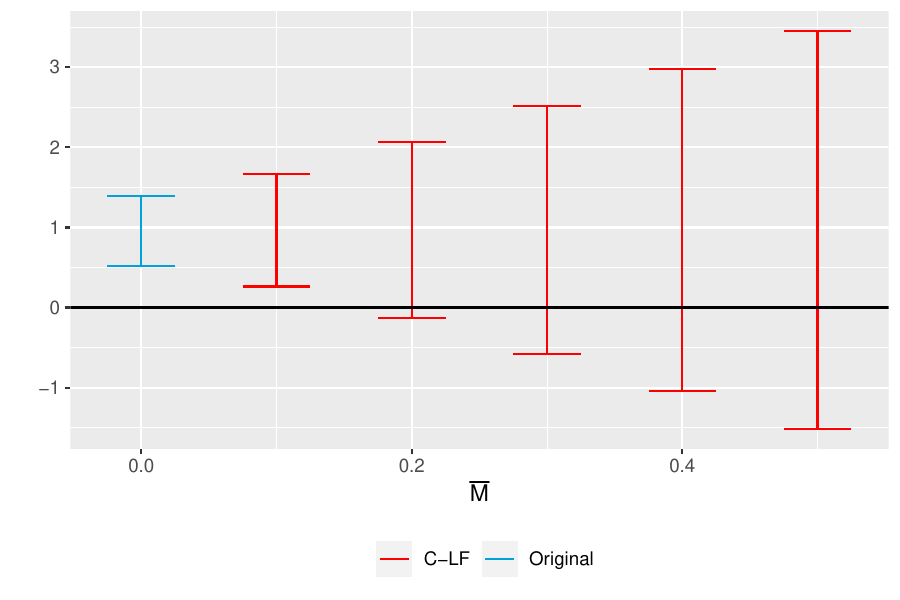}
\end{center}

And now we can interpret the TWFE as the average of the ACR parameters weighted by the population distribution of the dose Table \ref{tab:CTgaussian}.

\begin{table}\label{tab:CTgaussian}
\begingroup
\centering
\begin{tabular}{lc}
   \tabularnewline \midrule \midrule
   Dependent Variable:            & y\\  
   Model:                         & (1)\\  
   \midrule
   \emph{Variables}\\
   treatment $\times$ Post $=$ 1  & 1.436$^{***}$\\   
                                  & (0.2261)\\   
   \midrule
   \emph{Fixed-effects}\\
   time                           & Yes\\  
   firm                           & Yes\\  
   \midrule
   \emph{Fit statistics}\\
   Observations                   & 4,764\\  
   R$^2$                          & 0.31830\\  
   Within R$^2$                   & 0.05550\\  
   \midrule \midrule
   \multicolumn{2}{l}{\emph{Clustered (firm) standard-errors in parentheses}}\\
   \multicolumn{2}{l}{\emph{Signif. Codes: ***: 0.01, **: 0.05, *: 0.1}}\\
\end{tabular}
\par\endgroup 
\caption{TWFE with Gaussian continuous treatment}
\end{table}

We have a baseline value of $y$ of 1.147 before the treatment, so we can interpret this coefficient as the treatment leads to a $125\%$ increase in $y$ post-treatment. This is coherent with the version where we take the logarithm of $y$.

\subsubsection{Average Level effects: What was the effect of dose $d$?}

Under some the assumptions 1, 2, 3, and 5, the average treatment effect at $d$, ATE(d) is identified and is $\mathbb{E}\left[ \Delta y_t\vert D=d \right] - \mathbb{E}\left[ \Delta y_t\vert D=0 \right]$. Under these conditions, it is equal to ATT($d\vert d$), the average effect of dose $d$ on units that actually experienced dose $d$: $ATT(a \vert b) = \mathbb{E}\left[ y_t(a) - y_t(b) \vert D=b \right]$.

We know the true theoretical value as we constructed our data set:
\begin{equation}
ATE(d,t) = T_{D_{i,t}} + X_i \log\left[  1 + T_{D_{i,t}} \right]
\end{equation}

I will focus on the first period for simplicity: $t=1$, then I can display ATE($d$,1) as a function of $d$, it is almost linear and an ols can be fitted:
$ATE(d,1) = \alpha + \beta d + \epsilon_{d,1}$ and we find the results table \ref{tab:olsATE}.

\begin{table}\label{tab:olsATE}
\begin{center}
\begin{tabular}{lclc}
\toprule
\textbf{Dep. Variable:}    &       ATE        & \textbf{  R-squared:         } &     0.983   \\
\textbf{Model:}            &       OLS        & \textbf{  Adj. R-squared:    } &     0.982   \\
\textbf{Method:}           &  Least Squares   & \textbf{  F-statistic:       } &     5522.   \\
\textbf{Date:}             & Tue, 05 Sep 2023 & \textbf{  Prob (F-statistic):} &  5.50e-88   \\
\textbf{Time:}             &     10:18:55     & \textbf{  Log-Likelihood:    } &    87.719   \\
\textbf{No. Observations:} &         100      & \textbf{  AIC:               } &    -171.4   \\
\textbf{Df Residuals:}     &          98      & \textbf{  BIC:               } &    -166.2   \\
\textbf{Df Model:}         &           1      & \textbf{                     } &             \\
\textbf{Covariance Type:}  &    nonrobust     & \textbf{                     } &             \\
\bottomrule
\end{tabular}
\begin{tabular}{lcccccc}
                   & \textbf{coef} & \textbf{std err} & \textbf{t} & \textbf{P$> |$t$|$} & \textbf{[0.025} & \textbf{0.975]}  \\
\midrule
\textbf{const}     &      -0.0100  &        0.016     &    -0.626  &         0.532        &       -0.042    &        0.022     \\
\textbf{treatment} &       1.7264  &        0.023     &    74.312  &         0.000        &        1.680    &        1.772     \\
\bottomrule
\end{tabular}
\begin{tabular}{lclc}
\textbf{Omnibus:}       & 64.540 & \textbf{  Durbin-Watson:     } &    2.054  \\
\textbf{Prob(Omnibus):} &  0.000 & \textbf{  Jarque-Bera (JB):  } &  312.713  \\
\textbf{Skew:}          & -2.135 & \textbf{  Prob(JB):          } & 1.25e-68  \\
\textbf{Kurtosis:}      & 10.537 & \textbf{  Cond. No.          } &     3.04  \\
\bottomrule
\end{tabular}
\end{center}
\caption{ATE($d$,1) as a function of the treatment intensity $d$}
\end{table}

\subsubsection{Average Slope effects: What was the incremental effect of the $d^{th}$ dose unit?}

Under the same assumptions, the slope is identified, we can directly use the table \ref{tab:olsATE} and this is the estimated coefficient of the ols. ACRT($d\vert d$) is the average difference between potential
outcomes under dose $d$ compared to potential outcomes under a marginal change in the dose for the group of units who actually experience dose $d$.
\begin{equation*}
ACR(d) = ACRT(d\vert d) = \frac{\partial \mathbb{E}\left[ \Delta y_t\vert D=d \right] }{\partial d}
\end{equation*}

\section{R: difference-in-differences, replication of \citet{Bleakley2010}}\label{sec:bleakleyrep}

\citep{Bleakley2010} uses\footnote{Here we will approach this method with a replication code 20200305\_Bleakley2010 of the paper, inspired by Professor Pascaline Dupas' PSE lecture.} the malaria-eradication campaigns in the United States (circa 1920) and other areas\footnote{we focus on the United States} to measure how much childhood exposure to malaria depresses labor productivity as an adult. Considering that \begin{quote}
children are more susceptible to malaria than adults [\ldots]. Most of a person's human-capital and physiological development happens in childhood
\end{quote}

The study uses the rapid impact of the treatment campaign which makes it possible to compare at a given year $t > 1920$ between individuals that were more or less exposed to malaria during childhood.

This study uses the cross-area heterogeneity in terms of malaria intensity pre-1920 to measure the effect of childhood exposure to malaria on adulthood labor productivity.

\begin{quote}
Areas with high malaria infection rates had more to gain from eradication, but the non-malarious areas serve as a comparison group
\end{quote}

It is reasonable to consider the malaria-eradication campaigns to be exogenous as 
\begin{quote}
episodes of rapid progress against malaria were made possible by critical innovations to knowledge and spending, which are shocks that came from outside the studied areas
\end{quote}

\begin{quote}
Being born later and in an area with high pre-campaign malaria burden implies more exposure to the benefits of the eradication programs. To test this hypothesis, census micro-data samples of native-born males from all four countries are used to construct panel data of cohorts by birth year and birthplace. 
\end{quote}

\subsection{The disease}

\begin{quote}
Malaria is a parasitic disease that afflicts humans. Acute symptoms of infection include fever, headache, and nausea. An important chronic symptom is anaemia. Malaria results in death in some cases, but the strains prevalent in the Americas (\emph{vivax}, and to a lesser extent \emph{malariae}) have quite low case-fatality rates, especially compared with the predominantly African variety (\emph{falciparum}).
The parasite has a complicated life-cycle that is partly spent in a mosquito vector and partly in the human host. The disease is transmitted when a mosquito takes a blood meal from an infected person and, some time later, bites another person. Because of the crucial role played by mosquitoes in the transmission cycle, warm and wetter climates are more likely to sustain endemic malaria.
\end{quote}

\subsection{Research design}

For the United-States, we define:
\begin{itemize}
\item "young" as those born in 1920 or later
\item "old" as those born in 1899 or earlier
\end{itemize}
taking some margin for the definition of old, as we could technically set it in 1902:

\begin{center}
\includegraphics[scale=0.4]{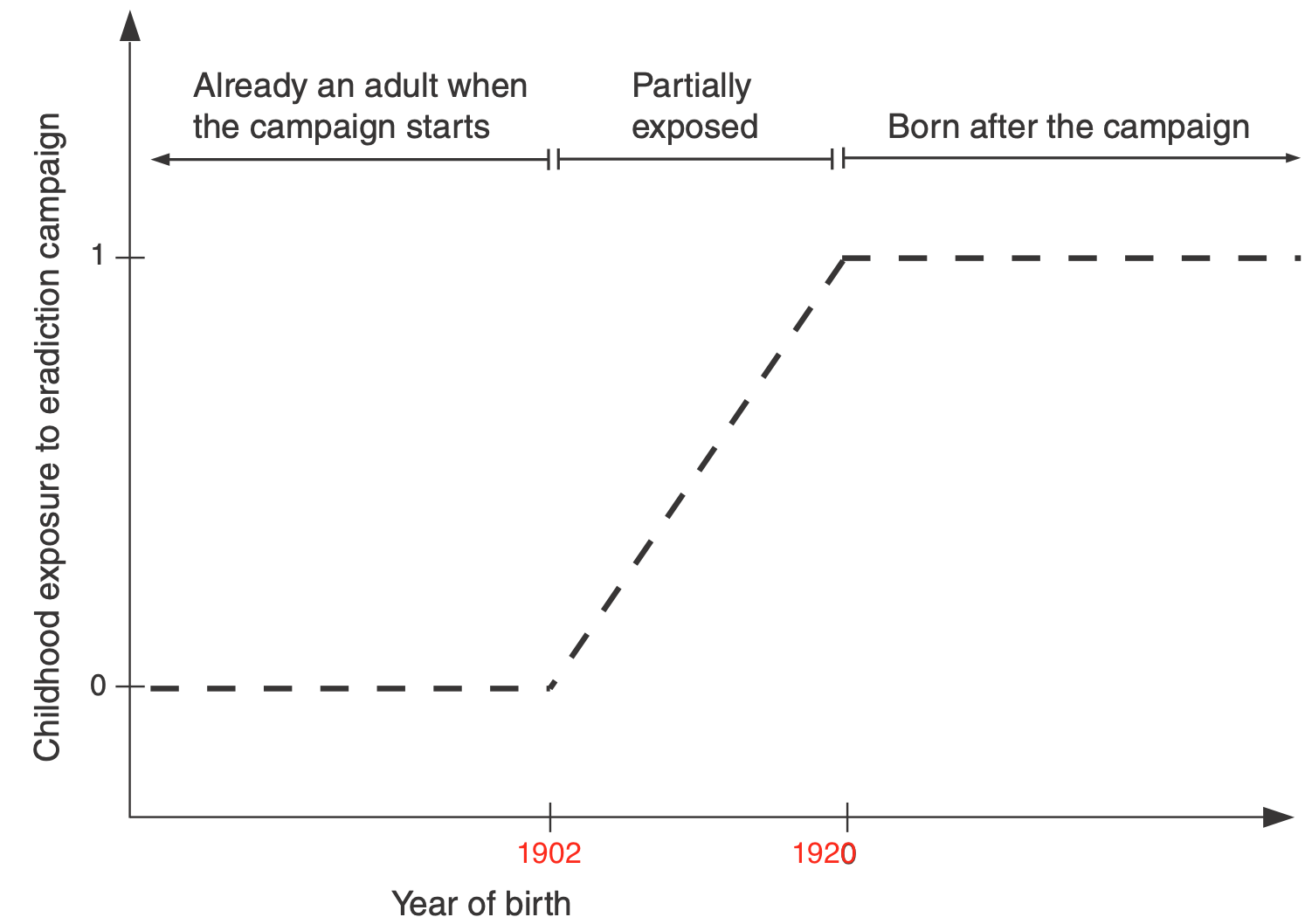}
\end{center}

We replicate Table 1 in Table \ref{tab:ols1} (omit to show the intercept for ease of reading) to measure the effect of malaria mortality per 10,000 people, 1890 of each area $j$ on these two differences. $\beta$ is the coefficient of interest in specification \ref{eq:malariaspec}. Column (3) and (4) Table \ref{tab:ols1} are computed with weighted least squares as in the paper using the square root of the cell sizes to account for the different precision with which cohort means are estimated.

\begin{equation}\label{eq:malariaspec}
\bar{Y}_{j,post} - \bar{Y}_{j,pre} = \alpha + \beta \mbox{malmort1890}_{j,pre} + \epsilon_{j, post}
\end{equation}

with:
\begin{itemize}
\item $\bar{Y}$ either the Duncan's socio-economic index (sei) or the occupation income score (occscore) averaged for a birthplace $j$ for people born well before ($<=1899$) or just after ($>=1920$) the treatment campaign
\item $j$ the area of birth
\item $\mbox{malmort1890}_{j,pre}$ is the malaria mortality per 10,000 people in the area $j$ in year 1890
\item "diffsei" for each area of birth $j$ the difference of the average of sei between the young and the old cohorts
\item "diffoccscore" for each area of birth $j$ the difference of average occscore between the yound and the old cohorts
\end{itemize}

\begin{table}[!htbp] \centering 
  \caption{} 
  \label{tab:ols1} 
\resizebox{\textwidth}{!}{
\begin{tabular}{@{\extracolsep{5pt}}lcccc} 
\\[-1.8ex]\hline 
\hline \\[-1.8ex] 
 & \multicolumn{4}{c}{\textit{Dependent variable:}} \\ 
\cline{2-5} 
\\[-1.8ex] & diffsei & diffoccscore & diffsei & diffoccscore \\ 
\\[-1.8ex] & (1) & (2) & (3) & (4)\\ 
\hline \\[-1.8ex] 
 malmort1890 & 0.118$^{**}$ & 0.117$^{***}$ & 0.131$^{***}$ & 0.134$^{***}$ \\ 
  & (0.053) & (0.028) & (0.003) & (0.002) \\ 
  & & & & \\ 
\hline \\[-1.8ex] 
Observations & 49 & 49 & 7,211 & 7,211 \\ 
R$^{2}$ & 0.095 & 0.274 & 0.191 & 0.410 \\ 
Adjusted R$^{2}$ & 0.076 & 0.259 & 0.191 & 0.410 \\ 
Residual Std. Error & 0.136 (df = 47) & 0.071 (df = 47) & 0.439 (df = 7209) & 0.260 (df = 7209) \\ 
F Statistic & 4.959$^{**}$ (df = 1; 47) & 17.742$^{***}$ (df = 1; 47) & 1,699.031$^{***}$ (df = 1; 7209) & 5,008.722$^{***}$ (df = 1; 7209) \\ 
\hline 
\hline \\[-1.8ex] 
\textit{Note:}  & \multicolumn{4}{r}{$^{*}$p$<$0.1; $^{**}$p$<$0.05; $^{***}$p$<$0.01} \\ 
\end{tabular} 
}
\end{table} 

\subsection{Difference-in-differences estimator of the effect of malaria eradication on the Duncan SEI index}

We are following\footnote{we use the notations from Teppei Yamamoto's lecture on difference-in-differences methods} several individuals that we can divide into young and old and area of birth that we can divide into low and high-malaria intensity. We can consider that old individuals will never get treated and that low-malaria-intensity area will not benefit from the treatment. We would like to measure the treatment effect, that is the difference between the outcome $Y$ (the revenue here) for agents that were treated and for the same agents if they had not been treated:
$$ DiD =  E\left( Y_{i1}(1) \vert G_i = 1\right) - E\left( Y_{i1}(0) \vert G_i = 1\right) $$
this is the average treatment effect of the treated (ATT) with:
\begin{itemize}
\item $Y_{ig}(j)$: potential outcome for unit $i$ in group $g$ when treated ($j=1$) or not treated ($j=0$).
\item in our set up:
\begin{itemize}
\item $i$ is the individual index;
\item $g$ is the young when treatment occurred dummy;
\item $j$ is the "in a treated area" dummy;
\end{itemize}
\end{itemize}

Obviously, we cannot know the outcome for agents of the treatment group if they had not been treated: $E\left( Y_{i1}(0) \vert G_i = 1\right)$, because in fact they have been treated $G_i = 1$ so we only observe $Y_{i1}(1)$ and not $Y_{i1}(0)$ when $G_i = 1$. We use a control strategy, called the difference-in-differences (DiD)$Y_{i1}(0)$  when $G_i = 0$, but then we need some additional assumptions that will need to be econometrically tested:

\begin{equation}
\mbox{DiD} = \left[ E\left( Y_{i1}(1) \vert G_i = 1\right) - E\left( Y_{i1}(0) \vert G_i = 0\right)\right] - \left[ E\left( Y_{i0}(1) \vert G_i = 1\right) - E\left( Y_{i0}(0) \vert G_i = 0\right)\right]
\end{equation}
assuming parallel trends:
\begin{equation}
E\left[ Y_{i1}(0) - Y_{i0}(0) \vert G_i = 0 \right] = E\left[ Y_{i1}(0)-Y_{i0}(0) \vert G_i = 1\right]
\end{equation}

We can manually compute the DiD estimator which is exactly what we find in our first DiD specification.

\quad

\begin{tabular}{c|c|c|}
\cline{2-3}
& \multicolumn{2}{ c| }{Cohort} \\ \cline{1-3}
Area intensity & old =1 ($g=0$) & young = 1 ($g=1$) \\ \cline{1-3}
highmalaria=1 & Untreated & Treated\\
 & $E\left( Y_{i0}(0) \vert G_i = 1\right)$  & $E\left( Y_{i1}(1) \vert G_i = 1\right)$ \\ \cline{1-3}
lowmalaria=1 & Untreated & Untreated\\
 & $E\left( Y_{i0}(0) \vert G_i = 0\right)$  & $E\left( Y_{i1}(0) \vert G_i = 0\right)$ \\ \cline{1-3}
\end{tabular}

\quad

We first observe that old group income are measured in the years 1970 to 1950 and the old group income are measured in the years 1960 to 1990. So when measuring the average group income, there will be the effect that over time incomes have grown in the United States. There is another effect, in our sampled income observation the old group has an average age around 45 and the young group has an average age of 43, so we can expect that there is also an experience effect where people aged 45 should earn more than people aged 42. In the data, we find that the income growth effect dominates.

We can think of the DiD effect as following:
\begin{itemize}
\item $Y_l$ \quad ($E\left( Y_{i1}(0) \vert G_i = 0\right)$) is the revenue for people that were born in low-intensity malaria area, they almost did not benefit from the treatment as the area was almost clean already from malaria, so we can consider that they were not treated, as the low;
\item $O_l$ \quad ($E\left( Y_{i0}(0) \vert G_i = 0\right)$) is the revenue for people that were born in low-intensity malaria area, the difference $Y_l - O_l > 0$ should be mainly due to the fact that over time incomes have grown in the United States, effect that dominates the experience effect as the old group average age is higher than the young group;
\item $Y_h$ \quad ($E\left( Y_{i1}(1) \vert G_i = 1\right)$) is the revenue for people that were born in high-intensity malaria area, they benefited from the treatment;
\item $Y'_h$ \quad ($E\left( Y_{i1}(0) \vert G_i = 0\right)$) is a fictitious revenue for people that would be born in high-intensity malaria area at the same time as the cohort used to compute the average $Y_h$ but those would not have benefited from the treatment (which in practice is not possible to measure), our DiD estimator is $Y'_h - Y_h$;
\item $O_h$ \quad ($E\left( Y_{i0}(0) \vert G_i = 1\right)$) is the revenue for people that were born in high-intensity malaria area and became adult before the treatment, so the difference $Y_h - O_h$ reflect (1) the income growth effect (positive sign expected), (2) the fact that as young cohort benefited from the treatment they are more productive once adult and can earn more (positive sign expected) and (3) the experience effect (negative sign expected). (2) is the DiD we want to isolate from other effects.
\end{itemize}

Note that the high-intensity dotted line and the line of the low-intensity are parallel, this is called the parallel trend assumption.
\begin{center}
\includegraphics[scale=0.5]{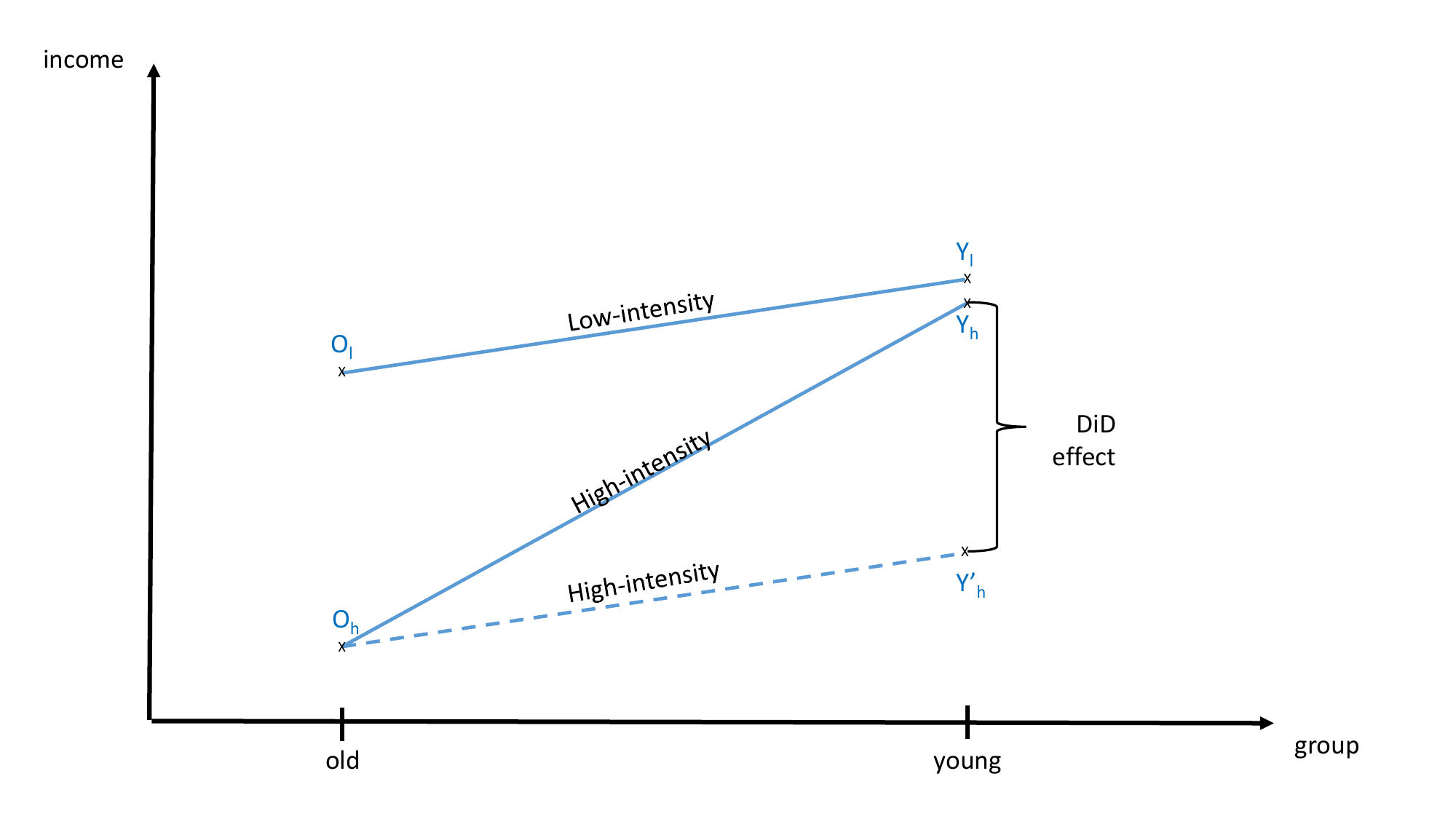}
\end{center}

We define the dummies:
\begin{itemize}
\item "high" malaria areas as those in the 90th percentile or greater in terms of "malmort1890"
\item "low" malaria areas as those in the 10th percentile or lower. 
\end{itemize}

Now we can estimate the DiD with a specification:

\begin{equation}\label{eq:dideq1}
\log(\mbox{Y}_{j,k,t}) = c + \alpha_k \mbox{highmalaria}_{j} + \alpha_j \mbox{young}_{k} + \beta \mbox{highmalaria}_{j} \mbox{young}_{k} + \epsilon_{j,k,t}
\end{equation}

We also test the following specifications with fixed effects, a two-way fixed effects (TWFE) regression, closer to the paper Table 4 and report the coefficients of interest in columns (3) and (4). \citet{roth2023s} shows that when the panel is balance, equation \ref{eq:dideq1} and \ref{eq:dideq2} are equivalent.
\begin{equation}\label{eq:dideq2}
\log(\mbox{Y}_{j,k,t}) = \delta_k + \delta_j + \delta_t  + \beta \mbox{highmalaria}_{j} \mbox{young}_{k} + \epsilon_{j,k,t}
\end{equation}

with:
\begin{itemize}
\item $Y$ either the Duncan's socio-economic index (sei) or the occupation income score (occscore)
\item $j$ the area of birth
\item $k$ the year of birth
\item $t$ the census year
\item $\mbox{young}_{k}$ is taken to be $1$ for "young" and $0$ otherwise
\item $\mbox{highmalaria}_{j}$ is taken to be $1$ for high intensity exposure to malaria in the area of birth (90\% quantile), $0$ otherwise, but NaN when there is a lack of malmort1890 information
\end{itemize}

\quad

$\beta$ is the DiD estimator.

\quad

We display the results in Table \ref{tab:did1} where we further restricted the data set to highmalaria or lowmalaria equal to $1$ and young or old equal to $1$ for the regressions.

\quad

\begin{table}[!htbp] \centering 
  \caption{} 
  \label{tab:did1} 
\resizebox{\textwidth}{!}{
\begin{tabular}{@{\extracolsep{5pt}}lcccc} 
\\[-1.8ex]\hline 
\hline \\[-1.8ex] 
 & \multicolumn{4}{c}{\textit{Dependent variable:}} \\ 
\cline{2-5} 
\\[-1.8ex] & sei & occscore & sei & occscore \\ 
\\[-1.8ex] & (1) & (2) & (3) & (4)\\ 
\hline \\[-1.8ex] 
 highmalaria & $-$0.215$^{***}$ & $-$0.171$^{***}$ &  &  \\ 
  & (0.064) & (0.044) &  &  \\ 
  & & & & \\ 
 young & 0.330$^{***}$ & 0.155$^{***}$ &  &  \\ 
  & (0.048) & (0.035) &  &  \\ 
  & & & & \\ 
 highmalaria:young & 0.159$^{***}$ & 0.143$^{***}$ & 0.151$^{**}$ & 0.133$^{***}$ \\ 
  & (0.057) & (0.038) & (0.070) & (0.045) \\ 
  & & & & \\ 
\hline \\[-1.8ex] 
Observations & 1,859 & 1,859 & 1,859 & 1,859 \\ 
R$^{2}$ & 0.396 & 0.360 & 0.535 & 0.511 \\ 
Adjusted R$^{2}$ & 0.395 & 0.359 & 0.500 & 0.474 \\ 
Residual Std. Error & 0.266 (df = 1855) & 0.172 (df = 1855) & 0.242 (df = 1728) & 0.156 (df = 1728) \\ 
\hline 
\hline \\[-1.8ex] 
\textit{Note:}  & \multicolumn{4}{r}{$^{*}$p$<$0.1; $^{**}$p$<$0.05; $^{***}$p$<$0.01} \\ 
\end{tabular} 
}
\end{table}

The standard errors are clustered at the region (bplg) level, as we can expect time series-type correlation for each region variables. To avoid the "Ashenfelter dip", we assume that there was no self selection of areas into the treatment which as explained in the paper is plausible. We also assume that the trend were the same before the treatment for each region, the so-called "parallel-trend" assumption.

\subsubsection{Robustness-checks}

We define the dummies:
\begin{itemize}
\item "veryyoung" as those born in 1940 or later
\item "veryold" as those born in 1860 or earlier
\item "kindofyoung" those young but not veryyoung
\item "kindofold" those old but not veryold
\end{itemize}

As robustness check, we can compare if these results are robust when comparing agents "very" exposed to the treatment ("veryyoung") to "kind-of-exposed to the treatment ("kindofyoung"), in Table \ref{tab:did2} where we limit our data set to high or low malaria intensity region and to young cohort. We also confirm that there is no significance in the DiD estimator, when comparing very old with kind of old where we limit our data set to high or low malaria intensity region and old cohort .

\begin{table}[!htbp] \centering 
  \caption{} 
  \label{tab:did2} 
\resizebox{\textwidth}{!}{
\begin{tabular}{@{\extracolsep{5pt}}lcccc} 
\\[-1.8ex]\hline 
\hline \\[-1.8ex] 
 & \multicolumn{4}{c}{\textit{Dependent variable:}} \\ 
\cline{2-5} 
\\[-1.8ex] & sei & occscore & sei & occscore \\ 
\\[-1.8ex] & (1) & (2) & (3) & (4)\\ 
\hline \\[-1.8ex] 
 highmalaria & $-$0.066$^{*}$ & $-$0.033$^{*}$ & $-$0.230$^{***}$ & $-$0.192$^{***}$ \\ 
  & (0.037) & (0.018) & (0.080) & (0.055) \\ 
  & & & & \\ 
 veryyoung & 0.044$^{***}$ & 0.019$^{***}$ &  &  \\ 
  & (0.015) & (0.005) &  &  \\ 
  & & & & \\ 
 highmalaria:veryyoung & 0.027$^{*}$ & 0.016$^{***}$ &  &  \\ 
  & (0.015) & (0.006) &  &  \\ 
  & & & & \\ 
 kindofold &  &  & 0.169$^{***}$ & 0.101$^{***}$ \\ 
  &  &  & (0.036) & (0.015) \\ 
  & & & & \\ 
 highmalaria:kindofold &  &  & 0.025 & 0.035 \\ 
  &  &  & (0.069) & (0.038) \\ 
  & & & & \\ 
\hline \\[-1.8ex] 
Observations & 640 & 640 & 1,219 & 1,219 \\ 
R$^{2}$ & 0.178 & 0.132 & 0.168 & 0.209 \\ 
Adjusted R$^{2}$ & 0.174 & 0.128 & 0.166 & 0.207 \\ 
Residual Std. Error & 0.086 (df = 636) & 0.050 (df = 636) & 0.310 (df = 1215) & 0.201 (df = 1215) \\ 
\hline 
\hline \\[-1.8ex] 
\textit{Note:}  & \multicolumn{4}{r}{$^{*}$p$<$0.1; $^{**}$p$<$0.05; $^{***}$p$<$0.01} \\ 
\end{tabular} 
}
\end{table}

\subsection{Figure 4 replication}

Before running multiple regressions, we compute for each year of birth the difference of average sei between high and low intensity malaria:
\begin{center}
\includegraphics[scale=0.2]{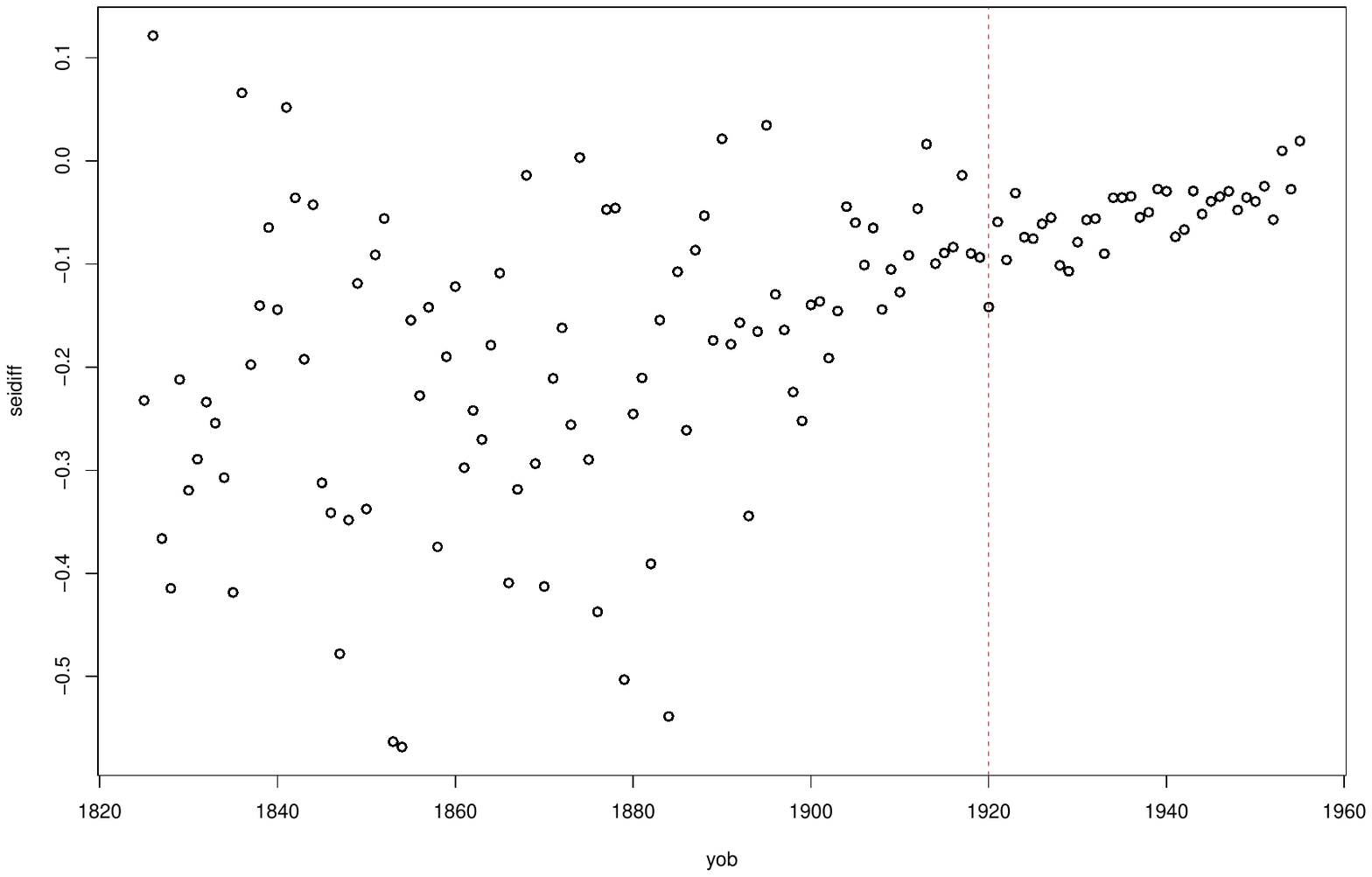}
\end{center}

We use the specification to extract the estimated $\beta_k$ per year of birth:
\begin{equation}
\log(\mbox{occscore}_{j,k,t}) = \beta_k \mbox{malariascaled}_{j} + \delta_j + \delta_k + \delta_t + \epsilon_{j,k,t}
\end{equation}

with:
\begin{itemize}
\item $j$ the area of birth
\item $k$ the year of birth
\item $t$ the census year
\item $\mbox{malariascalde}_{j}$ the malaria mortality per 10,000 people, 1890 scaled with the $95$ and $5$ quantile as $\frac{malmort1890}{q_{95}-q_{5}}$ 
\end{itemize}

We plot the estimated $\beta_k$ against the year of birth $k$:

\begin{center}
\includegraphics[scale=0.3]{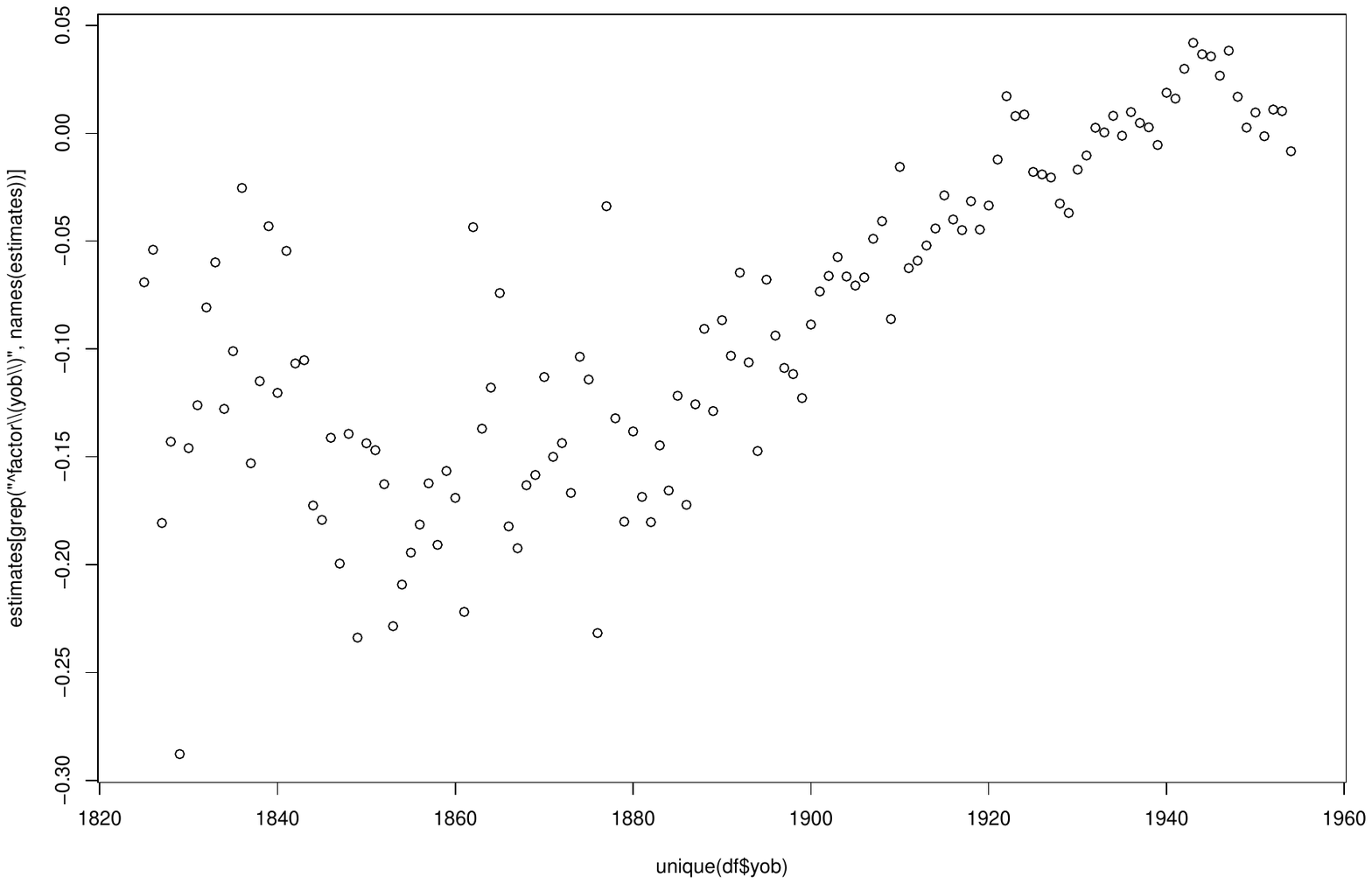}
\end{center}

\subsection{Difference-in-differences, pooled regression analysis}

Then we replicate Table 4 specifications of the first column for the four rows (omitting the fifth row that is computationally intensive) and display results in Table \ref{tab:table4}:
\begin{equation}
\log(\mbox{Y}_{j,k,t}) = \alpha + \beta \mbox{malmort1890}_{j} \mbox{exp\_mal}_{k} + \delta_k + \delta_j + \delta_t  + \epsilon_{j,k,t}
\end{equation}
we further test the specifications:
\begin{equation}
\log(\mbox{Y}_{j,k,t}) = \alpha + \beta \mbox{malmort1890}_{j} \mbox{exp\_mal}_{k} + \delta_k + \delta_j + \delta_t + \delta_j \times I_{t>1920}  + \beta_t \mbox{malmort1890}_{j} \mbox{exp\_mal}_k \times I_{t>1920} + \epsilon_{j,k,t}
\end{equation}

\begin{equation}
\log(\mbox{Y}_{j,k,t}) = \beta \mbox{malmort1890}_{j} \mbox{exp\_mal}_{k} + \delta_k \times \delta_j + \delta_k +  \delta_j + \delta_t   + \epsilon_{j,k,t}
\end{equation}

For the fourth row, we limit the data set to years after or equal to 1930.

We do not report the fifth row as it is too computationally intensive:
\begin{equation}
\log(\mbox{Y}_{j,k,t}) = \beta \mbox{malmort1890}_{j} \mbox{exp\_mal}_{k} + \delta_k \times \delta_j \times \delta_t + \epsilon_{j,k,t}
\end{equation}
and all possible combinations of $\delta_k$, $\delta_j$ and $\delta_t$.

with:
\begin{itemize}
\item $\mbox{exp\_mal}_{k}$ the fraction of population aged 0-20 exposed to malaria eradication campaign, which depends on your year of birth $k$
\end{itemize}

\begin{table}[!htbp] \centering 
  \caption{} 
  \label{tab:table4} 
\resizebox{\textwidth}{!}{
\begin{tabular}{@{\extracolsep{5pt}}lcccc} 
\\[-1.8ex]\hline 
\hline \\[-1.8ex] 
 & \multicolumn{4}{c}{\textit{Dependent variable:}} \\ 
\cline{2-5} 
\\[-1.8ex] & \multicolumn{4}{c}{sei} \\ 
\\[-1.8ex] & (1) & (2) & (3) & (4)\\ 
\hline \\[-1.8ex] 
 exp\_mal:malmort1890 & 0.111$^{***}$ & 0.098$^{***}$ & 0.084$^{***}$ & 0.124$^{***}$ \\ 
  & (0.013) & (0.003) & (0.010) & (0.034)  \\ 
  & & & & \\ 
\hline \\[-1.8ex] 
Observations & 11,165 & 11,165 & 11,165 & 7,123 \\ 
R$^{2}$ & 0.545 & 0.563 & 0.629 & 0.530 \\ 
Adjusted R$^{2}$ & 0.537 & 0.554 & 0.604 & 0.521 \\ 
Residual Std. Error & 0.238 (df = 10974) & 0.233 (df = 10926) & 0.220 (df = 10454) & 0.171 (df = 6987) \\ 
\hline 
\hline \\[-1.8ex] 
\textit{Note:}  & \multicolumn{4}{r}{$^{*}$p$<$0.1; $^{**}$p$<$0.05; $^{***}$p$<$0.01} \\ 
\end{tabular} 
}
\end{table} 

\subsection{Interpretation}

We run the specification replicating the first row of Table 5 in Table \ref{tab:table5} with standard error clustered at places of birth:
\begin{equation}
\log(\mbox{Y}_{j,k,t}) = \beta \mbox{malariascaled}_j \mbox{exp\_mal}_k + \delta_j + \delta_k + \delta_t + \epsilon_{j,k,t}
\end{equation}

\begin{table}[!htbp] \centering 
  \caption{} 
  \label{tab:table5} 
\begin{tabular}{@{\extracolsep{5pt}}lcc} 
\\[-1.8ex]\hline 
\hline \\[-1.8ex] 
 & \multicolumn{2}{c}{\textit{Dependent variable:}} \\ 
\cline{2-3} 
\\[-1.8ex] & sei & occscore \\ 
\\[-1.8ex] & (1) & (2)\\ 
\hline \\[-1.8ex] 
 exp\_mal:malmortscaled & 0.137$^{***}$ & 0.147$^{***}$ \\ 
  & (0.006) & (0.004) \\ 
  & & \\ 
  Maximal endemicity & .3 & .3 \\
  Income effect per probability of childhood infection & .46 & .49 \\
\hline \\[-1.8ex] 
Observations & 11,165 & 11,165 \\ 
R$^{2}$ & 0.865 & 0.839 \\ 
Adjusted R$^{2}$ & 0.862 & 0.836 \\ 
Residual Std. Error (df = 10974) & 0.419 & 0.265 \\ 
\hline 
\hline \\[-1.8ex] 
\textit{Note:}  & \multicolumn{2}{r}{$^{*}$p$<$0.1; $^{**}$p$<$0.05; $^{***}$p$<$0.01} \\ 
\end{tabular} 
\end{table} 


The DiD estimator is composed of two elements:
\begin{enumerate}
\item the effect on adult income of a given childhood malaria burden
\item the magnitude of decline of the malaria burden following the eradication campaigns
\end{enumerate}

The second element can be estimated to be $30\%$ in the United-States, which then allows to approximate the effect on adult income of childhood malaria exposure in units of infection rates.

We find an income effect per probability of childhood infection between $46$ and $49\%$ for the United States which is close to the $47$ and $60\%$ found in the original paper (in the original paper additional controls are used). As in the paper:
\begin{quote}
Being infected with malaria through childhood leads to a reduction in adult income of approximately 50 percent.
\end{quote}

\section{R: Introduction to GMM}

For this introduction to the Generalized Method of Moments, we follow the work of \citep{hamilton1994}, \citep{verbeek}, \citep{cochrane} and the package and examples suggest by \citep{gmmR} and suggest the codes \verb|GMM_Jondeau_test.R| and \verb|GMM_and_CAPM.R| with examples on CAPM and intertemporal CAPM tests.

\subsection{Introduction on the moments of a distribution: normality test}\label{sec:GMMWald}

If we consider that we have $T$ observations of a random variable $r_t$, we might want to test if as for a normal law its skewness and excess kurtosis\footnote{if kurtosis is $\kappa$, then excess kurtosis is $\kappa-3$} are nil. As in \citep{Jondeau2007} we define the four moments\footnote{the following are also called \emph{orthogonality conditions}}
\begin{align*}
e_{1,t}&=r_t - \mu\\
e_{2,t}&=(r_t-\mu)^2 - \sigma^2\\
e_{3,t}&=\left(\frac{r_t-\mu}{\sigma}\right)^3-s\\
e_{4,t}&=\left(\frac{r_t-\mu}{\sigma}\right)^4-\kappa-3
\end{align*}

We are searching for estimate of the vector $\theta = [\mu \quad \sigma \quad s \quad \kappa]'$ of dimension $N$ that satisfy the equalities above. Writing the vector-valued function $h(\theta,r_t) = [e_1 \quad e_2 \quad e_3 \quad e_4]'$ and the sample average $g(\theta,r_T) = \frac{1}{T} \sum_{t=1}^T h(\theta,r_t)$, we are searching for $\theta$ that minimizes the criterion function $g' W g$, where $W$ is a $N \times N$ positive definite symmetric weighing matrix reflecting the importance of matching each moment. If $\Gamma_\nu$ is the auto-covariance matrix of the vector-valued function\footnote{$\Gamma_\nu = E{[h(\theta_0,r_t)][h(\theta_0,r_{t-\nu})]'}$ with $\theta_0$ the true value of the moments} we can define $S = \sum_{\nu=-\infty}^{\nu=+\infty} \Gamma_\nu$, it can be demonstrated that $S$ is the asymptotic variance of the sample mean $S = \lim_{T \to \infty} T E\left([g(\theta_0,r_T)][g(\theta_0,r_{T})]'\right)$ and that the optimal weighing matrix is $S^{-1}$.

\quad

In practice, we start with an arbitrary weighing matrix, the identity $I$, this leads to a first estimate $\hat{\theta}_a$, from which can be estimated $\hat{S}_a^{-1}$, this is then taken as the new weighing matrix to establish a new estimate $\hat{\theta}_b$, until convergence.

As a numerical illustration\footnote{R code: GMM\_Jondeau\_test}, we follow the test suggested in \citep{Richardson:1993aa} following the steps and the data set in \citep{Jondeau2007} based on GMM and a Wald test, for which the null hypothesis of normality consist of nil skewness and excess kurtosis: $H_0: s = \kappa - 3 =0$, which can be written in matrix form with the constraint matrix $\Phi$:

\begin{equation*}
\Phi = 
\begin{bmatrix}
0 &0 &1 &0\\
0 &0 &0 &1
\end{bmatrix}
\end{equation*}

The test has the Wald statistic that under the null hypothesis is distributed as a $\chi^2(2)$:
\begin{equation}
T (\Phi \hat{\theta})' ( \Phi W_T \Phi)^{-1} (\Phi \hat{\theta})
\end{equation}

We use the data set suggested in \citep{Jondeau2007}, the raw version of the gmm function is not converging and we have to resort to defining lower and upper bound for the moments, as in \citep{gmmR} we also compute the gradient of the sample average function as it affects the Wald test statistic:
\begin{equation*}
\begin{bmatrix}
-1&0&0&0\\
0&-2\sigma&0&0\\
0&0&-1&0\\
0&0&0&-1
\end{bmatrix}
\end{equation*}

\subsection{Reminder on the standard error of the mean}

If we consider that we have $T$ observations of a random variable $U_t$ of mean $\mu$ and standard error $\sigma_U$, then we can compute the mean:
$$E_T(U_t) = \frac{1}{T} \sum_{t=1}^T U_t$$

The expectation of that mean is the actual mean of the random variable:
$$E\left(E_T(U_t)\right) = \frac{1}{T} \sum_{t=1}^T E(U_t) = \mu$$

However, the standard error $\sigma$ of that mean is:
$$\sigma^2 \left(E_T(U_t)\right) = \frac{1}{T^2} \sigma^2 \left( \sum_{t=1}^T U_t \right)$$

if $cov(U_t,U_{t-j}) = 0$ $\forall j \neq 0$:

$$\sigma \left(E_T(U_t)\right) = \frac{1}{\sqrt{T}} \sigma \left( U_t \right)$$

but in case of correlation:

$$\sigma^2 \left(E_T(U_t)\right) = \frac{1}{T}\left[ \sigma^2(U_t) + 2 \sum_{j=1}^T \frac{\vert T-j \vert}{T} cov(U_t,U_{t-j}) \right]$$

$$ \lim_{n\to\infty} \sigma^2 \left(E_T(U_t)\right) = \frac{1}{T}\left[ \sigma^2(U_t) + 2 \sum_{j=1}^{+\infty} cov(U_t,U_{t-j}) \right] = \frac{S}{T}$$

Hence, the heart of many problems in econometrics being correlation, we can work on $S$, let's now introduce GMM notations.

\subsection{GMM notations}

In a nutshell, if we suspect some issues on the exogeneity of the regressors and/or the circularity of the residuals, we might resort to the use of GMM based on assumptions on moment conditions. 

We consider a model:
$$E[f(w_t,z_t,b)] = 0$$
where $f$ is a vector function, $b$ is a vector containing all unknown parameters, $w_t$ is a vector of observable variables that could be endogenous or exogenous, and $z_t$ is the vector of instruments.

To estimate $\theta$ we consider:

$$g_T(\theta) = \frac{1}{T} \sum_{t=1}^T f(w_t,z_t,b)$$

The GMM finds the estimate $\hat{b}$ to the problem:
$$\min_b g_T(b)' W_T g_T(b)$$
with $W_T$ a positive definite matrix with plim $ W_T = W$.

\quad

\quad

If we work on the Intertemporal Asset Pricing Models as in \citep{verbeek} and apply the GMM to return data of 10 portfolio and consumption growth provided on the book companion website\footnote{R code: GMM\_and\_CAPM}.

We define consumption $C$, returns $r$ and some unknown coefficient $\delta$ and $\gamma$, the model can be derived to produce a set of conditions:

$$E \left(  \delta \left( \frac{C_{t+1}}{C_t}\right)^{-\gamma} (r_{j,t+1}-r_{f,t+1})  \right) = 0$$

with $\delta \left( \frac{C_{t+1}}{C_t}\right)^{-\gamma} $ called the stochastic discount factor.

\quad

We can rewrite with: $b \equiv [\delta \quad \gamma]'$, and $u$ the errors $u_{t+1}(b) = \delta \left( \frac{C_{t+1}}{C_t}\right)^{-\gamma} (r_{j,t+1}-r_{f,t+1})$.

The sample mean of the errors is: $g_T \equiv E_T[u_t(b)] = \frac{1}{T} \sum_{t=1}^T u_t(b)$.

The first stage estimate of $b$ with an arbitrary matrix $W$ (e.g. $W = I$) is: $$\hat{b}_1 = \min_b g_T(b)' W_T g_T(b)$$

Using $\hat{b}_1$, construct an estimate $\hat{S}$ of $S$: $$S \equiv \sum_{j=-\infty}^{+\infty} E [u_t(b) u_{t-j}(b)']$$

Second stage estimate of $b$, which is now a consistent, asymptotically normal, and asymptotically efficient estimate of $b$: $$\hat{b}_2 = \min_b g_T(b)' \hat{S}^{-1} g_T(b)$$

\subsection{Estimating the CAPM with GMM}

From section \ref{sec:CAPM}, we saw in equation \ref{eq:CAPM}: $R_i=R_f+\beta_i (R_M-R_f)$. In order to test the CAPM, we can run the OLS:
\begin{equation}\label{eq:CAPMtest}
R_i - R_f = \alpha +\beta_i (R_M-R_f) + \epsilon_t
\end{equation}
We follow \citep{MacKinlay:1991aa} test for the null hypothesis: $H_0: \alpha = 0$, using the moment conditions:
$$ E[\epsilon_t] = 0$$
$$E[\epsilon_t (R_{M,t}-R_{f,t})]=0$$

\subsection{Estimating the Risk Premia using GMM}

\href{https://www.kevinsheppard.com/teaching/python/notes/notebooks/example-gmm-estimation/}{Kevin Sheppard} suggest estimating the Risk Premia using GMM (in python).

\section{Demand and supply estimations}

\subsection{Prices and quantities, supply and demand}
\citet{Working:1927wf}:
\begin{quote}
The original observations of prices and corresponding quantities are the resultant of both supply and demand. [\ldots] The matter of correlation between shifts of the demand and supply curves is a more difficult problem to deal with. Every effort should be made to discover whether there is a tendency for the shifting of these to be interdependent. In case it is impossible to determine this, it should be carefully noted that the demand curve which is obtained is quite likely not to hold true for periods other than the one studied, and cannot be treated as corresponding to the demand curve of economic theory.
\end{quote}
\begin{center}
\includegraphics[scale=.5]{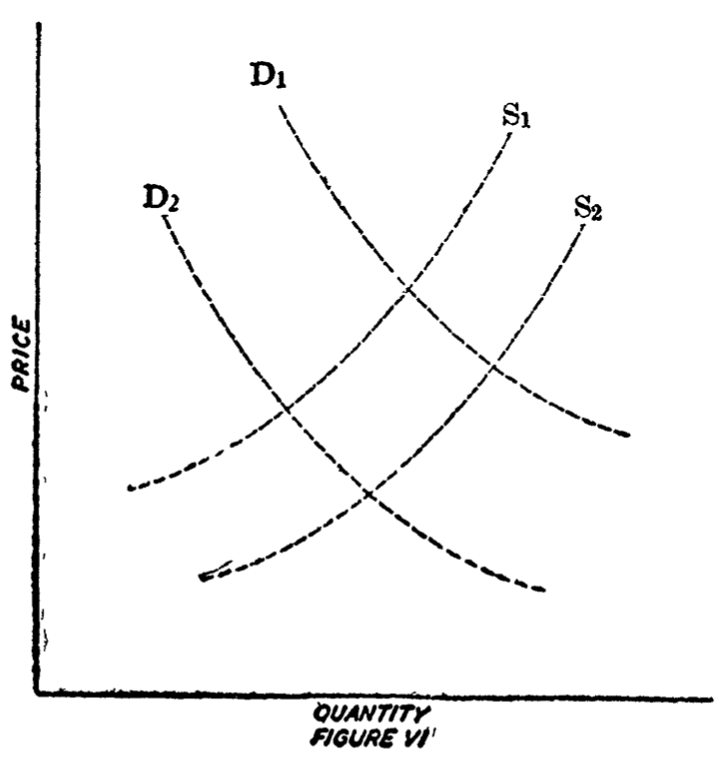}
\end{center}
This writes:
\begin{equation}
\begin{cases}
Q_t^d = \alpha_0 +\alpha_1 P_t + u_t\\
Q_t^s = \beta_0 +\beta_1 P_t + v_t\\
Q_t^d = Q_t^s
\end{cases}
\end{equation}

To deal with this simultaneity problem (simultaneous-equations bias), we would need an instrument that shift demand (and not limited to price, i.e. endogeneity also involves unobservables) without shifting supply (or vice-versa).

e.g. if you look at insurance supply and demand, you can use an instrument, variations in prudential cost to insurers, that should impact the supply but not the demand for insurance.

To summarize, the main issues we will face when studying supply and demand:
\begin{itemize}
\item prices are endogenous (strategically determined by firms);
\item supply and demand shifts occur simultaneously, we cannot isolate one shift with respect to the other, except if we find an instrument;
\item we have multiple products, with possible substitutions across products.
\end{itemize}

To make this more complicated to estimate for the econometrician, demand should be considered as not separable: $Q_t^d = f\left( X_t, P_t, u_t \right)$, with $X$ observable demand shifters and so $u$ unobservable demand shifters. $f$ is now a function we would need to estimate.

Nevertheless, we want to be able to estimate how choices respond to changes in prices, to advice on taxes or subsidies, the effect of company mergers, etc.

We would need:
\begin{itemize}
\item flexible functional forms;
\item exogenous variations;
\item unobserved individual-level heterogeneity in tastes.
\end{itemize}
While still being able to estimate such a model. We will drive through historical models and how they can be estimated.

\subsubsection{Prices and quantities, elasticities of commodities}
A commodity is, by definition, when all products are identical, regardless of which firm produced them (e.g. electricity). A differentiated product is, by definition, when the consumer can distinguish between products based on which company produced them (e.g. beer).

The price elasticity of demand (supply) measures the percentage change in the quantity of a commodity demanded (supplied) per unit of time resulting from a given percentage change in the price of the commodity:
\begin{equation}
e_{d\mbox{ or }s} =  \frac{\frac{ \Delta Q}{Q}}{\frac{\Delta P}{P} }= \frac{P \Delta Q}{Q \Delta P}
\end{equation}
Demand is elastic if $e_d<-1$, inelastic if $e_d>-1$. Supply is elastic if $e_s>1$ and inelastic if $e_s<1$

The income elasticity of demand $e_M$ measures the percentage change in the amount of a commodity purchased per unit time resulting from a given percentage change in the consumer's income $M$:
\begin{equation}
e_M=\frac{\frac{\Delta Q}{Q}}{\frac{\Delta M}{M}}
\end{equation}
If $e_M>1$, the good is a \textit{luxury} and if $e_M<0$ the good is inferior and if $0<e_M<1$ the good is a \textit{necessity}.

\subsubsection{Cross elasticity of demand}
The cross elasticity of demand between commodity $x$ and $y$ is:
\begin{equation}
e_{xy}=\frac{\frac{\Delta Q_x}{Q_x}}{\frac{\Delta P_y}{P_y}}
\end{equation}
if $e_{xy}>0$, $x$ and $y$ are substitutes, complements if negative.

\subsubsection{Engel curve}

Keeping $P_x$ and $P_y$ constant and consumer tastes constant, an Engel curve is showing the quantity consumed when the income $M$ varies.

\subsection{Looking at quantities consumed, income to estimate some elasticities}\label{sec:qelast}

We assume we have some aggregated data where for each goods $i$ (differentiated-product) at price $p_i$ a quantity $q_i$ is consumed by the population we are looking at (agents indexed by $k$) for a total consumption $M=\sum_{i=1}^n p_i q_i$. Writing the budget share of good $i$:
\begin{equation}\label{eq:budgetshare}
w_i = \frac{p_i q_i}{M}
\end{equation}

We are interested in the income elasticity, that is how much more do people consume of good $i$ when their income $M$ change:
\begin{equation}
\eta_i = \frac{\partial (\log q_i )}{\partial (\log M)}
\end{equation}
We have by definition $q_i=q_i(M,p_1, \dots,p_n)$, thus $$dq_i = \frac{\partial q_i}{\partial M} dM + \sum_{j=1}^n\frac{\partial q_i}{\partial p_j	}dp_j$$ with the Slutsky equation:
$$\frac{\partial q_i}{\partial p_j	} = s_{ij} - q_j  \frac{\partial q_i}{\partial M} $$
where $ s_{ij}$ is the substitution effect that holds real income constant. 
\subsubsection{Slutsky equation and Shephard's lemma}
To demonstrate the Slutsky equation, we need to define the Hicksian demand, which is the demand $h$ that minimizes expenditure $M$ at a given utility $u$: $h_i(p,u)=q_i\left(p, M(p,u)\right)$, then 
$$\frac{\partial h_i(p,u)}{\partial p_j} =\frac{\partial q_i}{\partial p_j} + \frac{\partial q_i}{\partial M}\frac{\partial M}{\partial p_j}$$
Now using the Shephard's lemma that given price and utility, the demand for a particular good equals the derivative of the expenditure function with respect to that price: $h_i(p,u)=\frac{\partial M(p,u)}{\partial p_i}$, then:
$$\frac{\partial q_i}{\partial p_j	} = \frac{\partial h_i(p,u)}{\partial p_j} - q_j  \frac{\partial q_i}{\partial M} $$

Hence:
$$dq_i = \frac{\partial q_i}{\partial M}\left(dM-\sum_{j=1}^n q_jdp_j\right) + \sum_{j=1}^ns_{ij} dp_j$$
using the identity $d(\log x)=\frac{dx}{x}$:
\begin{equation}\label{eq:diffeqlog}
d(\log q_i) = \frac{M}{q_i} \frac{\partial q_i}{\partial M} \left[ d(\log M) - \sum_{j=1}^n\frac{p_jq_j}{M} d(\log p_j) \right] + \sum_{j=1}^n\frac{p_j}{q_i}s_{ij} d(\log p_j)
\end{equation}

Hence the income elasticity writes:
\begin{equation}
\frac{M}{q_i} \frac{\partial q_i}{\partial M}
\end{equation}

\subsection{Working-Leser model}\label{sec:workingleser}

\citet{Working:1943wb} and \citet{Leser:1963vd} introduced a model for consumption analysis when survey data is available, they assume:
\begin{equation}\label{eq:WLequation}
w_i = \alpha_0+\beta_i \log M + \sum_j \beta_{ij} \log p_j +\sum_k \gamma_{ik} H_k + \epsilon_i
\end{equation}
where $ \epsilon_i$ are random disturbance terms and $H_k$ are fixed effects (information on the agents like age, family size, etc.).

We have from the budget share equation \ref{eq:budgetshare} that $$ \frac{\partial w_i}{\partial M} = \frac{Mp_i \frac{\partial q_i}{\partial M}-p_i q_i}{M^2} $$

and from the main assumption from Working-Leser, equation \ref{eq:WLequation} that $\frac{\partial w_i}{\partial M} = \frac{\beta_i}{M}$.

Rearranging, we find an expression income elasticity:
\begin{equation}
\eta_i = 1+\frac{\beta_i}{w_i}
\end{equation}
by definition the marginal share is $w_i \eta_i$, which here is $w_i + \beta_i$, which describe how a one unit increase in expenditure decomposes into each good $i$.

Without detailed survey, we have to work with aggregated data sets.

\subsection{Linear Expenditure Systems}
\citet{Klein:1947tz} introduced a Linear Expenditure Systems (LES) that was estimated by \citet{Stone:1954uk}. Now agents, to maximize their utility $U(q_1,\ldots,q_n)$, consume $ \sum_{j=1}^n p_j \gamma_j$ for subsistence with a budget constraint of total "income" $M=\sum_{i=1}^n p_i q_i$. The following assumption is made on the utility:
\begin{equation}\label{eq:LESutility}
U(q_1,\ldots,q_n)=\sum_{i=1}^n \beta_i \log\left(q_i - \gamma_i \right)
\end{equation}
with $q_i>\gamma_i$ and $\sum_{i=1}^n \beta_i=1$, we write the Lagrangian of the consumer maximizing his utility:
\begin{equation}
\sum_{i=1}^n \beta_i \log\left(q_i - \gamma_i \right) - \lambda \left[\sum_{i=1}^n p_i q_i - M \right]
\end{equation}
The first order conditions, $\frac{\partial .}{\partial q_i} = 0$ yield:
\begin{equation}
\lambda p_i = \frac{\beta_i}{q_i-\gamma_i}
\end{equation}
or
\begin{equation}
p_i q_i = p_i \gamma_i + \beta_i \frac{1}{\lambda}
\end{equation}
summing over $i$'s and using the assumption that $\sum_{i=1}^n \beta_i=1$:
\begin{equation}
M = \sum_{i=1}^n p_i \gamma_i + \frac{1}{\lambda}
\end{equation}
This yields the demand function:
\begin{equation}\label{eq:demandfunciton}
p_i q_i = p_i \gamma_i + \beta_i\left( M - \sum_{j=1}^n p_j \gamma_j \right)
\end{equation}
where $q_i > \gamma_i$.

Using equation \ref{eq:demandfunciton} to express $q_i$ as a function of $M$ and deriving using logarithmic differentiation:
\begin{equation}
\left( \ln (p_iq_i) \right)' = \frac{\left(p_i q_i \right)'}{p_i q_i} =  \left( \ln (p_i) \right)' + \left( \ln (p_i) \right)' = \partial \log(p_i) + \partial \log(q_i) = \frac{\gamma_i}{q_i} \partial \log(p_i) + \beta_i \frac{M}{p_iq_i}\partial \log(M)  + \ldots
\end{equation}
thus we find that in this model, $\beta_i w_i^{-1}$ is the income elasticity and  $0 < \beta_i < 1$ is the i\textsuperscript{th} marginal share, it represents the fraction of a one-dollar rise in of supernumerary\footnote{supernumerary in the sense that it is total expenditure less total committed expenditure} income that will be spent on good $i$. 

Commodities with income elasticities less than unity are called \textit{necessities}, those with elasticities greater than one are called \textit{luxuries}.

The model, to be estimated econometrically, rewrites:
\begin{equation}\label{eq:LES}
p_i q_i = \alpha_i + \beta_i M
\end{equation}
where $\alpha_i = \sum_{j=1}^n\left(\delta_{ij}-\beta_i\right) p_j \gamma_j$, with $\delta_{ij}$ the Kronecker delta\footnote{$\delta_{ij}=1 \mbox{ if } i=j \mbox{, } 0$ otherwise},  and $\sum_{i=1}^n\alpha_i=0$. Equation \ref{eq:LES} estimation has been discussed (\citet{Pollak:1969tv}, etc.) and remains a challenge, one of the main challenge lies in identifying $\gamma_j$.

It can be demonstrated that for this model, commodities are price inelastic and cross-price elasticities are negative (all pairs are gross complements).

\subsubsection{Introduction on UK data set on alcohol consumption}

We follow\footnote{code: vansteenberghe\_demand\_supply.R} \citet{selvanathan1995}, using UK alcohol data (1955-1976 beer, wine and spirits in liters per capita) to estimate marginal shares, income flexibility and income and price elasticities.

As in the original chapter, UK per capita consumption in 1985 was 109 liters of beer, 17 liters of wine and 4 liters of spirits.

\stepcounter{question}
\begin{itemize}
\item \textbf{Question \arabic{question}}
\begin{itemize}
\item Replicate figure 3.1 and 3.2 in \citet{selvanathan1995}:
\begin{itemize}
\item Alcohol consumption time series;
\item Alcohol Price Indexes time series.
\end{itemize}
\end{itemize}
\end{itemize}

We estimate equation \ref{eq:LES} and retrieve the income elasticities.

\subsection{The Rotterdam model}

Contrary to the LES that start with the consumer's utility function, the Rotterdam model starts with the demand function $q_i=q_i(M,p_1, \dots,p_n)$ and the utility function doesn't need to be specified (we do not use the assumptions of equation \ref{eq:LESutility}).

\citet{Clements:2015wj} present and review the Rotterdam demand model and compare it with some of its "competitors" like \citet{Deaton:1980vf}'s AIDS.

The consumer maximizes his utility (but we do not specify the functional form of his utility, so this is more general)\footnote{Nota bene: starting from a generic utility equation \ref{eq:URotterdam}, we can end up again with equation \ref{eq:diffeqlog}}
\begin{equation}\label{eq:URotterdam}
U(q_1,\ldots,q_n)
\end{equation}
so all the generic results from section \ref{sec:qelast} holds but we do not directly add a relationship between $w_i$ and $M$ as for the Working-Leser model, so it is more generic and we will need to find another approach to estimate the income elasticities.

We start by writing the \citet{divisia1925} index of the change in real income and prices: $d(\log Q) = \sum_i w_i d(\log q_i)$ and $d(\log P) = \sum_i w_i d(\log p_i)$, and following the budget constraint $d(\log M) = d(\log P) + d(\log Q)$, equation  \ref{eq:diffeqlog} can be simplified into:
\begin{equation}
d(\log q_i) = \eta_i d(\log Q) + \sum_j \eta_{ij} d(\log p_j)
\end{equation}
where $\eta_{ij}=\frac{p_j}{q_i} s_{ij}$ is the (i,j)\textsuperscript{th} price elasticity (income compensated). To apply this to time series, we replace finite change and take the elasticities to be constant:
\begin{equation}
Dq_{it} = \eta_i DQ_t + \sum_j \eta_{ij} Dp_{jt} + \epsilon_{it}
\end{equation}
where $Dx_t=\log x_t - \log x_{t-1}$ and  $\epsilon_{it}$ is a disturbance term. Multiplying both sides by an arithmetic average of the unconditional budget share of beverage $i$ is $\bar{w}_{it} = \frac{1}{2}\left(w_{it} + w_{it-1} \right)$
\begin{equation}\label{eq:rotterdammodel}
\bar{w}_{it} Dq_{it} = \theta_i DQ_t + \sum_j	\pi_{ij}Dp_{jt} + \mu_{it}
\end{equation}
which is the Rotterdam model. It describes the budget-share weighted change in the quantity consumed of good $i$, $\bar{w}_{it} Dq_{it}$, as a linear function of the change in real income $DQ_t $ and the change in each of the $n$ prices. 
This was further modified \citep{theil1980}
\begin{equation}\label{eq:rotterdam1}
\bar{w}_{it} Dq_{it} = \theta_i DQ_t + \Phi \theta_i ( Dp_{it}-DP'_t)
\end{equation}
with $DP'_t$ the overall Frisch price index: $DP'_t=\sum_j \theta_j Dp_{jt}$.

We compute the quantity and price log changes, $D_{qit}=\log q_{it}-\log q_{it-1}$ and $D_{pit}=\log p_{it}-\log p_{it-1}$.

\stepcounter{question}
\begin{itemize}
\item \textbf{Question \arabic{question}}
\begin{itemize}
\item Replicate figure 3.3, 3.4, 3.5 and 3.6 in \citet{selvanathan1995}:
\begin{itemize}
\item Log-change in Alcohol Consumption;
\item Log-change in Alcohol Prices;
\item Conditional Budget Shares of Beer, Wine and Spirits;
\item Unconditional Budget Shares of Beer, Wine and Spirits.
\end{itemize}
\end{itemize}
\end{itemize}

We define the  arithmetic average of the budget share of total alcohol:
$$\bar{w}'_{it} = \frac{\bar{w}_{it}}{\bar{W}_{gt}} $$ and the arithmetic average of the budget share of total alcohol $\bar{W}_{gt}=\sum_{i=1}^3 \bar{w}_{it}$.

\citet{selvanathan1995} demonstrates that under some conditions, equation \ref{eq:rotterdam1} can be rewritten into their equation 4.5:
\begin{equation}
\bar{W}_{gt} DQ_{gt} = \Theta_g \left[ DQ_t + \Phi \left( DP_{gt} - DP^*_t \right) \right]
\end{equation}

Before being able to estimate equation 4.5, the first equation to estimate in \citet{selvanathan1995}, equation 4.2
\begin{equation}
DQ_{gt}-DQ_t = \Phi\left( DP_{gt} - DP_t  \right)
\end{equation}
with the Divisia price $DP_{gt}=\sum_i \bar{w}'_{it} Dp_{it}$ and $DP_t=\sum_i \bar{w}_{it} DP_{it}$  and the Divisia volume $DQ_{gt}=\sum_i \bar{w}'_{it} Dq_{it}$ and $DQ_t$ the real income approximated with $DM_t-DP^*_t$ where $DM_t$ is the log-change in income (total expenditure) and $DP^*_t$ the CPI log-change. $DP^*_t$ proxies the Divisia price index $DP_t$.

Finally they estimate equation 4.11:
\begin{equation}\label{eq:rotterdam411}
\bar{w}_{it} Dq_{it} = \theta_i \left[ \bar{W}_{gt} DQ_{gt} + \Phi \Theta_g \left( Dp_{it} - DP_{gt} \right) \right]
\end{equation}

So now the marginal share is directly estimated: $ \theta_i$.

\citet{selvanathan1995} found that a one unit increase in income results in a $ \Theta_g =0.078$ increase in spending on alcohol. And a one unit increase in expenditure on alcohol decomposes into $\theta_1=.35$ for beer, $\theta_2=.2$ for wine and $\theta_3=.46$ for spirits.

The main criticisms of the Rotterdam model are (cf. \citet{Clements:2015wj} for a discussion and implications):
\begin{itemize}
\item constant marginal shares: the income elasticity is inversely proportional to the corresponding budget share;
\item constant Slutsky coefficients;
\item degenerate preferences.
\end{itemize}

\subsection{Almost Ideal Demand System}

\citet{Deaton:1980vf} introduced an \textbf{Almost Ideal Demand System} that satisfies
\begin{quote}
the axioms of choice exactly; aggregates perfectly over consumers without invoking parallel linear Engel curves; has a functional form which is consistent with known household-budget data; is simple to estimate, largely avoiding the need for non-linear estimation; and can be used to test the restrictions of homogeneity and symmetry through linear restrictions on fixed parameters.
\end{quote} and recommended their model to \begin{quote}
testing, extending, and improving conventional demand analysis.
\end{quote}
The main hypothesis of this model is that there is an expenditure function $c(u,p)$, with $u$ the utility and $p$ the price:
\begin{equation*}
\log\left(  c(u,p)\right) =  \alpha_0 + \sum_{i=1}^n \alpha_i \log(p_i) + \frac{1}{2}\sum_{i=1}^n\sum_{j=1}^n\gamma^*_{ij}\log(p_i)\log(p_j) + u \beta_0 \Pi_i p_i^{\beta_i}
\end{equation*} 
For a utility-maximizing consumer, using the Shephard's lemma:
\begin{equation}
w_i = \alpha_i + \beta_i \left[ \log(M) - \log(P) \right] + \sum_{j=1}^n \gamma_{ij} \log(p_j) + \epsilon_i
\end{equation}
with
\begin{equation*}
 \log(P) = \alpha_0 + \sum_{j=1}^n \alpha_j \log(p_j) + \frac{1}{2}\sum_{i=1}^n\sum_{j=1}^n\gamma_{ij}\log(p_i)\log(p_j)
\end{equation*}
and $\sum_{i=1}^n \alpha_i=1$, $\sum_{i=1}^n \beta_i=0$ and $\sum_{i=1}^n \gamma_{ij}=0, \forall j\in [1,n]$.

The income elasticity can be derived exactly as in section \ref{sec:workingleser} is $\eta_i= 1+\frac{\beta_i}{w_i}$

For homogeneity, we test whether $\sum_{j=1}^n \gamma_{ij} = 0, \forall i \in [1,n]$.
For symmetry, we test whether $\gamma_{ij}=\gamma{ji}, \forall i,j \in [1,n]\times [1,n]$.

\citet{Alley1992AnAI} review the AIDS model to study alcoholic beverage demand. We apply the AIDS model to the UK data set and again find that the spirits are a strong luxury item (elasticity greater than one).

Price is correlated with both unobserved product quality and unobserved demand shocks. \citet{Hausman:1994wo} introduced the Hausman instrument: prices in one city to instrument for prices in another.

\citet{Clements:2015wj} show that working with differentials, the AIDS model falls back to the Rotterdam estimation equation \ref{eq:rotterdam411}.

When number of products is large, estimations run into dimensionality problems, which is solved using logit demand model.

\subsection{Discrete choice model}
\subsubsection{Multinomial logit demand model}\label{sec:IIAlogit}
We follow \citet{McFadden1974} for a conditional logit analysis of qualitative choice behavior. An individual $i$ choose from a set of (discrete) options $J$ with probability $s_{ij}$ and receives utility $U_{ij}$ for choosing $j$:
\begin{equation}
s_{ij} = Prob\left(U_{ij}>U_{ik}, \forall j \neq k \right)
\end{equation}
we assume that ties occur with zero probability and that the utility can be separated into observed $V$ and unobserved $\epsilon$ components, $U_{ij} = V_{ij}+\epsilon_{ij}$ then the probability of $i$ choosing $j$ rewrites:
\begin{equation}
s_{ij}=Prob\left(\epsilon_{ij}-\epsilon_{ik}>V_{ik}-V_{ij}, \forall j \neq k \right)
\end{equation}
\citet{McFadden1974} demonstrates in his Lemma 1, that if $\epsilon$ are independent and identically distributed according to a Weibull distribution (p.d.f. of $e^{-e^{-\epsilon}}$), then the probability can be written:
\begin{equation}
s_{ij}=\frac{e^{V_{ij}}}{\sum_k e^{V_{ik}}}
\end{equation}
Three axioms are satisfied:
\begin{itemize}
\item \textbf{I}ndependence of \textbf{I}rrelevant \textbf{A}lternatives: the ratio of choice probabilities for $j$ and $k$ do not depend on an alternative $l$;
\item Positivity: $s_{ij}>0$;
\item Irrelevance of Alternative Set Effect, which is equation \ref{eq:axiom3}.
\end{itemize}

To identify the IIA logit, some normalization choices have to be made, usually, the utility of an outside option $0$ (e.g. not buying an insurance contract at all) is set to $0$:
\begin{equation}\label{eq:axiom3}
s_{ij}=\frac{e^{V_{ij}}}{1+\sum_k e^{V_{ik}}}
\end{equation}

Now if we assume that the price enters utility: $V_{ij}=\beta_p p_j + x_j \beta + \beta_z z_{ij}$, with $x_j$ observed attributes of good $j$ and $z_{ij}$ are also observed variables and $\beta_z$  is a vector of random terms with zero mean. Then:
$$\frac{\partial s_{ij}}{\partial p_j} = s_{ij}(1-s_{ij})\beta_p$$
and to evaluate the price elasticity:
\begin{equation}
\eta_j = \frac{\partial \log s_{ij}}{\partial \log p_j} = \beta_p p_j (1-s_{ij})
\end{equation}
Where we find the usual critics that the price elasticity is increasing in it own price and that the larger the share $s_{ij} \to 1$, the least elastic to price agent $i$ is.

\subsubsection{Nested logit}
IIA logit (or multinomial logit) presented section \ref{sec:IIAlogit} has i.i.d. unobserved components  $\epsilon$. This model is poor at explaining some observed substitution between goods. The Nested Logit relax part of this assumption by grouping consumers into categories $g$, $\epsilon_{ij}=\eta_{ig}+\tilde{\epsilon}_{ij}$. Demonstrations and derivations of this model is left out of this lecture.

In both presented logit demand model, we cite \citet{Berry:2021un} for criticisms that lead to allow $\beta$ and $\beta_p$ to vary for each consumer $i$, as these restrictions
\begin{quote}
imply that any two goods with the same (or similar) market shares---no matter how they differ in other respects--will have the same (or similar) own-price elasticities, equilibrium markups, and cross-price elasticities with respect to any third good. These are not only strong restrictions, but properties that are contrary to economic models of differentiated products, where, for example, goods that are more similar tend to have larger cross- price elasticities.
\end{quote}

\subsubsection{Random utility model - Random-coefficients logit model, a.k.a. mixed logit}

Random-coefficients logit model were introduced by \citet{Boyd:1980uf} and \citet{Cardell:1980wq}. We refer to \href{https://pyblp.readthedocs.io/en/stable/background.html}{pyblp} documentation for the model presentation. 

Such model have a wide range of applications, even in insurance \citep{Koijen:2016tw} and \citep{koijen2022}.

Estimations methods were developed, e.g. \citet{Berry:1995ud} (BLP) provide the following advantages:
\begin{itemize}
\item estimation is possible with only price and quantity data;
\item endogeneity of prices is taken into account;
\item elasticities are closer to economic analysis.
\end{itemize}

In this model, taste heterogeneity in the population is allowed and the parameters governing the distribution of this heterogeneity are estimated. To follow the last sections and compare, \citet{Miller2021} estimate a random coefficient nested logit on the U.S. beer market.

\citet{Nevo:2000wr} provide a practitioner's guide to BLP model and estimation methods are detailed in \citet{PyBLP}. The key estimation challenge is to find instruments as prices are endogenous and no close formula is possible.

In such a model, the utility of an agent $i$ now has random coefficients $\beta_i$, $\alpha_i$ to purchase product $j$ out of $N$ products:
\begin{equation}\label{eq:RClogit}
U_{ij} = \alpha_{i} p_j + x_j \beta_i+ \xi_j + \epsilon_{ij}
\end{equation}
with $\xi_j$ good $j$ unobserved (to the econometrician) characteristics\footnote{this can also be considered more generally as latent taste variation and latent product characteristics} which can be correlated with $p$. In this model, only $x_j$ are assumed exogenous. If $\epsilon_{ij}$ are i.i.d. draw from a standard type-1 extreme value distribution, then this yield  mixed\footnote{mixed in the sense that the coefficient $\alpha_i$ and $\beta_i$ are consumer heterogeneous} multinomial logit model.

We further assume that there is an outside option $0$ (not purchasing any good) and the indirect utility is normalized to $0$, $U_{i0}=0$.

We add some structure to the random coefficients:
\begin{equation}
\begin{bmatrix}
\alpha_i\\
\beta_i
\end{bmatrix}
= 
\begin{bmatrix}
\alpha\\
\beta
\end{bmatrix}
+ \Pi D_i + \Sigma \nu_i
\end{equation}
with $D_i$ a $d \times 1$ vector of demographic variables following a nonparametric distribution and $\nu_i$ a parametric distribution (taken from a standard normal law).

Following this assumptions, equation \ref{eq:RClogit} rewrites:
\begin{equation}
U_{ij} =\delta_{j}(x_j, p_j, \xi_j ; \theta_1) + \mu_{ij}(x_j, p_j, \nu_i, D_i ; \theta_2) + \epsilon_{ij}
\end{equation}
with the mean utility common to all consumers: $ \delta_{j} = x_j \beta + \alpha p_j +\xi_j$ and deviation from the mean utility capturing the effects of the random coefficients:
$$  \mu_{ij} = [-p_j, x_j] \left(\Pi D_i + \Sigma \nu_i\right) $$

We follow the notations in \citet{PyBLP}:
with $\theta_1$ the endogenous and exogneous \textit{linear} parameter (for $X_1$ characteristics, typically including the price) and $\theta_2$ \textit{nonlinear} parameter (typically for a subset of $X_1$ characteristics) that interact with the demographic characteristics. $\nu$ being the unobserved population heterogeneity that are drawn from standard normal distribution.

The market share of good $j$ is:
\begin{equation}
s_j(\delta, \tilde{\theta_2})  = \int \frac{\exp \left(  \delta_{j}+ \mu_{ij} \right)}{\sum_k \exp \left( \delta_{k}+ \mu_{ik} \right)} f(\mu_i \vert \tilde{\theta_2}) d\mu_i
\end{equation}
where $\tilde{\theta_2}$ is a subset of $\theta_2$ that governs heterogeneity and $f(\mu_i \vert \tilde{\theta_2}) $ denotes the mixing distribution over the heterogeneous types $i$.

On the supply side, the marginal cost for the firm is assumed to follow:
$ X_3 \gamma + \omega$ with $X_3$ observed supply-side characteristics and $\omega$ unobserved supply-side characteristics. The firm maximizes its profit:
\begin{equation}
\max \sum_j s_j (p_j - mc_j)
\end{equation}
which yields:
\begin{equation}\label{eq:equilibriumprice}
p = mc + \Delta (p)^{-1} s(p)
\end{equation}
where
$$-\Delta (p) = -H \odot \frac{\partial s}{\partial p}(p)$$
where $\odot $ is the Hadamard product and $H_{jk}=1$ if firm $k$ produces $j$, $0$ otherwise.

Now the \textbf{price elasticities} expression are not trivial and the reader can refer to \citet{Nevo:2000wr} page 525.
\begin{equation}
\eta_{jk} = \frac{\partial s_j p_k}{\partial p_k s_j} =     \begin{cases}
-\frac{p_j}{s_j}\int \alpha_i s_{ij}(1-s_{ij})d\ldots  ,& \text{if} j = k \\     
\frac{p_k}{s_j}\int \alpha_i s_{ij}s_{ik} d\ldots  & \text{otherwise}
    \end{cases}
\end{equation}

where the probability of individual $i$ purchasing good $j$ writes: $$s_{ij} = \frac{\exp\left( \delta_j + \mu_{ij} \right)}{1+\sum_k \exp\left( \delta_k + \mu_{ik} \right)}$$

With this, we need to estimate the equilibrium prices that satisfy equation \ref{eq:equilibriumprice}. For this, we approximate:
\begin{equation}
s_j = \frac{1}{ns} \sum_i s_{ij}
\end{equation}
where we take $ns$ draws for $\nu_i$ and $D_i$. We use the same method to approximate $\frac{\partial s_j}{\partial p_k}$ with the \citet{Morrow:2011vo} approach suggested in  \citet{PyBLP}.

\subsubsection{Mixed logit estimation}

The minimization program that comes to mind for this approach to estimate $\theta=[\beta,\Sigma,\Pi,\gamma]$ is:
\begin{equation}
\min_{\theta} \parallel s_{ij}(x,p,\xi,\theta) - S_{ij} \parallel
\end{equation}
with $s_{ij}$ the shares predicted by the theoretical model and $S_{ij}$ the observed market shares. But non-linearities and number of parameters render the estimation complicated.

A Generalized Method of Moments (GMM) estimator is constructed to estimate $\theta$, 
$$\min_\theta \hat{g}(\theta)\hat{W} \hat{g}(\theta)$$
with $W$ a weight matrice and $g(\theta)$ a vector of moments.

$\hat{\theta}$ is an instrument variable estimator if $$\hat{g}(\theta) = \frac{1}{N} \begin{bmatrix}
\sum_j Z'_{Dj}\xi_j\\
\sum_j Z'_{Sj} \omega_j
\end{bmatrix}$$
with $Z_{D}$ and $Z_{S}$ demand and supply-side instruments.

\citet{Berry:1995ud} uses three sets of instruments:
\begin{enumerate}
\item observed product characteristics 
\item sum of product characteristics for all models marketed by a single firm in a given market, excluding the considered model for the instrument
\item sum of product characteristics for all models in a given market, excluding the firm considered for the instrument
\end{enumerate}

\subsection{Discrete choice model estimations}
\subsubsection{Automobile prices in market equilibrium}

\citet{Berry:1995ud} specify the utility of agent $i$ for good $j$ in a market $t$, in logarithm $u_{ij}=\log U_{ij}$:
\begin{equation}
u_{ij}=\alpha \log\left(y_i - p_j \right) + x_j  \bar{\beta} + \xi_j + \sum_k \sigma_k x_{jk} \nu_{ik} + \epsilon_{ij}
\end{equation}
with $y_i$ the income of agent $i$. $x_j$ are observed product characteristics. $\xi$ reflects the difficult to quantify aspects of style, prestige, reputation, and pas experience as well as quantifiable characteristics of the car no available in the data set.

The marginal cost for the firm that produces good $j$ is assumed to be:
\begin{equation}\label{eq:mcBLP}
\ln \left(mc_j \right) = w_j \gamma + \omega_j
\end{equation}
where $w_j$ are observed cost shifters and $\omega_j$ unobserved cost shifters, the price $p_j$ is set by firms that maximize their profits: $\sum_{j} \left(p_j-mc_j\right) M s_j(p,x,\xi;\theta)$, first order condition yields:
$$mc_j = p_j-\Delta(p,x,\xi;\theta)^{-1} s_j(p,x,\xi;\theta)$$
where $\Delta_{jr}= -\frac{\partial s_r}{\partial p_j}$ if $r$ and $j$ are produced by the same firm, $0$ otherwise.

Equation \ref{eq:mcBLP} can be estimated if one assumes orthogonality conditions between $\omega$ and appropriate instruments. Furthermore, as $\xi_j$ is correlated with $p_j$ we need instruments $z$ for both the demand and pricing equations: $ E\left[\xi_j\vert z \right] = E\left[\omega_j\vert z \right] = 0$, with $z_j=\left[x_j,w_j \right]$.

\stepcounter{question}
\begin{itemize}
\item \textbf{Question \arabic{question}}
\begin{itemize}
\item Replicate\footnote{pyblp\_introduction.py} partially Table 1
\begin{itemize}
\item Per year (market), count models, and compute the weighed mean for:
\begin{itemize}
\item price, hpwt, space (size in the paper), air, MPG and MP dollar (mpd).
\end{itemize}
\end{itemize}
\end{itemize}
\end{itemize}

In the following exercise, we will follow \citet{berry1999} and assume that price sensitivity is inversely proportional to income, if income $y_i$ is drawn from a lognormal distribution, then:
\begin{equation}
u_{ij}=\alpha \frac{p_j}{y_i}+ x_j  \bar{\beta} + \xi_j + \sum_k \sigma_k x_{jk} \nu_{ik} + \epsilon_{ij}
\end{equation}

We first replicate the Logit estimates TABLE III in \citet{Berry:1995ud}:

\begin{tabular}{rll}
Variable & Estimate & Standard Errors\\
$p$ & -0.089 & 0.004\\
hpwt & -0.121 & 0.277\\
air & -0.035 & (0.073)\\
mpd & 0.263 & 0.043\\
space & 2.341 & 0.12\\
Constant & -10.068 & 0.253\\
\end{tabular}

Then we replicate TABLE IV:

\begin{tabular}{rll}
Variable & Estimate & Standard Errors\\
Constant &- 7.061 &0.941 \\
hpwt &2.883 &2.019 \\
air &1.521&0.891\\
mpd &-0.122& 0.320 \\
space &3.460&0.610\\
$\ln(y_i-p_j)$&43.501&6.427\\
\end{tabular}

And the cost side:

\begin{tabular}{rll}
Variable & Estimate & Standard Errors\\
Constant &0.952&0.194 \\
$\ln(hpwt)$ &0.477&0.056\\
air &0.619&0.038\\
$\ln(mpg)$ &-0.415&0.055 \\
$\ln(space)$ &-0.046&0.081\\
Trend&0.019&0.002\\
\end{tabular}

Then we extract some price elasticities for the market (year) 1990 for the Mazda323 and the BMW 735i of 6.358 and 3.515 respectively in TABLE V.

\quad

The results suggest that for the average consumer more space, more air conditioning increase the utility from the product.

\subsubsection{Automobile prices in market equilibrium - agent data}

\citet{Berry:1995ud} provided agent data, on their income distribution. They provided to the research community a file with draw for $\nu$ and most importantly, the weights $w_{it}$. In this integral approach, the observed share for good $j$ in market $t$ is approximated as: $s_{jt}=\sum_i w_{it} s_{ijt}$ and we know from the model that we can recover:
\begin{equation}
s_{ijt}=\frac{\exp\left(X_1 \beta + \xi_j + X_2 (\Sigma \nu' + \Pi d')_{ij} \right)}{1+\sum_k \exp\left(X_1 \beta + \xi_k + X_2 (\Sigma \nu' + \Pi d')_{ik} \right)}
\end{equation}

In fact, we do not observe the mean (across consumers) of the unobserved component of utility, so we cannot build $s_{ijt}$ and set $w_{it}$ . So we have to fall back to integration methods, as in \citet{PyBLP} Table 8.

 \section{python and R: copula application to risk measures}\label{sec:Rcopula}
The following\footnote{some elements can also be found \href{http://copula.r-forge.r-project.org/book/index.html}{here}} is inspired by \citep{copula2004}. The copula functions help to cope with the problems of specification of marginal univariate distribution. The copula functions model fat tails and tail dependence, separately. 
\subsection{Bivariate copula definition}

\begin{definition}
A two-dimension copula $C$ is a real function defined on $I \times I$ ($I=[0,1]$):
\begin{equation*}
C : I \times I \rightarrow I
\end{equation*}
\begin{enumerate}
\item grounded, i.e. $C(u,0) = C(0,v) = 0$
\item such that $C(u,1) = u$ and $C(1,v) = v$ $\forall (u,v) \in I \times I$
\item 2-increasing, i.e. for every rectangle $[u_1,u_2] \times [v_1,v_2]$ whose vertices lie in $I \times I$ such that $ u_1 \leq u_2, v_1 \leq v_2$: 
$$ C(u_2,v_2) - C(u_2,v_1) \geq C(u_1,v_2) - C(u_1,v_1) $$
\end{enumerate}
\end{definition}

\citep{sklar1959} wrote the theorem:
\begin{theorem}
Given any joint distribution $F(x_1,x_2)$ with continuous margins $F_1(x_1)$ and $F_2(x_2)$, there exists a \emph{unique} copula function $C : I \times I \rightarrow I$ such that:
\begin{equation}\label{eq:sklar}
F(x_1,x_2) = C\left(F_1(x_1),F_2(x_2)\right)
\end{equation}

Conversely\footnote{this is important because then "much of the study of joint distribution can be reduced to the study of copulas" \citep{schweizer1991}, copulas separate the marginal behaviors $F_i$ from their dependence.}, if $F_1(x_1)$ and $F_2(x_2)$ are continuous distribution functions, then \ref{eq:sklar} defines a bivariate distribution function with marginal distributions $F_1(x_1)$ and $F_2(x_2)$.
\end{theorem}

Since copulas are joint distribution functions of standard uniforms\footnote{from the probability integral transform, $u_i = Pr(X_i < x_i) \in [0,1]$, $u_i$ has a uniform distribution}, a copula computed at $F_1(x_1)$, $F_2(x_2)$ gives a joint distribution function at $(x_1,x_2)$:
\begin{equation}
\begin{cases}
C\left(F_1(x_1),F_2(x_2)\right) &= Pr\left( U_1 \leq F_1(x_1), U_2 \leq F_2(x_2)  \right)\\
&= Pr\left( F^{-1}_1(U_1) \leq x_1 , F^{-1}_2(U_2) \leq  x_2 \right)\\
&= Pr\left( X_1 \leq x_1 , X_2 \leq  x_2 \right)\\
&= F\left( x_1 ,  x_2 \right)\\
\end{cases}
\end{equation}

The copulas are bounded (Frechet bounds) by the minimum\footnote{$C^{-}(u_1,u_2)=\max (u_1 + u_2 - 1, 0)$, for perfect negative dependence} and the maximum\footnote{$C^{+}(u_1,u_2)=\min(u_1,u_2)$, for perfect positive dependence, also known as the \textit{comonotonic} copula} copulas:
\begin{theorem}
Copulas satisfy the following inequality $ \forall (u_1,u_2) \in I \times I$
\begin{equation}
C^{-}(u_1,u_2) \leq C(u_1,u_2) \leq C^{+}(u_1,u_2)
\end{equation}
\end{theorem}

An using the \citep{sklar1959} theorem, on can write the Frechet-Hoeffding inequality:
\begin{equation}
\max (F_1(x_1) + F_2(x_2) - 1, 0) \leq F(x_1,x_2) \leq \min(F_1(x_1), F_2(x_2))
\end{equation}

\begin{corollary}
Two continuous random variables $X_1$ and $X_2$ with distribution functions $F_1$ and $F_2$ are independent r.v.s iff they have the product\footnote{$C^{\perp}(u_1,u_2)=u_1 u_2$} copula: $F (x_1, x_2) = F_1(x_1)F_2(x_2)$.
\end{corollary}

\subsubsection{Copulas density}
When it exists, the density of a copula is defined as:
\begin{equation}
\frac{\delta^d C}{\delta u_1 \ldots \delta u_d}(u_1,\ldots,u_d)
\end{equation}

\subsubsection{Tail dependence}

Copulas are agile tools to model extreme events dependence between random variables, writing the lower tail dependence coefficient:
\begin{equation}
\lambda^l_{1,2} = \lim_{q \to 0} Pr\left(  U_1 < F_1^{-1}(q) \mid U_2 < F_2^{-1}(q) \right) = \lim_{q \to 0} q^{-1} C(q,q)
\end{equation}
and the upper tail dependence coefficient:
\begin{equation}
\lambda^u_{1,2} = \lim_{q \to 1} Pr\left( U_1 > F_1^{-1}(q) \mid U_2 > F_2^{-1}(q) \right) =  \lim_{q \to 1} \frac{1-2q+ C(q,q)}{1-q}
\end{equation}

which represents the conditional probability that variable $1$ takes a value in its lower (resp. upper) tail, given that $2$ takes a value in its lower (resp. upper) tail.

\subsection{Examples of copulas}

Some popular copulas are the Gaussian, Student t, Normal Mixture, Archimedean (such as Clayton or Gumbel), detailed table \ref{table:EXcopulas}.

\begin{table}[!h]
\centering
\begin{tabular}{| l | c | c |}
\hline
 Name &  generator  &  Acceptable value of $\theta$\\
 \hline
Clayton & $\frac{1}{\theta}\left(t^{-\theta} -1 \right)$ & $\theta>0$\\
Gumbel & $-\log(t)^\theta$&$\theta\geq 1$\\
Frank &  $-\log\left(\frac{e^{-\theta t}-1}{e^{-\theta}-1}\right)$&$\theta \in \mathbb R_{\ne 0}$\\
\hline
\end{tabular}
\caption {Archimedean copulas, some examples} 
\label{table:EXcopulas}
\end {table}

\subsection{Calibrating copulas with the Canonical Maximum Likelihood method (CML)}
The copula parameters may be estimated by the canonical maximum likelihood method (CML):
\begin{enumerate}
\item first the sample $(x_j^1,\ldots,x_j^d)$, $j=1,\ldots,n$ is transformed into uniform variates $(u_j^1,\ldots,u_j^d)$, $j=1,\ldots,n$,
\item estimate the marginals using the empirical distributions (without assumptions on the parametric form for each of them), i.e. $\hat{F}_i(u_j^i)$ with $j=1,\ldots,n$ .
\item estimate by maximum likelihood of the density:
\begin{equation}
\hat{\alpha} = ArgMax \sum_{j=1}^n \log\left( \frac{\delta C_\alpha}{\delta u^1 \ldots \delta u^d } (u_j^1,\ldots,u_j^d) \right)
\end{equation}
\end{enumerate}

As an illustration, for the Gumbel I copulas, the density is:
\begin{equation}
\begin{cases}
-a \left[ (-\log u)^a + (-\log v)^a \right]^{1/a-2} (-\log u)^{a-1}  (-\log v)^{a-1} (1/a - 1) exp\left[ - \left( (-\log u)^a + (-\log v)^a \right)^{1/a}\right] \frac{1}{uv} \\
+\\
\left[ (-\log u)^a + (-\log v)^a \right]^{2/a-2}(-\log u)^{a-1}  (-\log v)^{a-1} exp\left[ - \left( (-\log u)^a + (-\log v)^a \right)^{1/a}\right] \frac{1}{uv}
\end{cases}
\end{equation}

\subsection{Illustration with the VCAC and the CAC}

We use\footnote{vansteenberghe\_copula\_introduction.py} daily rates of the VCAC and the CAC from January 2000 to January 2017, as in Example II.6.4 of \citet{alexanderII}:
\begin{center}
\includegraphics[scale=0.5]{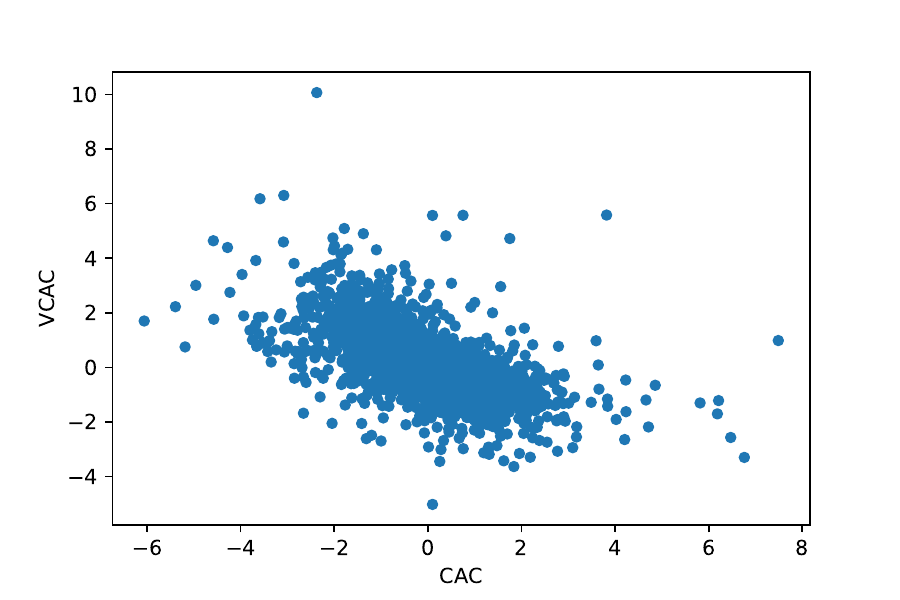}
\end{center}

We might assume that the marginals are Student-t distribution we want to fit on standardized observations. Once we have $\hat{F_1}$ and $\hat{F_2}$, we have the quasi immediate relationships:

If we fit a Gumbel copula, we know that from the Kendall $\tau$ there is the correspondence with the Gumbel's parameter:
\begin{equation}
\alpha = \frac{1}{1-\tau}
\end{equation}

and for the Clayton copula:
\begin{equation}
\alpha = \frac{2 \tau}{1-\tau}
\end{equation}

For bivariate normal copula and for the Student t copula, the correlation coefficient is $\sin\left( \tau \frac{\pi}{2} \right)$

\subsubsection{Copula estimation}

We observe realizations of random variables $x_1$ and $x_2$, we want to estimate the copula $C\left(F_1(x_1),F_2(x_2)\right)$, let's take the Frank copula as an example, we want thus to find the optimal parameter $\theta$ and we think of applying maximum likelihood techniques, but we do not know the distributions $F_1$ nor $F_2$ (the marginals). If we assume that the marginals and the copula are continuous and can be described by parameters, then we can differentiate $C\left(F_1(x_1),F_2(x_2)\right)$ and the density is:
\begin{equation}
c\left(F_1(x_1,\alpha_1),F_2(x_2,\alpha_2);\theta \right) \Pi_{i=1}^2 f_i(x_i, \alpha_i)
\end{equation}

Taking the log likelihood function:
\begin{equation}
\sum_{t=1}^T \ln\left( c\left(F_1(x_1,\alpha_1),F_2(x_2,\alpha_2);\theta \right) \right) + \sum_{i=1}^2 \sum_{t=1}^T  \ln\left(  f_i(x_i, \alpha_i) \right)
\end{equation}

We can maximize this in two steps:
\begin{enumerate}
\item estimate $\hat{\alpha}$ that maximizes $\sum_{i=1}^2 \sum_{t=1}^T  \ln\left(  f_i(x_i, \alpha_i) \right)$
\item estimate $\hat{\theta}$ that maximizes $\sum_{t=1}^T \ln\left( c\left(F_1(x_1,\hat{\alpha_1}),F_2(x_2,\hat{\alpha_2});\theta \right) \right)$
\end{enumerate}

For the bivariate t copula, the density\footnote{most densities can be found in the Wikipedia page of copula} is:
$$
c(u_1,u_2; \rho, \nu) = K \left(1 -\rho^2 \right)^{-.5} \left[1+ \left(\xi_1^2-2\rho\xi_1\xi_2+\xi_2^2 \right)(1-\rho^2)^{-1}\nu-{-1} \right]^{-\frac{\nu+2}{2}} \left[\left( 1+\nu^{-1}\xi_1^2 \right) \left( 1+\nu^{-1}\xi_2^2 \right)\right]^{\frac{\nu+1}{2}}
$$
with 
$$ K = \Gamma\left( \frac{\nu}{2}\right)^{n-1} \Gamma\left(\frac{\nu + 1}{2} \right)^{-n} \Gamma\left(  \frac{\nu+n}{2}\right)$$ 
and $\xi_i = t_\nu^{-1}(u_i)$

The ML estimate is closed to the estimate: $\hat{\rho} = \sin\left(\frac{\pi}{6} \rho_S\right)$ with $\rho_S$ the Spearman rank correlation coefficient, introduced section  \ref{sec:corronreturns}.

For the Frank copula density:
$$ \frac{\theta e^{\theta (u_1+u_2)}\left(e^\theta-1 \right)}{\left(e^\theta - e^{\theta u_1} -e^{\theta u_2} + e^{\theta(u_1+u_2)} \right)^2} $$

\subsection{Survival and joint survival copula}

Given two random variables with copula $C(.,.)$, the survival copula is defined as: 
\begin{equation}
\hat{C}(u,v) = u+v-1+C(1-u,1-v)
\end{equation}
and the joint survival function as:
\begin{equation}
\bar{C}(u,v) = \hat{C}(1-u,1-v) 
\end{equation}

\subsection{Extreme value copulas}

\begin{definition}
A copula $C$ is said to be an extreme-value copula if it verifies the following equality:
\begin{equation}
C(u_1^{\frac{1}{m}},\ldots,u_d^{\frac{1}{m}}) = \left[ C(u_1,\ldots,u_d) \right]^{\frac{1}{m}}
\end{equation}
with $m$ and integer satisfying $m\geq 1$
\end{definition}
We provide in table \ref{table:EVTcopulas} some extreme copulas which present the advantage to require only one parameter to be calibrated except for Marshall-Olkin.

\begin{table}[!h]
\centering
\begin{tabular}{| l | c | c |}
\hline
 Name &  $\alpha$  &  $C(u_1,...,u_d)$\\
 \hline
 $C^\bot$ &  & $u_1...u_d$  \\
Gumbel I &  $[1,\infty)$ & $\exp\big\{-(\tilde{u}^{\alpha}_1+...+\tilde{u}^{\alpha}_d)^{1/\alpha}\big\}$ \\
Gumbel II &  $[0,1]$ & $u_1...u_d\exp\Big\{\alpha\frac{\tilde{u}_1...\tilde{u}_d}{\tilde{u}_1+...+\tilde{u}_d}\Big\}$\\
Galambos &  $[0,\infty)$ & $u_1...u_d\exp\big\{(\tilde{u}^{-\alpha}_1+...+\tilde{u}^{-\alpha}_d)^{-1/\alpha}\big\}$\\
Marshall-Olkin &  $[0,1]^d$ & $u_1^{1-\alpha_1}...u_d^{1-\alpha_d}\min(u_1^{\alpha_1},...,u_d^{\alpha_d})$\\
$C^+$ &   & $\min(u_1,...,u_d)$\\
\hline
\end{tabular}
\caption {Writing $u_i$, $i=1,...,d$ marginals of uniform distributions and $\tilde{u}_i=-\log(u_i)$ $\forall i=1,...,d$.} 
\label{table:EVTcopulas}
\end {table}

\stepcounter{question}
\begin{itemize}
\item \textbf{Question \arabic{question}}
\begin{itemize}
\item Show that the Gumbel I (this is also true for the Clayton's) copula has asymmetric tail dependence ($\lambda^l_{1,2}\neq \lambda^u_{1,2}$)
\end{itemize}
\end{itemize}

We illustrate the copulas with the Gumbel I\footnote{it can be demonstrated that as $\alpha \to \infty$ the Gumbel copula converge to the maximum (product) copula} and Galambos:
\begin{center}
\includegraphics[scale=0.4]{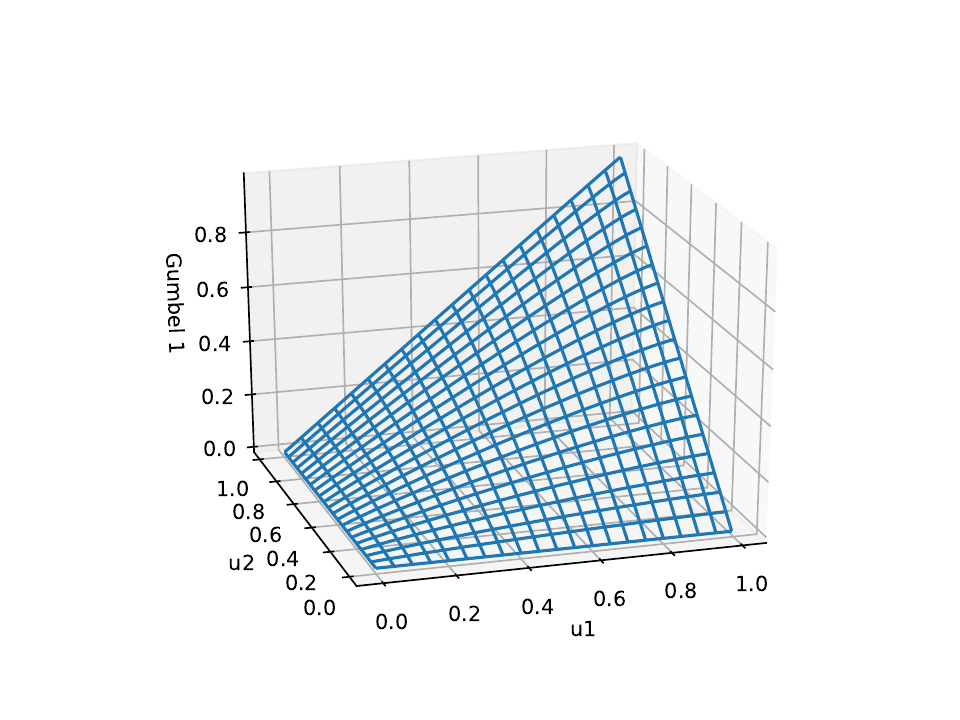}
\includegraphics[scale=0.4]{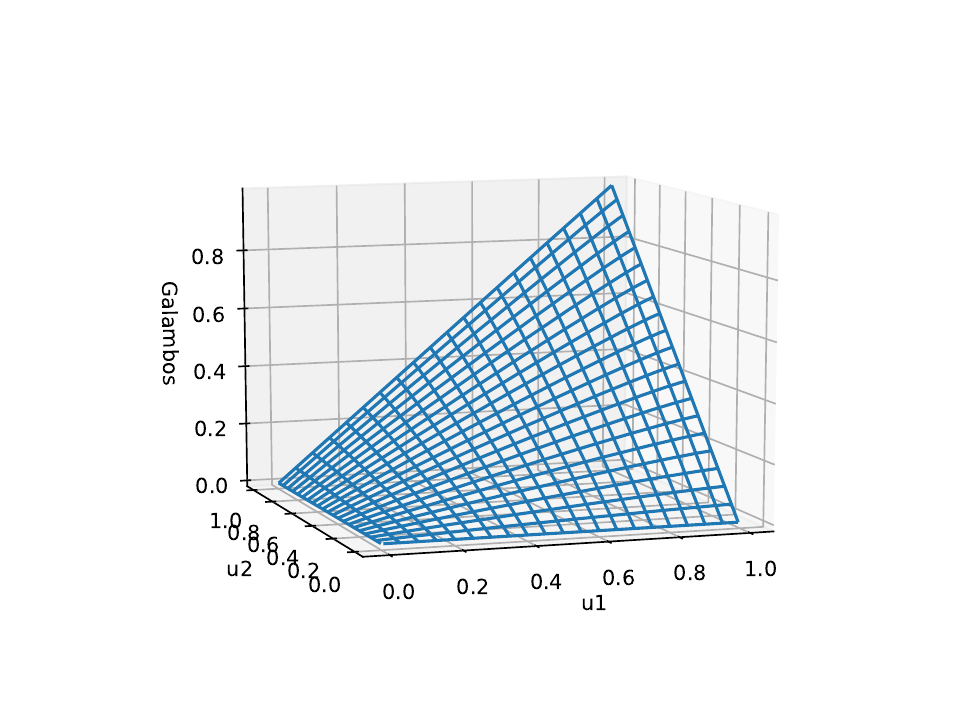}
\end{center}

\subsection{Domain of Attraction of Copula}

A copula $C$ belongs to the domain of attraction of an extreme value copula $C^*$ if for all $u$
\begin{equation}
\lim_{m \to \infty} C^m\left(u^{\frac{1}{m}}\right) = C^*(u)
\end{equation}

It can be demonstrated, (Marshall and Olkin, 1983), that the Gaussian copula (when the correlation matrix does not contain any 1 except in its diagonal) belongs to the domain of attraction of the independent copula.

\subsection{Kolmogorov-Smirnov adequacy test}
Kolmogorov-Smirnov theorem writes
\begin{theorem}
Given $(R_1,\ldots R_n)$ a n-sample following a law $\mu$ and $H_n$ its distribution function
\begin{equation}
H_n(R) = \frac{1}{n}\sum_{k=1}^n 1_{]-\infty,R]}(R_k)
\end{equation}
We assume that the distribution function $H$ of $\mu$ is continuous. Then $\sqrt{n}\|H_n-H\|_{\infty}$ converges in law to the Kolmogorov-Smirnov law $\mu_{KS}$ that can be characterized by the distribution function
\begin{equation}
H_{KS}(r)=1+2\sum_{k=1}^{\infty}(-1)^k\exp\{-2k^2r^2\}.
\end{equation}
\end{theorem}

The Kolmogorov-Smirnov test statistic is $$D_n=\sup_{u\in [0,1]^d}|\hat{C}_n(u)-C_{\hat{\theta_n}}(u)|$$ in practice this statistic is approximated by the following alerantive, chosing $N$ "high enough"
\begin{equation}
D_{n,obs}=\sup_{1\leq i_1,...,i_d\leq N} \Big|\hat{C}_n\Big(\frac{2i_1-1}{2N},...,\frac{2i_d-1}{2N}\Big)-C_{\theta_n}\Big(\frac{2i_1-1}{2N},...,\frac{2i_d-1}{2N}\Big)\Big|.
\end{equation}

\subsection{VaR uppder bound}
It can be demonstrated\footnote{\href{http://www.thierry-roncalli.com/download/Lecture-Notes-Copula-Ensai.pdf}{some elements here}}, that following a peak-over-threshold methodology then the VaR has an upper bound such that:
\begin{equation}
\mbox{VaR}_\alpha \left( \sum_{i=1}^d \nu_i R^i) \right) \leq K^{-1} (\alpha)
\end{equation}
with
\begin{equation}
 K^{-1} (\alpha) = \inf_{\left(\sum_{i=1}^d u_i - d + 1 \right)^+ = \alpha} \sum_{i=1}^d \nu_i F^{-1}_i(u_i)
\end{equation}

In practice, we create a grid of possible values for $u_i \in [0,1]$ $\forall i$ and select the combinations that equals $\alpha$ with a given tolerance\footnote{the more dense the grid and the lower the tolerance, the more precise our VaR measure will be}. We are then in a position to compute the VaR upper bound and compare it to the VaR measured with the product copula which is the one assuming independence between the random variables. We expect that $VaR_{\mbox{independent}} \leq VaR_{\mbox{upper}}$.

\section{Python and R: risk measures, correlation, clusters, an introduction}

We introduced the risk measure VaR section \ref{seq:VaR} and the Q-Q plot section \ref{sec:qqplot}. We are observing realizations $x_t$ from a random variable $X$, we drop the time consideration and reorder the observations so that $x_1 \leq x_2 \leq \ldots x_n$, we reconcile both notations writing:
\begin{equation*}
\widehat{VaR}_q = \widehat{Q}_n(q)  = x_{s}, \text{  with  }  s=[nq]+1
\end{equation*}
where $[nq]$ represents the integer part of $nq$. Intuitively, the more observations, the more precise our estimate of $Q$.

VaR is defined in the tail of the distribution, as for some data set the number of extreme events might be limited, extreme value theory is a tool to apply parametric methods to these data sets. As most parametric models assume that the rate of loss of assets are independently and identically normally distributed, the rate of loss time series have to be filtered before modelling the risk.

We present in this section that the financial returns are not independent nor identically distributed, thus first some dynamic models have to be applied to extract independent standardized innovations\footnote{some illustration can be found \href{https://cran.r-project.org/web/packages/qrmtools/vignettes/ARMA_GARCH_VaR.html}{here}}. Second the occurrences of extreme losses have to be model and third their dependence structure modelled before computing the risk measure VaR and performing a backtest to validate the approach.

\subsection{Independently and identically normally distributed returns tests}

To test for idd, we introduced the Ljung-Box test section \ref{sec:ljungbox}, the BDS test section \ref{sec:BDStest}. We further test for normality, 'heavy-tailness' and correlation breaks in the following sections.

\subsubsection{Richarson and Smith test (normality)}

We test the normality of the returns, following section \ref{sec:GMMWald} based on \citep{Richardson:1993aa} following the steps in \citep{Jondeau2007} based on GMM and a Wald test\footnote{R code: Richardson\_Smith\_test}. 

For those asset where the GMM converged\footnote{convergence code = 0}, we find the most asset returns satistics are above the $5.99$ threshold and fail the normality test. This test focuses on the skewness (symmetry on each side of the x=0 axis) and kurtosis (fat tails).

\subsubsection{Correlation Breakdown}\label{sec:correlationB}

We have time series of asset returns from 2006-12-01 until 2017-12-29 and we can split the data set in two periods of equal length (split at 2012-06-19). In the first half we have the direct aftermaths of the 2007-2009 financial crisis and in the second half occurs a recovery of the CAC 40 index. The average difference of the correlation between each asset returns is $8.6\%$, the average correlation being greater in the first half of the sample, during "turbulent time". As discussed in \citep{boyer1997}, we are aware of the selection bias and our correlation should be corrected before testing.

We perform\footnote{R code: assetmanager\_correlation\_matrices} a test on the correlation matrices as in \citep{jennrich:1970aa} with the null hypothesis $H_0$: both correlation matrices are equal. We strongly reject $H_0$.

\subsection{Dynamic models for financial time series: filtering volatility clusters}\label{sec:filter-GARCH}

We suspect that the empirical loss rates under study are not iid,  we suspect AR, MA, GARCH, therefore we need to first apply dynamic models to the time series before applying our risk measures.

\subsubsection{Univariate ARMA-GARCH model}

Most parametric models assume that the mean and volatility of the loss rate are constant over time and the loss rate are independent random variables. In practice, all three assumptions are violated for financial time series. To address these issues, \citet{McNeil:2000aa} suggest an ARMA-GARCH modelisation of the returns. Residuals can then be considered as iid and modelled using Extreme Value Theory (EVT).

As in section \ref{sec:uARMAGARCH}, we assume that we have autocorrelated financial loss rates modelled by the equation system known as the  $AR(1)-GARCH(1,1)$ and extract the innovations process $(Z_t)_t$. For such a process of rate of losses, $VaR$\footnote{estimation of the risk measure at date $t$ for a loss rate at date $t+1$}  knowing the filtration $\mathcal{F}_{t}$  can be written as:
\begin{equation*}
VaR^t_q(R_{t+1})= \mu_{t+1} + \sigma_{t+1} VaR^t_q(Z_{t+1})
\end{equation*}
As $(Z_t)_t$ is an iid series, we have that $VaR^t_q(Z_{t+1}) = VaR_q(Z)$ for all $t$. Thus we deduct that
\begin{equation*}\label{eq:VaRconditional}
VaR^t_q(R_{t+1})= \mu_{t+1} + \sigma_{t+1} VaR_q(Z)
\end{equation*}

We can compute the $VaR$ of the innovations as:
\begin{equation*}
VaR_q(Z) = G^{-1}(q) \text{ where $G$ is the distribution of the chosen law $g$ of the iid random variables $(Z_t)_t$}
\end{equation*}

\subsubsection{AR-EGARCH model}\label{sec:AREGARCH}

For financial times series negative correlation between returns and volatility is usually observed, with higher volatility in times of losses. \citep{Nelson:1991aa} introduced the AR-EGARCH model, used in \citep{Nguyen2019aa} to extract independent standardized innovations. For each asset $i$:
\begin{equation}
\begin{cases}
R_{t,i} &= \mu_{t,i} + \sigma_{t,i} Z_{t,i} \\
\mu_{t,i}  &= \mu_i + \sum_{k=1}^p \phi_k R_{t-k,i} \\
\log \left(\sigma^2_{t,i} \right) &= \omega_{i} + \sum_{j=1}^m \alpha_{j,i}\log \left(\sigma^2_{t-j,i} \right) + \sum_{j=1}^s \left[ \beta_{j,i}Z_{t-j,i} + \gamma_{j,i}\left( \vert Z_{t-j,i} \vert - E \vert Z_{t-j,i} \vert \right)  \right]
\end{cases}
\end{equation}

As in \citep{Nguyen2019aa}, the standardized innovations $Z_{t,i}$ are assumed to follow a univariate skewed Student-t distribution (sstd).

The asymmetry with respect to the returns sign in the volatility modelling is revealed by rearranging the equations. In the case of an EGARCH(1,1):
\begin{equation*}
\sigma^2_{t,i} = \sigma^{2\beta_{1,i}}_{t-1,i}e^{\omega_i}\begin{cases}
\exp\left[(\alpha_{1,i} + \gamma_{1,i}) Z_{t-1,i} \right] \quad \mbox{if } Z_{t-1,i}  \geq 0 \\
\exp\left[(\alpha_{1,i} - \gamma_{1,i}) Z_{t-1,i} \right] \quad \mbox{if } Z_{t-1,i}  < 0
\end{cases}
\end{equation*}

As expected, when fitting\footnote{assetmanager\_AR\_EGARCH.R} an AR(1)-EGARCH(1,1) on the loss rate, we find that loss rates levels are not predictable and the AR coefficient are not statistically different from $0$, we find that $\alpha_{1,i}$ and $\gamma_{1,i}$ are both positive and significant, meaning that we tend to observe higher volatility when losses are higher and be able to predict this phenomenon. We verify that for each model the weighted Ljung-Box\footnote{see section \ref{sec:ljungbox} for more detail on this test} Test on Standardized Residuals do not reject $H_0$ of independent distribution. We plot an example of the filtering:
\begin{center}
\includegraphics[scale=0.4]{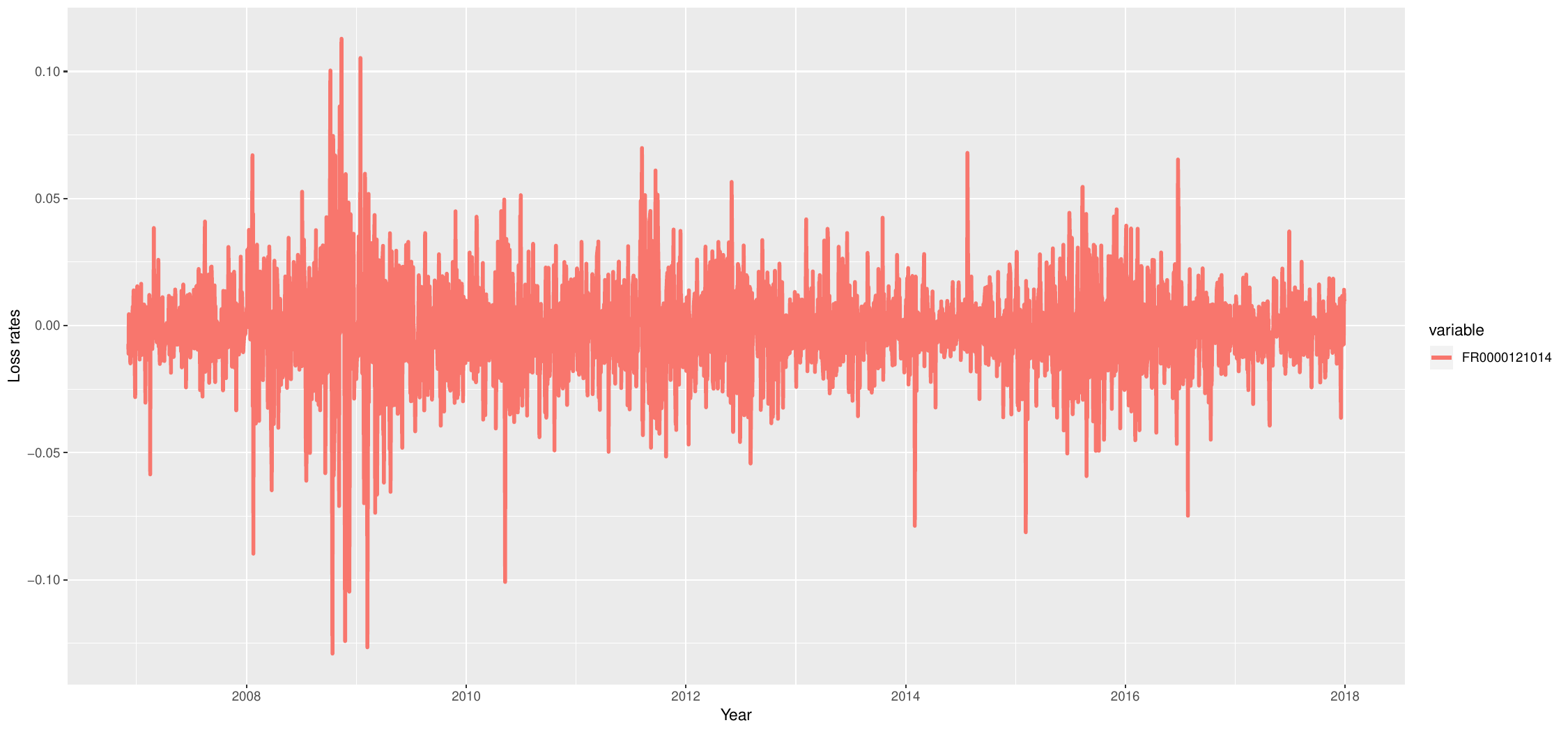}
\includegraphics[scale=0.4]{fig/eGARCH_innovations.pdf}
\end{center}

\subsubsection{DCC-GARCH(1,1) application to a portfolio}

We follow section \ref{sec:dcc} and apply a DCC-GARCH model to our returns. As this is demanding in terms of computing power, we do an example on the most weighed assets in the portfolio: FR0000120271 (Total), FR0000120578 (Sanofi) and FR0000121014 (LVMH). 
We have the following unconditional correlation for the sample:

\begin{tabular}{rlll}
              &FR0000120271 & FR0000120578  &FR0000121014\\
FR0000120271       &   1.00&          0.56 &         0.62\\
FR0000120578        &  0.56 &         1.00  &        0.51\\
FR0000121014         & 0.62  &        0.51   &       1.00\\
\end{tabular}

We plot the asset prices and returns evolution over time:
\begin{center}
\includegraphics[scale=0.4]{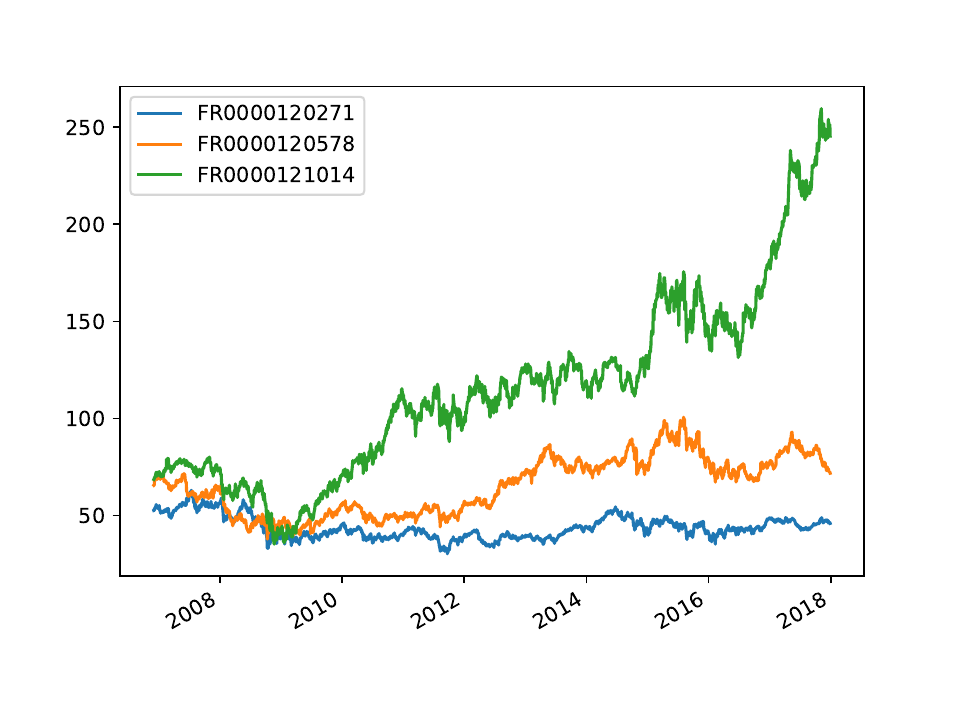}
\includegraphics[scale=0.4]{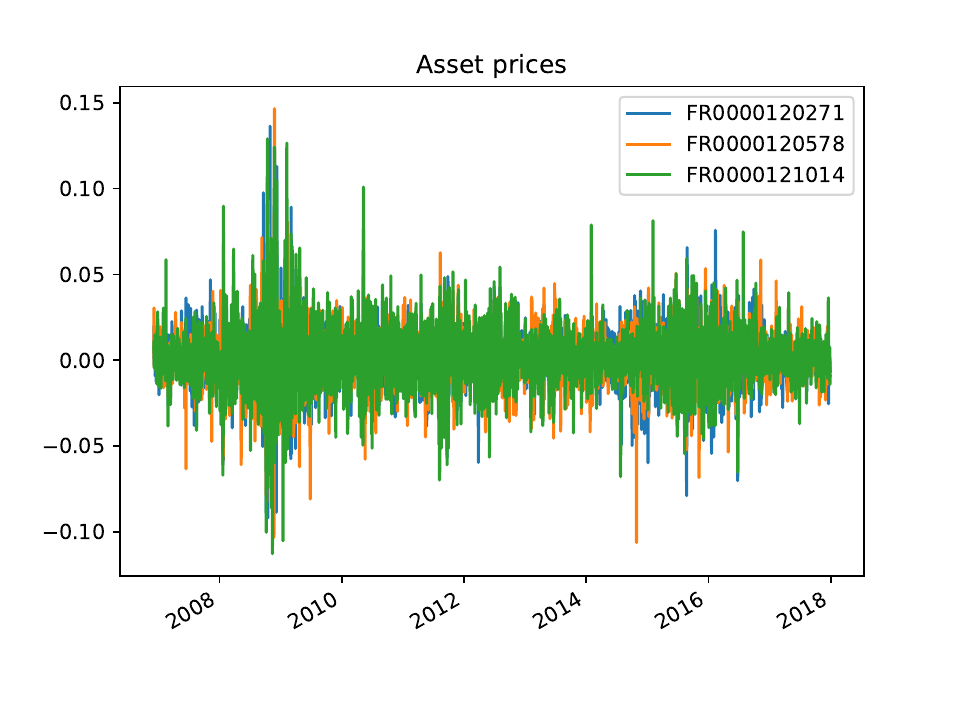}
\end{center}

We will be in a position to measure the unconditional VaR of the innovations and extract the volatility in equation \ref{eq:VaRconditional}.

We plot\footnote{assetmanager\_DCC\_M\_GARCH.R} the conditional correlations of the asset returns over time:
\begin{center}
\includegraphics[scale=0.45]{fig/DCC_conditional_correlations.pdf}
\end{center}
The conditional correlation is rather unstable and there is no clear pattern during the 2007-2009 financial crisis.

We can extract the standardized residuals from the DCC-model, the innovations, that we plot below the original returns to show how the model "filtered out" the volatility clusters:
\begin{center}
\includegraphics[scale=0.4]{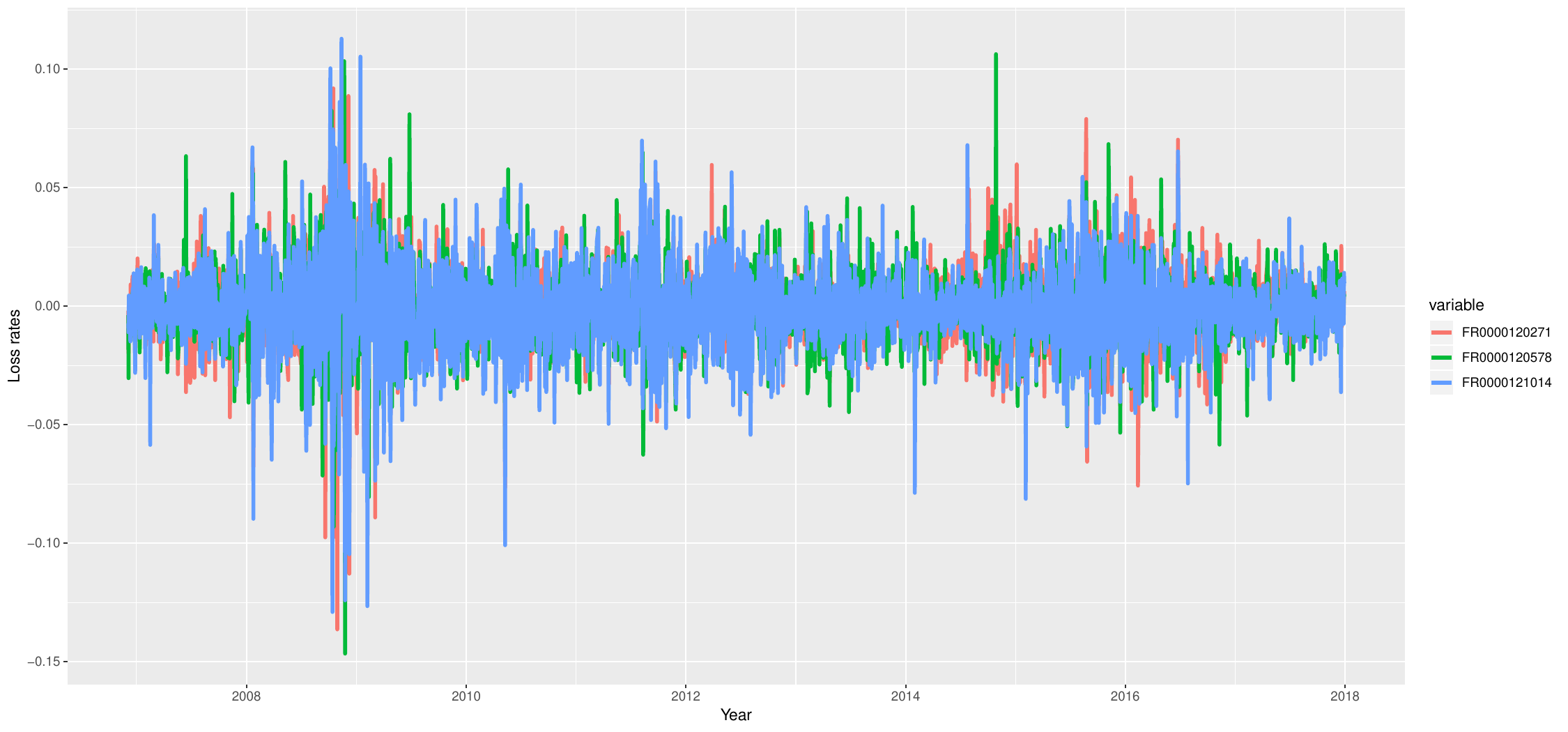}
\includegraphics[scale=0.4]{fig/DCC_innovations.pdf}
\end{center}

\subsubsection{Effects of negative news}

\citep{ENGLE:1993aa} present diagnostic tests which emphasize the asymmetry of the volatility response to news (good news and bad news have different predictability for future volatility) and praise the GJR-GARCH model from \citep{GLOSTEN:1993aa}, defining it in its simplest form:
\begin{equation}
h_t = w + \beta h_{t-1} +  \alpha \epsilon_{t-1}^2 +\gamma S_{t-1}^- \epsilon_{t-1}^2
\end{equation}
with the information set $F_{t-1}$, the expected return $m_t=E(y_t\vert F_{t-1})$, the conditional variance $h_t = Var(y_t\vert F_{t-1})$ and the news $\epsilon_t = y_t-m_t$. $S_{t}^-=1$ if $\epsilon_t<0$, $S_{t}^-=0$ otherwise. We can illustrate the asymmetric\footnote{vansteenberghe\_GJR\_leverage\_illustration.py} effect, respecting the conditions $0\leq \beta < 1$, $0\leq \alpha < 1$ and $\alpha + \gamma <1$ and $\gamma>0$:
\begin{center}
\includegraphics[scale=0.5]{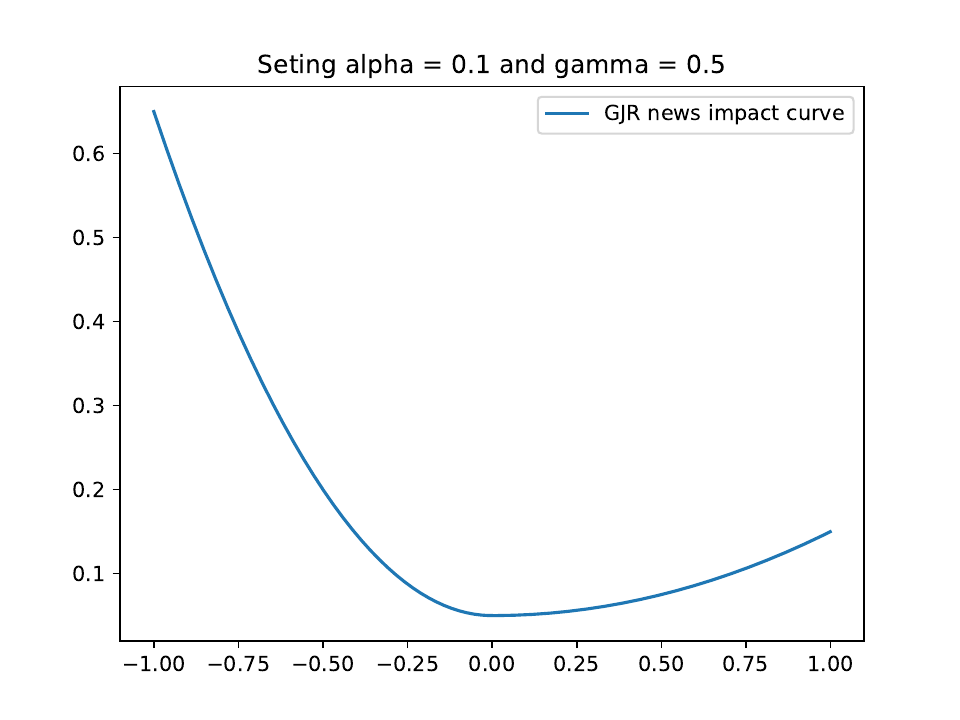}
\end{center}

In the above GJR model, $\gamma > 0$ and negative news will generate more future volatility then positive news, all else being equal.

\citep{ENGLE:1993aa} introduce three tests:
\begin{itemize}
\item Sign Bias Test;
\item Negative Size Bias Test, uses the variable $S_{t-1}^- \epsilon_{t-1}$
\item Positive Size Bias Test, uses the variable $(1-S_{t-1}^-) \epsilon_{t-1}$
\end{itemize}

These tests examine whether we can predict the squared normalized residual by some variables observed in the past which are not included in the volatility model being used. The sample size have to be significantly greater than $100$ for the results to be robust as demonstrated by the authors. 
The test consider the generic form of a GARCH model: $log h_t = log\left(h_{0t}(\delta_0' z_{0t})\right) + \delta_a' z_{at}$ with $h_{0t}(\delta_0' z_{0t})$ the model considered under the null. The null hypothesis: $H_0: \delta_a=0$ means that the model correctly capture the asymmetric effects of positive and/or/both negative news. In other terms, if $H_0$ is reject for a test, then the volatility model cannot capture all effects.

First we follow\footnote{GJR\_GARCH\_asset\_manager.R} the procedure in \citep{ENGLE:1993aa} to extract unpredictable part of the stock returns:
\begin{itemize}
\item regress $y_t$ on a constant and 5 day-of-week dummies, get the residuals $u_t$
\item regress $u_t$ on a constant and $u_{t-1}, \ldots, u_{t-6}$, get the residuals $\epsilon_t$
\end{itemize}

We apply a Ljung-Box\footnote{see section \ref{sec:ljungbox} for more detail on this test} test on the squared residuals $\epsilon^2_t$ and we reject the null hypothesis of independence, pointing to time-varying volatility in the returns.

Finally, applying the GJR model and preforming the sign bias test, we reject $H_0$ and believe our model is not well capturing the leverage, although as expected we have $\gamma > 0$. We should further investigate the GJR model before using it for our data.

\quad

Having "filtered out volatility clusters" we apply the Extreme Value Theory on the innovations as described in section \ref{sec:EVT}.

\section{Measure of tail risk}

\subsection{Measure of tail risk: VaR}

We estimate $F(x)$ for $x=u+y$ with $u$ some suitably chosen high threshold and $y>0$. Following assumption \ref{hyp1} we can write:
\begin{align*}
\bar{F}(x)=1-F(x)&=\mathbb{P}(R>x)\\
&=\mathbb{P}(R>u)\mathbb{P}(R>x | R>u) \\
&=\bar{F}(u)\mathbb{P}(R-u>x-u | R>u) \\
&=\bar{F}(u)\bar{F}_u(x-u) \\
&= \bar{F}(u)\Big(1+\xi\frac{x-u}{\beta}\Big)^{-1/\xi}
\end{align*} 
Thus for all $x=u+y$ with $y>0$:
\begin{equation}\label{FRE1}
F(x)= 1- \bar{F}(u)\Big(1+\xi\frac{x-u}{\beta}\Big)^{-1/\xi}.
\end{equation}
The equality \ref{FRE1} allows to obtain, when $F(u)$ is known, some high order quantile of the underlying distribution that we interpret as the $VaR$. Indeed, for some $q>F(u)$, the $VaR$ is defined as:
\begin{equation}\label{VaRu}
VaR_q=u+\frac{\beta}{\xi}\Big[\left(\frac{1-q}{\bar{F}(u)}\right)^{-\xi}-1\Big]
\end{equation}
To estimate the $VaR$, we replace $\xi$ and $\beta$ by their $ML$ estimates and $F(u)$ is replaced by its historical estimate $n^{-1} N_u $. We chose $u$ such that there are enough data points over $u$ and we have for all $x>u$:
\begin{equation}\label{VaRuE}
\widehat{VaR}_q=u+\frac{\widehat{\beta}}{\widehat{\xi}}\Big[\Big(\frac{n(1-q)}{N_u}\Big)^{-\widehat{\xi}}-1\Big]
\end{equation} 
The confidence intervals can be estimated using the Delta method described in Appendix.

Once the threshold $u$ has been chosen and the $VaR$ estimated, we have to backtest the results to validate the approach or else update the chosen threshold $u$.

\subsection{VaR Backtesting}\label{sec:varbacktest}

Backtesting is a process allowing us to accept or reject a risk model, we compare the modelled $VaR$ ex-ante with the historical loss rate observed ex-post with a statistical approach to test the validity of our measure.
We write $VaR_{q}^t(R_{t+1})$ the estimate of the $VaR$ for the period $t+1$,conditional on the information $\mathcal{F}_t$ available at date $t$ and for a coverage rate $p=1-q$. By definition of the conditional $VaR$ we have:
\begin{equation}
\mathbb{P}\big(R_{t+1} > VaR_{q}^t(R_{t+1}) \big)= p.
\end{equation}
We say that this is violation (or hit) when the observed loss rate ex-post, $R_{t+1}$, is higher than the ex-ante expected value $VaR_{q}^t(R_{t+1})$. We write $I_{t+1}(p)$ the random variable associated with a violation of the $VaR_{q}^t(R_{t+1})$ at date $t+1$:
\begin{equation*}
\begin{cases}
1 & \text{ if } R_{t+1} > VaR_{q}^t(R_{t+1})\\
0 & \text{ if } R_{t+1} \leq VaR_{q}^t(R_{t+1}).
\end{cases}
\end{equation*}
\citep{Christoffersen:1998aa} states that a $VaR$ model is valid if the violation series $(I_t(p))_{t=1,...,T}$ satisfy the two properties:
\begin{enumerate}
\item[a)]\textbf{The unconditional coverage (UC) hypothesis}

States that if the estimated $VaR$ is valid, the violation frequency ovserved over a period $T$, $\frac{1}{T}\underset{t=1}{\overset{T}{\sum}}I_t(p)$, should not be significantly different than the coverage rate $p$. If this frequency is higher than $p$ then our model under-estimate the risk and vice-versa. This property can be written as:
\begin{equation*}
\mathbb{E}(I_t(p))=\mathbb{P}\big(I_t(p)=1\big)=p.
\end{equation*}
\item[b)] \textbf{The independence (IND) hypothesis}

The violations of the $VaR$ have to be time independent, i.e. the random variable $I_t(p)$ has to be independent of $I_{t-k}(p)$ for all $1\leq k < t$. This property can be written as:
\begin{equation*}
\mathbb{P}\big(I_{t+1}(p)=1|\mathcal{F}_t\big)=\mathbb{E}\big(I_{t+1}(p)|\mathcal{F}_t\big)=\mathbb{P}\big(I_{t+1}(p)=1\big).
\end{equation*}
\end{enumerate}
A risk prevision model that satisfy both properties is said to be valid and has the conditional coverage property written as:
\begin{equation*}
\mathbb{P}\big(I_{t+1}(p)=1|\mathcal{F}_t\big)=\mathbb{E}\big(I_{t+1}(p)|\mathcal{F}_t\big)=p.
\end{equation*}
This implies that the series of violations of the $VaR$ is i.i.d following a Bernoulli law of parameter $p$.

\subsubsection{Unconditional coverage test: Kupiec (1995)}

As in  \citep{Kupiec:1995aa}, we test if the probability of an unconditional violation written $\pi$ is statistically different than the coverage rate $p$. The null hypothesis and its alternative are written:
\begin{equation*}
\begin{cases}
H_0:& \pi = p \\
H_1:& \pi \neq p.
\end{cases}
\end{equation*}
Assuming that the violations are independent, the likelihood under the null hypothesis and its alternative can be written respectively:
\begin{equation*}
\begin{cases}
L(p|I_1,...,I_T)=(1-p)^{T_0}p^{T_1} \\
L(\pi|I_1,...,I_T)=(1-\pi)^{T_0}\pi^{T_1}
\end{cases}
\end{equation*}
where $T_0$ and $T_1$ are the number of false and true in the violation sample. As the probability $\pi$ is unknown, we estimate it via the $ML$ methods and find $\pi = T_1/(T_0+T_1)$. 

We can perform the test with the statistics defined as the log-likelihood ratio:
\begin{equation*}
LR_{uc}=-2\log\Big(\frac{L(p|I_1,...,I_T)}{L(\pi|I_1,...,I_T)}\Big)\overset{d}{\Longrightarrow} \chi^2(1).
\end{equation*}
The higher this statistics, the less likely the null hypothesis. Under the null hypothesis, $LR_{uc}$ asymptotically follows a $\chi^2$ with one degree of freedom corresponding to the restriction $\pi=p$. The p-vlaue for this test is defined as $p-value = 1-F_{\chi^2(1)}(LR_{uc})$ where $F_{\chi^2(1)}$ is the distribution funciton of the $\chi^2$ with one degree of freedom.

\subsubsection{Regression based backtest: Engle and Manganelli (2004)}

We consider a linear regression model linking current violations to past violations and/or past information to test the conditionaly coverage hypothesis. We write the demeaned violation process $Hit_{t}(p)=I_t(p)-p$ and we have:
\begin{equation*}
Hit_{t+1}(p) =
\begin{cases}
1-p & \text{if } R_t>VaR_p^{t}(R_{t+1}) \\
-p  & \text{if } R_t\leq VaR_p^{t}(R_{t+1})
\end{cases}.
\end{equation*} 
\citep{Engle:2004aa} test the null hypothesis $H_0$ via the following linear regression:
\begin{align}\label{regressionEM}
Hit_{t+1}(p) &= a_0 + \sum_{k=1}^{K}a_k Hit_{t+1-k}(p) \nonumber\\
&+ \sum_{k=1}^{K}b_k g(Hit_{t+1-k}(p),Hit_{t-k}(p),...,\zeta_{t+1-k},\zeta_{t-k},...)+\epsilon_{t+1},
\end{align}
where the residuals $\epsilon_{t+1}$ satisfy:
\begin{equation*}
\epsilon_{t+1} =
\begin{cases}
1-p & \text{with probability } p  \\
-p  & \text{with probability } 1-p
\end{cases},
\end{equation*}
where $g(.)$ is a function of the past values of the hit variable $Hit$ and the variables $\zeta$ who belongs to the information set $\mathcal{F}_t$. The null hypothesis $H_0$ test the joint nullity of all regression coefficients $a_k$, $k=0,...,K$ et $b_k$, $k=1,...,K$. Equation \ref{regressionEM} implies:
\begin{align*}
\mathbb{E}\big(Hit_{t+1}(p)|\mathcal{F}_t\big) &= a_0 + \sum_{k=1}^{K}a_k Hit_{t+1-k}(p) \\
&+ \sum_{k=1}^{K}b_k g(Hit_{t+1-k}(p),Hit_{t-k}(p),...,\zeta_{t+1-k},\zeta_{t-k},...).
\end{align*}
Thus current violations are independent of past violations as long as $a_k=b_k=0$ for all $k=1,...,K$. If $a_0=0$, the unconditional coverage hypothesis is satisfied as $\mathbb{E}\big(Hit_{t+1}(p)|\mathcal{F}_t\big)=0$ as long as $a_0=a_k=b_k=0$ for all $k=1,...,K$. This implies that $\mathbb{P}(I_t(p)=1)=\mathbb{E}(I_t(p))=p$. Which is the same as the unconditional coverage test in \citep{Kupiec:1995aa}. Writing $\Phi = (a_0,a_k,...,a_K,b_1,...,b_K)'$ the $2K+1$ vector of the model parameters and $Z$ the matrix of explanatory variables of the model, the Wald statistics, written $DQ_{cc}$, associated with the jointn ullity of all coefficients verifies:
\begin{equation*}
DQ_{cc}=\frac{\widehat{\Phi'}Z'Z\widehat{\Phi}}{p(1-p)}\overset{d}{\Longrightarrow} \chi^2(2K+1).
\end{equation*}
The p-valule of the test is given by $p-value=1-F_{\chi^2(2K+1)}(DQ_{cc})$.

\subsection{EVT and VaR backtesting application to a portfolio}

We follow our example with the portfolio and the "filtered" innovations on which we apply the threshold exceedance method to fit a GPD. We compute the VaR at the $99\%$ confidence level. The optimal thresholds used oscillate in the vicinity of $1$.

\begin{table}[ht]
\centering
\begin{tabular}{rlrrrrrrr}
  \hline
 asset & GPD $\beta$ & GPD $\xi$ & VaRq & E. M. test p-value & original length & obs over u & u \\ 
  \hline
FR0000120271 & 0.57 & -0.00 & 2.95 & 0.15 & 2777 & 145 & 1.60 \\ 
FR0000120578 & 0.73 & 0.04 & 3.33 & 0.06 & 2777 & 129 & 1.60 \\ 
FR0000121014 & 0.53 & 0.08 & 2.90 & 0.13 & 2777 & 153 & 1.50 \\ 
   \hline
\end{tabular}
 \caption{VaR backtest for after the AR(1)-EGARCH(1,1) filtration}
\end{table}

\begin{table}[ht]
\centering
\begin{tabular}{rlrrrrrr}
  \hline
 asset & GPD $\beta$ & GPD $\xi$ & VaRq & E. M. test p-value & original length & obs over u & u \\ 
  \hline
  FR0000120271 & 0.51 & 0.09 & 2.98 & 0.79 & 2777 & 150 & 1.60 \\ 
FR0000120578 & 0.72 & 0.05 & 3.31 & 0.00 & 2777 & 130 & 1.60 \\ 
FR0000121014 & 0.52 & 0.12 & 2.91 & 1.00 & 2777 & 142 & 1.50 \\ 
   \hline
\end{tabular}
 \caption{VaR backtest for after the DCC-GARCH filtration}
\end{table}

And we compute\footnote{20190104\_assetmanager\_GPD.R} the conditional VaR at the confidence level of $99\%$ for each asset:
\begin{center}
\includegraphics[scale=0.4]{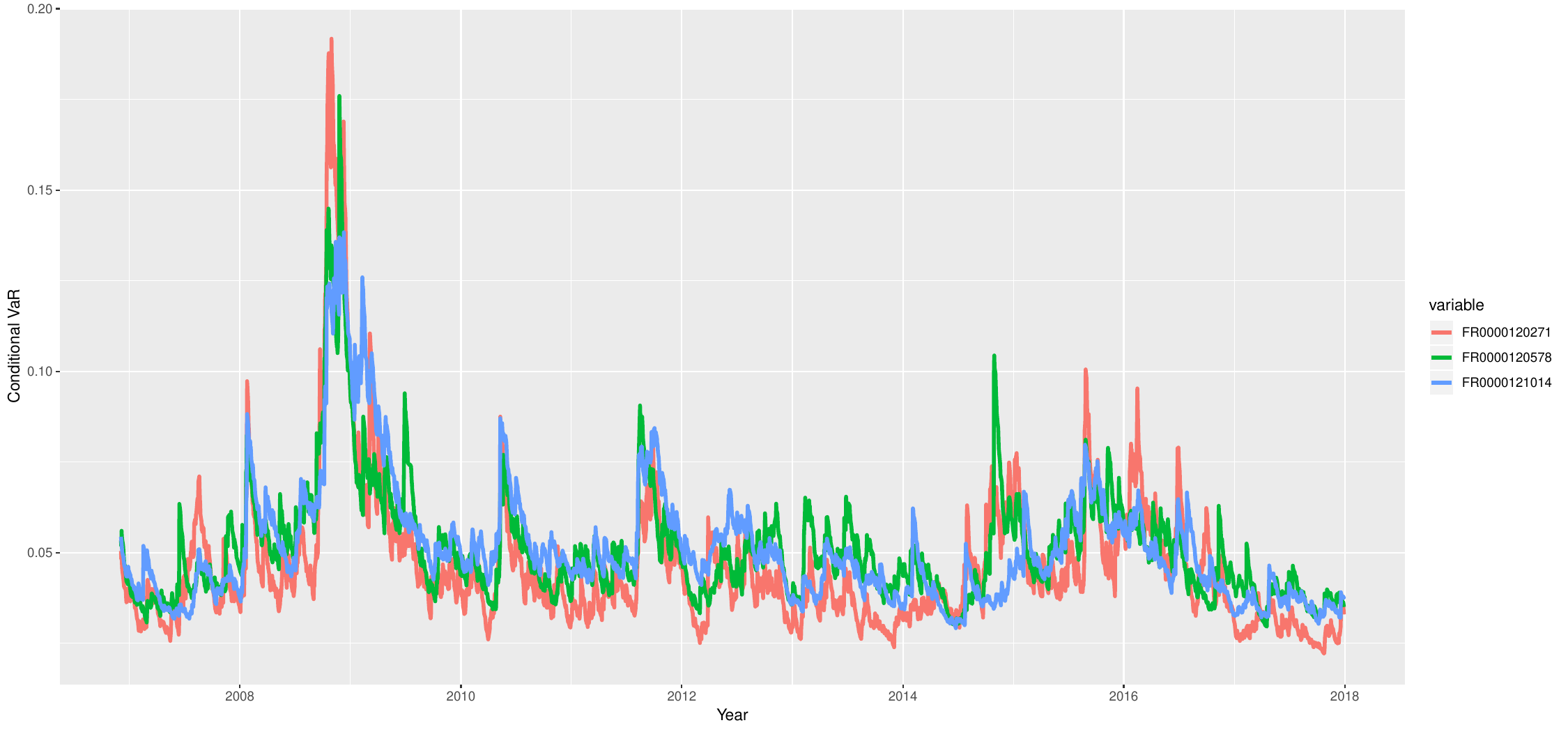}
\end{center}

\newpage
\bibliography{vansteenberghe_quantitative_methods_in_finance}

\end{document}